\documentclass[twoside,10pt]{./SetUps/kththesis}

\pdfoutput = 1
\usepackage{amsmath}
\usepackage{amssymb} 
\usepackage{amsfonts}
\usepackage{amsthm}
\usepackage{mathtools}
\usepackage{mathrsfs}
\usepackage{color}
\usepackage{longtable}
\usepackage{graphics} 
\usepackage{graphicx}
\usepackage{epsfig}   
\usepackage{epstopdf}
\usepackage{float}
\usepackage{rotating}
\usepackage{bbm}
\usepackage{bm}
\usepackage[table]{xcolor}
\usepackage{pgfplots}
\usepackage{tikz}
\usetikzlibrary{calc,positioning,shapes,shadows,arrows,fit}
\usepackage{hyperref}   
\usepackage{lipsum}     
\usepackage[config, font={small}, labelfont={bf}, margin={3mm}]{caption}
\usepackage{subcaption}
\usepackage{cleveref}
\usepackage{comment}
\usepackage{framed}
\usepackage{pifont}
\usepackage{lscape}
\usepackage{algorithm}
\usepackage[noend]{algpseudocode}
\usepackage{array}
\usepackage{url}
\usepackage{tabulary}
\usepackage{multirow} 
\usepackage{lettrine}
\usepackage[final]{microtype}
\usepackage{marvosym}
\usepackage{hyphenat}
\usepackage{afterpage}
\usepackage[numbers,sort&compress]{natbib}

\newtheorem{theorem}{Theorem}
\numberwithin{theorem}{chapter}
\newtheorem{lemma}{Lemma}
\numberwithin{lemma}{chapter}

\newtheorem{corollary}{Corollary}
\numberwithin{corollary}{chapter}
\theoremstyle{definition}
\newtheorem{proposition}{Proposition}
\numberwithin{proposition}{chapter}
\newtheorem{definition}{Definition}
\numberwithin{definition}{chapter}
\newtheorem{remark}{Remark}
\numberwithin{remark}{chapter}
\newtheorem{assumption}{Assumption}
\numberwithin{assumption}{chapter}
\newtheorem{example}{Example}
\numberwithin{example}{chapter}
\newtheorem{problem}{Problem}
\numberwithin{problem}{chapter}
\newtheorem{property}{Property}
\numberwithin{property}{chapter}

\newcommand{\scr}{\scriptscriptstyle}

\title{Planning and Control of Uncertain Cooperative Mobile Manipulator-Endowed Systems under Temporal-Logic Tasks}
\author{Christos K. Verginis}
\date{May 2020}
\defensedate{May 2020}
\thesistype{PhD Thesis} 
\imprint{Stockholm, Sweden, 2020}
\isbn{ISBN: 978-91-7873-503-7}
\trita{TRITA-EECS-AVL-2020:20}
\publisher{Universitetsservice US AB}
\defenseplace{}
\degree{}
\address{%
	KTH Royal Institute of Technology \\
	School of Electrical Engineering \\
	and Computer Science \\
	Division of Decision and Control Systems \\
	SE-100 44 Stockholm  \\
	Sweden}

\usepackage[bottom]{footmisc}
\usepackage{makeidx}
\makeindex

\begin{document}
\mainmatter

\maketitle

\addcontentsline{toc}{chapter}{Abstract}	
\thispagestyle{empty}
\begin{center}
\mbox{\Large \textbf{Abstract}}
\end{center}


Control and planning of multi-agent systems is an active and increasingly studied topic of research, with many practical applications such as rescue missions, security, surveillance, and transportation. This thesis addresses the planning and control of multi-agent systems under temporal logic tasks. The considered systems concern complex, robotic, manipulator-endowed systems, which can coordinate in order to execute complicated tasks, including object manipulation/transportation. Motivated by real life scenarios, we take into account high-order dynamics subject to model uncertainties and unknown disturbances. Our approach is based on the integration of tools from the areas of multi-agent systems, intelligent control theory, cooperative object manipulation, discrete abstraction design of multi-agent-object systems, and formal verification. 

The first part of the thesis is devoted to the design of continuous control protocols for the cooperative object manipulation/transportation by multiple robotic agents, and the relation of rigid cooperative manipulation schemes to multi-agent formation. We propose first a variety of centralized and decentralized control algorithms that do not employ force/torque information at the contact points and take into account both cases of rigid and rolling grasping points, dynamic uncertainties in the object's and agents' model, and potential constraint satisfaction, such as obstacle avoidance and input saturation. Next, we tackle the problem of robust formation control for a class of multi-agent systems and we analyze the relation between formation rigidity theory and rigid cooperative manipulation. 

In the second part of the thesis, we develop control schemes for the continuous coordination of multi-agent complex systems with uncertain dynamics. We first study the motion planning problem and propose novel adaptive control schemes for the collision-free navigation of single- and multi-agent spherical systems in obstacle-cluttered environments. Next, we focus on the leader-follower coordination problem of spherical multi-agent systems. More specifically, we design a robust adaptive decentralized control scheme for the successful navigation of a designated leader to a predefined point, while guaranteeing collision avoidance and connectivity maintenance properties. Finally, we design a closed-form robust barrier function-based control protocol for the collision avoidance of multiple $3$D ellipsoidal agents.  

The third part of the thesis is focused on the planning and control of multi-agent and multi-agent-object systems subject to complex tasks expressed as temporal logic formulas. We tackle first the case of local independent tasks for multi-agent systems, and by using previous results on multi-agent constrained navigation, we design a discrete abstraction of the agents' motion in the workspace and synthesize decentralized control policies that satisfy the agents' specifications. Next, in addition to the robotic agents, we take into account complex tasks to be satisfied by unactuated objects. We design a discrete abstraction that simulates the behavior of the agents and the objects in the workspace and we synthesize controllers for the agents that take into account both theirs and the objects' specifications.  


The fourth and final part of the thesis focuses on several extension schemes for
single-agent setups. Firstly, we consider the problem of single-agent motion planning under timed temporal tasks in an obstacle-cluttered environment. Using previous results on collision-free timed navigation, we develop a novel control policy that guarantees satisfaction of the agent's timed tasks as well as asymptotic optimality with respect to energy resources. Secondly, we tackle the motion planning problem for high-dimensional complex systems with uncertain dynamics in obstacle-cluttered environments.  We integrate intelligent control techniques with sampling-based motion planning algorithms to guarantee the safe navigation of the system to a predefined goal, while compensating for the model inaccuracies. Finally, we develop a novel control protocol that achieves asymptotic reference tracking for an unknown control affine system, while respecting at the same time funnel constraints. 

\thispagestyle{empty} 

\clearpage
	
\addcontentsline{toc}{chapter}{Sammanfattning}	
\thispagestyle{empty}
\begin{center}
\mbox{\Large \textbf{Sammanfattning}}
\end{center}

Reglering och planering av multiagent-system \"ar ett aktivt och v\"axande forskningsf\"alt med en rad praktiska till\"ampningar s\aa som r\"addningsuppdrag, \"overvakning, s\"akerhet och transport. Denna avhandling adresserar planering och reglering av multiagent-system med temporallogiska uppgifter. De ber\"orda systemen \"ar komplexa, robotiska, manipulatorbaserade system, vilka kan samarbeta f\"or att utf\"ora komplicerade uppgifter, bland annat manipulation och transport av objekt. Motiverade av verkliga scenarier tar vi h\"ansyn till h\"ogniv\aa\ dynamik med os\"akerheter och ok\"anda st\"orningar. V\aa rt angreppss\"att baseras p\aa\ integration av redskap fr\aa n f\"oljande omr\aa den: multiagent-system, intelligent reglerteknik, samarbetande objekt-manipulation, diskret abstrakt design av multi-agent-objekt system och formell verifiering. 

Avhandlingens f\"orsta del till\"agnas design av kontinuerliga protokoll f\"or samarbetande manipulering och transportering av objekt utf\"ord av flera robotagenter, och relationen av rigida samarbetskr\"avande manipulationsuppgifter. Vi f\"oresl\aa r f\"orst n\aa gra centraliserade och decentraliserade regleralgoritmer som saknar information om kraft och moment i kontaktpunkterna, men som tar h\"ansyn till b\aa de fasta och rullande greppunkter, dynamiska os\"akerheter i objektets och agentens modell, samt m\"ojlighet till att uppfylla villkor s\aa som att undvika hinder och m\"attning av insignaler. Som ett n\"asta steg behandlar vi reglering f\"or robust formering f\"or en klass av multi-agent system och vi analyserar relationen mellan teori f\"or formationsrigiditet och rigid samarbetskr\"avande manipulation.

I avhandlingens andra del utvecklar vi regleralgoritmer f\"or kontinuerlig koordinering av komplexa multi-agent system med os\"aker dynamik. Vi betraktar f\"orst r\"orelseplanering och f\"oresl\aa r nya adaptiva regleralgoritmer f\"or kollisionsfri navigering av enkel- och sf\"ariska- multiagent-system i hinderfyllda milj\"oer. Vi fokuserar sedan p\aa\ ledar-f\"oljare koordinering av sf\"ariska multiagent-system. Mer specifikt s\aa\ designar vi robust adaptiv decentraliserad reglering f\"or framg\aa ngsrik navigation av en utn\"amnd ledare till en f\"orutbest\"amt punkt,
samtidigt som vi garanterar att kollisioner kan undvikas och att sammankoppling uppr\"atth\aa lls.  Slutligen designar vi ett robust reglerprotokoll p\aa\ \aa terkopplad form baserat p\aa\ barri\"arfunktion f\"or kollisionsundvikande av multipla ellipsoidformade 3D-agenter.

Avhandlingens tredje del fokuserar p\aa\ planering och reglering av multiagent och multiagent-objekt system med komplexa uppgifter uttryckta med formler p\aa\ temporallogisk form. Vi behandlar f\"orst fallet med lokala oberoende uppgifter f\"or multiagent-system, och genom att anv\"anda tidigare resultat fr\aa n begr\"ansad navigering av multiagent-system designar vi en diskret abstraktion av agentens r\"orelse i arbetsytan och syntetiserar decentraliserade reglerpolicys som uppfyller agentens specifikationer.  F\"orutom robotagenterna tar vi sedan \"aven h\"ansyn till komplexa uppgifter som utf\"ors av op\aa verkade objekt. Vi designar en diskret abstraktion som simulerar agenternas beteenden och objekten i arbetsytan och vi syntetiserar regulatorer som tar h\"ansyn b\aa de till agenternas och objektens specifikationer. 


Den fj\"arde och sista delen av avhandlingen fokuserar p\aa\ flera utvidgningar f\"or singelagentfallet. F\"orst betraktar vi r\"orelseplanering f\"or singelagenter under temporala uppgifter i en hinderfylld milj\"o. Genom att anv\"anda tidigare resultat fr\aa n kollisionsfri tidsbegr\"ansad navigering utvecklar vi en ny reglerpolicy som garanterat uppfyller agentens tidsbegr\"ansningar  och \"ar asymptotiskt optimal med avseende p\aa\ energik\"allor. Sedan angriper vi r\"orelseplaneringproblemet f\"or m\aa ngdimensionella komplexa system med os\"aker dynamik i hinderfyllda milj\"oer. Vi integrerar intelligenta tekniker f\"or reglering med samplingsbaserade r\"orelseplaneringsalgoritmer f\"or att garantera s\"aker navigering av systemet till ett f\"orutbest\"amt m\aa l, samtidigt som vi kompenserar f\"or modellfel. Slutligen utvecklar vi nya regleringsprotokoll som uppn\aa r asymptotisk referensf\"oljning f\"or ett ok\"ant affint system, samtidigt som trattformade begr\"ansningar uppfylls.

\thispagestyle{empty} \mbox{}

\chapter*{Acknowledgements}
\addcontentsline{toc}{chapter}{Acknowledgements}

\thispagestyle{empty}

\vspace{3mm}
First and foremost, I would like to express my gratitude to my supervisor Prof. Dimos Dimarogonas for his valuable support, guidance and encouragement; his continuous feedback and inspiration has made this thesis possible, and it has been an excellent experience to work with him. I would also like to thank my co-supervisor Prof. Danica Kragic for her great insight, knowledge and assistance, and acknowledge the Knut och Alice Wallenberg Foundation, which funded partially my work.

Special thanks go to my advisor during my stay in Houston, Prof. Lydia Kavraki. She has been very kind and supportive, always providing significant research insights and motivation. The experience I gained is invaluable and I consider it a privilege to have worked with her group. 

I thank all my colleagues (current and former) at the Decision and Control Systems for creating a nice atmosphere, and for all the exciting, boring, difficult, happy, and sad moments we shared.
Special thanks go to my colleagues Alex, Lars, and Wences for our joint works and their continuous assistance and support, as well as to Pedro R., Pedro P., Pian, Peter, Sebastian, Antonio, Andrea, Xiao, Wei, Fei, Maria, Luis, Pierre-Jean, Leonardo, Dionysis, Christoforos, Shahab, Dimitris, Souleimane for our exciting group meetings and to Sofie for her MITL implementations. I would especially like to thank Dimitris Boskos for discussing with me all my strange research and notation questions and teaching me how to be technically rigorous. I am also grateful to Jana for conveying her excitement for temporal logics.  Big thanks go to Ziwei, Matteo, Yu, Imran, Cristina, Nicola, Francesca and Akash for our great collaboration during their master theses and for being patient with me. 
I would also like to thank all the professors and the administrative staff of the department. Special thanks to Silvia, Felicia, Anneli, Christer for their valuable help, and to Emmy for answering my vast amount of questions and making hence this thesis possible. 

Thank you David, for organizing the after work activities, and Mladen, Jezdimir, Rui, Rijad, Ines, and all the other kitchen regulars for our endless lunch discussions. Joana, thank you for pushing us to learn Swedish, our little group was successful while it lasted. I am grateful for Robert, Matias, Emma, and Elis for their positive spirit and for being great companions in our travels abroad. I am glad to have met the Italian attitude of Antonio, Riccardo, Valerio, Marco, and Demia. Thank you Peter for showing me stereograms and how to eat chestnut puree. Thank you Vahan for the khinkali. I would also like to thank my great roommates, Pedro Roque, Goncalo, Manne, Rong, and of course, Pedro Pereira, whose need for technical rigorousness motivated me and contributed to my compulsion. Thank you Dionisi, for always passing by to discuss random things, we have missed you. Thanks also to the SML people. Thank you Aldo for your positive attitude and inexplicably contagious laughter, Robin and Kuba, for refusing to give up on the robotic arms and my controllers, and Pedro, for upgrading the SML and pushing me (not very successfully though) to experiment with real robots.

I would further like to thank my Co4Robots partners and friends. Thank you Dimitri, Wei, Christos, and Pouria for your help in the project organization and implementation. Thank you Pedro, Sergio, Michali, Kosta Alevizo, Kosta Roditaki, Alexandre, Philipp, Alessandro, George, Claudio, and Meng for all our experiences during this project. I further thank the rest of the CSL people, Prof. Kyriakopoulos, Costas Vrohidis, and Panagiotis Vlantis for our collaboration during the project as well as my master thesis, and of course Babis, for exposing me to intelligent control systems and teaching me Lyapunov tricks. Babi, I wouldn't be here without you. 

I would like to thank my friends in Stockholm. Spyropoule and Charisi, thank you for all the weekend lunches, fikas, and drinks. Alex, thank you for being there for me when I started, for being a good friend, and for always having the answer in all my bureaucratic and application questions.
Pedro, thank you for being a great colleague, roommate, and friend. Thank you for co-piloting Co4robots and for the great time in all our trips abroad. Thank you for helping me with all my hardware and software questions, and in general, for putting up with me. Thank you for trying to put a band together and motivating me to play music again. Special thanks go to Wences and Oliv for all our shared moments and to Lars for co-organizing buffet Fridays and our fruitful discussions. 
I would also like to thank Dimitris, Vagelis, Psilos, Christoforos and Danilo for all the basketball moments we had so far and their incompetence in guarding me (please don't kick me out of the team). I hope Francisco is proud of us. 

I would further like to thank my friends in Greece and the rest of the world. Thank you Pano, Ilia, Tonia, Thanasi, and Faidona for everything we have been through together all these years. Thank you for your support and for being great friends. 

To my family, mother and brother, thank you from the bottom of my heart, for shaping my personality and always believing in me. You made me who I am and are always there for me. I am truly proud of you. 

\thispagestyle{empty} \mbox{}
\vspace{2mm}

Finally, I would like to express my utmost gratitude to my life partner, Stella, for constantly inspiring and motivating me all these years. And putting up with me. Stellou, this thesis could not have been written without your limitless moral support and unconditional love and it is yours as much as it is mine. 

%
%

\begin{flushright}
\emph{Christos Verginis} \\
Stockholm, Sweden \\
April 2020.
\end{flushright}

\thispagestyle{empty} \mbox{}
 
\afterpage{
	\pagestyle{empty}
	\newpage~\newpage
}
\newpage
\thispagestyle{empty} \mbox{} \vspace*{8cm}
\begin{flushright}
\emph{To my father}
\end{flushright}

\newpage
\thispagestyle{empty}
\mbox{}

\clearpage

\begin{KeepFromToc}
\tableofcontents
\end{KeepFromToc}

\chapter*{List of Abbreviations}
\addcontentsline{toc}{chapter}{List of Abbreviations}
 \rowcolors{2}{gray!25}{white}
\begin{tabulary}{\linewidth}{LL}
DNF & Decentralized Navigation Function \\
DoF & Degree of Freedom \\
FHOCP & Finite Horizon Optimal Control Problem \\
KRNF & Koditschek-Rimon navigation function  \\
LTL & Linear Temporal Logic \\
MILP & Mixed-Integer Linear Programming \\
MITL & Metric Interval Temporal Logic \\
MTL & Metric Temporal Logic \\
MPC & Model Predictive Control \\
MRNF & Multirobot Navigation Function \\
NMPC & Nonlinear Model Predictive Control \\
OCP & Optimal Control Problem \\ 
ODE & Ordinary Differential Equation \\
PPC & Prescribed Performance Control \\
RoI & Regions of Interest \\
RPF & Relation Proximity Function \\
RVF & Relation Verification Function \\
STL & Signal Temporal Logic \\
TBA & Timed B\"uchi Automata \\
TS & Transition System \\
TWTL & Time Window Temporal Logic \\
UAV & Unmanned Aerial Vehicle \\
WTS & Weighted Transition System \\
\end{tabulary}

\newpage

\newpage
\thispagestyle{empty} \mbox{}

\chapter*{List of Symbols}		
\addcontentsline{toc}{chapter}{List of Symbols}
\begin{longtable}{| p{.30\textwidth} | p{.70\textwidth} |} 
$\mathbb{N}$ & Set of natural numbers \\
$\mathbb{Q}$ & Set of rational numbers \\
$\mathbb{R}$ & Set of real numbers \\
$\mathbb{R}_{\geq 0}$ & Set of non-negative real numbers \\
$\mathbb{R}_{> 0}$ & Set of positive real numbers \\
$\mathbb{S}^{n-1} $ & Unit sphere in $\mathbb{R}^{n}$\\
$\mathbb{SO}(3)$ & Special orthogonal group in $3$ dimensions \\
$\mathbb{SE}(3)$ & Special Euclidean group in $3$ dimensions \\
$a \times b$ & Cross-product between two vectors $a,b\in\mathbb{R}^3$ \\
$S(x) \in \mathbb{R}^{3\times 3}$ & Skew-symmetric matrix of vector $x\in\mathbb{R}^3$ satisfying $S(a) b = a \times b $, for any vectors $a,b \in \mathbb{R}^3$ \\
$\lambda_\text{min}(A) \in \mathbb{R}$ & Minimum eigenvalue of a matrix $A\in\mathbb{R}^{n\times n}$\\
$\lambda_\text{max}(A) \in \mathbb{R}$ & Maximum eigenvalue of a matrix $A\in\mathbb{R}^{n\times n}$\\
$\sigma_\text{min}(A) \in \mathbb{R}$ & Minimum singular value of a matrix $A\in\mathbb{R}^{n\times n}$\\
$\partial \mathcal{A}$ & Boundary  of a set $\mathcal{A}$ \\
$\text{Int}(\mathcal{A})$ & Interior of a set $\mathcal{A}\subset \mathbb{R}^n$ \\
$\bar{\mathcal{A}}$ & Closure of a set $\mathcal{A}\subset \mathbb{R}^n$ \\
$x \succeq y$ &  Element-wise inequality for vectors $x, y\in\mathbb{R}^n$ \\
$(\alpha_1,\dots,\alpha_n)^\omega$ & Infinite sequence created by repeating $\alpha_1,\dots,\alpha_n$ \\
$\mathcal{B}(p, y) \subset \mathbb{R}^n$ & An open ball with center $p\in\mathbb{R}^n$ and radius $y \in \mathbb{R}_{>0}$ \\
$A \oplus B$ & Kronecker sum of the matrices $A\in \mathbb{R}^{n \times n}, B \in \mathbb{R}^{m \times m}$ \\
$A \otimes B$ & Kronecker product of the matrices $A \in \mathbb{R}^{m \times n} B \in \mathbb{R}^{p \times q}$ \\
$a \cdot b$ & Quaternion product of the quaternions $a, b \in \mathbb{S}^3$ \\
$a^+$ & Quaternion conjugate of $a\in\mathbb{S}^3$ \\
$\text{tr}(A)$ & Trace of a matrix $A \in \mathbb{R}^{n\times n}$ \\
$\det(A)$ & Determinant of a matrix $A \in \mathbb{R}^{n\times n}$ \\
$\| A\|_F$ & Frobenius norm of matrix $A \in \mathbb{R}^{n\times n}$ \\ 
$\text{span}(A)$ & Span of matrix $A\in\mathbb{R}^{n\times n}$ \\
$\text{rank}(A)$ & Rank of matrix $A \in \mathbb{R}^{n\times n}$ \\
$\text{null}(A)$ & Nullspace of matrix $A \in \mathbb{R}^{n\times n}$ \\
$\text{adj}(A)$ & Adjugate of matrix $A\in\mathbb{R}^{n\times n}$ \\
$A^\dagger $ & Moore-Penrose pseudo-inverse of matrix $A\in\mathbb{R}^{n\times n}$ \\
$\text{dim}(\mathbb{A})$ & Dimension of vector space $\mathbb{A}$  \\
$\nabla_x f(), \nabla^2_x f()$ & Gradient and Hessian, respectively, of function $f()\in\mathbb{R}$ with respect to $x \in\mathbb{R}^n$ \\
$\|x\|_1 \coloneqq |\sum_{i}^nx_i|$ & $\ell_1$ norm of vector $x = [x_1,\dots,x_n]^\top \in \mathbb{R}^n$\\
$\|x\| \coloneqq \sqrt{x^\top x}$ & Euclidean norm of vector $x = [x_1,\dots,x_n]^\top \in \mathbb{R}^n$\\
$\|A\| \coloneqq \sqrt{\lambda_{\text{max}}\left(A^\top A\right)}$ & Induced norm of matrix $A \in \mathbb{R}^{n\times n}$ \\
$\mathbbm{1}_n \in \mathbb{R}^n$ &  The column vector with all entries $1$ (subscript often omitted)\\
$I_n \in \mathbb{R}^{n \times n}$ & The unit matrix of dimension $n$ \\
$0_{m \times n} \in \mathbb{R}^{m \times n}$ & The $m \times n$ matrix with all entries zeros (subscript often omitted) \\
$\text{diag}(a_1,\dots,a_n)$ & Diagonal (block-diagonal) matrix with scalars (matrices) $a_i$ in the main diagonal (block diagonal) \\
$\mathbf{e}_{h} \in \mathbb{R}^3$ & Vector of one in the $h\in\{1,2,3\}$ element and zeros everywhere else \\ 
$\text{sgn}:\mathbb{R}\to \{-1,0,1\}$ & The sign function defined by $\text{sgn}(x)=1$, if $x>0$, $\text{sgn}(x)=-1$, if $x<0$, and $\text{sgn}(x)=0$, if $x=0$  \\
$\text{sgn}:\mathbb{R}^n\to\{-1,0,1\}^n$ &  $\text{sgn}(x)=[\text{sgn}(x_1),\dots,\text{sgn}(x_n)]^\top$ for $x=[x_1,\dots,x_n]^\top \in\mathbb{R}^n$\\
$\text{SGN}:\mathbb{R}\to[-1,1]$ & The set-valued sign function defined by $\text{SGN}(x)=1$, if $x>0$, $\text{SGN}(x)=-1$, if $x<0$, and $\text{SGN}(x)\in [0,1]$, if $x=0$ \\ 
$\text{SGN}:\mathbb{R}^n\to[-1,1]^n$ & $\text{SGN}(x)=[\text{SGN}(x_1),\dots,\text{SGN}(x_n)]^\top$ for $x=[x_1,\dots,x_n]^\top \in\mathbb{R}^n$ \\
$\text{sat}:\mathbb{R}\to[-1,1]$ & The saturation function $\text{sat}(x) = x$ if $|x|\leq 1$ and $\text{sat}(x) = \frac{x}{|x|} $ if $|x | > 1$ 
\end{longtable} 

\rowcolors{2}{}{}
\thispagestyle{empty} \mbox{}

\begin{tabulary}{\linewidth}{LL}

\end{tabulary} 
\rowcolors{2}{}{}
\thispagestyle{empty} \mbox{}

\clearpage	
\listoffigures

\chapter{Introduction}
\label{ch:Introduction}
The technological developments have been increasing exponentially during the last century, with an evident peak in the last few decades. The recent need for development of smart cities (including autonomy in industrial buildings, houses, highways, as well as automated rescue missions) calls for wider deployment of robots that must coordinate with each other to achieve a specific task. 
Additionally, noteworthy is the increasing evolution of wireless communication technology that results in the low-cost massive development of (internal and external) sensor devices. Along with the incapability of the corresponding computing units to process very large amounts of data in small amounts of time, this has given rise to a special case of systems that consist of multiple robots, namely multi-agent systems. Multi-agent systems consist of agents/robots that rely solely on local sensor information with respect to their neighboring robots to determine their actions, which is often called \textit{decentralized control}. 

During the last decade, decentralized control of multi-agent systems has gained a significant amount of attention due to the great variety of its applications, including  multi-robot systems, transportation, multi-point surveillance and biological systems. The main focus of multi-agent systems is the design of distributed control protocols in order to achieve global tasks, such as \emph{consensus} \cite{ren_beard_concensus, olfati_murray_concensus, jadbabaie_morse_coordination, tanner_flocking, dimos_rendezvous_problem}, in which all the agents are required to converge to a specific point, and \emph{formation} \cite{egerstedt_formation, oh_park_ahn_2015}, in which all the agents aim to form a predefined geometric shape. At the same time, the agents might need to fulfill certain transient  properties, such as \emph{network connectivity} \cite{magnus_2007_connectivity, zavlanos_2007_potential, zavlanos_2008_distributed} and/or \emph{collision avoidance} \cite{dimos_2006_automatica_nf}. 

A special case of multi-agent systems is cooperative robotic manipulators. In particular, when it comes to object manipulation/transportation, large/heavy payloads as well as complex maneuvers necessitate the deployment of more than one robot. The most common tasks consist of pick-and-place tasks and cooperative object transportation, while satisfying certain properties, such as collision- and singularity-avoidance. 

\begin{figure}[t!]
	\centering
	\includegraphics[scale = 0.50]{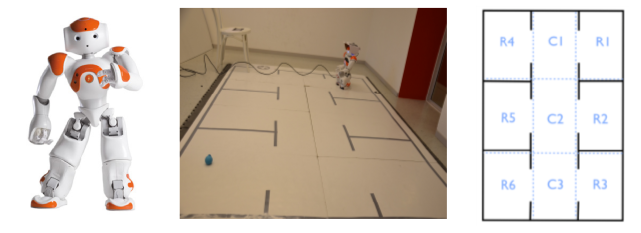}
	\caption{A humanoid robot moving to an environment consisting of $6$ rooms and $3$ corridor regions. In room $R6$ there exists a ball that the robot can grab.}
	\label{fig:intro_pic}
\end{figure}

Another topic that has troubled researchers the last decades is 
the control of multiple systems such that each agent/robot fulfills desired tasks given by high-level specifications expressed as temporal logic formulas. Temporal-logic based motion planning has gained a significant amount of attention over the last decade, since it provides a fully automated  correct-by-design controller synthesis approach for autonomous robots. Temporal logics, such as linear temporal logic (LTL), provide formal high-level languages that can describe planning objectives more complex than the well-studied navigation algorithms, and have been used extensively both in single- as well as in multi-agent setups. The objectives are given as a temporal logic formula with respect to a discretized abstraction of the system (usually a finite transition system), and then, a high-level discrete path is found by off-the-shelf model-checking algorithms, given the abstracted system and the task specification.   Consider, for instance, the robot in Figure \ref{fig:intro_pic} operating in a workspace which is partitioned into $6$ rooms and a corridor consisting of three regions. A high-level task for the robot might have the following form: ``Periodically visit rooms $R_1$, $R_4$, $R_6$, in this order, while avoiding rooms $R_2$, $R_3$ and $R_5$", or ``Grab the ball that lies in room $R_6$ and deliver it in room $R_3$ between 10 and 20 seconds". The aforementioned specifications include complex tasks where \emph{time} might play an important role. 

One of the main problems that arise when dealing with high-level tasks based on temporal-logic formulas is the construction of a discrete abstracted representation of the continuous system. More specifically, given a temporal-logic formula over a continuous workspace/state space, how does one partition this space into discrete state?s? Moreover, given a predefined partition, what are the control inputs of the agents that guarantee well-defined transitions among the discrete states? When multi-agent systems are concerned, the aforementioned specifications must also incorporate collision-avoidance as well as connectivity-maintenance properties among the robots, which brings the problem of abstraction to a new level of complexity.

Furthermore, consider a case where some unactuated objects must undergo a series of processes in a workspace with autonomous agents (e.g., car factories), expressed as temporal-logic high-level specifications. In such cases, the agents, except for satisfying their own motion specifications, are also responsible for coordinating with each other in order to transport the objects around the workspace. When the unactuated objects' specifications are expressed using temporal logics, then the motion- and task- planning of the agents' behavior becomes much more complex, since the discrete system abstraction has to also take into account the objects' goals.

The aforementioned problems become even more challenging when one takes into account system uncertainty. The dynamic model of real robotic systems cannot be accurately known by the user/designer, since it includes terms that might not be easy to identify, e.g., dynamic parameters (mass, inertia), friction, and other external disturbances. This becomes more apparent as the complexity of the considered systems increases (consider, e.g., a mobile robot vs a $6$-DoF robotic manipulator). These uncertainties are expected to affect the performance of the system, and since they cannot be accurately canceled by the control design, the latter must render the closed-loop system robust to them \cite{krstic1995nonlinear}.

Motivated by the above discussion, this thesis aims at solving the problem of decentralized motion- and task-planning of uncertain multi-agent and multi-agent-object systems under complex task specifications by integrating tools from the computer science and automatic control fields. 
The main contributions lie in the robust abstraction of the continuous coupled object-agents dynamics into a discrete representation of the system (transition systems) and the application of formal verification methodologies towards the satisfaction of temporal logic formulas.  More specifically, we break down the problem into three main subproblems. Motivated by the need of transition design for unactuated objects, we consider first the problem of cooperative object manipulation. We design control protocols for the centralized and decentralized cooperative manipulation of an object grasped by multiple robotic agents by means of rigid as well as rolling contacts, possibly subject to model uncertainties. Moreover, we study the relation of rigid cooperative manipulation with rigid formation control, and design a novel control algorithm for the latter problem. Secondly, again in view of transition design for multi-agent systems, we develop  numerous control protocols for the coordination of multi-agent systems, including multi-agent navigation and leader-follower coordination, subject to collision and connectivity constraints as well as model uncertainties. The third part draws from the previous ones to design well defined discrete abstractions for multi-agent and multi-agent-objects systems. In that way, we allow the expression of complex desired tasks as temporal logic specifications, for which we provide controller synthesis. Finally, we study some problems for single-agent setups, including timed temporal specifications in an obstacle-cluttered environment, integration of intelligent control protocols with sampling-based motion planning algorithms for complex uncertain systems, as well as asymptotic stability properties with funnel constraint satisfaction. 

The work developed in this thesis was supported by the research projects ``H2020 Research and Innovation Programme" under the Grant Agreements No. 644128 (AEROWORKS) and No. 731869 (Co4Robots), the0 H2020 ERC Starting Grant BUCOPHSYS, the Knut and Alice Wallenberg Foundation, the Swedish Research Council (VR), and the Swedish Foundation for Strategic Research. The next section presents the outline of this thesis.

\section{Thesis Outline and Contributions}
In this Section, we provide the outline of the thesis and indicate the contributions of each chapter. 
The thesis is divided into \emph{four main parts}, the first three of which aiming to solve the aforementioned problems, and the final one discussing single-agent extensions. 

\begin{itemize}
	\item The \emph{first part} consists of Chapters \ref{chapter:cooperative manip} and \ref{chapter:formation}. In this part, we first tackle the problem of cooperative manipulation of an object grasped by several robotic agents. We propose a variety of control algorithms, combining centralized and decentralized setups, rigid and rolling contacts, adaptive and Model Predictive control techniques, as well as incorporation of collision avoidance techniques with workspace obstacles. Next, motivated by its application to cooperative manipulation, we propose a novel control algorithm for the formation stabilization of a multi-agent team. Moreover, we explicitly study the relation of rigid formation control with robotic cooperative manipulation schemes through rigid contacts.	
	
	\item The \emph{second part} consists of Chapter \ref{chapter:synthesis}. In this part, we develop continuous control algorithms for multi-agent coordination under model uncertainties. More specifically, we propose first a novel adaptive control protocol for the single- and multi-agent collision-free navigation in an obstacle-cluttered workspace subject to uncertain dynamics and spherical shapes. Next, we tackle the leader-follower coordination problem. We develop an adaptive control algorithm for the leader navigation to a predefined goal while guaranteeing inter-agent collision avoidance and connectivity maintenance. Finally, motivated by real robotic structures, we present a closed form control protocol that achieves collision avoidance among \textit{ellipsoidal} agents, while compensating at the same time for the uncertain dynamics.
	
	\item The \emph{third part} consists of Chapter \ref{chapter:abstraction}. In this chapter, we design well-defined abstractions for multi-agent and multi-agent-object systems in discretized workspaces. This allows us to define complex tasks as temporal logic formulas and employ formal verification methodologies to synthesize control protocols. The discretized abstractions include both fully partitioned workspaces as well as discretization based on predefined regions of the workspace. We use control methodologies from the previous chapters as well as newly designed ones.  
	
	\item The \textit{fourth and final part}, consisting of Chapter \ref{chapter:single agent}, considers some challenging extensions for single-agent systems. Firstly, we tackle the problem of the optimal motion planning under \textit{timed} temporal logic specifications in an obstacle-cluttered environment. We use previous results on collision-free \textit{timed} navigation and we develop a novel reconfigurable framework that guarantees asymptotically optimal behavior. Secondly, we integrate adaptive control techniques with sampling-based algorithms for the motion planning problem of complex high-dimensional systems. We propose a two-layer approach that compensates for the system uncertainties and guarantees the collision-free navigation to the goal via a geometric path in an extended free space. Finally, we develop a novel control algorithm that guarantees asymptotic stability of a general class of uncertain systems subject to funnel constraints.
	
\end{itemize}

\subsubsection{Chapter \ref{chapter:cooperative manip} }

This chapter addresses the problem of cooperative manipulation of a single object by multiple robotic agents. We present first four control algorithm for the case of \textit{rigid} contact grasps. The first two are decentralized, adaptive closed-form techniques that aim to guarantee trajectory tracking by the object's center of mass while compensating for model uncertainties and external disturbances, and imposing predefined performance on the closed-loop system. Next, we design a centralized and a decentralized control protocol using the Nonlinear Model Predictive Control  methodology, which guarantee object transportation to a desired pose, while complying with other constraints, such as obstacle avoidance and input saturation. Finally, we consider the case of rolling contacts. We design a centralized control protocol that guarantees object trajectory tracking, robust to model uncertainties and center of mass location. Moreover, we propose a novel algorithm for contact maintenance of the agents with the object. By employing event-triggered agent communication, we extend the latter scheme to a decentralized version. It is noteworthy that none of the aforementioned approaches relies on force/torque sensor information.
The covered material is based on the following contributions \cite{verginisMastellaro,verginis2018TCST,alex_chris_med_2017,alex_chris_ecc_2018,verginisWencesACC,verginisWencesJournal}:

\begin{itemize}	
	\item C. K. Verginis, M. Mastellaro and D. V. Dimarogonas, ``Robust quaternion-based cooperative manipulation without force/torque information”, IFAC-PapersOnLine, 50(1), pp. 1754-1759, Toulouse, France, 2017.	
	\item C. K. Verginis, M. Mastellaro and D. V. Dimarogonas, ``Cooperative manipulation without force/torque measurements: Control design and experiments”, IEEE Transactions on Control Systems Technology, vol. 28, no. 3, pp. 713-729, 2020.
	\item A.  Nikou, C. K. Verginis and D. V. Dimarogonas, ``A nonlinear model predictive control scheme for cooperative manipulation with singularity and collision avoidance”, IEEE Mediterranean Conference on Control and Automation (MED), pp. 707-712, Valletta, Malta, 2017.
	\item C. K. Verginis, A. Nikou and D. V. Dimarogonas, ``Communication-based decentralized cooperative object transportation using nonlinear model predictive control”, IEEE European Control Conference (ECC), pp. 733-738, Limassol, Cyprus, 2018. 
	\item C. K. Verginis, W. S. Cortez and D. V. Dimarogonas, ``Adaptive Cooperative Manipulation with Rolling Contacts", to appear in the American Control Conference (ACC), Denver, Colorado, USA, 2020.	
	\item C. K. Verginis, W. S. Cortez and D. V. Dimarogonas, ``Decentralized adaptive Cooperative Manipulation with Rolling Contacts", under preparation.	
\end{itemize}

\subsubsection{Chapter \ref{chapter:formation} }

This chapter presents first a novel control protocol for the formation control of tree graphs in $\mathbb{SE}(3)$. The control laws are decentralized as well as robust to modeling uncertainties (parametric and structural) and external disturbances. The proposed methodology guarantees collision avoidance and connectivity maintenance among the initially connected agents and certain predefined functions characterize the transient and steady-state performance of the closed loop system. Next, we study the relation between rigid cooperative manipulation and rigid formations. By doing so, we provide novel conditions for the internal force-free cooperative manipulation based on the rigidity matrix of the underlying multi-agent system.  
The covered material is based on the following contributions \cite{verginis2019robust, verginisICRA20, verginis2019cooperative}:

\begin{itemize}		
	\item C. K. Verginis, A. Nikou and D. V. Dimarogonas, ``Robust formation control in $\mathbb{SE}(3)$ for tree-graph structures with prescribed transient and steady state performance”, Automatica 103 (2019): 538-548. 	
	\item C. K. Verginis and D. V. Dimarogonas: ``Energy-Optimal Cooperative Manipulation via Provable Internal-Force Regulation", to appear in the IEEE International Conference on Robotics and Automation (ICRA), Paris, France, 2020. 
	\item C. K. Verginis, D. Zelazo, and D. V. Dimarogonas: ``Cooperative Manipulation via Internal Force Regulation: A Rigidity Theory Perspective", Under Review. Arxiv Link: https://arxiv.org/pdf/1911.01297.pdf
\end{itemize}

\subsubsection{Chapter \ref{chapter:synthesis} }

This chapter tackles the problem of multi-agent coordination in the following ways. Firstly, we consider the problem of single- and multi-agent navigation in an obstacle-cluttered spherical environment under $2$nd-order uncertain dynamics. We propose an adaptive control scheme that guarantees the single-agent collision-free navigation to the goal from almost all initial conditions while compensating for the uncertain dynamics, which is then extended it to a decentralized priority-based multi-agent case. Secondly, we consider the leader-following coordination problem in the following sense. A leader agent aims at navigating to a pre-specified pose, while the entire team has to avoid collision with each other, as well as maintain connectivity. We develop a decentralized adaptive control protocol, compensating again for dynamic uncertainties, to guarantee accomplishment of the aforementioned specifications. The algorithms above consider spherical agents, which might be a conservative over-simplification when it comes to real robots. Therefore, we finally develop an adaptive control methodology that guarantees collision avoidance among \textit{ellipsoidal} agents. We propose a novel closed-form function that encodes collisions among $3$D ellipsoids and combine it with an adaptive control law that compensates for the model uncertainties. 
The covered material is based on the following contributions \cite{veginisAutomatica2020,verginisCDC19_LF,verginisLCSS}:
\begin{itemize}
	\item C. K. Verginis and D. V. Dimarogonas: ``Adaptive Robot Navigation with Collision Avoidance Subject	to 2nd-order Uncertain Dynamics", Under Review.
	\item C. K. Verginis and D. V. Dimarogonas, ``Adaptive  Leader-Follower  Coordination  of  Lagrangian  Multi-AgentSystems  under  Transient  Constraints”, IEEE Conference on Decision and Control (CDC), pp. 3833-3838, Nice, France, 2019.
	\item C. K. Verginis and D. V. Dimarogonas, ``Closed-Form Barrier Functions for Multi-Agent Ellipsoidal Systems with Uncertain Lagrangian Dynamics", IEEE Control System Letters,  pp. 727-732, 2019.
\end{itemize}

\subsubsection{Chapter \ref{chapter:abstraction}}

This chapter addresses the motion planning problem for multi-agent and multi-agent-object systems under high level complex tasks expressed as temporal logic formulas. We first focus on local temporal logic specifications for each agent individually. We use previous results to derive well-defined discrete abstractions based on pre-defined regions of interest in the workspace, possibly by accounting for collision and connectivity constraints. We use then standard formal verification techniques to derive paths that satisfy the independent tasks. Next, apart from the agents, we consider that unactuated objects have to satisfy certain temporal logic tasks. The robotic agents are now responsible for satisfying the objects' tasks, except for their own. We use again previous results to derive discrete abstractions of the coupled system's motion, based on both regions of interest as well as a complete workspace partition. We then apply the same formal verification-based strategy to obtain discrete paths that satisfy the agents' and the object's goals.
These results are based on \cite{verginisICRA2017, verginisCDC017,verginis2016distributed,Verginis_Auton_Robots,verginis_ifac17,Verginis_tase_subm}:

\begin{itemize}
	\item C. K. Verginis, Z. Xu and D. V. Dimarogonas, ``Decentralized motion planning with collision avoidance for a team of UAVs under high level goals”, IEEE International Conference on Robotics and Automation (ICRA), pp. 781-787, Singapore, 2017.
	\item C. K. Verginis and D. V. Dimarogonas, ``Robust decentralized abstractions for multiple mobile manipulators”, IEEE Conference on Decision and Control (CDC), pp. 2222-2227, Melbourne, Australia, 2017.
	\item C. K. Verginis and D. V. Dimarogonas, ``Distributed Cooperative Manipulation under Timed Temporal Specifications”, American Control Conference (ACC), pp. 1358-1363, Seattle, USA, 2017.
	\item C. K. Verginis and D. V. Dimarogonas, ``Timed abstractions for distributed cooperative manipulation”, Autonomous Robots, 42, no. 4 (2018): 781-799.
	\item C. K. Verginis and D. V. Dimarogonas, “Multi-agent motion planning and object transportation under high level goals”, IFAC-PapersOnLine, 50(1), pp. 15816-15821, Toulouse, France, 2017.
	\item C. K. Verginis, and D. V. Dimarogonas, ``Motion and cooperative transportation planning for multi-agent systems under temporal logic formulas”, BOSCH AI Conference, 2018.\\ Arxiv Link: https://arxiv.org/pdf/1803.01579.pdf.  	
\end{itemize}

\subsubsection{Chapter \ref{chapter:single agent}}

This chapter addresses some challenging extensions for single-agent setups. Firstly, we consider the problem of  motion planning under timed temporal tasks for a mobile robot in an obstacle-cluttered environment. We use previous results in collision-free timed navigation and develop a novel timed automata-based reconfiguration algorithm that achieves the satisfaction of the task in an asymptotically energy-optimal way. Secondly, we consider the motion planning problem for complex high-dimensional systems (e.g., robotic manipulators) with uncertain dynamics in obstacle-cluttered environments. We integrate in an innovative way sampling-based motion planning algorithms and adaptive control to provide a two layer framework that guarantees the safe navigation of the robot to its goal, while compensating for its uncertain dynamics. Finally, we consider the tracking problem for a class of uncertain nonlinear systems under funnel constraints. We develop a novel adaptive control protocol that achieves \textit{asymptotic} tracking while complying to the funnel specifications and without using any model information.
These results are based on \cite{verginis2019reconfigurable,verginisWAFR,verginis2019asymptotic,verginisTAC}:

\begin{itemize}
	\item C. K. Verginis, K. Vrohidis, C. P. Bechlioulis, K. J. Kyriakopoulos, and D. V. Dimarogonas, “Reconfigurable Motion Planning and Control in Obstacle Cluttered Environments under Timed Temporal Tasks”, IEEE International Conference on Robotics and Automation (ICRA), pp. 951-957, Montreal, Canada, 2019.
	\item C. K. Verginis, D. V. Dimarogonas, and L. E. Kavraki, ``Sampling-based Motion Planning for Uncertain High-dimensional Systems via Adaptive Control”, submitted to the Workshop on the Algorithmic Foundations of Robotics (WAFR), Oulu, Finland, 2020.
	\item C. K. Verginis and D. V. Dimarogonas, ``Asymptotic Stability of Uncertain Lagrangian Systems with Prescribed Transient Response”, IEEE Conference on Decision and Control, pp. 7037-7042, Nice, France, 2019.
	\item C. K. Verginis and D. V. Dimarogonas, ``Asymptotic Tracking of Second-order Nonsmooth Feedback Stabilizable Unknown Systems with Prescribed Transient Response”, under Review.	 	 
\end{itemize}

Finally, in Chapter \ref{chapter:conclusion}, conclusions of this thesis as well as future research directions are discussed.

\subsubsection{Contributions not included in this thesis}

The following publications are not covered in this thesis, but are related to the work presented here \cite{alex_chris_ppc_formation_ifac,alex_cdc_2017_formation,lindemann2017prescribed,alex_cdc_2017_timed_abstractions,wei2018asymptotic,verginisTianyang20,verginisNicola20}:

\begin{itemize}
	\item A. Nikou, C. K. Verginis and D. V. Dimarogonas, ``Robust distance-based formation control of multiple rigid bodies with orientation alignment”, IFAC-PapersOnLine, 50(1), pp. 15458-15463, Toulouse, France, 2017.	
	
	\item C. K. Verginis, A. Nikou and D. V. Dimarogonas, “Position and orientation based formation control of multiple rigid bodies with collision and avoidance and connectivity maintenance”, IEEE International Conference on Decision and Control (CDC), pp. 411-416, 2017, Melbourne, Australia.
	
	\item L. Lindemann, C. K. Verginis and D. V. Dimarogonas, ``Prescribed performance control for signal temporal logic specifications”, Proceedings of the IEEE Conference on Decision and Control (CDC), pp. 2997-3002, Melbourne, Australia, 2017.
	
	\item A. Nikou, C. K. Verginis, S. Heshmati-alamdari and D. V. Dimarogonas, ``Decentralized abstractions and timed constrained planning of a general class of coupled multi-agent systems”, Proceedings of the IEEE Conference on Decision and Control (CDC), pp. 990--995, Melbourne, Australia, 2017.
	
	\item J. Wei, C. K. Verginis, J. Wu, D. V. Dimarogonas, H. Sandberg, and K. H. Johansson, “Asymptotic and Finite-Time Almost Global Attitude Tracking: Representations Free Approach”, European Control Conference (ECC), pp. 3126-3131, Limassol, Cyprus, 2018. 
	
	\item T. Pan, C. K. Verginis, A. M. Wells, D. V. Dimarogonas, and L. E. Kavraki: “Augmenting Control Policies with Motion Planning for Robust and Safe Multi-robot Navigation", submitted to the IEEE International Conference on Intelligent Robots and Systems (IROS), Las Vegas, NV, USA, 2020.
	
	\item N. Lissandrini, C. K. Verginis, P. Roque, A. Cenedese, and D. V. Dimarogonas: “Decentralized Nonlinear MPC for Robust Cooperative Manipulation by Heterogeneous Aerial-Ground Robots", submitted to the IEEE International Conference on Intelligent Robots and Systems (IROS), Las Vegas, NV, USA, 2020.
\end{itemize}

   
\chapter{Cooperative Object Manipulation} \label{chapter:cooperative manip}
As mentioned in the previous chapter, cooperative manipulation of objects by autonomous robotic agents is of paramount importance in creating discrete representations of multi-object-robot systems as well as autonomizing item transportation tasks.

This chapter addresses the problem of cooperative manipulation of a single object by multiple robotic agents. We consider first the case where the agents grasp an object by means of \textit{rigid contacts}, and we present four novel control methodologies for the trajectory tracking by the object's center of mass, without the need for force/torque feedback at the grasping points. Firstly, we design an adaptive control protocol which employs quaternion-based feedback for the object orientation to avoid potential representation singularities. Secondly, we propose a control protocol that guarantees predefined transient and steady-state performance for the object trajectory. Both methodologies are decentralized, since the agents calculate their own signals without communicating with each other, as well as robust to external disturbances and model uncertainties. Load sharing coefficients are also introduced to account for potential differences in the agents' power capabilities. Thirdly, we turn to optimization techniques and use Nonlinear Model Predictive Control (NMPC) to guarantee convergence of the object's center of mass to a \textit{fixed} pose, both in a centralized and a communication-based decentralized framework. These approaches also guarantee collision avoidance properties among the robotic agents and potential workspace obstacles as well as avoidance of kinematic/representation singularities. 

Secondly, we consider the problem of object manipulation by means of \textit{rolling contacts}. We present a centralized control algorithm that achieves trajectory tracking by  the object as well as a decentralized extension using event-triggered communication, still without using force/torque feedback. Both schemes employ adaptive control ideas to compensate for potential uncertainties in the agents' and the object's dynamic parameters and do not use information regarding the object's center of mass, since the tracking concerns an observable point on the object. Contact slip avoidance is also guaranteed by novel optimization algorithms.
Simulation and experimental results support the theoretical findings.

\section{Introduction}

As highlighted in the previous chapter, multi-agent systems have gained significant attention the last years due to the numerous advantages they yield with respect to single-agent setups. In the case of robotic manipulation, heavy payloads and challenging maneuvers necessitate the employment of multiple robotic agents. Although collaborative manipulation of a single object, both in terms of transportation (regulation) and trajectory tracking, has been considered in the research community in the last decades, there still exist several challenges that need to be taken account by on-going research, both in control design as well as experimental evaluation. Moreover, along the lines of designing well-defined discretized abstractions for cooperative manipulation tasks, successful manipulation/transportation of objects plays a crucial role for the potential transitions between the states of the derived discrete system representation. 
In this chapter we model explicitly a system of multiple robotic agents grasping an object and develop control protocols for the pose and time trajectory tracking of the center of mass of the object. 

Early works develop control architectures where the robotic agents communicate and share information with each other, and completely decentralized schemes, where each agent uses only local information or observers, avoiding potential communication delays (see, indicatively, \cite{schneider1992object,sugar2002control,khatib1996decentralized,liu1996decentralized,liu1998decentralized,zribi1992adaptive,gudino2004control,wen1992motion,yoshikawa1993coordinated,kopf1989dynamic}). Impedance and hybrid force/position control is the most common methodology used in the related literature \cite{caccavale2000task,caccavale2008six,heck2013internal,erhart2013adaptive,erhart2013impedance,kume2007coordinated,szewczyk2002planning,tsiamis2015cooperative,ficuciello2014cartesian,ponce2016cooperative,gueaieb2007robust,Li_fuzzy2015,wen1992motion,yoshikawa1993coordinated,tzierakis2003independent,kopf1989dynamic,marino2017distributed}, where a desired impedance behavior is imposed potentially with force regulation. Most of the aforementioned works employ force/torque sensors to acquire feedback of the object-robots contact forces/torques, which however may result in a performance decline due to sensor noise or mounting difficulties. Recent technological advances allow manipulator grippers to grasp rigidly certain objects (see e.g., \cite{grasping2014}), which  can render the use of force/torque sensors unnecessary. Force/Torque sensor-free methodologies can be found in  \cite{zribi1992adaptive,wen1992motion,kume2007coordinated}, which have inspired the dynamic modeling in this work. Moreover, \cite{ficuciello2014cartesian} uses an external force estimator, without employing force sensors, \cite{liu1996decentralized} presents a force sensor-free control protocol with gain tuning, and \cite{caccavale2000task} considers the object regulation problem without force/torque feedback. Finally, force/torque sensor-free methodologies are developed in \cite{petitti2016decentralized}, where the robot dynamics are not taken into account, and in \cite{tzierakis2003independent}, where a linearization technique is employed.  

Another important characteristic is the representation of the agent and object orientation. The most commonly used tools for orientation representation consist of rotation matrices, Euler angles, and the pair angle-axis convention. Rotation matrices, however, are not commonly used in robotic manipulation tasks due to the difficulty of extracting an error vector from them. Moreover, the mapping from Euler angle/axis values to angular velocities exhibits singularities at certain points, rendering thus these representations incompetent. On the other hand, the representation using unit quaternions, which is employed in this work, constitutes a singularity-free orientation representation, without complicating the control design. In cooperative manipulation tasks, unit quaternions are employed in \cite{caccavale2008six,aghili2011self,caccavale2000task} as well as in \cite{erhart2016model}, where the interaction dynamics of cooperative manipulation are analyzed.

In addition, most works in the related literature consider known dynamic parameters regarding the object and the robotic agents. However, the accurate knowledge of such parameters, such as masses or moments of inertia, can be a challenging issue, especially for complex robotic manipulators; adaptive control protocols are proposed in  \cite{liu1998decentralized} with a gain tuning scheme, in \cite{caccavale2000task}, where the object regulation problem is considered, and in \cite{zribi1992adaptive}, \cite{ponce2016cooperative}. An estimation of parameters is included in \cite{petitti2016decentralized,marino2018two}, whereas \cite{gueaieb2007robust} and \cite{Li_fuzzy2015} employ fuzzy mechanisms to compensate for model uncertainties. In \cite{marino2017distributed,marino2017decentralized} the authors develop a task-oriented adaptive control protocol using observers. Kinematic uncertainties and joint limits are handled in \cite{aghili2011self}, \cite{erhart2013adaptive}, and \cite{ortenzi2018dual}, respectively. 

An internal force and load distribution analysis is performed in \cite{erhart2015internal}; \cite{tsiamis2015cooperative} employs a leader-follower scheme, and \cite{wang2015multi} develops a decentralized force consensus algorithm. Furthermore, \cite{chaimowicz2003hybrid} introduces hybrid modeling of cooperative manipulation schemes and \cite{murphey2008adaptive} includes intermittent contact; \cite{wang2016kinematic} proposes a kinematic-based multi-robot manipulation scheme, and \cite{bai2010cooperative,sieber2018human} address the problem from a formation-control point of view. In \cite{tanner2003nonholonomic} a navigation function-based approach is used, and object manipulation by aerial robots is considered in \cite{lippi2018cooperative,sanalitro2020full,gabellieri2019study}.

Another interesting direction regarding cooperative manipulation is the \textit{safe} transportation of an object in an obstacle-cluttered environment. 
In standard manipulation tasks, collision with obstacles of the environment has been dealt with only by exploiting the extra degrees of freedom that appear in over-actuated robotic agents. Potential field-based algorithms may suffer from local minima and navigation functions \cite{koditschek1990robot} cannot be extended to multi-agent second order dynamical systems in a trivial way. Moreover, these methods usually result in high control input values near obstacles that need to be avoided, which might conflict the saturation of the actual motor inputs. 

Other important properties that concern robotic manipulators are the
1) input saturation constraints, naturally characterizing real actuators, and 2)
singularities of the Jacobian matrix, which maps the joint velocities of the agent to a $6$D vector of generalized velocities. Such \textit{singular} \textit{kinematic} configurations, which indicate directions towards which the agents cannot move, must be always avoided, especially when dealing with task-space control in the end-effector \cite{siciliano2010robotics}. 
As already mentioned before, \textit{representation} singularities can also occur in the mapping from coordinate rates to angular velocities of a rigid body. Typical control schemes cannot guarantee satisfaction of a task while provably avoiding input saturations or singularities. 

The aforementioned properties can be considered as an instance of constrained-based control, which has always been of special interest to the automatic control/robotics community, due to the advantages it yields, by keeping variables of interest in specific compact sets, while achieving a primary task. A widely employed methodology in the last years is the methodology of Model Predictive Control (MPC) \cite{morrari_npmpc}, where a constrained optimization problem is solved for a finite horizon in the future, providing a prediction of the state evolution. For the design of a stabilizing feedback control law under such constraints, one would ideally look for a closed-loop solution for the feedback law satisfying the constraints while optimizing the performance. However, typically the optimal feedback law cannot be found analytically, even in the unconstrained case, since it involves the solution of the corresponding Hamilton-Jacobi-Bellman partial differential equations. One approach to circumvent this problem is the repeated solution of an open-loop finite-horizon optimal control problem for a given state. The first part of the resulting open-loop input signal is implemented and the whole process is repeated. Control approaches using this strategy are referred to as Nonlinear Model Predictive Control (NMPC) (see e.g. \cite{morrari_npmpc, frank_2003_nmpc_bible, frank_1998_quasi_infinite, frank_2003_towards_sampled-data-nmpc, fontes_2001_nmpc_stability, grune_2011_nonlinear_mpc, camacho_2007_nmpc, cannon_2001_nmpc, borrelli_2013_nmpc, fontes_2007_modified_barbalat}), which we use in this chapter for the problem of the constraint cooperative object manipulation. 

All the aforementioned approaches rely on the assumption that each robotic agent is \textit{rigidly} attached to the object, allowing it to apply any force/torque at the contact point. This rigidity assumption is highly restrictive as it only applies to objects on which a rigid grasp can be formed, excluding, e.g., objects with smooth surfaces or large boxes/spheres (e.g., packages), which cannot be rigidly grasped by a simple gripper. Non-rigid/rolling contacts, on the other hand, increase the number of objects
that can be grasped, increase the workspace of the system, and allow for modular manipulation scenarios in which robots can be swapped in/out to adjust the grasp online. Note that,
by employing rolling contacts, the cooperative manipulation problem becomes similar to robotic grasping \cite{ozawa2017grasp} albeit with moving “fingers.”

Rolling contacts complicate the problem as each contact may only apply a force that respects friction cone constraints to prevent slip, instead of an arbitrary wrench associated with rigid contacts \cite{Kerr1986}.  Early robotic grasping approaches required exact knowledge of the agent’s dynamics \cite{Kerr1986,cole1989kinematics}. Other recent techniques are robust to model uncertainties, but neglect rolling effects or dynamics \cite{fan2017robust,Caldas2015,Wimbock2012}, while other more sensor-deprived approaches assume the object is weightless \cite{Tahara2010,Kawamura2013}. The approach from \cite{ShawCortez2019} assumes a priori bounded
states, which does not apply to mobile manipulators that can be potentially considered. Adaptive control schemes that have also been developed require force and contact location sensing, and assume boundedness of the uncertain parameter estimates \cite{Ueki2008,Ueki2011}, or are limited to set-point (constant reference) manipulation \cite{Cheah1998}. 

Furthermore, for collaborative manipulation using rolling contacts, it is critical to
ensure the object does not slip. This is neglected by most of the aforementioned approaches, which assume either rigid grasps or simply no slip without guarantees. Methods of ensuring slip prevention are developed typically by solving an optimization problem online \cite{fan2017robust,ShawCortez2019,Fungtammasan2012}. However, \cite{fan2017robust,Fungtammasan2012} neglect the
dynamics of the system, which may perturb the system and
cause slip. The approach in \cite{ShawCortez2019} uses a conservative bound
on the dynamics, which overcompensates the amount of force
required to hold the object. Finally, most related works consider accurate knowledge of the
object center of mass, which can be difficult to obtain in
practice, especially in cases of complicated object shapes.

The contribution of this chapter consists of the following: Firstly, we introduce two novel close-form nonlinear control protocols for the trajectory tracking by the center of mass of an object that is \textit{rigidly} grasped by $N$ robotic agents, without using force/torque measurements at the grasping points. In particular, we develop first a decentralized control scheme that combines (i) adaptation laws to compensate for external disturbances and uncertainties of the agents' and the object's dynamic parameters, with (ii) quaternion modeling of the object's orientation that avoids undesired representation singularities. 
Then, we propose a decentralized model-free control scheme that guarantees \textit{predefined} transient and steady-state performance for the object's center of mass.

Secondly, we use NMPC to design control inputs for the navigation of the object to a final pose, while avoiding inter-agent collisions as well as collisions with obstacles. Moreover, we take into account constraints that emanate from control input saturation as well kinematic and representation singularities. We propose both a centralized and a decentralized methodology.

Thirdly, we propose an adaptive control protocol for the trajectory tracking by an observable point on an object that is manipulated by $N$ robotic agents in terms of \textit{rolling contacts}, also without using force/torque measurements at the grasping points. We develop a centralized as well as a decentralized event-triggered communication-based control scheme. Both schemes include the adaptive and quaternion modeling attributes of the rigid grasp schemes, and are robust to uncertainties of the object's center of mass pose, since the tracking concerns an observable a priori selected point on the object. Novel algorithms that guarantee contact slip avoidance are also developed. 
We provide detailed stability analyses for all the proposed schemes, whose validity is  verified by using simulation and experimental results.

\section{Rigid Contacts} \label{sec: Problem Formulation (TCST_coop_manip)}

\begin{figure}
	\centering
	\includegraphics[width = 0.5\textwidth]{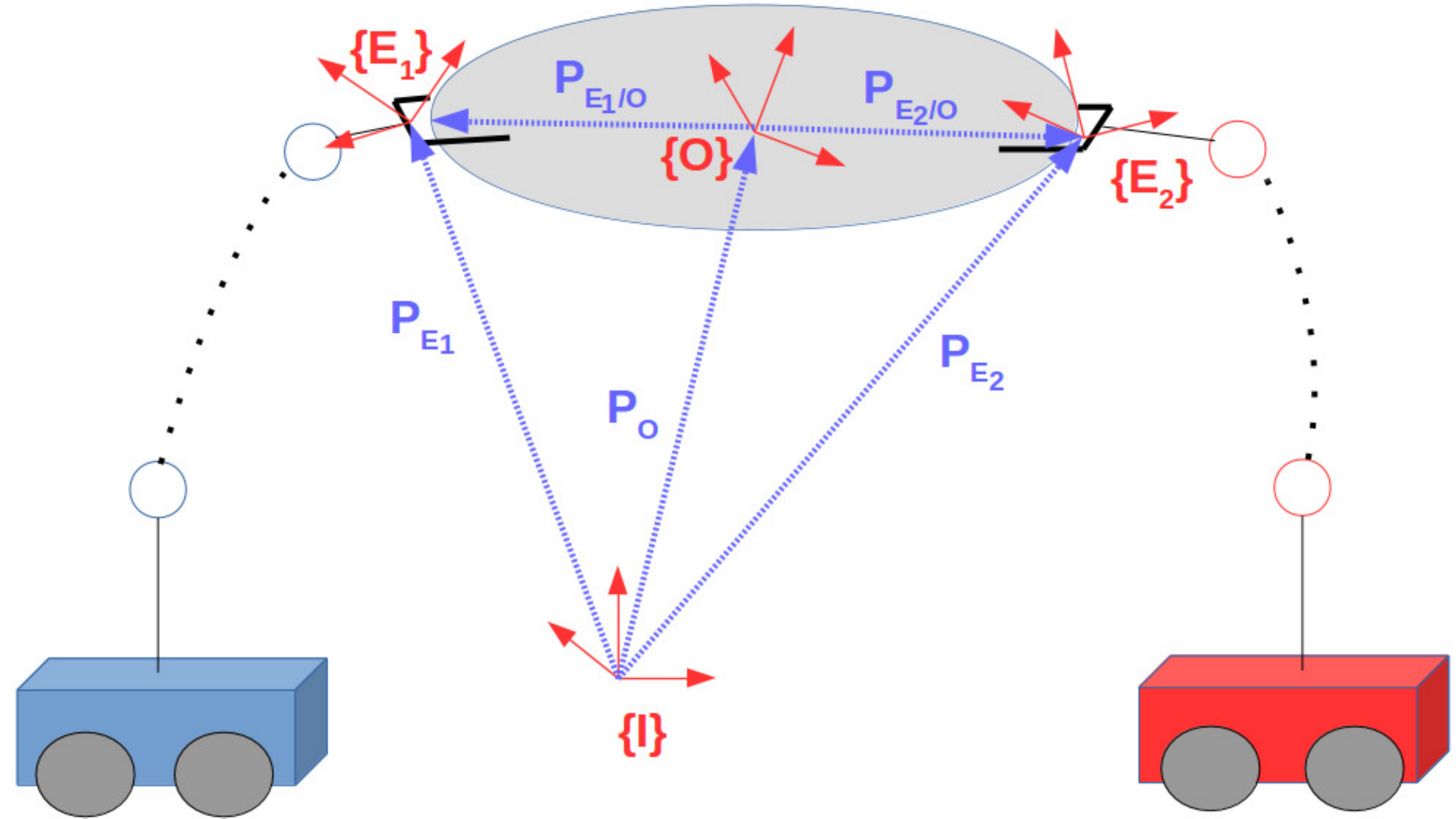}
	\caption{Two robotic agents rigidly grasping an object.\label{fig:Two-robotic-arms (TCST_coop_manip)}}
\end{figure}

Consider $N$ fully actuated robotic agents rigidly grasping an object  
(see Fig. \ref{fig:Two-robotic-arms (TCST_coop_manip)}). We denote by $\left\{ E_{i}\right\}$, $\left\{ O\right\}$
the end-effector and
object's center of mass frames, respectively; $\left\{ I\right\} $
corresponds to an inertial frame of reference. The rigidity assumption implies
that the agents can exert both forces and torques along all
directions to the object. In the following, we present the modeling of the coupled kinematics and dynamics of the object and the agents.

\subsection{System Model} \label{subsec:system model (TCST_coop_manip)}

We derive in this section the model of the system object-robots.


We denote by ${q}_i,\dot{{q}}_i \in\mathbb{R}^{n_i}$, with $n_i\in\mathbb{N}, \forall i\in\mathcal{N}\coloneqq\{1,\dots,N\}$, the generalized joint-space variables and their time derivatives of agent $i$, with ${q}_i \coloneqq [{q}_{i_1},\dots, {q}_{i_{n_i}}]$. The overall joint configuration is then ${q} \coloneqq [{q}^\top _1,\dots,{q}^\top _N]^\top , \dot{{q}} \coloneqq [\dot{{q}}^\top _1,\dots,\dot{{q}}^\top _N]^\top \in\mathbb{R}^{n}$, with $n \coloneqq \sum_{i\in\mathcal{N}}n_i$. In addition, the inertial position and orientation of the $i$th end-effector, denoted by ${p}_{\scriptscriptstyle E_i}$ and ${\eta}_{\scriptscriptstyle E_i}$, respectively, can be derived by the forward kinematics and are smooth functions of ${q}_i$, i.e. ${p}_{\scr E_i}\coloneqq p_{\scr E_i}(q_i):\mathbb{R}^{n_i}\to\mathbb{R}^3$, ${\eta}_{\scr E_i} \coloneqq \eta_{\scr E_i}(q_i):\mathbb{R}^{n_i}\to\mathbb{T}$, where $\mathbb{T}$ is an appropriate orientation space. 
The differential equation describing the dynamics of each agent is \cite{siciliano2010robotics}:
\begin{align}
{B}_i \ddot{{q}}_i+{C}_{{q}_i}\dot{{q}}_i+ {g}_{{q}_i} + {d}_{{q}_i}= {\tau}_i-{J}_i^\top {h}_i,\label{eq:manipulator joint_dynamics (TCST_coop_manip)}
\end{align}
where ${B}_i\coloneqq B_i(q_i):\mathbb{R}^{n_i}\to\mathbb{R}^{n_i\times n_i}$ is the positive definite inertia matrix, ${C}_{q_i} \coloneqq C_{q_i}(q_i,\dot{q}_i):\mathbb{R}^{2n_i}\to\mathbb{R}^{n_i\times n_i}$ is the Coriolis matrix, ${g}_{q_i} \coloneqq g_{q_i}(q_i):\mathbb{R}^{n_i}\to\mathbb{R}^{n_i}$ is the joint-space gravity term, ${d}_{q_i} \coloneqq d_{q_i}(q_i,\dot{q}_i,t):\mathbb{R}^{2n_i}\times\mathbb{R}_{\geq 0}\to\mathbb{R}^{n_i}$ is a bounded vector representing unmodeled friction, uncertainties and external disturbances, ${h}_i\in\mathbb{R}^{6}$ is the vector of generalized forces that agent $i$ exerts on the grasping point with the object and ${\tau}_i=[\tau_{i,1},\dots,\tau_{i,n_i}]^\top\in\mathbb{R}^{n_i}$ is the vector of joint torques, acting as control inputs, $\forall i\in\mathcal{N}$.   

The generalized velocity of each agent's end-effector ${v}_i \coloneqq [\dot{{p}}^\top_{\scr E_i},{\omega}^\top_{\scriptscriptstyle E_i}]^\top \in\mathbb{R}^6$, where $\omega_{\scriptscriptstyle E_i} \in \mathbb{R}^3$ is the respective angular velocity, can be considered as a transformed state through the differential kinematics ${v}_i = {J}_i\dot{{q}}_i$ \cite{siciliano2010robotics}, where ${J}_i\coloneqq J_i(q_i):\mathbb{R}^{n_i}\to\mathbb{R}^{6\times n_i}$ is a smooth function representing the geometric Jacobian matrix, $\forall i\in\mathcal{N}$ \cite{siciliano2010robotics}. The latter leads also to  
\begin{equation}
\dot{{v}}_i = {J}_i\ddot{{q}}_i + \dot{J}_i\dot{{q}}_i. \label{eq:accel (TCST_coop_manip)}
\end{equation}
We define also the sets $\mathsf{S}_i \coloneqq \{ q_i \in \mathbb{R}^{n_i} : \textup{det}(J_i(q_i)J_i(q_i)^\top) > 0\}$, which contains all the singularity-free configurations.
By employing the differential kinematics as well as \eqref{eq:accel (TCST_coop_manip)}, we obtain from \eqref{eq:manipulator joint_dynamics (TCST_coop_manip)} the transformed task space dynamics \cite{siciliano2010robotics}:
\begin{equation}
{M}_i\dot{{v}}_i+{C}_i{v}_i+ {g}_i + {d}_i ={u}_i-{h}_i,\label{eq:manipulator dynamics (TCST_coop_manip)}
\end{equation}
with the corresponding task space terms ${M}_i\coloneqq M_i(q_i):\mathsf{S}_i\to\mathbb{R}^{6\times6}$, ${C}_i\coloneqq C_i(q_i,\dot{q}_i):\mathsf{S}_i\times\mathbb{R}^{n_i}\to\mathbb{R}^{6\times6}$, ${g}_i\coloneqq g_i(q_i):\mathsf{S}_i\to\mathbb{R}^{6}$, ${d}_i\coloneqq d_i(q_i,\dot{q}_i,t):\mathsf{S}_i\times\mathbb{R}^{n_i}\times\mathbb{R}_{\geq 0}\to\mathbb{R}^6$ and ${u}_i=[u_{i,1},\dots,u_{i,6}]^\top\in\mathbb{R}^6$ being the task space wrench, related to ${\tau}_i$ via ${\tau}_i = {J}^\top _i{u}_i + ({I}_{n_i} - {J}^\top _i\widetilde{{J}}^\top _i){\tau}_{i0}$, where $\widetilde{{J}}_i$ is a generalized inverse of ${J}_i$ \cite{siciliano2010robotics}; ${\tau}_{i0}$ concerns redundant agents ($n_i > 6$) and does not contribute to end-effector forces. 


The agent task-space dynamics \eqref{eq:manipulator dynamics (TCST_coop_manip)} can be written in vector form as:
\begin{equation}
{M}\dot{{v}}+{{C}}v + {g} + {d} = {u}-{h},\label{eq:manipulator dynamics_vector_form (TCST_coop_manip)}
\end{equation}
where ${v} \coloneqq [{v}^\top_1,\dots,{v}^\top_N]\in\mathbb{R}^{6N}$, ${M} \coloneqq M(q) \coloneqq \textup{diag}\{[{M}_i]_{i\in\mathcal{N}}\}\in\mathbb{R}^{6N\times6N}$, ${C} \coloneqq C(q,\dot{q}) \coloneqq \textup{diag}\{[{C}_i]_{i\in\mathcal{N}}\}$ $\in\mathbb{R}^{6N\times6N}$, $h \coloneqq [{h}^\top_1,\dots,{h}^\top_N]^\top$, ${u} \coloneqq [{u}^\top_1,\dots$, ${u}^\top_N]^\top $, ${g} \coloneqq g(q) \coloneqq [{g}^\top_1,\dots,{g}^\top_N]^\top $, ${d} \coloneqq d(q,\dot{q},t) \coloneqq [{d}^\top_1,\dots,{d}^\top_N]^\top$ $\in\mathbb{R}^{6N}$.




Regarding the object, we denote by ${x}_{\scriptscriptstyle O}\coloneqq [{p}^\top _{\scriptscriptstyle O},{\eta}^\top _{\scriptscriptstyle O}]^\top \in\mathbb{M} \coloneqq \mathbb{R}^3\times \mathbb{T}$, ${v}_{\scriptscriptstyle O} \coloneqq [\dot{{p}}^\top_{\scriptscriptstyle O}, {\omega}^\top _{\scriptscriptstyle O}]^\top \in\mathbb{R}^{6}$ the pose and generalized velocity of its center of mass; $\eta_{\scr O}$ here denotes explicitly Euler angles $\eta_{\scr O} \coloneqq [\phi_{\scr O},\theta_{\scr O}, \psi_{\scr O}]^\top \in \mathbb{T} = \mathbb{R}^3$.
We consider the following second order dynamics, which can be derived based on the Newton-Euler formulation: 
\begin{subequations} \label{eq:object dynamics (TCST_coop_manip)}
	\begin{align}
	& \dot{{x}}_{\scriptscriptstyle O} = {J}_{\scriptscriptstyle O}{v}_{\scriptscriptstyle O}, \label{eq:object dynamics 1 (TCST_coop_manip)}\\
	& {M}_{\scriptscriptstyle O}\dot{{v}}_{\scriptscriptstyle O}+{C}_{\scriptscriptstyle O}{v}_{\scriptscriptstyle O}+{g}_{\scriptscriptstyle O}+ {d}_{\scriptscriptstyle O} = {h}_{\scriptscriptstyle O}, \label{eq:object dynamics 2 (TCST_coop_manip)}
	\end{align}
\end{subequations}
where ${M}_{\scriptscriptstyle O} \coloneqq M_{\scr O}(\eta_{\scr O}):\mathbb{T}\to\mathbb{R}^{6\times6}$ is the positive definite inertia matrix, ${C}_{\scriptscriptstyle O}\coloneqq C_{\scr O}(\eta_{\scr O},\omega_{\scr O}):\mathbb{T}\times\mathbb{R}^{6}\to\mathbb{R}^{6\times6}$ is the Coriolis matrix, ${g}_{\scriptscriptstyle O} \in \mathbb{R}^6$ is the gravity vector, ${d}_{\scriptscriptstyle O}\coloneqq d_{\scr O}(x_{\scr O},\dot{x}_{\scr O},t):\mathbb{M}\times\mathbb{R}^6\times\mathbb{R}_{\geq 0}\to\mathbb{R}^6$ a bounded vector representing modeling uncertainties and external disturbances, and ${h}_{\scriptscriptstyle O}\in\mathbb{R}^6$ is the vector of generalized forces acting on the object's center of mass. Moreover, ${J}_{\scriptscriptstyle O}\coloneqq J_{\scr O}(\eta_{\scr O}):\mathbb{T}\to\mathbb{R}^{6\times6}$ is the object representation Jacobian ${J}_{\scriptscriptstyle O}({\eta}_{\scriptscriptstyle O})\coloneqq\textup{diag}\{{I}_3, {J}_{\scriptscriptstyle O_{{\eta}}}\}$, where ${J}_{\scriptscriptstyle O_{\eta}} \coloneqq {J}_{\scriptscriptstyle O_{\eta}}(\eta_{\scr O}):\mathbb{T}\to\mathbb{R}^{3\times3}$:
\begin{equation}
{J}_{\scriptscriptstyle O_\eta} \coloneqq \begin{bmatrix}
1 & \sin(\phi_{\scriptscriptstyle O})\tan(\theta_{\scriptscriptstyle O}) & \cos(\phi_{\scriptscriptstyle O})\tan(\theta_{\scriptscriptstyle O}) \\
0 & \cos(\phi_{\scriptscriptstyle O}) & -\sin(\theta_{\scriptscriptstyle O}) \\
0 & \frac{\sin(\phi_{\scriptscriptstyle O})}{\cos(\theta_{\scriptscriptstyle O})} & \frac{\cos(\phi_{\scriptscriptstyle O})}{\cos(\theta_{\scriptscriptstyle O})}
\end{bmatrix}, \notag
\end{equation}
and is not well-defined when $\theta_{\scriptscriptstyle O}= \pm \tfrac{\pi}{2}$, which is referred to as \textit{representation singularity}. Moreover, it can be proved that 
\begin{subequations}\label{eq:J_norm (TCST_coop_manip)}
	\begin{align}
	& \lVert {J}_{\scriptscriptstyle O}(\eta_{\scr O})\rVert  = \sqrt{\tfrac{\lvert \sin(\theta_{\scriptscriptstyle O}) \rvert + 1}{1 - \sin^2(\theta_{\scriptscriptstyle O})}}, \\
	& \lVert {J}_{\scriptscriptstyle O}(\eta_{\scr O})^{-1}\rVert  = \sqrt{1+\sin(\theta_{\scr O})} \leq \sqrt{2},
	\end{align}
\end{subequations}
$\forall \eta_{\scr O}\in \mathbb{T}$.
We also denote by $R_{\scr O}\coloneqq R_{\scr O}(\eta_{\scr O}):\mathbb{T}\to\mathbb{SO}(3)$ the object's rotation matrix.

A possible way to avoid the aforementioned singularity is to transform the Euler angles to a unit quaternion representation for the orientation. Hence, the term ${\eta}_{\scriptscriptstyle O}$ can be transformed to the unit quaternion ${\zeta}_{\scriptscriptstyle O}=[\varphi_{\scriptscriptstyle O}, {\epsilon}^\top _{\scriptscriptstyle O}]^\top \in \mathbb{S}^3$, where $\varphi_{\scr O} \in [-1,1]$ and $\epsilon_{\scr O}\in \mathbb{R}^3$ are the scalar and vector parts, respectively
\cite{siciliano2010robotics}. The dynamics of $\zeta_{\scr O}$ can be proven to satisfy \cite{siciliano2010robotics}:
\begin{subequations} \label{eq:quat dynamics all (TCST_coop_manip)} 
\begin{align}
& \dot{{\zeta}}_{\scriptscriptstyle O} = \frac{1}{2}E({\zeta}_{\scriptscriptstyle O}){\omega}_{\scriptscriptstyle O} \label{eq:quat dynamics (TCST_coop_manip)}  \\
& {\omega}_{\scriptscriptstyle O} = 2 E({\zeta}_{\scriptscriptstyle O}) ^\top\dot{{\zeta}}_{\scriptscriptstyle O}, \label{eq:quat dynamics inverse (TCST_coop_manip)}
\end{align}
\end{subequations}
where $E:\mathbb{S}^3\to\mathbb{R}^{4\times3}$ is defined as:
\begin{equation*}
E(\zeta)=\left[\begin{array}{c}
-\epsilon^\top \\
\varphi I_3-S(\epsilon)
\end{array}\right], \forall \zeta=[\varphi,\epsilon^\top]^\top\in \mathbb{S}^3. \notag
\end{equation*}
and hence it holds that $E(\zeta)^\top E(\zeta) = I_3, \forall \zeta\in \mathbb{S}^3$.
It can be also shown that 
\begin{equation*}
\dot{\omega}_{\scriptscriptstyle O} = 2E(\zeta_{\scriptscriptstyle O})^\top \ddot{\zeta}_{\scriptscriptstyle O}. 
\end{equation*}



In view of Fig. \ref{fig:Two-robotic-arms (TCST_coop_manip)}, one concludes that the pose of the agents and the object's center of mass are related as  
\begin{subequations} \label{eq:coupled_kinematics (TCST_coop_manip)}
	\begin{align}
	{p}_{\scriptscriptstyle E_i}({q}_i) &= {p}_{\scriptscriptstyle O} +  {R}_{i}({q}_i) {p}^{\scriptscriptstyle E_i}_{\scriptscriptstyle E_{i}/O},\label{eq:coupled_kinematics_1 (TCST_coop_manip)} \\
	{\eta}_{\scriptscriptstyle E_i}({q}_i) &= {\eta}_{\scriptscriptstyle O} + {\eta}_{\scriptscriptstyle E_i/O}, \label{eq:coupled_kinematics_2 (TCST_coop_manip)}
	\end{align}
\end{subequations}
$\forall i\in\mathcal{N}$, where $R_{i} \coloneqq R_{i}(q_i):\mathbb{R}^{n_i}\to \mathbb{SO}(3)$ is the $i$'s end-effector rotation matrix, and ${p}^{\scriptscriptstyle E_i}_{\scriptscriptstyle E_i/O}$, ${\eta}_{\scriptscriptstyle E_i/O} \in \mathbb{R}^3$ are the \textit{constant} distance and orientation offset vectors between $\{O\}$ and $\{E_i\}$. 
Following \eqref{eq:coupled_kinematics (TCST_coop_manip)}, along with the
fact that, due to the grasping rigidity, it holds that ${\omega}_{\scriptscriptstyle E_i}={\omega}_{\scriptscriptstyle O}, \forall i\in\mathcal{N}$,
one obtains
\begin{equation}
{v}_i={J}_{\scriptscriptstyle O_i} {v}_{\scriptscriptstyle O}, \label{eq:J_o_i (TCST_coop_manip)}
\end{equation}
where ${J}_{\scriptscriptstyle O_i}\coloneqq J_{\scr O_i}({q}_i):\mathbb{R}^{n_i}\to\mathbb{R}^{6\times6}$ is the object-to-agent Jacobian matrix, with
\begin{equation}
{J}_{\scriptscriptstyle O_i}({x}) \coloneqq \left[\begin{array}{cc}
{I}_{ 3} & -{S}(R_i(x) p^{\scr E_i}_{\scr E_i/O})\\
0 & {I}_{3}
\end{array}\right], \forall {x}\in\mathbb{R}^{n_i}, \label{eq:J_o_i_def (TCST_coop_manip)}
\end{equation}
which is always full-rank. 
Moreover, from \eqref{eq:J_o_i (TCST_coop_manip)}, one obtains 
\begin{equation}
\dot{{v}}_i =  {J}_{\scriptscriptstyle O_i}\dot{{v}}_{\scriptscriptstyle O} + \dot{J}_{\scriptscriptstyle O_i}{v}_{\scriptscriptstyle O}. \label{eq:beta accel (TCST_coop_manip)}
\end{equation}
In addition, it can be proved for $J_{\scriptscriptstyle O_i}$ that
\begin{equation}
\| {J}_{\scriptscriptstyle O_i}({q_i}) \| \leq \left\| {p}^{\scriptscriptstyle E_i}_{\scriptscriptstyle O/E_i} \right \| + 1, \forall  {q_i}\in\mathbb{R}^{n_i}, i\in\mathcal{N}, \label{eq:J_O_i bound (TCST_coop_manip)}
\end{equation}
which will be used in the subsequent analysis.


The kineto-statics duality along with the grasp rigidity suggest that the force ${h}_{\scriptscriptstyle O}$ acting on the object's center of mass and the generalized forces ${h}_i,i\in\mathcal{N}$, exerted by the agents at the grasping points, are related through:
\begin{equation}
{h}_{\scriptscriptstyle O}={G}{{h}},\label{eq:grasp matrix (TCST_coop_manip)}
\end{equation}
where ${G}\coloneqq G(q):\mathbb{R}^{n}\to\mathbb{R}^{6\times 6N}$, with ${G}({{q}}) \coloneqq [{J}_{\scriptscriptstyle O_1}^\top,\dots,{J}_{\scriptscriptstyle O_N}^\top]$, 
is the full row-rank grasp matrix. 
By substituting \eqref{eq:manipulator dynamics_vector_form (TCST_coop_manip)} into \eqref{eq:grasp matrix (TCST_coop_manip)}, we obtain:
\begin{equation}
{h}_{\scriptscriptstyle O} = {G}\left({{u}} - {{M}}\dot{{{v}}} - {{C}}{v} - {{g}} - {d}  \right), \notag
\end{equation}
which, after substituting \eqref{eq:J_o_i (TCST_coop_manip)}, \eqref{eq:beta accel (TCST_coop_manip)} , \eqref{eq:object dynamics (TCST_coop_manip)}, and rearranging terms,
yields the overall system coupled dynamics: 
\begin{equation}
\widetilde{{M}}\dot{{v}}_{\scriptscriptstyle O}+\widetilde{{C}}{v}_{\scriptscriptstyle O}+\widetilde{{g}}+\widetilde{{d}}  = {G}u,\label{eq:coupled dynamics (TCST_coop_manip)}
\end{equation}
where 
\begin{subequations} \label{eq:coupled terms (TCST_coop_manip)}
	\begin{align}
	\widetilde{{M}}\coloneqq \widetilde{M}({x})  \coloneqq & {M}_{\scriptscriptstyle O}+{G} {M}{G}^\top \label{eq:coupled M (TCST_coop_manip)}\\
	\widetilde{{C}}\coloneqq\widetilde{{C}}({x})  \coloneqq & {C}_{\scriptscriptstyle O}+ {G}{C}{G}^\top + {G} {M}\dot{G}^\top \label{eq:coupled C (TCST_coop_manip)}\\
	\widetilde{{g}}\coloneqq\widetilde{{g}}({x})  \coloneqq & {g}_{\scriptscriptstyle O}+{G} {g}.\label{eq:coupled g (TCST_coop_manip)} \\ 
	\widetilde{{d}}\coloneqq \widetilde{{d}}({x},t)  \coloneqq & {d}_{\scriptscriptstyle O}+ {G} {d}\label{eq:coupled d (TCST_coop_manip)} 
	\end{align}
\end{subequations}
and ${x}$ is the overall state ${x}\coloneqq [{q}^\top,\dot{{q}}^\top,\eta^\top_{\scriptscriptstyle O},\omega_{\scriptscriptstyle O}^\top]^\top\in \mathbb{X} \coloneqq \mathsf{S}\times\mathbb{R}^{n+3}\times\mathbb{T}$, $\mathsf{S}\coloneqq\mathsf{S}_1\times\dots\times\mathsf{S}_N$.
Moreover, the following Lemma is necessary for the following analysis. 
\begin{lemma} \label{lem:coupled dynamics skew symmetry (TCST_coop_manip)}
	The matrix $\widetilde{{M}}({x})$ is symmetric and positive definite and the matrix $\dot{\widetilde{{M}}}({x}) - 2\widetilde{{C}}({x})$ is skew symmetric, i.e., 
	\begin{align}
	\Big[\dot{\widetilde{{M}}}({x})-2\widetilde{{C}}({x})\Big]^\top &= - \Big[\dot{\widetilde{{M}}}({x})-2\widetilde{{C}}({x})\Big], \forall {x}\in\mathbb{X} \notag \\
	{y}^\top\Big[\dot{\widetilde{{M}}}({x})-2\widetilde{{C}}({x})\Big]{y} &= 0, \ \ \forall {x}\in \mathbb{X},{y}\in\mathbb{R}^6. \notag
	\end{align}

\end{lemma}

\begin{proof}
	The matrices ${M}_{\scriptscriptstyle O}$ and ${M}_i$ are symmetric and positive definite, $\forall i\in\mathcal{N}$ and the matrices $\dot{{M}}_i - 2{C}_i$, ${M}_{\scriptscriptstyle O} - 2{C}_{\scriptscriptstyle O}$ are skew-symmetric, $\forall i\in\mathcal{N}$ \cite{siciliano2010robotics}, which leads to the skew-symmetry of $\dot{{M}} - 2{C}$.
	Therefore, since ${G}$ is full row-rank, we can conclude the symmetry and positive definiteness of $\widetilde{M}$.
	Regarding the skew symmetry of $\dot{\widetilde{{M}}}-2\widetilde{{C}}$, we define first $A\coloneqq{A}({x}) \coloneqq \dot{{G}} {M}{G}^\top$,
	and we have from \eqref{eq:coupled C (TCST_coop_manip)}:
	\begin{align}
	\dot{\widetilde{{M}}}-2\widetilde{{C}}=& \dot{{M}}_{\scriptscriptstyle O}-2{C}_{\scriptscriptstyle O} + {G} (\dot{{M}}-2{C}) {G}^\top + {A}-{A}^\top, \notag
	\end{align}
	which, by employing the skew-symmetry of ${M}_{\scriptscriptstyle O} - 2{C}_{\scriptscriptstyle O}$ and $\dot{{M}} - 2{C}$, leads to $[\dot{\widetilde{{M}}}-2\widetilde{{C}}]^\top  = -[ \dot{\widetilde{{M}}}-2\widetilde{{C}}]$, which completes the proof. 
\end{proof}
 The positive definiteness of $\widetilde{{M}}({x})$ leads to the property 
\begin{align}
\underline{m}{I}_6 \leq \widetilde{{M}}({x}) \leq \bar{{m}} I_6, \label{eq:dyn properties (TCST_coop_manip)}
\end{align}
$\forall {x}\in\mathbb{X}$, where $\underline{m}$ and $\bar{m}$ are positive unknown constants. 

\subsection{Problem Statement - Uncertain Model}

The general problem treated in this chapter is the tracking of a pose/trajectory by the object. We first assume that the object and robot models \eqref{eq:object dynamics (TCST_coop_manip)}, \eqref{eq:manipulator dynamics (TCST_coop_manip)} are uncertain, i.e., they are not fully available for feedback in the control design. 
Officially, the problem we are aiming to solve for the rigid contact case is the following:

\begin{problem}
	Given a desired bounded object smooth pose trajectory specified by $x_{\textup{d}} \coloneqq x_{\textup{d}}(t) \coloneqq [({p}_{\textup{d}})^\top, ({\eta}_{\textup{d}})^\top]^\top \coloneqq [({p}_{\textup{d}}(t))^\top, ({\eta}_{\textup{d}}(t))^\top]^\top :\mathbb{R}_{\geq 0}\to \mathbb{M}$, ${\eta}_\textup{d} \coloneqq [\varphi_\textup{d},\theta_\textup{d},\psi_\textup{d}] \coloneqq [\varphi_\textup{d}(t),\theta_\textup{d}(t),\psi_\textup{d}(t)]:\mathbb{R}_{\geq 0} \to \mathbb{T}$, with bounded first and second derivatives, 
	determine a continuous time-varying control law ${u}$ in \eqref{eq:coupled dynamics (TCST_coop_manip)} such that 
	\begin{equation*}
	\lim\limits_{t\rightarrow\infty}\left[\begin{array}{c}
	{p}_{\scriptscriptstyle O}(t)-{p}_\textup{d}(t)\\
	{\eta}_{\scriptscriptstyle O}(t)-{\eta}_\textup{d}(t)
	\end{array}\right]=0
	\end{equation*}
	\label{prob:problem1 (TCST_coop_manip)} 
\end{problem}

To solve the aforementioned problem, we need the following assumptions regarding the agent feedback and the kinematic singularities.

\begin{assumption} (Feedback) \label{ass:feedback (TCST_coop_manip)}
	Each agent $i\in\mathcal{N}$ has continuous feedback of its own state ${q}_i, \dot{{q}}_i$.
\end{assumption}
\begin{assumption} (Object geometry) \label{ass:object geometry (TCST_coop_manip)}
	Each agent $i\in\mathcal{N}$ knows the constant offsets ${p}^{\scriptscriptstyle E_i}_{\scriptscriptstyle E_i/O}$ and ${\eta}_{\scriptscriptstyle E_i/O}, \forall i\in\mathcal{N}$.
\end{assumption}
\begin{assumption} (Kinematic singularities) \label{ass:kinematic singularities (TCST_coop_manip)}
	The robotic agents operate away from kinematic singularities, i.e., ${q}_i(t)$ evolves in a closed subset of  
	$\mathsf{S}_i$, $\forall i\in\mathcal{N}$.
\end{assumption}

Assumption \ref{ass:feedback (TCST_coop_manip)} is realistic for real manipulation systems, since on-board sensors can provide accurately the measurements ${q}_i,\dot{{q}}_i$. The object geometrical characteristics in Assumption \ref{ass:object geometry (TCST_coop_manip)} can be obtained by on-board sensors, whose inaccuracies are not modeled here and constitute part of future work. Finally, Assumption \ref{ass:kinematic singularities (TCST_coop_manip)} states that the ${q}_i$ that achieve ${x}_{\scriptscriptstyle O}(t) = \bar{x}_{\textup{d}}(t), \forall t\in\mathbb{R}_{\geq 0}$ are sufficiently far from kinematic singular configurations.
Since each agent has feedback from its state ${q}_i, \dot{{q}}_i$, it can compute through the forward and differential kinematics the end-effector pose ${p}_{\scriptscriptstyle E_i}({q}_i), {\eta}_{\scriptscriptstyle E_i}({q}_i)$ and the velocity ${v}_i$, $\forall i\in\mathcal{N}$. Moreover, since it knows ${p}^{\scriptscriptstyle E_i}_{\scriptscriptstyle E_i/O}$ and ${\eta}_{\scriptscriptstyle E_i/O}$, it can compute ${J}_{\scriptscriptstyle O_i}({q}_i)$ from \eqref{eq:J_o_i_def (TCST_coop_manip)}, and ${x}_{\scriptscriptstyle O}$, ${v}_{\scriptscriptstyle O}$ by inverting \eqref{eq:coupled_kinematics (TCST_coop_manip)} and \eqref{eq:J_o_i (TCST_coop_manip)}, respectively. Consequently, each agent can then compute the object unit quaternion ${\zeta}_{\scriptscriptstyle O}$ as well as $\dot{{\zeta}}_{\scriptscriptstyle O}$.

{Note that, due to Assumption \ref{ass:object geometry (TCST_coop_manip)} and the grasp rigidity, the object-agents configuration is similar to a single closed-chain robot. The considered multi-agent setup, however, renders the problem more challenging, since the agents must calculate their own control signal in a decentralized manner, without communicating with each other. Moreover, each agent needs to compensate its own part of the (possibly uncertain/unknown) dynamics of the coupled dynamic equation \eqref{eq:coupled dynamics (TCST_coop_manip)}, while respecting the rigidity kinematic constraints. 


We present next two control schemes for the solution of Problem \ref{prob:problem1 (TCST_coop_manip)}. The proposed controllers are decentralized, in the sense that the agents calculate their control signal on their own, without communicating with each other, as well as robust, since they do not take into account the dynamic properties of the agents or the object (mass/inertia moments) or the uncertainties/external disturbances modeled by the function $\widetilde{{d}}({x},t)$ in \eqref{eq:coupled dynamics (TCST_coop_manip)}. The first control scheme is presented in Section \ref{subsec:Quaternion Controller (TCST_coop_manip)}, and is based on quaternion feedback and adaptation laws, while the second control scheme is given in Section \ref{subsec:PPC Controller (TCST_coop_manip)} and is inspired by the Prescribed Performance Control (PPC) methodology introduced in \cite{bechlioulis2014low}. 

\subsection{Adaptive Control with Quaternion Feedback} \label{subsec:Quaternion Controller (TCST_coop_manip)}
Firstly, we need the following assumption regarding the model uncertainties/external disturbances.
\begin{assumption} (Uncertainties/Disturbance parameterization) \label{ass:disturbance bound (TCST_coop_manip)}
	There exist constant \textit{unknown} vectors $\bar{d}_{\scriptscriptstyle O}\in\mathbb{R}^{\mu_{\scr O}}, \bar{d}_i\in\mathbb{R}^{\mu}$ and known functions $\delta_{\scriptscriptstyle O}\coloneqq \delta_{\scriptscriptstyle O}(x_{\scr O},\dot{x}_{\scr O},t):\mathbb{M}\times\mathbb{R}^6\times\mathbb{R}_{\geq 0}\to\mathbb{R}^{6\times\mu_{\scr O}}, \delta_i\coloneqq \delta_i(q_i,\dot{q}_i,t):\mathbb{R}^{2n_i}\times\mathbb{R}_{\geq 0}\to\mathbb{R}^{6\times\mu}$, such that $d_{\scriptscriptstyle O}(x_{\scriptscriptstyle O},\dot{x}_{\scriptscriptstyle O},t) = \delta_{\scriptscriptstyle O}(x_{\scriptscriptstyle O},\dot{x}_{\scriptscriptstyle O},t)\bar{d}_{\scriptscriptstyle O}$, 
	$d_i(q_i,\dot{q}_i,t) = \delta_i(q_i,\dot{q}_i,t)\bar{d}_i$,
	$\forall q_i,\dot{q}_i\in\mathbb{R}^{n_i}, x_{\scriptscriptstyle O}\in\mathbb{M}, \dot{x}_{\scriptscriptstyle O}\in\mathbb{R}^6, t\in\mathbb{R}_{\geq 0}, i\in\mathcal{N}$,
	{where $\delta_{\scriptscriptstyle  O}(x_{\scriptscriptstyle  O},\dot{x}_{\scriptscriptstyle  O},t)$ and  $\delta_i(q_i,\dot{q}_i,t)$ are continuous in $(x_{\scr O},\dot{x}_{\scr O})$ and $(q_i,\dot{q}_i)$, respectively, and uniformly bounded in $t$.}
\end{assumption} 
The aforementioned assumption is motivated by the use of Neural Networks for approximating unknown functions in compact sets \cite{lavretsky13adaptive}. More specifically, any continuous function $f(x):\mathbb{R}^n \to \mathbb{R}^m$ can be approximated on a known compact set $X\subset \mathbb{R}^n$ by a Neural Network equipped with $N$ Radial Basis Functions (RBFs) $\Phi(x)$ and using unknown ideal constant connection weights that are stored in a matrix $\Theta \in\mathbb{R}^{N\times m}$ as $f(x) = \Theta ^\top \Phi(x) + \varepsilon(x)$;
$\Theta ^\top \Phi(x)$ represents the parametric uncertainty and $\varepsilon(x)$ represents the unknown nonparametric uncertainty, which is bounded as $\|\varepsilon(x)\| \leq  \bar{\varepsilon}$ in $X$. In our case, the functions $\delta_{\scriptscriptstyle O}$, $\delta_i$ play the role of the known function $\Phi(x)$ and $\bar{d}_{\scriptscriptstyle O}$, $\bar{d}_i$ and $\mu$, $\mu_{\scriptscriptstyle O}$ represent the unknown constants $\Theta$ and the number of layers of the Neural Network, respectively. 
Nevertheless, in view of Neural Network approximation, Assumption $4$ implies that the nonparametric uncertainty is zero and that $d_{\scr O}$ and $d_i$ are \textit{known} functions of \textit{time}. These properties can be relaxed with non-zero bounded nonparametric uncertainties and \textit{unknown} but bounded time-dependent disturbances, i.e. $d_i(q_i,\dot{q}_i,t) = \delta_{i,q}(q_i,\dot{q}_i)\bar{d}_i + d_{i,t}(t) + \varepsilon_{i,q}(q_i,\dot{q}_i)$ and  
$d_{\scriptscriptstyle O}(x_{\scriptscriptstyle O},\dot{x}_{\scriptscriptstyle O},t) = \delta_{\scriptscriptstyle O, x}(x_{\scriptscriptstyle O},\dot{x}_{\scriptscriptstyle O})\bar{d}_{\scr O} + d_{\scriptscriptstyle O, t}(t) + \varepsilon_{\scriptscriptstyle O, x}(x_{\scriptscriptstyle O},\dot{x}_{\scriptscriptstyle O})$, where $d_{i,t}, d_{\scr O,t}$, $\varepsilon_{i,q}, \varepsilon_{\scr O,x}$ are bounded. In that case, instead of asymptotic convergence of the pose to the desired one, we can show convergence of the respective errors to a compact set around the origin. For more details on Neural Network approximation and adaptive control with illustrative examples, we refer the reader to \cite[Ch. 12]{lavretsky13adaptive}.

 The desired Euler angle orientation vector ${\eta}_\textup{d}:\mathbb{R}_{\geq 0}\to\mathbb{T}$ is transformed first to the unit quaternion ${\zeta}_\textup{d} \coloneqq {\zeta}_\textup{d}(t):\mathbb{R}_{\geq 0}\to \mathbb{S}^3$ \cite{siciliano2010robotics}. Then, we need to define the errors associated with the object pose and the desired pose trajectory. We first define the state that corresponds to the position error:
\begin{equation*}
{e}_p  \coloneqq {p}_{\scriptscriptstyle O}-{p}_\textup{d}. \label{eq:position error (TCST_coop_manip)}
\end{equation*}
Since unit quaternions do not form a vector space, they cannot be subtracted to form an orientation error; instead, we should use the properties of the quaternion group algebra. Let ${e}_\zeta = [e_{\varphi}, {e}^\top _{\epsilon}]^\top \in \mathbb{S}^3$ be the unit quaternion describing the orientation error. Then, it holds that \cite{siciliano2010robotics},
\begin{equation}
{e}_{\zeta}\coloneqq{\zeta}_\textup{d}\cdot{\zeta}_{\scriptscriptstyle O}^{+} = 
\begin{bmatrix}
\varphi_\textup{d} \\ {\epsilon}_\textup{d}
\end{bmatrix} \cdot
\begin{bmatrix}
\varphi_{\scriptscriptstyle O} \\ -{\epsilon}_{\scriptscriptstyle O}
\end{bmatrix}, \notag
\end{equation}
and yields
\begin{align*}
{e}_\zeta = \begin{bmatrix}
e_\varphi\\ {e}_\epsilon 
\end{bmatrix} \coloneqq
\begin{bmatrix}
\varphi_{\scriptscriptstyle O}\varphi_\textup{d}+{\epsilon}^\top _{\scriptscriptstyle O}{\epsilon}_\textup{d} \\
\varphi_{\scriptscriptstyle O}{\epsilon}_\textup{d}-\varphi_\textup{d}{\epsilon}_{\scriptscriptstyle O}+{S}({\epsilon}_{\scriptscriptstyle O}){\epsilon}_\textup{d} 
\end{bmatrix}.
\end{align*}
By employing the quaternion dynamics (see \eqref{eq:quat dynamics (TCST_coop_manip)}) and certain properties of skew-symmetric matrices \cite{campa2006kinematic}, it can be shown that the error dynamics of ${e}_p, e_\varphi$ are:
\begin{subequations} \label{eq:error_dynamics (TCST_coop_manip)}
	\begin{align}
	\dot{{e}}_{p}  = & \dot{{p}}_{\scriptscriptstyle O}-\dot{{p}}_\textup{d} \label{eq:position error dynamics (TCST_coop_manip)}\\
	\dot{e}_{\varphi}  = & \tfrac{1}{2}{e}^\top _{\epsilon}{e}_\omega \label{eq:eta_error dynamics (TCST_coop_manip)}  \\
	\dot{{e}}_{\epsilon} = & -\tfrac{1}{2}\left[e_\varphi {I}_3 + {S}({e}_{\epsilon}) \right]{e}_\omega - {S}({e}_{\epsilon}){\omega}_\textup{d},\label{eq:epsilon error dynamics (TCST_coop_manip)} 
	\end{align}
\end{subequations}
where ${e}_\omega \coloneqq {\omega}_{\scriptscriptstyle O} - {\omega}_\textup{d}$ is the angular velocity error, with ${\omega}_\textup{d} = 2{E} ({\zeta}_\textup{d})^\top \dot{{\zeta}}_\textup{d}$, as indicated by \eqref{eq:quat dynamics inverse (TCST_coop_manip)}. 

Due to the ambiguity of unit quaternions, when ${\zeta}_{\scriptscriptstyle O} = {\zeta}_\textup{d}$, then ${e}_{\zeta} = [1, {0}^\top_3]^\top \in \mathbb{S}^3$. If ${\zeta}_{\scriptscriptstyle O} = -{\zeta}_\textup{d}$, then ${e}_{\zeta} = [-1, {0}^\top_3]^\top \in \mathbb{S}^3$, which, however, represents the same orientation. Therefore,  the control objective established in Problem \ref{prob:problem1 (TCST_coop_manip)} is equivalent to 
\begin{equation*}
\lim\limits_{t\to\infty} 
\begin{bmatrix}
{e}_p(t) \\ \lvert e_\varphi(t) \rvert \\ {e}_\epsilon(t)
\end{bmatrix} = 
\begin{bmatrix}
0 \\ 1 \\ 0
\end{bmatrix}. 
\end{equation*}
The left hand side of \eqref{eq:manipulator dynamics (TCST_coop_manip)}, after employing \eqref{eq:J_o_i (TCST_coop_manip)} and \eqref{eq:beta accel (TCST_coop_manip)}, becomes
\begin{align}
& {M}_i \dot{{v}}_i + {C}_i {v}_i + {g}_i  + {d}_i =  {M}_i\Big({J}_{\scriptscriptstyle O_i}\dot{{v}}_{\scriptscriptstyle O} + \dot{J}_{\scriptscriptstyle O_i}{v}_{\scriptscriptstyle O}\Big)  + {C}_i {J}_{\scriptscriptstyle O_i} {v}_{\scriptscriptstyle O} + {g}_i + {d}_i. \notag
\end{align}
which, according to Assumption \ref{ass:disturbance bound (TCST_coop_manip)} and the fact that the manipulator dynamics can be linearly parameterized with respect to dynamic parameters \cite{Slotine_adaptive87}, becomes
	\begin{align*}
	& {M}_i{J}_{\scriptscriptstyle O_i} \dot{{v}}_{\scriptscriptstyle O} + \Big({M}_i \dot{J}_{\scriptscriptstyle O_i}+{C}_i{J}_{\scriptscriptstyle O_i}\Big){v}_{\scriptscriptstyle O} + {g}_i + {d}_i =  {Y}_i{\vartheta}_i + {\delta}_i \bar{d}_i, 
	\end{align*}
	$\forall i\in\mathcal{N}$, where ${\vartheta}_i \in\mathbb{R}^\ell,\ell\in\mathbb{N}$, are vectors of unknown but constant dynamic parameters of the agents, appearing in the terms ${M}_i,{C}_i,{g}_i$, and ${Y}_i \coloneqq {Y}_i({q}_i,\dot{{q}}_i,{v}_{\scriptscriptstyle O}, \dot{{v}}_{\scriptscriptstyle O}): \mathsf{S}
	\times\mathbb{R}^{n_i+12}\to\mathbb{R}^{6\times\ell}$ are known regressor matrices, independent of ${\vartheta}_i,i\in\mathcal{N}$. Without loss of generality, we assume here that the dimension of ${\vartheta}_i$ is the same, $\ell$ for all the agents. Similarly, the dynamical terms of the left hand side of \eqref{eq:object dynamics 2 (TCST_coop_manip)} can be written as 
	\begin{align*}
	& {M}_{\scriptscriptstyle O} \dot{{v}}_{\scriptscriptstyle O} + {C}_{\scriptscriptstyle O}{v}_{\scriptscriptstyle O} + {g}_{\scriptscriptstyle O} + 
	{d}_{\scriptscriptstyle O} =  {Y}_{\scriptscriptstyle O}{\vartheta}_{\scriptscriptstyle O} + 
	{\delta}_{\scriptscriptstyle O} \bar{d}_{\scriptscriptstyle O},
	\end{align*}
where ${\vartheta}_{\scriptscriptstyle O}\in\mathbb{R}^{\ell_{\scriptscriptstyle O}}, \ell_{\scriptscriptstyle O}\in\mathbb{N}$ is a vector of unknown but constant dynamic parameters of the object, appearing in the terms ${M}_{\scriptscriptstyle O},{C}_{\scriptscriptstyle O},{g}_{\scriptscriptstyle O}$, and 
${Y}_{\scriptscriptstyle O} \coloneqq {Y}_{\scriptscriptstyle O}({\eta}_{\scriptscriptstyle O},{{\omega}}_{\scriptscriptstyle O},{v}_{\scriptscriptstyle O}, \dot{{v}}_{\scriptscriptstyle O}):\mathbb{T}\times\mathbb{R}^{15}\to\mathbb{R}^{6\times\ell_{\scriptscriptstyle O}}$ is a known regressor matrix, independent of ${\vartheta}_{\scriptscriptstyle O}$. It is worth noting that
the choice for $\ell$ and $\ell_{\scriptscriptstyle O}$ is not unique. 
In view of the aforementioned expressions, the left-hand side of \eqref{eq:coupled dynamics (TCST_coop_manip)} can be written as: 	
\begin{align} \label{eq:parameter linearity 2 (TCST_coop_manip)}
& \widetilde{{M}}\dot{{v}}_{\scriptscriptstyle O}+ \widetilde{{C}}{v}_{\scriptscriptstyle O}+\widetilde{{g}} + \widetilde{{d}} = {Y}_{\scriptscriptstyle O}{\vartheta}_{\scriptscriptstyle O} + {\delta}_{\scriptscriptstyle O}\bar{d}_{\scriptscriptstyle O} + G\left(\widetilde{Y}{\vartheta} + \widetilde{\delta}\bar{d} \right)
\end{align}
where $\widetilde{Y} \coloneqq \widetilde{Y}({q},\dot{{q}},{v}_{\scriptscriptstyle O},\dot{{v}}_{\scriptscriptstyle O}) \coloneqq  \textup{diag}\{ [Y_i]_{i\in\mathcal{N}} \} \in \mathbb{R}^{6N \times N\ell}$, 
${\vartheta} \coloneqq [{\vartheta}^\top_1,\dots,{\vartheta}^\top_N]^\top \in\mathbb{R}^{N\ell}$, $\widetilde{\delta} \coloneqq \widetilde{\delta}(q,\dot{q},t) \coloneqq \textup{diag}\{[\delta_i]_{i\in\mathcal{N}}\}$ $\in\mathbb{R}^{6N \times N \mu}$, and $\bar{{d}} \coloneqq [\bar{d}_1^\top,\dots,\bar{d}^\top_N]^\top \in \mathbb{R}^{N\mu}$.

Let us now introduce the states $\hat{{\vartheta}}_{\scriptscriptstyle O}\in\mathbb{R}^{\ell_{\scriptscriptstyle O}}$ and $\hat{{\vartheta}}_i\in\mathbb{R}^{\ell}$ which represent the estimates of ${\vartheta}_{\scriptscriptstyle O}$ and ${\vartheta}_i$, respectively, by agent $i\in\mathcal{N}$, and the corresponding stack vector
$\hat{{\vartheta}} \coloneqq [\hat{{\vartheta}}_1^\top,\dots,\hat{{\vartheta}}_N^\top]^\top \in\mathbb{R}^{N\ell}$,
for which we formulate the associated errors as 
\begin{subequations} \label{eq:adaptation errors (TCST_coop_manip)}
	\begin{align}
	{e}_{\vartheta_{\scriptscriptstyle O}} \coloneqq & 
	{\vartheta}_{\scriptscriptstyle O} - \hat{{\vartheta}}_{\scriptscriptstyle O}    \label{eq:adaptation errors_1 (TCST_coop_manip)}\\
	{e}_{\vartheta} \coloneqq & \begin{bmatrix}
	{e}_{\vartheta_1} \\ \vdots \\ {e}_{\vartheta_N}
	\end{bmatrix}  \coloneqq 
	\begin{bmatrix}
	{\vartheta}_1- \hat{{\vartheta}}_1 \\ \vdots \\ {\vartheta}_N- \hat{{\vartheta}}_N
	\end{bmatrix} = {\vartheta}- \hat{{\vartheta}}.  \label{eq:adaptation errors_2 (TCST_coop_manip)}
	\end{align}
\end{subequations}
In the same vein, we introduce the states $\hat{d}_{\scriptscriptstyle O}\in\mathbb{R}^{\mu_{\scr O}}$ and $\hat{d}_i\in\mathbb{R}^\mu$ that correspond to the estimates of $\bar{d}_{\scriptscriptstyle O}$ and $\bar{d}_i$, respectively, by agent $i\in\mathcal{N}$, and the corresponding stack vector $\hat{{d}} \coloneqq [\hat{d}^\top_1,\dots,\hat{d}^\top_N]^\top\in\mathbb{R}^{N\mu}$, for which we also formulate the associated errors as
\begin{subequations} \label{eq:d bounde errors (TCST_coop_manip)}
	\begin{align}
	{e}_{d_{\scriptscriptstyle O}} \coloneqq & 
	\bar{d}_{\scriptscriptstyle O} - \hat{d}_{\scriptscriptstyle O} \in \mathbb{R}^{\mu_{\scr O}}	   \label{eq:d bounde errors_1 (TCST_coop_manip)}\\
	{e}_d \coloneqq & \begin{bmatrix}
	e_{d_1} \\ \vdots \\ e_{d_N}
	\end{bmatrix}  \coloneqq
	\begin{bmatrix}
	\bar{d}_1- \hat{d}_1 \\ \vdots \\ \bar{d}_N- \hat{d}_N
	\end{bmatrix} = \bar{{d}} - \hat{{d}} \in\mathbb{R}^{N\mu}. \label{eq:d bounde errors_2 (TCST_coop_manip)}
	\end{align}
\end{subequations}

Next, we design the reference velocity 
\begin{equation}
{v}_f \coloneqq {v}_\textup{d} - {K}_f {e} =
\begin{bmatrix}
\dot{{p}}_\textup{d} - k_p {e}_p \\ {\omega}_{\textup{d}} + k_\zeta {e}_\epsilon
\end{bmatrix} \label{eq:v_f (TCST_coop_manip)}
\end{equation} 
where $v_\textup{d} \coloneqq [\dot{p}^\top_\textup{d},\omega^\top_\textup{d}]^\top$, $e \coloneqq [{e}^\top_p, -{e}^\top_\epsilon]^\top\in\mathbb{R}^6$, and ${K}_f \coloneqq \textup{diag}\{k_p,k_\zeta\}$, with $k_p, k_\zeta$ positive control gains. 
We also introduce the respective velocity error ${e}_{v_f}$ as
\begin{equation}
{e}_{{v}_f} \coloneqq {v}_{\scriptscriptstyle O} - {v}_f, \label{eq:e_v_f (TCST_coop_manip)}
\end{equation} 
and 
design the adaptive control law ${u}_i$ in \eqref{eq:coupled dynamics (TCST_coop_manip)}, for each agent $i\in\mathcal{N}$, as $u_i: \mathbb{U}_{f_i} \times \mathbb{R}_{\geq 0} \to \mathbb{R}^6$ with
\begin{align}
{u}_i \coloneqq u_i(\chi_f,t) \coloneqq & {Y}_{f_i}\hat{{\vartheta}}_i + \delta_i\hat{d}_i +  J_{M_i}\bigg[ {Y}_{\scriptscriptstyle f_O}\hat{\vartheta}_{\scriptscriptstyle O}  - e - {K}_{v}{e}_{v_f} + {\delta}_{\scriptscriptstyle O}\hat{d}_{\scriptscriptstyle O} \bigg],	\label{eq:control laws adaptive quat (TCST_coop_manip)}
\end{align}
where $\mathbb{U}_f \coloneqq \mathsf{S}_i\times\mathbb{T}\times\mathbb{R}^{12+\ell+\ell_{\scr O}+\mu+\mu_{\scr O}}$, $\chi_f \coloneqq [q_i^\top,\eta_{\scr O}^\top,e^\top,e_{v_f}^\top,\hat{\vartheta}_i^\top,\hat{\vartheta}_{\scr O}^\top, \hat{d}_i,\hat{d}_{\scr O}]^\top$,
${Y}_{f_i} \coloneqq Y_i({q}_i,\dot{{q}}_i,{v}_f,\dot{v}_f)$, ${Y}_{\scriptscriptstyle f_O} \coloneqq Y_{\scr O}(\eta_{\scriptscriptstyle O},{\omega}_{\scriptscriptstyle O},{v}_f,\dot{v}_f ) $, ${K}_v$ is a positive definite gain matrix and $J_{M_i}\coloneqq J_{M_i}(q_i):\mathbb{R}^{n_i} \to \mathbb{R}^{6\times 6}$ are the matrices \cite{erhart2015internal}
\begin{equation} \label{eq:J_Hirche (TCST_coop_manip)}
	J_{M_i}(q_i) = \begin{bmatrix}
	m_i^\ast (m^\ast_{\scr O})^{-1} I_3 & m_i^\ast (J^\ast_{\scr O})^{-1} S\big(R_{\scr O}( \eta_{\scr E_i}(q_i) - \eta_{\scr E_i/O}\big)p^{\scr O}_{\scr O/E_i}) \\ 
	0 & J_i^\ast (J^\ast_{\scr O})^{-1}
	\end{bmatrix},
\end{equation}
for some positive coefficients $m_i^\ast\in\mathbb{R}_{>0}$ and positive definite matrices $J_i^\ast \in \mathbb{R}^{3\times 3}$, $\forall i\in\mathcal{N}$, satisfying 
\begin{align*}
	& m_{\scr O}^\ast = \sum_{i\in\mathcal{N}}m_i^\ast, \ \ \ \sum_{i\in\mathcal{N}} R_{i}(q_i)p^{\scr E_i}_{\scr E_i/O}m_i^\ast = 0 \\
	& J_{\scr O}^\ast = \sum_{i\in\mathcal{N}} J_i^\ast - \sum_{i\in\mathcal{N}}m_i^\ast S\big(R_{i}(q_i)p^{\scr E_i}_{\scr E_i/O}\big)^2.
\end{align*}


In addition, we design the following adaptation laws:
\begin{subequations} \label{eq:adaptation laws (TCST_coop_manip)}
	\begin{align}
	& \dot{\hat{{\vartheta}}}_i = -\gamma_i {Y}_{f_i}^\top  {J}_{\scriptscriptstyle O_i} {e}_{v_f}, \label{eq:adaptation laws_theta_i (TCST_coop_manip)}\\
	& \dot{\hat{d}}_i = -\beta_i {\delta}_i^\top{J}_{\scriptscriptstyle O_i}{e}_{v_f} \label{eq:adaptation laws_d_i (TCST_coop_manip)} \\
	& \dot{\hat{{\vartheta}}}_{\scriptscriptstyle O} = -\gamma_{\scriptscriptstyle O} Y_{\scriptscriptstyle f_O}^\top {e}_{v_f} \label{eq:adaptation laws_theta_o (TCST_coop_manip)}\\
	& \dot{\hat{d}}_{\scriptscriptstyle O_i} =  -\beta_{\scriptscriptstyle O} {\delta}_{\scriptscriptstyle O}^\top {e}_{v_f}, \label{eq:adaptation laws_d_o (TCST_coop_manip)}
	\end{align} 
\end{subequations}
with arbitrary bounded initial conditions, where $\beta_i, \beta_{\scriptscriptstyle O}, \gamma_i, \gamma_{\scriptscriptstyle O} \in\mathbb{R}_{> 0}$ are positive gains, $\forall i\in\mathcal{N}$. 

The control and adaptation laws can be written in vector form 
\begin{subequations} \label{eq:laws vector forms (TCST_coop_manip)}
	\begin{align}
	u =& \widetilde{Y}_f\hat{\vartheta} + \widetilde{\delta}\hat{d} + G^{+}_M\Big[Y_{\scriptscriptstyle f_O}\hat{\vartheta}_{\scriptscriptstyle O} - e  + \delta_{\scriptscriptstyle O}\hat{d}_{\scriptscriptstyle O}  - K_ve_{v_f} \Big] \label{eq:control laws adaptive quat vector form (TCST_coop_manip)} \\ 
	 \dot{\hat{\vartheta}} =& -\Gamma \widetilde{Y}_f^\top G^\top e_{v_f}  \label{eq:adaptation laws_theta_i vector form (TCST_coop_manip)} \\
	 \dot{\hat{d}} =& -B_g\widetilde{\delta}^\top G^\top e_{v_f} \label{eq:adaptation laws_d_i vector form (TCST_coop_manip)} \\
	\dot{\hat{\vartheta}}_{\scriptscriptstyle O} =& -\gamma_{\scriptscriptstyle O}Y_{\scriptscriptstyle f_O}^\top e_{v_f} \label{eq:adaptation laws_theta_o vector form (TCST_coop_manip)}\\
	\dot{\hat{d}}_{\scriptscriptstyle O} =&  -\beta_{\scriptscriptstyle O} \delta_{\scriptscriptstyle O} ^\top e_{v_f}, \label{eq:adaptation laws_d_o vector form (TCST_coop_manip)} 
	\end{align}
\end{subequations}
where $\widetilde{Y}_f \coloneqq \widetilde{Y}(q,\dot{q},v_{\scr f},\dot{v}_{\scr f})$, $G^{+}_M \coloneqq G^{+}_M(q)  \coloneqq [J^\top_{M_1},\dots,J^\top_{M_N}]^\top \in\mathbb{R}^{6N\times6}$, $B_g \coloneqq \textup{diag}\{[\beta_i I_\mu]_{i\in\mathcal{N}}\}$, and $\Gamma \coloneqq \textup{diag}\{[\gamma_iI_\ell]_{i\in\mathcal{N}}\}$. The matrix $G^{+}_M(q)$ was introduced in \cite{erhart2015internal}, where it was proved that it yields a load distribution that is free of internal forces. The parameters $m^\star_{\scr O}, m^\star_i$ are used to distribute the object's needed effort (the term that right multiplies $G^{+}_M(q)$ in \eqref{eq:control laws adaptive quat vector form (TCST_coop_manip)}) to the agents.

\begin{remark}[\textbf{Decentralized manner (adaptive controller)}]
	Notice from \eqref{eq:control laws adaptive quat (TCST_coop_manip)} and \eqref{eq:adaptation laws (TCST_coop_manip)} that the overall control protocol is decentralized in the sense that the agents calculate their own control signals without communicating with each other. In particular, the control gains and the desired trajectory can be transmitted off-line to the agents, which
	can compute the object's pose and velocity, and hence the signals $e$, $v_f$, $e_{v_f}$ from the inverse kinematics. For the computation of $J_{M_i}$, each agent needs knowledge of the offsets $p^{\scr E_i}_{\scr E_i/O}$, which can also be transmitted off-line to the agents. Moreover, by also transmitting off-line to the agents the initial conditions $\hat{\vartheta}_{\scr O}$, $\hat{d}_{\scr O}$, and via the adaptation laws \eqref{eq:adaptation laws_theta_o vector form (TCST_coop_manip)}, \eqref{eq:adaptation laws_d_o vector form (TCST_coop_manip)}, each agent has access to the adaptation signals $\hat{\vartheta}_{\scr O}(t), \hat{d}_{\scr O}(t)$, $\forall t\in\mathbb{R}_{\geq 0}$.
	Finally, the structure of the functions $\delta_i$, $\delta_{\scr O}$, $Y_i$, $Y_{\scr O}$, as well as the constants $m^\star_i$, $J^\star_i$ can be also known by the agents a priori.
\end{remark}

The following theorem summarizes the main results of this subsection.

\begin{theorem}
	Consider $N$ robotic agents rigidly grasping an object with coupled
	dynamics described by (\ref{eq:coupled dynamics (TCST_coop_manip)}) and unknown dynamic
	parameters. Then, under Assumptions \ref{ass:feedback (TCST_coop_manip)}-\ref{ass:disturbance bound (TCST_coop_manip)}, by applying the control protocol \eqref{eq:control laws adaptive quat (TCST_coop_manip)}
	with the adaptation laws \eqref{eq:adaptation laws (TCST_coop_manip)},  
	the object pose converges asymptotically to the desired pose trajectory. 
	Moreover, all closed loop signals are bounded. 
\end{theorem}

\begin{proof}
	Consider the nonnegative function 
	\begin{align}
 V \coloneqq& \tfrac{1}{2}e_p^\top e_p+2(1 - e_\varphi) + \tfrac{1}{2}e^\top_{v_f}\widetilde{M}e_{v_f} +\tfrac{1}{2}e^\top _\vartheta \Gamma^{-1} e_\vartheta +  \tfrac{1}{2\gamma_{\scr O}}e^\top _{\vartheta_{\scriptscriptstyle O}}e_{\vartheta_{\scriptscriptstyle O}} \notag\\  & + \tfrac{1}{2}e^\top _d B_g^{-1} e_d + \tfrac{1}{2\beta_{\scr O}}e^\top _{d_{\scriptscriptstyle O}}e_{d_{\scriptscriptstyle O}}, \label{eq:Lyap_f (TCST_coop_manip)}
	\end{align}
	By taking the derivative of $V$ and using \eqref{eq:e_v_f (TCST_coop_manip)}, \eqref{eq:v_f (TCST_coop_manip)}, \eqref{eq:parameter linearity 2 (TCST_coop_manip)}, and Lemma \ref{lem:coupled dynamics skew symmetry (TCST_coop_manip)}, we obtain
	\begin{align}
	\dot{V} =& -e^\top K_f e + e^\top_{v_f} [ G (u - \widetilde{Y}_f\vartheta -  \widetilde{\delta}\bar{d}) + e - \delta_{\scriptscriptstyle O}\bar{d}_{\scriptscriptstyle O}  
	- Y_{\scriptscriptstyle f_O}  \vartheta_{\scriptscriptstyle O}]  -e^\top_{\vartheta}\Gamma^{-1}\dot{\hat{\vartheta}} \notag \\ &- \tfrac{1}{\gamma_{\scr O}}e^\top_{\vartheta_{\scriptscriptstyle O}}\dot{\hat{\vartheta}}_{\scriptscriptstyle O}  - e^\top_d B_g^{-1}\dot{\hat{d}} - \tfrac{1}{\beta_{\scr O}}e^\top_{d_{\scriptscriptstyle O}} \dot{\hat{d}}_{\scriptscriptstyle O}, \notag
	\end{align}
	and after substituting the adaptive control and adaptation laws \eqref{eq:laws vector forms (TCST_coop_manip)} and using the fact that $G^\top G^{+}_M = I_6$, 
	\begin{align}
	\dot{V} =& -e^\top K_f e - e^\top_{v_f}K_{v}e_{v_f} - e^\top_{v_f} \Big[ G \Big( 
	\widetilde{Y}_f e_{\vartheta} + \widetilde{\delta} e_d\Big)  +    Y_{\scriptscriptstyle f_O} e_{\vartheta_{\scriptscriptstyle O}} + \delta_{\scriptscriptstyle O}e_{d_{\scriptscriptstyle O}}\Big] \notag \\
	&+ e^\top_{\vartheta} \widetilde{Y}_f^\top G^\top e_{v_f} + 
	  e^\top_d\widetilde{\delta}^\top G^\top e_{v_f} +  e^\top_{\vartheta_{\scriptscriptstyle O}} Y_{\scr f_O}^\top e_{v_f} + e^\top_{d_{\scriptscriptstyle O}} \delta_{\scr O}^\top e_{v_f}  \notag \\
	=& -k_p\|e_p \|^2 - k_\zeta\| e_\epsilon \|^2 - e^\top_{v_f}K_{v}e_{v_f},   \label{eq:V_dot adaptive final (TCST_coop_manip)}
	\end{align}
	which is non-positive. Note, however, that $\dot{V}$ is not negative definite, and we need to invoke invariance-like properties to conclude the asymptotic stability of $e_p, e_\epsilon, e_{v_f}$. Since the closed-loop system is non-autonomous (this can be verified by inspecting \eqref{eq:error_dynamics (TCST_coop_manip)}, the derivative of \eqref{eq:e_v_f (TCST_coop_manip)} and \eqref{eq:laws vector forms (TCST_coop_manip)}), LaSalle's invariance principle is not applicable, and we thus employ Barbalat's lemma 
	(Lemma \ref{lemma:barbalat (App_dynamical_systems)} of Appendix \ref{app:dynamical systems}). From \eqref{eq:V_dot adaptive final (TCST_coop_manip)} we conclude 
	the boundedness of $V$ and of $x$, which implies  the boundedness of the dynamic terms $\widetilde{M}(x), \widetilde{C}(x), \widetilde{g}(x)$. 
	Moreover, by invoking the boundedness of $p_\textup{d}(t), v_\textup{d}(t),\omega_\textup{d}(t), \dot{v}_\textup{d}(t), \dot{\omega}_\textup{d}(t)$, we conclude the boundedness of $v_f,v_{\scriptscriptstyle O},v_i,\hat{\vartheta}_{\scriptscriptstyle O}$, $\hat{\vartheta}$, $\hat{d}$, $\hat{d}_{\scriptscriptstyle O}$. By differentiating \eqref{eq:error_dynamics (TCST_coop_manip)}, we also conclude the boundedness of $\dot{v}_f$ and therefore, the boundedness of the control and adaptation laws \eqref{eq:control laws adaptive quat (TCST_coop_manip)} and \eqref{eq:adaptation laws (TCST_coop_manip)}. Thus, we can conclude the boundedness of the second derivative $\ddot{V}$ and {by invoking Corollary 8.1 of \cite{lavretsky13adaptive}}, the uniform continuity of $\dot{V}$. Therefore, according to Barbalat's lemma, we deduce that $\lim_{t\to\infty}\dot{V}(t) = 0$ and, consequently, that $\lim_{t\to\infty}e_p(t) = 0$, $\lim_{t\to\infty}e_{v_f}(t) = 0$, and $\lim_{t\to\infty} \| e_\epsilon(t) \|^2 = 0$, which, given that $e_\zeta$ is a unit quaternion, leads to the configuration $(e_p, e_{v_f}, e_\varphi, e_\epsilon) = (0,0,\pm 1,0)$. 

\end{proof}

\begin{remark}[\textbf{Unwinding}]
	Note that the two configurations where $e_\varphi = 1$ and $e_\varphi = -1$  represent the same orientation. The closed loop dynamics of $e_\varphi$, as given in \eqref{eq:eta_error dynamics (TCST_coop_manip)}, can be written, in view of \eqref{eq:v_f (TCST_coop_manip)}, as $\dot{e}_\varphi = k_\zeta\tfrac{1}{2}\|e_\epsilon\|^2 +  \tfrac{1}{2}[0^\top_3, e^\top_\epsilon]e_{v_f}$. Since the first term is always positive, we conclude that the equilibrium point $(e_p, e_{v_f}, e_\varphi, e_\epsilon) = (0,0,-1,0)$ is unstable. Therefore, there might be trajectories close to the configuration $e_\varphi = -1$ that will move away and approach $e_\varphi = 1$, i.e., a full rotation will be performed to reach the desired orientation (of course, if the system starts at the equilibrium $(e_p, e_{v_f}, e_\varphi,e_\epsilon) = (0,0,-1,0)$, it will stay there, which also corresponds to the desired orientation behavior). This is the so-called \textit{unwinding phenomenon} \cite{bhat2000topological}.
	Note, however, that the desired equilibrium point $(e_p, e_{v_f}, e_\varphi,e_\epsilon) = (0,0,1,0)$ is \textit{\textbf{eventually attractive}}, meaning that for each $\delta_\varepsilon >0$, there exist finite a time instant $T\geq 0$ such that $1 - e_\varphi(t) < \delta_\varepsilon, \forall t > T \geq 0$. A similar behavior is observed if we stabilize the point $e_\varphi = -1$ instead of $e_\varphi = 1$, by setting $e \coloneqq [e^\top_p, e^\top_\epsilon]^\top$ in \eqref{eq:v_f (TCST_coop_manip)} and considering the term $2(1 + e_\varphi)$ instead of $2(1 - e_\varphi)$ in the function \eqref{eq:Lyap_f (TCST_coop_manip)}. 
	
	In order to avoid the unwinding phenomenon, instead of the error $e = [e^\top_p, -e^\top _\epsilon]^\top$, we {can choose} $e = [e^\top_p, -e_\varphi e^\top _\epsilon]^\top$. Then by replacing the term $1 - e_\varphi$ with $1-e^2_\varphi$ in \eqref{eq:Lyap_f (TCST_coop_manip)} and using 
	\eqref{eq:laws vector forms (TCST_coop_manip)}, we conclude by proceeding with a similar analysis that 
	$(e_p, \|e_\epsilon\|e_\varphi, e_{v_f}) \rightarrow (0,0,0)$, which implies that the system is asymptotically driven to either the configuration $(e_p, e_{v_f}, e_\varphi,e_\epsilon) = (0,0,\pm 1,0)$, which is the desired one, or a configuration  $(e_p, e_{v_f}, e_\varphi,e_\epsilon) = (0,0,0,\widetilde{e}_\epsilon)$, where $\widetilde{e}_\epsilon\in \mathbb{S}^2$ is a unit vector. The latter represents a set of invariant undesired equilibrium points. The closed loop dynamics are $\dot{e}_\varphi = \frac{1}{2}e_\varphi\|e_\epsilon\|^2 + \frac{1}{2}[0^\top_3, e^\top_\epsilon]e_{v_f}$, and 
	$\frac{\partial}{\partial t}\|e_\epsilon\|^2 = -e_\varphi^2\|e_\epsilon\|^2  - e_\varphi[0^\top_3, e^\top_\epsilon]e_{v_f}$. 
	We can conclude from the term $[0^\top_3, e^\top_\epsilon]e_{v_f}$ that there exist trajectories that can bring the system close to the undesired equilibrium, rendering thus the point  $(e_p, e_{v_f}, e_\varphi,e_\epsilon) = (0,0,\pm 1,0)$ only locally asymptotically stable. 
	It has been proved that $e_\varphi = \pm 1$ cannot be globally stabilized with a purely continuous controller \cite{bhat2000topological}. Discontinuous control laws have also been proposed (e.g., \cite{mayhew2011quaternion}), whose  combination with adaptation techniques constitutes part of our future research directions. 
\end{remark}

\begin{remark} [\textbf{Robustness (adaptive controller)}]
	Notice also that the control protocol compensates the uncertain dynamic parameters and external disturbances through the adaptation laws \eqref{eq:adaptation laws (TCST_coop_manip)}, although the errors \eqref{eq:adaptation errors (TCST_coop_manip)}, \eqref{eq:d bounde errors (TCST_coop_manip)} do not converge to zero, but remain bounded. Finally, the control gains $k_p, k_\zeta, K_{v}$ can be tuned appropriately so that the proposed control inputs do not reach motor saturations in real scenarios.
\end{remark}


\begin{figure}[t]
	\centering	
	\begin{subfigure}[b]{\columnwidth}
		\centering
		\includegraphics[width=.65\columnwidth]{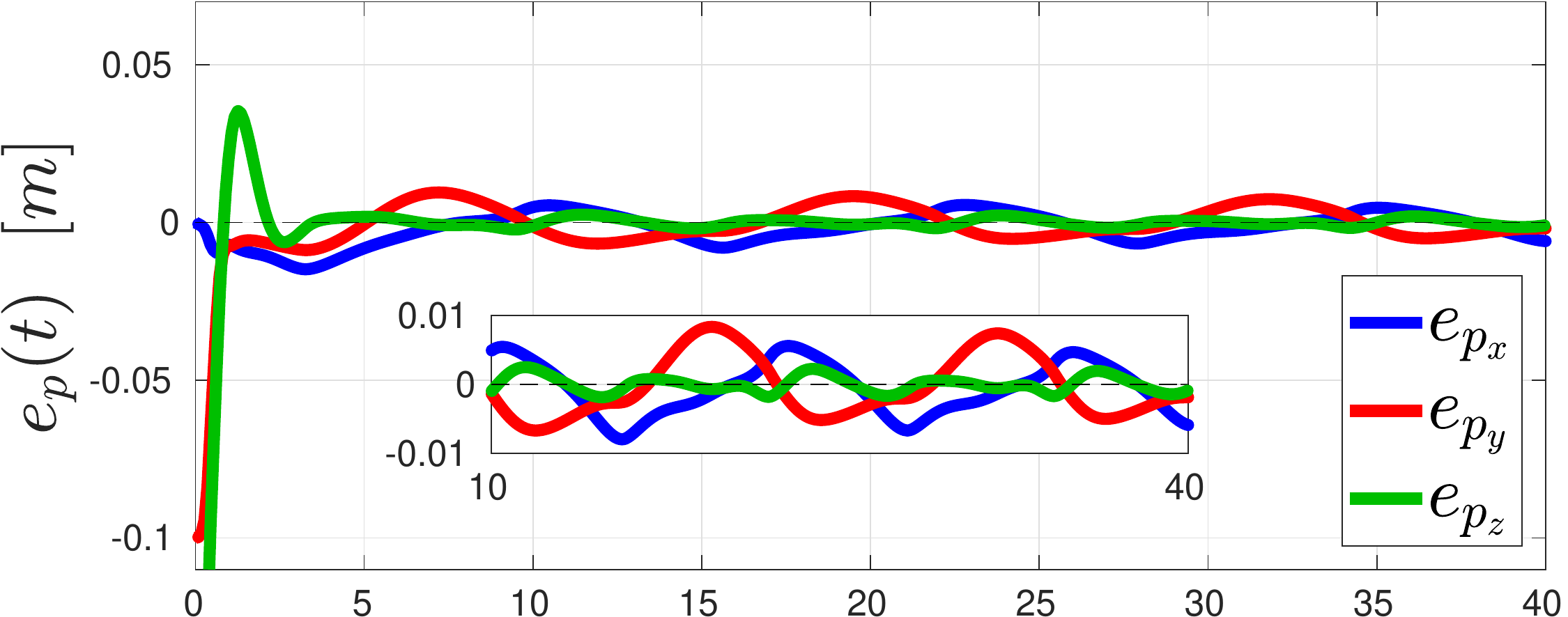}
		\caption{}
	\end{subfigure}		
	\begin{subfigure}[b]{\columnwidth}
		\centering
		\includegraphics[width=.65\columnwidth]{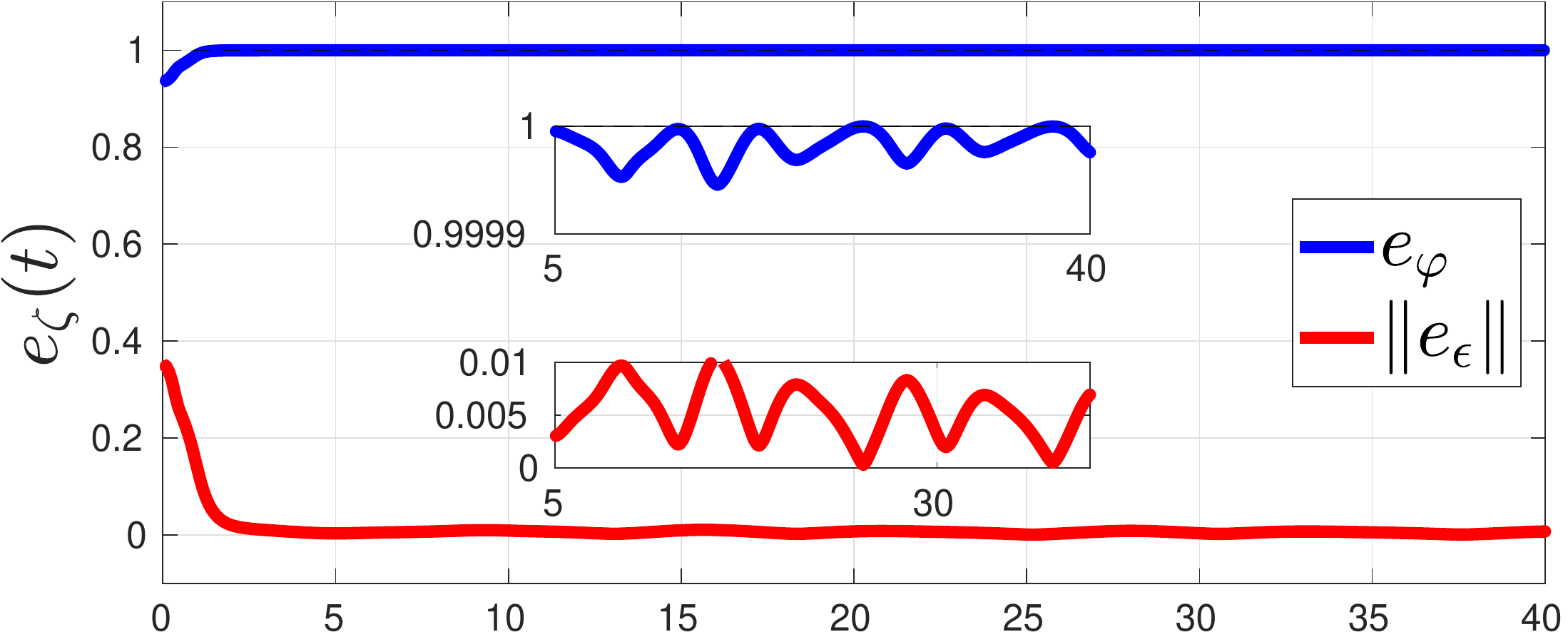}
		\caption{}
	\end{subfigure}
	
	\begin{subfigure}[b]{\columnwidth}
		\centering
		\includegraphics[width=.65\columnwidth]{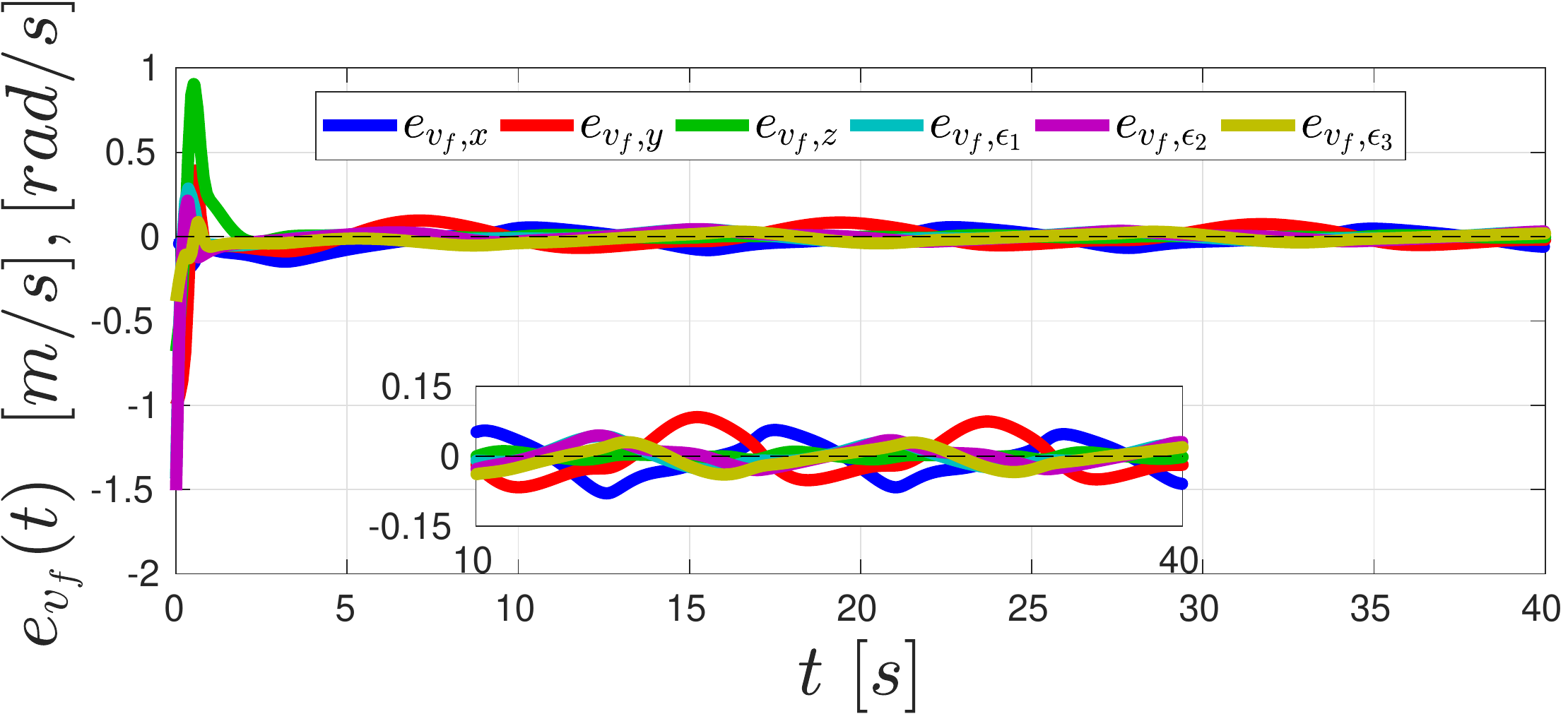}
		\caption{}
	\end{subfigure}
	
	\caption{Simulation results for the control scheme of Section \ref{subsec:Quaternion Controller (TCST_coop_manip)}; (a): The position errors $e_p(t)$; (b): The quaternion errors $e_\varphi(t)$, $\|e_\varepsilon(t)\|$; (c) The velocity errors $e_{v_f}(t)$, $\forall t\in[0,40]$. A zoomed version of the steady-state response has been included in all plots.} \label{fig:adapt_sim_errors (TCST_coop_manip)}
\end{figure}

\begin{figure}[t]
	\centering
	\begin{subfigure}[b]{\columnwidth}
		\centering
		\includegraphics[width=.65\columnwidth]{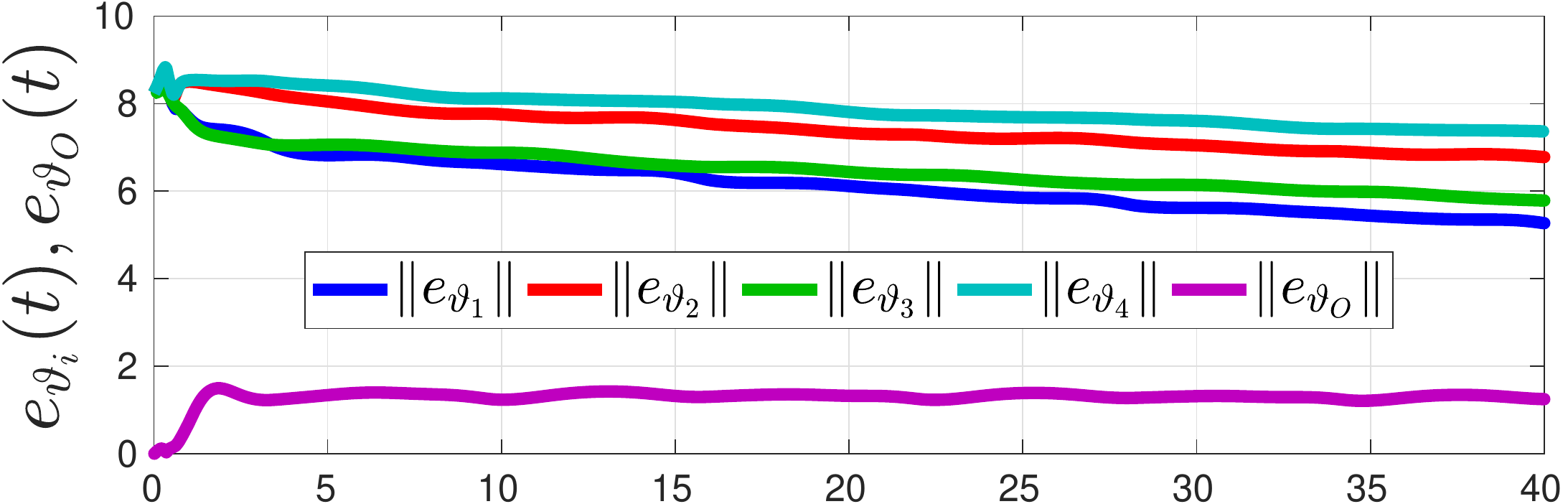}
		\caption{}
		\label{fig:Ng1 (TCST_coop_manip)} 
	\end{subfigure}
	
	\begin{subfigure}[b]{\columnwidth}
		\centering
		\includegraphics[width=.65\columnwidth]{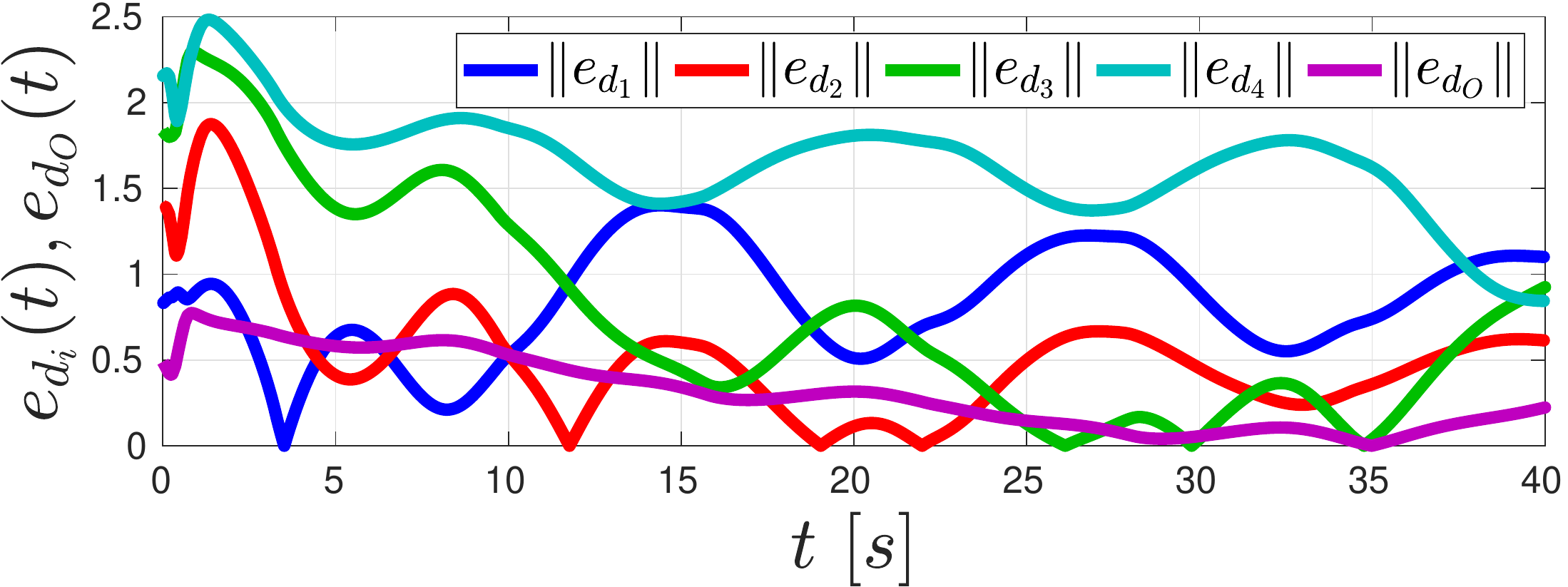}
		\caption{}
		\label{fig:Ng2 (TCST_coop_manip)}
	\end{subfigure}
	
	\caption{The adaptation error norms $\|e_{\vartheta_i}(t)\|$, $i\in\mathcal{N}$, $\|e_{\vartheta_{\scr O}}(t)\|$ (a), $\|e_{d_i}(t)\|$, $i\in\mathcal{N}$, $\|e_{d_{\scr O}}(t)\|$ (b), of the control scheme of Section \ref{subsec:Quaternion Controller (TCST_coop_manip)} $\forall t\in[0,40]$.} \label{fig:adapt_sim_theta_dis_errors (TCST_coop_manip)}
\end{figure}

\begin{figure}
	\centering
	\begin{subfigure}[b]{\columnwidth}
		\centering
		\includegraphics[width=.65\columnwidth]{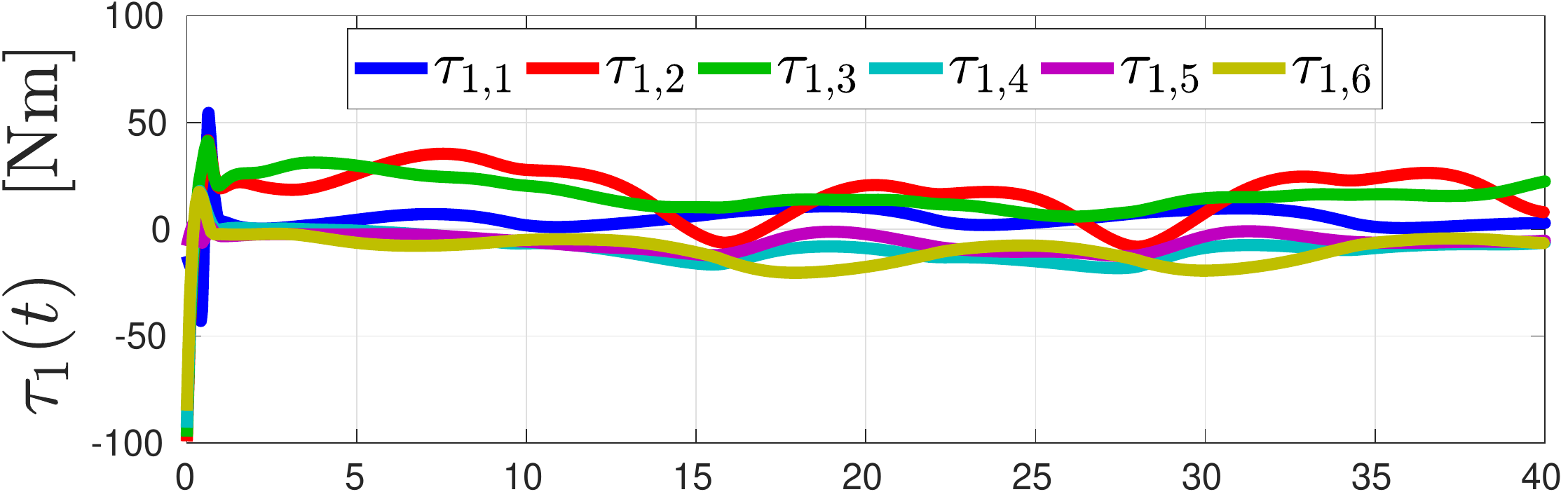}
		\caption{}
		\label{fig:Ng1 (TCST_coop_manip)} 
	\end{subfigure}
	
	\begin{subfigure}[b]{\columnwidth}
		\centering
		\includegraphics[width=.65\columnwidth]{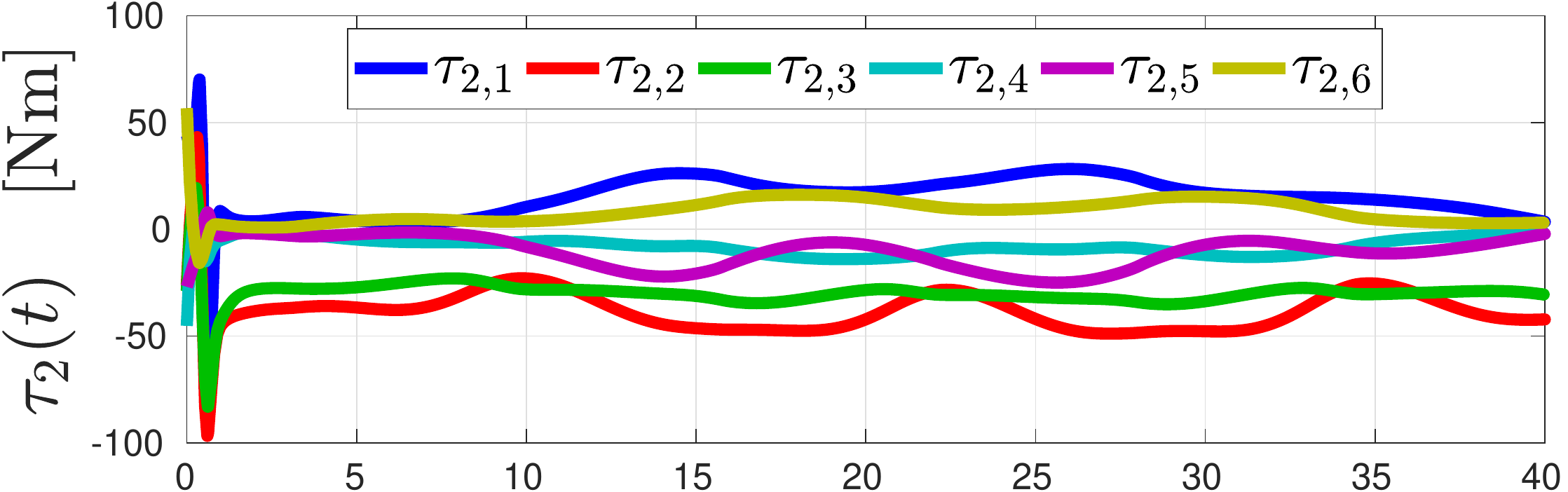}
		\caption{}
		\label{fig:Ng2 (TCST_coop_manip)}
	\end{subfigure}
	
	\begin{subfigure}[b]{\columnwidth}
		\centering
		\includegraphics[width=.65\columnwidth]{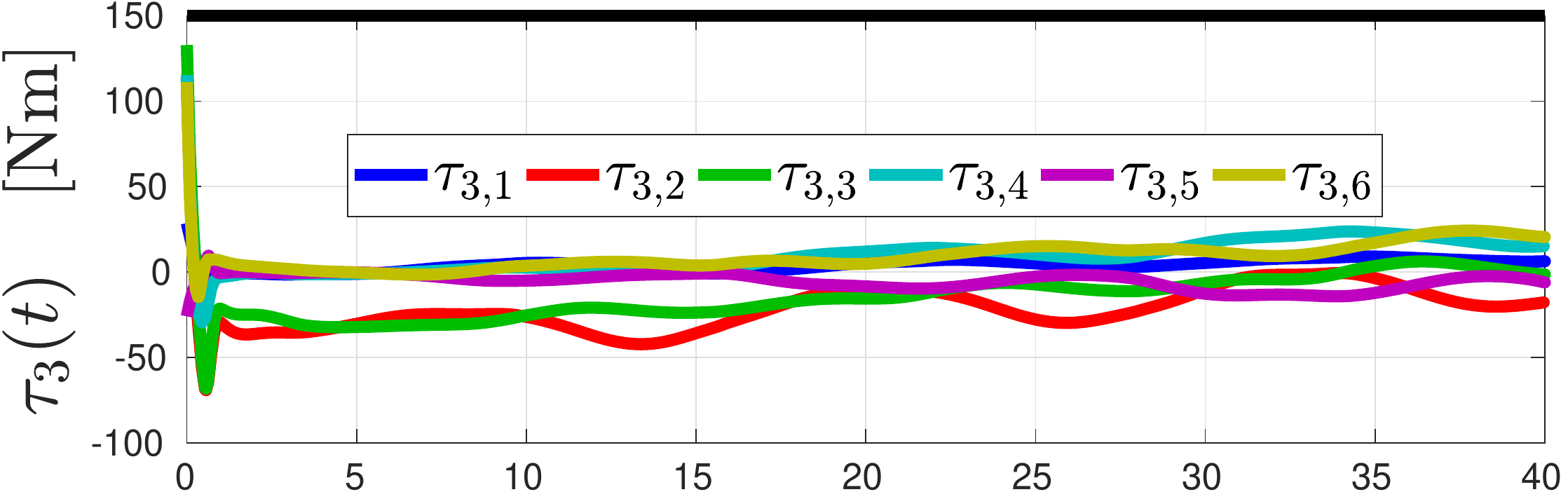}
		\caption{}
		\label{fig:Ng2 (TCST_coop_manip)}
	\end{subfigure}
	
	\begin{subfigure}[b]{\columnwidth}
		\centering
		\includegraphics[width=.65\columnwidth]{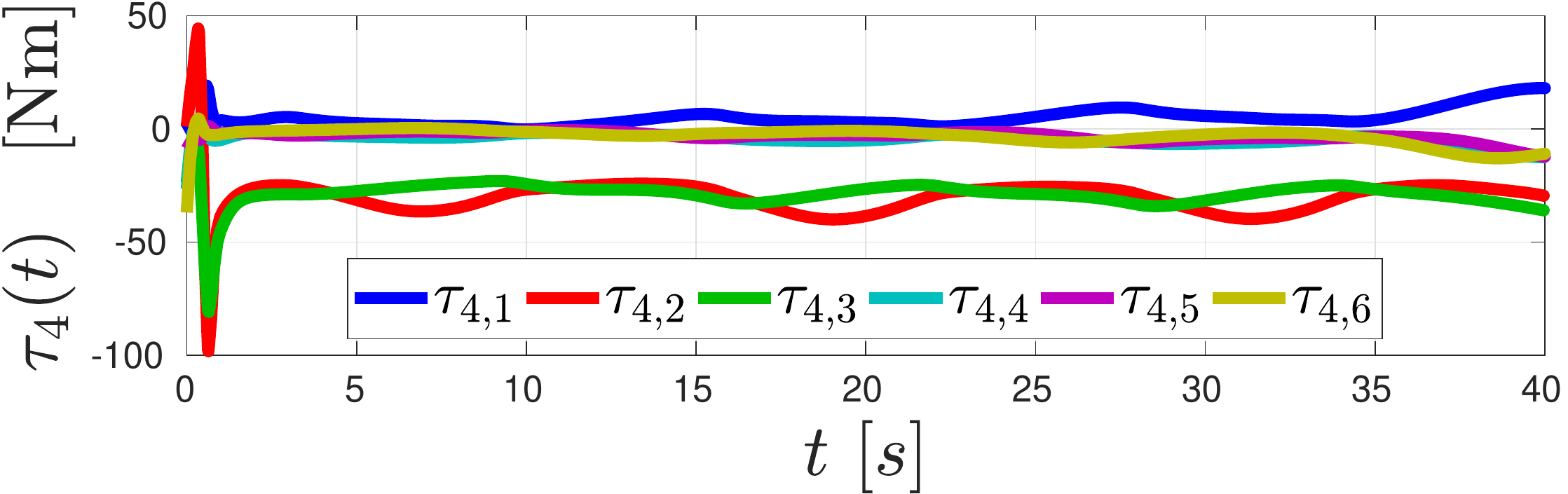}
		\caption{}
		\label{fig:Ng2 (TCST_coop_manip)}
	\end{subfigure}

	\caption{The agents' joint torques $\tau_i(t)$, $i\in\mathcal{N}$, (in (a)-(d), respectively) of the control scheme of Section \ref{subsec:Quaternion Controller (TCST_coop_manip)} $\forall t\in[0,40]$, and the motor saturation (with black), which has not been plotted in (a), (b), (d) for better visualization.} \label{fig:adapt_sim_tau (TCST_coop_manip)}
\end{figure}

\subsubsection{Simulation Results}
	
	We provide here simulation results for the developed control scheme.
	The tested scenario consists of four UR$5$ robotic manipulators rigidly grasping a rectangular object. The object's initial pose is $x_{\scr O}(0) = [-0.225,-0.612,0161$, $-\pi,\frac{\pi}{3},0]^\top$ $(\textup{[m]},\textup{[rad]})$ with respect to a chosen inertial frame and the desired trajectory is set as $p_\textup{d}(t) = [-0.225 + 0.1\sin(0.5t),-0.612 + 0.2\cos(0.5t),0.25 + 0.05\sin(0.5t)]^\top$, $\eta_\textup{d}(t)=[-\pi + 0.25\cos(0.5t),\frac{\pi}{3} + A_\theta\sin(0.25t),0.25\cos(0.5t)]^\top$,
	where $A_\theta = \frac{\pi}{6}$ (note that the desired pitch angle reaches the configuration of $\frac{\pi}{2}$, which yields a representation singularity in the Euler-angle formulation). In view of Assumption \ref{ass:disturbance bound (TCST_coop_manip)}, we set $d_i = (\|q_i\|\sin(\omega_{d_i}t + \phi_{d_i}) + \dot{q}_i)\bar{d}_i$ and $d_{\scriptscriptstyle O} = (\|\dot{x}_{\scriptscriptstyle  O}\|\sin(\omega_{d_{\scriptscriptstyle O}}t + \phi_{d_{\scriptscriptstyle O}}) + v_{\scriptscriptstyle O})\bar{d}_{\scriptscriptstyle  O}$, where the constants $\omega_{d_i}, \phi_{d_i}$, $\omega_{d_{\scriptscriptstyle O}}, \omega_{d_{\scriptscriptstyle O}}$ are randomly chosen in the interval $(0,1)$, $\forall i\in\mathcal{N}$. Regarding the force distribution matrix \eqref{eq:J_Hirche (TCST_coop_manip)}, we set $m_i^\star = 1$, $\forall i\in\mathcal{N}$, and $J_1^\star = 0.6I_3$, $J_2^\star = 0.4I_3$, $J_3^\star = 0.75I_3$, $J_4^\star = 0.25I_3$ to demonstrate a potential difference in the agents' power capabilities. In addition, we set an artificial saturation limit for the joint  motors as $\bar{\tau} = 150 \ \textup{Nm}$. 
	We set the control gains appearing in \eqref{eq:control laws adaptive quat (TCST_coop_manip)} and \eqref{eq:adaptation laws (TCST_coop_manip)} as $k_p = \textup{diag}\{[5,5,2]\},$ $k_\zeta = 3I_3$, $K_v = 400I_6$, $\gamma_i = \gamma_{\scr O} = \beta_i = \beta_{\scr O} = 1$, $\forall i\in\mathcal{N}$. The simulation results are depicted in Figs. \ref{fig:adapt_sim_errors (TCST_coop_manip)}-\ref{fig:adapt_sim_tau (TCST_coop_manip)} for $t\in[0,40]$ seconds. More specifically, Fig. \ref{fig:adapt_sim_errors (TCST_coop_manip)} shows the evolution of the pose and velocity errors $e_p(t), e_\zeta(t)$, $e_{v_f}(t)$, Fig. \ref{fig:adapt_sim_theta_dis_errors (TCST_coop_manip)} depicts the norms of the adaptation errors $e_{\vartheta_i}(t), e_{\vartheta_{\scr O}}(t)$,   $e_{d_i}(t), e_{d_{\scr O}}(t)$, and Fig. \ref{fig:adapt_sim_tau (TCST_coop_manip)} shows the resulting joint torques $\tau_i(t)$, $\forall i\in\{1,\dots,4\}$. Note that $e_p(t), e_\zeta(t)$ and $e_{v_f}(t)$ converge to the desired values and the adaptation errors are bounded, as predicted by the theoretical analysis.	
	
	One can conclude from the aforementioned figures that the simulation results verify the theoretical findings, since asymptotic stability is achieved. Moreover, the joint torques respect the saturation values we set. 
	The simulations were carried out in the MATLAB R2017a environment on a $i7$-$5600$ laptop computer at $2.6$Hz, with $8$GB of RAM.
	}

\begin{figure}[t]
	\centering
	\begin{subfigure}[b]{\columnwidth}
		\centering
		\includegraphics[trim =0cm 0cm 0cm 0cm, width=.65\columnwidth]{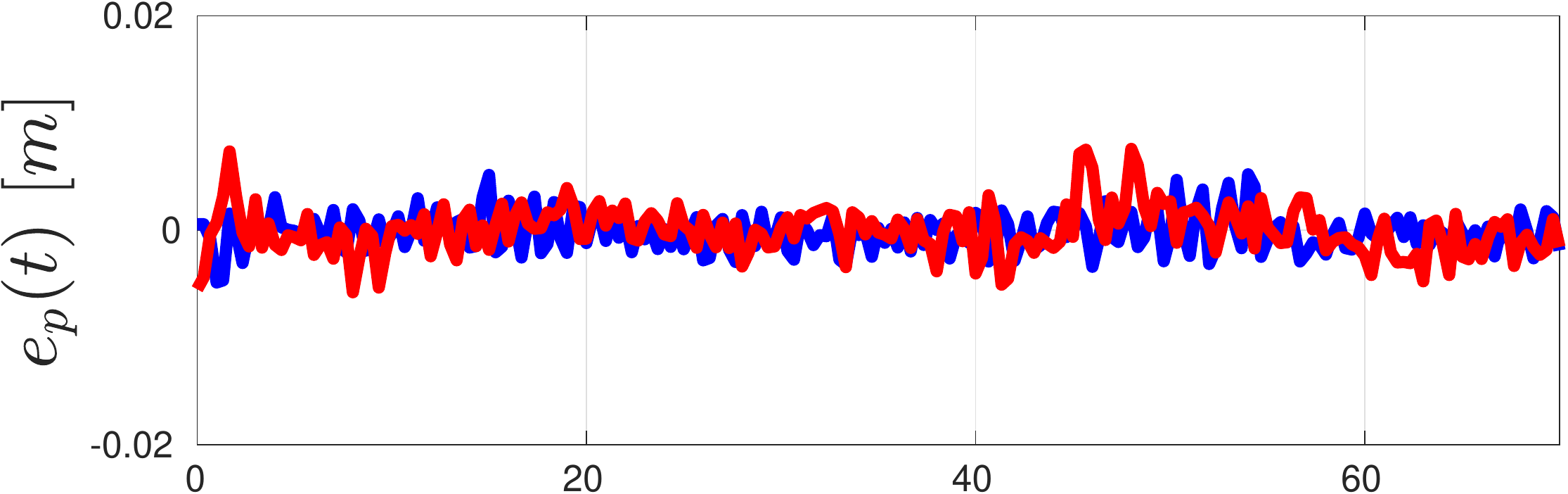}
		\caption{}
		\label{fig:Ng1 (TCST_coop_manip)} 
	\end{subfigure}
	
	\begin{subfigure}[b]{\columnwidth}
		\centering
		\includegraphics[width=.65\columnwidth]{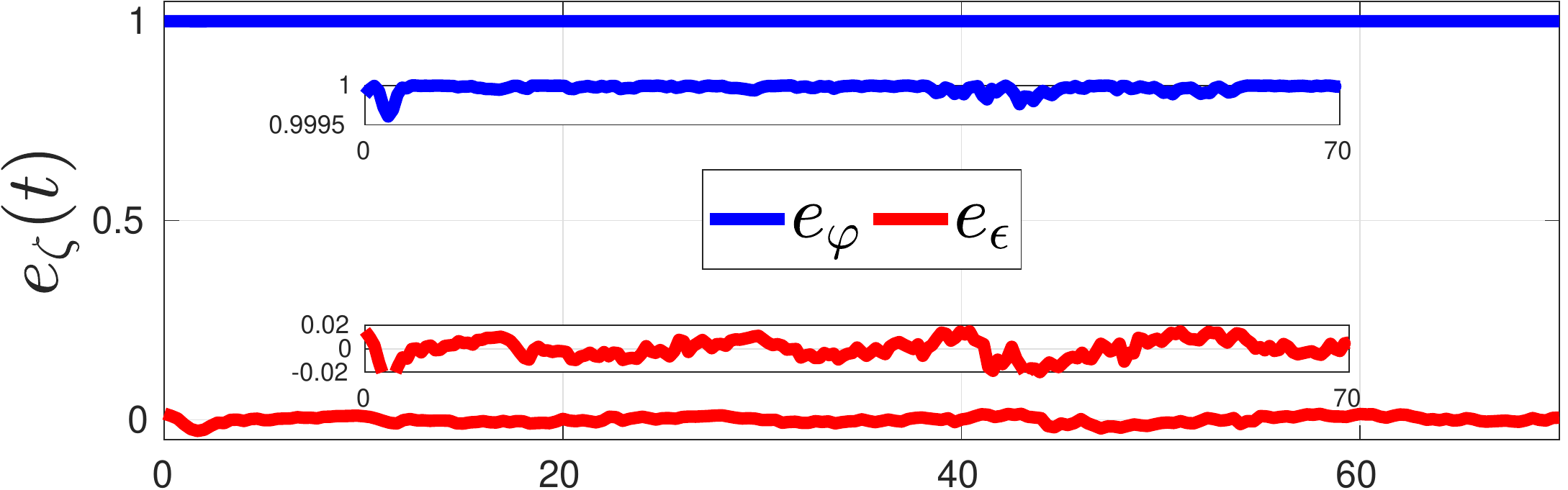}
		\caption{}
		\label{fig:Ng2 (TCST_coop_manip)}
	\end{subfigure}
	
	\begin{subfigure}[b]{\columnwidth}
		\centering
		\includegraphics[width=.65\columnwidth]{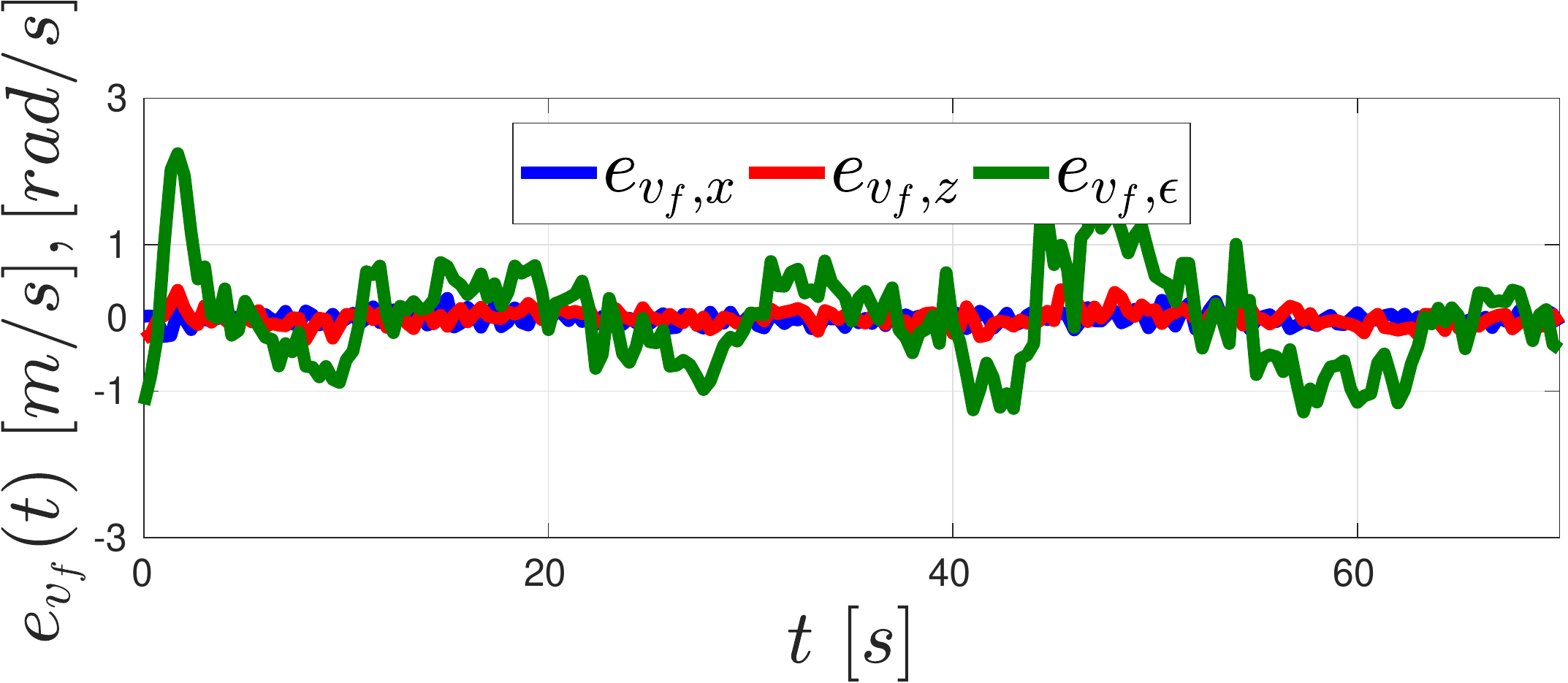}
		\caption{}
		\label{fig:Ng2 (TCST_coop_manip)}
	\end{subfigure}
	
	\caption{Experimental results for the control scheme of Section \ref{subsec:Quaternion Controller (TCST_coop_manip)}; (a): The position errors $e_p(t)$; (b): The quaternion errors $e_\varphi(t)$, $e_\varepsilon(t)$; (c) The velocity errors $e_{v_f}(t)$, $\forall t\in[0,70]$.} \label{fig:adapt_quat_errors (TCST_coop_manip)}
\end{figure}

\begin{figure}
	\centering
	\includegraphics[width=.65\columnwidth]{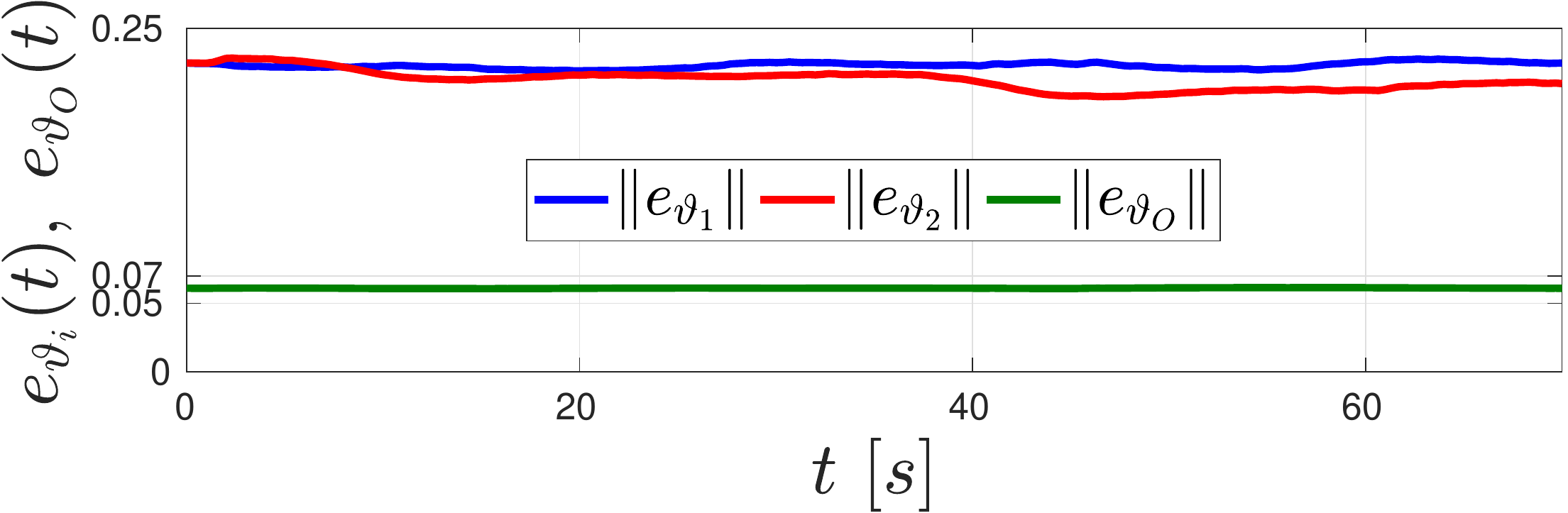} 
	\caption{The norms of the adaptation signals $e_{\vartheta_i}(t), \forall i\in\{1,2\}$ (left) and $e_{\vartheta_{\scriptscriptstyle O}}(t)$, (right) $\forall t\in[0,70]$ of the experiment of the controller in Section \ref{subsec:Quaternion Controller (TCST_coop_manip)}.}
	\label{fig:adaptive_exp_theta_adaptation_signals (TCST_coop_manip)}
\end{figure}

\begin{figure}
	\centering
	\includegraphics[width=.65\columnwidth]{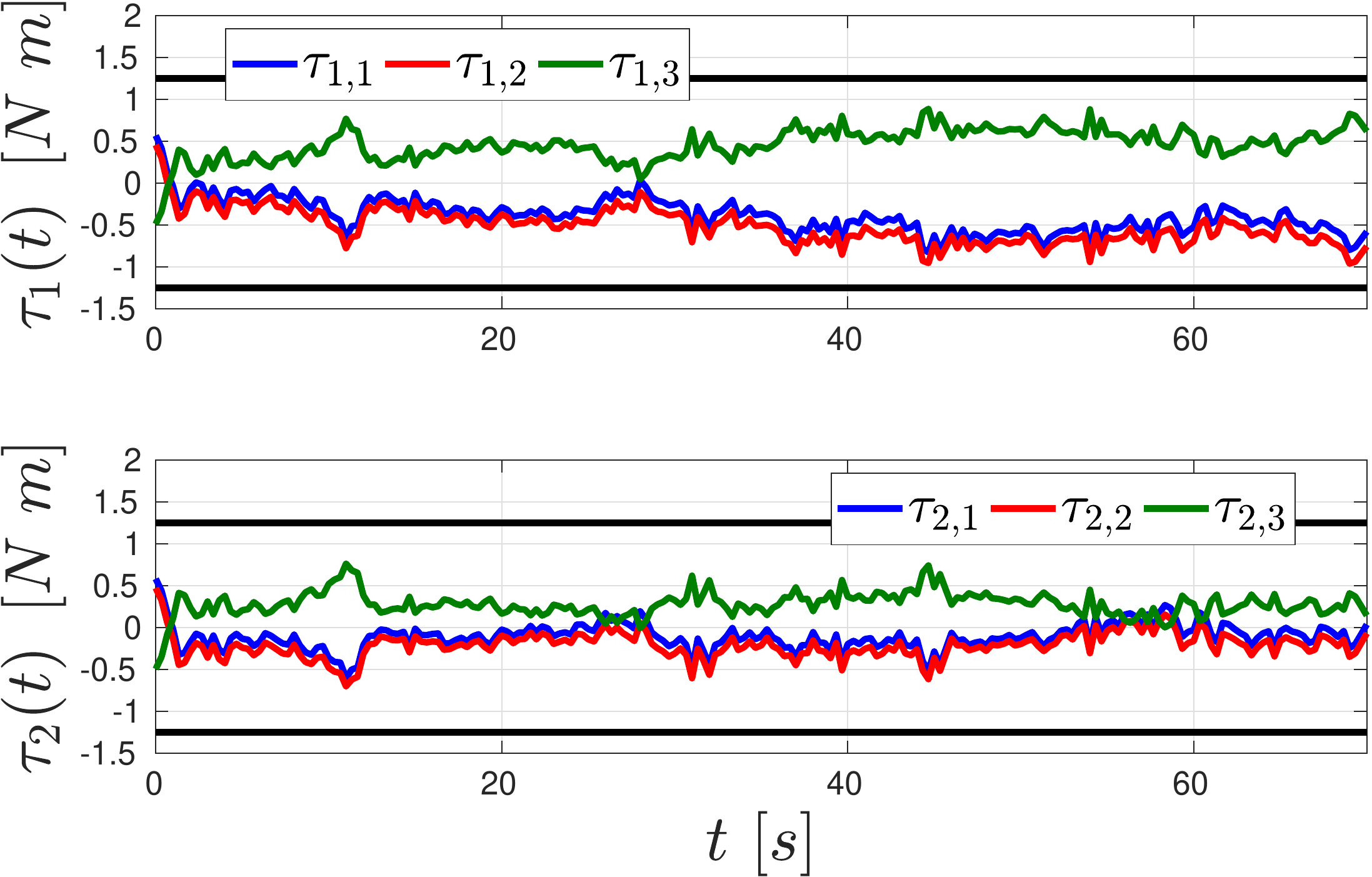} 
	\caption{The agents' joint torques of the experiment of the controller in Section \ref{subsec:Quaternion Controller (TCST_coop_manip)}, for $t\in[0,70]$, with their respective limits (with black).}
	\label{fig:adaptive_exp_theta_inputs (TCST_coop_manip)}
\end{figure}

\subsubsection{Experimental Results}

We further validate the developed control scheme through experimental results.
The tested scenario for the experimental setup consists of two WidowX Robot Arms rigidly grasping a wooden cuboid object of initial pose $x_{\scr O}(0) = [0.3, 0, 0.15, 0, 0, 0]^\top$ $(\textup{[m]},\textup{[rad]})$, which has to track a planar time trajectory $p_{\textup{d}}(t)= [0.3 + 0.05\sin(\frac{2\pi t}{35}), 0.15 - 0.05\cos(\frac{2\pi t}{35})]^\top$, $\eta_\textup{d}(t) 
= \frac{\pi}{20}\sin(\frac{5\pi t}{35})$.
For that purpose, we employ the three rotational -with respect to the $y$ axis - joints of the arms. The lower joint consists of a MX-$64$ Dynamixel Actuator, whereas each of the two upper joints consists of a MX-$28$ Dynamixel Actuator from the MX Series. Both actuators provide feedback of the joint angle and rate $q_i, \dot{q}_i$, $\forall i\in\{1,2\}$.
The micro-controller used for the actuators of each arm is the ArbotiX-M Robocontroller, which is serially connected to an i-$7$ desktop computer  with $4$ cores and $16$GB RAM. All the computations for the real-time experiments are performed at a frequency of $120$ [Hz]. Finally, we consider that the MX-$64$ motor can exert a maximum torque of $3$ [Nm], and the MX-$28$ motors can exert a maximum torque of $1.25$ [Nm], values that are slightly more conservative than the actual limits.  
The load distribution coefficients are set as $m_1^\star = m_2^\star = 1$, and $J^\star_1 = 0.75I_3$, $J^\star_2 = 0.25I_3$. 
For the adaptive quaternion-feedback control scheme, {we set $\delta_{\scriptscriptstyle O}(x_{\scriptscriptstyle O},\dot{x}_{\scriptscriptstyle O}, t)=0$, $\delta_i(q_i,\dot{q}_i,t) = 0$, $\forall i\in\mathcal{N}$, which essentially means that we do not model any external disturbances}. We also set the control gains appearing in  \eqref{eq:control laws adaptive quat (TCST_coop_manip)} and \eqref{eq:adaptation laws (TCST_coop_manip)} as $k_p = 50$, $k_\zeta = 80$, $K_v = \textup{diag}\{3.5,0.5,0.5\}$. 
The experimental results are depicted in Fig. \ref{fig:adapt_quat_errors (TCST_coop_manip)}-\ref{fig:adaptive_exp_theta_inputs (TCST_coop_manip)} for $t\in[0,70]$ seconds. More specifically, Fig. \ref{fig:adapt_quat_errors (TCST_coop_manip)} pictures the pose and velocity errors $e_p(t), e_\zeta(t), e_{v_f}(t)$, Fig. \ref{fig:adaptive_exp_theta_adaptation_signals (TCST_coop_manip)} depicts the norms of the adaptation errors $e_{\vartheta_i}(t)$, $e_{\vartheta_{\scr O}}(t)$, and Fig. \ref{fig:adaptive_exp_theta_inputs (TCST_coop_manip)} shows the joint torques  $\tau_1(t)$, $\tau_2(t)$ of the agents. {Although external disturbances and modeling uncertainties are not taken into account in the system model, they are indeed present during the experiment run time and one can observe that the errors converge to the desired values and the adaptation errors remain
bounded, verifying the theoretical findings.}
{A video illustrating the simulation and experimental results (along with the control scheme of the next section) can be found on \href{https://youtu.be/jJWeI5ZvQPY}{https://youtu.be/jJWeI5ZvQPY}.}

\subsection{Prescribed Performance Control} \label{subsec:PPC Controller (TCST_coop_manip)}

In this section, we adopt the concepts and techniques of prescribed performance control, proposed in \cite{bechlioulis2008robust}, in order to achieve predefined  transient and steady-state response for the derived error, as well as ensure that $\theta_{\scriptscriptstyle O}(t)\in(-\tfrac{\pi}{2},\tfrac{\pi}{2}), \forall t\in\mathbb{R}_{\geq 0}$. As stated in Appendix \ref{app:PPC}, prescribed performance characterizes the behavior where a signal evolves strictly within a predefined region that is bounded by absolutely decaying functions of time, called performance functions. This signal is represented by the object's pose error 
\begin{equation}\label{eq:ppc errors (TCST_coop_manip)}
e_s \coloneqq \begin{bmatrix}
e_{s_x}, e_{s_y}, e_{s_z}, e_{s_\phi}, e_{s_\theta}, e_{s_\psi}
\end{bmatrix}^\top
\coloneqq 
x_{\scriptscriptstyle O} - x_\textup{d}
\end{equation}
Similarly to the result of the previous subsection, the Euler angle Euclidean difference here does not represent a valid orientation distance metric. However, as also stated before, the desired equilibrium point will be rendered \textit{eventually attractive}, which stems from stabilization on the unit sphere. A PPC scheme based on a proper distance metric on $\mathbb{SO}(3)$ is introduced in the next chapter.

We now relax Assumption \ref{ass:disturbance bound (TCST_coop_manip)} and impose a controllability assumption on $\theta_\textup{d}$, given that Euler angles are used now:
\begin{assumption} [Uncertainties/Disturbances bound]  \label{ass:disturbance bound_ppc (TCST_coop_manip)}$\ $
	{The functions $d_{\scr O}(x_{\scr O},\dot{x}_{\scr O},t)$ and $d_{i}(q_i,\dot{q}_i,t)$ are continuous in $(x_{\scr O},\dot{x}_{\scr O})$ and $(q_i,\dot{q}_i)$, respectively, and bounded in $t$ by unknown positive constants $\bar{d}_{\scr O}$ and $\bar{d}_i$, respectively, $\forall i\in\mathcal{N}$.}
\end{assumption} 
\begin{assumption} \label{ass:theta des}
	It holds that $\theta_\textup{d}(t)\in[-\bar{\theta},\bar{\theta}]\subset(-\tfrac{\pi}{2},\tfrac{\pi}{2}), \forall t\in\mathbb{R}_{\geq 0}$.	 
\end{assumption}
More specifically, the requirement $\theta_\textup{d}(t)\in[-\bar{\theta},\bar{\theta}]\subset(-\tfrac{\pi}{2},\tfrac{\pi}{2}), \forall t\in\mathbb{R}_{\geq 0}$ is a necessary condition needed to ensure that tracking of $\theta_\textup{d}$ will not result in singular configurations of ${J}_{\scriptscriptstyle O}({\eta}_{\scriptscriptstyle O})$. The constant $\bar{\theta}\in[0,\tfrac{\pi}{2})$ can be taken arbitrarily close to $\tfrac{\pi}{2}$.

The mathematical expressions of prescribed performance are given by the following inequalities:
\begin{equation}
-\rho_{s_k}(t)< e_{s_k}(t) <\rho_{s_k}(t), \forall k\in\mathcal{K},  \label{eq:ppc (TCST_coop_manip)}
\end{equation}
where $\mathcal{K} \coloneqq \{x,y,z,\phi,\theta,\psi\}$ and $\rho_k:\mathbb{R}_{\geq 0}\rightarrow\mathbb{R}_{> 0}$, with 
\begin{equation}
\rho_{s_k} \coloneqq \rho_{s_k}(t) \coloneqq (\rho_{s_k,\scriptscriptstyle 0}-\rho_{s_k,\scriptscriptstyle\infty})\exp(-l_{s_k}t)+\rho_{s_k,\scriptscriptstyle \infty},  \ \forall k\in\mathcal{K},  \label{eq:rho (TCST_coop_manip)}
\end{equation}
are designer-specified, smooth, bounded and decreasing positive functions of time with $l_{s_k}, \rho_{s_k,\scriptscriptstyle \infty}, k\in\mathcal{K}$, positive parameters incorporating the desired transient and steady-state performance respectively. The terms $\rho_{s_k,\scriptscriptstyle \infty}$ can be set arbitrarily small, achieving thus practical convergence of the errors to zero. Next, we propose a state feedback control protocol that does not incorporate any information on the agents' or the object's dynamics or the external disturbances and guarantees \eqref{eq:ppc (TCST_coop_manip)} for all $t\in\mathbb{R}_{\geq 0}$. 
More specifically, given the errors \eqref{eq:ppc errors (TCST_coop_manip)}:\\
\textbf{Step I-a}. Select the functions $\rho_{s_k}$ as in \eqref{eq:rho (TCST_coop_manip)} with 
\begin{enumerate}[(i)]
	\item 	$\rho_{s_\theta,\scriptscriptstyle 0}  = \rho_{s_\theta}(0)= \theta^*, \rho_{s_k,\scriptscriptstyle 0} = \rho_{s_k}(0) > \lvert e_{s_k}(0) \rvert, \forall k\in\mathcal{K}\backslash\{\theta\}$, 
	\item $l_{s_k} \in\mathbb{R}_{>0}, \forall k\in\mathcal{K}$,
	\item $\rho_{s_k,\scriptscriptstyle \infty}\in(0,\rho_{s_k,0}), \forall k\in\mathcal{K}$,
\end{enumerate}
where $\theta^*$ is a positive constant satisfying $\theta^* + \bar{\theta} < \frac{\pi}{2}$.  \\
\textbf{Step I-b}. Introduce the normalized errors 
\begin{equation}
\xi_{s} \coloneqq \begin{bmatrix}
\xi_{s_x}, \dots, \xi_{s_\psi} \end{bmatrix}^\top
\coloneqq \rho_s^{-1}e_s,	\label{eq:ksi_s (TCST_coop_manip)}
\end{equation}
where $\rho_s \coloneqq \rho_s(t)\coloneqq\textup{diag}\{\left[\rho_{s_k}\right]_{k\in\mathcal{K}}\}\in\mathbb{R}^{6\times6}$, as well as the transformed state functions $\varepsilon_s:(-1,1)^6\to\mathbb{R}^6$, and signals $r_s:(-1,1)^6\to\mathbb{R}^{6\times 6}$, with  
\begin{align}
\varepsilon_s \coloneqq \varepsilon_s(\xi_s) &\coloneqq   \begin{bmatrix}
\varepsilon_{s_x}, \dots, \varepsilon_{s_\psi} \end{bmatrix}^\top 
\coloneqq 
\begin{bmatrix}
\ln\Big(\frac{1 + \xi_{s_x}}{1 - \xi_{s_x}} \Big), \dots, \ln\Big(\frac{1 + \xi_{s_\psi}}{1 - \xi_{s_\psi}}\Big) \end{bmatrix}^\top \label{eq:epsilon_s (TCST_coop_manip)}\\
r_s\coloneqq r_s(\xi_s) &\coloneqq 
\textup{diag}\{[r_{s_k}(\xi_{s_k})]_{k\in\mathcal{K}}\}
\coloneqq 
\textup{diag}\left\{ \left [\frac{\partial \varepsilon_{s_k}}{\partial \xi_{s_k}} \right]_{k\in\mathcal{K}} \right \} \notag \\
& = \textup{diag}\left\{ \left [\frac{2}{1-\xi^2_{s_k} } \right]_{k\in\mathcal{K}} \right \} \label{eq:r_s (TCST_coop_manip)},
\end{align} 
and design the reference velocity vector $v_r : (-1,1)^6\times \mathbb{R}_{\geq 0} \to \mathbb{R}^6$ with 
\begin{align}
& v_r \coloneqq  v_r(\xi_s,t) \coloneqq 
-g_s J_{\scriptscriptstyle O}\Big( \eta_\textup{d}(t) + \rho_{s_\eta}(t)\xi_{s_\eta} \Big)^{-1}\rho_s^{-1}r_s\varepsilon_{s}, \label{eq:v_r (TCST_coop_manip)}
\end{align}
where $\rho_{s_\eta} \coloneqq \rho_{s_\eta}(t) \coloneqq \textup{diag}\{\rho_{s_\phi},\rho_{s_\theta},\rho_{s_\psi}\}$, $\xi_{s_\eta}\coloneqq [\xi_{s_\phi},\xi_{s_\eta},\xi_{s_\phi}]^\top$, and we have further used the relation $\xi_s = \rho_s^{-1}(x_{\scriptscriptstyle O}- x_\textup{d})$ from \eqref{eq:ppc errors (TCST_coop_manip)} and \eqref{eq:ksi_s (TCST_coop_manip)}.\\
\textbf{Step II-a}. Define the velocity error vector 
\begin{equation}
e_v \coloneqq \begin{bmatrix}
e_{v_x}, \dots,  e_{v_\psi}  	
\end{bmatrix}^\top
\coloneqq v_{\scriptscriptstyle O} - v_r,  \label{eq:e_v_r (TCST_coop_manip)}
\end{equation} 
and select the corresponding positive performance functions $\rho_{v_k} \coloneqq \rho_{v_k}(t):\mathbb{R}_{\geq 0}\rightarrow\mathbb{R}_{>0}$ with $\rho_{v_k}(t) \coloneqq (\rho_{v_k, \scriptscriptstyle 0} - \rho_{v_k,\scriptscriptstyle \infty})\exp(-l_{v_k}t) + \rho_{v_k,\scriptscriptstyle \infty}$, such that $\rho_{v_k,\scriptscriptstyle 0}  = \lVert e_{v}(0) \rVert + \alpha, l_{v_k}>0$ and $\rho_{v_k,\scriptscriptstyle \infty}\in(0,\rho_{v_k,0}), \forall k\in\mathcal{K}$, where $\alpha$ is an arbitrary positive constant.\\
\textbf{Step II-b}. Define the normalized velocity error   
\begin{equation}
\xi_v \coloneqq \begin{bmatrix}
\xi_{v_x}, \dots, \xi_{v_\psi} \end{bmatrix}^\top
\coloneqq \rho_v^{-1}e_v,	\label{eq:ksi_v (TCST_coop_manip)}
\end{equation}	
where $\rho_v\coloneqq \rho_v(t)\coloneqq\textup{diag}\{\left[\rho_{v_k}\right]_{k\in\mathcal{K}}\}$, as well as the transformed states 
$\varepsilon_v:(-1,1)^6\to\mathbb{R}^6$ and signals $r_v:(-1,1)^6\to\mathbb{R}^{6\times6}$, with 
\begin{align}
\varepsilon_v \coloneqq \varepsilon_v(\xi_v) &\coloneqq  \begin{bmatrix}
\varepsilon_{v_x}, \dots, \varepsilon_{v_\psi} \end{bmatrix}^\top
\coloneqq 
\begin{bmatrix}
\ln\Big(\frac{1 + \xi_{v_x}}{1 - \xi_{v_x}} \Big), \dots, \ln\Big(\frac{1 + \xi_{v_\psi}}{1 - \xi_{v_\psi}}\Big) \end{bmatrix}^\top \notag \\  
r_v(\xi_v) &\coloneqq 
\textup{diag}\{[r_{v_k}(\xi_{v_k})]_{k\in\mathcal{K}}\}
\coloneqq 
\textup{diag}\left\{ \left [\frac{\partial \varepsilon_{v_k}}{\partial \xi_{v_k}} \right]_{k\in\mathcal{K}} \right \} \notag \\
& = \textup{diag}\left\{ \left[\frac{2}{1-\xi^2_{v_k} } \right]_{k\in\mathcal{K}} \right \} \label{eq:r_v (TCST_coop_manip)},
\end{align} 
and design the decentralized feedback control protocol for each agent $i\in\mathcal{N}$ as $u_i:\mathsf{S}_i\times(-1,1)^6\times\mathbb{R}_{\geq 0}$, with
\begin{equation}
u_i \coloneqq u_i(q_i,\xi_v,t) \coloneqq -g_v J_{M_i}(q_i) \rho_v^{-1}r_v\varepsilon_v, \label{eq:control_law_ppc (TCST_coop_manip)} 
\end{equation}
where $g_v$ is a positive constant gain and $J_{M_i}$ as defined in \eqref{eq:J_Hirche (TCST_coop_manip)}.
The control laws \eqref{eq:control_law_ppc (TCST_coop_manip)} can be written in vector form $u \coloneqq [u^\top_1,\dots,u^\top_N]^\top$, with:
\begin{align}
\hspace{-2mm}u = -g_v G^{+}_M(q)\rho_v^{-1}r_v\varepsilon_v. \label{eq:control_law_ppc_vector_form (TCST_coop_manip)}
\end{align}

\begin{remark} [\textbf{Decentralized manner and robustness (PPC)}]
	Similarly to \eqref{eq:laws vector forms (TCST_coop_manip)},  notice from \eqref{eq:control_law_ppc (TCST_coop_manip)} that each agent $i\in\mathcal{N}$ can calculate its own control signal, without communicating with the rest of the team, rendering thus the overall control scheme decentralized. The  terms $l_k$, $\rho_{k,0}$, $\rho_{k,\scriptscriptstyle \infty}$, $\alpha$, $l_{v_k}$, and $\rho_{v_k,\infty}$, $k\in\mathcal{K}$ needed for the calculation of the performance functions can be transmitted off-line to the agents. Moreover, the Prescribed Performance Control protocol is also robust to uncertainties of model uncertainties and external disturbances. In particular, note that the control laws do not even require the structure of the terms $\widetilde{M}, \widetilde{C}, \widetilde{g}, \widetilde{d}$, but only the positive definiteness of $\widetilde{M}$, as will be observed in the subsequent proof of Theorem \ref{th:thorem_ppc (TCST_coop_manip)}. It is worth noting that, in the case that one or more agent failed to participate in the task, then the remaining agents would need to appropriately update their control protocols (e.g., update $J_{M_i}$) to compensate for the failure.   
\end{remark}

The main results of this subsection are summarized in the following theorem. 
\begin{theorem}\label{th:thorem_ppc (TCST_coop_manip)}
	Consider $N$ agents rigidly grasping an object with unknown coupled dynamics \eqref{eq:coupled dynamics (TCST_coop_manip)}. Then, under Assumptions \ref{ass:feedback (TCST_coop_manip)}-\ref{ass:kinematic singularities (TCST_coop_manip)}, \ref{ass:disturbance bound_ppc (TCST_coop_manip)},   
	the decentralized control protocol \eqref{eq:ksi_s (TCST_coop_manip)}-\eqref{eq:control_law_ppc (TCST_coop_manip)} guarantees that $-\rho_{s_k}(t) < e_{s_k}(t) < \rho_{s_k}(t), \forall k\in\mathcal{K},t\in\mathbb{R}_{\geq 0}$ from all initial conditions satisfying $\lvert \theta_{\scr O}(0)-\theta_\textup{d}(0) \rvert < \theta^*$ (from \textbf{Step I-a} (i)), with
	all closed loop signals being bounded.
\end{theorem}
\begin{proof}
	{The proof consists of two main parts. Firstly, we prove that there exists a maximal solution $(\xi_s(t),\xi_v(t))\in(-1,1)^{12}$ for $t\in[0,\tau_{\max})$, where $\tau_{\max}>0$. Secondly, we prove that $(\xi_s(t), \xi_v(t))$ is contained in a compact subset of $(-1,1)^{12}$ and consequently, that $\tau_{\max} = \infty$. Without loss of generality, we assume that $v_{\scr O}(0) = 0$.}
	
	 \underline{Part A}: Consider the combined state $\sigma \coloneqq [q,\xi_s,\xi_v]\in \mathsf{S}\times\mathbb{R}^{12}$. Differentiation of $\sigma$ yields, in view of \eqref{eq:J_o_i (TCST_coop_manip)}, \eqref{eq:ksi_s (TCST_coop_manip)} and \eqref{eq:ksi_v (TCST_coop_manip)}
		\begin{align} \label{eq:sigma_dot_1 (TCST_coop_manip)}
		\dot{\sigma} = \begin{bmatrix}
		\widetilde{J} G^\top v_{\scr O} \\
		\rho_s^{-1}(\dot{x}_{\scriptscriptstyle O} - \dot{x}_\textup{d} - \dot{\rho}_s \xi_s) \\
		\rho_v^{-1}(\dot{v}_{\scr O} - \dot{v}_r - \dot{\rho}_v \xi_v),
		\end{bmatrix},
		\end{align}
		where $\widetilde{J}\coloneqq\widetilde{J}(q)\coloneqq \textup{diag}\{[J_i(q_i)^\top(J_i(q_i)J_i(q_i)^\top)^{-1}]_{i\in\mathcal{N}}\}\in\mathbb{R}^{n\times 6N}$ is well defined due to Assumption \ref{ass:kinematic singularities (TCST_coop_manip)}. Then, by employing \eqref{eq:object dynamics (TCST_coop_manip)}, \eqref{eq:ppc errors (TCST_coop_manip)}, \eqref{eq:ksi_s (TCST_coop_manip)}, and \eqref{eq:v_r (TCST_coop_manip)}-\eqref{eq:control_law_ppc_vector_form (TCST_coop_manip)} as well as $G G^{+}_M = I_6$, we can express the right-hand side of \eqref{eq:sigma_dot_1 (TCST_coop_manip)} as a function of $\sigma$ and $t$, i.e.,
		$$\dot{\sigma} = f_{\textup{cl}}(\sigma,t) \coloneqq 
		\begin{bmatrix}
		f_{\textup{cl},q}(\sigma,t) \\
		f_{\textup{cl},s}(\sigma,t) \\ 
		f_{\textup{cl},v}(\sigma,t) 
		] \end{bmatrix}, \notag $$
		with 
		\begin{align*}
			f_{\textup{cl},q}(\sigma,t) \coloneqq & \widetilde{J}(q) G(q)^\top \big( \rho_v(t) \xi_v + v_r(\xi_s,t) \big) \\
			f_{\textup{cl},s}(\sigma,t) \coloneqq & \rho_s(t)^{-1}\big[ J_{\scr O}(\eta_\textup{d}(t) + \rho_{s_\eta}(t)\xi_{s_\eta})\rho_v(t)\xi_v - \dot{\rho}_s(t)\xi_s \\
			&\hspace{35mm} -g_s \rho_s(t)^{-1}r_s(\xi_s)\varepsilon_s(\xi_s) - \dot{x}_\textup{d}(t)  \big] \\
			f_{\textup{cl},v}(\sigma,t) \coloneqq & -\rho_v(t)^{-1}\Bigg( 
			\widetilde{M}(x(\sigma,t)) \bigg[ \widetilde{C}(x(\sigma,t))\big(\rho_v(t)\xi_v + v_r(\xi_s,t) \big) \\
			& +\widetilde{g}(x(\sigma,t)) + \widetilde{d}(x(\sigma,t),t) + g_v\rho_v(t)^{-1} r_v(\xi_v) \varepsilon_v(\xi_v)  \bigg] - \dot{\rho}_v(t)\xi_v \\
			& + \frac{\partial v_r(\xi_s,t)}{\partial t} +  \frac{\partial v_r(\xi_s,t)}{\partial \xi_s} f_{\textup{cl},s}(\sigma,t)
			\Bigg),
		\end{align*}
		and we also express $x$ as a function of $\sigma$ and $t$ via 
		\begin{equation*}
			x(\sigma,t) = \begin{bmatrix}
			q \\ \dot{q} \\ \eta_{\scr O} \\ \omega_{\scr O}
			\end{bmatrix} = 
			\begin{bmatrix}
			q \\
			f_{\textup{cl},q}(\sigma,t)\\
			\eta_\textup{d}(t) + \rho_{s_\eta}(t)\xi_{s_\eta} \\ 
			\big(\rho_v(t)\xi_v + v_r(\xi_s,t) \big)_{3:6}
			\end{bmatrix}
		\end{equation*}		
	where $(\cdot)_{3:6}$ denotes the three last components of the vector.
	{Consider now the open and nonempty set $\Omega \coloneqq \mathsf{S}\times(-1,1)^{12}$. The choice of the parameters $\rho_{s_k,0}$ and $\rho_{v_k,0}, k\in\mathcal{K}$ in \textbf{Step I-a} and \textbf{Step II-a}, respectively, along with the fact that the initial conditions satisfy $|\theta_{\scriptscriptstyle O}(0) - \theta_\textup{d}(0) | < \theta^*$ imply that $| e_{s_k}(0) | < \rho_{s_k}(0), | e_{v_k}(0) | < \rho_{v_k}(0), \forall k\in\mathcal{K}$ and hence $[ \xi_s(0)^\top, \xi_v(0)^\top ]^\top\in(-1,1)^{12}$. 
		Moreover, it can be verified that $f_{\textup{cl}}: \Omega\times\mathbb{R}_{\geq 0} \to \mathbb{R}^{n+12}$ is locally Lipschitz in $\sigma$ over the set $\Omega$ and continuous and locally integrable in $t$ for each fixed $\sigma\in\Omega$. Therefore, the hypotheses of Theorem \ref{thm:ode solution (App_dynamical_systems)}  in Appendix \ref{app:dynamical systems} hold and the existence of a maximal solution $\sigma:[0,\tau_{\max})\to\Omega$, for $\tau_{\max} > 0$, is ensured. We thus conclude  
		\begin{align} \label{eq:ksi_bounded_open (TCST_coop_manip)}
		\xi_{s_k}(t), \  
		\xi_{v_k}(t)\in (-1,1)
		\end{align}
		$\forall k\in\mathcal{K}, t\in[0,\tau_{\max})$, which also implies that $\| \xi_s(t) \| < \sqrt{6}$, and $\|\xi_v(t) \| < \sqrt{6}, \forall t\in[0,\tau_{\max})$. In the following, we show the boundedness of all closed loop signals and $\tau_{\max} = \infty$. }
	
	{ \underline{Part B}:}
	{Note first from \eqref{eq:ksi_bounded_open (TCST_coop_manip)}, that  $\lvert \theta_{\scriptscriptstyle O}(t) - \theta_\textup{d}(t) \rvert < \rho_\theta(t) \leq \rho_{\theta}(0) = \theta^*$, which, since $\theta_\textup{d}(t)\in[-\bar{\theta}, \bar{\theta}],\forall t\in\mathbb{R}_{\geq 0}$, implies that $ | \theta_{\scriptscriptstyle O}(t) | \leq \widetilde{\theta}\coloneqq \bar{\theta} + \theta^* < \frac{\pi}{2}, \forall t\in[0,\tau_{\max})$. Therefore, by employing \eqref{eq:J_norm (TCST_coop_manip)}, one obtains that, $\forall t\in[0,\tau_{\max})$,
		\begin{equation}
		\| J_{\scriptscriptstyle O}(\eta_{\scriptscriptstyle O}(t)) \| \leq \bar{J}_{\scriptscriptstyle O} \coloneqq \sqrt{\frac{ |\sin(\widetilde{\theta}) |+1}{1-\sin^2(\widetilde{\theta})}} < \infty. \label{eq:J_O_bar (TCST_coop_manip)}
		\end{equation}
		Consider now the positive definite function $V_s \coloneqq \tfrac{1}{2}\| \varepsilon_s\|^2$. Differentiating $V_s$ along the solutions of the closed loop system yields $\dot{V}_s = \varepsilon_s^\top r_s\rho_s^{-1}\dot{\xi}_s$, which, in view of \eqref{eq:sigma_dot_1 (TCST_coop_manip)}, \eqref{eq:ksi_v (TCST_coop_manip)}, \eqref{eq:v_r (TCST_coop_manip)} and the fact that $\dot{x}_{\scr O} = J_{\scr O}(\eta_{\scr O})v_{\scr O} = J_{\scr O}(\eta_{\scr O})(v_r + e_v)$, becomes		
		\begin{align*} 
		\dot{V}_s =& - g_s \|\rho_s^{-1}r_s\varepsilon_s \|^2 - \varepsilon_s^\top r_s\rho_s^{-1}\Big(\dot{x}_\textup{d} +\dot{\rho}_s\xi_s   - J_{\scriptscriptstyle O}e_v \Big) \notag \\
		 \leq &  -g_s \|\rho_s^{-1}r_s\varepsilon_s \|^2 + \|\rho_s^{-1}r_s\varepsilon_s \| \Big(\|\dot{x}_\textup{d}\|    + \| J_{\scriptscriptstyle O}\rho_v\xi_v \|  +  \| \dot{\rho}_s\xi_s \| \Big).  
		\end{align*}		
		In view of \eqref{eq:J_O_bar (TCST_coop_manip)}, \eqref{eq:ksi_bounded_open (TCST_coop_manip)}, and the structure of $\rho_{s_k}, \rho_{v_k}, k\in\mathcal{K}$, as well as the fact that $v_{\scriptscriptstyle O}(0) = 0$ and the boundedness of $\dot{x}_\textup{d} $, the last inequality becomes  
		\begin{align*}
		\dot{V}_s \leq & -g_s \|\rho_s^{-1}r_s\varepsilon_s \|^2 +   \|\rho_s^{-1}r_s\varepsilon_s \| \bar{B}_s,   
		\end{align*}		
		$\forall t\in[0,\tau_{\max})$, with 
		\begin{equation*}
			\bar{B}_s \coloneqq \sqrt{6}\bar{J}_{\scr O}(\|v_r(0)\|+\alpha) + \sup_{t>0}\|\dot{x}_\textup{d}(t)\| + \sqrt{6} \max_{k\in\mathcal{K}}\{ l_k(\rho_{s_k,0}-\rho_{s_k,\infty}) \}, 
		\end{equation*}
		independent of $\tau_{\max}$.
		Therefore, $\dot{V}_s$ is negative when $\|\rho_s^{-1}r_s\varepsilon_s\| > \frac{\bar{B}_s}{g_s}$, which, by employing \eqref{eq:r_s (TCST_coop_manip)}, the decreasing property of $\rho_{s_k}, k\in\mathcal{K}$ as well as \eqref{eq:ksi_bounded_open (TCST_coop_manip)}, is satisfied when $\| \varepsilon_s \| > \frac{\max_{k\in\mathcal{K}}\{\rho_{s_k,0}\} \bar{B_s}}{2 g_s}$. Hence, by using Theorem \ref{th:uub_khalil (App_dynamical_systems)} of Appendix \ref{app:dynamical systems}, we conclude that
		\begin{align}
		\| \varepsilon_s(\xi_s(t)) \| \leq \bar{\varepsilon}_s \coloneqq \max\Bigg\{ \| \varepsilon_s(0) \|, \frac{\max\limits_{k\in\mathcal{K}}\{\rho_{s_k,0}\} \bar{B_s}}{2 g_s}   \Bigg\}, \label{eq:epsilon_s_bar (TCST_coop_manip)}
		\end{align}   
		$\forall t\in[0,\tau_{\max})$. Furthermore, since $| \varepsilon_{s_k}| \leq \|\varepsilon_s\|, \forall k\in\mathcal{K}$, taking the inverse logarithm function from \eqref{eq:epsilon_s (TCST_coop_manip)}, we obtain 		
		\begin{align}
		-1 < \frac{\exp(-\bar{\varepsilon}_s)-1}{\exp(-\bar{\varepsilon}_s)+1} =: -\bar{\xi}_{s} \leq  \xi_{s_k}(t) \leq \bar{\xi}_{s} \coloneqq \frac{\exp(\bar{\varepsilon}_s)-1}{\exp(\bar{\varepsilon}_s)+1} < 1, \label{eq:ksi_s_bar (TCST_coop_manip)}
		\end{align}		
		$\forall t\in[0,\tau_{\max})$. Hence, recalling \eqref{eq:r_s (TCST_coop_manip)} and \eqref{eq:v_r (TCST_coop_manip)},
		we obtain the boundedness of $r_s(\xi_s(t))$, $v_r(t)$, $\forall t\in [0,\tau_{\max})$, and in view of $v_{\scr O} = v_r + e_v$, \eqref{eq:e_v_r (TCST_coop_manip)}, \eqref{eq:ksi_bounded_open (TCST_coop_manip)},  \eqref{eq:J_o_i (TCST_coop_manip)} and \eqref{eq:J_O_i bound (TCST_coop_manip)}, the boundedness of $v_{\scr O}(t)$ and $v_i(t)$ as
		\begin{align}
			&\|r_s(\xi_s(t))\| \leq \bar{r}_s \coloneqq \frac{2}{1 - \bar{\xi}_s^2} = \frac{ (\exp(\bar{\varepsilon}_s) + 1)^2}{2\exp(\bar{\varepsilon}_s)}, \notag \\
			&\|v_r(t)\| \leq \bar{v}_r \coloneqq g_s\sqrt{2} \frac{\bar{\varepsilon}_s(\exp(\bar{\varepsilon}_s) + 1)^2}{2\min_{k\in\mathcal{K}}\{\rho_{s_k,\infty}\}\exp(\bar{\varepsilon}_s)} \notag \\
			& \|v_{\scr O}(t)\| \leq \bar{v}_{\scr O} \coloneqq \bar{v}_r + \sqrt{6} \max_{k\in\mathcal{K}}\{ \rho_{v_k,0}\} \notag \\
			& \|v_i(t) \| \leq \bar{v}_i \coloneqq \big(\| p^{\scr E_i}_{\scr O/E_i} + 1\big)\bar{v}_{\scr O}, \forall i\in\mathcal{N},	\label{eq:v_i_bar (TCST_coop_manip)}
		\end{align} 
		$\forall t\in [0,\tau_{\max})$, 
		From \eqref{eq:ksi_s_bar (TCST_coop_manip)}, \eqref{eq:object dynamics 1 (TCST_coop_manip)}, and \eqref{eq:ppc errors (TCST_coop_manip)} we also conclude the boundedness of $x_{\scr O}(t)$, $\dot{x}_{\scr O}(t)$, as 
		\begin{align*}
			\|x_{\scr O}(t) \| &\leq \bar{x}_{\scr O} \coloneqq \sup_{t>0}\|x_\textup{d}(t)\| + \sqrt{6} \xi_s \max_{k\in\mathcal{K}}\{\rho_{s_k,0}\}, \\
			\|\dot{x}_{\scr O}(t)\| &\leq \bar{J}_{\scr O}\bar{v}_{\scr O},
		\end{align*}
		$\forall t\in [0,\tau_{\max})$. The coupled kinematics \eqref{eq:coupled_kinematics (TCST_coop_manip)} and Assumption \ref{ass:kinematic singularities (TCST_coop_manip)} imply also the boundedness of $p_{\scr E_i}(t)$, $q_i(t)$, and $\dot{q}_i(t)$, $\forall i\in\mathcal{N}$, as 
		$\|q(t)\| \leq \bar{q}$, $\|\dot{q}(t)\| \leq \bar{J}\|v\|\leq \bar{J}\sum_{i\in\mathcal{N}}\bar{v}_i$ for a positive constant $\bar{J}$, $[0,\tau_{\max})$. Hence, we conclude that 
		\begin{equation*}
			\|x(t)\| \leq \bar{x} \coloneqq \bar{q} + \bar{J} \sum_{i\in\mathcal{N}} \bar{v}_i + \bar{x}_{\scr O} + \bar{J}_{\scr O} \bar{v}_{\scr O},
		\end{equation*}		
	    $[0,\tau_{\max})$. 
		In a similar vein, by differentiating the reference velocity \eqref{eq:v_r (TCST_coop_manip)} and using \eqref{eq:epsilon_s (TCST_coop_manip)}, \eqref{eq:r_s (TCST_coop_manip)}, and \eqref{eq:epsilon_s_bar (TCST_coop_manip)}, we also conclude the boundedness of $\dot{v}_r(t)$ by a positive constant $\bar{\dot{v}}_r$, $\forall t\in [0,\tau_{\max})$. }

	{Applying the aforementioned line of proof, we consider the positive definite function $V_v \coloneqq \tfrac{1}{2}\| \varepsilon_v\|^2$. By differentiating $V_v$ we obtain $\dot{V}_v = \varepsilon_v^\top r_v\rho_v^{-1}\dot{\xi}_v$, which, in view of \eqref{eq:sigma_dot_1 (TCST_coop_manip)}, \eqref{eq:e_v_r (TCST_coop_manip)}, \eqref{eq:coupled dynamics (TCST_coop_manip)}, becomes 
		\begin{align}
		 \dot{V}_v =& - g_v \varepsilon_v^\top r_v\rho_v^{-1}\widetilde{M} \rho_v^{-1}r_v\varepsilon_v + \varepsilon_v^\top r_v\rho_v^{-1}\Big( 
		-  \dot{\rho}_v \xi_v  - \widetilde{M}\Big[\widetilde{C}(\rho_v\xi_v +  v_r) + \widetilde{g}  \notag \\ 
		&+ \widetilde{d}\Big] -  \dot{v}_r \Big). \label{eq:V_v dot 1 (TCST_coop_manip)} 
		\end{align}
		Invoking Assumption \ref{ass:disturbance bound_ppc (TCST_coop_manip)} and the boundedness of $q_i(t)$, $\dot{q}_i(t)$, $x_{\scr O}(t)$, $\dot{x}_{\scr O}(t)$, $\forall t\in[0,\tau_{\max})$, we conclude the boundedness of  $d_{\scr O}(x_{\scr O}(t),\dot{x}_{\scr O}(t),t)$ and $d_i(q_i(t),\dot{q}_i(t),t)$ by positive finite constants  $\underline{d}'_{\scr O}$, $\underline{d}'_i$, $\forall i\in\mathcal{N}$, respectively, $\forall t\in[0,\tau_{\max})$. Hence, from  \eqref{eq:J_O_i bound (TCST_coop_manip)} and \eqref{eq:coupled dynamics (TCST_coop_manip)}, we also obtain the boundedness of $\widetilde{d}(x(t))$ as 
		\begin{equation*}
			\|\widetilde{d}(x(t)) \| \leq \underline{d} \coloneqq \underline{d}'_{\scr O} + \sum_{i\in\mathcal{N}}\{\| p^{\scr E_i}_{\scr O/E_i} \| + 1 \}\underline{d}'_i. 
		\end{equation*}
		In addition, the continuity of $\widetilde{C}(x), \widetilde{g}(x)$ implies the existence of positive and finite constant $\bar{c},\bar{g}$ such that $\|\widetilde{C}(x(t))\| \leq \bar{c}$, $\|\widetilde{g}(x(t))\| \leq \bar{g}$, $\forall t\in[0,\tau_{\max})$.}
		
	{Thus, by combining the aforementioned discussion with the boundedness of $\dot{v}_r$, the positive definitiveness and boundedness of $\widetilde{M}(x)$, \eqref{eq:dyn properties (TCST_coop_manip)} and \eqref{eq:ksi_bounded_open (TCST_coop_manip)}, we obtain from  \eqref{eq:V_v dot 1 (TCST_coop_manip)} 		
		\begin{align*}
		& \dot{V}_v \leq -g_v \underline{m} \| \rho_v^{-1}r_v\varepsilon_v \|^2  +   \| \rho_v^{-1}r_v\varepsilon_v \| \bar{B}_v,  
		\end{align*}		
		$\forall t\in[0,\tau_{\max})$, where 
		\begin{equation*}
			\bar{B}_v \coloneqq \sqrt{6} \max_{k\in\mathcal{K}}\{ l_{v_k}(\rho_{v_k,0} - \rho_{v_k,\infty}) \} + \bar{\dot{v}}_r + \bar{m}(\bar{g} + \underline{d} + \bar{c}(\bar{v}_r + \sqrt{6}(\|v_r(0)\| + \alpha )))
		\end{equation*}   is a positive and finite constant, independent of $\tau_{\max}$.}

	{By proceeding similarly as with $\dot{V}_s$, we conclude that 
		\begin{align}
		\|\varepsilon_v(\xi_v(t)) \| \leq \bar{\varepsilon}_v \coloneqq \max\Bigg\{\| \varepsilon_v(0)\|, \frac{\max\limits_{k\in\mathcal{K}}\{\rho_{v_k,0}\}\bar{B}_v}{2g_v\underline{m}} \Bigg\}, \label{eq:epsilon_v_bar (TCST_coop_manip)}
		\end{align}
		$\forall t\in[0,\tau_{\max})$, from which we obtain		
		\begin{align}
		-1 < \frac{\exp(-\bar{\varepsilon}_v)-1}{\exp(-\bar{\varepsilon}_v)+1} =: -\bar{\xi}_{v} \leq  \xi_{v_k}(t) \leq \bar{\xi}_{v} \coloneqq \frac{\exp(\bar{\varepsilon}_v)-1}{\exp(\bar{\varepsilon}_v)+1} < 1, \label{eq:ksi_v_bar (TCST_coop_manip)}
		\end{align}		
		$\forall t\in[0,\tau_{\max})$. 
		In view of \eqref{eq:r_v (TCST_coop_manip)}, \eqref{eq:control_law_ppc (TCST_coop_manip)}, this also implies 
		\begin{align}
			& \|r_v(\xi_v(t))\| \leq \bar{r}_v \coloneqq \frac{2}{1 - \bar{\xi}_v^2} = \frac{ (\exp(\bar{\varepsilon}_v) + 1)^2}{2\exp(\bar{\varepsilon}_v)}, \notag \\
			& \|u_i(t)\| \leq \bar{u}_i \coloneqq g_v \bar{J}_{M_i} \max_{k\in\mathcal{K}} \{ \rho_{v_k,\infty}^{-1} \}\bar{r}_v \bar{\varepsilon}_v, \label{eq:u_i_bar (TCST_coop_manip)}
		\end{align}
		$\forall t\in[0,\tau_{\max})$, where $\bar{J}_{M_i}$ is an upper bound of $\|J_{M_i}(q_i)\|$, which can be proven to be independent of $q$.
		
		What remains to be shown is that $\tau_{\max} = \infty$. We can conclude from the aforementioned analysis, Assumption \ref{ass:kinematic singularities (TCST_coop_manip)}, and \eqref{eq:ksi_s_bar (TCST_coop_manip)}, \eqref{eq:ksi_v_bar (TCST_coop_manip)} that the solution $\sigma(t)$ remains in a compact subset $\Omega'$ of $\Omega$, $\forall t\in[0,\tau_{\max})$, namely $\sigma(t) \in \Omega'$,
		$\forall t\in[0,\tau_{\max})$. Hence, according to Theorem \ref{thm:forward_completeness (App_dynamical_systems)} of Appendix \ref{app:dynamical systems}, it holds that $\tau_{\max} = \infty$. Thus, all closed loop signals remain bounded and moreover $\sigma(t)\in\Omega' \subset \Omega, \forall t\in\mathbb{R}_{\geq 0}$. Finally, by multiplying \eqref{eq:ksi_s_bar (TCST_coop_manip)} by $\rho_k(t), k\in\mathcal{K}$, we obtain
		\begin{equation}
		-\rho_{s_k}(t) < -\bar{\xi}_s\rho_{s_k}(t) \leq e_{s_k}(t) \leq \bar{\xi}_s\rho_{s_k}(t) < \rho_{s_k}(t), \label{eq:ppc_final (TCST_coop_manip)}
		\end{equation}
		$\forall t\in\mathbb{R}_{\geq 0}$, which leads to the conclusion of the proof. }
\end{proof}

\begin{remark} [\textbf{Prescribed Performance}]
	From the aforementioned proof it can be deduced that the Prescribed Performance Control scheme achieves its goal without resorting to the need of rendering the ultimate bounds $\bar{\varepsilon}_s,\bar{\varepsilon}_v$ of the modulated pose and velocity errors $\varepsilon_s, \varepsilon_v$ arbitrarily small by adopting extreme values of the control gains $g_s$ and $g_v$ (see \eqref{eq:epsilon_s_bar (TCST_coop_manip)} and \eqref{eq:epsilon_v_bar (TCST_coop_manip)}). More specifically, notice that \eqref{eq:ksi_s_bar (TCST_coop_manip)} and \eqref{eq:ksi_v_bar (TCST_coop_manip)} hold no matter how large the finite bounds $\bar{\varepsilon}_s, \bar{\varepsilon}_v$ are. In the same spirit, large uncertainties involved in the coupled model \eqref{eq:coupled dynamics (TCST_coop_manip)} can be compensated, as they affect only the size of $\varepsilon_v$ through $\bar{B}_v$, but leave unaltered the achieved stability properties. Hence, the actual performance given in \eqref{eq:ppc_final (TCST_coop_manip)}, which is solely determined by the designed-specified performance functions $\rho_{s_k}(t), \rho_{v_k}(t), k\in\mathcal{K}$, becomes isolated against model uncertainties, thus extending greatly the robustness of the proposed control scheme.
\end{remark}

\begin{remark}[\textbf{Control Input Bounds}]
	The aforementioned analysis of the Prescribed Performance Control methodology reveals the derivation of bounds for the velocity $v_i$ and control input $u_i$ of each agent. Note the explicit bounds $\bar{v}_i$  and $\bar{u}_i$ for $v_i$ and $u_i$ (see \eqref{eq:v_i_bar (TCST_coop_manip)}, \eqref{eq:u_i_bar (TCST_coop_manip)}), respectively, which depend on the control gains, the bounds of the dynamic terms, the desired trajectory, and the performance functions.
	Therefore, given desired bounds for the agents' velocity $\bar{v}_{i,b}$ and input $\bar{u}_{i,b}$ (derived from bounds on the joint velocities and torques $\dot{q}_i$, $\tau_i$, respectively) and that the upper bounds of the dynamic terms are known, we can tune appropriately the control gain $g_s$, $g_v$ as well as the parameters $\rho_{s_k,0}, \rho_{v_k,0}, \rho_{s_k,\infty}, \rho_{v_k,\infty}, l_{s_k}, l_{v_k}$ in order to achieve $\bar{v}_i \leq \bar{v}_{i,b}, \bar{u}_i \leq \bar{u}_{i,b}, \forall i\in\mathcal{N}$. 
	It is also worth noting that the selection of the control gains $g_s, g_v$ affects the evolution of the errors $e_s, e_v$ inside the corresponding performance envelopes.   
\end{remark}

{\begin{remark} [\textbf{Internal forces}]
		The internal forces were proven, in \cite{erhart2015internal}, to be regulated to zero using the distribution matrix $J_{M_i}$ from \eqref{eq:J_Hirche (TCST_coop_manip)}. That result, however, did not take into account the actual dynamic parameters of the robots. In the next chapter we analyze the internal forces in rigid cooperative manipulation and provide conditions that achieve their regulation to zero.		
	\end{remark}}

\subsubsection{Simulation Results}
We provide here simulation results for the developed control scheme.

\begin{figure*}
	\centering
	\includegraphics[trim = 3.5cm 0cm 0cm 0cm, scale=0.28]{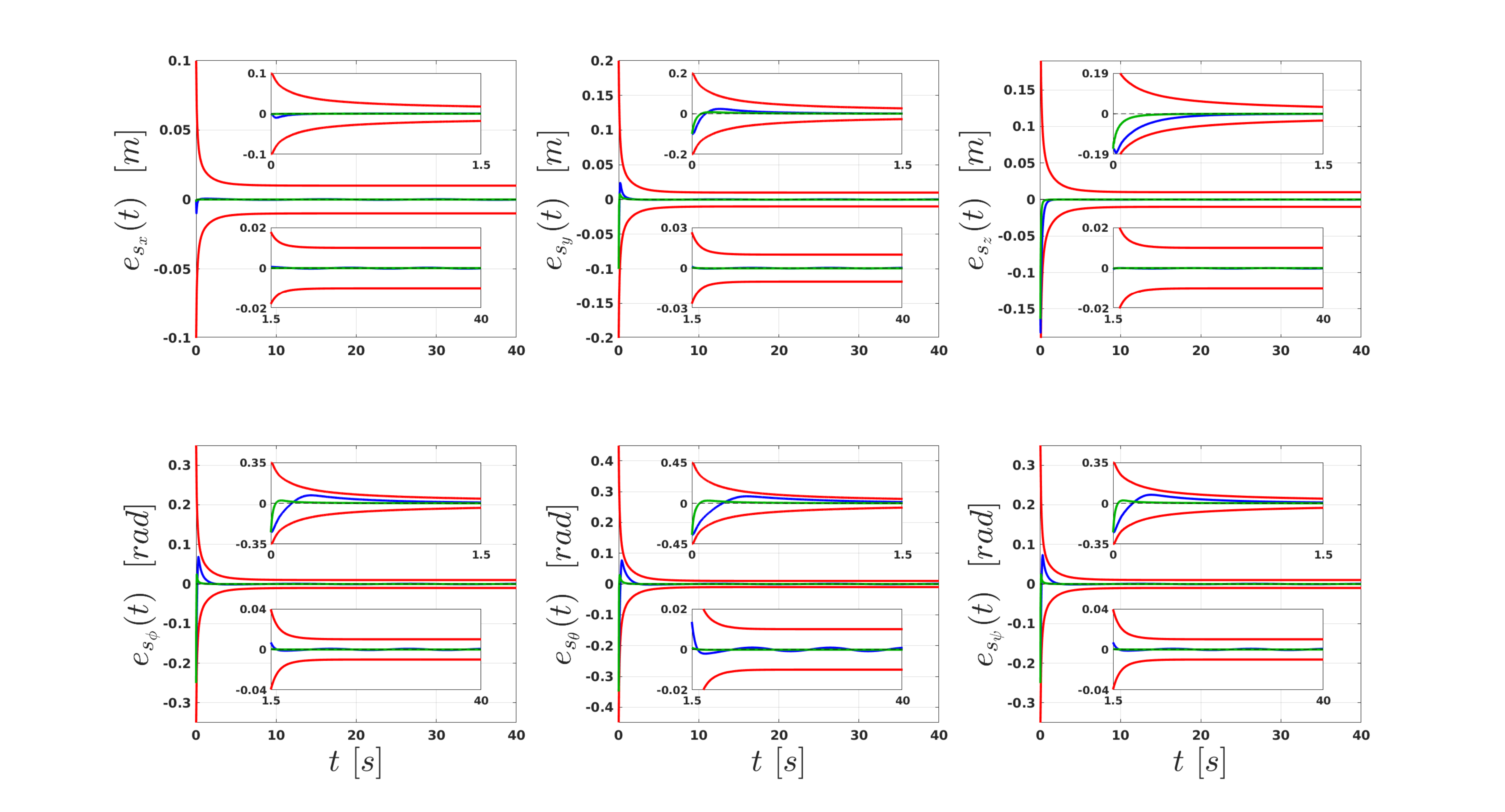} 
	\caption{Simulation results for the controller of Section \ref{subsec:PPC Controller (TCST_coop_manip)}, with (in blue) and without (in green) taking into account input constraints; Top: The position errors $e_{s_x}(t)$, $e_{s_y}(t)$, $e_{s_z}(t)$ (with blue {and green, respectively}) along with the respective performance functions (with red); Bottom: The orientation errors $e_{s_\phi}(t)$, $e_{s_\theta}(t)$, $e_{s_\psi}(t)$ (with blue {and green, respectively}) along with the respective performance functions (with red), $\forall t\in[0,40]$. Zoomed versions of the transient and steady-state response have been included for all plots.}
	\label{fig:ppc_pose_errors (TCST_coop_manip)}
\end{figure*}

\begin{figure}
	\centering
	\includegraphics[width=0.8\columnwidth]{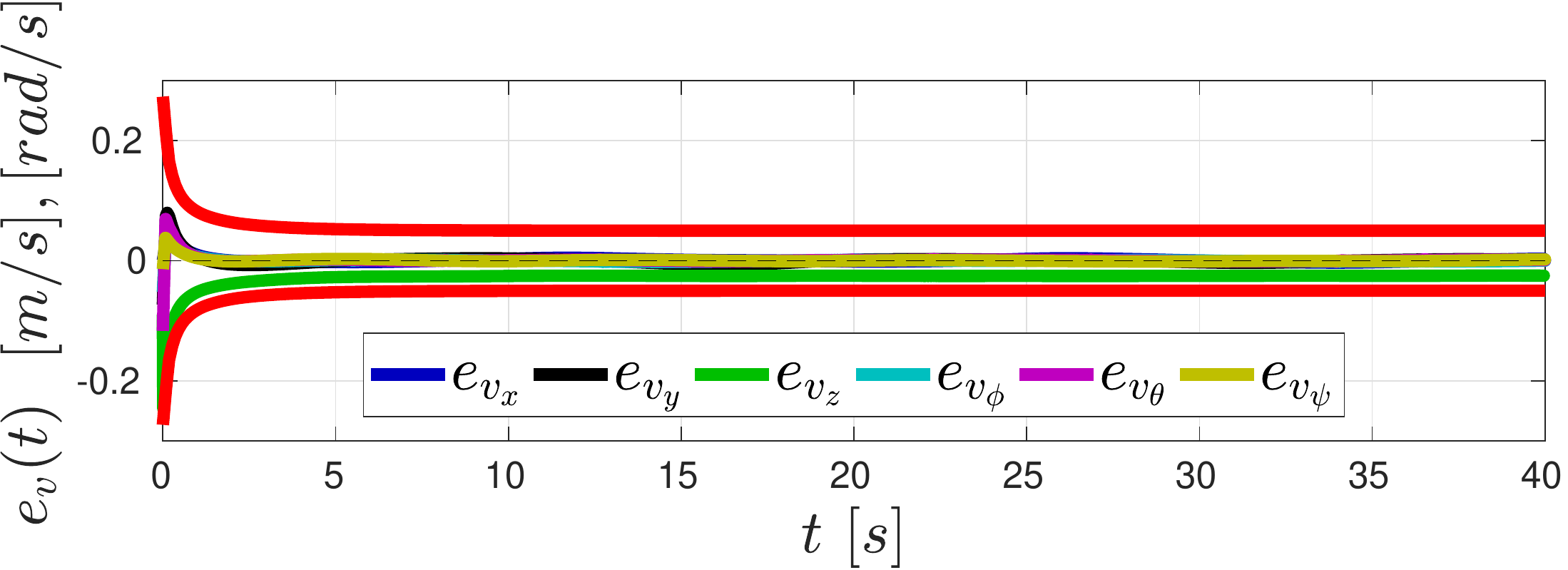} 
	\caption{The velocity errors $e_v(t)$ along with the respective performance functions (with red) for the controller of Section \ref{subsec:PPC Controller (TCST_coop_manip)}, $\forall t\in[0,40]$.}
	\label{fig:ppc_velocity_errors (TCST_coop_manip)}
\end{figure}

\begin{figure}
	\centering
	\begin{subfigure}[b]{\columnwidth}
		\centering
		\includegraphics[width=.65\columnwidth]{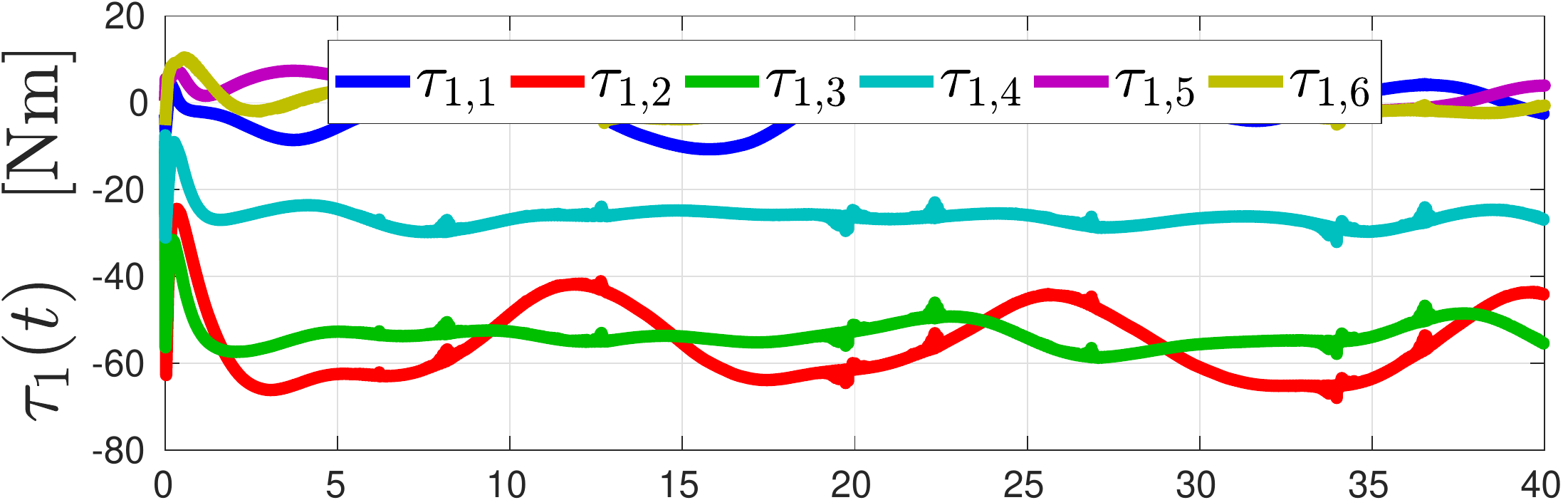}
		\caption{}
		\label{fig:Ng1 (TCST_coop_manip)} 
	\end{subfigure}
	
	\begin{subfigure}[b]{\columnwidth}
		\centering
		\includegraphics[width=.65\columnwidth]{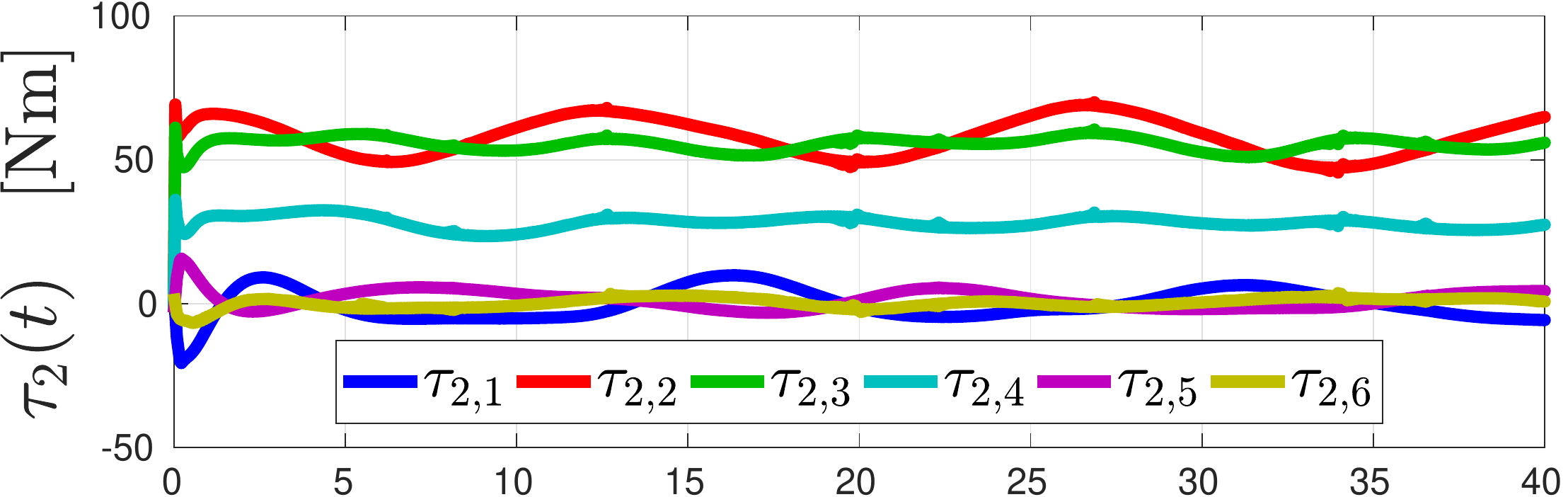}
		\caption{}
		\label{fig:Ng2 (TCST_coop_manip)}
	\end{subfigure}
	
	\begin{subfigure}[b]{\columnwidth}
		\centering
		\includegraphics[width=.65\columnwidth]{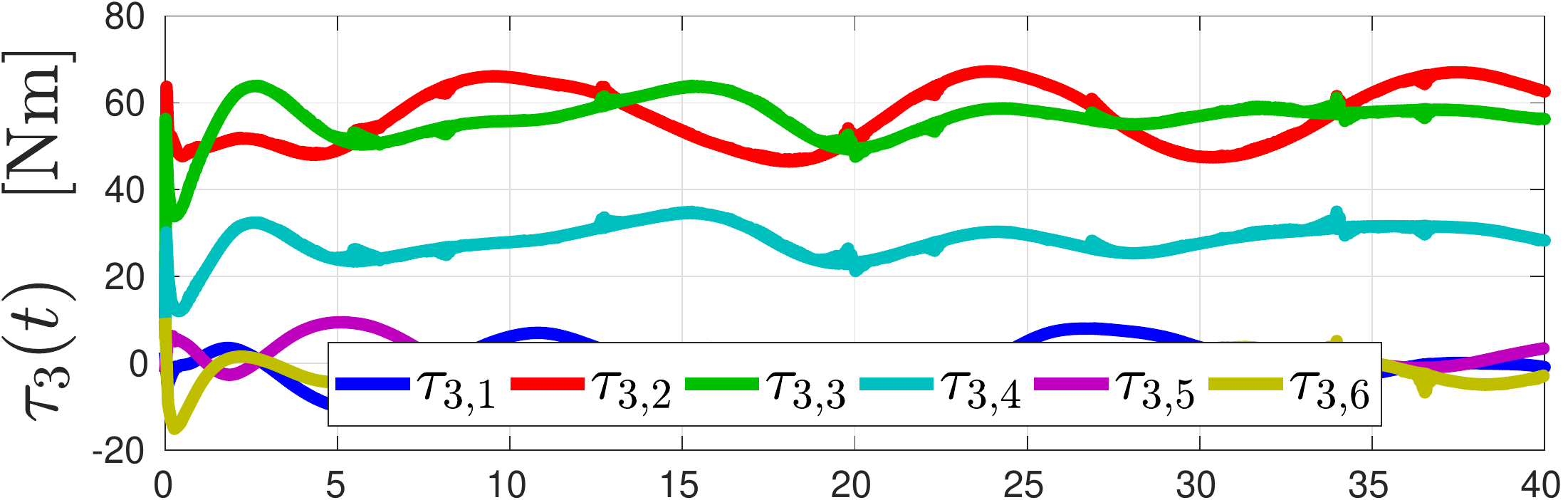}
		\caption{}
		\label{fig:Ng2 (TCST_coop_manip)}
	\end{subfigure}
	
	\begin{subfigure}[b]{\columnwidth}
		\centering
		\includegraphics[width=.65\columnwidth]{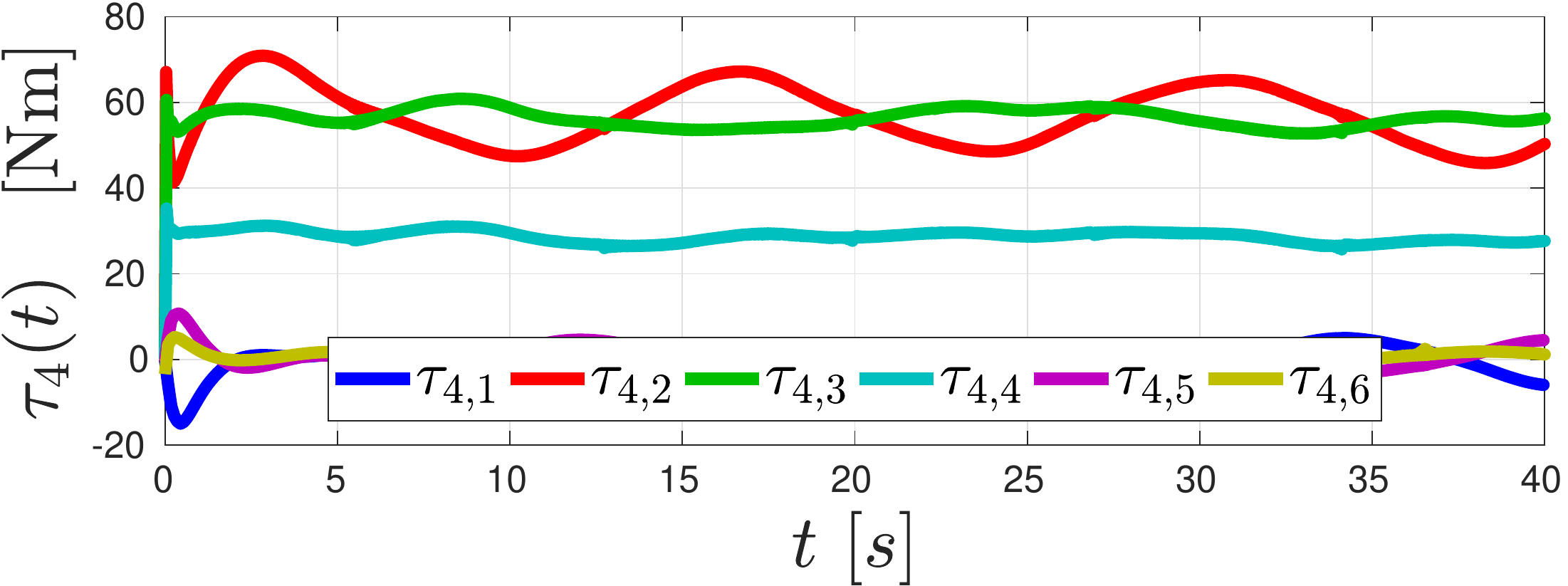}
		\caption{}
		\label{fig:Ng2 (TCST_coop_manip)}
	\end{subfigure}
	
	\caption{The agents' joint torques $\tau_i(t)$, $i\in\mathcal{N}$, (in (a)-(d), respectively) of the control scheme of Section \ref{subsec:PPC Controller (TCST_coop_manip)} $\forall t\in[0,40]$ by taking into account input constraints.} \label{fig:ppc_tau (TCST_coop_manip)}
\end{figure}

\begin{figure}[t]
	\centering
	\includegraphics[width=.95\columnwidth]{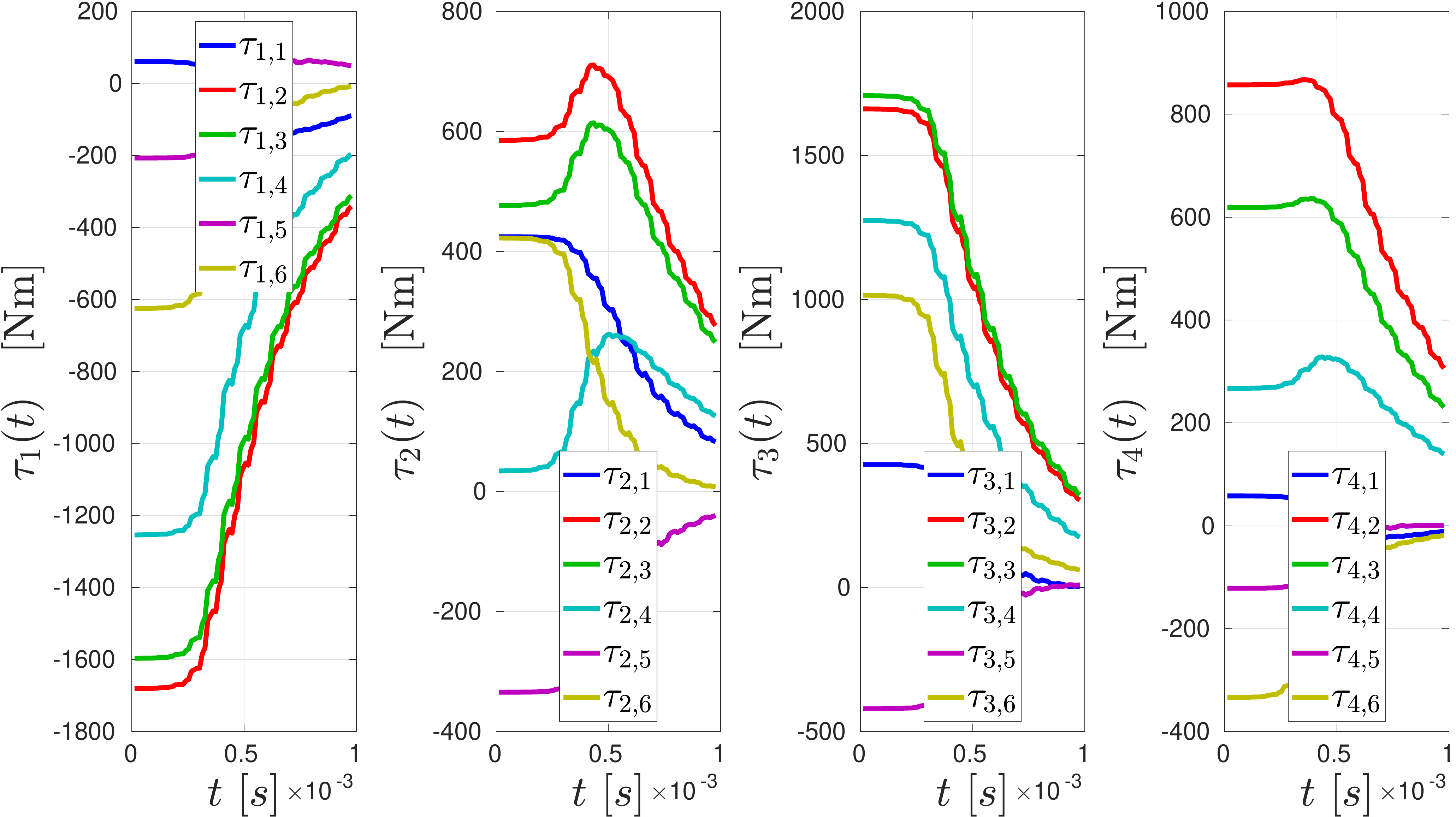} 
	\caption{The agents' joint torques $\tau_i(t)$, $i\in\mathcal{N}$, (in (a)-(d), respectively) of the control scheme of Section \ref{subsec:PPC Controller (TCST_coop_manip)} $\forall t\in[0,40]$ without taking into account input constraints, $\forall t\in[0,0.001]$.}
	\label{fig:ppc_tau_comp (TCST_coop_manip)}
\end{figure}

{The tested scenario is identical to the one used for the adaptive control scheme of Section \ref{subsec:Quaternion Controller (TCST_coop_manip)}, with the modification of $A_\theta = \frac{\pi}{9}$, in order to avoid $\theta_\textup{d}(t) = \pm \frac{\pi}{2}$.
We set the performance functions as $\rho_{s_k}(t)= (|e_{s_k}(0)| + 0.09)\exp(-0.5t)+0.01$, $\rho_{v_k}(t) =(|e_{v_k}(0)| + 0.95)\exp(-0.5t)+0.05$, $\forall k\in\mathcal{K}$, and the control gains of \eqref{eq:v_r (TCST_coop_manip)}, \eqref{eq:control_law_ppc (TCST_coop_manip)} as $g_s = 0.005$, $g_v = 10$, respectively, by following the bounds derived in the previous section and considering known dynamic bounds. The simulation results are depicted in Figs. \ref{fig:ppc_pose_errors (TCST_coop_manip)}-\ref{fig:ppc_tau (TCST_coop_manip)}, for $t\in[0,40]$ seconds. In particular, Fig. \ref{fig:ppc_pose_errors (TCST_coop_manip)} depicts the evolution of the pose errors $e_s(t)$ {(in blue)}, along with the respective performance functions $\rho_s(t)$ (in red), Fig. \ref{fig:ppc_velocity_errors (TCST_coop_manip)} depicts the evolution of the velocity errors $e_v(t)$, along with the respective performance functions $\rho_v(t)$, and Fig. \ref{fig:ppc_tau (TCST_coop_manip)} shows the resulting joint torques $\tau_i(t)$, $\forall i\in\{1,\dots,4\}$.	
One can conclude from the aforementioned figures that the simulation results verify the theoretical findings, since the errors $e_s(t)$, $e_v(t)$ stay confined in the performance function funnels. Moreover, the joint torques in respect the saturation values we set. {For comparison purposes, we also simulate the same system without taking into account any input constraints. In order to achieve good performance in terms of overshoot, rise, and settling time, we set the control gains as $g_s = 1$, $g_v = 200$. The resulting pose errors are depicted in Fig. \ref{fig:ppc_pose_errors (TCST_coop_manip)} for $t\in[0,40]$ seconds (with green) along with the performance functions (with red), and the resulting torques are depicted in Fig. \ref{fig:ppc_tau_comp (TCST_coop_manip)} for $t\in[0,0.001]$ seconds. This small time interval is sufficient to observe the high-value initial peaks of the torque inputs that do not satisfy the desired constraint of  $\bar{\tau} = 150 \ \textup{Nm}$, which can be attributed to the lack of gain calibration. Nevertheless, note also the better performance of the pose errors, in terms of overshoot, rise and settling time, as pictured in Fig. \ref{fig:ppc_pose_errors (TCST_coop_manip)}. Finally, note that any Prescribed Performance Control methodology would fail to solve Problem \ref{prob:problem1 (TCST_coop_manip)} with $\theta(0) = \frac{\pi}{2}$ or $\theta_d(t) = \frac{\pi}{2}$ for some $t\in\mathbb{R}_{\geq 0}$, in contrast to the adaptive quaternion-feedback control scheme of Section \ref{subsec:Quaternion Controller (TCST_coop_manip)}.} 
The simulations were carried out in the MATLAB R2017a environment on a $i7$-$5600$ laptop computer at $2.6$Hz, with $8$GB of RAM.}

\subsubsection{Experimental Results} \label{subsec:exps (TCST_coop_manip)}

\begin{figure*}
	\centering
	\includegraphics[scale = 0.53]{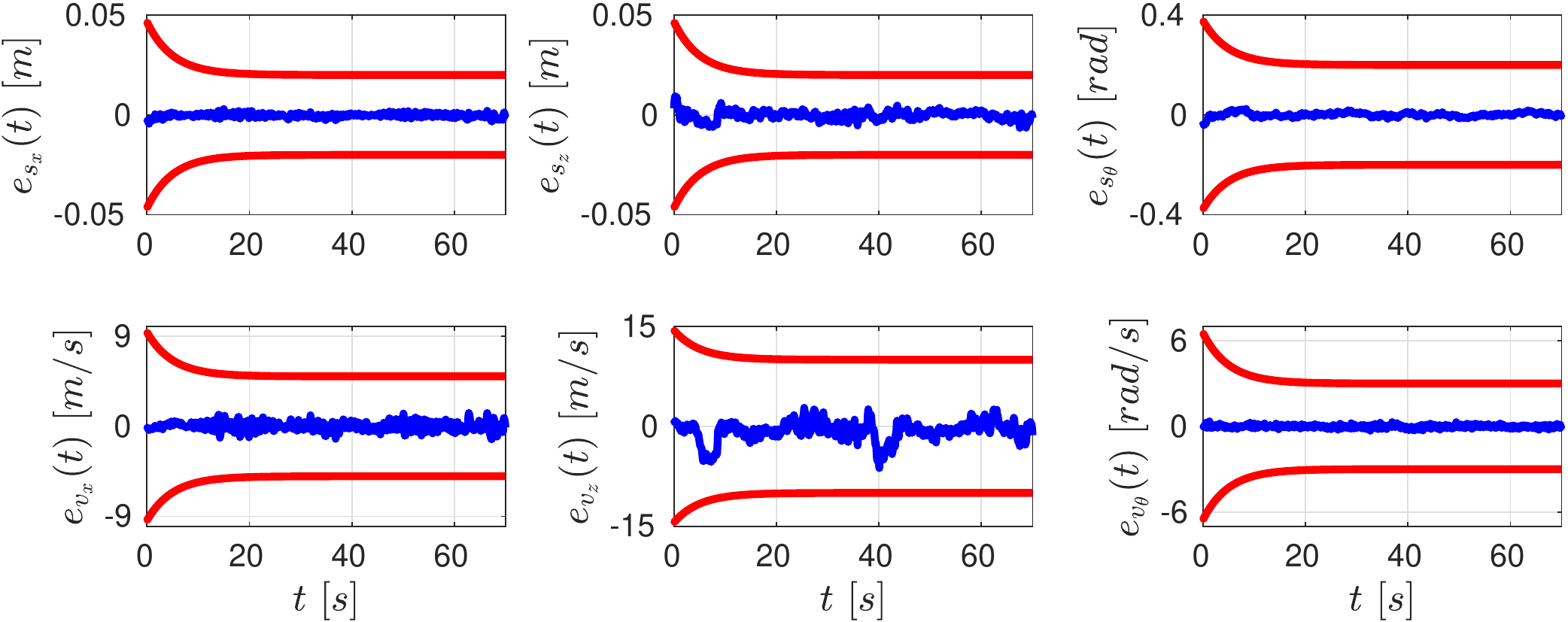} 
	\caption{Experimental results for the controller of Section \ref{subsec:PPC Controller (TCST_coop_manip)}; Top: the pose errors $e_{s_x}(t)$, $e_{s_z}(t)$, $e_{s_\theta}(t)$ (with blue) along with the respective performance functions (with red); Bottom: The velocity errors $e_{v_x}(t)$, $e_{v_z}(t)$, $e_{v_\theta}(t)$ (with blue) along with the respective performance functions (with red), $\forall t\in[0,70]$.}
	\label{fig:ppc_exp_theta_errors (TCST_coop_manip)}
\end{figure*}

\begin{figure}
	\centering
	\includegraphics[width=.65\columnwidth]{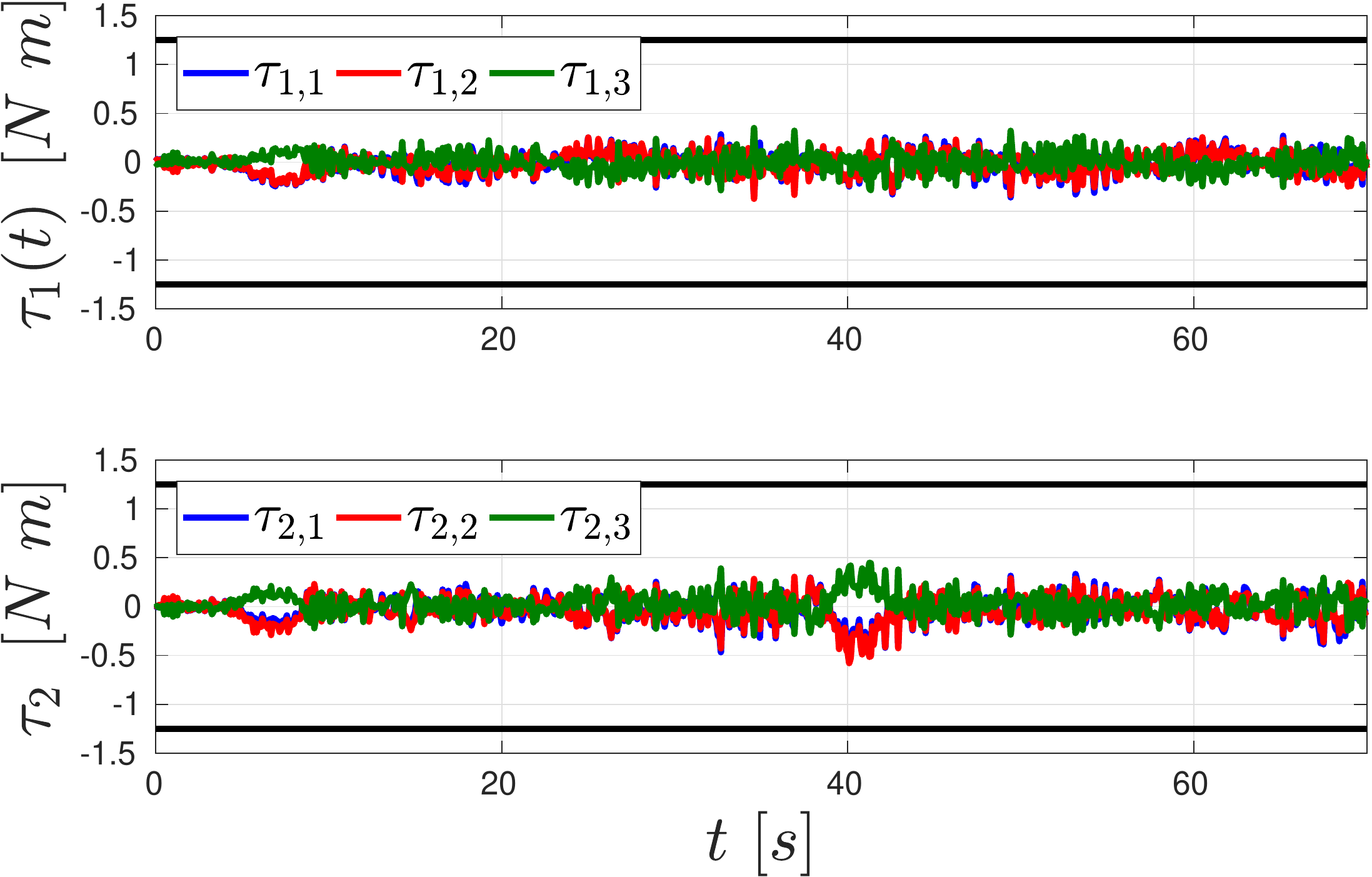} 
	\caption{The agents' joint torques of the experiment of the controller in Section \ref{subsec:PPC Controller (TCST_coop_manip)}, $\forall t\in[0,70]$, with their respective limits (with black).}
	\label{fig:ppc_exp_theta_inputs (TCST_coop_manip)}
\end{figure}

We provide here experimental results for the developed Prescribed Performance Control scheme. The scenario here is identical to the one used for Section \ref{subsec:Quaternion Controller (TCST_coop_manip)}.
We set the performance functions as  $\rho_{s_{x}}(t) = \rho_{s_{z}}(t) = 0.03\exp(-0.2t) + 0.02$ [m], $\rho_{s_{\theta}}(t) = 0.2\exp(-0.2t) + 0.2$ [rad], $\rho_{v_x}(t) = 5\exp(-0.2t) + 5$ [m/s], $\rho_{v_z}(t) = 5\exp(-0.2t) + 10$ [m/s], and $\rho_{v_{\theta}}(t) = 4\exp(-0.2t) + 3$ [m/s], and the control gains of \eqref{eq:v_r (TCST_coop_manip)} and \eqref{eq:control_law_ppc (TCST_coop_manip)} as $g_s = 0.05$ and $g_v=10$, respectively. The experimental results are depicted in Fig. \ref{fig:ppc_exp_theta_errors (TCST_coop_manip)}-\ref{fig:ppc_exp_theta_inputs (TCST_coop_manip)} for $t\in[0,70]$ seconds. In particular, Fig. \ref{fig:ppc_exp_theta_errors (TCST_coop_manip)} shows the pose and velocity errors $e_s(t)$, $e_v(t)$ along with the respective performance functions, and Fig. \ref{fig:ppc_exp_theta_inputs (TCST_coop_manip)} depicts the joint torques $\tau_1(t)$, $\tau_2(t)$ of the agents.
We can conclude that the experimental results verify the theoretical analysis, since the errors evolve strictly within the prespecified performance bounds. Note also that the joint torques respect the saturation limits. {A video illustrating the simulation and experimental results (along with the ones of the previous section's control scheme) can be found on \href{https://youtu.be/jJWeI5ZvQPY}{https://youtu.be/jJWeI5ZvQPY}.}

\subsection{Discussion} \label{subsec:Dissusion (TCST_coop_manip)}
{ In view of the aforementioned results, we mention some worth-noting differences between the two control schemes. Firstly, note that the PPC methodology allows for \textit{exponential} convergence of the errors to the set defined by the values $\rho_{s_k,\infty}$, $\rho_{v_k,\infty}$, achieving \textit{predefined} transient and steady-state performance, without the need to resort to tuning of the control gains. The adaptive quaternion-feedback methodology, however, can only guarantee that the errors converge asymptotically to zero as $t\to\infty$. This is verified by the simulation results, where the error trajectories $e_p(t), e_\zeta(t)$ and $e_v(t)$ show an oscillatory behavior. Improvement of such performance (in terms of overshoot, rise, and settling time) would require appropriate gain tuning. 
	Secondly, note that, as shown in the simulations section, the quaternion-feedback methodology allows for trajectories where the pitch angle of the object $(\theta_{\scriptscriptstyle O})$ can be $\pm 90$ degrees, in contrast to the PPC methodology, where that configuration is ill-posed, since the matrix $J_{\scriptscriptstyle O}(\eta_{\scriptscriptstyle O})$ is not defined. Finally, the adaptive quaternion-feedback methodology can be considered less robust to modeling uncertainties in real-time scenarios, since it accounts only for parametric uncertaintes (the unknown terms $\vartheta_i$, $\vartheta_{\scriptscriptstyle O}$, $d_i$, $d_{\scriptscriptstyle O}$), assuming a \textit{known} structure of the dynamic terms. The PPC methodology, however, does not require any information of the structure or the parameters of the dynamic model (note that the only requirements are the positive definiteness of the coupled inertia matrix, the locally Lipschitz and continuity properties of the dynamic terms and the boundedness - with respect to time -  of the disturbances $d_i, d_{\scriptscriptstyle O}$). In that sense, one would expect the PPC methodology to perform better in real-time experiments, where unmodeled dynamics are involved. The fact, however, that PPC is a control scheme that does not contain any information of the model structure makes it more difficult to tune (in terms of gain tuning) in order to achieve robot velocities and torques that respect specific bounds, especially when the bounds  of the dynamic terms are unknown. This has been noticed during both simulations and experiments.}

\subsection{Problem Statement - Constrained Transportation}

We deal here with a slightly different problem, that is, the problem of cooperatively transporting an object to a desired \textit{constant} pose, while complying to certain constraints. Such constraints consist of inter-robot collision avoidance, collision avoidance with obstacles, singularity avoidance, as well as robot velocity and torque saturation constraints.

Consider $Z \in \mathbb{N}$ obstacles $\mathcal{O}_z \subset \mathbb{R}^3$, $z \in\mathcal{Z} \coloneqq \{1,\dots,Z\}$ and denote by $\mathcal{A}_i(q_i) \subset \mathbb{R}^3$, $i\in\mathcal{N}$, $\mathcal{A}_{\scr O}(x_{\scr O}) \subset \mathbb{R}^3$ the physical volumes occupied by agent $i$, at state $q_i$, $i\in\mathcal{N}$, and the object, at state $x_{\scr O}$, respectively. 

\begin{remark}
	As mentioned before, since the geometric object parameters $p^{\scriptscriptstyle E_i}_{\scriptscriptstyle E_i/O}$ and $\eta_{\scriptscriptstyle E_i/O}$ are known, each agent can compute $p_{\scriptscriptstyle O},\eta_{\scriptscriptstyle O}$ and $v_{\scriptscriptstyle O}$ from the coupled kinematics and dynamics, respectively, without employing any sensory data. In the same vein, all agents can also compute the object's bounding ellipsoid $\mathcal{C}_{\scriptscriptstyle O}$, which depends on $q$.
\end{remark} 

We can now formulate the problem considered here:
\begin{problem} \label{problem: main problen (MED)}
	Consider $N$ robotic agents rigidly grasping an object, governed by the coupled dynamics \eqref{eq:coupled dynamics (TCST_coop_manip)}. Given a desired \textit{constant} pose $x_{\textup{d}} \coloneqq [({p}_{\textup{d}})^\top, ({\eta}_{\textup{d}})^\top]^\top$, ${p}_\textup{d}\in\mathbb{R}^3,
	{\eta}_\textup{d} \coloneqq [\varphi_\textup{d},\theta_\textup{d}, \psi_\textup{d}]\in\mathbb{T}$, with $\theta_\textup{d} \in [-\bar{\theta},\bar{\theta}] \subset \left(-\frac{\pi}{2},\frac{\pi}{2}\right)$, design the control input $u\in\mathbb{R}^{6N}$ such that $\lim\limits_{t\to\infty}x_{\scriptscriptstyle O}(t) = x_\textup{d}$, while ensuring the satisfaction of the following collision avoidance and singularity properties:
	\begin{enumerate}
		\item $\mathcal{A}_i(q_i(t))\cap\mathcal{O}_z = \emptyset, \forall i\in\mathcal{N}, z\in\mathcal{Z}$,
		\item $\mathcal{A}_{\scriptscriptstyle O}(x_{\scriptscriptstyle O}(t))\cap\mathcal{O}_z = \emptyset, \forall z \in \mathcal{Z}$,
		\item $\mathcal{A}_i(q_i(t))\cap\mathcal{A}_{j}(q_{j}(t)) = \emptyset, \forall i, j \in \mathcal{N}, i\neq j$,
		\item $-\tfrac{\pi}{2} < -\bar{\theta} \le  \theta_{\scriptscriptstyle O}(t) \le -\bar{\theta} < \tfrac{\pi}{2}$,
		\item $q_i(t) \in {\mathsf{S}}_i$,
	\end{enumerate} 
	$\forall t\in\mathbb{R}_{\geq 0}$, as well as the input and velocity magnitude constraints: $\lvert \tau_{i,k} \rvert \leq \bar{\tau}_i, \lvert \dot{q}_{i_k} \rvert \leq \bar{\dot{q}}_i, \forall k\in\{1,\dots,n_i\}, i\in\mathcal{N}$, for some positive constants $\bar{\tau}_i, \bar{\dot{q}}_i, i\in\mathcal{N}$.%
\end{problem}

In order to solve the aforementioned problem, we need the following reasonable assumption regarding the workspace, which implies that the collision-free space is connected:
\begin{assumption} \label{ass:feasility_assumption (MED+ECC)}
	(Problem feasibility) The set $\{(q,x_{\scr O}) \in \mathbb{R}^n\times\mathbb{M}: \mathcal{A}_i(q_i)\cap\mathcal{O}_z = \emptyset, \mathcal{A}_i(q_i)\cap\mathcal{A}_j(q_j) = \emptyset, \mathcal{A}_{\scr O}(x_{\scriptscriptstyle O})\cap\mathcal{O}_z = \emptyset, \forall i,j\in\mathcal{N}, i\neq j, z\in\mathcal{Z}\}$, is connected.	
\end{assumption}

 We also define the following sets:
\begin{align}
\mathsf{S}_{i, {\scriptscriptstyle O}} &\coloneqq \{q_i\in\mathbb{R}^{n_i} : \mathcal{A}_i(q_i)\cap\mathcal{O}_z = \emptyset, \forall z \in \mathcal{Z} \}, \ \ \forall i \in \mathcal{N} \notag \\
\mathsf{S}_{\scriptscriptstyle A} &\coloneqq \{q\in\mathbb{R}^{n} : \mathcal{A}_i(q_i)\cap\mathcal{A}_{j}(q_{j}) = \emptyset, \forall i,j \in \mathcal{N},i\neq j\}, \notag \\
\mathsf{S}_{\scriptscriptstyle O} &\coloneqq \{x_{\scriptscriptstyle O} \in \mathbb{M}: \mathcal{A}_{\scriptscriptstyle O}(x_{\scriptscriptstyle O}) \cap \mathcal{O}_z = \emptyset  \}. \notag
\end{align}
associated with the desired collision-avoidance properties.


We present next two control schemes, based on Nonlinear Model Predictive Control (NMPC), for the solution of Problem \ref{problem: main problen (MED)}. The first one is a \textit{centralized scheme}, where a central computer unit (e.g., on one of the robotic agents) has global feedback and computes the control input of the entire team. Secondly, we develop a \textit{decentralized} scheme, where each robotic agent computes its own control signal. The latter
is based on a leader-follower coordination as well as inter-agent communication.

We also assume that $d_i(\cdot) = d_{\scr O}(\cdot) = 0$, $\forall i\in\mathcal{N}$ in the dynamics \eqref{eq:manipulator dynamics (TCST_coop_manip)}, \eqref{eq:object dynamics (TCST_coop_manip)}, and that the system model is \textit{accurately known}. Potential uncertainties could be taken into account by using robust variations of NMPC, like, e.g., tube-based NMPC \cite{nikou2019decentralized}.

\subsection{Centralized NMPC} \label{sec:MPC (MED)}

In this section, a centralized systematic solution to Problem \ref{problem: main problen (MED)} is introduced. Our overall approach builds on designing a  Nonlinear Model Predictive control scheme for the system of the manipulators and the object. Nonlinear Model Predictive Control (see e.g. \cite{morrari_npmpc, frank_2003_nmpc_bible, frank_1998_quasi_infinite, frank_2003_towards_sampled-data-nmpc, fontes_2001_nmpc_stability, grune_2011_nonlinear_mpc, camacho_2007_nmpc, cannon_2001_nmpc, borrelli_2013_nmpc}) has been proven suitable for dealing with nonlinearities and state and input constraints.

The coupled agents-object \emph{nonlinear dynamics} can be written in compact form as follows:

\begin{equation} \label{eq:main_system (MED)}
\dot{x}_c = f_c(x_c,u) \coloneqq \begin{bmatrix}
f_{c_1}(x_c,u) \\
f_{c_2}(x_c,u) \\
f_{c_3}(x_c,u) 
\end{bmatrix}, x_{c0} \coloneqq x_c(0),
\end{equation}
where $x_c \coloneqq [x_{\scriptscriptstyle O}^\top, v_{\scriptscriptstyle O}^\top, q^\top]^\top \in \mathbb{M} \times \mathbb{R}^{n+6}, u \in \mathbb{R}^{6N}$ and $f_c : \mathbb{M}\times\mathsf{S}\times\mathbb{R}^{6N+6} \to \mathbb{R}^{n+12}$, with 
\begin{align}
f_{c_1}(x_c,u) &\coloneqq J_{\scriptscriptstyle O}(\eta_{\scriptscriptstyle O})v_{\scriptscriptstyle O}, \notag \\
f_{c_2}(x_c,u) &\coloneqq \widetilde{M}(x)^{-1}\left[G(q)u - \widetilde{C}(x)v_{\scriptscriptstyle O} - \widetilde{g}(x) \right],  \notag \\
f_{c_3}(x_c,u) &\coloneqq \widetilde{J}(q) G(q)^\top v_{\scriptscriptstyle O}, \notag 
\end{align}
where we have used the first equation of \eqref{eq:sigma_dot_1 (TCST_coop_manip)}.
Note that $f_c$ is \emph{locally Lipschitz continuous} in its domain since it is continuously differentiable there. Next, we define the respective errors:
\begin{align} \label{eq:error (MED)}
e_c &\coloneqq x_c - x_\textup{d} =
\begin{bmatrix}
x_{\scriptscriptstyle O} \\ 
v_{\scriptscriptstyle O} \\
q
\end{bmatrix}
-
\begin{bmatrix}
x_\textup{d} \\
\dot{x}_\textup{d} \\
q_\textup{d} 
\end{bmatrix} =
\begin{bmatrix}
x_{\scriptscriptstyle O}-x_\textup{d} \\ 
v_{\scriptscriptstyle O} \\
q - q_\textup{d} 
\end{bmatrix} \in \mathbb{M}\times\mathbb{R}^{6}\times\mathsf{S}, 
\end{align}
where $q_\textup{d}\coloneqq[q^\top_{1,\textup{d}},\dots,q^\top_{N,\textup{d}}]^\top \in \mathbb{R}^n$ is appropriately chosen to comply with the coupled kinematics \eqref{eq:coupled_kinematics (TCST_coop_manip)} and $x_\textup{d}$.
The error dynamics are then $\dot{e}_c(t) = f_c(x_c(t),u(t))$, which can be appropriately transformed to:
\begin{equation}
\dot{e}_c = f_e(e_c,u), \ \ e_{c0} \coloneqq e_c(0) =  x_c(0)-x_{\textup{d}}. \label{eq:error_dynamics (MED)}
\end{equation}
where $f_e \coloneqq f_c(e_c+x_{\textup{d}}, u)$. By ignoring over-actuated input terms, we have that $\tau_i = J_i^\top u_i$, which yields
\begin{align*}
\lVert \tau_i \rVert \leq \bar{\tau}_i \Leftrightarrow \sigma_{\min}(J_i^\top)\lVert u_i \rVert \leq \bar{\tau}_i,
\end{align*} 
where we have employed the property $\sigma_{\min}(J^\top_i)\lVert u_i\rVert \leq \lVert J^\top_i u_i \rVert$, with $\sigma_{\min}(J^\top_i)$ being  positive, if the constraint $q_i\in{\mathsf{S}}_i$ is always satisfied. Hence, the constraint $\lvert \tau_{i,k} \rvert \leq \bar{\tau}_i$ is equivalent to 
\begin{equation*}
\lVert u_i \rVert \leq \frac{\bar{\tau}_i}{\sigma_{\min}(J^\top_i)}, \forall i\in\mathcal{N}.
\end{equation*}

Let us now define the following compact set $U_c \subseteq \mathbb{R}^{6N}$:
\begin{equation} \label{eq:input constraint set (MED)}
U_c \coloneqq \left\{ u\in\mathbb{R}^{6N} :  \lVert u_i \rVert \leq \frac{\bar{\tau}_i}{\sigma_{\min}(J^\top_i)}, \forall i\in\mathcal{N}, k\in\{1,\dots,n_i\}\right\},
\end{equation}
as the set that captures the control input constraints of the error dynamics system \eqref{eq:error_dynamics (MED)}. By using \eqref{eq:main_system (MED)} to express $\dot{q}$ as a function of $v_{\scr O}$, 
we define also the set $X_c \subseteq \mathbb{R}^{n+12}$:
\begin{align*}
X_c \coloneqq \Big\{ &x_c \in \mathbb{R}^{n+12} : \theta_{\scriptscriptstyle O} \in [\bar{\theta}, \bar{\theta}], \| J_i(q_i)^\top(J_i(q_i)J_i(q_i)^\top)^{-1}J_{\scr O_i}(q_i)v_{\scr O}\| \leq \bar{\dot{q}}_i, \\ 
&i\in\mathcal{N},  q \in{\mathsf{S}} \cap \mathsf{S}_{\scriptscriptstyle A} \cap (\mathsf{S}_{1, {\scriptscriptstyle O}}\times\dots\times \mathsf{S}_{N, {\scriptscriptstyle O}}), x_{\scriptscriptstyle O}\in \mathbb{M} \cap \mathsf{S}_{\scriptscriptstyle O}(x_{\scriptscriptstyle O}) \Big\}.
\end{align*}
The set $X_c$ captures all the state constraint of the system dynamics \eqref{eq:main_system (MED)}. In view of \eqref{eq:error (MED)}, we define the set $E_c \subseteq \mathbb{R}^{n+12}$ as:
\begin{equation*}
E_c \coloneqq \{e_c \in \mathbb{R}^{n+12}: e_c \in X_c \oplus (-x_\textup{d}) \},
\end{equation*}
as the set that captures all the constraints of the error dynamics system \eqref{eq:error_dynamics (MED)}.

The problem in hand is the design of a control input $u(t)\in U_c$ such that $\lim_{t \to \infty} e_c(t) = 0$ while ensuring  $e_c(t) \in E_c, \forall t\in\mathbb{R}_{\geq 0}$.
The proposed Nonlinear Model Predictive scheme is presented hereafter.

Consider a sequence of sampling times $\{t_j\}$, $j \in \mathbb{N}$, with a constant sampling period $h_s \in (0,T_p)$, where $T_p$ is the prediction horizon, such that: 
\begin{equation*} 
t_{j+1} = t_j + h_s, \forall \ j \ge 0.
\end{equation*}
In the sampling-data NMPC, a finite-horizon open-loop optimal control problem (OCP) is solved at discrete sampling time instants $t_j$ based on the current state error information $e_c(t_j)$. The solution is an optimal control signal $\hat{u}(s)$, for $s \in [t_j,t_j+T_p]$. For more details, the reader is referred to \cite{frank_2003_nmpc_bible}. The open-loop input signal applied in between the sampling instants is given by the solution of the following Optimal Control Problem (OCP):
\begin{subequations}
	\begin{align}
	&\min\limits_{\hat{u}(\cdot)} J_c(e_c(t_j),\hat{u}(\cdot)) \coloneqq \min\limits_{\hat{u}(\cdot)} \left\{  V_c(\hat{e}_c(t_j+T_p)) + \int_{t_j}^{t_j+T_p}  F_c(\hat{e}_c(s), \hat{u}(s))  ds \right\}  \label{eq:mpc_minimazation (MED)} \\
	&\hspace{-4mm}\text{subject to:} \notag \\
	&\hspace{1mm} \dot{\hat{e}}_c(s) = f_e(\hat{e}_c(s),\hat{u}(s)), \hat{e}_c(t_j) = e_c(t_j), \label{eq:diff_mpc (MED)} \\
	&\hspace{1mm} \hat{e}_c(s) \in E_c, \hat{u}(s) \in U_c, s \in [t_j,t_j+T_p], \\
	&\hspace{1mm} \hat{e}_c(t_j+T_p)\in\mathcal{E}_{c_f}, \label{eq:mpc_terminal_set (MED)}
	\end{align}
\end{subequations}
where the hat $\hat{\cdot}$ denotes the predicted variables (internal to the controller), i.e. $\hat{e}_c(\cdot)$ is the solution of \eqref{eq:diff_mpc (MED)} driven by the control input $\hat{u}(\cdot): [t_j, t_j+T_p] \to U_c$ with initial condition $e_c(t_j)$. Note that the predicted values are not necessarily the same with the actual closed-loop values (see \cite{frank_2003_nmpc_bible}). The term $F_c: E_c \times U_c \to \mathbb{R}_{\ge 0}$, is the \emph{running cost}, and is chosen as:
\begin{equation*} 
F_c(e_c,u) \coloneqq e_c^\top Q_c e_c + u^\top R_c u.
\end{equation*}
The terms $V_c: E_c \to \mathbb{R}_{ > 0}$ and $\mathcal{E}_{c_f}$ are the \emph{terminal penalty cost} and \emph{bounded terminal set}, respectively, and are used to enforce the stability of the system. The terminal cost is given by $V_c(e_c)\coloneqq e_c^\top P_c e_c$;
$Q_c\in\mathbb{R}^{(n+12)\times(n+12)}$ is chosen as a diagonal positive semi-definite matrix, and $P_c,R_c \in\mathbb{R}^{(n+12)\times(n+12)}$ as diagonal positive definite matrices.  

The solution of the OCP  \eqref{eq:mpc_minimazation (MED)}-\eqref{eq:mpc_terminal_set (MED)} starting at time $t_j$ provides an optimal control input denoted by $\hat{u}^\star(s; e_c(t_j))$, for $s \in [t_j, t_j+T_p]$. It defines the open-loop input that is applied to the system until the next sampling instant $t_{j+1}$:
\begin{equation} \label{eq:control_input_star (MED)}
u(s; e_c(t_j)) = \hat{u}^\star(s; e_c(t_j)),s \in [t_j, t_{j+1}).
\end{equation} 
The corresponding \emph{optimal value function} is given by $J_c^\star(e_c(t_j), \hat{u}^\star(\cdot; e_c(t_j)))$.
where $J_c(\cdot)$ as is given in \eqref{eq:mpc_minimazation (MED)}. The control input $ u(s; e_c(t_j))$ is a feedback, since it is recalculated at each sampling instant using the new state information. The solution of \eqref{eq:error_dynamics (MED)} starting at time $t_j$ from an initial condition $e_c(t_j)$, applying a control input $u: [t_j, t_{j+1}] \to U_c$ is denoted by $e_c(s; u(\cdot), e_c(t_j)), s \in [t_j, t_{j+1}]$.

Through the following theorem, we guarantee the stability of the system which is the solution to Problem 1 (see also Theorem \eqref{th:main theorem (App_NMPC)} in Appendix \ref{app:NMPC}).

\begin{theorem} \label{th:main theorem MED}
	Let Assumption \ref{ass:feasility_assumption (MED+ECC)} hold. Suppose also that:
	\begin{enumerate}
		\item The OCP \eqref{eq:mpc_minimazation (MED)}-\eqref{eq:mpc_terminal_set (MED)} is feasible for the initial time $t = 0$.
		\item The terminal set $\mathcal{E}_{c_f} \subseteq E_c$ is closed, with $0 \in \mathcal{E}_{c_f}$.
		\item The terminal set and terminal cost are chosen such that there exists an admissible control input (according to Def. \ref{def:admissible_control_input (App_NMPC)} of Appendix \ref{app:NMPC}) $u_{c_f}: [0, h_s] \to U_c$ such that for all $e_c(s) \in \mathcal{E}_{c_f}$ it holds that:
		\begin{enumerate}
			\item $e_c(s) \in \mathcal{E}_{c_f}, \forall \ s \in [0, h_s]$.
			\item $\displaystyle \frac{\partial V_c}{\partial{e}_c} f_e(e_c(s), u_{c_f}(s)) + F_c(e_c(s), u_{c_f}(s)) \le 0, \forall \ s \in [0, h_s].$
		\end{enumerate}
	\end{enumerate}
	Then, the closed loop system \eqref{eq:error_dynamics (MED)}, under the control input \eqref{eq:control_input_star (MED)}, converges to the origin for $t \to \infty$, i.e., $\lim_{t\to\infty} e_c(t) = 0$.
\end{theorem}

\begin{proof}
	The proof is identical to the proof of Theorem 2.1 in \cite{frank_2003_nmpc_bible}.
\end{proof}

\subsubsection{Simulation Results}

To demonstrate the efficiency of the proposed control protocol, we consider the following simulation scenario.

\begin{figure}[t!]
	\centering
	\includegraphics[width = 0.6\textwidth]{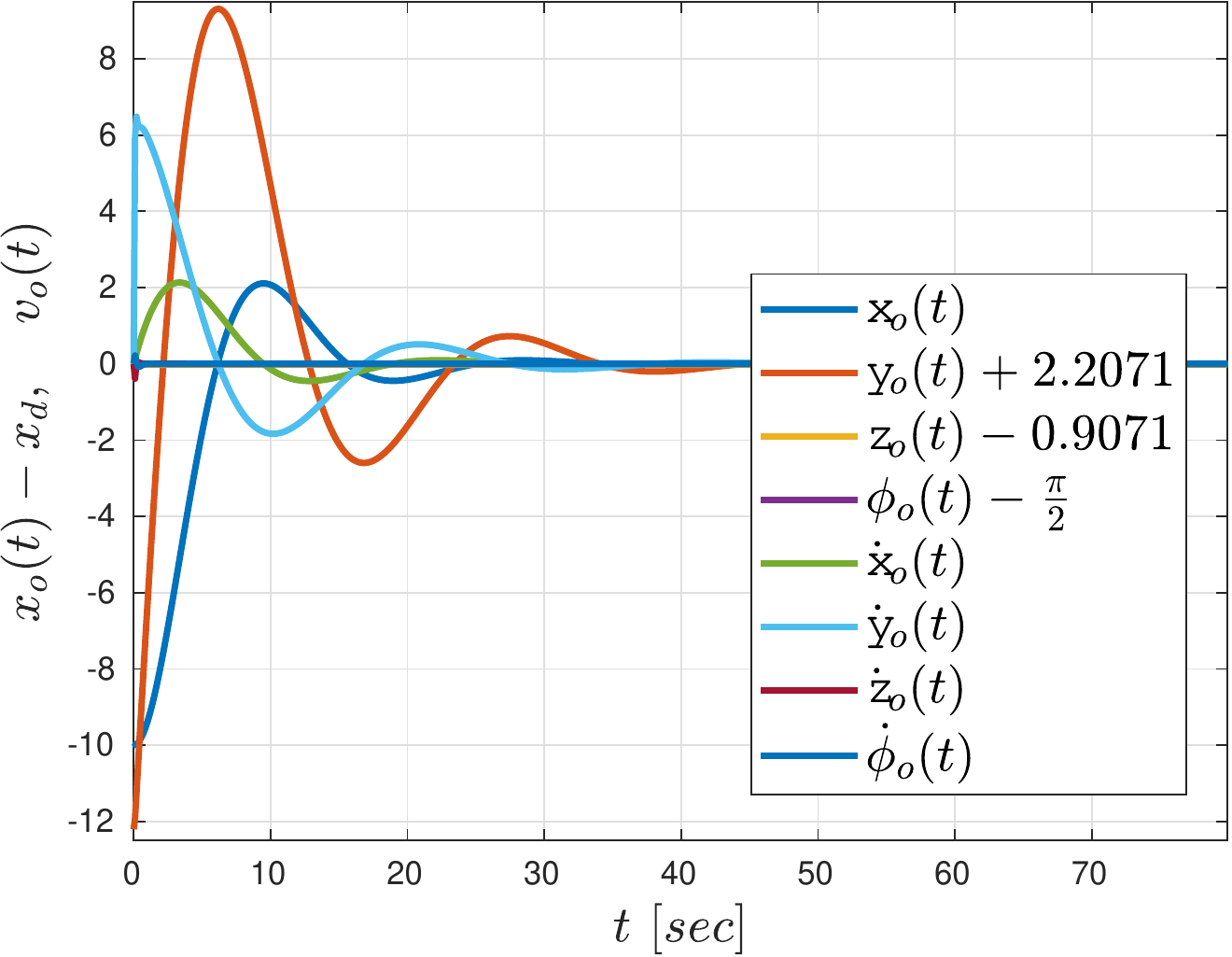}
	\caption{The errors of the object  for $t\in[0,80]$ seconds.\label{fig:sim9 (MED)}}
\end{figure}

\begin{figure}[t!]
	\centering
	\includegraphics[width = 0.6\textwidth]{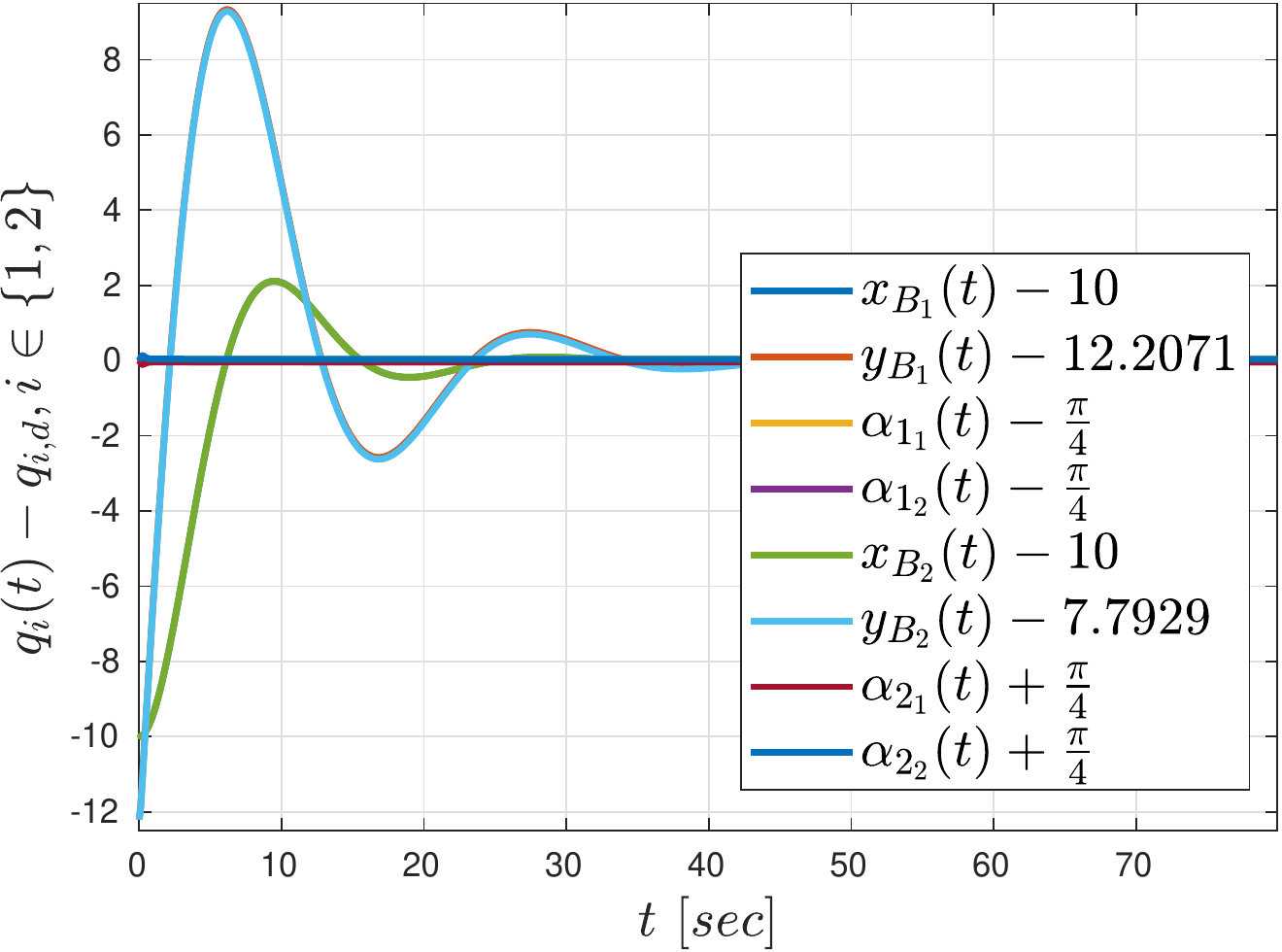}	
	\caption{The errors of robotic agents for $t\in[0,80]$ seconds.\label{fig:sim11 (MED)}}
\end{figure}

\begin{figure}[t!]
	\centering
	\includegraphics[width = 0.6\textwidth]{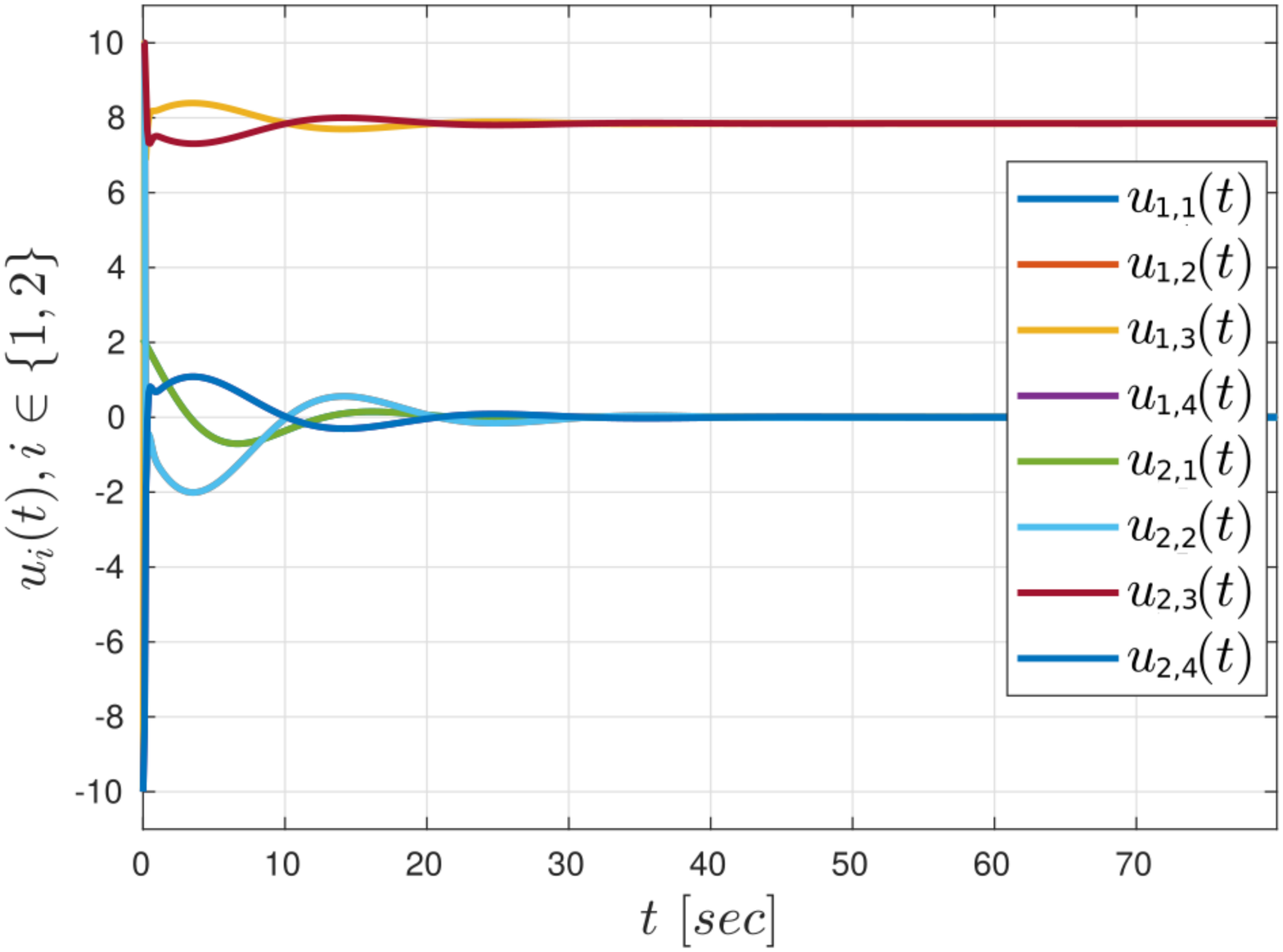}
	\caption{The control inputs of the actuators of the robotic agents $u_i(t)$, $\forall t\in[0, 80]$ seconds.\label{fig:sim16 (MED)}}
\end{figure}

Consider $N=2$ ground vehicles equipped with $2$ DOF manipulators, rigidly grasping an object with $n_1 = n_2 = 4, n = n_1+n_2 =8$. From \eqref{eq:main_system (MED)} we have that $x = [x_{\scriptscriptstyle O}^\top, v_{\scriptscriptstyle O}^\top, q^\top]^\top \in \mathbb{R}^{16}$, $u \in \mathbb{R}^{8}$, with $x_{\scriptscriptstyle O} = [p_{\scriptscriptstyle O}^\top, \phi_{\scriptscriptstyle O}]^\top \in \mathbb{R}^4$, $v_{\scriptscriptstyle O} = [\dot{p}_{\scriptscriptstyle O}^\top, \omega_{\scriptscriptstyle O}]^\top \in \mathbb{R}^4$, where $\omega_{\scr O} \in\mathbb{R}$ occurs with respect to only one axis. 
We also denote $p_{\scriptscriptstyle O} = [\mathsf{x}_{\scriptscriptstyle O}, \mathsf{y}_{\scriptscriptstyle O}, \mathsf{z}_{\scriptscriptstyle O}]^\top \in \mathbb{R}^3$, $q = [q_1^\top, q_2^\top]^\top \in \mathbb{R}^8$, $q_i = [p_{\scriptscriptstyle B_i}^\top, \alpha_i^\top]^\top \in \mathbb{R}^4$, $p_{\scriptscriptstyle B_i} = [x_{\scriptscriptstyle B_i}, y_{\scriptscriptstyle B_i}]^\top \in \mathbb{R}^2$, $\alpha_i = [\alpha_{i_1}, \alpha_{i_2}]^\top \in \mathbb{R}^2, i \in \{1,2\}$, where $p_{\scr B_i}$ are the vehicles' positions, and $\alpha_i$ the manipulator angles. The manipulators become singular when $\sin(\alpha_{i_1}) = 0, i \in \{1,2\}$, thus the state constraints for the manipulators are set to:
\begin{align}
\epsilon \leq \alpha_{1_1} \leq \frac{\pi}{2}- \epsilon, & -\frac{\pi}{2}+ \epsilon \leq \alpha_{1_2} \leq \frac{\pi}{2} - \epsilon, \notag \\
-\frac{\pi}{2} + \epsilon \leq \alpha_{2_1} \leq - \epsilon,& -\frac{\pi}{2}+\epsilon \leq \alpha_{2_2} \leq \frac{\pi}{2}-\epsilon. \notag
\end{align}
We also consider the input constraints:
\begin{equation*}
-10 \le u_{i,j}(t) \le 10, i \in \{1, 2\}, j \in \{1,\dots,4\}.
\end{equation*}
The initial conditions are set to:
\begin{align*}
x_{\scriptscriptstyle O}(0) &= \left[0, -2.2071, 0.9071, \frac{\pi}{2} \right]^\top, v_{\scriptscriptstyle O}(0) = \left[0, 0, 0, 0\right]^\top, \\
q_{1}(0) &= \left[0, 0, \frac{\pi}{4}, \frac{\pi}{4}\right]^\top, q_{2}(0) = \left[0, -4.4142, -\frac{\pi}{4}, -\frac{\pi}{4}\right]^\top,
\end{align*}
(in (m, rad), (m/s, rad/s), rad, rad/s, respectively). The desired goal states are set to: 
\begin{gather*}
x_\textup{d} = \left[10, 10, 0.9071, \frac{\pi}{2} \right]^\top, \\
q_{1, \textup{d}}  = \left[10, 12.2071, \frac{\pi}{4}, \frac{\pi}{4}\right]^\top, 
q_{2, \textup{d}} =  \left[10, 7.7929, -\frac{\pi}{4}, -\frac{\pi}{4}\right]^\top,
\end{gather*}
(in (m, rad), rad, respectively). We set a spherical obstacle between the initial and the desired pose of the object, with center $(5,5,1)$ m and radius $2$ m.
The sampling time is $h = 0.1$ seconds, the horizon is set to $T_p = 0.5$ seconds, and the total simulation time is $80$ seconds; The matrices $P_c, Q_c, R_c$ are set to:
\begin{equation}
P_c = Q_c = 0.5  I_{16}, R_c = 0.5 I_{8}. \notag 
\end{equation}
The terminal set is taken as a ball of radius $0.1$ m around $0$. 
The simulation results are depicted in Fig. \ref{fig:sim9 (MED)}-Fig. \ref{fig:sim16 (MED)},  which show that the states of the agents as well as the states of the object converge to the desired ones while guaranteeing that all state and input constraints are met. The simulation scenarios were carried out by using the NMPC toolbox given in \cite{grune_2011_nonlinear_mpc} and they took $23500$ seconds in MATLAB Environment on a desktop computer with $8$ cores, $3.60$ GHz CPU and $16$GB of RAM.

\subsection{Decentralized NMPC} \label{sec:MPC (ECC)}

In this section, in order to reduce the computational complexity of the NMPC, we develop a decentralized counterpart, where each robotic agent calculates its own control signal. 

We first decouple the dynamics \eqref{eq:coupled dynamics (TCST_coop_manip)} for each agent's MPC. We define $x_{\scr O_i}:\mathbb{R}^{n_i}\to\mathbb{M}$, $v_{\scr O_i}:\mathbb{R}^{2n_i}\to \mathbb{R}^6$ with $x_{\scr O_i}(q_i) \coloneqq [p_{\scr O_i}(q_i)^\top, \eta_{\scr O_i}(q_i)^\top]^\top \in \mathbb{M}$, 
\begin{subequations} \label{eq:coupled kinematics (MED+ECC)}
\begin{align}
	p_{\scr O_i}(q_i) \coloneqq&  p_{\scr E_i}(q_i) + R_{i}(q_i) p^{\scr E_i}_{\scr O/E_i} \\ 
	\eta_{\scr O_i}(q_i) \coloneqq& \eta_{\scr E_i}(q_i) + \eta_{\scr O/E_i} 	
\end{align}
\end{subequations}
$\forall i\in\mathcal{N}$, as well as 
\begin{align}
	v_{\scr O_i}(q_i,\dot{q}_i) \coloneqq [\dot{p}_{\scr O_i}(q_i)^\top, \omega_{\scr O_i}(q_i,\dot{q}_i)^\top]^\top \coloneqq J_{\scr i_O}(q_i) v_i(q_i,\dot{q}_i), \ \ \forall i\in\mathcal{N} \label{eq:object-end-effector jacobian (ECC)}, 
\end{align}
where $J_{\scr i_O}(q_i) \coloneqq J_{\scr O_i}(q_i)^{-1}$, $\forall i\in\mathcal{N}$, which are derived from \eqref{eq:coupled_kinematics (TCST_coop_manip)} and \eqref{eq:J_o_i (TCST_coop_manip)}, respectively; $x_{\scr O_i}$ and $v_{\scr O_i}$ are the pose and velocity of the object as computed by agent $i\in\mathcal{N}$.

Consider now the constants $c_i$, with $0< c_i < 1$ and $\sum\limits_{i\in\mathcal{N}}c_i = 1$ that play the role of load sharing coefficients for the agents. Then the object dynamics \eqref{eq:object dynamics (TCST_coop_manip)} can be written as:
\begin{align*}
\sum\limits_{i\in\mathcal{N}}c_i\Big\{M_{\scriptscriptstyle O}\big(\eta_{\scriptscriptstyle O_i}(q_i)\big)\dot{v}_{\scriptscriptstyle O_i}(q_i,\dot{q}_i) &  + C_{\scriptscriptstyle O}\big(\eta_{\scriptscriptstyle O_i}(q_i), \omega_{\scriptscriptstyle O_i}(q_i,\dot{q}_i)\big)v_{\scriptscriptstyle O_i}(q_i,\dot{q}_i) + g_{\scriptscriptstyle O} \Big\} =\\
&\sum\limits_{i\in\mathcal{N}}J_{\scriptscriptstyle O_i}(q_i)^\top h_i,
\end{align*} 
from which, by employing the grasp coupling (see \eqref{eq:grasp matrix (TCST_coop_manip)}), the differential kinematics of the agents, \eqref{eq:object-end-effector jacobian (ECC)}, and after straightforward algebraic manipulations, we obtain the coupled dynamics
\begin{align}
&\sum\limits_{i\in\mathcal{N}}\Big\{M_{D_i}(q_i)\ddot{q}_i + C_{D_i}(q_i,\dot{q}_i)\dot{q}_i + g_{D_i}(q_i) \Big\} = \sum\limits_{i\in\mathcal{N}}J_{\scriptscriptstyle O_i}(q_i)^\top u_i,
\label{eq:coupled dynamics 2 (ECC)}
\end{align}	
where:
\small 
\begin{align*}
	M_{D_i} \coloneqq M_{D_i}(q_i) \coloneqq& c_iM_{\scriptscriptstyle O}J_{i_{\scriptscriptstyle O}}J_i + J_{\scriptscriptstyle O_i}^\top M_iJ_i, \notag \\
 C_{D_i} \coloneqq C_{D_i}(q_i,\dot{q}_i) \coloneqq& J_{\scriptscriptstyle O_i}^\top \Big( M_i\dot{J}_i + C_iJ_i\Big) +   c_iM_{\scriptscriptstyle O}J_{i_{\scriptscriptstyle O}}\dot{J}_i +  c_iM_{\scriptscriptstyle O}\dot{J}_{i_{\scriptscriptstyle O}}J_i  + c_iC_{\scriptscriptstyle O}, \\ \notag
g_{D_i} \coloneqq g_{D_i}(q_i) \coloneqq& c_ig_{\scriptscriptstyle O} + J_{\scriptscriptstyle O_i}^\top g_i,
\end{align*}
\normalsize
$\forall i\in\mathcal{N}$.
Since the scheme developed here is decentralized, we need the following assumption regarding the agent communication:

\begin{assumption} \label{ass:sensing_assumption (ECC)}
	(Sensing and communication capabilities) Each agent $i\in\mathcal{N}$ is able to
	continuously communicate with the other agents $j\in\mathcal{N}\backslash\{i\}$ and transmit appropriate information. 
\end{assumption}

Along with the sets $\mathsf{S}_{i, {\scriptscriptstyle O}}$, $\mathsf{S}_{\scriptscriptstyle A}$ defined in the previous section, we also define  
\begin{align*}
	\mathsf{S}_{\scriptscriptstyle O_i}& \coloneqq \{ q_i\in\mathbb{R}^{n_i}: \mathcal{A}_{\scriptscriptstyle O}(x_{\scriptscriptstyle O_i}(q_i))\cap\mathcal{O}_z = \emptyset, \forall z\in\mathcal{Z}\}, \\
	\mathsf{S}_{i,\mathcal{A}}(q_{-i})& \coloneqq \{ q_i \in \mathbb{R}^{n_i}: \mathcal{A}_i(q_i) \cap \mathcal{A}_j(q_j) = \emptyset, \forall j\in\mathcal{N}\backslash\{i\}\},	
\end{align*}
where $q_{-i}\coloneqq[q_1^\top,\dots,q^\top_{i-1},q^\top_{i+1},\dots,q_N^\top]^\top$, $\forall i \in \mathcal{N}$.

To design a decentralized NMPC control scheme, we employ a leader-follower perspective. More specifically, as will be explained in the sequel, at each sampling time, a leader agent solves part of the coupled dynamics \eqref{eq:coupled dynamics 2 (ECC)} via an NMPC scheme, and transmits its predicted variables to the rest of the agents. Assume, without loss of generality, that the leader corresponds to agent $i=1$. Loosely speaking, the proposed solution proceeds as follows: agent $1$ solves, at each sampling time step, the receding horizon model predictive control subject to the forward dynamics:
\begin{equation}
M_{D_1}\ddot{q}_1 + {C}_{D_1}\dot{q}_1 + {g}_{D_1}  = J_{\scriptscriptstyle O_1}^\top u_1, \label{eq:dynamics_leader (ECC)}
\end{equation}
and a number of inequality constraints, as will be clarified later. After obtaining a control input sequence and a set of predicted variables for $q_1,\dot{q}_1$, denoted as $\hat{q}_1,\hat{\dot{q}}_1$, it transmits the corresponding predicted state for the object $x_{\scriptscriptstyle O_1}(\hat{q}_1), v_{\scriptscriptstyle O_1}(\hat{q}_1,\hat{\dot{q}}_1)$ for the control horizon to the other agents $\{2,\dots,N\}$. Then, the followers solve the receding horizon NMPC subject to the forward dynamics:
\begin{equation}
{M}_{D_i}\ddot{q}_i + {C}_{D_i}\dot{q}_i + g_{D_i}  = J_{\scriptscriptstyle O_i}^\top u_i, \label{eq:dynamics_followers (ECC)} 
\end{equation}
the state equality constraints: 
\begin{align} \label{eq:equality state constr 1 (ECC)}
& x_{\scriptscriptstyle O_i}(q_i) = x_{\scriptscriptstyle O_1}(\hat{q}_1), v_{\scriptscriptstyle O_i}(q_i,\dot{q}_i) = v_{\scriptscriptstyle O_1}(\hat{q}_1,\hat{\dot{q}}_1),
\end{align}
$i \in \{2,\dots,N\}$ as well as a number of inequality constraints that incorporate obstacle and inter-agent collision avoidance. More specifically, we consider that there is a priority sequence among the agents, which we assume, without loss of generality, that is defined by $\{1,\dots,N\}$. Each agent, after solving its optimization problem, transmits its calculated predicted variables to the agents of lower priority, which take them into account for collision avoidance. Note that the coupled object-agent dynamics are implicitly taken into account in equations \eqref{eq:dynamics_leader (ECC)}, \eqref{eq:dynamics_followers (ECC)} in the following sense. Although the coupled model \eqref{eq:coupled dynamics 2 (ECC)} does not imply that each one of these equations is satisfied, by forcing each agent to comply with the specific dynamics through the optimization procedure, we guarantee that \eqref{eq:coupled dynamics 2 (ECC)} is satisfied, since it's the result of the addition of \eqref{eq:dynamics_leader (ECC)} and \eqref{eq:dynamics_followers (ECC)}, for $i = 1$ and every $i \in \{2,\dots,N\}$, respectively. Intuitively, the leader agent is the one that determines the path that the object will navigate through, and the rest of the agents are the followers that contribute to the transportation.  Moreover, the equality constraints \eqref{eq:equality state constr 1 (ECC)} guarantee that the predicted variables of the agents $\{2,\dots,N\}$ will comply with the rigidity at the grasping points. 

By using the notation $x_{q_i} \coloneqq [x^\top_{q_i,1},x^\top_{q_i,2}]^\top\coloneqq[q^\top_i, \dot{q}^\top_i]^\top\in\mathbb{R}^{2n_i}$, $i\in\mathcal{N}$, the nonlinear dynamics of each agent can be written as:
\begin{equation} \label{eq:main_system (ECC)}
\dot{x}_{q_i} = f_{q_i}(x_{q_i},u_i) \coloneqq 	
\begin{bmatrix}
f_{q_i,1}(x_{q_i}) \\
f_{q_i,2}(x_{q_i},u_i) 		
\end{bmatrix}, 
\end{equation}
where ${f}_{q_i}: \mathsf{S}_i\times\mathbb{R}^{n_i+6}\to\mathbb{R}^{2n_i}$ is the locally Lipschitz function:
\begin{align*}
	{f}_{q_i,1}(x_{q_i}) \coloneqq& x_{q_i,2}, \\
	 {f}_{q_i,2}(x_{q_i},u_i) \coloneqq& \widehat{M}_{D_i}(q_i)\Big( J_{\scriptscriptstyle O_i}(q_i)^\top u_i - C_{D_i}(q_i,\dot{q}_i)\dot{q} - g_{D_i}(q_i)  \Big),
\end{align*} 
$\forall i \in \mathcal{N}$, where $\widehat{M}_{D_i}: \mathsf{S}_i\to\mathbb{R}^{n_i\times6}$, is the pseudo-inverse 
\begin{equation*}
	\widehat{M}_{D_i}(q_i) \coloneqq M_{D_i}(q_i)^\top\Big(M_{D_i}(q_i){M}_{D_i}(q_i)^\top\Big)^{-1}.
\end{equation*}
$\forall i\in\mathcal{N}$. We define now the error vector $e_{D_1}:\mathbb{R}^{2n_i}\to \times\mathbb{R}^{12}$, as: $$e_{D_1}(x_{q_1}) \coloneqq \begin{bmatrix}
x_{\scriptscriptstyle O_1}(q_1) - x_\textup{d} \\
v_{\scriptscriptstyle O_1}(q_1,\dot{q}_1),
\end{bmatrix}$$
which gives us the \emph{error dynamics}:
\begin{equation} \label{eq:error_dynamics (ECC)}
\dot{e}_{D_1} = f_{D_1}(x_{D_1},u_1),
\end{equation}
\small
with $f_{D_1}: \mathsf{S}_i \times\mathbb{R}^{n_i+6}\to\mathbb{R}^{12}$: 
\begin{align*}
f_{D_1}(x_{q_1},u_1) \coloneqq &\\
&\hspace{-5mm} \begin{bmatrix}	
J_{\scr O}(\eta_{\scr O_1}(q_1)) J_{1_{\scr O}}(q_1)J_1(q_1)\dot{q}_1 \\
J_{1_{\scr O}}(q_1) J_1(q_1) f_{q_1,2}(x_{q_1},u_1) + \Big(J_{1_{\scr O}}(q_1) \dot{J}_1(q_1) + \dot{J}_{1_{\scr O}}(q_1)J_1(q_1)\Big)\dot{q}_1,	
\end{bmatrix}
\end{align*}
\normalsize
where we employed \eqref{eq:error_dynamics (ECC)} and the object dynamics.
The input constraint sets are defined similarly to \eqref{eq:input constraint set (MED)} as
\begin{equation*}
	U_{D_i} \coloneqq \left\{ u_i\in\mathbb{R}^{6} :  \lVert u_i \rVert \leq \frac{\bar{\tau}_i}{\sigma_{\min}(J^\top_i)} \right\},
\end{equation*}
Define also the sets 
\begin{align*}
X_{D_1}(q_{-1}) \coloneqq \Bigg\{ &x_{q_1} \in \mathbb{R}^{2n_1} : \theta_{\scriptscriptstyle O_1}(q_1)\in [-\bar{\theta}, \bar{\theta}], \lvert \dot{q}_{1_k} \rvert \leq \bar{\dot{q}}_1, \forall k\in\{1,\dots,n_1\}, \\
& q_1 \in {\mathsf{S}}_1 \cap \mathsf{S}_{1,\mathcal{A}}(q_{-1}) \cap \mathsf{S}_{1, {\scriptscriptstyle O}} \cap {\mathsf{S}}_{\scriptscriptstyle O_1} \Bigg\}	\\
X_{D_i}(q_{-i})  \coloneqq \Bigg\{ &x_{q_i} \in \mathbb{R}^{2n_i} : \lvert \dot{q}_{i_k} \rvert \leq \bar{\dot{q}}_i, \forall k\in\{1,\dots,n_i\}, q_i \in {\mathsf{S}}_i \cap \mathsf{S}_{i,\mathcal{A}}\cap \mathsf{S}_{i, {\scriptscriptstyle O}}  \Bigg\},
\end{align*}
$\forall i\in\{2,\dots,N\}$. 
The sets $X_{D_i}$ capture all the state constraints of the system dynamics \eqref{eq:main_system (ECC)}, i.e., representation- and singularity-avoidance, collision avoidance among the agents and the obstacles, as well as collision avoidance of the object with the obstacles, which is assigned to the leader agent only. We further define the set
\begin{align*}
E_{D_1}(q_{-1}) \coloneqq \{ e_{D_1}(x_{q_1}) \in  \mathbb{R}^{12} : x_{q_1} \in X_1(q_{-1})\}, 
\end{align*} 
which now represents the constraints set for the NMPC scheme of the leader.

The main problem at hand is the design of a \emph{feedback control law} $u_1 \in U_{D_1}$ for agent $1$ which guarantees that the error signal $e_{D_1}$ with dynamics given in \eqref{eq:error_dynamics (ECC)}, satisfies $\lim_{t \to \infty} \|e_{D_1}(x_{q_1}(t))\| \to 0$, while ensuring singularity avoidance, collision avoidance between the leader, the object and the obstacles as well as collision avoidance between the leader and the followers in their current position.
The role of the followers $\{2,\dots,N\}$ is, through the load-sharing coefficients $c_2,\dots,c_N$ in \eqref{eq:coupled dynamics 2 (ECC)}, to contribute to the object trajectory execution, as derived by the leader agent $1$, while also avoiding collisions. In order to solve the aforementioned problem, we propose a NMPC scheme, that is presented hereafter.

Consider a sequence of sampling times $\{t_j\}$, $j \in \mathbb{N}$ as defined in the centralized scheme, with $t_{j+1} = t_j + h_s$, $h_s \in(0,T_p)$, and $T_p$ the respective horizon. For agent $1$, the open-loop input signal applied in between the sampling instants is given by the solution of the following FHOCP:
\begin{subequations}
	\begin{align}
	&\hspace{0mm}\min\limits_{\hat{u}_1(\cdot)} J_{D_1}(e_{D_1}(x_{q_1}(t_j)),\hat{u}_1(\cdot)) \coloneqq \min\limits_{\hat{u}_1(\cdot)} \Bigg\{  V_{D_1}(e_{D_1}(\hat{x}_{q_1}(t_j+T_p))) \notag \\
	&\hspace{19mm} + \int_{t_j}^{t_j+T_p} \Big[ F_{D_1}(e_{D_1}(\hat{x}_{q_1}(s)), \hat{u}_1(s)) \Big] ds \Bigg\}  \label{eq:mpc_minimazation (ECC)} \\
	&\hspace{0mm}\text{subject to:} \notag \\
	&\hspace{1mm} \dot{e}_{D_1}(\hat{x}_{q_1}(s)) = f_{D_1}(\hat{x}_{q_1}(s),\hat{u}_1(s)), \ e_{D_1}(\hat{x}_{q_1}(t_j)) = f_{D_1}(x_{q_1}(t_j)),   \label{eq:diff_mpc (ECC)} \\
	&\hspace{1mm} e_{D_1}(\hat{x}_{q_1}(s)) \in E_{D_1}(q_{-1}(t_j)), s \in [t_j,t_j+T_p], \label{eq:mpc_constrained_set_1 (ECC)} \\
	&\hspace{1mm} \hat{u}_1(s)\in U_{D_1}, s \in [t_j,t_j+T_p], \label{eq:mpc_constrained_set_2 (ECC)} \\
	&\hspace{1mm} e_{D_1}(\hat{x}_{q_1}(t_j+T_p)) \in \mathcal{E}_{D_1}. \label{eq:mpc_terminal_set (ECC)}
	\end{align}
\end{subequations}
At a generic time $t_j$ then, agent $1$ solves the aforementioned FHOCP. 
The functions $F_{D_1} : E_{D_1}(q_{-1}(t_j)) \times U_{D_1} \to \mathbb{R}_{\geq 0}$, $V_{D_1}: \mathcal{E}_{D_1}(q_{-1}(t_j)) \to \mathbb{R}_{\geq 0}$ stand for the \emph{running cost} and the \emph{terminal penalty cost}, respectively, and they are defined as: 
\begin{align*}
	F_{D_1} (e_{D_1}, u_1) \coloneqq& e_{D_1}^{\top} Q_{D_1} e_{D_1} + u_1^{\top} R_{D_1} u_1 \\
	V_{D_1} (e_{D_1}) \coloneqq& e_{D_1}^{\top} P_{D_1} e_{D_1},
\end{align*} 
where $R_{D_1} \in \mathbb{R}^{6 \times 6}$ and $P_{D_1} \in \mathbb{R}^{2n_1 \times 2n_1}$ are symmetric and positive definite gain matrices; $Q_{D_1} \in \mathbb{R}^{2n_1 \times 2n_1}$ is a symmetric and positive semi-definite controller gain matrix. The bounded \textit{terminal} set is defined here as $\mathcal{E}_{D_1}$, and we assume that $\mathcal{E}_{D_1} \subset \bigcap_{j\in\mathbb{N}}\{ E_{D_1}(q_{-1}(t_j)) \} \neq \emptyset$.


The solution to FHOCP \eqref{eq:mpc_minimazation (ECC)} - \eqref{eq:mpc_terminal_set (ECC)} starting at time $t_j$ provides an optimal control input, denoted by
$\hat{u}_1^{\star}(s;\ e_{D_1}(x_{q_1}(t_j)),x_q(t_j))$, $s \in [t_j, t_j + T_p]$, $x_q\coloneqq[x_{q_1}^\top,\dots,x_{q_N}^\top]^\top$. This control input is then applied to the system until the next sampling instant $t_{j+1}$:
\begin{align}
u_1\left(s;e_{D_1}(x_{q_1}(t_j),x_q(t_j))\right) = \hat{u}_1^{\star}\left(s; e_{D_1}(x_{D_1}(t_j)),x_q(t_j)\right), \ \ \forall s \in [t_j, t_{j+1}).
\label{eq:optimal_input (ECC)}
\end{align}
At time $t_{j+1} = t_{j}+h_s$ a new FHOCP is solved in the same manner, leading to a receding horizon approach. The control input $u_1(\cdot)$ is of feedback form, since it is recalculated at each sampling instant based on the then-current state. The solution of \eqref{eq:error_dynamics (ECC)} starting at time $t_j$, from an initial condition $x_{q}(t_j), e_{D_1}(x_{q_1}(t_j))$, by application of the control input $u_1 : [t_j, t_{j+1}] \to U_{D_1}$ is denoted by $e_{D_1}\big(x_{q_1}(s;u_1(\cdot), x_q(t_j), e_{D_1}(x_{q_1}(t_j)\big)$, $s \in [t_j, t_j+T_p]$. 

After the solution of the FHOCP and the calculation of the predicted states 
$\hat{x}_{q_1}(s)$, $s\in[t_j,t_{j+1})$, 
at each time instant $t_j$, agent $1$ transmits the values $\hat{q}_1(s)$, $\hat{\dot{q}}_1(s)$ as well as $x_{\scr O_1}(\hat{q}_1(s))$ and $v_{\scr O_1}(\hat{q}_1(s),\hat{\dot{q}}_1(s))$, as computed by \eqref{eq:coupled kinematics (MED+ECC)}, \eqref{eq:object-end-effector jacobian (ECC)}, $\forall s \in [t_j, t_j+T_p]$, to the rest of the agents $\{2,\dots,N\}$. The rest of the agents then proceed as follows. Each agent $i\in\{2,\dots,N\}$, solves the following FHOCP:
\begin{subequations}
	\begin{align}
	&\hspace{0mm}\min\limits_{\hat{u}_i(\cdot)} J_{D_i}(x_{q_i}(t_j)),\hat{u}_i(\cdot)) \label{eq:mpc_minimazation followers (ECC)} \\
	&\hspace{0mm}\text{subject to:} \notag \\
	&\hspace{1mm} \dot{x}_{q_i} = f_{q_i}(x_{q_i}(s),u_i(s)),  \label{eq:diff_mpc followers (ECC)} \\
	&\hspace{1mm} x_{q_i}(s) \in X_i\big(\hat{q}_1(s),\dots,\hat{q}_{i-1}(s),q_{i+1}(t_j),\dots,q_N(t_j)\big), \label{eq:mpc_constrained_set_1 followers (ECC)} \\
	&\hspace{1mm} x_{\scr O_i}(q_i(s)) = x_{\scr O_1}(\hat{q}_1(s)), s \in [t_j,t_j+T_p] \label{eq:equality constr_1 followers (ECC)} \\
	&\hspace{1mm} v_{\scr O_i}(q_i(s),\dot{q}_i(s)) = v_{\scr O_1}(\hat{q}_1(s), \hat{\dot{q}}_1(s)), s \in [t_j,t_j+T_p] \label{eq:equality constr_2 followers (ECC)} \\
	&\hspace{1mm} u_i(s)\in U_{D_i}, s \in [t_j,t_j+T_p], \label{eq:mpc_constrained_set_2 followers (ECC)} 
	\end{align}
\end{subequations}
at every sampling time $t_j$, where $J_{D_i}$ is an associated cost function. The constraint \eqref{eq:mpc_constrained_set_1 followers (ECC)} guarantees that agent $i$ will obtain a trajectory that does not collide with the predicted trajectories of the agents higher in priority, or the agents lower in priority at $t_j$.
Note that, through the equality constraints \eqref{eq:equality constr_1 followers (ECC)}, \eqref{eq:equality constr_2 followers (ECC)}, the follower agents must comply with the trajectory  computed by the leader $\hat{q}_1(s), \hat{\dot{q}}_1(s)$. 
This can be problematic in the sense that this trajectory might drive the followers to collide with an obstacle or among each other. Resolution of such cases, however, is not in the scope of this thesis. We state that with the following assumption:
\begin{assumption} \label{ass:follower feasibility (ECC)}
	The sets 
	$\{ q\in\mathbb{R}^n : x_{\scr O_i}(q_i(s)) = x_{\scr O_1}(\hat{q}_1(s))$, $v_{\scr O_i}(q_i(s),\dot{q}_i(s))$ $=$ $v_{\scr O_1}(\hat{q}_1(s)$, $\hat{\dot{q}}_1(s))$, $q_i \in X_i\big(\hat{q}_1(s),\dots,\hat{q}_{i-1}(s),q_{i+1}(t_j),\dots,q_N(t_j)\big)\}$  are nonempty, $\forall i\in\{2,\dots,N\}$, $\forall s\in [t_j, t_j+T_p]$, $j\in\mathbb{N}$.
\end{assumption}
Next, similarly to the leader agent, agent $i>1$ calculates the predicted states $\hat{q}_i(s),\hat{\dot{q}}_i(s), s\in[t_j,t_j+T_p]$, which then transmits to the agents $\{i+1,\dots,N\}$. In that way, at each time instant $t_j$, each agent $i\in\{2,\dots,N\}$ receives the other agents' states (as stated in Assumption \ref{ass:sensing_assumption (ECC)}), incorporates the constraint \eqref{eq:mpc_constrained_set_1 followers (ECC)} for the agents $\{i+1,\dots,N\}$, 
receives the predicted states $\hat{q}_\ell(s), \hat{\dot{q}}_\ell(s)$ from the agents $\ell\in\{2,\dots,i-1\}$ and incorporates the collision avoidance constraint \eqref{eq:mpc_constrained_set_1 followers (ECC)} for the entire horizon. Loosely speaking, we consider 
that each agent $i\in\mathcal{N}$ takes into account the first state of the next agents in priority ($q_\ell(t_j),\ell\in\{i+1,\dots,N\}$), as well as the transmitted predicted variables $\hat{q}_\ell(s), \ell\in\{1,\dots,i-1\}$ of the previous agents in priority, for collision avoidance.
Intuitively, the leader agent executes the planning for the followed trajectory of the object's center of mass (through the solution of the FHOCP \eqref{eq:mpc_minimazation (ECC)}-\eqref{eq:mpc_terminal_set (ECC)}), the follower agents contribute in executing this trajectory through the load sharing coefficients $c_i$ (as indicated in the coupled model \eqref{eq:coupled dynamics 2 (ECC)}), and the agents low in priority are responsible for collision avoidance with the agents of higher priority.
Moreover, the aforementioned equality constraints \eqref{eq:equality constr_1 followers (ECC)}, \eqref{eq:equality constr_2 followers (ECC)} as well as the forward dynamics \eqref{eq:mpc_minimazation followers (ECC)} guarantee the compliance of all the followers with the model \eqref{eq:coupled dynamics 2 (ECC)}. 

Therefore, given the constrained FHOCP \eqref{eq:mpc_minimazation followers (ECC)}-\eqref{eq:mpc_constrained_set_2 followers (ECC)}, the solution of the problem lies in the capability of the leader agent to produce a state trajectory that guarantees  $x_{\scr O_1}(q_1(t)) \to x_\text{des}$, by solving the FHOCP \eqref{eq:mpc_minimazation (ECC)}-\eqref{eq:mpc_terminal_set (ECC)}, which is discussed in Theorem \ref{th:main theorem (ECC)}.




\begin{theorem} \label{th:main theorem (ECC)}
	Suppose that Assumptions \ref{ass:feasility_assumption (MED+ECC)} - \ref{ass:follower feasibility (ECC)} hold as well as 
	\begin{itemize}
		\item The FHOCP \eqref{eq:mpc_minimazation (ECC)}-\eqref{eq:mpc_terminal_set (ECC)} is feasible for the initial time $t = 0$
		\item The terminal set $\mathcal{E}_{D_1}$ is closed, with $0\in\mathcal{E}_{D_1}$
		\item The terminal set and terminal cost are chosen such that, $\forall e_{D_1} \in \mathcal{E}_{D_1}$, there exists an admissible control input $u_{D_{1F}} : [0, h_s] \to U_{D_1}$ such that for all $e_{D_1}(x_{q_1}(s)) \in \mathcal{E}_{D_1}$, $\forall s\in[0,h_s]$  and 
		\begin{equation*}
			\dfrac{\partial V_{D_1}}{\partial e_{D_1}} f_{D_1}(e_{D_1}(x_{q_1}(s)),u_{D_{1F}}(s)) + F_{D_1}(e_{D_1}(x_{q_1}(s)),u_{D_{1F}}(s)) \leq 0
		\end{equation*}
	\end{itemize} 
	Then, the system \eqref{eq:error_dynamics (ECC)}, under the control input \eqref{eq:optimal_input (ECC)}, converges to the origin when $t \to \infty$, i.e. $\lim_{t\to\infty} e_{D_1}(x_{q_1}(t)) = 0$.
\end{theorem}
\begin{proof}
		The proof is identical to the proof of Theorem 2.1 in \cite{frank_2003_nmpc_bible}.
\end{proof}

\subsubsection{Simulation Results}

\begin{figure}[t!]
	\vspace{2mm}
	\centering
	\includegraphics[scale = 0.25]{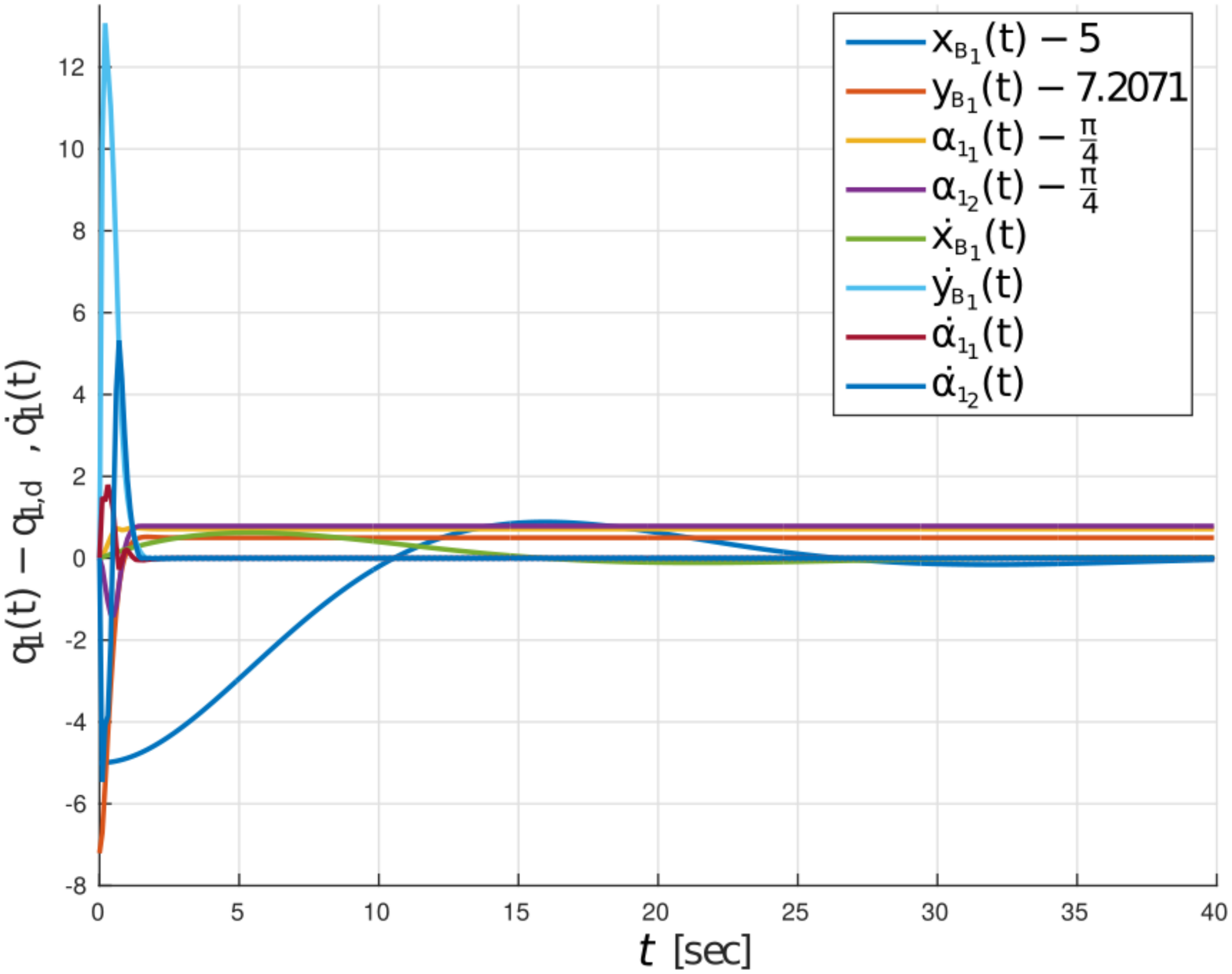}
	\caption{The error states of agent $1$.}
	\label{fig:error_ag_1 (ECC)}
\end{figure}

\begin{figure}[t!]
	\centering
	\includegraphics[scale = 0.25]{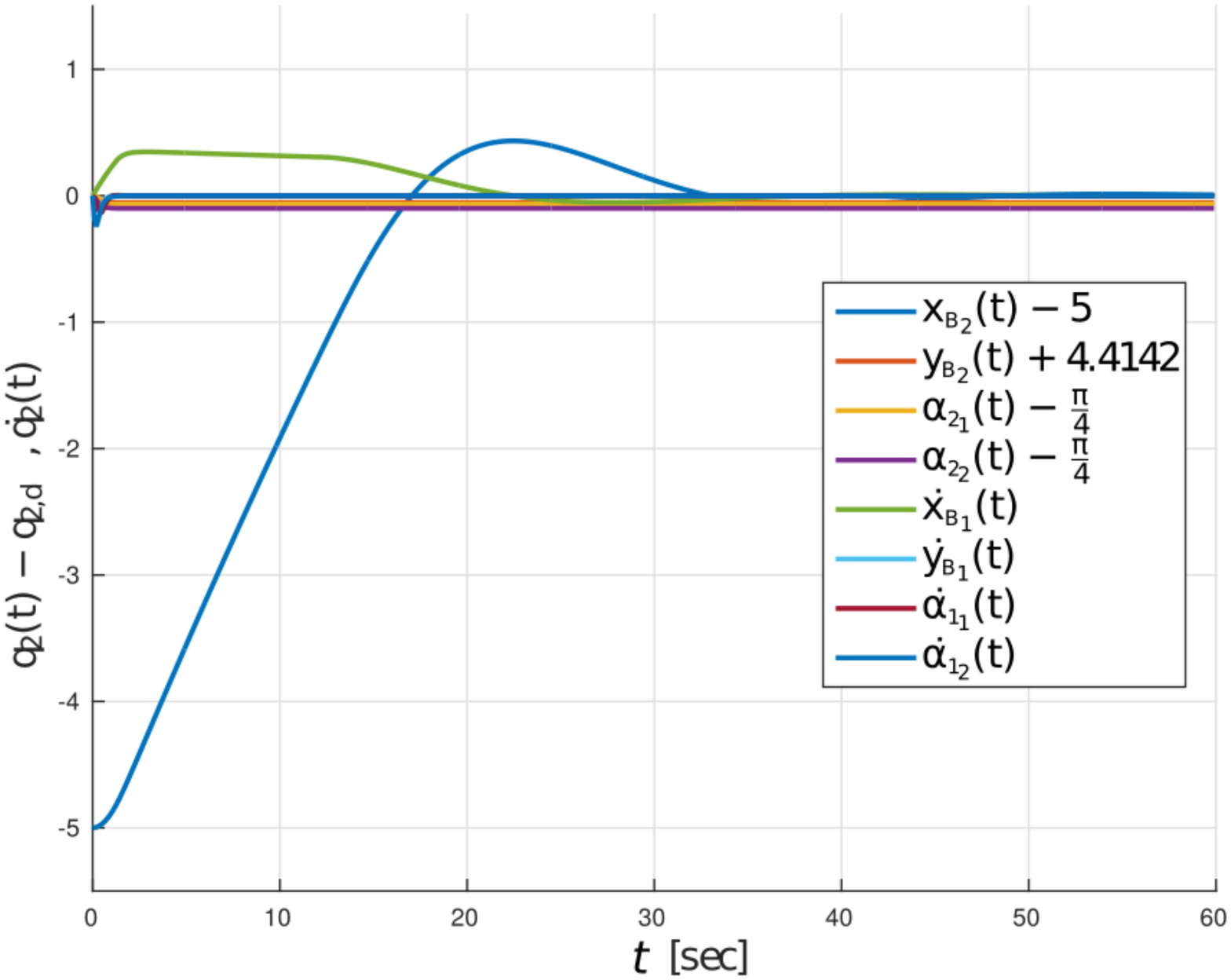}
	\caption{The error states of agent $2$.}
	\label{fig:errors_ag_2 (ECC)}
\end{figure}

\begin{figure}[t!]
	\centering
	\includegraphics[scale = 0.25]{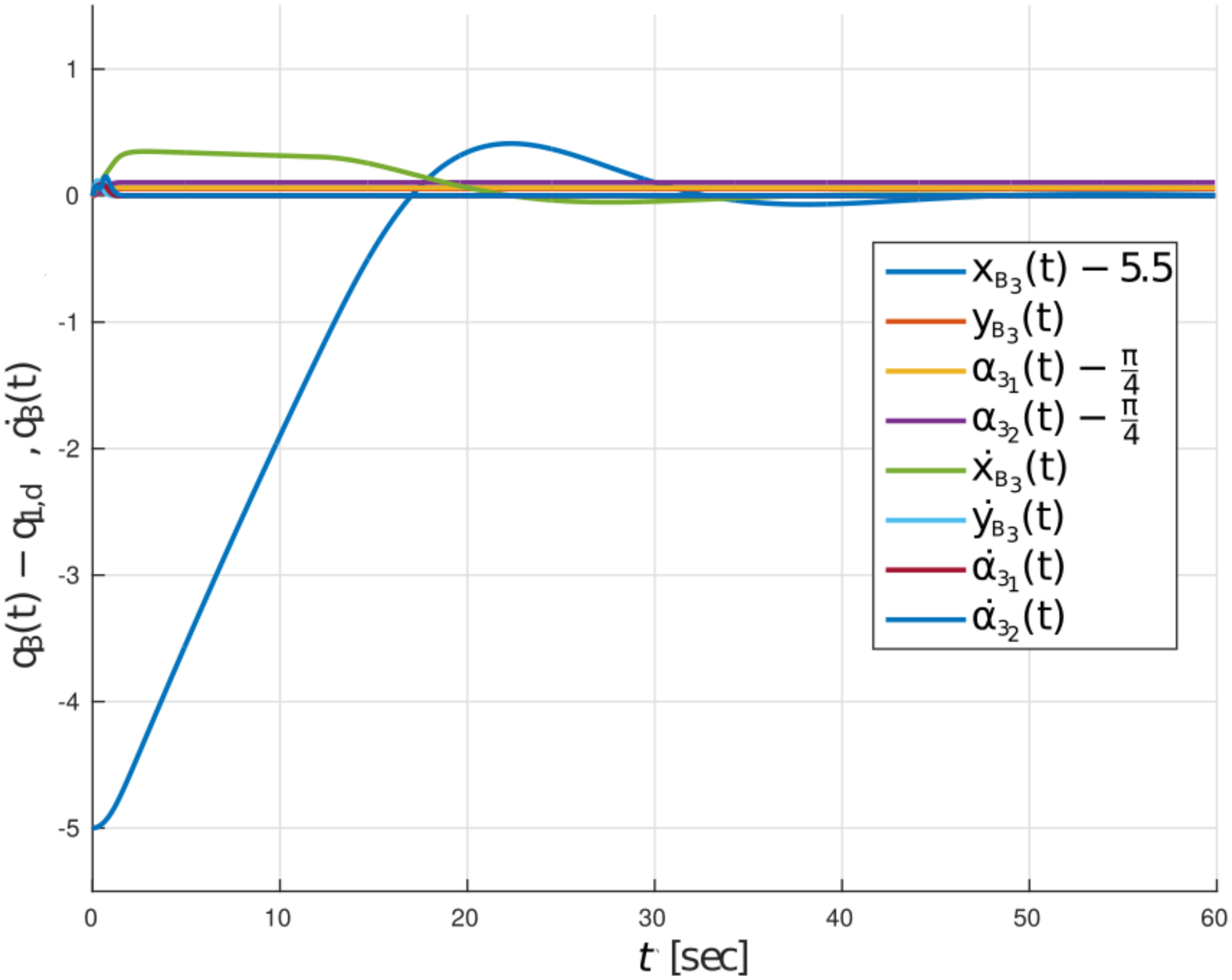}
	\caption{The error states of agent $3$.}
	\label{fig:errors_ag_3 (ECC)}
\end{figure}

\begin{figure}[t!]
	\centering
	\includegraphics[scale = 0.45]{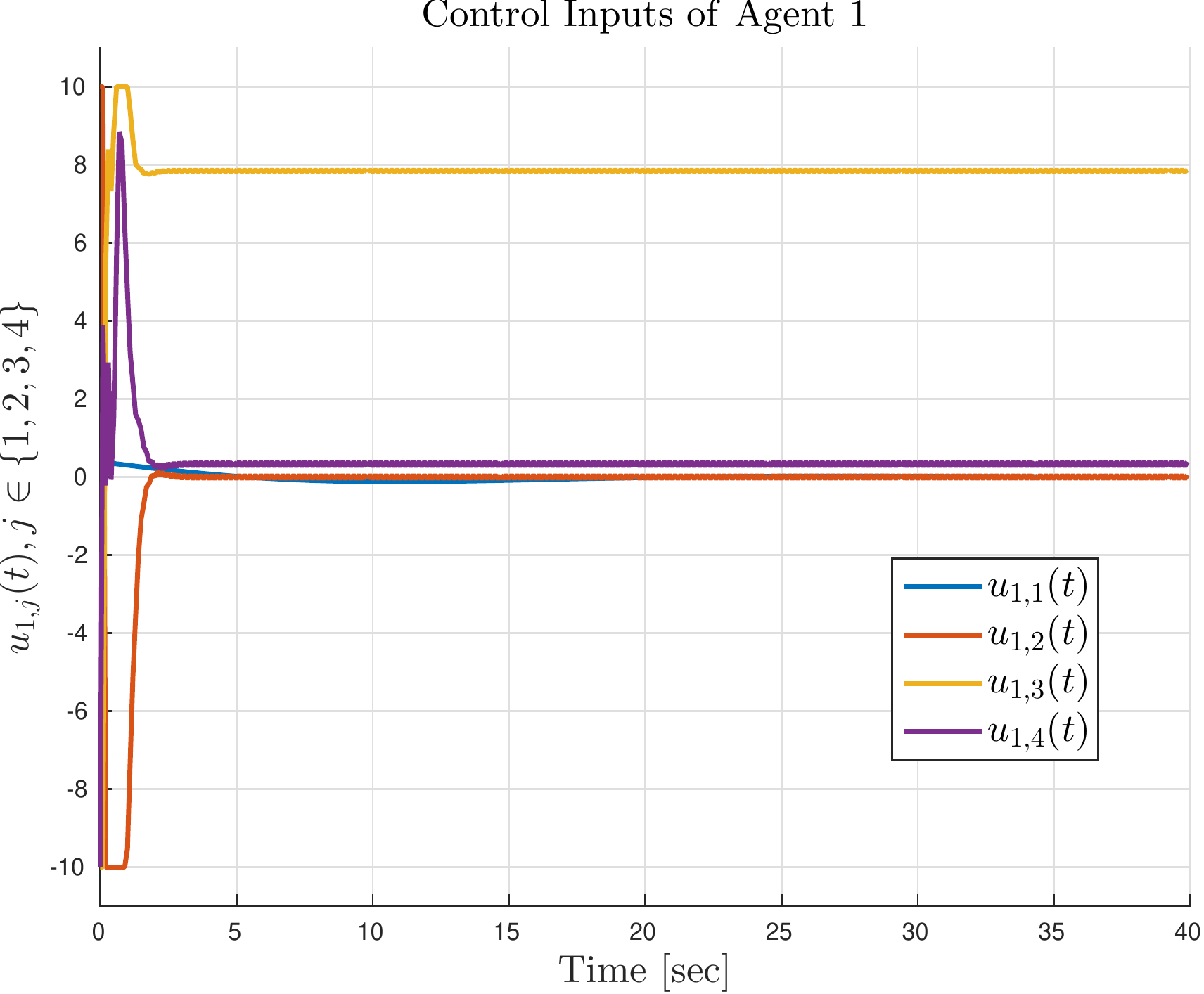}
	\caption{The control inputs of agent $1$ with $-10 \le u_{1,j}(t) \le 10$, $\forall j\in\{1,\dots,4\}$.}
	\label{fig:control_inputs_ag_1 (ECC)}
\end{figure}

\begin{figure}[t!]
	\centering
	\includegraphics[scale = 0.45]{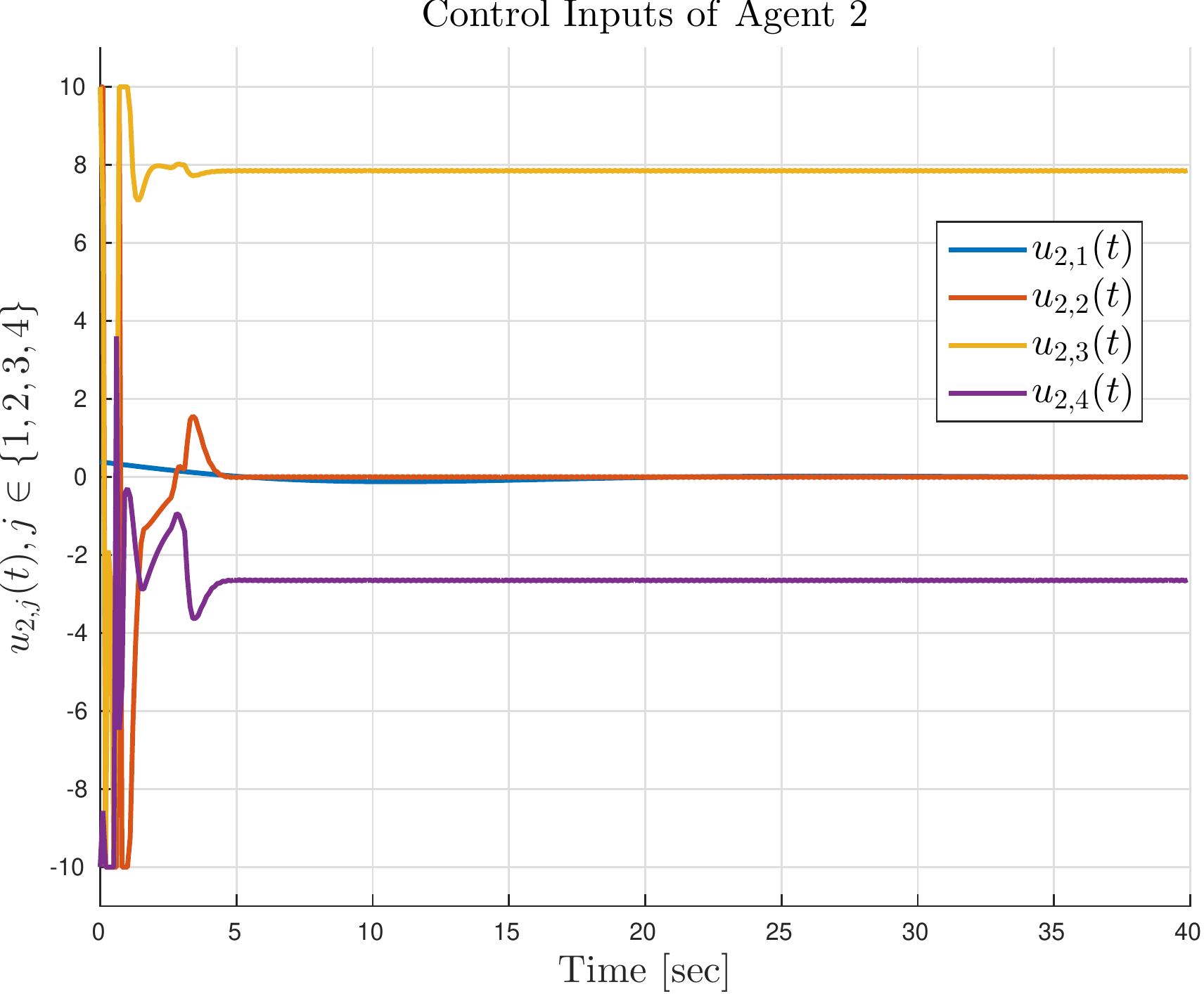}
	\caption{The control inputs of agent $2$ with $-10 \le u_{2,j}(t) \le 10$, $\forall j\in\{1,\dots,4\}$.}
	\label{fig:control_inputs_ag_2 (ECC)}
\end{figure}

\begin{figure}[t!]
	\centering
	\includegraphics[scale = 0.45]{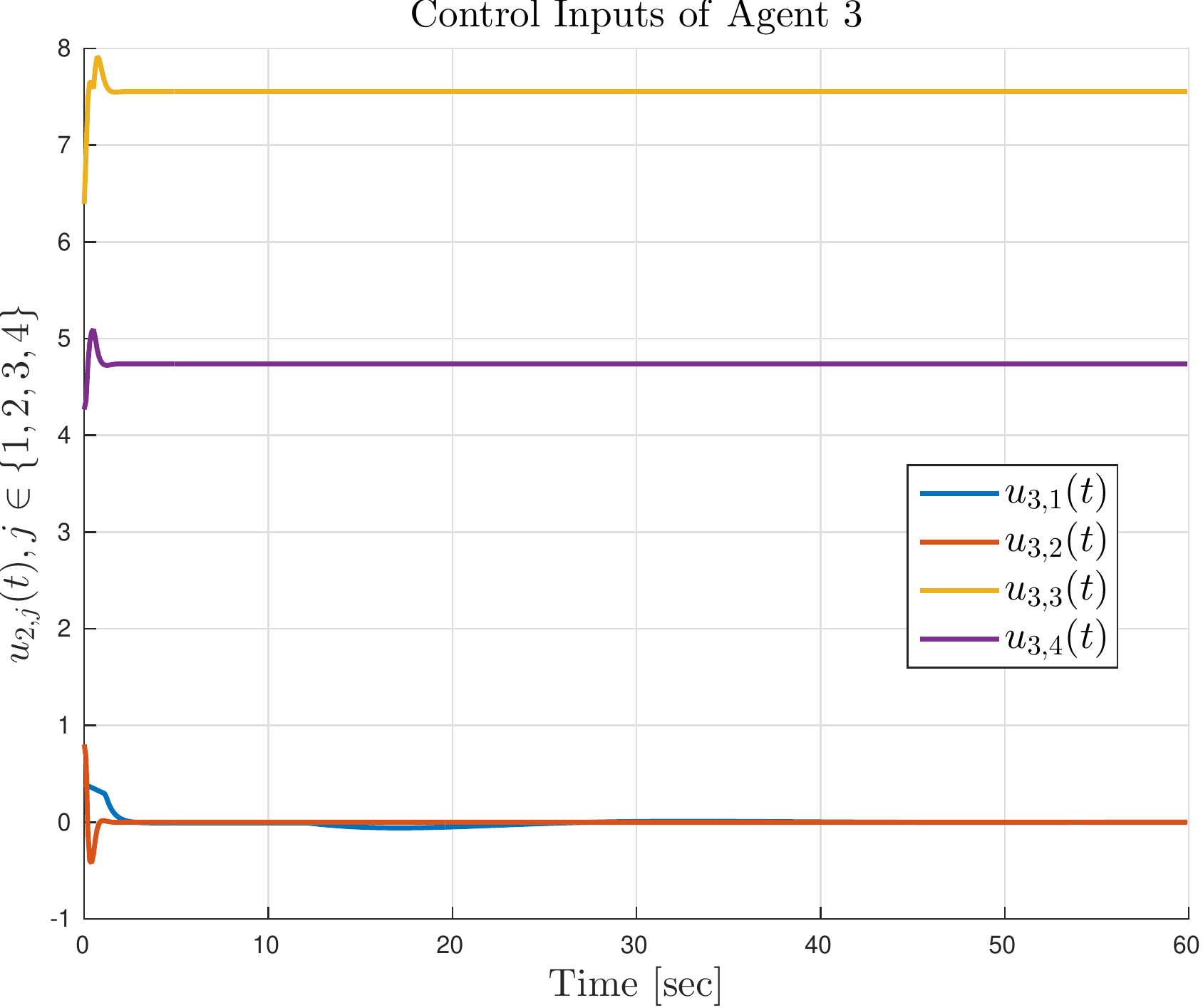}
	\caption{The control inputs of agent $3$ with $-10 \le u_{3,j}(t) \le 10$, $\forall j\in\{1,\dots,4\}$.}
	\label{fig:control_inputs_ag_3 (ECC)}
\end{figure}

To demonstrate the efficiency of the proposed control protocol, we consider a simulation example with $N=3$ ground vehicles equipped with $2$ DOF manipulators, rigidly grasping an object with $n_1 = n_2 = n_3 = 4$, $n = n_1+n_2+n_3= 12$. The states of the agents are given as: $q_i = [p_{\scriptscriptstyle B_i}^\top, \alpha_i^\top]^\top \in \mathbb{R}^4$, $p_{\scriptscriptstyle B_i} = [x_{\scriptscriptstyle B_i}, y_{\scriptscriptstyle B_i}]^\top \in \mathbb{R}^2$, $\alpha_i = [\alpha_{i_1}$, $\alpha_{i_2}]^\top \in \mathbb{R}^2$, $i \in \{1,2,3\}$. The state of the object is $x_{\scriptscriptstyle O} = [p_{\scriptscriptstyle O}^\top, \phi_{\scriptscriptstyle O}]^\top \in \mathbb{R}^4$ and it is calculated though the states of the agents. The singularity and input constraints are set as in the centralized case. The initial conditions of agents and the object are set to: 
\begin{gather*}
q_{1}(0) = [0.5, 0, \frac{\pi}{4}, \frac{\pi}{4}]^\top, q_{2}(0) = [0, -4.4142, -\frac{\pi}{4}, -\frac{\pi}{4}]^\top, \\
q_{3}(0) = [-0.50, -4.4142, -\frac{\pi}{4}, -\frac{\pi}{4}]^\top, \dot{q}_1(0) = \dot{q}_2(0) = \dot{q}_3(0) = [0, 0, 0, 0]^\top, \\
x_{\scriptscriptstyle O}(0) = [0, -2.2071, 0.9071, \frac{\pi}{2}]^\top, \dot{x}_{\scriptscriptstyle O}(0) = [0,0,0,0]^\top
\end{gather*} 
(in rad, rad/s (m, rad), (m/s, rad/s), respectively). The desired goal state the object is set to $$x_{\scriptscriptstyle O, \text{des}} = [5, -2.2071, 0.9071, \frac{\pi}{2}]^\top$$ (m, rad), which, due to the structure of the considered robots, corresponds uniquely to
\begin{gather*}
q_{1, \text{des}} = [5.5, 0, \frac{\pi}{4}, \frac{\pi}{4}]^\top, q_{2, \text{des}} = [5, -4.4142, -\frac{\pi}{4}, -\frac{\pi}{4}]^\top, \\
q_{3, \text{des}} = [4.5, 0, -\frac{\pi}{4}, -\frac{\pi}{4}]^\top, \dot{q}_{1, \text{des}} = \dot{q}_{2, \text{des}} = \dot{q}_{3, \text{des}} =  [0, 0, 0, 0]^\top
\end{gather*} 
(in rad and rad/s, respectively). We set an obstacle between the initial and the desired pose of the object. The obstacle is spherical with center $(2.5,-2.2071,1)$ m and radius $\sqrt{0.2}$ m. The sampling time is $h_s = 0.1$ seconds, the horizon is $T_p = 0.5$ seconds, and the total simulation time is $60$ seconds; The matrices $P_{D_i}$, $Q_{D_i}$, $R_{D_i}$ are set to: $P_{D_i} = Q_{D_i} = 0.5 I_{8}$, $R_{D_i} = 0.5 I_{4}$, $\forall i\in\{1,2,3\}$, and the load sharing coefficients as $c_1 = 0.3$, $c_2 = 0.5$, and $c_3 = 0.2$. The functions $J_{D_i}$ are chosen as simple quadratic functions of their arguments. The control input constraints are taken as in the previous section.
The simulation results are depicted in Figs. \ref{fig:error_ag_1 (ECC)}- \ref{fig:control_inputs_ag_3 (ECC)}; Figs. \ref{fig:error_ag_1 (ECC)},  \ref{fig:errors_ag_2 (ECC)} \ref{fig:errors_ag_3 (ECC)} show the error states of agent $1$, $2$ and $3$, respectively, which converge to $0$; Figs. \ref{fig:control_inputs_ag_1 (ECC)} - \ref{fig:control_inputs_ag_3 (ECC)} depict the control inputs of the three agents. Note that the different load-sharing coefficients produce slightly different inputs. The simulation was carried out by using the NMPC toolbox given in \cite{grune_2011_nonlinear_mpc} and it took $13450 \sec$ in MATLAB Environment on a desktop computer with $8$ cores, $3.60$ GHz CPU and $16$GB of RAM. Note the significant time difference with respect to the centralized case of the previous section. 
Finally, a video illustrating an implementation of the algorithm in real hardware can be found on \href{https://youtu.be/f\_95UCAAp6M}{https://youtu.be/f\_95UCAAp6M}.

\section{Rolling Contacts} \label{sec:rolling contacts (ACC)}

In this section we relax the assumption on the rigid grasping points. In particular, we assume that the robotic agents are connected to the object in terms of \textit{rolling contacts}. As discussed before, this more natural approach to cooperative manipulation allows for a wider class of objects to be manipulated, and allows for modular manipulation scenarios where robots can be swapped to adjust the grasp online. 

\begin{figure}
	\centering
	\includegraphics[width = 0.5\textwidth]{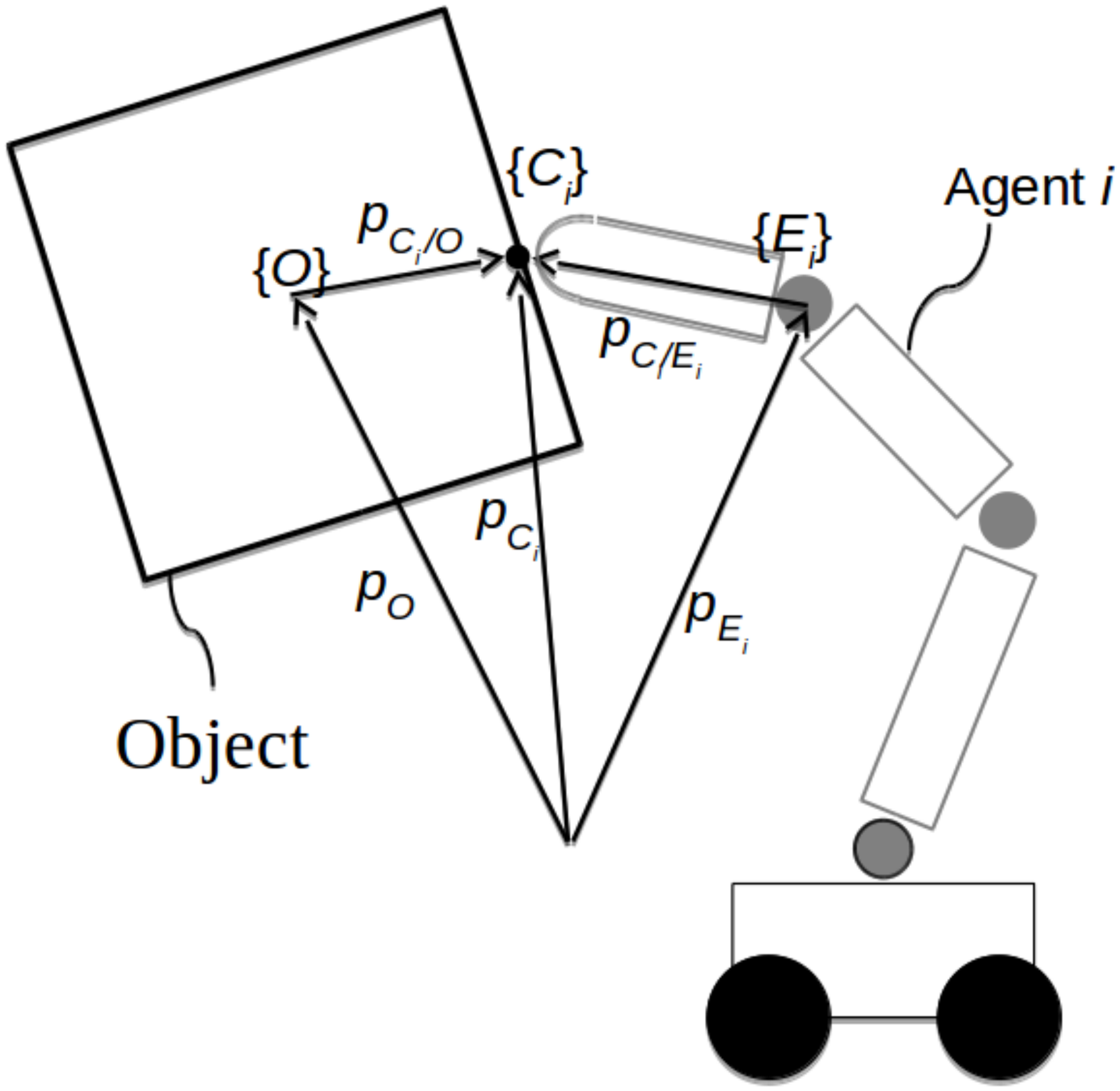}
	\caption{A robotic agent in contact with a rigid object via a rolling contact.\label{fig:Two-robotic-arms (ACC_rolling)}}
\end{figure}

\subsection{System Model} \label{sec:system model rolling (ACC_rolling)}

We provide here the model of the coupled system, which is slightly different with respect to that of Section \ref{subsec:system model (TCST_coop_manip)} to account for the rolling contact constraints. 
As before, we consider $N\in\mathbb{R}$ robotic agents grasping a rigid object in $3$D space, with generalized variables $q_i,\dot{q}_i \in\mathbb{R}^{n_i}$. We relax the assumption of fully actuated agents by requiring $n_i \geq 3$, $\forall i\in\mathcal{N}$. Each agent has a smooth, convex “fingertip” (i.e. passive end-effector) of high stiffness that is
in contact with an object via a smooth contact surface. In addition to the end-effector frames $\mathcal{E}_i$, we add the contact frames $\mathcal{C}_i$, located at $p_{\scr C_i} \in\mathbb{R}^3$, with respect to the inertial frame; $\mathcal{C}_i$ are defined as Gauss frames \cite{murray1994}, where one of the axes is defined orthonormal to the contact plane. We further define $p_{\scr C_i/E_i} \coloneqq p_{\scr C_i} - p_{\scr E_i}$, $\forall i\in\mathcal{N}$.  A visual representation of
the contact geometry for the $i$th agent is shown in Fig. \ref{fig:Two-robotic-arms (ACC_rolling)}.  The dynamics of the $i$th agent is given by \cite{murray1994}
\begin{equation} \label{eq:hand dynamics (ACC_rolling)}
B_i \ddot{q}_i + C_{q_i} \dot{q}_i + {g}_{q_i} = -J_{h_i}^\top {f}_{\scr C_i}  + {\tau}_i
\end{equation}
with the dynamics terms as in \eqref{eq:manipulator joint_dynamics (TCST_coop_manip)}, ${f}_{\scr C_i} \in \mathbb{R}^{3}$ is the contact force in \textit{three dimensions}, 
and $J_{h_i} \coloneqq J_{h_i}(q_i,p_{\scr C_i/E_i}) :\mathbb{R}^{n_i+3} \to \mathbb{R}^{3 \times n_i}$ is the  Jacobian matrix to the contact point,  defined by 
\begin{equation*}\label{eq:hand jacobian (ACC_rolling)}
J_{h_i}({q}_i,{p}_{\scr C_i/E_i}) {\coloneqq} \begin{bmatrix}
I_3 & - S({p}_{\scr C_i/E_i})  \end{bmatrix} J_i({q}_i), 
\end{equation*}
where $J_i$ is the manipulator Jacobian that maps $\dot{q}_i \mapsto {v}_i$, defined in the previous section. Note the difference of \eqref{eq:manipulator dynamics (TCST_coop_manip)} and \eqref{eq:hand dynamics (ACC_rolling)} due to the presence of the rolling contacts. Note also that disturbances are not taken into account here for simplicity.

The full hand Jacobian matrix is $J_h \coloneqq J_h(q,p_{\scr EC}) \coloneqq \text{diag}\{[J_{h_i}]_{i\in{N}}\} \in \mathbb{R}^{3N \times n}$, with $p_{\scr EC} \coloneqq [p_{\scr C_1/E_1}^\top,\dots, p_{\scr C_N/E_N}^\top]^\top$ $\in \mathbb{R}^{3N}$.
As before, we consider that the dynamical parameters (masses, moments of inertia) appearing in the terms $B_i$, $C_{q_i}$, ${g}_{q_i}$ are \textit{unknown}, $\forall i\in\mathcal{N}$. The dynamics \eqref{eq:hand dynamics (ACC_rolling)} can be written in  vector form as
\begin{equation} \label{eq:hand dynamics stack (ACC_rolling)}
B\ddot{q} + C_{q}\dot{q} + {g}_q = -J_h^\top f_{\scr C} + \tau,
\end{equation}
where $B \coloneqq \text{diag}\{[B_i]_{i\in\mathcal{N}}\}$, $C_q \coloneqq \text{diag}\{[C_{q_i}]_{i\in\mathcal{N}}\}$ $\in \mathbb{R}^{n\times n}$, ${g}_q \coloneqq  [{g}_{q_1}^\top,\dots$, ${g}_{q_N}^\top]$ $^\top$ $\in\mathbb{R}^{n}$ , ${f}_{\scr C} \coloneqq [{f}_{\scr C_1}^\top,\dots,{f}_{\scr C_N}^\top]^\top \in\mathbb{R}^{3N}$. With a slight abuse of notation, we assume that the set $\mathsf{S}$ ($\mathsf{S}_i$) contains the configurations $q$ ($q_i$) that yield a non-singular $J_h$ $(J_{h_i})$ (in contrast to just $J$ ($J_i$) of the case of rigid contacts)

A common assumption in the majority of the related literature is that the object center of mass is accurately known, which is typically not the case in practice. {We assume here tracking of a traceable point $p_o$ on the object surface instead of the center of mass $p_{\scr O}$, whose information is considered unknown. Note that appropriate sensor equipment, e.g., cameras and markers, can accurately track such points in practice. }
Hence, to remove the dependency on an unknown object center of mass, we perform a standard rigid body transformation to the conventional object dynamics as follows. Let 
${R}_{o} \coloneqq {R}_{o}({\eta}_{\scr o}) : \mathbb{T} \to  \mathbb{SO}(3)$ be the respective rotation matrix of a frame attached to ${p}_o$, and ${x}_{o} := [{p}_{o}^\top, {\eta}_{o}^\top]^\top \in \mathbb{M}$, ${v}_{o} := [\dot{{p}}_o^\top, {\omega}_{o}^\top]^\top {\in\mathbb{R}^6}$ denote the pose and generalized velocity of the object frame, with (without loss of generality) ${\eta}_{o} = \eta_{\scr O}$, $R_{o} = {R}_{\scr O}$, and ${\omega}_{o} = \omega_{\scr O}$. The position vector from ${p}_{o}$ to the respective contact point is ${p}_{\scr C_i/o} \coloneqq p_{\scr C_i} - p_{o} \in \mathbb{R}^3$, $\forall i \in \mathcal{N}$.
Moreover, define $p_{\scr C_i/O} \coloneqq p_{\scr C_i} - p_{\scr O}$, $\forall i\in\mathcal{N}$, and $p_{\scr OC} \coloneqq [p_{\scr C_1/O}^\top,\dots, p^\top_{\scr C_N/O}]^\top$, $p_{\scr oC} \coloneqq [p^\top_{\scr C_1/o},\dots, p^\top_{\scr C_N/o}]^\top$. 

Given the rolling contacts, the conventional object dynamics with respect to the object center of mass are given by the Newton-Euler formulation:
\begin{equation}\label{eq:object dynamics conventional (ACC_rolling)}
M_{\scr O} \dot{v}_{\scr O} + {C}_{\scr O}v_{\scr O} + {g}_{\scr O} = \bar{G}_\mathcal{R} f_{\scr C}
\end{equation}
with the dynamic terms as in \eqref{eq:object dynamics (TCST_coop_manip)}, and 
$\bar{G}_\mathcal{R}:\mathbb{R}^{3N} \to \mathbb{R}^{6 \times 3N}$ is the grasp map, 
defined by $\bar{G}_\mathcal{R} \coloneqq \bar{G}_\mathcal{R}(p_{\scr OC}) \coloneqq [\bar{G}_{\mathcal{R}_1}, ..., \bar{G}_{\mathcal{R}_N}]$ where $\bar{G}_{\mathcal{R}_i}\coloneqq \bar{G}_{\mathcal{R}_i}(p_{\scr C_i/O}):\mathbb{R}^3 \to \mathbb{R}^{6\times 3}$, with
\begin{equation*}\label{eq:grasp map}
\bar{G}_{\mathcal{R}_i}(p_{\scr C_i/O}) \coloneqq \left[ \begin{array}{c} I_3 \\ S({p}_{\scr C_i/O}) \end{array} \right].
\end{equation*}
Note the difference of $\bar{G}_\mathcal{R}$ with respect to the rigid contact-case \eqref{eq:J_o_i_def (TCST_coop_manip)}. 
We now perform a transformation of the aforementioned dynamics to account for ${p}_{o}$. 
Let $J_a \coloneqq J_a(\eta_{\scr O}):\mathbb{T} \to \mathbb{R}^{6\times 6}$ be defined as:
\begin{equation} \label{eq:J_a (ACC_rolling)}
{ J_a({\eta}_{\scr O}) \coloneqq} \begin{bmatrix}
I_3 & S({R}_{\scr O} {p}_{\scr Oo}^{\mathcal{O}} ) \\ {0} & I_3 \end{bmatrix}
\end{equation}
where ${p}_{\scr Oo}^{\mathcal{O}} \coloneqq R_{\scr O}^\top (p_{\scr O} - {p}_{o})$, such that ${v}_{\scr o} = J_a {v}_{\scr O}$.
Note that ${p}_{\scr Oo}^{\mathcal{O}}$ is constant.

Substitution of ${v}_{o} = J_a {v}_{\scr O}$ and left multiplication by $J_a^\top$ in \eqref{eq:object dynamics conventional (ACC_rolling)} yields the adjusted object dynamics {with respect to ${p}_{o}$}:
\begin{equation} \label{eq:object dynamics (ACC_rolling)}
{M}_{o}{\dot{v}}_{o} + {C}_{o}{v}_{o} + {g}_{o} = {G}_{\mathcal{R}} {f}_{\scr C},
\end{equation}
{where 
	\begin{align*}
		{M}_{o}\coloneqq & {M}_{o}(\eta_{\scr O}) \coloneqq J_a^\top M_{\scr O} J_a,\\
		{C}_{o}\coloneqq & {C}_{o}(\eta_{\scr O},\omega_{\scr O}) \coloneqq J_a^\top ( M_{\scr O} \dot{J}_a + C_{\scr O} J_a), \\
		{g}_{o}\coloneqq & {g}_{o}(\eta_{\scr O})\coloneqq  J_a^\top g_{\scr O},	\\
		{G}_{\mathcal{R}} \coloneqq & {G}_{\mathcal{R}}(p_{\scr oC}) \coloneqq [{G}_{\mathcal{R}_1},\dots,{G}_{\mathcal{R}_N}] \coloneqq   J_a^\top \bar{G}_{\mathcal{R}},
	\end{align*}
	 with ${G}_{\mathcal{R}_i} \coloneqq {G}_{\mathcal{R}_i}(p_{\scr C_i/o})$, and
	\begin{align*}
	{G}_{\mathcal{R}_i}(p_{\scr C_i/o}) \coloneqq J_a^\top \bar{G}_{\mathcal{R}_i} =&
	\begin{bmatrix}
	I_3  \\
	- S(R_{\scr O}{p}^{\mathcal{O}}_{\scr Oo}))+ S({p}_{\scr C_i/O})
	\end{bmatrix} 
	=
	\begin{bmatrix}
	I_3  \\
	S(p_{\scr C_i/o})
	\end{bmatrix},
	\end{align*}
	Note that $\bar{G}_\mathcal{R}$ does not depend on ${p}_{\scr O}$. Note also  by the relation ${p}_{\scr O} = {p}_{o} - {R}_{\scr O} {p}^{\mathcal{O}}_{\scr Oo}$,  that ${M}_{o}$, ${C}_{o}$, ${g}_{o}$ are functions of ${\eta}_{o} = \eta_{\scr O}$, ${\omega}_{o} = {\omega}_{\scr O}$ with dependency on the constant but unknown term ${p}^{\mathcal{O}}_{\scr Oo}$. {We also note the following relation that will be needed subsequently:
	\begin{align}\label{eq:G'vo relations (ACC_rolling)}
	\bar{G}_{\mathcal{R}_i}^\top {v}_{\scr O} =&
	\begin{bmatrix}
	I_3  \\
	S({p}_{\scr C_i} - p_{\scr O})
	\end{bmatrix}^\top \begin{bmatrix} I_3 & S(R_{\scr O} {p}_{\scr Oo}^{\mathcal{O}})\\ 0 & I_3 \end{bmatrix} {v}_{o}
	= {G}_{\mathcal{R}_i}^\top {v}_{o}
	\end{align}
}Similarly to the agents, the object dynamic parameters appearing in the terms ${M}_{o}$, ${C}_{o }$, ${g}_{o}$ are considered to be \textit{unknown}.

The more practical consideration of rolling contacts, as opposed to a rigid grasp, requires no slip to occur between the agents and object by ensuring that each contact force remains inside the friction cone defined by:
\begin{equation}\label{eq:friction cone (ACC_rolling)}
\mathcal{F}_{\scr C_i}(\mu_f) \coloneqq \{ {f}^{\scr C_i}_{\scr C_i} \in \mathbb{R}^3 : f_{\scr C,n_i} \mu_f \geq \sqrt{f_{\scr C,x_i}^2 + f_{\scr C,y_i}^2} \}
\end{equation}
where ${f}^{\mathcal{C}_i}_{c_i} \coloneqq R_{\scr C_i}^\top f_{\scr C_i} =:   (f_{\scr C,x_i}, f_{\scr C,y_i}, f_{\scr C,n_i}) $ is the $i$th contact force written in frame $\mathcal{C}_i$, whose orientation is described by $R_{\scr C_i}\coloneqq R_{\scr C_i}(\eta_{\scr C_i}):\mathbb{T}\to\mathbb{SO}(3)$, $\forall i\in\mathcal{N}$, ($\eta_{\scr C_i}\in\mathbb{T}$ being the respective Euler-angle orientation), with tangential force components $f_{\scr C,x_i}, f_{\scr C,y_i} \in \mathbb{R}$ and normal force component $f_{\scr C,n_i} \in \mathbb{R}$, $\mu_f \in \mathbb{R}_{>0}$ is the friction coefficient. The full friction cone is the Cartesian product of all the friction cones: $\mathcal{F}_{\scr C} \coloneqq \mathcal{F}_{\scr C_1} \times ... \times \mathcal{F}_{\scr C_n}$. 

In practice, it is common to approximate the friction cone by an inscribed pyramid with $l_f \in\mathbb{R}_{ > 0}$ sides. The set associated with this pyramid is defined as 
\begin{equation} \label{eq:pyramid (ACC_rolling)}
	\widetilde{\mathcal{F}}_{\scr C_i}(\mu_f) \coloneqq \{ f^{\scr C_i}_{\scr C_i} \in \mathbb{R}^3 : \Lambda_i(\mu_f) f^{\scr C_i}_{\scr C_i} \succeq 0 \},
\end{equation}
where $\Lambda_i(\mu_f) \in \mathbb{R}^{l_f\times 3}$. The overall friction pyramid is then $\widetilde{\mathcal{F}}_{\scr C}(\mu_f) \coloneqq \{ f^{\scr C}_{\scr C} \in\mathbb{R}^{3N} : \Lambda(\mu_f) f^{\scr C}_{\scr C} \succeq 0 \}$, where $f^{\scr C}_{\scr C} \coloneqq [(f^{\scr C_1}_{\scr C_1})^\top,\dots, (f^{\scr C_N}_{\scr C_N})^\top]^\top$, and $\Lambda \coloneqq \text{diag}\{ [\Lambda_i]_{i\in\mathcal{N}} \}$.

When the contact points do not slip, the grasp relation $J_h \dot{q} = \bar{G}_{\mathcal{R}}^\top {v}_{\scr O}$ holds \cite{cole1989kinematics}, which, after substituting \eqref{eq:G'vo relations (ACC_rolling)}, becomes:
\begin{equation} \label{eq:contact vel compact (ACC_rolling)}
{v}_{\scr C}  \coloneqq J_h \dot{q} = {G}_{\mathcal{R}}^\top {v}_{o},
\end{equation}
where ${v}_{\scr C} \coloneqq [{v}_{\scr C_1}^\top,\dots,{v}_{\scr C_N}^\top]^\top \in \mathbb{R}^{3n}$ is the vector of contact velocities. 

{As in Section \ref{subsec:Quaternion Controller (TCST_coop_manip)}, we use for the object orientation the unit quaternion choice ${\zeta}_{\scr O} \coloneqq [\varphi_{\scr O},{\epsilon}_{\scr O}^\top]^\top \in \mathbb{S}^3$}.
Let hence now a desired pose trajectory, {$p_{\textup{d}}:\mathbb{R}_{\geq 0} \to \mathbb{R}^3$}, ${\zeta}_{\textup{d}} \coloneqq [\varphi_{\textup{d}},\epsilon_{\textup{d}}^\top]^\top :\mathbb{R}_{\geq 0} \to \mathbb{S}^3$, to be tracked by ${x}_{o}$.
To that end, similar to Section \ref{subsec:Quaternion Controller (TCST_coop_manip)}, we define the position error ${e}_{p_o} \coloneqq {p}_{o}-{p}_{\textup{d}}$ as well as the quaternion product ${e}_\zeta \coloneqq {\zeta}_{\textup{d}} \cdot {\zeta}_{\scr O}^+$. The aim is then to regulate ${e}_{p_o}$ to zero and ${e}_\zeta$ to $[\pm 1,{0}^\top]^\top$. Moreover, we aim {at ensuring} that the agents are always in contact with the object and slipping is avoided.
{Formally, the problem is defined as follows.}
%
\begin{problem} \label{problem:main (ACC_rolling)}
	Given a desired bounded, smooth object pose trajectory defined by ${p}_{\textup{d}}:\mathbb{R}_{\geq 0}\to\mathbb{R}^{3}$, ${\zeta}_{\textup{d}}:\mathbb{R}_{\geq 0}\to\mathbb{S}^3$, with bounded first and second derivatives, as well as uncertain agent and object dynamic parameters involved in \eqref{eq:hand dynamics (ACC_rolling)} and \eqref{eq:object dynamics (ACC_rolling)}, respectively, determine a control law $\tau$ in \eqref{eq:hand dynamics stack (ACC_rolling)} such that the following conditions hold: 
	\begin{enumerate}
		\item $\lim_{t\to \infty} \left({e}_{p_o}(t), {e}_\zeta(t) \right) = \left({0}, [\pm 1, {0}^\top ]^\top \right)$		
		\item {${f}^{\scr C_i}_{\scr C_i}(t) \in \mathcal{F}_{\scr C_i}, \forall t > 0$, $i\in\mathcal{N}$.}
	\end{enumerate}
\end{problem}

In order to solve the aforementioned problem the following assumptions are made for the grasp:
\begin{assumption}\label{asm:full rank G (ACC_rolling)}
	The grasp consists of $N \geq 3$ agents with non-collinear contact points and $\textup{Null}(\bar{G}_{\mathcal{R}}) \bigcap {\textup{Int}(\mathcal{F}_{\scr C})} \neq \emptyset$.
\end{assumption}

\begin{assumption}\label{asm:nonsingular, no exessive rolling (ACC_rolling)}
	{The matrix $J_h(q)$ is non-singular, and the contact points do not exceed the fingertip surface.}
\end{assumption}

%

\begin{remark} \label{rm:Jh to slip (ACC_rolling)}
	{Note that $N\geq 3$ agents with non-collinear contact points ensures $\bar{G}_{\mathcal{R}}$ is full row rank \cite{murray1994}. The condition that $\textup{null}(\bar{G}_\mathcal{R}) \bigcap {\textup{Int}(\mathcal{F}_{\scr C})} \neq \emptyset$ ensures the existence of a contact force that lies within the friction cone and yields a desired object wrench, which is called the force-closure condition} \cite{cole1989kinematics}. Force-closure depends on the initial grasp, and can be ensured by existing high-level grasp planning methods \cite{Hang2016}. Moreover, by incorporating optimization techniques, as e.g. in \cite{shaw2018grasp}, we can enforce prevention of excessive rolling of the contacts and thus relax the respective part of Assumption \ref{asm:nonsingular, no exessive rolling (ACC_rolling)}.   {Finally, the non-singular condition of $J_h$ intuitively implies that tracking the desired reference trajectory does not force the agents through such singular configurations (such an assumption was also considered in the case of rigid grasps). 
	This can also be achieved by exploiting internal motions of redundant agents ($n_i > 3$).}
\end{remark}

We also assume that the contact vectors $R_{i}^\top {p}_{\scr C_i/E_i}$ and their derivatives are measured accurately online, $\forall i\in\mathcal{N}$. This can be achieved either by the use of appropriate tactile sensors {or forward simulation of the contact dynamics \cite{murray1994}}. By also assuming the geometry of the fingertips known, we can also compute $\eta_{\scr C_i}$ and hence $R_{\scr C_i}$ online, $\forall i\in\mathcal{N}$.
Finally, note that $B_i(\cdot)$ are positive definite, and $\dot{B}_i(\cdot) - 2C_{q_i}(\cdot)$ are skew-symmetric, $\forall i\in\mathcal{N}$, similarly to $M_i(\cdot)$, and $\dot{M}_i(\cdot) - 2C_i(\cdot)$. 

%

In the following, we present two \textit{adaptive} control schemes for the solution of Problem \ref{problem:main (ACC_rolling)}, a centralized one, where one computer unit (or a ``leader" agent) computes the input commands for the entire team, as well as a decentralized one, based on event-triggered inter-agent communication.

\subsection{Centralized Scheme}

This section presents the centralized proposed control scheme, which employs adaptive control techniques for the compensation of the dynamic uncertainties of the agent and object present in the problem setup.

Without loss of generality, we assume that $n_i =3$, $\forall i\in\mathcal{N}$, i.e., the agents are not redundant. The proposed solution can be trivially extended to redundant cases, e.g., by following the analysis of \cite[Chapter 6]{murray1994}.
By combining the agent and object dynamics \eqref{eq:hand dynamics stack (ACC_rolling)}, \eqref{eq:object dynamics (ACC_rolling)} as well as \eqref{eq:contact vel compact (ACC_rolling)}, we can obtain the coupled dynamics 
\begin{equation} \label{eq:coupled dynamics (ACC_rolling)}
{\widetilde{B}}{\dot{v}}_{o} + \widetilde{C}_q{v}_{o} + \widetilde{g}_q  = {G}_\mathcal{R} J_h^{-T}\tau,
\end{equation}
where 
\begin{align*}
	\widetilde{B} \coloneqq & \widetilde{B}(\check{x}) \coloneqq {M}_{o} + {G}_\mathcal{R} J_h^{-T} B J_h^{-1} {G}_\mathcal{R}^\top , \\
	\widetilde{C}_q \coloneqq & \widetilde{C}_q(\check{x},\dot{\check{x}}) \coloneqq {C}_{o} + {G}_{\mathcal{R}} J_h^{-T} \big( C_q J_h^{-1} {G}_\mathcal{R}^\top  + B \frac{d}{dt}(J_h^{-1}{G}_\mathcal{R}^\top ) \big), \\
	\widetilde{g}_q \coloneqq & \widetilde{g}_q(\check{x}) \coloneqq {g}_{o} + {G}_{\mathcal{R}} J_h^{-T} g_q,
\end{align*}
and $\check{x} \coloneqq [ \eta_{\scr O}^\top ,{q}^\top ,p_{\scr EC}^\top ,{p}_{\scr oC}^\top ]^\top  \in \mathbb{T} \times \mathsf{S}\times\mathbb{R}^{6N}$. 
{The following lemma states useful properties of \eqref{eq:coupled dynamics (ACC_rolling)}:}
\begin{lemma} \label{lem:coupled dynamics skew symmetry (ACC_rolling)}
	The matrix $\widetilde{B}$ is symmetric and positive-definite, and the matrix $\dot{\widetilde{B}} - 2 \widetilde{C}_q$ is skew-symmetric.
\end{lemma}
\begin{proof}
	The proof is similar to the one of Lemma \ref{lem:coupled dynamics skew symmetry (TCST_coop_manip)} and is omitted.
\end{proof}
Next, we proceed to parameterizing the dynamics with respect to constant but unknown dynamic parameters, similarly to the case of rigid contacts. 
In particular, the left-hand side of the object dynamics (with respect to ${p}_{o}$) is  parameterized as:
\begin{equation*}
{M}_{o}({\eta}_{o})\dot{{v}}_{o} + {C}_{o}({\eta}_{o},{\omega}_{o}){v}_{o} + g_{o} = Y_{\mathcal{R}_{o}}({\eta}_{o},{\omega}_{o},{v}_{o},\dot{{v}}_{o}){\vartheta}_{\mathcal{R}_{o}},
\end{equation*}
where ${\vartheta}_{\mathcal{R}_{o}} \in \mathbb{R}^{\ell_{\mathcal{R}_{o}}}$, ${\ell_{\mathcal{R}_{o}}} \in\mathbb{N}$, is a vector containing the unknown object dynamic parameters, similarly to $\vartheta_{\scr O}$ defined in Section \ref{subsec:Quaternion Controller (TCST_coop_manip)}, but also including the term ${p}^{\mathcal{O}}_{\scr Oo}$, introduced in \eqref{eq:J_a (ACC_rolling)}, and 
$Y_{\mathcal{R}_{o}} : \mathbb{T}\times \mathbb{R}^{15} \to \mathbb{R}^{6\times \ell_{\mathcal{R}_o}}$ is the respective (known) regressor matrix.
Similarly, the part of \eqref{eq:coupled dynamics (ACC_rolling)} that concerns the robotic agents can be linearly parameterized {as}:
\begin{align*}
	B_i J_{h_i}^{-1} {G}_{\mathcal{R}_i}^\top \dot{{v}}_{o} + \left(B_i\frac{\partial }{\partial t}\left(J_{h_i}^{-1} {G}_{\mathcal{R}_i}^\top \right)  +  C_{q_i} J_{h_i}^{-1} {G}_{\mathcal{R}_i}^\top \right){v}_{o} &+ g_{q_i} = \\
	& Y_{\mathcal{R}_i}(\check{x}_i, \dot{\check{x}}_i, {v}_{o}, \dot{{v}}_{o}) \vartheta_{\mathcal{R}_i},
\end{align*}
with $\check{x}_i \coloneqq [\eta_{o}^\top, q_i^\top, p^\top_{\scr C_i/E_i}, p^\top_{\scr C_i/o}]^\top \in \mathbb{T} \times \mathsf{S}_i \times \mathbb{R}^6$, $Y_{\mathcal{R}_i}: \mathbb{T} \times \mathsf{S}_i \times \mathbb{R}^{30} \to \mathbb{R}^{3\times\ell_\mathcal{R}}$ being agent $i$'s regressor matrix, and $\vartheta_{\mathcal{R}_i} \in \mathbb{R}^{\ell_{\mathcal{R}}}$, $\ell_{\mathcal{R}}\in\mathbb{N}$ the respective vector of unknown, constant parameters. The aforementioned parameterization is written in vector form:
\begin{align*}
& B J_h^{-1} {G}_{\mathcal{R}}^\top  \dot{{v}}_{o}  + \left(B\frac{\partial }{\partial t}( J_h^{-1} {G}_{\mathcal{R}}^\top ) +  C_q J_h^{-1} G_{\mathcal{R}}^\top  \right){v}_{o} + g_q = Y_\mathcal{R}(\check{x},\dot{\check{x}},{v}_{o},\dot{{v}}_{o}) {\vartheta_{\mathcal{R}}},
\end{align*}
where $Y_\mathcal{R} \coloneqq Y_\mathcal{R}(\check{x},\dot{\check{x}},{v}_{o},\dot{{v}}_{o})$ $\coloneqq$ $\textup{diag}\{ [Y_{\mathcal{R}_i}]_{i\in\mathcal{N}}\}$ $\in$ $\mathbb{R}^{3N \times \ell_{\mathcal{R}}}$, and $\vartheta_{\mathcal{R}}$ $\coloneqq$ $[\vartheta_{\mathcal{R}_1}^\top,\dots,\vartheta_{\mathcal{R}_N}^\top]^\top$ $\in$ $\mathbb{R}^{N\ell_{\mathcal{R}}}$.

Therefore, the left-hand side of the coupled dynamics \eqref{eq:coupled dynamics (ACC_rolling)} can be written as 
\begin{align} \label{eq:regressor dynamics (ACC_rolling)}
& \widetilde{B}\dot{{v}}_o +  \widetilde{C}_q {v}_o + \widetilde{g}_q =    Y_{\mathcal{R}_{o}}({\eta}_{o},{\omega}_{o},{v}_{o},\dot{{v}}_{o})\vartheta_{\mathcal{R}_o}+ {G}_{\mathcal{R}} J_h^{-T} Y_{\mathcal{R}} (\check{{x}},\dot{\check{{x}}},{v}_{o},\dot{{v}}_{o}) {\vartheta_{\mathcal{R}}}
\end{align}

Let now $\hat{\vartheta}_\mathcal{R} \in\mathbb{R}^{N\ell_{\mathcal{R}}}$, $\hat{\vartheta}_{\mathcal{R}_o} \in \mathbb{R}^{\ell_{\mathcal{R}_{o}}}$, be the estimates of ${\vartheta}_\mathcal{R}$ and $\vartheta_{\mathcal{R}_{o}}$, respectively, by the agents, and the respective errors 
$e_{\mathcal{R}_\vartheta} \coloneqq \hat{\vartheta}_\mathcal{R} - \vartheta_\mathcal{R}$, and ${e}_{\mathcal{R}_\vartheta,o} \coloneqq \hat{\vartheta}_{\mathcal{R}_{o}} -{\vartheta}_{\mathcal{R}_{o}}$.

We provide next the proposed control protocol. 
First, we design the reference velocity signal ${v}_{f_o}\in\mathbb{R}^6$ and the associated velocity error ${e}_{v_o}$ as 
\begin{subequations}
	\begin{align} \label{eq:v_f and e_v (ACC_rolling)}
	{v}_{f_o} &:= v_{\textup{d}} - K e_{\mathcal{R}} \coloneqq \begin{bmatrix}
	\dot{p}_{\text{d}}  \\ {\omega}_{\textup{d}} \end{bmatrix} -  \begin{bmatrix} {k}_p {e}_{p_o} \\ - {k}_\eta \displaystyle \frac{{e}_\epsilon}{e_\varphi^3}  \end{bmatrix} \\
	{e}_{v_o} & \coloneqq {v}_{o} - {v}_{f_o},
	\end{align}
\end{subequations}
where ${K} = \text{diag}\{{k}_p I_3, {k}_\eta I_3\}\in\mathbb{R}^3$ is the positive definite gain matrix used in \eqref{eq:v_f (TCST_coop_manip)},$e_\mathcal{R} \coloneqq [{e}^\top _{p_o}, -\frac{{e}^\top _\epsilon}{e_\varphi^3}]^\top $, and ${v}_{\text{d}} \coloneqq v_{\text{d}}(t) \coloneqq [\dot{p}_{\text{d}}^\top , {\omega}_{\text{d}}^\top ]^\top $. Note the difference in the definition of $e_\mathcal{R}$ and $e$ from \eqref{eq:v_f (TCST_coop_manip)}, which will account tot stabilizing the scalar quaternion error $e_\varphi$ to either $1$ or $-1$, depending on $e_\varphi(0)$, while guaranteeing that $e_\varphi(t) \neq 0$, $\forall t\geq 0$ (provided that $e_\varphi(0) \neq 0$), and rendering thus \eqref{eq:v_f and e_v (ACC_rolling)} well defined.

We design now the control protocol as $\tau: \mathcal{T}_\mathcal{R} \times \mathbb{R}_{\geq 0} \to \mathbb{R}^n$, with
\begin{align} \label{eq:control law stack (ACC_rolling)}
\tau\coloneqq \tau(\chi_\mathcal{R},t) =& Y_r{\hat{\vartheta}_\mathcal{R}} +  J_h^\top  ({G_\mathcal{R}^\dagger} {f}_{\text{d}} + {f}_{\text{int}} ), 
\end{align}
where $\chi_\mathcal{R}\coloneqq [\check{x}^\top,\dot{\check{x}}^\top, e_{\mathcal{R}}^\top, e_{v_o}^\top, \hat{\vartheta}_\mathcal{R}^\top, \hat{\vartheta}_{\mathcal{R}_o}^\top,\eta_{\scr C}^\top]^\top$, $\mathcal{T}_\mathcal{R} \coloneqq \{\chi_\mathcal{R}\in \mathbb{T}^{N+1} \times \mathsf{S} \times \mathbb{R}^{15N+15+ N\ell_\mathcal{R}+\ell_{\mathcal{R}_o}}:  e_\varphi \neq 0\}$, $\eta_{\scr C}\coloneqq [\eta_{\scr C_1}^\top,\dots,\eta_{\scr C_N}^\top]^\top$,
{$G_\mathcal{R}^\dagger$ is the Moore-Penrose pseudoinverse of $G_\mathcal{R}$,} ${f}_{\text{d}} \coloneqq Y_{o_r} \hat{\vartheta}_{\mathcal{R}_o} - {e_\mathcal{R}} - K_v {e}_{v_o}$ with $K_v\in\mathbb{R}^{6\times 6}$ the positive definite gain matrix used in \eqref{eq:control laws adaptive quat (TCST_coop_manip)}, $Y_r \coloneqq Y_\mathcal{R}(\check{{x}},\dot{\check{{x}}},{v}_{f_o}, \dot{v}_{f_o})$, $Y_{o_r} \coloneqq Y_{\mathcal{R}_o}({\eta}_{o},{\omega}_{o},{v}_{f_o}, \dot{v}_{f_o})$,
and ${f}_{\text{int}} \coloneqq f_{\text{int}}(q,\eta_{\scr C}): \mathsf{S}\times\mathbb{T}^N \to \mathbb{R}^{3N}$ is a term in the nullspace of $G_\mathcal{R}$ to prevent contact slip, {which will be designed later}. Moreover, we design the adaptation signals 
\begin{subequations} \label{eq:adaptation laws stack (ACC_rolling)}
	\begin{align}
	\dot{{\hat{\vartheta}}}_{\mathcal{R}} &= {\text{Proj}({\hat{\vartheta}_\mathcal{R}},{- \Gamma} Y_r^\top  J_h^{-1} G_\mathcal{R}^\top {e}_{v_o}),  }\\ 
	\dot{{\hat{\vartheta}}}_{\mathcal{R}_o} &= \text{Proj}({\hat{\vartheta}_{\mathcal{R}_o}}, {-\Gamma_o} Y_{o_r}^\top  {e}_{v_o}), 
	\end{align}
\end{subequations}
where {$\Gamma \in \mathbb{R}^{N\ell_\mathcal{R}\times N\ell_\mathcal{R}}, \Gamma_o \in\mathbb{R}^{\ell_{\mathcal{R}_o}\times \ell_{\mathcal{R}_o}}$} are constant positive definite gain matrices (as in \eqref{eq:adaptation laws (TCST_coop_manip)}), and {$\text{Proj}()$ is the projection operator, which satisfies}
\cite{lavretsky13adaptive}:
\begin{equation} \label{eq:proj property (ACC_rolling)}
(\hat{y} - {y})^\top  (W^{-1}\text{Proj}({y},W{z}) - {z}) \leq 0,
\end{equation}
for any symmetric positive definite $W\in \mathbb{R}^{\ell_z\times \ell_z}$, and  $\forall \hat{y}, {y}, {z} \in\mathbb{R}^{\ell_z}$, for some $\ell_z \in\mathbb{N}$. 
{Moreover, by appropriately choosing the initial conditions of the estimates ${\hat{\vartheta}_\mathcal{R}}(0)$, ${\hat{\vartheta}_{\mathcal{R}_o}}(0)$, 
	we guarantee via the projection operator that ${\hat{\vartheta}_\mathcal{R}}(t)$, ${\hat{\vartheta}_{\mathcal{R}_o}}(t)$ will stay uniformly bounded in predefined sets defined by finite constants $\bar{\vartheta}_\mathcal{R}$, $\bar{\vartheta}_{\mathcal{R}_o}$, i.e.,
	$\|\hat{\vartheta}_\mathcal{R}(t)\| \leq \bar{{\hat{\vartheta}}}_\mathcal{R}$, $\|\hat{\vartheta}_{\mathcal{R}_o}(t)\| \leq \bar{\hat{\vartheta}}_{\mathcal{R}_o}$, $\forall t\geq 0$. Hence, we can achieve the boundedness of the respective errors as} 
\begin{subequations}\label{eq:e_nu bars (ACC_rolling)}
	\begin{align} 
	\|{e}_{\vartheta_\mathcal{R}}(t) \| &\leq \bar{e}_{\vartheta_\mathcal{R}} \coloneqq  \bar{\vartheta}_\mathcal{R} + \|\vartheta_\mathcal{R}\| \\
	\|{e}_{\vartheta_{\mathcal{R},o}}(t) \| &\leq \bar{e}_{\vartheta_{\mathcal{R},o}} \coloneqq  \bar{\vartheta}_{\mathcal{R}_o} + \|\vartheta_{\mathcal{R}_o}\|.
	\end{align}
\end{subequations}
More details can be found in \cite[Chapter 11]{lavretsky13adaptive}.

{
	We design next the internal force component ${f}_{\text{int}}$ to guarantee slip prevention.
	Slip is addressed by ensuring the contact forces remain inside the friction cone as specified in \eqref{eq:friction cone (ACC_rolling)}. From \eqref{eq:pyramid (ACC_rolling)}, we have to guarantee that  $\Lambda_i(\mu_f) R_{\scr C_i}^\top  f_{\scr C_i} \succeq 0, \forall i\in\mathcal{N}$, or in vector form, 
	\begin{equation}\label{eq:linearized friction constraint (ACC_rolling)}
	\Lambda(\mu_f) R_{\scr C}^\top  {f}_{\scr C} \succeq 0,
	\end{equation}
	where $R_{\scr C}\coloneqq  R_{\scr C}(\eta_{\scr C}) \coloneqq \text{diag}\{[R_{\scr C_i}]_{i\in\mathcal{N}}\}$.
	
	The design of the internal force component, ${f}_{\text{int}}$, to ensure \eqref{eq:linearized friction constraint (ACC_rolling)} is performed as follows. First, ${f}_{\text{int}}$ must be in the nullspace of $G_\mathcal{R}$, i.e., $G_\mathcal{R} {f}_{\text{int}} = 0$. Second, the internal force must satisfy \eqref{eq:linearized friction constraint (ACC_rolling)}. Third, the normal component of the internal force with respect to the contact plane must always be positive (i.e. the manipulators cannot ``pull" on the contact point). To enforce this condition we design\footnote{We use the notation ${f}_{\text{int}} = [{f}_{\text{int}_1}^\top ,\dots,{f}_{\text{int}_N}^\top ]^\top$.} ${f}_{\text{int}_i} =   f'_{\text{int}} R_{\scr C_i} {\ell}_{\text{int},i}$, where ${\ell}_{\text{int},i}\coloneqq [\ell_{\text{int},i_x},\ell_{\text{int},i_y},\ell_{\text{int},i_z}]^\top $ is the internal force direction in the contact frame $\mathcal{C}_i$, $i \in \mathcal{N}$, and {$f'_{\text{int}} \in\mathbb{R}_{>0}$ is a gain parameter to be designed}. Without loss of generality let $\ell_{\text{int},i_z}$ be aligned with the normal direction of the contact frame such that  $\ell_{\text{int},i_z} > 0$, $i \in \mathcal{N}$, ensures that only pushing forces are applied at each contact. Satisfaction of the aforementioned conditions is done by solving the following convex quadratic program to define the internal force controller 
	\begin{subequations} \label{eq:quadratic program internal forces (ACC_rolling)}
		\begin{equation}
		{f}_{\text{int}} = f'_ {\text{int}} R_{\scr C} {\ell^*_{\textup{int}}}
		\end{equation}
		\begin{align} 
		\ell_{\textup{int}}^* =  & \text{argmin}_{{\ell}_{\textup{int}} } \left\{ \sum_{i\in\mathcal{N}} \ell_{\textup{int},i_x}^2 + \ell_{\textup{int},i_y}^2 + \ell_{\textup{int},i_z}^2 \right\} \\ 
		&\text{ s. t. } \\
		& G_\mathcal{R} R_{\scr C} \ell_{\textup{int}}= {0}, \\
		& \ell_{\textup{int},i_z} > 0, \ \ \ \ \   \forall i\in\mathcal{N}, \\
		& \Lambda_i(\mu_f) {\ell}_{\textup{int},i} \succ 0, \ \forall i\in\mathcal{N},
		\end{align}
	\end{subequations}
	where ${\ell}_\textup{int} \coloneqq [{\ell}_{\textup{int},1}^\top ,\dots, {\ell}_{\textup{int},N}^\top ]^\top $. Note that, since the contact points form a force-closure configuration, \eqref{eq:quadratic program internal forces (ACC_rolling)} always has a feasible solution. 
	
	Finally, to satisfy \eqref{eq:linearized friction constraint (ACC_rolling)}, ${f}_{\text{int}}$ must apply sufficient force inside the friction cone to reject perturbations that will arise during the manipulation motion that can push the contact force outside of the friction cone. Rejection of these perturbations is performed by designing the gain $f'_{\text{int}}$ as follows. For simplicity we define the terms ${k}_{\textup{int}} \coloneqq \Lambda(\mu_f) R_{\scr C}^\top  {G^\dagger_{\mathcal{R}}} {f}_{\text{d}}, {l}_{\textup{int}} = \Lambda(\mu_f) {\ell^*_{\text{int}}}$,
	and we denote by $k_{\textup{int},j}$ and $l_{\textup{int},j}$ the $j$th scalar element of ${k}_{\textup{int}}$ and ${l}_{\textup{int}}$ respectively for $j\in\{1,\dots,Nl_f\}$. 
	
	Noting that $\Lambda(\mu_f){\ell^*}_{\textup{int}} \succ 0$ from \eqref{eq:quadratic program internal forces (ACC_rolling)}, we define
	the \textit{decreasing} function $\kappa_{\textup{int}} : \mathbb{R} \to \mathbb{R}_{\geq 0}$ as 
	\begin{equation*}
	\kappa_\textup{int}(x) \coloneqq \begin{cases} 
	-x, \hspace{5mm} \text{if } x \leq -1, \\
	q_{\textup{int}}(x), \hspace{3.3mm} \text{if } -1 \leq x \leq 0,  \\
	0, \hspace{8mm} \text{if } x \geq 0
	\end{cases},
	\end{equation*}
	where $q_\textup{int}(x) \geq 0$, $\forall x\in[-1,0]$, is an appropriate polynomial that ensures continuous differentiability of $\kappa_{\textup{int}}$, for instance $q_{\textup{int}}(x) = x^3 + 2x^2$. Then one can verify that  $\kappa_{\textup{int}}(x) + 1 \geq -x$, $\forall x \in \mathbb{R}$.
	We now design the magnitude scaling for the internal forces as 
	\begin{equation} \label{eq:f_int scalar (ACC_rolling)}
	f'_{\text{int}} = \frac{\kappa(\min_{j}\{k_{\textup{int},j}\}) + 1 + \epsilon_f}{\min_j\{ l_{\textup{int},j} \}},
	\end{equation}
	where $\epsilon_f \in \mathbb{R}_{>0}$ is a tuning gain. The intuition behind \eqref{eq:f_int scalar (ACC_rolling)} is to upper bound elements of the control and the system dynamics to prevent either from pushing the contact force outside of the friction cone. The term $\kappa(\min_{j}\{k_j\}) + 1$ cancels out any effects from ${f}_\textup{d}$. The term $\epsilon_f$ handles the system dynamics, which is guaranteed to be bounded in the following theorem.
	
	{\begin{remark}
			The internal force control presented here accounts for the dynamics of the system by appropriately scaling $f'_{\text{int}}$, which rejects perturbations from causing slip. However, as opposed to \cite{ShawCortez2018b}, we relax the condition that $\epsilon_f$ must upper bound all of the dynamics terms by exploiting knowledge of the applied controller via the term $\kappa_{\textup{int}}(\min_{j}\{k_{\textup{int},j}\})$. This reduces the amount of squeezing force applied to prevent crushing the object.
		\end{remark}}
	}
	The stability and slip prevention guarantees of the proposed controller are presented in the following theorem.
	
	\begin{theorem} \label{th:main th (ACC_rolling)}
		Consider $N$ robotic agents in contact with an object, described by the dynamics \eqref{eq:hand dynamics stack (ACC_rolling)}, \eqref{eq:object dynamics (ACC_rolling)}, and suppose Assumptions \ref{asm:full rank G (ACC_rolling)} and \ref{asm:nonsingular, no exessive rolling (ACC_rolling)} hold. a the desired object pose $[{p}_{\textup{d}}^\top, {\eta}^\top_{\textup{d}}]^\top:\mathbb{R}_{\geq 0} \to \mathbb{R}^3\times \mathbb{S}^3$ be bounded with bounded first and second derivatives. Moreover, assume that $e_\varphi(0) \neq 0$ and ${f}^{\scr C_i}_{\scr C_i}(0) \in \textup{Int}(\mathcal{F}_{\scr C_i}(\mu_f))$, $\forall i\in\mathcal{N}$.
		Then, the control protocol \eqref{eq:v_f and e_v (ACC_rolling)}-\eqref{eq:f_int scalar (ACC_rolling)} guarantees that $ \text{lim}_{t\to \infty} \left({e}_{p_o}(t), {e}_\eta(t) \right) = \left({0}, [\pm 1, {0}^\top  ]^\top  \right)$,
		as well as boundedness of all closed-loop signals. Moreover, by choosing a sufficiently large $\epsilon_f$ in \eqref{eq:f_int scalar (ACC_rolling)}, it holds that ${f}^{\scr C_i}_{\scr C_i}(t) \in \mathcal{F}_{\scr C_i}, \forall t > 0$, $i\in\mathcal{N}$.
	\end{theorem}
	\begin{proof}
		
		Consider the stack vector state ${\chi} \coloneqq [{e}_{p_o}^\top , {e}_\epsilon^\top , {e}_{v_o}^\top , {e}_{\vartheta_{\mathcal{R}}}^\top , {e}_{\vartheta_{\mathcal{R}},o}^\top ]^\top  \in \mathcal{X} \coloneqq \mathbb{R}^{12+N\ell_\mathcal{R}+\ell_{\mathcal{R}_o}}$. Next, note by 
		\eqref{eq:hand dynamics stack (ACC_rolling)}, \eqref{eq:object dynamics conventional (ACC_rolling)}, and \eqref{eq:contact vel compact (ACC_rolling)} that, when ${f}_{\scr C_i}^{\scr C_i} \in \mathcal{F}_{\scr C_i}$, each ${f}_{\scr C_i}^{\scr C_i}$ can be written as a function of the stack state, i.e., ${f}_{\scr C_i}^{\scr C_i}= {f}_{\scr C_i}^{\scr C_i}({\chi})$, $\forall i\in\mathcal{N}$. 
		Consider also the set 
		\begin{align*}
		\mathcal{U} \coloneqq   \{ {\chi} & \in \mathcal{X} : \|{e}_\epsilon \| < \bar{e}_\epsilon, \|{e}_{p_o}\| < \bar{e}_{p_o}, \|{e}_{v_o}\| < \bar{e}_{v_o},   \|{e}_{\vartheta_\mathcal{R}}\| < \tilde{e}_{\vartheta_\mathcal{R}}, \\
		& \|{e}_{\vartheta_{\mathcal{R},o}}\| < \tilde{e}_{\vartheta_{\mathcal{R},o}},   {f}_{\scr C_i}^{\scr C_i}({\chi}) \in {\textup{Int}({\mathcal{F}}_{\scr C_i})}, \forall i\in\mathcal{N} \},
		\end{align*}
		{for some positive} constants $\bar{e}_\epsilon$, $\bar{e}_{v_o}$, $\bar{e}_{p_o}$ satisfying $\|e_\epsilon(0)\| \leq \bar{e}_\epsilon$, $\|{e}_{v_o}(0)\| < \bar{e}_{v_o}$, $\|{e}_{p_o}(0)\| < \bar{e}_{p_o}$, and $\tilde{e}_{\vartheta_\mathcal{R}}, \tilde{e}_{\mathcal{R},o}$ larger than $\bar{e}_{\vartheta_\mathcal{R}}, \bar{e}_{\vartheta_{\mathcal{R},o}}$, respectively, which were introduced in \eqref{eq:e_nu bars (ACC_rolling)}. Note that ${\chi}(0) \in \mathcal{U}$. Next,
		by using \eqref{eq:control law stack (ACC_rolling)} and \eqref{eq:adaptation laws stack (ACC_rolling)}, one obtains the closed-loop dynamics $\dot{{\chi}} = {h}_\chi({\chi},t)$, where ${h}_\chi : \mathcal{X} \times \mathbb{R}_{\geq 0} \to \mathcal{X}$ is a function that is continuous in $t$ and locally Lipschitz in ${\chi}$. 
		Then, according to {Theorem \ref{thm:ode solution (App_dynamical_systems)} } of Appendix \ref{app:dynamical systems}, there exists a positive time constant $t_{\max} > 0$ and a unique solution ${\chi}:[0,t_{\max}) \to \mathcal{U}$, i.e., defined for  $[0,t_{\max})$ and satisfying ${\chi}(t) \in \mathcal{U}$, $\forall t\in[0,t_{\max})$. Hence, slip is prevented and the dynamics \eqref{eq:coupled dynamics (ACC_rolling)} are well-defined, for $t\in[0,t_{\max})$.
		
		Let now the Lyapunov function 
		\begin{align} \label{eq:Lyapunov (ACC_rolling)}
		V_f \coloneqq \frac{1}{2}{e}_{p_o}^\top  {e}_{p_o} + \frac{2}{e_\varphi^2} + \frac{1}{2}{e}_{v_o}^\top  \widetilde{B} {e}_{v_o}
		+ \frac{1}{2} {e}_{\vartheta_\mathcal{R}}^\top \Gamma^{-1}  {e}_ {\vartheta_\mathcal{R}} +   \frac{1}{2} {e}_{\vartheta_{\mathcal{R},o}}^\top  \Gamma_o^{-1} {e}_{\vartheta_{\mathcal{R},o}}.
		\end{align}
		Since $e_\varphi(0) \neq 0$, it holds that $V_f(0) \leq \bar{V}_{f0}$ for a finite positive $\bar{V}_{f0}$. 
		Differentiation of $V_f$  results in:
		\begin{align*}
		\dot{V}_f 
		= &{e}_\mathcal{R}^\top  ({v}_o - {v}_{\text{d}}) + \frac{1}{2} {e}_{v_o}^\top  \dot{\widetilde{B}} {e}_{v_o} + {e}_{v_o}^\top  ( -\widetilde{C}_q {v}_o - \widetilde{g}_q 
		- \widetilde{B}\dot{v}_{f_o}   
		+ G_\mathcal{R} J_h^{-T} {\tau}) \\
		&+ {e}_{\vartheta_\mathcal{R}}^\top \Gamma^{-1} \dot{\vartheta}_\mathcal{R} + {e}_{\vartheta_{\mathcal{R},o}}^\top  \Gamma_o^{-1} \dot{\vartheta}_{\mathcal{R}_o}.
		\end{align*}
		Exploitation of the skew symmetry of $\dot{\widetilde{B}}-2\widetilde{C}_q$, use of ${v}_o = {e}_{v_o} + {v}_{f_o}$, use of \eqref{eq:regressor dynamics (ACC_rolling)}, and substitution of the control law \eqref{eq:control law stack (ACC_rolling)} results in:
		\begin{align*}
		\dot{V}_f  
		=& -{e_\mathcal{R}}^\top  K {e}_\mathcal{R} - {e}^\top_{v_o} K_v {e}_{v_o} + {e}_{v_o}^\top ( { Y_{o_r}}{e}_{\vartheta_{\mathcal{R},o}} {+ G_\mathcal{R}J_h^{-T} Y_r} {e}_{\vartheta_\mathcal{R}}) \\ 
		& + {e}_{\vartheta_{\mathcal{R}}}^\top \Gamma^{-1} \dot{\vartheta}_{\mathcal{R}}  + {e}_{\vartheta_{\mathcal{R},o}}^\top  \Gamma_o^{-1} \dot{\vartheta}_{\mathcal{R}_o},
		\end{align*}
		where we used the fact that $G_\mathcal{R} {f}_{\text{int}} = {0}$ {through \eqref{eq:quadratic program internal forces (ACC_rolling)}}. Finally, by substituting the adaptation laws \eqref{eq:adaptation laws stack (ACC_rolling)}, we obtain
		\begin{align*}
		\dot{V}_f =& -{e}_\mathcal{R}^\top  K {e}_\mathcal{R} - {e}^\top_{v_o} K_v {e}_{v_o} + {e}_{\vartheta_\mathcal{R}} ^\top \bigg (\Gamma^{-1}\text{Proj}({\vartheta_\mathcal{R}},  {-Y_r}^\top  J_h^{-1} G_\mathcal{R}^\top {e}_{v_o} )  \\
		& +\Gamma Y_r^\top  J_h^{-1} G_\mathcal{R}^\top  {e}_{v_o} \bigg)  + {e}_{\vartheta_{\mathcal{R}_o}}^\top  \bigg(\Gamma_o^{-1}\text{Proj}({\vartheta}_{\mathcal{R}_o}, {- Y_{o_r}}^\top  {e}_{v_o}) {+} \Gamma_o Y_{o_r}^\top  {e}_{v_o} \bigg) 
		\end{align*}
		which, by invoking the projection operator property \eqref{eq:proj property (ACC_rolling)}  becomes $\dot{V}_f \leq -{e}_\mathcal{R}^\top  K {e}_\mathcal{R} - {e}_{v_o}^\top  K_v {e}_{v_o}$.
		Thus $\dot{V}_f$ is negative semi-definite, and $V_f$ is bounded in a compact set as $V_f(t) \leq V_f(0)$, $\forall t\in[0,t_{\max})$. In addition, $e_\varphi(t) \neq 0$, $\forall t\in[0,t_{\max})$. Hence, the terms ${e}_{p_o}(t)$, ${e}_\epsilon(t)$, $e_\varphi(t)$ are bounded in a compact set defined by $V_f(0)$ and not dependent on $t_{\max}$, $\forall t\in[0,t_{\max})$. Therefore, since ${p}_{\text{d}}(t)$ and ${\eta}_{\text{d}}(t)$ are bounded and have bounded derivatives,  one concludes that ${p}_o(t)$, ${\eta}_o(t)$ ${v}_o(t)$, and ${v}_{f_o}(t)$, $\dot{v}_{f_o}(t)$ are also bounded in compact sets, $\forall t\in[0,t_{\max})$. This also implies boundedness of $\check{x},\dot{\check{x}}$, as introduced in \eqref{eq:coupled dynamics (ACC_rolling)}, which,
		along with Assumption \ref{asm:nonsingular, no exessive rolling (ACC_rolling)} and properties of Euler-Lagrange systems \cite{Ortega1998},
		implies that $Y_\mathcal{R}()$, $Y_r$, $Y_{\mathcal{R}_o}()$, $Y_{o_r}$ are also bounded in compact sets that are independent of $t_{\max}$, $\forall t\in[0,t_{\max})$. 
		We prove next the slip prevention using the design of the internal force component ${f}_{\text{int}}$. By using \eqref{eq:hand dynamics stack (ACC_rolling)}, \eqref{eq:object dynamics (ACC_rolling)} and \eqref{eq:contact vel compact (ACC_rolling)}, one obtains the following expression for the interaction forces:
		\begin{align}
		{f}_{\scr C} =& W_h^{-1} \bigg( J_h B^{-1}\left[ \tau - {g}_q - \left(C_qJ_h^{-1} G_\mathcal{R}^\top  + B \frac{d}{dt}(J_h^{-1} G_\mathcal{R}^\top )\right){v}_o \right] \notag \\
		& + G_\mathcal{R}^\top  M_o^{-1}(C_o{v}_o + {g}_o )\bigg), \label{eq:contact force (ACC_rolling)}
		\end{align}
		where $W_h \coloneqq  J_h B^{-1} J_h^\top  + G_\mathcal{R}^\top  M_o^{-1} G_\mathcal{R}$, which, by replacing $\tau$, using ${v}_{f_o} = {e}_{v_o} + {v}_o$ and \eqref{eq:regressor dynamics (ACC_rolling)}, adding and subtracting $W_h^{-1} G_\mathcal{R}^\top  M_o^{-1} {G_\mathcal{R}G_\mathcal{R}^\dagger} {f}_{\textup{d}}$ and adding $W_h^{-1} G_\mathcal{R}^\top  M_o^{-1} G_\mathcal{R} {f}_{\textup{int}} = {0}$, becomes 
		\begin{equation} \label{eq:f_c final (ACC_rolling)}
		{f}_{\scr C} = G_\mathcal{R}^\dagger {f}_{\textup{d}} + {f}_{\text{int}} + {h_f}
		\end{equation}
		where 
		\begin{align*}
		h_f \coloneqq & W_h^{-1} J_h B^{-1}(g_q - Y_\mathcal{R}(\check{{x}},\dot{\check{{x}}},{e}_{v_o},\dot{e}_{v_o})\vartheta_\mathcal{R} +  Y_r{e}_{\vartheta_\mathcal{R}}) + 
		W_h^{-1} G_\mathcal{R}^\top  M_o ( {e}_\mathcal{R} \\
		& + K_v {e}_{v_o}  +  Y_{\mathcal{R}_o}(\eta_o,{\omega}_o,{e}_{v_o},\dot{e}_{v_o}) \vartheta_{\mathcal{R}_o} - Y_{o_r} {e}_{\vartheta_{\mathcal{R},o}}- g_o).
		\end{align*}
		
		By combining the aforementioned expression with \eqref{eq:linearized friction constraint (ACC_rolling)}, one obtains the following condition for slip prevention: 
		\begin{align} \label{eq:slip prev}
		\Lambda(\mu_f) R_{\scr C}^\top  {f}_{\text{int}} \succeq {-\Lambda(\mu_f) R_{\scr C}^\top  G^\dagger {f}_{\text{d}} -  \Lambda(\mu_f) R_{\scr C}^\top  {h_f}.}
		\end{align}
		
		Note that due to the aforementioned Lyapunov analysis, as well as the adaptation laws \eqref{eq:adaptation laws stack (ACC_rolling)} through the projection operator, ${e}_\mathcal{R}(t)$, ${e}_{v_o}(t)$, $\dot{{e}}_{v_o}(t)$, ${e}_{\vartheta_\mathcal{R}}(t)$, ${e}_{\vartheta_{\mathcal{R},o}}(t)$ are bounded in compact set independent of $t_{\max}$, $\forall t\in[0,t_{\max})$. By combining this with the aforementioned analysis, we conclude 
		that ${h}_f$ is  bounded for all $\forall t\in[0,t_{\max})$ in a compact set, independent of $t_{\max}$. 
		Hence, {by} denoting $\varepsilon_h$ the maximum bound of the elements of {$\pm  \Lambda(\mu_f) R_{\scr C}^\top  {h}_f$ }and using the designed internal force component ${f}_{\text{int}} = f'_{\text{int}} R_{\scr C}{\ell_{\textup{int}}}$, a sufficient condition for \eqref{eq:slip prev} to hold is for the $j$th element  to satisfy
		\begin{align*}
		l_{\textup{int},j} f'_{\text{int}} \geq {-k_{\textup{int},j}} + \varepsilon_h ,
		\end{align*}
		$\forall j\in\{1,\dots, Nl_f\}$. By substituting \eqref{eq:f_int scalar (ACC_rolling)}, the left side satisfies
		\begin{align*}
		l_{\textup{int},j} \frac{\kappa(\min_{j}\{k_{\textup{int},j}\}) + 1 + \epsilon_f +\delta_f}{\min_j\{ l_{\textup{int},j} \}} &\geq \kappa(\min_{j}\{k_{\textup{int},j}\}) + 1 + \epsilon_f + \delta_f \\
		&\geq {- k_{\textup{int},j}} + \epsilon_f,
		\end{align*}
		where we use $\kappa_{\textup{int}}(x) \geq 0$, $\kappa_{\textup{int}}(x) + 1 \geq -x$, $\forall x\in\mathbb{R}$, and {$\kappa_{\textup{int}}(\min_j(k_{\textup{int},j})) \geq \kappa_{\textup{int}}(   k_{\textup{int},j})$, $\forall j\in\{1,\dots,Nl_f\}$}, since $\kappa_{\textup{int}}()$ is decreasing. Hence, by choosing a large enough $\epsilon_f$ we guarantee $\epsilon_f \geq \varepsilon_h$ and hence contact slip is avoided $\forall t\in[0,t_{\max})$. In fact, the internal forces analysis above and the fact that $\Lambda(\mu_f)$ defines pyramid constraints  imply that ${f}^{\scr {C}_i}_{\scr C_i} \in \bar{\mathcal{F}}_{\scr C_i}$,  where $ \bar{\mathcal{F}}_{\scr C_i}$ is a compact subset of $\text{Int}(\mathcal{F}_{\scr C_i})$, $\forall i\in\mathcal{N}$. Therefore, since ${e}_{\vartheta_\mathcal{R}}$ and ${e}_{\vartheta_\mathcal{R},o}$ are uniformly bounded through the projection operator by $\bar{e}_{\vartheta_\mathcal{R}}$ and $\bar{e}_{\vartheta_\mathcal{R},o}$, respectively, by choosing large enough $\bar{e}_{p_o}$, $\bar{e}_\epsilon$, and $\bar{e}_{v_o}$ in the definition of $\mathcal{U}$, $\chi(t)$ belongs to a compact subset $\bar{\mathcal{U}}$ of $\mathcal{U}$, $\forall t\in[0,t_{\max})$. Thus by invoking Theorem \ref{thm:forward_completeness (App_dynamical_systems)}  of Appendix \ref{app:dynamical systems}, it follows that $t_{\max} = \infty$.
		
		Note, finally,  that ${\tau}(\chi_\mathcal{R}(t),t)$, as designed in \eqref{eq:control law stack (ACC_rolling)}, is bounded, $\forall t\geq 0$. Therefore, one can conclude that
		$\dot{e}_{v_o}(t)$ and thus $\ddot{q}(t)$ is bounded, $\forall t\geq 0$.
		%
		%
		Hence, it follows that $\ddot{V}_f(t)$ is also bounded, $\forall t\geq 0$. 
		Thus by invoking 
		Barbalat's lemma (Lemma \ref{lemma:barbalat (App_dynamical_systems)} of Appendix \ref{app:dynamical systems}), it follows that $\lim_{t\to\infty}\dot{V}_f(t) = 0$ and so $\lim_{t\to\infty}{e}_\mathcal{R}(t) \to {0}$ and $\lim_{t\to\infty}{e}_{v_o}(t) \to 0$. This implies that $\lim_{t\to\infty}{e}_\epsilon(t) \to {0}$, which, given that ${e}_\eta$ is a unit quaternion and $e_\varphi(t) \neq 0$, $\forall t\geq 0$, ensures asymptotic stability of the pose error as $\lim_{t\to\infty}({e}_{p_o}(t), {e}_\zeta(t)) = ({0},[\text{sgn}(e_\varphi(0)), {0}^\top ]^\top )$.
		
	\end{proof}
	
	\begin{remark}
		Note that the bound $\varepsilon_h$ of ${h}_f$ in \eqref{eq:f_c final (ACC_rolling)} can be computed a priori. {In practice, the terms $\vartheta_\mathcal{R}$, $\vartheta_{\mathcal{R}_o}$, which concern masses and moments of inertia of the object and the agents, can be known a priori up to a certain accuracy, leading thus to respective bounds.
		Hence,} one can compute upper bounds for $V_f(0)$ and hence for ${e}_\mathcal{R}$, ${e}_{v_o}$, ${e}_{\vartheta_\mathcal{R}}$, and ${e}_{\vartheta_{\mathcal{R}},o}$. Since the structure of the dynamic terms is known, this can also lead to a bound of the terms $W_h^{-1}$, $B^{-1}$, $M_o$, $Y_\mathcal{R}(\cdot)$, $Y_{\mathcal{R}_o}(\cdot)$, $Y_r$, and $Y_{o_r}$ {that appear in ${h}_f$}. Hence, tuning of $\epsilon_f$ to overcome $\varepsilon_h$ can be performed off-line.  
	\end{remark}

\subsubsection{Simulation Results} 
The proposed control algorithm ensures asymptotic stability for cooperative manipulation with rolling contacts, as well as no slip, while being robust to dynamic uncertainties of the object-robot system. In this section, we implement the proposed control scheme on three 6 DoF mobile manipulators consisting of a 3 DoF, {$3$ kg} base (X-Y translation, rotation about Z) and a 3-DoF manipulator {with $3$ identical links of length $0.3$ m and mass of $0.5$ kg each, as depicted in Fig. \ref{fig.initial conf (ACC_rolling)}}. The objective is to transport a {$2$ kg} box along the desired reference trajectory defined by {${p}_{\text{d}}(t) \coloneqq [0.1 \sin(0.125t),0.1 \sin(0.125t),0.1 \sin(0.125t)]^\top$ m}, {${\eta}_{\text{d}}(t) \coloneqq [\cos(0.1\sin(.125t)), 0,0, \sin(0.1\sin(0.125t))]^\top$}. The control gains used are: $k_p = 1$, $k_\eta = .5$, $K_v = \text{diag}[5,5,5,2,2,2]$, $\epsilon_f = 0.1$, $\Gamma_o = 0.5 I_{\ell_{\mathcal{R}_o}}$, $\Gamma = 0.5 I_{N\ell_\mathcal{R}}$. The control is implemented with 30\% error in all uncertain parameter (including the object center of mass), and the projection operator enforces the following bounds on the uncertain terms: $\bar{\hat{\theta}}_{\mathcal{R}} = 2.25$, $\bar{\hat{\theta}}_{\mathcal{R}_o} = 1.5$. 

The simulation results are depicted in Figs.  \ref{fig.pos (ACC_rolling)}-\ref{fig.u (ACC_rolling)} for $50$ seconds. More specifically,
Figs. \ref{fig.pos (ACC_rolling)} and \ref{fig.ori (ACC_rolling)} show the resulting error trajectories of the object-agent system, which satisfy $\lim_{t\to\infty}{e}_{p_o}(t) = {0}$, $\lim_{t\to\infty}{e}_\epsilon(t) = {0}$, and $\lim_{t\to\infty}{e}_{\varphi}(t) = \text{sgn}({e}_\varphi(0)) = 1$ in the presence of rolling effects. {Fig. \ref{fig.nu (ACC_rolling)}} illustrates the boundedness of the uncertain parameters, ${\hat{\vartheta}}_\mathcal{R}, \hat{\vartheta}_{\mathcal{R}_o}$ that is enforced by the proposed control scheme. Fig. \ref{fig.mu (ACC_rolling)} shows the required friction, {$\mu_{ri} := \frac{\sqrt{f_{\scr C,x_i}^2 + f_{\scr C,y_i}^2}}{f_{\scr C,n_i}}$}, which denotes the minimum friction coefficient necessary to prevent slip throughout the motion \cite{ShawCortez2018b}. If the required friction surpasses the true coefficient, then the contact point will slip and the grasp is compromised. As shown in Fig. \ref{fig.mu (ACC_rolling)}, however, the required friction for each contact is below the true coefficient of $\mu_f = 0.9$, which indicates that slip is prevented as guaranteed by the proposed method. Finally, Fig. \ref{fig.u (ACC_rolling)} depicts the control inputs of the agents. As predicted by the theoretical analysis, asymptotic error stability as well as contact slip prevention are achieved.

\begin{figure}[hbtp] 
	\centering
	\includegraphics[scale=.32]{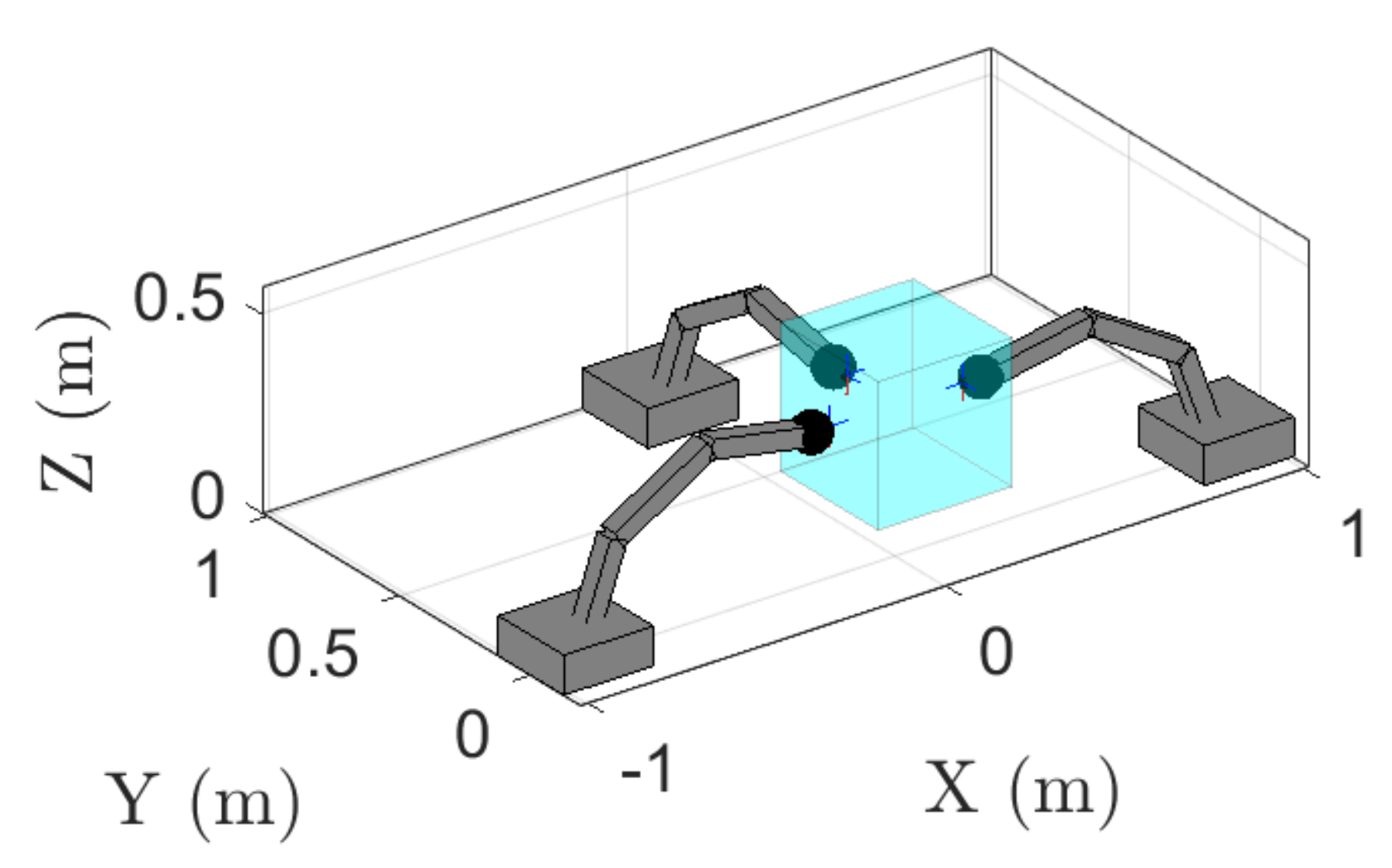}
	\caption{Initial configuration of the system that consists of three mobile manipulators and a rigid object.}\label{fig.initial conf (ACC_rolling)}
\end{figure}

\begin{figure}[hbtp] 
	\centering
	\includegraphics[scale=.45]{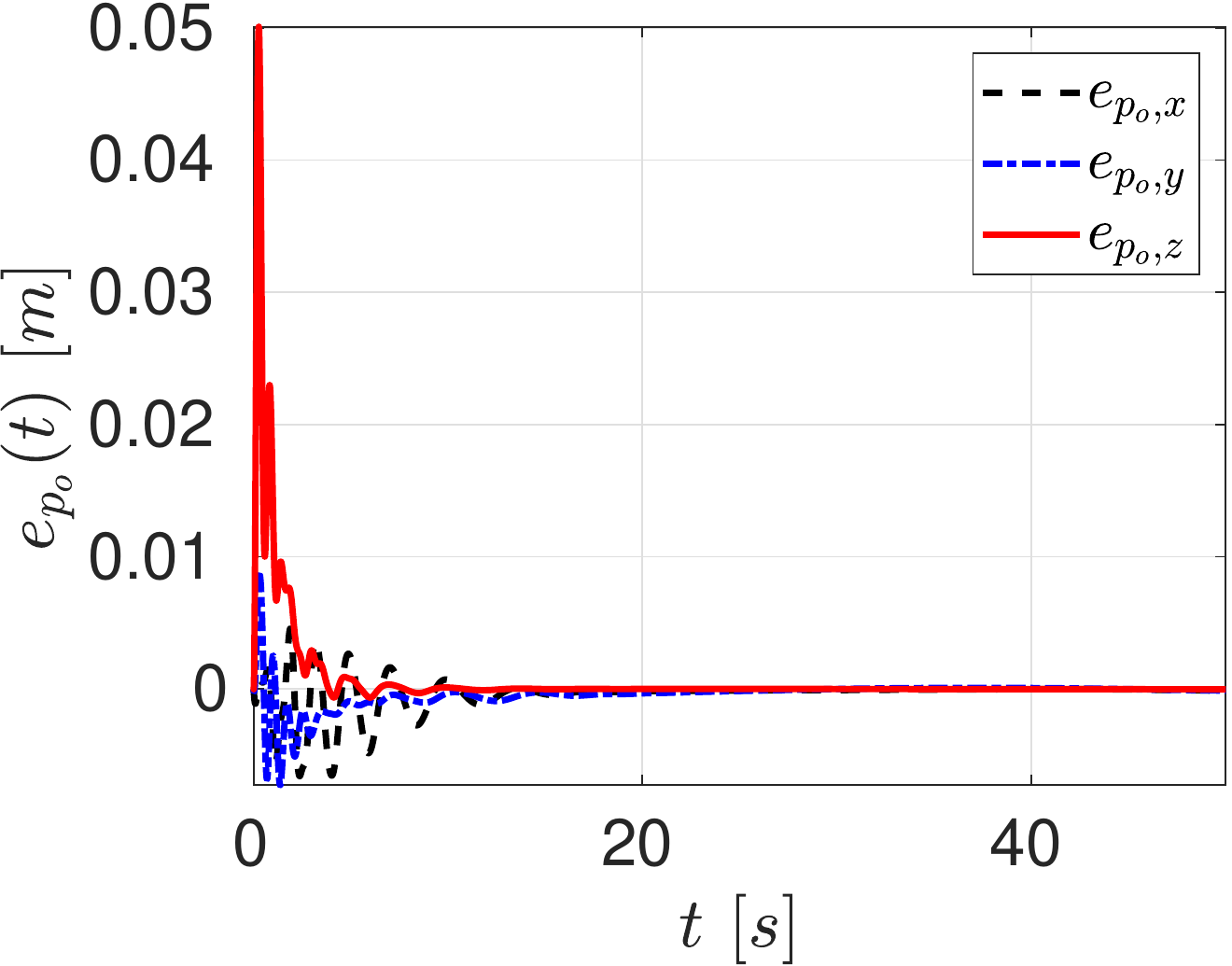}
	\caption{The evolution of the position error, ${e}_{p_o}(t)$, $\forall t\in[0,50]$.}\label{fig.pos (ACC_rolling)}
\end{figure}
\begin{figure}[hbtp] 
	\centering
	\includegraphics[scale=.45]{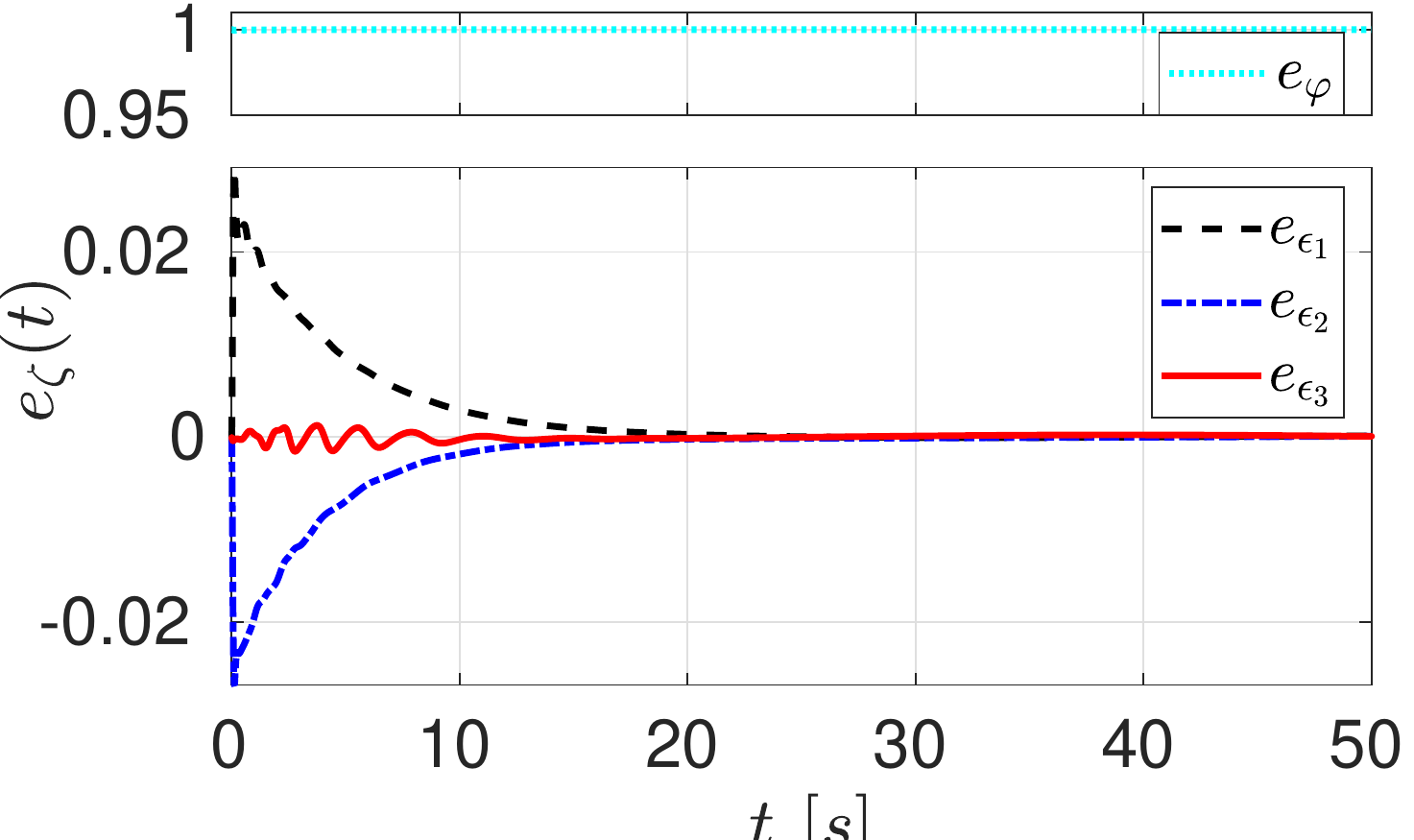}
	\caption{The evolution of $e_\varphi(t)$, ${e}_\epsilon(t)$, $\forall t\in[0,50]$.} \label{fig.ori (ACC_rolling)}  
\end{figure}

\begin{figure}
	\centering
	\includegraphics[scale=.35]{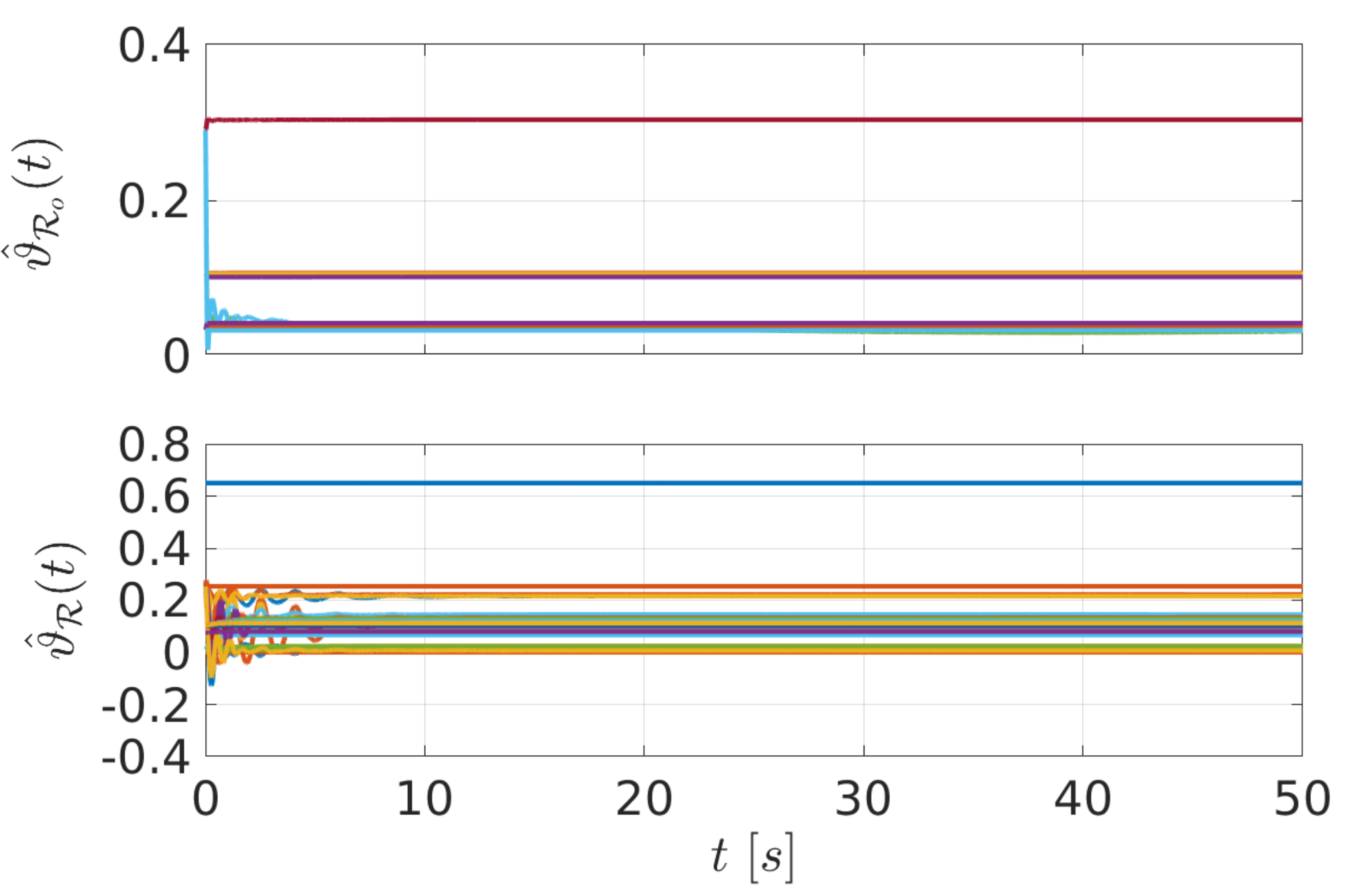}
	\caption{The evolution of $\hat{\vartheta}_{\mathcal{R}_o}(t)$, $\hat{\vartheta}_\mathcal{R}(t)$, $\forall t\in[0,50]$.} \label{fig.nu (ACC_rolling)}  
\end{figure}

\begin{figure}[hbtp] 
	\centering
	\includegraphics[scale=.45]{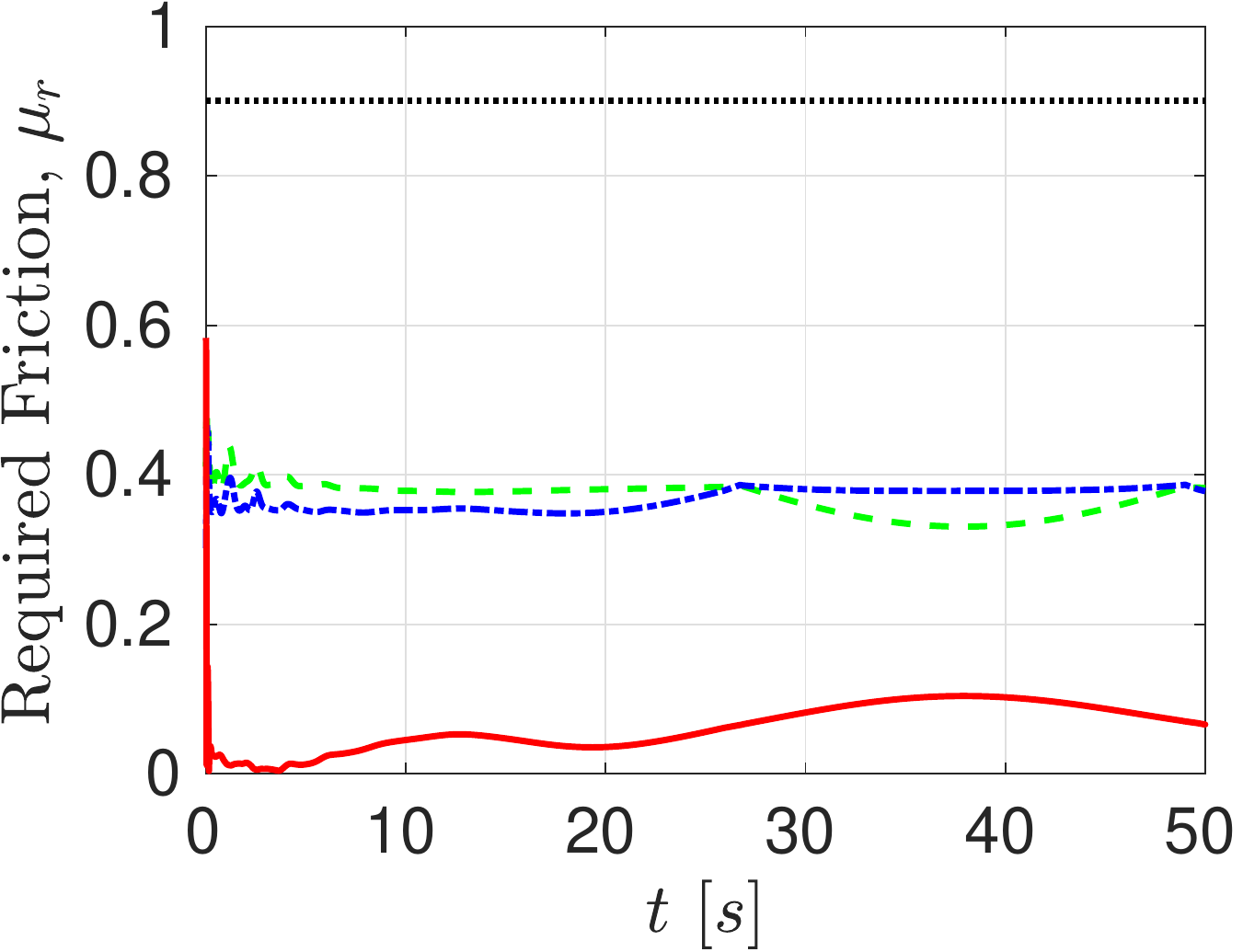}
	\caption{The required friction to prevent slip for the three agents. The black dashed line represents $\mu_f = 0.9$.}\label{fig.mu (ACC_rolling)}
\end{figure}

\begin{figure}
	\centering
	\includegraphics[scale=.45]{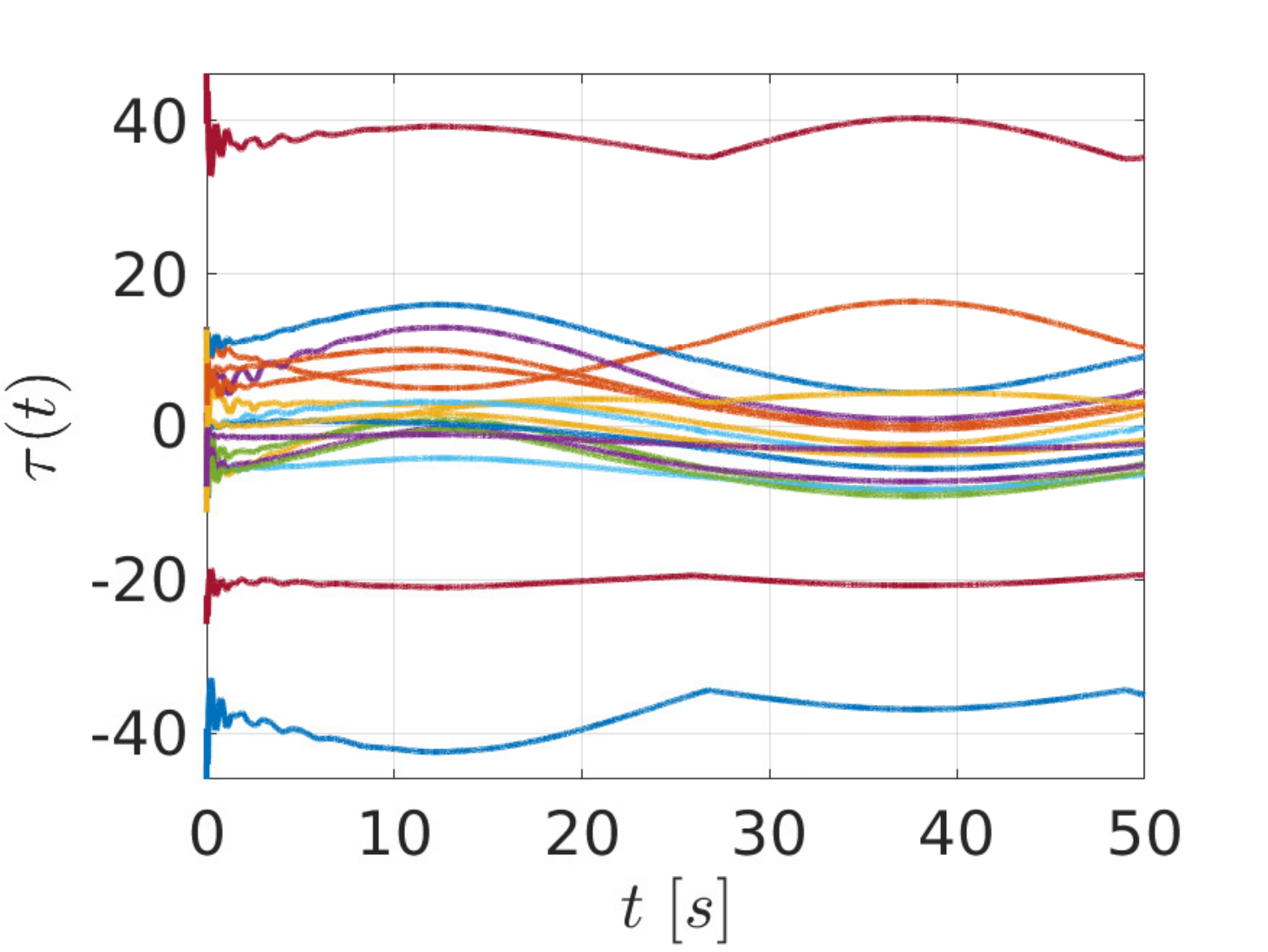}
	\caption{The resulting inputs ${\tau}(t)$ of the agents, $\forall t\in[0,50]$.} \label{fig.u (ACC_rolling)}  
\end{figure}

\subsection{Decentralized Scheme}

This section presents a decentralized extension of the aforementioned scheme via event-triggered communication among the agents. 
The event-triggered control requires an update law (to be designed) that updates relevant variables at each time $t_k \in \mathbb{R}_{>0}$ for $k \in \mathbb{N}$. We use the subscript with $k$ to denote a variable that is held constant over the time interval $[t_k, t_{k+1})$ and updated at each $t_k$. 
The variables communicated among the agents at time $t_k$ are $p_{\scr C_i/o}(t_k)$ and $p_{\scr C_i/E_i}(t_k)$, allowing all the agents to reconstruct $G_{\mathcal{R},k} \coloneqq [G_{\mathcal{R},k_1}, ..., G_{\mathcal{R},k_N}]$, as well as $R_{C,k} \coloneqq \text{diag}\{R_{\scr C,{k_i}}\}_{i \in \mathcal{N}}$, with $G_{\mathcal{R},k_i} \coloneqq G_{\mathcal{R}_i}(p_{\scr C_i/o}(t_k))$, $R_{\scr C,k_i} \coloneqq R_{\scr C_i}(t_k)$, $\forall i\in\mathcal{N}$. 
The event-triggered manipulation control law is defined as:
\begin{subequations} \label{eq:event triggered manipulation control (dec_rolling)}
	\begin{align} 
		{\tau}_{k_i}(\chi_{\mathcal{R}_i},t)  =& Y_{r_i}\hat{\vartheta}_{\mathcal{R}_i} +  J_{h_i}^\top ( {\lambda}_{k_i}  + {f}_{\text{int},{k_i}} ), \\
		{\lambda}_{k_i} \coloneqq& - G_{\mathcal{R}_i	}^\top K_v {e}_{v_o} + G^\ast_{\mathcal{R},k_i}(Y_{o_r} \hat{\vartheta}_{\mathcal{R}_o} - {e}_\mathcal{R})
	\end{align}
\end{subequations}
with $\chi_{\mathcal{R}_i} \coloneqq [\check{x}_i^\top,\dot{\check{x}}_i^\top,e^\top_\mathcal{R},e^\top_{v_o},\hat{\vartheta}_{\mathcal{R}_i}^\top, \hat{\vartheta}_{\mathcal{R}_o}^\top, \eta_{\scr C_i}^\top]^\top \in \mathcal{T}_{\mathcal{R}_i} \coloneqq \{\chi_{\mathcal{R}_i}\in \mathbb{T}^2\times \mathsf{S}_i \times \mathbb{R}^{30+\ell_{\mathcal{R}}+\ell_{\mathcal{R}_o}} \times \mathbb{T} : e_\varphi \neq 0\}$
$\forall i\in\mathcal{N}$, with the decentralized version of the adaptive update laws defined by \eqref{eq:adaptation laws stack (ACC_rolling)}: 
\begin{subequations} \label{eq:adaptation laws stack (dec_rolling)}
\begin{align} 
	\dot{{\hat{\vartheta}}}_{\mathcal{R}_i} &= {\text{Proj}(\hat{\vartheta}_{\mathcal{R}_i},{- \Gamma}_i Y_{r_i}^\top  J_{h_i}^{-1} G_{\mathcal{R}_i}^\top {e}_{v_o}),  }\\ 
	\dot{{\hat{\vartheta}}}_{\mathcal{R}_o} &= \text{Proj}({\hat{\vartheta}_{\mathcal{R}_o}}, {-\Gamma_o} Y_{o_r}^\top  {e}_{v_o}).
\end{align}
\end{subequations}
where $Y_{r_i} \coloneqq Y_{\mathcal{R}_i}(\check{x}_i,\dot{\check{x}}_i,v_{f_o},\dot{v}_{f_o})$, $\Gamma_i \in \mathbb{R}^{\ell_\mathcal{R}\times\ell_\mathcal{R}}$ is such that $\Gamma = \text{diag}\{ [\Gamma_i]_{i\in\mathcal{N}}\}$ from \eqref{eq:adaptation laws stack (ACC_rolling)}, and
${f}_{\text{int},{k_i}}  \in \mathbb{R}^3$ is the event-triggered internal force control yet to be designed. Similarly, $G^*_{\mathcal{R},k}$ is a generalized inverse of the grasp map at $t = t_k$ and we denote $G^*_{\mathcal{R},k} =: [G^{\ast^\top}_{\mathcal{R},k_1},...,G^{\ast^\top}_{\mathcal{R},k_N}]^\top$. Let $\Delta G_{\mathcal{R}} := G_{\mathcal{R}} - G_{\mathcal{R},k}$, $\Delta p_{\scr oC} \coloneqq [\Delta p_{\scr C_i/o}^\top, \dots, \Delta p_{\scr C_N/o}^\top]^\top \coloneqq p_{\scr oC} - p_{\scr oC}(t_k)$, $\Delta p_{\scr EC} \coloneqq [\Delta p_{\scr C_i/E_i}^\top, \dots, \Delta p_{\scr C_N/E_N}^\top]^\top \coloneqq p_{\scr EC} - p_{\scr EC}(t_k)$, $\Delta R_{\scr C} \coloneqq R_{\scr C} - R_{\scr C,k}$ denote the triggering errors. Note that \eqref{eq:event triggered manipulation control (dec_rolling)} is only dependent on the full grasp map, $G_{\mathcal{R},k}$, in the term $G_{\mathcal{R},k}^\ast$, whereas the adaptation laws \eqref{eq:adaptation laws stack (dec_rolling)} and remainder of the control depend on $G_{\mathcal{R}_i}$. 



Moreover, in order to ensure no slip, recall that the condition \eqref{eq:linearized friction constraint (ACC_rolling)} must hold. Notice that when there is no communication between agents, $R_{\scr C}$ and $G_\mathcal{R}$ are unknown as each agent only has knowledge of $R_{(dec_rolling) C,k}$ and $G_{\mathcal{R},k}$. Thus if the original internal force control \eqref{eq:quadratic program internal forces (ACC_rolling)} is implemented with $R_{\scr C,k}$, $G_{\mathcal{R},k}$ the errors $\Delta R_{\scr C}$ and $\Delta G_\mathcal{R}$ may induce slip. To account for this, we use a conservative $\mu_f' \in (0, \mu_f)$ that effectively shrinks the friction pyramid. The design of $\mu_f'$ is dependent on the allowable error that will result from triggering. This presents a trade-off where to reduce communication (i.e reduced triggering), a smaller more conservative $\mu'_f$ is required, and vice versa (i.e. larger $\mu_f'$ requires more communication between agents). We introduce the following Lemma to compute $\mu_f'$.

\begin{lemma}\label{lem:no slip perturbed_new (dec_rolling)}
	Let $\mu_f > 0$, and $W \in \mathbb{R}^{3\times 3}$ satisfying 
	$\|W\| \leq \delta_c$, where $\delta_c$ is a positive constant satisfying
	\begin{equation} \label{eq:delta_c (dec_rolling)}
	\delta_c < \frac{\sqrt{\mu_f^2+1}-1}{\sqrt{\mu_f^2+1}+1} < 1,
	\end{equation}
	and define 
	\begin{equation} \label{eq:mu_new (dec_rolling)}
	\mu_f' \coloneqq \tan\left(\tan^{-1}(\mu_f) - \cos^{-1}\left(\frac{1-\delta_c}{1+\delta_c}\right) \right).
	\end{equation}
	If ${y}\in \mathbb{R}^3$ satisfies ${y} \in \mathcal{F}_{\scr C_i}(\mu_f')$, then $(I_3 - W){y} \in \mathcal{F}_{\scr C_i}(\mu_f)$.
\end{lemma}
\begin{proof}
	Denote by $\theta_W$ the angle defined by ${y}$ and $(I_3 - W){y}$,  satisfying
	\begin{align} \label{eq:cos theta_C (dec_rolling)}
	\cos(\theta_W) = \frac{{y}^\top (I_3 - W){y}}{\|(I_3 - W){y}\| \|{y}\|} = \frac{{y}^\top (I_3 - W_{\text{sym}}){y}}{\|(I_3 - W){y}\| \|{y}\|},
	\end{align}
	where $W_{\text{sym}} \coloneqq \frac{W + W^\top}{2}$. Note that $\lambda_{\max}(W_{\text{sym}}) \leq \|W_{\text{sym}}\| \leq \|W\| \leq \delta_c < 1$ and hence
	$I_3 - W_{\text{sym}}$ has strictly positive eigenvalues, rendering $\cos(\theta_W)$ positive and $|\theta_W| < \frac{\pi}{2}$. 
	
	Moreover, it holds $\|(I_3 - W){y}\|\|{y}\| \leq  (1  + \|W\|)\|{y}\|^2$ as well as ${y}^\top (I_3 - W_{\text{sym}}){y} \geq \lambda_{\min}(I_3 - W_{\text{sym}})\|{y}\|^2 \geq (1 - \lambda_{\max}(W_{\text{sym}}))\|{y}\|^2  \geq (1 - \|W\|)\|{y}\|^2$. 
	Hence, by taking into account $\|W\| \leq \delta_c$ and \eqref{eq:delta_c (dec_rolling)}, \eqref{eq:cos theta_C (dec_rolling)} becomes
	\begin{align*}
	\cos(\theta_W) \geq \frac{1 - \|W\|}{1 + \|W\|} \geq  \frac{1 - \delta_c}{1 + \delta_c} > \frac{1}{\sqrt{\mu_f^2 + 1}} = \cos(\tan^{-1}(\mu_f)),
	\end{align*}
	implying 
	\begin{align} \label{eq:theta_W tmp (dec_rolling)}
	|\theta_W| \leq \cos^{-1}\left(\frac{1-\delta_c}{1+\delta_c}\right) < \tan^{-1}(\mu_f)
	\end{align}
	and rendering hence $\mu'_f$ positive.
	
	In order for $(I_3 - W){y}$ to belong to $\mathcal{F}_{c_i}(\mu_f)$, 
	${y}$ must lie in a new friction cone $\mathcal{F}_{\scr C_i}(\mu^\ast)$, whose angle $\tan^{-1}(\mu^\ast)$ must be reduced by $|\theta_W|$ from $\tan^{-1}(\mu_f)$, leading thus to $\mu^\ast \coloneqq \tan(\tan^{-1}(\mu_f) - |\theta_W|)$. In view of \eqref{eq:theta_W tmp (dec_rolling)}, it holds that $\mu_f' < \mu_f^\ast$ and hence $\mathcal{F}_{\scr C_i}(\mu_f') \subset \mathcal{F}_{\scr C_i}(\mu_f^\ast)$ and ${y} \in \mathcal{F}_{\scr C_i}(\mu_f') \Rightarrow {y} \in \mathcal{F}_{\scr C_i}(\mu_f^\ast) \Rightarrow (I_3 - W){y} \in \mathcal{F}_{\scr C_i}(\mu_f)$.  
\end{proof}

The event-triggered internal force controller is now defined as follows:
\begin{subequations} \label{eq:event triggered internal forces (dec_rolling)}
	\begin{equation}
	{f}_{\text{int},k} = f'_ {\text{int},k} R_{\scr C,k} {\ell}_{\text{int},k}^*
	\end{equation}
	
	\begin{align}  \label{eq:event triggered internal forces QP (dec_rolling)}
		{\ell}_{\text{int},k}^* =  & \text{argmin}_{{\ell}} \   {\ell}^\top {\ell}  \\ 
		&\text{ s. t. } \\
		&  G_{\mathcal{R},k} R_{\scr C,k} \ell = 0  \\
		& \Lambda_i(\mu'_f) {\ell}_i \succ {0}, \ \forall i\in\mathcal{N},
	\end{align}
\end{subequations}

\begin{equation} \label{eq:event triggered f_int scalar (dec_rolling)}
f'_{\text{int},k} \coloneqq \frac{\kappa(\min_{j}\{b_{k_j}\}) + 1 + \epsilon_f }{ \min_j\{ l_{k_j} \} - \varepsilon_d \delta_p \max_j\{ \ell^*_{\text{int},k_j} \}},
\end{equation}
\begin{equation*}
{b}_{k} \coloneqq \Lambda(\mu_f') R_{\scr C,k}^\top {\lambda}_{k}, {l}_k \coloneqq \Lambda(\mu'_f) {\ell}_{\text{int},k}^*
\end{equation*}
where $\lambda_k \coloneqq [\lambda_{k_1}^\top,\dots,\lambda_{k_N}^\top]^\top$, $b_{k_j}$, $\ell_i \in\mathbb{R}^3$ is the $i$th agent's part in vector $\ell$, $l_{k_j}$ and $\ell_{\text{int},k_j}$ are the $j$th scalar elements of ${b}_k$, ${l}_k$ and ${\ell}_{\text{int},k}$, respectively, $\varepsilon_d , \epsilon_f\in \mathbb{R}_{>0}$ are design parameters, and $\mu'_f$ is defined as in \eqref{eq:mu_new (dec_rolling)}.
Note that ${\ell}_{\text{int},k}^*$ is constant for $t \in [t_k, t_{k+1})$ such that it need only be computed at each $k$ update.

Now that the full control protocol is defined, the final step is to define the event-triggering condition to update ${G}_k$ and $R_{\scr C,k}$ which are:
\begin{subequations} \label{eq:triggering conditions (dec_rolling)}
\begin{align}
	& ||\Delta {p}_{\scr C_i/o}|| = \delta_p, \label{eq:triggering condition G 1 (dec_rolling)} \\
	& \delta_p := \text{min} \left\{ \frac{1}{\sum_i ||G_{\mathcal{R},k_i}^\ast||} \text{min} \{k_1 - c_{\mathcal{R}_2}, 2 k_2 - c_{\mathcal{R}_2} \}, \frac{\delta_c}{2 \varepsilon_c}, \frac{\min_j \{l_{k_j} \}}{\varepsilon_d \max_j \{\ell^*_{\text{int},k_j}\}}\right\} \label{eq:triggering condition G 2 (dec_rolling)} \\
	&|| \Delta {p}_{\scr C_i/E_i} || = \delta_r < \frac{\delta_c}{2 \varepsilon_c} \\
	\label{eq:triggering condition R (dec_rolling)} \\
	&{e}_{v_o}^\top  \begin{bmatrix} 0 \\ S(\Delta {p}_{\scr C_i/o})\end{bmatrix} \Big(  {f}_{\text{int},{k_i}}  + G_{\mathcal{R},k_i}^\ast Y_{o_r} \hat{\vartheta}_{\mathcal{R}_o} \Big) - c_{\mathcal{R}_2} \gamma_q  = 0,  \label{eq:triggering condition ev (dec_rolling)}
\end{align}
\end{subequations}
where $f_{\text{int},k_i}\in\mathbb{R}^3$ is the $i$th component of $f_{\text{int},k}$,
$k_1 \coloneqq \lambda_{\min}(K)$, $k_2 \coloneqq \lambda_{\min}(G_\mathcal{R} G_\mathcal{R}^\top K_v)$, $c_{\mathcal{R}_2}$, $\gamma_q$, $\delta_e$, $\varepsilon_c$, are design parameters. Note that $k_2 > 0$ due to the fact that $G_\mathcal{R}$ is full row rank. The time for which an event is triggered is when \eqref{eq:triggering condition G 1 (dec_rolling)}, \eqref{eq:triggering condition R (dec_rolling)}, or \eqref{eq:triggering condition ev (dec_rolling)} are satisfied, and formally defined as:
\begin{align}\label{eq:triggering condition (dec_rolling)}
	t(0) = 0, \ t_{k+1} = \text{inf}  \{ t \in \mathbb{R}: t > t_k \land ( \eqref{eq:triggering condition G 1 (dec_rolling)} \lor \eqref{eq:triggering condition R (dec_rolling)} \lor \eqref{eq:triggering condition ev (dec_rolling)}  ) \}
\end{align}
Note that the condition \eqref{eq:triggering condition (dec_rolling)} can be evaluated by each agent individually. When one agent identifies a triggering condition, the agent then signals to all agents that an update is required and all agents then \textit{only} communicate ${p}_{\scr C_i/o}$ and ${p}_{\scr C_i/f}$ for all $i \in \mathcal{N}$.

The proposed control is decentralized with aperiodic updates of \text{only} each agent's contact information. The event-triggered, decentralized control law ensures practical asymptotic stability of the origin as presented in the following theorem:
\begin{theorem} \label{th:event triggered th (dec_rolling)}
	Consider $N$ robotic agents in contact with an object, described by the dynamics \eqref{eq:hand dynamics stack (ACC_rolling)}, \eqref{eq:object dynamics (ACC_rolling)}, and suppose Assumptions \ref{asm:full rank G (ACC_rolling)} and \ref{asm:nonsingular, no exessive rolling (ACC_rolling)} hold. Let the desired object pose $[{p}^\top_{\textup{d}}, {\eta}^\top_{\textup{d}}]^\top:\mathbb{R}_{\geq 0} \to \mathbb{R}^3\times \mathsf{S}^3$ be bounded with bounded first and second derivatives. Moreover, assume that $e_\varphi(0) \neq 0$ and ${f}^{\scr C_i}_{\scr C_i}(0) \in \textup{Int}( \mathcal{F}_{\scr C_i})$, $\forall i\in\mathcal{N}$.
	Then, by choosing sufficiently large control gains $k_p$, $k_\eta$, $K_v$, the event-triggered control protocol \eqref{eq:event triggered manipulation control (dec_rolling)}, \eqref{eq:adaptation laws stack (dec_rolling)}, \eqref{eq:event triggered internal forces (dec_rolling)} with event-triggered mechanism \eqref{eq:triggering condition (dec_rolling)} guarantees ultimate boundedness of ${e}_\mathcal{R}$, ${e}_v$ in a set around the origin, and by choosing sufficiently large $\epsilon_f$, $\varepsilon_c$, $\varepsilon_d$, it holds that ${f}^{\scr C_i}_{\scr C_i}(t) \in \mathcal{F}_{\scr C_i}, \forall t > 0$, $i\in\mathcal{N}$.
\end{theorem}
\begin{proof}
	The proof is structured into 2 Cases. Case 1 addresses the system if no event is triggered. Case 2 addresses the triggering conditions and ensuring non-Zeno behavior for the time updates.
	
	Case 1: Here we address the case when no event is triggered such that $t_k = 0$ and $t_{k+1} = \infty$. We note from the proof of Theorem \ref{th:main th (ACC_rolling)} that the continuous control law, ${\tau}$ from \eqref{eq:control law stack (ACC_rolling)} ensures asymptotic stability of the system with the Lyapunov candidate function, $V_f$, defined in \eqref{eq:Lyapunov (ACC_rolling)}. We define the following compact set:
	\begin{equation*}
	\Omega_k :=\{ {\chi} \in \mathcal{X} : V_f({\chi}(t)) \leq V_k({\chi}) \}, k \in \mathbb{Z}_{\geq 0}
	\end{equation*}
	with $V_f$ as defined in \eqref{eq:Lyapunov (ACC_rolling)}, and $V_k := V_f(\chi(t = t_k))$.
	
	From ${f}^{\scr C_i}_{\scr C_i}(0) \in \text{Int}(\mathcal{F}^{\scr C_i}_{\scr C_i})$, the same analysis from Theorem \ref{th:main th (ACC_rolling)} applies here such that there exists a $t_{\max} \in \mathbb{R}_{>0}$ such that for $t \in [0, t_{\max})$, slip does not occur and the solution is unique. In the following we will apply the Lyapunov analysis over the time interval $[0, t_{\max})$. 
	
	After substitution of \eqref{eq:event triggered manipulation control (dec_rolling)}, $\dot{V}_f$ becomes 
	\begin{align*}
		\dot{V}_f =& - {e}_\mathcal{R}^\top K {e}_\mathcal{R} -{e}_{v_o}^\top G_\mathcal{R}G_{\mathcal{R}}^\top K_v {e}_{v_o} + {e}_{v_o}^\top G_\mathcal{R} G_{\mathcal{R},k}^\ast Y_{o_r} \hat{\vartheta}_{\mathcal{R}_o} - {e}_{v_o}^\top Y_{o_r} {\vartheta}_{\mathcal{R}_o} \\
		&- {e}_{v_o}^\top ( G_\mathcal{R}G^\ast_{\mathcal{R},k} - I_6){e_\mathcal{R}} + {e}_{v_o}^\top G_\mathcal{R} {f}_{\text{int}_k} + {e}_{\vartheta_{\mathcal{R},o}}^\top \Gamma^{-1}_o \dot{\hat{\vartheta}}_{\mathcal{R}_o}
	\end{align*}	
	From $\Delta G_\mathcal{R} = G_{\mathcal{R}} - G_{\mathcal{R},k}$ it follows that $G_\mathcal{R} G_{\mathcal{R},k}^\ast - I = \Delta G_\mathcal{R} G_{\mathcal{R},k}^\ast$ which yields, along with \eqref{eq:adaptation laws stack (dec_rolling)}, \eqref{eq:proj property (ACC_rolling)}, and the fact that $G_{\mathcal{R},k} {f}_{\text{int}_k} = {0}$:
	\begin{align*}
		\dot{V}_f \leq & - k_1\|{e}_\mathcal{R}\|^2 - k_2\|{e}_{v_o}\|^2 + {e}_{v_o}^\top \Delta G_\mathcal{R} G^\ast_{\mathcal{R},k} Y_{o_r} \hat{\vartheta}_{\mathcal{R}_o} - {e}_{v_o}^\top \Delta G_{\mathcal{R}} G^\ast_{\mathcal{R},k} {e}_\mathcal{R}  \\
		& + {e}_{v_o}^\top \Delta G_{\mathcal{R}} {f}_{\text{int}_k}.
	\end{align*}
	Note that $k_2$ can be increased by tuning  $K_v$. From $\Delta G_{\mathcal{R}_i}\coloneqq G_{\mathcal{R}_i} - G_{\mathcal{R},k_i} = \begin{bmatrix} 0 \\ S(\Delta {p}_{\scr C_i/o}) \end{bmatrix}$ and $\Delta G_\mathcal{R} = [\Delta G_{\mathcal{R}_1}, ..., \Delta G_{\mathcal{R}_N}]$, it follows that $\| \Delta G_{\mathcal{R}} G_{\mathcal{R},k}^\ast \| \leq ||\Delta G_{\mathcal{R}} ||||G_{\mathcal{R},k}^\ast|| \leq \sum_i ||\Delta {p}_{\scr C_i/o}|| ||G_{\mathcal{R},k_i}^\ast||$. From the triggering condition \eqref{eq:triggering condition (dec_rolling)}, it follows that $||\Delta {p}_{\scr C_i/o}|| \leq \delta_p$ for all $i \in \mathcal{N}$. We thus define $c_{\mathcal{R}_1} := \delta_p \sum_i ||G_{\mathcal{R},k_i}^\ast||$, which is constant between events, such that  $||\Delta G_\mathcal{R} G^\ast_{\mathcal{R},k} \| \leq c_{\mathcal{R}_1}$. Note that Assumptions \ref{asm:full rank G (ACC_rolling)} and \ref{asm:nonsingular, no exessive rolling (ACC_rolling)} as well as the fact that slip does not occur for $[0,t_{\max})$ imply that $||\Delta G_\mathcal{R} G^\ast_{\mathcal{R},k} \|$ is well defined and bounded, $\forall i\in\mathcal{N}$. 
	Hence $\dot{V}_f$ becomes 
	\begin{align*}
		\dot{V}_f \leq & - k_1\|{e}_\mathcal{R}\|^2 - k_2\|{e}_{v_o}\|^2 + {e}_{v_o}^\top \Delta G_\mathcal{R} G^\ast_{\mathcal{R},k} Y_{o_r} \hat{\vartheta}_{\mathcal{R}_o} + c_{\mathcal{R}_1} \| {e}_{v_o} \| \| {e}_\mathcal{R}\|  \\
		& + {e}_{v_o}^\top \Delta G_\mathcal{R} {f}_{\text{int}_k} \\ 
	\end{align*}	
	We then complete the squares such that $c_{\mathcal{R}_1} \|{e}_{v_o}\| \|{e}_\mathcal{R}\| \leq \frac{c_{\mathcal{R}_1}}{2} \|{e}_{v_o}\|^2 + \frac{c_{\mathcal{R}_1}}{2} \|{e}_\mathcal{R}\|^2 $ and hence $\dot{V}_f$ becomes
	\begin{align*}
		\dot{V}_f \leq & - \left(k_1 - \frac{c_{\mathcal{R}_1}}{2}  \right)\|{e}_\mathcal{R}\|^2 - \left(k_2 - \frac{c_{\mathcal{R}_1}}{2}  \right)  \|{e}_{v_o}\|^2   + {e}_{v_o}^\top \Delta G_\mathcal{R} G^\ast_{\mathcal{R},k} Y_{o_r} \hat{\vartheta}_{\mathcal{R}_o} \\
		&+ {e}_{v_o}^\top \Delta G_\mathcal{R} {f}_{\text{int}_k} \\
	\end{align*}
	Now we introduce $c_{\mathcal{R}_2} \in \mathbb{R}_{>0}$ such that:
	\begin{align*}
		\dot{V}_f \leq & - \left(k_1 - \frac{c_{\mathcal{R}_1}}{2} - c_{\mathcal{R}_2}\right)\|{e}_\mathcal{R}\|^2 - \left(k_2 - \frac{c_{\mathcal{R}_1}}{2}  - c_{\mathcal{R}_2}\right)  \|{e}_{v_o}\|^2   \\
		& - c_{\mathcal{R}_2} \|{e}_\mathcal{R}\|^2 - c_{\mathcal{R}_2} \|{e}_{v_o}\|^2 + {e}_{v_o}^\top \Delta G_\mathcal{R} G^\ast_{\mathcal{R},k} Y_{o_r} \hat{\vartheta}_{\mathcal{R}_o} + {e}_{v_o} \Delta G_\mathcal{R} {f}_{\text{int}_k} \\
		=: &-(k_e- c_{\mathcal{R}_2})\|{e}_\mathcal{R}\|^2 - (k_{e_v} - c_{\mathcal{R}_2})\|{e}_{v_o}\|^2 - c_{\mathcal{R}_2}\|{e}_\mathcal{R}\|^2 - c_{\mathcal{R}_2}\|{e}_{v_o}\|^2  \\
		& + {e}_{v_o}^\top \Delta G_\mathcal{R} G^\ast_{\mathcal{R},k} Y_{o_r} \hat{\vartheta}_{\mathcal{R}_o} + {e}_{v_o}^\top \Delta G_\mathcal{R} {f}_{\text{int}_k} 
	\end{align*}
	where $k_e \coloneqq k_1 - \frac{c_{\mathcal{R}_1}}{2} $ and $k_{e_v} \coloneqq k_2 - \frac{c_{\mathcal{R}_1}}{2}$. {By choosing large enough $k_p$, $k_\eta$, and $K_v$, we can achieve $k_e > c_{\mathcal{R}_2}$ and $k_{e_v} > c_{\mathcal{R}_2}$.} 
	
	Let now $ \mathcal{Q} \coloneqq \{ {\chi} \in \mathcal{X} : \| {e}_\mathcal{R} \|^2 + \| {e}_{v_o} \|^2 \leq \gamma_q \}$. Note that $\mathcal{Q}$ is compact since ${e}_{\vartheta_\mathcal{R}}$, ${e}_{\vartheta_{\mathcal{R},o}}$ are bounded as per \eqref{eq:e_nu bars (ACC_rolling)}. 
	Moreover, in $\mathcal{X} \backslash \mathcal{Q}$ it holds that $\|{e}_\mathcal{R}\|^2 + \|{e}_{v_o}\|^2 > \gamma_q$ and hence $c_{\mathcal{R}_2}\|{e}_\mathcal{R}\|^2 + c_{\mathcal{R}_2}\|{e}_{v_o}\|^2 > c_{\mathcal{R}_2}\gamma_q$, and $\dot{V}_f$ becomes
	\begin{align*}
		\dot{V}_f \leq & -(k_e- c_{\mathcal{R}_2})\|{e}_\mathcal{R}\|^2 - (k_{e_v} - c_{\mathcal{R}_2})\|{e}_{v_o}\|^2 \\
		& + {e}_{v_o}^\top \Delta G_\mathcal{R} G^\ast_{\mathcal{R},k} Y_{o_r} \hat{\vartheta}_{\mathcal{R}_o} + {e}_{v_o}^\top \Delta G_\mathcal{R} {f}_{\text{int}_k}  - c_{\mathcal{R}_2}\gamma_q 
	\end{align*}
	According to \eqref{eq:triggering condition ev (dec_rolling)}, it holds, between events, that 
	\begin{align*}
		&{e}_{v_o}^\top  \begin{bmatrix} 0 \\ S(\Delta {p}_{\scr C_i/o}) \end{bmatrix} \Big(  {f}_{\text{int},{k_i}}  + G_{\mathcal{R},k_i}^\ast Y_{o_r} \hat{\vartheta}_{\mathcal{R}_o} \Big) - c_{\mathcal{R}_2} \gamma_q \leq 0.
	\end{align*}
	By summing for all $i\in\mathcal{N}$, the latter becomes
	\begin{align*}
		{e}_{v_o}^\top \Delta G_\mathcal{R} G_{\mathcal{R},k}^\ast (Y_{o_r}\hat{\vartheta}_{\mathcal{R}_o} + {f}_{\text{int}_k}) - c_{\mathcal{R}_2} \gamma_q \leq 0,
	\end{align*}
	implying that  
	$\dot{V}_f \leq - (k_e- c_{\mathcal{R}_2}) \|{e}_\mathcal{R}\|^2 - (k_{e_v} - c_{\mathcal{R}_2})\|{e}_{v_o}\|^2 \leq 0$. By following Barbalat's Lemma, it can be shown that $\chi$ will enter the set $\mathcal{Q}$ in finite time. 
	
	By using \eqref{eq:e_nu bars (ACC_rolling)}, we now investigate $\dot{V}_f$ inside $\mathcal{Q}$ for which it holds $\|{e}_{v_o} \| \leq \sqrt{\gamma_q}$:
	\begin{align*}
		\dot{V}_f \leq & -k_e \|{e}_\mathcal{R}\|^2 - k_{e_v}\|{e}_{v_o}\|^2 + {e}_{v_o}^\top \Delta G_\mathcal{R} G^\ast_{\mathcal{R},k} Y_{o_r} \hat{\vartheta}_{\mathcal{R}_o} + {e}_{v_o} \Delta G_\mathcal{R} {f}_{\text{int}_k} \\
		\leq & -k_e \|{e}_\mathcal{R}\|^2 - k_{e_v}\|{e}_{v_o}\|^2 - \beta_\vartheta\| {e}_{\vartheta_\mathcal{R}} \|^2 - \beta_{\vartheta_o}\|{e}_{\vartheta_{\mathcal{R},o}}\|^2 + \beta_\vartheta \bar{e}_{\vartheta_\mathcal{R}}^2 \\
		& + \beta_{\vartheta_o}\bar{e}_{\vartheta_{\mathcal{R},o}}^2 +{e}_{v_o}^\top \Delta G_\mathcal{R} G^\ast_{\mathcal{R},k} Y_{o_r} {e}_{\vartheta_{\mathcal{R},o}} + {e}_{v_o}^\top \Delta G_\mathcal{R} G^\ast_{\mathcal{R},k} Y_{o_r} {\vartheta}_{\mathcal{R}_o} \\
		& + {e}_{v_o}^\top \Delta G_\mathcal{R} {f}_{\text{int}_k} \\
		\leq & -k_e \|{e}_\mathcal{R}\|^2 - k_{e_v}\|{e}_{v_o}\|^2 - \beta_\vartheta\| {e}_{\vartheta_{\mathcal{R}}} \|^2 - \beta_{\vartheta_o}\|{e}_{\vartheta_{\mathcal{R},o}}\|^2 + \beta_\vartheta \bar{e}_{\vartheta_\mathcal{R}}^2 \\
		& + \beta_{\vartheta_o}\bar{e}_{\vartheta_{\mathcal{R},o}}^2 + \sqrt{\gamma_q} \|\Delta G_\mathcal{R} G^\ast_{\mathcal{R},k}\| \|Y_{o_r}\| \bar{e}_{\vartheta_{\mathcal{R},o}}  \\
		& + \sqrt{\gamma_q} \| \Delta G_\mathcal{R} G^\ast_{\mathcal{R},k} \| \| Y_{o_r} \| \|{\vartheta}_{\mathcal{R}_o}\|  + \sqrt{\gamma_q}\|\Delta G_\mathcal{R}\| \|{f}_{\text{int}_k}\|, 
	\end{align*}
	where  $\beta_\vartheta$, $\beta_{\vartheta_o} \in \mathbb{R}_{>0}$ are positive constants. Since $\|\Delta G_\mathcal{R} G^\ast_{\mathcal{R},k} \|$ $\leq$ $c_{\mathcal{R}_1}$, it holds that $\|\Delta G_\mathcal{R}\| \leq \frac{c_{\mathcal{R}_1}}{\| G^\ast_{\mathcal{R},k}\|}$, which is bounded, since $\sigma_{\text{min}}(G^\ast_{\mathcal{R},k}) = \frac{1}{\sigma_{\max}(G_{\mathcal{R},k})}$, and $G_{\mathcal{R},k}$ is full row rank. Furthermore ${f}_{\text{int}_k}$ is constant between events. Thus in view of \eqref{eq:e_nu bars (ACC_rolling)} and 
	since $\chi$ lies in the compact set $\mathcal{Q}$, we can conclude  that there exists a $\bar{\delta}_k$ such that:
	\begin{align*}
		\bar{\delta }_k \geq & + \beta_\vartheta \bar{e}_{\vartheta_{\mathcal{R}}}^2 + \beta_{\vartheta_o}\bar{e}_{\vartheta_{\mathcal{R},o}}^2 + \sqrt{\gamma_q} \|\Delta G_\mathcal{R} G^\ast_{\mathcal{R},k}\| \|Y_{o_r}\| \bar{e}_{\vartheta_o}  \\
		&+ \sqrt{\gamma_q} \| \Delta G_\mathcal{R} G^\ast_{\mathcal{R},k} \| \| Y_{o_r} \| \|{\vartheta}_{\mathcal{R}_o}\|  + \sqrt{\gamma_q}\|\Delta G_\mathcal{R}\| \|{f}_{\text{int}_k}\| 
	\end{align*}
	Hence $\dot{V}$ becomes 
	\begin{align*}
		\dot{V} \leq -k_\chi \|{\chi}\|^2 + \bar{\delta}_k,
	\end{align*}
	where $k_\chi \coloneqq \min\{k_e, k_{e_v}, \beta_\vartheta, \beta_{\vartheta_o}\}$. Therefore, by invoking Lemma \ref{lemma:barbalat (App_dynamical_systems)} of Appendix \ref{app:dynamical systems},  we guarantee that ${\chi}$ is ultimately bounded in a compact set defined by $k_\chi$ and $\bar{\delta}_k$, for $t \in [0, t_{\max})$.
	
	Now we investigate the slip prevention properties, similar to that of Theorem \ref{th:main th (ACC_rolling)}. The same derivation of ${f}_C$ yields:
	\begin{align*}\label{eq:contact force event triggered (dec_rolling)}
		{f}_{\scr C} = &W_h^{-1} \bigg( J_h B^{-1}\left[ \tau_k - {g}_q - \left(C_qJ_h^{-1} G_\mathcal{R}^\top + B \frac{d}{dt}(J_h^{-1} G_\mathcal{R}^\top)\right){v}_o \right] \\   &+ G_\mathcal{R}^\top M_o^{-1}(C_o{v}_o + {g}_o )\bigg). 
	\end{align*}
	By following a similar procedure as with the previous section, we conclude that  
	\begin{equation*} 
	{f}_{\scr C} = {\lambda}_k + (I - W_h^{-1} G_\mathcal{R}^\top M_o^{-1} \Delta G_\mathcal{R}) {f}_{\text{int}_k} + {h}_{f_k}
	\end{equation*}
	\begin{align*}
		{h}_{f_k} \coloneqq  & W_h^{-1} J_h B^{-1}(g_q - Y_\mathcal{R}(\check{x},\dot{\check{x}},e_{v_o},\dot{e}_{v_o}) + Y_r e_{\vartheta_\mathcal{R}}) + \notag \\
		&   W_h^{-1} G_\mathcal{R}^\top M_o^{-1} \bigg( G_\mathcal{R}G_\mathcal{R}^\top K_v {e}_{v_o} + {e}_\mathcal{R} + Y_{\mathcal{R}_o}(\eta_o,\omega_o,e_{v_o},\dot{e}_{v_o}){\vartheta}_{\mathcal{R}_o} \notag \\
		&  - Y_{o_r}e_{\vartheta_{\mathcal{R},o}}   + \Delta G_\mathcal{R} G_{\mathcal{R},k}^\ast({e}_\mathcal{R} - Y_{o_r}\hat{\vartheta}_{\mathcal{R}_o}) - g_o \bigg)
	\end{align*}
	
	Substitution of ${f}_{\scr C}$ into \eqref{eq:linearized friction constraint (ACC_rolling)}, which  ensures slip prevention, yields the following condition to be satisfied:
	\begin{align*}
		\Lambda(\mu_f) R_{\scr C}^\top ( I_{3N} - W_h^{-1} G_\mathcal{R}^\top M_o^{-1} \Delta G_\mathcal{R}) {f}_{\text{int}_k} & \succeq - \Lambda(\mu_f) R_{\scr C,k}^\top {\lambda}_k - \Lambda(\mu_f) R_{\scr C,k}^\top {h}_{f_k} 
	\end{align*}
	
	From the boundedness of signals, we conclude that ${h}_{f_k}$ is bounded for all $\forall t \in [0,t_{\max})$ in a compact set, independent of $t_{\max}$. Now let $\varepsilon_{h_k}$ denote the maximum bound of the elements of $\pm  \Lambda(\mu_f') R_{\scr C,k}^\top {h}_{f_k}$ and substitute ${f}_{\text{int}_k} = f_{\text{int}_k}' R_{\scr C,k} {\ell}_{\text{int},k}^*$ with $W \coloneqq R_{\scr C}^\top(\Delta R_{\scr C} {+} W_h^{-1} G_\mathcal{R}^\top M_o^{-1} \Delta G_\mathcal{R} R_{\scr C,k})$ to re-write the sufficient condition for no slip as:
	\begin{align*} 
		f_{\text{int},k}'\Lambda(\mu_f)(I_3 - W) {\ell}_{\text{int},k}^* & \succeq - \Lambda(\mu_f) R_{\scr C,k}^\top {\lambda}_k+
		\varepsilon_{h_k} \mathbbm{1}
	\end{align*}
	or, for each agent separately, 
	\begin{align}\label{eq:slip prev event triggered each agent (dec_rolling)}
		f_{\text{int},k}'\Lambda_i(\mu_f)(I_3 - W_{ii}) {\ell}_{\text{int},k,i}^* & \succeq - \Lambda_i(\mu_f) R_{\scr C,k_i}^\top {\lambda}_{k_i}  \notag \\
		&\hspace{-10mm} + f_{\text{int},k}'\sum_i\sum_{j\neq i} D_{ij} {\ell}^*_{\text{int},k,j} +
		\varepsilon_h {\mathbbm{1}}
	\end{align}
	where ${\ell}_{\text{int},k,i}^*\in\mathbb{R}^3$ is the $i$th agent's part in $\ell_{\text{int},k}^*$ (as opposed to the scalar $\ell_{\text{int},k_i}^*$),
	$W_{ii}$ is $i$th block matrix of $W$'s diagonal and $D_{ij} \in \mathbb{R}^3$ is the $ij$-block matrix of $\Lambda(\mu_f) W$. Here we show that the triggering conditions \eqref{eq:triggering condition G 1 (dec_rolling)} and \eqref{eq:triggering condition G 2 (dec_rolling)} ensure that $\lambda_{\max}(W) \leq \delta_c < 1$. By boundedness of the system dynamics, for sufficiently large $\varepsilon_{c} \in \mathbb{R}_{>0}$ bounding the terms $W_h^{-1} G^\top_\mathcal{R}M_o^{-1}$, it follows that $\lambda_{\max}(W) \leq ||W|| \leq \varepsilon_{c}( \delta_r +  \delta_p)$. Since $\delta_r < \frac{\delta_c}{2 \varepsilon_{c}}$ and $\delta_p < \frac{\delta_c}{2 \varepsilon_{c}}$ it follows that $\lambda_{\max}(W_{ii}) \leq  \delta_c < \frac{\sqrt{\mu^2+1}-1}{\sqrt{\mu^2 + 1}+1} <1$, $\forall i \in \mathcal{N}$, and hence Lemma \ref{lem:no slip perturbed_new (dec_rolling)} dictates that \eqref{eq:slip prev event triggered each agent (dec_rolling)} is satisfied when 
	\begin{align*} 
		f_{\text{int},k}'  \Lambda_i(\mu'_f) {\ell}_{\text{int},k,i}^* \succeq - \Lambda_i(\mu_f) R_{\scr C,k_i}^\top {\lambda}_{k_i} + \sum_i\sum_{j\neq i} D_{ij} {\ell}^*_{\text{int},k,j} +
		\varepsilon_{h_k} \mathbbm{1},
	\end{align*}
	where $\mu'_f$ as given by \eqref{eq:mu_new (dec_rolling)}. 
	
	From the boundedness of signals, there exists a $\varepsilon_d \in \mathbb{R}_{>0}$ such that $||D_{ij}|| \leq \varepsilon_d \delta_p$  for all $i,j \in \mathcal{N}$. Furthermore, 
	we know ${\ell}_{\text{int},k}^*$ is constant between updates and known to all agents. Therefore at each update $\min_j \{l_{k_j}\}$ and $\max_j \{\ell_{\text{int},k_j}^*\}$ can be computed, and we note that $\max_j \{\ell_{\text{int},k_j}^*\}$ is bounded by Assumption \ref{asm:full rank G (ACC_rolling)}. Thus a sufficient condition for the above expression to hold is:
	\begin{align*}
		f'_{\text{int},k} (\min_j\{l_{k_j}\} - \varepsilon_d \delta_p \max_j\{\ell_{\text{int},k_j}^*\}) \succeq  -b_{k_j} +  \varepsilon_{h_k}
	\end{align*}
	$\forall j\in\{1,\dots, Nn_s\}$, which is feasible since
	$\delta_p < \dfrac{\min_j\{l_{k_j}\}}{\varepsilon_d \max_j\{\ell_{\text{int},k_j}^*\}}$ from \eqref{eq:triggering condition G 2 (dec_rolling)}.

	By substituting \eqref{eq:event triggered f_int scalar (dec_rolling)} with the choice of $\epsilon_f \geq \varepsilon_d$, the left side satisfies
	\begin{align*}
		\kappa(\min_{j}\{b_{k_j}\}) + 1 + \epsilon_f \geq - b_{k_j} + \varepsilon_{h_k},
	\end{align*}
	where we use $\kappa(x) \geq 0$, $\kappa(x) + 1 \geq -x$, $\forall x\in\mathbb{R}$, and $\kappa(\min_j(b_{k_j})) > \kappa(b_{k_j})$, $\forall j\in\{1,\dots,Nl_f\}$, since $\kappa()$ is decreasing. Hence, by choosing a large enough $\epsilon_f$ we guarantee $\epsilon_f \geq \varepsilon_{h_k}$ and hence contact slip is actively prevented $\forall t\in[0,t_{\max})$. Following the proof of \ref{th:main th (ACC_rolling)}, it follows that $t_{\max} = \infty$, and thus slip prevention is ensured for the entirety of the manipulation task. 
	
	Furthermore, since $t_{\max} = \infty$, the previous Lyapunov analysis ensures that $\dot{V}_f < 0$ when $\|{\chi}\| > \sqrt{\frac{\bar{\delta}_k}{k_\chi}}$, guaranteeing thus, in view of Lemma \ref{lemma:barbalat (App_dynamical_systems)} of Appendix \ref{app:dynamical systems}, that ${\chi}$ will be ultimately bounded in a compact set around the origin, rendering the closed-loop system practically asymptotically stable.

	Case 2: In the previous analysis, practical asymptotic stability is ensured when no triggering occurs. Here we show that indeed the event triggering preserves the results from Case 1 and that the system does not exhibit Zeno behavior. For any $\chi \in \mathcal{X}\setminus \mathcal{Q}$, it follows that $\dot{V}_f \leq 0$. Thus if any event triggers in $\mathcal{X}\setminus \mathcal{Q}$ at $t = t_k$, $\Delta G_{\mathcal{R},k} = 0$ and $ \Delta R_{C,k} = 0$, and  it is straightforward to see that $V_{k+1} \leq V_k$ such that $\Omega_{k+1} \subseteq \Omega_k$. Furthermore $\dot{V}_f \leq 0$ holds after the event occurs and ensures $\chi$ enters $\mathcal{Q}$ in finite time.
	
	For $\chi \in \mathcal{Q}$, the condition $\dot{V} \leq k_\chi \|\chi\|^2 + \bar{\delta}_k$ holds although $\bar{\delta}_k$ will change between events. However, since $\chi$ and ${f}_{\text{int}_k}$ are bounded in $\mathcal{Q}$, there exists a maximum $\bar{\delta} \geq \bar{\delta}_k$ for which $\chi$ is ultimately bounded, and practical stability is preserved.
	
	Now we show there exists a lower bound between each event time instant. Events \eqref{eq:triggering condition G 1 (dec_rolling)} and \eqref{eq:triggering condition R (dec_rolling)} are dependent on bounds $\delta_r$, and $\delta_p$, where $\delta_r > 0$ is fixed and $\delta_p > 0$ and will never tend to zero due to boundedness of ${p}_{\scr C_i/o}$. From the continuous differentiability of ${p}_{\scr C_i/o}$ and ${p}_{\scr C_i/E_i}$, let $L_p, L_r \in \mathbb{R}_{>0}$ denote their respective Lipschitz constants. It follows that there exist lower bounds on event times defined by $\Delta t_p = \delta_p/L_p$, $\Delta t_r = \delta_r/L_r$, respectively. 
	
	Similarly, the event defined by \eqref{eq:triggering condition ev (dec_rolling)} depends on the bound $c_{\mathcal{R}_2}\gamma_q$. Denote by ${e}_{v_o} = [{e}_{v_o,p}^\top, {e}_{v_o,\eta}^\top]^\top \in\mathbb{R}^3 \times \mathbb{R}^3$. Then  \eqref{eq:triggering condition ev (dec_rolling)} occurs when 
	\begin{equation*}
		{e}_{v_o,\eta}(t_{k+1})^\top S(\Delta {p}_{\scr C_i/o}(t_{k+1}))  \tilde{{h}}_{f,k_i}(t_{k+1}) = c_{\mathcal{R}_2} \gamma_q,
	\end{equation*}
	where $\tilde{{h}}_{f,k_i} \coloneqq {f}_{\text{int},{k_i}} + G^\ast_{\mathcal{R},k_i} Y_{o_r} \hat{\vartheta}_{\mathcal{R}_o}(t)$, $\forall i\in\mathcal{N}$, $\forall k\in\mathbb{N}$, with $t_1 = 0$. Therefore, since $\|\Delta {p}_{\scr C_i/o}\| \leq \delta_p$ from \eqref{eq:triggering condition G 2 (dec_rolling)} and ${e}_{v_o}$, $Y_{o_r}$, $\hat{\vartheta}_{\mathcal{R}_o}$, ${f}_{\text{int},k}$ are bounded in compact sets for $t\in[t_k,t_{k+1})$ from the previous analysis, there exist positive constants $\underline{e}$ and $\underline{h}_i$ such that $\| {e}_{v_o,\eta}(t_{k+1})\| \geq \underline{e}$ and $\|\tilde{{h}}_{f,k_i}(t_{k+1})\| \geq \underline{h}_i$, $\forall i\in\mathcal{N}$. 
	Hence, by taking into account \eqref{eq:triggering condition G 1 (dec_rolling)} it holds that $c_{\mathcal{R}_2} \gamma_q \leq {\delta}_p \|{e}_{v_o,\eta}(t_{k+1}) \| \|\tilde{{h}}_{k_i}(t_{k+1})\|$ $\Delta t_e$, with $\Delta t_e$ being the inter-sampling time between the updates defined by \eqref{eq:triggering condition ev (dec_rolling)}. We conclude then that  $\Delta t_e \geq \frac{c_{\mathcal{R}_2} \gamma_q}{ \underline{e} \underline{h}_i {\delta}_p}$. 
	
	Finally, as $t_k$ is defined by satisfaction of any events from \eqref{eq:triggering condition G 1 (dec_rolling)}, \eqref{eq:triggering condition R (dec_rolling)}, or \eqref{eq:triggering condition ev (dec_rolling)}, it follows that $\Delta t \coloneqq t_{k+1} - t_k = \min \{ \Delta t_p, \Delta t_r, \Delta t_e \}$ where $\Delta t > 0$ and lower bounded. 	
\end{proof}

\section{Conclusion} \label{sec:Conclusion and FW (TCST_coop_manip)}
This chapter presented novel control protocols for the cooperative manipulation of a single object by $N$ robotics agents without employing force sensing. Firstly, we focused on rigid grasps, by introducing two adaptive decentralized control schemes that used quaternion-feedback and prescribed performance control, respectively. Next, we incorporated collision avoidance by using nonlinear MPC, in a centralized and a communication-based decentralized scheme. 
Secondly, we considered the case of rolling contacts. We developed novel adaptive centralized and  decentralized control schemes that compensate for the object's and the agents' dynamic uncertainties and guarantee avoidance of contact loss at the contact points.

\chapter{Formation Control and Rigid Cooperative Manipulation} \label{chapter:formation}
As discussed in Chapter \ref{ch:Introduction}, an important problem associated with multi-agent coordination is formation control. On one hand, formation specifications can be imposed by a higher level planner associated with temporal tasks. Moreover, as we show here, a particular instance of multi-agent formations, namely rigid formation, is tightly associated to rigid cooperative manipulation presented in the previous chapter. More specifically, 
this chapter addresses the following two topics.

Firstly, it deals with 
the problem of distance- and orientation-based formation control of a class of second-order nonlinear multi-agent systems in $\mathbb{SE}(3)$, under static and undirected communication topologies. More specifically, we design a decentralized model-free control protocol in the sense that each agent uses only local information from its neighbors to calculate its own control signal, without incorporating any knowledge of the model nonlinearities and exogenous disturbances. Moreover, the transient and steady-state response is solely determined by certain designer-specified performance functions and is fully decoupled by the agents' dynamic model, the control gain selection, the underlying graph topology as well as the initial conditions. Additionally, by introducing certain inter-agent distance constraints, we guarantee collision avoidance and connectivity maintenance between neighboring agents. 

Secondly, we introduce a new notion of distance rigidity, namely distance- and bearing-rigidity in $\mathbb{SE}(3)$, and we connect it with rigid cooperative manipulation. More specifically, the nodes of a general rigid framework are associated to the robotic agents of rigid cooperative manipulation schemes and the  object-agent  interaction  forces are expressed by  using  the  rigidity matrix of the graph formed by the robots,  which  encodes  the  infinitesimal  rigid  body  motions  of the  system.  Moreover,  we  show  that  the  associated  cooperative manipulation grasp matrix is related to the rigidity matrix via a range-nullspace relation, based on which we provide novel results on the relation between the arising interaction and internal forces and  consequently  on  the  energy-optimal  force  distribution  on  a cooperative  manipulation  system.

\section{Introduction}

During the last decades, decentralized control of networked multi-agent systems has gained a significant amount of attention due to the great variety of its applications, including  multi-robot systems, transportation, multi-point surveillance and biological systems. The main focus of multi-agent systems is the design of distributed control protocols in order to achieve global tasks, such as consensus \cite{ren_beard_consensus, olfati_murray_concensus, jadbabaie_morse_coordination, tanner_flocking}, and at the same time fulfill certain properties, e.g., network connectivity \cite{egerstedt_formation, zavlanos_2008_distributed}. 

A particular multi-agent problem that has been considered in the literature is the formation control problem, where the agents represent robots that aim to form a prescribed geometrical shape, specified by a certain set of desired relative configurations between the agents. The main categories of formation control that have been studied in the related literature are (\cite{oh_park_ahn_2015}) position-based control, displacement-based control, distance-based control and orientation-based control.

In distance-based formation control, inter-agent distances are actively controlled to achieve a desired formation, dictated by desired inter-agent distances. Each agent is assumed to be able to sense the position of its neighboring agents. When orientation alignment is considered as a control design goal, the problem is known as orientation-based (or bearing-based) formation control. The orientation-based control steers the agents to configurations that achieve desired relative orientation angles. In this work, we aim to design a decentralized control protocol such that both distance- and orientation-based formation is achieved.

The literature in distance-based formation control is rich, and is traditionally categorized in single or double integrator agent dynamics and directed or undirected communication topologies (see e.g. \cite{oh_park_ahn_2015, olfati_murray_2002, smith_broucke_francis_2006, hendrickx_anderson_delvenne_blondel_2007, anderson-yu-dasgupta-morse_2007, anderson_yu_fidan_hendrickx_2008, dimos_kalle_2008, cao_anderson_morse_yu_2008, yu_anderson_dagsputa_fidan_2009, krick_broucke_francis_2009, dorfler_francis_2010, oh_ahn_2011e, cao_morse_yu_anderson_dagsputa_2011, summers_yu_dagsputa_anderson_2011_tac, park_oh_ahn_2012_gradient, belabbas2012robustness, oh_ahn_2014a}).
Orientation-based formation control has been addressed in \cite{basiri_2010_angle_formation, eren_2012_bearing_formation, oh_ahn_2014_angle_based_formation, zhao2016bearing}, whereas the authors in \cite{oh_ahn_2014_angle_based_formation, bishop_2015_distributed, fathian2016globally} have considered the combination of distance- and orientation-based formation.

In most of the aforementioned works in formation control, the two-dimensional case with simple dynamics and point-mass agents has been dominantly considered. In real applications, however, the engineering systems have nonlinear second order dynamics and are usually subject to exogenous disturbances and modeling errors. Other important issues concern the connectivity maintenance, the collision avoidance between the agents and the transient and steady-state response of the closed loop system, which have not been taken into account in the majority of related woks. Thus, taking all the above into consideration, the design of robust distributed control schemes for the multi-agent formation control problem becomes a challenging task. 

Another special instance of formation control that has practical relevance and numerous applications in robotics is that of \textit{rigid formations}. Two cases of rigid formation control have been widely studied in the literature, namely \textit{distance rigidity} and \textit{bearing rigidity}. The classic distance rigidity theory studies the problem of under what conditions can the geometric pattern of a network be uniquely determined if the length (distance) of each edge in the multi-agent team is fixed. It is a combinatorial theory for characterizing the ``stiffness" or ``flexibility" of structures formed by rigid bodies connected by flexible linkages or hinges, and it has been applied extensively in distance-based formation control and network localization \cite{de2016distributed,sun2018distributed,chen2017global,mou2016undirected,zelazo2015decentralized,tian2013global,oh2014distance,krick2009stabilisation,anderson2008rigid,eren2004,mao2007,james2006}. Bearing rigidity theory studies the fundamental problem of under what conditions can the geometric pattern of a multi-agent system be uniquely determined if the bearing of each edge is fixed \cite{zhao2019bearingRig}, and it has been used for bearing-based control and estimation problems \cite{Zhao16bearing,Tron15RigidComp,Eren12Formation,Bishop2011Stabilization}. Recent works have developed bearing rigidity theory on the manifolds of $\mathbb{SE}(2)$ \cite{bearingSE2} and $\mathbb{SE}(3)$ \cite{bearingSE3}. In this chapter, we introduce the notion of \textit{distance and bearing rigidity}, which studies under what conditions can the geometric pattern of a multi-agent system be uniquely determined if both the \textit{distance} and the \textit{bearing} of each edge is fixed. Moreover, we combine the latter with \textit{rigid} cooperative manipulation, i.e., configurations where a number of robotic agents are attached to a common object by means of rigid contact points.

As shown in the previous chapter, rigid cooperative manipulation by robotic agents (i.e., when the grasps are rigid) is an important and challenging topic, indispensable in cases of difficult maneuvers or heavy payloads.
An important property in rigid cooperative manipulation systems that has been studied thoroughly in the {related} literature and overlooked in the previous chapter, is the regulation of internal forces. Internal forces are forces exerted by the agents at the grasping points that do not contribute to the motion of the object. While a certain amount of such forces is required in many cases (e.g., to avoid contact loss in multi-fingered manipulation), they need to be minimized in order to prevent object damage and unnecessary effort of the agents. Most works in rigid cooperative manipulation assume a certain decomposition of the {interaction} forces in motion-inducing and internal ones, without explicitly showing that the actual internal forces will be indeed regulated to the desired ones (e.g., \cite{tsiamis2015cooperative,heck2013internal,caccavale2000task}); \cite{walker1991analysis,williams1993virtual,chung2005analysis,erhart2015internal} analyze specific load decompositions based on whether they provide internal force-free expressions, whereas \cite{erhart2016model} is concerned with the cooperative manipulation interaction dynamics.	
The decompositions in the aforementioned works, however, are based on the inter-agent distances and do not take into account the actual dynamics of the agents. The latter, as we show in this chapter, is tightly connected to the internal forces as well as their relation to the total force exerted by the agents at the grasping points.

This chapter deals with the following two topics:
\begin{enumerate}
	\item
	Firstly, we address the distance-based formation control problem with orientation alignment for a team of rigid bodies operating in $\mathbb{SE}(3)$, with unknown second-order nonlinear dynamics and external disturbances. We propose a purely decentralized control protocol that guarantees distance formation, orientation alignment as well as collision avoidance and connectivity maintenance between neighboring agents and in parallel ensures the satisfaction of prescribed transient and steady state performance. The prescribed performance control framework has been incorporated in multi-agent systems in \cite{babis_2014_formation} and \cite{mehdifar2019prescribed} for minimally rigid formations, 
	where first order dynamics have been considered without taking into account the problem of orientation alignment.
	\item 
	We integrate rigid cooperative manipulation with rigidity theory. Motivated by rigid cooperative manipulation systems, where the inter-agent distances \textit{and} bearings are fixed, we introduce the notion of \textit{distance and bearing rigidity} in the special Euclidean group $\mathbb{SE}(3)$. Based on recent results, we show next that the interaction forces in a rigid cooperative manipulation system depend on the distance and bearing rigidity matrix, a matrix that encodes the allowed coordinated motions of the multi-agent-object system. Moreover, we prove that the cooperative manipulation grasp matrix, which relates the object and agent velocities, is connected via a range-nullspace relation to the rigidity matrix. Furthermore, 
	we rely on the aforementioned findings to provide new results on the internal force-based rigid cooperative manipulation. We {derive}
	novel results on the relation between the arising interaction and internal forces in a cooperative manipulation system.  This leads to novel conditions on the internal force-free object-agents force distribution and consequently to optimal, in terms of energy resources, cooperative manipulation. 
\end{enumerate}

Finally, we verify all the theoretical findings through simulation results.

\section{Formation Control in $\mathbb{SE}(3)$} \label{sec:formation control}

\subsection{Problem Formulation} \label{sec:prob_formulation (formation)}
Consider a set of $N$ rigid bodies, with $\mathcal{N} = \{ 1,2, \ldots, N\}$, $N  \geq 2$, operating in a workspace $W\subseteq \mathbb{R}^3$. We consider that each agent occupies a ball $\mathcal{B}(p_i,r_i)$, where $p_i\in\mathbb{R}^3$ is the position of the agent's center of mass with respect to an inertial frame $\mathcal{F}_o$ and $r_i\in\mathbb{R}_{>0}$ is the agent's radius (see Fig. \ref{fig:agents_geometry (formation)}). We also denote by $R_i\in  \mathbb{SO}(3) $ the rotation matrix associated with the orientation of the $i$th rigid body.    {Moreover, we denote by $v_{i,L}\in\mathbb{R}^3$ and $\omega_i\in\mathbb{R}^3$ the linear and angular velocity of agent $i$ with respect to frame $\mathcal{F}_o$. The vectors $p_i$ are expressed in $\mathcal{F}_o$ coordinates, whereas $v_{i,L}$ and $\omega_i$ are expressed in a local frame $\mathcal{F}_i$ centered at each agent's center of mass. The position, though, of $\mathcal{F}_o$, is not required to be known by the agents, as will be shown later.}    
By defining $x_i \coloneqq (p_i, R_i) \in  \mathbb{SE}(3) $ and $v_i\coloneqq [v_{i,L}^\top, \omega^\top_i]^\top\in\mathbb{R}^6$, we model each agent's motion with the $2$nd order Newton-Euler dynamics:
\begin{subequations}\label{eq:system (formation)} 
	\begin{align} 
	&\hspace{-3mm} \dot{x}_i = (  {R_iv_{i,L}}, R_i S(\omega_i)) \in \mathbb{T}_{R_i}, \label{eq:system_1 (formation)} \\ 
	&\hspace{-3mm}   {u_i = M_i \dot{v}_i + C_i v_i+g_i + w_i},  \label{eq:system_2 (formation)} 
	\end{align}
\end{subequations}
where the matrix $M_i\in \mathbb{R}^{6\times6}$ is the constant positive definite inertia matrix, $C_i \coloneqq C_i(v_i): \mathbb{R}^6 \to  \mathbb{R}^{6\times6}$ is the Coriolis matrix, $g_i \coloneqq g_i(x_i): \mathbb{SE}(3) \to\mathbb{R}^6$ is the body-frame gravity vector,  $w_i \coloneqq w_i(x_i,v_i,t):\mathbb{SE}(3)\times\mathbb{R}^6\times\mathbb{R}_{\geq 0}  \to  \mathbb{R}^6$ is a bounded vector representing model uncertainties and external disturbances,  and $\mathbb{T}_{R_i} \coloneqq \mathbb{R}^3\times T_{R} \mathbb{SO}(3)$, where $T_{R} \mathbb{SO}(3)$ is the tangent space to $\mathbb{SO}(3)$ at $R$. Finally, $u_i\in\mathbb{R}^6$ is the control input vector representing the $6$D generalized force acting on agent $i$. The following properties hold for the aforementioned terms:
\begin{itemize}
	\item   {The terms $M_i, C_i(\cdot), g_i(\cdot)$ are \textit{unknown}, $C_i(\cdot), g_i(\cdot)$ are continuous and it holds that 
		\begin{subequations} \label{eq:M property (formation)}	
			\begin{align}
			& 0 < \underline{m}_i < \bar{m}_i < \infty \label{eq:M property 1 (formation)}\\
			& \|g_i(x_i)\| \leq \bar{g}_i, \forall x_i\in \mathbb{SE}(3) , \label{eq:M property 2 (formation)}
			\end{align}
		\end{subequations}
		$\forall i\in\mathcal{N}$, where $\bar{g}_i$ is a finite \textit{unknown} positive constant and $\underline{m}_i\coloneqq \lambda_{\min}(M_i)$, and $\bar{m}_i \coloneqq \lambda_{\max}(M_i)$, which are also \textit{uknown}}, $\forall i\in\mathcal{N}$. 
	\item   {The functions $w_i(x_i,v_i,t)$ are assumed to be continuous in $v_i\in\mathbb{R}^6$ and for each fixed $v_i\in\mathbb{R}^6$, the functions $(x_i,t) \to  w_i(x_i,v_i,t)$ are assumed to be bounded by \textit{unknown} positive finite constants $\bar{w}_i$, i.e.,  $\| w_i(x_i,v_i,t) \|\leq \bar{w}_i < \infty$, $\forall x_i\in \mathbb{SE}(3) ,t\in\mathbb{R}_{\geq 0}$, $i\in\mathcal{N}$. }
\end{itemize}



The dynamics \eqref{eq:system (formation)} can be written in a vector form representation as:
\begin{subequations} \label{eq:system_MAS (formation)} 
	\begin{align} 
	\dot{x} & =   {h_x}, \label{eq:system_1_MAS (formation)} \\ 
	u & = M \dot{v} + C v +g + w,  \label{eq:system_2_MAS (formation)}  
	\end{align}
\end{subequations}
where $x \coloneqq (x_1,\dots,x_N)\in \mathbb{SE}(3)^N$, $v \coloneqq [v_1^\top, \dots, v_N^\top]^\top$ $\in \mathbb{R}^{6N}$, $u \coloneqq [u_1^\top, \dots, u_N^\top]^\top \in \mathbb{R}^{6N}$, and
\begin{align*}
h_{x} \coloneqq h_{x}(x,v) \coloneqq& (h_{x_1}(x_1,v_1),\dots,h_{x_N}(x_N,v_N)) \\
\coloneqq& ((R_1v_{1,L}, R_1S(\omega_1)),\dots,(R_Nv_{N,L}, R_NS(\omega_N)) )\\ 
&\in \mathbb{T}_{R_1}\times\dots\times\mathbb{T}_{R_N}, \\
M \coloneqq& \text{diag}\{[M_i]_{i\in\mathcal{N}} \} \in\mathbb{R}^{6N\times 6N}, \\
C\coloneqq C(v) \coloneqq& \text{diag}\{[C_i(v_i)]_{i\in\mathcal{N}} \} \in\mathbb{R}^{6N\times 6N}, \\
g\coloneqq g(x) \coloneqq& [g_1(x_1)^\top,\dots, g_N(x_N)^\top]^\top \in\mathbb{R}^{6N},  \\ 
w\coloneqq w(x,v,t) \coloneqq& [w_1(x_1,v_1,t)^\top,\dots, w(x_N,v_N,t)^\top]^\top \in\mathbb{R}^{6N}.
\end{align*}

\begin{figure}[t!]
	\centering
	\begin{tikzpicture}[scale = 0.45]	
	\draw [color=black,thick,->,>=stealth'](-9, -5) to (-7, -5);
	\draw [color=black,thick,->,>=stealth'](-9, -5) to (-9, -3);
	\draw [color=black,thick,->,>=stealth'](-9, -5) to (-10, -6.5);
	\node at (-9.8, -5.0) {$\mathcal{F}_o$};
	
	\draw [color = blue, fill = blue!20] (-4.5,0) circle (2.5cm);
	\node at (-5.7, 0.0) {$\mathcal{F}_i$};
	\draw[green,thick,dashed] (-4.5,0) circle (5.0cm);
	\draw [color=black,thick,->,>=stealth'](-9, -5) to (-4.5, -0.1);
	\node at (-7.7, -3.0) {$p_i$};
	\draw [color=green,thick,dashed,->,>=stealth'](-4.5, 0.0) to (-8.93, 2.43);
	\node at (-7.3, 2.15) {$\varsigma_i$};
	\draw [color=black,thick,dashed,->,>=stealth'](-4.5, 0.0) to (-2.0, 0.0);
	\node at (-3.3, 0.3) {$r_i$};
	\node at (-4.5, 0.0) {$\bullet$};
	
	\draw [color = red, fill = red!20] (3.2, 0) circle (1.5cm);
	\node at (2.5, 0.3) {$\mathcal{F}_j$};
	\draw[orange,thick,dashed,] (3.2, 0) circle (4.1cm);
	\draw [color=black,thick,->,>=stealth'](-9, -5) to (3.2, -0.1);
	\node at (-5.0, -3.9) {$p_j$};
	\draw [color=orange,thick,dashed,->,>=stealth'](3.2, 0.0) to (3.2, -4.0);
	\node at (3.8, -2.7) {$\varsigma_j$};
	\draw [color=black,thick,dashed,->,>=stealth'](3.2, 0.0) to (4.7, 0.0);
	\node at (4.1, 0.3) {$r_j$};
	\node at (3.2, 0.0) {$\bullet$};
	\end{tikzpicture}
	\caption{Illustration of two agents $i$, $j \in \mathcal{N}$ in the workspace; $\mathcal{F}_o$ is the inertial frame, $\mathcal{F}_i$, $\mathcal{F}_j$ are the frames attached to the agents' center of mass, $p_i$, $p_j \in \mathbb{R}^3$ are the positions of the center of mass with respect to $\mathcal{F}_o$; $r_i$, $r_j$ are the radii of the agents and $\varsigma_i > \varsigma_j$ are their sensing ranges.}
	\label{fig:agents_geometry (formation)}
\end{figure}
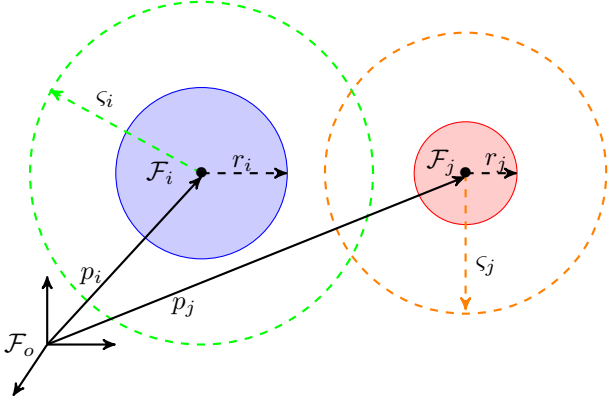

It is also further assumed that each agent has a limited sensing range of $\varsigma_i > \max_{i,j\in\mathcal{N}}\{r_i+r_j\}$. Therefore, by defining the set $\mathcal{N}_i:\mathbb{R}^{3N}\rightrightarrows\mathbb{N}$, with $\mathcal{N}_i(p) \coloneqq \{j\in\mathcal{N} : p_j\in\mathcal{B}(p_i,\varsigma_i)\}$, and $p\coloneqq [p^\top_1,\dots,p^\top_N]^\top\in\mathbb{R}^{3N}$, agent $i$ can measure the relative offset   {$R^\top_i(p_i - p_j)$ (i.e., expressed in $i$'s local frame)}, the distance $\|p_i - p_j\|$, as well as the relative orientation $R^\top_j R_i$ with respect to its neighbors $j\in\mathcal{N}_i(p)$. In addition, we consider that each agent can measure its own velocity subject to time- and state-varying bounded noise, i.e., agent $i$ has continuous feedback of $\widetilde{v}_i \coloneqq [  {\widetilde{v}_{i,L}^\top},\widetilde{\omega_i}^\top]^\top \coloneqq v_i + \mathsf{n}_i$, $\forall i\in\mathcal{N}$, where $\mathsf{n}_i\coloneqq \mathsf{n}_i(x_i,t):\mathbb{SE}(3) \times\mathbb{R}_{\geq 0}\to\mathbb{R}^6$ are vector fields  bounded by \textit{unknown} positive finite constants $\bar{\mathsf{n}}_i$, i.e., $\|\mathsf{n}_i(x_i,t) \| \leq \bar{\mathsf{n}}_i$, $\forall x_i\in \mathbb{SE}(3) , t\in\mathbb{R}_{\geq 0}$, $i\in\mathcal{N}$. Moreover, the vector fields  $\mathsf{n}_{i,d}\coloneqq \mathsf{n}_{i,d}(x_i,\dot{x}_i,t): \mathbb{SE}(3) \times\mathbb{T}_{R_i}\times\mathbb{R}_{\geq 0}\to\mathbb{R}^6$ with $\mathsf{n}_{i,d}(x_i,\dot{x}_i,t) \coloneqq \dot{\mathsf{n}}_i(x_i,\dot{x}_i) = \frac{\partial \mathsf{n}_i(x_i,t)}{\partial x_i}\dot{x}_i + \frac{\partial \mathsf{n}_i(x_i,t)}{\partial t}$ are assumed to be continuous in $\dot{x}_i\in \mathbb{T}_{R_i}$ and for each fixed $\dot{x}_i\in \mathbb{T}_{R_i}$, the functions $(x_i,t)\to \mathsf{n}_{i,d}(x_i,\dot{x}_i,t)$ are assumed to be bounded by \textit{unknown} positive finite constants $\bar{\mathsf{n}}_{i,d}$, i.e.,  $\|\mathsf{n}_{i,d}(x_i,\dot{x}_i,t)\| \leq \bar{\mathsf{n}}_{i,d}$, $\forall x_i\in \mathbb{SE}(3) , t\in\mathbb{R}_{\geq 0}$, $i\in\mathcal{N}$.}


\begin{remark} (\textbf{Local relative feedback})
	Note that the agents do not need to have information of any   {common} global inertial frame. The feedback they obtain is relative with respect to their neighboring agents (expressed in their local frames) and they are not required to perform transformations in order to obtain absolute positions/orientations.   {In the same vein, note also that the velocities $v_i$ are vectors expressed in the agents' local frames.}
	
\end{remark}

The topology of the multi-agent network is modeled through the   {\textit{undirected}} graph $\mathcal{G} \coloneqq (\mathcal{N},\mathcal{E})$, with $\mathcal{E}=\{(i,j)\in\mathcal{N}^2 : j\in\mathcal{N}_i(p(0)) \text{ and } i\in\mathcal{N}_j(p(0))\}$ (i.e., the initially connected agents), which is assumed to be nonempty and   {\textit{connected}}. We further denote $\mathcal{K} \coloneqq \{1,\dots,K\}$ where $K \coloneqq |\mathcal{E}|$. Given the $k$-th edge, we use the simplified notation $(k_1,k_2)$ for the function that assigns to edge $k$ the respective agents, with $k_1,k_2\in\mathcal{N}$, $\forall k\in\mathcal{K}$. 
{Since the agents are heterogeneous with respect to their sensing capabilities (different sensing radii $\varsigma_{i}$), the fact that the initial graph is nonempty, connected and undirected implies that}
\begin{equation*}
\lVert p_{k_2}(0)-p_{k_1}(0) \rVert < d_{k,\text{con}},
\end{equation*}
with $d_{k,\text{con}} \coloneqq \min\{\varsigma_{k_1},\varsigma_{k_2}\},\forall k\in \mathcal{K}$. 
We also consider that $\mathcal{G}$ is static in the sense that no edges are added to the graph. We do not exclude, however, edge removal through connectivity losses between initially neighboring agents, which we guarantee to avoid.   {That is, the proposed methodology guarantees that $\lVert p_{k_2}(t)-p_{k_1}(t) \rVert < d_{k,\text{con}}$, $\forall k\in\mathcal{K}$, $\forall t\in\mathbb{R}_{\geq 0}$.} 
It is also assumed that at $t=0$ the neighboring agents are at a collision-free configuration, i.e., $d_{k,\text{col}} < \lVert p_{k_2}(0)-p_{k_1}(0)\rVert, \forall k\in \mathcal{K}$, with $d_{k,\text{col}} \coloneqq r_{k_1}+r_{k_2}$. Hence, we conclude that 
\begin{equation}
d_{k,\text{col}} < \lVert p_{k_2}(0)-p_{k_1}(0)\rVert < d_{k,\text{con}}, \forall k\in \mathcal{K}. \label{eq:at t=0 (formation)}
\end{equation}

The desired formation is specified by the constants $d_{k,\text{des}}\in\mathbb{R}_{\geq 0}, R_{k,\text{des}}\in\mathbb{SO}(3), \forall k\in \mathcal{K}$, for which, the formation configuration is called \textit{feasible} if the set $\{x\in \mathbb{SE}(3)^N  : \lVert p_{k_2} - p_{k_1} \rVert = d_{k,\text{des}}, \ R^\top_{k_2}R_{k_1} = R_{k,\text{des}}, \forall k\in \mathcal{K}\}$ is nonempty.   
Due to the fact that the agents are not dimensionless and their communication capabilities are limited, the control protocol, except from achieving a desired inter-agent formation and maintaining connectivity, should also guarantee for all $t\in\mathbb{R}_{\geq 0}$ that the initially neighboring agents avoid collision with each other. Therefore, all pairs $(k_1,k_2)\in\mathcal{N}^2$ of agents that initially form an edge must remain within distance greater than $d_{k,\text{col}}$ and less than $d_{k,\text{con}}$. We also make the following assumptions that  on the graph topology: 
\begin{assumption} \label{assump:basic_assumption (formation)}
	The communication graph $\mathcal{G}$ is a tree.
\end{assumption}

Formally, the robust formation control problem under the aforementioned constraints is formulated as follows:
\begin{problem} \label{problem (formation)}
	Given $N$ agents governed by the dynamics \eqref{eq:system (formation)}, under Assumption \ref{assump:basic_assumption (formation)} and given the desired inter-agent configuration constants $d_{k,\text{des}}  {\in\mathbb{R}_{\geq 0}}$, $R_{k,\text{des}}  {\in \mathbb{SO}(3) }$, with $d_{k,\text{col}}<d_{k,\text{des}} < d_{k,\text{con}}$, $\forall k\in \mathcal{K}$, design decentralized control laws $u_i \in\mathbb{R}^6, i\in\mathcal{N}$ such that, $\forall \ k \in \mathcal{K}$, the following hold:
	\begin{enumerate}
		\item $\lim\limits_{t  \to  \infty} \|p_{k_2}(t)-p_{k_1}(t)\| = d_{k,\text{des}}$;
		\item $\lim\limits_{t  \to  \infty} [R_{k_2}(t)]^\top R_{k_1}(t) = R_{k,\text{des}}$; 
		\item $d_{k,\text{col}} < \|p_{k_2}(t)-p_{k_1}(t)\| < d_{k,\text{con}}, \forall \ t \in \mathbb{R}_{\geq 0}$.
	\end{enumerate}
\end{problem}

{The term ``robust" here refers to robustness of the proposed methodology with respect to the unknown dynamics and external disturbances in \eqref{eq:system (formation)} as well as the unknown noise $\mathsf{n}_i(\cdot)$ in the velocity feedback.}

\subsection{Problem Solution} \label{sec:solution (formation)}

Let us first introduce the distance and orientation errors:
\begin{subequations} \label{eq:errors (formation)}
	\begin{align}
	e_k &\coloneqq \left\| p_{k_2}-p_{k_1} \right\|^2-d_{k,\text{des}}^2 \ \   {\in\mathbb{R}},  \label{eq:error e_k (formation)}\\
	\psi_k &\coloneqq \frac{1}{2} \text{tr}\Big[I_{3} - R^\top_{k,\text{des}} R^\top_{k_2} R_{k_1}  \Big] \ \   {\in [0,2]}, \label{eq:error psi_k (formation)}
	\end{align}
\end{subequations}
$\forall k \in \mathcal{K}$,   {where we have used Proposition \ref{prop:R trace (app_useful_prop)}} of Appendix \ref{app:useful_prop}. Regarding $e_k$, our goal is to guarantee $\lim_{t\to\infty} e_k(t) \to 0$ from all initial conditions satisfying \eqref{eq:at t=0 (formation)}, while avoiding inter-agent collisions and connectivity losses among the initially connected agents specified by $\mathcal{E}$. Regarding $\psi_k$, we aim to guarantee the following:
\begin{enumerate}
	\item $\lim_{t\to\infty}\psi_k(t) \to 0$, which, according to Proposition \ref{prop:R trace (app_useful_prop)} of Appendix \ref{app:useful_prop} implies that $$\lim_{t\to\infty} R_{k_2}(t)^\top R_{k_1}(t) = R_{k,\text{des}}$$
	\item  $\psi_k(t) < 2$, $\forall t\in\mathbb{R}_{\geq 0}$, since the configuration $\psi_k = 2$ is an undesired equilibrium, as will be clarified later\footnote{It has been proved that topological obstructions do not allow global stabilization on $ \mathbb{SO}(3) $ with a continuous feedback control law (see \cite{lee2017attitude,lee10control,bhat2000topological})}.
\end{enumerate}

By using the properties of skew-symmetric matrices presented Appendix \ref{app:useful_prop}, we derive the following dynamics of the errors \eqref{eq:errors (formation)}:
\begin{subequations} \label{eq:errors derivative (formation)}
	\begin{align}
	\dot{e}_k &=   {2(p_{k_2}-p_{k_1})^\top(R_{k_2}v_{k_2,L}-R_{k_1}v_{k_1,L})} \notag \\
	& =2  { (R^\top_{k_1}\widetilde{p}_{k_2,k_1})^\top (R^\top_{k_1}R_{k_2}v_{k_2,L}-v_{k_1,L})}, \label{eq:error e_k dot (formation)} \\
	\dot{\psi}_k &= \frac{1}{2}e_{R_k}^\top (R^\top_{k_1} R_{k_2}\omega_{k_2} - \omega_{k_1}), \label{eq:error psi_k dot (formation)} 
	\end{align}
\end{subequations} 
where $\widetilde{p}_{k_2,k_1} \coloneqq p_{k_2} - p_{k_1}$ and $e_{R_k}\coloneqq S^{-1}(R^\top_{k_1}R_{k_2}R_{k,\text{des}}-R^\top_{k,\text{des}}R^\top_{k_2}R_{k_1})$, $\forall k\in \mathcal{K}$.

By employing Proposition \ref{prop:e_R frobenious (app_useful_prop)} of Appendix \ref{app:useful_prop}, we obtain $\|e_{R_k}\|^2 = \|R^\top_{k_2}R_{k_1}-R_{k,\text{des}}  \|^2_\text{F}(1 - \tfrac{1}{8} \|R^\top_{k_2}R_{k_1}-R_{k,\text{des}}\|^2_\text{F})$ as well as \\
\begin{align}
\|R^\top_{k_2}R_{k_1}-R_{k,\text{des}}\|^2_\text{F} & = \text{tr}\Big[ (R^\top_{k_2}R_{k_1}-R_{k,\text{des}})^\top(R^\top_{k_2}R_{k_1}-R_{k,\text{des}}) \Big] \notag \\
& = \text{tr}\left[2I_3 -2R^\top_{k,\text{des}}R^\top_{k_2}R_{k_1} \right] = 4\psi_k \notag.
\end{align}
Hence, it holds that:
\begin{equation} \label{eq:e_R_k and psi_k (formation)}
\|e_{R_k} \|^2 = 2\psi_k (2 - \psi_k),
\end{equation}
which implies that: $\|e_{R_k}\| = 0 \Rightarrow \psi_k = 0 \ \textit{or} \ \psi_k =2$, $\forall k\in\mathcal{M}$. The two configurations $\psi_k =0$ and $\psi_k = 2$ correspond to the desired and undesired equilibrium, respectively.  

The concepts and techniques of prescribed performance control (see Appendix \ref{app:PPC}) are adapted in this work in order to: a) achieve predefined transient and steady-state response for the distance and orientation errors $e_k$, $\psi_k$, $\forall k \in \mathcal{K}$, as well as ii) avoid the violation of the collision and connectivity constraints between initially neighboring agents, as presented in Section \ref{sec:prob_formulation (formation)}. The mathematical expressions of prescribed performance are given by the inequality objectives: 
\begin{subequations} \label{eq:ppc ineq (formation)}
	\begin{align}
	-C_{k,\text{col}} \rho_{e_k}(t) &< e_k(t) < C_{k,\text{con}} \rho_{e_k}(t), \\
	0 & \leq   \psi_{k}(t) < \rho_{\psi_k}(t) < 2, \label{eq:ppc ineq psi (formation)}
	\end{align}
\end{subequations}
$\forall k \in \mathcal{K}$, where $\rho_{e_k}\coloneqq \rho_{e_k}(t):\mathbb{R}_{\geq 0} \to  \left[\tfrac{\rho_{\scr e_k,\infty}} {\max\{C_{k,\text{con}}, C_{k,\text{col}} \}},1\right]$,  $\rho_{\psi_k}\coloneqq \rho_{\psi_k}(t):\mathbb{R}_{\geq 0}\to[\rho_{\scr \psi_k,\infty}, \rho_{\scr \psi_k,0}]$, with 
\begin{align*}
\rho_{e_k}(t) & \coloneqq \left[1 - \frac{\rho_{\scr e_k,\infty}} {\max\{C_{k,\text{con}}, C_{k,\text{col}} \}}\right] e^{-l_{e_k} t}  + \frac{\rho_{\scr e_k,\infty}} {\max\{C_{k,\text{con}}, C_{k,\text{col}} \} } , \\
\rho_{\psi_k}(t) &\coloneqq (\rho_{\scr \psi_k,0} - \rho_{\scr \psi_k,\infty})e^{-l_{\psi_k} t} + \rho_{\scr \psi_k,\infty}, 
\end{align*}
are designer-specified, smooth, bounded, and decreasing functions of time; the constants $l_{e_k}$, $l_{\psi_k}$ $\in\mathbb{R}_{\geq 0}$, and $\rho_{\scr e_k,\infty}\in(0,\max\{C_{k,\text{con}}, C_{k,\text{col}} \})$, $\rho_{\scr \psi_k,\infty}\in (0,\rho_{\scr \psi_k,0})$, $\forall k \in \mathcal{K}$, incorporate the desired transient and steady-state performance specifications respectively, as presented in Section \ref{app:PPC}, and $C_{k,\text{col}}$, $C_{k,\text{con}}\in\mathbb{R}_{>0},\forall k \in \mathcal{K}$, are associated with the collision and connectivity constraints. In particular, we select
\begin{subequations} \label{eq:C_k (formation)}
	\begin{align}
	C_{k,\text{col}} &\coloneqq d^2_{k, \text{des}} - d^2_{k,\text{col}}, \\
	C_{k,\text{con}} &\coloneqq d^2_{k,\text{con}} - d^2_{k, \text{des}} ,
	\end{align}
\end{subequations}
$\forall k \in \mathcal{K}$, which, since the desired formation is compatible with the collision and connectivity constraints (i.e., $d_{k,\text{col}}<d_{k,\text{des}}<d_{k,\text{con}}, \forall k \in \mathcal{K}$), ensures that $C_{k,\text{col}},C_{k,\text{con}}\in\mathbb{R}_{>0},\forall k \in \mathcal{K}$, and consequently, in view of \eqref{eq:at t=0 (formation)}, that:
\begin{subequations} \label{eq:ppc_time_0 (formation)}
	\begin{align}
	-C_{k,\text{col}} \rho_{e_k}(0) < e_k(0) <\rho_{e_k}(0) C_{k,\text{con}}, 
	\end{align}
	$\forall k \in \mathcal{K}$. Moreover, assuming that $\psi_k(0) < 2$, $\forall k\in \mathcal{K}$, by choosing 
	\begin{equation}
	\rho_{\scr \psi_k,0} = \rho_{\psi_k}(0) \in \Big(\psi_k(0),2\Big), \label{eq:rho_psi_0 (formation)}
	\end{equation}
	it is also guaranteed that:
	\begin{align}
	{0 \leq \psi_k(0) < \rho_{\psi_k}(0)} < 2,
	\end{align}
\end{subequations}
$\forall k \in \mathcal{K}$. Hence, if we guarantee prescribed performance via \eqref{eq:ppc ineq (formation)}, by setting the steady-state constants $\rho_{\scr e_k,\infty}, \rho_{\scr \psi_k,\infty}$ arbitrarily close to zero and by employing the decreasing property of $\rho_{e_k}(t),\rho_{\psi_k}(t),\forall k \in \mathcal{K}$, we guarantee practical convergence of the errors $e_k(t), \psi_k(t)$ to zero and we further obtain:
\begin{subequations} \label{eq:ppc goal (formation)}
	\begin{align} 
	-C_{k,\text{col}} &< e_k(t) < C_{k,\text{con}}, \\
	{0} &  {\leq \psi_{k}(t) < \rho_{\psi_k}(t)},
	\end{align}
\end{subequations}
$\forall t\in\mathbb{R}_{\geq 0}$, which, owing to \eqref{eq:C_k (formation)}, implies:
\begin{align*}
d_{k,\text{col}} < \lVert p_{k_2}(t)-p_{k_1}(t)\rVert < d_{k,\text{con}},
\end{align*}
$\forall k \in \mathcal{K}, t \in \mathbb{R}_{\geq 0}$, providing, therefore, a solution to problem \ref{problem (formation)}.   {Moreover, note that the choice of $\rho_{\scr \psi_k,0}$ along with \eqref{eq:ppc goal (formation)} guarantee that $\psi_k(t) < 2$, $\forall t\in\mathbb{R}_{\geq 0}$ and the avoidance of the singular equilibrium}.

In the sequel, we propose a decentralized control protocol that does not incorporate any information on the agents' dynamic model and guarantees \eqref{eq:ppc ineq (formation)} for all $t\in\mathbb{R}_{\geq 0}$. 

Given the errors $e_k, \psi_k$ defined in the previous section, we perform the following steps: 

\textbf{Step I-a}: Select the corresponding functions $\rho_{e_k}(\cdot), \rho_{\psi_k}(\cdot)$ and positive parameters $C_{k,\text{con}}$, $C_{k,\text{col}}$, $k \in \mathcal{K}$, following \eqref{eq:ppc ineq (formation)}, \eqref{eq:rho_psi_0 (formation)}, and \eqref{eq:C_k (formation)}, respectively, in order to incorporate the desired transient and steady-state performance specifications as well as the collision and connectivity constraints, and define the normalized errors, $\forall k \in \mathcal{K}$,
\begin{align}
\xi_{e_k} \coloneqq \frac{e_k}{\rho_{e_k}}, \xi_{\psi_k} \coloneqq  \frac{\psi_k}{\rho_{\psi_k}}. \label{eq:ksi_k (formation)}
\end{align}
\textbf{Step I-b}: Define the transformations $T_{e_k}:(-C_{k,\text{col}},C_{k,\text{con}})$ $\to\mathbb{R}$, $k \in \mathcal{K}$,  and $T_{\psi}:[0,1)\to[0,\infty)$ by
\begin{align*}
T_{e_k}(x) \coloneqq \ln\Bigg(\frac{1+\tfrac{x}{C_{k,\text{col}}}}{1-\tfrac{x}{C_{k,\text{con}}}}\Bigg), T_{\psi}(x) \coloneqq \ln\Big(\frac{1}{1 - x}\Big), 
\end{align*}
$\forall k \in \mathcal{K}$, and the transformed error states $\varepsilon_{e_k}\coloneqq \varepsilon_{e_k}(\xi_{e_k}) : (-1,1) \to \mathbb{R}$, $\varepsilon_{\psi_k}\coloneqq \varepsilon_{\psi_k}(\xi_{\psi_k}) : [0,1) \to \mathbb{R}_{\geq 0}$, $\forall k \in \mathcal{K}$, 
\begin{subequations} \label{eq:epsilon errors (formation)}
	\begin{align}
	\varepsilon_{e_k} &\coloneqq T_{e_k}(\xi_{e_k}), \label{eq:epsilon e_k (formation)}\\
	\varepsilon_{\psi_k} &\coloneqq T_{\psi}(\xi_{\psi_k}). \label{eq:epsilon psi_k (formation)} 
	\end{align}
\end{subequations}
Next, we design the decentralized reference velocity vector for each agent as
\begin{align} \label{eq:vel_i_des (formation)}
v_{i,\text{des}}  \coloneqq 
\begin{bmatrix}
v_{i,L\text{des}}\\ 
\omega_{i,\text{des}} 
\end{bmatrix}  \coloneqq -  {\delta_i}\begin{bmatrix}
{2\sum\limits_{k\in\mathcal{M}} \alpha_f\frac{r_{e_k}(\xi_{e_k})}{\rho_{e_k}}\varepsilon_{e_k} R^\top_{k_1}\widetilde{p}_{k_2,k_1} } \\
\sum\limits_{k\in\mathcal{K}}  \alpha_f\frac{r_{\psi}(\xi_{\psi_k})}{\rho_{\psi_k}}e_{R_k}
\end{bmatrix}, 
\end{align}
where $\delta_i\in\mathbb{R}_{>0}$ are positive gains, $\forall i\in\mathcal{N},$ $r_{e_k}:(-C_{k,\text{col}},C_{k,\text{con}})\to[1,\infty), r_{\psi}:[0,1)\to[1,\infty)$, with $r_{e_k}(x) \coloneqq \frac{\partial T_{e_k}(x)}{\partial x}$, $r_{\psi}(x)\coloneqq \frac{\partial T_{\psi}(x)}{\partial x}$, 
{and the function $\alpha_f\coloneqq \alpha_f(i,k,R_{k_1},R_{k_2})$ is defined as $\alpha_f(i,k,R_{k_1},R_{k_2}) = -I_3$, if $i$ is the tail of the $k$th edge ($i=k_1$), $\alpha_f(i,k,R_{k_1},R_{k_2}) = R^\top_{k_2} R_{k_1}$ if $i$ is the head of the $k$th edge ($i=k_2$), and $0$ otherwise (see Appendix \ref{app:Rigidity} for more details on graph edges). The assignment of the  head and tail in each edge can be done off-line according to the specified orientation of the graph.

\textbf{Step II-a}: Define for each agent the velocity errors $e_{v_i}$ $\coloneqq$ $[e_{v_i,1}^\top,\dots,e_{v_i,6}^\top]^\top$ $\coloneqq$ $\widetilde{v}_i - v_{i,\text{des}}$, $\forall i\in\mathcal{N}$, and design the decreasing performance functions as $\rho_{v_{i,\ell}}\coloneqq\rho_{v_{i,\ell}}(t):\mathbb{R}_{\geq 0} \to  [\rho_{\scriptscriptstyle v^{0}_{i,\ell}}, \rho_{\scr v^{\infty}_{i,\ell}}]$, with $\rho_{v_{i,\ell}}(  {t})\coloneqq (\rho_{\scriptscriptstyle v^{0}_{i,\ell}} - \rho_{\scriptscriptstyle v^{\infty}_{i,\ell}})\exp(-l_{v_{i,\ell}}t) + \rho_{\scriptscriptstyle v^{\infty}_{i,\ell}}$, where the constants $\rho_{\scriptscriptstyle v^{0}_{i,\ell}}, \rho_{\scriptscriptstyle v^{\infty}_{i,\ell}}, l_{v_{i,\ell}}$ incorporate the desired transient and steady-state specifications, with the design constraints $\rho_{\scriptscriptstyle v^{0}_{i,\ell}} > | e_{v_{i,\ell}}(0)|$, $ \rho_{\scriptscriptstyle v^{\infty}_{i,\ell}}\in(0,\rho_{\scriptscriptstyle v^{0}_{i,\ell}})$, $\forall \ell \in\{1,\dots,6\}$, $i\in\mathcal{N}$.   {The term $e_{v_{i,\ell}}(0)$ can be measured be each agent at $t=0$ directly after the calculation of $v_{i,\text{des}}(0)$.}

Moreover, define the normalized velocity errors
\begin{align}
\xi_{v_i} 
& \coloneqq 
\begin{bmatrix}
\xi_{v_i,1}, \dots, \xi_{v_i,6}
\end{bmatrix}^\top \coloneqq \rho_{v_i}^{-1} e_{v_i}, \label{eq:ksi_i_v (formation)}
\end{align}
where $\rho_{v_i}\coloneqq\rho_{v_i}(t)\coloneqq \text{diag}\{ [\rho_{v_{i,\ell}}]_{\ell \in \{1,\dots,6\}}\}$, $\forall i\in\mathcal{N}$.


\textbf{Step II-b}: Define the transformation $T_v:(-1,1)\to\mathbb{R}$ as 
\begin{equation*}
T_v(x)\coloneqq\ln\Big(\frac{1+x}{1-x}\Big), 
\end{equation*}
and the transformed error states $\varepsilon_{v_i}:(-1,1)^6 \to \mathbb{R}^6$ as
\begin{align}
\varepsilon_{v_i}(\xi_{v_i})\coloneqq\varepsilon_{v_i} \coloneqq \begin{bmatrix}
\varepsilon_{v_i,1} \\ \vdots \\ \varepsilon_{v_i,6}
\end{bmatrix}
\coloneqq 
\begin{bmatrix}
T_v(\xi_{v_i,1}) \\ \vdots \\ T_v(\xi_{v_i,6}) 
\end{bmatrix}. \label{eq:epsilon v (formation)}
\end{align}
Finally, 
design the decentralized control protocol for each agent $i\in\mathcal{N}$ as $u_i: (-1,1)^6\times \mathbb{R}_{\geq 0}$, with
\begin{equation}
u_i\coloneqq u_i(\xi_{v_i},t) \coloneqq -\gamma_i \rho_{v_i}(t)^{-1}\bar{r}_v(\xi_{v_i})\varepsilon_{v_i}(\xi_v), \label{eq:u_i (formation)}
\end{equation}
where $\bar{r}_v(\xi_{v_i}) \coloneqq \text{diag}\{ [r_v(\xi_{v_i,\ell})]_{\ell\in\{1,\dots,6\}}\}$ with $r_{v}:(-1,1)\to[1,\infty)$, $r_v(x)\coloneqq \frac{\partial T_v(x)}{\partial x}$, and  $\gamma_i \in\mathbb{R}_{>0}$ are positive gains, $\forall i\in\mathcal{N}$. 

\begin{remark}(\textbf{Control protocol intuition}) Note that the selection of $C_{k,\text{col}}, C_{k,\text{con}}$ according to \eqref{eq:C_k (formation)} and of $\rho_{\psi_k}(t),\rho_{v_i,\ell}(t)$ such that
	$\rho_{\scr \psi_k,0}=\rho_{\psi_k}(0) \in (\psi_k(0),2), \rho_{\scr v^0_{i,\ell}}=\rho_{v_{i,\ell}}(0) > \lvert e_{v_{i,\ell}}(0)\rvert$ along with \eqref{eq:at t=0 (formation)}, guarantee that $\xi_{e_k}(0)\in(C_{k,\text{col}},C_{k,\text{con}})$, $\psi_{k}(0)\in[0,2)$, $\xi_{v_{i,\ell}}(0)\in(-1,1)$, $\forall k \in \mathcal{K}$, $\ell\in\{1,\dots,6\}$, $i\in\mathcal{N}$. The prescribed performance control technique enforces these normalized 
	errors $\xi_{e_k}(t), \xi_{\psi_k}(t)$ and $\xi_{v_i,\ell}(t)$ to remain strictly within the sets $(
	-C_{k,\text{col}},C_{k,\text{con}}), [0,2)$, and $(-1,1)$, respectively, $\forall k \in \mathcal{K}, \ell \in \{1,\dots,6\}, i\in\mathcal{N},t\geq0$, guaranteeing thus a solution to Problem \ref{problem (formation)}. It can be verified that this can be achieved by maintaining the boundedness of the modulated errors $\varepsilon_{e_k}(t), \varepsilon_{\psi_k}(t)$ and $\varepsilon_{v_i}(t)$ in a compact set, $\forall t\geq0$.
\end{remark}


\begin{remark} (\textbf{Arbitrarily fast convergence to} $\psi_k = 0$)
	The configurations where $\|e_{R_k}\| = 0 \Leftrightarrow \psi_k = 0$ or $\psi_k = 2$ are equilibrium configurations that result in $\omega_{k_1,\text{des}} = \omega_{k_2,\text{des}} = 0$, $\forall k\in\mathcal{K}$. If $\psi_k(0) = 2$, which is a local minima, the orientation formation specification for edge $k$ cannot be met, since the system becomes uncontrollable. This is an inherent property of stabilization in $ \mathbb{SO}(3) $, and cannot be resolved with a purely continuous controller   {\cite{bhat2000topological}}. Moreover, initial configurations $\psi_k(0)$ starting arbitrarily close to $2$ might take infinitely long to be stabilized at $\psi_k = 0$ with common continuous methodologies \cite{mayhew2011quaternion}. Note however, that the proposed control law guarantees convergence to $\psi_k = 0$ arbitrarily fast, given that $\psi_k(0) < 2$. More specifically, given the initial configuration $\psi_k(0) < 2$, we can always choose $\rho_{\scr \psi_k,0}$  such that $\psi_k(0) < \rho_{\scr \psi_k,0} < 2$, regardless of how close $\psi_k(0)$ is  to $2$. Then, as proved in the next section, the proposed control algorithm guarantees \eqref{eq:ppc ineq psi (formation)} and the transient and steady-state performance of the evolution $\psi_k(t)$ is determined solely by $\rho_{\psi_k}(t)$ and more specifically, the rate of convergence is determined by the term $l_{\psi_k}$.  {It can be observed from the desired angular velocities designed $\omega_{i,\text{des}}$ in \eqref{eq:vel_i_des (formation)} that close to the configuration $\psi_k(0) = 2$, the term $e_{R_k}(0)$, which is close  to zero (since $\psi_k(0) = 2 \Rightarrow \|e_{R_k}(0)\| = 0$), is compensated by the term $r_\psi(\xi_{\psi_k}(0)) = \frac{1}{1 - \xi_{\psi_k}(0)}$, which attains large values (since $\xi_{\psi_k}(0) = \frac{\psi_k(0)}{\rho_{\scr \psi_k,0}}$ is close to 1). Moreover, potentially large values (but always bounded, as proved in the next section) for $\omega_{i,\text{des}}$ and hence $u_i$ due to the term $r_\psi(\xi_{\psi_k}(0))$ can be compensated by tuning the control gains $\delta_i$ and $\gamma_i$.} 	
\end{remark}

\begin{remark} (\textbf{Decentralized manner, relative feedback,   {and robustness}})
	Notice by \eqref{eq:vel_i_des (formation)} and \eqref{eq:u_i (formation)} that the proposed control protocols are distributed in the sense
	that each agent uses only local \textit{relative} information to calculate its
	own signal. In that respect, regarding every edge $k$, the parameters $\rho_{\scr e_k,\infty}, \rho_{\scr \psi_k,\infty}, l_{e_k}, l_{\psi_k}$, as well as the sensing radii $\varsigma_j,\forall j\in \mathcal{N}_i(p(0))$, which are needed for the calculation of the performance functions $\rho_{e_k}(t), \rho_{\psi_k}(t)$, can be transmitted off-line to the agents $k_1,k_2\in\mathcal{N}$. In the same vein, regarding $\rho_{v_{i,\ell}}(t)$, i.e., the constants $\rho_{\scr v^{\infty}_{i,\ell}}, l_{v_{i,\ell}}$ can be transmitted off-line to each agent $i$, which can also compute $\rho_{\scr v^0_{i,\ell}}$, given the initial velocity errors $e_{v_i}(0)$. Notice also from \eqref{eq:vel_i_des (formation)} that each agent $i$ uses only relative feedback with respect to its neighbors.   {In particular, for the calculation of $v_{i,L\text{des}}$, the tail of edge $k$, i.e., agent $k_1$, uses feedback of $R^\top_{k_1}(p_{k_2}-p_{k_1})$, and the head of edge $k$, i.e., agent $k_2$, uses feedback of $R^\top_{k_2}R_{k_1}R^\top_{k_1}(p_{k_2}-p_{k_1}) = R^\top_{k_2}(p_{k_2}-p_{k_1})$. Both of these terms are the relative inter-agent position difference expressed in the agents' local frames. For the calculation of $\omega_{i,\text{des}}$, agents $k_1$ and $k_2$ require feedback of the relative orientation $R^\top_{k_2}R_{k_1}$, as well as the signal $S^{-1}(R^\top_{k_1}R_{k_2}R_{k,\text{des}} - R^\top_{k,\text{des}}R^\top_{k_2}R_{k_1})$, which is a function of $R^\top_{k_2}R_{k_1}$}. The aforementioned signals encode information related to the relative pose of each agent with respect to its neighbors, without the need for knowledge of a common global inertial frame. 	
	It should also be noted that the proposed
	control protocol \eqref{eq:u_i (formation)} depends exclusively on
	the velocity of each agent and not on the velocity (expressed in a local frame) of its neighbors.
	Moreover, the proposed control law does not incorporate any prior knowledge of
	the model nonlinearities/disturbances, enhancing thus its robustness. Finally, the proposed
	methodology results in a low complexity. Notice that no hard
	calculations (neither analytic nor numerical) are required to output the
	proposed control signal.
\end{remark}

We provide now the main result of this section0, which is summarized in the following theorem.
\begin{theorem} \label{thm:main theorem (formation)}
	Consider the multi-agent system described by the dynamics \eqref{eq:system_MAS (formation)}, under a static tree communication graph $\mathcal{G}$, aiming at establishing a formation described by the desired offsets $d_{k,\text{des}}\in(d_{k,\text{col}},d_{k,\text{con}})$ and $R_{k,\text{des}}$, $\forall k\in\mathcal{K}$. Then, the control protocol \eqref{eq:ksi_k (formation)}-\eqref{eq:u_i (formation)} guarantees the prescribed transient and steady-state performance 
		\begin{align*}
		-C_{k,\text{col}} \rho_{e_k}(t) &< e_k(t) < C_{k,\text{con}} \rho_{e_k}(t), \\
		{0} &  {\leq \psi_{k}(t) < \rho_{\psi_k}(t)},
		\end{align*}
	$\forall k \in \mathcal{K}$, $t\in\mathbb{R}_{\geq 0}$, under all initial conditions satisfying  {$\psi_k(0) < 2$, $\forall k\in\mathcal{K}$} and \eqref{eq:at t=0 (formation)}, providing thus a solution to Problem \ref{problem (formation)}. 
\end{theorem}

\begin{proof}
	We start by defining some vector and matrix forms of the introduced signals and functions: 	
	\begin{center}
		\begin{tabular}{ l l}
			$e \coloneqq [e_1,\dots,e_K]^\top$ & $\psi \coloneqq [\psi_1,\dots,\psi_K]^\top$ \\
			$e_R \coloneqq [ e_{R_1}^\top, \dots, e_{R_K}^\top]^\top$  & 
			$\bar{e}_v \coloneqq [ e_{v_1}^\top, \dots, e_{v_N}^\top]^\top$ \\
			$\xi_a \coloneqq [\xi_{a_1},\dots,\xi_{a_K}]^\top$ & $\xi_v\coloneqq [\xi_{v_1}^\top,\dots, \xi_{v_N}^\top]^\top$ \\  
			$\varepsilon_e \coloneqq \varepsilon_e(\xi_e)\coloneqq [\varepsilon_{e_1},\dots,\varepsilon_{e_K}]^\top$ &  $\varepsilon_\psi \coloneqq \varepsilon_\psi(\xi_\psi) \coloneqq [\varepsilon_{\psi_1},\dots,\varepsilon_{\psi_K}]^\top$ \\
			$\varepsilon_v \coloneqq \varepsilon_v(\xi_v) \coloneqq [\varepsilon^\top_{v_1},\dots,\varepsilon^\top_{v_N}]^\top$  & $\widetilde{p} \coloneqq [\widetilde{p}^\top_{1_2,1_1},\dots,\widetilde{p}^\top_{K_2,K_1}]^\top$ \\ ${v_L\coloneqq [v_{1,L}^\top,\dots,v_{N,L}^\top}]^\top$ &
			$v_{L\text{des}}\coloneqq [v_{1,L\text{des}}^\top,\dots,v_{N,L\text{des}}^\top]^\top$ \\
			$\omega \coloneqq [\omega_1^\top,\dots,\omega_N^\top]^\top$ &$\omega_{\text{des}} \coloneqq [\omega_{1,\text{des}}^\top,\dots,\omega_{N,\text{des}}^\top]^\top$ \\
			$v_{\text{des}} \coloneqq [v_{1,\text{des}}^\top, \dots, v_{N,\text{des}}^\top]^\top$ & $\rho_a\coloneqq\rho_a(t) \coloneqq \text{diag}\{[\rho_{a_k}(t)]_{k\in \mathcal{K}}\}$ \\
			$\rho_v\coloneqq \rho_v(t) \coloneqq \text{diag}\{[\rho_{v_i}(t)]_{i\in\mathcal{N}}\}$ &
			$r_e(\xi_e) \coloneqq \text{diag}\{ [r_{e_k}(\xi_{e_k})]_{k\in \mathcal{K}}\}$ \\ $\Sigma_e\coloneqq \Sigma_e(\xi_e,t) \coloneqq r_e(\xi_e)\rho_e(t)^{-1}$ & $\widetilde{r}_\psi(\xi_\psi) \coloneqq \text{diag}\{ [r_{\psi}(\xi_{\psi_k})]_{k\in \mathcal{K}}\}$ \\ 
			$\Sigma_\psi\coloneqq \Sigma_\psi(\xi_\psi,t) \coloneqq \widetilde{r}_\psi(\xi_\psi)\rho_\psi(t)^{-1}$ & 
			$\widetilde{r}_v(\xi_v) \coloneqq \text{diag}\{ [\bar{r}_{v}(\xi_{v_i})]_{i\in\mathcal{N}}$ \\
			 $\Sigma_v\coloneqq \Sigma_v(\xi_v,t) \coloneqq \widetilde{r}_v(\xi_v)\rho_v(t)^{-1}$
		\end{tabular}
	\end{center}
	where $a\in\{e,\psi\}$.
	
	With the introduced notation, \eqref{eq:errors derivative (formation)} can be written in vector form as:
	\begin{subequations}  \label{eq:e_deriv (formation)}
		\begin{align}
		\dot{e}
		= 
		\begin{bmatrix}
		\dot{e}_1 \\
		\vdots \\
		\dot{e}_K
		\end{bmatrix} =&
		\begin{bmatrix}
		&   {2(R^\top_{1_1}\widetilde{p}_{1_2,1_1})^\top(R^\top_{1_1}R_{1_2}v_{1_2,L}-v_{1_1,L})} \notag \\
		\vdots \\
		&   {2(R^\top_{K_1}\widetilde{p}_{K_2,K_1})^\top(R^\top_{K_1}R_{K_2}v_{K_2,L}-v_{K_1,L})}
		\end{bmatrix} \notag \\
		=&
		2 \begin{bmatrix}
		\widetilde{p}^\top_{1_2,1_1}  & \dots & 0 \\
		\vdots & \ddots & \vdots\\
		0 &  \dots  & \widetilde{p}^\top_{K_2,K_1} \\
		\end{bmatrix}
		{\hat{R} D_R^\top v_L}	
		=:
		\mathbb{F}_p^\top   {\hat{R} D_R^\top v_L}, \label{eq:e_p_deriv (formation)} \\ \notag
		\dot{\psi} =
		\begin{bmatrix}
		\dot{\psi}_1 \\
		\vdots \\
		\dot{\psi}_K
		\end{bmatrix} =&  \frac{1}{2}
		\begin{bmatrix}
		e_{R_1}^\top (R^\top_{1_1} R_{1_2}\omega_{1_2}-\omega_{1_1})\\
		\vdots \\
		e_{R_K}^\top (R^\top_{K_1} R_{K_2}\omega_{1_2}-\omega_{K_1})
		\end{bmatrix} \notag \\
		=& \frac{1}{2}\begin{bmatrix}
		e_{R_1}^\top  & \dots & 0\\
		\vdots & \ddots & \vdots\\
		0 &  \dots  & e_{R_K}^\top \\
		\end{bmatrix}  {D_R^\top}\omega =: \mathbb{F}_{R}^\top D_R^\top\omega,
		\label{eq:e_q_deriv (formation)} 
		\end{align}
	\end{subequations}
	where   {$\hat{R} \coloneqq \text{diag}\{[R_{k_1}]_{k\in\mathcal{K}}\}\in\mathbb{R}^{3K\times 3K}$},
	\begin{align*}
	\mathbb{F}_p \coloneqq \mathbb{F}_p(\widetilde{p}) & \coloneqq
	2 \begin{bmatrix}
	\widetilde{p}_{1_2,1_1}  & \dots & 0 \\
	\vdots & \ddots & \vdots\\
	0 &  \dots  & \widetilde{p}_{K_2,K_1}
	\end{bmatrix} \in \mathbb{R}^{3K\times K},  \\
	\mathbb{F}_R\coloneqq \mathbb{F}_R(e_R) & \coloneqq
	\frac{1}{2}\begin{bmatrix}
	e_{R_1} & \dots & 0 \\
	\vdots & \ddots & \vdots\\
	0 &  \dots  & e_{R_K} \\
	\end{bmatrix} \in \mathbb{R}^{3K\times K}, 
	\end{align*}
	$D_R\coloneqq D_R(R,G)\in \mathbb{R}^{3N}\times\mathbb{R}^{3K}$ is the \emph{orientation incidence matrix} of the graph: 
	\begin{align} 
	&   {D_R(R,\mathcal{G}) \coloneqq  \bar{R}^\top\left[ D\otimes I_3 \right] \hat{R} }, \label{eq:incidence_D_R (formation)}
	\end{align}
	{with $\bar{R}\coloneqq\text{diag}\{[R_i]_{i\in\mathcal{N}}\}\in\mathbb{R}^{3N\times3N}$, and $D\coloneqq D(\mathcal{G})$ is the incidence matrix of the graph (see Section \ref{sec:graph_theory (app_ridigity)} of Appendix \ref{app:Rigidity})}. 
	The terms $\bar{R}$ and $\hat{R} $ in $D_R$ correspond to the block diagonal matrix with the agents' rotation matrices along the main block diagonal, and the block diagonal matrix with the rotation matrix of each edge's tail along the main block diagonal, respectively.   These two terms have motivated the incorporation of the terms $\alpha_f(\cdot)$ in the desired velocities $v_{i,\text{des}}$ designed in \eqref{eq:vel_i_des (formation)}, since,  as shown next, the vector form $v_{\text{des}}$ yields the orientation incidence matrix $D_{R}(R,\mathcal{G})$.

	The desired velocities \eqref{eq:vel_i_des (formation)} and control inputs \eqref{eq:u_i (formation)} can be written in vector form as 
	\begin{subequations} \label{eq:control design vectors (formation)}
		\begin{align}
		{v_{L\text{des}}} &= -  {\Delta D_R \hat{R} ^\top}\mathbb{F}_p\Sigma_e\varepsilon_e, \label{eq:control design p vectors (formation)}\\
		\omega_\text{des} &= -  {\Delta}D_R \left[ \Sigma_\psi \otimes I_3 \right]   {e_R}, \label{eq:control design R vectors (formation)}\\
		u &= -\Gamma \ \Sigma_v\varepsilon_v, \label{eq:control design u vectors (formation)}
		\end{align}
	\end{subequations}
	where   {$\Delta \coloneqq \text{diag}\{[\delta_iI_3]_{i\in\mathcal{N}}\}\in\mathbb{R}^{3N\times3N}$} and  $\Gamma \coloneqq \text{diag}\{[\gamma_iI_6]_{i\in\mathcal{N}} \}\in\mathbb{R}^{6N\times6N}$. Note from \eqref{eq:control design u vectors (formation)} and \eqref{eq:ksi_k (formation)}, \eqref{eq:ksi_i_v (formation)}, \eqref{eq:epsilon errors (formation)}, \eqref{eq:epsilon v (formation)} that $u$ can be expressed as a function of the states $x,v,t$.  Hence, the closed loop system can be written as
	
	\begin{align*}
	\dot{x} &=  h_x(x,v) \\
	\dot{v} &= -M^{-1}\Big\{ C(v)v + g(x) + w(x,v,t) - u(\cdot) \Big\} =: h_v(x,v,t).
	\end{align*}
	By defining $z \coloneqq (x,v)\in \mathbb{SE}(3) ^{N}\times\mathbb{R}^{6N}$, we can 
	write the closed loop system in vector form as 
	\begin{equation}
	\dot{z} = h_z(z,t) \coloneqq (h_x(z), h_v(z,t)). \label{eq:closed loop vector (formation)}
	\end{equation}
	
	Next, define the set 
	\begin{align*}
	\Omega \coloneqq \bigg\{  (x,v,t) & \in \mathbb{SE}(3) ^N\times\mathbb{R}^{6N}\times\mathbb{R}_{\geq 0} :
	 \xi_{e_k}(p_{k_1},p_{k_2},t) \in (-C_{k,\text{col}}, C_{k,\text{con}}), \\ &  {\xi_{\psi_k}(R_{k_1},R_{k_2},t) < 1}, \xi_{v_i}(x,v_i,t)\in(-1,1)^6, \forall k\in \mathcal{K} \bigg\},
	\end{align*}
	where we abuse the notation and express $\xi_{e_k}$, $\xi_{\psi_k}$, $\xi_{v_i}$ from \eqref{eq:ksi_k (formation)}, \eqref{eq:ksi_i_v (formation)} as a function of the states.   {It can be verified that the set $\Omega$ is open due to the continuity of the operators $\xi_{e_k}(\cdot), \xi_{\psi_k}(\cdot), \xi_{v_i}(\cdot)$ and nonempty, due to \eqref{eq:C_k (formation)}.}   {Our goal here is to prove first that \eqref{eq:closed loop vector (formation)} has a unique and maximal solution $(z(t),t)$ in $\Omega$ and then that this solution stays in a compact subset of $\Omega$.}
	
	It can be verified that the function $h:\Omega  \to  \mathbb{T}_{R_1}\times\cdots\times\mathbb{T}_{R_N} \times\mathbb{R}^{6N}$ is (a) continuous in $t$ for each fixed $(x,v)\in\{ (x,v)\in  \mathbb{SE}(3) ^N\times\mathbb{R}^{6N}: (x,v,t)\in\Omega\}$, and (b) continuous and locally lipschitz in $(x,v)$ for each fixed $t\in\mathbb{R}_{\geq 0}$. Therefore, the conditions of Theorem \ref{thm:ode solution (App_dynamical_systems)} of Appendix \ref{app:dynamical systems} are satisfied and hence, we conclude the existence of a unique and maximal solution of \eqref{eq:closed loop vector (formation)} for a timed interval $[0,t_{\max})$, with $t_{\max} >0$, such that $(z(t),t)\in\Omega$, $\forall t\in[0,t_{\max})$. This implies that 
	\begin{subequations} \label{eq:ksi tau_max (formation)}
		\begin{align}
		\xi_{e_k}(t) & = \frac{e_k(t)}{\rho_{e_k}(t)}\in(-1,1) \label{eq:ksi e tau_max (formation)}, \\
		\xi_{\psi_k}(t) & = \frac{\psi_k(t)}{\rho_{\psi_k}(t)} < 1, \label{eq:ksi psi tau_max (formation)} \\
		\xi_{v_i}(t) & = \rho_{v_i}(t)^{-1}e_{v_i}(t)\in (-1,1)^6, \label{eq:ksi v tau_max (formation)} 
		\end{align}
	\end{subequations}
	$\forall k\in \mathcal{K}$, $i\in\mathcal{N}$,  $t\in[0,t_{\max})$. Therefore, the signals $e_k(t), \psi_k(t), e_{v_i}(t)$ are bounded for all $t\in[0,t_{\max})$. In the following, we aim to show that the solution $(z(t),t)$ is bounded in a compact subset of $\Omega$ and hence,   {by employing Theorem \ref{thm:forward_completeness (App_dynamical_systems)} of Appendix \ref{app:dynamical systems},} that $t_{\max} = \infty$. 
	
	Consider the positive definite Lyapunov candidate $V_e$ $\coloneqq$ $V_e(\varepsilon_e):(-1,1)^K$ $\to\mathbb{R}_{\geq 0}$, with $V_e(\varepsilon_e) \coloneqq \tfrac{1}{2}\|\varepsilon_e\|^2$, which is well defined for $t\in[0,t_{\max})$, due to \eqref{eq:ksi e tau_max (formation)}. By differentiating $V_e$ and taking into account the dynamics $\dot{\xi}_e = \rho_e(t)^{-1} \left[ \dot{e} - \dot{\rho}_e(t)\xi_e \right]$, we obtain 
	\begin{equation*}
		\dot{V}_e = \left[\frac{\partial V_e}{\partial \varepsilon}\right] \dot{\varepsilon}_e = \varepsilon^\top_e \Sigma_e \left(  {\mathbb{F}_p^\top \hat{R} D_R^\top v_{L}} -  \dot{\rho}_e\xi_e \right),
	\end{equation*}
	which, by substituting ${v_L} =   {\widetilde{v}_L} - \mathsf{n}_{p} = e_{v_p} +   {v_{L\text{des}}} - \mathsf{n}_p$ and \eqref{eq:e_deriv (formation)},  becomes
	\begin{align} \label{eq:V_e 1 (formation)}
	\dot{V}_e = & -\varepsilon^\top_e \Sigma_e\mathbb{F}_p^\top   {\widetilde{D}}\mathbb{F}_p\Sigma_e\varepsilon_e  +  \varepsilon^\top_e \Sigma_e \Big[   {\mathbb{F}_p^\top \hat{R} D_R^\top(e_{v_p} - \mathsf{n}_p)} -\dot{\rho}_e\xi_e \Big],
	\end{align} 
	where   { $\widetilde{D}\coloneqq\widetilde{D}(\mathcal{G})$ $\coloneqq$ $\hat{R} D_R^\top D_R \hat{R} ^\top$ $=$ $D^\top\otimes I_3$ $\Delta$ $D$ $\otimes I_3 \in\mathbb{R}^{3K\times 3K}$ (by employing \eqref{eq:incidence_D_R (formation)})}, and $e_{v_p}$, $\mathsf{n}_p$ are the linear parts of $\bar{e}_v$ and $\mathsf{n}\coloneqq [\mathsf{n}_1^\top,\dots,\mathsf{n}_N^\top]^\top$   {(i.e., the stack vector of the first three components of every $e_{v_i} $, $\mathsf{n}_i$)}, respectively. Note first that, due to \eqref{eq:ksi v tau_max (formation)}, the function $e_{v_p}(t)$ is bounded for all $t\in[0,t_{\max})$. Moreover, note that \eqref{eq:ksi e tau_max (formation)} implies that $  {0<}d_{k,\text{col}} < \|p_{k_1}(t)-p_{k_2}(t) \| < d_{k,\text{con}}$, $\forall t\in[0,t_{\max})$. Therefore, it holds that $\text{rank} (\mathbb{F}_p(\widetilde{p}(t))) = K$, $\forall t\in[0,t_{\max})$. In addition, since $\mathcal{G}$ is a connected tree graph   {and $\delta_i\in\mathbb{R}_{>0}$, $\forall i\in\mathcal{N}$,} $\widetilde{D}$ is positive definite (see Lemma \ref{lemma:tree formation (App_Rigidity)} of Appendix \ref{app:Rigidity}) and  hence $\text{rank}(\widetilde{D}) = {3K}$. Hence, we conclude that $\text{rank}\Big( \mathbb{F}_p(\widetilde{p}(t))^\top\widetilde{D}\mathbb{F}_p(\widetilde{p}(t))\Big) = K$ and the positive definiteness of $\mathbb{F}_p(\widetilde{p}(t)) ^\top\widetilde{D}\mathbb{F}_p(p(t))$, $\forall t\in[0,t_{\max})$. 
	{In addition, since $\|p_{k_2}(t) - p_{k_1}(t)\| < d_{k,\text{con}}$, we also conclude that the term $\mathbb{F}_p^\top\hat{R} D_R^\top$ is upper bounded, $\forall t\in[0,t_{\max})$}. Finally, $\dot{\rho}_e$ and $\mathsf{n}_p$ are bounded by definition and assumption, respectively, $\forall x\in \mathbb{SE}(3) ^N, t\in\mathbb{R}_{\geq 0}$. Note that all the aforementioned bounds are independent of $t_{\max}$. We obtain now from \eqref{eq:V_e 1 (formation)}:
	\begin{align*}
	\dot{V}_e &\leq - \underline{\lambda}_{\scr \widetilde{D}} \| \Sigma_e \varepsilon_e \|^2 + \|\Sigma_e \varepsilon_e \| \bar{B}_e = - \underline{\lambda}_{\scr \widetilde{D}} \| \Sigma_e\varepsilon_e \| \left(\| \Sigma_e\varepsilon_e \| - \frac{\bar{B}_e}{\underline{\lambda}_{\scr \widetilde{D}}} \right),
	\end{align*}
	$  {\forall} t\in[0,t_{\max})$ where 
	\begin{align*}
	\underline{\lambda}_{\scr \widetilde{D}} & \coloneqq \inf\limits_{\scr p(t),t\in[t_0,  {t}_{\max})}\Big\{\lambda_{\min}\Big(\mathbb{F}_p(\widetilde{p}(t))^\top\widetilde{D}\mathbb{F}_p(\widetilde{p}(t))\Big)\Big\}  \geq d^2_{k,\text{col}}\lambda_{\min}(\widetilde{D}) > 0,
	\end{align*}
	and $\bar{B}_e$ is a positive constant, independent of $t_{\max}$, satisfying the following inequality:
	$\bar{B}_e \geq \|  {\mathbb{F}_p(\widetilde{p}(t))^\top\hat{R} D_R^\top} (e_{v_p}(t) - \mathsf{n}_p(x(t),t)) - \dot{\rho}_e(t)\xi_e(t) \|, \forall t\in[0,t_{\max})$.   {Note that, in view of the aforementioned discussion, $\bar{B}_e$ is finite}. 
	
	Hence, we conclude that $\dot{V}_e < 0 \Leftrightarrow \| \Sigma_e\varepsilon_e \| > \frac{\bar{B}_e}{\underline{\lambda}_{\scr \widetilde{D}}}$. By noting that 
	\begin{align*}
	r_{e_k}(x) = \frac{\partial T_{e_k}(x)}{\partial x} = \frac{\frac{1}{C_{k,\text{col}}}+\frac{1}{C_{k,\text{con}}}}{\left(1 + \frac{x}{C_{k,\text{col}}}\right)\left(1-\frac{x}{C_{k,\text{con}}}\right)}> \frac{1}{C_{k,\text{col}}} + \frac{1}{C_{k,\text{con}}},
	\end{align*}
	$\forall x\in(-C_{k,\text{col}},C_{k,\text{con}})$, as well as $\rho_{e_k}(t) \leq 1, \forall t\in\mathbb{R}_{\geq 0}$, $k\in \mathcal{K}$,   {we conclude that $\| \Sigma_e(\xi_e(t),t)\varepsilon_e(\xi_e(t)) \|= \sqrt{\sum_{k\in\mathcal{K}} \frac{r_{e_k}(\xi_{e_k}(t))^2}{\rho_{e_k}(t)^2}\varepsilon_{e_k}(\xi_e(t))^2} \geq \bar{C}_e\|\varepsilon_e(\xi_e(t))\|$, $\forall t\in[0,t_{\max})$, where $\bar{C}_e \coloneqq \max_{k\in\mathcal{K}}\left\{ \frac{1}{C_{k,\text{col}}} + \frac{1}{C_{k,\text{con}}} \right\}$. Hence, we conclude that 
		$\dot{V}_e < 0$, $\forall \|\varepsilon_e\| \geq 
		\tfrac{\bar{B}_e}{\underline{\lambda}_{\scr \widetilde{D}}\bar{C}_e}$, $\forall t\in[0,t_{\max})$} and therefore
	\begin{equation} \label{eq:bar epsilon e (formation)}
	\|\varepsilon_e(\xi_e(t)) \| \leq \bar{\varepsilon}_e \coloneqq \max\left\{ \varepsilon_e(\xi_e(0)), \tfrac{\bar{B}_e}{\underline{\lambda}_{\scr \widetilde{D}}\bar{C}_e} \right\},
	\end{equation}
	$t\in[0,t_{\max})$, and by taking the inverse logarithm function:
	\begin{equation} \label{eq:bar ksi e (formation)}
	- C_{k,\text{col}} < -\underline{\xi}_{e} \leq \xi_{e_k}(t) \leq \bar{\xi}_{e} <  C_{k,\text{con}},
	\end{equation}
	$\forall t\in[0,t_{\max})$, where $\bar{\xi}_e \coloneqq \tfrac{\exp(\bar{\varepsilon}_e)-1}{\exp(\bar{\varepsilon}_e)+1}C_{k,\text{con}}$, and $\underline{\xi}_e \coloneqq \tfrac{\exp(-\bar{\varepsilon}_e)-1}{\exp(-\bar{\varepsilon}_e)+1}C_{k,\text{col}}$.    Hence, \eqref{eq:bar epsilon e (formation)} and \eqref{eq:bar ksi e (formation)} imply the boundedness of $\varepsilon_{e_k}(\xi_{e_k}(t))$, $r_{e_k}(\xi_{e_k}(t))$, $\widetilde{p}(t)$, and $p(t)$ in compact sets, $\forall k\in \mathcal{K}$, and therefore, through \eqref{eq:vel_i_des (formation)}, the boundedness of $v_{i,L\text{des}}(t)$, $\forall i\in\mathcal{N}$, $t\in[0,t_{\max})$. 
	
	Similarly, consider the positive definite Lyapunov  candidate $V_\psi\coloneqq V_\psi(\xi_\psi):[0,1)^K\to\mathbb{R}_{\geq 0}$, with $V_\psi = 2\sum_{k\in \mathcal{K}}\varepsilon_{\psi_k}$. By differentiating $V_\psi$ and taking into account the dynamics $\dot{\xi}_{\psi_k} = \rho_{\psi_k}^{-1} \left[ \dot{\psi}_k - \dot{\rho}_{\psi_k}\xi_{\psi_k} \right]$, we obtain 
	\begin{equation*}
		\dot{V}_\psi \coloneqq \left[\frac{\partial V_\psi}{\partial \varepsilon_e }\right]\dot{\varepsilon}_\psi = 2\sum_{k\in \mathcal{K}} \frac{r_{\psi}(\xi_{\psi_k})}{\rho_{\psi_k}}(\dot{\psi}_{k} - \dot{\rho}_{\psi_k}\xi_{\psi_k}),
	\end{equation*}
	which, after substituting \eqref{eq:error psi_k dot (formation)}, \eqref{eq:e_deriv (formation)},  becomes 	
	\begin{align*}
	\dot{V}_\psi = \ & e_R^\top \left( \Sigma_\psi \otimes I_3 \right) D_R^\top\omega -2\sum\limits_{k\in \mathcal{K}}\frac{r_{\psi}(\xi_{\psi_k})}{\rho_{\psi_k}}\dot{\rho}_{\psi_k}\xi_{\psi_k}  \\
	= \ & e_R^\top \left( \Sigma_\psi\otimes I_3 \right) D_R^\top \left(  \omega_\text{des} + e_{v_R} - \mathsf{n}_R \right) -2\sum\limits_{k\in \mathcal{K}}\frac{r_{\psi}(\xi_{\psi_k})}{\rho_{\psi_k}}\dot{\rho}_{\psi_k}\xi_{\psi_k}, \\ 
	\end{align*}
	{where  $e_{v_R}$ and $\mathsf{n}_R$ are the angular parts of $\bar{e}_v$ and $\mathsf{n}$ (i.e., the stack vector of the last three components of every $e_{v_i}$, $\mathsf{n}_i$), respectively.} By substituting \eqref{eq:control design R vectors (formation)} and defining $\widetilde{\Sigma}_\psi \coloneqq \Sigma_\psi\otimes I_3\in\mathbb{R}^{3K \times 3K}$, $\widetilde{D}_R \coloneqq D_R^\top   {\Delta} D_R\in\mathbb{R}^{3K \times 3K}$, we obtain: 	
	\begin{align}
	\dot{V}_\psi = \ & -e_R^\top \widetilde{\Sigma}_\psi\widetilde{D}_R \widetilde{\Sigma}_\psi e_R   +e_R^\top\widetilde{\Sigma}_\psi D_R^\top \left(e_{v_R} - \mathsf{n}_R \right)  - 2\sum\limits_{k\in  {\mathcal{K}}}\frac{r_{\psi}(\xi_{\psi_k})}{\rho_{\psi_k}}\dot{\rho}_{\psi_k}\xi_{\psi_k}. \label{eq:V_psi 1 (formation)}
	\end{align}
	
	According to \eqref{eq:incidence_D_R (formation)},   {$D_R$ $=$ 
	$\bar{R}^\top$ $\left(D\otimes I_3 \right)$ $\hat{R} $. Since $\bar{R}$ and $\hat{R} $ are rotation (and thus unitary) matrices, the singular values of $D_R$ are identical to the ones of $D$, and hence $\lambda_{\min}(\widetilde{D}_R) = \lambda_{\min}(\widetilde{D}) > 0$. Indeed, let $D\otimes I_3 = U\Sigma_D V^\top$ be a singular value decomposition of $D\otimes I_3$, where $U$, $V$ are unitary matrices, and $\Sigma_D$ is a diagonal matrix containing the singular values of $D\otimes I_3$. Then 
	$ D_R = \bar{R}^\top U \Sigma_D V^\top \hat{R} 
	=  \widetilde{U}\Sigma_D \widetilde{V}^\top$
	where $\widetilde{U} \coloneqq \bar{R}^\top U$, and $\widetilde{V} = \hat{R} ^\top V$ are unitary matrices (being products of unitary matrices). Thus, $\widetilde{U}\Sigma_D\widetilde{V}^\top$ is the singular value decomposition of $D_R$, and hence its singular values are the diagonal values of $\Sigma_D$}. By further defining $\beta \coloneqq$ $[\beta_1^\top,\dots,\beta_K^\top]^\top$ $\coloneqq$ $D_R^\top (e_{v_R} - \mathsf{n}_R)\in\mathbb{R}^{3M}$, with $\beta_k\in\mathbb{R}^3$, $\forall k\in\mathcal{K}$, \eqref{eq:V_psi 1 (formation)} becomes	
	\begin{align*}
	\dot{V}_\psi\leq& -\lambda_{\min}( \widetilde{D} )\| \widetilde{\Sigma}_\psi e_R \|^2+ \sum\limits_{k\in\mathcal{K}}\frac{r_{\psi}(\xi_{\psi_k})}{\rho_{\psi_k}} e_{R_k}^\top \beta_k  - 2\sum\limits_{k\in\mathcal{K}}\frac{r_{\psi}(\xi_{\psi_k})}{\rho_{\psi_k}}\dot{\rho}_{\psi_k}\xi_{\psi_k}.
	\end{align*}
	Note that, by construction, $\xi_{\psi_k} \geq 0$, $\forall k\in\mathcal{K}$, and $r_\psi(x) = \frac{\partial T_\psi(x)}{\partial x} = \frac{1}{1-x} > 1, \forall x < 1$. Hence, in view of \eqref{eq:ksi psi tau_max (formation)}, we conclude that $r_\psi(\xi_{\psi_k}(t)) > 1$, $\forall t\in[0,t_{\max})$. By noting also that $\dot{\rho}_{\psi_k}(t) < 0,\forall t\in\mathbb{R}_{\geq 0}$, 
	$\dot{V}_\psi$ becomes
	\begin{align*}
	\dot{V}_\psi \leq & -\lambda_{\min}( \widetilde{D} )\sum\limits_{k\in \mathcal{K}}\Bigg[\frac{r_{\psi}(\xi_{\psi_k})}{\rho_{\psi_k}}\Bigg]^2\| e_{R_k} \|^2 + \bar{B}_{\psi_1}\sum\limits_{k\in \mathcal{K}}\frac{r_{\psi}(\xi_{\psi_k})}{\rho_{\psi_k}} \|e_{R_k}\| \\
	& + 2\max\limits_{k\in \mathcal{K}}\{ l_{\psi_k}(\rho_{\scr \psi_k,0} - \rho_{\scr \psi_k,\infty}) \}\sum\limits_{k\in \mathcal{K}}\frac{r_{\psi}(\xi_{\psi_k})}{\rho_{\psi_k}}\xi_{\psi_k},
	\end{align*}
	where $\bar{B}_{\psi_1}$ is a positive constant, independent of $t_{\max}$, satisfying  $\bar{B}_{\psi_1} \geq \max_{k\in \mathcal{K}}\{\| \beta_k(t) \|\}$, $\forall t\in[0,t_{\max})$.   {Note that $\bar{B}_{\psi_1}$ is finite, $\forall t\in[0,t_{\max})$, due to \eqref{eq:ksi psi tau_max (formation)} and the boundedness of the noise signals.}
	After substituting \eqref{eq:e_R_k and psi_k (formation)}, we obtain
	\small
	\begin{align}
	&\dot{V}_\psi \leq -2\lambda_{\min}( \widetilde{D} )\sum\limits_{k\in \mathcal{K}}\Bigg[\frac{r_{\psi}(\xi_{\psi_k})}{\rho_{\psi_k}}\Bigg]^2 \psi_k(2-\psi_k) \notag  \\
	& + \bar{B}_{\psi_1}\sum\limits_{k\in \mathcal{K}}\frac{r_{\psi}(\xi_{\psi_k})}{\rho_{\psi_k}} \sqrt{2\psi_k(2-\psi_k)}+ 2\max\limits_{k\in \mathcal{K}}\{ l_{\psi_k}(\rho_{\scr \psi_k,0} - \rho_{\scr \psi_k,\infty})\}\sum\limits_{k\in\mathcal{K}}\frac{r_{\psi}(\xi_{\psi_k})}{\rho_{\psi_k}}\xi_{\psi_k}. \label{eq:V_psi 2 (formation)}
	\end{align}	
	\normalsize
	From \eqref{eq:ksi psi tau_max (formation)} we conclude that $0\leq \psi_k(t) < \rho_{\psi_k}(t) \leq \rho_{\scr \psi_k,0} < 2$, and hence $2-\psi_k(t) \geq 2 - \rho_{\scr \psi_k,0}=: \underline{\rho}_k > 0$ $\forall t\in[0,t_{\max})$, $k \in \mathcal{K}$. Moreover, by noticing that $2-\psi_k\leq 2$, $\rho_{\psi_k}(t) \leq \rho_{\scr \psi_k,0}$, and $\psi_k = \xi_{\psi_k}\rho_{\psi_k}(t)$, $\forall k\in \mathcal{K}$, \eqref{eq:V_psi 2 (formation)} becomes	
	\begin{align*}
	&\dot{V}_\psi \leq -\widetilde{\mu} \sum\limits_{k\in \mathcal{K}} r_{\psi}(\xi_{\psi_k}) ^2 \xi_{\psi_k}  + \frac{2\bar{B}_{\psi_1}}{\max\limits_{k\in \mathcal{K}}\{\sqrt{\rho_{\scr \psi_k,0}}\}}\sum\limits_{k\in \mathcal{K}} r_{\psi}(\xi_{\psi_k}) \sqrt{\xi_{\psi_k}} \\
	& \hspace{8mm} + 2\max\limits_{k\in \mathcal{K}}\left\{ \frac{l_{\psi_k}(\rho_{\scr \psi_k,0} - \rho_{\scr \psi_k,\infty})}{\rho_{\scr \psi_k,0}} \right\} \sum\limits_{k\in \mathcal{K}}r_{\psi}(\xi_{\psi_k})\xi_{\psi_k},
	\end{align*} 
	where 
	\begin{equation*}
		\widetilde{\mu} \coloneqq \frac{2\lambda_{\min}(\widetilde{D})\min_{k\in \mathcal{K}}\{\underline{\rho}_k\}}{\max_{k\in \mathcal{K}}\{\rho_{\scr \psi_k,0}\}}.	
	\end{equation*}		
	{From \eqref{eq:ksi psi tau_max (formation)}, \eqref{eq:ksi_k (formation)}, and the fact that $\psi_k\in[0,2]$, it holds that $\xi_{\psi_k}(t) < \sqrt{\xi_{\psi_k}(t)}, \forall k\in \mathcal{K}$}. By also employing the property $$\sum_{k\in \mathcal{K}} r_{\psi_k}(\xi_{\psi_k})\sqrt{\xi_{\psi_k}} \leq \sqrt{K}\sqrt{\sum_{k\in \mathcal{K}} r_{\psi_k}(\xi_{\psi_k})^2\xi_{\psi_k}},$$ we obtain
	\begin{align*}
	&\dot{V}_\psi \leq \notag -\sqrt{\sum\limits_{k\in \mathcal{K}}  r_{\psi}(\xi_{\psi_k})^2 \xi_{\psi_k}} \left(\widetilde{\mu}\sqrt{\sum\limits_{k\in \mathcal{K}} r_{\psi_k}(\xi_{\psi_k})^2 \xi_{\psi_k}} - \bar{B}_\psi \right),
	\end{align*}
	where:
	\begin{align*}
	\bar{B}_\psi & \coloneqq 2\sqrt{K}\Bigg(\frac{\bar{B}_{\psi_1}}{\max\limits_{k\in \mathcal{K}}\{\sqrt{\rho_{\scr \psi_k,0}}\}} + \max\limits_{k\in \mathcal{K}}\left\{ \frac{l_{\psi_k}(\rho_{\scr \psi_k,0} - \rho_{\scr \psi_k,\infty})}{\rho_{\scr \psi_k,0}} \right\}\Bigg).
	\end{align*} 
	We conclude therefore that $\dot{V}_\psi < 0$ $\Leftrightarrow$ $\sqrt{\sum_{k\in \mathcal{K}} r_{\psi}(\xi_{\psi_k})^2 \xi_{\psi_k}} > \tfrac{\bar{B}_\psi}{\widetilde{\mu}}$. From \eqref{eq:epsilon psi_k (formation)}, given $y = T_\psi(x)$, we obtain: 
	\begin{align*}
	&r_{\psi}(x)^2 x = \left[\frac{\partial T_\psi(x)}{\partial x}\right]^2 T_\psi^{-1}(y) = \frac{1}{(1-x)^2}T_\psi^{-1}(y) \\ &=\frac{1}{\left[1-T_\psi^{-1}(y)\right]^2}T_\psi^{-1}(y) = \exp(y)\left(\exp(y)-1\right),
	\end{align*}
	$\forall x\in[0,1)$. Therefore,
	$ r_{\psi}(\xi_{\psi_k})^2\xi_{\psi_k}$ $= \exp(\varepsilon_{\psi_k})$ $\left(\exp(\varepsilon_{\psi_k})-1 \right)$, and
	according to Proposition \ref{prop: f(x) (app_useful_prop)} of Appendix \ref{app:useful_prop},   
	\begin{align*}
	\sqrt{\sum_{k\in \mathcal{K}}\left[r_{\psi}(\xi_{\psi_k})\right]^2 \xi_{\psi_k}} & = \sqrt{\sum_{k\in \mathcal{K}} \exp(\varepsilon_{\psi_k}) \left( \exp(\varepsilon_{\psi_k})-1 \right)}  \geq \sqrt{\sum_{k\in \mathcal{K}} \varepsilon^2_{\psi_k}} = \| \varepsilon_{\psi}\|.
	\end{align*}	
	Hence, we conclude that $\dot{V}_\psi < 0, \forall \| \varepsilon_{\psi}\| > \tfrac{\bar{B}_{\psi}}{\widetilde{\mu}}$. Therefore, 
	\begin{align} \label{eq:bar epsilon psi (formation)}
	\| \varepsilon_\psi(\xi_\psi(t)) \| \leq \bar{\varepsilon}_\psi \coloneqq \max\left\{\varepsilon_\psi(\xi_\psi(0)), \tfrac{\bar{B}_\psi}{\widetilde{\mu}}\right\},
	\end{align}
	and, by taking the inverse logarithm: 
	\begin{align} \label{eq:bar ksi psi (formation)}
	0 \leq -\underline{\xi}_\psi \leq \xi_{\psi_k}(t) \leq \bar{\xi}_\psi < 1,
	\end{align}
	where $\bar{\xi}_\psi \coloneqq \tfrac{\exp(\bar{\varepsilon}_\psi)-1}{\exp(\bar{\varepsilon}_\psi)}$ and $\underline{\xi}_\psi \coloneqq \tfrac{\exp(-\bar{\varepsilon}_\psi)-1}{\exp(-\bar{\varepsilon}_\psi)}$, $\forall k\in\mathcal{K}$.  Therefore, we conclude the boundedness of $\varepsilon_{\psi_k}(\psi_k(t)), r_{\psi_k}(\xi_{\psi_k}(t))$, $\bar{e}_v(t)$ in compact sets, $\forall k\in \mathcal{K}$, and therefore, through \eqref{eq:vel_i_des (formation)}, the boundedness of $\omega_{i,\text{des}}(t)$, $\forall i\in\mathcal{N}, t\in[0,t_{\max})$. From the proven boundedness of $p(t)$ and $p_{i,\text{des}}(t)$, we also conclude the boundedness of $\mathsf{n}(x(t),t)$ and invoking $\widetilde{v} = v + \mathsf{n}(x,t) = \bar{e}_v(t)-v_{\text{des}}(t)$ and \eqref{eq:ksi v tau_max (formation)}, the boundedness of $v(t)$ and $\dot{x}(t)$,  $\forall t\in[0,t_{\max})$. 
	Moreover, in view of \eqref{eq:bar epsilon e (formation)}, \eqref{eq:bar ksi e (formation)}, \eqref{eq:closed loop vector (formation)}, \eqref{eq:vel_i_des (formation)}, we also conclude the boundedness of $\dot{v}_{\text{des}}(t)$. 
	
	Proceeding along similar lines, we consider the positive definite Lyapunov candidate $V_v\coloneqq V_v(\varepsilon_v):(-1,1)^{6N}\to\mathbb{R}_{\geq 0}$ with $V_v = \tfrac{1}{2}\varepsilon^\top_v\Gamma\varepsilon_v$. By computing $\dot{V}_v = \left[ \frac{\partial V_v}{\partial \varepsilon_v}\right]\dot{\varepsilon}_v$ and using the dynamics $\dot{\xi}_v = \rho_v^{-1}(\dot{e}_v$ $- \dot{\rho}_v\xi_v)$, we obtain
	\begin{align}
	\dot{V}_v =&  \varepsilon^\top_v\Gamma\Sigma_v \left[\dot{v} + \dot{\mathsf{n}} \right]  - \varepsilon^\top_v\Gamma\Sigma_v\dot{v}_{\text{des}} - \varepsilon^\top_v\Gamma\Sigma_v\dot{\rho}_v\xi_v = - \varepsilon^\top_v \Sigma_v\Gamma M^{-1}\Gamma\Sigma_v\varepsilon_v \notag\\
	& -\varepsilon^\top_v \Sigma_v \Big\{\Gamma M^{-1}\Big[Cv + g + w \Big] - \dot{\mathsf{n}} + \dot{v}_{\text{des}} + \dot{\rho}_v\xi_v \Big\}.  \label{eq:V_v 1 (formation)}
	\end{align}
	Since we have proved the boundedness of $v(t)$ and $\dot{x}(t)$, $\forall t\in[0,t_{\max})$ the terms  $Cv$, $\dot{\mathsf{n}}$, and $w$ are also bounded, $t\in[0,t_{\max})$, due to the continuities of $C$, $w$, and $\dot{\mathsf{n}}$ in $v$, $\dot{x}$ and the boundedness of $w$ and $\dot{\mathsf{n}}$ in $x,t$. Moreover, $g$, $\xi_v$, and $\dot{\rho}_{v}$ are also bounded due to \eqref{eq:M property 2 (formation)}, \eqref{eq:ksi v tau_max (formation)}, and by construction, respectively. By also using \eqref{eq:M property 1 (formation)}, we obtain from \eqref{eq:V_v 1 (formation)}: 
	\begin{align*}
	&  \dot{V}_v \leq  - \underline{\lambda}_K \|\Sigma_v \varepsilon_v \|^2 + \|\Sigma_v \varepsilon_v \| \bar{B}_v,	 
	\end{align*}
	where $\bar{B}_v$ is {a} positive finite term, independent of $t_{\max}$, satisfying $\bar{B}_v$ $\geq$ $\Big\| \frac{\max_{i\in\mathcal{N}}\{\gamma_i\}}{\min_{i\in\mathcal{N}}\{\underline{m}_i\}} \Big[C(v(t))v + g(x(t)) + w(x(t),v(t),t) \Big] -\dot{\mathsf{n}}(x(t),t) + \dot{v}_{\text{des}}(t) + \dot{\rho}_v(t)\xi_v(t) \Big\|$, and $\underline{\lambda}_K \coloneqq \frac{\min_{i\in\mathcal{N}}\{\gamma_i\}^2}{\max_{i\in\mathcal{N}}\{\bar{m}\}} > 0$.
	Hence, $\dot{V}_v < 0 \Leftrightarrow \|\Sigma_v \varepsilon_v \| > \tfrac{\bar{B}_v}{\underline{\lambda}_{K}}$.   {By noting that 
		\begin{align*} 
		r_{v}(x) = \frac{\partial T_{v}(x)}{\partial x} = \frac{2}{(1+x)(1-x)} > 2 >1,  
		\end{align*}
		$\forall x\in(-1,1)$, as well as $\rho_{v_i,\ell}(t) \leq \rho_{v^0_i,\ell}$, $\forall \ell\in\{1,\dots,6\},t\in\mathbb{R}_{\geq 0}$, we conclude that $\|\Sigma_v \varepsilon_v(\xi_v(t)) \| = \sqrt{\sum_{i\in\mathcal{N}} \sum_{\ell\in\{1,\dots,6\}}\frac{r_v(\xi_{v_i,\ell}(t))^2}{\rho_{v_{i,\ell}}(t)^2} \varepsilon_{v_{i,\ell}}(\xi_{v_i,\ell}(t))^2}$ $\geq$ $\frac{1}{\widetilde{\rho}}\|\varepsilon_v(\xi_v(t))\|$, $\forall t\in[0,t_{\max})$, where $\widetilde{\rho} \coloneqq \max\limits_{\stackrel{i\in\mathcal{N}}{m\in\{1,\dots,6\}} }\{ \rho_{\scr v^0_{i,m}} \}$. 
		Hence, we conclude that $\dot{V}_v < 0, \forall \| \varepsilon_v \| \geq  \frac{\widetilde{\rho} \bar{B}_v}{\underline{\lambda}_K},  \forall t\in[0,t_{\max})$,
	} and consequently that 
	\begin{align*} 
	\| \varepsilon_v(\xi_v(t))\| \leq \bar{\varepsilon}_v \coloneqq \max\left\{ \varepsilon_v(\xi_v(0)), \frac{\widetilde{\rho} \bar{B}_v}{\underline{\lambda}_K} \frac{\max\limits_{i\in\mathcal{N}}\{\gamma_i\}}{\min\limits_{i\in\mathcal{N}}\{\gamma_i\}} \right\},
	\end{align*}
	$\forall t\in[0,t_{\max})$
	and by taking the inverse logarithm function: 
	\begin{equation} \label{eq:bar ksi v (formation)}
	-1 < -\bar{\xi}_v \leq \xi_{v_{i, \ell}}(t) \leq \bar{\xi}_v < 1,
	\end{equation}
	$\forall \ell \in\{1,\dots,6\}$, $t\in[0,t_{\max})$ where $\bar{\xi}_v \coloneqq \tfrac{\exp(\varepsilon_v)-1}{\exp(\varepsilon_v)+1} = -\tfrac{\exp(-\varepsilon_v)-1}{\exp(-\varepsilon_v)+1}$.   {Note that the term $\bar{B}_v$ is finite, $\forall t\in[0,t_{\max})$. Moreover, the term $\varepsilon_v(\xi_v(0))$ is finite due to the choice $\rho_{\scr v^0_{i,\ell}} > |e_{v_i,\ell}(0)|, \forall \ell \in\{1,\dots,6\},i\in\mathcal{N}$. Hence, since $\underline{\lambda}_K$ is strictly positive, the term $\bar{\varepsilon}_v$ is also finite.}
	Thus,  the term $\widetilde{r}_v(\xi_v(t)) $ and hence the control laws \eqref{eq:u_i (formation)} are also bounded in compact sets for all $t\in[0,t_{\max})$.
	
	{ What remains to be shown is that $t_{\max} = \infty$. Towards that end, suppose that $t_{\max}$ is finite, i.e., $t_{\max} < \infty$. Then, according to Theorem \ref{thm:forward_completeness (App_dynamical_systems)} of Appendix \ref{app:dynamical systems}, it holds that $\lim\limits_{t \to  t^{-}_{\max}}\Big(\| z(t) \| + \frac{1}{d_\mathcal{S}((z(t),t),\partial \Omega)} \Big) = \infty$, where $d_\mathcal{S}(\cdot)$ is the distance of $(z(t),t)$ to $\partial \Omega$. We first rewrite the condition in a more explicit form, in order to account for the matrix tuple $R\in \mathbb{SO}(3) ^N$. We define $z_{p,v} \coloneqq [p^\top,v^\top]^\top\in\mathbb{R}^{3N}\times\mathbb{R}^{6N}$, the projection sets $\Omega_R \coloneqq \{(R,t)\in \mathbb{SO}(3) ^N\times\mathbb{R}_{\geq 0} : (x,v,t)\in\Omega\}$ and $\Omega_{p,v} \coloneqq \{ (p,v,t)\in\mathbb{R}^{3N}\times\mathbb{R}^{6N}\times\mathbb{R}_{\geq 0} : (x,v,t)\in\Omega \}$ as well as the distance from a set $A\subset \mathbb{SO}(3) ^N\times\mathbb{R}_{\geq 0}$ as $d_{\mathcal{S}, \mathbb{SO}(3) }:  \mathbb{SO}(3) ^N\times\mathbb{R}_{\geq 0} \times 2^{ \mathbb{SO}(3) ^N\times\mathbb{R}_{\geq 0}}\to\mathbb{R}_{\geq 0}$ with $d_{\mathcal{S}, \mathbb{SO}(3) }((R,t),A) \coloneqq \inf\limits_{(R_A,t_A)\in A}\{\|R - R_A\|_T + t-t_A\}$, where $\| \cdot \|_T$ is the induced norm in $ \mathbb{SO}(3) ^N$ defined as $\|R\|_T \coloneqq \sum_{i\in\mathcal{N}}\|R_i\|_F$, for $R = (R_1,\dots,R_N)\in\mathbb{SO}(3)^N$. Therefore, the condition of Theorem \ref{thm:forward_completeness (App_dynamical_systems)} of Appendix \ref{app:dynamical systems} can now be stated as follows: Since $t_{\max} < \infty$, 
		it holds that
		\begin{align} 
		L \coloneqq& \lim\limits_{t \to  t^{-}_{\max}}\Big(\|p(t)\| + \|v(t)\| + \|R(t)\|_T +  \notag \\ 
		&\frac{1}{d_\mathcal{S}((z_{p,v}(t),t),\partial \Omega_{p,v}) + d_{\mathcal{S}, \mathbb{SO}(3) }((R(t),t),\partial \Omega_R) }  \Big) = \infty \label{eq:L t_max < infty (formation)}
		\end{align}
		which we aim to prove that is a contradiction. Firstly, it holds that 
		\begin{equation*}
			\|R(t)\|_T = \sum_{i\in\mathcal{N}}\|R_i(t)\|_F \leq N \sup_{t\in [0,t_{\max})}\left\{ \max_{i\in\mathcal{N}}\left\{R_i(t)\right\}\right\}.
		\end{equation*} 
		However, according to Proposition \ref{prop:R trace (app_useful_prop)}, it holds that $-1 \leq \text{tr}(R) \leq 3$ for any $R\in \mathbb{SO}(3) $. Hence, $\|R(t)\|_T \leq 3N, \forall t\in[0,t_{\max}]$. Moreover, from \eqref{eq:bar ksi v (formation)} and \eqref{eq:ksi_i_v (formation)} we obtain $\|\bar{e}_v(t)\| \leq \sqrt{6}\bar{\xi}_v\widetilde{\rho}$, $\forall t\in[0,t_{\max})$. By invoking \eqref{eq:bar epsilon e (formation)}, \eqref{eq:bar epsilon psi (formation)}, we can also conclude that there exists a finite $\bar{v}_{\text{des}}$ such that $\|v_{\text{des}}(t)\| \leq \bar{v}_{\text{des}}$, $\forall t\in[0,t_{\max})$. Hence, since $\|\mathsf{n}_i(x_i(t),t)\| \leq \bar{\mathsf{n}}_i$, $\forall t\in\mathbb{R}_{\geq 0}, i\in\mathcal{N}$, $v = \widetilde{v} - \mathsf{n} = \bar{e}_v + v_{\text{des}} - \mathsf{n}$ implies that there exists a finite $\bar{v}$ such that $\|v(t)\| \leq \bar{v}$, $\forall t\in[0,t_{\max})$. Hence, $\| p(t)\| = \| \int_{0}^{t_{\max}}\bar{R}(s)v(s)ds\| \leq \int_{0}^{t_{\max}}\|\bar{R}(s)v(s)\|ds = \int_{0}^{t_{\max}}\|v(s)\|ds \leq  \int_{0}^{t_{\max}}\bar{v}ds \Rightarrow \|p(t)\| \leq t_{\max}\bar{v}$, $\forall t\in[0,t_{\max})$, which proves the boundedness of $\|p(t)\|$, since $t_{\max}<\infty$. }
	
	{Next, note that $\partial \Omega_{p,v} = \{(p,v,t)\in \mathbb{R}^{3N}\times\mathbb{R}^{6N}\times\mathbb{R}_{\geq 0} : ( \exists k\in\mathcal{K} : \xi_{e_k}(p_{k_1},p_{k_2},t) = -C_{k,\text{col}} \text{ or } \xi_{e_k}(p_{k_1},p_{k_2},t) = C_{k,\text{con}} ) \text{ or } (\exists i\in\mathcal{N},\ell\in\{1$, $\dots$, $6\} : \xi_{v_{i,\ell}}(x,v_i,t) = -1 \text{ or }  \xi_{v_{i,\ell}}(x,v_i,t) = 1 ) \}$ and $\partial \Omega_R$ $ = $ $\{(R,t)\in \mathbb{SO}(3) ^N\times\mathbb{R}_{\geq 0}$ $: \exists k\in\mathcal{K}$ $:$ $\xi_{\psi_k}(R_{k_1}$,$R_{k_2}$,$t)$ $= 1 \}$. We have proved, however, from \eqref{eq:bar ksi e (formation)}, \eqref{eq:bar ksi psi (formation)}, and \eqref{eq:bar ksi v (formation)} that the maximal solution satisfies the strict inequalities $-C_{k,\text{col}} < -\underline{\xi}_e \leq \xi_{e_k}(p_{k_1}(t),p_{k_2}(t),t) \leq \bar{\xi}_e < C_{k,\text{con}}$, $\xi_{\psi_k}(R_{k_1}(t),R_{k_2}(t),t) \leq \bar{\xi}_{\psi} < 1$, and $|\xi_{v_{i,\ell}}(x(t),v_i(t),t)| \leq \bar{\xi}_v < 1 $, $\forall k\in\mathcal{K}$, $\ell\in\{1,\dots,6\}$, $i\in\mathcal{N}$, $t\in[0,t_{\max})$. Therefore, we conclude that there exist strictly positive constants $\epsilon_{p,v}$, $\epsilon_R$ $\in\mathbb{R}_{>0}$ such that $d_\mathcal{S}((z_{p,v}(t),t),\partial \Omega_{p,v}) \geq \epsilon_{p,v}$ and $d_{\mathcal{S}, \mathbb{SO}(3) }((R(t),t),\partial \Omega_R) \geq \epsilon_R$, $\forall t\in[0,t_{\max})$. Therefore, we have proved that 
		\begin{align*}
		&L \leq (t_{\max} + 1)\bar{v} + 3N + \frac{1}{\epsilon_{p,v}+\epsilon_R}  < \infty,
		\end{align*}
		since $t_{\max}$ is finite. This contradicts \eqref{eq:L t_max < infty (formation)} and hence, we conclude that $t_{\max} = \infty$.
	}
	
	We have proved the containment of the errors $e_k(t)$, $\psi_k(t)$ in the domain defined by the prescribed performance funnels:
	\begin{align*} 
	-C_{k,\text{col}}\rho_{e_k}(t) &< e_k(t) < C_{k,\text{con}}\rho_{e_k}(t), \\
	{0} &   {\leq \psi_k(t) < \rho_{\psi_k}(t)},
	\end{align*}
	$\forall k\in \mathcal{K}$, $t\in \mathbb{R}_{\geq 0}$, which also implies that 
	\begin{align*}
	d_{k,\text{col}} <& \| p_{k_1}(t) - p_{k_2}(t) \| < d_{k,\text{con}},\\
	0\leq &\psi_k(t) < 2,
	\end{align*}
	$\forall k\in \mathcal{K}$, $t\in\mathbb{R}_{\geq 0}$, i.e., avoidance of the singularity $\psi_k = 2$ and satisfaction of the collision and connectivity constraints for the initially connected edge set $\mathcal{E}$. 
\end{proof}

\begin{remark} 
		It is worth noting that the control scheme applies also to the case that the graph is minimally rigid, which implies similar properties regarding 	$\widetilde{D}$ \cite{mehdifar2019prescribed}.
\end{remark}

\subsection{Simulation Results} \label{sec: simulation_results (formation)}

We considered $N=4$ spherical agents with $\mathcal{N} = \{1,2,3,4\}$ and dynamics of the form \eqref{eq:system (formation)}, with $r_i=1 \text{m}$ and $\varsigma_i = 4 \text{m}$, $i\in\{1,\dots,4\}$. We selected the exogenous disturbances and measurement noise as $w_i = A_{w_i}\sin(\omega_{w,i}t)\dot{x}_i$, and $\mathsf{n}_i = A_{n_i}\sin(\omega_{\mathsf{n},i}t)\dot{x}_i$, where the parameters $A_{w_i}, A_{n_i}, \omega_{w,i}, \omega_{\mathsf{n},i}$ as well as the dynamic parameters (mass and moment of inertia) of the agents were randomly chosen in $[0,1]$, $\forall i\in\mathcal{N}$. The initial conditions were taken as: $p_1(0)=[0,0,0]^\top \ \text{m}$, $p_2(0)=[2,2,2]^\top \ \text{m}$, $p_3(0)=[2,4,4]^\top \ \text{m}$, $p_4(0)=[2,3,3]^\top \ \text{m}$, $R_1(0)=R_3(0)=R_4(0) = I_{3}$ and $$R_2(0) = \begin{bmatrix}
-0.3624 & 0.0000 & 0.9320 \\
0.6591 & 0.7071 & 0.2562 \\
-0.6591 & 0.7071 & -0.2562
\end{bmatrix},$$ $v_1(0) = v_2(0 = v_3(0) = v(4) = 0$, which give the edge set $\mathcal{E} = \{\{1,2\}, \{2,3\}, \{2,4\} \}$ and the incidence matrix:
\begin{equation*}
D(\mathcal{G}) = 
\begin{bmatrix}
-1 & 0 & 0 \\
1 & -1 & -1 \\
0 & 1 & 0 \\
0 & 0 & 1 
\end{bmatrix}.
\end{equation*}
The desired graph formation was defined by the constants $d_{k,\text{des}} = 2.5\text{m}$, 
\begin{equation*}
R_{k,\text{des}} =
\begin{bmatrix}
0.5000 & -0.8660 & 0.0000 \\
0.6124 & 0.3536 & -0.7071 \\
0.6124 & 0.3536 & 0.7071
\end{bmatrix}, \forall k\in\{1,2,3\}.
\end{equation*}
The definitions of $d_{k, \text{col}}$, $d_{k, \text{con}}$ yield: $d_{k, \text{col}} = 2$ and $d_{k, \text{con}} = 4$. Invoking \eqref{eq:C_k (formation)}, we have $C_{k, \text{col}} = 2.25$ and $C_{k, \text{con}} = 9.75$. Moreover, the parameters of the performance functions were chosen as $\rho_{\scr e_k,\infty} $ $= \rho_{\scr \psi_k,\infty} = 0.1$, $\rho_{\scr \psi_k,0} = 1.99 > \max\{\rho_{\scr \psi_1}(0), \rho_{\scr \psi_2}(0), \rho_{\scr \psi_3}(0)\}$ and $l_{e_k} = l_{\psi_k} = 0.7$. In addition, we chose $\rho_{\scriptscriptstyle v^{0}_{i,\ell}} = 2 | e_{v_i,\ell}(0)| + 0.5$, $l_{v_{i_\ell}} = 1.55$ and $\rho_{\scriptscriptstyle v^{\infty}_{i,\ell}} = 0.15$, for every $i \in \{1, \dots, 4\}$, $\ell \in \{1, \dots, 6\}$. Finally, the control gains were set to $\Gamma = 10 I_{24}$ and $\Delta = I_{24}$.

The simulation results are shown in Fig.\ref{fig:dist_error_1 (formation)}-\ref{fig:control_input_4 (formation)}. In particular, Fig. \ref{fig:dist_error_1 (formation)}-\ref{fig:dist_error_3 (formation)} and Fig. \ref{fig:orient_error_1 (formation)}-\ref{fig:orient_error_3 (formation)} show the distance error signals and the orientation error signals, respectively. All the errors remain within the predefined bounds and converge to $0$. Fig. \ref{fig:dist_agents (formation)} shows the distance between the agents. The connectivity is maintained for all times as well as the agents do not collide with each other. Finally, Fig. \ref{fig:control_input_1 (formation)}-Fig. \ref{fig:control_input_4 (formation)} depict the control input signals of the agents which remain bounded for all times. A video illustrating the simulation results can be found on \href{https://www.youtube.com/watch?v=Z4xLyO1twvk}{https://www.youtube.com/watch?v=Z4xLyO1twvk}.

\begin{figure}[t!]
	\centering
	\includegraphics[scale = 0.45]{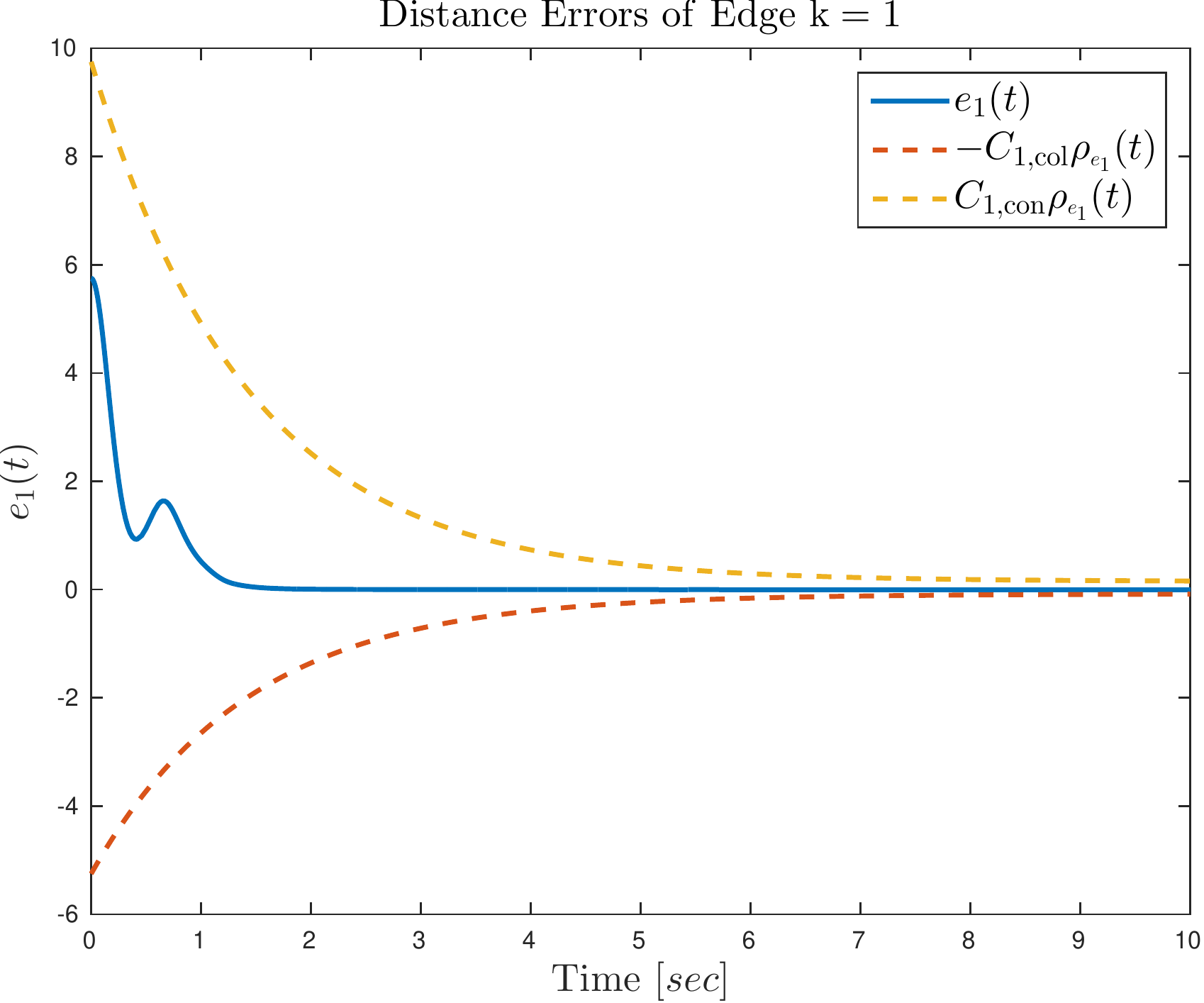}
	\caption{The distance error signal of the edge $(1,2)$.}
	\label{fig:dist_error_1 (formation)}
\end{figure}

\begin{figure}[t!]
	\centering
	\includegraphics[scale = 0.45]{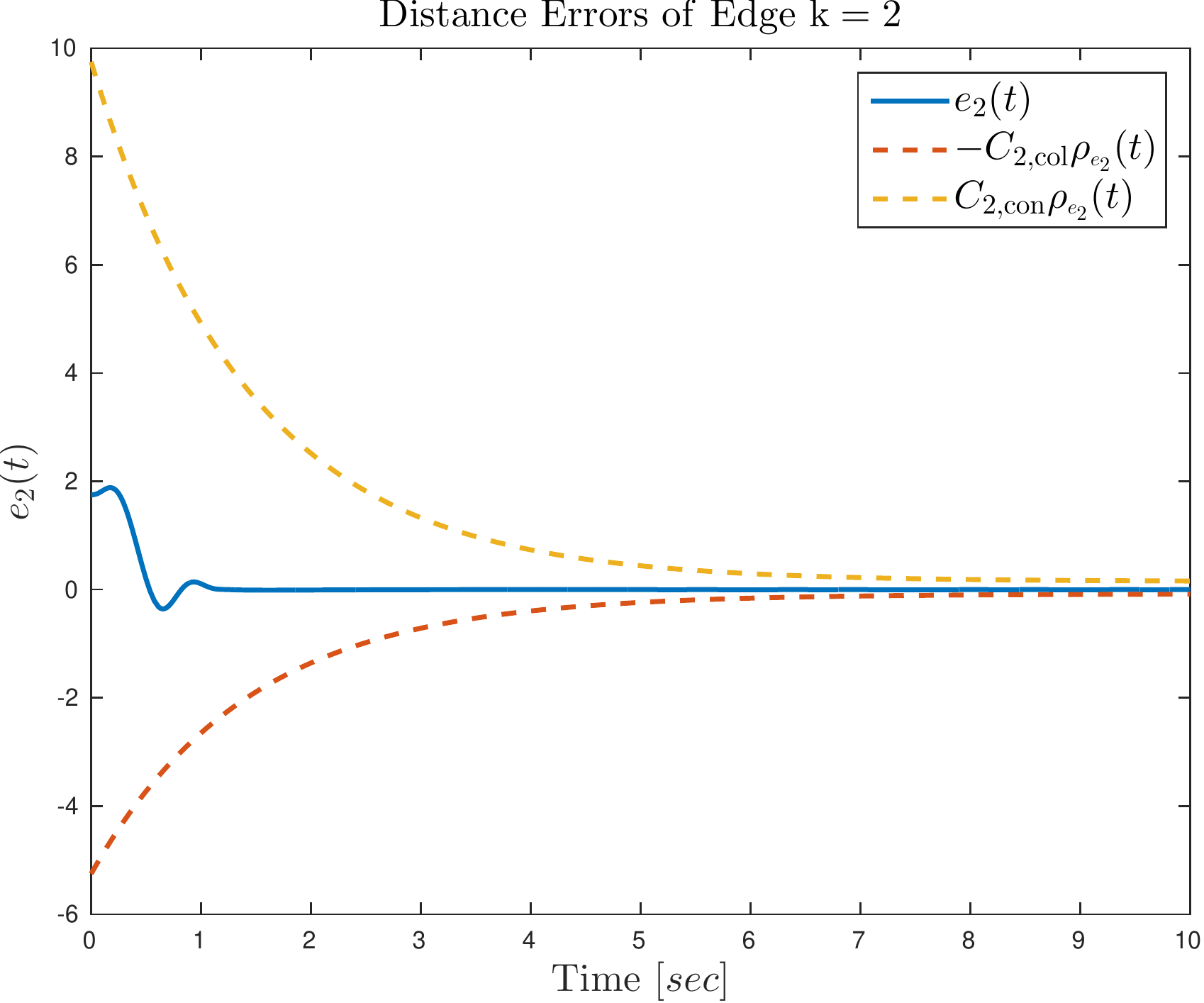}
	\caption{The distance error signal of the edge $(2,3)$.}
	\label{fig:dist_error_2 (formation)}
\end{figure}

\begin{figure}[t!]
	\centering
	\includegraphics[scale = 0.45]{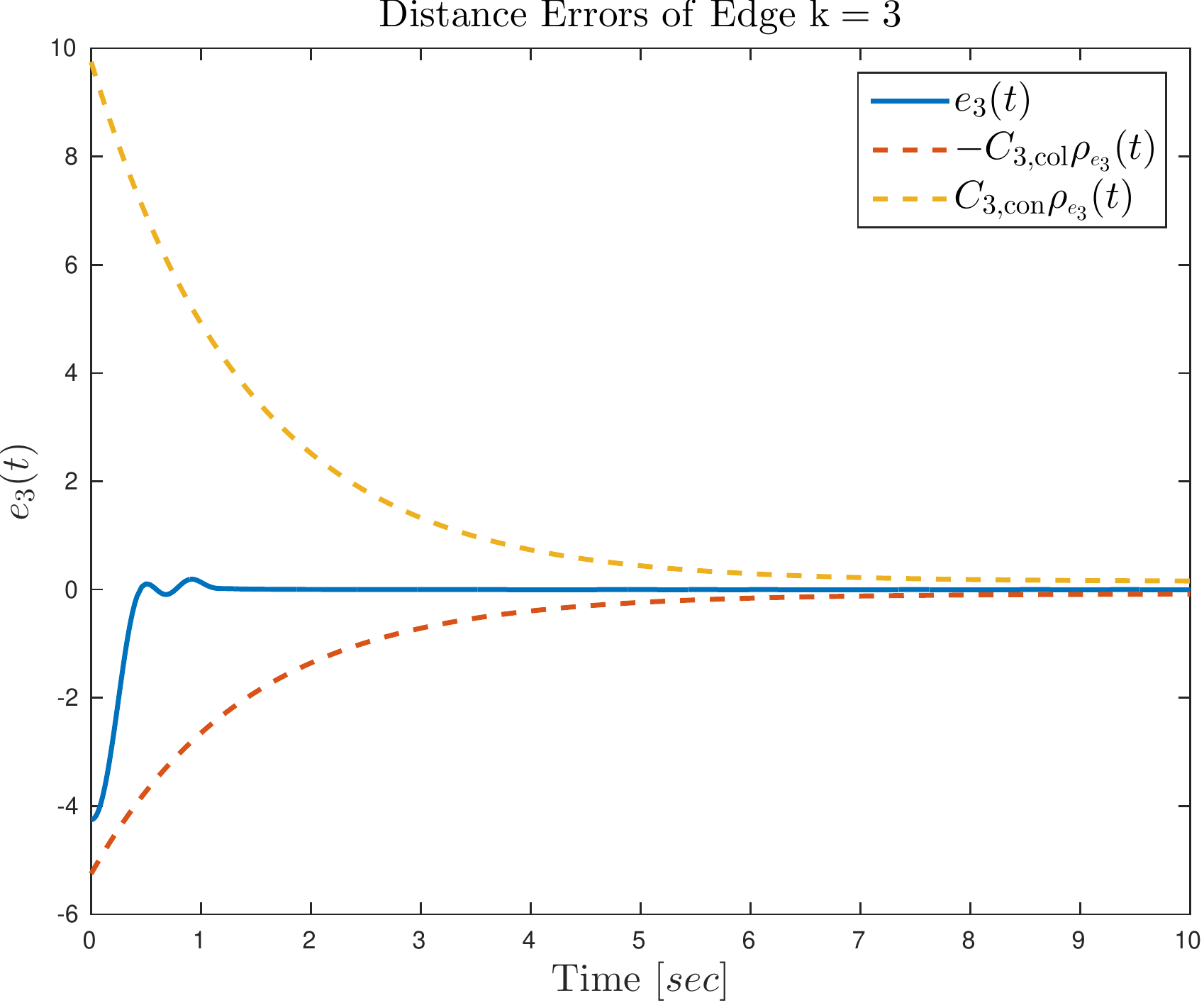}
	\caption{The distance error signal of the edge $(2,4)$.}
	\label{fig:dist_error_3 (formation)}
\end{figure}

\begin{figure}[t!]
	\centering
	\includegraphics[scale = 0.45]{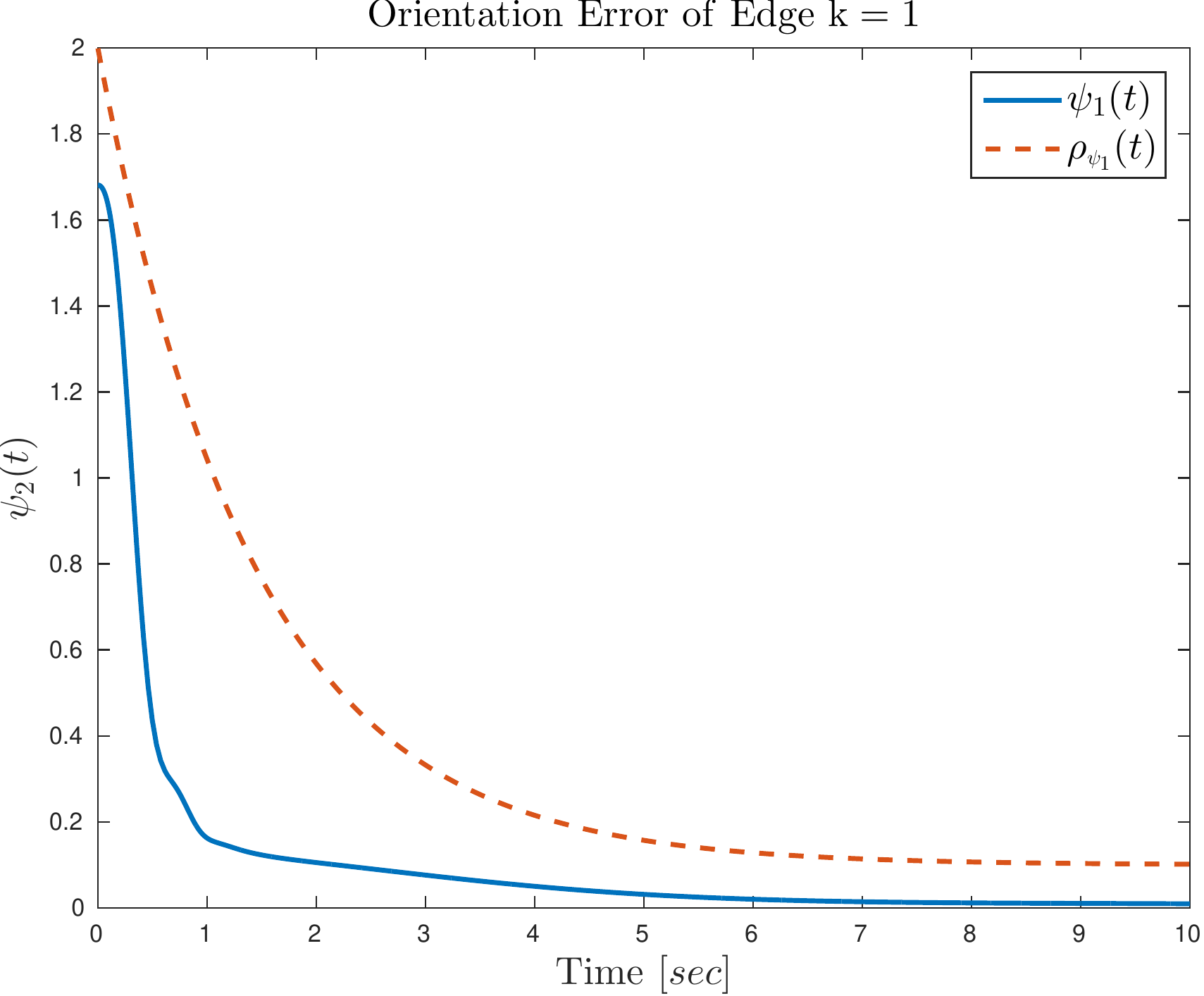}
	\caption{The orientation error signal of the edge $(1,2)$.}
	\label{fig:orient_error_1 (formation)}
\end{figure}

\begin{figure}[t!]
	\centering
	\includegraphics[scale = 0.45]{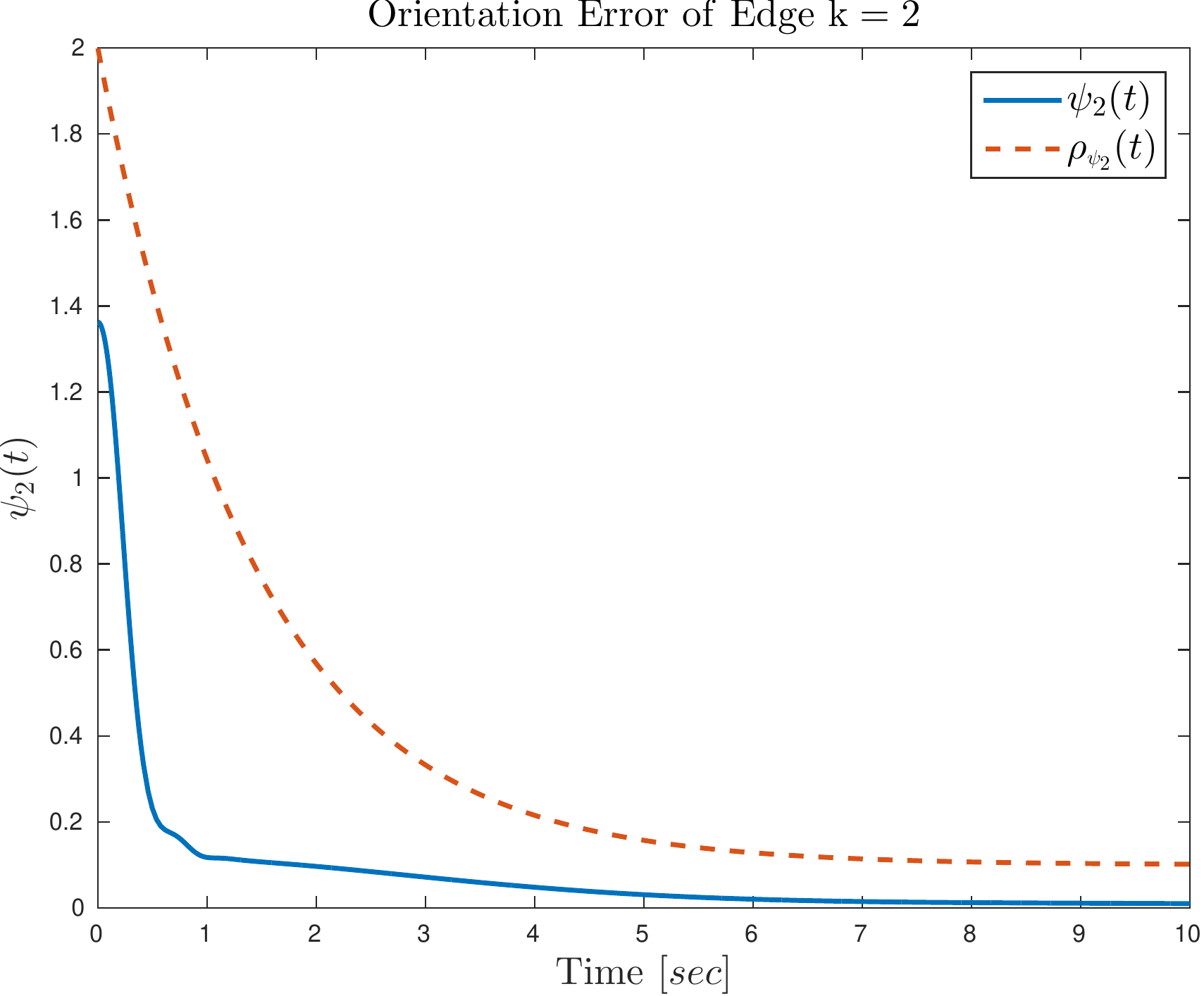}
	\caption{The orientation error signal of the edge $(2,3)$.}
	\label{fig:orient_error_2 (formation)}
\end{figure}

\begin{figure}[t!]
	\centering
	\includegraphics[scale = 0.45]{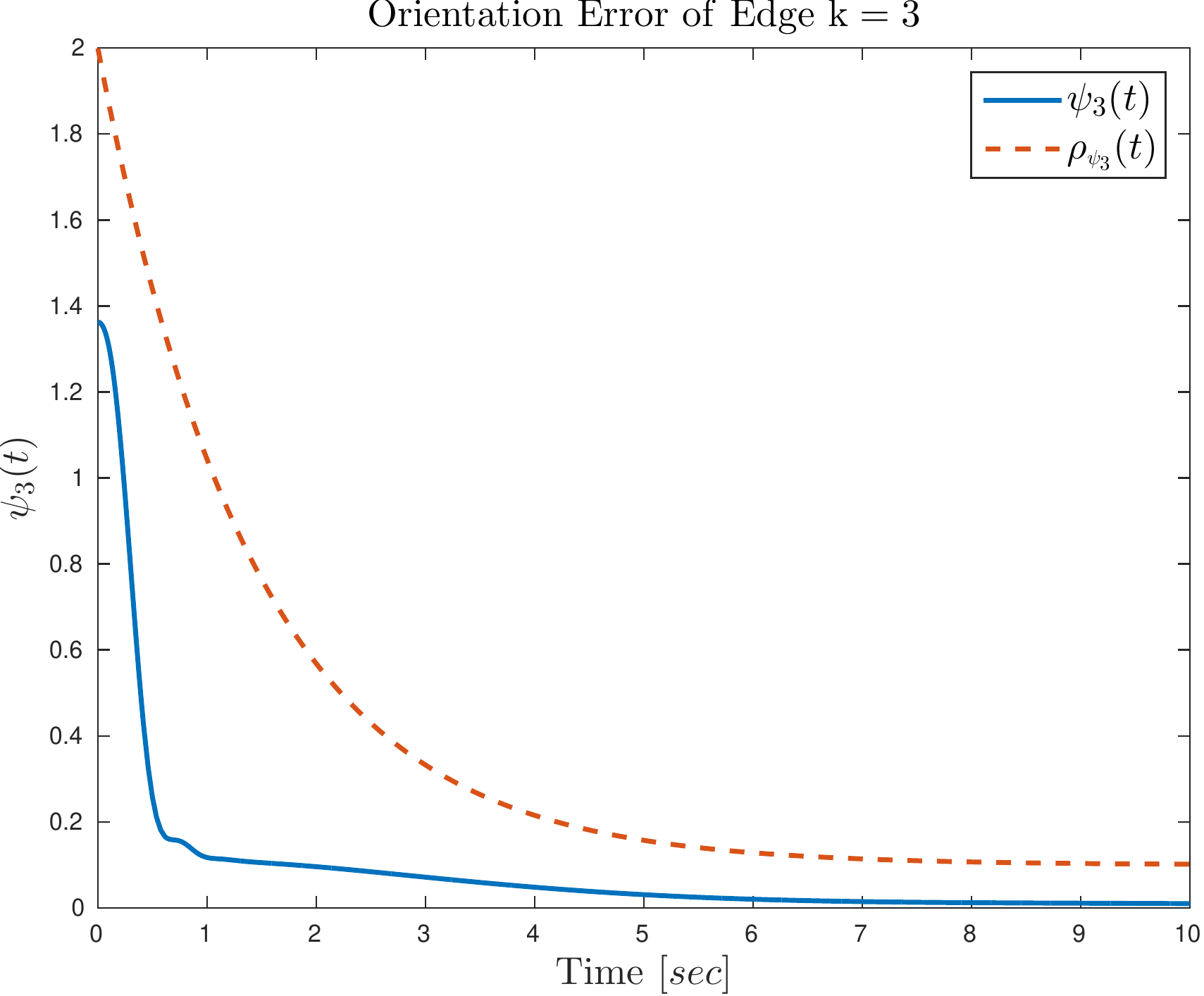}
	\caption{The orientation error signal of the edge $(2,4)$.}
	\label{fig:orient_error_3 (formation)}
\end{figure}

\begin{figure}[t!]
	\centering
	\includegraphics[scale = 0.45]{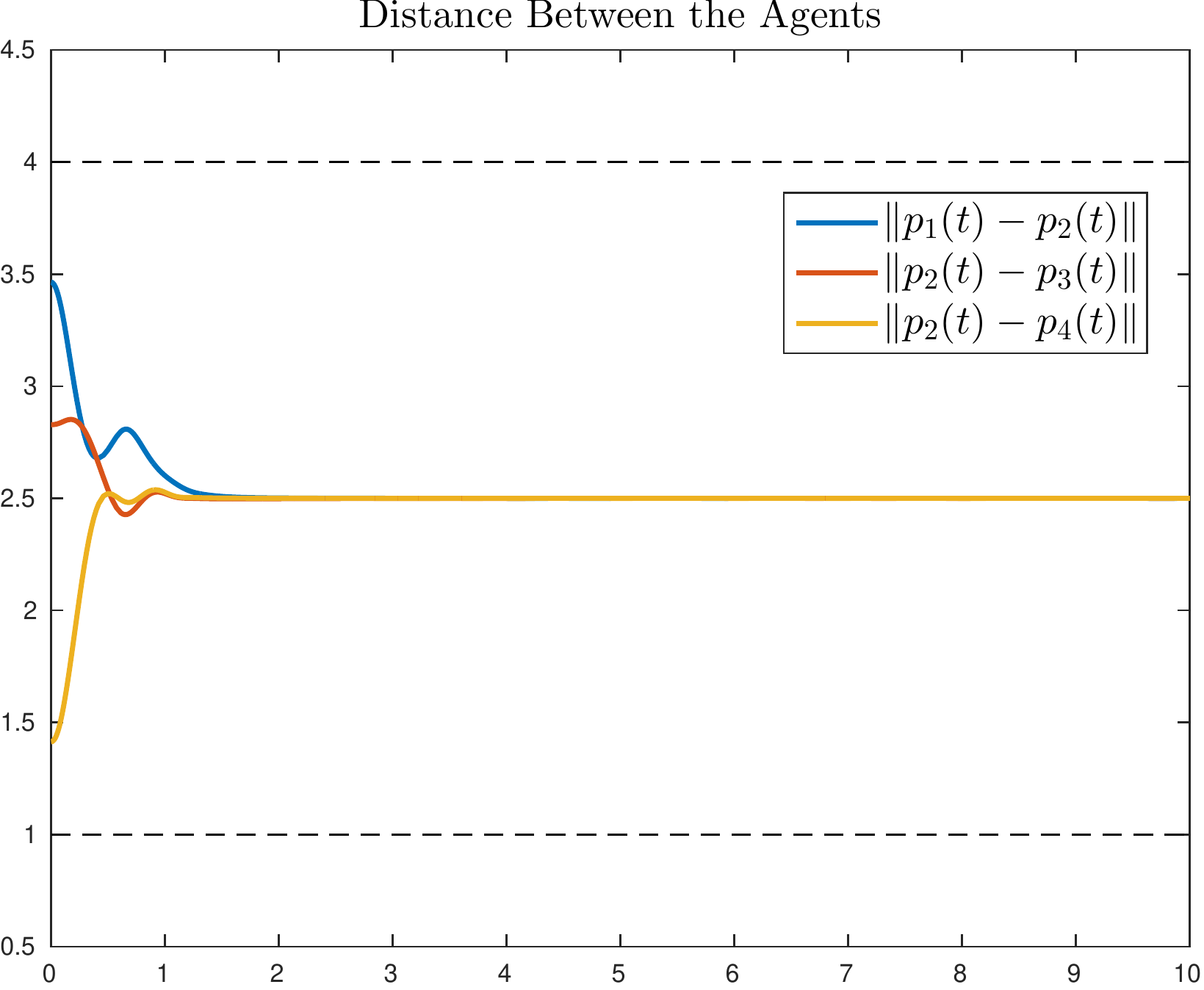}
	\caption{The distance between the agents.}
	\label{fig:dist_agents (formation)}
\end{figure}

\begin{figure}[t!]
	\centering
	\includegraphics[scale = 0.45]{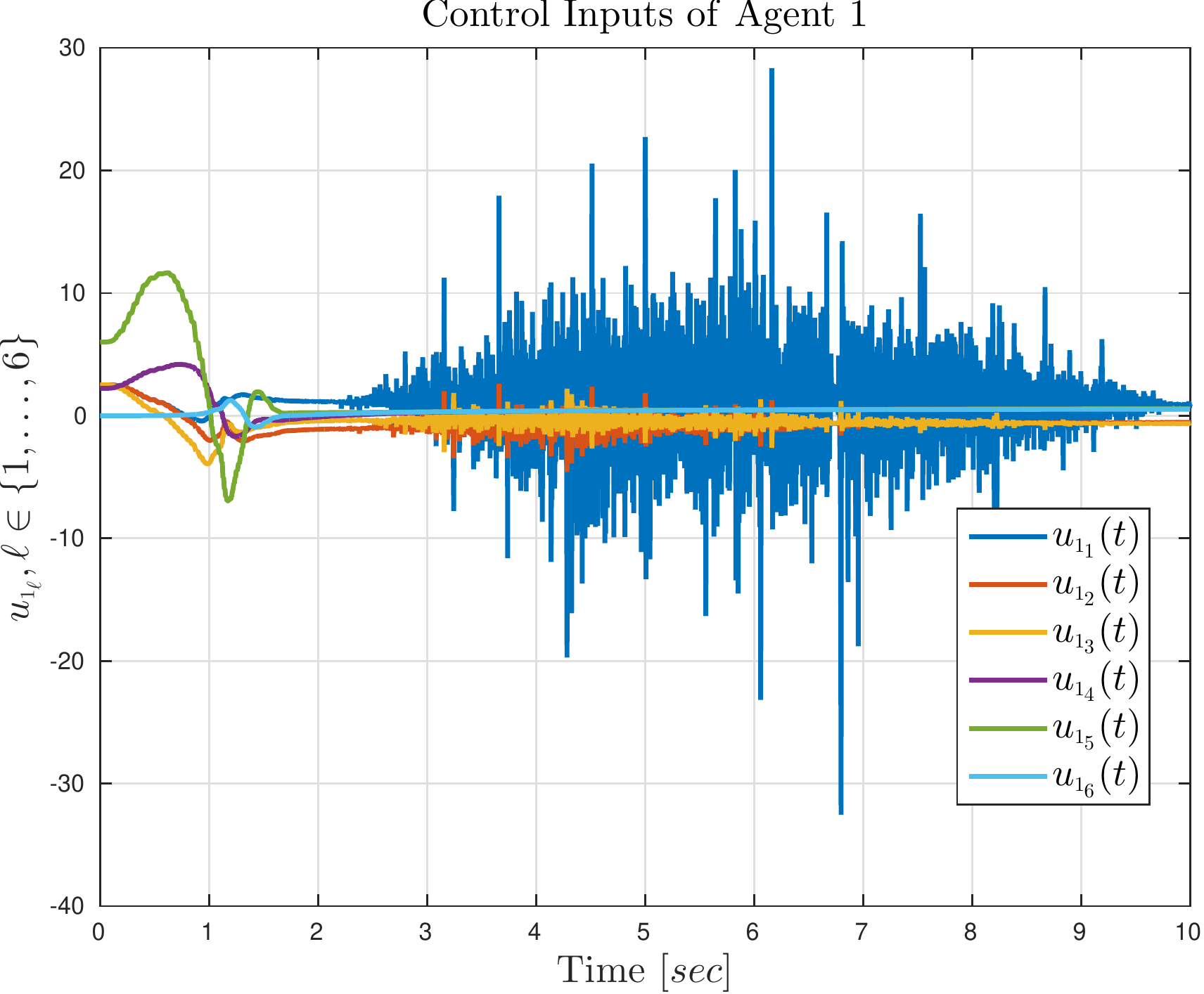}
	\caption{The control input signals of agent $1$.}
	\label{fig:control_input_1 (formation)}
\end{figure}

\begin{figure}[t!]
	\centering
	\includegraphics[scale = 0.45]{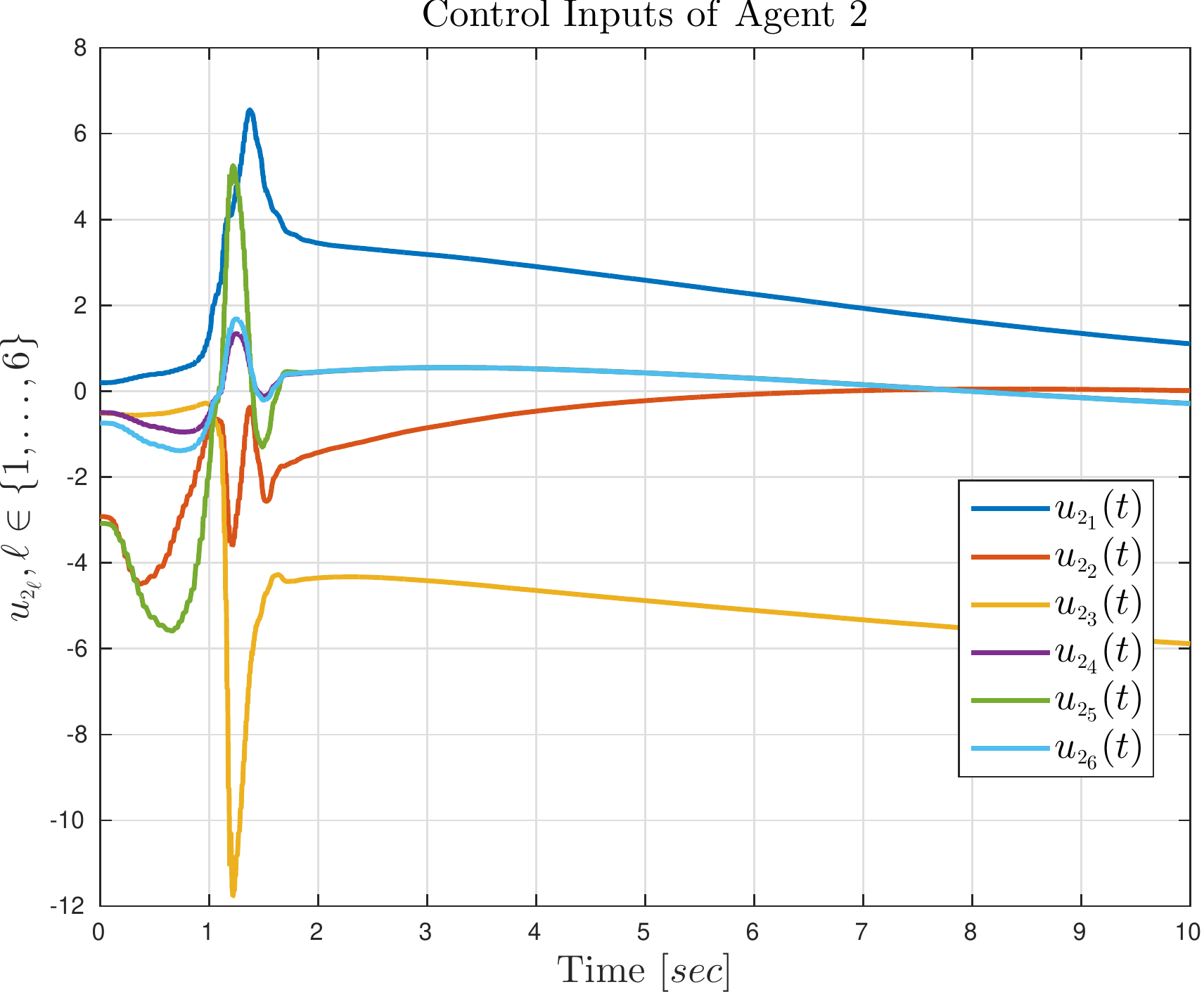}
	\caption{The control input signals of agent $2$.}
	\label{fig:control_input_2 (formation)}
\end{figure}

\begin{figure}[t!]
	\centering
	\includegraphics[scale = 0.45]{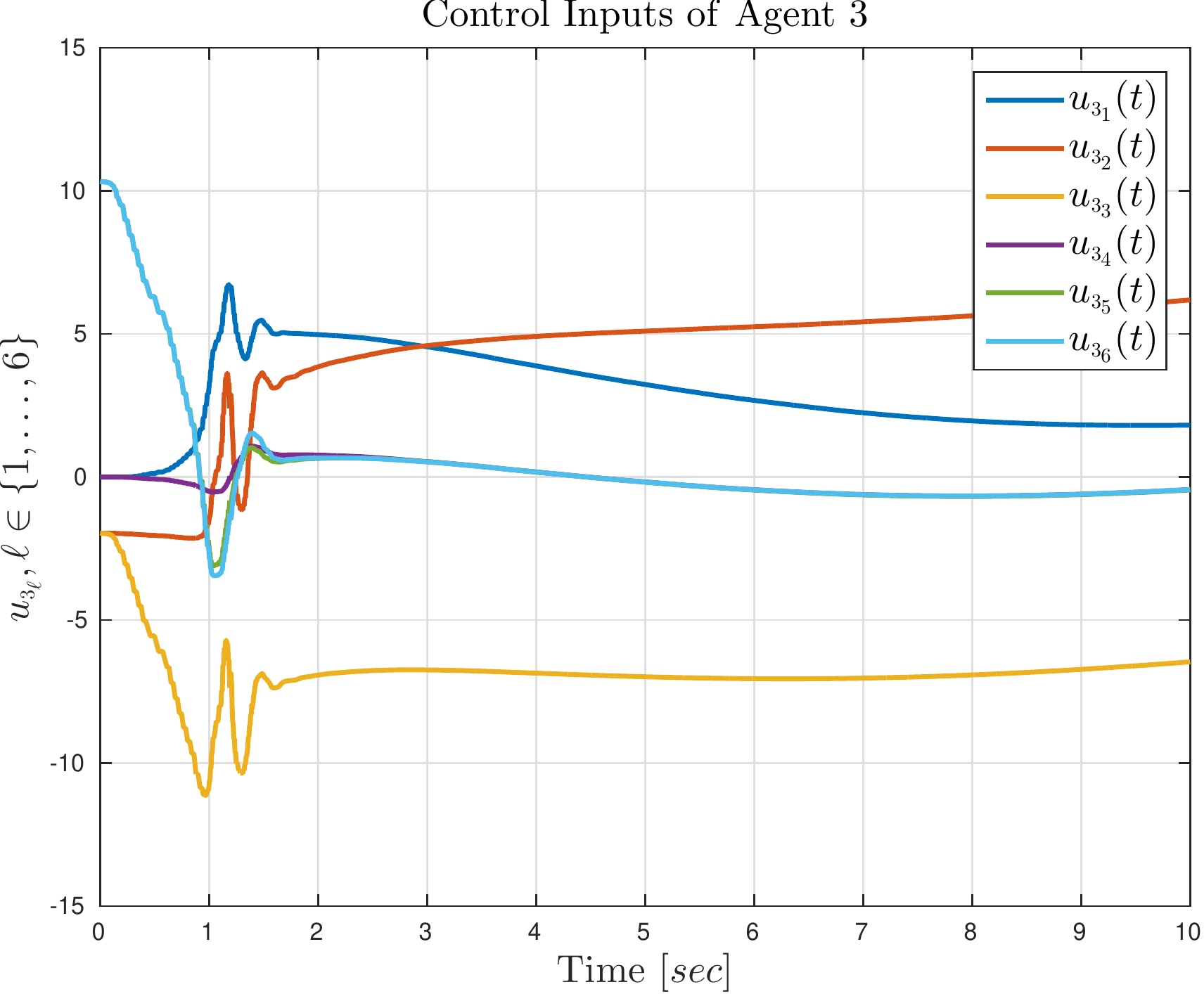}
	\caption{The control input signals of agent $3$.}
	\label{fig:control_input_3 (formation)}
\end{figure}

\begin{figure}[t!]
	\centering
	\includegraphics[scale = 0.45]{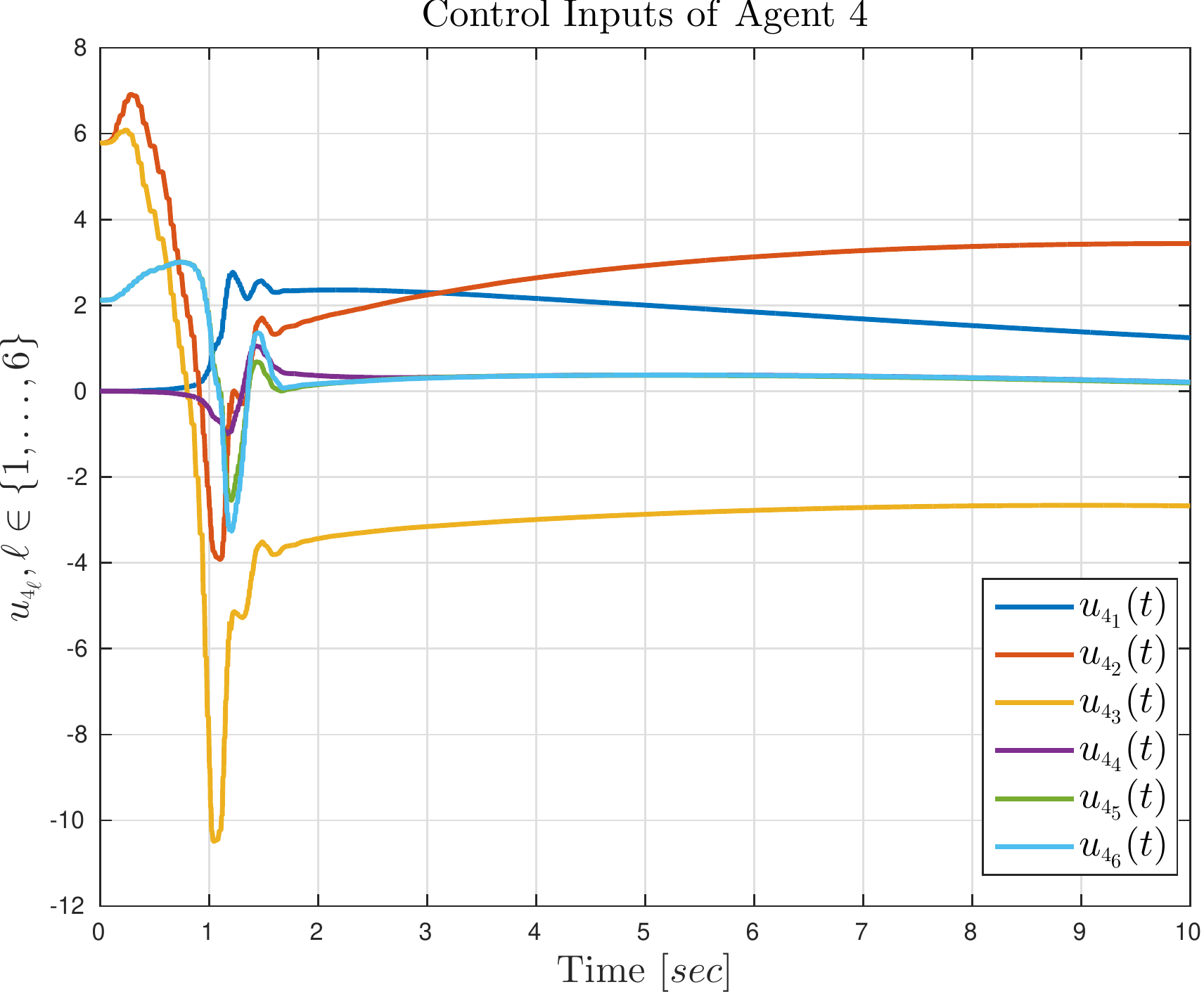}
	\caption{The control input signals of agent $4$.}
	\label{fig:control_input_4 (formation)}
\end{figure}

\section{Cooperative Manipulation via Internal Force Regulation: A Rigidity Theory Approach} \label{sec:rigidity + coop manip}

Except from the more classical formation control framework, connected to the previous section, formation control can be connected to cooperative manipulation, where the nodes of the formation graph are the robotic agents grasping the object.
In this section we associate classical \textit{rigidity} theory with rigid cooperative manipulation. In particular, motivated by the rigid grasps in a cooperative manipulation scheme, we first introduce the notion of distance and bearing rigidity of a graph in $\mathbb{SE}(3)$.  Next, we associate the nodes of the graph to the robotic agents in a cooperative manipulation scheme, and we provide new results on the interaction and internal forces as well as optimal cooperative manipulation. 


\subsection{Cooperative Manipulation Modeling}

Regarding rigid cooperative manipulation, we follow the notation and dynamics of the previous chapter. We provide here a brief recap.

Let $N$ robotic agents, with $\mathcal{N} \coloneqq \{1,\dots,N\}$, rigidly grasping an object, with $q_i\in\mathbb{R}^{n_i}$ their joint configurations, $q\coloneqq [q_1^\top,\dots,q_N^\top]^\top \in \mathbb{R}^n$, $n\coloneqq \sum_{i\in\mathcal{N}}n_i$, $p_{\scr E_i} \in \mathbb{R}^3, \eta_{\scr E_i} \in \mathbb{T}$, $R_i(\eta_i)\in\mathbb{SO}(3)$ and $v_i \coloneqq [\dot{p}_{\scr E_i}^\top, \omega_{\scr E_i}^\top]^\top \in \mathbb{R}^6$ the end-effector poses and velocities, and $x_i \coloneqq (p_{\scr E_i}, R_i) \in \mathbb{SE}(3)$, $x\coloneqq (x_1,\dots,x_N)\in\mathbb{SE}(3)^N$, $v\coloneqq [v_1^\top,\dots,v_N^\top]^\top\in\mathbb{R}^{6N}$. The stacked agent dynamics in joint- and task-space are (see \eqref{eq:manipulator joint_dynamics (TCST_coop_manip)} and \eqref{eq:manipulator dynamics_vector_form (TCST_coop_manip)}
\begin{subequations}	\label{eq:manipulator dynamics_vector_form (rigid+coopmanip)}
	\begin{align}
	{B}({q})\ddot{q}+C_q(q,\dot{{q}})\dot{q} + g_q(q)& = \tau - J(q)^\top h, \label{eq:manipulator dynamics_joint space vector_form (rigid+coopmanip)} \\
	{M}({q})\dot{v}+{C}({q},\dot{{q}})v + {g}({q}) &= {u}-h,\label{eq:manipulator dynamics_task space vector_form (rigid+coopmanip)}
	\end{align}
\end{subequations}
where $J\coloneqq \text{diag}\{[J_i]_{i\in\mathcal{N}}\}$, and we have removed the disturbance vector for simplicity. We remind the reader that the task-space terms are defined in $\mathsf{S}_i=\{ q_i\in\mathbb{R}^n : \det(J_i(q_i)J_i(q_i)^\top) > 0\}$. 
The object pose and velocity are denoted by $x_{\scr O} \coloneqq [p_{\scr O}^\top,\eta_{\scr O}^\top]^\top \in \mathbb{M}$, $R_{\scr O}(\eta_{\scr O})\in\mathbb{SO}(3)$, $v_{\scr O}\coloneqq [\dot{p}_{\scr O}^\top,\omega_{\scr O}^\top]^\top \in\mathbb{R}^6$, and dynamics
\begin{subequations} \label{eq:object dynamics (rigid+coopmanip)}
\begin{align}
	&\dot{R}_{\scr O} = S(\omega_{\scr O})R_{\scr O}, \label{eq:object dynamics 1 (rigid+coopmanip)} \\
	&M_{\scr O}(\eta_{\scr O})\dot{v}_{\scr O} + C_{\scr O}(\eta_{\scr O},\omega_{\scr O})v_{\scr O} + g_{\scr O} = h_{\scr O}.\label{eq:object dynamics 2 (rigid+coopmanip)}
\end{align}
\end{subequations} 

In view of Fig. \ref{fig:Two-robotic-arms (TCST_coop_manip)}, one obtains the coupled kinematics
\begin{align} \label{eq:J_o_i (rigid+coopmanip)}
	v_i = J_{\scr O_i}v_{\scr O},
\end{align}
where $J_{\scr O_i}\coloneqq J_{\scr O_i}(x_i):\mathbb{SE}(3) \to \mathbb{R}^{6\times 6}$ is the object-to-agent Jacobian introduced in \eqref{eq:J_o_i (TCST_coop_manip)}, redefined here as a function of $x_i$ instead of $q_i$, i.e.,
\begin{equation*}
	J_{\scr O_i}(x_i) \coloneqq \begin{bmatrix}
	I_3 & -S\big( R_i^\top p^{\scr E_i}_{\scr E_i/O} \big) \\
	0 & I_3
	\end{bmatrix},
\end{equation*}
which forms the respective grasp matrix 
\begin{equation} \label{eq:grasp matrix def. (rigid+coopmanip)}
G\coloneqq G(x) \coloneqq [J_{\scriptscriptstyle O_1}(x_1)^\top,\dots,J_{\scriptscriptstyle 	O_N}(x_N)^\top] \in \mathbb{R}^{6 \times 6N},
\end{equation}
and has full column-rank due to the rigidity of the grasping contacts; Note that \eqref{eq:J_o_i (rigid+coopmanip)} can now be written in stack vector form as 
\begin{equation} \label{eq:grasp matrix velocities (rigid+coopmanip)}
v = G^\top v_{\scr O}.
\end{equation}}
Next, we associate $h$ and $h_{\scr O}$ via $G$ (as in \eqref{eq:grasp matrix (TCST_coop_manip)}) to obtain
\begin{equation}
h_{\scriptscriptstyle O}=G h, \label{eq:grasp matrix (rigid+coopmanip)}
\end{equation}
which leads to the coupled dynamics (see \eqref{eq:coupled dynamics (TCST_coop_manip)})
\begin{equation}
	\widetilde{M}(\bar{x})\dot{v}_{\scriptscriptstyle O}+\widetilde{C}(\bar{x})v_{\scriptscriptstyle O}+\widetilde{g}(\bar{x})  = G u,\label{eq:coupled dynamics (rigid+coopmanip)}
\end{equation}
where we slightly change the notation with respect to the previous chapter as $\bar{x} \coloneqq [q^\top,\dot{q}^\top,\eta_{\scr O}^\top, \omega_{\scr O}^\top]^\top \in \mathbb{X}= \mathsf{S}\times\mathbb{R}^{n+6}\times\mathbb{T}$ (instead of $x$). 

The vector of interaction forces $h$ among the agents  and the object can be decoupled into motion-induced and internal forces 
\begin{equation} \label{eq:interaction forces (rigid+coopmanip)}
h = h_\text{m} + h_\text{int}.
\end{equation}
The internal forces $h_\text{int}$ are squeezing forces that the agents exert to the object and belong to the nullspace of $G(x)$ (i.e., $G h_\textup{int} = 0$). Hence, they do not contribute to the acceleration of the coupled system and result in internal stresses that might damage the object. A closed form analytic expression for $h_\text{m}$ and $h_\text{int}$ will be given in  the next section.

{Note from \eqref{eq:grasp matrix velocities (rigid+coopmanip)} that the agent velocities $v$ belong to the range space of $G^\top$. Therefore, since $G$ is a matrix that encodes rigidity constraints, this motivates the association of $G$ to the \emph{rigidity matrix} used in formation rigidity theory, and of the rigid cooperative manipulation scheme to a multi-agent rigid formation scheme. To this end, we introduce now the notion of Distance and Bearing Rigidity in $\mathbb{SE}(3)$.} 	

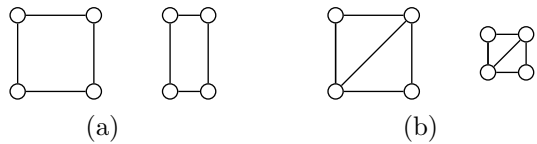
\begin{figure}[t!]
	\centering
	\begin{tikzpicture}[scale = 1]    
	\node(1) [line width = 0.5] at (0,0)[shape=circle,draw,inner sep=2pt]{};
	\node(2) [line width = 0.5] at (1,0)[shape=circle,draw,inner sep=2pt]{$ $};
	\node(3) [line width = 0.5] at (0,-1)[shape=circle,draw,inner sep=2pt]{$ $};
	\node(4) [line width = 0.5] at (1,-1)[shape=circle,draw,inner sep=2pt]{$ $};
	
	\path [-] [line width = 0.5]
	(1) edge node [above]      {}   (2)
	(1) edge node [above]      {}   (3)
	(3) edge node [above]      {}   (4)
	(2) edge node [above]      {}   (4);
	
	\node(5) [line width = 0.5] at (2,0)[shape=circle,draw,inner sep=2pt]{$ $};
	\node(6) [line width = 0.5] at (2.5,0)[shape=circle,draw,inner sep=2pt]{$ $};
	\node(7) [line width = 0.5] at (2,-1)[shape=circle,draw,inner sep=2pt]{$ $};
	\node(8) [line width = 0.5] at (2.5,-1)[shape=circle,draw,inner sep=2pt]{$ $};
	
	\path [-] [line width = 0.5]
	(5) edge node [above]      {}   (6)
	(5) edge node [above]      {}   (7)
	(6) edge node [above]      {}   (8)
	(7) edge node [above]      {}   (8);
	
	\node at (1.5,-1.1,1) {(a)};
	\end{tikzpicture}
	\qquad\hspace{5mm}
	\begin{tikzpicture}[scale = 1]    
	\node(1) [line width = 0.5] at (0,0)[shape=circle,draw,inner sep=2pt]{};
	\node(2) [line width = 0.5] at (1,0)[shape=circle,draw,inner sep=2pt]{$ $};
	\node(3) [line width = 0.5] at (0,-1)[shape=circle,draw,inner sep=2pt]{$ $};
	\node(4) [line width = 0.5] at (1,-1)[shape=circle,draw,inner sep=2pt]{$ $};
	
	\path [-] [line width = 0.5]
	(1) edge node [above]      {}   (2)
	(1) edge node [above]      {}   (3)
	(1) edge node [above]      {}   (3)
	(3) edge node [above]      {}   (4)
	(2) edge node [above]      {}   (4)
	(2) edge node [above]      {}   (3);
	
	\node(5) [line width = 0.5] at (2,-0.25)[shape=circle,draw,inner sep=2pt]{};
	\node(6) [line width = 0.5] at (2.5,-0.25)[shape=circle,draw,inner sep=2pt]{$ $};
	\node(7) [line width = 0.5] at (2,-0.75)[shape=circle,draw,inner sep=2pt]{$ $};
	\node(8) [line width = 0.5] at (2.5,-0.75)[shape=circle,draw,inner sep=2pt]{$ $};
	
	\path [-] [line width = 0.5]
	(5) edge node [above]      {}   (6)
	(5) edge node [above]      {}   (7)
	(5) edge node [above]      {}   (7)
	(7) edge node [above]      {}   (8)
	(6) edge node [above]      {}   (8)
	(6) edge node [above]      {}   (7);
	\node at (1.5,-1.1,1) {(b)};
	\end{tikzpicture}
	\caption{Illustration  of  bearing  rigidity.  The  networks  in  (a)  are  not  bearing rigid because the same inter-neighbor bearings may lead to different geometric patterns  of  the  networks,  for  example,  a  square  on  the  left  and  a  rectangle on  the  right.  The  networks  in  (b)  are  bearing  rigid  because  the  same  inter-neighbor bearings imply the same geometric pattern though the networks may differ in terms of translation and scale.}
	\label{fig:bearing rig example (rigid+coopmanip)}
\end{figure}

\subsection{Distance and Bearing Rigidity in $\mathbb{SE}$(3)}\label{subsec:d+b rigidity (rigid+coopmanip)}

	\textcolor{black}{
		We begin be recalling that the range space of the grasp matrix $G^T$ corresponds to the rigid body translations and rotations of the system.  While this matrix appears naturally in the context of dynamic modeling of rigid bodies, it is also indirectly related to the notion of structural rigidity in discrete geometry, which is a combinatorial theory for determining the flexibility of ensembles formed by rigid bodies connected by flexible linkages or hinges.}
	
	\textcolor{black}{	
		In the classical structural rigidity theory, one considers a collection of rigid bars connected by joints allowing free rotations around the joint axis - this is known as a \emph{bar-and-joint} framework.  One is then interested in understanding what are the allowable motions of the framework, i.e., those motions that preserve the lengths of the bars and their connections to the joints.  The so-called \emph{trivial motions} for these frameworks are precisely the rigid body translations and rotations of the system.  For some frameworks, there may be additional motions, known as \emph{flexes}, that also preserve the constraints.  This is captured by the notion of \emph{infinitesimal motions} of the framework and is characterized by the \emph{rigidity matrix} of the framework \cite{asimow1978rigidity}. 	
	}
	
	\textcolor{black}{Here we can consider frameworks that also encode the pose of the joints in addition to the lengths of the rigid bars connecting them, leading to a \emph{distance and bearing}-type framework.   Bearing rigidity has been recently explored in the context of formation control and studies the  problem  of  under what  conditions  the  geometric  pattern  of  a  network  can  be uniquely determined if the bearing of each edge in the network is fixed \cite{zhao2019bearingRig} (see Fig. \ref{fig:bearing rig example (rigid+coopmanip)}).  The bearing rigidity has also been extended to frameworks embedded in $\mathbb{SE}(2)$ and $\mathbb{SE}(3)$\cite{bearingSE2,bearingSE3}.  Both the bearing and distance rigidity theories have found many applications for multi-agent systems, in particular for formation control and localization \cite{zhao2019bearingRig, Ahn2020, eren2004, zhao2016localizability}.
	}	
	
	\textcolor{black}{In this chapter we introduce and formalize the concept of distance and bearing rigidity (abbreviated as D\&B Rigidity in the following).  This is motivated by the fusion of both distance and pose constraints in the cooperative grasping problem.  D\&B Rigidity in $\mathbb{SE}(3)$ aims at studying the problem of under what conditions the geometric pattern of a network can be uniquely determined if \textit{both} the bearing and the distance of each edge in the network is fixed.
	} 
	\textcolor{black}{In this direction, we focus on the notion of infinitesimal rigidity for D\&B frameworks. We first formally define a D\&B framework in $\mathbb{SE}(3)$, similarly to Section \ref{sec:Rigidity theory (App_Rigidity)} of Appendix \ref{app:Rigidity}:}
	\begin{definition} 
		A framework in $\mathbb{SE}(3)$ is a triple $(\mathcal{G},p_\mathcal{G},R_\mathcal{G})$, where $\mathcal{G}\coloneqq (\mathcal{N},\mathcal{E})$ is a graph, 
		$p_\mathcal{G}: \mathcal{N} \to \mathbb{R}^3$ is a function mapping each node to a position in $\mathbb{R}^3$, and $R_\mathcal{G} : \mathcal{N} \to \mathbb{SO}(3)$ is a function associating each node with an orientation element of $\mathbb{SO}(3)$ (both with respect to an inertial frame).
	\end{definition}
	
	As in the previous section concerning formation control, we employ the Special Orthogonal Group (rotation matrices) $\mathbb{SO}(3)$ to express the orientation of the agents. Moreover, we use the shorthand notation $p_i \coloneqq p_\mathcal{G}(i)$, $R_i \coloneqq R_\mathcal{G}(i)$, $p \coloneqq [p_1^\top,\dots,p_N^\top]^\top \in\mathbb{R}^{3N}$, $R \coloneqq (R_1,\dots,R_N)\in\mathbb{SO}(3)^N$, $x_i \coloneqq (p_i,R_i)\in\mathbb{SE}(3)$, and $x \coloneqq (x_1,\dots,x_N) \in \mathbb{SE}(3)^N$.
	{The distances and bearings in a framework can be summarized through the following \textit{SE(3) D\&B rigidity function}, $\gamma_\mathcal{G}$, that encodes the rigidity constraints in the framework. 
		Consider a directed graph $\mathcal{G}$, where $\mathcal{E} \subseteq \{ (i,j) \in \mathcal{N}^2 : i\neq j \}$, as well as its \textit{undirected part} $\mathcal{E} \supseteq \mathcal{E}_u \coloneqq \{(i,j)\in\mathcal{E} : i<j \}$. Then $\gamma_\mathcal{G}$ can be formed by considering the distance and bearing functions $\gamma_{e,d}:\mathbb{R}^3\times\mathbb{R}^3\to\mathbb{R}_{\geq 0}$, $\gamma_{e,b}:\mathbb{SE}(3)^2\to\mathbb{S}^2$, with 
		\begin{subequations} \label{eq:gamma 1 (rigid+coopmanip)}
			\begin{align}
			&\gamma_{e,d}(p_i,p_j) \coloneqq \frac{1}{2}\|p_i - p_j\|^2, \forall e=(i,j)\in\mathcal{E}_u, \\
			&\gamma_{e,b}(x_i,x_j) \coloneqq						
			R_i^\top \frac{p_j - p_i}{\|p_i-p_j\|},	
			\forall e=(i,j)\in\mathcal{E},
			\end{align}
		\end{subequations}
		which encodes the distance $\|p_i-p_j\|$ between two agents as well as the local bearing vector $R_i^\top \frac{p_j - p_i}{\|p_i-p_j\|}$, expressed in the frame of agent $i$. Note that the distance functions  are considered only for the undirected part of $\mathcal{G}$, since $\gamma_{(i,j),d} = \gamma_{(j,i),d}$. Now $\gamma_\mathcal{G}$ is formed by stacking the aforementioned distance and bearing functions, i.e., $\gamma_\mathcal{G}\coloneqq\gamma_\mathcal{G}(x): \mathbb{SE}(3)^N \to \mathbb{R}^{|\mathcal{E}_u|}\times\mathbb{S}^{2|\mathcal{E}|}$, with 
		\begin{equation} \label{eq:gamma G (rigid+coopmanip)}
		\gamma_\mathcal{G}	\coloneqq \begin{bmatrix} \gamma_d(p) \\ \gamma_b(x) \end{bmatrix} \coloneqq \begin{bmatrix}
		\gamma_{1,d} \\
		\vdots \\
		\gamma_{|\mathcal{E}_u|,d} \\
		\gamma_{1,b} \\ 
		\vdots\\
		\gamma_{|\mathcal{E}|,b}
		\end{bmatrix}.
		\end{equation} }	
	Note that the aforementioned expressions for $\gamma_{e,d}$, $\gamma_{e,b}$ are not unique and other choices that capture the rigidity constraints can also be made.  We also mention our slight abuse of notation, where the index $k$ in $\gamma_{k,d}$ and $\gamma_{k,b}$ refers to a labeled edge in $\mathcal{E}_u$ and $\mathcal{E}_b$.  
	
	{In this section, we are interested in the set of D\&B \textit{infinitesimal} motions of a framework in $\mathbb{SE}(3)$. \textcolor{black}{These can be thought as perturbations to a framework in $\mathbb{SE}(3)$ that leave $\gamma_\mathcal{G}$ unchanged. More information about the separate distance and bearing infinitesimal motions can be found in Section \ref{sec:Rigidity theory (App_Rigidity)} of Appendix \ref{app:Rigidity}.} The set of D\&B \textit{infinitesimal} motions is characterized by the nullspace of the Jacobian of the $\mathbb{SE}(3)$-D\&B rigidity function \textcolor{black}{arising from the Taylor series expansion of $\gamma_\mathcal{G}$.}  That is, the nullspace of the matrix $\nabla_{(p,R)}\gamma_\mathcal{G}$, that we term the $\mathbb{SE}(3)$-D\&B \emph{rigidity matrix}. This matrix is denoted as $\mathcal{R}_\mathcal{G}:\mathbb{SE}(3)^N \to \mathbb{R}^{(|\mathcal{E}_u|+3|\mathcal{E}|) \times 6N} \coloneqq \nabla_{(p,R)}\gamma_\mathcal{G}$, i.e., 	
		\begin{equation} \label{eq:rigidity matrix (rigid+coopmanip)}
		\mathcal{R}_\mathcal{G}(x)  = \begin{bmatrix}
		\frac{\partial \gamma_{1,d}}{\partial p_1} & \frac{\partial \gamma_{1,d}}{\partial R_1} & \dots & \frac{\partial \gamma_{1,d}}{\partial p_N} & \frac{\partial \gamma_{1,d}}{\partial R_N} \\
		\vdots & & \ddots & & \vdots \\ 
		\frac{\partial \gamma_{|\mathcal{E}_u|,d}}{\partial p_1} & \frac{\partial \gamma_{|\mathcal{E}_u|,d} }{\partial R_1} & \dots & \frac{\partial \gamma_{|\mathcal{E}_u|,d} }{\partial p_N} & \frac{\partial \gamma_{|\mathcal{E}_u|,d} }{\partial R_N} \\
		\frac{\partial \gamma_{1,b}}{\partial p_1} & \frac{\partial \gamma_{1,b}}{\partial R_1} & \dots & \frac{\partial \gamma_{1,b}}{\partial p_N} & \frac{\partial \gamma_{1,b}}{\partial R_N} \\
		\vdots & & \ddots & & \vdots \\ 
		\frac{\partial \gamma_{|\mathcal{E}|,b}}{\partial p_1} & \frac{\partial \gamma_{|\mathcal{E}|,b} }{\partial R_1} & \dots & \frac{\partial \gamma_{|\mathcal{E}|,b} }{\partial p_N} & \frac{\partial \gamma_{|\mathcal{E}|,b} }{\partial R_N}
		\end{bmatrix},
		\end{equation}	
		with
		\begin{align*}
		\frac{\partial \gamma_{e,d}}{\partial x_i} =& \begin{bmatrix}
		\frac{\partial \gamma_{e,d}}{\partial p_i}	& \frac{\partial \gamma_{e,d}}{\partial R_i}
		\end{bmatrix} = \displaystyle
		\begin{bmatrix}
		(p_i - p_j)^\top & 0_{1 \times 3} 
		\end{bmatrix}, \\    
		\frac{\partial \gamma_{e,d}}{\partial x_j} =& \begin{bmatrix}
		\frac{\partial \gamma_{e,d}}{\partial p_j}	& \frac{\partial \gamma_{e,d}}{\partial R_j}
		\end{bmatrix} = \displaystyle
		\begin{bmatrix}
		(p_j - p_i)^\top & 0_{1 \times 3} 
		\end{bmatrix}, 
		\\
		\frac{\partial \gamma_{e,b}}{\partial x_i} =& \begin{bmatrix}
		\frac{\partial \gamma_{e,b}}{\partial p_i}	& \frac{\partial \gamma_{e,b}}{\partial R_i}
		\end{bmatrix} = \displaystyle
		\begin{bmatrix}
		-\frac{P_r(\gamma_{e,b})}{\|p_j-p_i\|}R_i^\top &  S(\gamma_{e,b})R_i^\top
		\end{bmatrix}, \\ 
		\frac{\partial \gamma_{e,b}}{\partial x_j} =& \begin{bmatrix}
		\frac{\partial \gamma_{e,b}}{\partial p_j}	& \frac{\partial \gamma_{e,b}}{\partial R_j}
		\end{bmatrix} =
		\begin{bmatrix}
		\frac{P_r(\gamma_{e,b})}{\|p_j-p_i\|}R_i^\top & 0_{3\times 3}
		\end{bmatrix}. 
		\end{align*}
		{Here, $P_r(\gamma_{e,b})$ is defined as
			$$P_r(\gamma_{e,b}) \coloneqq I_3 - \frac{(p_j-p_i)(p_j-p_i)^T}{\|p_j-p_i\|^2} ,$$
			and projects vectors onto the orthogonal complement of $(p_j-p_i)$}.  See \cite{zhao2019bearingRig} for more discussion on this projection matrix and its use in the bearing rigidity theory.
		Infinitesimal motions, therefore, are motions $x(t)$ produced by velocities $v(t)$ that lie in the nullspace of $\mathcal{R}_\mathcal{G}$, for which it holds that $\dot{\gamma}_\mathcal{G} = \mathcal{R}_\mathcal{G}(x(t)) v(t) = 0$, where $v\coloneqq [\dot{p}_1^\top,\omega_1^\top,\dots,\dot{p}_N^\top,\omega_N^\top]^\top$, as defined before. \textcolor{black}{The infinitesimal motions therefore depend on the number of motion degrees of freedom the entire framework possesses.  This directly relates to the structure of the underlying graph.  }
		Motions that preserve the distances and bearings of the framework for \emph{any} underlying graph are called $D\& B$ \textit{trivial motions}. This leads to the definition of \emph{infinitesimal rigidity}, stated below.
		\begin{definition} \label{def:inf rig (rigid+coopmanip)}
			A framework $(\mathcal{G}, p_\mathcal{G}, R_\mathcal{G})$ is D\&B infinitesimally rigid in $\mathbb{SE}(3)$ if every D\&B infinitesimal motion is a D\&B trivial motion. 
		\end{definition}
	}

	\textcolor{black}{
		We now aim to identify precisely what the trivial motions of a D\&B framework are, and to determine conditions for a framework to be infinitesimally rigid based on properties of the D\&B rigidity matrix.  Before we proceed, we first note that the D\&B rigidity function in $\mathbb{SE}(3)$ can be seen as a superposition of the rigidity functions associated with the {classic} distance rigidity theory \cite{asimow1978rigidity} and the $\mathbb{SE}(3)$ \textcolor{black}{bearing} rigidity theory \cite{bearingSE3}.  In particular, we note that $\mathcal{R}_{\mathcal{G},d}: \mathbb{R}^{3N} \to \mathbb{R}^{|\mathcal{E}_u|\times 3N}\coloneqq \nabla_p \gamma_{d}$ is the well-studied (distance) rigidity matrix, while $\mathcal{R}_{\mathcal{G},b}:\mathbb{SE}^{3N}\to\mathbb{R}^{3\mathcal{E}\times6N} \coloneqq \nabla_{(p,R)}\gamma_{\mathcal{G},b}$ is the $\mathbb{SE}(3)$ \textcolor{black}{bearing} rigidity matrix.  Note that the distance rigidity matrix is associated with the framework $(\mathcal{G},p_\mathcal{G})$, which is the projection of $(\mathcal{G},p_\mathcal{G},R_\mathcal{G})$ {to} $\mathbb{R}^3$. With an appropriate permutation, $P_R$, of the columns of $\mathcal{R}_{\mathcal{G}}$, we have that 
		\begin{align} \label{eq:R_G tilde (rigid+coopmanip)}
		\widetilde{\mathcal{R}}_\mathcal{G} \coloneqq &  \mathcal{R}_\mathcal{G} P_R \notag \\
		=& \begin{bmatrix}
		\frac{\partial \gamma_{1,d}}{\partial p_1} & \dots & \frac{\partial \gamma_{1,d}}{\partial p_N} & \frac{\partial \gamma_{1,d}}{\partial R_1} & \dots & \frac{\partial \gamma_{1,d}}{\partial R_N} \\	
		\vdots & & \ddots & & \vdots \\ 
		\frac{\partial \gamma_{M_\mathcal{G},d}}{\partial p_1} & \dots & \frac{\partial \gamma_{M_\mathcal{G},d}}{\partial p_N} & \frac{\partial \gamma_{M_\mathcal{G},d}}{\partial R_1} & \dots & \frac{\partial \gamma_{M_\mathcal{G},d}}{\partial R_N} \\    
		\frac{\partial \gamma_{1,b}}{\partial p_1} & \dots & \frac{\partial \gamma_{1,b}}{\partial p_N} & \frac{\partial \gamma_{1,b}}{\partial R_1} & \dots & \frac{\partial \gamma_{1,b}}{\partial R_N} \\
		\vdots & & \ddots & & \vdots \\ 
		\frac{\partial \gamma_{M_\mathcal{G},b}}{\partial p_1} & \dots & \frac{\partial \gamma_{M_\mathcal{G},b}}{\partial p_N} & \frac{\partial \gamma_{M_\mathcal{G},b}}{\partial R_1} & \dots & \frac{\partial \gamma_{M_\mathcal{G},b}}{\partial R_N} 
		\end{bmatrix},
		\end{align}
		which is equal to 
		\begin{equation*}
		\widetilde{\mathcal{R}}_\mathcal{G} = \begin{bmatrix} \begin{bmatrix}\mathcal{R}_{\mathcal{G},d} &  0_{|\mathcal{E}_u|\times 3N} \end{bmatrix} \\	
		\mathcal{R}_{\mathcal{G},b} 
		\end{bmatrix} =: \begin{bmatrix} \bar{\mathcal{R}}_{\mathcal{G},d} \\	
		\mathcal{R}_{\mathcal{G},b} 
		\end{bmatrix}.
		\end{equation*}
		The nullspace of $\widetilde{\mathcal{R}}_\mathcal{G}$, therefore, is the intersection of the nullspaces of $\bar{\mathcal{R}}_{\mathcal{G},d}$ and $\mathcal{R}_{\mathcal{G},b}$.
	}
	
	\textcolor{black}{With the above interpretation, we can now understand the trivial motions to be the intersection of trivial motions associated to distance rigidity with those associated to $\mathbb{SE}(3)$ bearing rigidity.  In particular, let 
		$$\mathcal{S}_d \coloneqq \mathrm{span} \left\{ \mathbbm{1}_N \otimes I_3, \mathcal{L}^{\circlearrowright}_{\mathbb{R}^3}(\mathcal{G})\right\},$$
		denote the trivial motions associated to a distance framework {\cite{asimow1978rigidity}}.  That is, $\mathbbm{1}_N \otimes I_3$ represents translations of the entire framework, and $\mathcal{L}^{\circlearrowright}_{\mathbb{R}^3}(\mathcal{G})$ is the rotational subspace induced by the graph $\mathcal{G}$ in $\mathbb{R}^3$, i.e., 
		$$\mathcal{L}^{\circlearrowright}_{\mathbb{R}^3}(\mathcal{G}) = \mathrm{span}\left\{ 
		\left(  I_3 \otimes S(\mathbf{e}_{h})\right)p_\mathcal{G}, h=1,2,3\right\}.$$
		These motions can be produced by {the} \emph{linear} velocities of the agents.  It is known that $\mathcal{S}_d \subseteq \mathrm{null}(\mathcal{R}_{\mathcal{G},d})$ for any underlying graph $\mathcal{G}$ \cite{asimow1978rigidity}. For the matrix $\bar{\mathcal{R}}_{\mathcal{G},d}$, we can define the corresponding set
		$$\bar{S}_d \coloneqq \mathrm{span}\left\{\begin{bmatrix}
		\mathbbm{1}_N \otimes I_3 \\ \star 
		\end{bmatrix}, \begin{bmatrix}
		\mathcal{L}^{\circlearrowright}_{\mathbb{R}^3}(\mathcal{G}) \\ \star 
		\end{bmatrix}\right\}\subseteq \mathrm{null}(\bar{\mathcal{R}}_{\mathcal{G},d}). $$
		Note that the distance rigidity does not explicitly depend on the orientation of the nodes when expressed as a point in $\mathbb{SE}(3)$.  This accounts for the free $\star$ entry in the subspace $\bar{S}_d$ corresponding to the rotations.}  Thus, the set of trivial motions in $\mathbb{R}^3$ can be seen as the projection of $\bar{S}_d$ in $\mathbb{R}^3$.

	Similarly, for an $\mathbb{SE}(3)$ \textcolor{black}{bearing} framework one can define the subspace \cite{bearingSE3}
		$$\mathcal{S}_b \coloneqq \mathrm{span}\left\{ \begin{bmatrix}
		\mathbbm{1}_N \otimes I_3 \\ 0_{3N \times 3}, 
		\end{bmatrix}, \begin{bmatrix}
		p_\mathcal{G} \\ 0_{3N}, 
		\end{bmatrix}, \mathcal{L}^{\circlearrowright}_{\mathbb{SE}(3)}(\mathcal{G})\right\},$$
		where the vector $[p_\mathcal{G}^T, 0_{3N}^T]^T$ represents a scaling of the framework. The space $\mathcal{L}^{\circlearrowright}_{\mathbb{SE}(3)}(\mathcal{G})$ is the rotational subspace induced by $\mathcal{G}$, in $\mathbb{SE}(3)$,
		\begin{equation} \label{eq:rot subspace SE3 (rigid+coopmanip)}
		\mathcal{L}^{\circlearrowright}_{\mathbb{SE}(3)}(\mathcal{G})=\mathsf{span}{\left\{ \begin{bmatrix}			
				\left(  I_3 \otimes S(\mathbf{e}_{h}) \right) p_{\mathcal{G}} \\
				\mathbbm{1}_n\otimes \mathbf{e}_h \\
		\end{bmatrix} , {h=1,2,3} \right\} }. 
		\end{equation}
		It is also known that $\mathcal{S}_b \subseteq \mathrm{null}(\mathcal{R}_{\mathcal{G},b})$. Thus $\mathcal{S}_b$ describes the trivial motions of an $\mathbb{SE}(3)$ \textcolor{black}{bearing} framework \cite{bearingSE3}.

	\textcolor{black}{
		The above discussion immediately leads to the following proposition.
		\begin{proposition}\label{prop.trivialmotions (rigid+coopmanip)}
			The trivial motions of a D\&B framework are characterized by the set
			$$\mathcal{S}_{db} \coloneqq \bar{\mathcal{S}}_d \cap \mathcal{S}_b = \mathrm{span}\left\{ \begin{bmatrix} \mathbbm{1}_N \otimes I_3\\0_{3N\times 3} \end{bmatrix} , \mathcal{L}^{\circlearrowright}_{\mathbb{SE}(3)}(\mathcal{G}) \right\}.$$
			Furthermore, it follows that $\mathcal{S}_{db} \subseteq \mathrm{null}({\widetilde{\mathcal{R}}_{\mathcal{G}}})$.
		\end{proposition}
	}
	\textcolor{black}{Having characterized the trivial motions, it now follows from Definition \ref{def:inf rig (rigid+coopmanip)} that for infinitesimal rigidity, we require that $\mathrm{null}({\widetilde{\mathcal{R}}_{\mathcal{G}}}) = \mathcal{S}_{db}$.  This is summarized in the following proposition. 
		\begin{proposition} \label{prop:ranks (rigid+coopmanip)}
			The framework $(\mathcal{G},p_\mathcal{G},R_\mathcal{G})$ is D\&B infinitesimally rigid in $\mathbb{SE}(3)$ if and only if
			\begin{align*}
			\textup{null}(\widetilde{\mathcal{R}}_\mathcal{G}) &=   \textup{null}(\bar{\mathcal{R}}_{\mathcal{G},d})\cap\textup{null}(\mathcal{R}_{\mathcal{G},b}) \notag \\ 
			&= \textup{span}\left\{ \begin{bmatrix}
			\mathbbm{1}_N \otimes I_3 \\
			0_{3N\times 3}, 
			\end{bmatrix}, \mathcal{L}^{\circlearrowright}_{\mathbb{SE}(3)}(\mathcal{G}) \right\}=\mathcal{S}_{db}.
			\end{align*}
			Equivalently, the D\&B framework is infinitesimally rigid in $\mathbb{SE}(3)$ if and only if
			\begin{align*}
			\textup{rank}(\widetilde{\mathcal{R}}_\mathcal{G}) &= \textup{dim}(\widetilde{\mathcal{R}}_\mathcal{G}) - \textup{dim}(\textup{null}(\widetilde{\mathcal{R}}_\mathcal{G})) = 6N - 6.
			\end{align*}
		\end{proposition}
	}
	{Hence, all the motions produced by the nullspace of $\widetilde{\mathcal{R}}_\mathcal{G}$ for an infinitesimally rigid framework must correspond to trivial motions, i.e., coordinated translations and rotations.} 
	{Moreover, given \eqref{eq:R_G tilde (rigid+coopmanip)}, it follows that $(\mathcal{G},p_\mathcal{G},R_\mathcal{G})$ is D\&B infinitesimally rigid in $\mathbb{SE}(3)$ if and only if
		\begin{align} \label{eq:R_G nullspace (rigid+coopmanip)}
		\textup{null}(\mathcal{R}_\mathcal{G}) = \{x=P_Ry \in \mathbb{SE}(3)^N : y \in \textup{null}(\widetilde{\mathcal{R}}_\mathcal{G}) \},
		\end{align}
		i.e., {the nullspace of $\mathcal{R}_\mathcal{G}$ consists of} the vectors of $\textup{null}(\widetilde{\mathcal{R}}_\mathcal{G})$ whose elements are permutated by $P_R$. }
	
	{It is worth noting that the aforementioned results are not valid if the rigidity matrix loses rank, i.e., $\textup{rank}(\mathcal{R}_\mathcal{G}) < \max\{\textup{rank}(\mathcal{R}_\mathcal{G}(x)), x\in\mathbb{SE}(3)\}$. These are degenerate cases that correspond, for example, to when all agents are aligned along a direction $\mathbf{v} \in \mathbb{S}^2$.  For more discussion on these degenerate cases, the reader is referred to \cite{Pasquetti2019}.} 
	
	{As a last remark, we observe that frameworks over the complete graph, $(\mathcal{K}_N, p_{\mathcal{K}_N},R_{\mathcal{K}_N})$, are (except for the degenerate configurations), infinitesimally rigid.  That is, $\textup{rank}(\widetilde{\mathcal{R}}_{\mathcal{K}_N}) =6N - 6$. This leads to the following corollary.
		\begin{corollary} \label{cor:R_g_tilde null}
			Consider the D\&B frameworks $(\mathcal{G},p_\mathcal{G},R_\mathcal{G})$ and $(\mathcal{K}_N,p_\mathcal{G},R_\mathcal{G})$ for nondegenrate configurations $(p_\mathcal{G},R_{\mathcal{G}})$.  Then $(\mathcal{G},p_\mathcal{G},R_\mathcal{G})$ is D\&B infinitesimally rigid if and only if
			$$ \textup{rank}(\widetilde{\mathcal{R}}_{\mathcal{G}})= \textup{rank}(\widetilde{\mathcal{R}}_{\mathcal{K}_N}) = 6N-6.$$
		\end{corollary}
	}
	
	{In the next section, we use the aforementioned results to link the D\&B rigidity matrix of a complete graph to the forces $h_\text{m}$ and $h_\text{int}$ of \eqref{eq:interaction forces (rigid+coopmanip)}}. 

	\subsection{Interaction Forces based on the D\&B Rigidity Matrix and Internal-Force based Optimal Cooperative Manipulation} \label{sec:Main results (rigid+coopmanip)}
	
	We provide here the main results of this section. Firstly, we give a closed form expression for the interaction and internal forces of the coupled system object-robots. Next, we connect these forces with the D\&B rigidity matrix introduced before. After that, we use these results to provide a novel relation between  the arising interaction and internal forces and we give conditions on the agent force distribution for cooperative manipulation free from internal forces.
	For the rest of the chapter, we use the following notation for the cooperative object-manipulation system:
	\begin{center}
	\begin{tabular}{ l l}
	$\widetilde{x} \coloneqq [\dot{q}^\top, v_{\scr O}^\top]^\top$ &
	$\bar{B}\coloneqq\bar{B}(\bar{x}) \coloneqq \textup{diag}\{ B(q), M_{\scr O}(\eta_{\scr O})  \}$ \\ 
	$\bar{\tau} \coloneqq [\tau^\top, 0_{1\times 6}]^\top$ & 
	$\bar{C}_q\coloneqq\bar{C}_q(\bar{x}) \coloneqq \textup{diag}\{ C_q(q,\dot{q}), C_{\scr O}(\eta_{\scr O},{\omega}_{\scr O})\}$ \\
	$\bar{g}_q \coloneqq \bar{g}_q(\bar{x})\coloneqq [g_q(q)^\top, g_{\scr O}^\top]^\top$ & $\bar{M}\coloneqq \bar{M}(\bar{x}) \coloneqq \textup{diag}\{M(q),M_{\scr O}(\eta_{\scr O})\}$ \\
	$\bar{v} \coloneqq [v^\top, v_{\scr O}^\top]^\top$ & 
	$\bar{C}\coloneqq\bar{C}(\bar{x}) \coloneqq \textup{diag}\{ C(q,\dot{q}), C_{\scr O}(\eta_{\scr O},\omega_{\scr O}) \} $ \\
	$\bar{g}\coloneqq\bar{g}(\bar{x}) \coloneqq [g(q)^\top, g_{\scr O}^\top]^\top$ & $\bar{u} \coloneqq [u^\top, 0_{1\times 6}]^\top$\\
	$\bar{J}\coloneqq\bar{J}(q) \coloneqq \textup{diag}\{J(q),I_6\}$
	\end{tabular}
	\end{center}

	\subsubsection{Interaction Forces Based on the D\&B Rigidity Matrix}
	
	{We provide here closed form expressions for the interaction forces of the coupled object-agents system and link them to the D\&B rigidity matrix notion introduced in the previous section. In particular, we consider that the robotic agents and the object form a graph {that will be defined in the sequel}. Note that, due to the rigidity of the grasping points, the forces exerted by an agent influence, not only the object, but all the other agents as well. Hence, since there exists interaction among all the pairs of agents as well as the agents and the object, 
		we model their connection as a complete graph, as described rigorously below. Moreover, as will be clarified later, the rigidity matrix of this graph encodes the constraints of the agents-object system, imposed by the rigidity of the grasping points, and plays an important role in the expression of the agents-object interaction forces. }
	
	{Let the robotic agents form a  framework $(\mathcal{G},p_\mathcal{G},R_\mathcal{G})$ in $\mathbb{SE}(3)$, where $\mathcal{G}\coloneqq (\mathcal{N},\mathcal{E})$ is the complete graph, i.e., $\mathcal{E} = \{(i,j) \in \mathcal{N}^2 : i \neq j \}$, and $p_\mathcal{G} \coloneqq [p_1^\top,\dots,p_N^\top]^\top$, $R_\mathcal{G} \coloneqq (R_1,\dots,R_N)$. Consider also the undirected part $\mathcal{E}_u = \{(i,j)\in\mathcal{E} : i<j\}$ of $\mathcal{E}$, as also described in the previous section. Since the graph is complete, we conclude that $|\mathcal{E}| = N(N-1)$ and $|\mathcal{E}_u| = \frac{N(N-1)}{2}$. Moreover, consider the extended framework $(\bar{\mathcal{G}},p_{\bar{\mathcal{G}}}, R_{\bar{\mathcal{G}}})$ of the robotic agents and the object, i.e., where the object is considered as the $(N+1)$th agent; $\bar{\mathcal{G}}$ is the complete graph $\bar{\mathcal{G}} \coloneqq (\mathcal{\bar{N}},\bar{\mathcal{E}})$, where $\mathcal{\bar{N}} \coloneqq \{1,\dots,\bar{N}\}$, $\bar{N} \coloneqq N+1$, and $\bar{\mathcal{E}} \coloneqq \{ (i,j) \in \bar{\mathcal{N}}^2 : i \neq j\}$, with $|\bar{\mathcal{E}}| = \bar{N}(\bar{N}-1)$. Let also $\bar{\mathcal{E}}_u\coloneqq \{(i,j)\in\bar{\mathcal{N}}: i < j\}$ be the undirected edge part, with $|\bar{\mathcal{E}}_u| = \frac{\bar{N}(\bar{N}-1)}{2}$.  }
	
	{
		{Consider} now the rigidity functions
		$\gamma_{e,d}:\mathbb{R}^3\times\mathbb{R}^3\to\mathbb{R}_{\geq 0}$, $\forall e \in \bar{\mathcal{E}}_u$ and $\gamma_{e,b}:\mathbb{SE}(3)^2\to\mathbb{S}^2$, $\forall e \in \bar{\mathcal{E}}$, as given in \eqref{eq:gamma 1 (rigid+coopmanip)}, as well as the stack vector $\gamma_{\bar{\mathcal{G}}}: \mathbb{SE}(3)^{\bar{N}} \to \mathbb{R}^{\frac{\bar{N}(\bar{N}-1)}{2}}\times\mathbb{S}^{2\bar{N}(\bar{N}-1)}$ as given in \eqref{eq:gamma G (rigid+coopmanip)}.	 
		The rigidity constraints of the framework are encoded in the constraint $\gamma_{\bar{\mathcal{G}}} = \text{const.}$. Since the rigidity of the framework stems from the rigidity of the grasping points, these constraints encode also the rigidity constraints of the object-agent cooperative manipulation.
		{By differentiating $\gamma_{\bar{\mathcal{G}}}=\text{const.}$, one obtains }
		\begin{align*}
		&	\mathcal{R}_{\bar{\mathcal{G}}} \bar{v} = 0 \Leftrightarrow \mathcal{R}_{\bar{\mathcal{G}}} \bar{J}
		\widetilde{x} = 0 \Rightarrow 	
		\mathcal{R}_{\bar{\mathcal{G}}} \bar{J}
		\dot{\widetilde{x}}  = 
		- \big( \dot{\mathcal{R}}_{\bar{\mathcal{G}}}\bar{J} + \mathcal{R}_{\bar{\mathcal{G}}}\dot{\bar{J}} \big)
		\widetilde{x},
		\end{align*}
		where $\mathcal{R}_{\bar{\mathcal{G}}}\coloneqq \mathcal{R}_{\bar{\mathcal{G}}}(x, x_{\scr O} ):\mathbb{SE}(3)^{\bar{N}}\to \mathbb{R}^{\frac{7\bar{N}(\bar{N}-1)}{2}\times(6\bar{N})}$ is the rigidity matrix associated to $\bar{\mathcal{G}}$ and has the form  \eqref{eq:rigidity matrix (rigid+coopmanip)}. We {now} write the aforementioned equations as	
		\begin{align*}		
		&A
		\dot{\widetilde{x}}  = b,			
		\end{align*}
		where		
		\begin{subequations}     \label{eq:A_total, b_total (rigid+coopmanip)}
			\begin{align}
			&A\coloneqq  A(\bar{x},t) \coloneqq \mathcal{R}_{\bar{\mathcal{G}}}(x,x_{\scr O})\bar{J}(q) \\
			&b\coloneqq b(\bar{x},t) \coloneqq - \big( \dot{\mathcal{R}}_{\bar{\mathcal{G}}}(x,x_{\scr O})\bar{J}(q) + \mathcal{R}_{\bar{\mathcal{G}}}(x,x_{\scr O})\dot{\bar{J}}(q) \big)
			\widetilde{x}.
			\end{align}
		\end{subequations}    
		One can verify that the motion of the cooperative object-agents manipulation system that is enforced by the aforementioned constraints corresponds to rigid body motions (coordinated translations and rotations of the system). Hence, since $\bar{\mathcal{G}}$ is complete, the analysis of the previous section dictates 
		that these motions are the infinitesimal motions of the framework and are the ones produced by the nullspace of  {$\mathcal{R}_{\bar{\mathcal{G}}}(x, x_{\scr O} )$}}.
	
	{Next, we turn to the main focus of our results, which is the case of internal forces and we consider the framework comprising only of the robotic agents $(\mathcal{G},p_\mathcal{G},R_\mathcal{G})$.	
		The inter-agent rigidity constraints are expressed by the D\&B rigidity functions
		$\gamma_{e,d}:\mathbb{R}^3\times\mathbb{R}^3\to\mathbb{R}_{\geq 0}$, $\forall e \in \mathcal{E}_u$ and $\gamma_{e,b}:\mathbb{SE}(3)^2\to\mathbb{S}^2$, $\forall e \in \mathcal{E}$, as given in \eqref{eq:gamma 1 (rigid+coopmanip)}, as well as the stack vector $\gamma_{\mathcal{G}}: \mathbb{SE}(3)^{N} \to \mathbb{R}^{\frac{N(N-1)}{2}}\times\mathbb{S}^{2N(N-1)}$ as given in \eqref{eq:gamma G (rigid+coopmanip)}. Differentiation of $\gamma_\mathcal{G}(x(q)) = \text{const.}$, which encodes the rigidity constraints of the system comprised by the robotic agents,
		yields    
		\begin{align*}
		&  	\mathcal{R}_\mathcal{G} v = 0 \Leftrightarrow \mathcal{R}_\mathcal{G} J\dot{q} = 0  \Rightarrow  \mathcal{R}_\mathcal{G} J\ddot{q} = -\big( \mathcal{R}_\mathcal{G}\dot{J} + \dot{\mathcal{R}}_\mathcal{G}J\big) \dot{q},
		\end{align*}
		written more compactly as 
		\begin{align*}
		A_\textup{int}\ddot{q} =  b_\textup{int},
		\end{align*}
		where
		\begin{subequations}     \label{eq:A_int, b_int (rigid+coopmanip)}
			\begin{align}   
			&  A_\textup{int}\coloneqq  A_\textup{int}(q,\dot{q},t) \coloneqq \mathcal{R}_\mathcal{G}(x(q)) J(q), \label{eq:A_int} \\   & b_\textup{int}\coloneqq b_\textup{int}(q,\dot{q},t) \coloneqq -\big( \mathcal{R}_\mathcal{G}(x(q))\dot{J}(q) + \dot{\mathcal{R}}_\mathcal{G}(x(q))J(q)\big) \dot{q}.       
			\end{align}\label{eq:b_int}
		\end{subequations}		
		Similarly to the case of $\bar{\mathcal{G}}$, we conclude that the agent motions produced by the aforementioned constraints correspond to rigid body motions, which are the infinitesimal motions produced by the nullspace of $\mathcal{R}_\mathcal{G}$.}

	{After giving the rigidity constraints in the cooperative manipulation system, we are now ready to derive the expressions for the interaction forces, $h$, in terms of the aforementioned rigidity matrices. We follow the same methodology as in \cite{erhart2015internal}. Consider first \eqref{eq:manipulator dynamics_joint space vector_form (rigid+coopmanip)} and \eqref{eq:object dynamics 2 (rigid+coopmanip)} written in vector form as 
		\begin{equation*} 
		\bar{B}\dot{\widetilde{x}} + \bar{C}_q\widetilde{x} + \bar{g} = \bar{\tau} + \begin{bmatrix}
		-J^\top h \\ h_{\scr O}
		\end{bmatrix},
		\end{equation*}	
		{with the barred terms as introduced in the beginning of this section.}	
		We use Gauss' principle \cite{Udwadia93Constrained} to derive closed form expressions for $J^\top h$ and $h_{\scr O}$. Let the \textit{unconstrained} coupled object-robots system be
		\begin{equation*}
		\bar{B}\alpha
		\coloneqq \bar{\tau} - \bar{C}_q
		\widetilde{x} - \bar{g},
		\end{equation*}	
		where $\alpha \coloneqq \alpha(\bar{x}):\mathbb{R}^{2n+6}\times\mathbb{T}\to\mathbb{R}^{n+6}$ is the unconstrained acceleration, i.e., the acceleration the system would have if the agents did not grasp the object. According to Gauss's principle \cite{Udwadia93Constrained}, the actual accelerations $\dot{\widetilde{x}}$ of the system are the closest ones to $\alpha(\bar{x})$, while satisfying the rigidity constraints. More rigorously, $\dot{\widetilde{x}}$ is the solution of the constrained minimization problem 
		\begin{align*}
		&\min_{z} \ \ \big( z - \alpha(\bar{x}) \big)^\top \bar{B}(\bar{x}) \big( z - \alpha(\bar{x}) \big)  \\ 		
		&\hspace{1mm} \textup{s.t.} \ \ \ A(\bar{x},t) z = b(\bar{x},t).
		\end{align*}	
		The solution to this problem is obtained by using the 
		Karush-Kuhn-Tucker conditions \cite{boyd2004convex} and has a closed-form expression. It can be shown that it satisfies 
		\begin{align*}
		\bar{B}z = \alpha +  A^\top\big( A\bar{B}^{-1} A^\top \big)^\dagger \big(b- A\alpha \big),
		\end{align*}
		where $^\dagger$ denotes the Moore-Penrose inverse.
		The aforementioned expression is compliant with the one in \cite{Udwadia92NewPerspective},
		\begin{equation*}
		\bar{B}z = \alpha + \bar{B}^{\frac{1}{2}} \big( A \bar{B}^{-\frac{1}{2}} \big)^\dagger \big( b - A\alpha\big),
		\end{equation*}
		since it holds that $A^\top (A \bar{B}^{-1} A^\top)^\dagger = \bar{B}^{\frac{1}{2}} (A\bar{B}^{-\frac{1}{2}}) ^\dagger$. Indeed, according to Theorem 3.8 of \cite{albert72Pseudoinverse}, it holds that $H^\dagger = H^\top (H H^\top)^\dagger$, for any $H\in\mathbb{R}^{x\times y}$. Then the aforementioned equality is obtained by setting $H = A\bar{B}^{-\frac{1}{2}}$.} 
	
	{Therefore, the forces, projected onto the joint-space of the agents, have the form
		\begin{subequations}	\label{eq:forces 1 (rigid+coopmanip)}
			\begin{align} 
			\begin{bmatrix}
			-J^\top h \\ 
			h_{\scr O}
			\end{bmatrix}
			&= A^\top(A \bar{B}^{-1} A^\top)^\dagger(b  - A\alpha) \\
			&= \bar{B}^{\frac{1}{2}} \big( A \bar{B}^{-\frac{1}{2}} \big)^\dagger \big( b - A\alpha\big).		
			\end{align}
		\end{subequations}
		Consider now that $h_{\scr O} = h_\textup{m} = 0_6 \Leftrightarrow h =  h_\textup{int}$, i.e., the agents produce only internal forces, without inducing object acceleration. Then, the agent dynamics are 
		\begin{align*} 
		B\ddot{q} + C_q\dot{q}+ g_{q}  = \tau - J^\top h_{\text{int}},
		\end{align*}
		and the respective \textit{unconstrained} acceleration $\alpha_{\textup{int}} \coloneqq \alpha_{\textup{int}}(q,\dot{q}) :\mathbb{R}^{2n} \to \mathbb{R}^n$ {is given by}
		\begin{equation*}
		B \alpha_\textup{int} \coloneqq \tau - C_q\dot{q} - g_q.
		\end{equation*}
		Hence, by proceeding in a similar fashion as for $\dot{\widetilde{x}}$, we derive an expression for the \textit{internal} forces as
		\begin{subequations}	\label{eq:internal forces 1}
			\begin{align} 
			-J^\top h_\textup{int} &= A_\textup{int}^\top \big(A_\textup{int} B^{-1} A_\textup{int}^\top \big)^\dagger \big(b_\textup{int} - A_\textup{int} \alpha_\textup{int}\big) \label{eq:internal forces 1_1 (rigid+coopmanip)} \\
			&= B^{\frac{1}{2}} \big( A_\textup{int} B^{-\frac{1}{2}} \big)^\dagger \big(b_\textup{int} - A_\textup{int} \alpha_\textup{int}\big). \label{eq:internal forces 1_2}
			\end{align}
		\end{subequations}
		with $A_\textup{int}$, $b_\text{int}$ as defined in \eqref{eq:A_int, b_int (rigid+coopmanip)}.}

	Therefore, one concludes that when the unconstrained motion of the system does not satisfy the constraints (i.e., when $b_\textup{int} \neq A_\textup{int} \alpha_\textup{int}$), then the actual accelerations of the system are modified in a manner directly proportional to the extent to which these constraints are violated. Moreover, it is evident from the aforementioned expression that the internal forces depend, not only on the relative distances $p_i - p_j$, but also on the closed loop dynamics and the inertia of the unconstrained system (see the dependence on $\alpha_\textup{int}$ and $B$). Therefore, given a desired force $h_{\scr O,\text{d}}$ to be applied to the object, an internal force-free distribution to agent forces $h_{i,\text{d}}$ at the grasping points cannot be independent of the system dynamics. This is clearly illustrated in the following example. 
	\begin{figure}
		\centering
		\includegraphics[width = 0.3\textwidth]{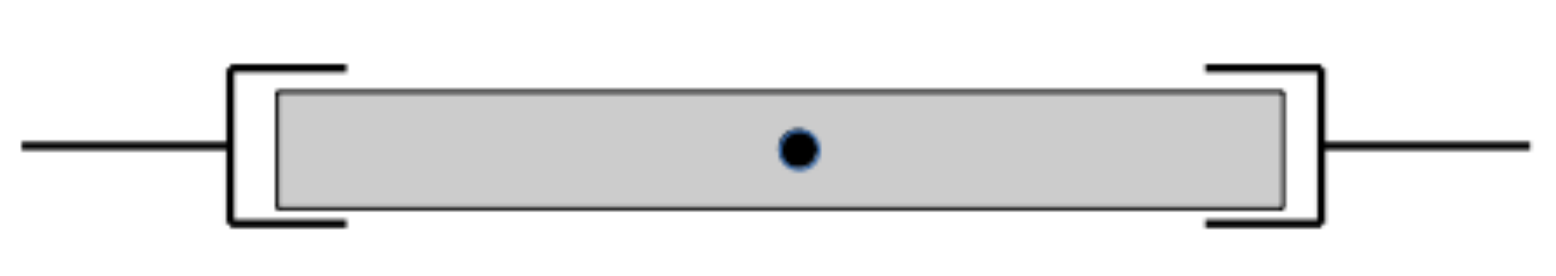}
		\caption{Two agents rigidly grasping an object in a $1$D scenario.\label{fig:1d_grasp}}
	\end{figure}

	\begin{example} \label{ex:1d}
		Consider a simplified $1$D scenario, with two agents rigidly grasping an object (see Fig. \ref{fig:1d_grasp}) subject to the dynamics 
		\begin{align*}
		& m_i \ddot{p}_i = u_i - h_i, \ \ i\in\{1,2\} \\
		& m_{\scr O} \ddot{p}_{\scr O} = h_{\scr O} = h_1 + h_2,
		\end{align*}
		with the Jacobian matrices being $J_1 = J_2 = 1$.
		The inter-agent constraints here are simply $\dot{p}_1=\dot{p}_2 \Rightarrow \begin{bmatrix}
		1 & -1
		\end{bmatrix}
		\begin{bmatrix}
		\ddot{p}_1^\top & \ddot{p}_2^\top
		\end{bmatrix}^\top = 0$, which gives $A = \begin{bmatrix}
		1 & -1
		\end{bmatrix}$, and $b = 0$. In view of \eqref{eq:internal forces 1}, one can conclude that in this simplified scenario internal forces appear when 
		\begin{equation*}
		\frac{u_1}{m_1} \neq \frac{u_2}{m_2},
		\end{equation*}
		which depends on the masses of the agents.
	\end{example}

	Note that, as dictated in the previous section, the rigidity matrix $\mathcal{R}_\mathcal{G}$ is not unique, since different choices of $\gamma_\mathcal{G}$ that encode the rigidity constraints can be made.  Hence, one might think that different expressions of $\mathcal{R}_\mathcal{G}$ will result in different rigidity constraints of the form \eqref{eq:A_int, b_int (rigid+coopmanip)} and hence different interaction and internal forces - which is unreasonable. Nevertheless, note that all different expressions of the rigidity matrix $\mathcal{R}_\mathcal{G}$ have the same nullspace (the coordinated translations and rotations of the framework),  and that suffices to prove that this is not the case, as illustrated in Corollary \ref{corol:different R same h_int (rigid+coopmanip)}.
	
	\begin{corollary} \label{corol:different R same h_int (rigid+coopmanip)}
		Let $\mathcal{R}_{\mathcal{G},1}$ and $\mathcal{R}_{\mathcal{G},2}$ such that $\textup{null}(\mathcal{R}_{\mathcal{G},1})=\textup{null}(\mathcal{R}_{\mathcal{G},2})$ and let
		\begin{align*}
		J^\top h_{\textup{int},i} \coloneqq& B^{\frac{1}{2}}\left( \mathcal{R}_{\mathcal{G},i}J B^{-\frac{1}{2}} \right)^\dagger \bigg( \big( \mathcal{R}_{\mathcal{G},i}\dot{J} + \dot{\mathcal{R}}_{\mathcal{G},i}J\big) \dot{q} +  \mathcal{R}_{\mathcal{G},i} J B^{-1}\alpha_\textup{int} \bigg),
		\end{align*}    	
		$\forall i\in\{1,2\}$, where we have used \eqref{eq:internal forces 1} and \eqref{eq:A_int, b_int (rigid+coopmanip)}. Then $h_{\textup{int},1} = h_{\textup{int},2}$. 
	\end{corollary}    
	\begin{proof}
		The poses and velocities in the terms $(\mathcal{R}_{\mathcal{G},i}\dot{J} + \dot{\mathcal{R}}_{\mathcal{G},i}J)\dot{q}$ are the actual ones resulting from the coupled system dynamics and hence they respect the rigidity constraints imposed by $R_{\mathcal{G},i}J\ddot{q} = (\mathcal{R}_{\mathcal{G},i}\dot{J} + \dot{\mathcal{R}}_{\mathcal{G},i}J)\dot{q}$, $\forall i\in\{1,2\}$. Therefore, exploiting the positive definiteness of $B$, we need to prove that $( \mathcal{R}_{\mathcal{G},1} J B^{-\frac{1}{2}})^\dagger \mathcal{R}_{\mathcal{G},1} J = ( \mathcal{R}_{\mathcal{G},2} J B^{-\frac{1}{2}} )^\dagger \mathcal{R}_{\mathcal{G},2} J$. In view of Definition \ref{def:left equiv. (app_useful_prop)} and Proposition \ref{prop:left equiv. and nullspace (app_useful_prop)} in Appendix \ref{app:useful_prop}, since $\mathcal{R}_{\mathcal{G},1}$ and $\mathcal{R}_{\mathcal{G},2}$ have the same nullspace, they are left equivalent matrices and there exists an invertible matrix $P$ such that $\mathcal{R}_{\mathcal{G},1} = P \mathcal{R}_{\mathcal{G},2}$. Hence, it holds that
		\begin{align*}
		 \left( \mathcal{R}_{\mathcal{G},2} J B^{-\frac{1}{2}} \right)^\dagger \mathcal{R}_{\mathcal{G},2}J -
		\left( \mathcal{R}_{\mathcal{G},1} J B^{-\frac{1}{2}} \right)^\dagger \mathcal{R}_{\mathcal{G},1}J & = \\
		&\hspace{-60mm}\bigg( \left( \mathcal{R}_{\mathcal{G},2} J B^{-\frac{1}{2}} \right)^\dagger \mathcal{R}_{\mathcal{G},2} J B^{-\frac{1}{2}} - 
		\left( P \mathcal{R}_{\mathcal{G},2} J B^{-\frac{1}{2}} \right)^\dagger P \mathcal{R}_{\mathcal{G},2} J B^{-\frac{1}{2}}  \bigg) B^{\frac{1}{2}}, 
		\end{align*}
		which is {equal to} $0$, according to Proposition \ref{prop: pseudoinv. of A, KA (app_useful_prop)} of Appendix \ref{app:useful_prop} and the positive definiteness of $B$.
	\end{proof}
	One can verify that a similar argument holds for the interaction forces $\begin{bmatrix}
	-J^\top h \\ h_{\scr O}
	\end{bmatrix}$ and $\mathcal{R}_{\bar{\mathcal{G}}}$ as well.

	The aforementioned expressions concern the forces in the joint-space of the robotic agents. 
	The next Corollary gives the expression of the forces in task-space:
	\begin{corollary} \label{corol:h task space (rigid+coopmanip)}
		The internal forces $h_\textup{int}$ are given by 
		\begin{align} \label{eq:internal forces task-space (rigid+coopmanip)}
		h_\textup{int} = \mathcal{R}_\mathcal{G}^\top\left(\mathcal{R}_\mathcal{G} M^{-1} \mathcal{R}_\mathcal{G}^\top\right)^\dagger \left( \dot{\mathcal{R}}_\mathcal{G} v + \mathcal{R}_\mathcal{G} \alpha^\textup{ts}_{\textup{int}}  \right), 
		\end{align}
		where $\alpha^\textup{ts}_\textup{int}\coloneqq \alpha^\textup{ts}_\textup{int}(q,\dot{q}):\mathsf{S}\times\mathbb{R}^n\to \mathbb{R}^{6N}$ is the acceleration vector of the task-space unconstrained system 
		\begin{align*}
		M \alpha^\textup{ts}_{\textup{int}} \coloneqq u -Cv - g,
		\end{align*}
		and the forces $h, h_{\scr O}$ are given by 
		\begin{equation} \label{eq:forces task-space (rigid+coopmanip)}
		\begin{bmatrix}
		-h \\ 
		h_{\scr O}
		\end{bmatrix} = -\mathcal{R}_{\bar{\mathcal{G}}}^\top \left(\mathcal{R}_{\bar{\mathcal{G}}} \bar{M}^{-1} \mathcal{R}_{\bar{\mathcal{G}}}^\top\right)^\dagger \left( \dot{\mathcal{R}}_{\bar{\mathcal{G}}} \bar{v} + \mathcal{R}_{\bar{\mathcal{G}}} \alpha^\textup{ts}  \right), 
		\end{equation}
		where $\alpha^\textup{ts} \coloneqq \alpha^\textup{ts}(\bar{x}):\mathbb{X}\to\mathbb{R}^{6N+6}$ is the acceleration vector of the task-space unconstrained system 
		\begin{align*}
		\bar{M} \alpha^\textup{ts} \coloneqq \bar{u} - \bar{C}
		\bar{v} - \bar{g}.
		\end{align*}
	\end{corollary}	
	\begin{proof}
		By using the expressions of $M(q), C(q,\dot{q})v, g(q)$ from \eqref{eq:manipulator dynamics_vector_form (rigid+coopmanip)} to expand \eqref{eq:internal forces task-space (rigid+coopmanip)}, one can conclude that $J^\top h_\textup{int}$,  with $h_\textup{int}$ given by \eqref{eq:internal forces task-space (rigid+coopmanip)} and in view of \eqref{eq:A_int, b_int (rigid+coopmanip)}, is equal to \eqref{eq:internal forces 1_1 (rigid+coopmanip)}.  Similarly, by expanding the dynamic terms of \eqref{eq:forces task-space (rigid+coopmanip)} and using \eqref{eq:A_total, b_total (rigid+coopmanip)}, one can verify that the vector $[-(J^\top h)^\top, h_{\scr O}^\top]^\top$, with $[-h^\top,h_{\scr O}^\top]^\top$ given by \eqref{eq:forces task-space (rigid+coopmanip)}, is equal to \eqref{eq:forces 1 (rigid+coopmanip)}.
	\end{proof}	
	
	We also show later that the derived forces \eqref{eq:forces task-space (rigid+coopmanip)} are consistent with the relation $h_{\scr O} = G(x)h$ (see \eqref{eq:grasp matrix (rigid+coopmanip)}).
	
	We now give a more explicit expression for $h$. One can verify that, by appropriately arranging the rows of $\gamma_{\mathcal{G}}$, it holds that 
	\begin{equation} \label{eq:R_g_bar (rigid+coopmanip)}
	\mathcal{R}_{\bar{\mathcal{G}}} \coloneqq \begin{bmatrix}
	\mathcal{R}_\mathcal{G} & 0_{\frac{7N(N-1)}{2} \times 6} \\
	\mathcal{R}_{\scr O_1} & \mathcal{R}_{\scr O_2} 
	\end{bmatrix} \in \mathbb{R}^{\frac{7\bar{N}(\bar{N}-1)}{2} \times (6N+6)},
	\end{equation} 
	where $\mathcal{R}_{\scr O_1} \in \mathbb{R}^{7N\times6N}$ and $\mathcal{R}_{\scr O_2}\in\mathbb{R}^{7N\times 6}$ are the matrices
	\begin{align*}
	\mathcal{R}_{\scr O_1} & \coloneqq \begin{bmatrix}
	(p_1 - p_{\scr O})^\top & 0_{1\times 3} & \dots & 0_{1\times 3} & 0_{1\times 3}\\
	\vdots & 
	\vdots & \dots & \ddots & \vdots \\
	0_{1\times 3} & 0_{1\times 3} & \dots & (p_N - p_{\scr O})^\top & 0_{1\times 3} \\
	\frac{\partial \gamma_{e_{1\scr O},b}}{\partial p_1} & \frac{\partial \gamma_{e_{1\scr O},b}}{\partial R_1} & \dots & 0_{3\times 3} & 0_{3\times 3} \\
	\frac{\partial \gamma_{e_{\scr O1},b}}{\partial p_1} & \frac{\partial \gamma_{e_{\scr O1},b}}{\partial R_1}  & \dots & 0_{3\times 3} & 0_{3\times 3} \\		
	\vdots & \dots & \ddots & \vdots & \vdots \\
	0_{3\times 3} & 0_{3\times 3} & \dots & \frac{\partial \gamma_{e_{N \scr O},b}}{\partial p_N} & \frac{\partial \gamma_{e_{N \scr O},b}}{\partial R_N}  \\
	0_{3\times 3} & 0_{3\times 3} & \dots &  \frac{\partial \gamma_{e_{\scr ON},b}}{\partial p_N} & \frac{\partial \gamma_{e_{\scr ON},b}}{\partial R_N} 
	\end{bmatrix}  \\
	\end{align*}
	\begin{align*}
	\mathcal{R}_{\scr O_2} &\coloneqq \begin{bmatrix}
	-(p_1-p_{\scr O})^\top & 0_{1\times 3} \\
	\vdots & \vdots \\
	-(p_N-p_{\scr O})^\top & 0_{1\times 3} \\
	\frac{\partial \gamma_{e_{1 \scr O},b}}{\partial p_{\scr O}} & \frac{\partial \gamma_{e_{1 \scr O},b}}{\partial R_{\scr O}} \\
	\frac{\partial \gamma_{e_{\scr O1},b}}{\partial p_{\scr O}} & \frac{\partial \gamma_{e_{\scr O1},b}}{\partial R_{\scr O}} \\
	\vdots & \vdots \\
	\frac{\partial \gamma_{e_{N \scr O},b}}{\partial p_{\scr O}} & \frac{\partial \gamma_{e_{N \scr O},b}}{\partial R_{\scr O}} \\
	\frac{\partial \gamma_{e_{\scr ON},b}}{\partial p_{\scr O}} & \frac{\partial \gamma_{e_{\scr ON},b}}{\partial R_{\scr O}} 
	\end{bmatrix},
	\end{align*}
	where $e_{i\scr O} \coloneqq (i,\bar{N})$, $e_{\scr Oi} \coloneqq (\bar{N},i) \in \bar{\mathcal{E}}$ corresponding to the edge among the $i$th agent and the object, $\forall i\in\mathcal{N}$. 
	Therefore, \eqref{eq:forces task-space (rigid+coopmanip)} can be written as
	\begin{subequations} \label{eq:h + h_o (rigid+coopmanip)}	
		\begin{align} 
		&h = \begin{bmatrix} \mathcal{R}_\mathcal{G}^\top & \mathcal{R}^\top_{\scr O_1}\end{bmatrix} \left(\mathcal{R}_{\bar{\mathcal{G}}} \bar{M}^{-1} \mathcal{R}_{\bar{\mathcal{G}}}^\top\right)^\dagger(\dot{\mathcal{R}}_{\bar{\mathcal{G}}}\bar{v} + \mathcal{R}_{\bar{\mathcal{G}}}\alpha^{\textup{ts}}) \label{eq:h + h_o:h (rigid+coopmanip)} \\		
		&h_{\scr O} = -\begin{bmatrix} 0 & \mathcal{R}^\top_{\scr O_2} \end{bmatrix} 
		\left(\mathcal{R}_{\bar{\mathcal{G}}} \bar{M}^{-1} \mathcal{R}_{\bar{\mathcal{G}}}^\top\right)^\dagger(\dot{\mathcal{R}}_{\bar{\mathcal{G}}}\bar{v} + \mathcal{R}_{\bar{\mathcal{G}}}\alpha^{\textup{ts}}). \label{eq:h + h_o:h_o (rigid+coopmanip)}
		\end{align}
	\end{subequations}
	Note also that 
	\begin{equation} \label{eq:GR_o1' = -R_o2' (rigid+coopmanip)}
	G \mathcal{R}_{\scr O_1}^\top = - \mathcal{R}_{\scr O_2}^\top,
	\end{equation} 
	which will be used in the analysis to follow.
	
	Another expression for the interaction forces $h$ can be obtained by differentiating \eqref{eq:grasp matrix velocities (rigid+coopmanip)}, which, after using \eqref{eq:manipulator dynamics_vector_form (rigid+coopmanip)} and \eqref{eq:object dynamics (rigid+coopmanip)} yields after straightforward manipulations (similarly to  \eqref{eq:contact force (ACC_rolling)})
	\begin{align} \label{eq:h 1 (rigid+coopmanip)}
	h =& \left(M^{-1} + G^\top M_{\scr O}^{-1} G\right)^{-1} \bigg( M^{-1}(u - g - C v) - \dot{G}^\top v_{\scr O}  + \notag \\
	&G^\top M_{\scr O}^{-1}(C_{\scr O}v_{\scr O} + g_{\scr O}) \bigg).
	\end{align}
	In order to show the consistency of our results, we prove next that \eqref{eq:h + h_o:h (rigid+coopmanip)} and \eqref{eq:h 1 (rigid+coopmanip)} are identical.
	\begin{corollary} \label{corol:h1=h2 (rigid+coopmanip)}
		Let $h_a$ be given by \eqref{eq:h + h_o:h (rigid+coopmanip)} and $h_b$ be given by \eqref{eq:h 1 (rigid+coopmanip)}. Then $h_a = h_b$.
	\end{corollary}
	\begin{proof}
		By using \eqref{eq:R_g_bar (rigid+coopmanip)}, \eqref{eq:h + h_o:h (rigid+coopmanip)} is expanded as 
		\begin{align*}
		h_a = & \mathcal{R}_{\scr O_1}^\top\left(\mathcal{R}_{\scr O_1}M^{-1}\mathcal{R}_{\scr O_1}^\top + \mathcal{R}_{\scr O_2}M_{\scr O}^{-1}\mathcal{R}_{\scr O_2}^\top\right)^\dagger ( \dot{\mathcal{R}}_{\scr O_1} v + \dot{\mathcal{R}}_{\scr O_2} v_{\scr O} \\ & + \mathcal{R}_{\scr O_1} M^{-1}(u - g - Cv) 
		- \mathcal{R}_{\scr O_2} M_{\scr O}^{-1}(g_{\scr O}+C_{\scr O}v_{\scr O}) ) 			
		\end{align*}
		which, after using $\eqref{eq:GR_o1' = -R_o2' (rigid+coopmanip)}$ and $v = G^\top v_{\scr O}$, becomes
		\begin{align*}
		h_a =& \mathcal{R}_{\scr O_1}^\top\left(\mathcal{R}_{\scr O_1}(M^{-1} + G^\top M_{\scr O}^{-1} G)\mathcal{R}_{\scr O_1}^\top\right)^\dagger ( \dot{\mathcal{R}}_{\scr O_1} v -  \dot{\mathcal{R}}_{\scr O_1}G^\top v_{\scr O} - \mathcal{R}_{\scr O_1}\dot{G}^\top v_{\scr O} \\
		& + \mathcal{R}_{\scr O_2}M^{-1}(u - g - Cv) 
		+ \mathcal{R}_{\scr O_1}G^\top M_{\scr O}^{-1}(g_{\scr O}+C_{\scr O}v_{\scr O}) ) \\
		=& \mathcal{R}_{\scr O_1}^\top\left(\mathcal{R}_{\scr O_1}(M^{-1} + G^\top M_{\scr O}^{-1} G)\mathcal{R}_{\scr O_1}^\top\right)^\dagger \mathcal{R}_{\scr O_1}\bigg(  - \dot{G}^\top v_{\scr O} +M^{-1}(u - g - C v) \\
		& + M_{\scr O}^{-1} (g_{\scr O} + C_{\scr O} v_{\scr O}) \bigg).
		\end{align*}
		Denote now for convenience $M_G \coloneqq M^{-1} + G^\top M_{\scr O}^{-1} G$. According to Theorem 3.8 of \cite{albert72Pseudoinverse}, it holds that $\mathcal{R}_{\scr O_1}^\top\left(\mathcal{R}_{\scr O_1}M_G\mathcal{R}_{\scr O_1}^\top\right)^\dagger \mathcal{R}_{\scr O_1} = M_G^{-\frac{1}{2}}\left(\mathcal{R}_{\scr O_1} M_G^{\frac{1}{2}}\right)^\dagger \mathcal{R}_{\scr O_1}$. 	
		Next, note that $\mathcal{R}_{\scr O_1}$ has linearly independent columns and hence 
		\begin{equation*}
			\left(\mathcal{R}_{\scr O_1} M_G^{\frac{1}{2}}\right)^\dagger = \left(M_G^{\frac{1}{2}}\mathcal{R}_{\scr O_1}^\top\mathcal{R}_{\scr O_1}M_G^{\frac{1}{2}}\right)^{-1} M_G^{\frac{1}{2}}\mathcal{R}_{\scr O_1}^\top = M_G^{-\frac{1}{2}}\left(\mathcal{R}_{\scr O_1}^\top \mathcal{R}_{\scr O_1}\right)^{-1}\mathcal{R}_{\scr O_1}^\top,
		\end{equation*}
		since $M_G$ is symmetric and positive definite.	
		Therefore, we conclude that $M_G^{-\frac{1}{2}}\left(\mathcal{R}_{\scr O_1} M_G^{\frac{1}{2}}\right)^\dagger \mathcal{R}_{\scr O_1} = M_G^{-1}$, and hence $h_a = h_b$.
	\end{proof}
	
	\begin{remark}	
		According to Theorem 3.8 of \cite{albert72Pseudoinverse}, the task-space internal forces can also be written as 
		\begin{equation} \label{eq:internal forces task-space 2}
		h_\textup{int} = M^{\frac{1}{2}} \left(\mathcal{R}_\mathcal{G} M^{-\frac{1}{2}} \right)^\dagger \left( \dot{\mathcal{R}}_\mathcal{G} v + \mathcal{R}_\mathcal{G} \alpha^\textup{ts}_{\textup{int}}  \right),
		\end{equation}
		which is compliant with the result in \cite{erhart2015internal}.
	\end{remark}
	
	One concludes, therefore, that in order to obtain internal force-free trajectories, the term 
	$\dot{\mathcal{R}}_\mathcal{G} v + \mathcal{R}_\mathcal{G} \alpha^\textup{ts}_{\textup{int}} = \dot{\mathcal{R}}_\mathcal{G}v  + \mathcal{R}_\mathcal{G} M^{-1}(u - C\dot{v} -g)$ must belong to the nullspace of $M^{\frac{1}{2}} \left(\mathcal{R}_\mathcal{G} M^{-\frac{1}{2}}\right)^\dagger$. The latter, however, is identical to the nullspace of $\mathcal{R}_\mathcal{G}$, since it holds that $\textup{null}(\mathcal{R}_\mathcal{G} M^{-1/2})^\dagger = \textup{null}(M^{-\frac{1}{2}}\mathcal{R}_\mathcal{G}^\top)$ and $M$ is  positive definite. 
	This result is summarized in the following corollary. 
	
	\begin{corollary} \label{coroll:A and b (rigid+coopmanip)}
		The cooperative manipulation system is free of internal forces, i.e., {$h_{\textup{int}} =  0$}, if and only if 
		\begin{equation*}
		\dot{\mathcal{R}}_\mathcal{G}v  + \mathcal{R}_\mathcal{G} M^{-1}(u - C\dot{v} -g) \in \textup{null}(\mathcal{R}_\mathcal{G}^\top)
		\end{equation*}
	\end{corollary}
	
	In cooperative manipulation schemes, the most energy-efficient way of transporting an object is to exploit the full potential of the cooperating robotic agents, i.e., each agent does not exert less effort at the expense of other agents, which might then potentially exert more effort than necessary. For instance, consider a rigid cooperative manipulation scheme, with 
	only one agent (a leader) working towards bringing the object to a desired location, whereas the other agents have zero inputs. Since the grasps are rigid, if the leader has sufficient power, it will achieve the task by ``dragging" the rest of the agents, compensating for their dynamics, and creating non-negligible internal forces. In such cases, when
	the cooperative manipulation system is rigid (i.e., the grasps are considered to be rigid), the optimal strategy of transporting an object 
	is achieved by regulating the internal forces to zero. 	Therefore, from a control perspective, the goal of a rigid cooperative manipulation system is to design a control protocol that achieves a desired cooperative manipulation task, while guaranteeing that the internal forces {remain} zero.

	\subsubsection{Cooperative Manipulation via Internal Force Regulation}
	
	{We derive here a new relation between the interaction and internal forces $h$ and $h_{\text{int}}$, respectively. Moreover, we	derive novel sufficient and necessary conditions on the agent force distribution for the provable regulation of the internal forces to zero, according to \eqref{eq:internal forces task-space (rigid+coopmanip)}, and we show its application in a standard inverse-dynamics control law that guarantees trajectory tracking of the object's center of mass. This is based on the following {main theorem}, which links the complete agent graph rigidity matrix $\mathcal{R}_\mathcal{G}$ to the grasp matrix $G$: }
	
	\begin{theorem} \label{th:null G range R_T (rigid+coopmanip)}
		Let $N$ robotic agents, with configuration $x = (p,R)\in\mathbb{SE}(3)^N$, rigidly grasping an object and associated with a grasp matrix $G(x)$, as in \eqref{eq:grasp matrix def. (rigid+coopmanip)}. Let also the agents be 		
		modeled by a framework on the complete graph $(\mathcal{K}_N, p_{\mathcal{K}_N}, R_{\mathcal{K}_N}) = (\mathcal{K}_N, p, R)$ in $\mathbb{SE}(3)$, which is associated with a rigidity matrix $\mathcal{R}_{\mathcal{K}_N}$. 
		Let also $x$ be such that $\textup{rank}(\mathcal{R}_{\mathcal{K}_N}(x)) = {\max_{y \in \mathbb{SE}(3)^N}\{ \textup{rank}(\mathcal{R}_{\mathcal{K}_N}(y))  \}}$.
		Then it holds that 
		\begin{equation*} 
		\textup{null}(G(x)) = \textup{range}(\mathcal{R}_{\mathcal{K}_N}(x)^\top).
		\end{equation*}
	\end{theorem}
	\begin{proof}
		Since $\mathcal{R}_{\mathcal{K}_N}$ is associated to the complete graph and $\textup{rank}(\mathcal{R}_{\mathcal{K}_N}(x))$ $=$ $\max_{y \in \mathbb{SE}(3)^N}\{ \textup{rank}(\mathcal{R}_{\mathcal{K}_N}(y))  \}$, the framework $(\mathcal{K}_N, p, R)$ is infinitesimally rigid. 
		Hence, the nullspace of $\mathcal{R}_{\mathcal{K}_N}$ consists only of the infinitesimal motions of the framework, i.e., coordinated translations and rotations, as defined in Proposition \ref{prop.trivialmotions (rigid+coopmanip)}. 		
		In particular, in view of \eqref{eq:R_G nullspace (rigid+coopmanip)}, {Proposition \ref{prop:ranks (rigid+coopmanip)}}, and \eqref{eq:rot subspace SE3 (rigid+coopmanip)}, one concludes that $\textup{null}(\mathcal{R}_{\mathcal{K}_N})$ is the linear span of $1_N\otimes \begin{bmatrix}
		I_3 \\ 0_{3 \times 3}
		\end{bmatrix}$ and the vector space
		$[\chi_1^\top, \dots, \chi_N^\top]^\top \in \mathbb{SE}(3)^N$, with $\chi_i\coloneqq [\chi_{i,p}^\top, \chi_{i,R}^\top]^\top\in\mathbb{SE}(3)$, satisfying
		\begin{subequations} \label{eq:rot subsp (rigid+coopmanip)}
			\begin{align}
			& \chi_{i,p} - \chi_{j,p} = -S(p_i - p_j) \chi_{i,R} \\
			& \chi_{i,R} = \chi_{j,R},
			\end{align}
		\end{subequations}
		where $p_i \coloneqq p_{\mathcal{K}_N}(i)$, $p_j \coloneqq p_{\mathcal{K}_N}(j)$, $\forall i,j \in\mathcal{N}$, with $i\neq j$. In view of \eqref{eq:grasp matrix velocities (rigid+coopmanip)}, one obtains $v = G^\top v_{\scr O}$, where
		\begin{align*}
		G^\top = \begin{bmatrix}
		I_3 & -S(p_{1\scr O})\\
		0  & I_3 \\ 
		\vdots & \vdots \\
		I_3 & -S(p_{N \scr O}) \\
		0  & I_3 
		\end{bmatrix}.
		\end{align*}			
		The first $3$ columns of $G^\top$ form the space $1_N\otimes \begin{bmatrix}
		I_3 \\ 0_{3 \times 3}
		\end{bmatrix}$ whereas the last $3$ columns $G^\top$ span the aforementioned rotation vector space. Indeed, for any  $\dot{p}_{\scr O}$, $\omega_{\scr O} \in \mathbb{R}^6$ the range of these columns is
		\begin{align*}
		\begin{bmatrix}
		-S(p_{1\scr O}) \dot{p}_{\scr O} \\
		\omega_{\scr O} \\
		\vdots \\
		-S(p_{N \scr O}) \dot{p}_{\scr O} \\
		\omega_{\scr O}
		\end{bmatrix},
		\end{align*}
		for which it is straightforward to verify that \eqref{eq:rot subsp (rigid+coopmanip)} holds.	Hence, $\textup{null}(\mathcal{R}_{\mathcal{K}_N}) = \textup{range}(G^\top)$ and by using the rank-nullity theorem the result follows.
	\end{proof}
	
	Hence, {since the internal forces belong to $\text{null}(G)$}, one concludes that they are comprised of all the vectors $z$ for which there exists a $y$ such that $z = \mathcal{R}_\mathcal{G}^\top y$. This can also be verified by inspecting \eqref{eq:internal forces task-space 2}; one can prove that $\textup{range}( M^{\frac{1}{2}} (\mathcal{R}_\mathcal{G} M^{-\frac{1}{2}})^\dagger ) = \textup{range}(\mathcal{R}_\mathcal{G}^\top)$. The aforementioned result provides significant insight regarding the  control of the motion of the coupled cooperative manipulation system. In particular,  
	by using \eqref{eq:internal forces task-space 2} and Theorem \ref{th:null G range R_T (rigid+coopmanip)}, we provide next {new conditions on the agent force distribution for provable avoidance of internal forces. We first derive a novel relation between the agent forces $h$ and the internal forces $h_\text{int}$. }
	
	{
		In many works in the related literature, the force $h$ is decomposed as 
		\begin{equation} \label{eq:force decomp}
		h = h_\textup{m} + h_\textup{int} =  G^\ast G h + (I - G^\ast G)h,
		\end{equation}
		where $G^\ast$ is a right inverse of $G$. The term $G^\ast G h$ is a projection of $h$ on the range space of $G^\top$, whereas the term $(I - G^\ast G)h$ is a projection of $h$ on the null space of $G$. A common choice is the Moore-Penrose inverse $G^\ast = G^\dagger$, which equals to $G^\top (GG^\top)^{-1}$. This specific choice yields the vector $G^\ast G h=G^\dagger G h \in \textup{range}(G^\top)$ that is closest to $h$, i.e., $\|h - G^\dagger G h\| \leq \|h - y\|$, $\forall y\in\textup{range}(G^\top)$. However, as the next theorem states, if the second term of \eqref{eq:force decomp} must equal $h_\text{int}$, as this is defined in \eqref{eq:internal forces task-space 2}, $G^\ast$ must {actually} be the weighted pseudo inverse $MG^\top(GMG^\top)^{-1}$.}
	
	{
		\begin{theorem} \label{th:internal forces (rigid+coopmanip)}
			Consider $N$ robotic agents rigidly grasping an object with coupled dynamics \eqref{eq:coupled dynamics (rigid+coopmanip)}. Let $h \in \mathbb{R}^{6N}$ be the stacked vector of agent forces exerted at the grasping points. Then the agent forces $h$ and the internal forces $h_\text{int}$ are related as:
			\begin{equation*}
			h_\textup{int} =  (I_{6N} - MG^\top(GMG^\top)^{-1}G) h.
			\end{equation*}
		\end{theorem}		
		In order to prove Theorem \ref{th:internal forces (rigid+coopmanip)}, we first need the following preliminary result.
		\begin{proposition} \label{prop:forces 2 (rigid+coopmanip)}
			Consider the grasp and rigidity matrices $G$, $\mathcal{R}_\mathcal{G}$, respectively, of the cooperative manipulation system. Then it holds that 
			\begin{equation*} 
			MG^\top\left(G M G^\top\right)^{-1} G + M^\frac{1}{2}\left(\mathcal{R}_\mathcal{G}M^{-\frac{1}{2}}\right)^\dagger \mathcal{R}_\mathcal{G} M^{-1} = I.
			\end{equation*}		
		\end{proposition}
		\begin{proof}
			Let $A_f \coloneqq  \mathcal{R}_\mathcal{G} M^{-\frac{1}{2}} $ and $B_f \coloneqq G M^{\frac{1}{2}}$. Then $\textup{range}(A_f^\top) = \textup{null}(B_f)$. Indeed, according to Theorem \ref{th:null G range R_T (rigid+coopmanip)}, it holds that if $z = \mathcal{R}^\top_\mathcal{G} y$, for some $y\in\mathbb{R}^6$, then $G z =0$. By multiplying by $M^{-\frac{1}{2}}$, we obtain $M^{-\frac{1}{2}} z = M^{-\frac{1}{2}} \mathcal{R}^\top_\mathcal{G}y$, which implies that $\hat{z}\coloneqq M^{-\frac{1}{2}}z \in \textup{range}( (\mathcal{R}_\mathcal{G}M^{\frac{1}{2}})^\top)$. It also holds that $B_f \hat{z} =  GM^{\frac{1}{2}}\hat{z} = Gz = 0$, and hence $\hat{z}\in\textup{null}(B_f)$. Therefore, in view of Proposition \ref{prop: pseudo-inverse property identity (app_useful_prop)} of Appendix \ref{app:useful_prop}, Theorem 3.8 of \cite{albert72Pseudoinverse}, {according  to which $G^\top(GMG^\top)^\dagger = M^{-\frac{1}{2}}(GM^{\frac{1}{2}})^\dagger $}, and the fact that $GMG^\top$ is invertible, we conclude that 
			\begin{align*}
			&\left(G M^{\frac{1}{2}}\right)^\dagger GM^{\frac{1}{2}} + \left(\mathcal{R}_\mathcal{G}M^{-\frac{1}{2}}\right)^\dagger\mathcal{R}_\mathcal{G}M^{-\frac{1}{2}} = I \Leftrightarrow \\
			&M^\frac{1}{2}G^\top(G M G^\top)^\dagger GM^\frac{1}{2} + \left(\mathcal{R}_\mathcal{G}M^{-\frac{1}{2}}\right)^\dagger\mathcal{R}_\mathcal{G}M^{-\frac{1}{2}} = I,
			\end{align*}
			and by left and right multiplication by $M^{\frac{1}{2}}$ and $M^{-\frac{1}{2}}$, respectively, the result follows.
		\end{proof}
		We are now ready to prove Theorem \ref{th:internal forces (rigid+coopmanip)}.

		\begin{proof}					
			We first show that
			\begin{align*}		
			&	 (I - MG^\top(GMG^\top)^{-1}G)\left(M^{-1} + G^\top M_{\scr O}^{-1}G\right)^{-1} =  M^\frac{1}{2}\left(\mathcal{R}_\mathcal{G}M^{-\frac{1}{2}}\right)^\dagger \mathcal{R}_\mathcal{G}.
			\end{align*}
			Indeed, since $\left(M^{-1} + G^\top M_{\scr O}^{-1}G\right)^{-1}$ has full rank, it suffices to show that
			\begin{align*}
			&(I - MG^\top(GMG^\top)^{-1}G) =  M^\frac{1}{2}\left(\mathcal{R}_\mathcal{G}M^{-\frac{1}{2}}\right)^\dagger \mathcal{R}_\mathcal{G} \left(M^{-1} + G^\top M_{\scr O}^{-1}G\right), 
			\end{align*}
			which can be concluded from the fact that $\mathcal{R}_\mathcal{G} G^\top = 0$ (due to Theorem \ref{th:null G range R_T (rigid+coopmanip)}) and Proposition \ref{prop:forces 2 (rigid+coopmanip)}.
			Therefore, in view of \eqref{eq:h 1 (rigid+coopmanip)}, it holds that 
			\begin{align*}
			&(I - MG^\top(GMG^\top)^{-1}G)h = (I - MG^\top(GMG^\top)^{-1}G)(M^{-1} + \\& G^\top M_{\scr O}^{-1} G)^{-1}
			\bigg(- \dot{G}^\top v_{\scr O} 
			+M^{-1}(u - g - C v) + G^\top M_{\scr O}^{-1}(C_{\scr O}v_{\scr O} + g_{\scr O}) \bigg) = \\
			&M^\frac{1}{2}\left(\mathcal{R}_\mathcal{G} M^{-\frac{1}{2}}\right)^\dagger\mathcal{R}_\mathcal{G}\big(M^{-1}(u - g - C v) + G^\top M_{\scr O}^{-1}(C_{\scr O}v_{\scr O} + g_{\scr O}\big)- \dot{G}^\top v_{\scr O} ,
			\end{align*}
			which, in view of the facts that $\mathcal{R}_\mathcal{G} G^\top = 0$, and hence by differentiation $-\mathcal{R}_\mathcal{G} \dot{G}^\top = \dot{\mathcal{R}}_\mathcal{G} G^\top$, as well as $G^\top v_{\scr O} = v$, becomes
			\begin{align*}
			M^\frac{1}{2}\left(\mathcal{R}_\mathcal{G} M^{-\frac{1}{2}}\right)^\dagger\big(\dot{\mathcal{R}_\mathcal{G}}v + \mathcal{R}_\mathcal{G}M^{-1}(u - g - C v)\big) = h_\text{int}.
			\end{align*}
		\end{proof}} 
		
		{
			Based on Theorem \ref{th:internal forces (rigid+coopmanip)}, we provide in the next theorem new results on the optimal distribution of a force to the robotic agents, i.e., a distribution that provably yields zero internal forces.
		}
		{
			\begin{theorem} \label{th:optimal internal force distribution (rigid+coopmanip)}
				Consider $N$ robotic agents rigidly grasping an object, with coupled dynamics \eqref{eq:coupled dynamics (rigid+coopmanip)}. Let a desired force to be applied to the object $h_{\scr O,\textup{d}} \in \mathbb{R}^6$, which is distributed to the agents' desired forces as $h_\textup{d} = G^\ast h_{\scr O,\textup{d}}$, and where $G^\ast$ is a right inverse of $G$, i.e., $GG^\ast = I_6$. Then there are no internal forces, i.e., $h_\textup{int} = 0$, if and only if 
				\begin{equation*}
				G^\ast =   MG^\top(GMG^\top)^{-1}.
				\end{equation*}
			\end{theorem}}
			
			\begin{proof}
				{
					According to Theorem \ref{th:internal forces (rigid+coopmanip)}, the derivation of $h_\text{d}$ that yields zero internal forces can be formulated as a quadratic minimization problem:			
					\begin{align*}
					\text{QP}: \hspace{2mm }&\min_{h_\text{d}} \hspace{5mm} \|h_\textup{int}\|^2 =  h_\text{d}^\top H h_\textup{d} 	 \notag \\
					&\text{s.t. }  \hspace{6.5mm} Gh_\textup{d} = h_{\scr O,\textup{d}},
					\end{align*}
					where $H \coloneqq (I_{6N} - MG^\top(GMG^\top)^{-1}G)^\top(I_{6N} - MG^\top(GMG^\top)^{-1}G)$. }
				{
					Firstly, note that the choice $G^\ast = MG^\top(GMG^\top)^{-1}h_{\scr O,\textup{d}}$ is a minimizer of QP, since $GG^\ast = I_6$, and $H G^\ast h_{\scr O,\textup{d}} = 0$, and therefore sufficiency is proved. }
				
				{In order to prove necessity, we prove next that $G^\ast$ is a strict minimizer, i.e., there is no other right inverse of $G$ that is a solution  of QP. Note first that $G\in\mathbb{R}^{6\times6N}$ has full row rank, which implies that the dimension of its nullspace is $6N-6$. Let $Z \coloneqq [z_1,\dots,z_{6N-6}] \in \mathbb{R}^{6\times(6N-6)}$ be the matrix formed by the vectors $z_1,\dots,z_{6N-6}\in \mathbb{R}^{6N}$ that span the nullspace of $G$. It follows that $\textup{rank}(Z) = 6N-6$ and $GZ = 0$. Let now the matrix $H' \coloneqq Z^\top H Z \in\mathbb{R}^{(6N-6)\times(6N-6)}$. Since $GZ = 0 \Rightarrow Z^\top G^\top = 0$, it follows that $H' = Z^\top Z$. Hence, $\textup{rank}(H') = \textup{rank}(Z) = 6N-6$, which implies that $H'$ is positive definite. Therefore, according to  \cite[Theorem $1.1$]{gould1985practical}, QP has a strong minimizer. 
				}		
			\end{proof}
			
			{
				The aforementioned theorem provides novel necessary and sufficient conditions for provable minimization of internal forces in a cooperative manipulation scheme. As discussed before, this is crucial for achieving energy-optimal cooperative manipulation, where the agents do not have to ``waste" control input and hence energy resources that do not contribute to object motion. 
				Related works that focus on deriving internal force-free distributions $G^\ast$, e.g., \cite{walker1991analysis,chung2005analysis,williams1993virtual,erhart2015internal}, are solely based on the inter-agent distances, neglecting the actual dynamics of the agents and the object. The expression \eqref{eq:internal forces task-space (rigid+coopmanip)}, however, gives new insight on the topic and suggests that the dynamic terms of the system play a significant role in the arising internal forces, as also indicated by Corollary \ref{coroll:A and b (rigid+coopmanip)}. This is further exploited by Theorem \ref{th:optimal internal force distribution (rigid+coopmanip)} to derive a right-inverse that depends on the inertia of the system. Note also that, as explained in \cite{erhart2015internal} and illustrated in Example \ref{ex:1d}, the internal forces depend on the acceleration of the robotic agents and hence the incorporation of $M$ in $G^\ast$ is something to be expected.						
			}
			
			{
				The forces $h$, however, are not the actual control input of the robotic agents, and hence we cannot simply set $h = h_\text{d} = MG^\top(GMG^\top)^{-1}Gh_{\scr O,\text{d}}$ for a given $h_{\scr O,\text{d}}$. Therefore, we design next a standard inverse-dynamics control algorithm controller that guarantees tracking of a desired trajectory by the object center of mass while provably achieving regulation of the internal forces to zero.				 
			}	
			
			Let a desired position trajectory for the object center of mass be $p_\textup{d}:\mathbb{R}_{\geq 0}\to\mathbb{R}^3$, and $e_p \coloneqq p_{\scr O} - p_\text{d}$. Let also a desired object orientation be expressed in terms of a desired rotation matrix $R_\textup{d}:\mathbb{R}_{\geq 0} \to \mathbb{SO}(3)$, with $\dot{R}_\textup{d} = S(\omega_\textup{d})R_\textup{d}$, where $\omega_\textup{d}:\mathbb{R}_{\geq 0} \to\mathbb{R}^3$ is the desired angular velocity. Then an orientation error metric that was also used in the previous formation-control section is 
			\begin{equation} \label{eq:e_O rot mat}
			e_{\scr O} \coloneqq \frac{1}{2}\textup{tr}\left(I_3 - R_\textup{d}^\top R_{\scr O} \right) \ \ \in [0,2],
			\end{equation}
			which, after differentiation and by using \eqref{eq:object dynamics 1 (rigid+coopmanip)} becomes (see also  \eqref{eq:error psi_k dot (formation)})
			\begin{equation} \label{eq:e_O_dot rot mat}
			\dot{e}_{\scr O} = \frac{1}{2}e_R^\top R^\top_{\scr O}\left( \omega_{\scr O} - \omega_\textup{d} \right),
			\end{equation}
			where $e_R \coloneqq S^{-1}\left( R_\textup{d}^\top R_{\scr O} - R^\top_{\scr O} R_\textup{d} \right) \in\mathbb{R}^3$. It holds that 
			\begin{equation*} 
			e_R = 0 \Leftrightarrow \begin{cases}
			e_{\scr O} = 0 \Leftrightarrow \textup{tr}(R_\textup{d}^\top R_{\scr O}) = 3 \Leftrightarrow R_{\scr O} = R_\textup{d}
			\\
			e_{\scr O} = 2 \Leftrightarrow \textup{tr}(R_\textup{d}^\top R_{\scr O}) = -1 \Leftrightarrow R_{\scr O} \neq R_\textup{d}
			\end{cases}			.
			\end{equation*}
			The second case represents an undesired equilibrium, where the desired and the actual orientation differ by $180$ degrees. This issue is caused by topological obstructions on $\mathbb{SO}(3)$ and it has been proven that no continuous controller can achieve \textit{global} stabilization \cite{mayhew2011quaternion}. {The following control design guarantees that $e_{\scr O}(t) < 2$, $\forall t\in\mathbb{R}_{\geq 0}$, from all initial conditions satisfying $e_{\scr O}(0)<2$.}%
			
			{The next corollary shows that a standard inverse-dynamics control protocol guarantees convergence of $p(t)-p_\textup{d}(t)$, $e_{\scr O}(t)$ to zero while avoiding internal forces, provided that the right inverse $G^\ast = MG^\top(GMG^\top)^{-1}$ is used.}

			{
				\begin{corollary} \label{corol:u design rot mat}
					Consider $N$ robotic agents rigidly grasping an object with coupled dynamics \eqref{eq:coupled dynamics (rigid+coopmanip)}. Let a desired trajectory be defined by $p_\textup{d}:\mathbb{R}_{\geq 0}\to\mathbb{R}^{3}$, $R_\textup{d}:\mathbb{R}_{\geq 0} \to \mathbb{SO}(3)$, {$\dot{p}_\textup{d}, \omega_\textup{d}\in\mathbb{R}^3$}, and assume that $e_{\scr O}(0) < 2$, with $e_{\scr O}$ as defined in \eqref{eq:e_O rot mat}. Consider the inverse-dynamics control law
					\begin{align} 
					u &= g + \left( C G^\top + M \dot{G}^\top \right) v_{\scr O} +  G^\ast\left( g_{\scr O} + C_{\scr O} v_{\scr O} \right) \notag \\
					&  \hspace{20mm}+ \left(M G^\top + G^\ast M_{\scr O}\right)\left( \dot{v}_\textup{d} - K_d e_v - K_p e_x  \right), \label{eq:u rot mat (rigid+coopmanip)}
					\end{align}
					where $e_v \coloneqq v_{\scr O} - v_\textup{d}$, $v_\textup{d} \coloneqq [\dot{p}_\textup{d}^\top, \omega_\textup{d}^\top]^\top \in\mathbb{R}^6$, $e_x \coloneqq [e_p^\top, \frac{1}{2(2-e_{\scr O})^2}e_R^\top R_{\scr O}^\top]^\top$,  $K_p\coloneqq \textup{diag}\{ K_{p_1}, k_{p_2} I_3 \}$, where $K_{p_1} \in\mathbb{R}^{3\times 3}, K_d\in\mathbb{R}^{6\times 6}$ are positive definite matrices, and $k_{p_2}\in\mathbb{R}_{>0}$ is a positive constant.
					Then the solution of the closed-loop coupled system satisfies the following:
					\begin{enumerate}			
						\item $e_{\scr O}(t) < 2$, $\forall t\in\mathbb{R}_{\geq 0}$
						\item $\lim_{t\to\infty}(p_{\scr O}(t) - p_\textup{d}(t)) = 0$,  $\lim_{t\to\infty} R_\textup{d}(t)^\top R_{\scr O}(t) =  I_3$
						\item There are no internal forces, i.e., $h_\textup{int}(t) = 0$, $\forall t\in\mathbb{R}_{\geq 0}$, if and only if 			
						\begin{equation*}
						G^\ast = MG^\top(GMG^\top)^{-1}.
						\end{equation*}
					\end{enumerate}   
				\end{corollary}
			}
			\begin{proof}
				\begin{enumerate}
					\item By substituting \eqref{eq:u rot mat (rigid+coopmanip)} in \eqref{eq:coupled dynamics (rigid+coopmanip)} and using $G G^\ast = I_6$, we obtain, in view of \eqref{eq:coupled M (TCST_coop_manip)}-\eqref{eq:coupled g (TCST_coop_manip)} and the positive definiteness of $\widetilde{M}$:
					\begin{align}
					&\widetilde{M}\left( \dot{e}_v + K_d e_v + K_p e_x \right) = 0_6 \ \Rightarrow \dot{e}_v = -K_d e_v - K_p e_x. \label{eq:dot_e_v closed loop}
					\end{align}			
					{Consider now the function}
					\begin{equation*}
					V \coloneqq \frac{1}{2}e_p^\top K_{p_1} e_p + \frac{k_{p_2}}{2-e_{\scr O}} + \frac{1}{2}e_v^\top e_v,
					\end{equation*}
					for which it holds $V(0) < \infty$, since $e_{\scr O}(0) < 2$. By differentiating $V$, and using \eqref{eq:e_O_dot rot mat} and \eqref{eq:dot_e_v closed loop}, one obtains 
					\begin{align*}
					\dot{V} =& \begin{bmatrix}
					e_p^\top K_{p_1} & \frac{k_{p_2}}{2(2-e_{\scr O})^2}e_R^\top R_{\scr O}^\top
					\end{bmatrix} e_v - e_v^\top\left( K_d e_v + K_p e_x \right) 
					= -e_v^\top K_d e_v \leq 0
					\end{align*}
					Hence, it holds that $V(t) \leq V(0) < \infty$, which implies that $\frac{k_{p_2}}{2-e_{\scr O}(t)}$ is bounded and consequently $e_{\scr O}(t) < 2$.
					\item Since $V(t) \leq V(0) < \infty$, the errors $e_p$, $e_v$ are bounded, 	which, given the boundedness of the desired trajectories $p_\textup{d}$, $R_\textup{d}$ and their derivatives, implies the boundedness of the control law $u$. Hence, it can be proved that $\ddot{V}$ is bounded which implies the uniform continuity of $\dot{V}$. Therefore, according to
					Barbalat's lemma (Lemma \ref{lemma:barbalat (App_dynamical_systems)} of Appendix \ref{app:dynamical systems}), we deduce that $\lim_{t\to\infty} \dot{V}(t) = 0 \Rightarrow \lim_{t\to\infty} e_v(t) = 0$. Since $e_x(t)$ is also bounded, it can be proved by using the same arguments that $\lim_{t\to\infty} \dot{e}_v(t) = 0$ and hence \eqref{eq:dot_e_v closed loop} implies that $\lim_{t\to\infty}e_x(t) = 0$.			
					\item {Let the desired object force be 
						\begin{equation} \label{eq:h O des}
						h_{\scr O,\text{d}} = C_{\scr O}v_{\scr O} + g_{\scr O}  + M_{\scr O}\alpha_\textup{d},
						\end{equation}
						where $\alpha_\textup{d} \coloneqq \dot{v}_\textup{d} - K_d e_v - K_p e_x$, which implies that \eqref{eq:u rot mat (rigid+coopmanip)} becomes
						\begin{equation*}
						u= g + (CG^\top + M\dot{G}^\top)v_{\scr O} + MG^\top\alpha_\text{d} + G^\ast h_{\scr O,\text{d}}
						\end{equation*}
						In view of Theorem \ref{th:optimal internal force distribution (rigid+coopmanip)}, it suffices to prove $h = h_{\text{d}} = G^\ast h_{\scr O,\text{d}}$. By substituting \eqref{eq:u rot mat (rigid+coopmanip)} in the expression \eqref{eq:h 1 (rigid+coopmanip)} and canceling terms, we obtain
						\small
						\begin{align*}
						h =& (M^{-1} + G^\top M_{\scr O}^{-1}G)^{-1}\big(M^{-1}G^\ast h_{\scr O,\textup{d}} + G^\top \alpha_\textup{d} +  G^\top M_{\scr O}^{-1}(C_{\scr O}v_{\scr O} + g_{\scr O}) \big).
						\end{align*}
						\normalsize
						Next, we add and subtract the term $G^\top M_{\scr O} GG^\ast h_{\scr O,\text{d}}$ to obtain 
						\small
						\begin{align*}
						h =& (M^{-1} + G^\top M_{\scr O}^{-1} G)^{-1}(M^{-1} + G^\top M_{\scr O}^{-1} G)G^\ast h_{\scr O,\text{d}} + \\
						& (M^{-1} + G^\top M_{\scr O}^{-1} G)^{-1}\big(G^\top M_{\scr O}^{-1}(M_{\scr O}\alpha_\text{d} + C_{\scr O}v_{\scr O} + g_{\scr O} - G^\top M_{\scr O} h_{\scr O,\text{d}} ) \big),
						\end{align*}
						\normalsize
						which, in view of \eqref{eq:h O des}, becomes $h = G^\ast h_{\scr O,\text{d}}$.}					
				\end{enumerate}
			\end{proof}

			\begin{remark}[\textbf{Uncertain dynamics and force sensing}]
				{Note that the employed inverse dynamics controller requires knowledge of the agent and object dynamics. In case of dynamic parameter uncertainty}, standard adaptive control schemes that attempt to estimate potential uncertainties in the model (see, e.g., \cite{marino2017distributed} or the previous chapter) would intrinsically create internal forces, since the dynamics of the system would not be accurately compensated. The same holds for schemes that employ force/torque sensors that provide the respective measurements at the grasp points (e.g., \cite{tsiamis2015cooperative,heck2013internal}) in periodic time instants. Since the interaction forces depend explicitly on the control input, such measurements will unavoidably correspond to the interaction forces of the previous time instants due to causality reasons, creating thus small disturbances in the dynamic model. 
			\end{remark}
			\begin{remark}[\textbf{Load-sharing}]
				{Finally, note that $G^\ast = MG^\top(GMG^\top)^{-1}$ induces an \textit{implicit} and natural load-sharing scheme via the incorporation of $M$. More specifically, note that the force distribution to the robotic agents via $G^\ast h_{\scr O,\text{d}}$ yields for each agent $M_i J_{\scr O_i} (\sum_{i\in\mathcal{N}} J_{\scr O_i}^\top M_i J_{\scr O_i})^{-1}$, $\forall i\in\mathcal{N}$. Hence, larger values of $M_i$ will produce larger inputs for agent $i$, implying that agents with larger inertia characteristics will take on a larger share of the object load.	
					Note that this is also a \textit{desired} load-sharing scheme, since larger dynamic values usually imply more powerful robotic agents.
				}
			\end{remark}

			In case it is required to achieve a \textit{desired} internal force $h_\textup{int,d}$, one can add in \eqref{eq:u rot mat (rigid+coopmanip)} a term of the form described in the next corollary.			
			\begin{corollary} \label{corol:h int d (rigid+coopmanip)}
				Let $h_\textup{int,d} \in \textup{null}(G)$ be a desired internal force to be achieved. Then adding the extra term $u_\textup{int,d} \coloneqq (I_{6N} - MG^\top (GM G^\top)^{-1}) h_\textup{int,d}$  			
				in \eqref{eq:u rot mat (rigid+coopmanip)} achieves $h_\textup{int} = h_\textup{int,d}$.
			\end{corollary}
			\begin{proof}
				Since $h_{\text{int,d}} \in \textup{null}(G) = \textup{range}(\mathcal{R}_\mathcal{G}^\top)$, it holds that $M^{-\frac{1}{2}}h_{\text{int,d}} \in \textup{range}(M^{-\frac{1}{2}}\mathcal{R}_\mathcal{G}^\top) = \textup{range}(\mathcal{R}_\mathcal{G}M^{-\frac{1}{2}})^\dagger$. Therefore, it holds that 
				\begin{align} \label{eq:des int 1}
				(\mathcal{R}_\mathcal{G}M^{-\frac{1}{2}})^\dagger \mathcal{R}_\mathcal{G} M^{-1}h_{\text{int,d}} = 
				(\mathcal{R}_\mathcal{G}M^{-\frac{1}{2}})^\dagger \mathcal{R}_\mathcal{G} M^{-\frac{1}{2}} (M^{-\frac{1}{2}}h_{\text{int,d}}) =& M^{-\frac{1}{2}}h_{\text{int,d}}.
				\end{align}
				Hence, \eqref{eq:internal forces task-space 2} yields the resulting internal forces
				\begin{align*}
				h_{\text{int}} =& M^{\frac{1}{2}} (\mathcal{R}_\mathcal{G} M^{-\frac{1}{2}})^\dagger \mathcal{R}_\mathcal{G} M^{-1}(I - MG^\top (GMG^\top)^{-1})h_{\text{int,d}} \\
				=& M^{\frac{1}{2}} (\mathcal{R}_\mathcal{G}M^{-\frac{1}{2}})^\dagger \mathcal{R}_\mathcal{G}M^{-1}h_{\text{int,d}} \\
				=& M^{\frac{1}{2}} M^{-\frac{1}{2}} h_{\text{int,d}} = h_{\text{int,d}},
				\end{align*}
				where we have used \eqref{eq:des int 1} and the fact that $\mathcal{R}_\mathcal{G}G^\top = 0$ from Theorem \ref{th:null G range R_T (rigid+coopmanip)}.
			\end{proof}
			
			Finally, in view of Theorem \ref{th:null G range R_T (rigid+coopmanip)}, one can also verify the consistency of the expressions of $h, h_{\scr O}$ in \eqref{eq:forces task-space (rigid+coopmanip)} with the grasp-matrix rigidity constraint $h_{\scr O} = G(x)h$ (see \eqref{eq:grasp matrix (rigid+coopmanip)}). 		
			Indeed, Theorem \ref{th:null G range R_T (rigid+coopmanip)} dictates that $G\mathcal{R}_\mathcal{G}^\top = 0$. Therefore, by combining \eqref{eq:GR_o1' = -R_o2' (rigid+coopmanip)} and \eqref{eq:h + h_o (rigid+coopmanip)} we conclude that $h_{\scr O} = G(x) h$. Note also that, in view of Corollary \ref{corol:different R same h_int (rigid+coopmanip)}, the result is still valid if different $\gamma_{\bar{\mathcal{G}}}$ and $\mathcal{R}_{\bar{\mathcal{G}}}$ are chosen.
			
			\begin{figure}
				\centering
				\includegraphics[width = 0.45\textwidth]{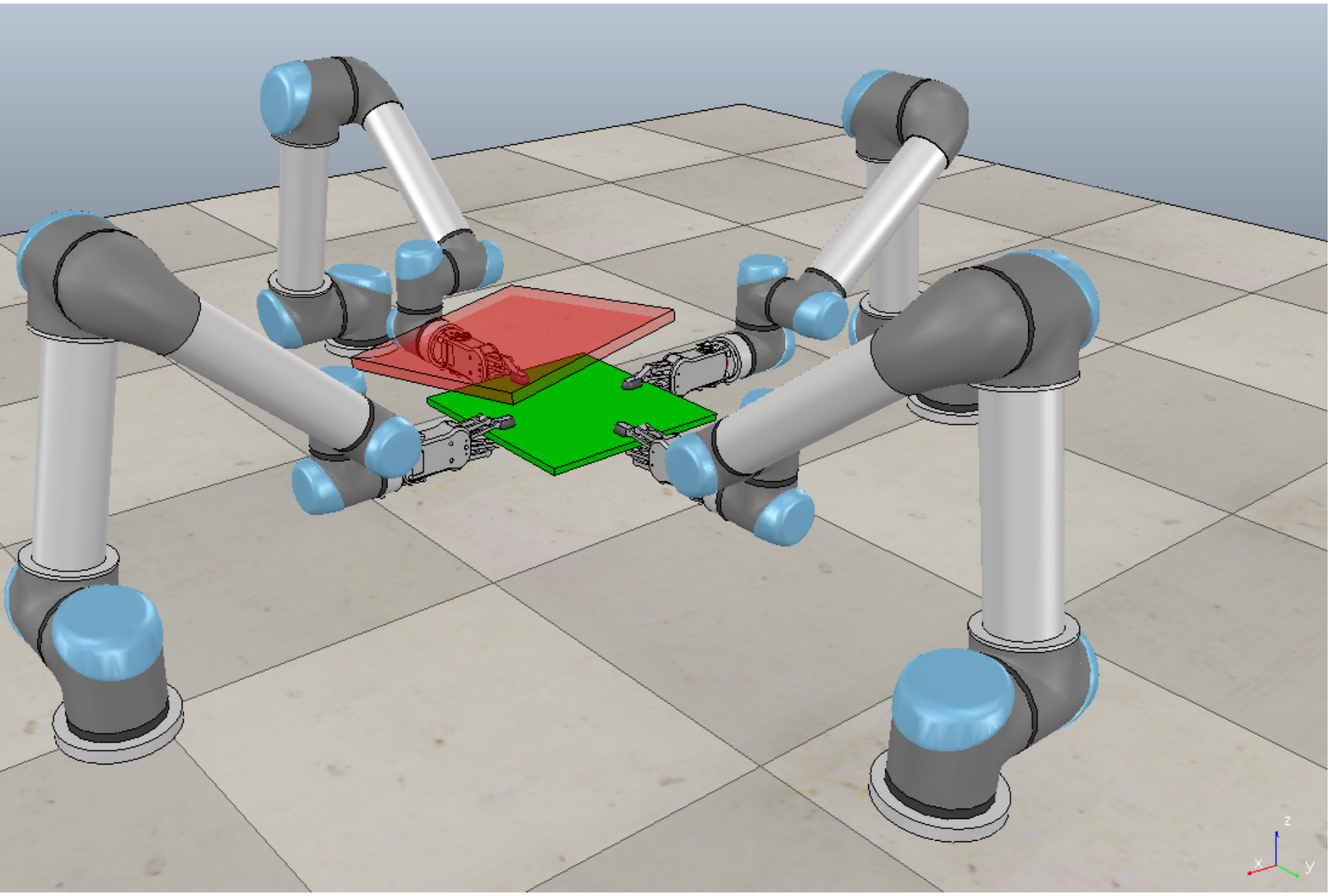}
				\caption{Four UR$5$ robotic arms rigidly grasping an object. The red counterpart represents a desired object pose at $t=0$.\label{fig:vrep_initial (rigid+coopmanip)}}
			\end{figure}
			
			\subsection{Simulation Results} \label{sec:Sim/Exp results (rigid+coopmanip)}
			{This section provides simulation results using $4$ identical UR5 robotic manipulators in the realistic dynamic environment V-REP \cite{Vrep}. The $4$ agents are rigidly grasping an object of $40$ kg in an initial configuration as shown in Fig. \ref{fig:vrep_initial (rigid+coopmanip)}. In order to verify the theoretical findings of the previous sections, we apply the controller \eqref{eq:u rot mat (rigid+coopmanip)} to achieve tracking of a desired trajectory by the object's center of mass. We simulate the closed loop system for two cases of $G^\ast$, namely the proposed one $G^\ast_1 = MG^\top(GMG^\top)^{-1}$ as well as the more standard choice $G^\ast_2 = G^\top(GG^\top)^{-1}$. Moreover, we show for $G^\ast_1$ the validity of Theorems \ref{th:internal forces (rigid+coopmanip)} and \ref{th:optimal internal force distribution (rigid+coopmanip)} by plotting the arising internal forces, and we also illustrate the achievement of a desired nonzero internal force. 	}
			
			{The initial pose of the object is set as $p_{\scr O}(0) = [-0.225,-0.612,0.161]^\top$, $\eta_{\scr O}(0) = [0,0,0]^\top$ and the desired trajectory as $p_\text{d}(t) = p_{\scr O}(0) + [0.2\sin(w_p{d}t + \varphi_\text{d}), 0.2\cos(w_pt+\varphi_\text{d}), 0.09+0.1\sin(w_pt + \varphi_\text{d})]^\top$, $\eta_\text{d}(t) = [0.15\sin(w_\phi t + \varphi_\text{d}),0.15\sin(w_\theta t + \varphi_\text{d}), 0.15\sin(w_\psi t + \varphi_\text{d})]^\top$ (in m and rad, respectively), where $\varphi_\text{d} = \frac{\pi}{6}$, $w_p = w_\phi = w_\psi = 1$, $w_\theta = 0.5$, and $\eta_\text{d}(t)$ is transformed to the respective $R_\text{d}(t)$. The control gains are set as $K_{p_1} = 15$, $k_{p_2} = 75$, and $K_d = 40 I_6$.}
			
\begin{figure}
				\centering
				\includegraphics[width = .9\textwidth]{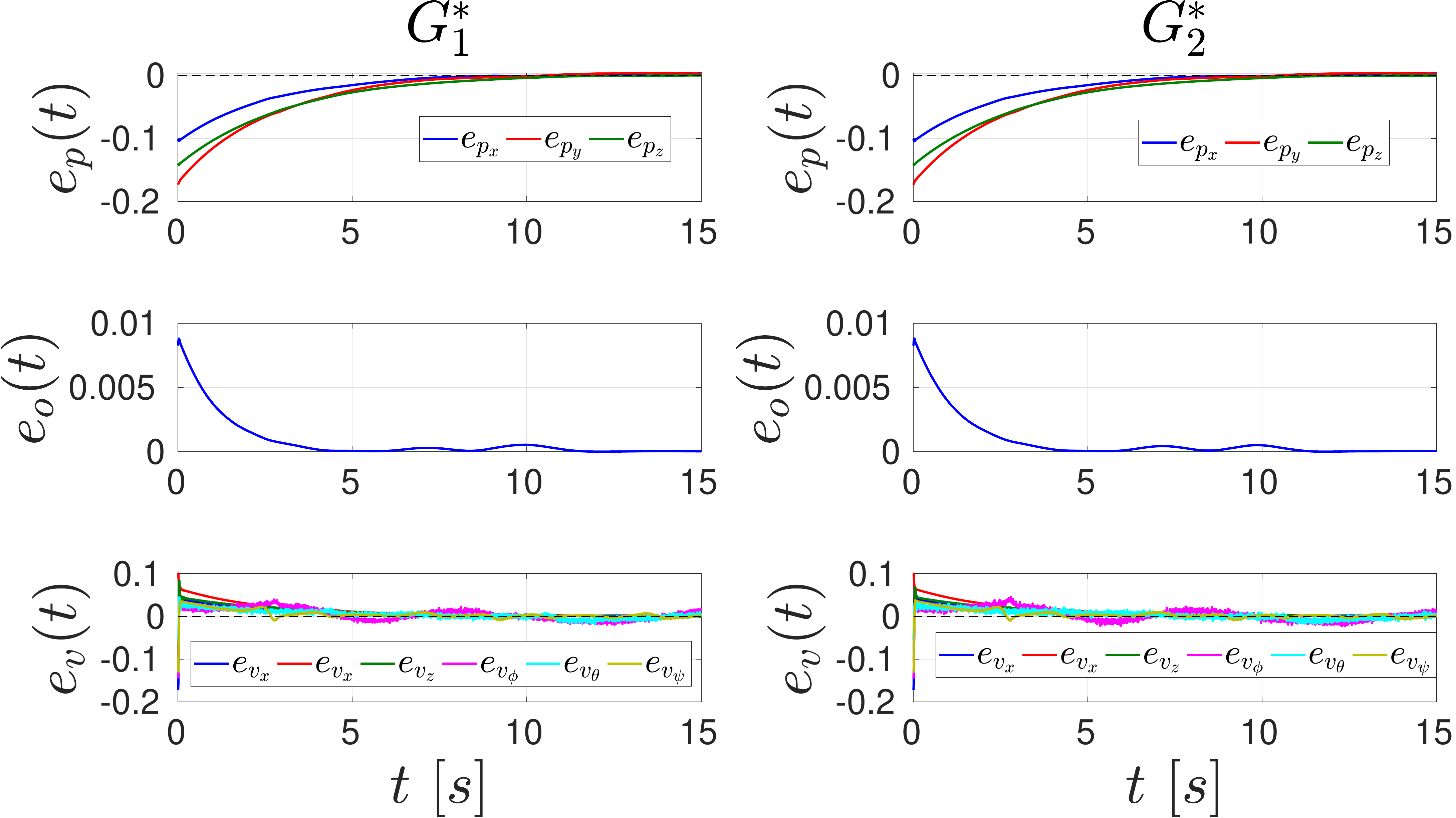}
				\caption{The error metrics $e_p(t)$, $e_{\scr O}(t)$, $e_v(t)$, respectively, top to bottom, for the two choices $G^\ast_1$ and $G^\ast_2$ and $t\in[0,15]$ seconds.\label{fig:errors (rigid+coopmanip)}}	
			\end{figure}

\begin{figure}
				\centering
				\includegraphics[width = .9\textwidth]{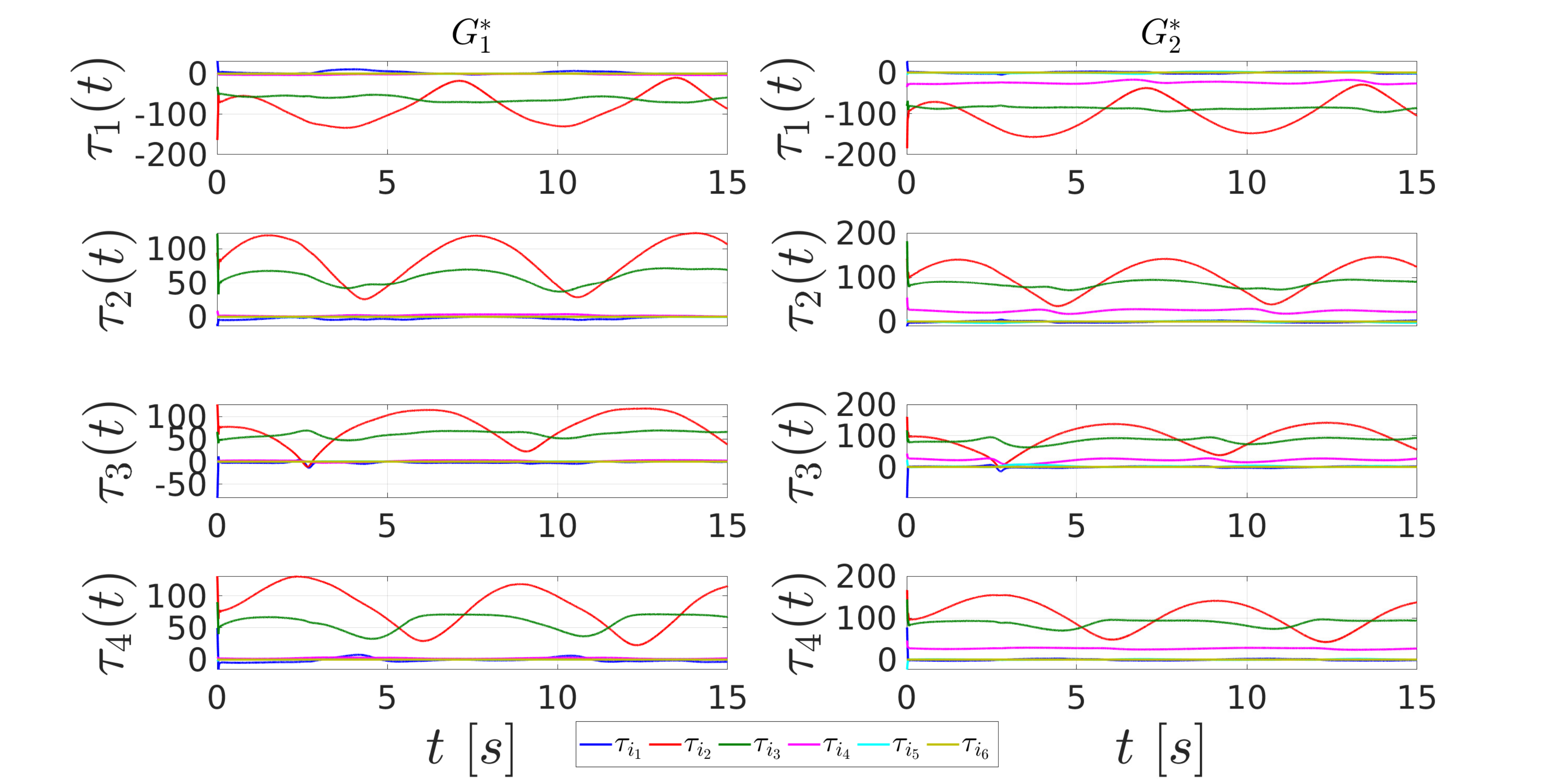}
				\caption{The resulting control inputs $\tau_i(t)$ for $G^\ast_1$ (left) and $G^\ast_2$ (right), $\forall i\in\{1,\dots,4\}$ and $t\in[0,15]$ seconds.\label{fig:inputs (rigid+coopmanip)}}
			\end{figure}

\begin{figure}
				\centering
				\includegraphics[width = .75\textwidth]{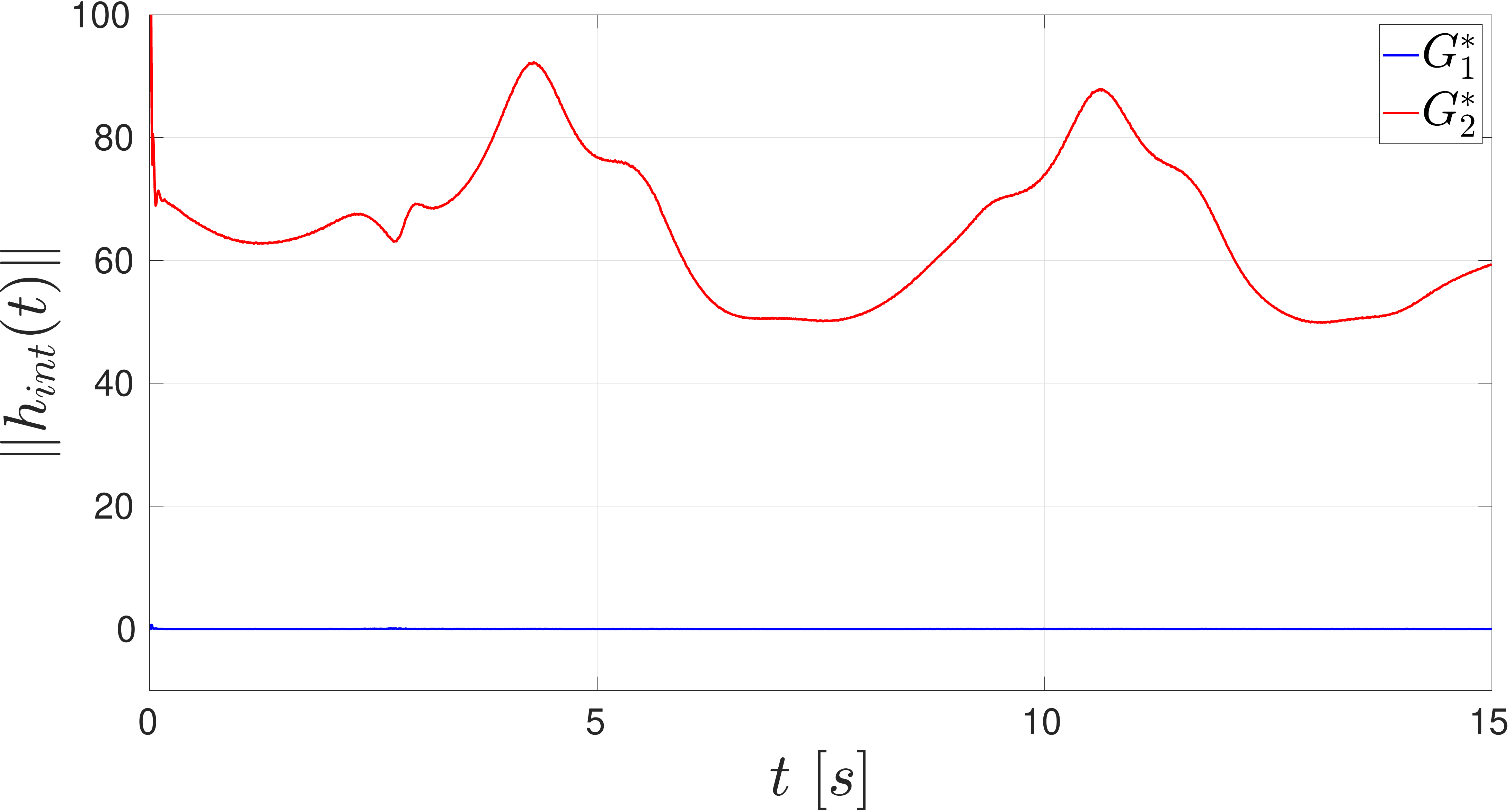}
				\caption{The norm of the internal forces $\|h_\text{int}(t)\|$ (as computed via \eqref{eq:internal forces task-space (rigid+coopmanip)}) for the two cases of $G^\ast$ and $t\in[0,15]$ seconds.\label{fig:h_int (rigid+coopmanip)}}
			\end{figure}

\begin{figure}
				\centering
				\includegraphics[width = .75\textwidth]{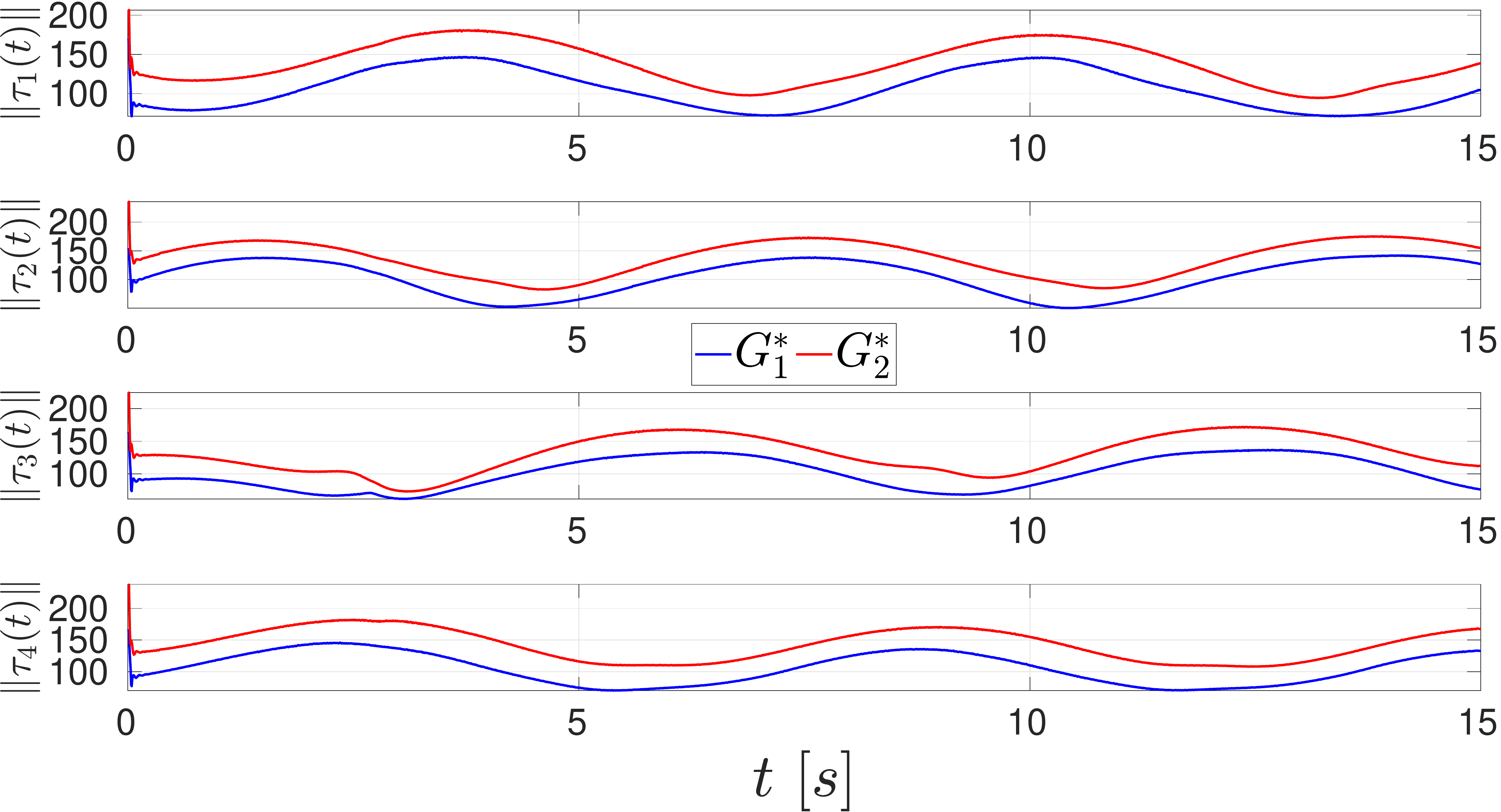}
				\caption{The norms of the resulting control inputs, $\|\tau_i(t)\|$ for $G^\ast_1$ (with blue) and $G^\ast_2$ (with red), $\forall i\in\{1,\dots,4\}$, and $t\in[0,15]$ seconds.\label{fig:inputs_norm (rigid+coopmanip)}}
			\end{figure}

			{
				The results are given in {Figs. \ref{fig:errors (rigid+coopmanip)}-\ref{fig:inputs_norm (rigid+coopmanip)}} for $15$ seconds. Fig. \ref{fig:errors (rigid+coopmanip)} depicts the pose and velocity errors $e_p(t)$, $e_{\scr O}(t)$, $e_v(t)$, which are shown to converge to zero for both choices of $G^\ast$, as expected. The control inputs $\tau_i(t)$ of the agents are shown in Fig. \ref{fig:inputs (rigid+coopmanip)}. Moreover,		
				the norm of the internal forces, $\|h_{\text{int}}(t)\|$, is computed via \eqref{eq:internal forces task-space (rigid+coopmanip)} and shown in Fig. \ref{fig:h_int (rigid+coopmanip)}. It is clear that $G^\ast_2$ yields significantly large internal forces, whereas  $G^\ast_1$ keeps them very close to zero, as proven in the theoretical analysis. The larger internal forces in the case of $G^\ast_2$ are associated with the larger control inputs $\tau_i$. This can be concluded from Fig. \ref{fig:inputs (rigid+coopmanip)} and is also more clearly visualized in Fig. \ref{fig:inputs_norm (rigid+coopmanip)}, which depicts the norms $\|\tau_i(t)\|$ for the two choices of $G^\ast$, $\forall i\in\{1,\dots,4\}$. It is clear that inputs of larger magnitude occur in the case of $G^\ast_2$, which create internal forces (in the nullspace of $G$). {A video illustrating the aforementioned simulations can be found on \href{https://youtu.be/a31LTBBkE-Q}{https://youtu.be/a31LTBBkE-Q}.} 
			}
			
\begin{figure}
				\centering
				\includegraphics[width = 0.75\textwidth]{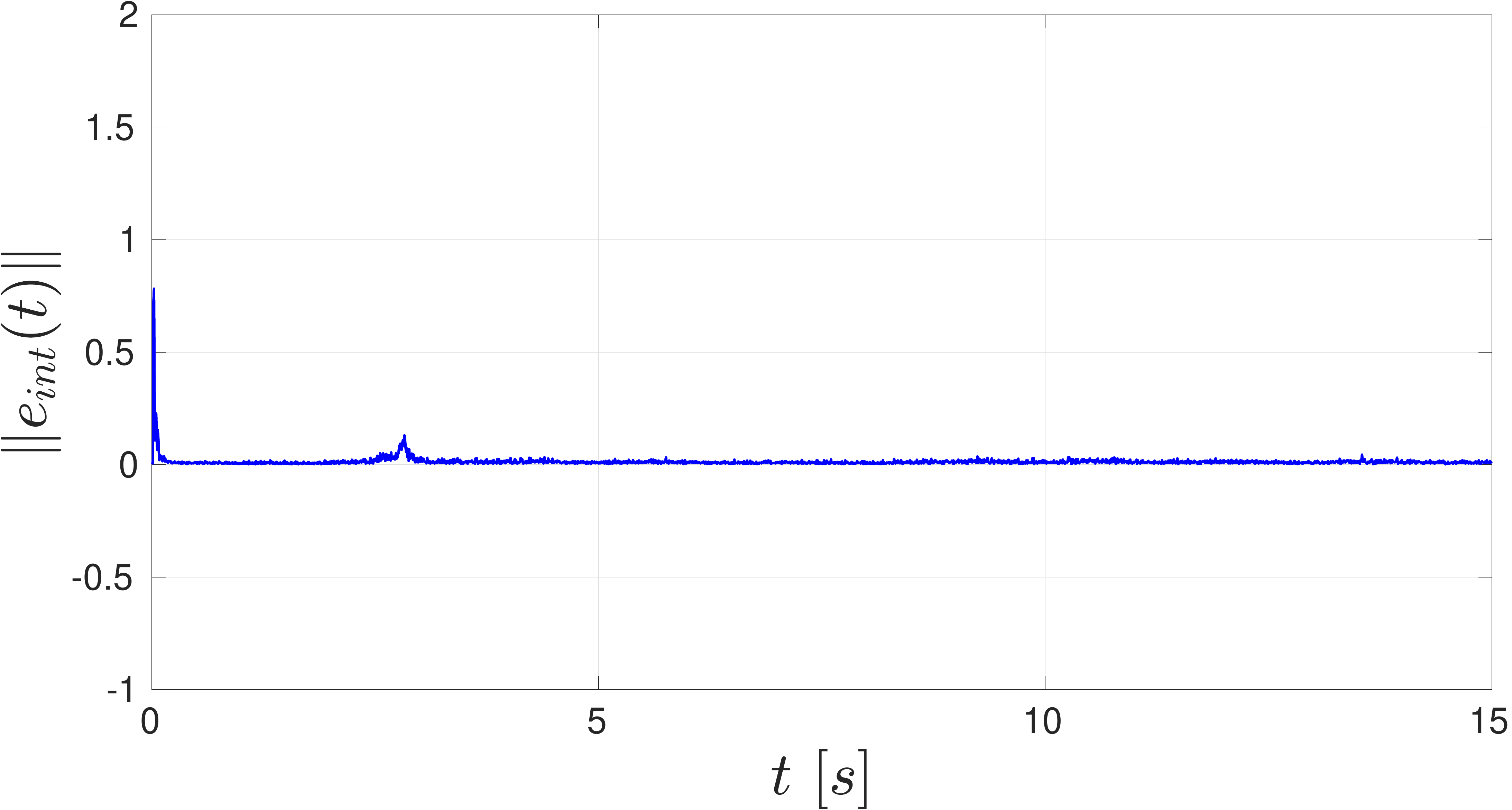}
				\caption{The norm of the internal force error $\|e_{\text{int}}(t)\| $, when using $G^\ast_1$ and for $t\in[0,15]$ seconds.\label{fig:e_int (rigid+coopmanip)}}
			\end{figure}

			{Finally, we set a random force vector $h_{\text{int,d}}$ in the nullspace of $G$ and we simulate the control law \eqref{eq:u rot mat (rigid+coopmanip)} with the extra component $u_{\text{int,d}}$ (see Corollary \ref{corol:h int d (rigid+coopmanip)}). Fig. \ref{fig:e_int (rigid+coopmanip)} illustrates the error norm $\|e_{\text{int}}(t)\|\coloneqq \|h_{\text{int,d}}(t) - h_{\text{int}}(t)\|$, which evolves close to zero. The minor observed deviations can be attributed to model uncertainties and hence the imperfect cancellation of the respective dynamics via \eqref{eq:u rot mat (rigid+coopmanip)}. }
			
\section{Conclusion} \label{sec:concl_formation}

This chapter focused on multi-agent formation control design as well as its connection to rigid cooperative manipulation. Firstly, we developed a model-free decentralized control protocol for distance- and orientation-based formation control for a class of multi-agent systems modeled by Newton-Euler dynamics. Collision avoidance as well as connectivity maintenance was guaranteed to be satisfied by the proposed feedback control scheme. Secondly, we linked rigidity theory to rigid cooperative manipulation, by relating the former's rigidity matrix to the latter's grasp matrix. Moreover, we provided novel conditions for the internal force-free rigid cooperative manipulation.

\chapter{Continuous Coordination of Multi-Agent Systems} \label{chapter:synthesis}
As discussed in Chapter \ref{ch:Introduction}, in order to be able to express complex tasks as temporal logic formulas, we need to have well-defined discrete representations of the continuous multi-agent system. Intuitively, this implies an appropriate discretization of the multi-agent state space, as well as the design of control schemes to navigate the agents among the points of this discretization. At the same time and since we are mainly interested in physical robotic agents, we need to guarantee safe multi-agent behavior, i.e., guarantee collision avoidance among the robotic agents and with potential workspace obstacles. On the same vein, we are interested in developing decentralized schemes, where the agents have local feedback only with respect to the neighbors. Therefore, multi-agent connectivity maintenance is another critical property we impose. Finally, as mentioned before, real robotic agents' equations of motion cannot be accurately known (model uncertainties) and are also subject to external disturbances. Hence, the control design needs to be robust and compensate appropriately for this partial model information (see the previous chapters, where adaptive control and PPC were used). 
 
Motivated by the above, this chapter presents continuous control schemes for the coordination of multi-agent robotic systems. More specifically, we address the following three problems. Firstly, we develop an algorithm for the single- as well as multi-agent go-to-goal and collision-avoidance problem for robotic agents with uncertain dynamics. Secondly, we develop a novel leader-follower scheme for the navigation of a leader to a predefined point subject to model uncertainties and collision avoidance and connectivity maintenance constraints. Since the aforementioned algorithms consider mainly spherical robotic agents, we finally introduce a control scheme that guarantees collision avoidance between robotic agents of ellipsoidal shape.

\section{Introduction} \label{sec:intro (cont_MAS)}

As mentioned in the previous chapters, multi-agent systems have received a large amount of attention lately, due to the advantages they bring with respect to single-agent setups. Apart from cooperative robotic manipulation and formation control, important multi-agent tasks, applicable to real robotic systems and studied in this chapter, consist of multi-robot navigation and leader-follower coordination. Moreover, we impose certain transient properties on the multi-agent system, such as collision avoidance \cite{wang2017safety,sabattini2015decentralized,claes2018multi,cai2007collision,li2013finite}, and/or connectivity maintenance \cite{zavlanos2011graph,sabattini2013distributed,ji2007distributed,sun2018robust,kantaros2016global,turpin2014capt, zavlanos2008distributed, zavlanos2007potential}, both crucial properties for real robotic systems. At the same time, we aim at developing control schemes that compensate for potentially uncertain dynamics of the robotic agents.

Multi-robot navigation with collision avoidance, possibly also with works-pace obstacles, is a special instance of the motion planning problem \cite{latombe2012robot,choset2005principles}. Several techniques have been developed in the related literature for robot motion planning with obstacle avoidance, such as discretization of the continuous space and employment of discrete algorithms (e.g., Dijkstra, $A^\star$), probabilistic roadmaps, sampling-based motion planning, and feedback-based motion planning \cite{lavalle2006planning}. The latter offers closed-form analytic solutions by usually evaluating appropriately designed artificial potential fields, avoiding thus the potential complexity of employing discrete algorithms or discretizing the robot workspace. At the same time, feedback-based methods provide a solution to the control aspect of the motion planning problem, i.e., the correctness based on the solution of the closed-loop differential equation that describes the robot model. 

Feedback-based motion planning has been receiving attention for more than two decades. Early works established the Koditschek-Rimon navigation function (KRNF) \cite{koditschek1990robot,rimon1992exact}, where the robot successfully converges to its goal while avoiding all obstacles from almost all initial conditions (in the sense of a measure-zero set), if the control gain of the goal term is chosen greater than a predefined constant. At the same time, an artificial potential fields based on harmonic functions and the panel method was proposed in \cite{kim1992real}. KRNFs were extended to more general workspaces and adaptive gain controllers  \cite{filippidis2011adjustable,filippidis2012navigation}, to multi-robot systems \cite{loizou2002closed,dimarogonas2006feedback,roussos2013decentralized}, and more recently, to convex potentials and obstacles \cite{paternain2017navigation}. The idea of gain tuning has been also employed to an alternative KRNF in \cite{tanner2005towards}. 

Tuning-free constructions of artificial potential fields have also been developed in the related literature; \cite{loizou2003closed} considers dynamic obstacles and non-smooth controllers, \cite{panagou2017distributed} tackles nonholonomic multi-robot systems, and in \cite{loizou2011closed,loizou2017navigation,vlantis2018robot} harmonic functions are combined with adaptive controllers for the goal gain to achieve almost global safe navigation. Harmonic functions are also used in \cite{waydo2003vehicle,szulczynski2011real}. A transformation of arbitrarily shaped worlds to points worlds, which facilitates the motion planning problem, is also considered in \cite{loizou2017navigation,vlantis2018robot} and in \cite{loizou2014multi} for multi-robot systems. The recent works of \cite{loizou2017navigation} and \cite{vrohidis2018prescribed} guarantee also safe navigation in a predefined \textit{time}.

Barrier functions for multi-robot collision avoidance are employed in \cite{wang2017safety} and optimization-based techniques via model predictive control (MPC) can be found in \cite{filotheou2018,mendes2017real}; \cite{van2011reciprocal} and \cite{roelofsen2017collision} propose reciprocal collision obstacle by local decision making for the desired velocity of the robot(s). Sensing uncertainties are taken into account in \cite{rodriguez2011collision}. 
A recent prescribed performance methodology for ellipsoidal obstacles is proposed in \cite{stavridis2017dynamical} and \cite{Grush18obstacle} extends a given potential field to $2$nd-order systems. A similar idea is used in \cite{montenbruck2015navigation}, where the effects of an unknown drift term in the dynamics are examined. Workspace decomposition methodologies with hybrid controllers are employed in \cite{arslan2016exact} for single- and \cite{arslan2016coordinated} for multi-robot systems, respectively; A hybrid controller is also designed in the recent work \cite{berkane2019hybrid}; \cite{mujahed2017admissible} employs reactive collision avoidance using admissible gaps, and \cite{slotine19avoidance} employs a contraction-based methodology that can also tackle the case of moving obstacles.

A common assumption that most of the aforementioned works consider is the simplified robot dynamics, i.e., single integrators/unicycle kinematics, without taking into account any robot dynamic parameters and where the control input is the robot velocity. Hence, indirectly, the schemes depend on an embedded internal system that converts the desired signal to the actual robot actuation command. The above imply that the actual robot trajectory might deviate from the desired one, jeopardizing its safety and possibly resulting in collisions. 

Second-order realistic robot models are considered in MPC-schemes, like \cite{mendes2017real,filotheou2018}. Such optimization techniques, however, might result in computationally expensive solutions for large horizons. Moreover, regarding model uncertainties, a global upper bound is required, which is used to enlarge the obstacle boundaries and might yield infeasible solutions. A $2$nd-order model is considered in \cite{stavridis2017dynamical}, without, however, considering any unknown dynamic terms. The same holds for \cite{Grush18obstacle}, where an already given potential function is extended to $2$nd-order systems. The works \cite{koditschek1991control,dimarogonas2006feedback,loizou2011closed,arslan2017smooth} consider simplified $2$nd-order systems with \textit{known} dynamic terms (and in particular, inertia and gravitational terms that are assumed to be successfully compensated); \cite{montenbruck2015navigation} guarantees the asymptotic stability of $2$nd-order systems with a class of unknown drift terms to the critical points of a given potential function. However, there is no characterization of the region of attraction of the goal by analyzing the equilibrium points of the whole closed-loop system. 

Another important feature of multi-agent systems is their coordination under leader-follower architectures, where an assigned leader aims at executing a task, and the rest of the team is concerned with secondary tasks, such as staying connected with the leader, forming a desired formation, or performing consensus protocols \cite{hu2010distributed,zhang2015leader,li2013distributed,liu2017adaptive,guo2016communication,mei2011distributed,gustavi2010sufficient}. When robotic teams are concerned, such schemes resemble cases where a leader agent contains information regarding a task, and the followers need to comply with certain specifications to aid the leader. 

Most leader-follower schemes in the related literature consider the follower consensus problem with fixed or time varying communication graphs, where the followers' states converge to the leader's one, which is assumed to have bounded velocity/acceleration \cite{hu2010distributed,zhang2015leader,li2013distributed,liu2017adaptive,guo2016communication,mei2011distributed}. Moreover, connectivity maintenance in the transient state is also taken into account in a variety of leader-follower works (e.g., \cite{ji2007distributed, gustavi2010sufficient,guo2016communication}).
Such schemes cannot be extended to multi-robot systems though, since collision avoidance is of uttermost importance and it is unreasonable to consider the convergence of the agents' states (e.g., positions) to the same value. Vehicular platoons are special cases of leader-follower structures where collision avoidance is taken into account \cite{verginis2015decentralized,verginis2018platoon,sadraddini2017provably}, restricted, however, to the longitudinal platoon-type sensing/communication graph. 

Moreover, as discussed before, many of the multi-agent works in the related literature 
consider simplified/known dynamics  (\cite{ren2005survey,oh2015survey,kantaros2016global,turpin2014capt,egerstedt2001formation, tanner2005towards, zavlanos2008distributed, panagou2017distributed,dimarogonas2006feedback,li2013distributed,ji2007distributed,zavlanos2011graph,gustavi2010sufficient,guo2016communication,hu2010distributed,sabattini2013distributed,sabattini2011arbitrarily,mastellone2008formation,zavlanos2007potential}), which can have crucial effects on the actual behavior of real robotic systems, whose dynamics are described accurately by Lagrangian models, jeopardizing their performance/safety. More complex/uncertain dynamics are taken into account in \cite{zhang2015leader,sun2018robust,liu2017adaptive}, without considering collision specifications; \cite{li2013finite} integrates collision avoidance  with finite boundedness of the inter-agent distances, and \cite{cai2007collision,claes2018multi,van2011reciprocal} deal with the multi-robot collision avoidance problem, without, however, providing theoretical guarantees with respect to the robot dynamics. Gain tuning is also performed in several works to cancel unknown nonlinearities, which are assumed to be uniformly bounded.
An MPC methodology is developed in \cite{filotheou2018}, which can be computationally infeasible in real-time when complex dynamics are considered.

As discussed before, collision avoidance is considered to be a crucial property in real robotic systems, and is tackled in a large variety of multi-robot works. The majority of the related works, however, considers spherical agents, which provide a straightforward metric for the inter-agent or the agent-to-obstacle distances. 
However, since the shapes of real robotic vehicles can be far from {spherical} (e.g., robotic manipulators), that approach can be too conservative and may prevent the agents from fulfilling their primary objectives. Ellipsoids, on the other hand, can approximate more accurately the volume of autonomous agents.

{The authors in \cite{rimon1992exact,loizou2017navigation,vlantis2018robot} employ diffeomorphisms to transform arbitrarily-shaped obstacles, including ellipsoids, to points. This methodology, however, is not straightforwardly extendable to the case of moving obstacles (i.e., multiple autonomous agents). A point-world transformation of multi-agent systems {was} taken into account in \cite{tanner2003nonholonomic}. 
As described in \cite{tanner2003nonholonomic} though, each agent's transformation deforms the other agents into shapes whose implicit closed-form equation (and hence a suitable distance metric) is not trivial to obtain.
The methodology of \cite{loizou2017navigation} provides useful insight, where the volume of each agent is ``absorbed" to the other agents via Minkowski sums. The closed-form implicit equation of the resulting shapes, however, although possible to obtain \cite{yan2015closed}, cannot be used to derive an appropriate distance metric in a straightforward way; \cite{rimon1997obstacle} derives a conservative inter-ellipsoid distance by employing ellipsoid-to-sphere transformations and eigenvalue computations. An arithmetic algorithm that produces velocities for inter-agent elliptical agents is derived in \cite{best2016real}, without, however, theoretical guarantees.
Optimization-based techniques (e.g., Model Predictive Control), which can be employed for collision avoidance of convex-shaped agents (like e.g., in Chapter \ref{chapter:cooperative manip}), can be too complex to solve, especially in cases where the control must be decentralized and/or complex dynamics are considered. The latter property constitutes another important issue regarding the related literature. In particular, most related works consider simplified single- or double-integrator models, which deviate from the actual dynamics and can lead to performance decline and safety jeopardy. 

Barrier functions constitute a suitable tool for expressing objectives like collision avoidance. Originated in optimization, they are continuous functions that diverge to infinity as their argument approaches the boundary of a desired/feasibly region. Barrier Lyapunov-like functions for general control systems can be found in \cite{romdlony2016stabilization,xu2018constrained}, and in \cite{wang2017safety,panagou2016distributed,lindemann2019control} for multi-agent systems, for obstacle avoidance with {spherical} obstacles/agents and time-dependent tasks. 

This chapter deals with the following three problems. 
Firstly, we consider the robot navigation in an obstacle-cluttered environment under $2$nd-order uncertain robot dynamics. The considered uncertainties consist of (i) unknown friction/drag terms, which are hard to model accurately, and (ii) unknown mass and unmodeled dynamics, motivated by transportation of objects of unknown mass or fuel consumption along a robot task, or cases where the robot is enhanced with other parts (e.g., robotic manipulators), whose dynamics are not known.  We design a novel $2$nd-order smooth navigation function which is integrated with adaptive control laws that compensate for the uncertain terms. Extensive analysis of the equilibrium points of the closed-loop system shows convergence to the desired goal from almost all initial conditions while avoiding obstacle collisions with the workspace boundary and spherical obstacles. The proposed scheme is then extended to star-worlds, i.e., workspaces with star-shaped obstacles \cite{rimon1992exact}. Finally, using the single-robot methodology, we propose a \textit{decentralized} hybrid coordination algorithm for the navigation of a multi-robot system in an environment cluttered with spherical obstacles.

Secondly, we propose a decentralized control protocol for the coordination of a multi-agent system with $2$nd order uncertain Lagrangian dynamics, subject to collision avoidance and connectivity maintenance. In particular, we consider that a leader agent has to navigate to a desired pose, inter-agent collisions must be avoided, and some of the initially connected agents have to remain connected. {We are mainly motivated by cases where a cooperative task (e.g., cooperative pick-and-place tasks) is assigned to a multi-agent system, but the details are given only to a leader agent, which has to lead the entire team along the desired task. 
{By using certain properties of the incidence matrix, we avoid issues of local minima and we relax the assumptions on the connectivity of the graph (as opposed to, e.g., \cite{gustavi2010sufficient,li2013distributed}) as well as the access of the leader's velocity by the followers. Moreover,	
we consider uncertain terms and unknown external disturbances in the dynamic model, which we cope with by using adaptive and discontinuous control laws. 

Finally, we design smooth closed-form barrier functions for the collision avoidance of ellipsoidal agents. By employing results from the computer graphics field, we 
derive a novel closed-form expression that represents a distance metric of two ellipsoids in $3$D space. Moreover, we use the latter to design a control protocol that guarantees the collision avoidance of a multi-agent system that aims to achieve a primary objective,
subject to uncertain $2$nd-order Lagrangian dynamics. The derived control law is (i) decentralized, in the sense that each agent calculates its control signal based on local information, (ii) discontinuous and adaptive, in order to compensate for the uncertainties and external disturbances.

\section{Adaptive Robot Navigation with Collision Avoidance Subject to $2$nd-order Uncertain Dynamics} \label{sec:adaptive nav (Automatica_adaptive_nav)}

We first consider the problem of single-robot navigation in a workspace cluttered with obstacles, subject to $2$nd-order dynamics, whose analysis is necessary for the extension to the more general multi-robot problem.

\subsection{Problem Statement}  \label{sec:PF (Automatica_adaptive_nav)}

Consider a spherical robot operating in a bounded workspace $\mathcal{W}$, characterized by its {position vector} $x \in \mathbb{R}^n$, $n\in\{2,3\}$ and radius $r > 0$, and subject to the dynamics:
\begin{subequations} \label{eq:dynamics (Automatica_adaptive_nav)}	
	\begin{align}
	& \dot{x}= v \\
	& m \dot{v} + f(x,v) + mg = u,
	\end{align}
\end{subequations}
where $m > 0$ is the \textit{unknown} mass, $g \in \mathbb{R}^n$ is the constant gravity vector, $u\in\mathbb{R}^n$ is the input vector, and $f:\mathbb{R}^{2n}\to \mathbb{R}^n$ is a friction-like function, satisfying the following assumption:
\begin{assumption} \label{ass:f (Automatica_adaptive_nav)}
	The function $f:\mathbb{R}^{2n}\to \mathbb{R}^n$ is analytic and satisfies 	
	\begin{equation*}
	\|f(x,v)\| \leq \alpha \|v\|,
	\end{equation*} 
	$\forall x,v \in \mathbb{R}^{2n}$, where $\alpha \in \mathbb{R}_{\geq 0}$ is an unknown positive constant.
\end{assumption}	
{The aforementioned assumption is a standard condition concerning friction-like terms, which are bounded by the robot velocity \cite{de1995new,makkar2005new}. Constant unknown friction terms could be also included in the dynamics (e.g., incorporated in the gravity vector).}
Note also that $\|f(x,v)\| \leq \alpha \|v\|$ implies $f(x,0) = 0$, and $\frac{\partial f(x,v)}{\partial x} \Big|_{v=0} = 0$.
The workspace is assumed to be an open ball centered at the origin 
\begin{equation} \label{eq:workspace (Automatica_adaptive_nav)}
\mathcal{W} \coloneqq \mathcal{B}(0,r_\mathcal{W}) = \{z \in \mathbb{R}^n : \|z\| < r_{\mathcal{W}} \},
\end{equation}
where $r_{\mathcal{W}} > 0$ is the workspace radius. The workspace contains $M\in\mathbb{N}$ closed sets $\mathcal{O}_j$, $j\in\mathcal{J} \coloneqq \{1,\dots,M\}$, corresponding to obstacles. Each obstacle is a closed ball centered at $c_j\in\mathbb{R}^3$, with radius $r_{o_j} > 0$:
\begin{equation*} 
\mathcal{O}_j \coloneqq \bar{\mathcal{B}}(c_j,r_{o_j}) = \{z\in\mathcal{W} : \|z-c_j\|\leq r_{o_j} \}, \ \ \forall j\in\mathcal{J}.
\end{equation*} 
The analysis that follows will be based on the transformed workspace: 
\begin{equation} \label{eq:transf. workspace (Automatica_adaptive_nav)}
\bar{\mathcal{W}} \coloneqq \{z \in \mathbb{R}^n : \|z\| < \bar{r}_\mathcal{W} \coloneqq r_\mathcal{W} - r\},
\end{equation}  
and the set of obstacles
\begin{equation*}
\bar{\mathcal{O}}_j \coloneqq \{z\in\mathcal{W} : \|z-c_j\|\leq \bar{r}_{o_j}\coloneqq r_{o_j} + r \}, \ \ \forall j\in\mathcal{J}.
\end{equation*}
and the robot is reduced to the point $x$.
The free space is defined as 
\begin{equation} \label{eq:sphere world (Automatica_adaptive_nav)}
\mathcal{F} \coloneqq \bar{\mathcal{W}} \backslash \bigcup_{j\in\mathcal{J}} \bar{\mathcal{O}}_j,
\end{equation}
also known as a \textit{sphere world} {\cite{koditschek1990robot}}.
We consider the following {common feasibility assumption \cite{koditschek1990robot}} for $\mathcal{F}$:
\begin{assumption} \label{ass: workspace (Automatica_adaptive_nav)}
	The workspace $\mathcal{W}$ {and} the obstacles $\mathcal{O}_j$ satisfy:
	\begin{align*}
	& \|c_i - c_j\| > r_{o_i} + r_{o_j} + 2r \\
	& r_\mathcal{W} - \|c_j\| > r_{o_j} + 2r.
	\end{align*}
\end{assumption}
The aforementioned assumption implies essentially that there is enough space among the obstacles and the workspace boundary and the obstacles for the robot to navigate, or equivalently, $\bar{\mathcal{O}}_j \subset \bar{\mathcal{W}}$ and $\bar{\mathcal{O}}_i \cap \bar{\mathcal{O}}_j = \emptyset$, $\forall i,j \in \mathcal{J}$, with $i\neq j$.

Moreover, Assumption \ref{ass: workspace (Automatica_adaptive_nav)} implies that we can find some $\bar{r} > 0$ such that 
\begin{subequations} \label{eq:r_bar (Automatica_adaptive_nav)}
	\begin{align}
	& \|c_i - c_j\| > r_{o_i} + r_{o_j} + 2r + 2\bar{r}, \ \ \forall i,j \in \mathcal{J}, i\neq j, \\
	& r_\mathcal{W} - \|c_j\| > r_{o_j} + 2r + 2\bar{r}, \ \ \forall j\in \mathcal{J}
	\end{align}
\end{subequations}

\begin{figure}[!ht]
	\centering
	\includegraphics[trim = 0cm 0cm 0cm -0cm,width = 0.55\textwidth]{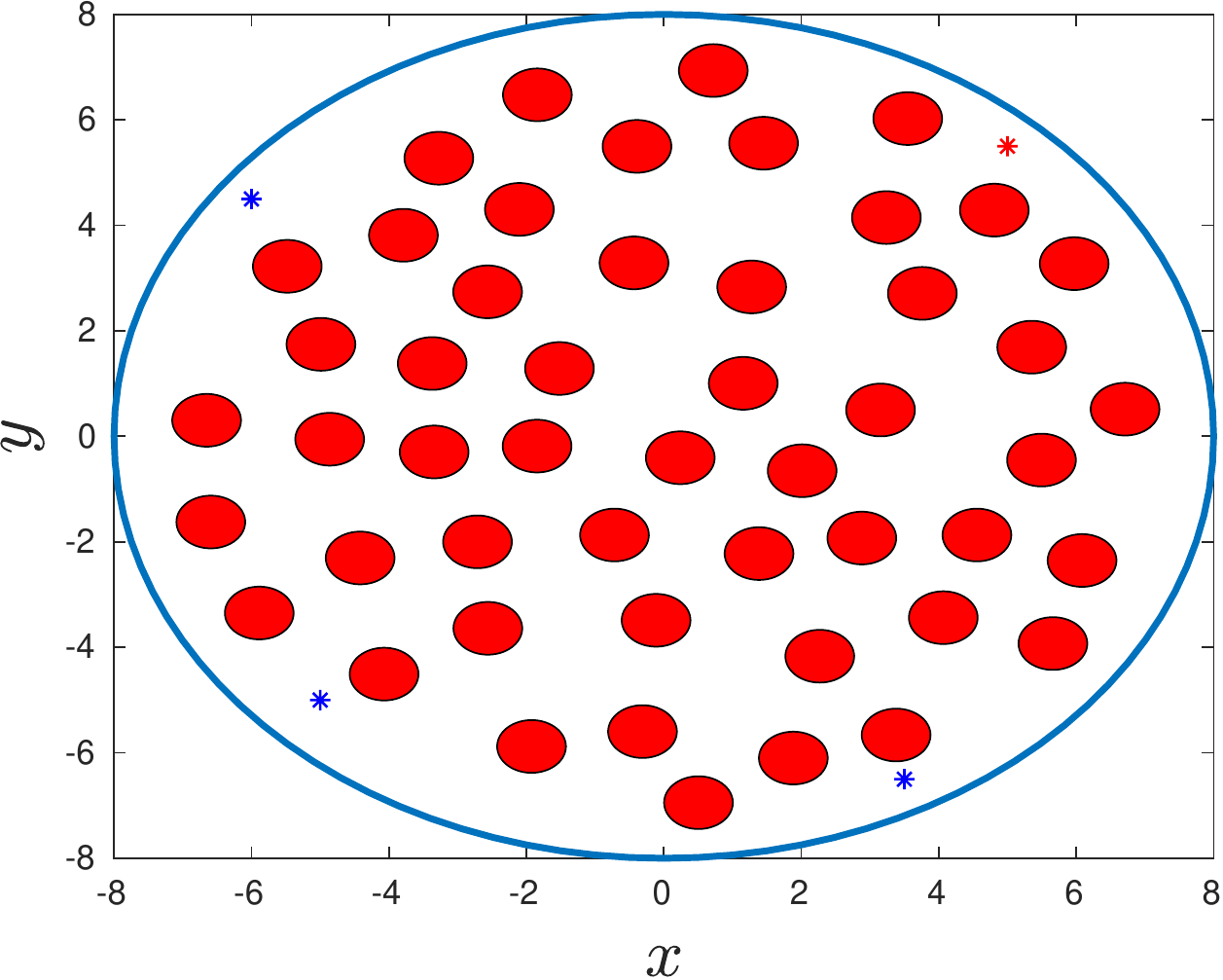}\\
	\caption{A $2$D example of the workspace $\bar{\mathcal{W}}$ with $50$ obstacles $\bar{\mathcal{O}}_j$, $j\in\{1,\dots,50\}$. The blue asterisks indicate potential initial configurations of the robot and the obstacles have been enlarged with the robot radius $r$. The red asterisk indicates a potential goal robot position.}\label{fig:2d_ws_example (Automatica_adaptive_nav)}
\end{figure}

This section treats the problem of navigating the robot to a destination $x_{\text{d}}$ while avoiding the obstacles and the workspace boundary, formally stated as follows:
\begin{problem} \label{prob 1 (Automatica_adaptive_nav)}
	Consider a robot subject to the \textit{uncertain} dynamics \eqref{eq:dynamics (Automatica_adaptive_nav)}, operating in the aforementioned sphere world, with $(x(t_0), v(t_0)) \in \mathcal{F}\times\mathbb{R}^n$. Given a destination $x_{\text{d}}\in \mathcal{F}$, design a control protocol $u$ such that 
	\begin{align*}
	& x(t) \in \mathcal{F}, \ \ t \geq t_0 \\
	& \lim_{t\to\infty}(x(t),v(t))  = (x_{\text{d}},0)
	\end{align*} 
\end{problem}
An illustration of the considered workspace is provided in Fig. \ref{fig:2d_ws_example (Automatica_adaptive_nav)}.

\subsection{Single-Agent Solution} \label{sec:main (Automatica_adaptive_nav)}

We provide in this section our methodology for solving Problem \ref{prob 1 (Automatica_adaptive_nav)}. Define first  the set $\bar{\mathcal{J}} \coloneqq \{0\}\cup\mathcal{J}$ as well as the distances $d_j\coloneqq d_j(x): \mathcal{F} \to \mathbb{R}_{\geq 0}$, $j\in\bar{\mathcal{J}}$, with $d_j(x) \coloneqq \|x-c_j\|^2 - \bar{r}_{o_j}^2$, $\forall j\in\mathcal{J}$, and $d_0\coloneqq d_0(x) \coloneqq \bar{r}_{\mathcal{W}}^2 - \|x\|^2$. Note that, by keeping $d_j(x) > 0$, $d_0(x) > 0$, we guarantee that $x \in \mathcal{F}$\footnote{A safety margin can also be included, which needs, however, to be incorporated in the constant $\bar{r}$ of \eqref{eq:r_bar (Automatica_adaptive_nav)}.}.

We introduce first the notion of the \textit{$2$nd-order navigation function}: 
\begin{definition} \label{def:2nd nf (Automatica_adaptive_nav)}
	A \textit{$2$nd-order navigation function} is a function $\phi \coloneqq \phi(x): \mathcal{F} \to \mathbb{R}_{\geq 0}$ of the form 
	\begin{equation} \label{eq:potential function (Automatica_adaptive_nav)}
	\phi(x) \coloneqq k_1\|x - x_{\text{d}}\|^2 + k_2\sum_{j\in\bar{\mathcal{J}}}\beta(d_j(x)),
	\end{equation}
	where $\beta:\mathbb{R}_{>0}\to\mathbb{R}_{\geq 0}$ is a (at least) twice contin. differentiable function and $k_1, k_2$ are positive constants, with the followings properties:
	\begin{enumerate}
		\item $\beta((0,\tau])$ is strictly decreasing,   				
		$\lim_{z \to 0} \beta(z) = \infty$, and $\beta(z) = \beta(\tau)$, $\forall z \geq \tau$, $j\in\bar{\mathcal{J}}$, for some $\tau > 0$, 
		\item $\phi(x)$ has a global minimum  at $x = x_{\text{d}} \in \text{int}(\mathcal{F})$ where $\phi(x_{\text{d}}) = 0$,			
		\item if $\beta'(d_k(x)) \neq 0$ {and $\beta''(d_k(x)) \neq 0$} for some $k\in\bar{\mathcal{J}}$, then $\beta'(d_j(x)) = \beta''(d_j(x)) = 0$, for all $j\in\bar{\mathcal{J}} \backslash\{k\}$, where $'$ and $''$ denote function derivatives.
		
		\item The function $\widetilde{\beta}:(0,\tau) \to \mathbb{R}_{\geq 0}$, with
		\begin{equation*}
		\widetilde{\beta}(z) \coloneqq \beta''(z) z\sqrt{z} 
		\end{equation*}
		is strictly decreasing.
	\end{enumerate}	
\end{definition}

By using the first property we will guarantee that, by keeping $\beta(d_j(x))$ bounded, there are no collisions with the obstacles or the free space boundary. Property $2$ will be used for the asymptotic stability of the desired point $x=x_{\text{d}}$. Property $3$ places the rest of the critical points of $\phi$  (which are proven to be saddle points) close to the obstacles, and the last property is used to guarantee that these are non-degenerate.

Examples for the function $\beta$ that satisfy {properties 1) and 4)} are 
\begin{equation*}
\beta(z) \coloneqq 
\begin{cases} \displaystyle
\bar{\beta}\frac{\exp\left(-\frac{1}{z}\right) + \exp\left(-\frac{1}{\tau - z}\right)}{\exp\left(-\frac{1}{z}\right)}, &  z \leq \tau \\
\bar{\beta}, & z \geq \tau,
\end{cases}	 
\end{equation*}
for any positive $\bar{\beta}$ and sufficiently small $\tau$, or the functions
\begin{align}
\beta(z) \coloneqq& 
\begin{cases} \displaystyle
\frac{1}{6z^5 - 15z^4 + 10z^3}, &  z \leq 1\\
1, & z \geq 1,
\end{cases} \label{eq:beta_exps (Automatica_adaptive_nav)} \\	 
\beta(z) \coloneqq& 
\begin{cases} \displaystyle
\ln^4\left(\frac{z}{\tau}\right), &  z \leq \tau\\
0, & z \geq \tau.
\end{cases} \notag
\end{align}	
Note that $\beta'(z) = \beta''(z) = 0$, for $z \geq \tau$. 	
We define also the constant 
\begin{equation} \label{eq:bar_r_d (Automatica_adaptive_nav)}
\bar{r}_{\text{d}} \coloneqq \min \left\{\bar{r}_{\mathcal{W}}^2-\|x_{\text{d}}\|^2, \min_{j\in\mathcal{J}}\left\{ \|x_{\text{d}} - c_j\|^2 - \bar{r}^2_{o_j} \right\} \right\}	
\end{equation}
as the minimum distance of the goal to the obstacles/workspace boundary.

We prove next that, by appropriately choosing $\tau$, only one $\beta(d_j(x))$, $j\in\bar{\mathcal{J}}$ affects the robotic agent for each $x\in\mathcal{F}$, and furthermore that {$\beta'(d_j(x_\text{d})) = \beta''(d_j(x_\text{d})) = 0$.  Hence, properties 2) and 3) of Def. \ref{def:2nd nf (Automatica_adaptive_nav)} are satisfied}.
\begin{proposition} \label{prop:tau (Automatica_adaptive_nav)}
	By choosing $\tau$ as 
	\begin{equation} \label{eq:tau choice (Automatica_adaptive_nav)}
	\tau \in (0,\min\{\bar{r}^2, \bar{r}_{\text{d}}\}),
	\end{equation}  
where $\bar{r}, \bar{r}_{\text{d}}$ were introduced in \eqref{eq:r_bar (Automatica_adaptive_nav)} and \eqref{eq:bar_r_d (Automatica_adaptive_nav)}, respectively,  we guarantee that at each $x\in \mathcal{F}$ there is not more than one $j\in\bar{\mathcal{J}}$ such that $d_j \leq \tau$, implying that $\beta'(d_j(x))$ and $\beta''(d_j(x))$ are non-zero. 
\end{proposition}	
\begin{proof}
	Assume that $d_j(x) \leq \tau$ for some $j\in\mathcal{J}$, $x\in\mathcal{F}$. Then, in view of \eqref{eq:r_bar (Automatica_adaptive_nav)}, it holds that 
	\begin{align*}
	& \|x - c_j\|^2  < \bar{r}^2 + \bar{r}^2_{o_j} \Rightarrow  \\
	& \|x - c_j\| < \bar{r} + \bar{r}_{o_j} =  \bar{r} + r + r_{o_j} < \|c_j - c_k\|    
	\end{align*}
	$\forall k\in\mathcal{J}\backslash\{j\}$, and hence, 
	\begin{align*}
	\|x - c_k\| &= \|x - c_j + c_j - c_k\|  \\
	& \geq \|c_j - c_k\| - \|x - c_j\|>  r_{o_k} + r + \bar{r} \Rightarrow \\
	\|x - c_k\|^2 &> (r_{o_k} + r + \bar{r})^2 > (r_{o_k} + r)^2 + \bar{r}^2,
	\end{align*}
	and hence $ d_k(x) > \bar{r}^2 > \tau$, $\forall k\in\mathcal{J}\backslash\{j\}$. Moreover, in view of \eqref{eq:r_bar (Automatica_adaptive_nav)}, it holds that 
	\begin{align*}
	\|x\| &\leq \|x-c_j\| + \|c_j\| \pm r_\mathcal{W} \Rightarrow \\
	\|x\| &< r_\mathcal{W} - r -\bar{r} \Rightarrow  (r_{\mathcal{W}}-r)^2  \geq (\|x\| + \bar{r})^2 \Rightarrow \\
	\bar{r}_\mathcal{W}^2 &\geq \|x\|^2 + \bar{r}^2  \Rightarrow  \bar{r}^2_\mathcal{W} - \|x\|^2 > \bar{r}^2, 
	\end{align*}
	and hence $d_o(x) > \tau$. Similarly, we conclude by contradiction that $d_o(x) \leq \tau \Rightarrow d_j > \tau$, $\forall j\in\mathcal{J}$. 
\end{proof}
Moreover, it holds for the desired equilibrium that 
\begin{align*}
&x = x_\text{d} \Leftrightarrow d_j(x) = \|x_\text{d} - c_j\|^2 - \bar{r}_{j}^2 \geq \bar{r}_\text{d} > \tau, 
\end{align*}	
and 
\begin{align*}
& x = x_\text{d} \Leftrightarrow d_0(x) = \bar{r}_\mathcal{W}^2 - \|x_\text{d}\|^2 \geq \bar{r}_\text{d} > \tau,
\end{align*}
and hence $\beta'(d_j(x_\text{d})) = \beta''(d_j(x_\text{d})) = 0$, $\forall j\in\bar{\mathcal{J}}$.

Intuitively, the obstacles and the workspace boundary  have a local region of influence defined by the constant $\tau$, which will play a significant role in determining the stability of the overall scheme later. Moreover, it encompasses also the potential \textit{local} sensing capabilities of the robot, since it takes into account the presence of the obstacles and the workspace boundary only when it is ``$\tau$-close" to them. Similar techniques have been used in the literature, e.g., \cite{arslan2016exact,vrohidis2018prescribed}.
The expressions for the gradient and the Hessian of $\phi$, which will be needed later, are the following:
\begin{subequations} \label{eq:grad + hessian (Automatica_adaptive_nav)}
	\begin{align}
	\nabla_x \phi(x) =& 2k_1(x - x_{\text{d}}) + 2k_2\sum_{j\in\mathcal{J}} \beta'(d_j)(x - c_j)  - 2k_2\beta'(d_0)x \\
	\nabla_x^2 \phi(x) =& 2 \left(k_1 - k_2\beta'(d_0) + k_2\sum_{j\in\mathcal{J}} \beta'(d_j)\right) I_n - 2k_2 \beta''(d_0)xx^\top \notag \\&  + 
	2k_2\sum_{j\in\mathcal{J}}\beta''(d_j)(x - c_j)(x - c_j)^\top.
	\end{align}
\end{subequations}
Given the aforementioned definitions, we design a reference signal $v_{\text{d}}\coloneqq v_{\text{d}}(x):\mathcal{F}\to\mathbb{R}^n$ for the robot velocity $v$ as 
\begin{equation} \label{eq:v_d (Automatica_adaptive_nav)}
v_{\text{d}}(x) = -\nabla_x \phi(x).
\end{equation}
Next, we will design the control input $u$ to guarantee tracking of the aforementioned reference velocity  as well as compensation of the unknown terms $m$ and $f(x,v)$. More specifically, we define the signals $\hat{m}\in\mathbb{R}$ and $\hat{\alpha}\in\mathbb{R}$ as the estimation terms of $m$ and $\alpha$ (see Assumption \ref{ass:f (Automatica_adaptive_nav)}), respectively, and the respective errors $\tilde{m} \coloneqq \hat{m} - m$, $\widetilde{\alpha} \coloneqq \hat{\alpha} - \alpha$. We design now the control law $u:\mathcal{F}\times\mathbb{R}^{n+2}\to\mathbb{R}^n$ as 
\begin{align} \label{eq:u (Automatica_adaptive_nav)}
u \coloneqq u(x,v,\hat{m},\hat{\alpha}) &\coloneqq  -k_\phi \nabla_x \phi(x) + \hat{m}(\dot{v}_{\text{d}} + g) - \left(k_v + \frac{3}{2}\hat{\alpha}\right)e_v,
\end{align}
where $e_v \coloneqq v - v_{\text{d}}$, and $k_v$, $k_\phi$ are positive gain constants. Moreover, we design the adaptation laws for the estimation signals as 
\begin{subequations} \label{eq:adaptation laws (Automatica_adaptive_nav)}
	\begin{align}
	\dot{\hat{m}} \coloneqq& -k_m e_v^\top( \dot{v}_{\text{d}}+g)  \\
	\dot{\hat{\alpha}} \coloneqq & k_\alpha \|e_v\|^2, 
	\end{align}
\end{subequations}
with $k_m$, $k_\alpha$ positive gain constants, $\hat{\alpha}(t_0) \geq 0$, and arbitrary finite initial condition $\hat{m}(t_0)$.
The correctness of the proposed control protocol is established in the following theorem: 
\begin{theorem}\label{th:single robot (Automatica_adaptive_nav)}
	Consider a robot operating in $\mathcal{W}$, subject to the uncertain $2$nd-order dynamics \eqref{eq:dynamics (Automatica_adaptive_nav)}. Given $x_{\textup{d}} \in \mathcal{F}$, the control protocol \eqref{eq:v_d (Automatica_adaptive_nav)}-\eqref{eq:adaptation laws (Automatica_adaptive_nav)}  guarantees the collision-free navigation to $x_{\textup{d}}$ from almost all initial conditions $(x(t_0),v(t_0),\hat{m}(t_0),\hat{\alpha}(t_0)) \in \mathcal{F}\times\mathbb{R}^{{n+1}}\times\mathbb{R}_{\geq 0}$, given a sufficiently small $\tau$ and that $k_\phi > \frac{\alpha}{2}$. Moreover, all closed loop signals remain bounded, $\forall t \geq t_0$.
\end{theorem}  
\begin{remark}
	Note that the proposed potential function \eqref{eq:potential function (Automatica_adaptive_nav)} is, in a sense, equivalent to the one designed in {\cite{koditschek1990robot}}, since the critical points are ``pushed" arbitrarily close to the obstacles and, as shown in the proof of Theorem \ref{th:single robot (Automatica_adaptive_nav)}, they are also non-degenerate by choosing $\tau$ small enough. 
	In contrast to {\cite{koditschek1990robot}}, however, as well as other related works (e.g., {\cite{loizou2011closed,vlantis2018robot,dimarogonas2006feedback}}), we do not require large goal gains (the gain $k_1$ here) in order to establish the correctness of the propose scheme. 
\end{remark}
\begin{remark}
	The proposed scheme can be also extended to \textit{unknown} environments, where the amount and location of the spherical obstacles is unknown a priori, and these are sensed locally on-line. In particular, by having a large enough sensing neighborhood, the location (and possibly the radius) of each obstacle $j\in\mathcal{J}$ can be sensed when $d_j > \tau$, and hence the respective term (which will be zero, since $\beta'(d_j) = 0$, for $d_j > \tau$) can be smoothly incorporated in $\nabla_x \phi(x)$.
\end{remark}
\begin{proof}[Proof of Theorem \ref{th:single robot (Automatica_adaptive_nav)}]
	Consider the Lyapunov candidate function 
	\begin{equation} \label{eq:V Lyap (Automatica_adaptive_nav)}
	V \coloneqq k_{\phi}\phi + \frac{m}{2}\|e_v\|^2 + \frac{3}{4k_\alpha}\widetilde{\alpha}^2 + \frac{1}{2k_m}\widetilde{m}^2.
	\end{equation}		
	Since $x(t_0) \in \mathcal{F}$, there exists a constant $\bar{d}_j$ such that $d_j(x(t_0)) \geq \bar{d}_j > 0$, $j\in\bar{\mathcal{J}}$, which implies the existence of a finite positive constant $\bar{V}_0$ such that $V(t_0) \leq \bar{V}_0$.
	By considering the time derivative of $V$ and using $v = e_v + v_{\text{d}}$ and Assumption \ref{ass:f (Automatica_adaptive_nav)}, we obtain after substituting \eqref{eq:adaptation laws (Automatica_adaptive_nav)}:
	\begin{align*}
	\dot{V} =& k_{\phi} \nabla_x \phi(x)^\top (e_v + v_{\text{d}}) + e_v^\top(u - mg - f(x,v) - m\dot{v}_{\text{d}} ) + \frac{3}{2}\widetilde{\alpha}\|e_v\|^2  \\
	&-\widetilde{m}e_v^\top (\dot{v}_{\text{d}} + g) \\
	\leq & -k_\phi \|\nabla_x \phi(x) \|^2 + e_v^\top(k_{\phi}\nabla_x \phi(x) + u - m(g +\dot{v}_{\text{d}}))  +\alpha\|e_v\|\|v\| \\
	& + \frac{3\widetilde{\alpha}}{2}\|e_v\|^2  - \widetilde{m}e_v^\top (\dot{v}_{\text{d}} + g), 		
	\end{align*}
	which, by substituting \eqref{eq:u (Automatica_adaptive_nav)} and using $\alpha \|e_v\|\|v\| \leq {\alpha}\|e_v\|^2 + \frac{\alpha}{2}\|\nabla_x \phi(x)\|^2 + \frac{\alpha}{2}\|e_v\|^2$, becomes
	\begin{align*}
	\dot{V} \leq & -\left(k_\phi - \frac{\alpha}{2}\right) \|\nabla_x \phi(x) \|^2 - k_v\|e_v\|^2 - \frac{3}{2}\hat{\alpha}\|e_v\|^2  + 
	\frac{3}{2}\alpha\|e_v\|^2  
	+ \widetilde{m}e_v^\top (g+ \\& \dot{v}_{\text{d}}) 
	+  \frac{3\widetilde{\alpha}}{2}\|e_v\|^2 - \widetilde{m}e_v^\top (\dot{v}_{\text{d}}+g) \\
	= & -\left(k_\phi - \frac{\alpha}{2}\right) \|\nabla_x \phi(x) \|^2 - k_v\|e_v\|^2 \leq 0.
	\end{align*}
	Hence, we conclude that $V(t)$ is non-increasing, and hence $\beta(d_j(x(t))) \leq V(t) \leq V(t_0) \leq \bar{V}_0$, $\forall t\geq t_0$, which implies that collisions with the obstacles and the workspace boundary are avoided, i.e., $x(t) \in \bar{\mathcal{F}} \coloneqq \left\{ x\in \mathcal{F}: \beta(d_j(x))\right.$ $\leq$ $\left.\bar{V}_0, \forall j\in\bar{\mathcal{J}}\right\}$, $\forall t \geq t_0$. Moreover,
	\eqref{eq:grad + hessian (Automatica_adaptive_nav)} implies also the boundedness of $\nabla_x \phi(x)|_{x(t)}$, $\forall t\geq t_0$.
	In addition, the boundedness of $V(t)$ implies also the boundedness of $x(t)$, $e_v(t)$, $\widetilde{m}(t)$, $\widetilde{\alpha}(t)$, $\widetilde{g}(t)$ and hence of $v(t)$,  $\hat{m}(t)$, $\hat{\alpha}(t)$, $\forall t\geq t_0$. More specifically, 
	by letting  $s \coloneqq [x^\top, v^\top, \widetilde{\alpha}$, $\widetilde{m}]^\top$, we conclude that $s(t)\in \bar{S}$, $\forall t\geq t_0$, with
	\begin{align*}
	\bar{S} \coloneqq& \bigg\{ s\in\bar{\mathcal{F}}\times \mathbb{R}^{n+2} : |\widetilde{\alpha}| \leq \sqrt{\frac{4}{3}k_\alpha\bar{V}_0}, 
	|\widetilde{m}| \leq \sqrt{2k_m\bar{V}_0}, \notag \\ 
	& \|v\| \leq \sqrt{2m\bar{V}_0} + \sup_{x\in\bar{\mathcal{F}}}\|\nabla_x\phi(x)\| \bigg\}
	\end{align*}
	Therefore, by invoking LaSalle's invariance principle (Theorem \ref{theorem:LaSalle (App_dynamical_systems)} of Appendix \ref{app:dynamical systems}), we conclude that the solution $s(t)$ will converge to the largest invariant set in $ S \coloneqq \{s\in\bar{S}:  \dot{V} = 0\}$, which, in view of \eqref{eq:v_d (Automatica_adaptive_nav)}, becomes $S \coloneqq \{s\in\bar{S} : {\nabla_x \phi(x) = 0, v = 0}\}$. Consider now the {closed-loop} dynamics for $s$:
	\begin{subequations} \label{eq:cl system (Automatica_adaptive_nav)}
		\begin{align}
		\dot{x} =& v \\
		\dot{v} =& \frac{1}{m}(\widetilde{m}g + \hat{m}\dot{v}_{\text{d}}- k_{\phi}\nabla_x \phi(x)  - \left(k_v + \frac{3}{2}\hat{\alpha}\right)(v + \nabla_x\phi(x)) - f(x,v) )  \label{eq:cl system v_dot (Automatica_adaptive_nav)} \\ 
		\dot{\widetilde{m}} =& -k_m (v + \nabla_x\phi(x))^\top (\dot{v}_{\text{d}}+ g) \\ 			 
		\dot{\widetilde{\alpha}} =& k_\alpha \|v + \nabla_x\phi(x)\|^2.
		\end{align}
	\end{subequations}
	Note that, in view of the aforementioned discussion and the continuous differentiability of $f(x,v)$, the right-hand side of \eqref{eq:cl system v_dot (Automatica_adaptive_nav)} is bounded in $\bar{S}$. Note also that \eqref{eq:grad + hessian (Automatica_adaptive_nav)} implies the boundedness of $\nabla^2_x\phi(x)$ in $\bar{\mathcal{F}}$. Moreover, by differentiating $\dot{v}$, using the closed loop dynamics \eqref{eq:cl system (Automatica_adaptive_nav)} and \eqref{eq:grad + hessian (Automatica_adaptive_nav)}, we conclude the boundedness of $\ddot{v}$ and the uniform continuity of $\dot{v}(t)$ in $\bar{S}$. Hence, since $\lim_{t\to\infty}v(t) = 0$, we invoke Barbalat's Lemma (Lemma \ref{lemma:barbalat (App_dynamical_systems)} in Appendix \ref{app:dynamical systems}) to conclude that $\lim_{t\to\infty}\dot{v}(t) = 0$.
	
	Therefore, the set $S$ consists of the points where $\dot{v} = v= \nabla_x \phi(x) = 0$, $\dot{v}_{\text{d}} = \nabla_x^2\phi(x) v = 0$, and by also using the property $f(x,0) = 0$ we obtain $\lim_{t\to\infty}\widetilde{m}(t) = 0$ and $\lim_{t\to\infty}\dot{s}(t) = 0$.
	Note also that $\hat{\alpha}:[t_0,\infty)\to\mathbb{R}_{\geq 0}$ is a monotonically increasing function and it converges thus to some constant positive value $\hat{\alpha}^\star > 0$, since {$\hat{\alpha}(t_0) \geq 0$, and $\lim_{t\to\infty}\dot{\hat{\alpha}}(t) = \lim_{t\to\infty}\dot{\widetilde{\alpha}}(t) = 0$}.
	Therefore, we conclude that the system will converge to an equilibrium $s^\star \coloneqq [(x^\star)^\top, 0^\top, 0, \hat{\alpha}^\star]$ satisfying $\nabla_x \phi(x)|_{x^\star} = 0$.
	
	Since $\lim_{t\to\infty} \nabla_x \phi(x)|_{x(t)} = \lim_{t\to\infty} v(t) = 0$, the system converges to the critical points of $\phi(x)$, i.e., we obtain from \eqref{eq:grad + hessian (Automatica_adaptive_nav)} that at steady-state: 
	\begin{equation} \label{eq:grad at equil (Automatica_adaptive_nav)}
	2k_1 (x^\star - x_{\text{d}}) = -k_2 \sum_{j\in\bar{J}} \beta'(d_j^\star)(x^\star-c_j),
	\end{equation}
	where $d_j^\star \coloneqq d_j(x^\star)$, $\forall j\in\bar{\mathcal{J}}$. 
	According to the choice of $\tau$ in \eqref{eq:tau choice (Automatica_adaptive_nav)}, $x^\star=x_{\text{d}}$ implies that $\beta'(d_j^\star) = 0$, $\forall j\in\bar{\mathcal{J}}$, and hence the desired equilibrium $x^\star=x_{\text{d}}$ satisfies \eqref{eq:grad at equil (Automatica_adaptive_nav)}. Other \textit{undesired} critical points of $\phi(x)$ consist of cases where the two sides of \eqref{eq:grad at equil (Automatica_adaptive_nav)} cancel each other out. However, as already proved, only one $\beta'_j$ can be nonzero for each $x\in\mathcal{F}$. Hence, the undesired critical points satisfy one of the following expressions:  
	\begin{subequations} \label{eq:grad at equil k (Automatica_adaptive_nav)}	
		\begin{align} 
		k_1 (x^\star - x_{\text{d}}) =& -k_2 \beta'(d_k^\star)(x^\star-c_k), \label{eq:grad at equil k in J (Automatica_adaptive_nav)}\\
		k_1 (x^\star - x_{\text{d}}) =& k_2 \beta'(d_0^\star)x^\star, \label{eq:grad at equil k=0 (Automatica_adaptive_nav)}
		\end{align}  
	\end{subequations}
	for some $k\in\bar{\mathcal{J}}$. In the case of \eqref{eq:grad at equil k=0 (Automatica_adaptive_nav)}, $x^\star$ is collinear with the origin and $x_{\text{d}}$. However, 	
	the choice of $\tau < \bar{r}_\mathcal{W}^2 - \|x_{\text{d}}\|^2$ in \eqref{eq:tau choice (Automatica_adaptive_nav)} implies that 
	\begin{align*}
	d_0^\star = \bar{r}^2_\mathcal{W} - \|x^\star\|^2 \leq \tau < \bar{r}^2_\mathcal{W} - \|x_\text{d}\|^2 \Leftrightarrow \|x^\star\| \geq \|x_\text{d}\|,
	\end{align*}
	and hence $x^\star - x_{\text{d}}$ and $x^\star$ have the same direction. Therefore, since $\beta'(d_j) < 0$, for $d_j < \tau$, $\forall j\in\bar{\mathcal{J}}$, \eqref{eq:grad at equil k=0 (Automatica_adaptive_nav)} is not feasible.
	
	Moreover, in the case of \eqref{eq:grad at equil k in J (Automatica_adaptive_nav)}, since $\beta'(d^\star_k) \leq 0$, $x^\star - x_{\text{d}}$ and $x^\star - c_k$ point to the same direction. Hence, the respective critical points $x^\star$ are on the $1$D line connecting $x_{\text{d}}$ and $c_k$. Moreover, since $\tau < \bar{r}_{\text{d}} \leq \|x_\text{d} - c_k\|^2 - \bar{r}_{o_k}^2$, as chosen in \eqref{eq:tau choice (Automatica_adaptive_nav)}, it holds that 
	\begin{align*}
	& d_k^\star = \|x^\star - c_k\|^2 - \bar{r}_{o_k}^2 < \|x_\text{d} - c_k\|^2 - \bar{r}_{o_k}^2 \Leftrightarrow  \|x^\star - x_{\text{d}}\| > \|x^\star - c_k\|.
	\end{align*}		
	We proceed now by showing that the critical points satisfying \eqref{eq:grad at equil k in J (Automatica_adaptive_nav)} are saddle points, which have a lower dimension stable manifold. 
	Consider, therefore, the error $e_x = x - x^\star$, where $x^\star \neq x_{\text{d}}$ represents the potential \textit{undesired} equilibrium point that satisfies \eqref{eq:grad at equil k in J (Automatica_adaptive_nav)}. {Let also $s_e \coloneqq [s_x^\top,\widetilde{\alpha}^\top]^\top$, where $s_x \coloneqq [e_x^\top, v^\top, \widetilde{m}]^\top$}, whose linearization around zero yields, after using \eqref{eq:cl system (Automatica_adaptive_nav)} and $\frac{\partial f(x,v)}{\partial x}\Big|_{v = 0} = 0$, 
	\begin{equation} \label{eq:s_e dot (Automatica_adaptive_nav)}
	\dot{s}_e = \bar{A}_s s_e,
	\end{equation}
	where 		
	\begin{align*}
	\bar{A}_s \coloneqq& \begin{bmatrix}
	A_s & 0 \\ 
	0^\top & 0
	\end{bmatrix} \\
	A_s \coloneqq& \begin{bmatrix}
	0_{n\times n} & I_n & 0\\
	A_{s,21} & A_{s,22} & g \\			
	-k_m g^\top(\nabla_x^2 \phi(x))^\top|_{x^\star} & -k_m g^\top & 0 
	\end{bmatrix},
	\end{align*}
	and
	\begin{align*}
	A_{s,21} \coloneqq& -\frac{1}{m}\left(k_{\phi} + k_v + \frac{3}{2}\hat{\alpha}^\star \right) \nabla_x^2\phi(x)\big|_{x^\star} \\
	A_{s,22} \coloneqq& -\nabla_x^2\phi(x)\big|_{x^\star} - \left(k_v + \frac{3}{2}\hat{\alpha}^\star\right)I_n - \frac{3}{2}\frac{\partial f(x,v)}{\partial v}\bigg|_{s^\star}.
	\end{align*}		
	We aim to prove that the equilibrium {$s_x^\star \coloneqq [0^\top,0^\top,0]^\top$} has at least one positive eigenvalue. To this end, consider a vector $\bar{\nu} \coloneqq [\mu \nu^\top, \nu^\top,0]^\top$	, where $\mu > 0$ is a positive constant, and $\nu \in\mathbb{R}^n$ is an orthogonal vector to $(x^\star - c_k)$, i.e. $\nu^\top(x^\star - c_k) = 0$. Then the respective quadratic form yields
	\begin{align*}
	\bar{\nu}^\top A_s \bar{\nu} =& \begin{bmatrix}
	\nu^\top A_{s,21} & \mu \nu^\top + \nu^\top A_{s,22} & \nu^\top g
	\end{bmatrix} \begin{bmatrix}
	\mu \nu \\ 
	\nu \\ 
	0
	\end{bmatrix} = \mu \nu^\top A_{s,21} \nu + \mu \|\nu\|^2 \\
	&\hspace{65mm}+ \nu^\top A_{s,22} \nu,		
	\end{align*}
	which, after employing \eqref{eq:grad + hessian (Automatica_adaptive_nav)} with $\beta'(d_j^\star) = 0$, $\forall j\in\mathcal{J}\backslash\{k\}$ and $\nu^\top(x^\star - c_k) = 0$, becomes
	\begin{align*}
	\bar{\nu}^\top A_s \bar{\nu} =& -\frac{2\mu k_1}{m}(k_{\phi} + k_v + \frac{3}{2}\hat{\alpha}^\star) \left(1 + \frac{k_2}{k_1} \beta'(d_k^\star) \right)\|\nu\|^2 + \mu\|\nu\|^2 \\
	&- 2k_1 \left(1 + \frac{k_2}{k_1} \beta'(d_k^\star) \right)\|\nu\|^2 - \left(k_v + \frac{3}{2}\hat{\alpha}^\star\right)\|\nu\|^2 -\nu^\top \frac{\partial f(x,v)}{\partial v}\bigg|_{s^\star} \nu.
	\end{align*}
	From \eqref{eq:grad at equil k in J (Automatica_adaptive_nav)}, by recalling that $\beta'(d_k) \leq 0$, we obtain that
	\begin{equation} \label{eq:at critical point (Automatica_adaptive_nav)}
	\frac{k_2}{k_1}\beta'(d_k^\star) = -\frac{\|x^\star - x_{\text{d}}\|}{\|x^\star - c_k\|}  < -1.
	\end{equation}
	Therefore by defining $c^\star \coloneqq -\frac{k_2}{k_1}\beta'(d_k^\star) -1 > 0$, we obtain
	\begin{align*}
	\bar{\nu}^\top A_s \bar{\nu} =& \Bigg(\frac{2\mu k_1}{m}k_{\phi}c^\star  +  \left(\frac{2\mu k_1}{m}c^\star -1 \right)\left(k_v + \frac{3}{2}\hat{\alpha}^\star \right) + \mu  + 2k_1c^\star \Bigg)\|\nu\|^2 \notag \\ & -\nu^\top \frac{\partial f(x,v)}{\partial v}\bigg|_{s^\star} \nu,
	\end{align*}
	which is rendered positive by choosing a sufficiently large $\mu$. Hence, $A_s$ has at least one positive eigenvalue. Next, we prove that $A_s$ has no zero eigenvalues by proving that its determinant is nonzero. For the determinant of $\nabla^2_x\phi(x)|_{x^\star}$, it holds in view of \eqref{eq:grad + hessian (Automatica_adaptive_nav)} that 
	\small
	\begin{align*}
	\det\left(\nabla_x^2\phi(x)|_{x^\star}\right) =& \det \Bigg(2\left(k_1 + k_2\beta'(d^\star_k)\right)I_n + 2k_2 \beta''(d^\star_k) (x^\star-c_k)(x^\star-c_k)^\top \Bigg).
	\end{align*} 
	\normalsize
	By using the property $\det(A + uv^\top) = (1+v^\top A^{-1}u) \det(A)$, for any invertible matrix $A$ and vectors $u,v$, we obtain 
	\small
	\begin{align} \label{eq:det hessian 1 (Automatica_adaptive_nav)}
	\det\left(\nabla_x^2\phi(x)|_{x^\star}\right) =& 2^n \big(k_1 + k_2\beta'(d_k^\star) \big)^n \Bigg(1 + \frac{k_2}{k_1\left(1 + \frac{k_2}{k_1}\beta'(d_k^\star)\right)}\beta''(d_k^\star)\|x^\star-c_k\|^2 \Bigg).
	\end{align}
	\normalsize
	In view of \eqref{eq:at critical point (Automatica_adaptive_nav)} and by using $\|x^\star -x_{\text{d}}\| - \|x^\star - c_k\| = \|x_{\text{d}} - c_k\|$ since $x^\star$, $c_k$ and $x_{\text{d}}$ are collinear,  \eqref{eq:det hessian 1 (Automatica_adaptive_nav)} becomes
	\begin{align*} 
	\det\left(\nabla_x^2\phi(x)|_{x^\star}\right) =& 2^n \big(k_1 + k_2\beta'(d_k^\star)\big)^n \Bigg(1 -\frac{k_2}{k_1 \|x_{\text{d}} - c_k\|}\beta''(d_k^\star)\|x^\star-c_k\|^3 \Bigg).
	\end{align*}
	Note that, since $\lim_{d_j \to 0}\beta(d_j) = \infty$ and $\beta(d_j)$ decreases to $\beta(d_j) = \beta(\tau)$, $\forall d_j \geq \tau$, the derivatives $\beta'(d_j)$ satisfy $\lim_{d_j\to 0}\beta'(d_j) = -\infty$ and increase to $\beta'(d_j) = 0$, $\forall d_j \geq \tau$. Hence, we conclude that $\beta''(d_j) > 0$, $\forall d_j \in (0,\tau)$. Therefore, in order for the critical point to be non-degenerate, we must guarantee that 
	\begin{equation} \label{eq:det hessian 3 (Automatica_adaptive_nav)}
	\frac{k_2}{k_1 \|x_{\text{d}} - c_k\|}\beta''(d_k^\star) \|x^\star-c_k\|^3 > 1.
	\end{equation}	
	By expressing $\|x^\star - c_k\|^3 = (d_k^\star + \bar{r}_{o_k}^2 )\sqrt{d_k^\star + \bar{r}_{o_k}^2}$, considering that $\|x_{\text{d}} - c_k\| \leq 2\bar{r}_{\mathcal{W}}$ and setting $\underline{r} \coloneqq \min_{j\in\mathcal{J}}\{\bar{r}_{o_j}\}$, a lower bound for the left-hand side of \eqref{eq:det hessian 3 (Automatica_adaptive_nav)} is 
	\begin{equation} \label{eq:det hessian 4 (Automatica_adaptive_nav)}
	f_\ell(d_k^\star) \coloneqq \frac{k_2}{2k_1 \bar{r}_{\mathcal{W}} } \beta''(d_k^\star)(d_k(x^\star) + \underline{r}^2 )\sqrt{d_k(x^\star) + \underline{r}^2}.
	\end{equation}
	According to Property $4$ of Definition \ref{def:2nd nf (Automatica_adaptive_nav)}, \eqref{eq:det hessian 4 (Automatica_adaptive_nav)} is a decreasing function of $d_k^\star$, for $d_k^\star \in (0,\tau)$, with $f_\ell(\tau) = 0$ and $\lim_{d_k^\star \to 0} f_\ell(d_k^\star) = \infty$. Therefore, there exists a positive $d_k^{\star\star} > 0$, such that $f_\ell(d_k^\star) > 1$, $\forall d_k^\star < d_k^{\star\star}$. Hence, by setting $\tau < d_k^{\star\star}$, we achieve $d_k^\star < \tau < d_k^{\star\star}$ and guarantee that $f_\ell(d_k^\star) > 1$.
		
	Next, by defining $A_{2ns} \coloneqq \begin{bmatrix}
	0_{n\times n} & I_n \\A_{s,21} & A_{s,22}
	\end{bmatrix}$, it holds that  
	\begin{align*}
	\det(A_{2ns}) =& \det(A_{s,21}) = (-1)^n\frac{1}{m^n}\left(k_{\phi} + k_v + \frac{3}{2}\hat{\alpha}^\star\right)^n \det(\nabla^2_x \phi(x)|_{x^\star}) \neq 0,
	\end{align*} 
	and 
	\begin{equation*}
	A_{2ns}^{-1} = \begin{bmatrix}
	\star & A_{s,21}^{-1} \\
	\star & 0_{n\times n}
	\end{bmatrix}
	\end{equation*}
	and therefore we obtain that
	\begin{align*}
	\det(A_s) =& \det(A_{s,21})\begin{bmatrix}
	k_m g^\top(\nabla_x^2 \phi(x))^\top|_{x^\star} & k_m g^\top
	\end{bmatrix}  A_{2ns}^{-1} \begin{bmatrix}
	0 \\ g 
	\end{bmatrix}  = \\
	&\det(A_{s,21}) \begin{bmatrix}
	k_m g^\top(\nabla_x^2 \phi(x))^\top|_{x^\star} & k_m g^\top
	\end{bmatrix}  \begin{bmatrix}
	A_{s,21}^{-1} \ g \\ 0 
	\end{bmatrix}  = \\
	&\det(A_{s,21})
	k_m g^\top(\nabla_x^2 \phi(x))^\top|_{x^\star}
	A_{s,21}^{-1} \ g = \\
	&k_m g^\top(\nabla_x^2 \phi(x))^\top|_{x^\star}
	\text{adj}(A_{s,21}) \ g,		 	
	\end{align*}
	which is non-zero, since $g \neq 0$,  $$\det\bigg(\nabla_x^2(\phi(x))^\top|_{x^\star}\text{adj}(A_{s,21})\bigg) = \det(\nabla_x^2(\phi(x))^\top|_{x^\star}) \det(A_{s,21})^{n-1} \neq 0$$ and hence the matrix that forms the latter quadratic form is nonsingular.
	
	Therefore, we conclude that $A_s$ is non-degenerate and has at least one positive eigenvalue. Note that {$\bar{A}_s$ has the same eigenvalues as $A_s$ and an extra zero eigenvalue}. According to the Reduction Principle (Theorem \ref{theorem:reduction pr (App_dynamical_systems)} of Appendix \ref{app:dynamical systems}), \eqref{eq:s_e dot (Automatica_adaptive_nav)} is locally topologically equivalent near the origin to the system
	\begin{align*}			
	\dot{\hat{\alpha}} &= k_\alpha \big\| v_\alpha(\hat{\alpha}) + \nabla_x\phi(x)|_{x_\alpha(\hat{\alpha})}\big\|^2 \\
	\dot{s}_x &= A_s s_x,		
	\end{align*}
	where 
	$v_\alpha(\hat{\alpha})$, $\nabla_x\phi(x)|_{x_\alpha(\hat{\alpha})}$ are the restrictions of $v$ and $\nabla_x\phi(x)$ to the center manifold of $\hat{\alpha}$ {(see Theorem \ref{theorem:reduction pr (App_dynamical_systems)} of Appendix \ref{app:dynamical systems})}. Regarding the trajectories of $s_x$, since $A_s$ is a non-degenerate saddle (it has at least one positive eigenvalue)  its stable manifold has dimension lower than $2n+1$ and is thus a set of zero measure.
	Therefore, all the initial conditions $(x(t_0),v(t_0),\widetilde{m}(t_0)) \in \mathcal{F}\times\mathbb{R}^{n+1}$, except for the aforementioned lower-dimensional manifold, converge to the desired equilibrium $(x_{\text{d}},0,0)$.		
\end{proof}

\begin{figure}[!ht]
	\centering
	\includegraphics[trim = 0cm 0cm 0cm -0cm,width = 0.45\textwidth]{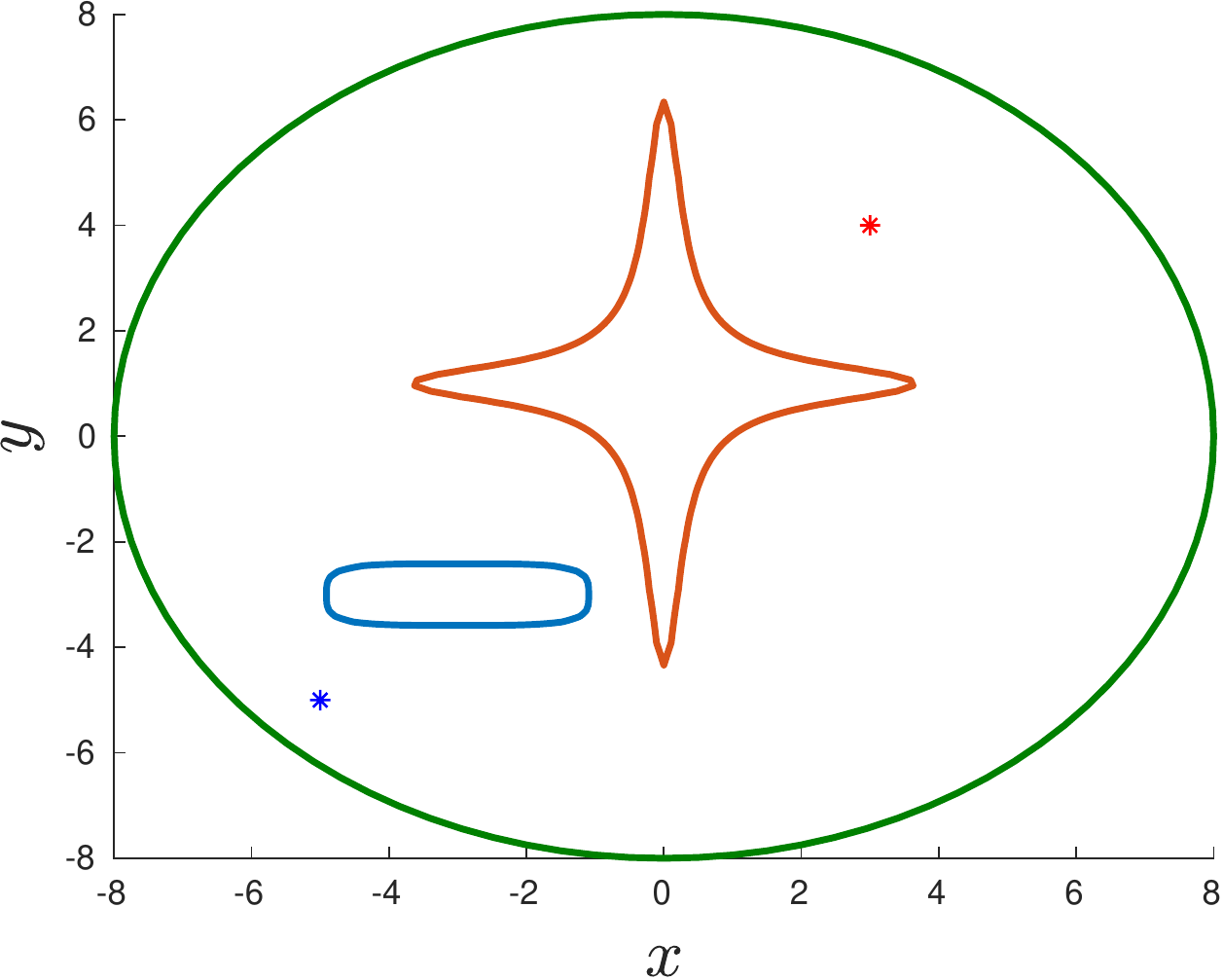}\\
	\caption{A workspace with two star-shaped obstacles. The blue asterisk indicates the center of the robot and the obstacles have been enlarged with the robot radius $r$ {through the Minkowski sum}.
		The red asterisk indicates a potential goal robot position.}\label{fig:2d_stars_initial (Automatica_adaptive_nav)}
\end{figure}

\begin{remark}
	The proof can be trivially extended to the $2$D case on the horizontal plane where there is no gravity, i.e., $g = 0$. In addition, it is worth noting that in that case, the estimation parameter $\hat{m}(t)$ will converge to a constant value different than the mass $m$, as {revealed by} a careful inspection of the closed-loop system.
	Moreover, the condition $k_\phi > \frac{\alpha}{2}$ of Theorem \ref{th:single robot (Automatica_adaptive_nav)} is only sufficient and not necessary, as will be shown in the simulation results.
\end{remark}

{\subsection{Dynamic Disturbance Addition}} \label{subsec:disturbance (Automatica_adaptive_nav)}

{Except for the already considered dynamic uncertainties, we can add to the right-hand side of \eqref{eq:dynamics (Automatica_adaptive_nav)} an unknown disturbance vector $d\coloneqq d(x,v,t):\mathbb{R}^{2n}\times\mathbb{R}_{\geq 0} \to \mathbb{R}^n$, i.e.,
		\begin{align*}
		& \dot{x}= v \\
		& m \dot{v} + f(x,v) + mg + d(x,v,t)= u,
		\end{align*}
	subject to a uniform boundedness condition $\|d(x,v,t)\| \ \leq \bar{d}$, $\forall x,v,t\in\mathbb{R}^{2n}\times\mathbb{R}_{\geq 0}$. In this case, by slightly modifying the control scheme, we still guarantee collision avoidance with the workspace obstacles and boundary. In addition, we 
	achieve uniform ultimate boundedness of the error signals as well as the gradient of $\phi$, as the analysis in this section shows.}

{ 
	The control scheme of the previous section is appropriately enhanced to incorporate the $\sigma$- modification \cite{lavretsky13adaptive}, a common technique in adaptive control. More specifically, the adaptation laws \eqref{eq:adaptation laws (Automatica_adaptive_nav)} are modified according to 
	\begin{align*}
	\dot{\hat{m}} \coloneqq& -k_m e_v^\top( \dot{v}_{\text{d}}+g) - \sigma_m\hat{m}  \\
	\dot{\hat{\alpha}} \coloneqq & k_\alpha \|e_v\|^2 - \sigma_\alpha\hat{\alpha}, 
	\end{align*}
	where $\sigma_m$, $\sigma_\alpha$ are positive gain constants, {to be appropriately tuned as per the analysis below}. }

{
	Consider now the function $V$ as defined \eqref{eq:V Lyap (Automatica_adaptive_nav)}. In view of the analysis of the previous section, the incorporation of $d(x,v,t)$, as well as the modification of the adaptation laws, the derivative of $V$ becomes 
	\begin{align*}
	\dot{V} \leq& -\left(k_\phi - \frac{\alpha}{2}\right)\|\nabla_x \phi(x)\|^2 - k_v\|e_v\|^2 + \|e_v\|\bar{d} - \frac{3}{2}\sigma_\alpha \widetilde{\alpha}\hat{\alpha} - \sigma_m\widetilde{m}\hat{m},
	\end{align*} 
	which, by using $\hat{\alpha} = \widetilde{\alpha} + \alpha$, $\hat{m} = \widetilde{m} + m$, as well as the properties $-ab = -\frac{1}{2}(a+b)^2 + \frac{a^2}{2} + \frac{b^2}{2}$, $ab = -\frac{1}{2}(a-b)^2 + \frac{a^2}{2} + \frac{b^2}{2}$, $\forall a,b \in \mathbb{R}$, becomes
	\begin{align*}
	\dot{V} \leq&  -\left(k_\phi - \frac{\alpha}{2}\right)\|\nabla_x \phi(x)\|^2 - \frac{k_v}{2}\|e_v\|^2 + \frac{\bar{d}^2}{2k_v}- 3\sigma_\alpha\frac{\widetilde{\alpha}^2}{4} - \sigma_m\frac{\widetilde{m}^2}{2} \\ & + 3\sigma_\alpha\frac{\alpha^2}{4} + \sigma_m\frac{m^2}{2} \\ 
	\leq & -k_\xi \|\xi\|^2 + d_\xi, 		
	\end{align*}
	where $\xi \coloneqq [\nabla_x\phi(x)^\top, e_v^\top, \widetilde{m}, \widetilde{\alpha}]^\top \in \mathbb{R}^{2n+2}$, $k_\xi \coloneqq \min\left\{k_\phi-\frac{\alpha}{2}, k_v, \frac{\sigma_m}{2},\frac{3\sigma_\alpha}{4}\right\}$, and
	$d_\xi \coloneqq \frac{\bar{d}^2}{2k_v} + 3\sigma_\alpha\frac{\alpha^2}{4} + \sigma_m\frac{m^2}{2}$. Therefore, $\dot{V}$ is negative when $\|\xi\| > \sqrt{\frac{d_\xi}{k_\xi}}$, which implies uniform ultimate boundedness of $\|\xi(t)\| $ in a set around zero, whose size is proportional to $d_\xi$, which can be shrunk by gain tuning (see Theorem \ref{th:uub_khalil (App_dynamical_systems)} of Appendix \ref{app:dynamical systems}). In addition, $\dot{V}$ is sign indefinite only in the set defined by $\|\xi \| < \sqrt{\frac{d_\xi}{k_\xi}}$ and negative otherwise, which implies that $V(t)$ remains bounded, $\forall t\geq 0$, and hence collisions with the workspace obstacles and boundary are avoided.  }

	Note that the aforementioned analysis guarantees that $\nabla_x\phi(x)|_{x(t)}$ will be ultimately bounded in a set close to zero. This point, however, might be a critical point of $\phi$ and it is not guaranteed that $x(t)$ will be bounded close to the goal configuration $x_\text{d}$. Nevertheless, intuition suggests that if the disturbance vector $d(x,v,t)$ does not behave adversarially, the agent will converge close to the goal configuration. This is also verified by the simulation results of Section \ref{sec:sim (Automatica_adaptive_nav)}.

\subsection{Extension to Star Worlds} \label{sec:Star (Automatica_adaptive_nav)}

Before moving to the multi-robot case, we discuss in this section how the proposed control scheme can be extended to generalized sphere worlds, a wider class of configuration spaces than the spheres worlds described in the previous section. In particular, we consider \textit{star worlds}, which are configuration spaces diffeomorphic to sphere worlds. In particular, a star world is a set of the form 
\begin{equation*}		
\mathcal{T} \coloneqq \bar{\mathcal{W}} \backslash \bigcup_{j\in\mathcal{J}} \bar{O}_{\mathcal{T}_j},
\end{equation*} 
where $\bar{\mathcal{W}}$ is a workspace of the form \eqref{eq:transf. workspace (Automatica_adaptive_nav)} and $\bar{O}_{\mathcal{T}_j}$ are $M$ disjoint star-shaped obstacles (indexed by $\mathcal{J} = \{1,\dots,M\}$). The latter are sets characterized by the  property that all rays emanating from a center point cross their boundary only once {\cite{rimon1992exact}} (see Fig. \ref{fig:2d_stars_initial (Automatica_adaptive_nav)}).

One can design a diffeomorphic mapping $H:\mathcal{T} \to \mathcal{F}$, where $\mathcal{F}$ is a sphere world of the type \eqref{eq:sphere world (Automatica_adaptive_nav)}. More specifically, $H$ maps the boundary of $\mathcal{T}$ to the boundary of $\mathcal{F}$. 

The control scheme of the previous section is modified now to account for the transformation $H$ as follows. The desired robot velocity is set to $v_{\text{d}}: \mathcal{T} \to \mathbb{R}^n$, with
\begin{equation} \label{eq:v_d star (Automatica_adaptive_nav)}
v_{\text{d}}(x) \coloneqq - J_H(x)^{-1} \nabla_{H(x)}\phi(H(x)),
\end{equation}
where $J_H(x) \coloneqq \frac{\partial H(x)}{\partial x}$ is the nonsingular Jacobian matrix of $H$. Next, by letting $e_v \coloneqq v - v_{\text{d}}$, the control law is designed as $u:\mathcal{T}\times\mathbb{R}^{n+2} \to \mathbb{R}^n$, with
\small
\begin{align} \label{eq:control law star (Automatica_adaptive_nav}
u\coloneqq u(x,v,\hat{m},\hat{\alpha}) \coloneqq& -k_\phi J_h(x)^\top \nabla_{H(x)} \phi(H(x)) +  \hat{m}(\dot{v}_{\text{d}} + g) - \left(k_v + \frac{3}{2}\hat{\alpha}\right)e_v,
\end{align}
\normalsize
where $\hat{m}$ and $\hat{\alpha}$ evolve according to the respective expressions in \eqref{eq:adaptation laws (Automatica_adaptive_nav)}. The next theorem gives the main result of this section. 
\begin{theorem}\label{th:single robot star worlds (Automatica_adaptive_nav)}
	Consider a robot operating in $\mathcal{W}$, subject to the uncertain $2$nd-order dynamics \eqref{eq:dynamics (Automatica_adaptive_nav)}. Given $x_{\textup{d}} \in \mathcal{T}$, the control protocol \eqref{eq:adaptation laws (Automatica_adaptive_nav)}, \eqref{eq:v_d star (Automatica_adaptive_nav)}, \eqref{eq:control law star (Automatica_adaptive_nav}  guarantees the collision-free navigation to $x_{\textup{d}}$ from almost all initial conditions $(x(t_0),v(t_0),\hat{m}(t_0),\hat{\alpha}(t_0)) \in \mathcal{T}\times\mathbb{R}^{n_1}\times\mathbb{R}_{\geq 0}$, given a sufficiently small $\tau$ and that $k_\phi > \frac{\alpha}{2}$. Moreover, all closed loop signals remain bounded, $\forall t \geq t_0$.
\end{theorem}  
\begin{proof}
	Following similar steps as in the proof of Theorem \ref{th:single robot (Automatica_adaptive_nav)}, we consider the Lyapunov candidate function 
	\begin{equation*}
	V \coloneqq k_{\phi}\phi(H(x)) + \frac{m}{2}\|e_v\|^2 + \frac{1}{2k_\alpha}\widetilde{\alpha}^2 + \frac{3}{4k_m}\widetilde{m}^2,
	\end{equation*}	
	whose derivative along the solutions of the closed loop system can be {proven} to satisfy 
	\begin{equation*}
	\dot{V} \leq -\left(k_\phi - \frac{\alpha}{2}\right) \|\nabla_{H(x)} \phi(H(x)) \|^2 - k_v\|e_v\|^2 \leq 0,
	\end{equation*}
	which proves the boundedness of the obstacle functions $\beta(d_j(H(x(t))))$, $\forall j\in\mathcal{J}, t\geq t_0$. Since the boundaries $\partial \bar{\mathcal{O}}_j$ are mapped to $\partial \bar{\mathcal{O}}_{\mathcal{T}_j}$ through $H(x)$, we conclude that $x(t) \in \mathcal{T}$, $t\geq t_0$ and no collisions occur. Next, by following similar arguments as in the proof of Theorem \ref{th:single robot (Automatica_adaptive_nav)}, we conclude that the solution will converge to a critical point of $\phi(H(x))$. By choosing a sufficiently small $\tau$ for the obstacle functions $\beta(d_j(H(x(t))))$, the critical points consist of the desired equilibrium, where $\beta'(d_j(H(x_{\text{d}}))) = 0$, $\forall j\in\mathcal{J}$, or undesired critical points $x^\star$ satisfying 
	\begin{equation} \label{eq:critical point star world (Automatica_adaptive_nav)}
	k_1(H(x^\star) - H(x_{\text{d}})) = -k_2 \beta'(d_{H_k}^\star)(H(x^\star) - H(c_k)),
	\end{equation}
	for some $k\in\mathcal{J}$, where we define $d_{H_j}^\star \coloneqq d_j(H(x^\star))$, $\forall j\in\mathcal{J}$. The respective linearization matrix $\bar{A}_s$ from \eqref{eq:s_e dot (Automatica_adaptive_nav)} becomes now	
	\begin{align*}
	\bar{A}_s \coloneqq& \begin{bmatrix}
	A_s & 0 \\ 
	0^\top & 0
	\end{bmatrix} \\
	A_s \coloneqq& \begin{bmatrix}
	0_{n\times n} & I_n & 0 \\
	A_{s,21} & A_{s,22} & g \\			
	A_{s,31} & -k_m g^\top & 0 
	\end{bmatrix},
	\end{align*}
	with 
	\begin{align*}
	A_{s,21} \coloneqq& -\frac{1}{m}\big( k_{\phi}J_H(x^\star)^\top + \left(k_v + \frac{3}{2}\hat{\alpha}^\star\right)J_H(x^\star)^{-1} \big) \nabla^2 \phi^\star J_H(x^\star) \\
	A_{s,22} \coloneqq& -J_H(x^\star)^{-1} \nabla^2 \phi^\star J_H(x^\star) - \left(k_v + \frac{3}{2}\hat{\alpha}^\star\right)I_n - \frac{\partial f(x,v)}{\partial v}\bigg|_{s^\star}, \\ 
	A_{s,31} \coloneqq & -k_m g^\top \big(J_H(x^\star)^{-1} \nabla^2 \phi^\star J_H(x^\star)\big)^\top
	\end{align*}
	and $\nabla^2 \phi^\star \coloneqq \nabla^2_{H(x)}\phi(H(x))|_{x^\star}$,
	around $x = x^\star, v = 0, \widetilde{m} = 0, \widetilde{\alpha} = \widetilde{\alpha}^\star$. Next, similarly to the proof of Theorem \ref{th:single robot star worlds (Automatica_adaptive_nav)}, we prove that $\bar{\nu}^\top A_s \bar{\nu} > 0$, for $\bar{\nu} \coloneqq [\mu \nu^\top, \nu^\top, 0]^\top$, where $\mu >0$ is a positive constant and $\nu \coloneqq J_H(x^\star)^{-1}\hat{\nu}$, with
	$\hat{\nu}\in\mathbb{R}^n$ a vector orthogonal to $(H(x^\star) - H(c_k))$.
	The respective quadratic form yields, after employing \eqref{eq:critical point star world (Automatica_adaptive_nav)} and defining $c^\star \coloneqq -\left(1 +  \frac{k_2}{k_1}\beta'(d^\star_{H_k}) \right) > 0$:
	\begin{align*}
	\bar{\nu}^\top A_s \bar{\nu} =& 	\hat{\nu}^\top \Bigg[ \frac{2k_1k_{\phi}\mu c^\star}{m} I_n + J_H(x)^{-\top} \Bigg( \bigg(\frac{2k_1 c^\star\left(k_v + \frac{3}{2}\hat{\alpha}^\star\right)}{m}  + \\
	&\hspace{-5mm} \mu - \left(k_v + \frac{3}{2}\hat{\alpha}^\star\right) + 2k_1 c^\star \bigg) I_n - \frac{\partial f(x,v)}{\partial v } \bigg|_{s^\star}   \Bigg) J_H(x)^{-1} 
	\Bigg] \hat{\nu},
	\end{align*}
	which can be rendered positive for sufficiently large $\mu$.

	Moreover, at a critical point $x^{\star,1}$ of $\phi(H(x))$, it holds that (see the proof of Prop. 2.6 in {\cite{koditschek1990robot}}),
	\begin{equation*}
	\nabla^2_{H(x)} \phi(H(x))|_{x^{\star,1}} = J_H(x^{\star,1})^\top \nabla^2_x \phi(x) |_{x^{\star,2}} J_H(x^{\star,1}),
	\end{equation*} 
	where $x^{\star,2} \coloneqq H(x^{\star,1})$ is a critical point of $\phi(x)$. Since $J_H(x)$ is nonsingular, it holds that $x^{\star,1}$ is non-degenerate if and only if $x^{\star,2}$ is non-degenerate. As already shown in the proof of Theorem \ref{th:single robot (Automatica_adaptive_nav)}, by choosing $\tau$ sufficiently small, we render the critical points of $\phi(x)$ that are close to the obstacles non-degenerate. 
	Hence, we conclude that the respective critical points of $\phi(H(x))$ are also non-degenerate and $\det(\nabla^2 \phi^\star) \neq 0$.
	
	Next, in order to prove that the critical point $(x^\star,0,0)$ is non-degenerate, we calculate the determinant of $A_s$. Following the proof of Theorem \ref{th:single robot (Automatica_adaptive_nav)}, we obtain that 
	\begin{align*}
	\det(A_s) &= \det(A_{s,21}) k_m g^\top \big(J_H(x^\star)^{-1} \nabla^2 \phi^\star J_H(x^\star)\big)^\top A_{s,21}^{-1}g \\
	&= k_m g^\top \big(J_H(x^\star)^{-1} \nabla^2 \phi^\star J_H(x^\star)\big)^\top \text{adj}(A_{s,21})g
	\end{align*}
	where 
	\begin{align*}
	\det(A_{s,21}) =&(-1)^n\Bigg( \frac{k^n_{\phi}}{m^n}\det\left(J_H(x^\star)\right) + \\
	& \hspace{-10mm} \left(k_v+\frac{3}{2}\hat{\alpha}^\star\right)^n \frac{1}{\det(J_H(x^\star))} \Bigg)\det(\nabla^2\phi^\star)\det(J_H(x^\star)),
	\end{align*}
	which is not zero, since $\det(\nabla^2 \phi^\star) \neq 0$ and $J_H(x^\star) \neq 0$. Hence, we conclude that the aforementioned quadratic form is also not zero and hence the non-degeneracy of the critical points under consideration. Hence, by following similar arguments as in the proof of Theorem \ref{th:single robot (Automatica_adaptive_nav)}, we conclude that the initial conditions that converge to these critical saddle points form a set of measure zero. 
\end{proof}

\subsection{Extension to Multi-Robot Systems} \label{sec:MAS (Automatica_adaptive_nav)}

This section is devoted to extending the results of Section \ref{sec:main (Automatica_adaptive_nav)} to Multi-Robot systems. Consider, therefore, $N\in\mathbb{N}$ spherical robots operating in a workspace $\mathcal{W}$ of the form \eqref{eq:workspace (Automatica_adaptive_nav)}, characterized by their position vectors $x_i\in\mathbb{R}^n$, as well as their radii $r_i > 0$, $i\in\mathcal{N} \coloneqq \{1,\dots,N\}$, and obeying the second-order uncertain dynamics \eqref{eq:dynamics (Automatica_adaptive_nav)}, i.e.,
\begin{subequations} \label{eq:dynamics MAS (Automatica_adaptive_nav)}
	\begin{align} 
	& \dot{x_i}= v_i \\
	& m_i \dot{v_i} + f_i(x_i,v_i) + m_ig = u_i,
	\end{align}
\end{subequations}
with $f_i(\cdot)$ satisfying $\|f_i(x_i,v_i)\| \leq \alpha_i\|v_i\|$, for unknown positive constants $\alpha_i$, $\forall i\in\mathcal{N}$. We also denote $x\coloneqq [x_1^\top,\dots,x_N^\top]^\top$, $v\coloneqq [v_1^\top,\dots,v_N^\top]^\top \in\mathbb{R}^{Nn}$.
The robots desire to navigate to their destination configurations $x_{\textup{d}_i}$, $i\in\mathcal{N}$.
The proposed multi-robot scheme is based on a prioritized leader-follower coordination. {Prioritization in multi-agent systems for {navigation-type} objectives has been employed in \cite{roussos2013decentralized} and \cite{guo2016communication}, where KRNF gain tuning-type methodologies for single integrator agents are developed. Moreover, \cite{guo2016communication} does not consider inter-agent collision avoidance and \cite{roussos2013decentralized} {does not assume} static obstacles.}

Intuitively, in the proposed prioritized leader-follower methodology, the leader robot, by appropriately choosing the offset $\tau$, ``sees" the other robots as static obstacles and hence the overall scheme reduces to the one of Section \ref{sec:main (Automatica_adaptive_nav)}.
The workspace is assumed to satisfy Assumption \ref{ass: workspace (Automatica_adaptive_nav)} and we further impose extra conditions on the initial states and destinations:

\begin{assumption} \label{ass:MAS workspace (Automatica_adaptive_nav)}
	The workspace $\mathcal{W}$, obstacles $\mathcal{O}_j$, $j\in\mathcal{J}$, and destinations {$x_{\text{d}_i}$, $i\in\mathcal{N}$,} satisfy:	
	\begin{align*}			
	&\|c_j - x_{\text{d}_i}\| > r_{o_j} + r_i + 2r_M + \varepsilon, \forall i,j\in\mathcal{N}\times\mathcal{J}  \\
	&\|x_{\text{d}_i} - x_{\text{d}_j}\| > r_i + r_j + 2r_M + 2\varepsilon, \forall i,j\in\mathcal{N}, i\neq j \\		
	&r_\mathcal{W} - \|x_{\text{d}_i}\| > r_i + 2r_M + \varepsilon, \forall i\in\mathcal{N}
	\end{align*}	
	whereas the initial positions satisfy:
	\begin{align*}
	&\|c_j - x_i(t_0)\| > r_{o_j} + r_i + 2r_M, \forall i,j\in\mathcal{N}\times\mathcal{J}  \\
	&r_\mathcal{W} - \|x_i(t_0)\| > r_i + 2r_M, \forall i\in\mathcal{N}  \\
	&\|x_{\text{d}_i} - x_j(t_0) \| > r_i + r_j + 2r_M + \varepsilon, \forall i,j\in\mathcal{N}, i \neq j, 
	\end{align*}
	for an arbitrarily small positive constant  $\varepsilon$, $\forall i\in\mathcal{N}, j\in\mathcal{J}$, where $r_M \coloneqq \max_{i\in\mathcal{N}}\{r_i\}$. 
\end{assumption}

Loosely speaking, the aforementioned assumption states that the pairwise distances among obstacles, workspace boundary, initial conditions and final destinations are large enough so that one robot can always navigate between them. {Since} the convergence of the agents to the their destinations is asymptotic, we incorporate the threshold $\varepsilon$, which is the desired proximity we want to achieve to the destination, as will be clarified in the sequel. 
{Intuitively, since we cannot achieve $x_i = x_{\text{d}_i}$ in finite time, the high-priority agents will stop once $\|x_i-x_{\text{d}_i}\|= \varepsilon$, which is  included in the aforementioned conditions to guarantee the feasibility of the collision-free navigation for the lower-priority agents.}

Similarly to the single-agent case, we can find a positive constant $\bar{r}$ such that \eqref{eq:r_bar (Automatica_adaptive_nav)} hold as well as

\begin{subequations} \label{eq:r_bar_mas (Automatica_adaptive_nav)}
	\begin{align}			
	&\|c_j - x_i(t_0)\| > r_{o_j} + r_i + 2r_M + 2\bar{r}, \forall i,j\in\mathcal{N}\times\mathcal{J}  \\
	&r_\mathcal{W} - \|x_i(t_0)\| > r_i + 2r_M + 2\bar{r}, \forall i\in\mathcal{N} \\
	&\|c_j - x_{\text{d}_i}\| > r_{o_j} + r_i + 2r_M + \varepsilon + 2\bar{r}, \forall i,j\in\mathcal{N}\times\mathcal{J}  \\
	&\|x_{\text{d}_i} - x_{\text{d}_j}\| > r_i + r_j + 2r_M + 2\varepsilon + 2\bar{r}, \forall i,j\in\mathcal{N}, i\neq j \\
	&\|x_{\text{d}_i} - x_j(t_0) \| > r_i + r_j + 2r_M + \varepsilon + 2\bar{r}, \forall i,j\in\mathcal{N}, i \neq j, \\
	&r_\mathcal{W} - \|x_{\text{d}_i}\| > r_i + 2r_M + \varepsilon + 2\bar{r}, \forall i\in\mathcal{N}
	\end{align}
\end{subequations}

We consider that the agents have a limited sensing range, defined by a radius ${\varsigma_i} > 0$, $i\in\mathcal{N}$, and we assume that each agent $i$ can sense the state of its neighbors, as stated next. 
\begin{assumption} \label{ass:sensing (Automatica_adaptive_nav)}
	Each agent $i\in\mathcal{N}$ has a limited sensing radius $\varsigma_i$, satisfying $\varsigma_i > \sqrt{\min(\bar{r}^2,\bar{r}_\text{d})} + r_i + r_j +2r_M + 2\bar{r}$, with $\bar{r}_\text{d}$ as defined in \eqref{eq:bar_r_d (Automatica_adaptive_nav)}, and has access to $(x_j,v_j)$, $\forall j\in\{ j\in\mathcal{N} : \|x_i-x_j\| \leq \varsigma_i \}$.
\end{assumption} 

Moreover, we consider {that} the destinations, $x_{\text{d}_i}$, $i\in\mathcal{N}$, {as well as the radii, $r_i$}, are transmitted off-line to all the agents{\footnote{{This implies that the agents can compute $r_M$ offline.}}}.
Consider now a prioritization of the agents, possibly based on some desired metric (e.g., distance to their destinations), which can be performed off-line and transmitted to all the agents.
Our proposed scheme is based on the following algorithm. The agent with the highest priority is designated as the leader of the multi-agent system, indexed by $i_\mathcal{L}$, whereas the rest of the agents are considered as the followers, defined by the index set $\mathcal{N}_\mathcal{F} \coloneqq \mathcal{N} \backslash \{i_\mathcal{L}\}$. The followers and leader employ a control protocol that has the same structure as the one of Section \ref{sec:main (Automatica_adaptive_nav)}. The key difference here lies in the definition of the free space for followers and leaders. We define first the sets
\begin{align*}
&\bar{\mathcal{W}}_{i_\mathcal{L}} \coloneqq \{z = [z_1^\top,\dots,z_N^\top]^\top \in\mathbb{R}^{Nn}: \|z_{i_\mathcal{L}}\| < r_\mathcal{W} - r_{i_\mathcal{L}} \},  \\
&\bar{\mathcal{O}}_{i_\mathcal{L},j} \coloneqq \{ z = [z_1^\top,\dots,z_N^\top]^\top \in \bar{\mathcal{W}}_{i_\mathcal{L}} : \|z_i-c_j\| \leq r_{o_j} + r_i \}, \forall j\in\mathcal{J} \\
& \mathcal{C}_{i_\mathcal{L}} \coloneqq \{ z = [z_1^\top,\dots,z_N^\top]^\top \in \bar{\mathcal{W}}_{i_\mathcal{L}}: \|z_{i_\mathcal{L}} - z_j \| \leq r_{i_\mathcal{L}} + r_j, \forall j\in\mathcal{N} \backslash \{i_\mathcal{L}\}  \},  
\end{align*}
which correspond to the leader agent, as well as the follower sets
\begin{align*}
&\bar{\mathcal{W}}_i \coloneqq \{z = [z_1^\top,\dots,z_N^\top]^\top \in \mathbb{R}^{Nn}: \|z_i\| < r_\mathcal{W} - r_i - 2r_M - 2\bar{r} \} \\
&\bar{\mathcal{O}}_{i,j} \coloneqq \{ z = [z_1^\top,\dots,z_N^\top]^\top \in \bar{\mathcal{W}}_i : \|z_i-c_j\| \leq r_{o_j} + r_i + 2r_M + 2\bar{r} \}, \forall j\in\mathcal{J} \\
& \mathcal{C}_i \coloneqq \{ z = [z_1^\top,\dots,z_N^\top]^\top \in \bar{\mathcal{W}}_i: \|z_i - z_{i_\mathcal{L}} \| \leq r_i + r_{i_\mathcal{L}},  \\
&\hspace{12mm}  \|z_i - z_j \| \leq r_i + r_j + 2r_M + 2\bar{r}, \forall j\in\mathcal{N} \backslash \{i_\mathcal{L}{,i}\}, \\
&\hspace{12mm}  \|z_i-x_{\text{d}_j}\| \leq r_i + r_j + 2r_M + 2\bar{r} + \varepsilon, \forall j\in\mathcal{N}\backslash\{i\}  \}, 
\end{align*}
$\forall i\in \mathcal{N}_\mathcal{F}$. The free space for the agents is defined then as 
\begin{align*}
\mathcal{F}_i \coloneqq \bar{\mathcal{W}}_i \bigg\backslash \left\{ \left( \bigcup_{j\in\mathcal{J}} \bar{\mathcal{O}}_{i,j} \right)   \bigcup  \mathcal{C}_i  \right\}, \forall i\in\mathcal{N}.
\end{align*}
It can be verified that, in view of \eqref{eq:r_bar_mas (Automatica_adaptive_nav)}, the sets $\mathcal{F}_i$ are nonempty and  $x(t_0) \in \mathcal{F}_M\coloneqq \bigcap_{i\in\mathcal{N}}\mathcal{F}_i$. 

The main difference lies in the fact that the follower agents aim to keep a larger distance from each other, the obstacles, and the workspace boundary than the leader agent, and in particular, a distance enhanced by $2r_M + 2\bar{r}$. In that way, the leader agent will be able to choose an  appropriate constant $\tau$ (as in the single-agent case of Section \ref{sec:main (Automatica_adaptive_nav)}) so that  it is influenced at each time instant only by one of the obstacles/followers, and will be also able to navigate among the obstacles/followers. Note that the followers are required to stay away also from other agents' destinations, since a potential local {minimum} in such configurations can prevent the leader agent from reaching its goal. 
We provide next the mathematical details of the aforementioned reasoning.

Consider the leader distances $d_{i_\mathcal{L},o_k}$, $d_{i_\mathcal{L},j}$, $d_{i_\mathcal{L},o_0}:\mathcal{F}_{i_\mathcal{L}} \to \mathbb{R}_{\geq 0}$ as 
\begin{align*}
& d_{i_\mathcal{L},o_k}\coloneqq d_{i_\mathcal{L},o_k}(x) \coloneqq \|x_{i_\mathcal{L}} - c_k \|^2 - (r_{i_\mathcal{L}} + r_{o_k})^2, \forall k\in\mathcal{J} \\
& d_{i_\mathcal{L},j}\coloneqq d_{i_\mathcal{L},j}(x) \coloneqq \|x_{i_\mathcal{L}} - x_j \|^2 - (r_{i_\mathcal{L}} + r_j)^2, \forall j\in\mathcal{N}_\mathcal{F} \\
& d_{i_\mathcal{L},o_0}\coloneqq d_{i_\mathcal{L},o_0}(x) \coloneqq (r_\mathcal{W}+r_{i_\mathcal{L}})^2 -  \|x_{i_\mathcal{L}}\|^2
\end{align*}
and the follower distances $d_{i,o_k}$, $d_{i,i_\mathcal{L}}$, $d_{i,j}$, $d_{i,\text{d}_j}$ $d_{i,o_0}:\mathcal{F}_i \to \mathbb{R}_{\geq 0}$ as 
\begin{align*}
& d_{i,o_k}\coloneqq d_{i,o_k}(x) \coloneqq \|x_i - c_k \|^2 - (r_i + r_{o_k} + 2r_M + 2\bar{r})^2, \forall k\in\mathcal{J} \\
& d_{i,i_\mathcal{L}}\coloneqq d_{i,i_\mathcal{L}}(x) \coloneqq \|x_i - x_{i_\mathcal{L}} \|^2 - (r_i + r_{i_\mathcal{L}})^2 = d_{i_\mathcal{L},i}(x) \\
& d_{i,j}\coloneqq d_{i,j}(x) \coloneqq \|x_i - x_j \|^2 - (r_i + r_j + 2r_M + 2\bar{r})^2, \forall j\in\mathcal{N}_\mathcal{F} \\
& d_{i,\text{d}_j}\coloneqq d_{i,\text{d}_j}(x) \coloneqq \|x_i - x_{\text{d}_j} \|^2 - (r_i + r_j + 2r_M + 2\bar{r} + \varepsilon)^2,  \forall j\in\mathcal{N}\backslash\{i\} \\
& d_{i,o_0}\coloneqq d_{i,o_0}(x) \coloneqq (r_\mathcal{W} - r_i - 2r_M - 2\bar{r})^2 -  \|x_i\|^2,
\end{align*}
$\forall i\in\mathcal{N}_\mathcal{F}$. Note that $d_{i,j}(x) = d_{j,i}(x)$, $\forall i,j\in\mathcal{N}_\mathcal{F}$, with $i \neq j$ and also that $x\in\mathcal{F}_M$ is equivalent to all the aforementioned distances being positive.

Let now functions $\beta$, $\beta_i$, $i\in\mathcal{N}$, that satisfy the properties of Definition \ref{def:2nd nf (Automatica_adaptive_nav)}, as well as the respective constants $\tau$, $\tau_i$, such that $\beta'(z)=\beta''(z) =0$, $\forall z\geq \tau$, $\beta'_i(z) = \beta''_i(z) = 0$, $\forall z \geq \tau_i$, $i\in\mathcal{N}$. The $2$nd-order navigation functions for the agents {are} now defined as $\phi_i\coloneqq \phi_i(x):\mathcal{F}_i \to \mathbb{R}_{\geq 0}$, $\forall i\in\mathcal{N}$, with
\begin{align*}
\phi_i(x) &\coloneqq k_{1_i}\|x_i - x_{\text{d}_i}\|^2 + k_{2_i}\bigg(b_{1_i}(x) + b_{2_i}(x) + k_{f_i}b_{3_i}(x)\bigg) \\
b_{1_i}&\coloneqq b_{1_i}(x)\coloneqq \sum_{j\in\bar{\mathcal{J}}} \beta_i( d_{i,o_j}(x) ) \\
b_{2_i} &\coloneqq b_{2_i}(x) \coloneqq \sum_{j\in\mathcal{N}\backslash\{i\}}  \beta(d_{i,j}(x)) \\
b_{3_i}&\coloneqq b_{3_i}(x) \coloneqq \sum_{j\in\mathcal{N}\backslash\{i\}}  \beta_i(d_{i,{\text{d}_j}}(x)),
\end{align*}
and $k_{f_{i_\mathcal{L}}} = 0$, $k_{f_i} = 1$, $\forall i\in\mathcal{N}_\mathcal{F}$. Note that the robotic agents can choose independently their $\tau_i$, $i\in\mathcal{N}$, that concerns the collision avoidance with the obstacles and the workspace boundary. The pair-wise inter-agent distances, however, are required to be the same and hence the same $\beta$ (and hence $\tau$) is chosen (see the terms $b_{2_i}(x)$ in $\phi_i(x)$), {which can, nevertheless, be done off-line.}
To achieve convergence of the leader to its destination, we choose $\tau$ and $\tau_{i_\mathcal{L}}$ as in Section \ref{sec:main (Automatica_adaptive_nav)}, i.e., $\tau, \tau_{i_\mathcal{L}} \in (0, \min\{\bar{r}^2,\bar{r}_\text{d}\})$. Regarding the ability of the agents to sense each other when $d_{i,j}(x) < \tau$,  it holds that
\begin{align*}
&d_{i,j}(x) < \tau \Leftrightarrow \|x_i - x_j\|^2 \leq \tau + (r_i + r_j + 2r_M + 2\bar{r})^2 \Rightarrow \\
&\|x_i - x_j\| \leq \sqrt{\tau} + r_i + r_j + 2r_M + 2\bar{r} \Rightarrow \\
&\|x_i - x_j\| \leq \sqrt{\min\{\bar{r}^2,\bar{r}_\text{d}\}} + r_i + r_j + 2r_M + 2\bar{r} < \varsigma_i,
\end{align*}
$\forall i,j\in\mathcal{N}$, $i\neq j$, as dictated by Assumption \ref{ass:sensing (Automatica_adaptive_nav)}. 

The control protocol follows the same structure as the single-agent case presented in Section \ref{sec:main (Automatica_adaptive_nav)}. In particular, we define the reference velocity for each agent as $v_{\text{d}_i}:\mathcal{F}_i\to\mathbb{R}^n$, with
\begin{equation} \label{eq:v_d MAS (Automatica_adaptive_nav)}
v_{\text{d}_i}\coloneqq v_{\text{d}_i}(x) \coloneqq -\nabla_{x_i}\widetilde{\phi}_i(x),
\end{equation}
where $\widetilde{\phi}_i:\mathcal{F}_i\to\mathbb{R}_{\geq 0}$ is the slightly modified function:
\begin{align*}
\widetilde{\phi}_i(x) &\coloneqq k_{1_i}\|x_i - x_{\text{d}_i}\|^2 + k_{2_i}\bigg(b_{1_i}(x) + 2b_{2_i}(x) + k_{f_i}b_{3_i}(x)\bigg).
\end{align*}
{The need for modification of $\phi_i$ to $\widetilde{\phi}_i$ stems from the  differentiation of the terms $b_{2_i}$, as will be clarified in the subsequent analysis.}

The control law is now designed as $u_i:\mathcal{F}_i\times\mathbb{R}^{Nn+2} \to \mathbb{R}^n$, with
\begin{align} \label{eq:u_MAS (Automatica_adaptive_nav)}
u_i \coloneqq &u_i(x,v,\hat{m}_i,\hat{\alpha}_i) \coloneqq  -k_{\phi_i} \nabla_{x_i}\widetilde{\phi}_i(x) + \hat{m}_i(\dot{v}_{\text{d}_i} + g) - \left(k_{v_i} + \frac{3}{2}\hat{\alpha}_i \right)e_{v_i}, 
\end{align}
$\forall i\in\mathcal{N}$; $k_{\phi_i}$, $k_{v_i}$ are positive constants, $e_{v_i}$ are the velocity errors $e_{v_i} \coloneqq v_i - v_{\text{d}_i}$, and $\hat{m}_i$, $\hat{\alpha}_i$ denote the estimates of $m_i$ and $\alpha_i$, respectively, by agent $i$, evolving according to 
\begin{subequations} \label{eq:adaptation laws MAS (Automatica_adaptive_nav)}
	\begin{align}
	\dot{\hat{m}}_i \coloneqq& -k_{m_i} e_{v_i}^\top( \dot{v}_{\text{d}_i}+g)  \\
	\dot{\hat{\alpha}}_i \coloneqq & k_{\alpha_i} \|e_{v_i}\|^2, 
	\end{align}
\end{subequations}
with $k_{m_i}$, $k_{\alpha_i}$ positive gain constants, $\hat{\alpha}_i(t_0) \geq 0$, and arbitrary initial conditions $\hat{m}_i(t_0)$, $\forall i\in\mathcal{N}$. We further denote $\hat{m} \coloneqq [\hat{m}_1,\dots,\hat{m}_N]^\top$, $\hat{\alpha} \coloneqq [\hat{\alpha}_1,\dots,\hat{\alpha}_N]^\top \in \mathbb{R}^{N}$.

As presented below, the leader agent will converge to its destination from almost all initial conditions that satisfy Assumption \ref{ass:MAS workspace (Automatica_adaptive_nav)}, whereas the followers might get stuck in local minima. Once the leader reaches $\varepsilon$-close to its destination (where $\varepsilon$ was introduced in Assumption \ref{ass:MAS workspace (Automatica_adaptive_nav)}), it switches off its control and the next robotic agent in the priority list becomes the leader. We assume that once an agent reaches its goal, it can broadcast this information for the next agent in priority.
This occurs iteratively until all the robotic agents reach their destinations. The following theorem considers the convergence of a leader to its destination. 

\begin{theorem}\label{th:MAS (Automatica_adaptive_nav)}
	Consider $N$ robots operating in $\mathcal{W}$, subject to the uncertain $2$nd-order dynamics \eqref{eq:dynamics MAS (Automatica_adaptive_nav)}, and a leader $i_\mathcal{L}$.  
	Under Assumptions \ref{ass:f (Automatica_adaptive_nav)}-\ref{ass:sensing (Automatica_adaptive_nav)}, the control protocol \eqref{eq:v_d MAS (Automatica_adaptive_nav)}-\eqref{eq:adaptation laws MAS (Automatica_adaptive_nav)} guarantees collision avoidance between the agents and the agents and obstacles/workspace boundary as well as convergence of $x_{i_\mathcal{L}}$ to $x_{\textup{d}_{i_\mathcal{L}}}$ from almost all initial conditions $(x(t_0),v(t_0),\hat{m}(t_0),\hat{\alpha}(t_0))$ $\in$ $\mathcal{F}_M\times\mathbb{R}^{N(n+1)}\times\mathbb{R}^N_{\geq 0}$, given sufficiently small $\tau$, $\tau_{i_\mathcal{L}}$, and that $k_{\phi_i} > \frac{\alpha_i}{2}$, $i\in\mathcal{N}$. Moreover, all closed loop signals remain bounded, $\forall t \geq t_0$.
\end{theorem} 


\begin{proof} 
	We prove first the avoidance of collisions by considering the function
	\begin{equation*}
	V_M \coloneqq \sum_{i\in\mathcal{N}}\bigg\{k_{\phi_i}\phi_i + \frac{m_i}{2}\|e_{v_i}\|^2 + \frac{3}{4k_{\alpha_i}}\widetilde{\alpha}_i^2 + \frac{1}{2k_{m_i}}\widetilde{m}_i^2 \bigg\}.
	\end{equation*} 
	Since $x(t_0) \in {F}_M$, $V_M(t_0)$ is bounded. 
	Differentiation of $V_M$ yields, after using the property $$\sum_{i\in\mathcal{N}}\sum_{j\in\mathcal{N}\backslash\{i\}}(x_i-x_j)^\top(v_i - v_j) = 2\sum_{i\in\mathcal{N}}\sum_{j\in\mathcal{N}\backslash\{i\}}(x_i-x_j)^\top v_i,$$
	\begin{align*}
	\dot{V}_M =& \sum_{i\in\mathcal{N}}\Bigg\{ 2k_{\phi_i}k_{1_i}(x_i-x_{\text{d}_i}) - 2k_{\phi_i}k_{2_i}\bigg(\beta_i'(d_{i,o_0})x - \sum_{k\in\mathcal{J}}\beta_i'(d_{i,o_k})(x_i - c_k) \\
	& - 2\sum_{j\in\mathcal{N}\backslash\{i\}}\beta'(d_{i,j})(x_i - x_j) - \sum_{j\in\mathcal{N}\backslash\{i\}}k_{f_i} \beta_i'(d_{i_{\textup{d}_j}})(x_i - x_{\textup{d}_j}) \bigg)^\top v_i \\
	& + e_{v_i}^\top(u_i - f_i(x_i,v_i) - m_ig - m_i\dot{v}_{\text{d}_i} )  +  \frac{3}{2k_{\alpha_i}}\widetilde{\alpha}_i\dot{\hat{\alpha}}_i + \frac{1}{2k_{m_i}}\widetilde{m}_i\dot{\hat{m}}_i \Bigg\}  \\
	\leq & \sum_{i\in\mathcal{N}}\bigg\{ k_{\phi_i} \nabla_{x_i}\widetilde{\phi}_i(x)^\top v_i + e_{v_i}^\top(u_i -m_i(g + \dot{v}_{\text{d}_i}) ) + \frac{3}{2}\alpha_i\|e_{v_i}\|\|v_i\|  \\
	&+ \frac{3}{2}\widetilde{\alpha}_i\|e_{v_i}\|^2  - \widetilde{m}_ie_{v_i}^\top(\dot{v}_{\text{d}_i} + g)  \bigg\},
	\end{align*}
	which, by using $v_i = e_{v_i} + v_{\text{d}_i}$ and substituting the control and adaptation laws \eqref{eq:u (Automatica_adaptive_nav)},\eqref{eq:adaptation laws MAS (Automatica_adaptive_nav)}, becomes
	\begin{align*}
	\dot{V}_M \leq -\sum_{i\in\mathcal{N}}\bigg\{ \left(k_{\phi_i}-\frac{\alpha_i}{2}\right)\|\nabla_{x_i}\widetilde{\phi}_i(x)\|^2 + k_{v_i}\|e_{v_i}\|^2 \bigg\} \leq  0,	
	\end{align*}
	and hence, $V_M(t) \leq V(t_0)$, which implies the boundedness of all {closed-loop} signals as well as that collisions between the agents and the agents and obstacles/workspace boundary are avoided $\forall t \geq t_0$. Moreover, following similar arguments as in the proof of Theorem \ref{th:single robot (Automatica_adaptive_nav)}, we conclude that $\lim_{t\to\infty}\|\nabla_{x_i}\widetilde{\phi}_i(x(t))\| = \lim_{t\to\infty}\|e_{v_i}(t)\|$ $=$ $\lim_{t\to\infty}\|v_i(t)\|$ $=$ $\lim_{t\to\infty}\|\dot{v}_i(t)\|$ $=$ $0$, $\forall i\in\mathcal{N}$. For the followers $\mathcal{N}_\mathcal{F}$, depending on the choice of $\tau_i$, $i\in\mathcal{N}_\mathcal{F}$, the critical point $\nabla_{x_i}\widetilde{\phi}_i(x(t)) = 0$ might either correspond to their destination $x_{\text{d}_i}$ or a local minimum. In any case, it holds that $x(t)\in\mathcal{F}_M$, $\forall t \geq t_0$, and hence, for all the followers $i\in\mathcal{N}_\mathcal{F}$,
	\begin{subequations} \label{eq:d_j static obs (Automatica_adaptive_nav)}
		\begin{align}
		& \|x_i(t) - c_k\| > r_i + r_{o_k} + 2r_M + 2\bar{r},\forall k\in\mathcal{J} \label{eq:d_j static obs 1 (Automatica_adaptive_nav)}\\
		&\|x_i(t) - x_j(t)\| > r_i + r_j + 2r_M + 2\bar{r}, 
		\forall j\in\mathcal{N}_\mathcal{F}\backslash\{i\} \label{eq:d_j static obs 2 (Automatica_adaptive_nav)}\\
		& r_\mathcal{W} - \|x_i\|  > r_i + 2r_M + 2\bar{r}, \label{eq:d_j static obs 3 (Automatica_adaptive_nav)}\\		
		& \|x_i(t) - x_{\text{d}_j} \| > r_i + r_j + 2r_M + 2\bar{r} +\varepsilon, 
		\forall j\in\mathcal{N}\backslash\{i\}, \label{eq:d_j static obs 4 (Automatica_adaptive_nav)}
		\end{align}
	\end{subequations}
	$\forall t > t_0$. Therefore, since $\lim_{t\to\infty}\|v_i(t)\| = \lim_{t\to\infty}\|\dot{v}_i(t)\| = 0$, $\forall i\in\mathcal{N}$, the multi-robot case reduces to the single-robot case of Section \ref{sec:main (Automatica_adaptive_nav)}, where the followers resemble static obstacles. Note that the obstacle constraints \eqref{eq:r_bar (Automatica_adaptive_nav)} are always satisfied by the followers (see \eqref{eq:d_j static obs 1 (Automatica_adaptive_nav)}-\eqref{eq:d_j static obs 3 (Automatica_adaptive_nav)}); \eqref{eq:d_j static obs 4 (Automatica_adaptive_nav)} implies that the configuration that corresponds to the leader destination, i.e., $[x_1^\top,\dots,x_{i_\mathcal{L}-1}^\top,x_{\text{d}_{i_\mathcal{L}}}^\top,x_{i_\mathcal{L}+1}^\top,\dots,x_N^\top]^\top$, belongs always in its free space $\mathcal{F}_{i_\mathcal{L}}$.
	Hence, by choosing sufficiently small $\tau, \tau_{i_\mathcal{L}}$ in the interval $(0,\min(\bar{r}^2,\bar{r}_\text{d}))$, with $\bar{r}_\text{d}$ as defined in \eqref{eq:bar_r_d (Automatica_adaptive_nav)}, we guarantee the safe navigation of $x_{i_\mathcal{L}}$ to $x_{\text{d}_{i_\mathcal{L}}}$ from almost all initial conditions, as in Section \ref{sec:main (Automatica_adaptive_nav)}.
\end{proof}

When the current leader $i_\mathcal{L}$ reaches $\varepsilon$-close to its goal, at a time instant $t_{i_{\mathcal{L}}}$\footnote{Note that the proven asymptotic stability of Theorem \ref{th:MAS (Automatica_adaptive_nav)} guarantees that this will occur in finite time.}, it broadcasts this information to the other agents, switches off its control and remains immobilized, considered hence as a static obstacle with center $c_{M+1}\coloneqq x_{i_\mathcal{L}}(t_{i_{\mathcal{L}}})$ and radius $r_{M+1}$  by the rest of the team. Note that $\|c_{M+1} - x_{\text{d}_{i_\mathcal{L}}}\| \leq \varepsilon$ and hence, in view of \eqref{eq:r_bar_mas (Automatica_adaptive_nav)}, $\|c_j - c_{M+1}\| > r_{o_j} + r_{i_\mathcal{L}} + 2r_M + 2\bar{r}$, $\forall j\in\mathcal{J}$, and $r_\mathcal{W} - \|c_{M+1}\| > r_{i_\mathcal{L}}+2r_M + 2\bar{r}$, satisfying the obstacle spacing properties \eqref{eq:r_bar (Automatica_adaptive_nav)}. 	
The next agent $i'_{\mathcal{L}}\in \widetilde{\mathcal{N}}\coloneqq \mathcal{N}\backslash \{i_{\mathcal{L}}\}$  in priority is then assigned as a leader for navigation, and we redefine the sets 
\begin{align*}
&\widetilde{\bar{\mathcal{O}}}_{i'_\mathcal{L},j} \coloneqq \{ q \in {\bar{\mathcal{W}}_{i'_\mathcal{L}}} : \|q_i-c_j\| \leq r_{o_j} + r_i \}, \forall j\in\widetilde{\mathcal{J}}\\
& \widetilde{\mathcal{C}}_{i'_\mathcal{L}} \coloneqq \{ q \in {\bar{\mathcal{W}}_{i'_\mathcal{L}}}: \|q_{i'_\mathcal{L}} - q_j \| \leq r_{i'_\mathcal{L}} + r_j, \forall j\in \widetilde{\mathcal{N}} \backslash \{i'_\mathcal{L}\}  \},  \\
&\widetilde{\bar{\mathcal{O}}}_{i,j} \coloneqq \{ q \in \bar{\mathcal{W}}_i : \|q_i-c_j\| \leq r_{o_j} + r_i + 2r_M + 2\bar{r} \}, \forall j\in\widetilde{\mathcal{J}} \\
&\widetilde{\mathcal{C}}_i \coloneqq \{ q \in \bar{\mathcal{W}}_i: \|q_i - q_{i'_\mathcal{L}} \| \leq r_i + r_{i'_\mathcal{L}},  \\
&\hspace{12mm}  \|q_i - q_j \| \leq r_i + r_j + 2r_M + 2\bar{r}, \forall j\in\widetilde{\mathcal{N}} \backslash \{i'_\mathcal{L}{,i}\}, \\
&\hspace{12mm}  \|q_i-x_{\textup{d}_j}\| \leq r_i + r_j + 2r_M + 2\bar{r} + \varepsilon, \forall j\in\widetilde{\mathcal{N}}\backslash\{i\}  \},  
\end{align*}
$\forall i\in\widetilde{\mathcal{N}}\backslash\{i'_\mathcal{L}\}$, where $\widetilde{\mathcal{J}} \coloneqq \mathcal{J}\cup\{M+1\}$, to account for the new obstacle $M+1$. The new free space is  $$\widetilde{\mathcal{F}}_i \coloneqq \bar{\mathcal{W}}_i \backslash \left\{ \left( \bigcup_{j\in\widetilde{\mathcal{J}}} \widetilde{\bar{\mathcal{O}}}_{i,j} \right) \cup  \widetilde{\mathcal{C}}_i \right\}, \forall i\in\widetilde{\mathcal{N}}$$ and, in view of \eqref{eq:d_j static obs (Automatica_adaptive_nav)}, one can conclude that $x_{i'_{\mathcal{L}}}(t_{i_{\mathcal{L}}}) \in \widetilde{\mathcal{F}}_{i'_{\mathcal{L}}}$, $x_i(t_{i_{\mathcal{L}}}) \in \widetilde{\mathcal{F}}_i$ $\forall i\in\widetilde{\mathcal{N}}\backslash\{i'_{\mathcal{L}}\}$. Therefore, the application of Theorem \ref{th:MAS (Automatica_adaptive_nav)} with $t_{i_{\mathcal{L}}}$ as $t_0$ and agent $i'_{\mathcal{L}}$ as leader guarantees its navigation $\varepsilon$-close to $x_{\text{d}_{i'_{\mathcal{L}}}}$. Applying iteratively the aforementioned reasoning, we guarantee the successful navigation of all the agents.}
More specifically, we initially set off-line the priorities of the agents based on a desired metric (e.g., distance to the goal), and set $i_\mathcal{L}$ as the top priority agent. 
Then the following procedure is iterated until all agents have reached their goals. The agents apply the control protocol \eqref{eq:v_d MAS (Automatica_adaptive_nav)}-\eqref{eq:adaptation laws MAS (Automatica_adaptive_nav)}. When the leader agent satisfies $\|x_{i_{\mathcal{L}}} - x_{\text{d}_{i_\mathcal{L}}}\| \leq \varepsilon$, it switches off its control, broadcasts this information to all other agents, and the next leader is chosen as the next agent in priority. Therefore, in view of Theorem \ref{th:MAS (Automatica_adaptive_nav)}, all agents will eventually reach $\varepsilon$-close to their destinations. {This is illustrated in Algorithm \ref{alg:1 (Automatica_adaptive_nav)}, which is run by each agent separately. The algorithm receives as input the agent index and destination $i$, $x_{\text{d}_i}$, respectively, as well as the priority vector $\mathsf{Pr}$, which have been set a priori. Next, depending on the priority (lines 3, 4), agent $i$ applies the control algorithm \eqref{eq:v_d MAS (Automatica_adaptive_nav)}-\eqref{eq:adaptation laws MAS (Automatica_adaptive_nav)} (line 8). In  case agent $i$ has the top priority and reaches its goal, it broadcasts to the other agents that it has arrived and exits the loop (lines 5-7). Finally, the agents are equipped with a callback function $\mathsf{Receive}$ that continuously checks whether some agent $j\in\mathcal{N}\backslash\{i\}$ broadcasts the arrival to its destination, so that they update accordingly the priority vector $\mathsf{Pr}$ (lines 9, 10). Note that the latter is a synchronous procedure and the priority variable $\mathsf{Pr}$ is always the same for all agents.}

\begin{algorithm}	
	\caption{Hybrid Control Strategy for Agent $i$}\label{alg:1 (Automatica_adaptive_nav)}
	\begin{algorithmic}[1]
		\State \textbf{Input}: $i$, $x_{\text{d}_i}$, $\mathsf{Pr} $
		\While { $\mathsf{True}$ }		
		\If {$\mathsf{Pr}[i] > \mathsf{Pr}[j], \forall j\in\mathcal{N}\backslash\{i\} $}
		\State $i_\mathcal{L} \leftarrow i$;
		\If{$\|x_i - x_{\text{d}_i}\| \leq \varepsilon$}
		\State $\mathsf{Broadcast}(``\mathsf{arrived}")$;
		\State $\mathsf{break}$;
		\EndIf
		\EndIf
		\State Apply \eqref{eq:v_d MAS (Automatica_adaptive_nav)}-\eqref{eq:adaptation laws MAS (Automatica_adaptive_nav)} 
		\If {$\mathsf{Receive}(``\mathsf{Arrived}",j)$ }
		\State $\mathsf{Pr} \leftarrow \mathsf{Update}(\mathsf{Pr},j)$;
		\EndIf
		\EndWhile
		%
	\end{algorithmic}
\end{algorithm}


{As a final remark, note that $\varepsilon$ can be arbitrarily small, achieving thus practical convergence of the agents to {their} destinations $x_{\text{d}_i}$, $i\in\mathcal{N}$}.


\subsection{Simulation Results} \label{sec:sim (Automatica_adaptive_nav)}

This section verifies the theoretical findings of Sections \ref{sec:main (Automatica_adaptive_nav)}-\ref{sec:MAS (Automatica_adaptive_nav)} via computer simulations. 

\subsubsection{Sphere worlds} \label{sec:Sim simple (Automatica_adaptive_nav)}

We consider first a $2$D workspace on the horizontal plane with $r_\mathcal{W} = 8$, populated with $M = 50$ randomly placed obstacles, whose radius, enlarged by the robot radius, is $\bar{r}_{o_j} = 0.5$, $\forall j\in\mathcal{J}$, as depicted in Fig. \ref{fig:2d_ws_example (Automatica_adaptive_nav)}. The mass, and function $f(x,v)$, both \textit{unknown} to the robotic agent, are taken as $m=1$, and $f(x,v) = \frac{\alpha}{16} \sin(0.5(x_\mathsf{x} + x_\mathsf{y}))F(v)v$, with $F(v) = \text{diag}\{ [\exp(-\text{sgn}(v_i)v_i) + 1]_{i\in\{\mathsf{x},\mathsf{y}\}} \}$,
and $\alpha=10$, where we denote $(x_\mathsf{x},x_\mathsf{y}) = x$, $(v_\mathsf{x},v_\mathsf{y}) = v$.
We choose the goal position as $x_\text{d} = (5,5)$, which the robot aims to converge to from $3$ different initial positions, namely $x(0) = -(5,5), (-6,4.5)$, and $(3.5,-7)$. We choose a variation of \eqref{eq:beta_exps (Automatica_adaptive_nav)} for $\beta$ with $\tau = \bar{r}^2$ and $\bar{\beta}$ = 100. The control gains are chosen as $k_1 = 0.04$, $k_2=5$, $k_v = 20$, $k_\phi = 1$, and $k_m = k_\alpha = 0.01$. The results for $t\in[0,100]$ seconds are depicted in Figs. \ref{fig:2d_traj_1_2_3 (Automatica_adaptive_nav)}, \ref{fig:2d_u_1_2_3 (Automatica_adaptive_nav)}; \ref{fig:2d_traj_1_2_3 (Automatica_adaptive_nav)} (left) shows that the robot navigates to its destination without any collisions, and \ref{fig:2d_u_1_2_3 (Automatica_adaptive_nav)} depicts the input and adaptation signals $u(t)$, $\hat{\alpha}(t)$, $\hat{m}(t)$. In addition, note that the fact that $\alpha > 2$ does not affect the performance of the proposed control protocol and hence we can verify that the condition $k_\phi > \frac{\alpha}{2}$ is only sufficient and not necessary. {Moreover, in order to verify the results of Section \ref{subsec:disturbance (Automatica_adaptive_nav)}, we add a bounded time-varying disturbance vector $d(x,v,t) = d(t) \coloneqq 2\left[\sin(0.5t+\frac{pi}{3}), \cos(0.4t-\frac{\pi}{4})\right]^\top \in \mathbb{R}^2$ and we choose the extra control gains as $\sigma_m = \sigma_\alpha = 0.1$. The results are depicted in Fig. \ref{fig:2d_traj_1_2_3 (Automatica_adaptive_nav)} (right), which shows the collision-free navigation of the agent to a set close to $x_\text{d}$, and Fig. \ref{fig:2d_u_1_2_3 (Automatica_adaptive_nav)}, which shows the input and adaptation signals $u(t)$, $\hat{\alpha}(t)$, $\hat{m}(t)$.}

\begin{figure}[!ht]
	\centering
	\includegraphics[trim = 0cm 0cm 0cm -0cm,width = \textwidth]{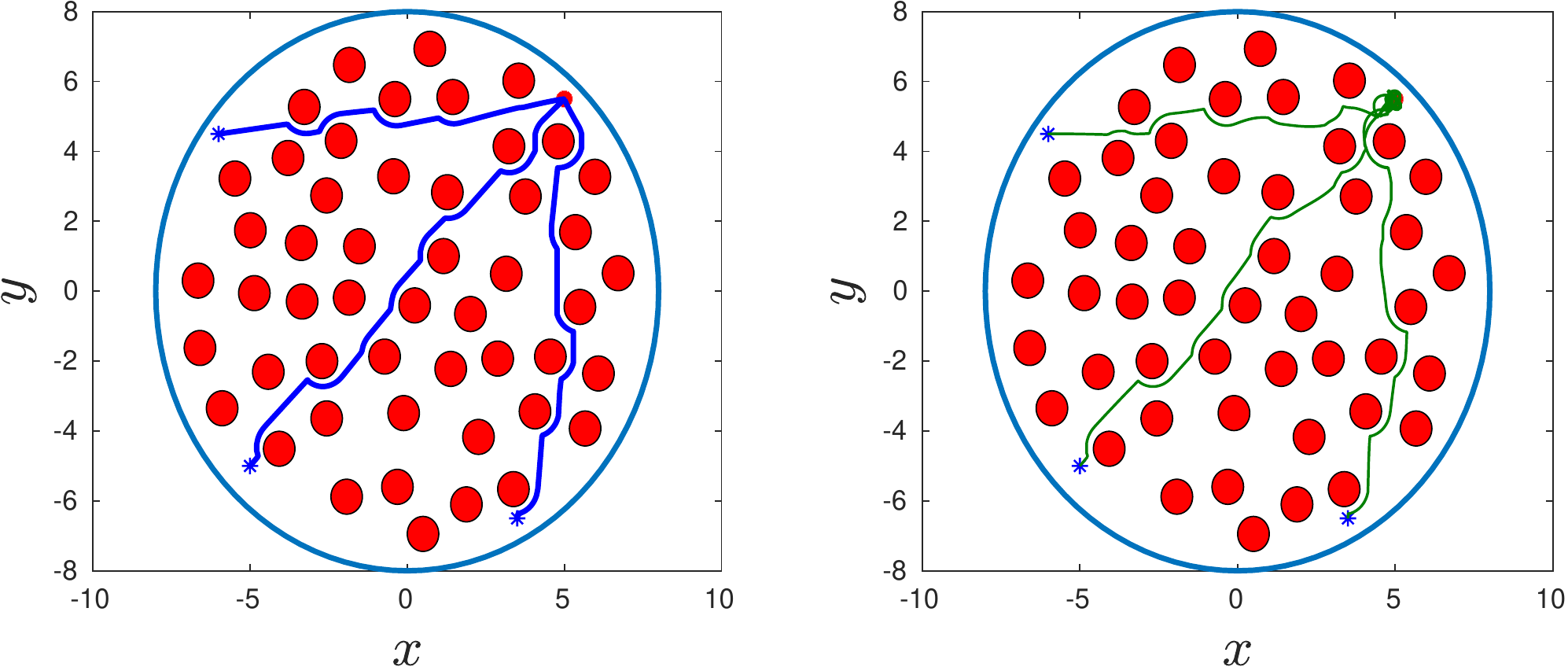}\\
	\caption{The resulting trajectories $x(t)$, $t\in[0,100]$ seconds, from the initial points $-(5,5), (-6,4.5)$, and $(3.5,-7)$ to the destination $(5,5)$. Left: without any disturbances. Right: with bounded disturbance $d(x,v,t)$. }\label{fig:2d_traj_1_2_3 (Automatica_adaptive_nav)}
\end{figure}

\begin{figure}[!ht]
	\centering
	\includegraphics[trim = 0cm 0cm 0cm -0cm,width = 0.9\textwidth]{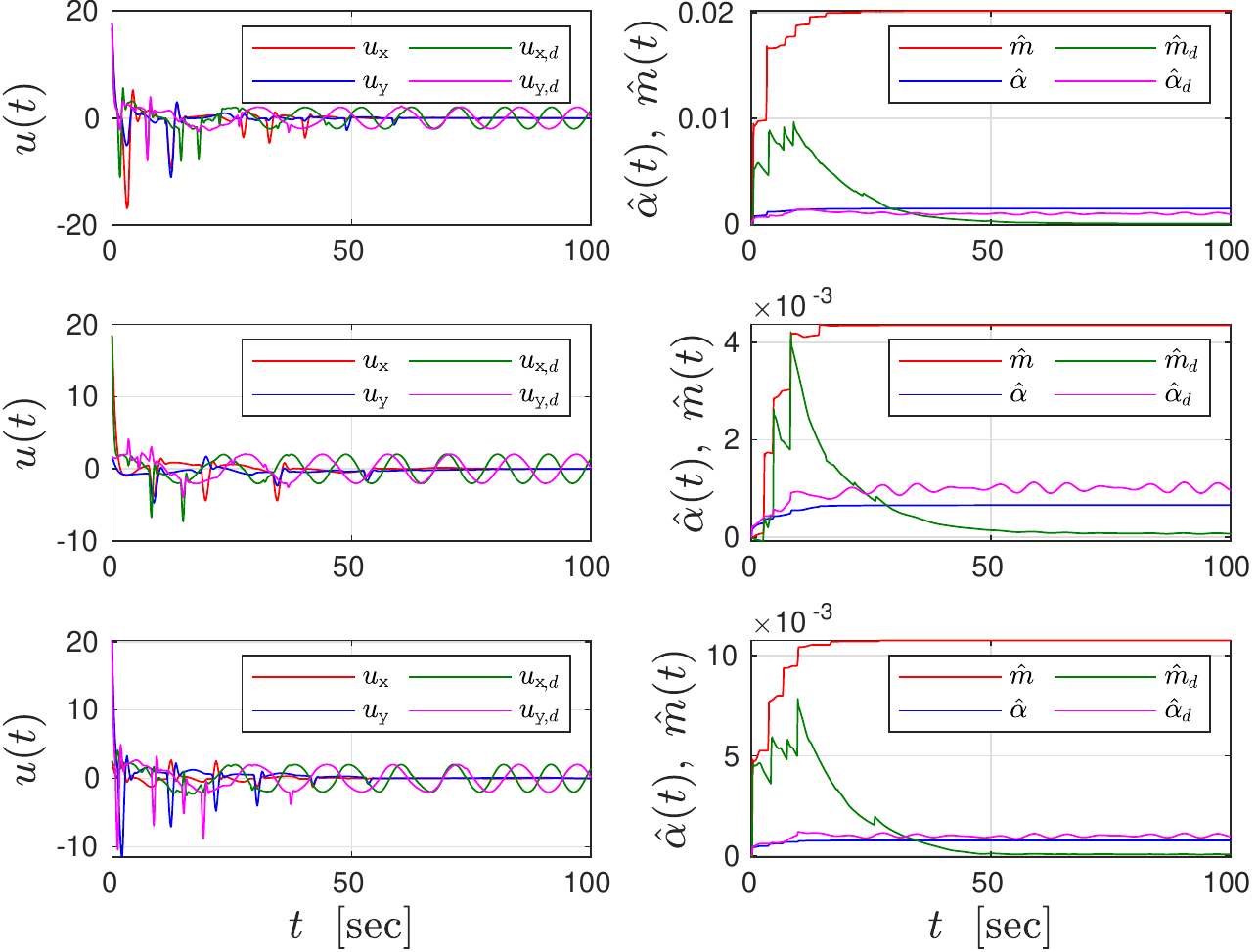}\\
	\caption{Left: The resulting input signals $u(t)=(u_\mathsf{x}(t),u_{\mathsf{y}}(t))$, $t\in[0,100]$ seconds, for the $2$D trajectories of Fig. \ref{fig:2d_traj_1_2_3 (Automatica_adaptive_nav)}. Right: The resulting adaptation signals $\hat{\alpha}(t)$, $\hat{m}(t)$, $t\in[0,100]$ seconds, for the $2$D trajectories of Fig. \ref{fig:2d_traj_1_2_3 (Automatica_adaptive_nav)}. The extra subscript $d$ corresponds to the model where a bounded disturbance vector $d(x,v,t)$ was included. }\label{fig:2d_u_1_2_3 (Automatica_adaptive_nav)}
\end{figure}

Next, we consider a $3$D workspace with $r_\mathcal{W} = 8$, populated with $M = 150$ randomly placed obstacles, whose radius, enlarged by the robot radius, is $\bar{r}_{o_j} = 0.5$, $\forall j\in\mathcal{J}$; $f(x,v)$ amd $m$ as well as the $\beta$ functions and control gains are chosen as in the $2$D scenario. 
We choose the goal position as $x_\text{d} = (4,4,4)$, which the robot aims to converge to from $3$ different initial positions, namely $x(0) = -(4,4,4), (-4,4,-4)$, and $(-4,-4,4)$. The parameter $\bar{r}$ is chosen as $\bar{r} = 0.75$. The robot navigation as well as the input and adaptation signals $u(t)$, $\hat{\alpha}(t)$, $\hat{m}(t)$ are depicted in Figs. \ref{fig:3d_traj_1_2_3 (Automatica_adaptive_nav)}, and \ref{fig:3d_u_1_2_3 (Automatica_adaptive_nav)} for $t\in[0,100]$ seconds. Note that the robot navigates to its destination without any collisions and that $\hat{m}$ converges to $m$, as predicted by the theoretical results.

\begin{figure}[!ht]
	\centering
	\includegraphics[trim = 0cm 0cm 0cm -0cm,width = 0.7\textwidth]{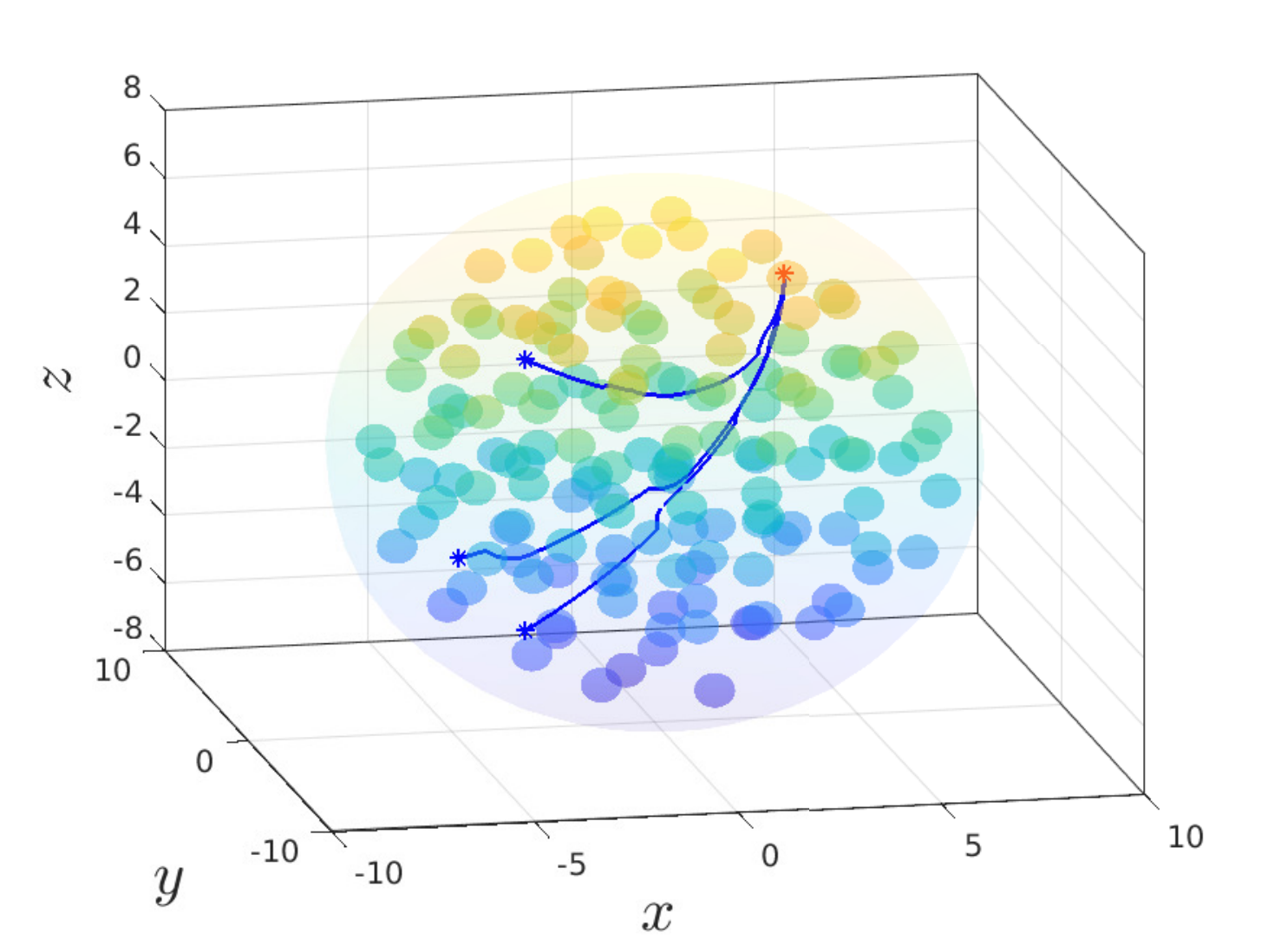}\\
	\caption{The resulting trajectories $x(t)$, $t\in[0,100]$ seconds, from the initial points $-(5,5), (-6,4.5)$, and $(3.5,-7)$ to the destination $(5,5)$.}\label{fig:3d_traj_1_2_3 (Automatica_adaptive_nav)}
\end{figure}

\begin{figure}[!ht]
	\centering
	\includegraphics[trim = .5cm 0cm 0cm -0cm,width = 0.9\textwidth]{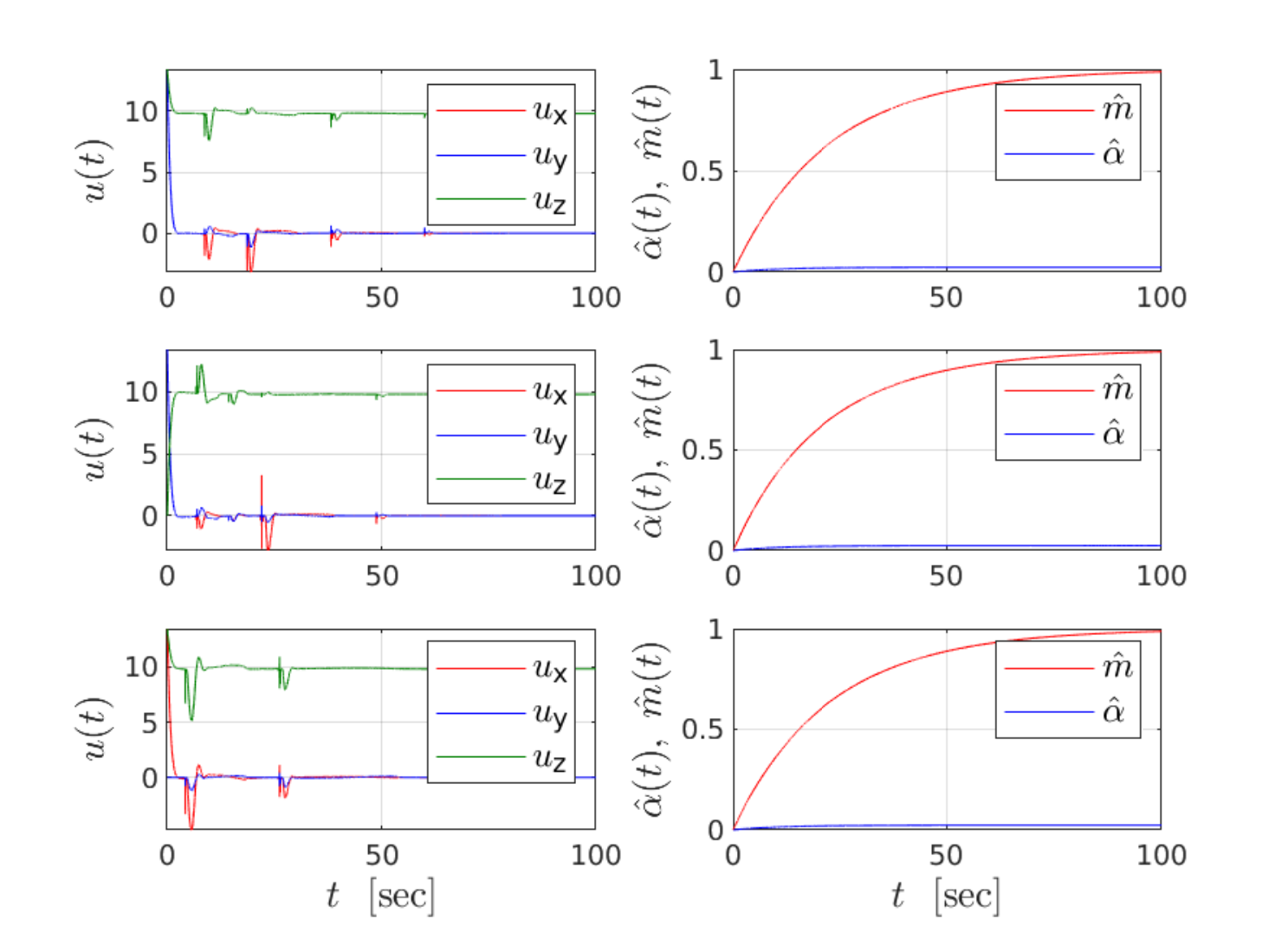}\\
	\caption{Left: The resulting input signals $u(t)=(u_\mathsf{x}(t),u_\mathsf{y}(t),u_\mathsf{z}(t))$, $t\in[0,100]$ seconds, for the $3$D trajectories of Fig. \ref{fig:3d_traj_1_2_3 (Automatica_adaptive_nav)}. Right: The resulting adaptation signals $\hat{\alpha}(t)$, $\hat{m}(t)$, $t\in[0,100]$ seconds, for the $3$D trajectories of Fig. \ref{fig:3d_traj_1_2_3 (Automatica_adaptive_nav)}.}\label{fig:3d_u_1_2_3 (Automatica_adaptive_nav)}
\end{figure}
%

\subsubsection{Star worlds} \label{sec:Sim star (Automatica_adaptive_nav)}

{
	Next, we illustrate the performance of the control protocol of Section \ref{sec:Star (Automatica_adaptive_nav)} in a $2$D and a $3$D star-world. We first consider the $2$D
	workspace shown in Fig. \ref{fig:2d_stars_initial (Automatica_adaptive_nav)}, with $r_\mathcal{W} = 8$, which contains $2$ star-shaped obstacles, centered at $(-3,-3)$ and $(0,1)$, respectively. The mass $m$ and function $f(x,v)$ are {given} as in the sphere-world case, with $\alpha = 1$.  
	In order to transform the workspace to a sphere world, we employ the transformation proposed in \cite{rimon1992exact}. In the transformed sphere world, we choose  $\bar{r}=4$ and $\bar{r}_{o_j}=0.5$, whereas the function $\beta$ is chosen as in the sphere-world case. The initial and goal position are selected as $x(0) = (-5,-5)$ and $x_\text{d} = (3,4)$, respectively, and the control gains as $k_1 = 0.04$, $k_2=.2$, $k_v = 20$, $k_\phi = 1$, and $k_m = k_\alpha = 0.01$. The results are depicted in Figs. \ref{fig:2d_stars_traj (Automatica_adaptive_nav)} and \ref{fig:u_alpha_2d_stars (Automatica_adaptive_nav)}, for $t\in[0,500]$ seconds. More specifically, \ref{fig:2d_stars_traj (Automatica_adaptive_nav)} shows the resulting trajectory, both in the original star world as well as in the transformed sphere world, and {Fig. \ref{fig:u_alpha_2d_stars (Automatica_adaptive_nav)}} depicts the resulting control input $u(t)$ and the adaptation signals $\hat{\alpha}(t)$, $\hat{m}(t)$.}

{
	Next, we consider a $3$D workspace, with $2$ star-shaped obstacles, centered at $(-4,-4,-2)$, $(1,2,2)$, similar to the previous $2$D star-shaped workspace, and $r_\mathcal{W} = 12$. By setting the initial and goal configurations at $(-5.1,-5.2,-5)$, and $(3,4,4)$, respectively, and all the parameters and control gains as in the $2$D counterpart, we obtain the results shown in Figs. \ref{fig:3d_stars_traj (Automatica_adaptive_nav)} and \ref{fig:u_alpha_3d_stars (Automatica_adaptive_nav)}, for $200$ seconds; \ref{fig:3d_stars_traj (Automatica_adaptive_nav)} shows the safe robot navigation to the goal and Fig. \ref{fig:u_alpha_3d_stars (Automatica_adaptive_nav)} depicts the evolution of the control and adaptation signals $u(t)$, $\hat{\alpha}(t)$, and $\hat{m}(t)$.}

{
	\subsubsection{Multi-Agent case}
	Finally, we use the control scheme of Section \ref{sec:MAS (Automatica_adaptive_nav)} in a multi-agent scenario. We consider $20$ agents in a $2$D workspace of $r_\mathcal{W} = 120$, populated with $70$ obstacles. The agents and obstacles are randomly initialized to satisfy the conditions of the free space of Section \ref{sec:MAS (Automatica_adaptive_nav)} (see Fig. \ref{fig:mas_initial (Automatica_adaptive_nav)}). The radius of the agents and the obstacles is chosen as $r_i=r_{o_j} = 2$, $\forall i\in\mathcal{N},j\in\mathcal{J}$, and the sensing radius of the agents is taken as $\varsigma_i = 20$, $\forall i\in\mathcal{N}$. The functions $\beta$, $\beta_i$ are chosen as in the previous subsections, and we also choose $\bar{r} = 4$, $\varepsilon = 0.1$. The results are depicted in Figs. \ref{fig:gamma_MAS (Automatica_adaptive_nav)}-\ref{fig:dis_MAS (Automatica_adaptive_nav)} for $870$ seconds. More specifically, Fig. \ref{fig:gamma_MAS (Automatica_adaptive_nav)} shows the convergence of the distance errors $\|x_i(t) - x_{\text{d}_i}\|$ to zero, $\forall i\in\mathcal{N}, t\in[0,870]$, and Fig. \ref{fig:traj_MAS (Automatica_adaptive_nav)} depicts the trajectories $x_i(t)$ of the agents in the workspace, $\forall i\in\mathcal{N}, t\in[0,870]$, from which it is clear that there {is} no collision with the workspace boundary. Finally, Fig. \ref{fig:dis_MAS (Automatica_adaptive_nav)} shows the minimum of the distances $\|x_i(t) - x_j(t)\| - 2r$, $\forall i,j\in\mathcal{N}$, $i\neq j$, and $\|x_i(t) - c_j\| - 2r$, $\forall i\in\mathcal{N}, j\in\mathcal{J}$, defined as 
	\small	
		\begin{align*}
		\beta_{\min}(t) \coloneqq &\min \bigg\{\min_{i,j\in\mathcal{N},i\neq j}\big\{\|x_i(t)-x_j(t)\|-2r\big\}, \min_{(i,j)\in\mathcal{N}\times\mathcal{J}}\big\{ \|x_i(t)-c_j\|-2r \big\} \bigg\},
		\end{align*}
	\normalsize
	which stays strictly positive, $\forall t\in[0,870]$, implying that collisions are avoided. A video illustrating the multi-robot case can be found on\\ \href{https://vimeo.com/393443782}{https://vimeo.com/393443782}. }

\begin{figure}[!ht]
	\centering
	\includegraphics[trim = 0cm 0cm 0cm -0cm,width = 0.75\textwidth]{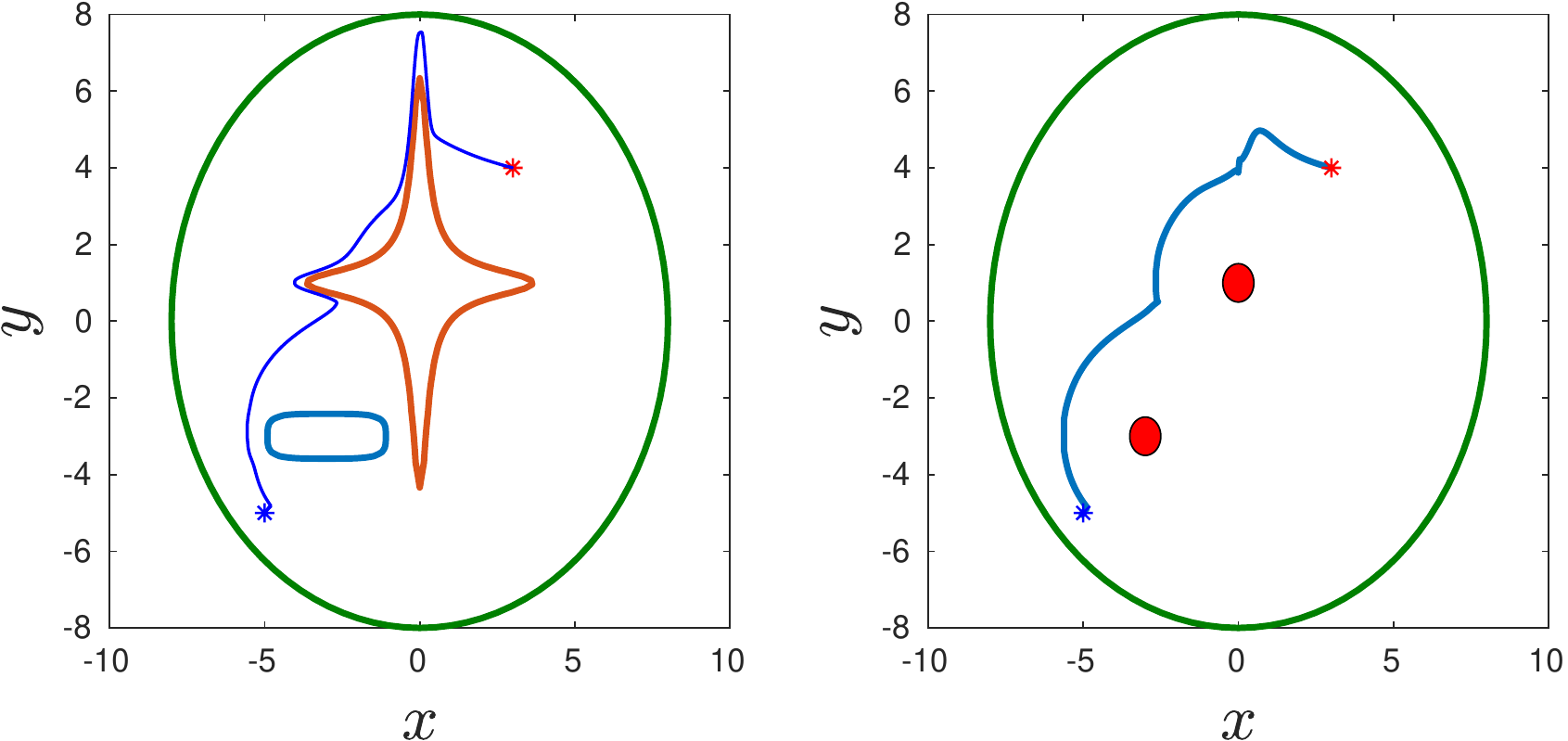}\\
	\caption{Left: The resulting trajectory $x(t)$, $t\in[0,500]$ seconds, from the initial points $-(5,5)$ to the destination $(3,4)$, in the $2$D star world workspace. Right: The respective trajectory in the transformed sphere world.}\label{fig:2d_stars_traj (Automatica_adaptive_nav)}
\end{figure}

\begin{figure}[!ht]
	\centering
	\subcaptionbox{\label{fig:u_alpha_2d_stars (Automatica_adaptive_nav)}}
	{\includegraphics[width = 0.45\textwidth]{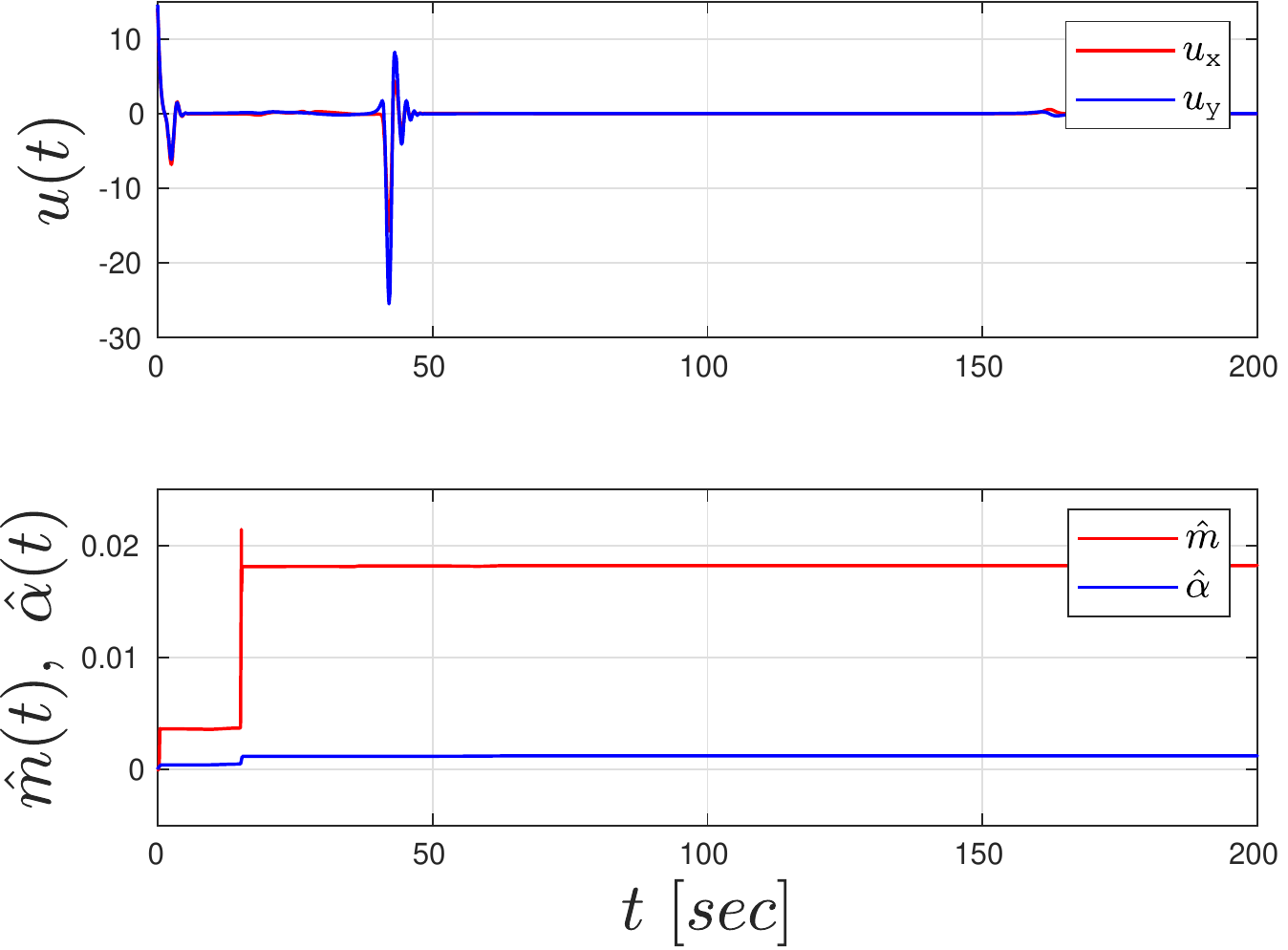}}
	\subcaptionbox{ \label{fig:u_alpha_3d_stars (Automatica_adaptive_nav)} }
	{\includegraphics[width = 0.45\textwidth]{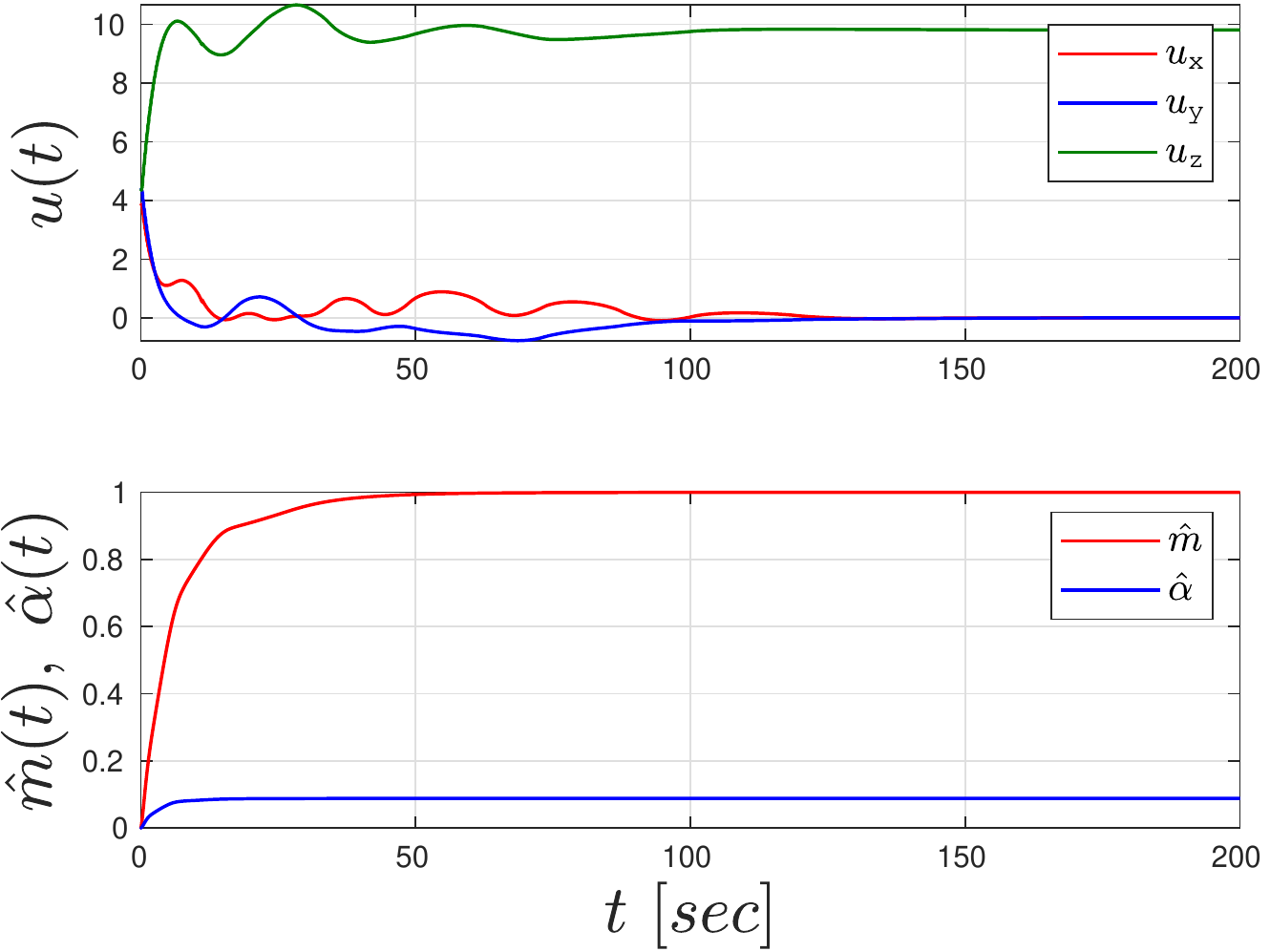}}
	\caption{The input and adaptations signals $u(t)$, $\hat{\alpha}(t)$, $\hat{m}(t)$, for the $2$D (a) and $3$D (b) star world workspaces, for $[0,500]$ and $[0,200]$ seconds, respectively.} \label{fig.simulation_states (Automatica_adaptive_nav)}
\end{figure}

%

\begin{figure}[!ht]
	\centering
	\includegraphics[trim = 0cm 0cm 0cm -0cm,width = 0.85\textwidth]{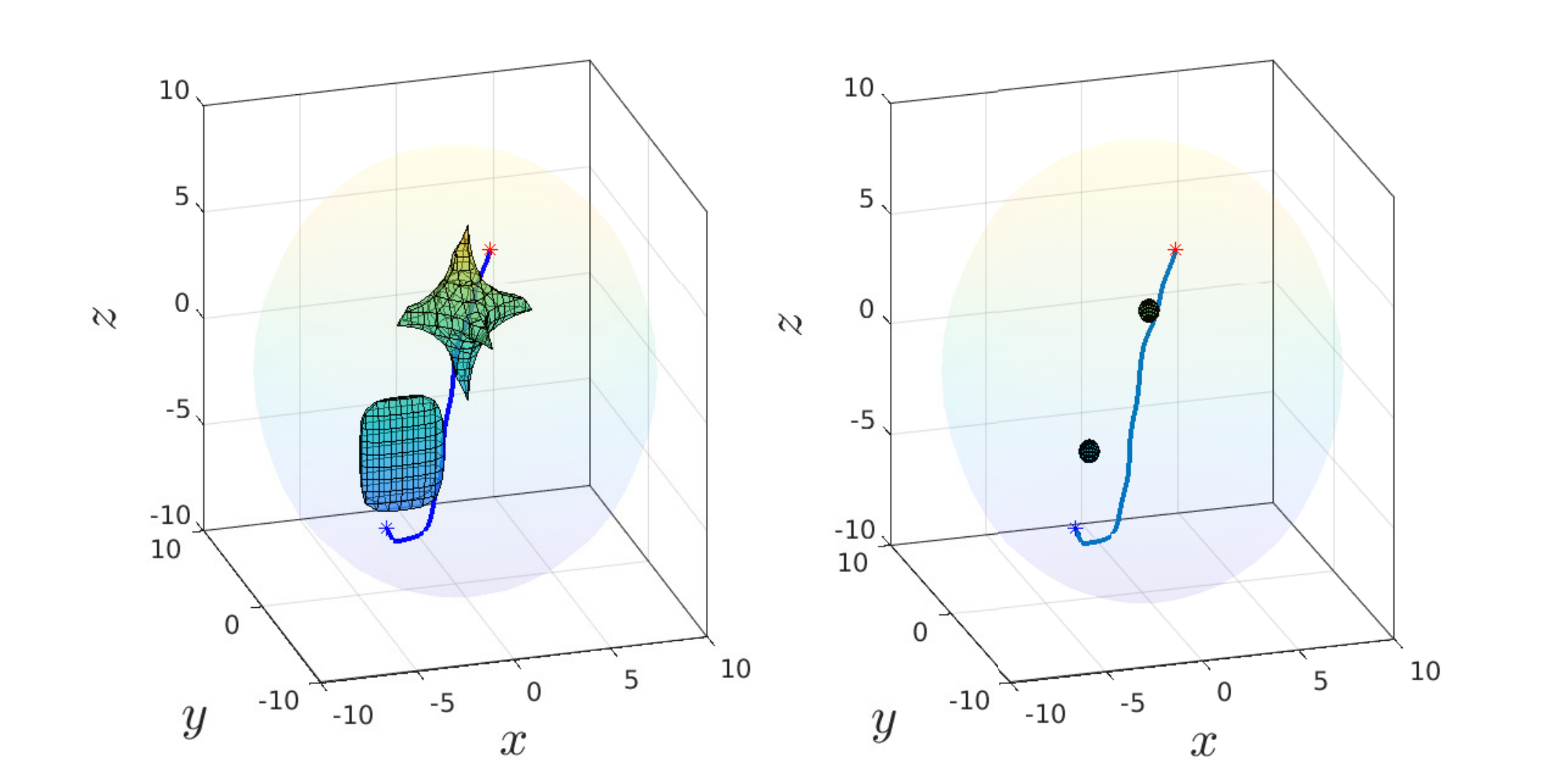}\\
	\caption{Left: The resulting trajectory $x(t)$, $t\in[0,200]$ seconds, from the initial points $-(4,4,2)$ to the destination $(1,2,2)$, in the $3$D star world workspace. Right: The respective trajectory in the transformed sphere world.}\label{fig:3d_stars_traj (Automatica_adaptive_nav)}
\end{figure}

\begin{figure}[!ht]
	\centering
	\includegraphics[trim = 0cm 0cm 0cm -0cm,width = 0.65\textwidth]{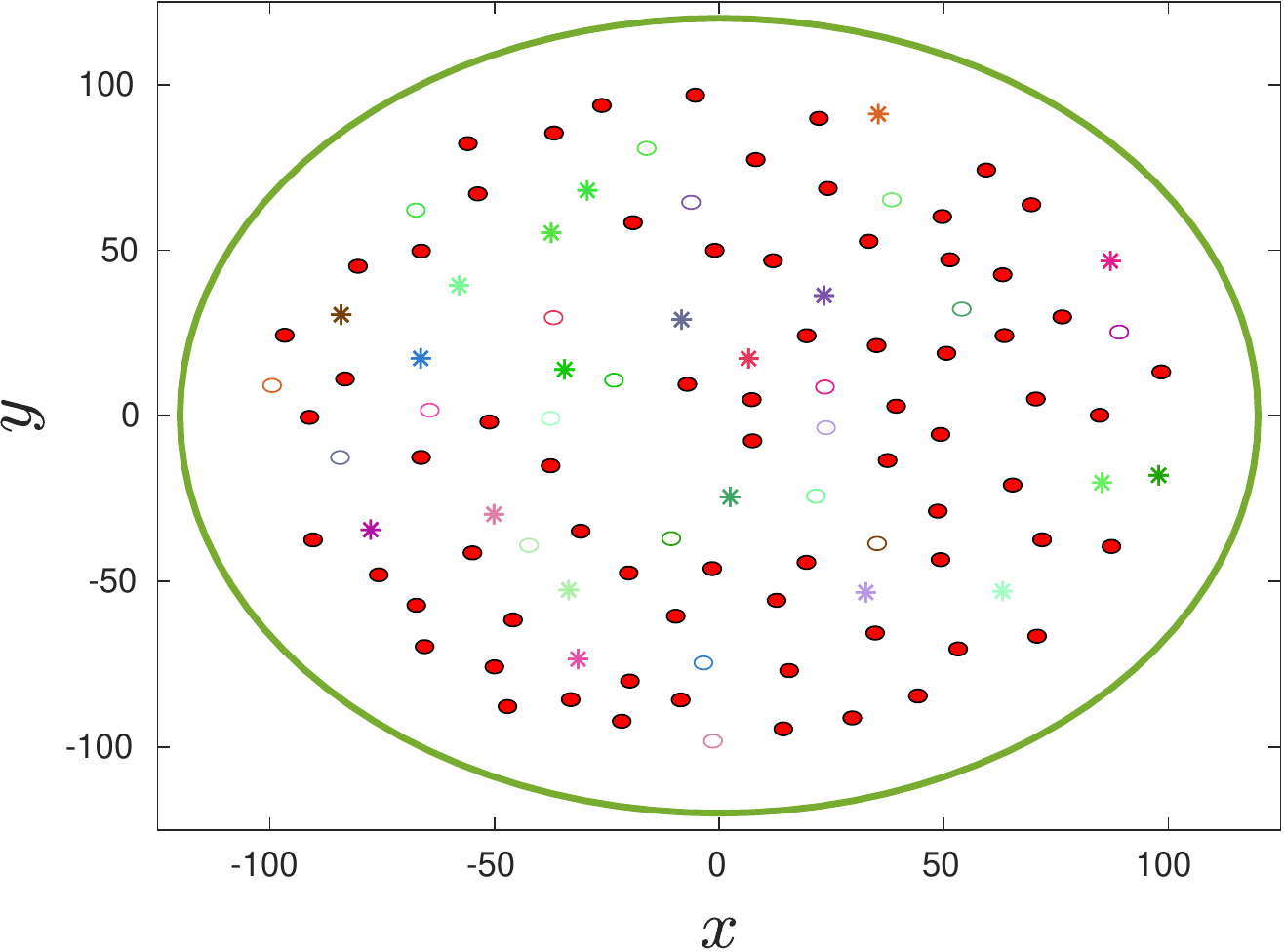}\\
	\caption{The initial configurations of the multi-agent scenario. The obstacles  are depicted as filled red disks whereas the agents as circles. The destinations are shown with asterisk.}\label{fig:mas_initial (Automatica_adaptive_nav)}
\end{figure}

\begin{figure}[!ht]
	\centering
	\includegraphics[trim = 0cm 0cm 0cm -0cm,width = 0.7\textwidth]{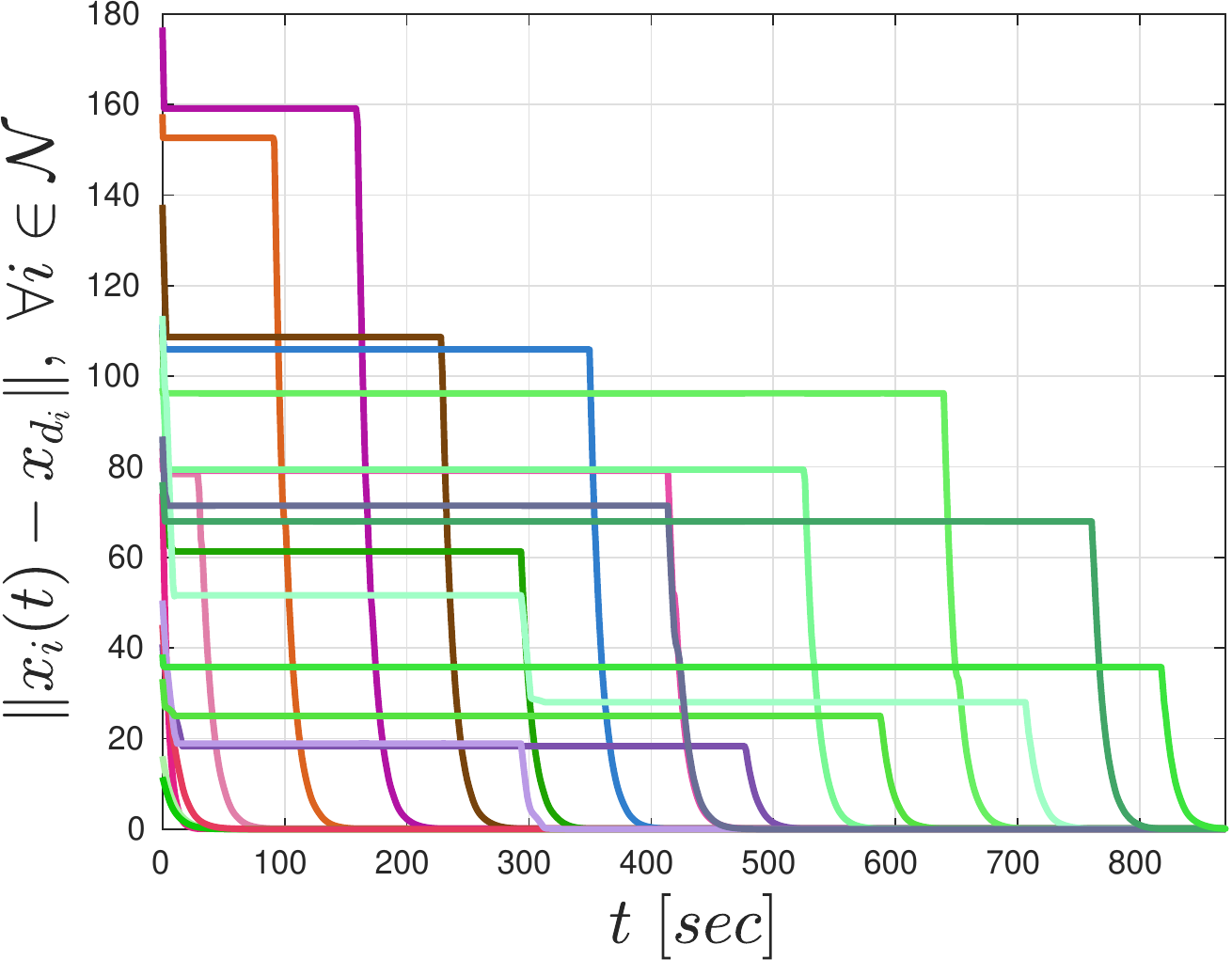}\\
	\caption{The resulting signals $\|x_i(t) - x_{\text{d}_i}\|$, $\forall i\in\mathcal{N}$, shown to converge to zero for the multi-agent scenario.}\label{fig:gamma_MAS (Automatica_adaptive_nav)}
\end{figure}

\begin{figure}[!ht]
	\centering
	\includegraphics[trim = 0cm 0cm 0cm -0cm,width = 0.65\textwidth]{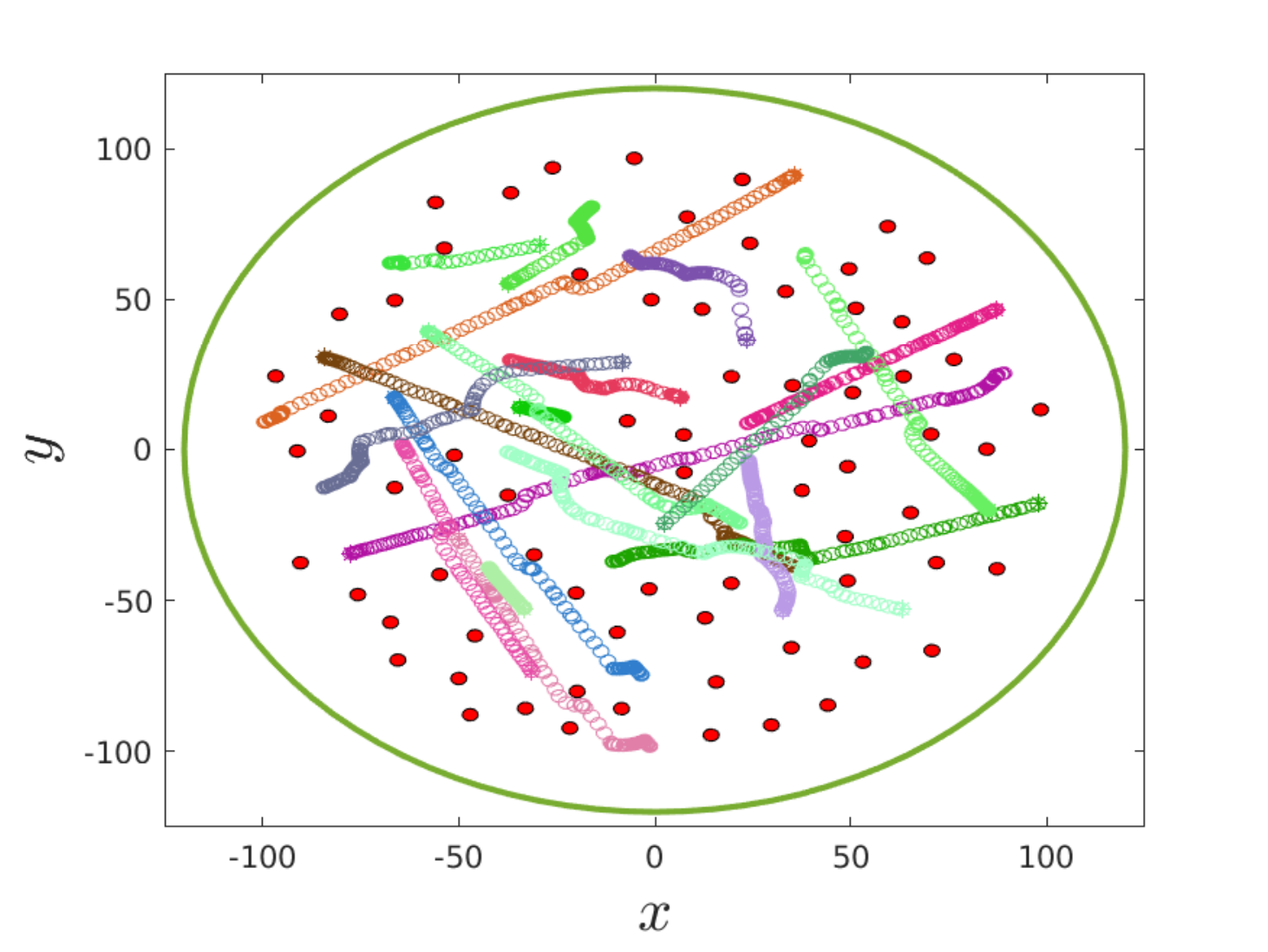}\\
	\caption{The resulting trajectories of the agents $x_i(t)$ in the $2$D workspace, $\forall i\in\mathcal{N}, t\in[0,870]$ seconds, for the multi-agent scenario. }\label{fig:traj_MAS (Automatica_adaptive_nav)}
\end{figure}

\begin{figure}[!ht]
	\centering
	\includegraphics[trim = 0cm 0cm 0cm -0cm,width = 0.7\textwidth]{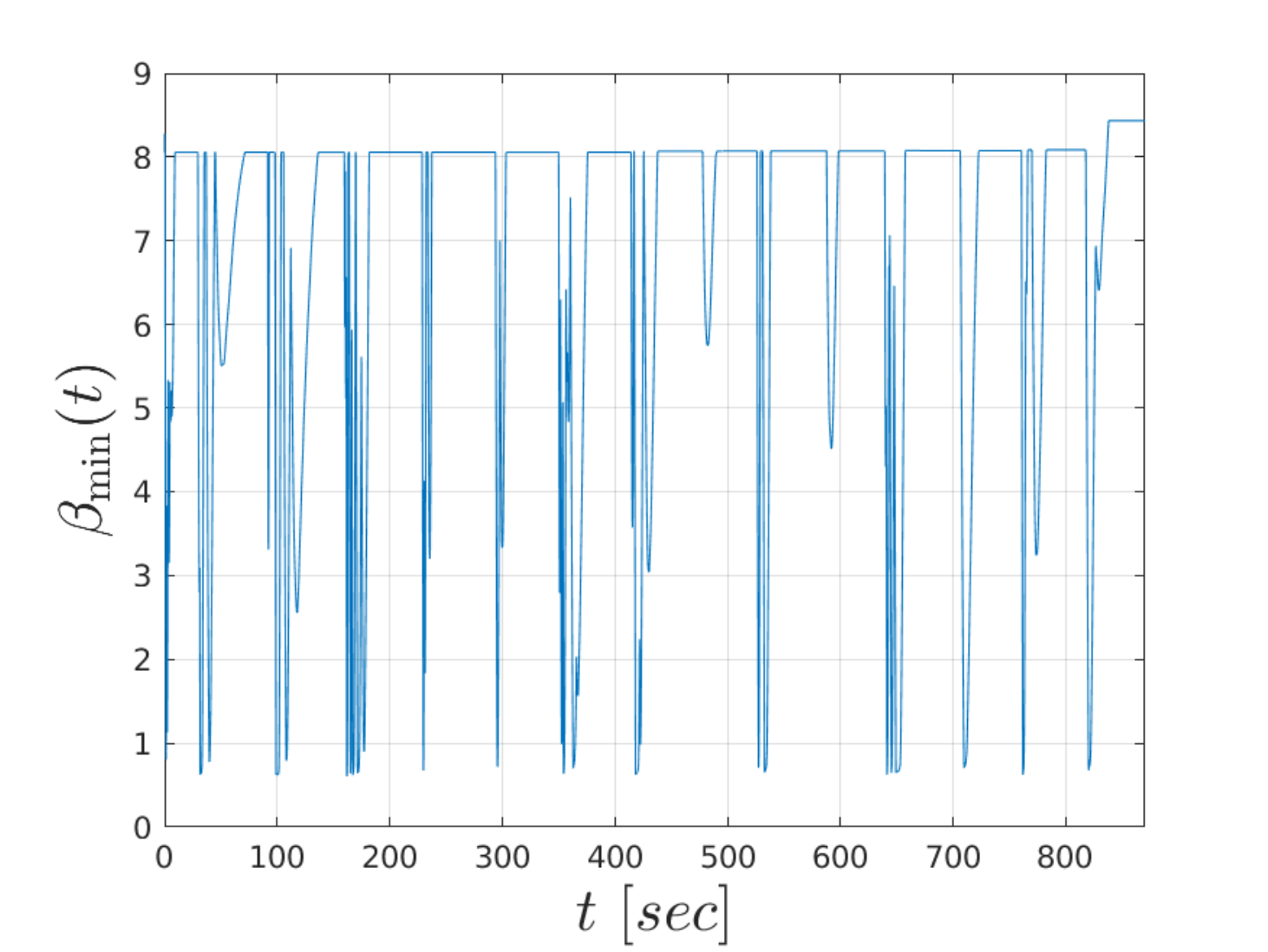}\\
	\caption{The signal $\beta_{\min}(t)$, which stays strictly positive, for all $t\in[0,870]$, implying that inter-agent collisions and agent-obstacle collisions are avoided.  }\label{fig:dis_MAS (Automatica_adaptive_nav)}
\end{figure}

\section{Adaptive Leader-Follower Coordination with Transient Constraints} \label{sec:LF}

We study next the problem of leader-follower coordination of a multi-agent system in an obstacle-free workspace. In particular, we consider that a leader agent aims to navigate to a predefined position, subject to collision and connectivity constraints, as well as uncertain $2$nd-order dynamics. We use innovatively a useful property of the graph's incidence matrix (see Appendix \ref{app:Rigidity}) to obtain the desired results.

\subsection{Problem Formulation} \label{sec:PF (CDC_LF)}
Consider $N>1$ autonomous robotic agents, with $\mathcal{N} \coloneqq \{1,\dots,N\}$, operating in $\mathbb{R}^n$ and described by the spheres $\mathcal{A}_i(x_i) \coloneqq \bar{\mathcal{B}}_i(x_i,r_i) = \{y\in\mathbb{R}^n : \|x_i-y\| \leq r_i \}$, with $x_i \in\mathbb{R}^n$ being agent $i$'s center, and $r_i\in\mathbb{R}_{> 0}$ its bounding radius. 
In contrast to the previous section, we consider now the more general Lagrangian dynamics for the agents (see Chapter \ref{chapter:cooperative manip}):
\begin{subequations} \label{eq:dynamics (CDC_LF)}
\begin{align}	
&\dot{x}_i = v_i \\
&M_i(x_i)\dot{v}_i + C_i(x_i,v_i)\dot{x}_i + g_i(x_i) + f_i(x_i,v_i) + d_i(t)  = u_i,
\end{align}
\end{subequations}
where $M_i\coloneqq M_i(x_i):\mathbb{R}^n\to\mathbb{R}^{n\times n}$ are positive definite  inertia matrices, with the standard property (see \eqref{eq:dyn properties (TCST_coop_manip)})
$$0<\underline{m} I_n \leq M_i(x) \leq \bar{m} I_n,$$ $\forall x\in\mathbb{R}^n,i\in\mathcal{N}$, for positive constants $\underline{m}, \bar{m}$, $C_i\coloneqq C_i(x_i,v_i):\mathbb{R}^{2n}\to\mathbb{R}^{n\times n}$ are the Coriolis terms, $g_i\coloneqq g_i(x_i):\mathbb{R}^n\to\mathbb{R}^n$ are the gravity vectors, $f_i\coloneqq f_i(x_i,v_i):\mathbb{R}^{2n}\to\mathbb{R}^n$ are unknown vector fields that represent friction-like terms (as in \eqref{eq:dynamics MAS (Automatica_adaptive_nav)}), $d_i\coloneqq d_i(t):\mathbb{R}_{\geq 0}\to\mathbb{R}^n$ are unknown external  disturbances and modeling uncertainties, and $u_i\in\mathbb{R}^n$ are the agents' control inputs, $\forall i\in\mathcal{N}$. The terms $M_i$, $C_i$ and $g_i$ are continuous in their arguments, the terms $f_i$ are Lebesgue measurable and locally bounded, and $d_i$ are uniformly bounded. Note that here we do not require $f_i$ and $d_i$ to be continuous everywhere, since we will employ non-smooth analysis for the stability of the closed-loop system. 
Moreover, as in Chapter \ref{chapter:cooperative manip}, we consider that the dynamic terms $M_i$, $C_i$, and $g_i$ include unknown constant dynamic parameters of the agents (e.g., masses, moments of inertia), denoted by the vectors $\theta_i\in \mathbb{R}^{\ell}$, $\ell \in \mathbb{N}$, $\forall i\in\mathcal{N}$.
The Lagrangian system \eqref{eq:dynamics (CDC_LF)} satisfies the following well-known properties (as in Chapter \ref{chapter:cooperative manip}): 
\begin{property}	\label{ass:skew-symm (CDC_LF)}
	The terms $\dot{M}_i(x)-2C_i(x,z)$ are skew-symmetric, i.e., $(\dot{M}_i(x)-2C_i(x,z))^\top=2C_i(x,z)-\dot{M}_i(x)$ and $y^\top(\dot{M}_i(x)-2C_i(x,z))y = 0$, $\forall x,y,z\in\mathbb{R}^n$, $i\in\mathcal{N}$.
\end{property}
\begin{property}  \label{ass:dynamics factorization (CDC_LF)}
	The dynamic terms of \eqref{eq:dynamics (CDC_LF)} can be linearly parameterized with respect to the agents' dynamic parameters. That is, for any vectors $x,y,z,w \in \mathbb{R}^n$, it holds that
	$M_i(x)y + C_i(x,z)w + g_i(x) = Y_i(x,z,w,y)\theta_i$, $\forall x,y,z,w \in\mathbb{R}^n$, where $Y_i:\mathbb{R}^{4n}\to\mathbb{R}^{n\times \ell}$ are known regressor matrices, and $\theta_i\in \mathbb{R}^{\ell}$,  $\ell\in\mathbb{N}$, are vectors of constant but unknown dynamic parameters of the agents, $\forall i\in\mathcal{N}$.
\end{property}

Moreover, we impose the following assumptions on the system \eqref{eq:dynamics (CDC_LF)}, which encapsulate standard properties of friction-terms and external disturbances, similar to Assumption \ref{ass:f (Automatica_adaptive_nav)}:
\begin{assumption} \label{ass:f_i+d_i (CDC_LF)}
	It holds that $\|f_i({x}_i,v_i)\|_1 \leq \alpha_i\|v_i\|_1$, $\|d_i(t)\|_1 \leq d_{b_i}$, $\forall x_i,v_i\in\mathbb{R}^{2n}, t\in\mathbb{R}_{\geq 0}$, where $\alpha_i$, $d_{b_i}$ are \textit{unknown} positive constants, $i\in\mathcal{N}$.
\end{assumption}
We aim to compensate $f_i$ and $d_i$ by using discontinuous adaptive control.
Without loss of generality, we assume that agent $i=1$ corresponds to the team leader, whereas $i>1$ are the followers, which belong to the set $\mathcal{N}_\mathcal{F}\coloneqq \{2,\dots,N\}$. The task of the leader is to navigate to a desired pose $x_{\textup{d}}\in\mathbb{R}^n$, and the entire team is responsible for guaranteeing collision avoidance as well as connectivity maintenance properties.	

In addition, as in Sections \ref{sec:formation control} and \ref{sec:MAS (Automatica_adaptive_nav)}, we consider that each agent has a limited sensing radius $\varsigma_i\in\mathbb{R}_{>0}$, with $\varsigma_i > \max_{j\in\mathcal{N}}\{r_i+r_j\}$, which implies that the agents can sense each other without colliding.
Based on this, we model the topology of the multi-agent network through the undirected graph $\mathcal{G}(x) \coloneqq (\mathcal{N},\mathcal{E}(x))$, with  $\mathcal{E}(x) \coloneqq \{(i,j)\in\mathcal{N}^2 : \|x_i - x_j \| \leq \min\{\varsigma_i, \varsigma_j \} \}$, where $x\coloneqq [x_1^\top,\dots, x_N^\top]^\top \in \mathbb{R}^{nN}$. We further denote $K(x)\coloneqq |\mathcal{E}(x)|$. Given the $k$-th edge in the edge set $\mathcal{E}(x)$, we use the notation $(k_1,k_2)\in\mathcal{N}^2$ that gives the agent indices that form edge $k\in\mathcal{K}(x)$, where $k_1$ is the tail and $k_2$ is the head of edge $k$, and $\mathcal{K}(x)\coloneqq\{1,\dots,K(x)\}$ is an arbitrary numbering of the edges $\mathcal{E}(x)$. 


As discussed before, the leader agent $i=1$ aims at navigating to $x_\text{d}$. We also need to guarantee that inter-agent collisions are avoided for all times, and that some initial edges, denoted by $\mathcal{E}_0 \subset \mathcal{E}(x(0))$, are preserved. The motivation for that is mainly potential cooperative tasks that the agents have to accomplish, whose details are provided only to a leader agent. Then, the leader has to guide the entire team to the points of interest, which is guaranteed via graph connectivity (all agents are part of an edge). There exist, nevertheless, more sophisticated and less conservative ways to maintain graph connectivity than just maintaining part of the initial edges \cite{zavlanos2011graph,sabattini2013distributed,turpin2014capt}. Such schemes are not included in the current framework. The results in this section are more general, in the sense that neither the graph connectivity of $\mathcal{G}(x(0))$, $\mathcal{G}_0\coloneqq (\mathcal{N},\mathcal{E}_0)$, nor the connectivity to the leader are \textit{technical} requirements of the analysis, as shown below.
Formally, the problem treated in this section is the following:

\begin{problem} \label{prob: 1 (CDC_LF)}
	Consider $N$ spherical autonomous robotic agents with dynamics \eqref{eq:dynamics (CDC_LF)}. Given Properties \ref{ass:skew-symm (CDC_LF)}-\ref{ass:dynamics factorization (CDC_LF)}  and Assumption \ref{ass:f_i+d_i (CDC_LF)}, develop a decentralized control strategy that guarantees $1$) achievement of the leader's task, $2$) inter-agent collision avoidance, and $3$) connectivity maintenance between a subset of the initially connected agents, i.e.,
	\begin{enumerate}
		\item $\lim\limits_{t\to\infty} (x_1(t) - x_{\text{d}}) = 0$, 
		\item $\mathcal{A}_i(x_i(t))\cap\mathcal{A}_j(x_j(t)) = \emptyset$, $\forall t\in\mathbb{R}_{\geq 0},i,j\in\mathcal{N}$, $i\neq j$,
		\item $\|x_{k_1}(t) - x_{k_2}(t) \| \leq \min\{\varsigma_{m_1},\varsigma_{m_2}\}$, $\forall t\in\mathbb{R}_{\geq 0}$, $k\in \mathcal{K}_0 \subset\mathcal{K}(x(0))$,
	\end{enumerate} 
	where $\mathcal{K}_0\coloneqq \{1,\dots, K_0\}$ is an edge numbering for the edge set $\mathcal{E}_0$, with $K_0 \coloneqq |\mathcal{E}_0|$.
\end{problem} 

\subsection{Problem Solution}\label{sec:main results (CDC_LF)}

In this section we propose a decentralized control protocol for the solution of Problem \ref{prob: 1 (CDC_LF)}.

Besides the edge set $\mathcal{E}_0$, with edge numbering $\mathcal{K}_0$ and $K_0$ edges, which needs to remain connected, consider also
the complete graph $\bar{\mathcal{G}} \coloneqq (\mathcal{N},\bar{\mathcal{E}})$, with $\bar{\mathcal{E}}\coloneqq \{ (i,j), \forall i,j\in\mathcal{N}, i < j\}$, $\bar{K}\coloneqq |\bar{\mathcal{E}}| = \frac{N(N-1)}{2}$, and the edge numbering $\bar{\mathcal{K}}\coloneqq  \{1,\dots,K_0,K_0+1,\dots,\bar{K}\}$, where $\{K_0+1,\dots,\bar{K}\}$ corresponds to the edges in  $\bar{\mathcal{E}}\backslash \mathcal{E}_0$. Moreover, denote by $D_0$ and $\bar{D}$ the incidence matrices of $\mathcal{G}_0$ and $\bar{\mathcal{G}}$, respectively (see Section \ref{sec:graph_theory (app_ridigity)} of Appendix \ref{app:Rigidity}).

We construct now the local collision and connectivity functions for all edges $\bar{\mathcal{K}}$ and $\mathcal{K}_0$, respectively. Given positive constants $\bar{\beta}_{\mathfrak{c}}$ and $\bar{\beta}_{\mathfrak{n}}$,  let  $\beta_{\mathfrak{c},k}:\mathbb{R}_{\geq 0}\to[0,\bar{\beta}_{\mathfrak{c}}]$ and $\beta_{\mathfrak{n},l}:\mathbb{R}_{\geq 0}\to[0,\bar{\beta}_{\mathfrak{n}}]$, with
\begin{align*}
\beta_{\mathfrak{c},k}(x) &\coloneqq \left\{ \begin{matrix}
\vartheta_{\mathfrak{c},k}(x) & 0 \leq x < \bar{d}_{\mathfrak{c},k}, \\
\bar{\beta}_\mathfrak{c} &	 \bar{d}_{\mathfrak{c},k} \leq x
\end{matrix} \right., \\
\beta_{\mathfrak{n},l}(x) &\coloneqq \left\{ \begin{matrix}
\vartheta_{\mathfrak{n},l}(x) & 0 \leq x < \underline{d}^2_{\mathfrak{n},l} \\
\bar{\beta}_\mathfrak{n}&	 \underline{d}^2_{\mathfrak{n},l} \leq x \\
\end{matrix} \right.,
\end{align*}
$\forall k\in\bar{\mathcal{K}}$, $l\in\mathcal{K}_0$, where $\vartheta_{\mathfrak{c},k}:\mathbb{R}_{\geq 0}\to[0,\bar{\beta}_\mathfrak{c}]$, $\vartheta_{\mathfrak{n},l}:$ $\mathbb{R}_{\geq 0}$ $\to$ $[0,\bar{\beta}_\mathfrak{n}]$ are polynomials that guarantee that $\beta_{\mathfrak{c},k}$ and $\beta_{\mathfrak{n},l}$, respectively, are twice continuously differentiable, $\forall k\in\bar{\mathcal{K}}$, $l\in\mathcal{K}_0$. The aforementioned functions are smooth switches, similar the the one used in Section \ref{sec:adaptive nav (Automatica_adaptive_nav)}. 
Then, we choose  
\begin{center}
	\begin{tabular}{ l l}
	$\beta_{\mathfrak{c},k} \coloneqq \beta_{\mathfrak{c},k}(\iota_k)$, & $\iota_k\coloneqq\iota_k(x_{k_1},x_{k_2})\coloneqq\|x_{k_1}-x_{k_2}\|^2 - (r_{k_1}+r_{k_2})^2$ \\
	$\beta_{\mathfrak{n},l} \coloneqq \beta_{\mathfrak{n},l}(\nu_l)$, & $\nu_l\coloneqq\nu_l(x_{l_1},x_{l_2})\coloneqq \underline{d}^2_{\mathfrak{n},l} - \|x_{l_1}-x_{l_2}\|^2$
	\end{tabular}
\end{center}
with $\underline{d}_{\mathfrak{n},k} \coloneqq \min\{\varsigma_{k_1},\varsigma_{k_2}\}$ and we also set $\bar{d}_{\mathfrak{c},k}\coloneqq \underline{d}^2_{\mathfrak{n},k}- (r_{k_1}+r_{k_2})^2$, $\forall k\in\bar{\mathcal{K}},l\in\mathcal{K}_0$.  The terms $\bar{\beta}_\mathfrak{c}$, $\bar{\beta}_\mathfrak{n}$ can be any positive constants.
Note that $\beta_{\mathfrak{c},k}$ and $\beta_{\mathfrak{n},l}$ take into account the limited sensing capabilities of the agents and their derivatives vanish at collisions and connectivity breaks, respectively, of the respective edges. All the parameters for the construction of $\beta_{\mathfrak{c},k}$, $\beta_{\mathfrak{n},l}$ can be transmitted off-line to the agents. 

Regarding the uncertain terms of \eqref{eq:dynamics (CDC_LF)}, note that  $\theta_i\in\mathbb{R}^{\ell}$, $\alpha_i\in\mathbb{R}$, and $d_{b_i}\in\mathbb{R}$ from Properties \ref{ass:skew-symm (CDC_LF)}, \ref{ass:dynamics factorization (CDC_LF)} and Assumption \ref{ass:f_i+d_i (CDC_LF)} are unknown to the agents. Hence, we define the estimations of these terms $\hat{\theta}_i\in\mathbb{R}^{\ell}$, $\hat{\alpha}_i\in\mathbb{R}$, $\hat{d}_{b_i}\in\mathbb{R}$, $\forall i\in\mathcal{N}$, with the respective errors $\widetilde{\theta}_i \coloneqq \hat{\theta}_i - \theta_i$, $\widetilde{\alpha}_i \coloneqq \hat{\alpha}_i - \alpha_i$, $\widetilde{d}_{b_i} \coloneqq \hat{d}_{b_i} - d_{b_i}$, $\forall i\in\mathcal{N}$. By using adaptive control techniques, we prove in the following that these estimations compensate appropriately for the unknown terms, without necessarily converging  to them. In addition, we define the leader error signal $s_e \coloneqq x_1 - x_{\text{d}}$
and $\alpha^\mathfrak{c}_{i,k}$ and $\alpha^\mathfrak{n}_{i,l}$ as:
\begin{alignat*}{2}
& \begin{aligned} 
&\alpha^\mathfrak{c}_{i,k} \coloneqq  \begin{cases}
-1,\ i=k_1\\
1, \hspace{3.5mm} i=k_2\\
0, \hspace{3mm} \text{ otherw.} \\
\end{cases}\\
\end{aligned}
& \hskip 0.01em &
\begin{aligned}
&\alpha^\mathfrak{n}_{i,l} \coloneqq  \begin{cases}
-1,\ i=l_1\\
1, \hspace{3.5mm} i=l_2\\
0, \hspace{3mm} \text{ otherw.} \\
\end{cases} \\
\end{aligned}
\end{alignat*} 
$\forall k\in\bar{\mathcal{K}}$, $l\in\mathcal{K}_0$, $i\in\mathcal{N}$, which provide boolean values depending on whether agent $i$ is part (head or tail) of edge $m$ and $l$ (as in \eqref{eq:control design vectors (formation)}). Finally, we define, $\forall k\in\bar{\mathcal{K}}$, $l\in\mathcal{K}_0$, the terms 
\begin{alignat*}{2}
&\begin{aligned}
&\beta'_{\mathfrak{c},k} \coloneqq \frac{\partial }{\partial \iota_k}\left(\frac{1}{\beta_{\mathfrak{c},k}(\iota_k)}\right), 
\end{aligned}
&\hskip 0.2em &
\begin{aligned} 
&\beta'_{\mathfrak{n},l} \coloneqq \frac{\partial }{\partial \nu_l}\left(\frac{1}{\beta_{\mathfrak{n},l}(\nu_l)}\right),
\end{aligned}
\end{alignat*} 
which diverge to infinity in a collision and a connectivity break of the agents $k_1,k_2$ and $l_1,l_2$, respectively.
We propose now the following decentralized adaptive control protocol.
Choose  the agents' desired velocity  as 
\begin{subequations} \label{eq:v des i (CDC_LF)}
	\begin{align} 
	& v_{d_1} = -\gamma_e s_e +  \sum\limits_{k\in\bar{\mathcal{K}}}\alpha^\mathfrak{c}_{1,k} \beta'_{\mathfrak{c},k}\frac{\partial \iota_k}{\partial x_{k_1}}+ \sum\limits_{l\in\mathcal{K}_0}\alpha^\mathfrak{n}_{1,l}  \beta'_{\mathfrak{n},l}\frac{\partial \nu_l}{\partial x_{l_1}}  \\		
	& v_{d_i} = k_i \left(\sum\limits_{k\in\bar{\mathcal{K}}}\alpha^\mathfrak{c}_{i,k} \beta'_{\mathfrak{c},k}\frac{\partial \iota_k}{\partial x_{k_1}} + \sum\limits_{l\in\mathcal{K}_0}\alpha^\mathfrak{n}_{i,l}  \beta'_{\mathfrak{n},l}\frac{\partial \nu_l}{\partial x_{l_1}}\right),  \forall i\in\mathcal{N}_\mathcal{F} 
	\end{align}
\end{subequations}
that concerns the collision avoidance and connectivity maintenance properties, with the extra term $\gamma_e s_e$ for the leader to guarantee the navigation to $x_\text{d}$. The terms 
$\gamma_e$, $k_i$ are positive constants, $\forall i\in\mathcal{N}_\mathcal{F}$. 
Since $v_{d_i}$ is not the actual velocity of the agents, we define the errors $e_{v_i} \coloneqq v_i - v_{d_i}$, $\forall i\in\mathcal{N}$, and design the decentralized control laws $u_i: \mathcal{X}_i \to \mathbb{R}^6$
\begin{align}
u_i \coloneqq u_i(\chi_i) =& \sum\limits_{k\in\bar{\mathcal{M}}}\alpha^\mathfrak{c}_{i,k} \beta'_{\mathfrak{c},k}\frac{\partial \iota_k}{\partial x_{k_1}} +\sum\limits_{l\in\mathcal{K}_0}\alpha^\mathfrak{n}_{i,l} \beta'_{\mathfrak{n},l}\frac{\partial \nu_l}{\partial x_{l_1}} - k_{v_i}e_{v_i}  - \widetilde{s}_{e_i} + Y_{r_i}\hat{\theta}_i\notag \\
& - \text{sgn}(e_{v_i})\|\dot{x}_i\|_1 \hat{\alpha}_i- \text{sgn}(e_{v_i})\hat{d}_{b_i},  \label{eq:control law (CDC_LF)}
\end{align}
$\forall i\in\mathcal{N}$, where $\chi_i \coloneqq [x^\top,v^\top,\hat{\theta}_i^\top,\hat{\alpha}_i,\hat{d}_{b_i},]^\top$, $v = [v_1^\top,\dots,v_N^\top]^\top$, $$\mathcal{X}_i\coloneqq \{ \chi_i \in \mathbb{R}^{2Nn+\ell+2} : \iota_k(x_{k_1},x_{k_2}) > 0, \nu_l(x_{l_1},x_{l_2}) > 0, \forall k\in\bar{\mathcal{K}}, l\in\mathcal{K}_0 \},$$
$\widetilde{s}_{e_1}= \gamma_e s_e$, $\widetilde{s}_{e_i} = 0,\forall i\in \mathcal{N}_\mathcal{F}$, $Y_{r_i} \coloneqq Y_i(x_i,v_i,v_{d_i},\dot{v}_{d_i})$, and $k_{v_i}$ are positive gains. Moreover, we design the adaptation signals
\begin{align} \label{eq:adaptation laws (CDC_LF)}
&\left.
\begin{matrix*}[l]
\dot{\hat{d}}_{b_i} = \gamma_{i,d} \|e_{v_i}\|_1, \\
\dot{\hat{\alpha}}_i = \gamma_{i,f} \|e_{v_i} \|_1 \|v_i\|_1, \\
\dot{\hat{\theta}}_i = -\gamma_{i,\theta}Y_{r_i}^\top e_{v_i}
\end{matrix*} \right \} i\in\mathcal{N},
\end{align}
with arbitrary bounded initial conditions, and positive constants $\gamma_{i,d}$, $\gamma_{i,f}$, $\gamma_{i,\theta}$, $\forall i\in\mathcal{N}$. 
Note from \eqref{eq:control law (CDC_LF)} that, unlike the usual case in the related literature, the leader contributes to the collision avoidance and connectivity maintenance properties, apart from just guaranteeing achievement of its task. Regarding the rest of the terms, $Y_i(\cdot)\hat{\theta}$, $\text{sgn}(e_{v_i})\|\dot{x}_i\|_1\hat{\alpha}_i$, and $\text{sgn}(e_{v_i})\hat{d}_{b_i}$ compensate for the unknown terms $\theta_i$, $f_{b_i}$, and $d_{b_i}$, respectively, and $e_{v_i}$ is a dissipative velocity term that ensures closed-loop stability.
The  main results of this section are summarized in the following theorem.

\begin{theorem} \label{th:main th (CDC_LF)}
	Consider a multi-agent team $\mathcal{N}$, described by the dynamics \eqref{eq:dynamics (CDC_LF)} subject to Properties \ref{ass:skew-symm (CDC_LF)}, \ref{ass:dynamics factorization (CDC_LF)} and Assumption \ref{ass:f_i+d_i (CDC_LF)}. Then, application of the control and adaptation laws \eqref{eq:control law (CDC_LF)}, \eqref{eq:adaptation laws (CDC_LF)} 
	guarantees: 1) navigation of the leader agent to $x_\textup{d}$, 2) connectivity maintenance of the subset $\mathcal{E}_0$ of the initial edges, 3) inter-agent collision avoidance, and 4) boundedness of all closed loop signals, from all collision-free initial configurations, i.e., $\mathcal{A}_i(x_i(0))\cap\mathcal{A}_j(x_j(0)) = \emptyset$, $\forall i,j\in\mathcal{N}$, with $i\neq j$, providing thus a solution to Problem \ref{prob: 1 (CDC_LF)}. Moreover, it holds that $\lim_{t\to\infty}v_i(t) = 0$, $\forall i\in\mathcal{N}$.
\end{theorem}

\begin{proof}
	{By employing  \eqref{eq:dynamics (CDC_LF)}, \eqref{eq:control law (CDC_LF)}, \eqref{eq:adaptation laws (CDC_LF)}, we can write the {closed-loop} system as
		\begin{subequations} \label{eq:system closed loop (CDC_LF)}
			\begin{align}
			&\dot{x}_i = v_i \\
			&	\dot{v}_i = -M_i(x_i)^{-1}\bigg(C_i(x_i,{v}_i){v}_i + g_i(x_i) + \mathsf{K}[f_i](x_i,v_i)  + d_i(t)-\mathsf{K}[u_i] \bigg) \label{eq:system closed loop x (CDC_LF)} \\		
			&\dot{\hat{d}}_{b_i} = \gamma_{i,d} \|e_{v_i}\|_1 \\
			&\dot{\hat{\alpha}}_i = \gamma_{i,f} \|e_{v_i} \|_1 \|{v}_i\|_1 \\
			&\dot{\hat{\theta}}_i = -\gamma_{i,\theta} Y_i(x_i,v_i,v_{d_i},\dot{v}_{d_i})^\top e_{v_i}		
			\end{align}
		\end{subequations}
		$\forall i\in\mathcal{N}$, where $\mathsf{K}[f_i]$ and $\mathsf{K}[u_i]$ are the Filippov regularizations of $f_i$ and $u_i$, respectively, $\forall i\in\mathcal{N}$. In particular, $\mathsf{K}[u_i]$ is formed by substituting $\text{sgn}(e_{v_i})$ with $\text{SGN}(e_{v_i})$ in \eqref{eq:control law (CDC_LF)}. Let now $\hat{d}_b \coloneqq [\hat{d}_{b_1},\dots,\hat{d}_{b_N}]^\top$, $\hat{\alpha} \coloneqq [\hat{\alpha}_1,\dots,\hat{\alpha}_N]^\top$, $\hat{\theta} \coloneqq [\hat{\theta}_1^\top,\dots,\hat{\theta}_N^\top]^\top$, $\chi \coloneqq [x^\top,v^\top,\hat{d}_b^\top,\hat{\alpha}^\top,\hat{\theta}^\top]^\top$  and consider the set 		
		\begin{align*}
		\mathcal{X} \coloneqq \{\chi \in\mathbb{R}^{2Nn+2N+\ell N}:\chi_i \in \mathcal{X}_i,\forall i\in\mathcal{N}\}.
		\end{align*}		
		Since, initially the agents do not collide and $\mathcal{E}_0$ is a subset of the initially connected agents $\mathcal{E}(x(0))$, it holds that $\chi(0)\in\mathcal{X}$. The right hand side of \eqref{eq:system closed loop (CDC_LF)} is measurable in $t$ over $\mathbb{R}_{\geq 0}$ and Lebesgue measurable and locally bounded in $\chi$ on $\mathcal{X}$. Therefore, by invoking Prop. \ref{prop:Filippov exist (app_dynamical_systems)} of Appendix \ref{app:dynamical systems}, there exists at least a Filippov solution $\mu_{\scr LF}:[0,t_{\max}) \to \mathcal{X}$ {for some} $t_{\max} > 0$. }
	Consider now the function 
	\begin{align} \label{eq:V_1 (CDC_LF)}
	V_1 \coloneqq & \frac{\gamma_e}{2}\left\|s_e\right\|^2  + \sum_{k\in\bar{\mathcal{K}}}  \frac{1}{\beta_{\mathfrak{c},k}}  
	+\sum_{l\in\mathcal{K}_0}\frac{1}{\beta_{\mathfrak{n},l}} 
	\end{align}	
	which is well defined when $\mu_{\scr LF}\in\mathcal{X}$. 
	By considering
	the time derivative of $V_1$, and
	taking into account that $\frac{\partial \iota_k}{\partial x_{k_1}} =- \frac{\partial \iota_k}{\partial x_{k_2}}$, $\forall k\in\bar{\mathcal{K}}$, $\frac{\partial \nu_l}{\partial x_{l_1}} = -\frac{\partial \nu_l}{\partial x_{l_2}}$, $\forall l\in\mathcal{K}_0$, we obtain 
	\begin{align}
	\dot{V}_1 
	=& \gamma_e s_e^\top v_1 - \beta^\top (\widetilde{D}\otimes I_n)^\top v, \label{eq:V_1_dot (CDC_LF)} 
	\end{align}
	where $\beta$ $\coloneqq$ $[\beta_{\mathfrak{c}}^\top, \beta_{\mathfrak{n}}^\top]^\top$ $\in$ $\mathbb{R}^{\bar{K}+K_0}$, $\beta_{\mathfrak{c}}$ $\coloneqq$ $[\beta'_{\mathfrak{c},1}\frac{\partial \iota_1 }{\partial x_{1_1}}$, $\dots$, $\beta'_{\mathfrak{c},\bar{K}}\frac{\partial \iota_{\bar{K}} }{\partial x_{\bar{K}_1}} ]^\top$ $\in$ $\mathbb{R}^{\bar{K}}$,
	$\beta_{\mathfrak{n}}$ $\coloneqq$ $[\beta'_{\mathfrak{n},1}\frac{\partial \nu_1 }{\partial x_{1_1}}$, $\dots$, $\beta'_{\mathfrak{n},K_0}\frac{\partial \nu_{K_0} }{\partial x_{(K_0)_1}} ]^\top\in\mathbb{R}^{K_0}$,
	and $\widetilde{D} \coloneqq [\bar{D}, D_0]\in \mathbb{R}^{N\times(\bar{K}+K_0)}$, where $\bar{D}$ and $D_0$ are the incidence matrices corresponding to $\bar{\mathcal{E}}$ and $\mathcal{E}_0$, respectively. Let now $\widetilde{d}_i^\top \in \mathbb{R}^{\bar{K}+K_0}$, $i\in\mathcal{N}$, be the rows of $\widetilde{D}$, i.e., $\widetilde{D} = [\widetilde{d}_1,\dots,\widetilde{d}_N]^\top$. 
	Then, \eqref{eq:V_1_dot (CDC_LF)} can be written as 
	\begin{equation*}
		\dot{V}_1 \coloneqq \gamma_e s_e^\top v_1 - \sum_{i\in\mathcal{N}} \beta^\top (\widetilde{d}_i\otimes I_n){v}_i = (\gamma_e s_e^\top - \beta^\top (\widetilde{d}_1 \otimes I_n))v_1 - \sum_{i\in\mathcal{N}_\mathcal{F}} \beta^\top (\widetilde{d}_i\otimes I_n){v}_i
	\end{equation*}
	and \eqref{eq:v des i (CDC_LF)} and \eqref{eq:control law (CDC_LF)} as
	\begin{subequations} \label{eq:v_d_i,u_i rewr (CDC_LF)}
		\begin{align}
		v_{d_1} =& - \gamma_e s_e + (\widetilde{d}_1 \otimes I_n)^\top \beta \\	
		v_{d_i} =& k_i(\widetilde{d}_i\otimes I_n)^\top \beta, \hspace{10mm} \forall i\in\mathcal{N}_\mathcal{F} \\
		u_i =& (\widetilde{d}_i\otimes I_n)^\top\beta
		- \widetilde{s}_{e_i} + Y_i(x_i,v_i,v_{d_i},\dot{v}_{d_i})\hat{\theta}_i - k_{v_i}e_{v_i}   -\text{sgn}(e_{v_i})\|v_i\|_1 \hat{\alpha}_i \notag\\
		& - \text{sgn}(e_{v_i})\hat{d}_{b_i},  \hspace{10mm} \forall i\in\mathcal{N}. \label{eq:u_i rewr (CDC_LF)}
		\end{align}
	\end{subequations}
	Achievement of the desired velocities, i.e., $v_i = v_{d_i}$, $\forall i\in\mathcal{N}$, would imply that 
	\begin{align*}
	&\dot{V}_1 = -\|\gamma_e s_e - (\widetilde{d}_1\otimes I_n)^\top \beta \|^2 
	-\sum_{i\in \mathcal{N}_\mathcal{F}} k_i \|(\widetilde{d}_i\otimes I_n)^\top\beta \|^2.
	\end{align*}	
	The actual velocities of the agents, however, are not necessarily equal to the desired ones $v_{d_i}$, and therefore we use a backstepping-like technique to proceed. Consider the vector $\mathsf{z}\in \mathcal{Z}$, with 
	\small
	\begin{align*}
	\mathsf{z} \coloneqq \left[ 
	s_e^\top, \left(\frac{1}{\beta_{\mathfrak{c},1}} \right)^\frac{1}{2},\dots,\left(\frac{1}{\beta_{\mathfrak{c},\bar{K}}}\right)^\frac{1}{2}, \left(\frac{1}{\beta_{\mathfrak{n},1}}\right)^\frac{1}{2},\dots, \left(\frac{1}{\beta_{\mathfrak{n},K_0}}\right)^{\frac{1}{2}}, e_v^\top, \widetilde{d}_b^\top, \widetilde{\alpha}^\top, \widetilde{\theta}^\top \right]^\top,
	\end{align*}	
	\normalsize
	where $e_v \coloneqq [e_{v_1}^\top,\dots,e_{v_N}^\top]^\top\in\mathbb{R}^{nN}$, $\widetilde{d}_b \coloneqq [\widetilde{d}_{b_1},\dots, \widetilde{d}_{b_N}]^\top\in\mathbb{R}^N$, $\widetilde{\alpha} \coloneqq [\widetilde{\alpha}_1,\dots,\widetilde{\alpha}_N]^\top\in\mathbb{R}^N$, $\widetilde{\theta} \coloneqq [\widetilde{\theta}_1^\top,\dots,\widetilde{\theta}_N^\top]\in\mathbb{R}^{\ell N}$, and $\mathcal{Z} \coloneqq \mathbb{R}^{n+(2+n+\ell)N+\bar{K}+K_0}$. Similar to \eqref{eq:system closed loop (CDC_LF)}, we guarantee the existence of a Filippov solution $\mathsf{z}:[{0},t_{\max})\to\mathcal{Z}$ for the respective {closed-loop} system obtained by differentiating $\mathsf{z}$.
	We aim to prove that $\mathsf{z}(t)$ remains in a compact subset of $\mathcal{Z}$, which implies that $\chi$ remains in a compact subset of $\mathcal{X}$.	
	Define the barrier-like function $V_{\scr LF}\coloneqq V_{\scr LF}(\mathsf{z},t):\mathcal{Z}\times[0,t_{\max})\to\mathbb{R}_{\geq 0}$, with
	\small
	\begin{align*} 
	V_{\scr LF}(\mathsf{z},t) \coloneqq &  V_1(\mathsf{z}) + \sum_{i\in\mathcal{N}}\bigg\{ \frac{1}{2}e^\top_{v_i} M_i(x_i(t))e_{v_i} + \frac{1}{2\gamma_{i,d}}\widetilde{d}_{b_i}^2 +\frac{1}{2\gamma_{i,f}}\widetilde{\alpha}_i^2 + \frac{1}{2\gamma_{i,\theta}}\|\widetilde{\theta}_i\|^2  \bigg\},
	\end{align*}
	\normalsize
	for which, by using the fact $\underline{m} \leq M_i(x) \leq \bar{m}$, $\forall x\in\mathbb{R}^n$, $i\in\mathcal{N}$, it holds that $W_1(\mathsf{z}) \leq V_{\scr LF}(\mathsf{z},t) \leq W_2(\mathsf{z})$, where $W_1, W_2:\mathcal{Z}\to\mathbb{R}_{\geq 0}$ are positive definite functions. 
	Since initially the agents do not collide and $\mathcal{E}_0$ is a subset of the initially connected agents $\mathcal{E}(x(0))$, $V_1$, as defined in \eqref{eq:V_1 (CDC_LF)}, is well-defined, and hence $V_{\scr LF}(\mathsf{z}(0),0)$, $\frac{1}{\beta_{\mathfrak{c},k}(\iota_m(0))}$, $\frac{1}{\beta_{\mathfrak{n},l}(\nu_l(0))}$ are bounded, $\forall k\in\bar{\mathcal{K}}, l\in\mathcal{K}_0$, i.e., $V_{\scr LF}(\mathsf{z}(0),0) \leq \bar{V}$ for a finite constant $\bar{V}$. By taking the derivative of $V_{\scr LF}$, 
	and in view of Lemma \ref{lem:Chain rule (App_dynamical_systems)} of Appendix \ref{app:dynamical systems},	
	one obtains $\dot{V}_{\scr LF}(\mathsf{z}(t),t) \overset{a.e.}{\in} \dot{\widetilde{V}}_{\scr LF}(\mathsf{z}(t),t)$, where $\dot{\widetilde{V}}_{\scr LF}(\mathsf{z}(t),t)$ is the intersection of the inner products of the all generalized gradients of $V_{\scr LF}$ with the right-hand size of \eqref{eq:system closed loop (CDC_LF)}.
	Since $V_{\scr LF}(\mathsf{z},t)$ is continuously differentiable, the generalized gradient reduces to the standard gradient and one obtains	
	\begin{align*}
	\dot{\widetilde{V}}_{\scr LF} \subset& \ \dot{V}_1 + \sum_{i\in\mathcal{N}} \bigg\{ \frac{1}{2}e_{v_i}^\top\dot{M}_ie_{v_i} + e_{v_i}^\top \big(u_i - C_i\dot{x}_i - g_i - f_i  - d_i \big) - e_{v_i}M_i\dot{v}_{d_i} \\
	& +  \frac{1}{\gamma_{i,f}}\widetilde{\alpha}_i\dot{\hat{\alpha}}_i +  \frac{1}{\gamma_{i,d}}\widetilde{d}_{b_i}\dot{\hat{d}}_{b_i} + \frac{1}{\gamma_{i,\theta}}\widetilde{\theta}_i^\top \dot{\hat{\theta}}_i  \bigg\}.
	\end{align*}	
	By substituting $v_i = e_{v_i} + v_{d_i}$ in $C_iv_i$ and \eqref{eq:V_1_dot (CDC_LF)}, and using Properties \ref{ass:skew-symm (CDC_LF)}, \ref{ass:dynamics factorization (CDC_LF)}, we obtain
	\small
	\begin{align*}
	\dot{\widetilde{V}}_{\scr LF}  \subset &  -\|\gamma_e s_e - (\widetilde{d}_1\otimes I_n)^\top \beta \|^2 
	-\sum_{i\in \mathcal{N}_\mathcal{F} } k_i \|(\widetilde{d}_i\otimes I_n)^\top\beta \|^2 + \gamma_e s_e^\top e_{v_1} + \\
	& \sum_{i\in\mathcal{N}} \bigg\{ e_{v_i}^\top \big( u_i   -Y_{r_i}\theta_i  -f_i - d_i  - (\widetilde{d}_i\otimes I_n)^\top \beta \big) +  \frac{1}{\gamma_{i,f}}\widetilde{\alpha}_i\dot{\hat{\alpha}}_i + \frac{1}{\gamma_{i,d}}\widetilde{d}_{b_i}\dot{\hat{d}}_{b_i} \\
	& + \frac{1}{\gamma_{i,\theta}}\widetilde{\theta}_i^\top \dot{\hat{\theta}}_i   \bigg\}.
	\end{align*}
	\normalsize
	Next, by substituting the control laws \eqref{eq:v_d_i,u_i rewr (CDC_LF)}, the right-hand side becomes
	\small
	\begin{align*} 
	\dot{\widetilde{V}}_{\scr LF} \subset &-\|\gamma_e s_e - (\widetilde{d}_1\otimes I_n)^\top \beta \|^2 - \sum_{i\in\mathcal{N}_\mathcal{F}}k_i \|(\widetilde{d}_i\otimes I_n)^\top\beta \|^2   
	+ \sum_{i\in\mathcal{N}} \bigg\{ e_{v_i}^\top \bigg(  Y_{r_i}\widetilde{\theta}_i \\
	& - k_{v_i}e_{v_i}  - \text{SGN}(e_{v_i})\big(\left\|v_i\right\|_1 \hat{\alpha}_i + \hat{d}_{b_i}\big)   - f_i - d_i  \bigg)  +  \frac{1}{\gamma_{i,f}}\widetilde{\alpha}_i\dot{\hat{\alpha}}_i  + \frac{1}{\gamma_{i,d}}\widetilde{d}_{b_i}\dot{\hat{d}}_{b_i}  \\
	&+ \frac{1}{\gamma_{i,\theta}}\widetilde{\theta}_i^\top \dot{\hat{\theta}}_i   \bigg\}. 
	\end{align*}	
	\normalsize
	By employing the property $x^\top\text{sgn}(x) = \|x\|_1$,  $\forall x\in\mathbb{R}^n$ (which also implies that $x^\top \text{SGN}(x) = \|x\|_1$, since $x^\top \text{SGN}(x) = \{0\}$ when $x=0$), as well as \eqref{eq:v_d_i,u_i rewr (CDC_LF)} and Assumption \ref{ass:f_i+d_i (CDC_LF)}, 
	we obtain	
	\begin{align*}
	\max_{z\in \dot{\widetilde{V}}_{\scr LF}}\{ z\} \leq & -W_\zeta(\mathsf{z})  + \sum_{i\in\mathcal{N}} \bigg\{ e_{v_i}^\top Y_{r_i}\widetilde{\theta}_i  + \alpha_i\|e_{v_i}\|_1\|v_i\|_1 + d_{b_i} \|e_{v_i}\|_1 -k_{v_i}\|e_{v_i}\|^2 \\
	& - \left\|e_{v_i}\right\|_1\big(\left\|v_i\right\|_1 \hat{\alpha}_i + \hat{d}_{b_i}\big)    +  \frac{1}{\gamma_{i,f}}\widetilde{\alpha}_i\dot{\hat{\alpha}}_i  + \frac{1}{\gamma_{i,d}}\widetilde{d}_{b_i}\dot{\hat{d}}_{b_i}  + \frac{1}{\gamma_{i,\theta}}\widetilde{\theta}_i^\top \dot{\hat{\theta}}_i   \bigg\},
	\end{align*}	
	where $W_\zeta:\mathcal{Z}\to\mathbb{R}_{\geq 0}$, with $$W_\zeta(\mathsf{z}) \coloneqq  \|\gamma_e s_e - (\widetilde{d}_1\otimes I_n)^\top \beta \|^2 +
	\sum_{i\in \mathcal{N}_\mathcal{F} } k_i \|(\widetilde{d}_i\otimes I_n)^\top\beta \|^2.$$
	Finally, by substituting $\alpha_i = \hat{\alpha}_i-\widetilde{\alpha}_i$, $d_{b_i} = \hat{d}_{b_i} - \widetilde{d}_{b_i}$, $\forall i\in\mathcal{N}$, as well as the adaptation laws \eqref{eq:adaptation laws (CDC_LF)}, we obtain 
	\begin{align*}
	\max_{z\in \dot{\widetilde{V}}_{\scr LF}}\{ z\} \leq -W_\zeta(\mathsf{z})  - \sum_{i\in\mathcal{N}} k_{v_i} \|e_{v_i}\|^2 =: -W(\mathsf{z}).
	\end{align*} 
	Therefore, we conclude that $z \leq -W(\mathsf{z})$, $\forall z\in \dot{\widetilde{V}}_{\scr LF}(\mathsf{z}(t),t)$, $t\in[0,t_{\max})$, $\mathsf{z}\in\mathcal{Z}$, where $W:\mathcal{Z}\to\mathbb{R}_{\geq 0}$ is a positive semi-definite function defined on $\mathcal{Z}$. Hence, the conditions of Theorem 
	\ref{th:nonsmooth LaSalle (App_dynamical_systems)} of Appendix \ref{app:dynamical systems} hold, according to which we conclude that all Filippov solutions starting in $\mathsf{z}(0)\in\bar{\mathcal{Z}} \coloneqq \{\mathsf{z} \in \mathcal{B}(0,r_\zeta) : W_2(\mathsf{z}) < \min_{\|\mathsf{z}\| = r_\zeta} W_1(\mathsf{z})\}$ are extended to $t_{\max} = \infty$,
	satisfy $\mathsf{z}(t)\in\bar{\mathcal{Z}}$ for all $t\in\mathbb{R}_{\geq 0}$ and any positive $r_\zeta$, and $\lim_{t\to\infty} W(\mathsf{z}(t)) = 0$. Thus, the terms $\beta_{\mathfrak{c},k}(\iota_k(t))$, $\beta_{\mathfrak{n},l}(\nu_l(t))$ are bounded, $\forall t\in\mathbb{R}_{\geq 0}$, $k\in\bar{K}$, $l\in\mathcal{K}_0$, which implies that connectivity breaks of the set $\mathcal{E}_0$ and inter-agent collisions are avoided, $\forall t\in\mathbb{R}_{\geq 0}$.  
	
	In addition, it holds that $\lim_{t\to\infty}e_{v_i}(t) = 0$, $\forall i\in\mathcal{N}$, $\lim_{t\to\infty}(\widetilde{d}_i\otimes I_n)^\top\beta(t) = 0$, $\forall i\in\mathcal{N}_\mathcal{F}$, as well as $\lim_{t\to\infty}\|\gamma_e s_e - (\widetilde{d}_1\otimes I_n)^\top \beta\| = 0$. We employ now Property \ref{prop:D (App_dynamical_systems)} of Appendix \ref{app:dynamical systems} for incidence matrices, which dictates that $\sum_{i\in\mathcal{N}} \widetilde{d}_i = 0$. Hence, it holds that 
	\begin{equation*}
		\lim_{t\to\infty}(\widetilde{d}_1\otimes I_n)^\top \beta = -  \lim_{t\to\infty} \sum_{i\in\mathcal{N}_\mathcal{F}}(\widetilde{d}_i\otimes I_n)^\top \beta = 0,
	\end{equation*}
	which implies that $\lim_{t\to\infty}\|\gamma_e s_e - (\widetilde{d}_1\otimes I_n)^\top \beta\| = 0 \Rightarrow \lim_{t\to\infty} s_e = 0$, meaning that the leader agent will converge to its destination. Moreover, one concludes that	
	$\lim_{t\to\infty}v_i(t)=0$, $\forall i\in\mathcal{N}$ due to \eqref{eq:v_d_i,u_i rewr (CDC_LF)}. Note that $r_\zeta$ can be any positive constant and hence the result is global with respect to $\mathsf{z}$, i.e., all initial configurations that are collision-free and satisfy $\mathcal{E}_0\subset \mathcal{E}(x(0))$. Moreover, {the} fact that $t_{\max} = \infty$ implies that there is no Zeno behavior due to the discontinuous nature of the controller.
\end{proof}

\begin{remark}
	Inspection of the closed loop dynamics \eqref{eq:system closed loop x (CDC_LF)} reveals that, if $\lim_{t\to\infty}\dot{v}_i(t) = 0$, then the follower agents will converge to an invariant set where $d_i(t) = Y_{r_i}\widetilde{\theta}_i$. 
\end{remark}

\begin{remark}	
	Note that initial connectivity of the graphs $\mathcal{G}(x(0))$, $\mathcal{G}_0$ and connectivity to the leader are not technical requirements, as is usually the case in the related literature (e.g., \cite{gustavi2010sufficient,li2013distributed}). In such cases, the leader will still converge to $x_\text{d}$, inter-agent collisions will not occur, and the edges of $\mathcal{E}_0$, will be preserved. Regarding the unknown terms $f_i$, $d_i$, $\theta_i$, note from Theorem \ref{th:main th (CDC_LF)} and its proof that these are successfully compensated, without the need of convergence of the respective errors to zero. Finally, observe that the framework can be also applied to the multi-robot navigation problem, via alternating between leaders and followers and appropriate prioritization.
\end{remark}
\begin{figure}[t!]
	\centering
	\includegraphics[trim = 0cm 0cm 0cm 0cm, width = 0.55\textwidth]{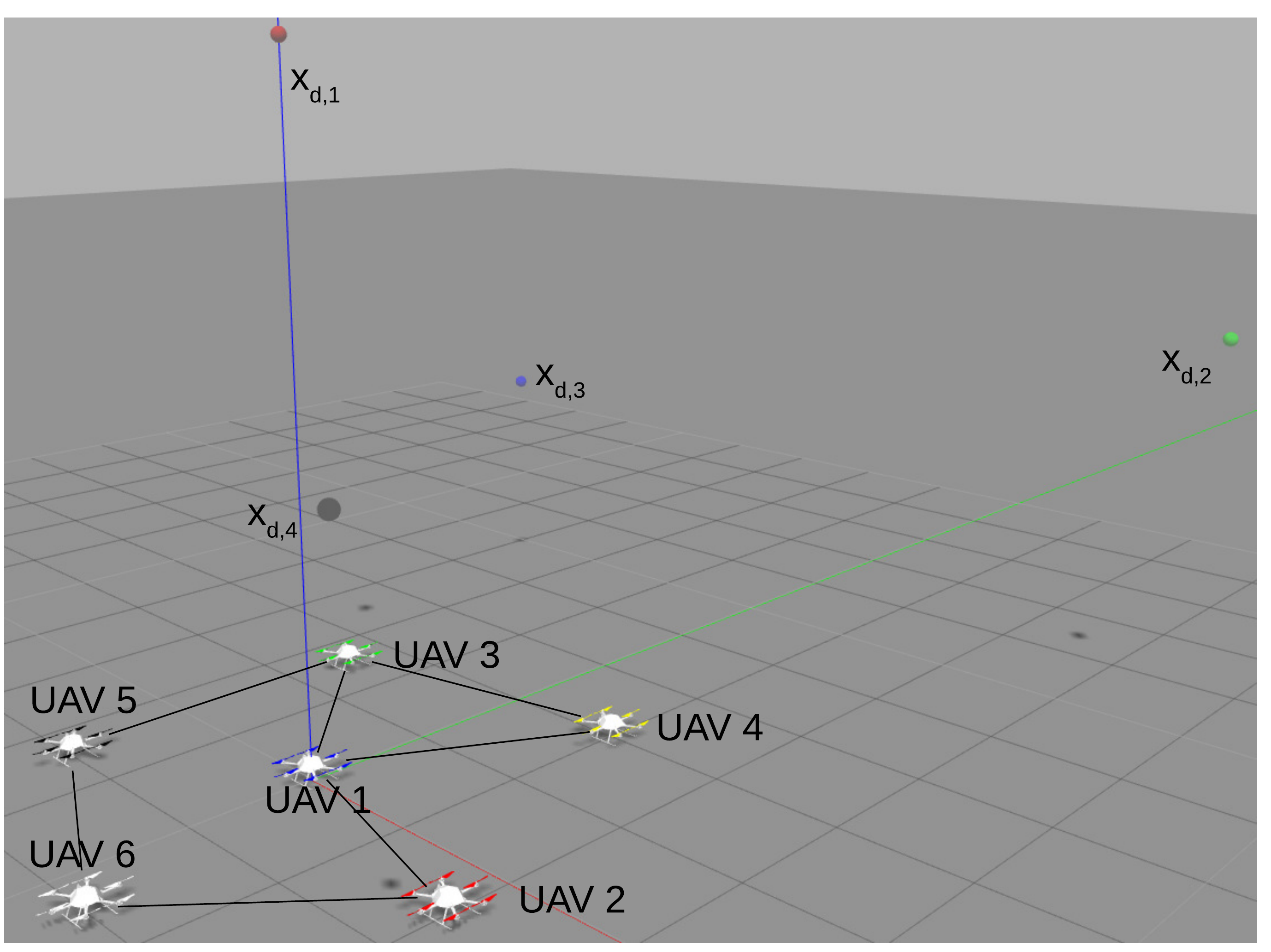}
	\caption{The initial positions of the $6$ UAVs, along with the desired leader goals $x_{\text{d},k}$, $k\in\{1,\dots,4\}$, and the edge set $\mathcal{E}_0$.}
	\label{fig:sim_init (CDC_LF)}
\end{figure}

\begin{figure*}[!tbp]
	\centering		
	\subcaptionbox{\label{fig:e_leader (CDC_LF)}}
	{\includegraphics[trim = 0cm 0cm 0cm 0cm,width=0.32\textwidth, height=0.17\textheight]{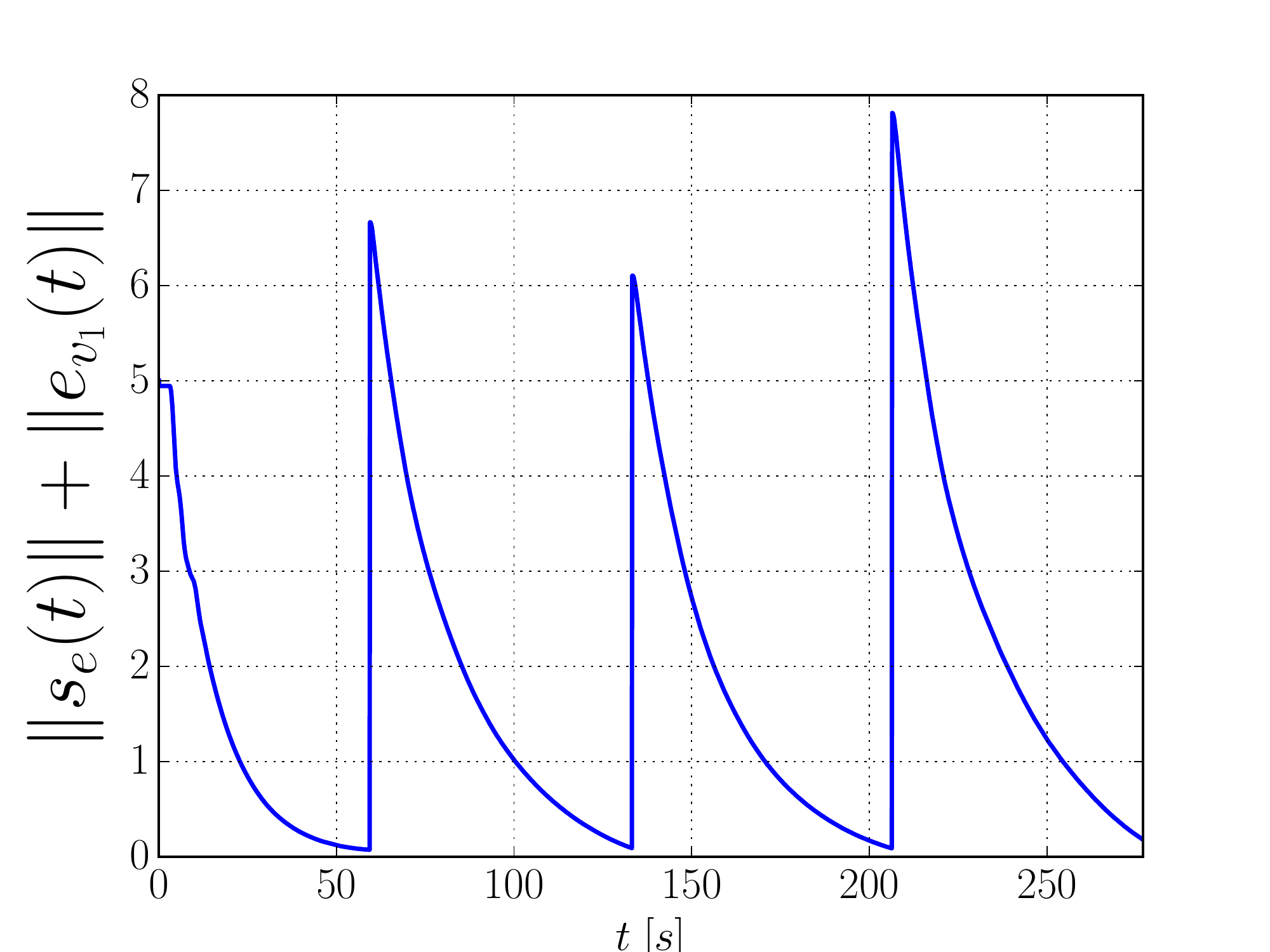}}
	\subcaptionbox{\label{fig:betas (CDC_LF)}}
	{\includegraphics[trim = 0cm 1.5cm 0cm 0cm,width=0.32\textwidth,height=0.17\textheight]{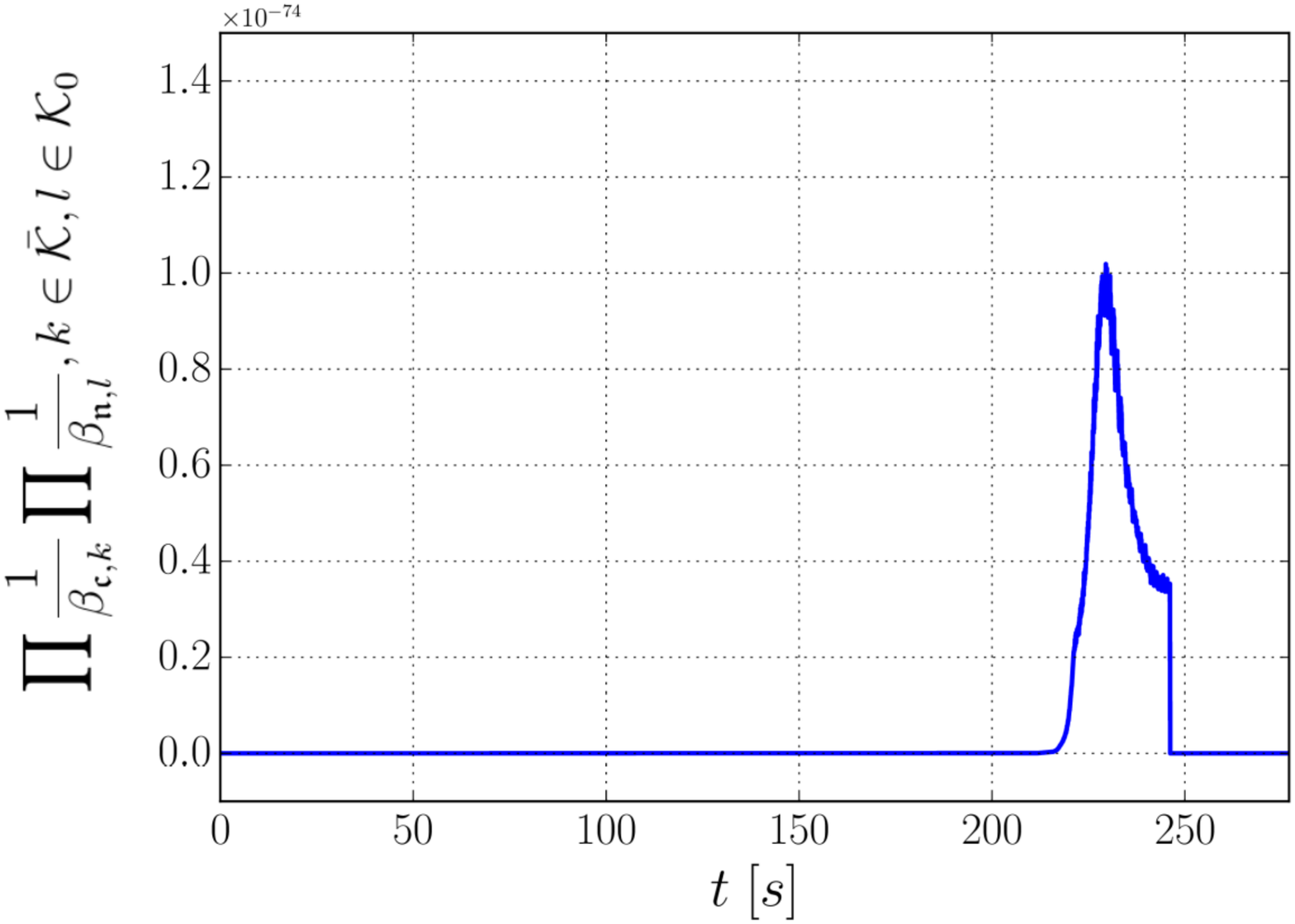}}	
	\subcaptionbox{\label{fig:adaptation laws (CDC_LF)}}{\includegraphics[trim = 0cm .5cm 0cm 0cm,width=0.32\textwidth,height=0.16\textheight]{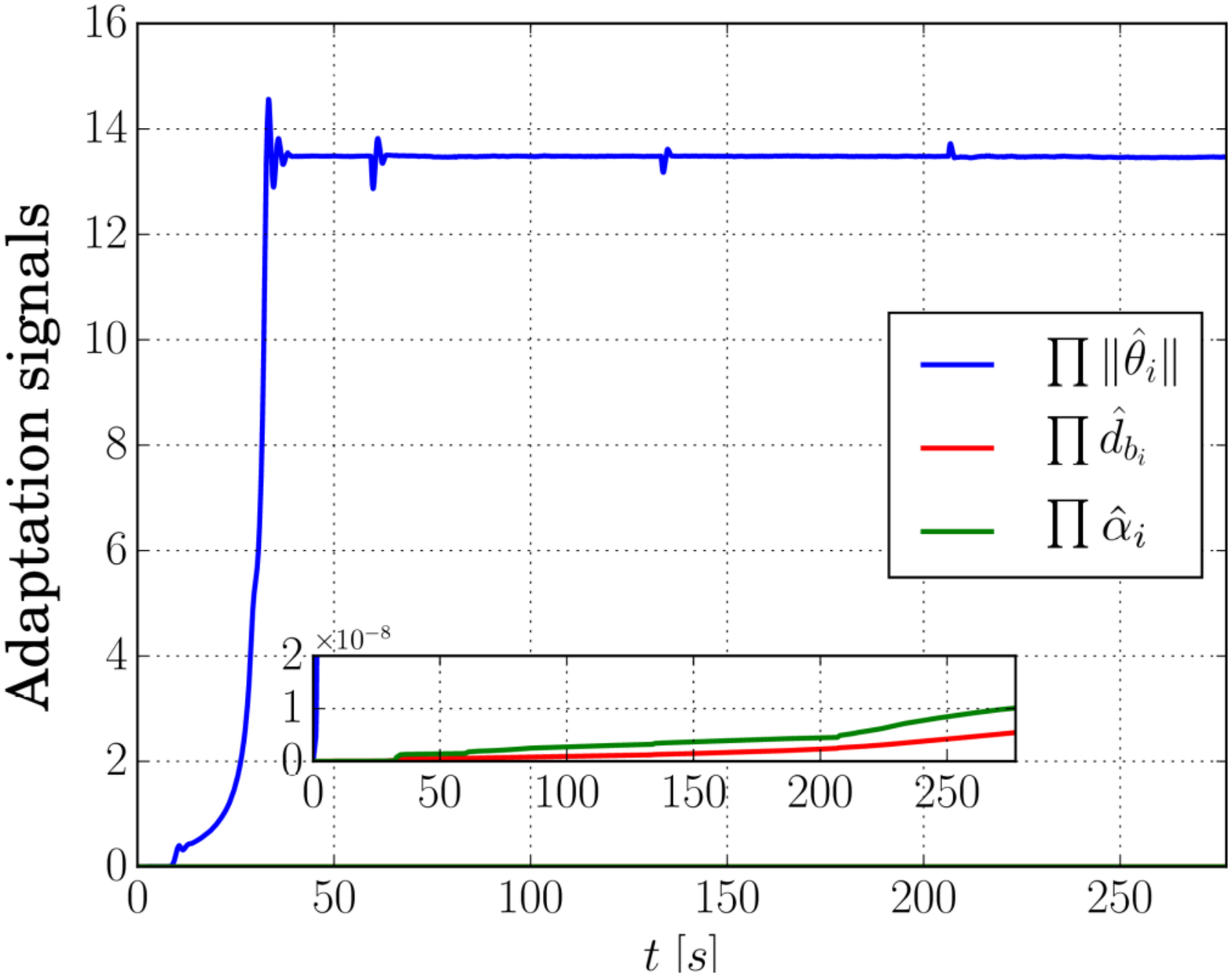}}	
	\caption{(a): The leader signal $\|s_e(t)\|+\|e_{v_1}(t)\|$, which converges to zero for every navigation objective; (b) the product $\prod_{k\in\bar{\mathcal{K}}}\frac{1}{\beta_{\mathfrak{c},k}(\iota_k(t))}$ $\prod_{l\in\mathcal{K}_0}\frac{1}{\beta_{\mathfrak{n},l}(\nu_l(t))}$, which remains bounded, proving thus the collision and connectivity properties (the zero values stem from the computer's lower numerical limits); (c) the adaptation signals $\prod_{i\in\{1,\dots,6\}}\|\hat{\theta}_i(t)\|$, $\prod_{i\in\{1,\dots,6\}} \hat{d}_{b_i}(t)$, $\prod_{i\in\{1,\dots,6\}} \hat{\alpha}_i(t)$, which remain bounded, $\forall t\in[0,277]$ s. }\label{fig:data (CDC_LF)}
\end{figure*}

\subsection{Simulation Results}\label{sec:Simulation (CDC_LF)}
We conducted simulations with $N=6$ UAVs in $\mathbb{R}^3$ using the realistic robotic simulator Gazebo \cite{koenig2004design}. We considered bounding radii $r_i=0.35 \text{m}$, sensing ranges $\varsigma_i = 3\text{m}$, $\forall i\in\mathcal{N}$, and initial positions $x_1(0) = [0,0,0.1]^\top$, $x_2(0) = [2,-0.5,0.1]^\top$, $x_3(0) = [-1.5,1.5,0.1]^\top$, $x_4(0) = [1,2,0.1]^\top$, $x_5(0) = [-1.5,-1,0.1]^\top$, and $x_6(0) = [0.5,-1.5,0.1]^\top$ m,  (see Fig. \ref{fig:sim_init (CDC_LF)}). We also considered that the leader has $4$ navigation objectives, that is, to sequentially navigate to the points $x_\text{d,1} = [0,0,5]^\top$, $x_\text{d,2} = [4,5,3]^\top $, $x_\text{d,3} = [-2,4,2]^\top$, $x_\text{d,4} = [3,-2,3]^\top$ m (pictured as small spheres in Fig. \ref{fig:sim_init (CDC_LF)}). Since this work provides asymptotic results with respect to the error $s_e$, the leader switches navigation goal each time it gets closer than $0.075\text{m}$ to the current goal, i.e., $\|s_e\| \leq 0.075\text{m}$.
We also considered $$\mathcal{E}_0=\{(1,2),(1,3),(1,4),(3,4),(3,5),(5,6),(2,6)\},$$ as shown in Fig. \ref{fig:sim_init (CDC_LF)} via straight black lines. 
The unknown parameters $\theta_i$ concerned the UAVs' mass and the gravity constant. The control gains and parameters were set as $\gamma_e = 0.7$, $k_i = 5$, $\forall i\in\{2,\dots,6\}$, and $\gamma_{i,\theta} = 0.1$, $\gamma_{i,d} = 0.01$, $\gamma_{i,f} = 0.1$, $k_{v_i} = 2$, $\forall i\in\{1,\dots,6\} $. The simulation results are shown in Figs. \ref{fig:data (CDC_LF)}-\ref{fig:control inputs (CDC_LF)} for $t\in[0,277]$ s. More specifically, Fig. \ref{fig:data (CDC_LF)} shows (a) the evolution of the signal $\|s_e(t)\|+\|e_{v_1}(t)\|$, which converges to zero for each navigation objective, (b) the evolution of the product $\prod_{k\in\bar{\mathcal{K}}}\frac{1}{\beta_{\mathfrak{c},k}(\iota_k(t))}\prod_{l\in\mathcal{K}_0}\frac{1}{\beta_{\mathfrak{n},l}(\nu_l(t))}$, which remains bounded, verifying thus the collision avoidance and connectivity maintenance properties, and (c) the evolution of the products of the adaptation signals $\prod_{i\in\{1,\dots,6\}}\|\hat{\theta}_i(t)\|$, $\prod_{i\in\{1,\dots,6\}} \hat{d}_{b_i}(t)$, $\prod_{i\in\{1,\dots,6\}} \hat{\alpha}_i(t)$, which remain bounded, verifying thus the boundedness of the individual signals. Moreover, Fig. \ref{fig:path (CDC_LF)} depicts the evolution of the multi-agent system along the $4$ navigation objectives, with the connectivity of $\mathcal{E}_0$ (straight black lines), and Fig. \ref{fig:control inputs (CDC_LF)} shows the control inputs of the UAVs.
The simulations were carried out in a ROS-Python interface of an i$7$-$8750$H laptop computer with $12$ cores at $2.2$GHz and $16$GB of RAM and an illustrating video can be found in \url{https://youtu.be/bzzXC-v2hEM}.

\begin{figure*}[!tbp]
	\centering			
	\subcaptionbox{\label{fig:path 1 (CDC_LF)}}{\includegraphics[trim = 0cm 0cm 0cm 0cm,width=0.45\textwidth]{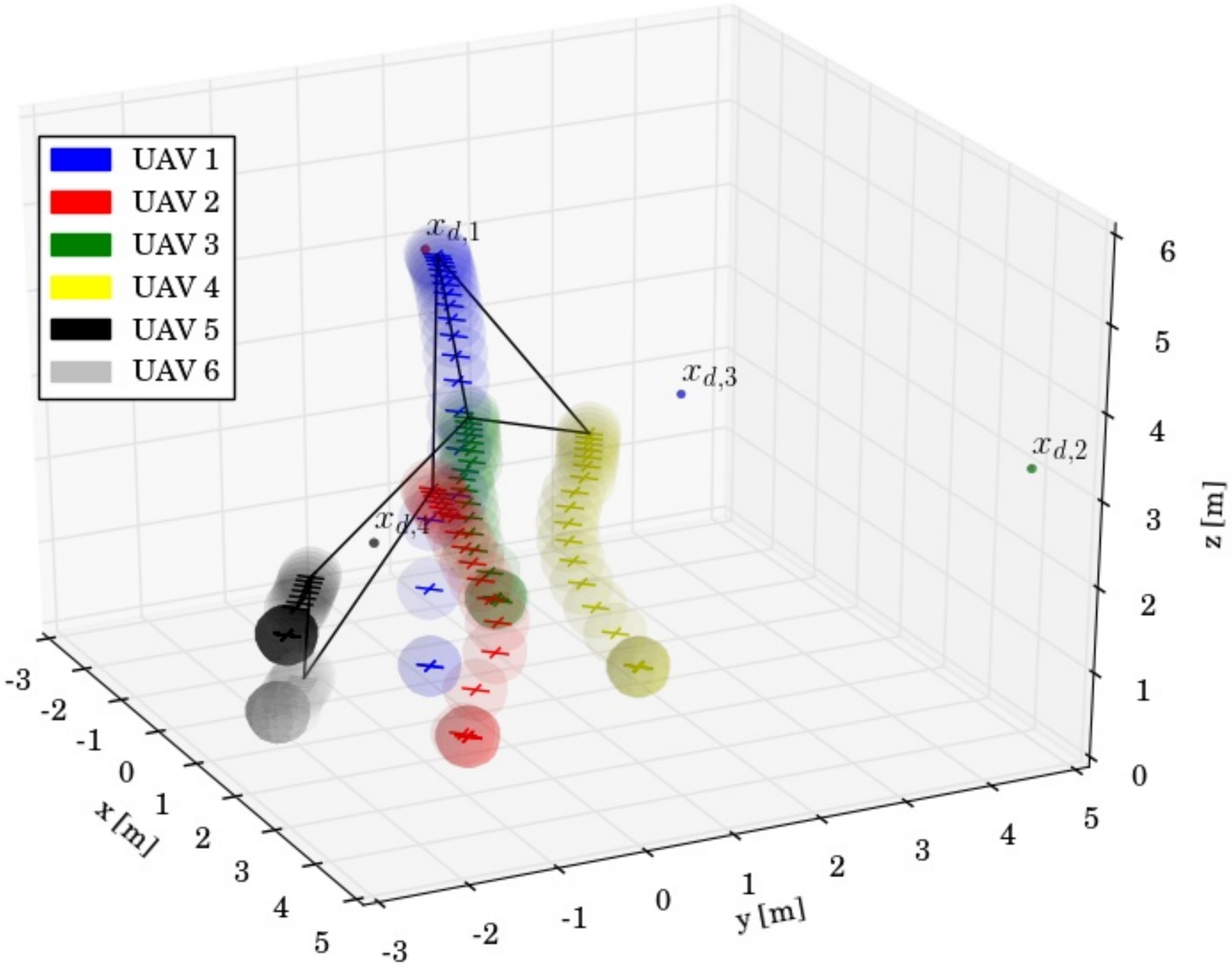}}
	\subcaptionbox{\label{fig:path 2 (CDC_LF)}}
	{\includegraphics[trim = 0cm 0cm 0cm 0cm,width=0.45\textwidth]{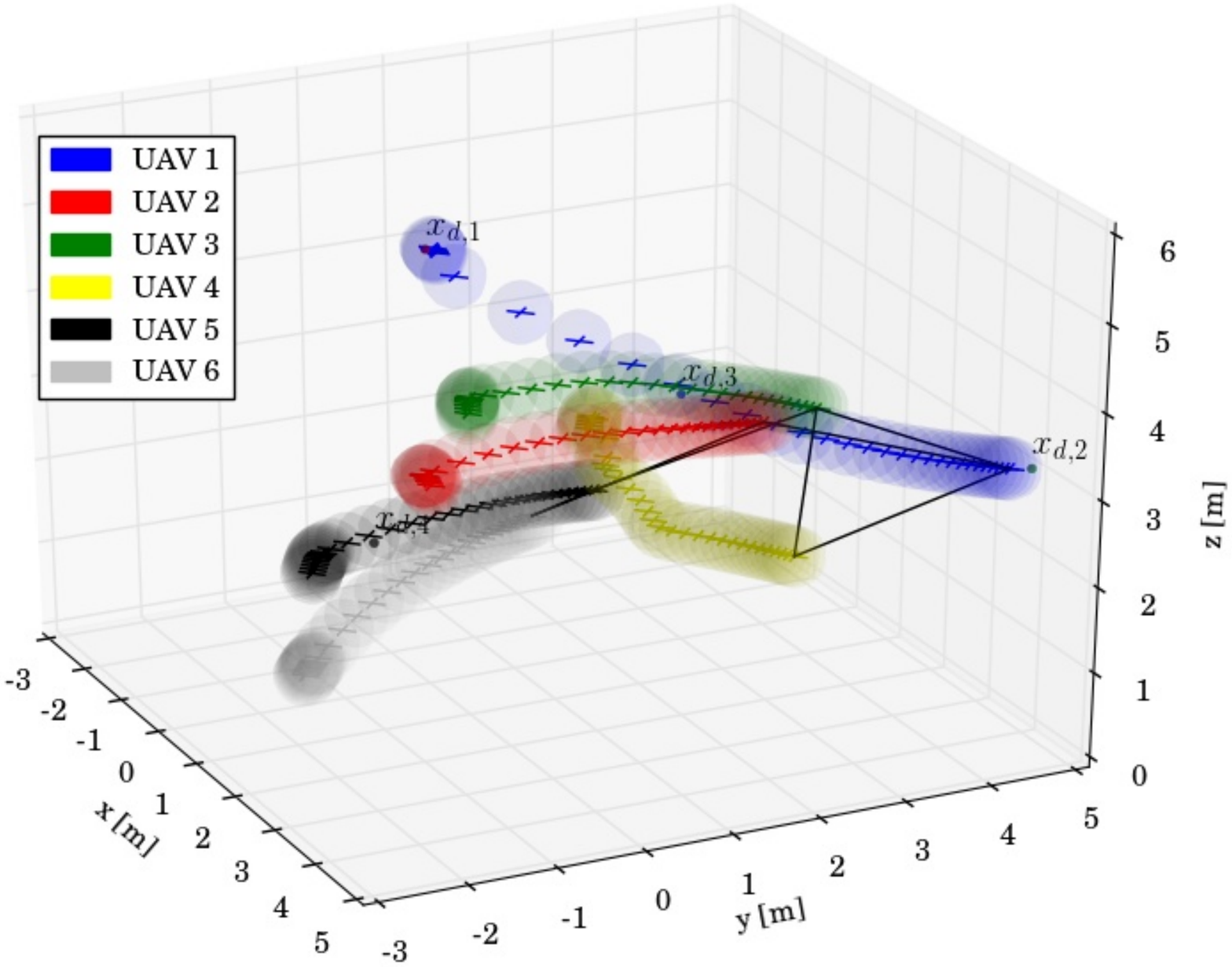}}	
	\subcaptionbox{\label{fig:path 3 (CDC_LF)}}{\includegraphics[trim = 0cm .75cm 0cm 0cm,width=0.45\textwidth]{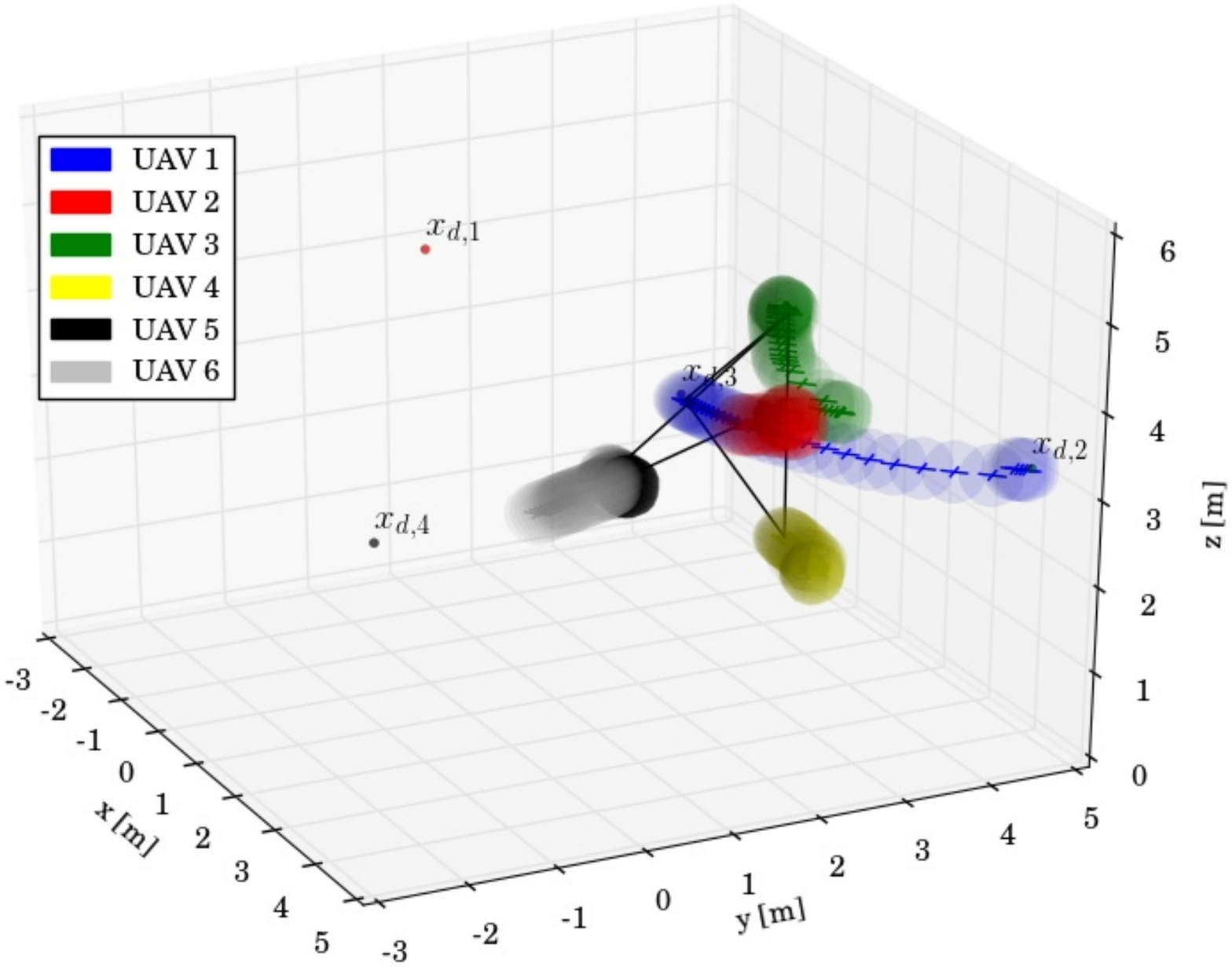}}
	\subcaptionbox{\label{fig:path 4 (CDC_LF)}}
	{\includegraphics[trim = 0cm 0.75cm 0cm 0cm,width=0.45\textwidth]{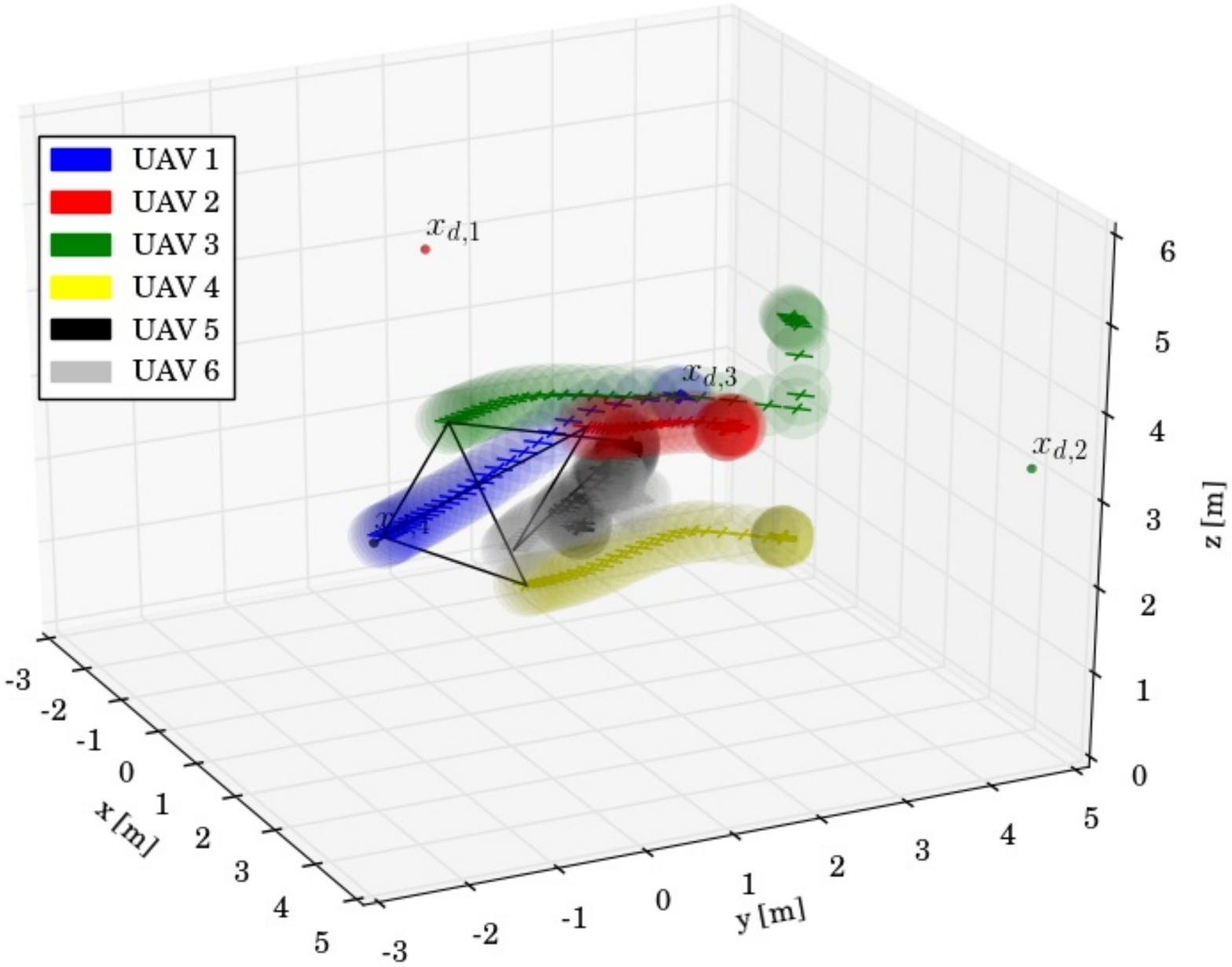}}
	\caption{The motion of the multi-agent system as the leader navigates to $x_{\text{d},1}$ (a), $\dots$, $x_{\text{d},4}$ (d). The connectivity of $\mathcal{E}_0$ is also pictured via straight lines.}\label{fig:path (CDC_LF)}
\end{figure*}
\begin{figure}[t!]
	\centering
	\includegraphics[trim = 0cm 0cm 0cm 0cm,width = 0.85\textwidth]{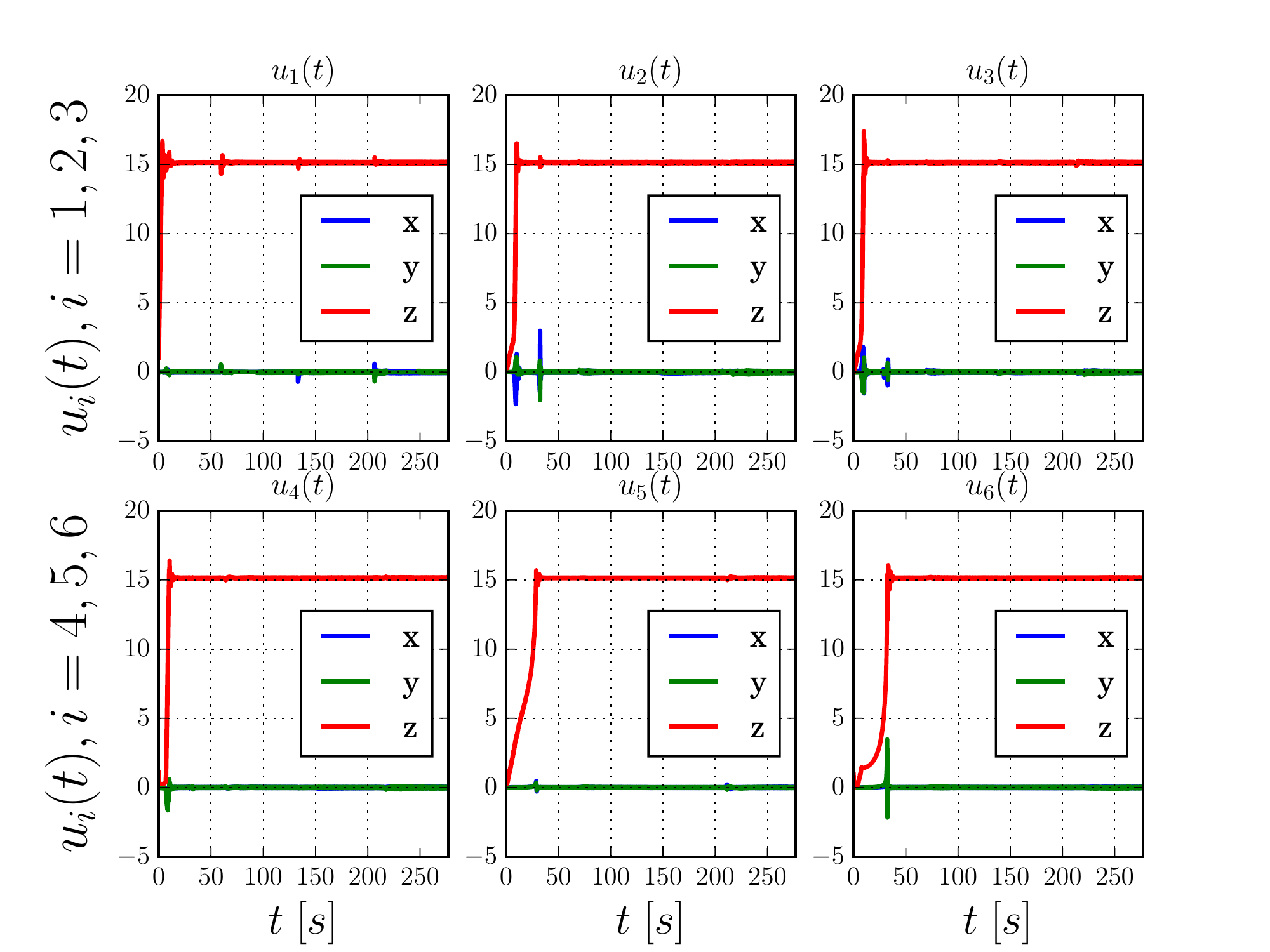}
	\caption{The resulting control inputs $u_i(t)$, $i\in\{1,\dots,6\}$, $t\in[0,277]$ s.}
	\label{fig:control inputs (CDC_LF)}
\end{figure}

\section{Closed-Form Collision Avoidance of Ellipsoidal Multi-Agent Systems} \label{sec:Ellipsoids}

The previous sections, as well as Section \ref{sec:formation control} of the previous chapter, considered spherical agents, which is a common assumption also in the related literature. In this section, we turn our attention to robotic agents whose volume is approximated as an ellipsoid in $\mathbb{R}^3$, since such an approximation is more realistic for robotic agents (see e.g., Fig. \ref{fig:ellips motivation (LCSS_EL)}). 
We develop a class of \textit{closed-form} barrier functions that approximate the distance between two such ellipsoids and design an adaptive control scheme for the collision avoidance of the multi-agent system, subject to some primary task and $2$nd-order uncertain dynamics, like in the previous section.

\begin{figure}[!ht]
	\centering
	\includegraphics[width = 0.5\textwidth]{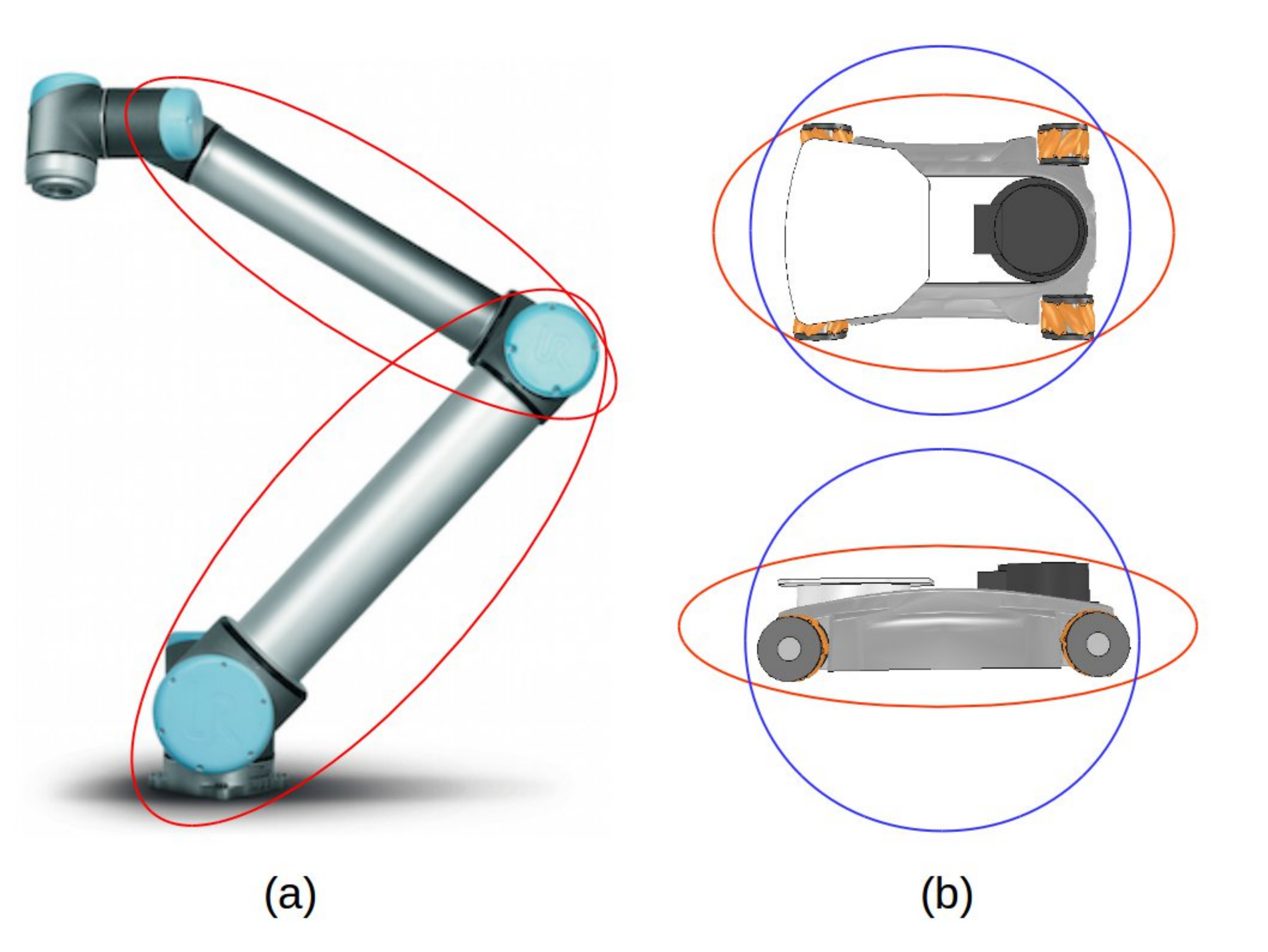}\\
	\caption{Ellipsoid approximation of (a) the rigid links of a robotic manipulator, (b) a mobile robot (top and front view).}\label{fig:ellips motivation (LCSS_EL)}
\end{figure}

\subsection{Problem Formulation} \label{sec:problem form (LCSS_EL)}
Consider $N>1$ ellipsoidal autonomous agents, with $\mathcal{N} \coloneqq \{1,\dots,N\}$, operating in $\mathbb{SE}(3)$, and described now by the ellipsoids $$\mathcal{A}_i(x_{s_i}) \coloneqq \{y\in\mathbb{R}^4 : y^\top A_i(x_{s_i})y \leq 0 \};$$ $x_{s_i} \coloneqq [p_i^\top, \zeta_i^\top]^\top \in\mathbb{M}\coloneqq \mathbb{R}^3\times \mathbb{S}^3$ is the $i$th agent's center of mass pose, where $p_i\in\mathbb{R}^3$  is its inertial position and $\zeta_i \coloneqq [\varphi_i, \epsilon_i^\top]^\top\in \mathbb{S}^3$ its unit quaternion-based orientation,
with $\varphi_i\in\mathbb{R}$, $\epsilon_i\in\mathbb{R}^3$ its scalar and vector parts, respectively, subject to $\|\zeta_i\| = 1$;
$A_i(x_{s_i}) \coloneqq T_i^{-\top}(x_{s_i}) \hat{A}_i T_i^{-1}(x_{s_i})$, with $\hat{A}_i\coloneqq \text{diag}\{ l^{-2}_{x,i}, l^{-2}_{y,i}, l^{-2}_{z,i}, -1 \}$, corresponding to the principal axis lengths $l_{x,i}, l_{y,i}, l_{z,i}\in\mathbb{R}_{> 0}$ of agent $i$'s ellipsoid, and $T_i\in \mathbb{SE}(3)$ is the transformation matrix describing the translation and orientation of agent $i$'s center of mass, $\forall i\in\mathcal{N}$.
The agents' motion follows the  standard Lagrangian dynamics (similar to \eqref{eq:dynamics (CDC_LF)}):
\small
\begin{subequations} \label{eq:dynamics (LCSS_EL)}
	\begin{align}	
	& \dot{x}_{s_i} = \bar{E}_\zeta(\zeta_i)v_i \\ 
	& M_i(x_{s_i})\dot{v}_i + C_i(x_{s_i},v_i)v_i + g_i(x_{s_i}) + f_i(x_{s_i},v_i) + d_i(t)  = u_i, 
	\end{align}
\end{subequations}
\normalsize
where $v_i \coloneqq [\dot{p}_i^\top, \omega_i^\top]^\top$ is agent $i$'s velocity, with $\omega_i\in\mathbb{R}^3$ being its angular velocity,  $\bar{E}_\zeta: \mathbb{S}^3 \to \mathbb{R}^{7\times6}$ is the matrix mapping the quaternion rates to velocities, defined as $\bar{E}_\zeta \coloneqq \text{diag}\{I_3, E(\zeta)\}$ and $E(\zeta)$ as defined in Section \ref{subsec:system model (TCST_coop_manip)};
The rest of the terms are the same as in \eqref{eq:dynamics (CDC_LF)}, with unknown dynamic parameters in $M_i$, $C_i$, $g_i$, and unknown $f_i$, $d_i$, $\forall i\in\mathcal{N}$.
Here we also consider that $u_i$ is decomposed as $u_i = u_{f,i} + u_{s,i}$, where $u_{f_i}$ is a bounded term that is responsible for some (potentially cooperative) task, and $u_{s,i}$ is a control term to be designed in order to achieve multi-agent decentralized collision avoidance, $\forall i\in\mathcal{N}$.
More specifically, we consider that 
$\phi_s(x_s)\in\mathbb{R}_{\geq 0}$ is a term that corresponds to the cooperative task dictated by $u_{f,i}$, with $$u_{f,i}=\bar{E}_\zeta(\zeta_i)^\top\frac{\partial \phi_s(x_s)}{\partial x_{s_i}},$$ $\forall i\in\mathcal{N}$, $\mathsf{c}_1(x_s) \leq \phi_s(x_s) \leq \mathsf{c}_2(x_s)$, for continuous positive definite functions $\mathsf{c}_1,\mathsf{c}_2$, and nonempty sets 
$\{ x_s \in \mathcal{X}_s : x_s =\phi_s^{-1}(y)\}$, $\forall y\in\mathbb{R}_{\geq 0}$, where $x_s\coloneqq[x_{s_1}^\top,\dots,x_{s_N}^\top]^\top$, and $$\mathcal{X}_s \coloneqq \{ x_s\in\mathbb{M}^N: \mathcal{A}_i(x_{s_i})\cap\mathcal{A}_j(x_{s_j}) = \emptyset, \forall i,j\in\mathcal{N},i\neq j\};$$ {$\phi_s$ can be also a function of $\widetilde{x}_s \coloneqq [p_1^\top-p_2^\top,\dots,p_N^\top-p_{N-1}^\top,\zeta_1^\top,\dots,\zeta_N^\top]^\top$ that concerns potential formation control objectives. Then $\mathcal{X}_s$ becomes $$\{ \widetilde{x}_s\in\mathbb{R}^{\frac{3N(N-1)}{2}}\times \mathbb{S}^{3N} : \mathcal{A}_i(x_{s_i})\cap\mathcal{A}_j(x_{s_j}) = \emptyset, \forall i,j\in\mathcal{N}, i\neq j\}.$$ The conditions for $\phi$ are satisfied by standard quadratic functions, e.g., $$\phi_s(x_s) = \sum_{i\in\mathcal{N}}\{\|p_i-p_{\textup{d}_i}\|^2 + e_{\zeta_i}^2\}$$ (for multi-agent navigation) or $$\phi_s(\widetilde{x}) = \sum_{(i,j)\in\mathsf{F}}\{\|p_i-p_j-p_{\textup{d}_{i,j}}\|^2 + e_{\zeta_{i,j}}^2\}$$ (for formation) for sufficiently distant $p_{\textup{d}_i}, p_{\textup{d}_{i,j}}$, where $\mathsf{F}$ is a potential formation set and $e_{\zeta_i}, e_{\zeta_{i,j}}$ represent appropriate quaternion errors (see Section \ref{subsec:Quaternion Controller (TCST_coop_manip)}). Note that $\phi_s$ and $u_{f,i}$ are \textit{not} responsible for collision avoidance or compensating model uncertainties.}

The terms $M_i$ and $C_i$ satisfy Property \ref{ass:skew-symm (CDC_LF)}, $\forall i\in\mathcal{N}$, as well as a slightly modified version of Property \ref{ass:dynamics factorization (CDC_LF)} that accounts only for $g$:
\begin{property}  \label{ass:dynamics factorization (LCSS_EL)}
	The gravity terms of \eqref{eq:dynamics (LCSS_EL)} can be written as 
	$g_i(z) = Y_{g_i}(z)\theta_{g_i}$, $\forall z\in\mathbb{M}, i\in\mathcal{N}$, where $Y_{g_i}:\mathbb{M}\to\mathbb{R}^{6\times \ell}$ are known continuous matrices, and $\theta_{g_i}\in \mathbb{R}^{\ell}$,  $\ell\in\mathbb{N}$, are  constant but unknown dynamic parameters of the agents, $\forall i\in\mathcal{N}$.
\end{property}
Moreover, the unknown disturbances $d_i$ satisfy Assumption \ref{ass:f_i+d_i (CDC_LF)} for unknown $d_{b_i}$, $\forall i\in\mathcal{N}$, whereas we impose a stronger assumption on $f_i$ for simplicity:

\begin{property}{\cite{kelly2006control}} \label{ass:friction (LCSS_EL)}
	The friction terms are dissipative, i.e., $v_i^\top f_i(x_{s_i},v_i) > 0$, $\forall x_{s_i}\in\mathbb{M}, v_i\neq 0, i\in\mathcal{N}$.
\end{property}
	
As before, we consider that each robot has a limited sensing radius $\varsigma_i\in\mathbb{R}_{>0}$, with the restriction now $\varsigma_i > \max\{l_{x,i},l_{y,i},l_{z,i}
\} + \max_{j\in\mathcal{N}}\big\{\max\{l_{x,j},\\l_{y,j},l_{z,j}\}\big\}+ \varepsilon$ for an arbitrarily small positive constant $\varepsilon$, which implies that the agents can sense each other without colliding.
Based on this, the undirected time-varying graph that models the topology of the multi-robot network becomes now $\mathcal{G}(p) \coloneqq (\mathcal{N},\mathcal{E}(p))$, with  $\mathcal{E}(p) \coloneqq \{(i,j)\in\mathcal{N}^2 : \|p_i - p_j \| \leq \min\{\varsigma_i, \varsigma_j\} \}$, $p\coloneqq[p_1^\top,\dots,p_N^\top]^\top$, and we further define the agent time-varying neighborhood $\mathcal{N}_i(p)\coloneqq \{j\in\mathcal{N}: \|p_i-p_j\| < \varsigma_i\} $, $\forall i\in\mathcal{N}$. Moreover, we consider again the complete graph $\bar{\mathcal{G}} \coloneqq (\mathcal{N},\bar{\mathcal{E}})$, with $\bar{\mathcal{E}}\coloneqq \{ (i,j), \forall i,j\in\mathcal{N}, i < j\}$, $\bar{K}\coloneqq |\bar{\mathcal{E}}| = \frac{N(N-1)}{2}$  and an edge numbering set $\bar{\mathcal{K}}\coloneqq  \{1,\dots,\bar{K}\}$. Finally, we use the same notation for $(k_1,k_2)$ that give the robot indices that form edge $k$.

As discussed before, the agents need to avoid collisions with each other, while executing their task, dictated by $u_{f,i}$. To that end, we aim to design \textit{closed-form} barrier functions and decentralized feedback control laws $u_{s,i}$ that guarantee collision avoidance among the ellipsoidal agents, while compensating appropriately for the model uncertainties and the external disturbances. 
Formally, the treated problem is the following:	
	
\begin{problem} \label{prob: 1 (LCSS_EL)}
		Given $N$ $3$D ellipsoidal autonomous agents with the uncertain Lagrangian dynamics \eqref{eq:dynamics (LCSS_EL)} executing tasks dictated by $u_{f,i}$, design 
		\begin{enumerate}
			\item closed-form barrier functions that encode collision avoidance of the agents,
			\item {decentralized control laws in $u_{s,i}$ that guarantee inter-agent collision avoidance, i.e., $\mathcal{A}_i(x_{s_i}(t))\cap\mathcal{A}_j(x_{s_j}(t)) = \emptyset$, $\forall i,j\in\mathcal{N}$, $i\neq j$, as well as boundedness of all closed loop signals}. 
		\end{enumerate} 
	\end{problem}

\subsection{Problem Solution} \label{sec:main results (LCSS_EL)}
	
	This section describes the proposed solution to Problem \ref{prob: 1 (LCSS_EL)}. In order to deal with the ellipsoidal collision avoidance, we employ results from computer graphics that are related to detection of ellipsoid collision and we build appropriate barrier functions whose boundedness implies the collision-free trajectories. Moreover, we use adaptive and discontinuous control laws to appropriately compensate for the uncertainties and external disturbances of  \eqref{eq:dynamics (LCSS_EL)}.
	
	We employ first the results described in Proposition \ref{prop:ellipsoids (app_useful_prop)} of Appendix \ref{app:useful_prop} to build an appropriate ellipsoidal barrier function. Note, however, that these results concern planar ellipsoids and cannot be straightforwardly extended to the $3$D case, which is the case of the considered multi-agent system. For that reason, we consider the respective planar projections.
	For an ellipsoid $\mathcal{A}_i, i\in\mathcal{N}$, we denote as $\mathcal{A}^{xy}_i,\mathcal{A}^{xz}_i, \mathcal{A}^{yz}_i$ its projections on the planes $x$-$y$, $x$-$z$ and $y$-$z$, respectively, with corresponding matrix terms $A^{xy}_i,A^{xz}_i, A^{yz}_i$, i.e., $$\mathcal{A}^s_i(x_{s_i}) \coloneqq \{y\in\mathbb{R}^3 : y^\top A^s_i(x_{s_i})y \leq 0 \}, \forall s \in \{xy,xz,yz\}.$$ Note that in order for $\mathcal{A}_i, \mathcal{A}_j$ to collide (touch externally), all their projections on the three planes must also collide, i.e.,
	\begin{align*}		
	&\mathcal{A}_i(x_{s_i})\cap\mathcal{A}_j(x_{s_j}) \neq \emptyset \land 
	\textup{Int}(\mathcal{A}_i(x_{s_i}))\cap \textup{Int}({\mathcal{A}}_j(x_{s_j})) = \emptyset
	\Leftrightarrow \\
	&\mathcal{A}^s_i(x_{s_i})\cap\mathcal{A}^s_j(x_{s_j}) \neq \emptyset \land  \textup{Int}(\mathcal{A}^s_i(x_{s_i}))\cap \textup{Int}(\mathcal{A}^s_j(x_{s_j})) = \emptyset, \forall s\in\{xy,xz,yz\}.
	\end{align*}
	Therefore, {$\mathcal{A}_i$ and $\mathcal{A}_j$  do not collide if and only if $\mathcal{A}^s_i(x_{s_i})\cap\mathcal{A}^s_j(x_{s_j})=\emptyset$} for some  $s\in\{xy,xz,yz\}$. In view of Proposition \ref{prop:ellipsoids (app_useful_prop)} of Appendix \ref{app:useful_prop}, that means that the characteristic equations $$f^{s}_{i,j}(\lambda) \coloneqq \det(\lambda A^s_i(x_{s_i}) - A^s_j(x_{s_j}))=0$$ must always have one positive real root and two negative distinct roots for at least one $s\in\{xy,xz,yz\}$. Hence, by denoting the discriminant of $f^s_{i,j}(\lambda) = 0$ as $\Delta^s_{i,j}(x_{s_i},x_{s_j})$, Proposition \ref{prop:cubic (app_useful_prop)} of Appendix \ref{app:useful_prop} suggests that $\Delta^s_{i,j}(x_{s_i},x_{s_j})$ must remain always positive for at least one $s\in\{xy,xz,yz\}$, since a collision would imply $\Delta^s_{i,j}(x_{s_i},x_{s_j})=0$, $\forall s\in\{xy,xz,yz\}$. Therefore, by defining the smooth function \cite{loizou2017navigation}  
	\begin{equation} \label{eq:sigma smooth switch}
		\sigma(z) \coloneqq 
		\begin{cases}
			\exp(-\frac{1}{z}), & z > 0 \\
			0, & z \leq 0
		\end{cases}
	\end{equation}
	we conclude that {$\mathcal{A}_i$ and $\mathcal{A}_j$ do not collide if and only if $$\sigma(\Delta^{xy}_{i,j}(x_{s_i},x_{s_j})) + \sigma(\Delta^{xz}_{i,j}(x_{s_i},x_{s_j})) + \sigma(\Delta^{yz}_{i,j}(x_{s_i},x_{s_j})) > 0,$$} since a collision would result in $\Delta^s_{i,j}(x_{s_i},x_{s_j}) = 0 \Leftrightarrow \sigma(\Delta^s_{i,j}(x_{s_i},x_{s_j})) = 0, \forall s\in\{xy,xz,yz\}$. We aim now at defining a decentralized continuously differentiable function for each edge $k\in\bar{\mathcal{K}}$ that incorporates the collision avoidance property of agents $k_1,k_2$. We need first the following result regarding the discriminant of $f^s_{i,j}(\lambda) = 0$:
	\begin{proposition} \label{prop:eq discr equal (LCSS_EL)}
		Let $\Delta_1$, $\Delta_2$ be the discriminants of $f_1(\lambda)\coloneqq\det(\lambda A - B) = 0$, $f_2(\lambda)\coloneqq\det(\lambda B - A) = 0$, respectively, where $A, B\in\mathbb{R}^{3\times3}$. Then  $\Delta_1 = \Delta_2$.
	\end{proposition}
	\begin{proof}
		Let $$\det(\lambda A - B) = 0 \Leftrightarrow f_1(\lambda) \coloneqq  c_3\lambda^3 + c_2\lambda^2 + c_1\lambda + c_0 = 0,$$ with $c_\ell$ $\in\mathbb{R}$, $\forall \ell\in\{0,\dots,3\}$. It can be verified that $$\det(\lambda B - A) = 0 \Leftrightarrow f_2(\lambda) = -c_0\lambda^3 - c_1\lambda^2 - c_2\lambda - c_3 = 0.$$ 
		Let $\lambda_1,\lambda_2,\lambda_3$ be the solutions of $f_1(\lambda) = 0$, i.e. $f_1(\lambda_1)=f_1(\lambda_2)=f_1(\lambda_3)=0$, and $\lambda_1\lambda_2\lambda_3 = -\frac{c_0}{c_3}$. By substituting $\frac{1}{\lambda_\ell}$ in $f_2(\lambda)$, $\ell\in\{1,2,3\}$, we obtain $$	-c_0\lambda_\ell^{-3} - c_1\lambda_\ell^{-2} - c_2\lambda_\ell^{-1} - c_3  = -(c_3\lambda_\ell^3 + c_2\lambda_\ell^2 + c_1\lambda_\ell + c_0) 
		= -f_1(\lambda_\ell) = 0.$$
		Hence, $\frac{1}{\lambda_1},\frac{1}{\lambda_2},\frac{1}{\lambda_3}$ are the solutions of $f_2(\lambda) = 0$. The discriminants of $f_1(\lambda) = 0$ and $f_2(\lambda) = 0$ are 
		\begin{align*}
			\Delta_1 = c_3^4 (\lambda_1-\lambda_2)^2 (\lambda_1-\lambda_3)^2 (\lambda_2-\lambda_3)^2
		\end{align*} 
		and 		
		\begin{align*}
			\Delta_2 =& (-c_0)^4 \left(\lambda_1^{-1}-\lambda_2^{-1}\right)^2 \left(\lambda_1^{-1}- \lambda_3^{-1}\right)^2 \left(\lambda_2^{-1}-\lambda_3^{-1}\right)^2  \\
			=& c_0^4(\lambda_1\lambda_2\lambda_3)^{-4} (\lambda_2-\lambda_1)^2 (\lambda_3-\lambda_1)^2 (\lambda_3-\lambda_2)^2,
		\end{align*}
		respectively, 
		which, by substituting $c_0 = -c_3\lambda_1\lambda_2\lambda_3$, becomes $\Delta_2 = \Delta_1$.
	\end{proof}
	
	Therefore, we conclude that the discriminants $\Delta^s_{i,j}()$ and $\Delta^s_{j,i}()$ of $\det(\lambda A^s_i(x_{s_i}) - A^s_j(x_{s_j})) = 0$ and $\det(\lambda A^s_j(x_{s_j}) - A^s_i(x_{s_i})) = 0$, respectively, are the same, for all $s\in\{xy,xz,yz\}$. Hence, we can define uniquely for each edge $k\in\bar{\mathcal{K}}$ the continuously differentiable function $\Delta_k:\mathbb{K}^2 \to \mathbb{R}_{\geq 0}$, with
	\begin{align} \label{eq:Delta_m (LCSS_EL)}
	&\Delta_k(x_{s_{k_1}},x_{s_{k_2}}) \coloneqq \sigma(\Delta^{xy}_{k_1,k_2}(x_{s_{k_1}},x_{s_{k_2}})) + \sigma(\Delta^{xz}_{k_1,k_2}(x_{s_{k_1}},x_{s_{k_2}})) \notag \\
	&+ \sigma(\Delta^{yz}_{k_1,k_2}(x_{s_{k_1}},x_{s_{k_2}})),
	\end{align} 
	which needs to remain positive for all times in order to achieve the collision avoidance property, i.e., $\Delta_k(x_{s_{k_1}}(t),x_{s_{k_2}}(t)) > 0$, $\forall t\in\mathbb{R}_{\geq 0}, k\in\bar{\mathcal{K}}$. Note that, in view of Proposition \ref{prop:eq discr equal (LCSS_EL)}, the agents $k_1$ and $k_2$ can calculate \eqref{eq:Delta_m (LCSS_EL)} based on $\Delta^s_{k_1,k_2}(x_{s_{k_1}},x_{s_{k_2}})$ and  $\Delta^s_{k_2,k_1}(x_{s_{k_2}},x_{s_{k_1}})$, respectively, $\forall s\in\{xy,xz,yz\}, k\in\bar{\mathcal{K}}$.
	
	We still need to incorporate the fact the that agents have a limited sensing radius, and that agent $i$ does not have access to the functions $\Delta^s_{i,j}(x_{s_i},x_{s_j})$, when $j\notin\mathcal{N}_i(p)$. To that end, we define first the greatest lower bound of $\Delta_k$ when both agents $k_1,k_2$ are in each other's sensing radius, i.e.,  
	\begin{equation} \label{eq:tilde delta (LCSS_EL)}
	\widetilde{{\Delta}}_k \coloneqq \inf_{\substack{(x_{s_{k_1}},x_{s_{k_2}})\in\mathbb{M}^2\\\lVert p_{k_1} - p_{k_2} \rVert \leq \min\{\varsigma_{k_1}, \varsigma_{k_2} \}}}\{\Delta_k(x_{s_{k_1}},x_{s_{k_2}}) \}, \forall k\in\bar{\mathcal{K}}.
	\end{equation}
	Since $\varsigma_i > \max\{l_{x,i},l_{y,i},l_{z,i}
	\} + \max_{j\in\mathcal{N}}\big\{\max\{l_{x,j},l_{y,j},l_{z,j}\}\big\}+ \varepsilon$, $\forall i\in\mathcal{N}$, it follows that there exists a positive constant $\varepsilon_\Delta$ such that $\widetilde{\Delta}_{k} \geq \varepsilon_\Delta > 0, \forall k\in\bar{\mathcal{K}}$. Next, we define the smooth switching functions $\beta_k:\mathbb{R}_{\geq 0} \to [0,\bar{\beta}_{m}]$, with \cite{loizou2017navigation}
	\begin{align} \label{eq:beta_m (LCSS_EL)}
	{\beta_k(z)} = \bar{\beta}_k\frac{\sigma(z)}{\sigma(z) + \sigma\left(\bar{\Delta}_k-z\right)},
	\end{align}
	where $\bar{\Delta}_k$ is a positive constant satisfying $\bar{\Delta}_k < \widetilde{\Delta}_k$, $\forall k\in\bar{\mathcal{K}}$. 
	Then, by choosing $\beta_{k} \coloneqq \beta_{k}(\gamma_\sigma \Delta_k(x_{s_{k_1}},x_{s_{k_2}}))$, where $\gamma_\sigma$ is a positive scaling constant, we incorporate the limited sensing radius of the agents in the collision avoidance scheme, since $\frac{\partial \beta_k(z)}{\partial z}$ vanishes when $k_1\notin \mathcal{N}_{k_2}(p)$ or $k_2\notin \mathcal{N}_{k_1}(p)$, i.e., when at least one of the agents that form edge $k$ lies outside  the sensing range of the other agent. Note that $\beta_k$ are similar to the switches defined in Section \ref{sec:main results (CDC_LF)}.  The terms $\bar{\beta}_k$ can be any positive constants, $\forall k\in\bar{\mathcal{K}}$. 
	All the  necessary information for the construction of the functions $\beta_{k}$, $\Delta_{k}$, i.e., the constants $\bar{\Delta}_k$, $\bar{\beta}_k$ and the lengths $l_{x,i}$, $l_{y,i}$, $l_{z,i}$, $i\in\mathcal{N}$, can be transmitted off-line to the agents. 
	
	We can now define a suitable barrier function for each edge $k\in\bar{\mathcal{K}}$ as any continuously differentiable function $b_k : \mathbb{R}_{\geq 0} \to \mathbb{R}_{\geq 0}$ with the property $\lim_{z\to0} b_k(z) = \infty$, e.g., $b_k(z) = \frac{1}{z}$, $k\in\bar{\mathcal{K}}$. The barrier function for edge $k$ is then $b_k \coloneqq b_k(\beta_k)$, $\forall k\in\bar{\mathcal{K}}$. 
	
	We propose now a decentralized feedback control law for the solution of Problem \ref{prob: 1 (LCSS_EL)}. Firstly, as in Section \ref{sec:main results (CDC_LF)}, we define the estimations of the unknown terms
	$\theta_{g_i}\in\mathbb{R}^{\ell}$ and $d_{b_i}\in\mathbb{R}$  as $\hat{\theta}_{g_i}\in\mathbb{R}^{\ell}$ and $\hat{d}_{b_i}\in\mathbb{R}$, with the respective errors $\widetilde{\theta}_{g_i} \coloneqq \hat{\theta}_{g_i} - \theta_{g_i}$ and $\widetilde{d}_{b_i} \coloneqq \hat{d}_{b_i} - d_{b_i}$, $\forall i\in\mathcal{N}$. 
	By using adaptive and discontinuous control techniques, we prove in the following that these estimations compensate appropriately for the unknown terms, without necessarily converging  to them. In particular, we design the feedback control laws for $u_{s,i}: \mathcal{X}_{s_i} \to \mathbb{R}^6$ as 
	\begin{align} \label{eq:control law (LCSS_EL)}
	u_{s,i} \coloneqq u_{s,i}(\chi_{s_i}) = & \sum_{k\in\bar{\mathcal{K}}} \alpha_{i,k} \kappa_k \bar{E}_\zeta(\zeta_i)^\top \frac{\partial \Delta_k}{\partial x_{s_i}} + Y_i(x_{s_i})\hat{\theta}_{g_i} - k_{v_i}v_i - \hat{d}_{b_i}\text{sgn}(v_i),
	\end{align}
	where $\chi_{s_i} \coloneqq [x_s^\top,v_i^\top,\hat{\theta}_{g_i}^\top,\hat{d}_{b_i}]^\top$, $\mathcal{X}_{s_i} \coloneqq \mathcal{X}_s \times \mathbb{R}^{7+\ell}$, with $\mathcal{X}_s$ as defined in Section \ref{sec:problem form (LCSS_EL)}. Moreover, $\alpha_{i,k} = -1$ if agent $i$ is part of edge $k$, and $\alpha_{i,k} = 0$ otherwise, $\forall i\in\mathcal{N}$, $k\in\bar{\mathcal{K}}$, $\kappa_k \coloneqq  \frac{\partial b_k(\beta_k)}{\partial \beta_k} \frac{\partial \beta_k(\Delta_k)}{\partial \Delta_k}$, $\forall k\in\bar{\mathcal{K}}$, and $k_{v_i}$ are positive constant gains. Finally, we design the associated adaptation laws
	\begin{align} \label{eq:adaptation laws (LCSS_EL)}
	\left.\begin{matrix}
	\dot{\hat{\theta}}_{g_i} &\coloneqq& -\gamma_{i,\theta} Y_i(x_{s_i})^\top v_i \\
	\dot{\hat{d}}_{b_i} &\coloneqq& \gamma_{i,d} \|v_i\|_1
	\end{matrix} \ \ \ \right\} \forall i\in\mathcal{N},
	\end{align}
	{with arbitrary bounded initial conditions}, where $\gamma_{\theta,i}$ and $\gamma_{d,i}$ are positive gains, $\forall i\in\mathcal{N}$. The correctness of \eqref{eq:control law (LCSS_EL)}-\eqref{eq:adaptation laws (LCSS_EL)} is shown in the following theorem:
	\begin{theorem}	
		Consider a multi-agent system comprised of $3$D ellipsoidal agents and subject to the dynamics \eqref{eq:dynamics (LCSS_EL)} at a collision-free initial configuration, i.e., $\mathcal{A}_i(x_{s_i}(0)) \cap \mathcal{A}_j(x_{s_j}(0)) = \emptyset$, $\forall i,j\in\mathcal{N}$ with $i\neq j$. Then, application of the control and adaptation laws \eqref{eq:control law (LCSS_EL)}, \eqref{eq:adaptation laws (LCSS_EL)} guarantees that the agents avoid collisions for all times, i.e., $\mathcal{A}_i(x_{s_i}(t)) \cap \mathcal{A}_j(x_{s_j}(t)) = \emptyset$, $\forall i,j\in\mathcal{N}$ with $i\neq j$, $t\in \mathbb{R}_{\geq 0}$, with all closed loop signals being bounded. Moreover, 
		$\lim_{t\to\infty}v_i(t) = 0, \forall i\in\mathcal{N}$.
	\end{theorem}
	\begin{proof}
		Consider the vector $\chi_s$ $\coloneqq$ $\big[x_s^\top,  v^\top, \widetilde{\theta}_g^\top, \widetilde{d}_b^\top \big]^\top$ $\in \widetilde{\mathcal{X}}_s \coloneqq \mathcal{X}_s\times\mathbb{R}^{7N+\ell N}$, 
		$v \coloneqq [v_1^\top,\dots,v_N^\top]^\top\in\mathbb{R}^{6N}$, $\widetilde{d}_b \coloneqq [\widetilde{d}_{b_1},\dots, \widetilde{d}_{b_N}]^\top\in\mathbb{R}^N$,  $\widetilde{\theta}_g \coloneqq [\widetilde{\theta}_{g_1}^\top,\dots,\widetilde{\theta}_{g_N}^\top]\in\mathbb{R}^{\ell N}$. Since the initial configuration is collision-free, it holds that $\chi_s(0)\in \widetilde{\mathcal{X}}_s$. 
		By combining \eqref{eq:dynamics (LCSS_EL)}, \eqref{eq:control law (LCSS_EL)}, and \eqref{eq:adaptation laws (LCSS_EL)}, it can be verified that the conditions of Prop. 3 of Prop. \ref{prop:Filippov exist (app_dynamical_systems)} in Appendix \ref{app:dynamical systems} are satisfied and hence we conclude that {at least} one Filippov solution exists and any {such solution} satisfies $\chi_s: [0,t_{\max}) \to \widetilde{\mathcal{X}}_s$ for a positive $t_{\max}$. 
		Define  $$\mathsf{z}_s \coloneqq \big[\phi_s, b_1, \dots, b_{\bar{M}},  v^\top, \widetilde{\theta}_g^\top, \widetilde{d}_b^\top \big]^\top \in \mathcal{Z}_s\coloneqq \mathbb{R}^{\bar{K}+7N+\ell N+1},$$
		{where $\phi_s$ is the cooperative term defined in Section \ref{sec:problem form (LCSS_EL)}. Note that $\mathsf{z}_s(0) \in \mathcal{Z}_s$ and, for any finite $r_s$, $\mathsf{z}_s \in \bar{\mathcal{B}}(0,r_s) \subset \mathcal{Z}_s \Leftrightarrow \chi_s \in \widetilde{\mathcal{X}}_s$, which we prove in the following.} 
		Define the function \small $$V_s\coloneqq V_s(\mathsf{z}_s) \coloneqq \phi_s + \sum_{k\in\bar{\mathcal{K}}}b_k +  \sum_{i\in\mathcal{N}}
		\left\{ \frac{1}{2}v_i^\top M_i(x_{s_i})v_i + \frac{1}{2\gamma_{i,d}}\widetilde{d}_{b_i}^2 + \frac{1}{2\gamma_{i,\theta}}\|\widetilde{\theta}_{g_i}\|^2 \right\},$$	
		\normalsize
		for which it holds that $W_{s_1}(\mathsf{z}_s) \leq V_s(\mathsf{z}_s) \leq W_{s_2}(\mathsf{z}_s)$ for positive definite functions $W_{s_1}, W_{s_2}$ on $\mathcal{Z}_s$. Since $\mathsf{z}_s(0)\in\mathcal{Z}_s$, we conclude that $V_s(\mathsf{z}_s(0))$ is well defined, and hence there exists a finite constant $\bar{V}_s$ such $V_s(\mathsf{z}_s(0)) \leq \bar{V}_s$ and $b_k(0) \leq \bar{V}_s$, $\forall k\in\bar{\mathcal{K}}$.
		By differentiating $V_s$ along the solutions of the closed loop system and in view of Lemma \ref{lem:Chain rule (App_dynamical_systems)} we obtain $\dot{V}_s \in \dot{\widetilde{V}}_s \coloneqq \cap_{\xi\in\partial V_s(\mathsf{z}_s)}\xi^\top \mathsf{K}[\dot{\mathsf{z}}_s]$. Since $V$ is continuously differentiable, the generalized gradient reduces to the standard gradient and therefore, after using Properties \ref{ass:skew-symm (CDC_LF)}, \ref{ass:dynamics factorization (LCSS_EL)}, and grouping terms, we obtain	
		\begin{align*}
		\max_{z\in\dot{\widetilde{V}}_s} \{z\} \leq&  \sum_{i\in\mathcal{N}} \Bigg\{  \sum_{k\in\bar{\mathcal{K}}}\left[\alpha_{i,k} \kappa_k \frac{\partial \Delta_k}{\partial x_{s_i}}^\top \bar{E}_\zeta(\zeta_i) \right]v_i +  \|v_i\|_1 \|d_i(t)\|_1  +  v_i^\top \bigg( u_i \\
		&\hspace{-10mm} - Y_i(x_{s_i})\theta_{g_i} + \bar{E}_\zeta(\zeta_i)^\top \frac{\partial \phi(x_s)}{\partial x_{s_i}}  \bigg) 
		 {-v_i^\top f_i(v_i)} +  \frac{1}{\gamma_{i,d}}\widetilde{d}_{b_i}\dot{\hat{d}}_{b_i} +
		\frac{1}{\gamma_{i,\theta}}\widetilde{\theta}_{g_i}^\top\dot{\hat{\theta}}_{g_i} \Bigg\},
		\end{align*}	
		By also using Property \ref{ass:friction (LCSS_EL)} and Assumption 
		\ref{ass:f_i+d_i (CDC_LF)},
		substituting $u_i = u_{f,i} + u_{s,i}$ with $u_{f,i} = \bar{E}_\zeta(\zeta_i)^\top \frac{\partial \phi(x_s)}{\partial x_{s_i}}$ and \eqref{eq:control law (LCSS_EL)}, the adaptation laws \eqref{eq:adaptation laws (LCSS_EL)}, and using $\widetilde{d}_{b_i} = \hat{d}_{b_i} - d_{b_i}$, $\widetilde{\theta}_{g_i} = \hat{\theta}_{g_i} - \theta_{g_i}$ and the property $z^\top\text{sign}(z) = \|z\|_1$, $\forall z\in\mathbb{R}^n$, we obtain $$\max_{z\in \dot{\widetilde{V}}_s} \{z\} \leq -\sum_{i\in\mathcal{N}} k_{v_i}\|v_i\|^2 =: W_s(\mathsf{z}_s).$$
		Therefore, $z \leq -W_s(\mathsf{z}_s(t))$, $\forall z\in\dot{\widetilde{V}}_s(\mathsf{z}_s(t))$, $t\in[0,t_{\max})$, where $W_s:\mathcal{Z}_s \to \mathbb{R}_{\geq 0}$ is a positive semi-definite function defined on $\mathcal{Z}_s$. Hence, by applying Theorem \ref{th:nonsmooth LaSalle (App_dynamical_systems)} of Appendix \ref{app:dynamical systems}, we conclude that $t_{\max} = \infty$, $\mathsf{z}_s(t)$ is bounded in the compact set $\{\mathsf{z}_s \in \bar{\mathcal{B}}(0,r_s):W_{s_2}(\mathsf{z}_s)\leq c\}$, $\forall t\in\mathbb{R}_{\geq 0}$ for any $r_s$ and $c$ satisfying $\bar{\mathcal{B}}(0,r_s)\subset \mathcal{Z}_s$, $c < \min_{\|x\| = r_s} W_{s_1}(\mathsf{z}_s)$, and $\lim_{t\to\infty} W_s(\mathsf{z}_s(t)) = 0 \Rightarrow \lim_{t\to\infty}v(t) = 0$. {Note that, since the sets $\{x_s\in\mathcal{X}_s: x_s = \phi_s^{-1}(y)\}$ are nonempty, $r_s$ can be chosen arbitrarily large, corresponding to all collision-free initial configurations}. Therefore, 
		inter-agent collisions are avoided, and the adaptation signals $\hat{\theta}_{g_i}$, $\hat{d}_{b_i}$, remain bounded, $\forall i\in\mathcal{N}$, $t\in\mathbb{R}_{\geq 0}$. The continuity of the terms $Y_i(\cdot)$ implies also their boundedness and hence  the boundedness of the control  signals \eqref{eq:control law (LCSS_EL)}, \eqref{eq:adaptation laws (LCSS_EL)}, $t\in\mathbb{R}_{\geq 0}$. 
	\end{proof}
	\begin{remark}
		It can be verified that $\det(\lambda A^s_{k_1}(x_{s_{k_1}}) - A^s_{k_2}(x_{s_{k_2}})$, and hence $b_k$, are functions of $p_{k_1}-p_{k_2}$, $\zeta_{k_1}$, $\zeta_{k_2}$. Therefore, if $\phi_s$ is a function of $\widetilde{x}$, the aforementioned analysis still holds by setting $\mathcal{X}_s = \{ \widetilde{x} \in\mathbb{R}^{\frac{3N(N-1)}{2}}\times\mathbb{S}^3 : \mathcal{A}_i(x_{s_i})\cap\mathcal{A}_j(x_{s_j}) = \emptyset, \forall i,j\in\mathcal{N}, i\neq j\}$. 
		Moreover, note that achievement of the objectives expressed by $\phi_s$ is not pursued here and may not be necessarily guaranteed due 	to the 	potentially counteracting terms of $u_i$. 
	\end{remark}
	
	\begin{remark}
		Since $\Delta^s_{i,j} = \Delta^s_{j,i}$ (due to Proposition \ref{prop:eq discr equal (LCSS_EL)}), $\forall i,j\in\mathcal{N}$, $i\neq j$, the control scheme can be extended to directed communication graphs, by setting for the $i$th agent $b_{i,j} = b_{i,j}(\beta_{i_j}(\Delta_{i,j}(x_{s_i},x_{s_j})))$, $\forall j\in\mathcal{N}\backslash\{i\}$, with $\Delta_{i,j}(x_{s_i},x_{s_j})$ as in \eqref{eq:Delta_m (LCSS_EL)} and $\beta_{i_j}$ as in \eqref{eq:beta_m (LCSS_EL)}, $\widetilde{{\Delta}}_{i,j}$ as in \eqref{eq:tilde delta (LCSS_EL)},
		and appropriately modifying the control law. Similarly, collision avoidance with static environment obstacles could be incorporated in the overall scheme.
	\end{remark}

	\subsection{Simulation Results} \label{sec:sim (LCSS_EL)}
	We consider a simulation example with $N = 8$ rigid bodies in $\mathbb{SE}(3)$, described by ellipsoids with axes lengths $l_{x,i} = 0.5$m, $l_{y,i} = 0.3$m, $l_{z,i} = 0.2$m, $\forall i\in\mathcal{N}$. {The initial poses are (in m)
		
	\begin{center}
		\begin{tabular}{ l l}
			 $p_1=[3,3,0]^\top$, & $p_2 = -[3,3,0]^\top$ \\
			 $p_3=[3,-3,0]^\top$, & $p_4 = [-3,3,0]^\top$ \\
			 $p_5 = [3,3,3]^\top$, & $p_6 = -[3,3,3]^\top$ \\
			 $p_7= [3,-3,3]^\top$, & $p_8 = [-3,3,-3]^\top$ 
		\end{tabular}
	\end{center}
	\begin{center}
		\begin{tabular}{ l}
			 $\zeta_1= \zeta_8= [0.769,0.1696,0.6153,0.0358]^\top$ \\
			 $\zeta_2= \zeta_6= [0.8488,-0.3913,-0.0598,-0.3505]^\top$ \\
			 $\zeta_3= \zeta_5= [0.7638,-0.5283,-0.3275,-0.1738]^\top$\\
			 $\zeta_4= \zeta_7= [0.7257,0.3081,0.3714,0.4904]^\top$ 
		\end{tabular}
	\end{center}
		We consider that $\phi_s(x_s)$ describes an independent multi-agent navigation objective, with desired configurations as 
		\begin{align*}
			&p_{1_\text{d}} = p_2, p_{2_\text{d}}=p_1,p_{3_\text{d}}=p_4,p_{4_\text{d}}=p_3, \\ 
			&p_{5_\text{d}}= p_6, p_{6_\text{d}}= p_5, p_{7_\text{d}}=p_8,p_{8_\text{d}}=p_7,
		\end{align*}
		and $\zeta_{i_\text{d}}$ $=$ $[1$,$0$,$0$,$0]^\top$, $\forall i\in\mathcal{N}$. We set the errors $e_{p_i}\coloneqq p_i - p_{i_\text{d}}$ and $e_{\zeta_i} \coloneqq [e_{\varphi_i}, e_{\epsilon_i}^\top]^\top \coloneqq \zeta_{i_\text{d}} \cdot {\zeta}^+_i$, and $e_{\varphi_i}, e_{\epsilon_i}$ are the scalar and vector parts, respectively, of the quaternion error (see Section \ref{sec: Problem Formulation (TCST_coop_manip)} ). The desired quaternion configuration is achieved when $e_{\zeta_i} = [\pm 1,0,0,0]^\top$ and hence the function  $\phi_s(x_s)$ is chosen as 
		$$\phi_s  = \sum_{i\in\mathcal{N}}\left(\frac{1}{2}\|p_i - p_{i_\text{d}} \|^2  + 1 - e_{\varphi_i}^2  \right),$$ with $$\dot{\phi}_s = \sum_{i\in\mathcal{N}}\left( (p_i - p_{i_\text{d}})^\top\dot{p}_i - e_{\varphi_i}e_{\epsilon_i}^\top\omega_i  \right).$$ The control inputs $u_{f,i}$ are therefore chosen as $$u_{f,i} = [p_{i_\text{d}}^\top-p_i^\top, e_{\varphi_i}e_{\epsilon_i}^\top]^\top,$$ $\forall i\in\mathcal{N}$. The agent masses and moments of inertia are chosen randomly in the interval $(0,0.2]$. We also set $f_i(x_i,v_i)=m_{f_i}\sin(w_{f_i}t + \phi_{f_i})v_i$, $d_i(t) = (1/m_{f_i})\sin(w_{f_i}t + \phi_{f_i})$,  $\forall i\in\mathcal{N}$, with
		the terms $m_{f_i}$, $\omega_{f_i}$, and $\phi_{f_i}$ chosen randomly in the interval $(0,5]$, $\forall i\in\mathcal{N}$.
		We choose $b_k=\frac{1}{\beta_k}$, with $\bar{\beta}_k=1$, $\bar{\Delta}_k = 10^4$,  $\gamma_\sigma = 10^{-40}$, $\forall k\in\bar{\mathcal{K}}$, and $\hat{\theta}_{g_i}(0) = 0.1$, $\hat{d}_{b_i}(0) = 0.2$, $k_{v_i} = 1$, $\forall i\in\mathcal{N}$. The expressions for $\Delta_k(x_{s_{k_1}},x_{s_{k_2}})$ were derived by using the symbolic toolbox of MATLAB.} Fig. \ref{fig:3dplot (LCSS_EL)} shows a $3$D plot of the agent trajectories, and Fig. \ref{fig:b_m+gamma_i (LCSS_EL)} shows the minimum of the barrier functions $\min_{k\in\bar{\mathcal{K}}}\{\beta_k(t)\}$ (left), which is always positive, and the signals $\gamma_i(t) \coloneqq \|p_i-p_{i_\text{d}}\|^2 + 1 - e_{\varphi_i}^2$ and $v_i(t)$ (right), $\forall i\in\mathcal{N}$, $t\in[0,15]$. Finally, Fig. \ref{fig:u_i (LCSS_EL)} depicts the control inputs of the agents. 
	A short video that demonstrates the aforementioned simulation example can be found in {\url{https://youtu.be/IAni7zIMM7k}}.
	
	\begin{figure}[!ht]
		\centering
		\includegraphics[width = 0.75\textwidth]{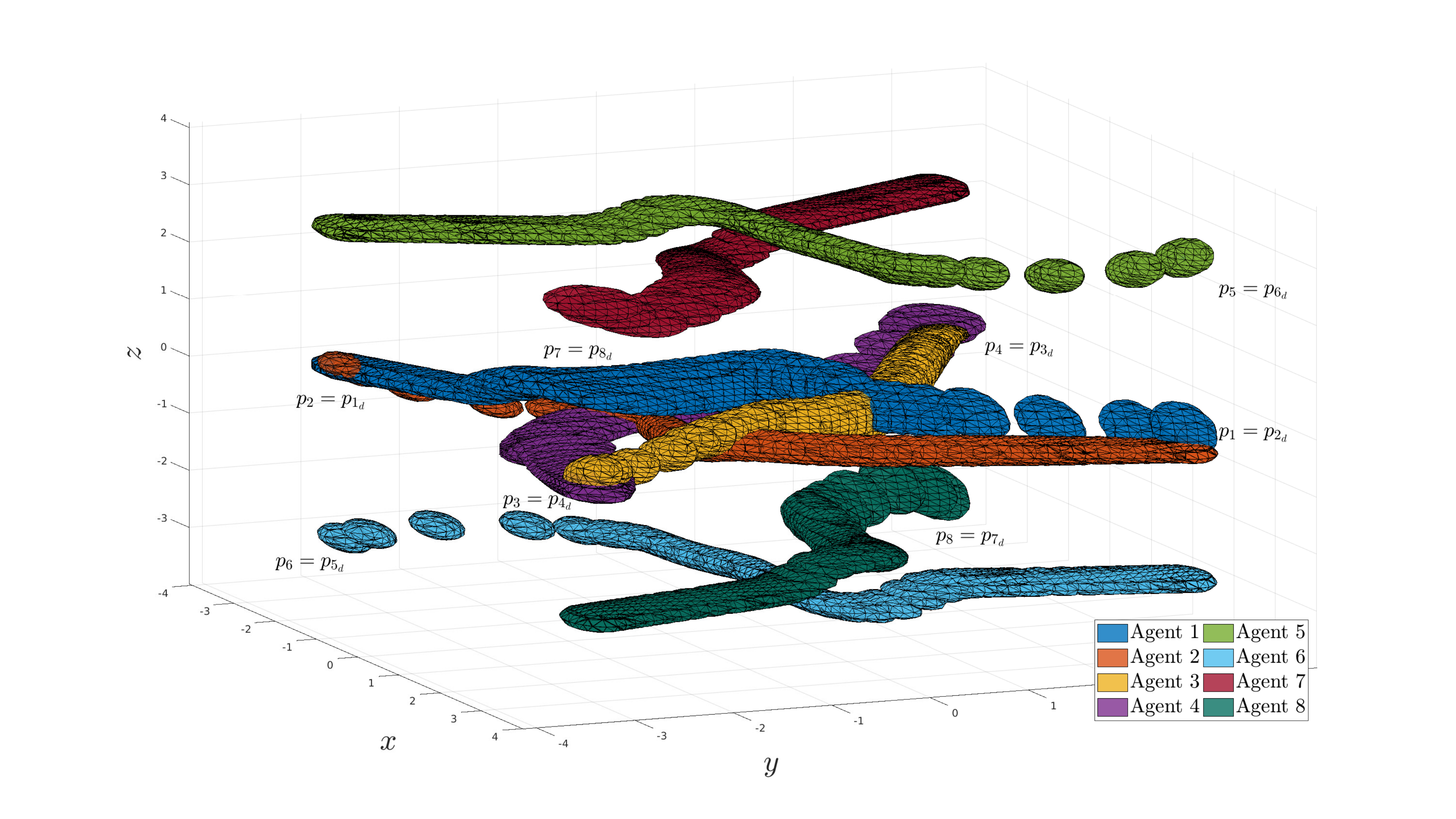}	
		\caption{The evolution of agent trajectories $\forall t\in[0,20]$ sec.}\label{fig:3dplot (LCSS_EL)}
	\end{figure}
	
	\begin{figure}[!ht]
		\centering		
		\begin{subfigure}{0.75\textwidth} 
		\includegraphics[trim =0cm 0cm 0cm 0cm, width= 0.8\textwidth]{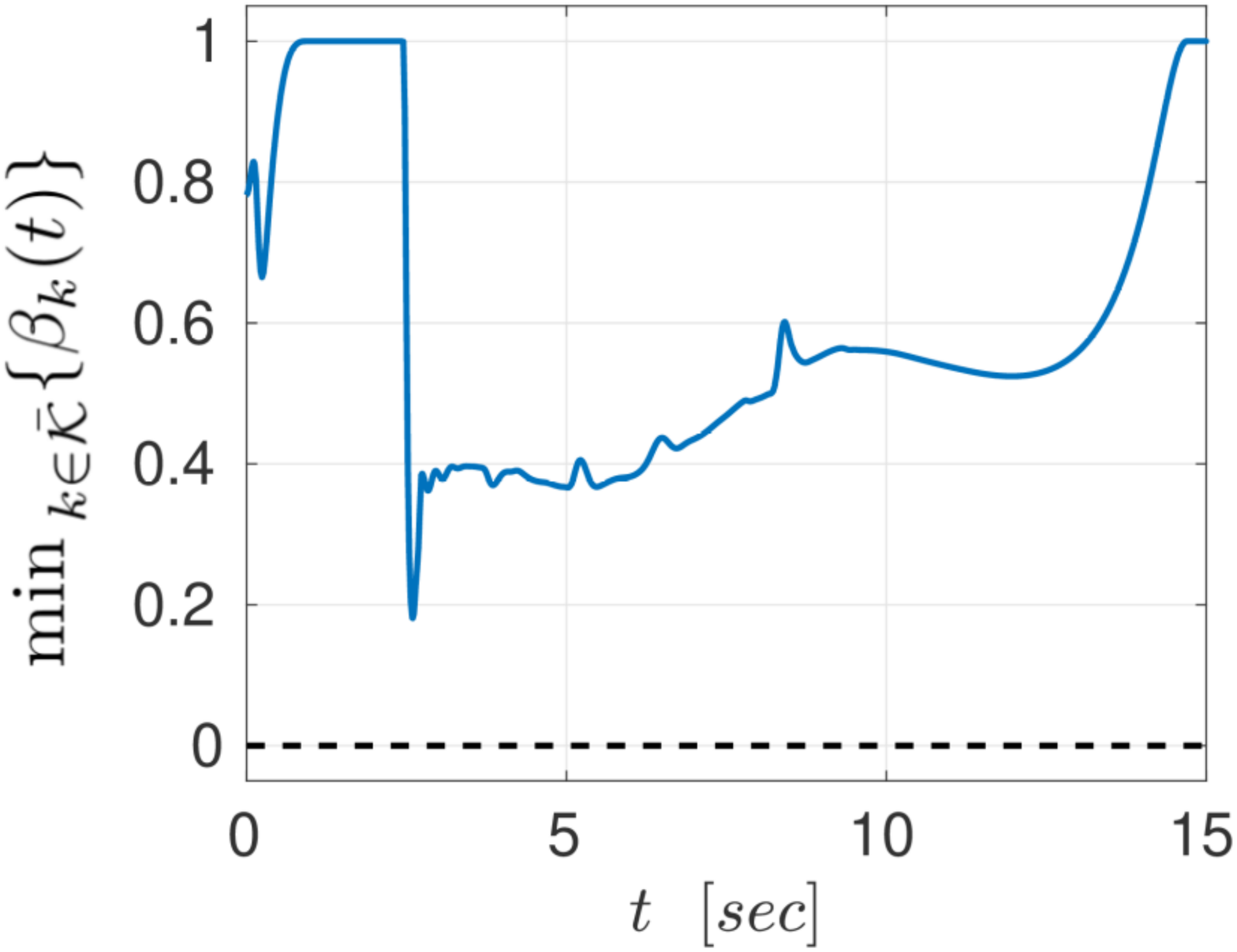}
		\end{subfigure}	
		\hfill
		\begin{subfigure}{0.75\textwidth} 	
		\includegraphics[trim =0cm 0cm 0cm 0cm, width = 0.8\textwidth]{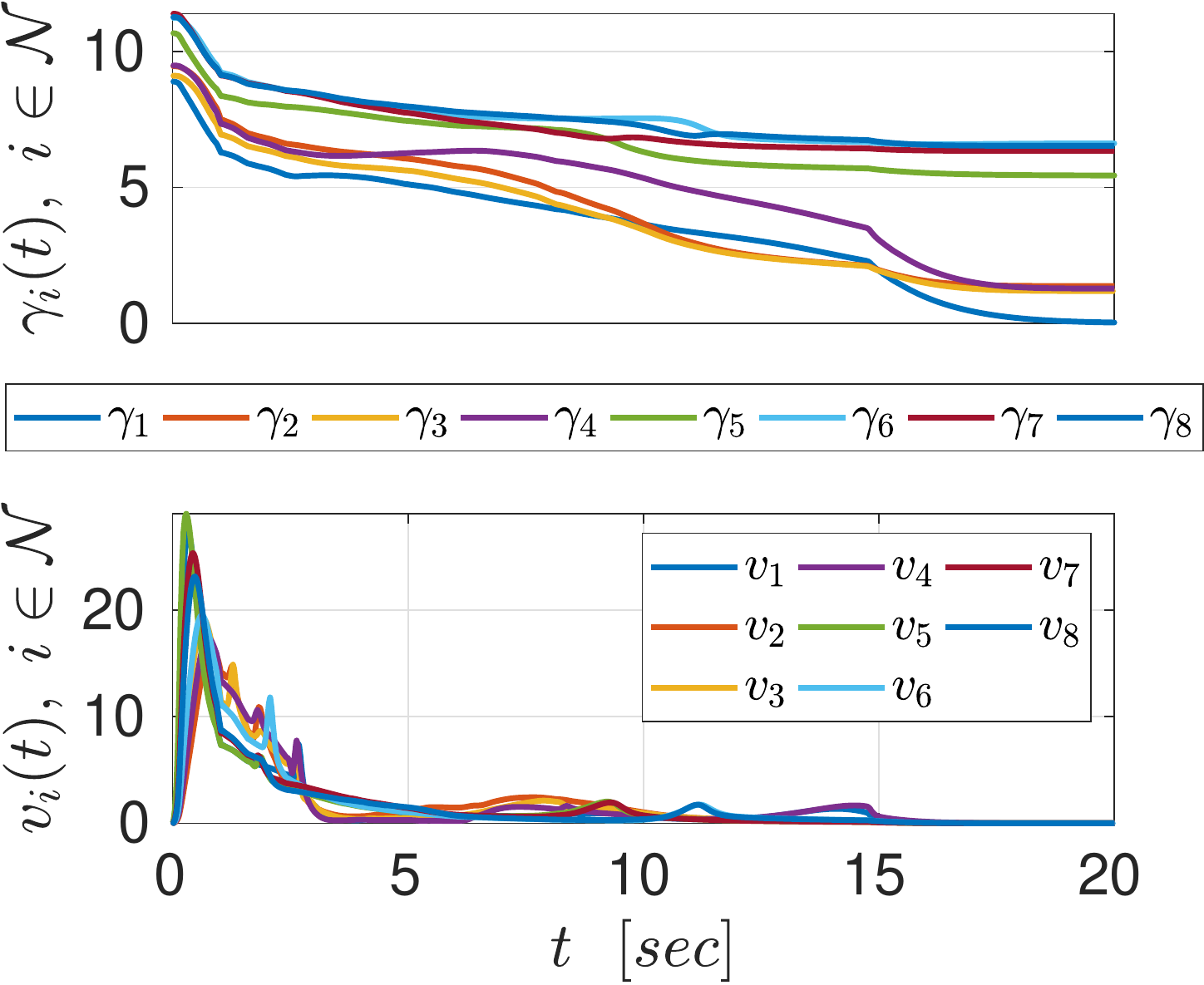}
		\end{subfigure}	
		\caption{Top: The evolution of the minimum of the functions $\min_{k\in\bar{\mathcal{K}}}\{\beta_k(t)\}$. Bottom: The evolution of the signals $\gamma_i(t)$ and $v_i(t)$, $\forall i\in \mathcal{N}$, $\forall t\in[0,20]$ sec.}\label{fig:b_m+gamma_i (LCSS_EL)}
	\end{figure}

	\begin{figure}[!ht]
		\centering
		\includegraphics[width = 0.9\textwidth]{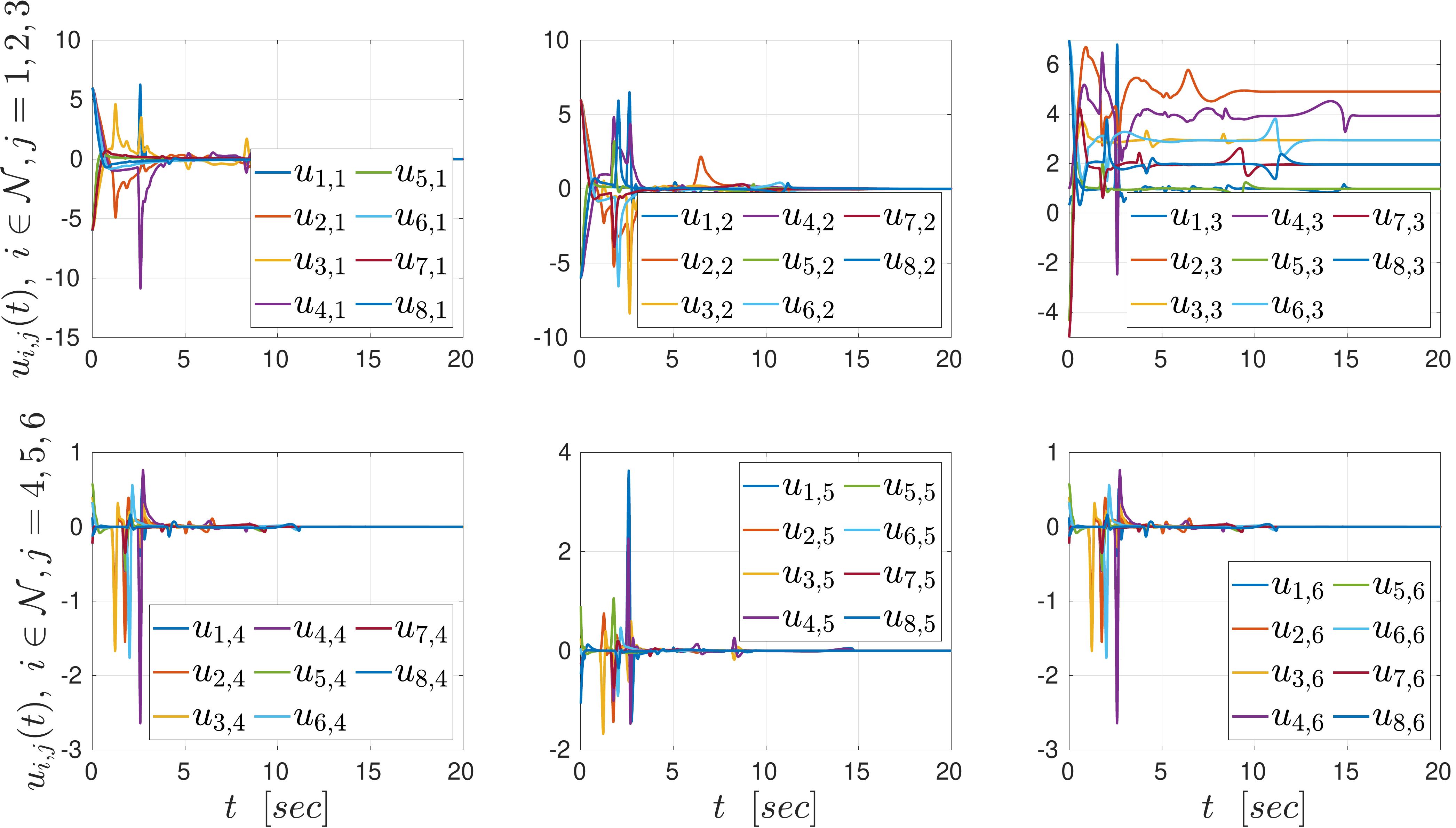}	
		\caption{The control inputs of the agents $u_i(t)$, $\forall t\in[0,20]$ sec, $i\in\mathcal{N}$.}\label{fig:u_i (LCSS_EL)}
	\end{figure} 
	
\section{Conclusion} 

This chapter presented several continuous control algorithms for multi-agent coordination of systems with uncertain dynamics. Firstly, we develop an adaptive control scheme for the almost global navigation of a single robotic agent in a workspace with obstacles, which is then extended to a decentralized multi-robot scheme. Secondly, we develop a decentralized adaptive multi-agent algorithm for the leader-follower coordination: A leader agent converges to a predefined goal point while the entire team avoids collision with each other and maintains connectivity. Finally, we develop a closed-form barrier function that encodes the distance between $3$D ellipsoids, and design a decentralized collision avoidance control scheme for a team of ellipsoidal robotic agents, while compensating for the dynamic uncertainties.

\chapter{Abstractions of Multi-Agent and Multi-Agent-Object Systems} \label{chapter:abstraction}
After designing continuous solutions to multi-agent problems, such as cooperative manipulation, formation, and navigation, we are ready to transit to the problem of multi-agent planning under temporal logic tasks. The content of the previous chapters can act as a means to obtain well-defined discrete representations (abstractions) of the continuous systems in hand. Therefore, this chapter addresses the motion and task planning of multi-agent \textit{and} multi-agent-object systems (systems comprised of multiple robotic agents and objects) subject to temporal logic constraints, focusing both on the abstraction technique as well as the control synthesis for the accomplishment of the tasks. 

More specifically, this chapter can be divided in two main parts. The first part tackles the motion planning of multi-robot teams under \textit{local} linear temporal tasks, i.e., when each robotic agent has its own task. The second part addresses the case where \textit{unactuated} objects of the workspace must satisfy a certain temporal logic task, with the robotic agents being responsible for their accomplishment.

\section{Introduction}\label{sec:intro (abstactions)}

Temporal-logic-based motion planning has gained significant attention in recent years, as it provides a fully automated correct-by-design controller synthesis approach for autonomous robots. Temporal logics such as linear temporal logic (LTL) and metric interval temporal logic (MITL) provide formal high-level languages that can describe complex planning objectives. 
As already discussed in the previous chapters, standard control problems are restricted to point-to-point navigation, multi-agent formation control, or consensus. 
Ultimately, however, we would like the robotic agents to execute more complex high-level tasks, involving combinations of safety ("never enter a dangerous regions"), surveillance ("keep visiting regions $A$ and $B$ infinitely often") or sequencing ("collect data in region $C$ and upload it in region $D$") properties. 
Temporal logic languages offer a means to express the aforementioned specifications, since they can describe complex planning objectives in a more efficient way than the well-studied navigation algorithms. 
The task specification is given as a temporal logic formula with respect to the discretized abstraction of the robot motion modeled as a finite transition system \cite{fainekos_planning,Belta2007,Bhatia2010,belta_2010_product_system}. Then a high-level discrete plan is found by off-the-shelf model-checking algorithms, given the finite transition system and the task specification \cite{baier2008principles}.
Temporal logics have been extensively used in the related literature for both single-  and multi-agent systems, e.g., \cite{Quottrup_ICRA2004,Loizou_CDC2005, Filippidis_CDC2012, guo_2015_reconfiguration,Nenchev_ECC2016,Feyzabadi_ICRA2016,Guo_CDC2015,Fainekos_Automatica2009,Kloetzer_ICNSC2006,ulusoy2013optimality,zhang2016motion, aksaray2016dynamic, Belta2007,Bhatia2010,Bhatia2011,Cowlagi2016,diaz2015correct}. 

A special and important class of autonomous robotic systems is the class of unmanned aerial vehicles (UAV), which can provide efficient multi-agent solutions in several problems, e.g., coverage or inspection. Control of aerial vehicles in a multi-agent setting has been well studied in the related literature. 
The standard problem of formation control for a team of aerial vehicles is addressed in \cite{Liu_ICIA2015,Yu_CCC2013, Koksal_ECC2015, Hou_ECC2016, Pereira_MED2016, Dong_ICST2015}, whereas \cite{Mercado_ECC2013,Roldao_ECC2013,Ulrigh_ACC2016,Hao_ACC2016,Ghamry_ICUAS2015} consider leader-follower formation approaches, where the latter also treats the problem of collision avoidance with static obstacles in the environment; \cite{Sunberg_ICRA2016}, \cite{Eskandarpour_ICROM2014,mansouri2015distributed} and \cite{Zhou_CDC2015} employ dynamic programming, Model Predictive Control and reachable set algorithms, respectively, for inter-agent collision avoidance, which is tackled also in \cite{alonso2015collision}. In \cite{Pierson_ICRA2016} the cooperative evader pursuit problem is treated. Aerial vehicles and temporal logic-based planning is considered in \cite{Karaman_CDC2008}, which addresses the vehicle routing problem using MTL specifications and in \cite{Karaman_ACC2008}, which approaches the LTL motion planning using MILP optimization techniques, both in a centralized manner. Markov Decision Processes are used for the LTL planning in \cite{Xiaoting_CCC2016}.  The aforementioned works, however, consider discrete agent models and do not take into account their continuous dynamics. 

The discretization of a multi-agent system to an abstracted finite transition system necessitates the design of appropriate continuous-time controllers for the transition of the agents among the states of the transition system \cite{baier2008principles}. Most works in the related literature, however, including the aforementioned ones, either assume that there \textit{exist} such continuous controllers or adopt single- and double-integrator models, ignoring the actual dynamics of the agents. Discretized abstractions, including design of the discrete state space and/or continuous-time controllers, have been considered in \cite{Belta2005,belta2006controlling,reissig2011computing,tiwari2008abstractions,rungger2015state} for general systems and \cite{boskos2015decentralized, belta2004abstraction} for multi-agent systems.

Another drawback of the majority of works in the related literature of temporal logic-based motion planning is the  point-agent assumption (as, e.g. in \cite{guo_2015_reconfiguration,Kloetzer_ICNSC2006,Fainekos_Automatica2009}), which does not take into account potential collisions between the robotic agents. The latter is a crucial safety property in real-time scenarios, where actual vehicles are used in the motion planning framework.

Furthermore, most works in the related literature consider temporal logic-based motion planning for fully actuated, autonomous agents.
Consider, however, cases where some \textit{unactuated} objects must undergo a series of processes in a workspace with autonomous agents (e.g., car factories). In such cases, the agents, except for
satisfying their own motion specifications, are also responsible for coordinating with each other in order to transport the objects around the workspace. When the unactuated objects’
specifications are expressed using temporal logics, then the abstraction of the agents’ behavior becomes much more complex, since it has to take into account the objects’ goals.
More specifically, we are here interested in complex tasks, possibly including time, such as ``never take the object to dangerous regions" or ``keep moving the object from region A to B within a predefined time interval" which must be executed via the control actions of the robotic agents.  \textit{Time constraints} can be incorporated in the motion planning temporal logic-based problem via specific logics, such as Metric and Metric Interval Temporal Logic (MTL, MITL) \cite{alur1996benefits,d2007expressiveness,alur1994theory}, as well as Time Window Temporal Logic (TWTL), or Signal Temporal Logic (STL). Such languages have been for multi-agent motion planning in several works (e.g., \cite{Alex16,karaman2011linear,aksaray2016dynamic,lindemann2018decentralized}).

This chapter addresses the motion planning problem of multi-agent systems as well multi-agent-object systems subject to complex tasks, expressed as temporal logic specifications. 
Firstly, we develop decentralized control protocols for the navigation of a multi-robot team among predefined regions or interest in the workspace, while taking into collision and/or connectivity properties. We consider separately the cases of (i) aerial vehicles, and (ii) mobile robotic manipulators. This allows us to abstract the continuous multi-agent dynamics as discrete transition systems (abstractions), which then can be used to obtain a path that satisfies the given \textit{local} LTL specifications, by employing formal method-based methodologies.

Secondly, we provide, similar to the first case, appropriate discrete abstractions for multi-robot-object systems, encoding the behavior of the robots as well as the \textit{unactuated objects} in the workspace. The proposed abstraction design involves both multi-robot \textit{safe} navigation as well as cooperative object transportation. The abstracted systems are then used to derive paths that satisfy the robotic agents' and the objects' (possibly timed) temporal goals. 

Although the proposed control schemes from the previous chapters can be used, we provide new control ideas as alternatives for the derivation of the discrete abstractions.

\section{Decentralized Motion Planning with Collision Avoidance for a Team of UAVs under High Level Goals}\label{sec:icra}

We first describe a decentralized hybrid control algorithm for the motion planning of aerial vehicles subject to Linear Temporal Logic (LTL) specifications.

\subsection{Problem Formulation} \label{sec:System and PF (ICRA 17)}
Consider $N$ aerial agents operating in a static workspace that is bounded by a large sphere in $3$-D space $\mathcal{W}\coloneqq \mathcal{B}(p_0,r_0)$, where ${p_0}\in \mathbb{R}^3$ and $r_0 \in\mathbb{R}_{>0}$ are the center and radius of $\mathcal{W}$. Within $\mathcal{W}$ there exist $K$ smaller spheres around points of interest, which are described by $\mathcal{\pi}_k \coloneqq \bar{\mathcal{B}}({p_{\pi_k}},r_{\pi_k})\subset \mathcal{W}$, where ${p_{\pi_k}}\in \mathbb{R}^3, r_{\pi_k}\in\mathbb{R}_{>0}$ are the central point and radius, respectively, of $\pi_k$. We denote the set of all $\pi_k$ as $\Pi=\{\pi_1,\dots,\pi_K \}$.  
Moreover, we introduce a set of atomic propositions $\Psi_i$ for each agent $i\in\{1,\dots,N\}$ that indicates certain properties of interest of agent $i$  in $\Pi$ and are expressed as boolean variables. The properties satisfied at each region $\pi_k$ are provided by the labeling function $\mathcal{L}_i:\Pi\rightarrow 2^{\Psi_i}$, which assigns to each region $\pi_k, k\in \mathcal{K}_\mathcal{R}\coloneqq\{1,\dots,K\}$ the subset of the atomic propositions $\Psi_i$ that are true in that region.  

\begin{figure}[!btp]
	\centering
	\includegraphics[scale = 0.5]{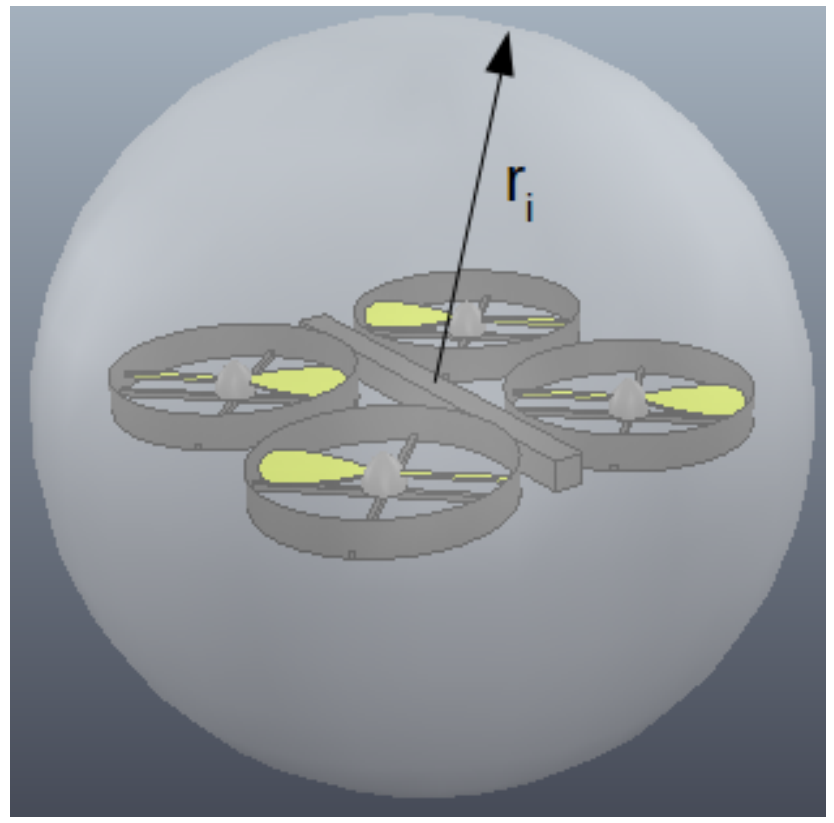}
	
	\caption{Bounding sphere of an aerial vehicle.\label{fig:quad (ICRA 17)}}
\end{figure} 

Each agent $i\in \mathcal{N} \coloneqq \{1,\dots,N\}$ occupies a bounding sphere $\bar{\mathcal{B}}({p_i},r_i)$, where ${p_i} \in \mathbb{R}^3$ is the center and $r_i \in\mathbb{R}_{>0}$ the radius of the sphere (Fig. \ref{fig:quad (ICRA 17)}). We also consider that $r_i < r_{\pi_k}, \forall i\in\mathcal{N},k\in\mathcal{K}_\mathcal{R}$, i.e., the regions of interest are larger than the aerial vehicles. The motion of each agent is controlled via its centroid ${p_i}$ through the single integrator dynamics: 
\begin{equation}
{\dot{p}_i} = {u_i}, i\in \mathcal{N}. \label{eq:dynamics (ICRA 17)}
\end{equation}
Moreover, similar to the previous chapter, we consider that agent $i$ has a limited sensing range of $\varsigma_i > \max_{i,j\in\mathcal{N}}$ $(r_i+r_j)$. Therefore, by defining the neighboring set $\mathcal{N}_i \coloneqq \{ j\in\mathcal{N}, \text{ s.t. } \Vert {p}_i - {p}_j \rVert \leq \varsigma_i \}$, agent $i$ knows at each configuration the position of all ${p_j}, \forall j\in\mathcal{N}_i$ as well as its own position ${p}_i$. The workspace is assumed to be perfectly known, i.e., ${p_{\pi_k}},r_{\pi_k}$ are known to all agents, for all $k\in\mathcal{K}_\mathcal{R}$. 

With the above ingredients, we provide the following definitions:
\begin{definition} \label{def:agent in region (ICRA 17)}
	An agent $i\in\mathcal{N}$ is in a region $\pi_k,k\in\mathcal{K}_\mathcal{R}$ at a configuration ${p_i}$, denoted as $\mathcal{A}_i({p_i})\in\pi_k$, if and only if $\bar{\mathcal{B}}({p_i},{r_i}) \subseteq \bar{\mathcal{B}}({p_{\pi_k}},r_{\pi_k})$.
\end{definition}
\begin{definition} \label{def:agent_transition (ICRA 17)}
	Assume that $\mathcal{A}_i({p_i}(t_0))\in\pi_k, i\in \mathcal{N},k\in\mathcal{K}_\mathcal{R}$ for some $t_0\geq 0$. Then there exists a transition for agent $i$ from region $\pi_k$ to region $\pi_{k'},k'\in\mathcal{K}_\mathcal{R}$, denoted as $\pi_k\rightarrow_i\pi_{k'}$, if and only if there exists a finite $t_f\geq 0$ such that 
	\begin{enumerate}
		\item $\mathcal{A}_i({p_i}(t_f))\in\pi_{k'}$, 
		\item $\bar{\mathcal{B}}(p_i(t),r_i) \subset \mathcal{W}$, 
		\item $\bar{\mathcal{B}}({p_i}(t),{r_i})\cap\bar{\mathcal{B}}({p_{\pi_m}},r_{\pi_m}) = \emptyset$, 
		\item $\bar{\mathcal{B}}({p_i}(t),{r_i})\cap\bar{\mathcal{B}}({p_{i'}}(t),{r_{i'}}) = \emptyset, \forall m\in\mathcal{K}_\mathcal{R}$ with $m\neq k,k', \forall i'\in \mathcal{N}$ with $i'\neq i$ and $t\in\left[0,t_f \right]$.
	\end{enumerate}
\end{definition}

Loosely speaking, an agent $i$ can transit between two regions of interest $\pi_k$ and $\pi_{k'}$, if there exists a bounded control trajectory ${u_i}$ in (\ref{eq:dynamics (ICRA 17)}) that takes agent $i$ from $\pi_k$ to $\pi_{k'}$ while avoiding entering all other regions, colliding with the other agents, or exiting the workspace boundary.

Our goal is to control the multi-agent system subject to (\ref{eq:dynamics (ICRA 17)}) so that each agent's behavior obeys a given specification over its atomic propositions $\Psi_i$. 

\begin{definition} 
Given a trajectory ${p_i}(t)$ of agent $i$, its corresponding \textit{behavior} is given by the infinite sequence $\mathfrak{b}_i(\mathsf{\breve{\psi}}_i) \coloneqq ({p_{i_1}},\mathsf{\breve{\psi}}_{i_1})({p_{i_2}},\mathsf{\breve{\psi}}_{i_2})\dots$, with $\mathbb{\breve{\psi}}_{i_m}\in2^{\Psi_i}$ and $\mathcal{A}_i({p_{i_m}})\in\pi_{k_m},  \mathsf{\breve{\psi}}_{i_m}\in\mathcal{L}_i(\pi_{k_m}),k_m\in\mathcal{K}_\mathcal{R},\forall m\in\mathbb{N}$. 
\end{definition}

The satisfaction of a LTL formula is provided by the following definition (see Appendix \ref{app:Logics} for more details on LTL formulas).

\begin{definition}
	The behavior $\mathfrak{b}_i(\breve{\psi}_i)$ satisfies an LTL formula $\mathsf{\Phi}$ if and only if $\breve{\psi}_i \models \mathsf{\Phi}$.
\end{definition}

The control objectives are given for each agent separately as LTL formulas $\mathsf{\Phi}_i$ over $\Psi_i, i\in\mathcal{N}$. An LTL formula is satisfied if there exists a behavior $\mathfrak{b}_i(\breve{\psi}_i)$ of agent $i$ that satisfies $\mathsf{\Phi}_i$. Formally, the problem treated in this section is the following:
\begin{problem} \label{problem (ICRA 17)}
	Given a set of aerial vehicles $N$ subject to the dynamics (\ref{eq:dynamics (ICRA 17)}) and $N$ LTL formulas $\mathsf{\Phi}_i,$ over the respective atomic propositions $\Psi_i, i\in\mathcal{N}$, achieve behaviors $\mathfrak{b}_i$ that (i) yield satisfaction of $\mathsf{\Phi}_i,\forall i\in\mathcal{N}$ and (ii) guarantee inter-agent collision avoidance.
\end{problem}

\subsection{Problem Solution } \label{sec:control strategy (ICRA 17)}

We provide here the proposed solution to Problem \ref{problem (ICRA 17)}, which consists of two main layers, that is, the design of a continuous control scheme, and the derivation of a high level path that satisfies $\mathsf{\Phi}_i$.

\subsubsection{Continuous Control Design} \label{subsec:Continuous Control (ICRA 17)}

The first ingredient of our solution is the development of a decentralized feedback control law that establishes a transition relation $\pi_k\rightarrow_i\pi_{k'}, \forall k,k'\in\mathcal{K}_\mathcal{R}$ according to Def. \ref{def:agent_transition (ICRA 17)}. The proposed approach is based on the concept of \textit{Decentralized Navigation Functions}, introduced in \cite{dimarogonas2007decentralized}, for which an overview can be found in Appendix \ref{app:NF}. 
More specifically, given that $\mathcal{A}_i({p_i}(t_0))$ for some $t_0 \geq 0$, we propose a decentralized control law ${u_i}$ for the transition $\pi_k\rightarrow_i\pi_{k'}$, as defined in Def. \ref{def:agent_transition (ICRA 17)}.

Initially, we define the set of ``undesired" regions as $\Pi_{k,k'} \coloneqq \{\pi_m\in\Pi, m\in\mathcal{K}_\mathcal{R} \backslash \{k ,k'\} \}$ and the corresponding free space $\mathcal{F}_{i_{k,k'}} \coloneqq \{p\in \mathcal{W}^N : \bar{\mathcal{B}}(p_i,r_i)\cap\bar{\mathcal{B}}(p_j,r_j) = \emptyset, \forall j \in \mathcal{N}\backslash\{j\}, \bar{\mathcal{B}}(p_i,r_i)\cap \pi = \emptyset, \forall \pi\in \Pi_{k,k'}\}$, with $p\coloneqq [p_1^\top,\dots,p_N^\top]^\top$. 
As the goal configuration we consider the centroid ${p_{\pi_{k'}}}$ of $\pi_{k'}$ and we construct the function $\gamma_{i_{k'}}:\mathbb{R}^3\rightarrow\mathbb{R}_{\geq 0}$ with $\gamma_{i_{k'}}({p_i}) \coloneqq \lVert {p_i} - {p_{\pi_{k'}}} \rVert^2$. For the collision avoidance between the agents, we employ the function $G_i:\mathcal{F}_{i_{k,k'}}\rightarrow\mathbb{R}$ as defined in \cite{dimarogonas2007decentralized}, which encodes the distances among the agents.

Moreover, we need some extra terms that guarantee that agent $i$ will avoid the rest of the regions as well as the workspace boundary. To this end, we construct the function $\alpha_{i_{k,k'}}:\mathbb{R}^3\rightarrow\mathbb{R}$ with $\alpha_{i_{k,k'}}({p_i}) \coloneqq \alpha_{i,0}({p_i})\prod_{ m\in{\Pi}_{k,k'}}\alpha_{i,m}({p_i})$, where the function $\alpha_{i,0}:\mathbb{R}^3\rightarrow\mathbb{R}$ is a measure of the distance of agent $i$ from the workspace boundary $\alpha_{i,0}(p_i) \coloneqq (r_0 - r_i)^2 - \lVert {p_i} - {p_0} \rVert^2$ and the function $\alpha_{i,m}:\mathbb{R}^3\rightarrow\mathbb{R}$ is a measure of the distance of agent $i$ from the undesired regions $\alpha_{i,m}(p_i) \coloneqq \lVert {p_i} - {p_m} \rVert^2 - (r_i + r_m)^2$. 

With the above ingredients, we construct the following navigation function $\varphi_{i_{k,k'}}:\mathcal{F}_{i_{k,k'}}\rightarrow[0,1]$:
\begin{equation*}
\varphi_{i_{k,k'}}({p}) \coloneqq \dfrac{\gamma_{i_{k'}}({p_i}) + f_{G_i}(G_i)}{ ( \gamma_{i_{k'}}^{\lambda_i}({p_i}) + G_i({p})\alpha_{i_{k,k'}}({p_i}) )^{1/\lambda_i}} 
\end{equation*}
for agent $i$, with $\lambda_i > 0$ and the following vector field: 
\begin{equation}
{c_{i_{k,k'}}}(p) \coloneqq \left\{ \begin{array}{cc}  -k_{g_i}\dfrac{\partial \varphi_{i_{k,k'}}({p})}{\partial{p_i}}, & \mbox{if }  \pi_{k} \not \equiv  \pi_{k'} \\  
0 &  \mbox{if } \pi_{k} \equiv  \pi_{k'} 
\end{array} \right. \label{eq:feedback_contr (ICRA 17)} 
\end{equation}
for all $t \geq t_0$, with $k_{g_i}>0$ and $f_{G_i}(G_i)$, defined in \cite{dimarogonas2007decentralized}, is a term that handles inter-agent collisions when an agent has reached its destination.

The navigation field (\ref{eq:feedback_contr (ICRA 17)}) guarantees that agent $i$ will not enter the undesired regions or collide with the other agents and $\lim_{t\rightarrow\infty}{p_i}(t) = {p_{\pi_{k'}}}$. The latter property of asymptotic convergence along with the assumption that $r_i < r_{\pi_k}, \forall i\in\mathcal{N},k\in\mathcal{K}_\mathcal{R}$, implies that there exists a finite time instant $t^{\scriptscriptstyle f}_{i,k'} \geq t_0$ such that ${p_i}(t^{\scriptscriptstyle f}_{i,k'}) \in \bar{\mathcal{B}}({p_{\pi_{k'}}},{r_{\pi_{k'}}})$ and more specifically that  $\mathcal{A}_i({p_i}(t^{\scriptscriptstyle f}_{i,k'}))\in\pi_{k'}$, which is the desired behavior. 
The time instant $t^{\scriptscriptstyle f}_{i,k'}$ can be chosen from the set $\{t\geq t_0, \mathcal{A}_i({p_i}(t))\in\pi_{k'}\}$. 

Note, however, that once agent $i$ leaves region $\pi_k$, there is no guarantee that it will not enter that region again (note that $\mathcal{F}_{i_{k,k'}}$ includes $\pi_k$), which might be undesirable. Therefore, we define the set $\Pi_{\emptyset,k'} \coloneqq \{\pi_m\in\Pi, m\in\mathcal{K}_\mathcal{R} \backslash \{k'\}\}$ and the corresponding free space
$\mathcal{F}_{i_{\emptyset,k'}} \coloneqq \{p\in \mathcal{W}^N : \bar{\mathcal{B}}(p_i,r_i)\cap\bar{\mathcal{B}}(p_j,r_j) = \emptyset, \forall j \in \mathcal{N}\backslash\{j\}, \bar{\mathcal{B}}(p_i,r_i)\cap \pi = \emptyset, \forall \pi\in \Pi_{\emptyset,k'}\}$, and we construct the function $\varphi_{i_{\emptyset,k'}}:\mathcal{F}_{i_{\emptyset,k'}}\rightarrow[0,1]$:
\begin{equation*}
\varphi_{i_{\emptyset,k'}}({p}) \coloneqq \dfrac{\gamma_{i_{k'}}({p_i})+f_{G_i}(G_i)}{ ( \gamma_{i_{k'}}^{\lambda_i}({p_i}) + G_i({p})\alpha_{i_{\emptyset,k'}}({p_i}) )^{1/\lambda_i}} 
\end{equation*}
where $\alpha_{i_{\emptyset,k'}}(p_i)\coloneqq\alpha_{i,0}({p_i})\prod_{ m\in\Pi_{\emptyset,k'}}\alpha_{i,m}({p_i})$, with corresponding vector field: 
\begin{equation}
{c_{i_{\emptyset,k'}}}(p) \coloneqq  -k_{g_i}\dfrac{\partial \varphi_{i_{\emptyset,k'}}({p})}{\partial{p_i}}, \label{eq:feedback_contr_2 (ICRA 17)}
\end{equation}
which guarantees that region $\pi_k$ will be also avoided. Therefore, we develop a switching control protocol that employs (\ref{eq:feedback_contr (ICRA 17)}) until agent $i$ is out of region $\pi_k$ and then switches to (\ref{eq:feedback_contr_2 (ICRA 17)}) until $t=t^{\scriptscriptstyle f}_{i,k'}$. Consider the following switching function:
\begin{equation*}
s_\mathsf{sat}(x) \coloneqq \dfrac{1}{2}(\text{sat}(2x-1)+1)
\end{equation*} 
and the time instant $t'_{i,k}$ that represents the moment that agent $i$ is out of region $\pi_k$, i.e., $t'_{i,k} \coloneqq \min \{t\geq t_0,  \bar{\mathcal{B}}({p_i}(t),{r_i})\cap \bar{\mathcal{B}}({p_{\pi_k}},{r_{\pi_k}})=\emptyset \}$. 
Then, we propose the following switching control protocol ${u_i}: \mathcal{F}_{i_{k,k'}}\cup\mathcal{F}_{i_{\emptyset,k}} \rightarrow \mathbb{R}^3$:
\begin{equation}
u_i\coloneqq {u_i}(p) = \left\{ \begin{array}{cc} {c_{i_{k,k'}}}(p), &  t\in [t_0,t'_{i,k}) \\
(1-s_\mathsf{sat}(\iota_{i,k})){c_{i_{k,k'}}}(p) + s_\mathsf{sat}(\iota_{i,k}){c_{i_{\emptyset,k'}}}(p), &  t\in [t'_{i,k},t^{\scriptscriptstyle f}_{i,k'})
\end{array} \right.
\label{eq:switch_controller (ICRA 17)}
\end{equation}
where $\iota_{i,k} \coloneqq \dfrac{t-t'_{i,k}}{\nu_i}$, and $\nu_i$ is a design parameter indicating the time period of the switching process, with $t^{\scriptscriptstyle f}_{i,k'}-t'_{i,k}>\nu_i > 0$. Invoking the continuity of ${p_i}(t)$, we obtain $\bar{\mathcal{B}}({p_i}(t^{\scriptscriptstyle f}_{i,k'}),r_i)\subset\bar{\mathcal{B}}({p_{\pi_{k'}}},{r_{\pi_{k'}}})$ and hence the control protocol (\ref{eq:switch_controller (ICRA 17)}) guarantees, for sufficiently small $\nu_i$, that agent $i$ will navigate from $\pi_k$ to $\pi_{k'}$ in finite time without entering any other regions or colliding with other agents and therefore establishes a transition $\pi_k\rightarrow_i\pi_{k'}$.

\subsubsection{High-Level Plan Generation} \label{subsec:High level plan (ICRA 17)}

The next step of our solution is the high-level plan, which can be generated using standard techniques inspired by automata-based formal verification methodologies. In Section \ref{subsec:Continuous Control (ICRA 17)}, we proposed a continuous control law that allows the agents to transit between any $\pi_k, \pi_{k'}\in\Pi$ in the given workspace $\mathcal{W}$, without colliding with each other. Thanks to this and to our definition of LTL semantics over the sequence of atomic propositions, we can abstract the motion capabilities of each agent as a finite transition system $\mathcal{T}_i$ as follows \cite{baier2008principles}:
\begin{definition}
	The motion of each agent $i\in\mathcal{N}$ in $\mathcal{W}$ is modeled by the following Transition System (TS):
	\begin{equation*}
	\mathcal{T}_i=(\Pi_i,\Pi^{\text{init}}_i, \rightarrow_i, \Psi_i, \mathcal{L}_i), 
	\end{equation*}
	where $\Pi_i\subseteq\Pi$ is the set of states represented by the regions of interest that the agent can be at, according to Def. \ref{def:agent in region (ICRA 17)}, $\Pi^{\text{init}}_i \subseteq \Pi_i $ is the set of initial states that agent $i$ can start from, $\rightarrow_i\subseteq\Pi_i\times\Pi_i$ is the transition relation established in Section \ref{subsec:Continuous Control (ICRA 17)}, and $\Psi_i, \mathcal{L}_i$ are the atomic propositions and labeling function respectively, as defined in Section \ref{sec:System and PF (ICRA 17)}.
\end{definition}

After the definition of $\mathcal{T}_i$, we translate each given LTL formula $\mathsf{\Phi}_i, i\in\mathcal{N}$ into a Büchi automaton $\mathcal{C}_i$ and we form the product $\widetilde{\mathcal{T}}_i=\mathcal{T}_i\times\mathcal{C}_i$. The accepting runs of $\widetilde{\mathcal{T}}_i$ satisfy $\mathsf{\Phi}_i$ and are directly projected to a sequence of waypoints to be visited, providing therefore a desired path for agent $i$. Although the semantics of LTL is defined over infinite sequences of atomic propositions, it can be proven that there always exists a high-level plan that takes a form of a finite state sequence followed by an infinite repetition of another finite state sequence. For more details on the followed technique, we kindly refer the reader to the related literature, e.g., \cite{baier2008principles}.

Following the aforementioned methodology, we obtain a high-level plan for each agent as sequences of regions and atomic propositions $r_{\scr \mathcal{T},i} \coloneqq \pi_{i_1} \pi_{i_2} \dots$ and $\breve{\psi}_i \coloneqq \breve{\psi}_{i_1} \breve{\psi}_{i_2}\dots$ with $i_m\in\mathcal{K}_\mathcal{R}, \breve{\psi}_{i_m}\in2^{\Psi_i}, \breve{\psi}_{i_m}\in\mathcal{L}_i(\pi_{i_m}), \forall m\in\mathbb{N}$ and $\breve{\psi}_i\models\mathsf{\Phi}_i,\forall i\in\mathcal{N}$. 

The execution of $r_{\mathcal{T},i},\breve{\psi}_i$ produces a trajectory ${p_i}(t)$ that corresponds to the behavior $\mathfrak{b}_i(\breve{\psi}_i) = ({p_{i_1}}(t),\breve{\psi}_{i_1})({p_{i_2}}(t),\breve{\psi}_{i_2})\dots$, with $\mathcal{A}_i({p_{i_m}})\in\pi_{i_m}$ and $\breve{\psi}_{i_m}\in\mathcal{L}_i(\pi_{i_m})$, $\forall m\in\mathbb{N}$. Therefore, since $\breve{\psi}_i\models \mathsf{\Phi}_i$, the behavior $\mathfrak{b}_i$ yields satisfaction of the formula $\mathsf{\Phi}_i$. Moreover, the property of inter-agent collision avoidance is inherent in the transition relations of $\mathcal{T}_i$ and guaranteed by the navigation control algorithm of Section \ref{subsec:Continuous Control (ICRA 17)}. 

\begin{remark}
	The proposed control algorithm is decentralized in the sense that each agent derives and executes its own plan without communicating with the rest of the team. The only  information that each agent has is the position of its neighboring agents that lie in its limited sensing radius. It is worth mentioning, nevertheless, that the workspace boundary and regions of interest have to satisfy certain assumptions, such as having a sufficient distance from each other or being sufficiently sparse.
\end{remark}

\begin{figure}[!btp]
	\centering
	\includegraphics[scale =0.45]{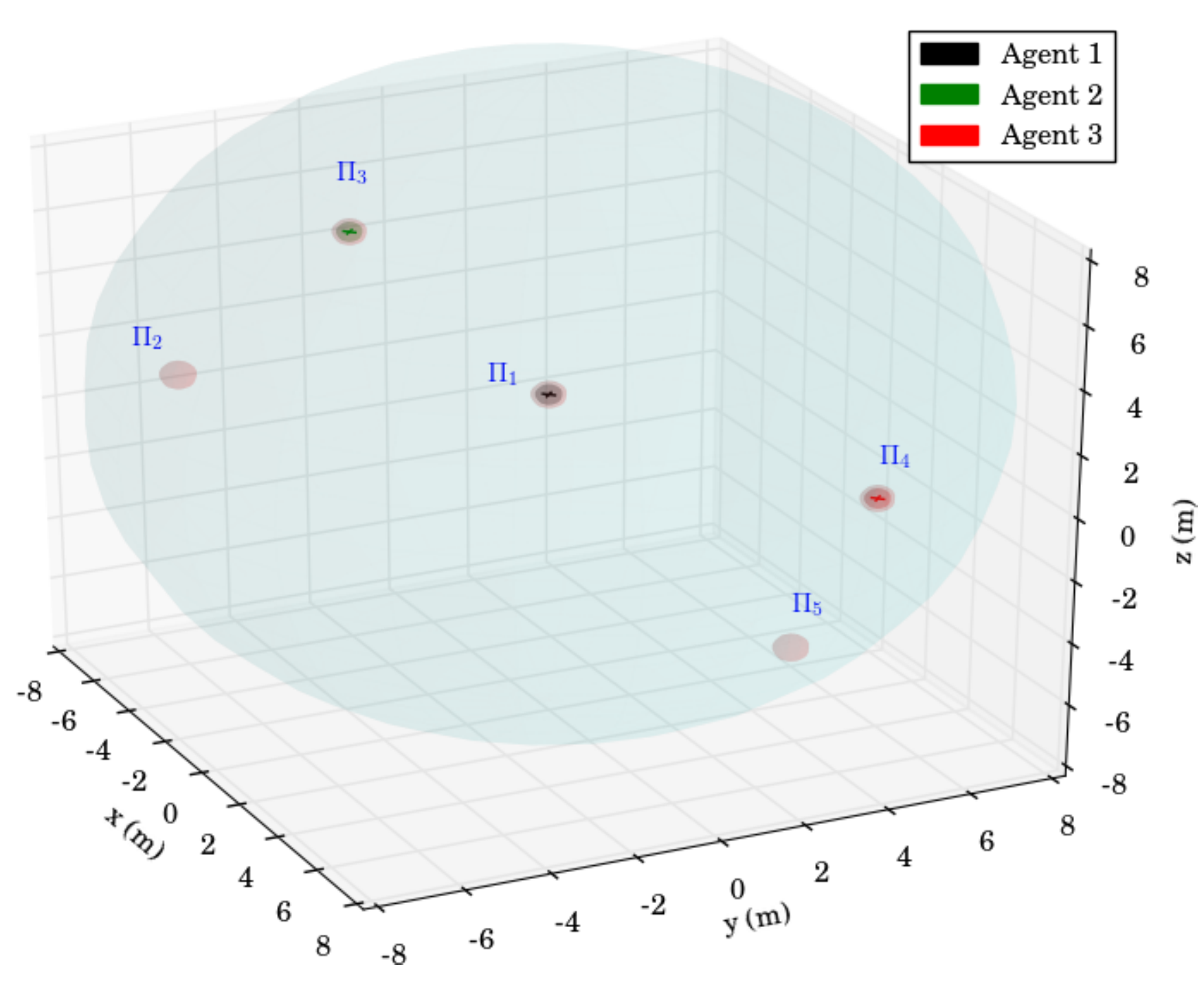}
	
	\caption{Initial workspace of the simulation studies. The grey spheres represent the regions of interest while the black, green and red crosses represent agents 1,2 and 3, respectively, along with their bounding spheres. \label{fig:Initial_workspace (ICRA 17)}}
\end{figure}

\begin{figure}[!btp]
	\centering
	\includegraphics[scale =0.9]{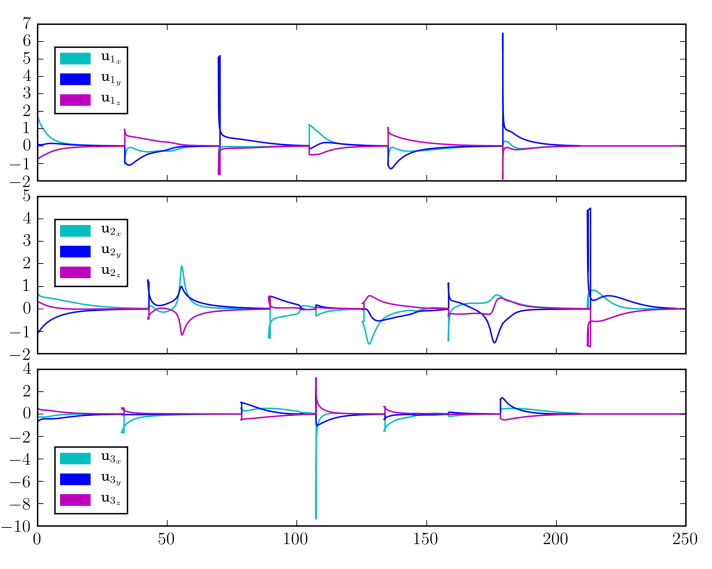}
	
	\caption{The resulting $3$-dimensional control signals of the $3$ agents for the simulation studies. Top: agent $1$, middle: agent $2$, bottom: agent $3$.\label{fig:sim_vel (ICRA 17)}}
\end{figure}

\subsection{Simulation and Experimental Results} \label{sec:simulation results (ICRA 17)}
To demonstrate the efficiency of the proposed algorithm, we consider $N=3$ aerial vehicles  with $r_i = 0.3$m, $\varsigma_i = 0.65$m,  $\forall i=\{1,2,3\}$, operating in a workspace $\mathcal{W}$ with $r_0 = 10$m and ${p_0} = [0,0,0]^\top$m. Moreover, we consider $K=5$ spherical regions of interest with $r_{\pi_k} = 0.4$m, $\forall k=\{1,\dots,5\}$ and ${p}_{\pi_1} = [0,0,2]^\top$m, ${p}_{\pi_2} = [1,-9,5]^\top$m, ${p}_{\pi_3} = [-8,-1,4]^\top$m, ${p}_{\pi_4} = [2,7,-2]^\top$m and ${p}_{\pi_5} = [7.5,2,-3]^\top$m.  The initial configurations of the agents are taken as ${p_1}(0) = {p}_{\pi_1}, {p_2}(0) = {p}_{\pi_3}, {p_3}(0) = {p}_{\pi_4}$ and therefore, $\mathcal{A}_1({p_1}(0))\in\pi_1, \mathcal{A}_2({p_2}(0))\in\pi_3$ and  $\mathcal{A}_3({p_3}(0))\in\pi_4$. An illustration of the described workspace is depicted in Fig. \ref{fig:Initial_workspace (ICRA 17)}.

\begin{figure}
	\begin{minipage}[b]{.5\linewidth}		
		\centering
		\includegraphics[width=0.7\textwidth,height=0.6\textwidth]{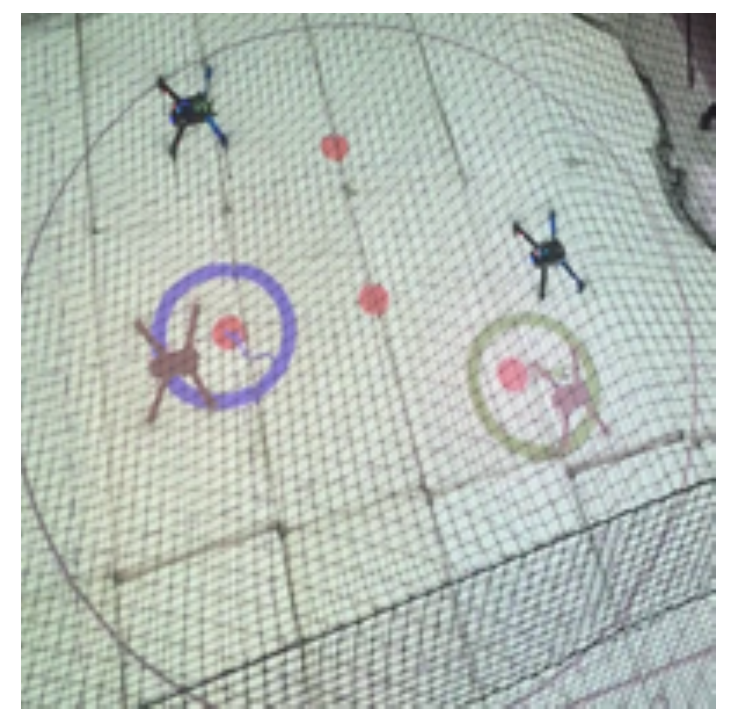}
		\subcaption{}
		\label{fig:Initial_workspace_exp_SML (ICRA 17)}	
	\end{minipage}
	\begin{minipage}[b]{.5\linewidth}
		\includegraphics[trim=-1cm 1.25cm 1cm -1.25cm, width=0.85\textwidth, height=0.75\textwidth]{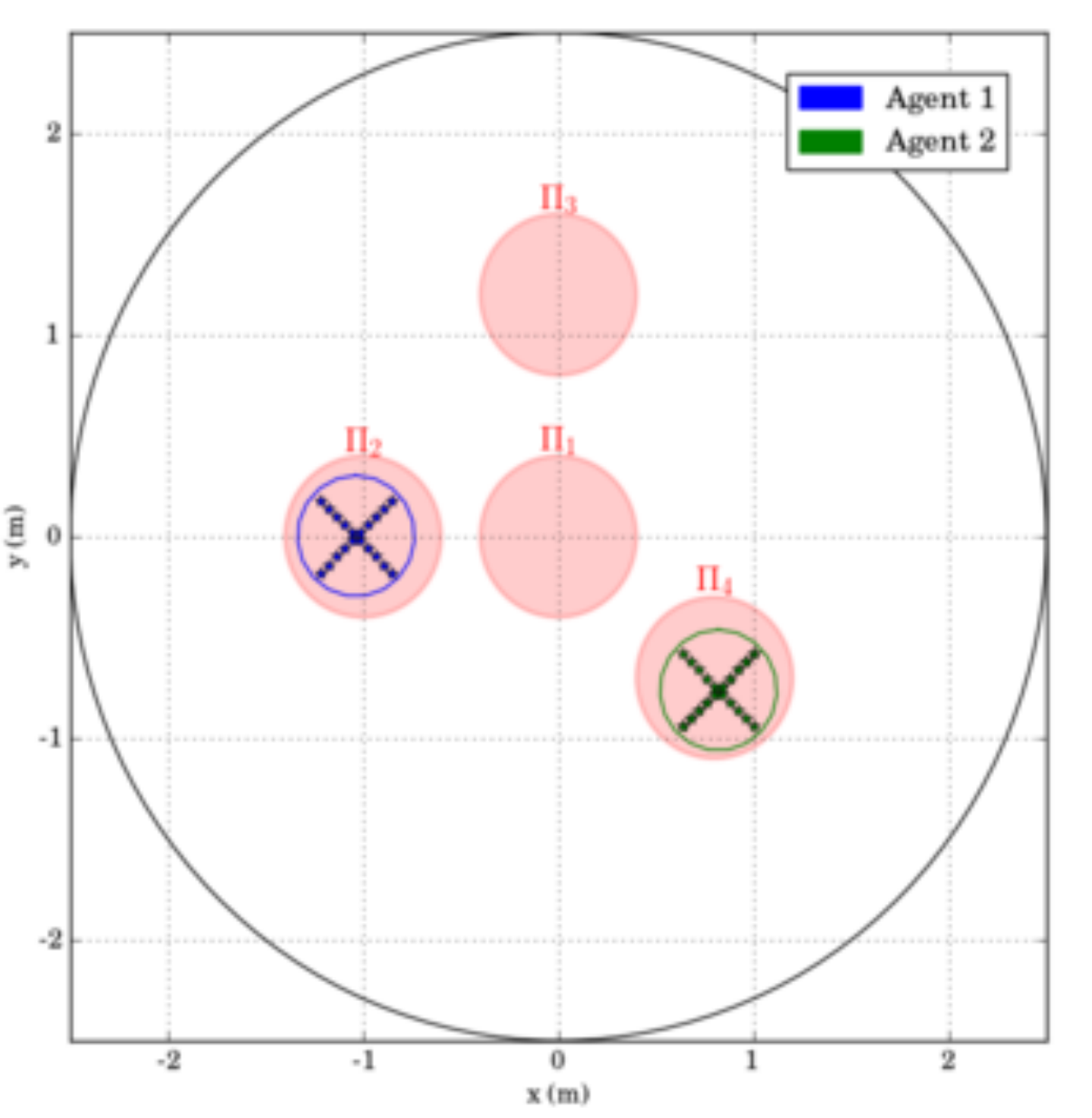}
		\subcaption{}
		\label{Initial_workspace_exp_simulated (ICRA 17)}
	\end{minipage}
	\caption{Initial workspace for the first real experimental scenario. (a): The UAVs with the projection of their bounding spheres, (with blue and green), and the centroids of the regions of interest (with red). (b): Top view of the described workspace. The UAVs are represented by the blue and green circled X's and the regions of interest by the red disks $\pi_1,\dots,\pi_4$.}\label{fig:Initial_workspace_exp_1 (ICRA 17)}
\end{figure}

We consider that agent $2$ is assigned with inspection tasks and has the  atomic propositions $\Psi_2 = \{ ``\text{ins}_{\text{a}}",``\text{ins}_{\text{b}}",``\text{ins}_{\text{c}}",``\text{ins}_{\text{d}}",``\text{obs}" \}$ with $\mathcal{L}_{2}(\pi_1) = \{``\text{obs}"\}$, $\mathcal{L}_{2}(\pi_2) = \{``\text{ins}_{\text{a}}"\}, \mathcal{L}_{2}(\pi_3) = \{``\text{ins}_{\text{b}}"\}, \mathcal{L}_{2}(\pi_4) = \{``\text{ins}_{\text{c}}"\}$ and $\mathcal{L}_{2}(\pi_5) = \{``\text{ins}_{\text{d}}"\}$, where we have considered that region $\pi_1$ is an undesired (``obstacle") region for this agent. More specifically, the task for agent $2$ is the continuous inspection of the workspace while avoiding region $\pi_1$. The corresponding LTL specification is $\mathsf{\Phi}_2 =  (\square \neg ``\text{obs}")\land\square(\lozenge``\text{ins}_{\text{a}}"\land\lozenge``\text{ins}_{\text{b}}"\land\lozenge``\text{ins}_{\text{c}}"\land\lozenge``\text{ins}_{\text{d}}")$. 
Agents $1$ and $3$ are interested in moving around resources scattered in the workspace and have propositions $\Psi_1 = \Psi_3 = \{ ``\text{res}_{\text{a}}",``\text{res}_{\text{b}}",``\text{res}_{\text{c}}",``\text{res}_{\text{d}}", ``\text{res}_{\text{e}}" \}$ with $\mathcal{L}_1(\pi_1)=\mathcal{L}_3(\pi_1)=\{\text{res}_{\text{a}}\}, \mathcal{L}_1(\pi_2)=\mathcal{L}_3(\pi_2)=\{\text{res}_{\text{b}}\}, \mathcal{L}_1(\pi_3)=\mathcal{L}_3(\pi_3)=\{\text{res}_{\text{c}}\}, \mathcal{L}_1(\pi_4)=\mathcal{L}_3(\pi_4)=\{\text{res}_{\text{d}}\}$ and $\mathcal{L}_1(\pi_5)=\mathcal{L}_3(\pi_5)=\{\text{res}_{\text{e}}\}$. We assume that $``\text{res}_{\text{a}}"$ is shared between the two agents whereas $``\text{res}_{\text{b}}"$ and $``\text{res}_{\text{e}}"$ have to be accessed only by agent $1$ and $``\text{res}_{\text{c}}"$ and $``\text{res}_{\text{d}}"$ only by agent $3$. The corresponding specifications are $\mathsf{\Phi}_1 = \square \neg (``\text{res}_{\text{c}}"\lor``\text{res}_{\text{d}}")\land \square\lozenge(``\text{res}_{\text{a}}" \bigcirc``\text{res}_{\text{e}}"\bigcirc``\text{res}_{\text{b}}")$ and $\mathsf{\Phi}_3 = \square \neg (``\text{res}_{\text{b}}"\lor``\text{res}_{\text{e}}")\land 	\square\lozenge(``\text{res}_{\text{a}}" \bigcirc``\text{res}_{\text{c}}"\bigcirc``\text{res}_{\text{d}}")$, where we have also included a specific order for the access of the resources. Next, we employ the off-the-shelf tool LTL2BA \cite{LTL2BA} to create the Büchi automata $\mathcal{C}_i, i=\{1,2,3\}$ and by following the procedure described in Section \ref{subsec:High level plan (ICRA 17)}, we derive the paths $p_1 = (\pi_1\pi_5\pi_2)^{\omega}, p_2 = (\pi_3\pi_2\pi_5\pi_4)^{\omega}, p_3 = (\pi_4\pi_1\pi_3)^{\omega}$, whose execution satisfies $\mathsf{\Phi}_1,\mathsf{\Phi}_2,\mathsf{\Phi}_3$. Regarding the continuous control protocol, we chose $k_{g_i} = 15, \lambda_i = 5, \forall i\in\{1,2,3\}$ in (\ref{eq:feedback_contr (ICRA 17)}), (\ref{eq:feedback_contr_2 (ICRA 17)}) and the switching duration in (\ref{eq:switch_controller (ICRA 17)}) was calculated online as $\nu_i = 0.1t'_{i,k}$, where we assume that the large distance between the regions $\pi_k$ (see Fig. \ref{fig:Initial_workspace (ICRA 17)}) implies that $t^{\scriptscriptstyle f}_{i,k'} > 1.1t'_{i,k}$ and thus, $\nu_i < t^{\scriptscriptstyle f}_{i,k'} - t'_{i,k}$. The simulation results are depicted in Fig. \ref{fig:sim_vel (ICRA 17)} and \ref{fig:path (ICRA 17)}. In particular, Fig. \ref{fig:path (ICRA 17)} illustrates the execution of the paths $(\pi_1\pi_5\pi_2)^2\pi_1, (\pi_3\pi_2\pi_5\pi_4)^2\pi_3\pi_2\pi_5$ and $(\pi_4\pi_1\pi_3)^2\pi_4$ by agents $1,2$ and $3$ respectively, where the superscript $2$ here denotes that the corresponding paths are executed twice. Fig. \ref{fig:sim_vel (ICRA 17)} depicts the resulting control inputs $u_i, \forall i\in\{1,2,3\}$. The figures demonstrate the successful execution of the agents' paths and therefore, satisfaction of the respective formulas with inter-agent collision avoidance.

\begin{figure*}
	\begin{minipage}[b]{.5\linewidth}	
		\centering
		\includegraphics[trim =0cm 0cm 0cm -1cm, width=0.7\textwidth, height=0.18\textheight]{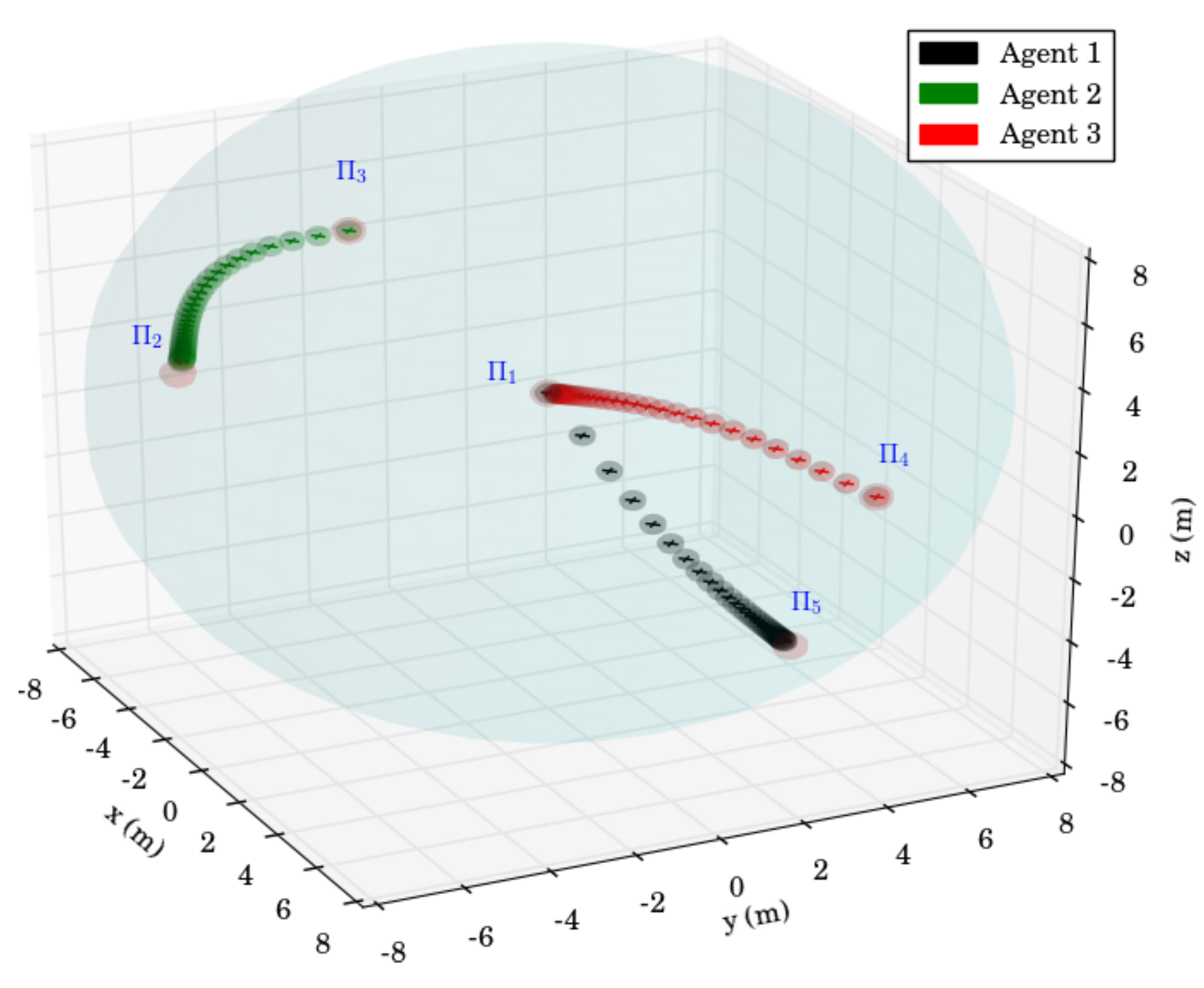}
		\label{fig:path:1 (ICRA 17)}
		\subcaption{}
	\end{minipage}
	\begin{minipage}[b]{.5\linewidth}
		\includegraphics[trim =-3.5cm 0cm 3.5cm -1cm, width=0.7\textwidth, height=0.18\textheight]{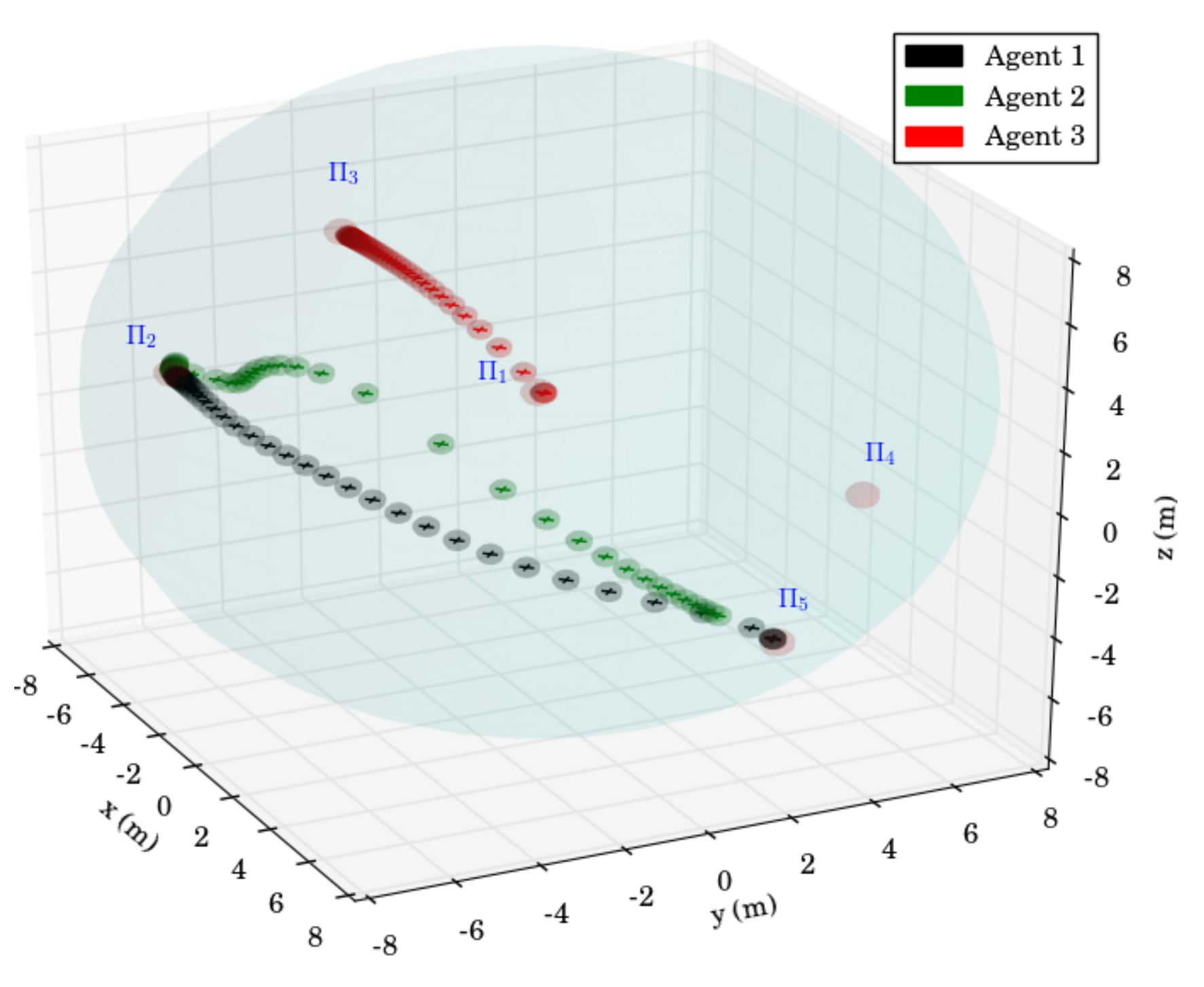}
		\label{fig:path:2 (ICRA 17)}
		\subcaption{}
	\end{minipage}
	\begin{minipage}[b]{.5\linewidth}
		\centering
		\includegraphics[trim =0cm 0cm 0cm -1cm, width=0.7\textwidth, height=0.18\textheight]{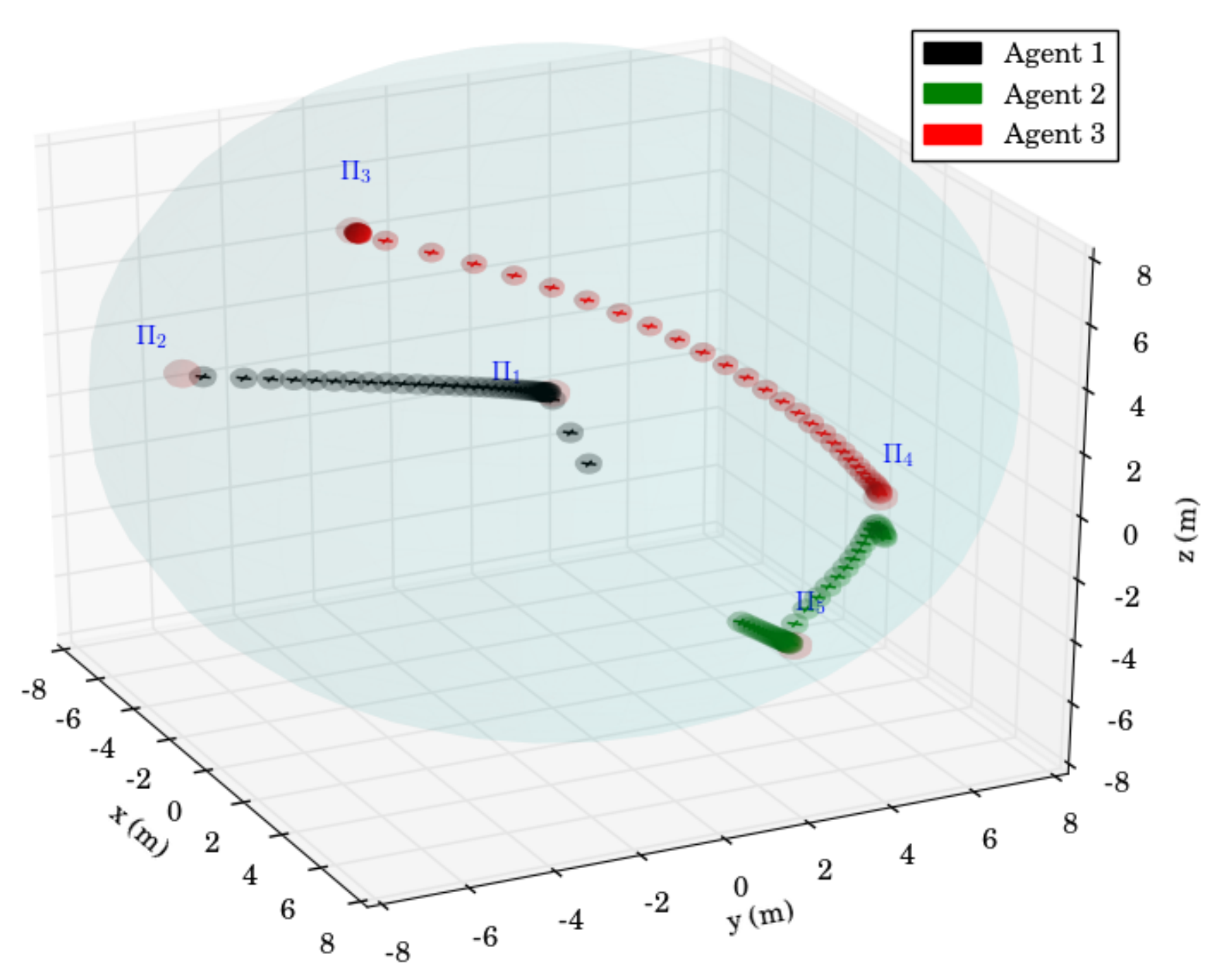}
		\label{fig:path:3 (ICRA 17)}
		\subcaption{}
	\end{minipage}
	\begin{minipage}[b]{.5\linewidth}
		\centering
		\includegraphics[trim =0cm 0cm 0cm -1cm, width=0.7\textwidth, height=0.18\textheight]{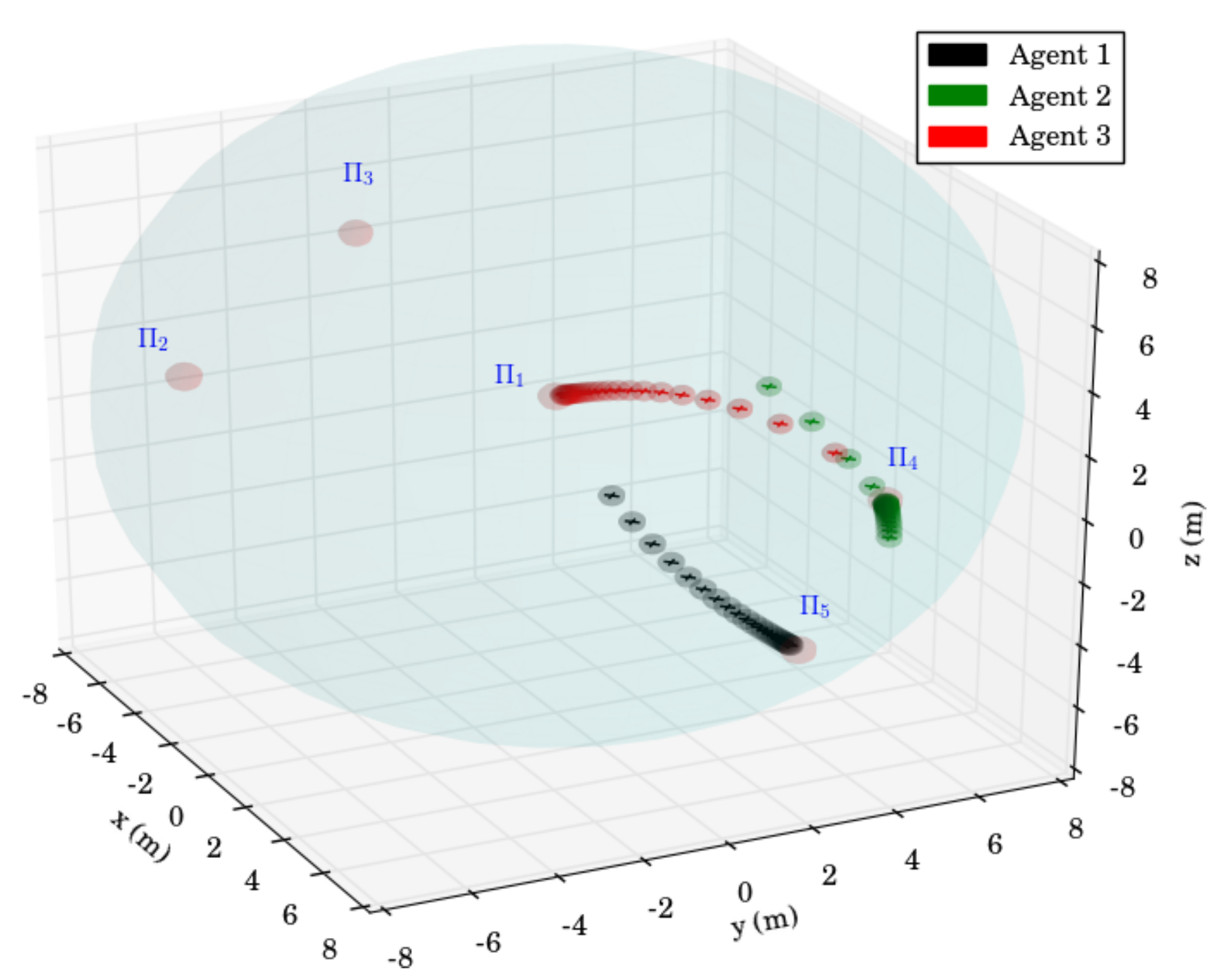}
		\label{fig:path:4 (ICRA 17)}
		\subcaption{}
	\end{minipage}
	\begin{minipage}[b]{.5\linewidth}
		\centering
		\includegraphics[trim =0cm 0cm 0cm -1cm, width=0.7\textwidth, height=0.18\textheight]{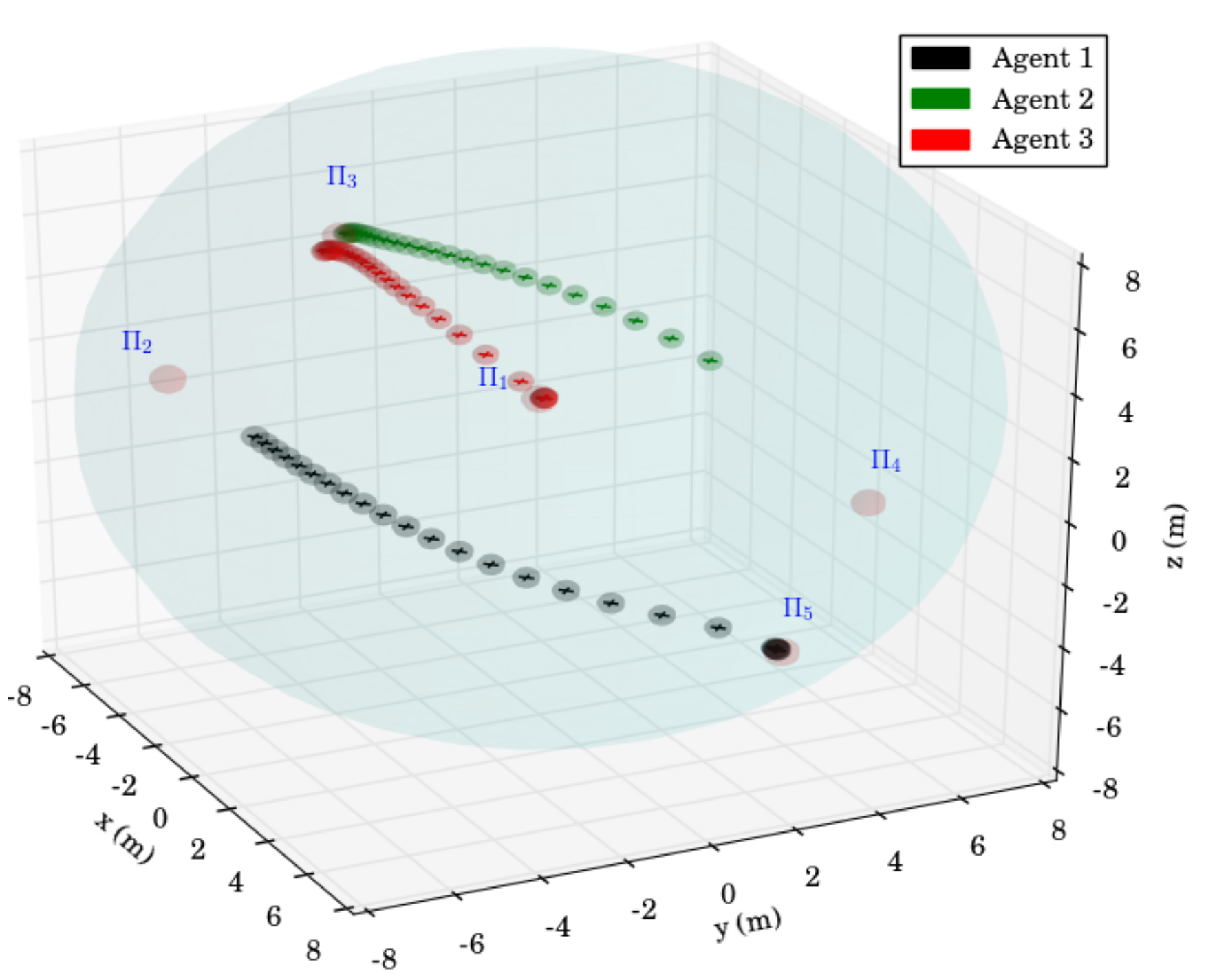}
		\label{fig:path:5 (ICRA 17)}
		\subcaption{}
	\end{minipage}
	\begin{minipage}[b]{.5\linewidth}
		\centering
		\includegraphics[trim =0cm 0cm 0cm -1cm, width=0.7\textwidth, height=0.18\textheight]{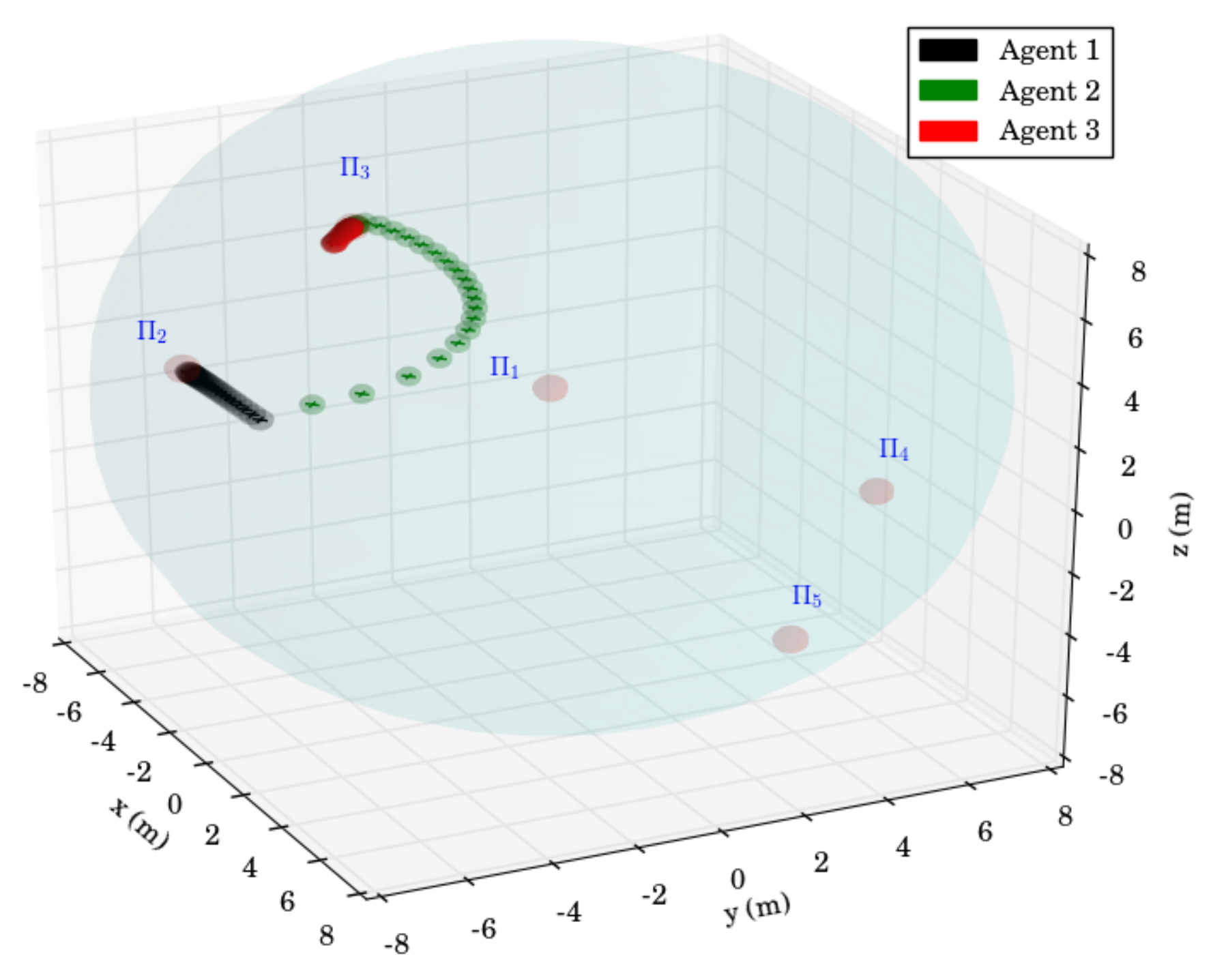}
		\label{fig:path:6 (ICRA 17)}
		\subcaption{}
	\end{minipage}
	\begin{minipage}[b]{.5\linewidth}
		\centering	
		\includegraphics[trim =0cm 0cm 0cm -1cm, width=0.7\textwidth, height=0.18\textheight]{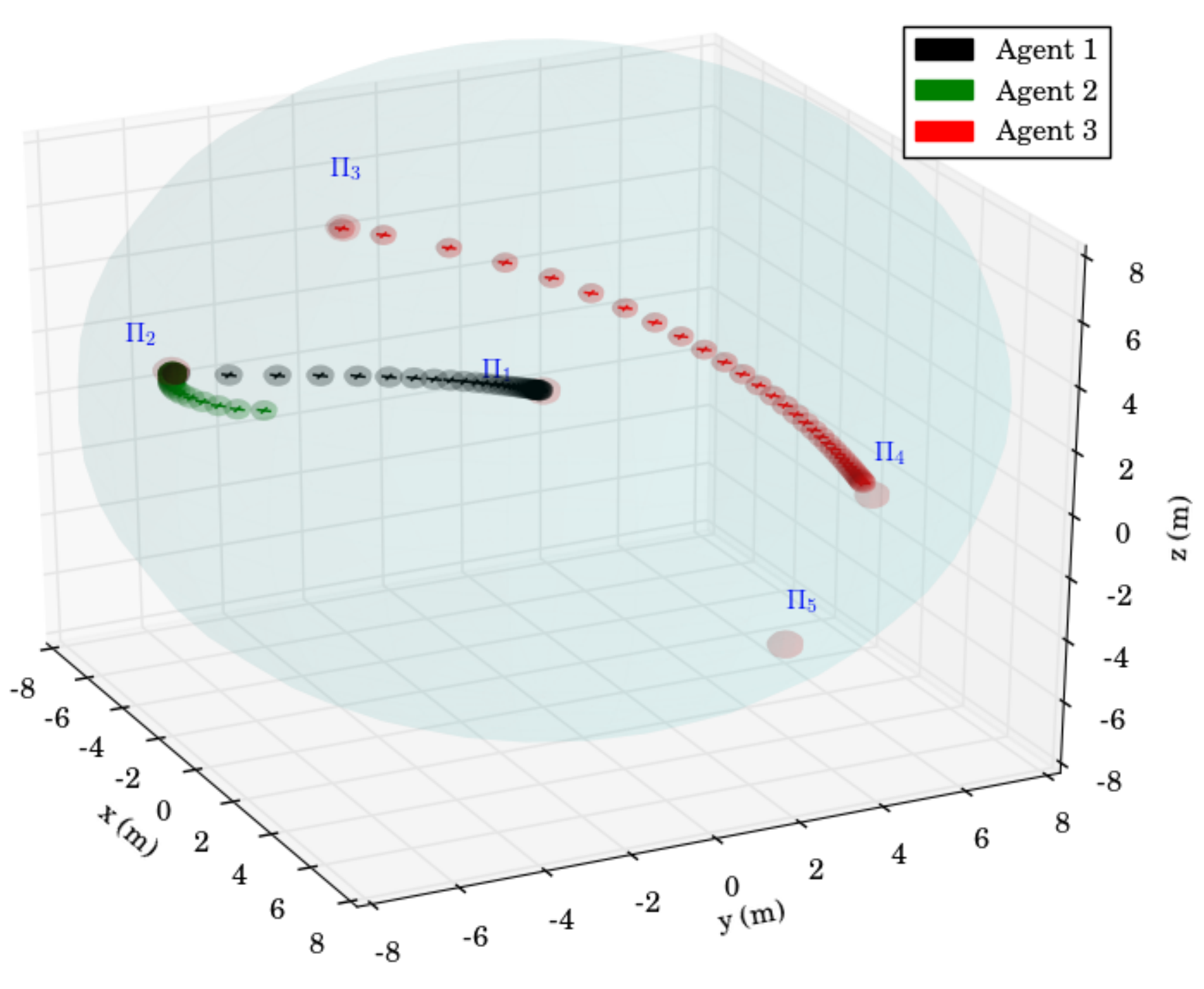}
		\label{fig:path:7 (ICRA 17)}
		\subcaption{}
	\end{minipage}
	\begin{minipage}[b]{.5\linewidth}
		\centering
		\includegraphics[trim =0cm 0cm 0cm -1cm, width=0.7\textwidth, height=0.18\textheight]{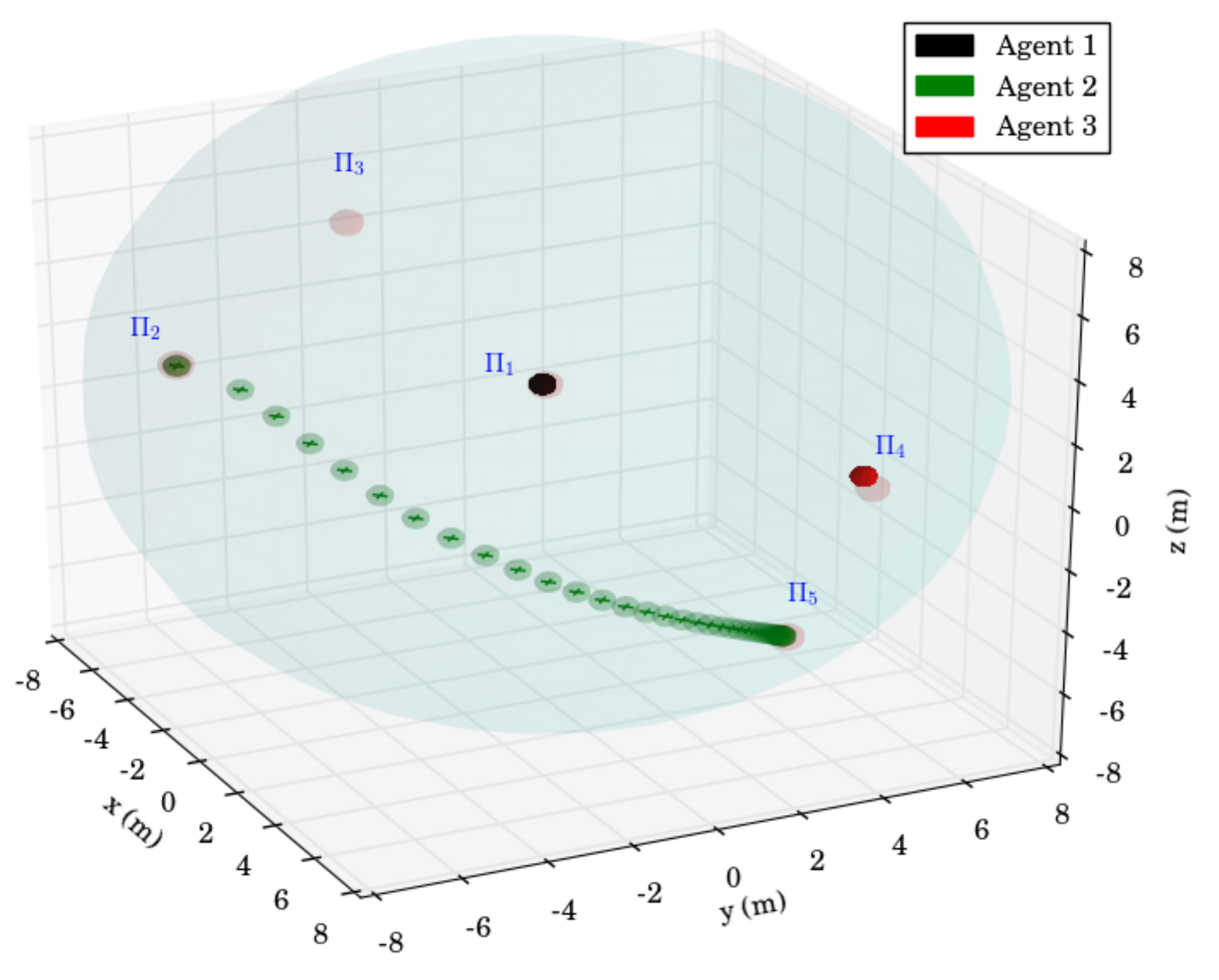}
		\label{fig:path:8 (ICRA 17)}
		\subcaption{}
	\end{minipage}	
	\caption{Execution of the paths $(\pi_1\pi_5\pi_2)^2\pi_1, (\pi_3\pi_2\pi_5\pi_4)^2\pi_3\pi_2\pi_5$ and $(\pi_4$ $\pi_1\pi_3)^2\pi_4$ by agents $1,2$ and $3$, respectively, for the simulation studies. }\label{fig:path (ICRA 17)}
\end{figure*}
\begin{figure*}[!htb]
	\begin{minipage}{.33\linewidth}	
		\centering
		\includegraphics[trim = 0cm 0cm 0cm 0cm,width=0.75\textwidth,height = 0.6\textwidth]{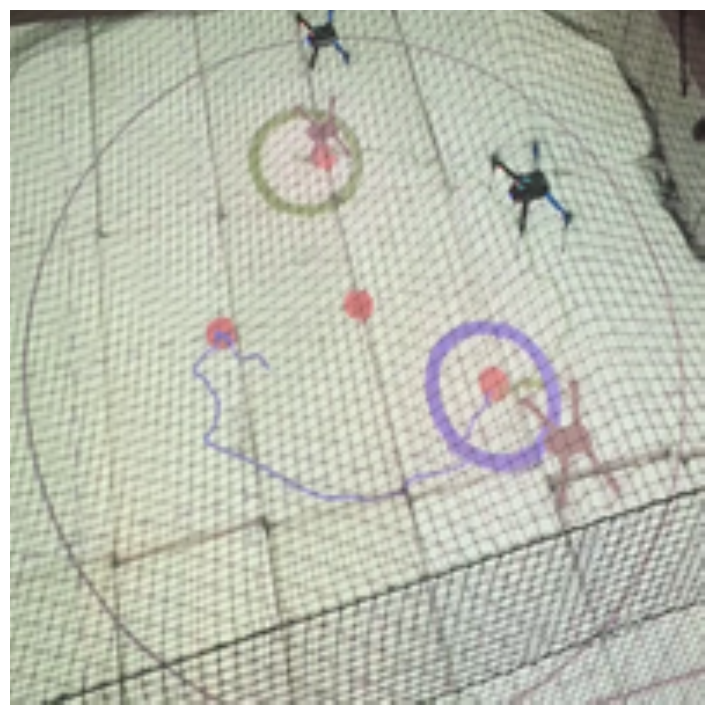}\label{fig:exp_1_path:1 (ICRA 17)}
		\subcaption{}
	\end{minipage}\hfill
	\begin{minipage}{.33\linewidth}
		\centering
		\includegraphics[trim = 0cm 0cm 0cm 0cm,width=0.75\textwidth,height = 0.6\textwidth]{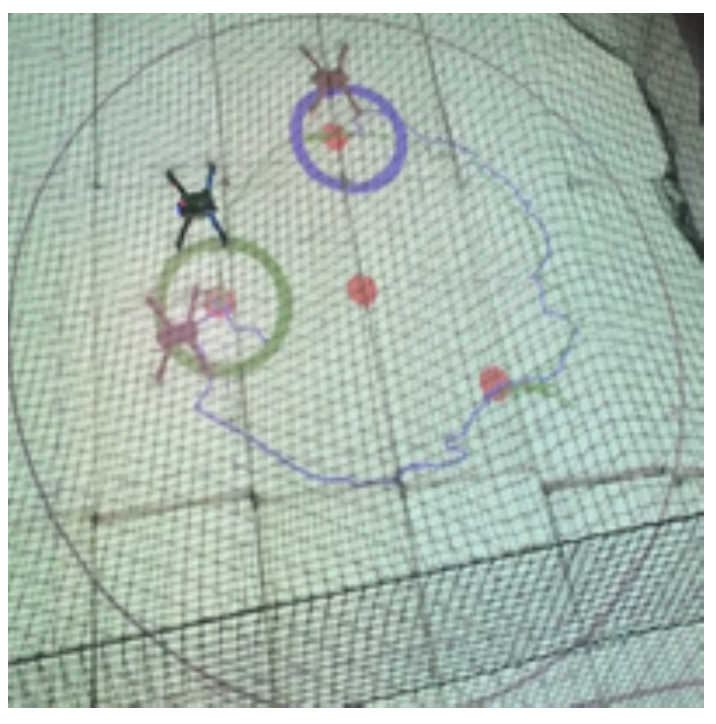}\label{fig:exp_1_path:2 (ICRA 17)}
		\subcaption{}
	\end{minipage}\hfill
	\begin{minipage}{.33\linewidth}
		\centering
		\includegraphics[trim = 0cm 0cm 0cm 0cm,width=0.75\textwidth,height = 0.6\textwidth]{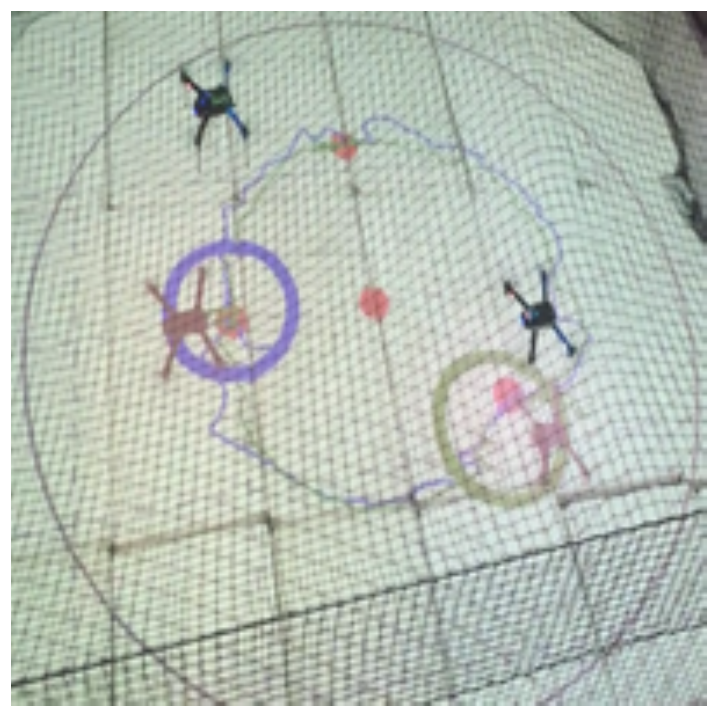}\label{fig:exp_1_path:3 (ICRA 17)}
		\subcaption{}		
	\end{minipage}
	\begin{minipage}{.33\linewidth}
		\centering
		\includegraphics[trim = 0cm 1cm 0cm -1cm,width=0.95\textwidth,height = 0.8\textwidth]{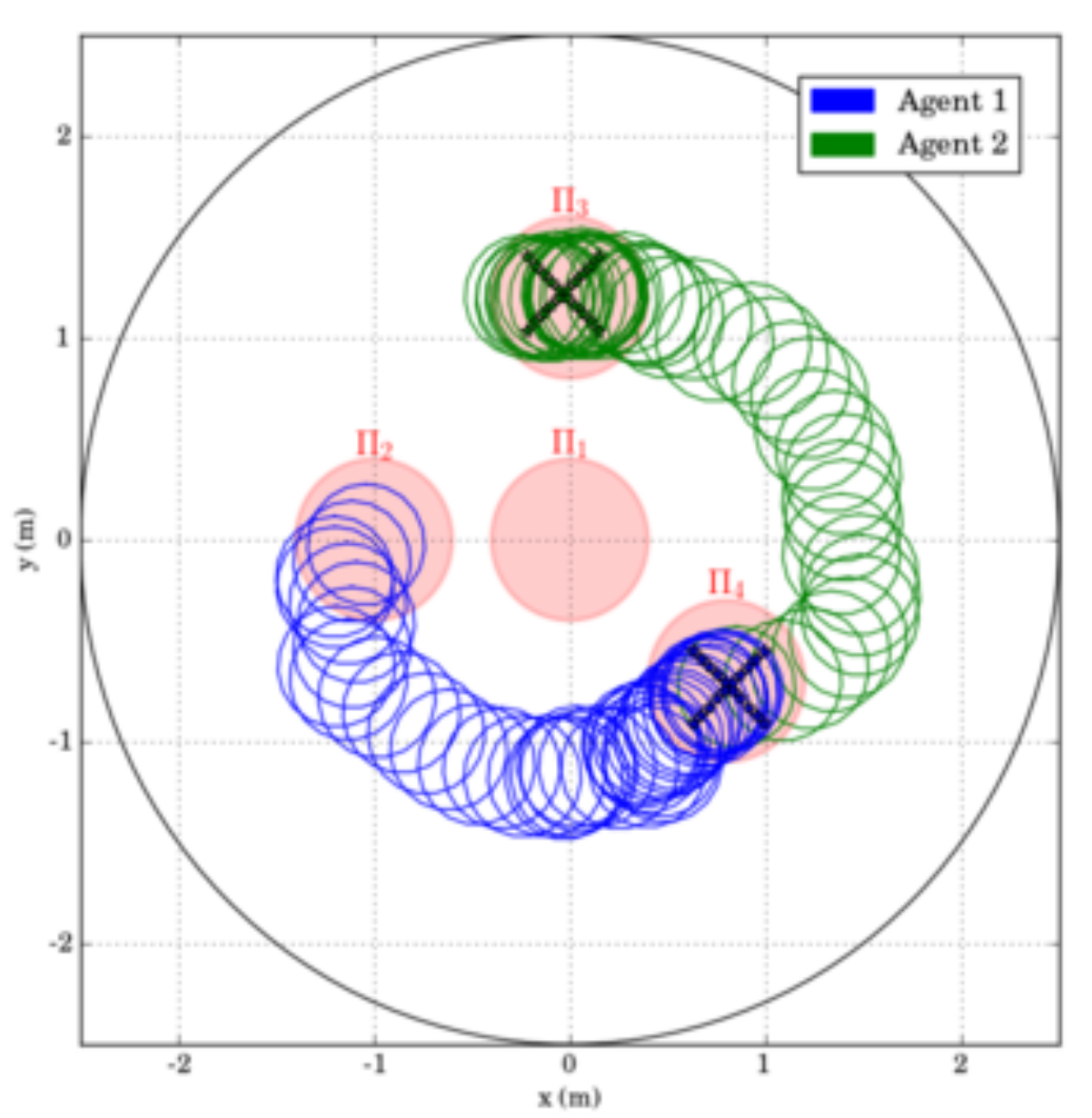}\label{fig:exp_1_path:1_sim (ICRA 17)}
		\subcaption{}
	\end{minipage}\hfill
	\begin{minipage}{.33\linewidth}
		\centering
		\includegraphics[trim = 0cm 1cm 0cm -1cm,width=0.95\textwidth,height = 0.8\textwidth]{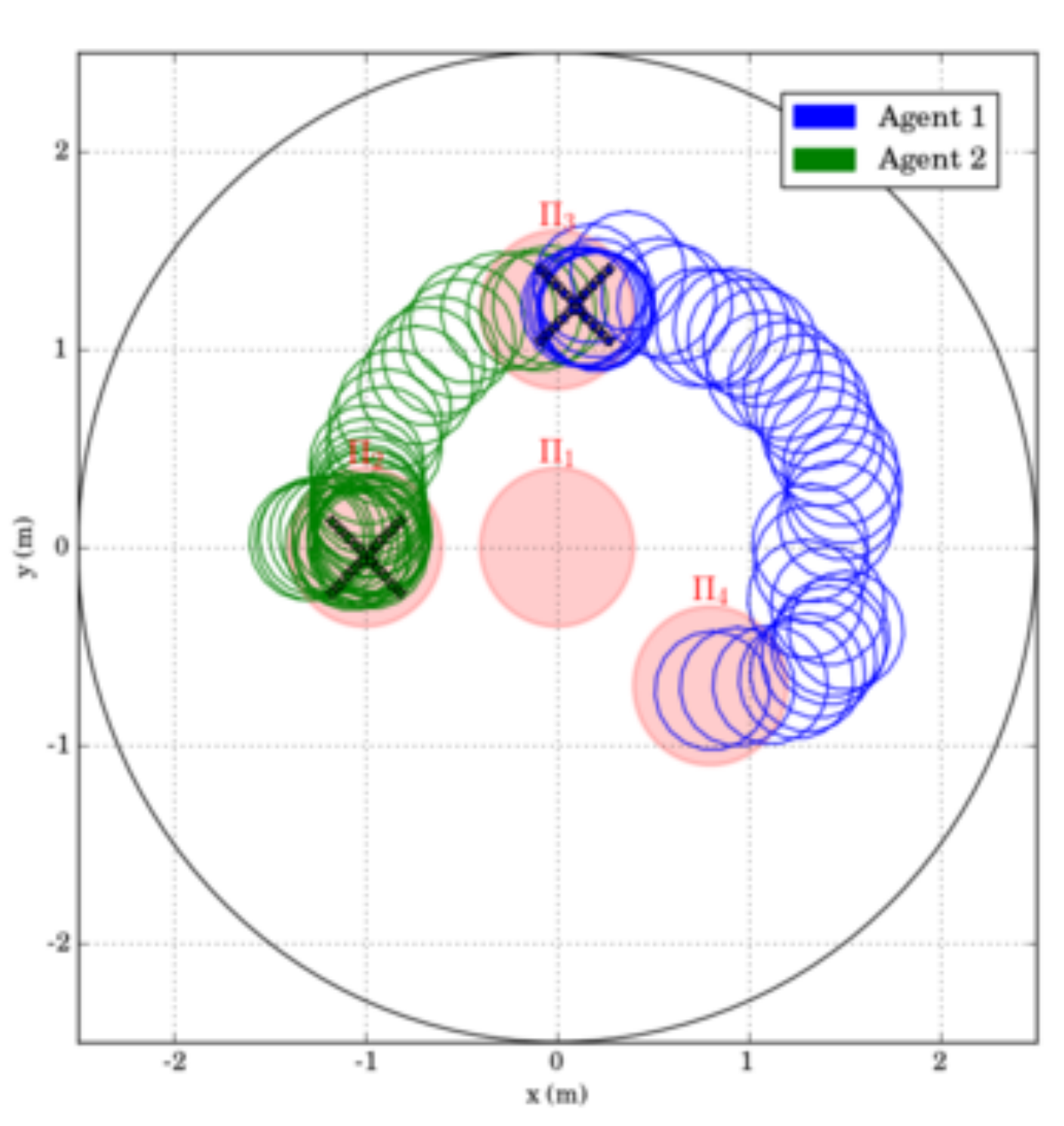}\label{fig:exp_1_path:2_sim (ICRA 17)}
		\subcaption{}
	\end{minipage}\hfill
	\begin{minipage}{.33\linewidth}
		\centering
		\includegraphics[trim = 0cm 1cm 0cm -1cm,width=0.95\textwidth,height = 0.8\textwidth]{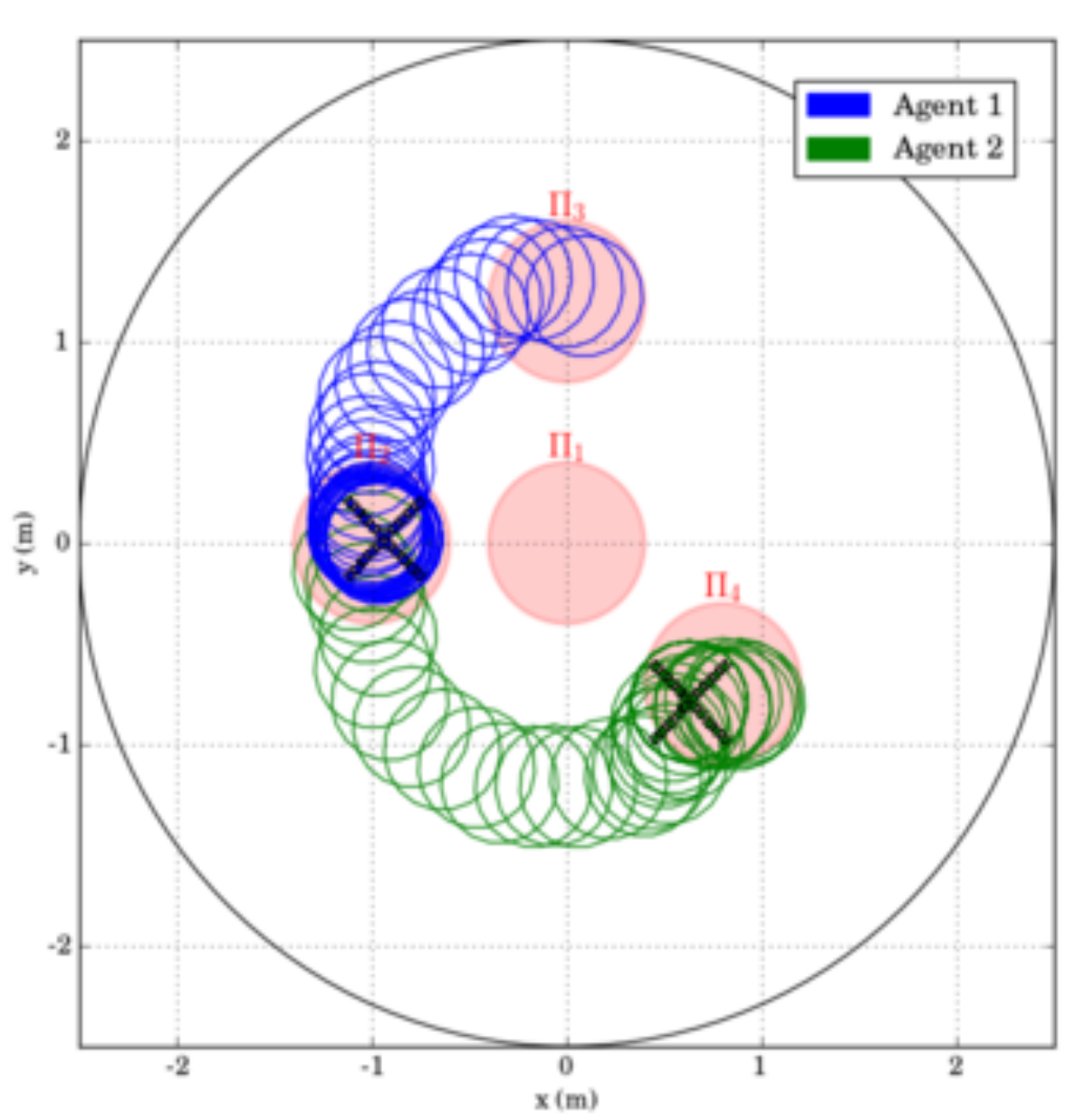}\label{fig:exp_1_path:3_sim (ICRA 17)}
		\subcaption{}
	\end{minipage}
	\caption{Execution of the paths $(\pi_2\pi_4\pi_3)^1$ and $(\pi_4\pi_3\pi_2)^1$ by agents $1$ and $2$, respectively for the first experimental scenario. (a), (d): $\pi_2\rightarrow_1\pi_4, \pi_4\rightarrow_2\pi_3$, (b), (e): $\pi_4\rightarrow_1\pi_3, \pi_3\rightarrow_2\pi_2$, (c), (f):$\pi_3\rightarrow_1\pi_2, \pi_2\rightarrow_2\pi_4$.  }\label{fig:exp_1_path (ICRA 17)}
\end{figure*}

\begin{figure}[!btp]
	\centering
	\includegraphics[trim = 0cm 0cm 0cm -0.5cm,width = 0.8\textwidth]{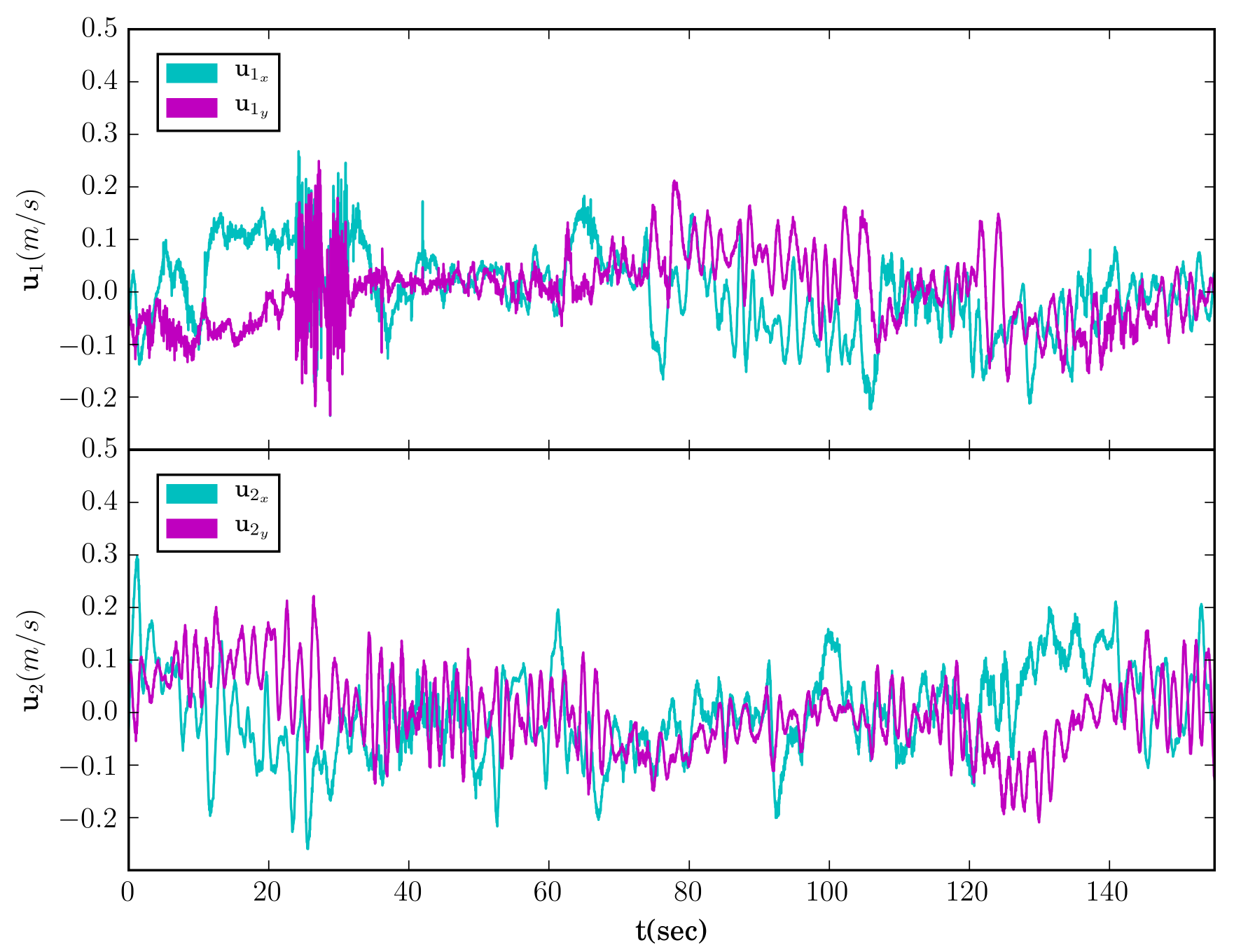}
	
	\caption{The resulting $2$-dimensional control signals of the $2$ agents for the first experimental scenario. Top: agent 1, bottom: agent 2. \label{fig:exp_1_vel (ICRA 17)}}
\end{figure}


The validity and efficiency of the proposed solution was also verified through real-time experiments. The experimental setup involved two remotely controlled \textit{IRIS+} quadrotors from $3$D Robotics, which we consider to have sensing range $\varsigma_i = 0.65$m, upper control input bound $\lvert {u}_m \rvert \leq 1$m/s, $m\in\{x,y,z\}$, and bounding spheres with radius $r_i = 0.3$m, $\forall i\in\{1,2\}$. We considered two $2$-dimensional scenarios in a workspace $\mathcal{W}$ with ${p_0} = [0,0]^\top$ and $r_0 = 2.5$m. 

The first scenario included $4$ regions of interest $\Pi=\{\pi_1,\dots,\pi_4\}$ in $\mathcal{W}$, with $r_{\pi_k} = 0.4,\forall k\in\{1,\dots,4\}$ and ${p_{\pi_1}} = [0,0]^\top$m, ${p_{\pi_2}} = [-1,0]^\top$m, ${p_{\pi_3}} = [0,1.25]^\top$m and ${p_{\pi_4}} = [0.8,-0.7]^\top$m. The initial positions of the agents were taken such that $\mathcal{A}_1({p_1}(0)) \in\pi_2$ and $\mathcal{A}_2({p_2}(0))\in\pi_4$ (see Fig. \ref{fig:Initial_workspace_exp_1 (ICRA 17)}). We also defined the atomic propositions $\Psi_1 = \Psi_2 = \{``\text{obs}",``a",``b",``c"\}$ with $L_1(\pi_1) = L_2(\pi_1)=\{``\text{obs}"\}, L_1(\pi_2) = L_2(\pi_2) = \{``a"\}, L_1(\pi_3) = L_2(\pi_3)=\{``b"\}, L_1(\pi_4) = L_2(\pi_4)=\{``c"\}$. In this scenario, we were interested in area inspection while avoiding the ``obstacle" region, and thus, we defined the individual specifications with the following LTL formulas: $\mathsf{\Phi}_1 = \mathsf{\Phi}_2 = \square\neg``\text{obs}"\land\square\lozenge(``a"\bigcirc``c"\bigcirc``b")$. By following the procedure described in Section \ref{subsec:High level plan (ICRA 17)}, we obtained the paths $p_1 = (\pi_2\pi_4\pi_3)^{\omega}, p_2 = (\pi_4\pi_2\pi_3)^{\omega}$. Fig. \ref{fig:exp_1_path (ICRA 17)} depicts the execution of the paths $(\pi_2\pi_4\pi_3)^1$ and $(\pi_4\pi_2\pi_3)^1$ by agents $1$ and $2$, respectively, and Fig. \ref{fig:exp_1_vel (ICRA 17)} shows the corresponding input signals, which do not exceed the control bounds $1$m/s. It can be deduced by the figures that the agents successfully satisfy their individual formulas, without colliding with each other. 

The second experimental scenario included $3$ regions of interest $\Pi=\{\pi_1,\dots,\pi_3\}$ in $\mathcal{W}$, with $r_{\pi_k} = 0.4,\forall k\in\{1,\dots,3\}$ and ${p_{\pi_1}} = [-1,-1.7]^\top$m, ${p_{\pi_2}} = [-1.3,1.3]^\top$m and ${p_{\pi_3}} = [1.2,0]^\top$m. The initial positions of the agents were taken such that $\mathcal{A}_1({p_1}(0)) \in\pi_1$ and $\mathcal{A}_2({p_2}(0)) \in\pi_2$ (see Fig. \ref{fig:Initial_workspace_exp_2 (ICRA 17)}). We also defined the atomic propositions $\Psi_1 = \Psi_2 = \{``\text{res}_\text{a}",``\text{res}_\text{b}",``\text{base}"\}$, corresponding to a base and several resources in the workspace, with $L_1(\pi_1) = L_2(\pi_1)=\{``\text{res}_\text{a}"\}, L_1(\pi_2) = L_2(\pi_2)= \{``\text{base}"\}, L_1(\pi_3) = L_2(\pi_3)=\{``\text{res}_\text{b}"\}$. We considered that the agents had to transfer the resources to the ``base" in $\pi_2$; both agents were responsible for $``\text{res}_\text{a}"$ but only agent $1$ should access $``\text{res}_\text{b}"$. The specifications were translated to the formulas $\mathsf{\Phi}_1 = \square(\lozenge(``\text{res}_\text{a}"\bigcirc``\text{base}")\land\lozenge(``\text{res}_\text{b}"\bigcirc``\text{base}")), \mathsf{\Phi}_2 = \square\neg``\text{res}_\text{b}"\land\square\lozenge(``\text{res}_\text{a}"\bigcirc``\text{base}")$ and the derived paths were $p_1 = (\pi_1\pi_2\pi_3\pi_2)^\omega$ and $p_2 = (\pi_1\pi_2)^\omega$. The execution of the paths $(\pi_1\pi_2\pi_3\pi_2)^1$ and $(\pi_2\pi_1)^2$ by agents 1 and 2, respectively, are depicted in Fig. \ref{fig:exp_2_path (ICRA 17)}, and the corresponding control inputs are shown in Fig. \ref{fig:exp_2_vel (ICRA 17)}. The figures demonstrate the successful execution and satisfaction of the paths and formulas, respectively, and the compliance with the control input bounds. 

Regarding the continuous control protocol in the aforementioned experiments, we chose $k_{g_i} = 3, \lambda_i = 2$ in (\ref{eq:feedback_contr (ICRA 17)}), (\ref{eq:feedback_contr_2 (ICRA 17)}) and the switching duration in (\ref{eq:switch_controller (ICRA 17)}) as $\nu_i = 0.1t'_{i,k}, \forall i\in\{1,2\}$.

\begin{figure}
	\begin{minipage}{0.5\linewidth}	
		\centering
		\includegraphics[trim = 0cm -0.2cm 0cm -1cm,width=0.8\textwidth,height=0.8\textwidth]{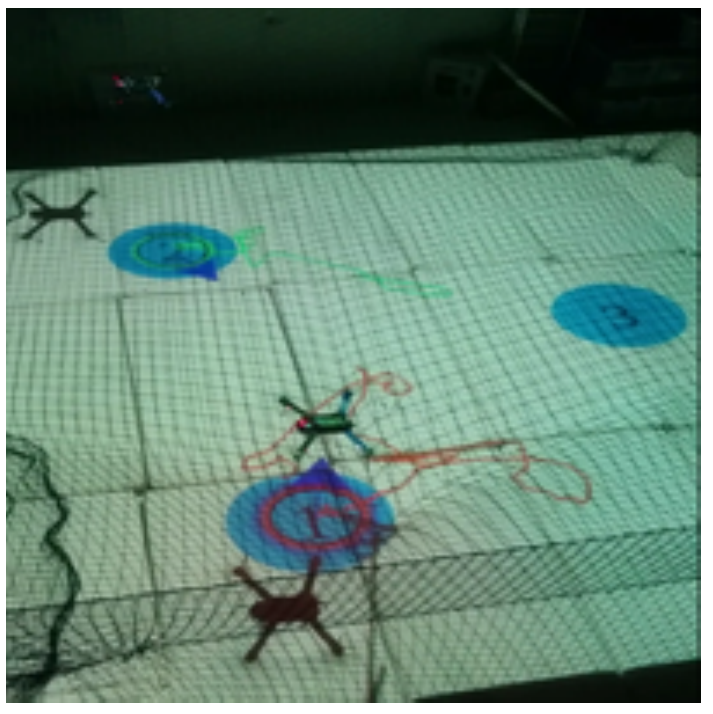}\label{fig:Initial_workspace_exp2_SML (ICRA 17)}
		\subcaption{}
	\end{minipage}
	\begin{minipage}{0.5\linewidth}
		\centering
		\includegraphics[trim = 0cm 0.5cm 0cm -0.5cm,width=0.95\textwidth, height=0.85\textwidth]{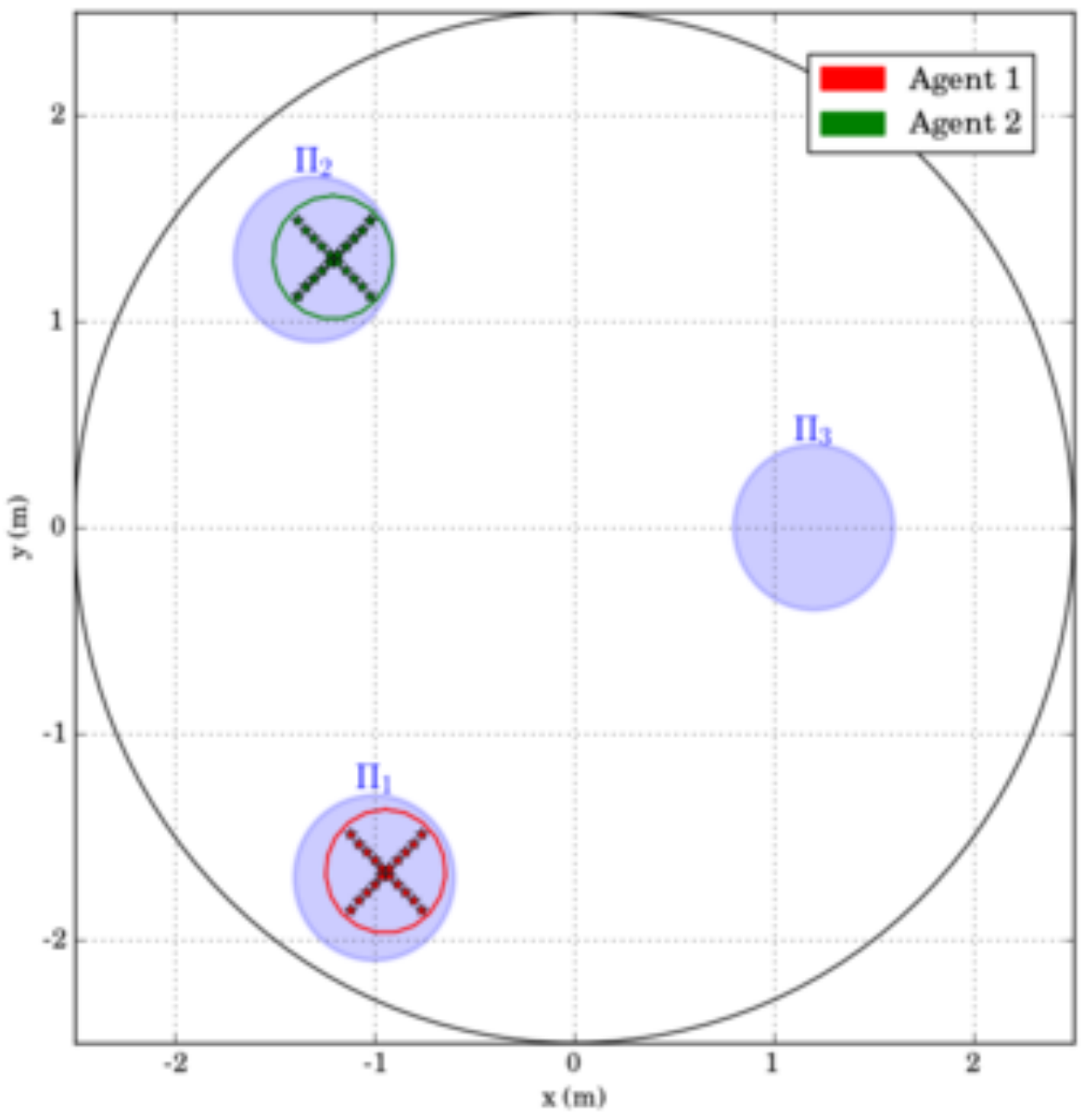}\label{Initial_workspace_exp2_simulated (ICRA 17)}
		\subcaption{}
	\end{minipage}
	\caption{Initial workspace for the second experimental scenario. (a): The UAVs with the projection of their bounding spheres, (with red and green), and the regions of interest (blue disks). (b): Top view of the described workspace. The UAVs are represented by the red and green circled X's and the regions of interest by the blue disks $\pi_1,\dots,\pi_3$.}\label{fig:Initial_workspace_exp_2 (ICRA 17)}
\end{figure}
\begin{figure}[!btp]
	\centering
	\includegraphics[trim = 0cm 0cm 0cm -0.5cm,width = 0.8\textwidth]{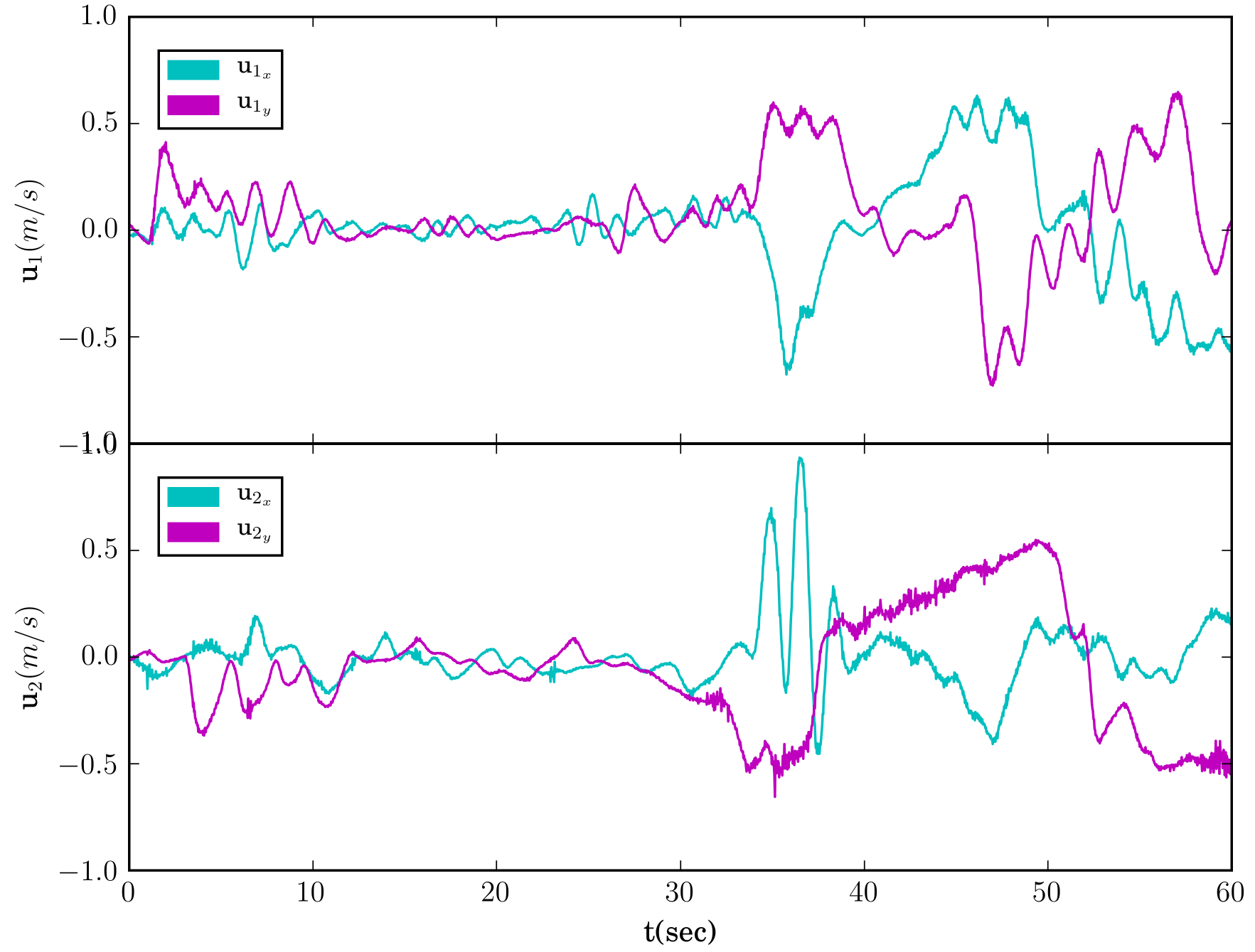}
	
	\caption{The resulting $2$-dimensional control signals of the $2$ agents for the second experimental scenario. Top: agent 1, bottom: agent 2. \label{fig:exp_2_vel (ICRA 17)}}
\end{figure} 

\begin{figure*}[!tbp]
	\begin{minipage}{0.33\linewidth}	
		\centering
		\includegraphics[trim = 0cm 0cm 0cm 0cm,width=0.75\textwidth,height = 0.6\textwidth]{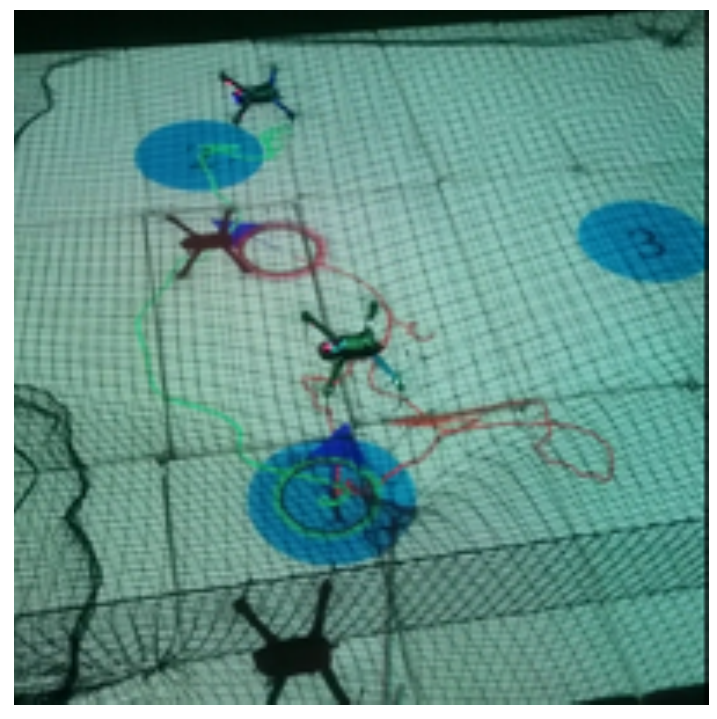}\label{fig:exp_2_path:1 (ICRA 17)}
		\subcaption{}
	\end{minipage}\hfill
	\begin{minipage}{0.33\linewidth}
		\centering	
		\includegraphics[trim = 0cm 0cm 0cm 0cm,width=0.75\textwidth,height = 0.6\textwidth]{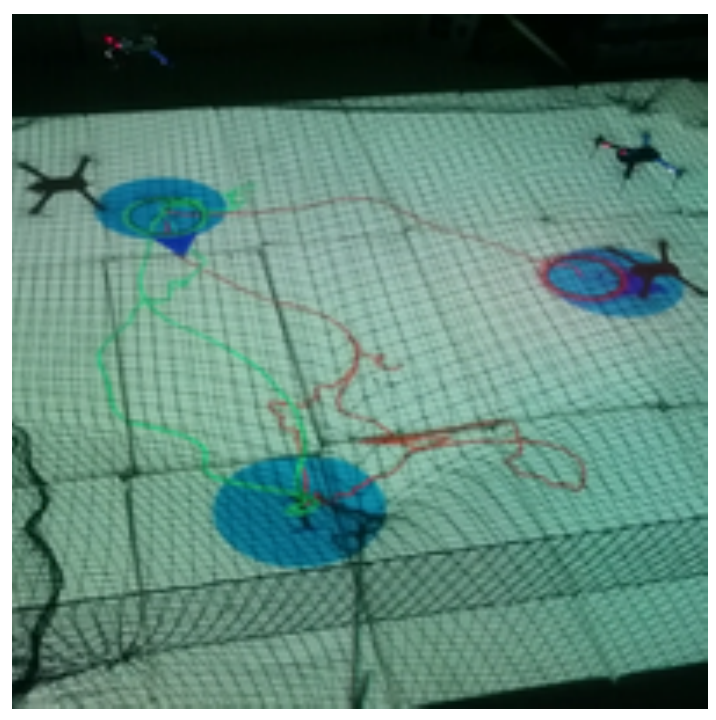}\label{fig:exp_2_path:2 (ICRA 17)}
		\subcaption{}
	\end{minipage}\hfill
	\begin{minipage}{0.33\linewidth}	
		\centering
		\includegraphics[trim = 0cm 0cm 0cm 0cm,width=0.75\textwidth,height = 0.6\textwidth]{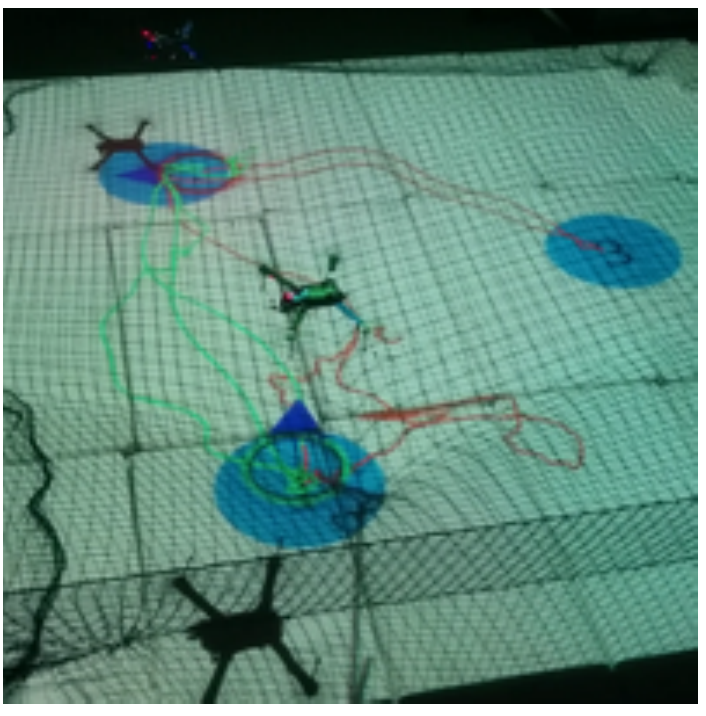}\label{fig:exp_2_path:3 (ICRA 17)}
		\subcaption{}
	\end{minipage}
	\begin{minipage}{0.33\linewidth}	
		\centering
		\includegraphics[trim = 0cm 0cm 0cm -1cm,width=0.95\textwidth,height = 0.8\textwidth]{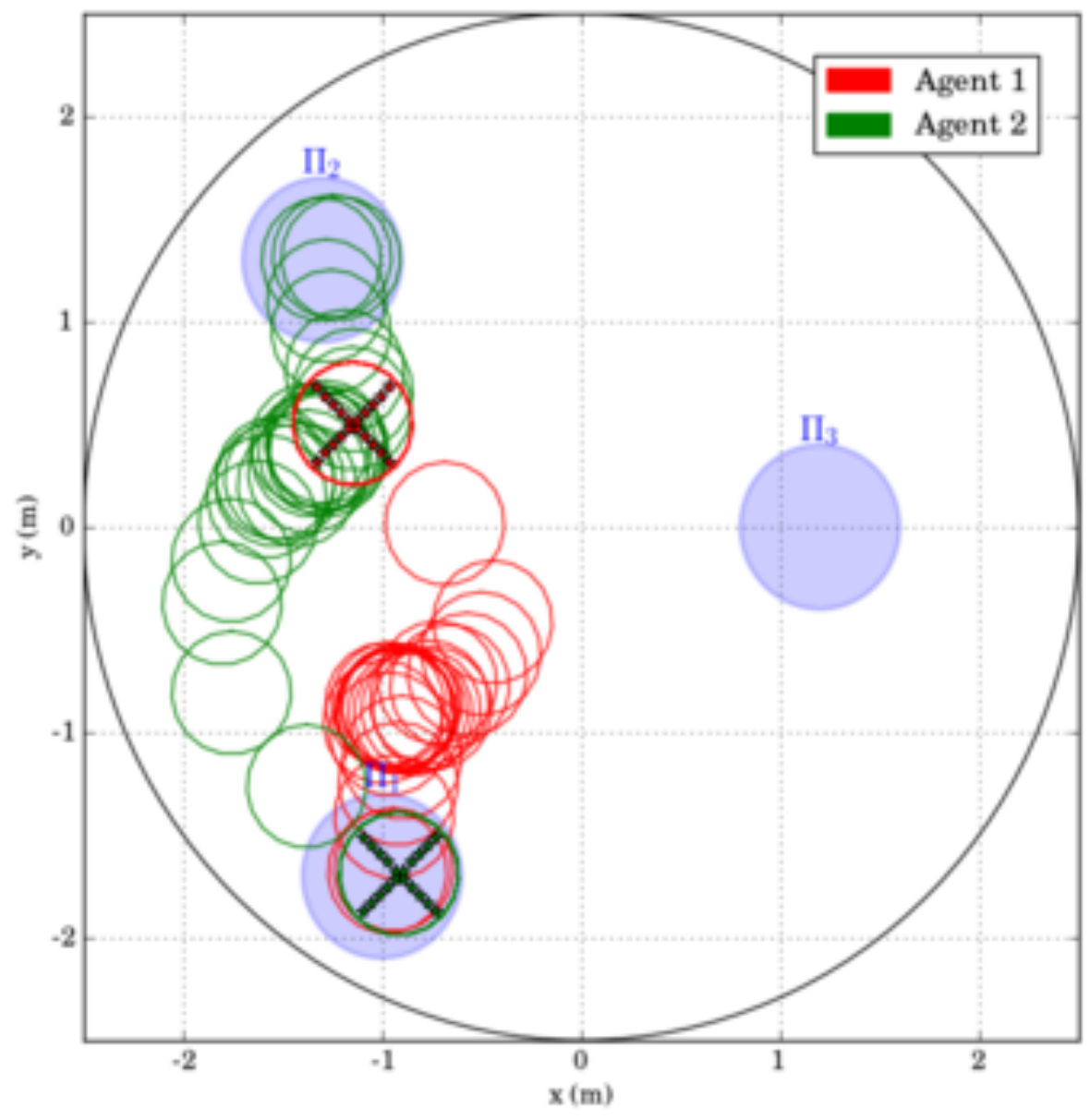}\label{fig:exp_2_path:1_sim (ICRA 17)}
		\subcaption{}
	\end{minipage}\hfill
	\begin{minipage}{0.33\linewidth}	
		\centering
		\includegraphics[trim = 0cm 0cm 0cm -1cm,width=0.95\textwidth,height = 0.8\textwidth]{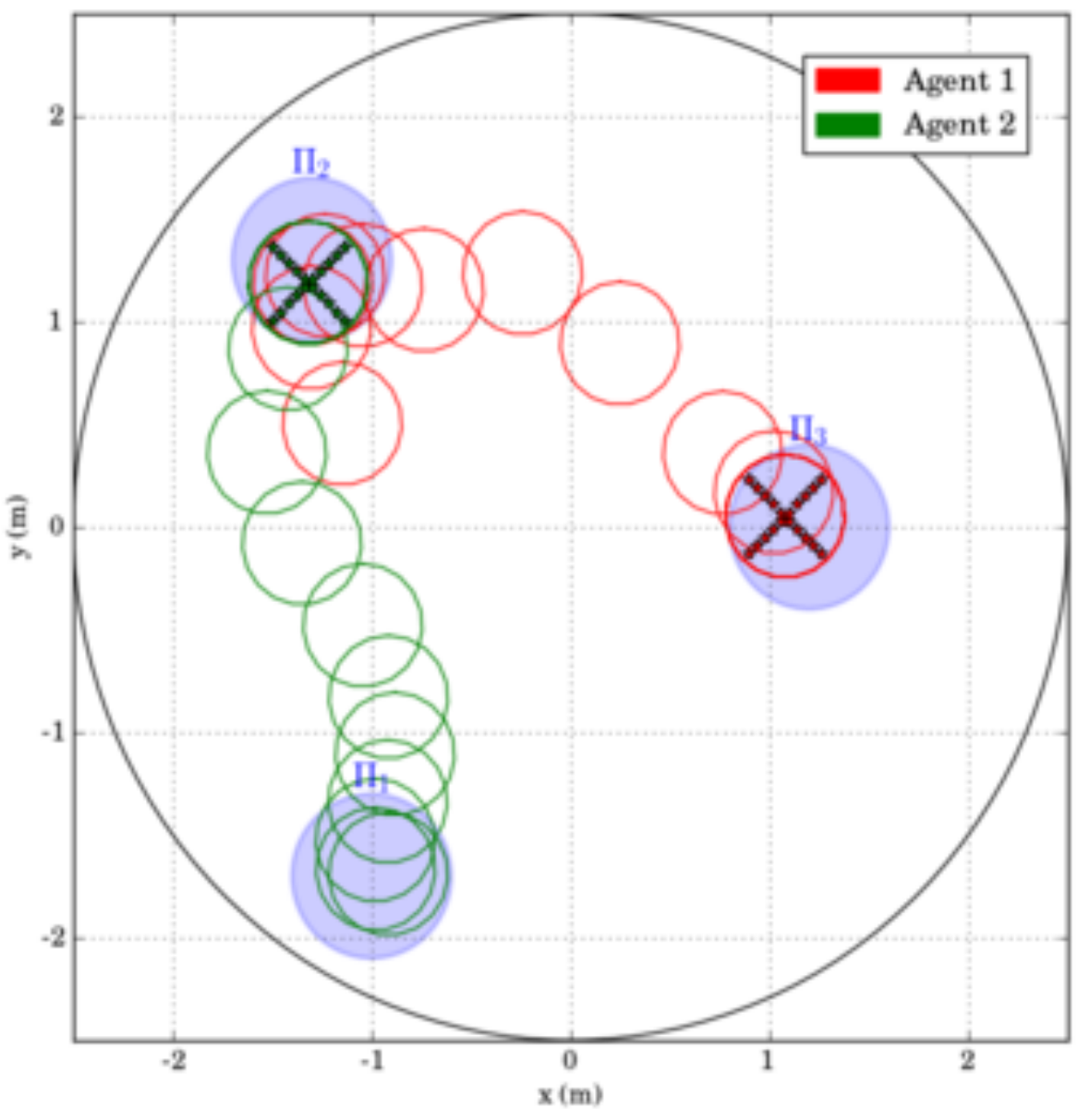}\label{fig:exp_2_path:2_sim (ICRA 17)}
		\subcaption{}
	\end{minipage}\hfill
	\begin{minipage}{0.33\linewidth}	
		\centering
		\includegraphics[trim = 0cm 0cm 0cm -1cm,width=0.95\textwidth,height = 0.8\textwidth]{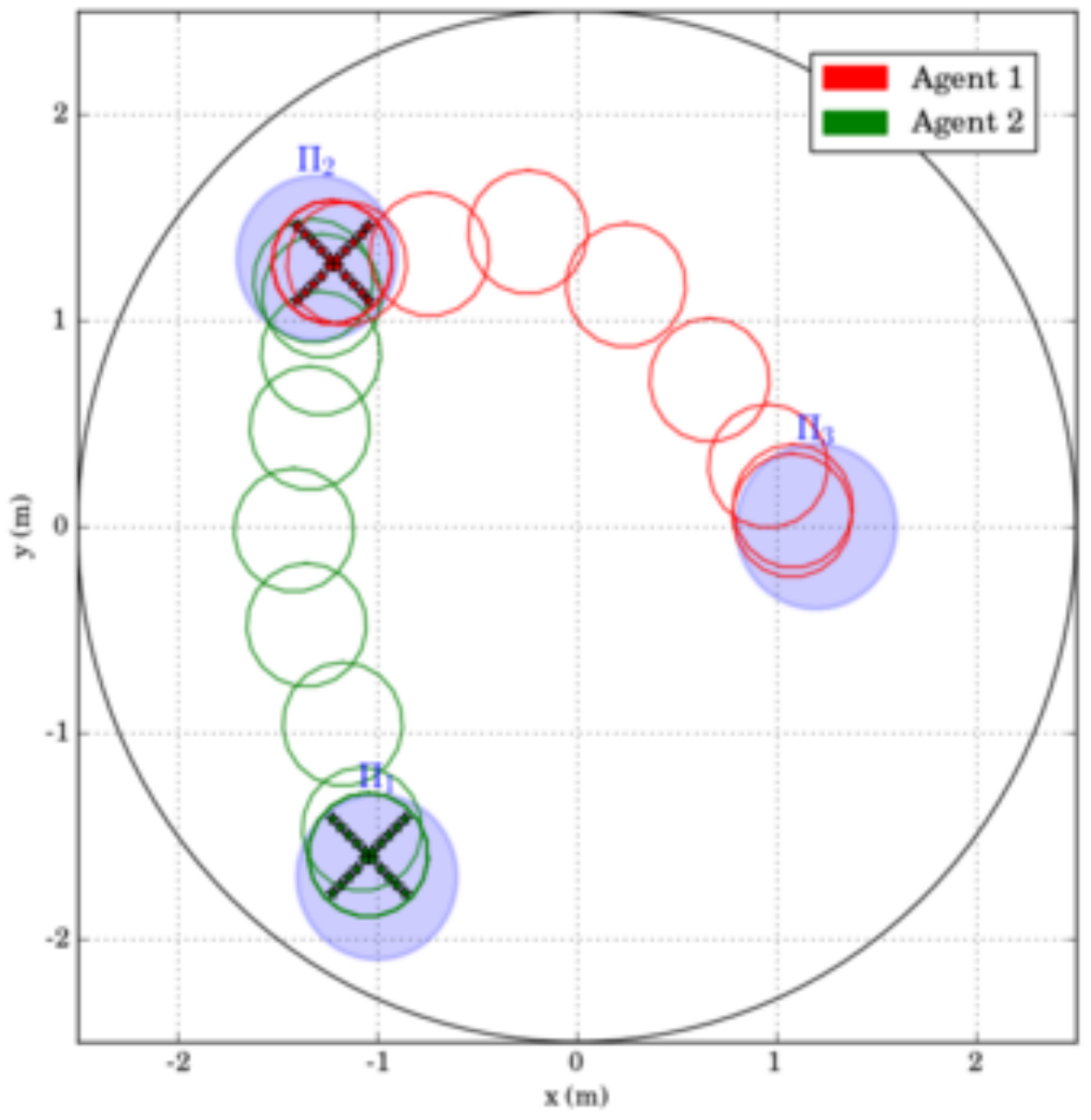}\label{fig:exp_2_path:3_sim (ICRA 17)}
		\subcaption{}
	\end{minipage}
	\caption{Execution of the paths $(\pi_1\pi_2\pi_3\pi_2)^1$ and $(\pi_2\pi_1)^2$ by agents $1$ and $2$, respectively for the second experimental scenario. (a), (d): $\pi_1\rightarrow_1\pi_2, \pi_2\rightarrow_2\pi_1$, (b), (e): $\pi_2\rightarrow_1\pi_3, \pi_1\rightarrow_2\pi_2$, (c), (f): $\pi_3\rightarrow_1\pi_2, \pi_2\rightarrow_2\pi_1$.}\label{fig:exp_2_path (ICRA 17)}
\end{figure*}


The simulations and experiments were conducted in Python environment using an Intel Core i7 2.4 GHz personal computer with 4 GB of RAM, and are clearly demonstrated in the video  found in \href{https://youtu.be/dO77ZYEFHlE}{https://youtu.be/dO77ZYEFHlE}.

\section{Robust Decentralized Abstractions for Multiple Mobile Manipulators}\label{sec:cdc}

We now turn our attention to a class of more complex systems, that is, mobile manipulators, which, unlike the previous section, have more complex and uncertain dynamics. In fact, we provide a more explicit dynamics formulation than just a sphere/ellipsoid, which was done so far. 
We describe a decentralized control algorithm that allows the derivation of a discrete abstraction of the multi-agent dynamics.

\subsection{Problem Formulation} \label{sec:PF (CDC 17)}

\begin{figure}[]
	\vspace{0.4cm}
	\centering
	\includegraphics[trim = 0 0 0 0,scale=0.25]{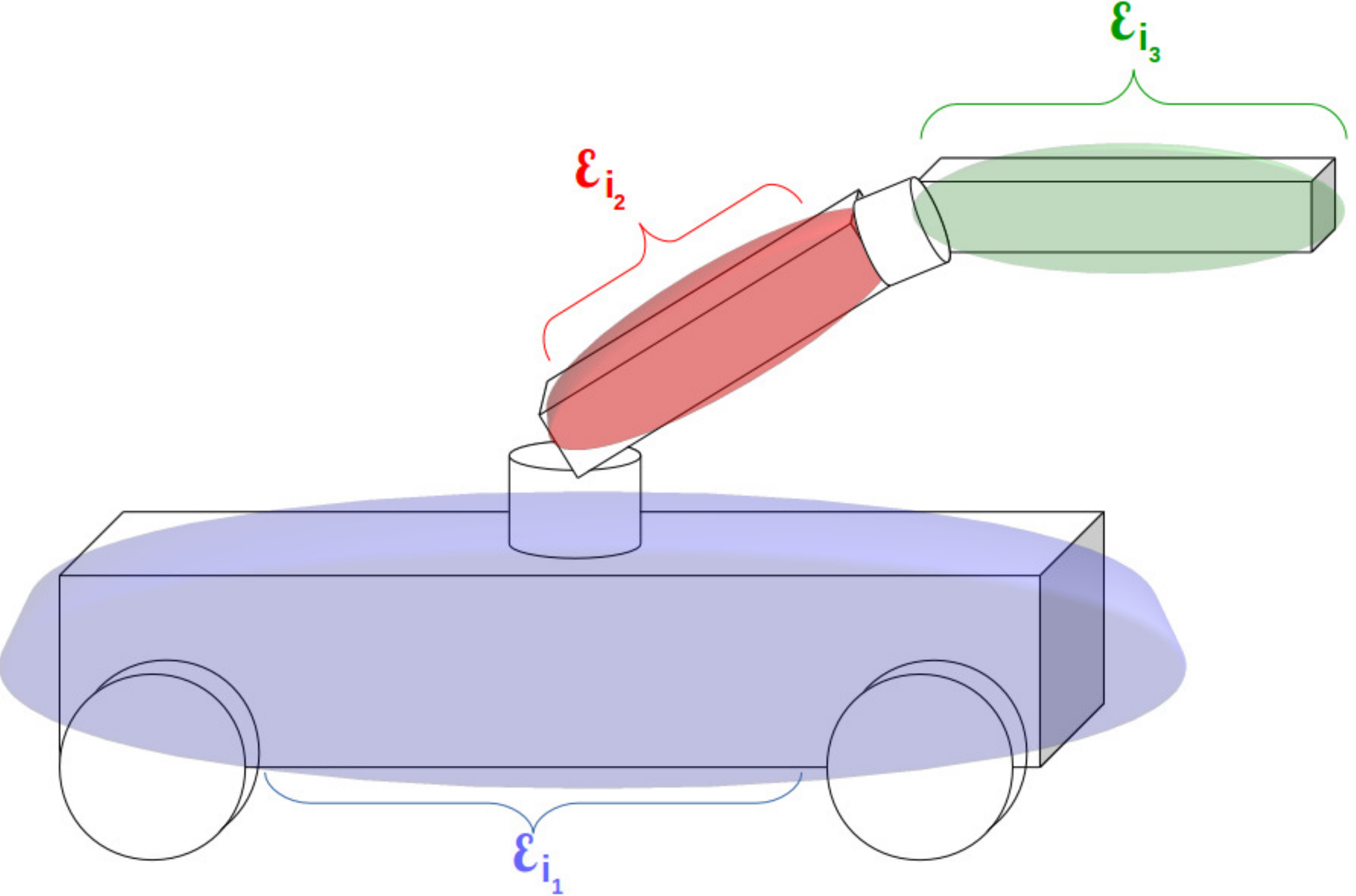}
	\caption{An agent that consists of $\ell_i = 3$ rigid links.\label{fig:Agent (CDC 17)}}
\end{figure}

Like before, consider $N\in\mathbb{N}$ fully actuated agents, with index set $\mathcal{N}$, composed by a robotic arm mounted on an
omnidirectional mobile base, operating in a static workspace $\mathcal{W}$ that is bounded by a large sphere in $3$D space, i.e. $\mathcal{W} = {\mathcal{B}}(p_0,r_0) = \{p\in\mathbb{R}^3 \text{ s.t. } \lVert p - p_0 \rVert < r_0\}$, where $p_0\in\mathbb{R}^3$ is the center of $\mathcal{W}$, and $r_0\in\mathbb{R}_{> 0}$ is its radius. Without loss of generality, we consider that $p_0 = 0$, corresponding to an inertial frame of reference. As in the previous section, we consider that within $\mathcal{W}$ there exist $K$ disjoint spheres around points of interest, which are described by $\pi_k = \bar{\mathcal{B}}(p_{\pi_k},r_{\pi_k}), k\in\mathcal{K}_\mathcal{R}$, where $p_k\in\mathbb{R}^3$ and $r_{\pi_k}\in\mathbb{R}_{>0}$ are the center and radius of the $k$th region, respectively. The regions of interest can be equivalently described by $\pi_k = \{z\in\mathbb{R}^4 \text{ s.t. } z^\top T_{\pi_k}z \leq 0 \}$, where $z=[p^\top, 1]^\top$ is the vector of homogeneous coordinates of $p\in\mathbb{R}^3$, and
\begin{equation*} 
T_{\pi_k} = \begin{bmatrix}
I_3 & p_{\pi_k} \\ 0_{1\times 3} & -r^2_{\pi_k}
\end{bmatrix}, \forall k\in\mathcal{K}_\mathcal{R}.
\end{equation*}
The dynamic model of each agent is given by the second-order Lagrangian dynamics (see \eqref{eq:manipulator joint_dynamics (TCST_coop_manip)} of Chapter \ref{chapter:cooperative manip}):
\begin{equation}
B_i(q_i)\ddot{q}_i + C_{q_i}(q_i,\dot{q}_i)\dot{q}_i + g_{q_i}(q_i) + f_i(q_i, \dot{q}_i) = \tau_i, \label{eq:dynamics (CDC 17)}
\end{equation}
$\forall i\in\mathcal{N}$, where $q_i\in\mathbb{R}^{n_i}$ is the vector of generalized  coordinates (e.g., pose of mobile base and joint coordinates of the arms), with $q\coloneqq [q_1^\top,\dots,q_N^\top]^\top$, and the rest of the terms as in \eqref{eq:manipulator joint_dynamics (TCST_coop_manip)} with a slight change of notation; $f_i(\cdot)$ here represents friction-like terms. Without loss of generality, we assume that $n_i = n\in\mathbb{N},\forall i\in\mathcal{N}$. In addition, we denote by $p_{\scriptscriptstyle B_i}\coloneqq p_{\scriptscriptstyle B_i}(q_i):\mathbb{R}^{n}\to\mathbb{R}^3$ the inertial position of the mobile base of agent $i$. 
Moreover, the matrix $\dot{B}_i - 2C_{q_i}$ is skew-symmetric \cite{siciliano2008springer}, and we further make the following assumption, similar to Assumption \ref{ass:f (Automatica_adaptive_nav)}:
\begin{assumption} \label{ass:f (CDC 17)}
	There exist positive constants $\alpha_i$ such that $\lVert f_i(q_i,\dot{q}_i) \rVert$ $\leq$  $\alpha_i\lVert q_i \rVert \lVert \dot{q}_i \rVert$, $\forall (q_i,\dot{q}_i)$ $\in\mathbb{R}^{n}\times\mathbb{R}^{n}$, $i\in\mathcal{N}$.
\end{assumption}

For the inter-agent collisions, we will use the ideas of Section \ref{sec:Ellipsoids} and ellipsoid collision.
We consider hence that each agent is composed by $\ell_i$ rigid links (see Fig. \ref{fig:Agent (CDC 17)}) with $\mathcal{Q}_i = \{1,\dots,\ell_i\}$ the corresponding index set. Each link of agent $i$ is approximated by the ellipsoid set \cite{choi2009continuous} $\mathcal{E}_{i_m}(q_i) = \{z\in\mathbb{R}^4 \text{ s.t. } z^\top E_{i_m}(q_i)z \leq 0\}$; $z=[p^\top, 1]^\top$ is the homogeneous coordinates of $p\in\mathbb{R}^3$, and $E_{i_m}:\mathbb{R}^n\to\mathbb{R}^{4\times4}$ is defined as $E_{i_m}(q_i) = T^{-T}_{i_m}(q_i)\hat{E}_{i_m}T^{-1}_{i_m}(q_i)$, where $\hat{E}_{i_m} = \text{diag}\{l^{-2}_{i_m,x},l^{-2}_{i_m,y},l^{-2}_{i_m,z},-1\}$ corresponds to the positive lengths $l_{i_m,x},l_{i_m,y},l_{i_m,z}$ of the principal axes of the ellipsoid, and $T_{i_m}:\mathbb{R}^n\to\mathbb{R}^{4\times4}$ is the transformation matrix for the coordinate frame $\{i_m\}$ placed at the center of mass of the $m$-th link of agent $i$, aligned with the principal axes of $\mathcal{E}_{i_m}$:
\begin{equation}
T_{i_m}(q_i) = \begin{bmatrix}
R_{i_m}(q_i) & p_{i_m}(q_i) \\ 0^\top_{1\times 3} & 1
\end{bmatrix}, \notag
\end{equation}
with $R_{i_m}:\mathbb{R}^n\to\mathbb{SO}^{3}$ being the rotation matrix of the link, $\forall m\in\mathcal{Q}_i,i\in\mathcal{N}$.
For an ellipsoid $\mathcal{E}_{i_m}, i\in\mathcal{N},m\in\mathcal{Q}_i$, we denote by $\mathcal{E}^{xy}_{i_m},\mathcal{E}^{xz}_{i_m}, \mathcal{E}^{yz}_{i_m}$ its projections on the planes $x$-$y$, $x$-$z$ and $y$-$z$, respectively, with corresponding matrix terms $E^{xy}_{i_m},E^{xz}_{i_m}, E^{yz}_{i_m}$. 

By following the procedure of Section \ref{sec:Ellipsoids}, we conclude that a sufficient condition for $\mathcal{E}_{i_m}$ and $\mathcal{E}_{j_l}$ not to collide is $\sigma(\Delta^{xy}_{i_m,j_l}) + \sigma(\Delta^{xz}_{i_m,j_l}) + \sigma(\Delta^{yz}_{i_m,j_l}) > 0$, with $\sigma()$ as defined in \eqref{eq:sigma smooth switch} and $\Delta^s_{i_m,j_l}$ is the discriminant of the equation $\det(\lambda E^s_{i_m}(q_i) - E^s_{j_l}(q_j))=0$, $\forall i,j\in\mathcal{N}$, $i\neq j$, $m\in \mathcal{Q}_i$, $l\in\mathcal{Q}_j$, where the subscript $s\in\{xy,yz,xz\}$ stands for the planar ellipsoid matrices.

Next, we define the constant $\bar{d}_{\scriptscriptstyle B_i}$, which is the maximum distance of the base to a point in the agent's volume over all possible configurations, i.e. $\bar{d}_{\scriptscriptstyle B_i} \coloneqq \sup_{\substack{q_i\in\mathbb{R}^n\\ p_i\in\bigcup_{m\in\mathcal{Q}_i} \mathcal{E}_{i_m}(q_i) }}\{ \lVert p_{\scriptscriptstyle B_i}(q_i) - p_i\rVert\}$. We also denote $\bar{d}_{\scriptscriptstyle B}$ $=$ $[\bar{d}_{\scriptscriptstyle B_1},\dots,\bar{d}_{\scriptscriptstyle B_N}]^\top$. 
Moreover, we consider that each agent has a sensor located at the center of its mobile base $p_{\scriptscriptstyle B_i}$ with a sensing radius $\varsigma_i \geq 2\max_{i\in\mathcal{N}}\{ \bar{d}_{\scriptscriptstyle B_i} \} + \varepsilon_d$, where $\varepsilon_d$ is an arbitrarily small positive constant. Hence, each agent has the sensing sphere $\mathcal{D}_i(q_i) \coloneqq \{p\in\mathbb{R}^3 \text{ s.t. } \lVert p - p_{\scriptscriptstyle B_i}(q_i) \rVert \leq \varsigma_i  \}$ and its neighborhood set at each time instant is defined as $\mathcal{N}_i(q_i) \coloneqq  \{j\in\mathcal{N}\backslash\{i\} \text{ s.t. } \lVert p_{\scriptscriptstyle B_i}(q_i) - p_{\scriptscriptstyle B_j}(q_j)\rVert \leq \varsigma_i \}$. 

As in Section \ref{sec:System and PF (ICRA 17)}, we are interested in defining transition systems for the motion of the agents in the workspace in order to be able to assign complex high level goals through logic formulas. Moreover, since many applications necessitate the cooperation of the agents in order to execute some task (e.g. transport an object), we consider that a nonempty subset $\widetilde{\mathcal{N}}_i \subseteq \mathcal{N}_i(q_i(0)), i\in\mathcal{N}$, of the initial neighbors of the agents must stay connected through their motion in the workspace, similarly to Section \ref{sec:LF}. In addition, it follows that the transition system of each agent must contain information regarding the current position of its neighbors. The problem in hand is equivalent to designing decentralized control laws $\tau_i,i\in\mathcal{N}$, for the appropriate transitions of the agents among the predefined regions of interest in the workspace.

Next, we provide the equivalent definitions to Def. \ref{def:agent in region (ICRA 17)} and \ref{def:agent_transition (ICRA 17)}. 
\begin{definition} \label{def:in region (CDC 17)}
	An agent $i\in\mathcal{N}$ is in region $k \in\mathcal{K}_\mathcal{R}$ at a configuration $q_i\in\mathbb{R}^n$, denoted as $\mathcal{A}_i(q_i)\in\pi_k$, if and only if 
	$\mathcal{E}_{i_m}(q_i) \subset \pi_k$, $\forall m\in\mathcal{Q}_i$.
\end{definition}

\begin{definition} 
	Agents $i,j\in\mathcal{N}$, with $i\neq j$, are in \textit{collision-free} configurations $q_i,q_j\in\mathbb{R}^n$, denoted as $\mathcal{A}_i(q_i)\not \equiv\mathcal{A}_j(q_j)$, if and only if $\mathcal{E}_{i_m}(q_i)\cap\mathcal{E}_{j_l}(q_j)=\emptyset, \forall m\in\mathcal{Q}_i,l\in\mathcal{Q}_j$.
\end{definition}

Given the aforementioned discussion, we make the following assumptions regarding the agents and the validity of the workspace 
\begin{assumption} \label{ass:validity (CDC 17)}	
	The regions of interest are
	\begin{enumerate}[(i)]
		\item  large enough such that all the agents can fit, i.e., given a specific $k\in\mathcal{K}_\mathcal{R}$, there exist $q_i, i\in\mathcal{N}$, such that 
		$\mathcal{A}_i(q_i)\in\pi_k$, $\forall i\in\mathcal{N}$, with $\mathcal{A}_i(q_i)\not\equiv\mathcal{A}_j(q_j)$, $\forall i,j\in\mathcal{N}$, with $i\neq j$. 
		\item sufficiently far from each other and the obstacle workspace, i.e., 
		\begin{align*}
		& \lVert p_{\pi_k} - p_{{\pi_k}'} \rVert \geq \max\limits_{i\in\mathcal{N}}\{ 2\bar{d}_{\scriptscriptstyle B_i}\} + r_{\pi_k} + r_{\pi_k'} + \epsilon_{\pi}, \notag \\
		& r_0 - \| p_k \| \geq \max\limits_{i\in\mathcal{N}}\{ 2\bar{d}_{\scriptscriptstyle B_i}\},  
		\end{align*}
		$\forall k,k'\in\mathcal{K}_\mathcal{R}, k\neq k'$, where $\epsilon_{\pi}$ is an arbitrarily small positive constant.
	\end{enumerate}
\end{assumption} 

Next, in order to proceed, we need the following definition.

\begin{definition}\label{def:agent transition (CDC 17)}
	Assume that $\mathcal{A}_i(q_i(t_0))\in\pi_k, i\in\mathcal{N}$, for some $t_0\in\mathbb{R}_{\geq 0},k\in\mathcal{K}_\mathcal{R}$, with $\mathcal{A}_i(q_i(t_0))\not \equiv\mathcal{A}_j(q_j(t_0)), \forall j\in\mathcal{N}\backslash\{i\}$. There exists a transition for agent $i$ between $\pi_k$ and $\pi_{k'}, k'\in\mathcal{K}_\mathcal{R}$, denoted as $(\pi_k,t_0)\xrightarrow{i}(\pi_{k'},t_f)$, if and only if there exists a finite time $t_f\geq t_0$, such that 
	\begin{itemize}
		\item $\mathcal{A}_i(q_i(t_f))\in\pi_{k'}$
		\item $\mathcal{A}_i(q_i(t))\not \equiv\mathcal{A}_j(q_j(t))$, $\forall j\in\mathcal{N}\backslash \{i\}$, 
		\item $\mathcal{E}_{i_m}(q_i(t))\cap\mathcal{E}_{i_\ell}(q_i(t))$, $\forall m,\ell\in\mathcal{Q}_i, m\neq \ell$,
		\item $\mathcal{E}_{i_m}(q_i(t))\cap\pi_{z} = \emptyset$, $\forall m\in\mathcal{Q}_i, z\in\mathcal{K}_\mathcal{R}\backslash\{k,k'\}$,
		\item $\mathcal{E}_{i_m}(q_i(t)) \subset \mathcal{W}$, $\forall m\in\mathcal{Q}_i$,
	\end{itemize} 
	$\forall t\in[t_0,t_f]$.
\end{definition}

Given the aforementioned definitions, the treated problem is the design of decentralized control laws for the transitions of the agents between two regions of interest in the workspace, while preventing collisions of the agents with each other, the workspace boundary, and the remaining regions of interest. More specifically, we aim to design a finite transition system for each agent of the form \cite{baier2008principles}
\begin{equation}
\mathcal{T}_i = (\Pi, \Pi_{i,0}, \xrightarrow{i}, \Psi_i, \mathcal{L}_i, \mathcal{H}_i), \label{eq:TS (CDC 17)}
\end{equation}
where $\Pi = \{\pi_1,\dots,\pi_K\}$ is the set of regions of interest that the agents can be at, according to Def. \ref{def:in region (CDC 17)}, $\Pi_{i,0}\subseteq \Pi$ is a set of initial regions that each agent can start from, $\xrightarrow{i}\subset(\Pi\times\mathbb{R}_{\geq 0})^2$ is the transition relation of Def. \ref{def:agent transition (CDC 17)}, $\Psi_i$ is a set of given atomic propositions, represented as boolean variables, that hold in the regions of interest, $\mathcal{L}_i:\Pi\to2^{\Psi_i}$ is a labeling function, and $\mathcal{H}_i:\Pi\to\Pi^{\lvert \widetilde{\mathcal{N}}_i \rvert}$ is a function that maps the region that agent $i$ occupies to the regions the initial neighbors $\widetilde{\mathcal{N}}_i$ of agent $i$ are at. Therefore, the treated problem is the design of bounded controllers $\tau_i$ for the establishment of the transitions $\xrightarrow{i}$.
Moreover, as discussed before, the control protocol should also guarantee the connectivity maintenance of a subset of the initial neighbors $\widetilde{\mathcal{N}	}_i,\forall i\in\mathcal{N}$. Another desired property important in applications involving robotic manipulators, is the nonsingularity of the Jacobian matrix $J_i:\mathbb{R}^n\to\mathbb{R}^{6\times n}$, that transforms the generalized coordinate rates of agent $i\in\mathcal{N}$ to generalized velocities \cite{siciliano2008springer} (also defined in Chapter \ref{chapter:cooperative manip}). That is, the agents should always remain in the sets $\mathsf{S}_i = \{q_i\in\mathbb{R}^{n} \text{ s.t. } \det(J_i(q_i)J_i(q_i)^\top) > 0\}$, $\forall i\in\mathcal{N}$.  

Formally, we define the problem treated in this section as follows:

\begin{problem} \label{Problem (CDC 17)}
	Consider $N$ mobile manipulators with dynamics \eqref{eq:dynamics (CDC 17)} and $K$ regions of interest $\pi_k,k\in\mathcal{K}_\mathcal{R}$, with $\dot{q}_i(t_0) < \infty, \mathcal{A}_i(q_i(t_0))\in \pi_{k_i}, k_i\in\mathcal{K}_\mathcal{R}$, $q_i(t_0)\in\mathsf{S}_i$, $\forall i\in\mathcal{N}$ and $\mathcal{A}_i(q_i(t_0))\not\equiv\mathcal{A}_j(q_j(t_0)), \mathcal{E}_{i_m}(q_i(t_0))\cap\mathcal{E}_{i_\ell}(q_i(t_0)) =\emptyset, \forall i,j \in\mathcal{N},i\neq j, m,\ell\in\mathcal{Q}_i,m\neq\ell$. 
	
	Given nonempty subsets of the initial edge sets $\widetilde{\mathcal{N}}_{i}\subseteq\mathcal{N}_i(q_i(0))\subseteq\mathcal{N}, \forall i\in\mathcal{N}$, as well as the indices $k'_{i}\in\mathcal{K}_\mathcal{R},i\in\mathcal{N}$, such that $\lVert p_{k'_i} - p_{k'_j}\rVert + r_{\pi_{k'_i}} + r_{\pi_{k'_j}} \leq \varsigma_i, \forall j\in\widetilde{\mathcal{N}}_i,i\in\mathcal{N}$, design decentralized controllers $\tau_i$ such that, for all $i\in\mathcal{N}$: 
	\begin{enumerate}
		\item $(\pi_{k_i},t_0) \xrightarrow{i} (\pi_{k'_i},t_{f_i})$, for some $t_{f_i}\geq t_0$,
		\item $r_0 - (\lVert p_{\scriptscriptstyle B_i}(t) \rVert  + \bar{d}_{\scriptscriptstyle B_i}) > 0,\forall t\in [t_0,t_{f_i}]$,
		\item $j_i^*\in{\mathcal{N}}_i(q_i(t)), \forall j_i^*\in\widetilde{\mathcal{N}}_i, t\in [t_0,t_{f_i}]$,
		\item $q_i(t) \in \mathsf{S}_i, \forall t\in [t_0,t_{f_i}]$.
	\end{enumerate} 
\end{problem}

The aforementioned specifications concern 1) the agent transitions according to Def. \ref{def:agent transition (CDC 17)}, 2) the confinement of the agents in $\mathcal{W}$, 3) the connectivity maintenance between a subset of initially connected agents and 4) the agent singularity avoidance. Moreover, the fact that the initial edge sets $\widetilde{\mathcal{N}}_i$ are nonempty implies that the sensing radius of each agent $i$ covers the regions $\pi_{k_j}$ of the agents in the neighboring set $\widetilde{\mathcal{N}}_i$. Similarly, the condition $\lVert p_{k'_i} - p_{k'_j}\rVert + r_{\pi_{k'_i}} + r_{\pi_{k'_j}} \leq \varsigma_i, \forall j\in\widetilde{\mathcal{N}}_i$, is a feasibility condition for the goal regions, since otherwise it would be impossible for two initially connected agents to stay connected. Intuitively, the sensing radii $\varsigma_i$ should be large enough to allow transitions of the multi-agent system to the entire workspace. 

\subsection{Problem Solution}  \label{sec:main results (CDC 17)}


To solve Problem \ref{Problem (CDC 17)}, we use the concept of potential fields, as done in  Section \ref{sec:icra}. Nevertheless, we do not provide an explicit closed-form expression of the potential function, but provide appropriate conditions. 

Let $\varphi_i$ be a \emph{decentralized potential function}, with the following properties: 
\begin{enumerate}[(i)]
	\item The function $\varphi_i(q)$ is not defined, i.e., $\varphi_i(q) = \infty$, $\forall i\in\mathcal{N}$, when a collision or a connectivity break occurs,
	\item The critical points of $\varphi_i$ where the vector field $\nabla_{q_i}\varphi_i(q)$ vanishes, i.e., the points where $\nabla_{q_i}\varphi_i(q) = 0$, consist of the goal configurations and a set of configurations whose region of attraction (by following the negated vector field curves) is a set of measure zero.	
	\item It holds that $\nabla_{q_i}\varphi_i(q) + \sum_{j\in\mathcal{N}_i(q_i)}\nabla_{q_i}\varphi_j(q) = 0$ $\Leftrightarrow$ $\nabla_{q_i}\varphi_i(q) = 0$ and \\ $\sum_{j\in\mathcal{N}_i(q_i)}\nabla_{q_i}\varphi_j(q) = 0$, $\forall i\in\mathcal{N},q\in \mathbb{R}^{Nn} $.
\end{enumerate}
More specifically, $\varphi_i(q)$ is a function of two main terms, a  \emph{goal function} $\gamma_i:\mathbb{R}^{n}\to\mathbb{R}_{\geq 0}$, which should vanish when $\mathcal{A}_i(q_i)\in\pi_{k'_i}$, and an \emph{obstacle function}, 
$\beta_i:\mathbb{R}^{Nn}\to\mathbb{R}_{\geq 0}$ that encodes inter-agent collisions, collisions between the agents and the obstacle boundary/undesired regions of interest, connectivity losses between initially connected agents and singularities of the Jacobian matrix $J_i(q_i)$; 
Next, we provide an analytic construction of the goal and obstacle terms. However, the construction of the function $\varphi_i$ is not taken into account. 

The control objective of agent $i$, i.e., reaching the region of interest $\pi_{k'_i}$, is encoded in the function $\gamma_i\coloneqq \gamma_i(q_i):\mathbb{R}^n\to\mathbb{R}_{\geq 0}$, defined as 
\begin{equation*} 
\gamma_i(q_i) \coloneqq \lVert q_i - q_{k'_i} \rVert^2,
\end{equation*}
where $q_{k'_i}$ is a configuration such that $r_{\pi_{k'_i}} - \|p_{\scriptscriptstyle B_i}(q_{k'_i}) - p_{k'_i} \| \leq \bar{d}_{\scriptscriptstyle B_i} - \epsilon_q$, for an arbitrarily small positive constant $\epsilon_q$, which implies $\mathcal{A}_i(q_{k'_i})\in \pi_{k'_i}$, $\forall i\in\mathcal{N}$. In case that multiple agents have the same target, i.e., there exists at least one $j\in\mathcal{N}\backslash\{i\}$ such that $\pi_{k'_j} = \pi_{k'_i}$, then we assume that $\mathcal{A}_i(q_{k'_i})\not\equiv \mathcal{A}_j(q_{k'_j})$. 


Inter-agent collisions, collisions with the boundary of the workspace and the undesired regions of interest, connectivity between initially connected agents and singularities of the Jacobian matrix $J_i(q_i),\forall i\in\mathcal{N}$ are encoded by a function $\beta_i$, defined next.

As mentioned before, a sufficient condition for ellipsoids $\mathcal{E}_{i_m}$ and $\mathcal{E}_{j_l}$ not to collide, is $\Delta_{i_m,j_l}(q_i,q_j) = \sigma(\Delta^{xy}_{i_m,j_l}(q_i,q_j) + \sigma(\Delta^{xz}_{i_m,j_l}(q_i,q_j)) + \sigma(\Delta^{yz}_{i_m,j_l}(q_i,q_j)$ $>$ $0$, $\forall m\in\mathcal{Q}_i,l\in\mathcal{Q}_j,i,j,\in\mathcal{N}$, and $\sigma$ as defined in
\eqref{eq:sigma smooth switch}.

Additionally, we define the greatest lower bound of the $\Delta_{i_m,j_l}$ when the point $p_{j_l}$ is on the boundary of the sensing radius $\partial D_i(q_i)$ of agent $i$, as $\widetilde{{\Delta}}_{i_m,j_l} = \inf_{\substack{(q_i,q_j)\in\mathbb{R}^{2n}\\\lVert p_{\scriptscriptstyle B_i}(q_i) - p_{j_l}(q_j) \rVert = \varsigma_i}}\{\Delta_{i_m,j_l}(q_i,q_j)\}, \forall m\in\mathcal{Q}_i,l\in\mathcal{Q}_j,i,j\in\mathcal{N}$. Since $\varsigma_i > 2\max_{i\in\mathcal{N}}\{\bar{d}_{\scriptscriptstyle B_i}\} + \epsilon_d$, it follows that there exists a positive constant $\epsilon_\Delta$ such that $\widetilde{\Delta}_{i_m,j_l} \geq \epsilon_\Delta > 0,\forall m\in\mathcal{Q}_i,l\in\mathcal{Q}_j,i,j\in\mathcal{N}, i\neq j$.


We further define the function $\zeta_{ij}\coloneqq\zeta_{ij}(q_i,q_j):\mathbb{R}^{n}\times\mathbb{R}^{n}\to\mathbb{R}$, with $\zeta_{ij}(q_i,q_j) \coloneqq \varsigma^2_i-\lVert p_{\scriptscriptstyle B_i}(q_i) - p_{\scriptscriptstyle B_j}(q_j)\rVert^2$, and the distance functions $\beta_{\mathfrak{c},i_m,j_l}\coloneqq \beta_{\mathfrak{c},i_m,j_l}(\Delta_{i_m,j_l}):\mathbb{R}_{\geq 0}\to\mathbb{R}_{\geq 0}, \beta_{\mathfrak{n},ij}\coloneqq \beta_{\mathfrak{n},ij}(\zeta_{ij}):\mathbb{R}\to\mathbb{R}_{\geq 0}$, $\beta_{iw}\coloneqq \beta_{iw}(\|p_{\scr B_i}\|^2):\mathbb{R}_{\geq 0}\to\mathbb{R}$ as 
\begin{align}
\beta_{\mathfrak{c},i_m,j_l}(\Delta_{i_m,j_l}) &\coloneqq 
\begin{cases}
\vartheta_{\mathfrak{c},i}(\Delta_{i_m,j_l}), &  0 \leq \Delta_{i_m,j_l} < \bar{\Delta}_{i_m,j_l}, \\
\bar{\Delta}_{i_m,j_l}, &  \bar{\Delta}_{i_m,j_l} \leq \Delta_{i_m,j_l}, \\
\end{cases} \notag \\
\beta_{\mathfrak{n},ij}(\zeta_{ij}) &\coloneqq 
\begin{cases}
0, &  \eta_{ij,c} < 0,\\
\vartheta_{\mathfrak{n},i}(\zeta_{ij}), &   0 \le \eta_{ij,c} < \varsigma^2_i, \\
d^2_{\text{con}_i}, &  \varsigma^2_i \le \zeta_{ij},  
\end{cases} \notag \\
\beta_{iw}(\lVert p_{\scriptscriptstyle B_i} \rVert^2) &=  (r_0 - \bar{d}_{\scriptscriptstyle B_i})^2 - \lVert p_{\scriptscriptstyle B_i} \rVert^2, \notag 
\end{align}
where $\bar{\Delta}_{i_m,j_l}$ is a constant satisfying $0 < \bar{\Delta}_{i_m,j_l} \leq \widetilde{\Delta}_{i_m,j_l}, \forall m\in\mathcal{Q}_i,l\in\mathcal{Q}_j,i,j\in\mathcal{N},i\neq j$, and $\vartheta_{\mathfrak{c},i}, \vartheta_{\mathfrak{n},i}$ are \textit{strictly increasing} polynomials appropriately selected to guarantee that the functions $\beta_{\mathfrak{c},i_m,j_l}$, and $\beta_{\mathfrak{n},ij}$, respectively, are twice continuously differentiable everywhere, with $\vartheta_{\mathfrak{c},i}(0) = \vartheta_{\mathfrak{n},i}(0) = 0, \forall i\in\mathcal{N}$. 
Note that the functions defined above use only local information in the sensing range $\varsigma_i$ of agent $i$. Similarly, $\beta_{iw}$ encodes the collision of agent $i$ with the workspace boundary. 

Finally, we choose the function $\beta_i\coloneqq \beta_i(q):\mathbb{R}^{Nn}\to\mathbb{R}_{\geq 0}$ as 

\begin{align*}
\beta_i(q) =& (\det(J_i(q_i)J_i(q_i)^\top))^2\beta_{iw}(\lVert p_{\scriptscriptstyle B_i} \rVert^2)\prod\limits_{j\in\widetilde{\mathcal{N}}_i}\beta_{\mathfrak{n},ij}(\zeta_{ij}) \notag\\ &\prod\limits_{(m,j,l)\in\widetilde{T}}\beta_{\mathfrak{c},i_m,j_l}(\Delta_{i_m,j_l})\prod\limits_{ (m,k)\in\widetilde{L}} \Delta_{i_m,\pi_k}(q_i), 
\end{align*}
$\forall i\in\mathcal{N}$, where $\widetilde{T} \coloneqq \mathcal{Q}_i\times\mathcal{N}\times\mathcal{Q}_j, \widetilde{L} \coloneqq \mathcal{Q}_i\times(\mathcal{K}_\mathcal{R}\backslash\{k_i,k'_i\})$.  Note that we have included the term $(\det(J_iJ^\top_i))^2$ to also account for singularities of $J_i, \forall i\in\mathcal{N}$ and the term $\prod_{(m,j,l)\in\widetilde{T}}\beta_{\mathfrak{c},i_m,j_l}(\Delta_{i_m,j_l})$ takes into account also the collisions between the ellipsoidal rigid bodies of agent $i$.

With the introduced notation, the properties of the functions $\varphi_i$ are: 
\begin{enumerate}[(i)]
	\item $\beta_i(q)\to 0 \Leftrightarrow (\varphi_i(q) \to \infty), \forall i\in\mathcal{N}$,
	\item $ \nabla_{q_i}\varphi_i(q)|_{q_i=q^\star_i} = 0,  \forall q^\star_i\in\mathbb{R}^n \text{ s.t. } \gamma_i(q^\star_i) = 0$ and the regions of attraction of the points $\{q \in\mathbb{R}^{Nn}: \nabla_{q_i}\varphi_i(q)|_{q_i=\widetilde{q}_i} = 0, \gamma_i(\widetilde{q}_i) \neq 0 \}, i\in\mathcal{N}$, are sets of measure zero.
\end{enumerate}

By further denoting $\mathbb{D}_i = \{q\in\mathbb{R}^{Nn}: \beta_i(q) > 0 \}$, we are ready to state the main theorem of this section:

\begin{theorem}
	Under the Assumptions \ref{ass:f (CDC 17)}-\ref{ass:validity (CDC 17)}, the decentralized control laws $\tau_i\coloneqq \tau_i(q,\dot{q}_i,\hat{\alpha}_i): \mathbb{D}_i\times\mathbb{R}^{n+1}\to \mathbb{R}^n$, with	
	\begin{align}
	& \tau_i(q,\dot{q}_i,\hat{\alpha}_i) = g_i(q_i) - \nabla_{q_i}\varphi_i(q) - \sum\limits_{j\in\mathcal{N}_i(q_i)}\nabla_{q_i}\varphi_j(q)  - \hat{\alpha}_i\lVert q_i \rVert \dot{q}_i - k_{v_i} \dot{q}_i, \label{eq:control law (CDC 17)}
	\end{align}
	with $k_{v_i}$ positive gain constants, $\forall i\in\mathcal{N}$, along with the adaptation laws 
	\begin{equation}
	\dot{\hat{\alpha}}_i = k_{\alpha_i}\lVert \dot{q}_i\rVert^2\lVert q_i\rVert, \label{eq:adaptation law (CDC 17)} 
	\end{equation}
	with $\hat{\alpha}_i(t_0) < \infty, k_{\alpha_i} \in\mathbb{R}_{\geq 0}$ positive gain constants, $\forall i\in\mathcal{N}$, guarantee the transitions $(\pi_{k_i},t_0)\xrightarrow{i}(\pi_{k'_i},t_{f_i})$ for finite $t_{f_i},i\in\mathcal{N}$ for almost all initial conditions, while ensuring $\beta_i > 0,\forall i\in\mathcal{N}$, as well as the boundedness of all closed loop signals, providing, therefore, a solution to Problem \ref{Problem (CDC 17)}.
\end{theorem}
\begin{proof}
	
	The closed loop system of \eqref{eq:dynamics (CDC 17)} is written as:
	\begin{align}
	M_i(q_i)\ddot{q}_i + N_i(q_i,\dot{q}_i)\dot{q}_i + f_i(q_i,\dot{q}_i) =& -\nabla_{q_i}\varphi_i(q_i) - k_{v_i}\dot{q}_i 
	- \hat{\alpha}\lVert q_i \rVert\dot{q}_i  -\notag  \\
	& \sum\limits_{j\in\mathcal{N}_i(q_i)}\nabla_{q_i}\varphi_j(q), \label{eq:closed loop dynamics (CDC 17)}
	\end{align}	
	$\forall i\in\mathcal{N}$.
	Due to Assumption \ref{ass:validity (CDC 17)}, the domain where the functions $\varphi_i(q)$ are well-defined (i.e., where $\beta_i > 0$) is connected. Hence,
	consider the Lyapunov-like function $V\coloneqq V(\varphi, \dot{q}, \widetilde{\alpha},q):\mathbb{R}^{N}\times\mathbb{R}^{Nn}\times\mathbb{R}^N\times\mathbb{D}_1\times\dots\times\mathbb{D}_N\to\mathbb{R}_{\geq 0}$, with
	\begin{align}
	V \coloneqq & \sum\limits_{i\in\mathcal{N}} \varphi_i(q) + \frac{1}{2}\dot{q}^\top_iM_i(q_i)\dot{q}_i + \frac{1}{2k_{\alpha_i}}\widetilde{\alpha}_i^2 \notag
	\end{align}
	where $\varphi$ and $\widetilde{\alpha}$ are the stack vectors containing all $\varphi_i$ and $\widetilde{\alpha}_i$, respectively, $i\in\mathcal{N}$, and $\widetilde{\alpha}_i \coloneqq \hat{\alpha}_i - \alpha_i, \forall i\in\mathcal{N}$. Note that, since there are no collision or singularities at $t_0$, the functions $\beta_i(q), i\in\mathcal{N}$, are strictly positive at $t_0$ which implies the boundedness of $V$ at $t_0$. Therefore, since $\dot{q}_i(t_0)<\infty$ and $\hat{\alpha}_i(t_0)<\infty, \forall i\in\mathcal{N}$, there exists a positive and finite constant $M<\infty$ such that $V_0\coloneqq V(t_0) \leq M$.
	
	By differentiating $V$, substituting the dynamics \eqref{eq:dynamics (CDC 17)}, employing the skew symmetry of $\dot{M}_i - 2N_i$ as well as  $\sum_{i\in\mathcal{N}} ( \nabla_{q_i}\varphi_i(q)^\top\dot{q}_i + \sum_{j\in\mathcal{N}_i(q_i)} \nabla_{q_j}\varphi_i(q)^\top\dot{q}_j)$ $=$ $\sum_{i\in\mathcal{N}}( \nabla_{q_i}\varphi_i(q)^\top +  \sum_{j\in\mathcal{N}_i(q_i)} \nabla_{q_i}\varphi_j(q)^\top) \dot{q}_i$, we obtain
	\begin{align*}
	\dot{V} =& \sum\limits_{i\in\mathcal{N}}\Bigg\{\dot{q}^\top_i \Bigg( \nabla_{q_i}\varphi_i(q) + \sum\limits_{j\in\mathcal{N}_i(q_i)} \nabla_{q_i}\varphi_j(q) + \tau_i - g_i(q_i)\Bigg) -\dot{q}_i^\top f_i(q_i,\dot{q}_i) \\& + \frac{1}{k_{\alpha_i}}\widetilde{\alpha}_i\dot{\hat{\alpha}}_i  \Bigg\},
	\end{align*}
	which, by substituting the control and adaptation laws \eqref{eq:control law (CDC 17)} and \eqref{eq:adaptation law (CDC 17)}, becomes:
	\begin{align*}
	\dot{V} =& \sum\limits_{i\in\mathcal{N}}\left\{ -k_{v_i}\lVert \dot{q}_i \rVert^2 - \hat{\alpha}_i\lVert \dot{q}_i \rVert^2 \lVert q_i \rVert  - \dot{q}_i^\top f_i(q_i,\dot{q}_i) + \widetilde{\alpha}_i\lVert \dot{q}_i \rVert^2 \lVert q_i \rVert\right\}  \notag\\ 
	\leq &  \sum\limits_{i\in\mathcal{N}}\left\{ -k_{v_i}\lVert \dot{q}_i \rVert^2 - (\hat{\alpha}_i-\alpha_i-\widetilde{\alpha}_i)\lVert \dot{q}_i \rVert^2 \lVert q_i \rVert  \right\} 
	\end{align*}
	where we have used the property $\lVert f_i(q_i,\dot{q}_i) \rVert \leq \alpha_i\lVert q_i \rVert \lVert \dot{q}_i\rVert$. Since $\widetilde{\alpha}_i = \hat{\alpha}_i - \alpha_i$, we obtain $\dot{V} \leq - \sum_{i\in\mathcal{N}} k_{v_i} \lVert \dot{q}_i\rVert^2$,
	which implies that $V$ is non-increasing along the trajectories of the closed loop system. Hence, we conclude that $V(t)\leq V_0 \leq M$, as well as the boundedness of $\widetilde{\alpha}_i, \varphi_i, \dot{q}_i$ and hence of $\hat{\alpha}_i, \forall i\in\mathcal{N}, t\geq t_0$. Therefore, we conclude that $\beta_i(q(t)) > 0, \forall t\geq t_0, i\in\mathcal{N}$. 
	
	Hence, inter-agent collisions, collision with the undesired regions and the obstacle boundary, connectivity losses between the subsets of the initially connected agents and singularity configurations are avoided.
	
	Moreover, by invoking LaSalle's Invariance Principle, the system converges to the largest invariant set contained in
	\begin{equation*}
	S_C \coloneqq \{(q,\dot{q})\in \mathbb{D}_1\times\dots\times\mathbb{D}_N\times\mathbb{R}^{Nn}\text{ s.t. } \dot{q} = 0_{Nn\times 1}\}. 
	\end{equation*}
	One can easily conclude that $\ddot{q}_i = 0, \forall i\in\mathcal{N}$, in $S_C$ and thus we conclude for the closed loop system \eqref{eq:closed loop dynamics (CDC 17)}  that $\nabla_{q_i}\varphi_i(q) = 0, \forall i\in\mathcal{N}$,
	since $\| f_i(q_i,0) \| \leq 0, \forall q_i\in\mathbb{R}^{n}$, in view of Assumption \ref{ass:f (CDC 17)}. Therefore, by invoking the properties of $\varphi_i(q)$, each agent $i\in\mathcal{N}$ will converge to a critical point of $\varphi_i$, i.e., all the configurations where $\nabla_{q_i}\varphi_i(q) = 0, \forall i\in\mathcal{N}$. However, due to properties of $\varphi_i(q)$, the initial conditions that lead to configurations $\widetilde{q}_i$ such that  $\nabla_{q_i}\varphi_i(q)|_{q_i=\widetilde{q}_i} = 0$ and $\gamma_i(\widetilde{q}_i) \neq 0$ are sets of measure zero in the configuration space \cite{koditschek1992robot}. Hence, the agents will converge to the configurations where $\gamma_i(q_i) = 0$ from almost all initial conditions, i.e., $\lim\limits_{t\to\infty}\gamma_i(q_i(t)) = 0$. Therefore, since $r_{\pi_{k'_i}} - \|p_{\scriptscriptstyle B_i}(q_{k'_i}) - p_{k'_i} \| \leq \bar{d}_{\scriptscriptstyle B_i} - \epsilon_q$, it can be concluded that there exists a finite time instance $t_{f_i}$ such that $\mathcal{A}_i(q_i(t_{f_i}))\in\pi_{k'_i}$, $\forall i\in\mathcal{N}$ and hence, each agent $i$ will be at its goal region $\pi_{k'_i}$ at time $t_{f_i}, \forall i\in\mathcal{N}$. In addition, the boundedness of $q_i,\dot{q}_i$ implies the boundedness of the adaptation laws $\dot{\hat{\alpha}}_i, \forall i\in\mathcal{N}$. Hence, the control laws \eqref{eq:control law (CDC 17)} are also bounded.
	
	\begin{figure*}
		\begin{minipage}{0.33\linewidth}		
			\centering
			\includegraphics[trim = 0cm 2cm 0cm 0cm,scale = 0.22]{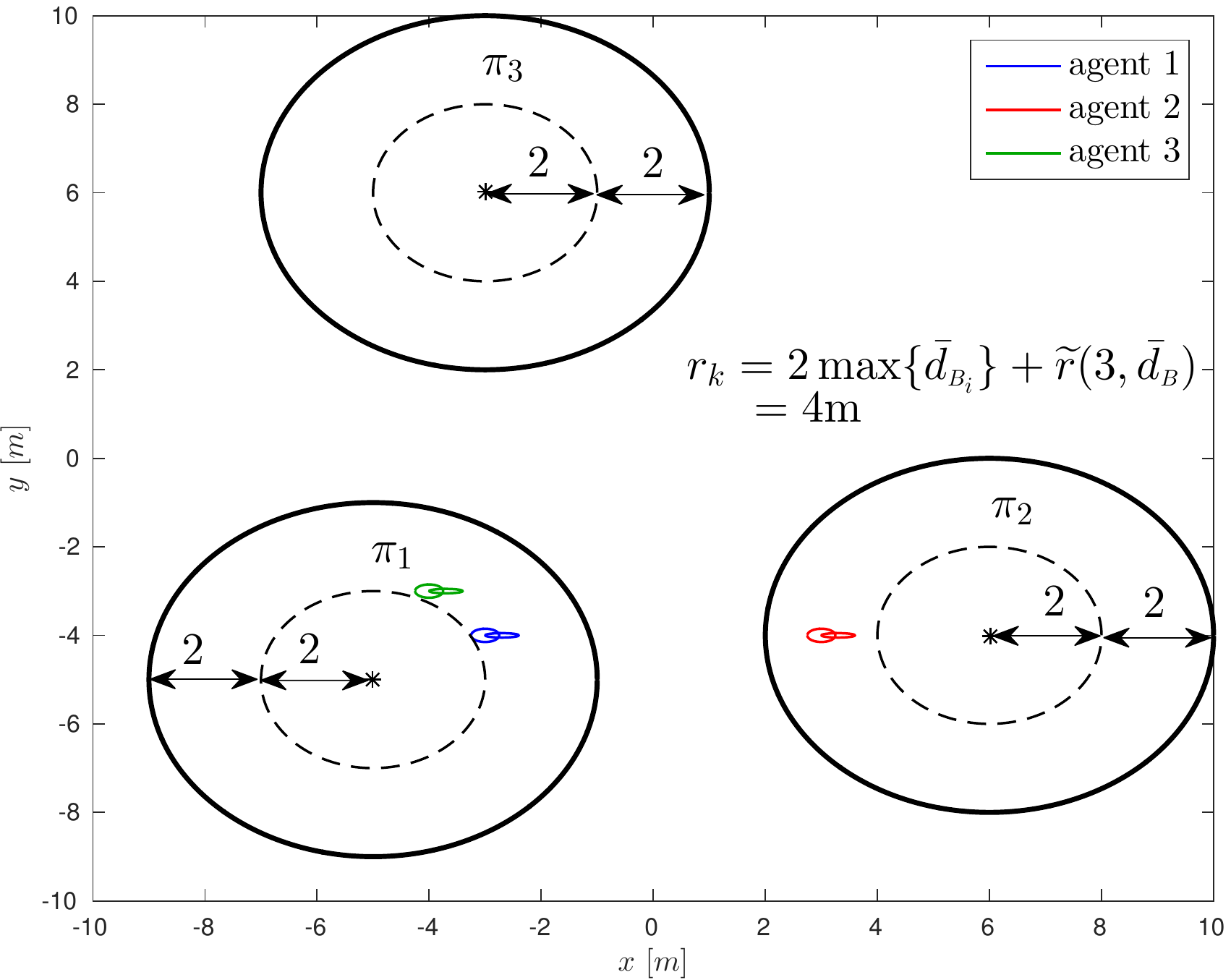}	
			\label{fig:navigation initial (CDC 17)}
			\subcaption{}
		\end{minipage}\hfill
		\begin{minipage}{0.33\linewidth}				
			\centering
			\includegraphics[scale =0.22]{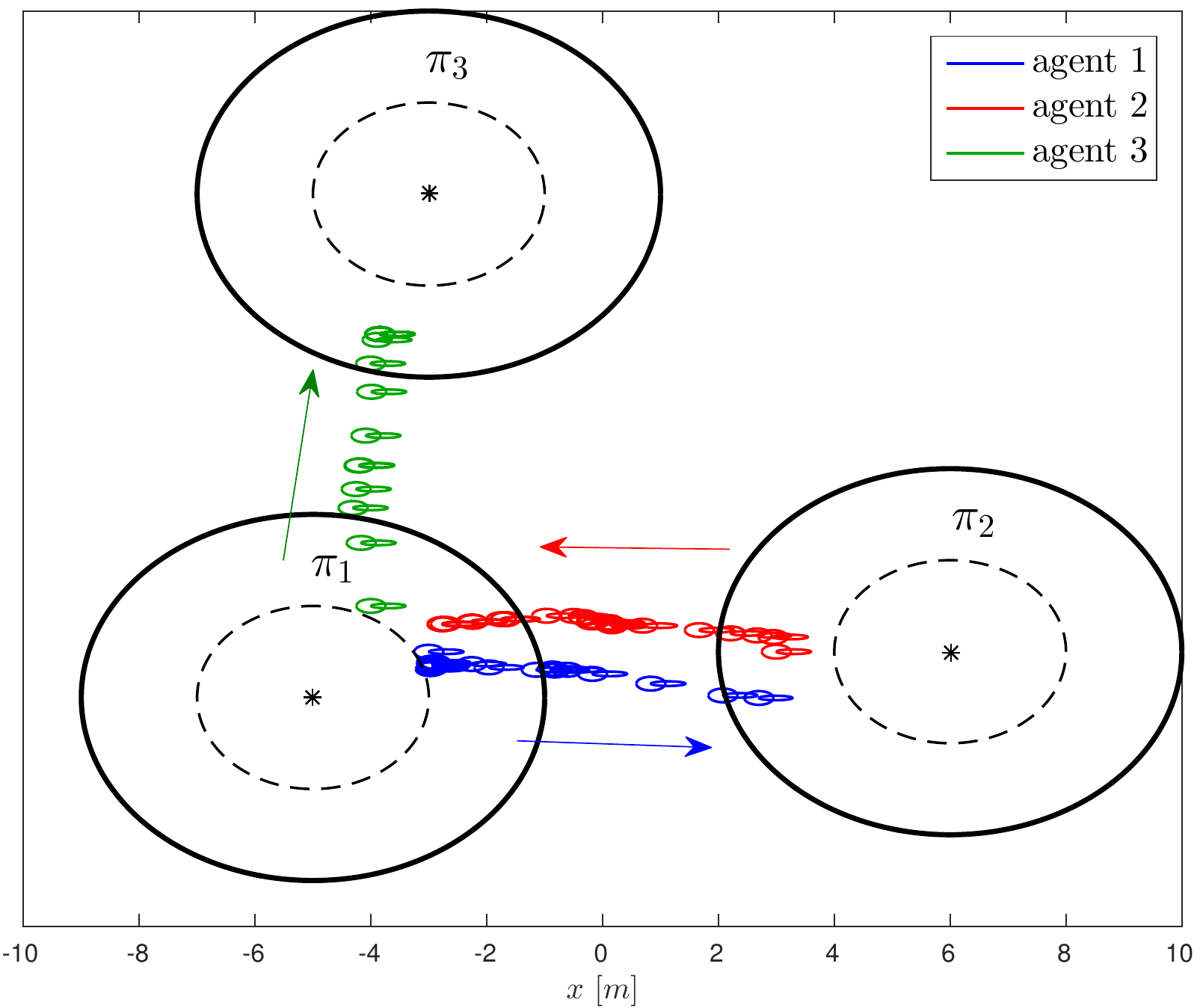}	
			\label{fig:navigation final (CDC 17)}
			\subcaption{}
		\end{minipage}\hfill
		\begin{minipage}{0.33\linewidth}				
			\centering
			\includegraphics[scale = 0.22]{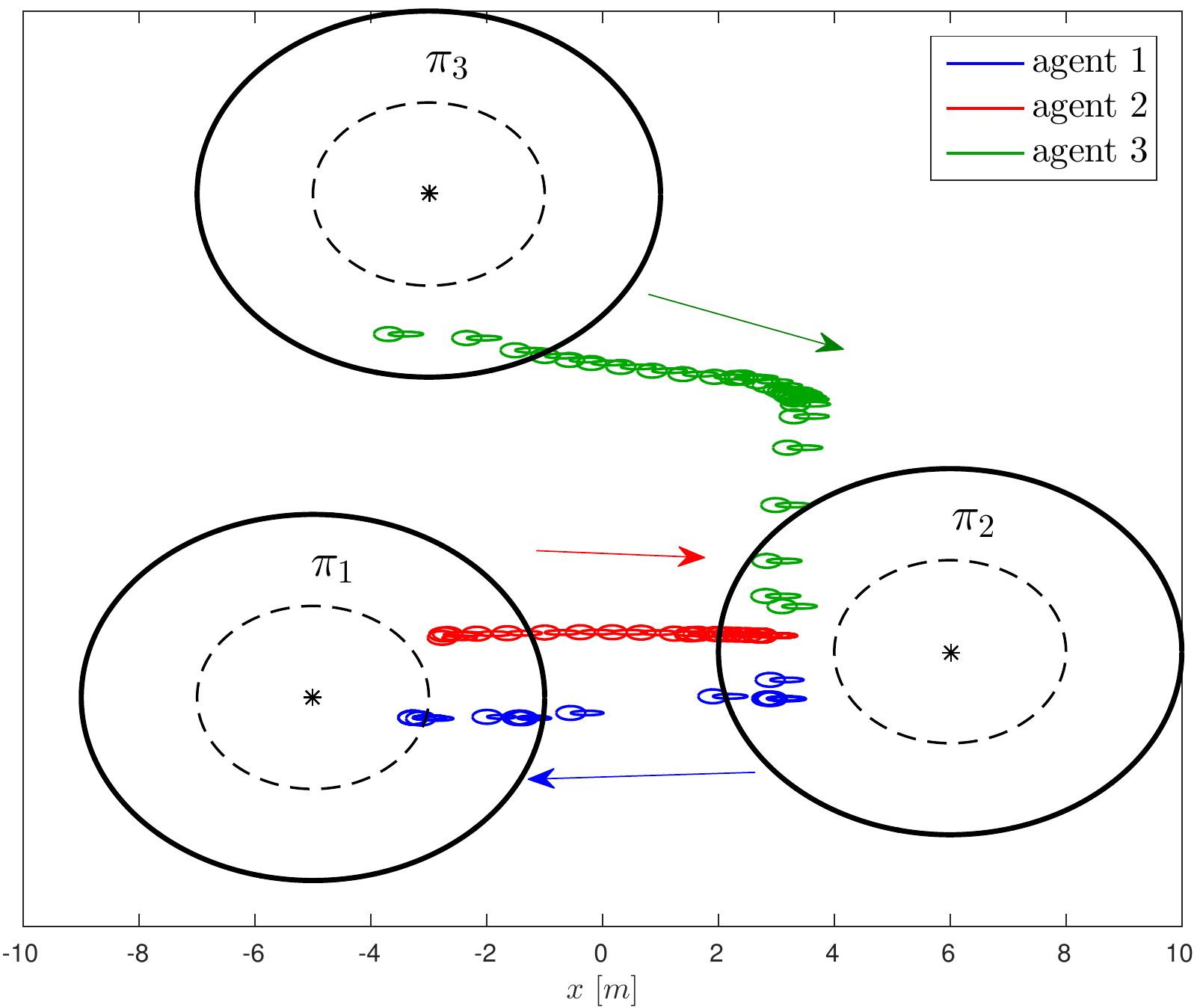}	
			\label{fig:navigation final2 (CDC 17)}
			\subcaption{}
		\end{minipage}
		\caption{(a): The initial position of the agents in the workspace of the simulation example. (b): The first transition of the agents in the workspace. Agent $1$ transits from $\pi_1$ to $\pi_2$, agent $2$ from $\pi_2$ to $\pi_1$, and agent $3$ from $\pi_{1}$ to $\pi_3$. (c): The second transition of the agents in the workspace. Agent $1$ transits from $\pi_2$ to $\pi_1$, agent $2$ from $\pi_1$ to $\pi_2$, and agent $3$ from $\pi_{3}$ to $\pi_2$.} \label{fig:navigation (CDC 17)}
	\end{figure*}			
	
\end{proof}

%
%

\subsubsection{Hybrid Control Framework} \label{subsec:hybrid (CDC 17)}
Due to the proposed continuous control protocol, the transitions $(\pi_{k_i},t_0)\xrightarrow{i}(\pi_{k'_i},t_{f_i})$ of Problem \ref{Problem (CDC 17)} are well-defined, according to Def. \ref{def:agent transition (CDC 17)}. Moreover, since all the agents $i\in\mathcal{N}$ remain connected with the subset of their initial neighbors $\widetilde{\mathcal{N}}_i$ and there exist finite constants $t_{f_i}$, such that $\mathcal{A}_i(q_i(t_{f_i}))\in\pi_{k'_i},\forall i\in\mathcal{N}$, all the agents are aware of their neighbors state, when a transition is performed. Hence, the transition system \eqref{eq:TS (CDC 17)} is well defined, $\forall i\in\mathcal{N}$.    
Consider, therefore, that $\mathcal{A}_i(q_i(0))\in\pi_{k_{i,0}}, k_{i,0}\in\mathcal{K}_\mathcal{R}, \forall i\in\mathcal{N}$, as well as a given desired path for each agent, that does not violate the connectivity condition of Problem \ref{Problem (CDC 17)}. Then, the iterative application of the control protocol \eqref{eq:control law (CDC 17)} for each transition of the desired path of agent $i$  guarantees the successful execution of the desired paths, with all the closed loop signals being bounded. 

\begin{remark}
	Note that, according to the aforementioned analysis, we implicitly assume that the agents start executing their respective transitions at the same time (we do not take into account individual control jumps in the Lyapunov analysis, i.e., it is valid only for one transition). Intuition suggests that if the regions of interest are sufficiently far from each other, then the agents will be able to perform the sequence of their transitions independently.     
	Detailed technical analysis of such cases is part of future research.
\end{remark}

\subsection{Simulation Results}\label{sec:simulations (CDC 17)}
To demonstrate the validity of the proposed methodology, we consider the simplified example of three agents in a workspace with $r_0 = 12$m and three regions of interest, with $r_{\pi_k} = 4$m, $\forall k\in\{1,2,3\}$. Each agent consists of a mobile base and a rigid link connected with a rotational joint, with $\bar{d}_{\scriptscriptstyle B_i} = 1$m, $\forall i\in\{1,2,3\}$. We also choose $p_1 = [-5,-5]^\top$m, $p_2 = [6,-4]^\top$m, $p_3 = [-3,6]^\top$m. 
The initial base positions are taken as $p_{\scriptscriptstyle B_1} = [-3,-4]^\top\text{m}, p_{\scriptscriptstyle B_2} = [3,-4]^\top\text{m}, p_{\scriptscriptstyle B_3} = [-4,-5]^\top\text{m}$, which imply that $\mathcal{A}_1(q_1(0)),\mathcal{A}_3(q_3(0))\in\pi_1$ and $\mathcal{A}_2(q_2(0))\in\pi_2$ (see Fig. \ref{fig:navigation (CDC 17)}(a). The control inputs for the agents are the $2$D force acting on the mobile base, and the joint torque of the link. We also consider a sensing radius of $d_{\text{con}_i} = 8\text{m}$ and the subsets of initial neighbors as $\widetilde{\mathcal{N}}_1 = \{2\}, \widetilde{\mathcal{N}}_2 = \{1,3\}$, and $\widetilde{\mathcal{N}}_3 = \{2\}$, i.e., agent $1$ has to stay connected with agent $2$, agent $2$ has to stay connected with agents $1$ and $3$ and agent $3$ has to stay connected with agent $2$. The agents are required to perform two transitions. Regarding the first transition, we choose $\pi_{k'_1} = \pi_2$ for agent $1, \pi_{k'_2} = \pi_1$ for agent $2$, and $\pi_{k'_3} = \pi_3$, for agent $3$. Regarding the second transition, we choose $\pi_{k'_1} = \pi_1, \pi_{k'_2} = \pi_2$, and $\pi_{k'_3} = \pi_2$. The control parameters and gains are chosen as $k_i = 5, k_{v_i} =10$, and $k_{\alpha_i} = 0.01, \forall i\in\{1,2,3\}$. We employ the potential field from \cite{dimarogonas2007decentralized}.
The simulation results are depicted in Fig. \ref{fig:navigation (CDC 17)}-\ref{fig:c tilde (CDC 17)}. In particular, Fig. \ref{fig:navigation (CDC 17)}(b) and \ref{fig:navigation (CDC 17)}(c) illustrate the two consecutive transitions of the agents. Fig. \ref{fig:gamma beta (CDC 17)} depicts the obstacle functions $\beta_i$ which are strictly positive, $\forall i\in\{1,2,3\}$. Finally, the control inputs are given in Fig. \ref{fig:inputs (CDC 17)} and the parameter errors $\widetilde{\alpha}$ are shown in Fig. \ref{fig:c tilde (CDC 17)}, which indicates their boundedness. As proven in the theoretical analysis, the transitions are successfully performed while satisfying all the desired specifications.

\begin{figure}[]
	\vspace{0.4cm}
	\centering
	\includegraphics[scale=0.35]{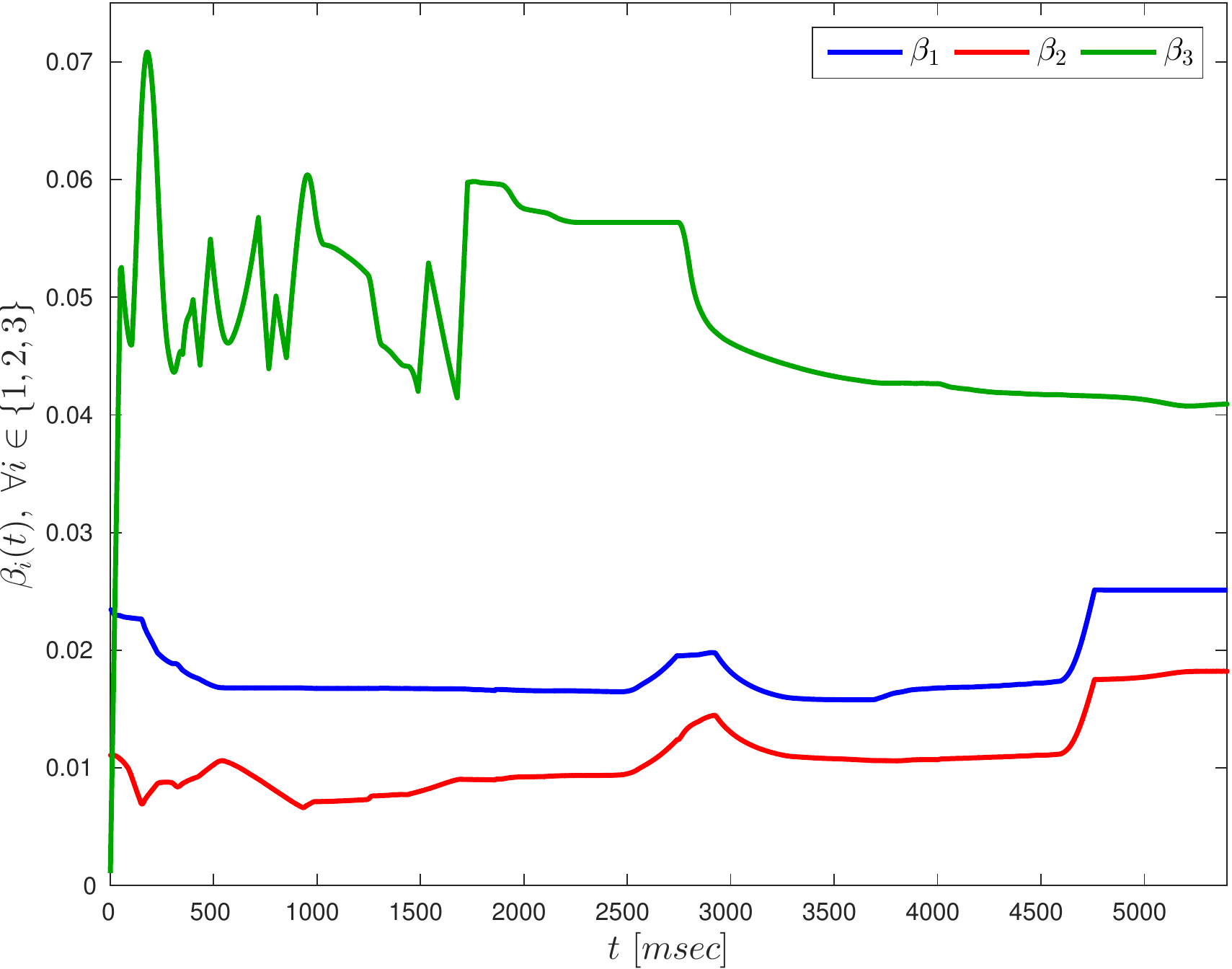}
	\caption{The obstacle functions $\beta_i, i\in\{1,2,3\}$, which remain strictly positive.  \label{fig:gamma beta (CDC 17)}}
\end{figure}

\begin{figure}[]
	\centering
	\includegraphics[scale=0.35]{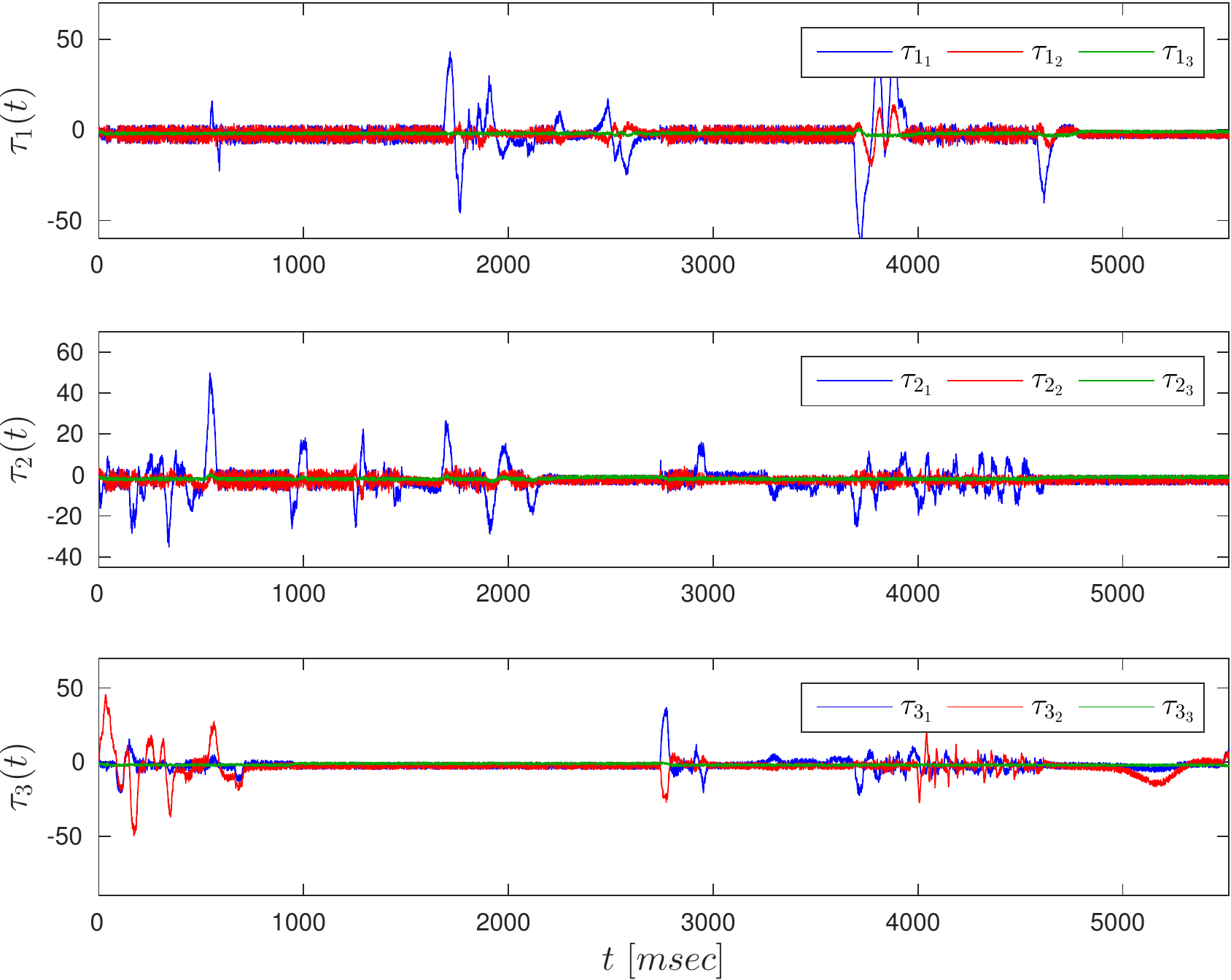}
	\caption{The resulting control inputs $\tau_i,\forall i\in\{1,2,3\}$ for the two transitions.\label{fig:inputs (CDC 17)}}
\end{figure}

%

\begin{figure}[]
	\vspace{0.4cm}
	\centering
	\includegraphics[scale=0.2]{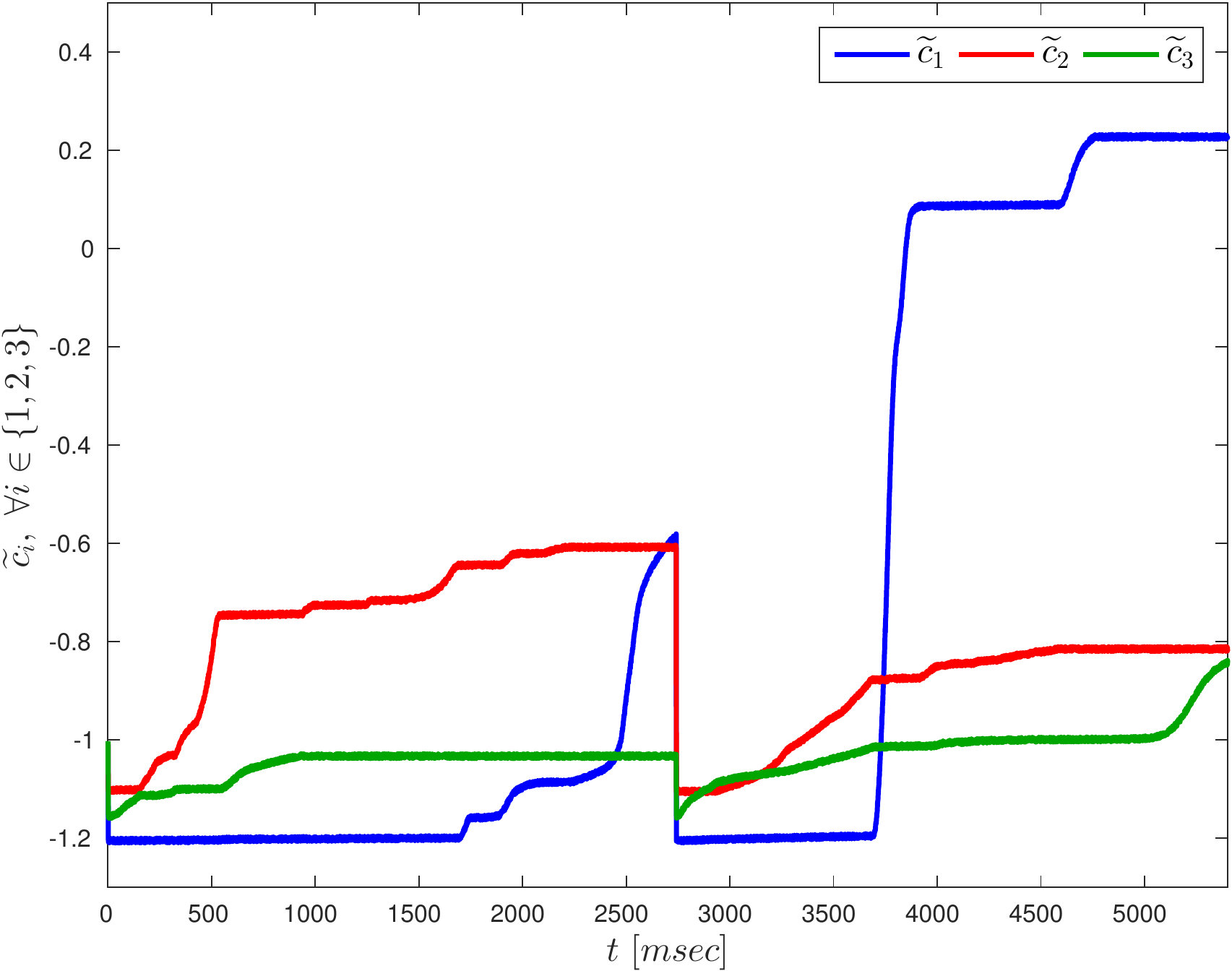}
	\caption{The parameter deviations $\widetilde{\alpha}_i,\forall i\in\{1,2,3\}$, which are shown to be bounded.\label{fig:c tilde (CDC 17)}}
\end{figure}

\section{Timed Abstractions for Distributed Cooperative Manipulation} \label{sec:AR}

We now switch our attention to multi-agent-object systems. Such systems include, except for a number of robotic agents, a certain number of \textit{unactuated} objects (items in the environment). We consider that these objects, along with the agents, have themselves some local tasks to complete, expressed as temporal specifications with respect to their locations. 
This section considers the problem of motion planning for one unactuated object, which is grasped by a number of robotic agents, under \textit{timed} temporal specifications, and in particular, Metric Interval Temporal Logic (MITL) specifications. In particular, we design appropriate well-defined \textit{timed} abstractions for a cooperatively manipulated object that allows us to express and solve the object motion planning problem under MITL formulas.

\subsection{Problem Formulation} 
\label{sec:Problem-Formulation (AR)}
Consider a bounded workspace $\mathcal{W} \subset \mathbb{R}^{3}$ containing $N$ robotic agents rigidly grasping an object, similar to what is shown in Fig. \ref{fig:Two-robotic-arms (TCST_coop_manip)}. 
The agents are considered to be fully actuated and they consist of a base that is able to move around the workspace (e.g., mobile or aerial vehicle) and a robotic arm.
The setup considered here is the same as in Section \ref{sec: Problem Formulation (TCST_coop_manip)}, which we briefly recap.
Each agent $i$ knows only its own state, position and velocity, as well as its own and the object's geometric parameters. More specifically, we assume that each agent $i$ knows the distance from its grasping point $\{E_i\}$ to the object's center of mass $\{O\}$ as well as the relative orientation offset between the two frames $\{E_i\}$ and $\{O\}$. This information can be either  retrieved on-line via appropriate sensors or transmitted off-line to the agents, without the need of inter-agent on-line communication. Finally, no interaction force/torque measurements are required and the dynamic model of the object and the agents is considered unknown. 

The dynamics of the agents are (see eq. \eqref{eq:manipulator dynamics (TCST_coop_manip)})
\begin{align*}
M_{i}(q_i)\dot{v}_i+C_{i}(q_i,\dot{q}_i)v_i+g_{i}(q_i) 
+ d_i(q_i,\dot{q}_i,t) =  u_{i} - h_{i},   
\end{align*}
whereas the object's 
	\begin{align*}
	& \dot{x}_{\scriptscriptstyle O} = J_{\scriptscriptstyle O}(\eta_{\scriptscriptstyle O})v_{\scriptscriptstyle O} \\ 
	& M_{\scriptscriptstyle O}(\eta_{\scriptscriptstyle O})\dot{v}_{{\scriptscriptstyle O}}+C_{{\scriptscriptstyle O}}(\eta_{\scriptscriptstyle O},\omega_{\scriptscriptstyle O})v_{{\scriptscriptstyle O}}+g_{\scriptscriptstyle O}+d_{\scriptscriptstyle O}(x_{\scr O},\dot{x}_{\scr O},t)  =  h_{\scriptscriptstyle O}, 
	\end{align*}
and the coupled dynamics by 
\begin{equation}
\widetilde{M}(x)\dot{v}_{{\scriptscriptstyle O}}+\widetilde{C}(x)v_{{\scriptscriptstyle O}}+\widetilde{g}(x) + \widetilde{d}(x,t) =G(q)^\top\bar{u},\label{eq:coupled dynamics 2 (AR)}
\end{equation}
with the coupling terms as in \eqref{eq:coupled terms (TCST_coop_manip)} and $x = [q^\top,\dot{q}^\top,\eta_{\scr O}^\top, \omega_{\scr O}^\top]$.

\begin{figure}
	\centering
	\includegraphics[width=0.6\textwidth]{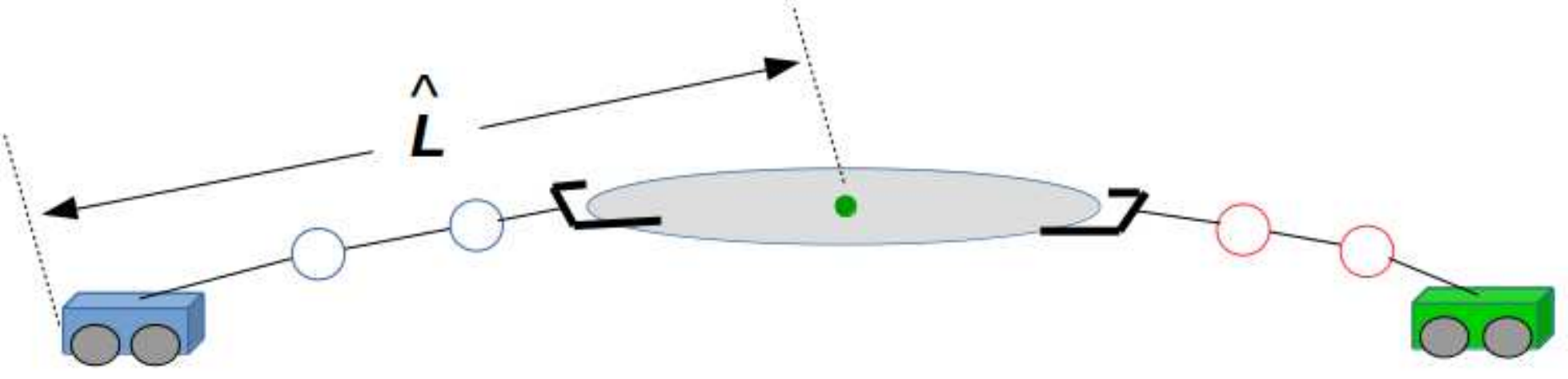}
	\caption{An example of the system shown in Fig. \ref{fig:Two-robotic-arms (TCST_coop_manip)}  in the configuration that produces $\hat{L}$. \label{fig:L0 (AR)}}
\end{figure}

\begin{figure}
	\centering
	\includegraphics[width=0.6\textwidth]{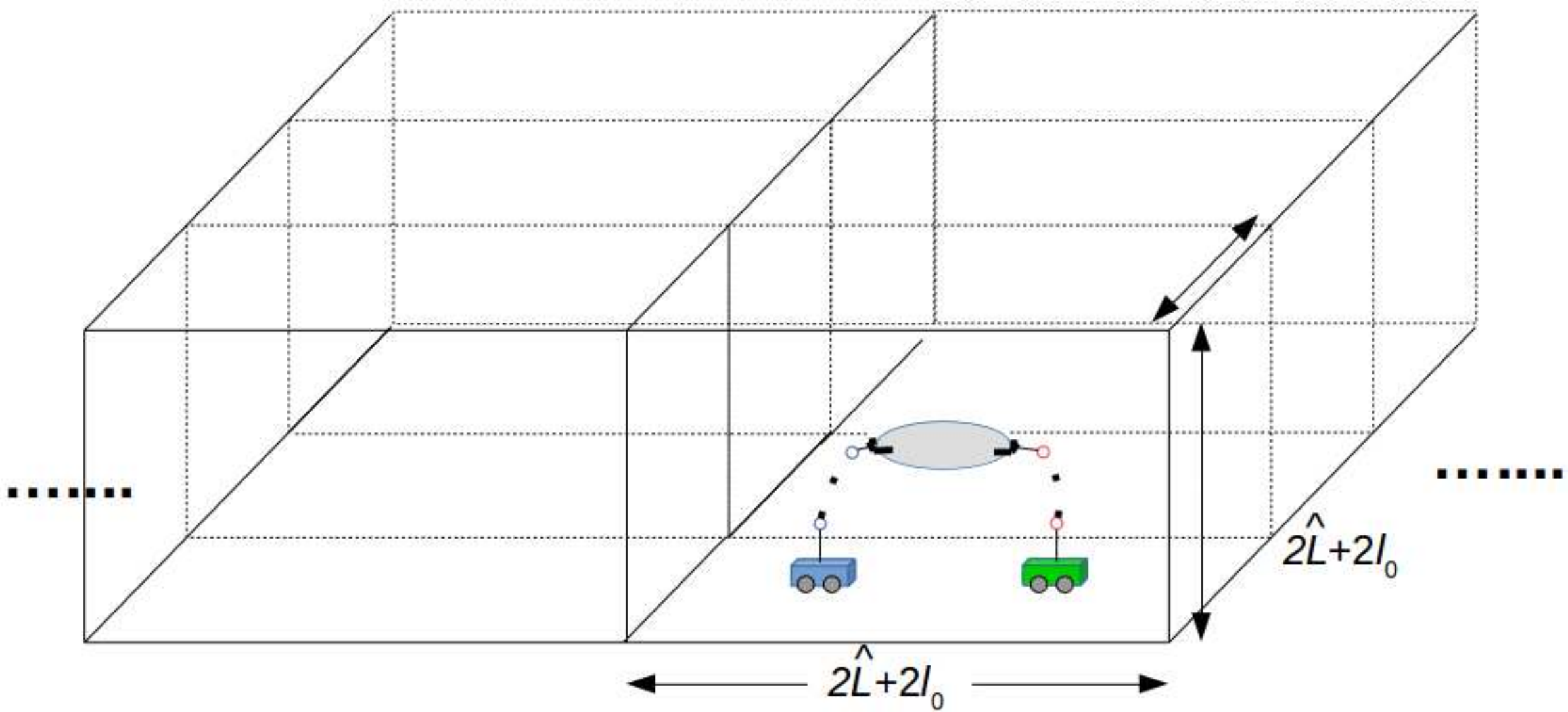}
	\caption{The workspace partition according to the bounding box of the coupled system.\label{fig:workspace_discretization (AR)}}
\end{figure}

\subsubsection{Workspace Partition} 
\label{subsec:wsp discret (AR)}
As already mentioned, we are interested in designing a well-defined abstraction of the coupled object-agents system, so that we can define MITL formulas over certain properties in a discrete set of regions of the workspace. Therefore, we provide now a partition of $\mathcal{W}$ into cell regions. We denote by $\mathcal{S}_{q}$ the set  that consists of all points $p_s\in\mathcal{W}$ that physically belong to the coupled system, i.e., they consist part of either the volume of the agents or the volume of the object. Note that these points depend on the actual value of $q$. We further define the constant $\hat{L} \geq \sup_{\substack{q\in\mathbb{R}^{{n}}\\p_s\in\mathcal{S}_{q}}} \|p_s - p_{\scriptscriptstyle O}(q) \|$, where, with a slight abuse of notation and in view of the coupled object-agents kinematics and the forward kinematics of the agents, we express $p_{\scriptscriptstyle O}$ as a function of $q$. Note that, although the explicit computation of $\mathcal{S}_{q}$ may not be possible, $\hat{L}$ is an upper bound of the maximum distance between the object center of mass and a point in the coupled system's volume over all possible configurations $q$, and thus, it can be measured. For instance, Fig. \ref{fig:L0 (AR)} 
shows $\hat{L}$ for the system of Fig. \ref{fig:Two-robotic-arms (TCST_coop_manip)}. It is straightforward to conclude that
\begin{equation}
\mathcal{S}_{q} \subset \mathcal{B}(p_{\scriptscriptstyle O}(q),\hat{L}),\forall q\in\mathbb{R}^{{n}}. \label{eq:p_s in L_hat (AR)}
\end{equation}
Next, we partition the workspace $\mathcal{W}$ into $R$ equally sized rectangular regions $\Pi=\left\{\pi_{1},\dots,\pi_{R}\right\}$, whose geometric centers are denoted by $p^c_{\pi_j}\in\mathcal{W}, j\in\{1,\dots,R\}$. The length of the region sides is set to $D = 2\hat{L}+2l_0$, where $l_0$ is an arbitrary positive constant. Hence, each region $\pi_j$ can be formally defined as follows: 
\begin{align}
\pi_j  \coloneqq &\{ p\in\mathcal{W} \text{ s.t. } (p)_k\in [ (p^c_{\pi_j})_k - \hat{L}-l_0, (p^c_{\pi_j})_k + \hat{L}+l_0 ), \forall k\in\{x,y,z\} \},  \label{eq:discretization 1 (AR)} \notag
\end{align}
with $\|p^c_{\pi_{j+1}} - p^c_{\pi_{j}}\| = \ (2\hat{L} + 2l_0), \forall j\in\{1,\dots,R-1\}$, and $\ (p^c_{\pi_j})_z$ $\coloneqq \hat{L} + l_0, \forall j\in\{1,\dots,R\}$; $(\cdot)_k, k\in\{x,y,z\}$, denotes the $k$-th coordinate. An illustration of the aforementioned partition is depicted in Fig. \ref{fig:workspace_discretization (AR)}.

Note that each $\pi_j$ is a uniformly bounded and convex set and also $\pi_j \cap \pi_{j'} = \emptyset, \forall j,j'\in\{1,\dots,R\}$ with $j\neq j'$. We also define the neighborhood $\mathcal{D}$ of region $\pi_j$ as the set of its adjacent regions, i.e., $\mathcal{D}(\pi_j) \coloneqq \{\pi_{j'}\in\Pi \text{ s.t. }$ $\|p^c_{\pi_j} - p^c_{\pi_{j'}}\| = (2\hat{L} + 2l_0) \}$, which is symmetric, i.e., $\pi_{j'}\in\mathcal{D}(\pi_j) \Leftrightarrow \pi_{j}\in\mathcal{D}(\pi_{j'})$.

To proceed we need the following definitions regarding the timed transition of the coupled system between two regions $\pi_j,\pi_{j'}$: 
\begin{definition} \label{def:system in region (AR)}
	The coupled object-agents system is in region $\pi_j$ at a configuration $q$, denoted as $\mathcal{A}(q)\in\pi_j$, if and only if the following hold:
	\begin{enumerate}
		\item $\mathcal{S}_{q} \subset \pi_j$ 
		\item $\|p_{\scriptscriptstyle O}(q)-p^c_{\pi_j}\| < l_0$.
	\end{enumerate}
\end{definition}

\begin{definition} \label{def:transition (AR)} 
	Assume that $\mathcal{A}(q(t_0))\in\pi_j, j\in\{1,\dots,R\}$, for some $t_0\in\mathbb{R}_{\geq 0}$. Then, there exists a transition for the coupled object-agents system from $\pi_j$ to $\pi_{j'}, j'\in\{1,\dots,R\}$ with time duration $\delta t_{j,j'}\in\mathbb{R}_{\geq 0}$, denoted as $\pi_j \xrightarrow{\mathcal{T}} \pi_{j'}$, if and only if 
	\begin{enumerate}
		\item $\mathcal{A}(q(t_0 + \delta t_{j,j'}))\in\pi_{j'}$,
		\item $\mathcal{S}_{q(t)} \subset \pi_i \cup \pi_j$,
		$\forall t\in[t_0,t_0 + \delta t_{j,j'}]$.
	\end{enumerate}
\end{definition}
Note that the entire system object-agents must remain in $\pi_j,\pi_{j'}$ during the transition and therefore the requirement $\pi_{j'}\in\mathcal{D}(\pi_j)$ is implicit in  Definition \ref{def:transition (AR)}. 

\subsubsection{Specification} \label{subsec:specf (AR)}
Given the workspace partition, we can introduce a set of atomic propositions $\Psi$ for the object, which are expressed as Boolean variables that correspond to  properties of interest in the regions of the workspace (e.g., ``Obstacle region", ``Goal region"). Formally, the labeling function $\mathcal{L}:\Pi\rightarrow2^{\Psi}$ assigns to each region $\pi_j$ the subset of the atomic propositions $\Psi$ that are true in $\pi_j$. We next provide the timed behavior, similar to Section \ref{sec:icra}
\begin{definition} \label{def:specification (AR)} 
	Given a time trajectory $q(t), t\geq0$, a \textit{timed sequence} of $q$ is the infinite sequence $\mathfrak{s}_t \coloneqq (q(t_1),t_1)(q(t_2),t_2)\dots$, with $t_m\in\mathbb{R}_{\geq 0}, t_{m+1} > t_m$ and $\mathcal{A}(q(t_m))\in\pi_{j_m}, j_m\in\{1,\dots,R\}, \forall m\in\mathbb{N}$.
	The \textit{timed behavior} of $\mathfrak{s}_t$ is the infinite sequence $\mathfrak{b}_t \coloneqq (\breve{\psi}_1,t_1)(\breve{\psi}_2,t_2)\dots$, with $\breve{\psi}_m\in 2^{\Psi}, \breve{\psi}_m \in\mathcal{L}(\pi_{j_m})$ for  $\mathcal{A}(q(t_m))\in\pi_{j_m}, j_m\in\{1,\dots,R\}$, $\forall m\in\mathbb{N}$, i.e., the set of atomic propositions that are true when $\mathcal{A}(q(t_m))\in\pi_{j_m}$. 
\end{definition}

The satisfaction of a MITL formula is provided by the following definition (see Appendix \ref{app:Logics} for more details on MITL formulas).

\begin{definition}
	The timed sequence $\mathfrak{s}_t$ satisfies a MITL formula $\mathsf{\Phi}$ if and only if $\mathfrak{b}_t \models \mathsf{\Phi}$. 
\end{definition}

We are now ready to state the problem treated in this section.

\begin{problem} \label{problem 1 (AR)}
	Given $N$ agents rigidly grasping an object in $\mathcal{W}$ subject to the coupled dynamics \eqref{eq:coupled dynamics 2 (AR)}, the workspace partition $\Pi$ such that $\mathcal{A}(q(0))\in\pi_{j_0},j_0\in\{1,\dots,R\}$, a MITL formula $\mathsf{\Phi}$ over $\Psi$ and the labeling function $\mathcal{L}$, derive a control strategy that achieves a timed sequence $\mathfrak{s}_t$ which yields the satisfaction of $\mathsf{\Phi}$.   
\end{problem}

\subsection{Main Results} \label{sec:main results (AR)}

\begin{figure}
	\centering
	\includegraphics[width=0.6\textwidth]{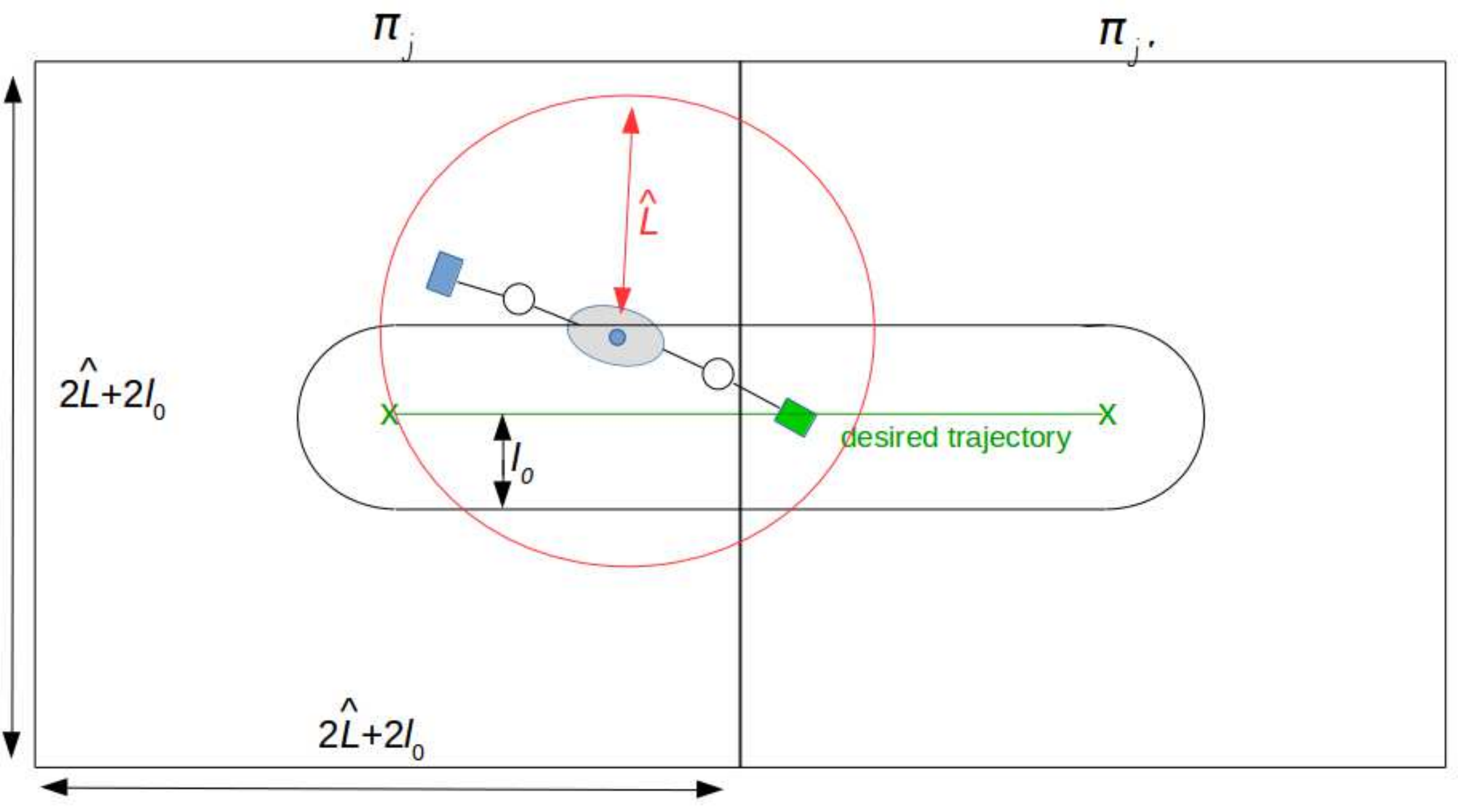}
	\caption{Top view of a transition between two adjacent regions $\mathsf{\pi}_j$ and $\mathsf{\pi}_{j'}$. Since $p_{\scriptscriptstyle O}\in\mathcal{B}(p_{j,j'}(t),l_0)$, we conclude that $\mathcal{S}_{q} \subset \mathcal{B}(p_{\scriptscriptstyle O},\hat{L}) \subset \mathcal{B}(p_{j,j'}(t),l_0+\hat{L})\subset \pi_j\cup\pi_{j'}$. \label{fig:grid3 (AR)}}
\end{figure}

\subsubsection{Control Design} \label{subsec:control design (AR)}
The first ingredient of the proposed solution is the design of a decentralized control protocol $u$ such that a transition relation between two adjacent regions according to Definition \ref{def:transition (AR)} is established. Assume, therefore, that $\mathcal{A}(q(t_0))\in\pi_j,j\in\{1,\dots,R\}$ for some $t_0\in\mathbb{R}_{\geq 0}$. We aim to find a bounded $\bar{u}$, such that $\mathcal{A}(q(t_0+\delta t_{j,j'}))\in\pi_{j'}$, with $\pi_{j'}\in\mathcal{D}(\pi_j)$, and  $\mathcal{S}_{q(t)} \subset \pi_j\cup\pi_{j'}, \forall  t\in[t_0,t_0+\delta t_{j,j'}]$, for a predefined arbitrary constant $\delta t_{j,j'}$ $\in\mathbb{R}_{\geq 0}
$ corresponding to the transition $\pi_j \xrightarrow{\mathcal{T}} \pi_{j'}$. 

The first step is to associate to the transition a smooth and bounded trajectory with bounded time derivative, defined by the line segment that connects $p^c_{\pi_j}$ and $p^c_{\pi_{j'}}$, i.e. define $p_{j,j'}:[t_0,\infty)\rightarrow\mathbb{R}^3$, such that $p_{j,j'}(t_0) =  p^c_{\pi_j}, p_{j,j'}(t) = p^c_{\pi_{j'}}, \forall t\geq t_0+\delta t_{j,j'}$ and
\begin{equation}
\mathcal{B}(p_{j,j'}(t), \hat{L}+l_0)\subset\pi_j\cup\pi_{j'}, \ \ \forall t\geq t_0.  \label{eq:desired_tr (AR)}
\end{equation}
An example of $p_{j,j'}$ is 
\begin{equation}
p_{j,j'}(t) = \left\{ \begin{matrix} \dfrac{p^c_{\pi_{j'}}- p^c_{\pi_{j}}}{\delta t_{j,j'}}t  +  \dfrac{p^c_{\pi_j}(\delta t_{j,j'}-1) - p^c_{\pi_{j'}}}{\delta t_{j,j'}}t_0, & \ \   t\in [t_0,t_0+\delta t_{j,j'}) \\
p^c_{\pi_{j'}}, & \ \ t\in [t_0+\delta t_{j,j'},\infty)  \end{matrix} \right. \label{eq:example of p_j (AR)}
\end{equation} 
The intuition behind the solution of Problem \ref{problem 1 (AR)} via the definition of $p_{j,j'}$ is the following:
if we guarantee that the object's center of mass stays $l_0$-close to $p_{j,j'}$, i.e., $\|p_{\scriptscriptstyle O}(t)-p_{j,j'}(t)\|<l_0,\forall t\geq t_0$, then $\|p_{\scriptscriptstyle O}(t_0+\delta t_{j,j'}) - p^c_{\pi_{j'}} \| < l_0$ and, by invoking \eqref{eq:p_s in L_hat (AR)} and \eqref{eq:desired_tr (AR)}, we obtain $\mathcal{S}_{q(t)} \subset \mathcal{B}(p_{\scriptscriptstyle O}(t),\hat{L})\subset\mathcal{B}(p_{j,j'}(t),\hat{L}+l_0)\subset\pi_j\cup\pi_{j'}, \forall t\geq t_0$ (and therefore $t\in[t_0,t_0+\delta t_{j,j'}]$), and thus the requirements of Definition \ref{def:transition (AR)} for the transition relation are met. Fig. \ref{fig:grid3 (AR)} illustrates the aforementioned reasoning.

Along with $p_{j,j'}$, we consider that the object has to comply with certain specifications associated with its orientation. Therefore, we also define a smooth and bounded orientation trajectory $\eta_{j,j'} \coloneqq [\phi_{j,j'},\theta_{j,j'},\psi_{j,j'}]^\top:[t_0,\infty)\rightarrow \mathbb{T}$ with bounded time derivative, that has to be tracked by the object's center of mass. We choose $\theta_{j,j'}(t)\in[-\bar{\theta},\bar{\theta}]\subset(-\tfrac{\pi}{2}, \tfrac{\pi}{2})$, $\forall t\in\mathbb{R}_{\geq 0}$, with $\bar{\theta}\in(0,\tfrac{\pi}{2})$, so as to ensure the singularity avoidance of $J_{\scriptscriptstyle O}(\eta_{\scriptscriptstyle O})$. 
We form, therefore, the desired pose trajectory $x_{j,j'}:[t_0,\infty)\rightarrow\mathbb{M}$, with $x_{j,j'}(t) \coloneqq [p_{j,j'}(t)^\top, \eta_{j,j'}(t)^\top]^\top$. In case of multiple consecutive transitions $\dots\pi_{h} \xrightarrow{\mathcal{T}} \pi_j \xrightarrow{\mathcal{T}} \pi_{j'} \xrightarrow{\mathcal{T}} \pi_{h'}\dots$ over the intervals $\ \ \dots$ ,$\delta t_{h,j}$, $\delta t_{j,j'}$, $\delta t_{j',h'}$,$\dots$, the desired orientation trajectories 
$\dots$, $\eta_{h,j}(t)$, $\eta_{j,j'}(t)$, $\eta_{j',h'}(t)$, $\dots$ 
must be continuous at the transition points, i.e., $\eta_{h,j}(t_0) = \eta_{j,j'}(t_0)$ and $\eta_{j,j'}(t_0 + \delta t_{j,j'})$ $= \eta_{j',h'}(t_0 + \delta t_{j,j'})$.

Therefore, Problem \ref{problem 1 (AR)} is equivalent to a problem of trajectory tracking within certain bounds. 

A suitable methodology for the control design in hand is that of prescribed performance control, which was used for the cooperative manipulation problem in Section \ref{subsec:PPC Controller (TCST_coop_manip)}. We describe it briefly here and associate it with the abstraction problem.

We consider first the associated position and orientation error as in \eqref{eq:ppc errors (TCST_coop_manip)}:
\begin{equation}
e_s \coloneqq \begin{bmatrix}
e_{s_x}, e_{s_y},  e_{s_z},  e_{s_\phi}, e_{s_\theta},  e_{s_\psi}  \end{bmatrix}^\top
\coloneqq x_{\scriptscriptstyle O} - x_{j,j'}(t).  \label{eq:e_pose_compact (AR)}
\end{equation}
Following that section as well as Appendix \ref{app:PPC}, the mathematical expressions of prescribed performance are given by the inequalities: 
	\begin{align}	\label{eq:ppc (AR)}
	-\rho_{s_k}(t)&< e_{s_k}(t)<\rho_{s_k}(t), \ \ \forall k\in\mathcal{K},  
	\end{align}
$\forall t\in[t_0,\infty)$, where $\mathcal{K} = \{x,y,z,\phi,\theta,\psi\}$, $\rho_{s_k}\coloneqq \rho_{s_k}(t):[t_0,\infty)\rightarrow\mathbb{R}_{> 0}$ with 
	\begin{align}	\label{eq:rho (AR)}
	\rho_{s_k}(t)&=(\rho_{s_k,0} - \rho_{s_k,\scr \infty})\exp(-l_{s_k}t) + \rho_{s_k,\scr \infty},  \ \ \forall k\in\mathcal{K},  
	\end{align}
as in \eqref{eq:rho (TCST_coop_manip)}.

The proposed prescribed performance control protocol does not incorporate any information on the agents' or the object's dynamics or the external disturbances and guarantees \eqref{eq:ppc (AR)} for all $t\in[t_0,\infty)$ and hence $[t_0,t_0+\delta t_{j,j'}]$, which,
by appropriately selecting $\rho_{s_k}(t), k\in\mathcal{K}$ and given that $\mathcal{A}(q(t_0))\in\pi_j$, guarantees a representation singularity-free (i.e., $\theta_{\scriptscriptstyle O}(t)\neq \tfrac{\pi}{2},t\in[t_0,\infty)$) transition $\pi_j \xrightarrow{\mathcal{T}} \pi_{j'}$ with time duration of $\delta t_{j,j'}$, as will be clarified in the sequel. 

As in Section \ref{subsec:PPC Controller (TCST_coop_manip)}, consider the following steps:
\textbf{Step I-a}. Select the corresponding functions $\rho_{s_k}$ as in \eqref{eq:rho (AR)} with 
\begin{enumerate}[(i)]
	\item 	$\rho_{s_\theta,\scriptscriptstyle 0}  = \rho_{s_\theta}(t_0)= \theta^*, \rho_{s_k,\scriptscriptstyle 0} = \rho_{s_k}(t_0) = l_0, \forall k\in\{x,y,z\}$ $\rho_{s_k,\scriptscriptstyle 0} = \rho_{s_k}(t_0) > \| e_{s_k}(t_0) \|, \forall k\in\{\phi,\psi\}$, 
	\item $l_{s_k} \in\mathbb{R}_{>0}, \forall k\in\mathcal{K}$,
	\item $\rho_{s_k,\scriptscriptstyle \infty}\in(0,\rho_{s_k,0}), \forall k\in\mathcal{K}$,
\end{enumerate}
where $\theta^*$ is a positive constant satisfying $\theta^* + \bar{\theta} < \frac{\pi}{2}$.

\textbf{Step I-b}. Introduce the normalized errors 
\begin{equation*}
\xi_{s} \coloneqq \begin{bmatrix}
\xi_{s_x}, \dots, \xi_{s_\psi} \end{bmatrix}^\top
\coloneqq \rho_s^{-1}e_s,	
\end{equation*}
where $\rho_s \coloneqq \rho_s(t)\coloneqq\textup{diag}\{\left[\rho_{s_k}\right]_{k\in\mathcal{K}}\}\in\mathbb{R}^{6\times6}$, as well as the transformed state functions $\varepsilon_s:(-1,1)^6\to\mathbb{R}^6$, and signals $r_s:(-1,1)^6\to\mathbb{R}^{6\times 6}$, with  
\begin{align*}
& \varepsilon_s \coloneqq \varepsilon_s(\xi_s) \coloneqq   \begin{bmatrix}
\varepsilon_{s_x}, \dots, \varepsilon_{s_\psi} \end{bmatrix}^\top 
\coloneqq 
\begin{bmatrix}
\ln\Big(\frac{1 + \xi_{s_x}}{1 - \xi_{s_x}} \Big), \dots, \ln\Big(\frac{1 + \xi_{s_\psi}}{1 - \xi_{s_\psi}}\Big) \end{bmatrix}^\top \\ 
&r_s\coloneqq r_s(\xi_s) \coloneqq 
\textup{diag}\{[r_{s_k}(\xi_{s_k})]_{k\in\mathcal{K}}\}
\coloneqq 
\textup{diag}\Big\{ \Big [\frac{\partial \varepsilon_{s_k}}{\partial \xi_{s_k}} \Big]_{k\in\mathcal{K}} \Big \} \notag \\
&\hspace{10mm} = \textup{diag}\Big\{ \Big [\frac{2}{1-\xi^2_{s_k} } \Big]_{k\in\mathcal{K}} \Big \} 
\end{align*} 
and design the reference velocity vector $v_r : (-1,1)^6\times \mathbb{R}_{\geq 0} \to \mathbb{R}^6$ with 
\begin{align*}
& v_r \coloneqq  v_r(\xi_s,t) \coloneqq 
-g_s J_{\scriptscriptstyle O}\Big( \eta_\textup{d}(t) + \rho_{s_\eta}(t)\xi_{s_\eta} \Big)^{-1}\rho_s(t)^{-1}r_s(\xi_s)\varepsilon_{s}, 
\end{align*}
where $\rho_{s_\eta} \coloneqq \rho_{s_\eta}(t) \coloneqq \textup{diag}\{\rho_{s_\phi},\rho_{s_\theta},\rho_{s_\psi}\}$, $\xi_{s_\eta}\coloneqq [\xi_{s_\phi},\xi_{s_\eta},\xi_{s_\phi}]^\top$, and we have further used the relation $\xi_s = \rho_s^{-1}(x_{\scriptscriptstyle O}- x_\textup{d})$ from \eqref{eq:ppc errors (TCST_coop_manip)} and \eqref{eq:ksi_s (TCST_coop_manip)}.\\
\textbf{Step II-a}. Define the velocity error vector 
\begin{equation*}
e_v \coloneqq \begin{bmatrix}
e_{v_x}, \dots,  e_{v_\psi}  	
\end{bmatrix}^\top
\coloneqq v_{\scriptscriptstyle O} - v_r,  
\end{equation*} 
and select the corresponding positive performance functions $\rho_{v_k} \coloneqq \rho_{v_k}(t):\mathbb{R}_{\geq 0}\rightarrow\mathbb{R}_{>0}$ with $\rho_{v_k}(t) \coloneqq (\rho_{v_k, \scriptscriptstyle 0} - \rho_{v_k,\scriptscriptstyle \infty})\exp(-l_{v_k}t) + \rho_{v_k,\scriptscriptstyle \infty}$, such that $\rho_{v_k,\scriptscriptstyle 0}  = \lVert e_{v}(0) \rVert + \alpha_b, l_{v_k}>0$ and $\rho_{v_k,\scriptscriptstyle \infty}\in(0,\rho_{v_k,0}), \forall k\in\mathcal{K}$, where $\alpha_b$ is an arbitrary positive constant.\\
\textbf{Step II-b}. Define the normalized velocity error   
\begin{equation*}
\xi_v \coloneqq \begin{bmatrix}
\xi_{v_x}, \dots, \xi_{v_\psi} \end{bmatrix}^\top
\coloneqq \rho_v^{-1}e_v,	
\end{equation*}	
where $\rho_v\coloneqq \rho_v(t)\coloneqq\textup{diag}\{\left[\rho_{v_k}\right]_{k\in\mathcal{K}}\}$, as well as the transformed states 
$\varepsilon_v:(-1,1)^6\to\mathbb{R}^6$ and signals $r_v:(-1,1)^6\to\mathbb{R}^{6\times6}$, with 
\begin{align}
&\varepsilon_v \coloneqq \varepsilon_v(\xi_v) \coloneqq  \begin{bmatrix}
\varepsilon_{v_x}, \dots, \varepsilon_{v_\psi} \end{bmatrix}^\top
\coloneqq 
\begin{bmatrix}
\ln\Big(\frac{1 + \xi_{v_x}}{1 - \xi_{v_x}} \Big), \dots, \ln\Big(\frac{1 + \xi_{v_\psi}}{1 - \xi_{v_\psi}}\Big) \end{bmatrix}^\top \notag \\  
&r_v(\xi_v) \coloneqq 
\textup{diag}\{[r_{v_k}(\xi_{v_k})]_{k\in\mathcal{K}}\}
\coloneqq 
\textup{diag}\Big\{ \Big [\frac{\partial \varepsilon_{v_k}}{\partial \xi_{v_k}} \Big]_{k\in\mathcal{K}} \Big \} \notag \\
&\hspace{10mm} = \textup{diag}\Big\{ \Big [\frac{2}{1-\xi^2_{v_k} } \Big]_{k\in\mathcal{K}} \Big \} \label{eq:r_v (AR)},
\end{align} 
and design the decentralized feedback control protocol for each agent $i\in\mathcal{N}$ as $u_i:\mathsf{S}_i\times(-1,1)^6\times\mathbb{R}_{\geq 0}$, with
\begin{equation}
u_i \coloneqq u_i(q_i,\xi_v,t) \coloneqq -g_v J_{M_i}(q_i) \rho_v^{-1}r_v(\xi_v)\varepsilon_v(\xi_v), \label{eq:control_law_ppc (AR)} 
\end{equation}
where $g_v$ is a positive constant gain and $J_{M_i}$ as defined in \eqref{eq:J_Hirche (TCST_coop_manip)}.


The control law \eqref{eq:control_law_ppc (AR)} can be written in vector form:
\begin{align} 
u = 
U^{j'}_{j} \coloneqq  -g_v G^{+}_M(q)\rho_v^{-1}r_v(\xi_v)\varepsilon_v(\xi_v), \label{eq:control_law_ppc_vector_form (AR)}
\end{align}
where $G^{+}_M(q)$ as in \eqref{eq:control_law_ppc_vector_form (TCST_coop_manip)}, and the notation $U^{j'}_{j}$ stands for the transition from $\pi_j$ to $\pi_{j'}$.	

The next theorem summarizes the results of this section. 
\begin{theorem}
	Consider $N$ agents rigidly grasping an object with unknown coupled dynamics \eqref{eq:coupled dynamics 2 (AR)} and $\mathcal{A}(q(t_0))\in\pi_j,j\in\{1,\dots,R\}$ as well as $\lvert \theta(t_0)-\theta_{j,j'}(t_0) \rvert < \theta^*$. Then, the distributed control protocol \eqref{eq:e_pose_compact (AR)}-\eqref{eq:r_v (AR)} guarantees that $\pi_j \xrightarrow{\mathcal{T}} \pi_{j'}$ with time duration $\delta t_{j,j'}$ and all closed loop signals being bounded, and thus establishes a transition relation between $\pi_j$ and $\pi_{j'}$ for the coupled object-agents system, according to Definition \ref{def:transition (AR)}. 
\end{theorem}

\begin{proof}
	By following the proof of Theorem \ref{th:thorem_ppc (TCST_coop_manip)}, we conclude that $ \xi_s(t)\in (-1,1)^6$, $\xi_v(t)\in(-1,1)^6$, $\forall t\in\mathbb{R}_{\geq 0}$. Therefore, it holds that $\lvert e_{s_k}(t) \rvert < \rho_{s_k}(t), \forall k\in\mathcal{K}$ and thus $\lvert e_{s_k}(t) \rvert < l_0,\forall k\in\{x,y,z\}$, $t\in[t_0,\infty)$, since $\rho_{s_k,0}=l_0, \forall k\in\{x,y,z\}$. Therefore, $p_{\scriptscriptstyle O}(q(t))$ $\in\mathcal{B}(p_{j,j'}(t),l_0), \forall t\geq t_0$ and, consequently, $p_{\scriptscriptstyle O}(q(t_0+\delta t_{j,j'}))$ $\in\mathcal{B}(p^c_{\pi_{j'}},l_0)$, since $p_{j,j'}(t_0+\delta t_{j,j'}) = p^c_{\pi_{j'}}$. Moreover, since $p_{\scriptscriptstyle O}(q(t))\in\mathcal{B}(p_{j,j'}(t),l_0)$, we deduce that $\mathcal{B}(p_{\scriptscriptstyle O}(q(t)),\hat{L})\subset\mathcal{B}(p_{j,j'}(t),l_0+\hat{L})$ and invoking (\ref{eq:p_s in L_hat (AR)}) and (\ref{eq:desired_tr (AR)}), we conclude that $\mathcal{S}_{q(t)}\subset \pi_j\cup\pi_{j'}, \forall t\in[t_0,t_0+\delta t_{j,j'}]\subset[t_0,\infty)$, and therefore a transition relation with time duration $\delta t_{j,j'}$ is successfully established. Finally, according to  the proof of Theorem \ref{th:thorem_ppc (TCST_coop_manip)}, it holds $|\theta_{\scr O}(t)| < \frac{\pi}{2}$, $\forall t \geq t_0$ and hence representation singularities are provably avoided.
\end{proof}

\subsubsection{High-Level Timed Plan Generation} 
\label{subsec:High level (AR)}
The second part of the proposed solution is the derivation of a high-level plan that satisfies the given MITL formula $\mathsf{\Phi}$ and can be generated using standard techniques from automata-based formal verification methodologies. Thanks to our proposed control law that allows the transition $\pi_j \xrightarrow{\mathcal{T}} \pi_{j'}$ for all $\pi_j\in\Pi$ with $\pi_{j'}\in\mathcal{D}(\pi_j)$ in a predefined time interval $\delta t_{j,j'}$, we can abstract the motion of the coupled object-agents system  as a finite Weighted Transition System (WTS) \cite{baier2008principles}
\begin{equation*}
\mathcal{T} = \{\Pi, \Pi_0, \xrightarrow{\mathcal{T}}, \Psi, \mathcal{L},  \gamma_\mathcal{T} \},
\end{equation*}
where 
\begin{itemize}
	\item $\Pi$ is the set of states defined in Section \ref{subsec:wsp discret (AR)},
	\item $\Pi_0\subset\Pi$ is a set of initial states,
	\item $\xrightarrow{\mathcal{T}} \subseteq\Pi\times\Pi$ is a transition relation according to Definition \ref{def:transition (AR)}.
	\item $\Psi$ and $\mathcal{L}$ are the atomic propositions and the labeling function, respectively, as defined in Section \ref{subsec:specf (AR)}, and 
	\item $\gamma_\mathcal{T}:(\xrightarrow{\mathcal{T}})\rightarrow\mathbb{R}_{\geq 0}$ is a map that assigns to each transition its time duration, i.e., $\gamma_\mathcal{T}(\pi_j \xrightarrow{\mathcal{T}} \pi_{j'}) = \delta t_{j,j'}$. 
\end{itemize}
Therefore, by designing the switching protocol $U^{r_{j+1}}_{r_j}(t)$ from (\ref{eq:control_law_ppc_vector_form (AR)}):
\begin{align*}
U^{r_{j+1}}_{r_j}(t) = -g_v G^{+}_M(q(t))\rho_v(t)^{-1}r_v(\xi_v(t))\varepsilon_v(\xi_v(t)), \forall t\in[t_j, t_j+\delta t_{r_j,r_{j+1}}),  
\end{align*}
$j\in\mathbb{N}$, with (i) $t_1 = 0$, (ii) $t_{j+1} = t_{j} + \delta t_{r_j,r_{j+1}}$ and (iii) $r_j\in\{1,\dots,R\}$, $\forall j\in\mathbb{N}$, 
we can define the \textit{timed run} of the WTS as the infinite sequence $r_{\mathcal{WTS}} \coloneqq (\pi_{r_1},t_1)(\pi_{r_2},t_2)\dots$, where $\pi_{r_1}\in\Pi_0$ with $\mathcal{A}(q(0))\in\pi_{r_1}, \pi_{r_j}\in\Pi, r_j\in\{1,\dots,R\}$ and $t_j$ are the corresponding time stamps such that $\mathcal{A}(q(t_j))\in\pi_{r_j}, \forall j\in\mathbb{N}$. Every timed run $r$ generates the \textit{timed word} $w_{\scr \mathcal{WT}}(r) \coloneqq (\mathcal{L}(\pi_{r_1}), t_1)(\mathcal{L}(\pi_{r_2}),t_2)\dots$ over $\Psi$ where $\mathcal{L}(\pi_{r_j}), j\in\mathbb{N}$, is the subset of the  atomic propositions $\Psi$ that are true when $\mathcal{A}(q(t_j))\in\pi_{r_j}$.

\begin{figure}
	\centering
	\includegraphics[width=0.45\textwidth]{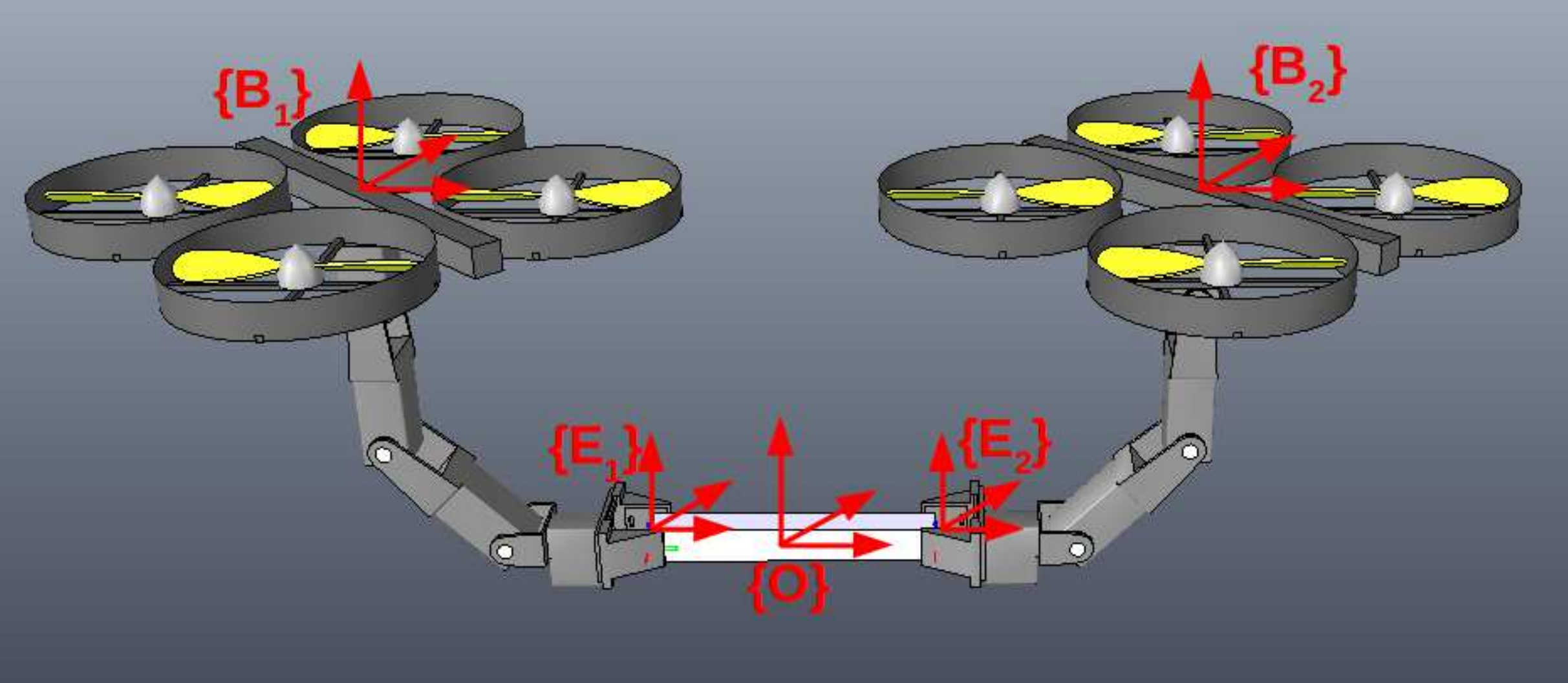}
	\caption{The aerial robots employed in the simulation rigidly grasping an object. \label{fig:v_rep_quads (AR)}}
\end{figure}

\begin{figure*}
	\begin{minipage}{0.5\linewidth}
		\centering
		\includegraphics[width=0.75\textwidth,height=0.2\textheight]{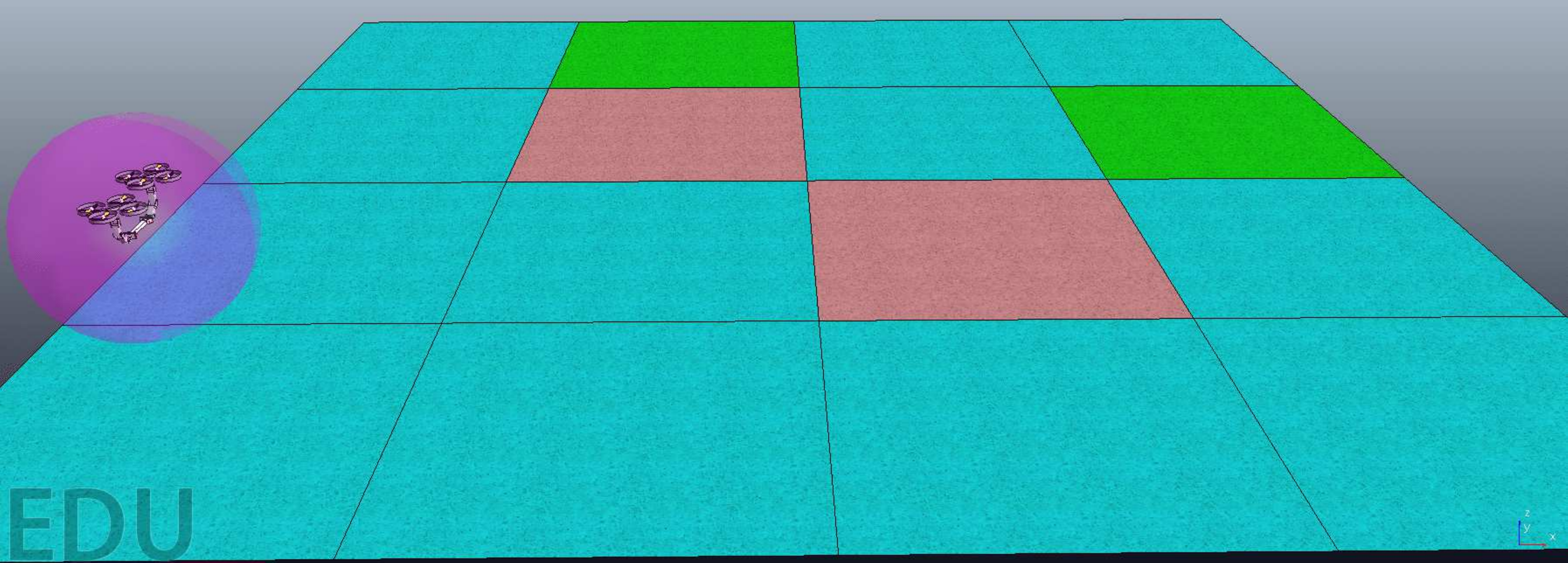}
		\subcaption{}
	\end{minipage}\hfill
	\begin{minipage}{0.5\linewidth}	
		\centering
		\includegraphics[trim = 0cm -2cm 0cm 1cm, width=0.75\textwidth, height=0.24\textheight]{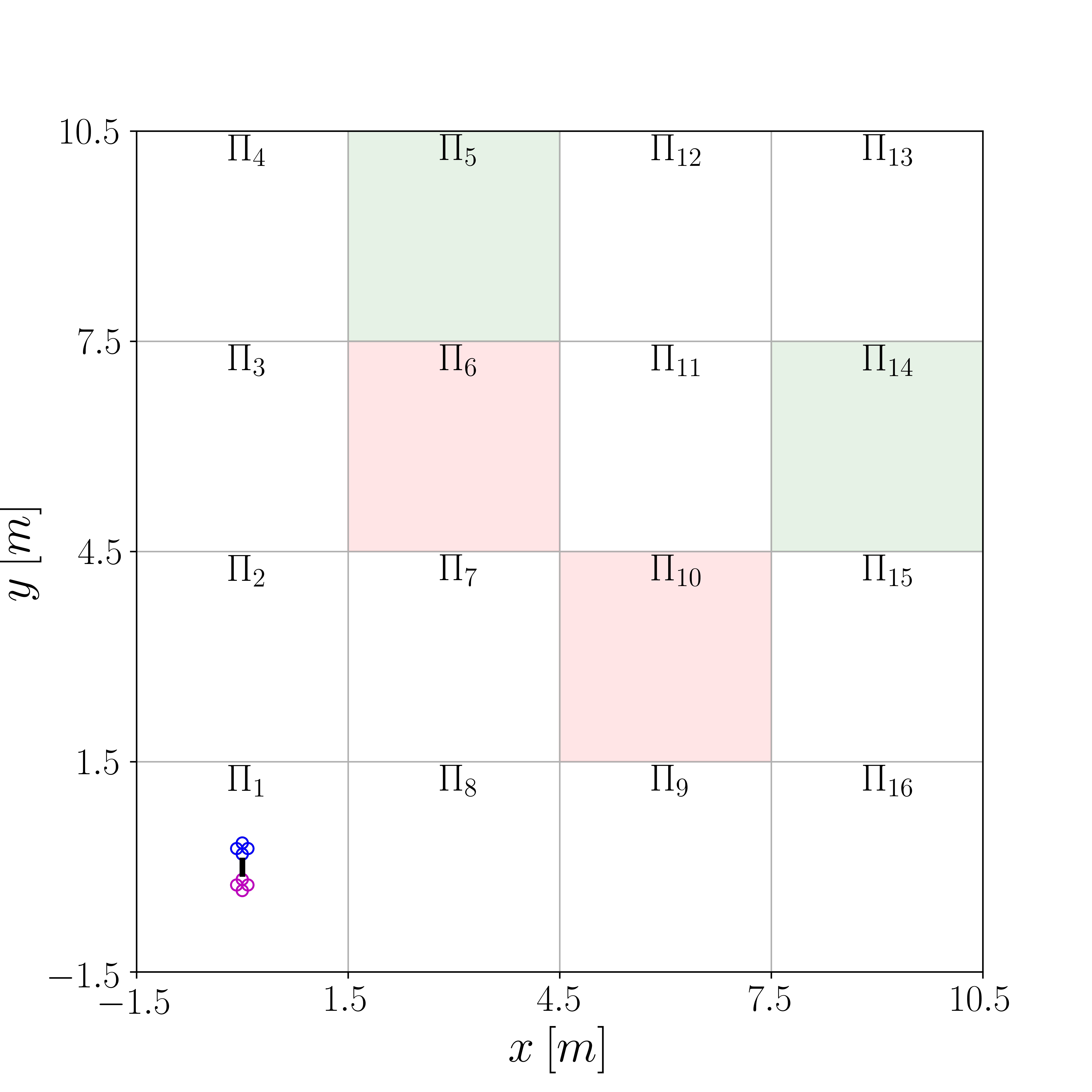}
		\subcaption{}
	\end{minipage}	
	\caption{Illustration of the initial workspace and pose of the system object-agents in the V-REP environment (a) and in top view (b). The red cells imply obstacle regions whereas the green cells are the goal ones.   \label{fig:init_wp (AR)}}
\end{figure*}

The given MITL formula $\mathsf{\Phi}$ is translated into a \textit{Timed B\"{u}chi Automaton} $\mathcal{A}^t_{\phi}$ \cite{alur1994theory} and the product $\mathcal{A}_{p}=\mathcal{T}\otimes\mathcal{A}^t_{\phi}$ is built \cite{baier2008principles}. The projection of the accepting runs of $\mathcal{A}_p$ onto $\mathcal{T}$ provides a \textit{timed run} $r_{\scr \mathcal{WT}}$ of $\mathcal{T}$ that satisfies $\phi$; $r_{\scr \mathcal{WT}}$ has the form $r_{\scr \mathcal{WT}} = (\pi_{r_1},t_1)(\pi_{r_2},t_2)\dots$, i.e., an infinite\footnote{It can be proven that if such a run exists, then there also exists a run that can be always represented as a finite prefix followed by infinite repetitions of a finite suffix \cite{baier2008principles}.} sequence of regions $\pi_{r_j}$ to be visited at specific time instants $t_j$ (i.e., $\mathcal{A}(q(t_j))\in\pi_{r_j}$) with $t_1 = 0$ and $t_{j+1} = t_j + \delta t_{r_j,r_{j+1}}, r_j\in\{1,\dots,R\}, \forall j\in\mathbb{N}$. More details on the technique can be found in \cite{baier2008principles,Alex16,alur1994theory}. 	

The execution of $r_{\scr \mathcal{WT}} = (\pi_{r_1},t_1)(\pi_{r_2},t_2)\dots$ produces a trajectory $q(t), t\in\mathbb{R}_{\geq 0}$, with timed sequence $\mathfrak{s}_t = (q(t_1),t_1)(q(t_2),t_2)\dots$, with $\mathcal{A}(q(t_j))\in\pi_{r_j}, \forall j\in\mathbb{N}$. Following Definition \ref{def:specification (AR)}, $\mathfrak{s}_t$ has the timed behavior $\mathfrak{b}_t = (\breve{\psi}_1,t_1)(\breve{\psi}_1,t_2)\dots$ with $\breve{\psi}_j\in\mathcal{L}(\pi_{r_j})$, for $\mathcal{A}(q(t_j))\in\pi_{r_j}, \forall j\in\mathbb{N}$. The latter implies that $\mathfrak{s}_t \models \mathsf{\Phi}$ and therefore that $\mathfrak{b}_t$ satisfies $\mathsf{\Phi}$. The aforementioned discussion is summarized as follows:
\begin{theorem}
	The execution of $r_{\scr \mathcal{WT}}=(\pi_{r_1},t_1)(\pi_{r_2},t_2)\dots$ of $\mathcal{T}$ that satisfies $\mathsf{\Phi}$ guarantees a timed behavior $\mathfrak{s}_t$ of the coupled object-agents system  that yields the satisfaction of $\mathsf{\Phi}$ and provides, therefore, a solution to Problem \ref{problem 1 (AR)}. 
\end{theorem}

\subsection{Simulation Results}\label{sec:Simulation Results (AR)}
The validity of the proposed framework is verified through a simulation study in the Virtual Robot Experimentation Platform (V-REP) \cite{Vrep}. We consider a rectangular rigid body of dimensions $0.025\times0.2\times0.025 \ \text{m}^3$ representing the object that is rigidly grasped by two agents. Each agent $i\in\mathcal{N}=\{1,2\}$ consists of a quadrotor base $\{B_i\}$ and a robotic arm of two revolute degrees of freedom as depicted in Fig. \ref{fig:v_rep_quads (AR)}. 
We consider that the quadrotor is fully actuated, as mentioned in Section \ref{sec:Problem-Formulation (AR)}, and there exists an embedded algorithm that translates the generalized force of the quadrotor base to the actual motor inputs.

The initial conditions of the system are taken as $p_{\scriptscriptstyle O}(0) = [0,0,1.5]^\top\text{m}$, $\eta_{\scriptscriptstyle O}(0)$ $=$ $[0,0,0]^\top\text{rad}$. The workspace is partitioned into $R=16$ regions, with $\hat{L} = 0.75 \ \text{m}$ and $l_0 = 0.5 \ \text{m}$. Fig. \ref{fig:init_wp (AR)} illustrates the aforementioned setup at $t=0$, from which it can be deduced that $\mathcal{A}(q(0))\in\pi_1$. We further define the atomic propositions $\Psi = \{``\text{green}_1", ``\text{green}_2", ``\text{red}", ``\text{obs}"\}$, representing goal ($``\text{green}_1", ``\text{green}_2"$) and obstacle ($`\text{obs}"$) regions with $\mathcal{L}(\pi_5) = \{ ``\text{green}_1"\}, \mathcal{L}(\pi_{14})$ $= \{ ``\text{green}_2"\}, \mathcal{L}(\pi_6) = \mathcal{L}(\pi_{10}) = \{ ``\text{obs}"\}$ and $\mathcal{L}(\pi_j) = \emptyset$, for the remaining regions.	

We consider the MITL formula $$\mathsf{\Phi} = (\square_{[0,\infty)} \neg ``\text{obs}" )\land \lozenge_{[0,60]}( ``\text{green}_1" \land ``\text{green}_2"),$$  which describes the following behavior: the coupled system
\begin{enumerate}
	\item must always avoid the obstacle regions, 
	\item must visit the greens region in the first $60$ seconds.  
\end{enumerate}
By following the procedure described in Section \ref{subsec:High level (AR)}, we obtain the accepting timed run 
\begin{align}
r_{\scr \mathcal{WT}} =& (\pi_{r_1},t_1)(\pi_{r_2},t_2)\dots  = (\pi_1,0)(\pi_2,6)(\pi_3,12)(\pi_4,18)(\pi_5,24)(\pi_{12},30)\notag \\ &
(\pi_{13},36)(\pi_{14},42)(\pi_{11},48)(\pi_{12},54)(\pi_5,60). \notag
\end{align}
Regarding each transition $\pi_{r_j} \xrightarrow{\mathcal{T}} \pi_{r_{j+1}}, j\in\{1,\dots,10\}$, we choose $\delta t_{r_j,r_{j'}} = 6 \ \text{s}$,  $p_{r_j,r_{j'}}(t)$ as in \eqref{eq:example of p_j (AR)} and
$\eta_{r_j,r_{j'}}(t) = [0,0,\tfrac{\pi}{4}\sin(\tfrac{\pi}{3}(t-t_{r_j}))]^\top$, where $t_{r_j} = j\delta t_{r_j,r_{j'}} = 6j$ plays the role of $t_0$ for each transition. Regarding the performance function parameters, we choose $\rho_{s_k,0} = \rho_{s_k}(t_{r_j}) = l_0 = 0.5 [\text{m}]$, $l_{s_k} = 0.5$, $\rho_{s_k,\scr \infty} = \lim\limits_{t\to\infty}\rho_{s_k}(t) = 0.1 \ [\text{m}], \forall k\in\{x,y,z\}$, $\rho_{s_k,0} = \rho_{s_k}(t_{r_j}) = \tfrac{\pi}{2} \ [\text{rad}]$, $l_{s_k} = 0.5$, $\rho_{s,k,\scr \infty} = \lim\limits_{t\to\infty}\rho_{s_k}(t) = \tfrac{\pi}{12} \ \text{r}$, $\forall k\in\{\phi,\theta, \psi\}$, $\rho_{v_k,0} = \rho_{v_k}(t_{r_j}) = 2\rvert e_{v_k}(t_{r_j}) \lvert + 0.5$, $l_{v_k} = 0.5$ and $\rho_{v_k,\scr \infty} = \lim\limits_{t\to\infty}\rho_{v_k}(t) = 0.1$, $k\in\mathcal{K}, j\in\{1,\dots,10\}$. The control gains are chosen as $g_s = 1$, $g_v = 10$, and the agents are set to contribute equally to the object motion.

The simulation results are depicted in Figs. \ref{fig:final_wp (AR)}-\ref{fig:tau (AR)}. More specifically, Fig. \ref{fig:final_wp (AR)} depicts the timed transitions of the coupled object-agents system, from which it can be deduced that $p_{\scriptscriptstyle O}(t)\in \mathcal{B}(p_{r_j,r_{j'}},l_0)$ and therefore $\mathcal{S}_q(t)\subset \pi_{r_j}\cup\pi_{r_{j'}}$, $\forall j\in\{1,\dots,10\}$. Moreover, Fig. \ref{fig:e_p (AR)} and \ref{fig:e_v (AR)} illustrate the errors $e_s(t)$ and $e_v(t)$ along with the performance functions $\rho_{s}(t), \rho_v(t)$, respectively, for all the transitions $\pi_{r_j}\to\pi_{r_{j'}}, j\in\{1,\dots,10\}$. Finally, the resulted control inputs $\tau_1, \tau_2$ for the two agents are shown in Fig. \ref{fig:tau (AR)}. A video showing the aforementioned simulation paradigm can be found on \href{https://youtu.be/AiAt9NqL1jo}{https://youtu.be/AiAt9NqL1jo}.

\begin{figure*}[t]
	\centering	
	\begin{minipage}{0.8\linewidth}
		\includegraphics[trim= -50cm 0cm 75cm 0cm,width=0.5\textwidth]{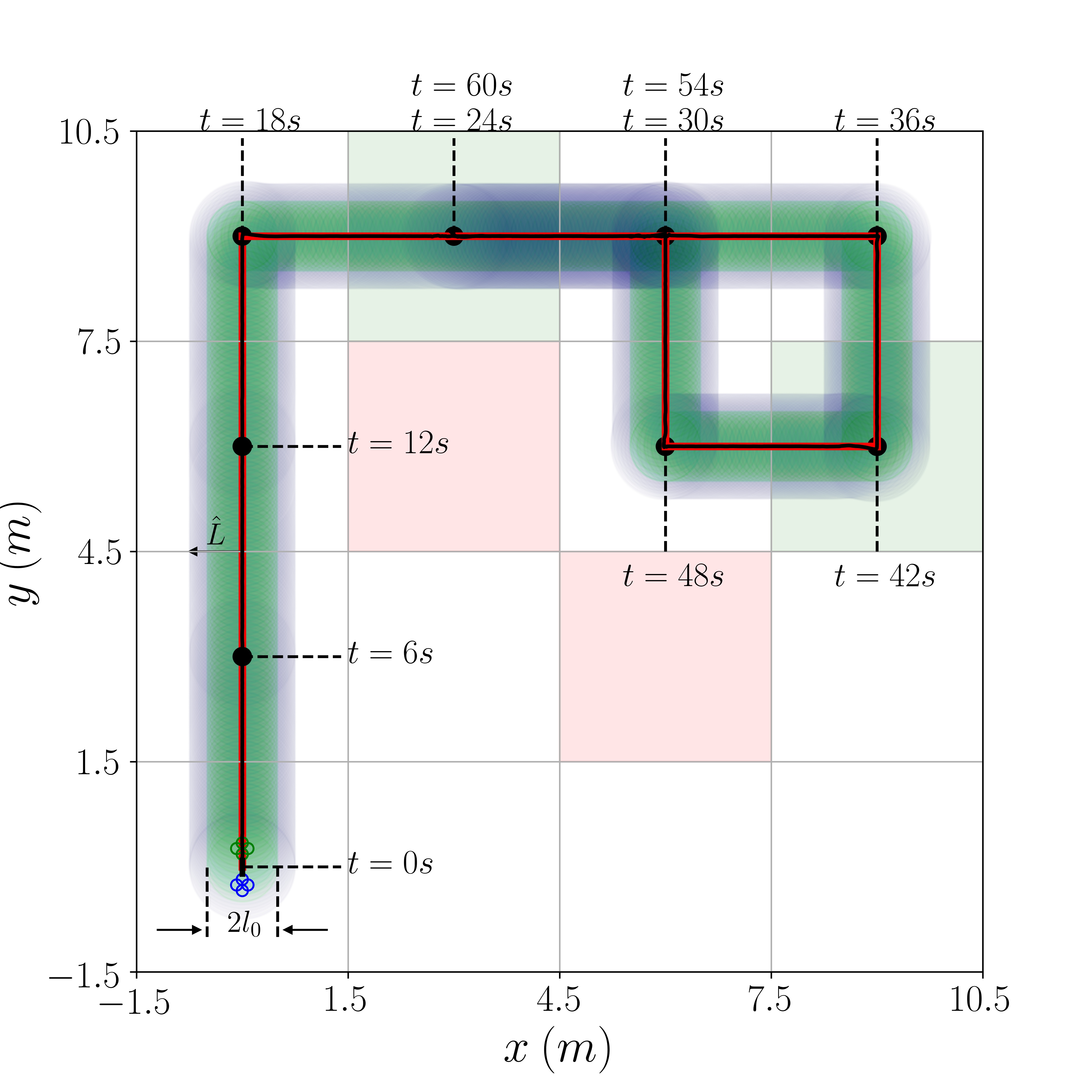}
		\subcaption{}	
	\end{minipage}
	\begin{minipage}{0.5\linewidth}
		\centering
		\includegraphics[trim = 0cm 0cm 0cm 0cm,width=0.75\textwidth]{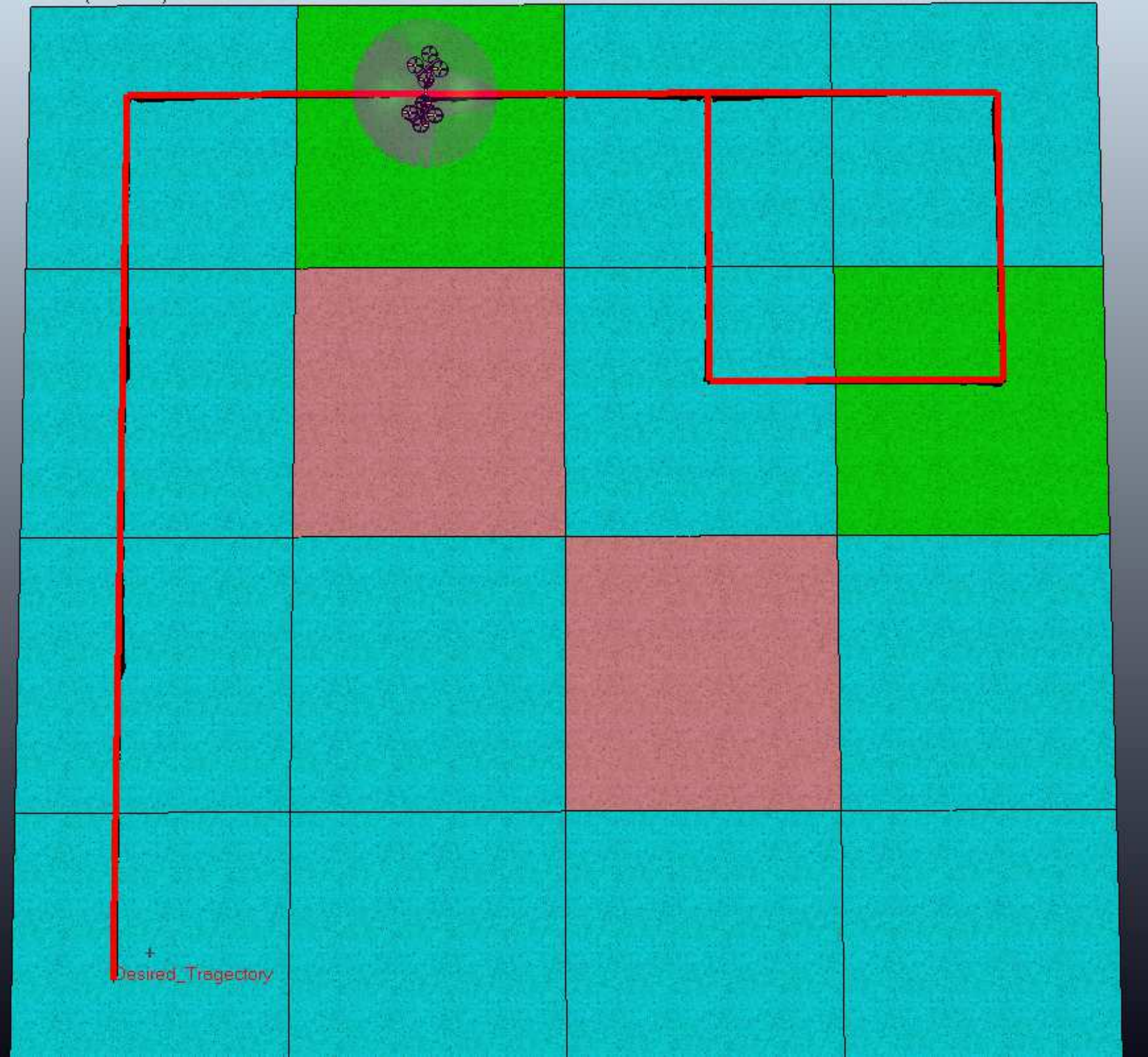}
		\subcaption{}
	\end{minipage}\hfill
	\begin{minipage}{0.5\linewidth}
		\centering
		\includegraphics[width=0.75\textwidth, height=0.23\textheight]{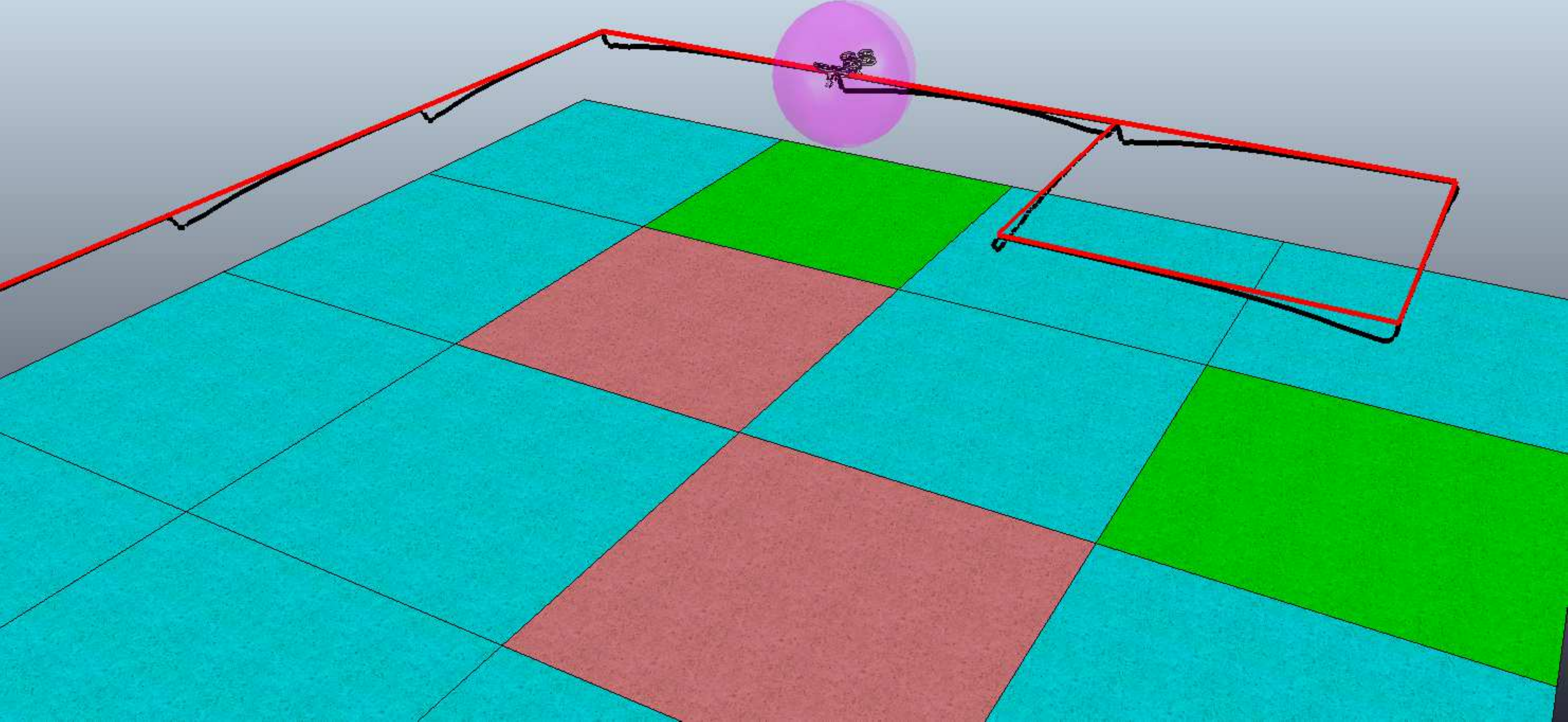}
		\subcaption{}
	\end{minipage}
	\caption{(a): The overall desired object trajectory (with red), the actual object trajectory (with black), the domain specified by $\mathcal{B}(p_{r_j,r_{j'}}(t),l_0), \forall j\in\{1,\dots,10\}$ (with green), and the domain specified by $\mathcal{B}(p_{\scriptscriptstyle O}(t),\hat{L})$ (with blue), for $t\in[0,60]$ s. (b), (c): Illustration of the system at the final region at $t=60$s in the V-REP environment along with the ball $\mathcal{B}(p_{\scriptscriptstyle O}(60),\hat{L})$. Since $p_{\scriptscriptstyle O}\in\mathcal{B}(p_{r_j,r_{j'}}(t),l_0)$, the desired timed run is successfully executed. \label{fig:final_wp (AR)}}
\end{figure*}

\begin{figure*}[t]
	\centering
	\includegraphics[width=0.9\textwidth]{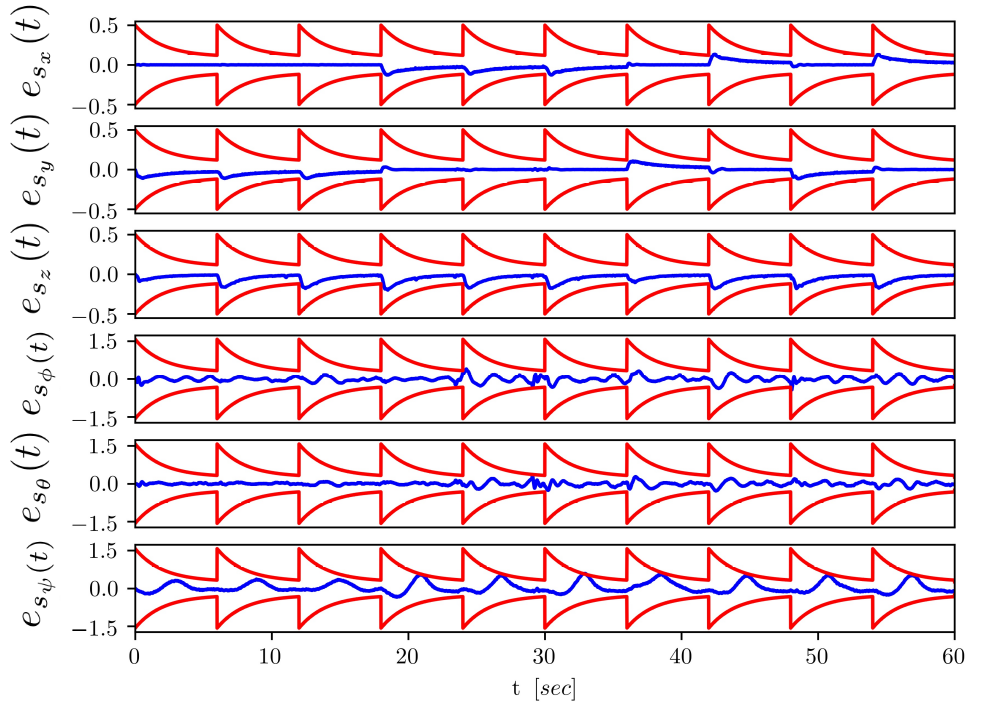}
	\caption{The pose errors $e_s(t)$ (with blue) along with the performance functions $\rho_s(t)$ (with red). \label{fig:e_p (AR)}}
\end{figure*}

\begin{figure*}[t]
	\centering
	\includegraphics[width=0.9\textwidth]{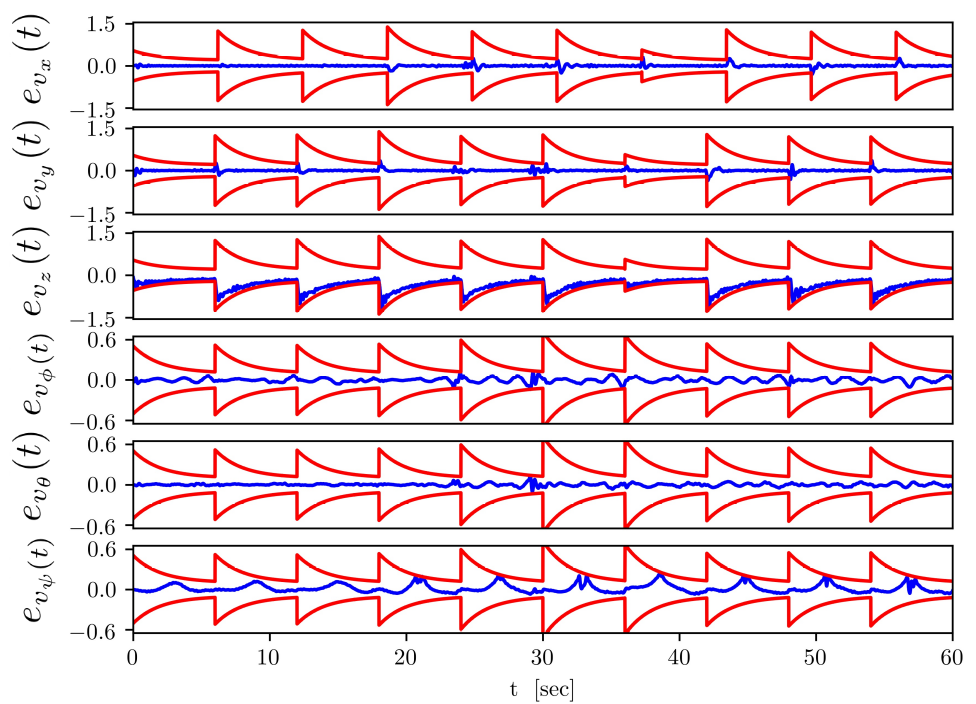}
	\caption{The velocity errors $e_v(t)$ (with blue) along with the performance functions $\rho_v(t)$ (with red). \label{fig:e_v (AR)}}
\end{figure*}

\begin{figure}
		\centering
		\includegraphics[trim=0cm 0cm 0cm 0cm,width=0.75\textwidth]{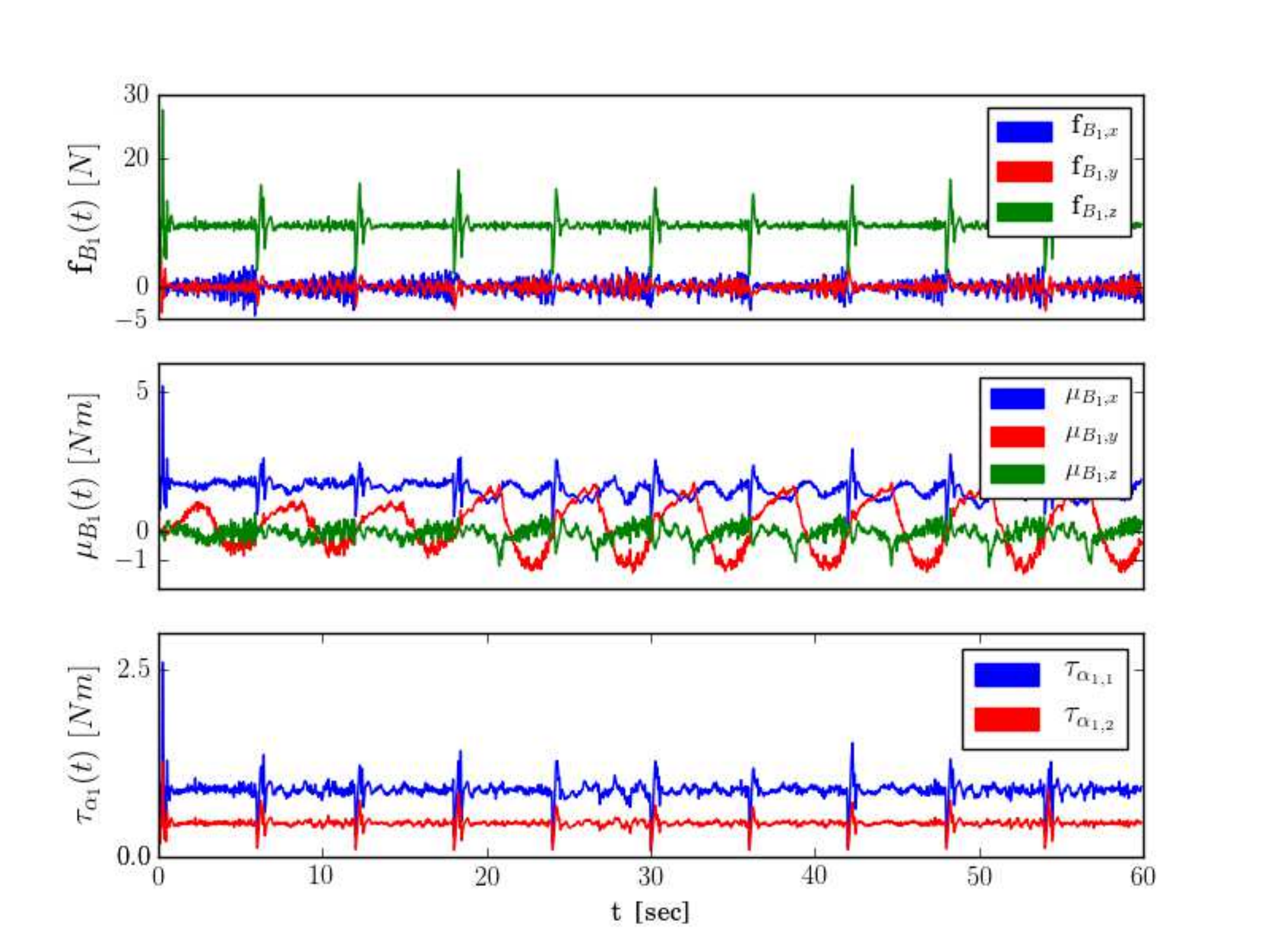}
		\includegraphics[trim=0cm 0cm 0cm 0cm, width=0.75\textwidth]{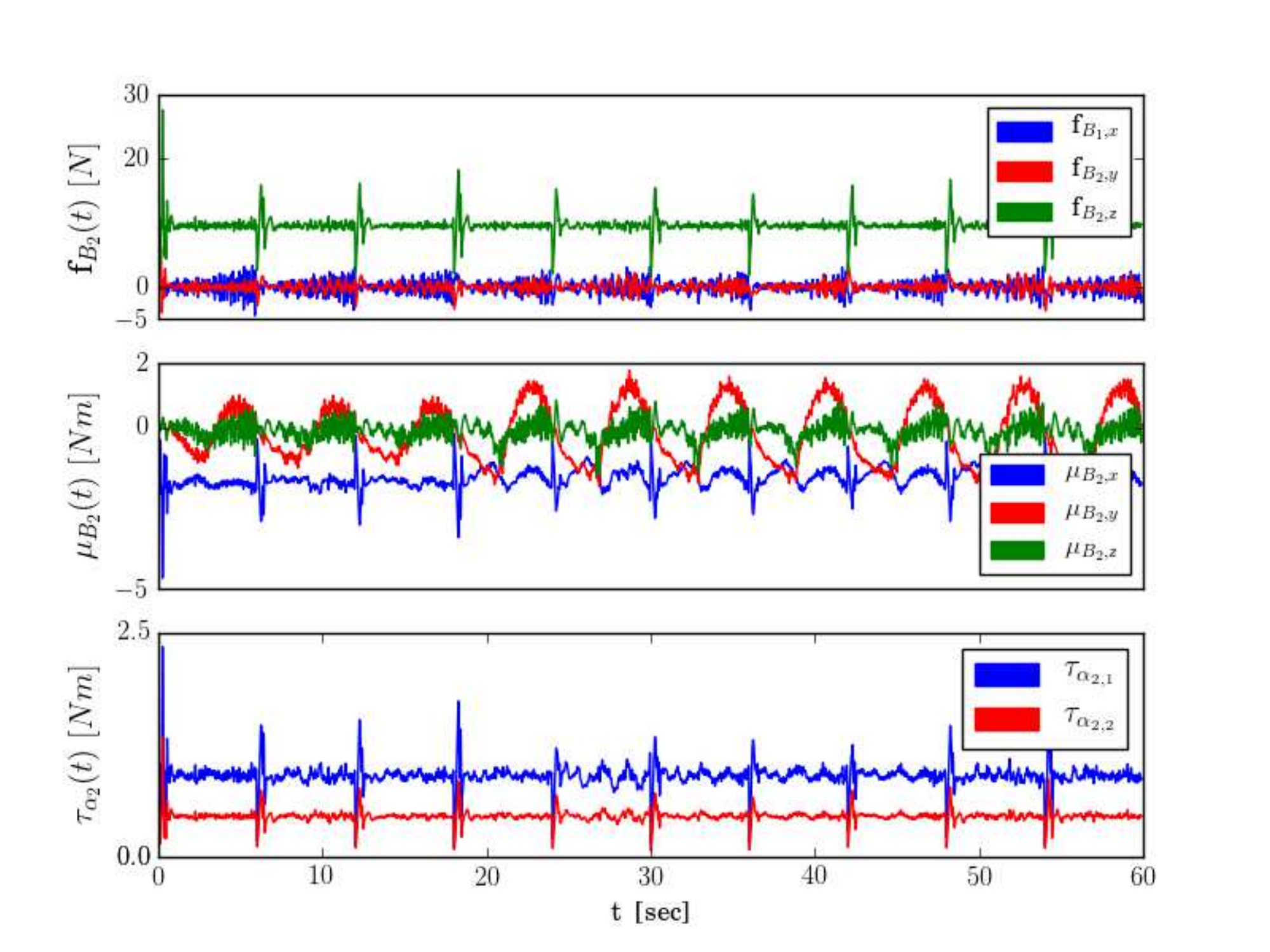}
	\caption{The resulting control inputs $\tau_i = [f^\top_{\scriptscriptstyle B_i}, \mu^\top_{\scriptscriptstyle B_i}, \tau_{\alpha_{i,1}}, \tau_{\alpha_{i,2}}]$ for $i=1$ and $i=2$; $f_{\scr B_i}, \mu_{\scr B_i}, \tau_{\alpha_i}$ are the quadrotor base forces and torques and the manipulator torque commands, respectively.} 
	\label{fig:tau (AR)}
\end{figure}

\section{Planning and Control for Multi-Robot-Object Systems under Temporal Logic Formulas} \label{sec:TASE}

The final section of this chapter considers the general case of a multi-robot-object system, with $N >1$ agents and $M>1$ objects. Unlike the previous section, the objects are now not assumed to be grasped in the starting configuration. Moreover, temporal logic specifications are imposed both to the robotic agents and the objects, whose behavior depends on the agent actions.

\subsection{Problem Formulation} \label{sec:Model and PF (TASE)}
Consider $N>1$ robotic agents operating in a workspace $\mathcal{W}$ with $M>0$ objects; $\mathcal{W}$ is a bounded open ball in $3$D space, i.e., $\mathcal{W}\coloneqq {\mathcal{B}}({0},r_0)$, where $r_0\in\mathbb{R}_{>0}$ is the radius  of $\mathcal{W}$. The objects are represented by rigid bodies whereas the robotic agents are fully actuated and consist of a fully actuated holonomic moving part (i.e., mobile base) and a robotic arm, having, therefore, access to the entire workspace.  Within $\mathcal{W}$ there exist $K>1$ smaller spheres around points of interest, which are described by $\mathcal{\pi}_k \coloneqq\bar{\mathcal{B}}({p}_{\pi_k},r_{\pi_k})$, where ${p}_{\pi_k}\in \mathbb{R}^3$ is the center and $r_{\pi_k} \in\mathbb{R}_{>0}$ the radius of $\pi_k$. 
We denote the set of all $\pi_k$ as $\Pi \coloneqq \{\pi_1,\dots,\pi_K \}$ and $\mathcal{K}_\mathcal{R} \coloneqq \{1,\dots,K\}$. 
Moreover, we introduce disjoint sets of atomic propositions $\Psi_i, \Psi^{\scriptscriptstyle O}_j$, expressed as boolean variables, that represent services provided to agent $i\in\mathcal{N}$ and object $j\in\mathcal{M}\coloneqq \{1,\dots,M\}$ in $\Pi$. The services provided at each region $\pi_k$ are given by the labeling functions $\mathcal{L}_i:\Pi\rightarrow2^{\Psi_i}, \mathcal{L}^{\scriptscriptstyle O}_j:\Pi\rightarrow2^{\Psi^{\scriptscriptstyle O}_j}$, which assign to each region $\pi_k, k\in\mathcal{K}_\mathcal{R}$, the subset of services $\Psi_i$ and $\Psi^{\scriptscriptstyle O}_j$, respectively, that can be provided in that region to agent $i\in\mathcal{N}$ and object $j\in\mathcal{M}$, respectively. In addition, we consider that the agents and the object are initially ($t=0$) in the regions of interest $\pi_{init(i)}, \pi_{init_{\scriptscriptstyle O}(j)}$, where the functions $init:\mathcal{N}\to\mathcal{K}_\mathcal{R}$, $init_{\scriptscriptstyle O}:\mathcal{M}\to\mathcal{K}_\mathcal{R}$ specify the initial region indices. 
%


The notation and modeling is identical to the one of the previous section and of Chapter \ref{chapter:cooperative manip}, which we briefly recap. For this section, 
we denote by $[z_i]_{i\in \mathcal{A}}$ the stack column vector of the vectors/scalars $z_i$, $i\in\mathcal{A}$, where $\mathcal{A}$ is an index set. 

We denote by ${q}_i,\dot{{q}}_i \in\mathbb{R}^{n_i}$, with $n_i\in\mathbb{N}, \forall i\in\mathcal{N}$, the generalized joint-space variables and their time derivatives for agent $i$. The overall joint configuration is then ${q} \coloneqq [{q}^\top_1,\dots,{q}^\top _N]^\top$, $\dot{{q}} \coloneqq [\dot{{q}}^\top _1,\dots,\dot{{q}}^\top _N]^\top \in\mathbb{R}^{n}$, with $n \coloneqq \sum_{i\in\mathcal{N}}n_i$. In addition, the inertial position and Euler-angle orientation of the $i$th end-effector, denoted by ${p}_{\scr E_i}=p_{\scr E_i}(q_i):\mathbb{R}^{n_i} \to \mathbb{R}^3$ and ${\eta}_{\scr E_i}\coloneqq \eta_{\scr E_i}(q_i):\mathbb{R}^{n_i}\to \mathbb{T}$, respectively, expressed in an inertial reference frame, can be derived by the forward kinematics. The generalized velocity of each agent's end-effector ${v}_i = [\dot{{p}}^\top_{\scr E_i},{\omega}^\top_{\scr E_i}]^\top \in\mathbb{R}^6$ is given by ${v}_i = {J}_i({q}_i)\dot{{q}}_i$, where ${J}_i = J_i(q_i):\mathbb{R}^{n_i}\to\mathbb{R}^{6\times n_i}$ is the geometric Jacobian matrix, $\forall i\in\mathcal{N}$. The matrix inverse of ${J}_i$ is well defined in the set away from \textit{kinematic singularities}, $\mathsf{S}_i \coloneqq \{{q}_i\in\mathbb{R}^{n_i}: \det({J}_i({q}_i) {J}_i({q}_i)^\top) > 0\}$, $\forall i\in\mathcal{N}$.  
The joint- and task-space dynamics of each agent are
\begin{subequations}\label{eq:manipulator dynamics (TASE)}
	\begin{align}
	&{B}_i({q}_i)\ddot{{q}}_i+{C}_{q_i}({q}_i,\dot{{q}}_i)\dot{q}_i+ {g}_{q_i}({q}_i) = {\tau}_i-J_i(q_i)^\top{h}_i \label{eq:manipulator dynamics_joint_space (TASE)}\\
	&{M}_i({q}_i)\dot{{v}}_i+{C}_i({q}_i,\dot{{q}}_i){v}_i+ {g}_i({q}_i) = {u}_i-{h}_i,\label{eq:manipulator dynamics_task_space (TASE)}
	\end{align}
\end{subequations}
with the standard dynamic terms (see previous chapters). Note again that the terms of \eqref{eq:manipulator dynamics_task_space (TASE)} are only defined in $\mathsf{S}_i$, away from singular configurations.
Avoidance of such configurations is not explicitly taken account here. Note, however, that the agents' tasks consist of navigating as well as cooperatively transporting the objects to predefined points in the workspace. This along with the fact that the agents consist of fully actuated moving bases imposes a kinematic redundancy, which can be exploited to avoid kinematic singularities.  

We consider that each agent $i$, for a given ${q}_i$, covers a spherical region $\mathcal{A}_i \coloneqq \bar{\mathcal{B}}_i(c_i(q_i),r_i) \subset \mathbb{R}^3$ of constant radius $r_i\in\mathbb{R}_{>0}$ that bounds its volume for that given ${q}_i$,  where ${c}_i\coloneqq c_i(q_i):\mathbb{R}^{n_i}\to\mathbb{R}^3$ is the center of the spherical region (a point on the robotic arm), $\forall i\in\mathcal{N}$; $\mathcal{A}_i$ can be obtained by considering the smallest sphere that covers the workspace of the robotic arm, extended with the mobile base part.
Moreover, we consider that the agents have specific power capabilities, which for simplicity, we match to positive integers $\mathfrak{c}_i > 0$, $i\in\mathcal{N}$, via an analogous relation.


Regarding the objects, we slightly change the notation with respect to the previous chapters and denote by ${x}^{\scriptscriptstyle O}_j\coloneqq [({p}^{\scriptscriptstyle O}_j)^\top,({\eta}^{\scriptscriptstyle O}_j)^\top]^\top \in\mathbb{M}$, ${v}^{\scriptscriptstyle O}_j \coloneqq [(\dot{{p}}^{\scriptscriptstyle O}_j)^\top, ({\omega}^{\scriptscriptstyle O}_j)^\top]^\top \in\mathbb{R}^{12}$, $\forall j\in\mathcal{M}$, the pose and generalized velocity of the $j$th object's center of mass. The object dynamic equations are given by the standard Newton-Euler form:
\begin{subequations} \label{eq:object dynamics (TASE)}
	\begin{align}
	& \dot{{x}}^{\scriptscriptstyle O}_j = {J}^{\scriptscriptstyle O}_j({x}^{\scriptscriptstyle O}_j){v}^{\scriptscriptstyle O}_j, \label{eq:object dynamics 1 (TASE)}\\
	& {M}_{\scriptscriptstyle O}({\eta}^{\scriptscriptstyle O}_j)\dot{{v}}^{\scriptscriptstyle O}_j+{C}_{\scriptscriptstyle O}({\eta}^{\scriptscriptstyle O}_j,{\omega}^{\scriptscriptstyle O}_j){v}^{\scriptscriptstyle O}_j+{g}_{\scriptscriptstyle O} = {h}^{\scriptscriptstyle O}_j. \label{eq:object dynamics 2 (TASE)}
	\end{align}
\end{subequations}

Similarly to the agents, each object's volume is represented by the spherical set $\mathcal{O}_j\coloneqq \mathcal{O}_j(p^{\scr O}_j) \coloneqq \bar{\mathcal{B}}_j(p^{\scr O}_j,r^{\scr O}_j) \subset \mathbb{R}^3$ of a constant radius $r^{\scriptscriptstyle O}_j \in\mathbb{R}_{>0}$, $\forall j\in\mathcal{M}$\footnote{Different center might be considered to obtain less conservative spherical volume.}. 

Similarly to \eqref{eq:coupled dynamics (TCST_coop_manip)}, the coupled dynamics between an object $j\in\mathcal{M}$ and a subset $\mathcal{V}\subseteq\mathcal{N}$ of agents that grasp it rigidly is given by 
\begin{align}
&\widetilde{{M}}_{\scriptscriptstyle \mathcal{V},j}\dot{{v}}^{\scriptscriptstyle O}_j+\widetilde{{C}}_{\scriptscriptstyle \mathcal{V},j}{v}^{\scriptscriptstyle O}_j+\widetilde{{g}}_{\scriptscriptstyle \mathcal{V},j}  = {G}_{\scriptscriptstyle \mathcal{V},j} {u}_{\scriptscriptstyle \mathcal{V}},\label{eq:coupled dynamics (TASE)}
\end{align}
where 
\begin{align*}
&\widetilde{{M}}_{\scriptscriptstyle \mathcal{V},j}\coloneqq \widetilde{{M}}_{\scriptscriptstyle \mathcal{V},j}({x}_{\scriptscriptstyle \mathcal{V},j})  \coloneqq  {M}_{\scriptscriptstyle O}+ {G}_{\scriptscriptstyle \mathcal{V},j} {M}_{\scriptscriptstyle \mathcal{V}}{G}_{\scriptscriptstyle \mathcal{V},j}^\top\\
&\widetilde{{C}}_{\scriptscriptstyle \mathcal{V},j}\coloneqq \widetilde{{C}}_{\scriptscriptstyle \mathcal{V},j}(x_{\scriptscriptstyle \mathcal{V},j})  \coloneqq  {C}_{\scriptscriptstyle O}+ {G}_{\scriptscriptstyle \mathcal{V},j} {M}_{\scriptscriptstyle \mathcal{V}}\dot{{G}}_{\scriptscriptstyle \mathcal{V},j}^\top
+ {G}_{\scriptscriptstyle \mathcal{V},j} {C}_{\scriptscriptstyle \mathcal{V}}{G}_{\scriptscriptstyle \mathcal{V},j}^\top \\
&\widetilde{{g}}_{\scriptscriptstyle \mathcal{V},j}\coloneqq \widetilde{{g}}_{\scriptscriptstyle \mathcal{V},j}({x}_{\scriptscriptstyle \mathcal{V},j})  \coloneqq  {g}_{\scriptscriptstyle O}+{G}_{\scriptscriptstyle \mathcal{V},j} {g}_{\scriptscriptstyle \mathcal{V}} \\ 
&G_{\mathcal{V},j} \coloneqq [(J^{\scr O}_{1,j})^\top,\dots, (J^{\scr O}_{|\mathcal{V}|,j})^\top], 
\end{align*}
$J^{\scr O}_{i,j} \in \mathbb{R}^{6\times 6}$, $i\in\mathcal{V}$, is the object-to-agent Jacobian matrix (see \eqref{eq:J_o_i_def (TCST_coop_manip)}), 
and ${x}_{\scriptscriptstyle \mathcal{V},j}$ is the overall state ${x}_{\scriptscriptstyle \mathcal{V},j} \coloneqq [{q}_{\scriptscriptstyle \mathcal{V}}^\top,\dot{{q}}_{\scriptscriptstyle \mathcal{V}}^\top, ({x}^{\scriptscriptstyle O}_j)^\top, ({{v}}^{\scriptscriptstyle O}_j)^\top]^\top\in \mathsf{S}_\mathcal{V}\times\mathbb{R}^{n_{\scriptscriptstyle \mathcal{V}}+6}\times\mathbb{M}$, where $\mathsf{S}_\mathcal{V} \coloneqq \prod_{i\in\mathcal{V}}\mathsf{S}_i$, and $n_\mathcal{V} \coloneqq |\mathcal{V}|$. The subscript $\mathcal{V}$ here corresponds to the agents of the set $\mathcal{V}$. We also use the following Lemma from Chapter \ref{chapter:cooperative manip} that is necessary for the following analysis. 
\begin{lemma} \label{lem:coupled dynamics skew symmetry (TASE)}
	The matrices ${B}_i({q}_i)$ and $\widetilde{{M}}_{\scriptscriptstyle \mathcal{V},j}$ are symmetric and positive definite and the matrices $\dot{{B}}_i - 2{C}_{q_i}$ and $\dot{\widetilde{{M}}}_{\scriptscriptstyle \mathcal{V},j} - 2\widetilde{{C}}_{\scriptscriptstyle \mathcal{V},j}$ are skew symmetric, $\forall i\in\mathcal{N}, j\in\mathcal{M}, \mathcal{V} \subseteq\mathcal{N}$.
\end{lemma}

Regarding the volume of the coupled agents-object system, we denote by $\mathcal{AO}_{\mathcal{V},j} \coloneqq \mathcal{AO}_{\mathcal{V},j}(p^{\scriptscriptstyle O}_j) \coloneqq \bar{\mathcal{B}}(p^{\scriptscriptstyle O}_j,r_{\scriptscriptstyle \mathcal{V},j})\subset\mathbb{R}^3$ the sphere centered at ${p}^{\scriptscriptstyle O}_j$ with constant radius $r_{\scriptscriptstyle \mathcal{V},j} \in\mathbb{R}_{>0}$, which is large enough to cover the volume of the coupled system in all configurations ${q}_{\scriptscriptstyle \mathcal{V}}$\footnote{$r_{\scriptscriptstyle \mathcal{V},j}$ can be chosen as the largest distance of the object's center of mass to a point in the agents' volume over all possible ${q}_{\scriptscriptstyle \mathcal{V}}$ (see previous section).}. This conservative formulation emanates from the sphere-world restriction of the multi-agent navigation function framework \cite{Loizou2006,koditschek1990robot}. In order to take into account other spaces, ideas from \cite{Loizou-RSS-14} could be employed or extensions of the respective works of \cite{koditschek1992robot}, \cite{loizou2017navigation} to the multi-agent case could be developed.

Moreover, in order to take into account the introduced agents' power capabilities $\mathfrak{c}_i$, $i\in\mathcal{N}$, we consider a function $\Lambda\in\{\mathsf{True},\mathsf{False}\}$ that outputs whether the agents that grasp an object are able to transport the object, based on their power capabilities. For instance, $\Lambda(m^{\scriptscriptstyle O}_j, \mathfrak{c}_{\mathcal{V}}) = \mathsf{True}$, where $m^{\scriptscriptstyle O}_j \in\mathbb{R}_{>0}$ is the mass of object $j$ and $\mathfrak{c}_{\mathcal{V}} \coloneqq [\mathfrak{c}_i]_{i\in\mathcal{V}}$, implies that the agents $\mathcal{V}$ have sufficient power capabilities to cooperatively transport object $j$. 

Next, we define the boolean functions $\mathcal{AG}_{i,j}:\mathbb{R}^{n_i}\times\mathbb{M}\to\{\mathsf{True},\mathsf{False}\}, i\in\mathcal{N},j\in\mathcal{M}$, to denote whether agent $i\in\mathcal{N}$ rigidly grasps an object $j\in\mathcal{M}$ at a given configuration ${q}_i,{x}^{\scriptscriptstyle O}_j$; We also define $\mathcal{AG}_{i,0}: \mathbb{R}^{n_i}\times\mathbb{M}^M\to\{\mathsf{True},\mathsf{False}\}$, to denote that agent $i$ does not grasp any objects, i.e., $\mathcal{AG}_{i,j}({q}_i,{x}^{\scriptscriptstyle O}_j) =  \mathsf{False}, \forall j\in\mathcal{M} \Leftrightarrow \mathcal{AG}_{i,0}({q}_i, {x}^{\scriptscriptstyle O}) = \mathsf{True}$, $\forall i\in\mathcal{N}$, where ${x}^{\scriptscriptstyle O} \coloneqq [ {x}^{\scriptscriptstyle O}_j]_{j\in\mathcal{M}}\in\mathbb{M}^M$.
Note also that $\mathcal{AG}_{i,\ell}({q}_i,{x}^{\scriptscriptstyle O}_\ell) = \mathsf{True}, \ell\in\mathcal{M} \Leftrightarrow \mathcal{AG}_{i,j}({q}_i,{x}^{\scriptscriptstyle O}_j)= \mathsf{False}, \forall j\in \mathcal{M}\backslash\{\ell\}$, i.e., agent $i$ can grasp at most one object at a time. 


We also assume the  existence of a procedure ${\mathcal{P}}_s$ that outputs whether or not a set of non-intersecting spheres fits in a larger sphere as well as possible positions of the spheres in the case they fit. More specifically, given a region of interest $\pi_k$ and a number $\widetilde{N}\in\mathbb{N}$ of sphere radii (of agents and/or objects) the procedure can be seen as a function ${\mathcal{P}}_s \coloneqq [\mathcal{P}_{s,0}, {\mathcal{P}}^\top_{s,1}]^\top$, where $\mathcal{P}_{s,0}:\mathbb{R}^{\widetilde{N}+1}_{\geq 0}\to\{\mathsf{True}, \mathsf{False}\}$ outputs whether the spheres fit in the region $\pi_k$ whereas  ${\mathcal{P}}_{s,1}$ provides possible configurations of the agents and the objects 
or ${0}$ in case the spheres do not fit.
For instance, $P_{s,0}(r_{\pi_2},r_1,r_3,r^{\scriptscriptstyle O}_1,r^{\scriptscriptstyle O}_5)$ determines whether the agents $1,3$ and the objects $1,5$ fit in region $\pi_2$, without colliding with each other; $({q}_1,{q}_3,{x}^{\scriptscriptstyle O}_1, {x}^{\scriptscriptstyle O}_5)={P}_{s,1}(r_{\pi_2},r_1,r_3,r^{\scriptscriptstyle O}_1,r^{\scriptscriptstyle O}_5)$ provides a set of configurations such that $\mathcal{A}_1({q}_1), \mathcal{A}_3({q}_3), \mathcal{O}_1({x}^{\scriptscriptstyle O}_1), \mathcal{O}_{5}({x}^{\scriptscriptstyle O}_5) \subset \pi_2$ and the pairwise intersections of the respective sets are empty.
The problem of finding an algorithm $\mathcal{P}_s$ is a special case of the sphere packing problem \cite{chen2008sphere}. Note, however, that we are not interested in finding the maximum number of spheres that can be packed in a larger sphere but, rather, in the simpler problem of determining whether a set of spheres can be packed in a larger sphere. 

The following definitions address the transitions of the agents and the objects between the regions of interest, as in the previous sections.

\begin{definition}(\textbf{Transition})  \label{def:agent transition (TASE)}
	Consider that $\mathcal{A}_i({q}_i(t_0)) \subset \pi_k$, for some $i\in\mathcal{N}, k\in\mathcal{K}_\mathcal{R}, t_0\in\mathbb{R}_{\geq 0}$, and
	\begin{align} \label{eq:transition conditions (TASE)}
	    \mathcal{A}_i(q_i(t_0)) \cap \bigg( \mathcal{A}_l(q_l(t_0)) \cup  \mathcal{O}_j(p^{\scr O}_j(t_0)) \cup \mathcal{AO}_{\mathcal{V},\ell}(p^{\scr O}_\ell(t_0))  \bigg) = \emptyset,
	\end{align}
	for all $l\in\mathcal{N}\backslash\{i\}$, $j\in\mathcal{M}$, and any $\mathcal{V} \subseteq \mathcal{N} \backslash\{i\}$, $\ell \in \mathcal{M}$ such that $\mathcal{AG}_{h,\ell}(q_h(t_0),x^{\scr O}_\ell(t_0)) = \mathsf{True}$, $\forall h\in\mathcal{V}$.	
	Then, there exists a transition for agent $i$ from region $\pi_k$ to $\pi_{k'},k'\in\mathcal{K}_\mathcal{R}$, denoted as $\pi_k\rightarrow_i\pi_{k'}$, if there exists a finite $t_f\geq t_0$ such that $\mathcal{A}_i({q}_i(t_f))\subset\pi_{k'}$, $\mathcal{A}_i(q_i(t)) \subset \mathcal{W}$, $\mathcal{A}_i(q_i(t)) \cap \pi_m =\emptyset$, $\forall t\in[t_0,t_f], m \in \mathcal{K}_\mathcal{R}\backslash\{k,k'\}$	
	and \eqref{eq:transition conditions (TASE)} holds for all $t\in[t_0,t_f]$.
\end{definition}

\begin{definition}(\textbf{Grasping}) \label{def:grasping (TASE)}
	Consider that $\mathcal{A}_i({q}_i(t_0))\subset\pi_k$, $\mathcal{O}_j({x}^{\scriptscriptstyle O}_j(t_0))\subset\pi_k$, $k\in\mathcal{K}_\mathcal{R}$ for some $i\in\mathcal{N}$, $j\in\mathcal{M}$, $t_0\in\mathbb{R}_{\geq 0}$, and \eqref{eq:transition conditions (TASE)} holds.  
	Then, agent $i$ \textit{grasps} object $j$, denoted as $i\xrightarrow{g}j$, if there exists a finite $t_f\geq t_0$ such that $\mathcal{AG}_{i,j}({q}_i(t_f),{p}^{\scriptscriptstyle O}_j(t_f)) = \mathsf{True}$, $\mathcal{A}_i({q}_i(t))\subset\pi_k$,   $\mathcal{O}_j({p}^{\scriptscriptstyle O}_j(t))\subset\pi_k$, $k\in\mathcal{K}_\mathcal{R}$, $\forall t\in[t_0,t_f]$, and \eqref{eq:transition conditions (TASE)} holds for all objects except for $j$ and all $t\in[t_0,t_f]$.
\end{definition}

\begin{definition}(\textbf{Releasing}) \label{def:releasing (TASE)}
	Consider that $\mathcal{A}_i({q}_i(t_0))\subset\pi_k$, $\mathcal{O}_j({p}^{\scriptscriptstyle O}_j(t_0))\subset\pi_k$, $k\in\mathcal{K}_\mathcal{R}$ for some $i\in\mathcal{N}$, $j\in\mathcal{M}$, $t_0\in\mathbb{R}_{\geq 0}$, with $\mathcal{AG}_{i,j}({q}_i(t_0),{x}^{\scriptscriptstyle O}_j(t_0)) = \mathsf{True}$, and \eqref{eq:transition conditions (TASE)} holding for all objects except for $j$. 
	Then, agent $i$ \textit{releases} object $j$, denoted as $i\xrightarrow{r}j$, if there exists a finite $t_f\geq t_0$ such that $\mathcal{AG}_{i,0}({q}_i(t_f),{x}^{\scriptscriptstyle O}(t_f)) = \mathsf{True}$, $\mathcal{A}_i({q}_i(t))\subset\pi_k$,  $\mathcal{O}_j({p}^{\scriptscriptstyle O}_j(t))\subset\pi_k$, $k\in\mathcal{K}_\mathcal{R}$, $t\in[t_0,t_f]$, and \eqref{eq:transition conditions (TASE)} holding for all objects except for $j$ and all $t\in[t_0,t_f]$.
\end{definition}

\begin{definition}(\textbf{Transportation}) \label{def:agent-object transition (TASE)}
	Consider a nonempty subset of agents $\mathcal{V}\subseteq\mathcal{N}$ and an object $j\in\mathcal{M}$ such that $\mathcal{AG}_{i,j}({q}_i(t_0), {x}^{\scriptscriptstyle O}_j(t_0)) = \mathsf{True}$, $\forall i\in\mathcal{V}$ and $\mathcal{AO}_{\mathcal{V},j}(p^{\scr O}_j(t_0)) \subset \pi_k$ for some $k\in\mathcal{K}_\mathcal{R}$, $t_0\geq 0$, with 	
	\begin{equation} \label{eq:transportation condition (TASE)}
		\mathcal{AO}_{\mathcal{V},j}(p^{\scr O}_j(t_0)) \cap \bigg( \mathcal{A}_l(q_l(t_0)) \cup \mathcal{O}_\ell(p^{\scr O}_\ell(t_0)) \cap \mathcal{AO}_{\mathcal{V}',j'}(p^{\scr O}_{j'}(t_0))  \bigg) = \emptyset,
	\end{equation}
	for all $l\in\mathcal{N}\backslash \mathcal{V}$, $\ell \in\mathcal{M}\backslash\{j\}$, and any $\mathcal{V'} \subseteq \mathcal{N}\backslash\mathcal{V}$, $j'\in\mathcal{M} \backslash\{j\}$ such that $\mathcal{AG}_{h,j'}(q_h(t_0),x^{\scr O}_{h}(t_0)) = \mathsf{True}$, $\forall h\in\mathcal{V}'$.
	Then, the team of agents $\mathcal{V}$ transports the object $j$ from region $\pi_k$ to region $\pi_{k'}, k'\in\mathcal{K}_\mathcal{R}$, denoted as $\pi_k \xrightarrow{T}_{\mathcal{V},j} \pi_{k'}$, if there exists a finite $t_f \geq t_0$  such that $\mathcal{AO}_{\mathcal{V},j}(p^{\scr O}_j(t_f))\subset\pi_{k'}$, $\mathcal{AO}_{\mathcal{V},j}(p^{\scr O}_j(t)) \subset \mathcal{W}$, $\mathcal{AG}_{i,j}({q}_i(t),{x}^{\scriptscriptstyle O}_j(t)) = \mathsf{True}$, $\forall i\in\mathcal{V}$, $\mathcal{AO}_{\mathcal{V},j}(p^{\scr O}_j(t)) \cap \pi_m = \emptyset$, $\forall m \in \mathcal{K}_\mathcal{R}\backslash\{k,k'\}$, $t\in[t_0,t_f]$, and \eqref{eq:transportation condition (TASE)} holding for all $t\in[t_0,t_f]$.
\end{definition}
Loosely speaking, the aforementioned definitions correspond to specific actions of the agents, namely \textit{transition}, \textit{grasp}, \textit{release}, and \textit{transport}. We do not define these actions explicitly though, since we will employ directly designed continuous control inputs $\tau_i$, $u_i$, as will be seen later. Moreover, in the \textit{grasping/releasing} definitions, we have not incorporated explicitly collisions between the agent and the object to be grasped/released other than the grasping point. Such collisions are assumed to be avoided.    


Our goal is to control the multi-agent system such that the agents and the objects obey a given specification over their atomic propositions $\Psi_i, \Psi^{\scriptscriptstyle O}_j, \forall i\in\mathcal{N},j\in\mathcal{M}$. 
Given the trajectories ${q}_i(t), {x}^{\scriptscriptstyle O}_j(t), t\in\mathbb{R}_{\geq 0}$, of agent $i$ and object $j$, respectively, their corresponding \textit{behaviors} are given by the infinite sequences 
\begin{align*}
& \mathfrak{b}_i \coloneqq ({q}_i(t),\breve{\psi}_i) \coloneqq ({q}_i(t_{i,1}),\breve{\psi}_{i,1})({q}_i(t_{i,2}),\breve{\psi}_{i,2})\dots, \\ 
& \mathfrak{b}^{\scriptscriptstyle O}_j \coloneqq ({x}^{\scriptscriptstyle O}_j(t),\breve{\psi}^{\scriptscriptstyle O}_j) \coloneqq ({x}^{\scriptscriptstyle O}_j(t^{\scriptscriptstyle O}_{j,1}),\breve{\psi}^{\scriptscriptstyle O}_{j,1}) ({x}^{\scriptscriptstyle O}_j(t^{\scriptscriptstyle O}_{j,2}),\breve{\psi}^{\scriptscriptstyle O}_{j,2})\dots,
\end{align*}
with $t_{i,\ell+1} > t_{i,\ell} \geq 0, t^{\scriptscriptstyle O}_{j,\ell+1} > t^{\scriptscriptstyle O}_{j,\ell} \geq 0, \forall \ell\in\mathbb{N}$, representing specific time stamps. The sequences $\breve{\psi}_i, \breve{\psi}^{\scriptscriptstyle O}_j$ are the services provided to the agent and the object, respectively, over their trajectories, i.e., $\breve{\psi}_{i,\ell}\in 2^{\Psi_i}, \breve{\psi}^{\scriptscriptstyle O}_{j,l}\in 2^{\Psi^{\scriptscriptstyle O}_j}$ with $\mathcal{A}_i({q}_i(t_{i,\ell})) \subset \pi_{k_{i,\ell}}, \breve{\psi}_{i,\ell} \in \mathcal{L}_i(\pi_{k_{i,\ell}})$ and $\mathcal{O}_j({x}^{\scriptscriptstyle O}_j(t^{\scriptscriptstyle O}_{j,l})) \subset \pi_{k^{\scriptscriptstyle O}_{j,l}}, \breve{\psi}^{\scriptscriptstyle O}_{j,l} \in \mathcal{L}^{\scriptscriptstyle O}_j(\pi_{k^{\scriptscriptstyle O}_{j,l}}), k_{i,\ell}, k^{\scriptscriptstyle O}_{j,l} \in\mathcal{K}_\mathcal{R}, \forall \ell,l\in\mathbb{N}, i\in\mathcal{N}, j\in\mathcal{M}$, where $\mathcal{L}_i$ and $\mathcal{L}^{\scriptscriptstyle O}_j$ are the previously defined labeling functions. The following Lemma then follows: 
\begin{lemma}
	The behaviors $\mathfrak{b}_i, \mathfrak{b}^{\scriptscriptstyle O}_j$ satisfy formulas $\mathsf{\Phi}_i, \mathsf{\Phi}_{\scriptscriptstyle O_j}$ if $\breve{\psi}_i \models \mathsf{\Phi}_i$ and $\breve{\psi}^{\scriptscriptstyle O}_j \models \mathsf{\Phi}^{\scriptscriptstyle O}_j$, respectively.
\end{lemma}

The control objectives are given as LTL formulas $\mathsf{\Phi}_i, \mathsf{\Phi}^{\scriptscriptstyle O}_j$ over $\Psi_i, \Psi^{\scriptscriptstyle O}_j$, respectively, $\forall i\in\mathcal{N},j\in\mathcal{M}$. The LTL formulas $\mathsf{\Phi}_i, \mathsf{\Phi}^{\scriptscriptstyle O}_j$ are satisfied if there exist behaviors $\mathfrak{b}_i,\mathfrak{b}^{\scriptscriptstyle O}_j$ of agent $i$ and object $j$ that satisfy  $\mathsf{\Phi}_i, \mathsf{\Phi}^{\scriptscriptstyle O}_j$. We are now ready to give a formal problem statement consider in this section:
\begin{problem} \label{problem (TASE)}
	Consider $N$ robotic agents and $M$ objects in $\mathcal{W}$ subject to the dynamics \eqref{eq:manipulator dynamics (TASE)} and \eqref{eq:object dynamics (TASE)}, respectively, not colliding at $t=0$, and
	\begin{equation*}
		\dot{{q}}_i(0) = {0}, {v}^{\scriptscriptstyle O}_{j} = {0}, \mathcal{A}_i({q}_i(0))\subset \pi_{\text{init}(i)}, \mathcal{O}_j({x}^{\scriptscriptstyle O}_j(0)) \subset\pi_{\text{init}_{\scriptscriptstyle O}(j)}, \forall i\in\mathcal{N},j\in\mathcal{M},
	\end{equation*}
	Given the disjoint sets $\Psi_i,\Psi^{\scriptscriptstyle O}_j$, $N$ LTL formulas $\mathsf{\Phi}_i$ over $\Psi_i$ and $M$ LTL formulas $\mathsf{\Phi}^{\scriptscriptstyle O}_j$ over $\Psi^{\scriptscriptstyle O}_j$, develop a control strategy that achieves behaviors $\mathfrak{b}_i, \mathfrak{b}^{\scriptscriptstyle O}_j$ which yield the satisfaction of 
	$\mathsf{\Phi}_i, \mathsf{\Phi}^{\scriptscriptstyle O}_j, \forall i\in\mathcal{N},j\in\mathcal{M}$.
\end{problem}
Note that it is implicit in the problem statement the fact that the agents/objects starting in the same region can actually fit without colliding with each other. Technically, it holds that $\mathcal{P}_{s,0}(r_{\pi_k},[r_i]_{i\in \{i\in\mathcal{N}: \text{init}(i)= k\} }$, $[r^{\scriptscriptstyle O}_j]_{j\in \{ j\in\mathcal{M}: \text{init}_{\scriptscriptstyle O}(j) = k\} }) = \mathsf{True}$, $\forall k\in\mathcal{K}_\mathcal{R}$.

\subsection{Problem Solution} \label{sec:main results (TASE)}

\subsubsection{Continuous Control Design}
The first ingredient of our solution is the development of feedback control laws that establish agent transitions and object transportations as defined in Def. \ref{def:agent transition (TASE)} and \ref{def:agent-object transition (TASE)}, respectively. Although the control protocols of Sections \ref{sec:MPC (MED)}, \ref{sec:MPC (ECC)}, \ref{sec:MAS (Automatica_adaptive_nav)} can be applied, we focus on an alternative design that follows the concept of multi-robot navigation functions (see Appendix \ref{app:NF}). 
Moreover, we do not focus on the grasping/releasing actions of Def. \ref{def:grasping (TASE)}, \ref{def:releasing (TASE)} and we refer to some existing methodologies that can derive the corresponding control laws (e.g., \cite{Cutkosky2012},\cite{Reis2015}). 

Assume that the conditions of Problem \ref{problem (TASE)} hold for some $t_0 \in\mathbb{R}_{\geq 0}$, i.e., all agents and objects are located in regions of interest with zero velocity.
We design a control law such that a subset of agents performs a transition between two regions of interest and another subset of agents performs cooperative object transportation, according to Def. \ref{def:agent transition (TASE)} and \ref{def:agent-object transition (TASE)}, respectively. 
Let $\mathcal{Z}, \mathcal{V}, \mathcal{G}, \mathcal{R}\subseteq \mathcal{N}$ denote disjoint sets of agents corresponding to transition, transportation, grasping and releasing actions, respectively, with $\lvert \mathcal{Z} \rvert + |\mathcal{V}| + \lvert \mathcal{G} \rvert + \lvert \mathcal{R} \rvert  \leq \lvert \mathcal{N} \rvert$ and $\mathcal{A}_z({q}_z(t_0))\subset \pi_{k_z}$, $\mathcal{A}_{\nu}({q}_\nu(t_0))\subset \pi_{k_\nu}$, $\mathcal{A}_{g}({q}_g(t_0))\subset \pi_{k_g}$, $\mathcal{A}_\rho({q}_\rho(t_0))\subset \pi_{k_\rho}$, where $k_z,k_\nu,k_g,k_\rho\in\mathcal{K}_\mathcal{R}$, $\forall z\in\mathcal{Z}, \nu\in\mathcal{V},g\in\mathcal{G},\rho\in\mathcal{R}$. Note that there might be idle agents in some regions, not performing any actions, i.e., the set $\mathcal{N}\backslash(\mathcal{Z}\cup\mathcal{V}\cup\mathcal{G}\cup\mathcal{R})$ might not be empty.

More specifically, regarding the transportation actions, we consider that the set $\mathcal{V}$ consists of $\bar{T}$ disjoint teams of agents, with each team consisting of agents that are in the same region of interest and aim to collaboratively transport an object, i.e. $\mathcal{V} = \mathcal{V}_1\cup\mathcal{V}_2\cup\dots\mathcal{V}_{\bar{T}}$, and $\mathcal{A}_{\nu}({q}_{\nu}(t_0))\subset \pi_{k_{\mathcal{V}_m}}, \forall \nu\in\mathcal{V}_m, m\in\{1,\dots,\bar{T}\}$, where $k_{\mathcal{V}_m}\in\mathcal{K}_\mathcal{R}, \forall m\in\{1,\dots,\bar{T}\}$.
Let also $\mathcal{S}\coloneqq\{s_{\mathcal{V}_1},s_{\mathcal{V}_2},\dots, s_{\mathcal{V}_{\bar{T}}}\}, \mathcal{X}\coloneqq\{[x_g]_{g\in\mathcal{G}}\}, \mathcal{Y}\coloneqq\{[y_\rho]_{\rho\in\mathcal{R}}\}\subseteq\mathcal{M}$ be disjoint sets of objects to be transported, grasped, and released, respectively. More specifically, each team $\mathcal{V}_m$ in the set $\mathcal{V}$ will transport cooperatively object $s_{\mathcal{V}_m}$, $m\in\{1,\dots,\bar{T}\}$, each agent $g\in\mathcal{G}$ will grasp object $x_g\in\mathcal{X}$ and each agent $\rho\in\mathcal{R}$ will release object $y_\rho\in\mathcal{Y}$. 
Then, suppose that the following conditions also hold at $t_0$:
\begin{itemize}	
	\item $\mathcal{AG}_{\rho,y_\rho}({q}_\rho(t_0),{x}^{\scriptscriptstyle O}_{y_\rho}(t_0)) = \mathsf{True}, \forall \rho\in\mathcal{R}$, 
	\item $\mathcal{AG}_{z,0}({q}_z(t_0),{x}^{\scriptscriptstyle O}(t_0)) = \mathsf{True}$, $\forall z\in\mathcal{Z}$, 
	\item $\mathcal{AG}_{g,0}({q}_g(t_0),{x}^{\scriptscriptstyle O}(t_0))=\mathsf{True}, \forall g\in\mathcal{G}$, 
	\item $\mathcal{AG}_{\nu,  s_{\mathcal{V}_m}}({q}_\nu(t_0),{x}^{\scriptscriptstyle O}_{s_{\mathcal{V}_m}}(t_0))  = \mathsf{True}$, $\forall \nu\in\mathcal{V}_m, m\in\{1,\dots,\bar{T}\}$,
	\item $\mathcal{AO}_{\mathcal{V}_m,s_{\mathcal{V}_m}}(p^{\scr O}_{s_{\mathcal{V}_m}}(t_0)) \subset \pi_{k_{\mathcal{V}_m}}$, $\forall m\in\{1,\dots,\bar{T}\}$,
	\item $\mathcal{O}_{x_g}({p}^{\scriptscriptstyle O}_{x_g}(t_0))\subset \pi_{k_g}, \forall g\in\mathcal{G}$, 
	\item $\mathcal{O}_{y_\rho}({p}^{\scriptscriptstyle O}_{y_\rho}(t_0))\subset \pi_{k_\rho}, \forall \rho\in\mathcal{R}$,
\end{itemize}
which mean, intuitively, that the objects $s_{\mathcal{V}_m}$, $x_g, y_\rho$ to be transported, grasped, released, are in the regions $\pi_{k_{\mathcal{V}_m}}$, $\pi_{k_g}$, $\pi_{k_\rho}$, respectively, and there is also grasping compliance with the corresponding agents.  
By also assuming that the agents do not collide with each other or with the objects (except for the transportation/releasing task agents), we guarantee that the conditions of Def. \ref{def:agent transition (TASE)}-\ref{def:agent-object transition (TASE)} hold. 

In the following, we design ${\tau}_{z}$ and ${u}_{\nu}$ such that $\pi_{k_z}\rightarrow_z\pi_{k'_z}$ and $\pi_{k_{\mathcal{V}_m}}\xrightarrow{T}_{\mathcal{V}_m,s_{\mathcal{V}_m}} \pi_{k'_{\mathcal{V}_m}}$, with $k'_z, k'_{\nu_m} \in\mathcal{K}_\mathcal{R}, \forall z\in\mathcal{Z}, m\in\{1,\dots,\bar{T}\}$, assuming that (i) there exist appropriate ${u}_g$ and ${u}_\rho$ that guarantee $g\xrightarrow{g}x_g$ and $\rho\xrightarrow{r}y_\rho$ in $\pi_{k_g}, \pi_{k_\rho}$, respectively, $\forall g\in\mathcal{G},\rho\in\mathcal{R}$ and (ii) that the agents and objects fit in their respective goal regions, i.e., 
\begin{align}
&\mathcal{P}_{s,0}\Big(r_{\pi_k}, [r_z]_{z\in\mathcal{Q}_{\mathcal{Z},k}}, [r_g]_{g\in\mathcal{Q}_{\mathcal{G},k}}, [r_\rho]_{\rho\in\mathcal{Q}_{\mathcal{R},k}}, [r_{\scriptscriptstyle \mathcal{V}_m,s_{\mathcal{V}_m}}]_{m\in\mathcal{Q}_{\mathcal{V},k}}, 
\notag \\
&\hspace{55mm} [r^{\scriptscriptstyle O}_{x_g}]_{g\in\mathcal{Q}_{\mathcal{G},k}}, [r^{\scriptscriptstyle O}_{y_\rho}]_{\rho\in\mathcal{Q}_{R,k}} \Big) = \mathsf{True} \label{eq:fit in goal regions (TASE)}
\end{align}
$\forall k\in\mathcal{K}_\mathcal{R}$, where we define the sets: $\mathcal{Q}_{\mathcal{Z},k} \coloneqq \{z\in\mathcal{Z}: k'_z = k \}, \mathcal{Q}_{\mathcal{G},k} \coloneqq \{g\in\mathcal{G}: k_g = k \}, \mathcal{Q}_{\mathcal{R},k} \coloneqq \{\rho\in\mathcal{R}: k_r = k \}, \mathcal{Q}_{\mathcal{V},k} \coloneqq \{m\in\{1,\dots,\bar{T}\}: k'_{\mathcal{V}_m} = k \}$, that correspond to the indices of the agents and objects that are in region $k\in\mathcal{K}_\mathcal{R}$.

\begin{example}
	As an example, consider $N = 6$ agents, $\mathcal{N} = \{1,\dots,6\}$, $M = 3$ objects, $\mathcal{M} = \{1,2,3\}$ in a workspace that contains $K = 4$ regions of interest, $\mathcal{K}_\mathcal{R} = \{1,\dots,4\}$. Let $t_0 = 0$ and, according to Problem \ref{problem (TASE)}, take $\text{init}(1) = \text{init}(5) = 1, \text{init}(2) = 2, \text{init}(3) = \text{init}(4) = 3$, and $\text{init}(6) = 4$, i.e., agents $1$ and $5$ are in region $\pi_{\text{init}(1)}=\pi_{\text{init}(5)} = \pi_1$, agent $2$ is in region $\pi_{\text{init}(2)} = \pi_2$, agents $3$ and $4$ are in region $\pi_{\text{init}(3)} = \pi_{\text{init}(4)}=\pi_3$ and agent $6$ is in region $\pi_{\text{init}(6)} = \pi_4$. We also consider $\text{init}_{\scriptscriptstyle O}(1) = 1, \text{init}_{\scriptscriptstyle O}(2) = 2, \text{init}_{\scriptscriptstyle O}(3) = 3$ implying that the $3$ objects are in regions $\pi_1,\pi_2$ and $\pi_3$, respectively. We assume that agents $1,5$ grasp objet $1$, and agents $3,4$ grasp object $3$, i.e., $\mathcal{AG}_{1,1}({q}_1(0),{x}^{\scriptscriptstyle O}_1(0))$ $=$ $\mathcal{AG}_{5,1}({q}_5(0),{x}^{\scriptscriptstyle O}_1(0))$ $=$ $\mathcal{AG}_{3,3}({q}_3(0),{x}^{\scriptscriptstyle O}_3(0))$ $=$ $\mathcal{AG}_{4,3}({q}_4(0),{x}^{\scriptscriptstyle O}_4(0))$ $=$ $\mathcal{AG}_{2,0}({q}_2(0),{x}^{\scriptscriptstyle O}(0))$ $=$ $\mathcal{AG}_{6,0}({q}_6(0),{x}^{\scriptscriptstyle O}(0))$ $=$ $\mathsf{True}$. Agents $1$ and $5$ aim to cooperatively transport object $1$ to $\pi_4$, agent $2$ aims to grasp object $2$, agents $3$ and $4$ aim to cooperatively transport object $3$ to $\pi_1$ and agent $6$ aims to perform a transition to region $\pi_2$. Therefore, $\mathcal{Z} = \{6\}, \bar{T} =2, \mathcal{V}_1 = \{1,5\}, \mathcal{V}_2 = \{3,4\}$, $\mathcal{V} = \mathcal{V}_1\cup\mathcal{V}_2 = \{1,5,4,3\}$, $\mathcal{G} = \{2\}, \mathcal{R} = \emptyset$, $s_{\mathcal{V}_1} = 1$, $s_{\mathcal{V}_2} = 2$, $\mathcal{S} = \{s_{\mathcal{V}_1},s_{\mathcal{V}_2}\} = \{1,2\}$, $\mathcal{X} = \{x_2\} = \{2\}, \mathcal{Y} = \emptyset$. Moreover, the region indices $k_z,k_\nu,k_g,k_r,k_{\mathcal{V}_m}, k'_z, k'_{\mathcal{V}_m}, z\in\mathcal{Z}=\{6\},\nu\in\mathcal{V}=\{1,5,4,3\}, g\in\mathcal{G}=\{2\},r\in\mathcal{R}=\emptyset,m\in\{1,2\}$, take the form $k_6 = 4, k_1=k_5 = 1, k_2 = 2, k_3=k_4 = 3, k_{\mathcal{V}_1} = 1, k_{\mathcal{V}_3} = 3, k'_6 = 2, k'_{\mathcal{V}_1} = 4, k'_{\mathcal{V}_2} = 1$. Finally, the actions that need to be performed by the agents are $\pi_1 \xrightarrow{T}_{\mathcal{V}_{1,1}} \pi_4$, $2 \xrightarrow{g} 2$, $\pi_3 \xrightarrow{T}_{\mathcal{V}_{2,3}} \pi_1$ and $\pi_4 \rightarrow \pi_2$.
\end{example} 
Next, for each region $\pi_k$, we compute from ${\mathcal{P}}_s$ a set of configurations for the agents and objects in this region. More specifically, 
\begin{align*}
&([{q}^\star_z]_{z\in\mathcal{Q}_{\mathcal{Z},k}}, [{q}^\star_g]_{g\in\mathcal{Q}_{\mathcal{G},k}}, [{q}^\star_\rho]_{\rho\in\mathcal{Q}_{\mathcal{R},k}}, [{x}^{\scriptscriptstyle O\star}_{s_{\mathcal{V}_m}}]_{m\in\mathcal{Q}_{\mathcal{V},k}}, [{x}^{\scriptscriptstyle O\star}_{x_g}]_{g\in\mathcal{Q}_{\mathcal{G},k}},     [{x}^{\scriptscriptstyle O\star}_{y_\rho}]_{\rho\in\mathcal{Q}_{\mathcal{R},k}}) = \notag \\ 
&{\mathcal{P}}_{s,1}\Big(r_{\pi_k}, [r_z]_{z\in\mathcal{Q}_{\mathcal{Z},k}}, [r_g]_{g\in\mathcal{Q}_{\mathcal{G},k}}, [r_\rho]_{\rho\in\mathcal{Q}_{\mathcal{R},k}}, [r_{\scriptscriptstyle \mathcal{V}_m,s_{\mathcal{V}_m}}]_{m\in\mathcal{Q}_{\mathcal{V},k}}, [r^{\scriptscriptstyle O}_{x_g}]_{g\in\mathcal{Q}_{\mathcal{G},k}}, \\
& \hspace{80mm}[r^{\scriptscriptstyle O}_{y_\rho}]_{\rho\in\mathcal{Q}_{R,k}} \Big),
\end{align*}
where we have used the notation of \eqref{eq:fit in goal regions (TASE)}. Hence, we now have the goal configurations for the agents $\mathcal{Z}$ performing the transitions as well as agents $\mathcal{V}$ performing the cooperative transportations. 

Following Section \ref{subsec:MAS NF (App:NF)}, we define the error functions  $\gamma_{z}:\mathbb{R}^{n_z}\rightarrow \mathbb{R}_{\geq 0}$ with $\gamma_{z}({q}_z) \coloneqq \lVert {q}_z - {q}^\star_z \rVert^2$, $\forall z\in\mathcal{Z}$, and $\gamma_{\scriptscriptstyle \mathcal{V}_m}:\mathbb{M}\rightarrow \mathbb{R}_{\geq 0}$ as $\gamma_{\scriptscriptstyle \mathcal{V}_m}({x}^{\scriptscriptstyle O}_{s_{\mathcal{V}_m}}) \coloneqq  \lVert {p}^{\scriptscriptstyle O}_{s_{\mathcal{V}_m}} - {p}^{\scriptscriptstyle O^\star}_{s_{\mathcal{V}_m}}\rVert^2$, where ${p}^{\scriptscriptstyle O^\star}_{s_{\mathcal{V}_m}}$ is the position part of ${x}^{\scriptscriptstyle O^\star}_{s_{\mathcal{V}_m}}$.  

Regarding the grasping agents $g \in\mathcal{G}$, these are assumed to operate in the sphere with the fixed center $c_g(q_g)$ and radius $r_g$. Regarding the releasing agent $\rho \in \mathcal{R}$ \textit{and} the respective objects $y_\rho$, $\rho$, these are assumed to operate in the sphere with the fixed center $c_\rho(q_\rho)$ and radius $r_\rho$.

Based on the above, we define the following collision functions: 
\small
\begin{align*}
&\beta_{i,l}({q}_i,{q}_l) \coloneqq  \|{c}_i({q}_i) - {c}_l({q}_l) \|^2 - (r_i + r_l)^2, 
\forall i,l\in\mathcal{N}\backslash\mathcal{V}, i\neq l, \notag \\
&\beta_{i,\scriptscriptstyle O_j}({q}_i) \coloneqq  \|{c}_i({q}_i) - {p}^{\scriptscriptstyle O}_j\|^2 - (r_i + r^{\scriptscriptstyle O}_j)^2, \forall i\in\mathcal{N}\backslash\mathcal{V}, j \in \mathcal{M}\backslash(\mathcal{S}\cup\mathcal{Y}) \notag \\
&\beta_{i,\scriptscriptstyle \mathcal{V}_m}({q}_i,{x}^{\scriptscriptstyle O}_{s_{\mathcal{V}_m}}) \coloneqq \|{c}_i({q}_i) - {p}^{\scriptscriptstyle O}_{s_{\mathcal{V}_m}} \|^2 - (r_i + r_{\scriptscriptstyle \mathcal{V}_m, s_{\mathcal{V}_m}})^2, \\
&\hspace{28mm} \forall i\in\mathcal{N}\backslash\mathcal{V}, m\in\{1,\dots,\bar{T}\}, \notag \\
&\beta_{\scriptscriptstyle \mathcal{V}_m, \mathcal{V}_\ell }({x}^{\scriptscriptstyle O}_{s_{\mathcal{V}_m}}, {x}^{\scriptscriptstyle O}_{s_{\mathcal{V}_\ell}}) \coloneqq \|{p}^{\scriptscriptstyle O}_{s_{\mathcal{V}_m}} - {p}^{\scriptscriptstyle O}_{s_{\mathcal{V}_\ell}} \|^2 - (r_{\scriptscriptstyle \mathcal{V}_m, s_{\mathcal{V}_m}} + r_{\scriptscriptstyle \mathcal{V}_\ell, s_{\mathcal{V}_\ell}})^2, \\
&\hspace{32mm}\forall m,\ell\in\{1,\dots,\bar{T}\}, m \neq \ell, \notag \\
&\beta_{\scriptscriptstyle \mathcal{V}_m, O_j}({x}^{\scriptscriptstyle O}_{s_{\mathcal{V}_m}}) \coloneqq \|{p}^{\scriptscriptstyle O}_{s_{\mathcal{V}_m}} - {p}^{\scriptscriptstyle O}_j \|^2 - (r_{\scriptscriptstyle \mathcal{V}_m, s_{\mathcal{V}_m}} + r^{\scriptscriptstyle O}_j)^2, \notag \\
&\hspace{25mm}\forall m\in\{1,\dots,\bar{T}\}, j\in\mathcal{M}\backslash(\mathcal{S}\cup\mathcal{Y}),\\
& \beta_{i,\pi_k}({q}_i) \coloneqq \| {c}_i({q}_i) - {p}_{\pi_{k} } \|^2 - (r_i + r_{\pi_k})^2,\forall i\in\mathcal{Z},k\in\mathcal{K}_\mathcal{R}\backslash\{k_z,k'_z\}, \notag \\
&\beta_{\scriptscriptstyle \mathcal{V}_m,\pi_k}({x}^{\scriptscriptstyle O}_{s_{\mathcal{V}_m}}) \coloneqq \| {p}^{\scriptscriptstyle O}_{s_{\mathcal{V}_m}} - {p}_{\pi_{k} } \|^2 - (r_{\scriptscriptstyle \mathcal{V}_m, s_{\mathcal{V}_m}}  + r_{\pi_k})^2, \\
&\hspace{25mm} \forall m\in\{1,\dots,\bar{T}\}, k\in\mathcal{K}_\mathcal{R}\backslash\{k_{\mathcal{V}_m},k'_{\mathcal{V}_m}\}, \notag \\
& \beta_{i,\scriptscriptstyle \mathcal{W}}({q}_i) \coloneqq (r_0 - r_i)^2 - \|{c}_i({q}_i) \|^2, \forall i\in\mathcal{N}\backslash\mathcal{V} \notag \\
& \beta_{\scriptscriptstyle \mathcal{V}_m,\mathcal{W}}({x}^{\scriptscriptstyle O}_{s_{\mathcal{V}_m}}) \coloneqq (r_0 - r_{\scriptscriptstyle \mathcal{V}_m, s_{\mathcal{V}_m}})^2 - \|{p}^{\scriptscriptstyle O}_{s_{\mathcal{V}_m}} \|^2, \forall m\in\{1,\dots,\bar{T}\},
\end{align*}
\normalsize
that incorporate collisions among the navigating agents, the navigating agents and the objects, the transportation agents, the transportation agents and the objects, the navigating agents and the undesired regions, the transportation agents and the undesired regions, the navigating agents and the workspace boundary, and the transportation agents and the workspace boundary, respectively.
Therefore, by following the procedure described in Section \ref{subsec:MAS NF (App:NF)}, we can form the total obstacle function $G:\mathbb{R}^{n_\mathcal{Z}}\times\mathbb{M}^{|\mathcal{S}| }  \to\mathbb{R}_{\geq 0}$, $n_\mathcal{Z}\coloneqq \sum_{z\in\mathcal{Z}}n_z$, and thus, define the navigation function \cite{koditschek1992robot,Loizou2006} $\varphi:\mathcal{F} \to [0,1]$ as 
\begin{equation*}
\varphi({q}_{\scriptscriptstyle \mathcal{Z}},{x}^{\scriptscriptstyle O}_{\scriptscriptstyle \mathcal{S}}) \coloneqq \frac{ \gamma({q}_{\scriptscriptstyle \mathcal{Z}},{x}^{\scriptscriptstyle O}_{\scriptscriptstyle \mathcal{S}}) }{ \Big( \gamma({q}_{\scriptscriptstyle \mathcal{Z}},{x}^{\scriptscriptstyle O}_{\scriptscriptstyle \mathcal{S}})^\kappa + G({q}_{\scriptscriptstyle \mathcal{Z}},{x}^{\scriptscriptstyle O}_{\scriptscriptstyle \mathcal{S}}) \Big)^{\frac{1}{\kappa}}},
\end{equation*}
where ${q}_{\scriptscriptstyle \mathcal{Z}} \coloneqq [q_z]_{z\in\mathcal{Z}}$, ${x}^{\scriptscriptstyle O}_{\scriptscriptstyle \mathcal{S}}$ $\coloneqq$ $[ {x}^{\scriptscriptstyle O}_{s_{\scriptscriptstyle \mathcal{V}_m}}  ]_{m\in\{1,\dots,\bar{T}\}}$ $\in \mathbb{M}^{|\mathcal{S}|}$, $\gamma({q}_{\scriptscriptstyle \mathcal{Z}},{x}^{\scriptscriptstyle O}_{\scriptscriptstyle \mathcal{S}}) \coloneqq \sum_{z\in\mathcal{Z}} \gamma_z({q}_z)$ $+$ $\sum_{m\in\{1,\dots,\bar{T}\}}$ $\gamma_{\scriptscriptstyle \mathcal{V}_m}({x}^{\scriptscriptstyle O}_{\scriptscriptstyle s_{\mathcal{V}_m}})$,
$\mathcal{F}$ is a subset of $\mathbb{R}^{n_\mathcal{Z}} \times \mathbb{M}^M$ where the collision functions are positive, and $\kappa > 0$ is a positive gain used to derive the proof correctness of $\varphi$ \cite{koditschek1992robot,Loizou2006}.  Note that, a sufficient condition for avoidance of the undesired regions and avoidance of collisions and singularities is $\varphi({q}_{\scriptscriptstyle \mathcal{Z}},{x}^{\scriptscriptstyle O}_{\scriptscriptstyle \mathcal{S}}) < 1$.

Next, we design the feedback control protocols $\tau_z:
\mathcal{F} \times \mathbb{R}^{n_z}\to \mathbb{R}^6, {u}_\ell:  \mathcal{F} \times \mathsf{S}_\ell \times\mathbb{R}^6$, $\forall z\in\mathcal{Z}, \ell\in\mathcal{V}_m, m\in\{1,\dots,\bar{T}\}$ as follows:
\begin{subequations} \label{eq:control_protocol (TASE)}
	\begin{align}
	&\tau_z = \tau_z(q_{\scr \mathcal{Z}},{x}^{\scriptscriptstyle O}_\mathcal{S},\dot{{q}}_z) \coloneqq {g}_{q_z} -  \nabla_{{q}_z}\varphi({q}_{\scriptscriptstyle \mathcal{Z}},{x}^{\scriptscriptstyle O}_{\scriptscriptstyle \mathcal{S}}) - {K}_z\dot{{q}}_z, 	\label{eq:control_protocol_transition (TASE)} \\
	&u_\ell = {u}_\ell(q_\mathcal{Z},{x}^{\scriptscriptstyle O}_\mathcal{S},q_\ell, {v}^{\scriptscriptstyle O}_{s_{\mathcal{V}_m}} ) \coloneqq  \left({J}^{\scriptscriptstyle O}_{\ell,s_{\mathcal{V}_m}}\right)^{-\top} \Big\{  \Big({g}_{\scriptscriptstyle O} -   \left({J}^{\scriptscriptstyle O}_{s_{\mathcal{V}_m}}\right)^\top \nabla_{\scriptscriptstyle {x}^{\scriptscriptstyle O}_{s_{\mathcal{V}_m}}} \varphi({q}_{\scriptscriptstyle \mathcal{Z}},{x}^{\scriptscriptstyle O}_{\scriptscriptstyle \mathcal{S}}) \notag \\
	&\hspace{43mm}- {v}^{\scriptscriptstyle O}_{s_{\mathcal{V}_m}} \Big) \Big\} + {g}_{\ell},  	\label{eq:control_protocol_transportation (TASE)}
	\end{align}
\end{subequations}
where ${K}_z = \text{diag}\{k_z\}\in\mathbb{R}^{n_z \times n_z}$, with $k_z > 0, \forall z\in\mathcal{Z}$, is a constant positive definite gain matrix. 
To characterize the solutions of the closed-loop system, we consider the function 
\begin{align*}
V \coloneqq &
\varphi({q}_{\scriptscriptstyle \mathcal{Z}},{x}^{\scriptscriptstyle O}_{\scriptscriptstyle \mathcal{S}}) + \frac{1}{2}\sum_{z\in\mathcal{Z}}\dot{{q}}^\top_z  {B}_z({q}_z) \dot{{q}}_z  +\frac{1}{2}\sum_{m\in\{1,\dots,\bar{T}\}} \left({v}^{\scriptscriptstyle O}_{s_{\mathcal{V}_m}}\right)^\top \widetilde{{M}}_{\scriptscriptstyle \mathcal{V}_m, s_{\mathcal{V}_m}}{v}^{\scriptscriptstyle O}_{s_{\mathcal{V}_m}}.
\end{align*}
Since no collisions occur and the robots and objects have zero velocity at $t_0$, we conclude that $V_0 \coloneqq V(t_0) =  \varphi({q}_{\scriptscriptstyle \mathcal{Z}}(t_0),{x}^{\scriptscriptstyle O}_{\scriptscriptstyle \mathcal{S}}(t_0)) =: \varphi_0 < 1$. 
	By differentiating $V$ and substituting \eqref{eq:manipulator dynamics (TASE)}, \eqref{eq:coupled dynamics (TASE)}, we obtain 
	\small
	\begin{align}
	&\dot{V} =  \notag\sum\limits_{z\in\mathcal{Z}}\Big\{ \nabla_{{q}_z}\varphi({q}_{\scriptscriptstyle \mathcal{Z}},{x}^{\scriptscriptstyle O}_{\scriptscriptstyle \mathcal{S}})^\top \dot{{q}}_z + \dot{{q}}^\top_z\Big( {\tau}_z - {C}_{q_z}\dot{{q}}_z -
	{g}_{q_z}\Big) +  \frac{1}{2}\dot{{q}}_z^\top\dot{{M}}_z \dot{{q}}_z \Big\} \notag \\
	& + \sum\limits_{m\in\{1,\dots,\bar{T}\}} \Bigg\{ \nabla_{{x}^{\scriptscriptstyle O}_{\scriptscriptstyle s_{\mathcal{V}_m}}} \varphi({q}_{\scriptscriptstyle \mathcal{Z}},{x}^{\scriptscriptstyle O}_{\scriptscriptstyle \mathcal{S}})^\top  \dot{{x}}^{\scriptscriptstyle O}_{\scriptscriptstyle s_{\mathcal{V}_m}}  +   \left({v}^{\scriptscriptstyle O}_{\scriptscriptstyle s_{\mathcal{V}_m}}\right)^\top \Big( \sum\limits_{\ell\in\mathcal{V}_m} [{J}^{\scriptscriptstyle O}_{\ell,\scriptscriptstyle s_{\mathcal{V}_m}}]^\top {u}_\ell - {g}_{\scriptscriptstyle O}  \notag \\
	&- \sum\limits_{\ell\in\mathcal{V}_m} [{J}^{\scriptscriptstyle O}_{\scriptscriptstyle \ell,s_{\mathcal{V}_m}}]^\top {g}_\ell - \widetilde{{C}}_{\scriptscriptstyle \mathcal{V}_m, s_{\mathcal{V}_m}}
	\Big) +   \frac{1}{2} \left({v}^{\scriptscriptstyle O}_{s_{\mathcal{V}_m}}\right)^\top \dot{\widetilde{{M}}}_{\scriptscriptstyle \mathcal{V}_m,s_{\mathcal{V}_m}} {v}^{\scriptscriptstyle O}_{\scriptscriptstyle s_{\mathcal{V}_m}} \Bigg\}, \notag
	\end{align}
	\normalsize
where we have also used the fact that ${f}_z = 0,\forall z\in\mathcal{Z}$, since the agents performing transportation actions are not in contact with any objects.   
	By employing Lemma \ref{lem:coupled dynamics skew symmetry (TASE)} as well as \eqref{eq:object dynamics 1 (TASE)}, $\dot{V}$ becomes:
	\small
	\begin{align}
	&\dot{V} =  \notag\sum\limits_{z\in\mathcal{Z}}\dot{{q}}^\top_z\Big(\nabla_{{q}_z}\varphi({q}_{\scriptscriptstyle \mathcal{Z}},{x}^{\scriptscriptstyle O}_{\scriptscriptstyle \mathcal{S}}) + {\tau}_z - {g}_{q_z} \Big)  + \\
	&  \sum\limits_{m\in \{1,\dots,\bar{T}\}} \left({v}^{\scriptscriptstyle O}_{\scriptscriptstyle s_{\mathcal{V}_m}}\right)^\top \Big( \sum\limits_{\ell\in\mathcal{V}_m} [{J}^{\scriptscriptstyle O}_{\ell,\scriptscriptstyle s_{\mathcal{V}_m}}]^\top ( {u}_\ell - {g}_\ell) - 
	{g}_{\scriptscriptstyle O} + [{J}^{\scriptscriptstyle O}_{\scriptscriptstyle s_{\mathcal{V}_m}}]^\top \nabla_{{x}^{\scriptscriptstyle O}_{\scriptscriptstyle s_{\mathcal{V}_m}}} \varphi({q}_{\scriptscriptstyle \mathcal{Z}},{x}^{\scriptscriptstyle O}_{\scriptscriptstyle \mathcal{S}}) 
	\Big), \notag
	\end{align}
	\normalsize
	and after substituting \eqref{eq:control_protocol (TASE)}:
	$$\dot{V} = -\sum_{z\in\mathcal{Z}}\dot{{q}}_z{K}_z\dot{{q}}_z - \sum_{m\in\{1,\dots,\bar{T}\}}\| {v}^{\scriptscriptstyle O}_{\scriptscriptstyle s_{\mathcal{V}_m}} \|^2,$$
	which is strictly negative unless $\dot{{q}}_z = {0}$, ${v}^{\scriptscriptstyle O}_{\scriptscriptstyle s_{\mathcal{V}_m}} = {0}, \forall z\in\mathcal{Z},m\in\{1,\dots,\bar{T}\}$. Since ${J}^{\scriptscriptstyle O}_{\scriptscriptstyle \ell,s_{\mathcal{V}_m}}$ is always non-singular, and ${J}_\ell({q}_\ell(t))$ has full-rank by assumption for the maximal solution, $\forall \ell\in\mathcal{V}_m, m\in\{1,\dots,\bar{T}\}$, the latter implies also that $\dot{{q}}_\ell = {0}$, $\forall \ell\in\mathcal{V}_m,m\in\widetilde{\mathcal{V}}$. Hence, $V(t) \leq V_0 < 1$, $\forall t\in[t_0,t_{\max})$, which suggests that $\varphi({q}_{\scriptscriptstyle \mathcal{Z}}(t),{x}^{\scriptscriptstyle O}_{\scriptscriptstyle \mathcal{S}}(t)) \leq \varphi_0 < 1$ $\forall t\geq t_0$. Moreover, according to La Salle's Invariance Principle \cite{khalil_nonlinear_systems}, the system will converge to the largest invariant set contained in the set where $\dot{{q}}_z = {0}, {v}^{\scriptscriptstyle O}_{\scriptscriptstyle s_{\mathcal{V}_m}}={0}, \forall z\in\mathcal{Z}, m\in\{1,\dots,\bar{T}\}$.
	We can also conclude that  $\lim_{t\to\infty}\ddot{{{q}}}_z(t) = {0}$, $\dot{{v}}^{\scriptscriptstyle O}_{\scriptscriptstyle s_{\mathcal{V}_m}} = {0}$, which, by employing \eqref{eq:control_protocol (TASE)}, \eqref{eq:manipulator dynamics (TASE)}, \eqref{eq:coupled dynamics (TASE)}, and the assumption of non-singular ${J}^{\scriptscriptstyle O}_{\scriptscriptstyle s_{\mathcal{V}_m}}$, $\forall t\in\mathbb{R}_{\geq 0}$, implies that $\nabla_{{q}_z}\varphi({q}_{\scriptscriptstyle \mathcal{Z}},{x}^{\scriptscriptstyle O}_{\scriptscriptstyle \mathcal{S}}) = {0}$, $\nabla_{{x}^{\scriptscriptstyle O}_{\scriptscriptstyle s_{\mathcal{V}_m}}} \varphi({q}_{\scriptscriptstyle \mathcal{Z}},{x}^{\scriptscriptstyle O}_{\scriptscriptstyle \mathcal{S}}) = {0}$, $\forall z\in\mathcal{Z}, m\in\{1,\dots,\bar{T}\}$. Since $\varphi$ is a navigation function \cite{Loizou2006}, by setting $\kappa$ large enough,  this condition is true only at the destination configurations (i.e., where $\gamma({q}_{\scriptscriptstyle \mathcal{Z}},{x}^{\scriptscriptstyle O}_{\scriptscriptstyle \mathcal{S}}) = 0$) and a set of isolated saddle points, whose region of attraction is a set of measure zero \cite{koditschek1992robot,koditschek1990robot}. Thus, the system converges to the destination configuration from almost everywhere, i.e., $\lVert {q}_z(t) - {q}^\star_z\rVert \rightarrow 0$ and $\lVert {p}^{\scriptscriptstyle O}_{\scriptscriptstyle s_{\mathcal{V}_m}}(t) - {p}^{\scriptscriptstyle O^\star}_{\scriptscriptstyle s_{\mathcal{V}_m}}\rVert \rightarrow 0$. Therefore, there exist finite time instants $t_{f_z},t_{f_m} > t_0$, such that $\mathcal{A}_z({q}_z(t_{f_z}))\subset \pi_{k'_z}$ and
	$\mathcal{AO}_{\mathcal{V}_m,s_{\mathcal{V}_m}}(p^{\scr O}_{s_{\mathcal{V}_m}}(t_{f_m}))$ $\subset \pi_{k'_{\mathcal{V}_m}}$, with inter-agent collision avoidance, $\forall z\in\mathcal{Z}, m\in\{1,\dots,\bar{T}\}$. Since the actions $g\xrightarrow{g}x_g$, $\rho \xrightarrow{r} y_\rho$ are also performed, we denote as $t_{f_g}, t_{f_\rho}$ the times that these actions have been completed, $g\in\mathcal{G}, \rho\in\mathcal{R}$. Hence, by setting $t_f \coloneqq \max\{ \max\limits_{z\in\mathcal{Z}}t_{f_z},\max\limits_{m\in\{1,\dots,\bar{T}\}}t_{f_m}, \max\limits_{g\in\mathcal{G}}t_{f_g}, \max\limits_{\rho\in\mathcal{R}}t_{f_\rho}\}$, all the actions of all agents will be completed at $t_f$.
	
	It should be noted that \cite{Loizou2006} does not take into account static obstacles. Since, however, the results are an extension of \cite{koditschek1990robot}, intuition suggests that the results are valid for sufficiently distant obstacles (in our case, the regions of interest).
	



\subsubsection{High-Level Plan Generation} \label{sec:high level plan (TASE)}

The second part of the solution is the derivation of a high-level plan that satisfies the given LTL formulas $\mathsf{\Phi}_i$ and $\mathsf{\Phi}^{\scriptscriptstyle O}_j$ and can be generated by using standard techniques from automata-based formal verification methodologies. Thanks to (i) the proposed control laws that allow agent transitions and object transportations $\pi_k\rightarrow_i\pi_{k'}$ and $\pi_k\xrightarrow{T}_{\mathcal{V},j}\pi_{k'}$, respectively, and (ii) the off-the-self control laws that guarantee grasp and release actions $i\xrightarrow{g}j$ and $i\xrightarrow{r}j$, we can abstract the behavior of the agents using a finite transition system as presented in the sequel.

\begin{definition} \label{def:TS objects all agents (TASE)}
	The coupled behavior of the overall system of all the $N$ agents and $M$ objects is modeled by the transition system $\mathcal{TS} = (\Pi_s,\Pi^{\text{init}}_s, \rightarrow_{s},\mathcal{AG}, \Psi, \mathcal{L}, \Lambda, {P}_s,\chi)$,
	where 
	\begin{enumerate}
		\item $\Pi_s\subset \bar{\Pi}\times\bar{\Pi}^{\scriptscriptstyle O}\times\bar{\mathcal{AG}}$ is the set of states; 
		$\bar{\Pi}\coloneqq\Pi_1\times\cdots\times\Pi_N$ and $\bar{\Pi}^{\scriptscriptstyle O}\coloneqq\Pi^{\scriptscriptstyle O}_1\times\cdots\times\Pi^{\scriptscriptstyle O}_M$ are the set of states-regions that the agents and the objects can be at, with $\Pi_i  = \Pi^{\scriptscriptstyle O}_j = \Pi,\forall i\in\mathcal{N},j\in\mathcal{M}$; 
		$\mathcal{AG} \coloneqq \mathcal{AG}_1\times\cdots\times\mathcal{AG}_N$ is the set of boolean grasping variables introduced in Section \ref{sec:Model and PF (TASE)}, with
		$\mathcal{AG}_i \coloneqq \{\mathcal{AG}_{i,0}\}\cup\{[\mathcal{AG}_{i,j}]_{j\in\mathcal{M}}\}, \forall i\in\mathcal{N}$. 
		By defining 
		$\bar{\pi} \coloneqq \left(\pi_{k_1},\cdots,\pi_{k_N}\right),\bar{\pi}_{\scriptscriptstyle O}  \coloneqq (\pi_{\scriptscriptstyle k^{\scriptscriptstyle O}_1},\cdots,\pi_{\scriptscriptstyle k^{\scriptscriptstyle O}_M}), \bar{w}=\left(w_1,\cdots,w_N\right)$, with $\pi_{k_i},\pi_{k^{\scriptscriptstyle O}_j}\in\Pi$ (i.e., $k_i,k^{\scriptscriptstyle O}_j\in\mathcal{K}_\mathcal{R},\forall i\in\mathcal{N},j\in\mathcal{M}$) and $w_i\in\mathcal{AG}_i, \forall i\in\mathcal{N}$, then the coupled state $\pi_s \coloneqq (\bar{\pi},\bar{\pi}_{\scriptscriptstyle O},\bar{w})$ belongs to $\Pi_s$, i.e., $(\bar{\pi},\bar{\pi}_{\scriptscriptstyle O},\bar{w})\in\Pi_s$ if
		\begin{enumerate}
			\item $\mathcal{P}_{s,0}\Big(r_{\pi_k}, [r_i]_{i\in\{ i\in\mathcal{N}: k_i = k \} }, [r^{\scriptscriptstyle O}_j]_{j\in\{ j\in\mathcal{M}:k^{\scriptscriptstyle O}_j= k \} }\Big) = \mathsf{True}$, i.e., the respective agents and objects fit in the region, $\forall k\in\mathcal{K}_\mathcal{R}$, 
			\item $k_i = k^{\scriptscriptstyle O}_j$ for all $i\in\mathcal{N}, j\in\mathcal{M}$ such that $w_i = \mathcal{AG}_{i,j} = \mathsf{True}$, i.e., an agent must be in the same region with the object it grasps,
		\end{enumerate} 
		\item $\Pi^{\text{init}}_s\subset\Pi_s$ is the initial set of states at $t=0$, which, owing to \textbf{(i)}, satisfies the conditions of Problem \ref{problem (TASE)},\\
		\item $\rightarrow_s\subset \Pi_s\times\Pi_s$ is a transition relation defined as follows: given the states $\pi_s, \widetilde{\pi}_s\in\Pi$, with
		\small
		\begin{align}
		\pi_s \coloneqq & (\bar{\pi},\bar{\pi}_{\scriptscriptstyle O},\bar{w}) \coloneqq (\pi_{k_1},\dots,\pi_{k_N}, \pi_{k^{\scriptscriptstyle O}_1},\dots,\pi_{k^{\scriptscriptstyle O}_M},w_1,\dots,w_N), \notag \\
		\widetilde{\pi}_s \coloneqq & ( \widetilde{\bar{\pi}},\widetilde{\bar{\pi}}_{\scriptscriptstyle O},\widetilde{\bar{w}}) \coloneqq (\pi_{\widetilde{k}_1},\dots,\pi_{\widetilde{k}_N}, \pi_{\widetilde{k}^{\scriptscriptstyle O}_1},\dots,\pi_{\widetilde{k}^{\scriptscriptstyle O}_1},\widetilde{w}_1,\dots,\widetilde{w}_N), \label{eq:pi_s (TASE)}
		\end{align}
		\normalsize
		a transition $\pi_s \rightarrow_s \widetilde{\pi}_s $ occurs if all the following hold:
		
		\begin{enumerate}
			\item $\nexists i\in\mathcal{N}, j\in\mathcal{M}$ such that $w_i = \mathcal{AG}_{i,j} = \mathsf{True}$, $\widetilde{w}_i = \mathcal{AG}_{i,0} = \mathsf{True}$, (or $w_i = \mathcal{AG}_{i,0} = \mathsf{True}$, $\widetilde{w}_i = \mathcal{AG}_{i,j} = \mathsf{True}$) and $k_i \neq \widetilde{k}_i$, i.e., there are no simultaneous grasp/release and navigation actions,
			\item $\nexists i\in\mathcal{N}, j\in\mathcal{M}$ such that $w_i = \mathcal{AG}_{i,j} = \mathsf{True}$, $\widetilde{w}_i = \mathcal{AG}_{i,0} = \mathsf{True}$, (or $w_i = \mathcal{AG}_{i,0} = \mathsf{True}$, $\widetilde{w}_i = \mathcal{AG}_{i,j} = \mathsf{True}$) and $k_i = k^{\scriptscriptstyle O}_j \neq \widetilde{k}_i=\widetilde{k}^{\scriptscriptstyle O}_j$, i.e., there are no simultaneous grasp/release and transportation actions,
			\item $\nexists i\in\mathcal{N}, j,j'\in\mathcal{M}$, with $j\neq j'$, such that $w_i = \mathcal{AG}_{i,j} = \mathsf{True}$ and $\widetilde{w}_i = \mathcal{AG}_{i,j'} = \mathsf{True}$ ($w_i = \mathcal{AG}_{i,j'} = \mathsf{True}$ and $\widetilde{w}_i = \mathcal{AG}_{i,j'} = \mathsf{True}$), i.e., there are no simultaneous grasp and release actions,
			\item $\nexists j\in\mathcal{M}$ such that $k^{\scriptscriptstyle O}_j \neq \widetilde{k}^{\scriptscriptstyle O}_j$ and $w_i \neq \mathcal{AG}_{i,j}, \forall i\in\mathcal{N}$ ( or $\widetilde{w}_i\neq \mathcal{AG}_{i,j}, \forall i\in\mathcal{N}$), i.e., there is no transportation of a non-grasped object,
			\item $\nexists j\in\mathcal{M},\mathcal{V}\subseteq \mathcal{N}$ such that $k^{\scriptscriptstyle O}_j \neq \widetilde{k}^{\scriptscriptstyle O}_j$ and $\Lambda(m^{\scriptscriptstyle O}_j, \mathfrak{c}_{\mathcal{V}}) = \mathsf{False}$, where $w_i = \widetilde{w}_i = \mathcal{AG}_{i,j} = \mathsf{True} \Leftrightarrow i\in\mathcal{V}$, i.e., the agents grasping an object are powerful enough to transfer it,
		\end{enumerate}

		\item $\Psi \coloneqq \bar{\Psi}\cup\bar{\Psi}^{\scriptscriptstyle O}$ with $\bar{\Psi}=\bigcup_{i\in\mathcal{N}}\Psi_{i}$ and $\bar{\Psi}^{\scriptscriptstyle O} = \bigcup_{j\in\mathcal{M}}\Psi^{\scriptscriptstyle O}_j$, are the atomic propositions of the agents and objects, respectively, as defined in Section \ref{sec:Model and PF (TASE)}.
		
		\item $\mathcal{L}:\Pi_s \rightarrow 2^\Psi$ is a labeling function defined as follows: Given a state $\pi_s$ as in \eqref{eq:pi_s (TASE)} and $\breve{\psi}_s \coloneqq \Big( \bigcup_{i\in\mathcal{N}}\breve{\psi}_i\Big)\bigcup \Big(\bigcup_{j\in\mathcal{M}}\breve{\psi}^{\scriptscriptstyle O}_j\Big)$ with $\breve{\psi}_i\in2^{\Psi_i},\breve{\psi}^{\scriptscriptstyle O}_j\in2^{\Psi^{\scriptscriptstyle O}_j}$, then $\breve{\psi}_s\in\mathcal{L}(\pi_s)$ if  $\breve{\psi}_i\in\mathcal{L}_i(\pi_{k_i})$ and $\breve{\psi}^{\scriptscriptstyle O}_j\in\mathcal{L}^{\scriptscriptstyle O}_j(\pi_{k^{\scriptscriptstyle O}_j}), \forall i\in\mathcal{N},j\in\mathcal{M}$.
		
		\item $\Lambda$ and ${P}_s$ as defined in Section \ref{sec:Model and PF (TASE)}.
		
		\item $\chi: (\to_s) \to \mathbb{R}_{\geq 0}$ is a function that assigns a cost to each transition $\pi_s \to_s \widetilde{\pi}_s$. This cost might be related to the distance of the agents' regions in $\pi_s$ to the ones in $\widetilde{\pi}_s$, combined with the cost efficiency of the agents involved in transport tasks (according to $\mathfrak{c}_i, i\in\mathcal{N}$).
	\end{enumerate} 
\end{definition}

Next, we form the global LTL formula $\mathsf{\Phi} \coloneqq (\land_{i\in\mathcal{N}}\mathsf{\Phi}_i)\land(\land_{j\in\mathcal{M}}\mathsf{\Phi}^{\scriptscriptstyle O}_j)$ over the set $\Psi$. Then, we translate $\mathsf{\Phi}$ to a B\"uchi Automaton $\mathcal{BA}$ and 
we build the product $\widetilde{\mathcal{TS}} \coloneqq \mathcal{TS}\times\mathcal{BA}$. Using basic graph-search theory, we can find the accepting runs of $\widetilde{\mathcal{TS}}$ that satisfy $\mathsf{\Phi}$ and minimize the total cost $\chi$. These runs are directly projected to a sequence of desired states to be visited in the $\mathcal{TS}$. Although the semantics of LTL are defined over infinite sequences of services, it can be proven that there always exists a high-level plan that takes the form of a finite state sequence followed by an infinite repetition of another finite state sequence. For more details on the followed technique, the reader is referred to the related literature, e.g., \cite{baier2008principles}.

Following the aforementioned methodology, we obtain a high-level plan as sequences of states and atomic propositions $\pi_\text{pl} \coloneqq \pi_{s,1} \pi_{s,2}\dots$ and $\check{\psi}_\text{pl} \coloneqq \check{\psi}_{s,1} \check{\psi}_{s,1}\dots \models \mathsf{\Phi}$, which minimizes the cost $\chi$, with 
\begin{align*}
& \pi_{s,\ell} \coloneqq (\bar{\pi}_\ell,\bar{\pi}_{\scriptscriptstyle O,\ell},\bar{w}_\ell ) \in\Pi_s, \forall \ell\in\mathbb{N}, \notag \\
& \check{\psi}_{s,\ell} \coloneqq \Big(\bigcup\limits_{i\in\mathcal{N}}\check{\psi}_{i,\ell} \Big)\bigcup\Big(\bigcup\limits_{j\in\mathcal{M}}\check{\psi}^{\scriptscriptstyle O}_{j,\ell} \Big) \in 2^{\Psi}, \mathcal{L}(\pi_{s,\ell}), \forall \ell\in\mathbb{N},
\end{align*}
where 
\begin{itemize}
	\item $\bar{\pi}_\ell \coloneqq \pi_{k_{1,\ell}}, \dots, \pi_{k_{N,\ell}}$, with $k_{i,\ell} \in\mathcal{K}_\mathcal{R},\forall i\in\mathcal{N}$, 
	\item $\bar{\pi}_{\scriptscriptstyle O,\ell} \coloneqq \pi_{k^{\scriptscriptstyle O}_{1,\ell}},\dots, \pi_{k^{\scriptscriptstyle O}_{N,\ell}}$, with $k^{\scriptscriptstyle O}_{j,\ell} \in\mathcal{K}_\mathcal{R},\forall j\in\mathcal{M}$, 
	\item $\bar{w}_\ell \coloneqq w_{1,\ell}, \dots, w_{N,\ell}$, with $w_{i,\ell} \in\mathcal{AG}_i, \forall i\in\mathcal{N}$,
	\item $\check{\psi}_{i,\ell} \in 2^{\Psi_i}, \mathcal{L}_i(\pi_{k_{i,\ell}}), \forall i\in\mathcal{N}$, 
	\item $\check{\psi}^{\scriptscriptstyle O}_{j,\ell}\in 2^{\Psi^{\scriptscriptstyle O}_j}, \mathcal{L}^{\scriptscriptstyle O}_j(\pi_{k^{\scriptscriptstyle O}_{j,\ell}}), \forall j\in\mathcal{M}$.
\end{itemize}

The path $\pi_\text{pl}$ is then projected to the individual sequences of the regions $\pi_{k^{\scriptscriptstyle O}_{j,1}} \pi_{k^{\scriptscriptstyle O}_{j,2}} \dots$ for each object $j\in\mathcal{M}$, as well as to the individual sequences of the regions $\pi_{k_{i,1}} \pi_{k_{i,2}} \dots$ and the boolean grasping variables $w_{i,1} w_{i,2}\dots$ for each agent $i\in\mathcal{N}$. The aforementioned sequences determine the behavior of agent $i\in\mathcal{N}$, i.e., the sequence of actions (transition, transportation, grasp, release or stay idle) it must take.

By the definition of $\mathcal{L}$ in Def. \ref{def:TS objects all agents (TASE)}, we obtain that $\check{\psi}_{i,\ell} \in\mathcal{L}_i(\pi_{k_{i,\ell}}), \check{\psi}^{\scriptscriptstyle O}_{j,\ell} \in\mathcal{L}^{\scriptscriptstyle O}_j( \pi_{k^{\scriptscriptstyle O}_{j,\ell}}), \forall i\in\mathcal{N},j\in\mathcal{M}, \ell\in\mathbb{N}$. Therefore, since $\mathsf{\Phi} = (\land_{i\in\mathcal{N}}\mathsf{\Phi}_i)\land(\land_{j\in\mathcal{M}}\mathsf{\Phi}_{\scriptscriptstyle O_j})$ is satisfied by $\check{\psi}_\text{pl}$, we conclude that $\check{\psi}_{i,1}\check{\psi}_{i,2}\dots \models \mathsf{\Phi}_i$ and $\check{\psi}^{\scriptscriptstyle O}_{j,1}\check{\psi}^{\scriptscriptstyle O}_{j,2}\dots\models \mathsf{\Phi}^{\scriptscriptstyle O}_j, \forall i\in\mathcal{N}, j\in\mathcal{M}$.

The sequences $\pi_{k_{i,1}}\pi_{k_{i,2}}\dots$, $\check{\psi}_{i,1}\psi_{i,2}\dots$ and $\pi_{k^{\scriptscriptstyle O}_{j,1}}\pi_{k^{\scriptscriptstyle O}_{j,2}}\dots, \check{\psi}^{\scriptscriptstyle O}_{j,1}\check{\psi}^{\scriptscriptstyle O}_{j,2}\dots$ over $\Pi, 2^{\Psi_i}$ and $\Pi, 2^{\Psi^{\scriptscriptstyle O}_j}$, respectively, produce the trajectories ${q}_i(t)$ and ${x}^{\scriptscriptstyle O}_j(t), \forall i\in\mathcal{N},j\in\mathcal{M}$. The corresponding behaviors are 
\begin{align*}
&\mathfrak{b}_i = ({q}_i(t),\check{\psi}_i) = ({q}_i(t_{i,1}),\check{\psi}_{i,1})({q}_i(t_{i,2}),\check{\psi}_{i,2})\dots \\
& \mathfrak{b}^{\scriptscriptstyle O}_j= ({x}^{\scriptscriptstyle O}_j(t),\check{\psi}^{\scriptscriptstyle O}_j)=({x}^{\scriptscriptstyle O}_j(t^{\scriptscriptstyle O}_{j,1}),\check{\psi}^{\scriptscriptstyle O}_{j,1}) ({x}^{\scriptscriptstyle O}_j(t^{\scriptscriptstyle O}_{j,2}),\check{\psi}^{\scriptscriptstyle O}_{j,2})\dots,
\end{align*} 
respectively, according to Section \ref{sec:Model and PF (TASE)}, with $\mathcal{A}_i({q}_i(t_{i,\ell}))\subset \pi_{k_{i,\ell}}, \check{\psi}_{i,\ell}\in\mathcal{L}_i(\pi_{k_{i,\ell}})$ and $\mathcal{O}_j({x}^{\scr O}_{j}(t^{\scriptscriptstyle O}_{j,m}))\in\pi_{k^{\scriptscriptstyle O}_{j,\ell}}, \check{\psi}^{\scriptscriptstyle O}_{j,\ell} \in\mathcal{L}^{\scriptscriptstyle O}_j(\pi_{k^{\scriptscriptstyle O}_{j,\ell}})$. Thus, it is guaranteed that $\check{\psi}_i \models \mathsf{\Phi}_i,\check{\psi}^{\scriptscriptstyle O}_j \models \mathsf{\Phi}^{\scriptscriptstyle O}_j$ and consequently, the behaviors $\mathfrak{b}_i$ and $\mathfrak{b}^{\scriptscriptstyle O}_j$ satisfy the formulas $\mathsf{\Phi}_i$ and $\mathsf{\Phi}^{\scriptscriptstyle O}_j$, respectively, $\forall i\in\mathcal{N},j\in\mathcal{M}$. The aforementioned reasoning is summarized in the next theorem:
\begin{theorem}
	The execution of the path $(\pi_{\text{pl}},\psi_{\textup{pl}})$ of $\mathcal{TS}$ guarantees behaviors $\mathfrak{b}_i,\mathfrak{b}^{\scriptscriptstyle O}_j$ that yield the satisfaction of $\mathsf{\Phi}_i$ and $\mathsf{\Phi}^{\scriptscriptstyle O}_j$, respectively, $\forall i\in\mathcal{N},j\in\mathcal{M}$, providing, therefore, a solution to Problem \ref{problem (TASE)}.  
\end{theorem}

\begin{remark}
	Note that although the overall set of states of $\mathcal{TS}$ increases exponentially with respect to the number of agents/objects/regions, some states are not reachable, due to our constraints for the object transportation and the size of the regions, reducing thus the state complexity.
\end{remark}

\begin{figure}	
	\centering
	\includegraphics[scale=0.55,trim = 0cm 0cm 0cm 0cm]{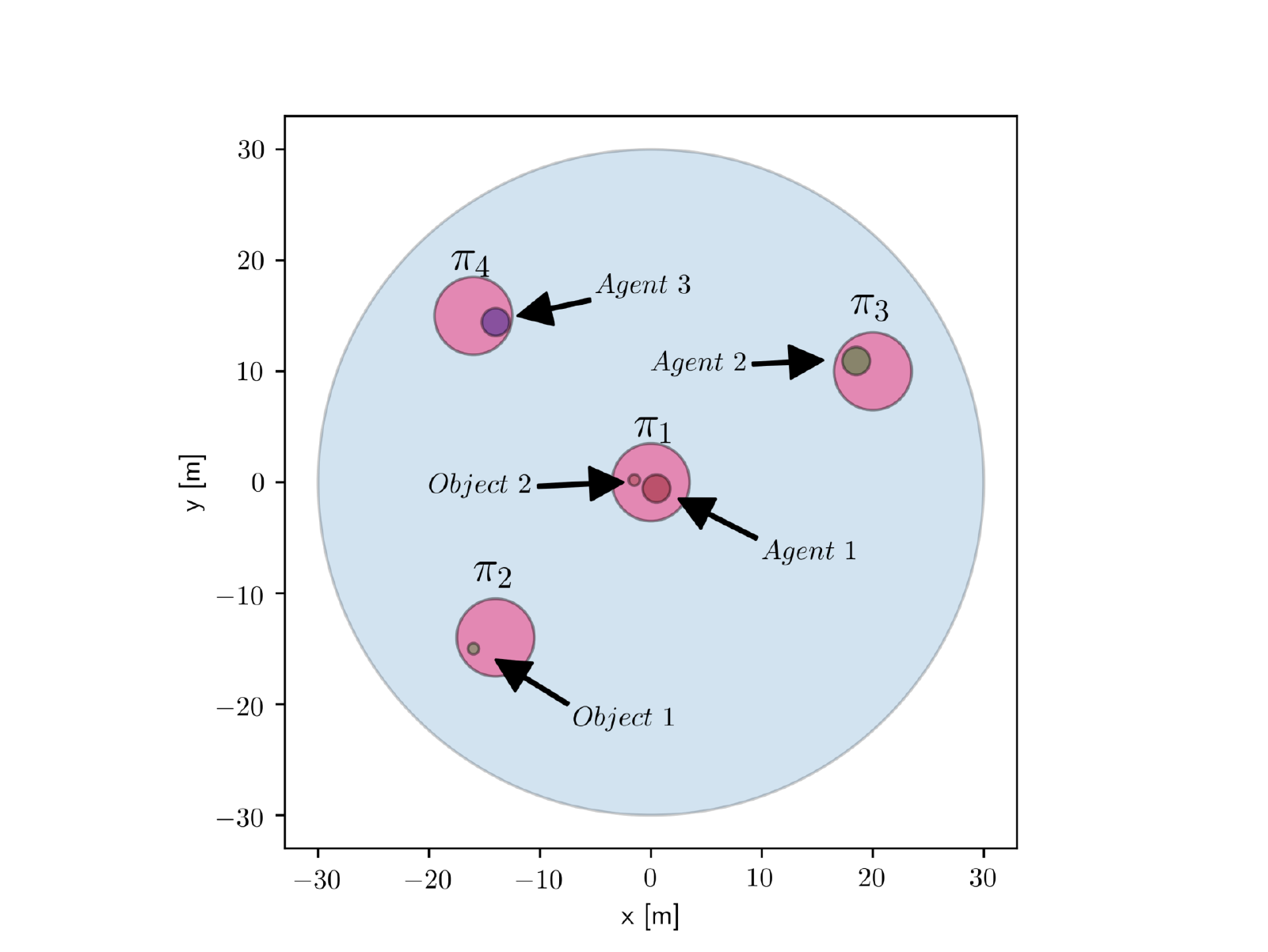}
	\caption{The initial workspace of the second simulation example, consisting of $3$ agents and $2$ objects. The agents and the objects are indicated via their corresponding radii. \label{fig:initial workspace (TASE)}}
\end{figure}

\subsection{Simulation Results} \label{sec:simulations (TASE)}

In this section we demonstrate our approach with computer simulations. We consider a workspace of radius $r_0 = 30 \text{m}$, with $K = 4$ regions of interest or radius $r_{\pi_k} = 3.5\text{m}$, $\forall k\in\mathcal{K}_\mathcal{R}$, centered at ${p}_{\pi_1} = [0,0,0]^\top, {p}_{\pi_2} = [-14,-14,0]^\top$m, ${p}_{\pi_3} = [20,-10,0]^\top$m, ${p}_{\pi_4} = [-16,15,0]^\top$, respectively (see Fig. \ref{fig:initial workspace (TASE)}). Moreover, we consider two cuboid objects of bounding radius $r^{\scriptscriptstyle O}_j = 0.5 \text{m}$, and mass $m^{\scriptscriptstyle O}_j = 0.5 \text{kg}$, $\forall j\in\{1,2\}$, initiated at  ${x}^{\scriptscriptstyle O}_{1}(0) = [-16,15,0.5,0,0,0]^\top$ (m,rad), ${x}^{\scriptscriptstyle O}_{2}(0) =$ $[-1.5$, $0.2$, $0.5$, $0,0,0]^\top$ (m,rad), which implies that $\mathcal{O}_1({x}^{\scriptscriptstyle O}_1(0)) \subset \pi_2$, and $\mathcal{O}_2({x}^{\scriptscriptstyle O}_1(0)) \subset \pi_1$. The considered agents consist of a mobile base and a $2$-dof rotational robotic arm. The mobile base is rectangular with dimensions $0.5\times 0.5\times 0.2 \ \text{m}^3$ and mass $0.5\text{kg}$, and the two arm links have length $1\text{m}$ and mass $0.5\text{kg}$ each.
The state vectors of the agents are ${q}_i = [\mathsf{x}_{c_i}, \mathsf{y}_{c_i}, q_{i_1}, q_{i_2}]^\top\in\mathbb{R}^4, \dot{{q}} = [\dot{\mathsf{x}}_{c_i}, \dot{\mathsf{y}}_{c_i}, \dot{q}_{i_1}, \dot{q}_{i_2}]^\top\in\mathbb{R}^4$, where $\mathsf{x}_{c_i}, \mathsf{y}_{c_i}$ are the planar position of the bases' center of mass, and $q_{i_1}, q_{i_2}$ the angles of the arms' joints. The geometric characteristics of the considered agents lead to a bounding radius of $r_i = 1.25 \text{m}$, $\forall i\in\mathcal{N}$. The atomic propositions are $\Psi_i =  \{ ``i\text{-}\pi_1",\dots,``i\text{-}\pi_4"\}$, $\forall i\in\mathcal{N}$, and $\Psi^{\scriptscriptstyle O} = \{ ``O_j\text{-}\pi_1",\dots,``O_j\text{-}\pi_4"\}$, $\forall j\in\mathcal{M}$, indicating whether the agents/objects are in the corresponding regions. The labeling functions are, therefore, $\mathcal{L}_i(\pi_k) = \{ ``i\text{-}\pi_k" \}$, $\mathcal{L}^{\scriptscriptstyle O}_j(\pi_k) = \{ ``O_j\text{-}\pi_k" \}$, $\forall k\in\mathcal{K}_\mathcal{R},i\in\mathcal{N},j\in\mathcal{M}$.
We test two scenarios with $N=2, 3$ agents, respectively. We generate the optimal high-level plan for these scenarios and present two indicative transitions of the continuous execution for the second case. The simulations were carried out using Python environment on a laptop computer with $4$ cores at $2.6$GHz CPU and $8$GB of RAM memory.

\begin{table}[t]
	\caption{The agent actions for the discrete path of the first simulation example}
	\label{table:Path 1 (TASE)}
	\centering
	\begin{tabular}{||c| c| c |c | ||} 
		\hline
		$\pi_{s,\ell} $ & Actions & $\pi_{s,\ell} $ & Actions \\ [0.5ex] 
		\hline\hline 		
		$\pi_{s,1}$ & ($-$) & $\pi_{s,14}$ & ($\pi_1 \xrightarrow{T}_{\{1,2\},2} \pi_2) $  \\ [0.5ex] 
		\hline
		$\pi_{s,2}$ & ($-$, $\pi_3 \to_2 \pi_1$)  & $\pi_{s,15}$ & ($1 \xrightarrow{r} 2$, $2 \xrightarrow{r} 2$)   \\ [0.5ex] 
		\hline
		$\pi_{s,3}$ & ($1 \xrightarrow{g} 2$, $2 \xrightarrow{g} 2$)   & $\pi_{s,16}$ & ($\pi_2 \to_2 \pi_4$, $\pi_2 \to_2 \pi_4$)   \\ [0.5ex]
		\hline
		$\pi_{s,4}$ &  ($\pi_1 \xrightarrow{T}_{\{1,2\},2} \pi_4 $)  &  $\pi_{s,17}$ & ($1 \xrightarrow{g} 1$, $2 \xrightarrow{g} 1$)  \\ [0.5ex]
		\hline
		$\pi_{s,5}$ &  ($\pi_4 \xrightarrow{T}_{\{1,2\},2} \pi_1 $)  & $\pi_{s,18}$ & ($\pi_4 \xrightarrow{T}_{\{1,2\},1} \pi_1 $)  \\ [0.5ex]
		\hline
		$\pi_{s,6}$ & ($1 \xrightarrow{r} 2$, $2 \xrightarrow{r} 2$)  & $\pi_{s,19}$ & ($\pi_1 \xrightarrow{T}_{\{1,2\},1} \pi_4 $)  \\ [0.5ex]
		\hline
		$\pi_{s,7}$ & ($\pi_1 \to_1 \pi_2$, $\pi_1 \to_2 \pi_2$)   & $\pi^\star_{s,20}$ & ($-$, $2 \xrightarrow{r} 1$)   \\ [0.5ex]
		\hline 
		$\pi_{s,8}$ & ($1 \xrightarrow{g} 1$, $2 \xrightarrow{g} 1$)  & $\pi^\star_{s,21}$ & ($-$, $\pi_4 \to_2 \pi_3$)   \\ [0.5ex]
		\hline 
		$\pi_{s,9}$ & ($\pi_2 \xrightarrow{T}_{\{1,2\},1} \pi_4 $)  & $\pi^\star_{s,22}$ & ($-$, $\pi_3 \to_2 \pi_4$)   \\ [0.5ex]
		\hline
		$\pi_{s,10}$ & ($1 \xrightarrow{r} 1$, $2 \xrightarrow{r} 1$)  & $\pi^\star_{s,23}$ & ($-$, $2 \xrightarrow{g} 1$)   \\ [0.5ex]
		\hline 
		$\pi_{s,11}$ & ($-$, $\pi_4 \to_2 \pi_3$)  & $\pi^\star_{s,24}$ & ($\pi_4 \xrightarrow{T}_{\{1,2\},1} \pi_1 $)  \\ [0.5ex]
		\hline
		$\pi_{s,12}$ & ($\pi_4 \to_2 \pi_1$, $\pi_3 \to_2 \pi_1$)  & 	$\pi^\star_{s,25}$ & ($\pi_1 \xrightarrow{T}_{\{1,2\},1} \pi_4 $)   \\ [0.5ex]	
		\hline 
		$\pi_{s,13}$ & ($1 \xrightarrow{g} 2$, $2 \xrightarrow{g} 2$) & &  \\ [0.5ex]
		\hline
	\end{tabular}
\end{table}

\begin{enumerate}
	\item  We consider $N=2$ agents with initial conditions ${q}_1(0)$ $=$ $[0.5\text{m}$, $0$, $\frac{\pi}{4}\text{rad}$, $\frac{\pi}{4}\text{rad}]^\top$,  ${q}_2(0) = [18.5\text{m},11.5\text{m},\frac{\pi}{4}\text{rad},\frac{\pi}{4}\text{rad}]^\top$, $\dot{{q}}_i(0) = [0,0,0,0]^\top, \forall i\in\{1,2\}$ which imply that $\mathcal{A}_1({q}_1(0)) \subset \pi_1$, $\mathcal{A}_2({q}_2(0)) \subset \pi_3$, and that no collisions occur at $t=0$. 
	We also assume that $\mathcal{AG}_{i,0}({q}_i(0),{x}^{\scriptscriptstyle O}(0)) = \top, \forall i\in\{1,2\}$.  We represent the agents' power capabilities with the scalars $\mathfrak{c}_1 = 2, \mathfrak{c}_2 = 4$ and construct the functions $\Lambda(m^{\scriptscriptstyle O}_1, \mathfrak{c}_\mathcal{V}) = \top $ if and only if $\sum_{\ell\in\mathcal{V}} \mathfrak{c}_\ell \geq 5$, with $\mathcal{AG}_{\ell,1} = \top \Leftrightarrow \ell\in\mathcal{V}$, and $\Lambda(m^{\scriptscriptstyle O}_2, \mathfrak{c}_\mathcal{V}) = \top$ if and only if $\sum_{\ell\in\mathcal{V}} \mathfrak{c}_\ell \geq 6$, with $\mathcal{AG}_{\ell,2} = \top \Leftrightarrow \ell\in\mathcal{V}$, i.e., the objects can be transported only if the agents that grasp them have a sum of capability scalars no less than $5$ and $6$, respectively. Regarding the cost $\chi$, we simply choose the sum of the distances of the transition and transportation regions, i.e., given $\pi_s, \widetilde{\pi}_s$ as in \eqref{eq:pi_s (TASE)} such that $\pi_s \to_s \widetilde{\pi}_s$, we have that $$\chi = \sum_{i\in\{1,2\}} \{ \|{p}_{\pi_{k_i}} - {p}_{\pi_{\widetilde{k}_i}} \|^2 \} + \sum_{j\in\{1,2\}}\|{p}_{\pi_{k^{\scriptscriptstyle O}_j}} - {p}_{\pi_{\widetilde{k}^{\scriptscriptstyle O}_j}} \|^2 \}.$$ The LTL formula is taken as
	\begin{multline*}
	(\square \neg ``1\text{-}\pi_3" ) \ \land \ ( \square\lozenge ``2\text{-}\pi_3" ) \ \land \ ( \square \lozenge ``O_1\text{-}\pi_1" ) \ \land \\ \square(``O_1\text{-}\pi_1" \to \bigcirc ``O_1\text{-}\pi_4") \ \land \ (\lozenge ``O_2\text{-}\pi_4"),
	\end{multline*} 
	which represents the following behavior. Agent $1$ must never go to region $\pi_3$, which must be visited by agent $2$ infinitely many times, object $1$ must be taken infinitely often to region $\pi_1$, always followed by a visit in region $\pi_4$, and object $2$ must be eventually taken to region $\pi_4$.
	
	The resulting transition system $\mathcal{TS}$ consists of $560$ reachable states and $7680$ transitions and it was created in $3.19 \sec$. The B\"uchi automaton $\mathcal{BA}$ contains $7$ states and $29$ transitions  and the product $\widetilde{\mathcal{TS}}$ contains $3920$ states and $50976$ transitions. Table \ref{table:Path 1 (TASE)} shows the actions of the agents for the derived path, which is the sequence of states $\pi_{s,1}\pi_{s,2}\dots...(\pi^\star_{s,20},\dots,\pi^\star_{s,25})^\omega$, where the states with ($^\star$) constitute the suffix that is run infinitely many times.  
	Loosely speaking, the derived path describes the following behavior: Agent $2$ goes first to $\pi_1$ to grasp and transfer object $2$ to $\pi_4$ and back to $\pi_1$ with agent $1$. The two agents then navigate to $\pi_2$ to take object $1$ to $\pi_4$. In the following, after agent $2$ goes to $\pi_3$,  they both go to $\pi_1$ to transfer object $2$ to $\pi_2$. Then, they navigate to $\pi_4$ to transfer object $1$ to $\pi_1$ and back.  Finally, the actions that are run infinitely many times consist of agent $2$ going to from $\pi_4$ to $\pi_3$ and back, and transferring object $1$ to $\pi_1$ and $\pi_4$ with agent $1$.  
	One can verify that the resulting path satisfies the LTL formula. Note also that the regions are not large enough to contain both agents and objects in a grasping configuration, which played an important role in the derivation of the plan.
	The time taken for the construction of the product $\widetilde{\mathcal{TS}}$ and the derivation of the path was $2.79 \sec$.

	\item We now consider $N=3$ agents with ${q}_1(0)$, ${q}_2(0)$, as in the first case, ${q}_3(0) = [-14, 15,\frac{\pi}{4}, \frac{\pi}{4}]^\top ([\text{m},\text{rad}])$ implying $\mathcal{A}_3({q}_3(0))$ $\in$ $\pi_4$, with $\mathcal{AG}_{3,0}({q}_i(0),{x}^{\scriptscriptstyle O}(0)) = \top$, $\mathfrak{c}_3 = 3$, and no collisions occurring at $t=0$. The functions $\Lambda$ and $\chi$ are the same as in the first case. The formula in this scenario is
	\begin{multline*}
	(\square \neg ``1\text{-}\pi_3") \ \land \ (\square\lozenge ``2\text{-}\pi_3") \ \land \ (\square\lozenge ``O_1\text{-}\pi_1") \ \land \\ \square(``O_1\text{-}\pi_1" \to \lozenge  ``O_1\text{-}\pi_4") \  \land \ (\square\lozenge ``O_2\text{-}\pi_3"),	
	\end{multline*}
	 which represents the following behavior. Agent $1$ must never visit region $\pi_3$, which must be visited infinitely many times by agent $2$, object $1$ must be taken infinitely many times to region $\pi_1$, eventually followed by a visit in region $\pi_4$, and object $2$ must be taken infinitely many times to region $\pi_2$.
	
	
	The resulting transition system $\mathcal{TS}$ consists of $3112$ reachable states and $154960$ transitions and it was created in $100.74 \sec$. The B\"uchi automaton $\mathcal{BA}$ contains $9$ states and $49$ transitions  and the product $\widetilde{\mathcal{TS}}$ contains $28008$ states and $1890625$ transitions. Table \ref{table:Path 2 actions (TASE)} shows the agent actions for the derived path as the sequence of states $\pi_{s,1}\pi_{s,2}\dots...(\pi^\star_{s,10},\pi^\star_{s,11})^\omega$. 
	In this case, the three agents navigate first to regions $\pi_2,\pi_1$, and $\pi_1$, respectively, and agents $2$ and $3$ take object $2$ to $\pi_3$. Next, agent $3$ goes to $\pi_2$ to transfer object $1$ to $\pi_1$ and then $\pi_4$ with agent $1$. The latter transportations occur infinitely often.	 	 	
	The time taken for the construction of the product $\widetilde{\mathcal{TS}}$ and the derivation of the path was $4573.89 \sec$. It is worth noting the exponential increase of the computation time with the simple addition of just one agent, which can be attributed to the centralized manner of the proposed methodology. The necessity, therefore, of less computational, decentralized schemes is evident and constitutes the main focus of our future directions.
\end{enumerate}

Next, we present the continuous execution of the transitions $\pi_{s,1} \to_s \pi_{s,2}$, and $\pi_{s,3}\to_s\pi_{s,4}$ for the second simulation scenario. More specifically, Fig. \ref{fig:transition_1 (TASE)} depicts the navigation of the three agents $\pi_1\to_1\pi_2$, $\pi_3\to_2\pi_1$, and $\pi_4\to_3\pi_1$, that corresponds to $\pi_{s,1}\to_s \pi_{s,2}$, with gains ${K}_z = \text{diag}\{0.01,0.01,0.01\}$, $\forall z\in\{1,2,3\}$. Moreover, Fig. \ref{fig:transition_2 (TASE)} depicts the transportation of object $2$ by agents $2$ and $3$, i.e., $\pi_1 \xrightarrow{T}_{\{2,3\}} \pi_3$, that corresponds to $\pi_{s,3}\to_s\pi_{s,4}$.

\begin{figure}	
	\centering
	\includegraphics[scale=0.55,trim = 0cm 0cm 0cm 0cm]{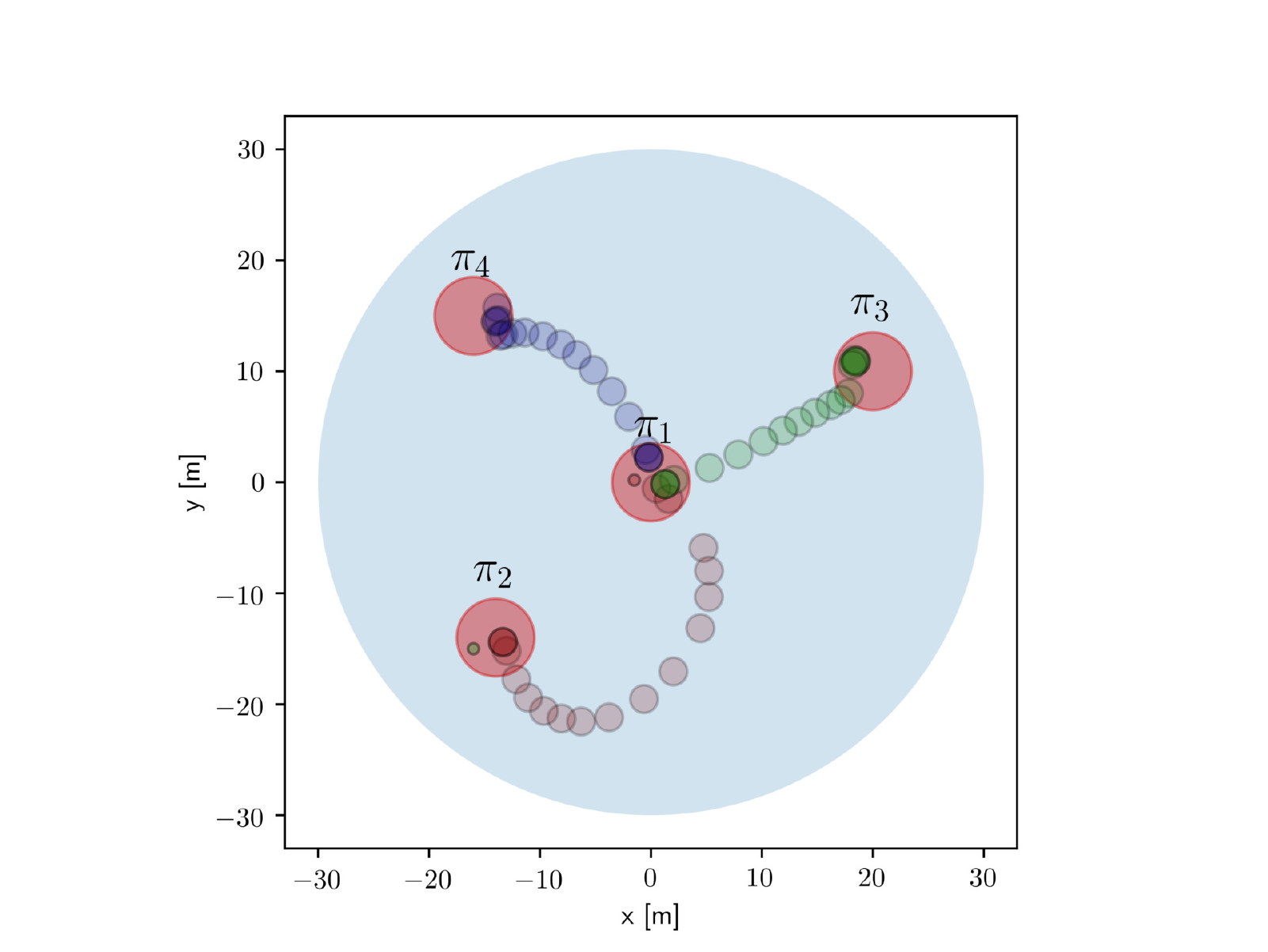}
	\caption{The transition $\pi_{s,1}\to_s \pi_{s,2}$ (a), that corresponds to the navigation of the agents $\pi_1\to_1\pi_2$, $\pi_3\to_2\pi_1$, $\pi_4\to_3\pi_1$. \label{fig:transition_1 (TASE)}}
\end{figure}

\begin{figure}	
	\centering
	\includegraphics[scale=0.55,trim = 0cm 0cm 0cm 0cm]{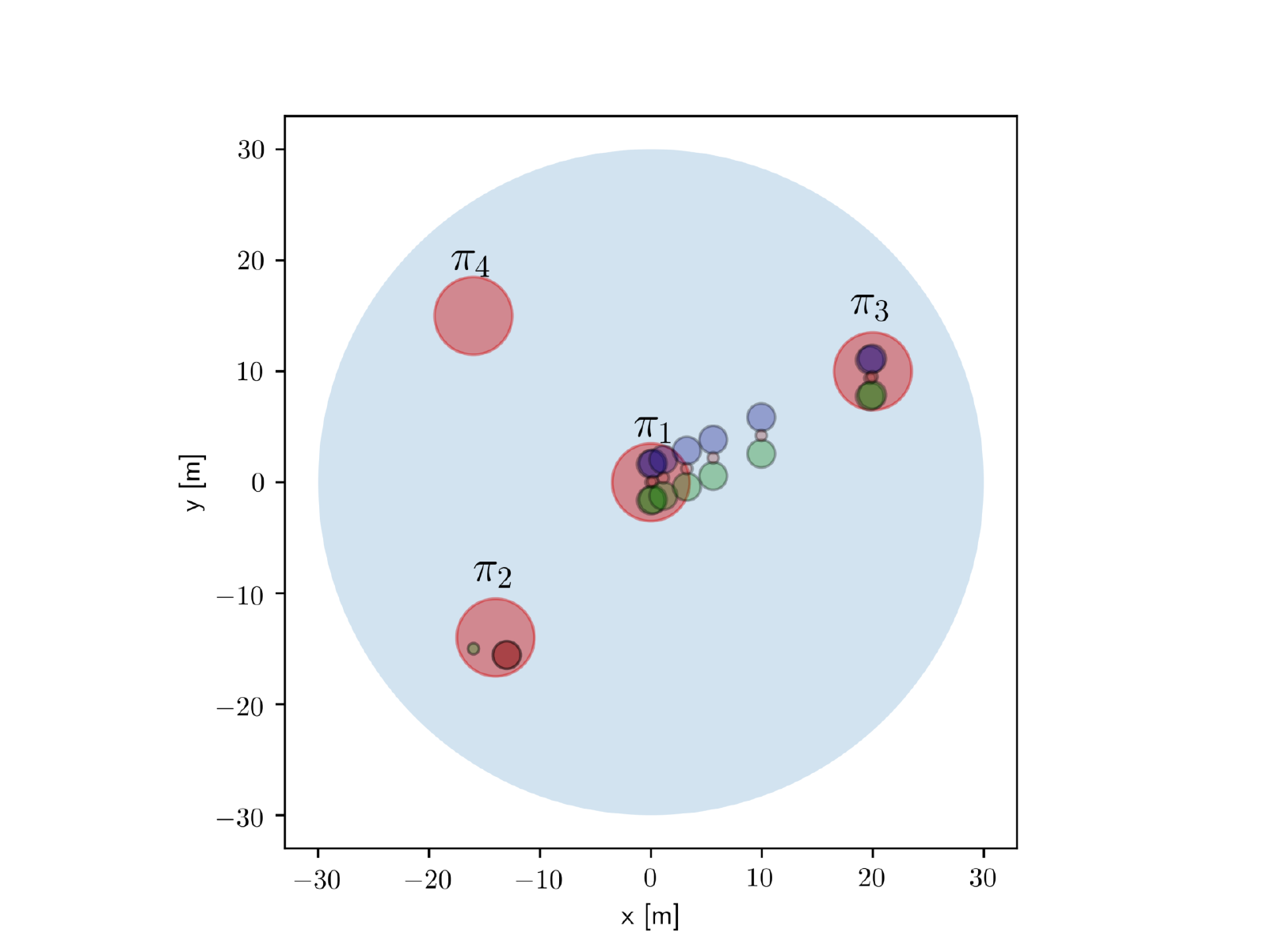}
	\caption{The transition $\pi_{s,3}\to_s \pi_{s,4}$ (b), that corresponds to the transportation $\pi_1 \xrightarrow{T}_{\{2,3\}} \pi_3$. \label{fig:transition_2 (TASE)}}
\end{figure}


\begin{table}[t]
	\caption{The agent actions for the discrete path of the second simulation example}
	\label{table:Path 2 actions (TASE)}
	\centering
	\begin{tabular}{||c| c ||} 
		\hline
		$\pi_{s,\ell} $ &  Actions\\ [0.5ex] 
		\hline\hline 		
		$\pi_{s,1}$ & ($-$) \\ [0.5ex] 	
		\hline 
		$\pi_{s,2}$ & ($\pi_1 \to_1 \pi_2, \pi_3 \to_2 \pi_1 , \pi_4\to_3\pi_1$) \\ [0.5ex] 	
		\hline 
		$\pi_{s,3}$ & ($-, 2\xrightarrow{g}1, 3\xrightarrow{g}2 $) \\ [0.5ex] 
		\hline 
		$\pi_{s,4}$ & ($-,\pi_1 \xrightarrow{T}_{\{2,3\},2} \pi_3,$) \\ [0.5ex] 
		\hline 
		$\pi_{s,5}$ & ($-,-, 3\xrightarrow{r}2$) \\ [0.5ex]
		\hline 
		$\pi_{s,6}$ & ($-,-,\pi_3\to_3\pi_2$) \\ [0.5ex]
		\hline 
		$\pi_{s,7}$ & ($1\xrightarrow{g}1, 3\xrightarrow{g}1$) \\ [0.5ex]
		\hline 
		$\pi_{s,8}$ & ($\pi_2 \xrightarrow{T}_{\{1,3\},1} \pi_1,-$) \\ [0.5ex] 
		\hline 
		$\pi_{s,9}$ & ($\pi_1 \xrightarrow{T}_{\{1,3\},1} \pi_4,-$) \\ [0.5ex] 
		\hline 
		$\pi^\star_{s,10}$ & ($\pi_4 \xrightarrow{T}_{\{1,3\},1} \pi_1,-$) \\ [0.5ex] 
		\hline 
		$\pi^\star_{s,11}$ & ($\pi_1 \xrightarrow{T}_{\{1,3\},1} \pi_4,-$) \\ [0.5ex] 
		\hline
	\end{tabular}
\end{table}

\section{Conclusion} \label{sec:conclusion (abstractions)}

This chapter presented hybrid control strategies for multi-agent systems and multi-agent-object systems under complex specifications expressed as temporal logic formulas. We considered firstly multi-agent teams of aerial vehicles and mobile manipulators with uncertain dynamics, by providing local agent abstractions as well local paths that satisfy the respective agents' LTL formulas. Secondly, we incorporated specifications of unactuated objects of the environment. Based on the previous chapters, we designed appropriate timed abstractions for a single object grasped by two agents in a partitioned workspace, and designed a timed path for it to follow, satisfying its timed specifications, expressed as MITL formulas. Next, we devised a hybrid control scheme for a system comprised of multiple robotic agents and objects that have local LTL formulas over a set of regions of interest in the workspace. We provided	a multi-agent-object abstraction as well as a path that satisfies the local specifications.

\chapter{Single-Agent Extensions} \label{chapter:single agent}   
This final chapter considers some additional problems for single-agent systems. Firstly, we consider the optimal motion planning of a single robot in a workspace with obstacles, under time temporal constraints. Unlike Section \ref{sec:AR}, we do not resort to a complete partition of the workspace, making thus the proposed algorithm more efficient. Moreover, a novel reconfiguration scheme guarantees that the obtained path is asymptotically optimal.

Secondly, we integrate adaptive control methodologies with sampling-based motion planning for high-dimensional complex systems, such as robotic manipulators. In particular, a standard adaptive control scheme is developed that compensates for the uncertain Lagrangian dynamics of the system and allows tracking of a predefined trajectory within certain bounds. These bounds are then passed to a RRT-variant planner that outputs a feasible collision-free geometric path to follow.  

Finally, we develop an extension of the standard Prescribed Performance Control methodology (see Appendix \ref{app:PPC}) that guarantees compliance with funnel constrains as well as \textit{asymptotic stability}. The developed scheme applies for control-affine $2$nd-order systems with \textit{completely unknown} dynamic terms.

\section{Introduction} \label{sec:Intro (single_agent)}

The first part of this chapter deals with robot motion planning under timed temporal constraints in an obstacle-cluttered workspace. As already discussed in the previous chapters, temporal logic-based motion planning has gained significant attention in recent years, since it provides a fully automated correct-by-design control synthesis approach for autonomous robots. An attribute that makes the problem both more interesting and challenging is the incorporation of time constraints in the temporal specification, as done in Section \ref{sec:AR}. 

In this case, however, we do not fully partition the workspace and take into account the environment obstacles via a continuous feedback control scheme proposed in \cite{vrohidis2018prescribed}. The latter guarantees \textit{timed} collision-free navigation and allows thus the discretization of the robot motion as a weighted transition system among a predefined set of regions of interest, as in Sections \ref{sec:icra} and \ref{sec:TASE}. Subsequently, we employ formal verification techniques to derive a plan that satisfies the \textit{untimed} specification and recast the assignment of the transition times as a convex optimization problem thereby achieving satisfaction of the timed specification. The transition times are recalculated after each transition, incorporating newly acquired information, and resulting in decreased control effort. In that sense, the proposed scheme is asymptotically optimal with respect to the robot control effort. \\

The second part of the chapter focuses on the motion planning problem of complex high-dimensional systems (e.g., robotic manipulators) with dynamic uncertainties in  obstacle-cluttered environments. In particular, we integrate sampling-based motion planning and adaptive control techniques to provide a computationally efficient framework that navigates the system to a desired goal while provably avoiding obstacles and compensating for the uncertain dynamics. 

For complex systems in high-dimensional spaces, closed-form feedback control fails to guarantee global solutions, and randomized planning has been introduced to overcome the respective scalability issues of standard motion planners (e.g., $A^\ast$); {\cite{kavraki1996probabilistic,hsu1997path, lavalle1998rapidly}} introduce the notions of probabilistic roadmaps (PRM) {and random trees (RRT, EST)}, respectively, which constitute efficient and probabilistically complete solutions to multiple- and single-query, respectively, high-dimensional motion planning problems.
The intuition behind these algorithms is the addition of random sampled states of the free space to a discrete graph/tree, promoting the search of the unexplored free space. 

Furthermore, although the initial works derive geometric solutions in the configuration space, {trees} have been extended to kinodynamic planning, where the {robot dynamics} $\dot{x} = f(x,t,u)$ are taken into account {\cite{hsu1997path,lavalle2001randomized,csucan2009kinodynamic,vidal2019online}}. In these algorithms, the robot dynamics are simulated forward in time, possibly by randomly sampling inputs, in order to find a feasible path. Except for the randomized inputs, the incremental step as well as the duration of this forward simulation are often also chosen randomly. {In high dimensional spaces, this randomness might require excessive tuning of the aforementioned parameters in order to find a solution in a reasonable amount of time.}
Along the lines of PRM, \cite{tedrake2010lqr} and \cite{reist2016feedback} introduce the notion of LQR-trees, which constitute trees of trajectories that probabilistically cover the state space. In that way, every controllable initial condition belongs to the region  of attraction (funnel) of a trajectory and is thus driven to the goal via local linearization of the dynamics and optimal feedback control. Dynamics linearization and reachability sets were also recently used to develop an optimal kinodynamic algorithm, namely R3T \cite{wur3t}.

A potential drawback of the aforementioned algorithms on kinodynamic motion planning is their {strong} dependence on the robot dynamics, which {in general} might be uncertain/unknown. The accurate identification of the dynamic models of real robots is a very tedious and often ineffective procedure. Therefore, the robot model used in standard forward simulation-based kinodynamic algorithms might deviate from the actual dynamics, outputting hence paths that might be colliding with obstacles or difficult to be realized by the actual robotic system.
Similar to the LQR-trees, \cite{majumdar2017funnel} proposes an algorithm that builds trees of funnels based on the (known) bounds of model disturbances, restricted however to polynomial robot dynamics. Planning under uncertainty has been also considered in a stochastic framework and via belief trees \cite{du2011probabilistic,pairet2018uncertainty,bry2011rapidly,agha2014firm}.
These approaches, however, usually deal with linearized dynamics, and/or propagate the uncertainties on the planning horizon, constraining thus excessively the free space.

In this chapter, we propose a two-layer framework that integrates ``intelligent" feedback control protocols with geometric motion planning for high-dimensional Lagrangian holonomic systems (e.g., robotic manipulators). {Firstly, motivated by the difficulty of measuring accurately the robotic system's dynamical parameters (like masses, and moments of inertia) as well as potential external disturbances, we design a feedback control scheme that does not use any information on these parameters/disturbances. The control scheme is a variation of standard adaptive control design, and aims at achieving tracking of a given trajectory for the robot, which is assumed to obey $2$nd-order dynamics. The tracking of the trajectory is achieved within certain bounds that stem from the aforementioned uncertainties/disturbances.} These bounds create an implicit funnel around the trajectory, which can be further shrunk by appropriate tuning of the control parameters, {the latter being a standard procedure in adaptive control design}. This funnel is then incorporated in a RRT-like algorithm, which outputs a path connecting an initial configuration to the goal. The construction of the RRT and the {employed control protocol} guarantee that the robot will follow the derived path without colliding with the workspace obstacles. 
In that way, by using appropriate feedback control, the proposed methodology ``relieves" the sampling-based motion planner of the robot dynamics and their uncertainties, {hence the problem of constructing a path becomes purely geometrical}. {The motion planner relies only on the performance of the control layer, encoded in the aforementioned bounds}. Similar ideas were pursued in \cite{le2012sequential} and \cite{luders2010bounds}; {\cite{le2012sequential}, however, just provides a general idea of interfacing the planning and control layers, without elaborating on a particular systematic control technique}, {while} \cite{luders2010bounds} considers mainly predictive controllers for linear systems, without avoiding the forward simulation of the available system model.
The proposed framework exhibits the following important characteristics: 1) The robot dynamics are not {forward} simulated and hence they are decoupled from the motion planner. Consequently, even though a $2$nd-order system is considered, the motion planner is purely geometrical and depends on the geometry of the configuration space as well as the bounds of the robot uncertainties. 2) We do not resort to linearization of the dynamics and computation of basins of attraction around the output trajectories, since the designed feedback control protocol applies directly to the nonlinear model. Finally,
{the proposed algorithm is expected, in practice, to exhibit lower complexity than standard kinodynamic planning algorithms, since it is purely geometrical and does not simulate any differential equations}. 
The proposed methodology is validated using a UR5 robotic manipulator in V-REP environment \cite{Vrep}. \\

The third part of this chapter deals with the problem of \textit{asymptotic stability} subject to funnel constraints for a class of $2$nd-order uncertain systems, which have been mostly studied through robust and adaptive control, as well as neural network/fuzzy logic control \cite{krstic1995nonlinear,farrell2006adaptive}.  There exists a variety of works achieving both asymptotic and ``practical" (ultimately bounded errors) stability under the presence of model uncertainties (e.g.,  \cite{wu2018adaptive,chen2008robust,xian2004continuous,wang2015stabilization,grip2010parameter,astolfi2003immersion,zhang2015continuous,bidikli2013asymptotically,marino1999adaptive,ge2002adaptive,wang2002adaptive,yang2007adaptive,wang2010adaptive}). The majority of the related works that achieve asymptotic stability assume parametric uncertainty of the underlying dynamics, and employ standard adaptive control techniques to compensate for them. Neural network approximations and fuzzy logic controllers have been also extensively used  (e.g., \cite{ge2002adaptive,wang2002adaptive,yang2007adaptive,wang2010adaptive}),
being valid, however, only in certain compact sets of the state space, and possibly yielding complex structures. Asymptotic stability subject to parametric and structural uncertainties is guaranteed in \cite{xian2004continuous} under a set of initial conditions, where gain tuning and growth conditions on the unknown terms are assumed. The same property is achieved in \cite{bidikli2013asymptotically}, where the controller uses partial information of the input matrix, as well as gain tuning.

A well-studied special instance of adaptive control is funnel control, where the output of the system is confined to a predefined funnel \cite{bechlioulis2010prescribed,theodorakopoulos2016low,berger2018funnel}. It is a model-free control scheme of high-gain type, with numerous applications {during} the last years. Examples include chemical reactors \cite{ilchmann2004input}, robotic manipulation \cite{karayiannidis2012model} (and Section \ref{subsec:PPC Controller (TCST_coop_manip)}), vehicle platooning \cite{verginis2018platoon,verginis2015decentralized}, temporal logic planning (see Section \ref{sec:AR}), {and} multi-agent systems \cite{hashim2017adaptive,bechlioulis2017decentralized,macellari2017multi} (and also Section \ref{sec:formation control}). The intuition behind funnel control is the incorporation of an adaptive gain in the control scheme, which increases (in absolute value) 
as the system's output reaches the funnel's boundary. In that way, the system's output is ``pushed" to always remain inside the funnel.  
Funnel control has been developed for both linear (e.g., \cite{ilchmann2006asymptotic}) and nonlinear systems (e.g., \cite{ilchmann2007tracking,bechlioulis2010prescribed,berger2018funnel}), involving parametric (e.g., \cite{karayiannidis2012model}) as well as structural (e.g., \cite{berger2018funnel}) dynamic uncertainties, for a wide class of systems. A funnel bang-bang controller for SISO systems was developed in \cite{liberzon2013bang}. 

An important property that most related funnel-control works fail to achieve is that of asymptotic stability subject to unknown nonlinear dynamics. Traditional funnel control guarantees only confinement of the system output in a prespecified funnel, and thus the closest property to asymptotic stability that can be achieved is that of ``practical stability", where the funnel converges arbitrarily close to zero. 
The latter, however, might yield undesired large inputs due to the small funnel values, and can be problematic in real-time systems. On the other hand, with potential guarantees of asymptotic stability, the funnel is not needed to converge close to zero, and can be used in order to encode just transient constraints for the system. Asymptotic tracking subject to transient constraints has been considered in several works \cite{ilchmann2006asymptotic,dresscher2017prescribing,macellari2017multi,karayiannidis2012model}; \cite{ilchmann2006asymptotic,dresscher2017prescribing,macellari2017multi} consider linear systems (LTI and double integrators), whereas \cite{karayiannidis2012model} assumes known model structure, with the uncertainties being only parametric; 
Along with the funnel confinement objective, finite-time stability has been also considered in \cite{han2017prescribed} for a Lagrangian-dynamics model. 
One can conclude that the aforementioned works cannot be extended in a straightforward manner to nonlinear systems where the dynamic terms have both \textit{parametric and structural} uncertainties. In addition, a class of systems for which funnel control has not been taken into account in the related works is the non-smooth type, i.e., systems with discontinuous right-hand side. Such models are motivated by real-time systems, where several dynamic terms (e.g., friction) can be accurately modeled by discontinuous functions of the state.

The third part of this chapter considers the asymptotic tracking control problem subject to transient constraints imposed by a predefined funnel for a class of  MIMO systems satisfying a loose set of assumptions. The control design combines adaptive and discontinuous control techniques and its region of attraction is independent of the system (unknown) dynamics, and relies on the initial funnel condition. If the latter is a design parameter, the results can be rendered global. It is worth noting that asymptotic stability has not been guaranteed in the related literature for such systems under the mild considered assumptions.


\section{Reconfigurable  Motion Planning and Control in Obstacle Cluttered Environments under Timed Temporal Tasks} \label{sec:reconf ... (icra19)}

We first tackle the problem of single-robot motion planning in workspace with obstacles under time temporal constraints. We develop a novel reconfigurable control scheme that achieves asymptotic optimality of the derived paths. 

\subsection{Problem Formulation\label{sec:Problem-Formulation (icra19)}}

Consider a robotic agent operating in an open bounded subset $\mathcal{W}$ of the $2$-dimensional Euclidean space. 
In addition, the workspace is populated with $m\in\mathbb{N}$ connected, closed sets $ \{O_i\}_{i\in\mathcal{J}}$, indexed by the set $\mathcal{J} \coloneqq \{1,\dots,m\}$, representing obstacles.
Accordingly, we define the free space as 
\begin{equation*}
\mathcal{F} \coloneqq \mathcal{W}\backslash \bigcup\limits_{i\in\mathcal{J}}O_i,
\end{equation*}	
\begin{remark}
	To facilitate the exposition, we assume that all the data describing the workspace are known \textit{a priori}. The analysis remains the same for the case of initially unknown workspaces where obstacles are discovered along the way.
\end{remark}
The agent is assumed to be a point\footnote{Treating a robot with volume can be achieved by initially ``transferring'' its volume to the other workspace entities (\textit{e.g.,} obstacles) and subsequently  considering it as a point.} described by the position variable $x\in\mathbb{R}^2$ which is governed by the single integrator dynamics,
\begin{equation}
\dot{x} = u,\quad u\in\mathbb{R}^2. \label{eq:dynamics (icra19)}
\end{equation}
Moreover, similarly to the previous sections, we consider that there exist $K$ points of interest in the free space, denoted by $c_{\pi_k}\in \mathcal{F}$,  for every $k\in\mathcal{K}_\mathcal{R}\coloneqq \{1,\dots,K\}$, with $\Pi \coloneqq \{c_{\pi_1},\dots,c_{\pi_K}\}$, that correspond to certain properties of interest (\textit{e.g.,} gas station, obstacle region, repairing area, etc.)
These properties of interest are expressed as boolean variables via the finite set of atomic propositions $\Psi$. The properties satisfied at each point are provided by the labeling function $\mathcal{L}: \Pi \to 2^{\Psi}$, which assigns to each point $c_{\pi_k}, k\in\mathcal{K}_\mathcal{R}$, the subset of the atomic propositions that hold true in that point. 

Since, in practice, the aforementioned properties shared by a point of interest are naturally inherited to some neighborhood of that point we define for each $k\in\mathcal{K}_\mathcal{R}$, the \emph{region of interest} $\pi_k$ corresponding to the point of interest $c_{\pi_k}$ as the set
\[
\pi_k  \coloneqq \bar{\mathcal{B}}(c_{\pi_k}, r_{\pi_k})\cap \mathcal{F},\quad r_{\pi_k}\in\mathbb{R}_{>0}.
\]
We also let  $\pi_{\mathcal{W}} \coloneqq \mathcal{F}\backslash(\cup_{k\in\mathcal{K}_\mathcal{R}} \pi_k) $ be the subset of the free space outside the regions of interest. We define thus the set $\widetilde{\Pi} \coloneqq \{\pi_{k}\}_{k\in\mathcal{K}_\mathcal{R}}\cup\{\pi_{\mathcal{W}}\}$ as well as the corresponding  labeling function as $\widetilde{\mathcal{L}}: \widetilde{\Pi} \to 2^{\Psi}$, with $\mathcal{L}(c_{\pi_k}) = \{p\} \Leftrightarrow \widetilde{\mathcal{L}}(\pi_{k}) = \{p\}$, $\forall k\in\mathcal{K}_\mathcal{R}$, and $\widetilde{\mathcal{L}}(\pi_{\mathcal{W}}) = \emptyset$.  
The agent is assumed to be in a region $\pi_k$, $k\in\mathcal{K}_\mathcal{R}$, in $\pi_{\mathcal{W}}$, simply when $x \in \pi_k$ and $x\in\pi_{\mathcal{W}}$, respectively.  
We assume that, for all $k\in\mathcal{K}_\mathcal{R}$, the location of the points $c_{\pi_k}$ as well as the radii $r_{\pi_k}$ are known.

We make the following standard assumptions \cite{rimon1992exact} regarding the geometry of the workspace and the regions of interest.
\begin{assumption}\label{ass:SphereWorld (icra19)}
	The collection of sets comprised of all obstacles and regions of interest is pairwise disjoint.
\end{assumption}
The aforementioned assumption simply states that the obstacles/regions of interest are sufficiently away from each other as well as the workspace boundary.

As already mentioned, we are interested in defining timed temporal formulas over the atomic propositions $\Psi$, and hence, over the regions of interest $\Pi$ of $\mathcal{F}$. To that end, we need to discretize the system using a finite set of states. We will achieve that by guaranteeing timed transitions between the regions of interest in $\Pi$ and by building a well-defined timed transition system among them. We first  need the following definition regarding the transitions of the agent.

\begin{definition} \label{def:transition (icra19)}
	Assume that $x(t_k)\in \mathcal{F}$, for a $t_k\in\mathbb{R}_{\geq 0}$, \textit{i.e.,} the agent is either in a region $\pi_k$, for some $k\in\mathcal{K}_\mathcal{R}$, or in $\pi_{\mathcal{W}}$. Then, given $\delta\in\mathbb{R}_{>0}$, there exists a \textit{timed transition} to $\pi_\ell$, $\ell\in\mathcal{K}_\mathcal{R}$, 
	denoted as $\pi_k \to \pi_\ell$ (or $\pi_{\mathcal{W}} \to \pi_\ell$), if there exists a time-varying feedback control law $u:\mathcal{F}\times[t_k,t_\ell]\to\mathbb{R}^2$, with $t_\ell\geq t_k + \delta$, such that the solution $x$ of the closed loop system \eqref{eq:dynamics (icra19)} satisfies the following:
	\begin{enumerate}
		\item $x(t)\in\pi_l$,\quad for all $t\in[t_k+\delta,t_l)$,
		\item $x(t) \in \mathcal{F},\quad \text{for all}~ t \in [t_k,t_\ell]$,
		\item $x(t) \not\in \pi_{m},\quad \text{for all}~ m\in \mathcal{M}_s,~ t \in [t_k,t_\ell]$,
	\end{enumerate}
	where $\mathcal{M}_s \coloneqq \mathcal{K}_\mathcal{R}\backslash\{k,\ell\}$ if $x(t_k)\in\pi_k$ and $\mathcal{M}_s \coloneqq \mathcal{K}_\mathcal{R}\backslash\{\ell\}$ if $x(t_k)\in\pi_{\mathcal{W}}$.
\end{definition}
Intuitively, according to \autoref{def:transition (icra19)}, the agent has to transit between two regions $\pi_k, \pi_\ell$ (or $\pi_{\mathcal{W}}$ and $\pi_\ell$), while avoiding all other regions of interest, obstacles, as well as the workspace boundary. In what follows, we sometimes use $\pi_k \xrightarrow{\delta} \pi_\ell$ instead of $\pi_k 
\to \pi_\ell$ to emphasize the transition time $\delta$. We have included the space outside the regions $\pi_{\mathcal{W}}$ to account for initial conditions that might satisfy $x(t_k) \notin \cup_{k\in\mathcal{K}_\mathcal{R}}\pi_k$.    
Next, we define the behavior of the agent, in order to formulate the problem of timed specifications.

\begin{definition} \label{def:specification (icra19)}
	Consider an agent trajectory $x:[t_0, \infty) \to\mathcal{F}$ of \eqref{eq:dynamics (icra19)}, where $t_0\in\mathbb{R}_{\geq 0}$. Then, a \textit{timed behavior} of $x$ is the infinite sequence $\mathfrak{b}\coloneqq (x(t_1),\breve{\psi}_1,t_1)(x(t_2),\breve{\psi}_0,t_2)\dots$, where $t_1t_2\dots$ is a time sequence according to Def. \ref{def:time sequence (App_logics)} of Appendix \ref{app:Logics}, $x(t_0) \in \widetilde{\Pi}$, 
	$x(t_{l})\in \pi_{j_l}$, $j_l\in\mathcal{K}_\mathcal{R}, \forall l\in\mathbb{N}$, and $\breve{\psi}_l=\mathcal{L}(\pi_{j_l})\subseteq 2^{\Psi}$, \textit{i.e.,} the subset of atomic propositions that are true when $x(t_j)\in\pi_{j_l}, \forall l\in\mathbb{N}$.	
\end{definition}

The specifications in this section are expressed via a Metric Interval Temporal Logic (MITL) formula $\mathsf{\Phi}$ (see Appendix \ref{app:Logics} for more details), although other timed variants could be used. 
The timed behavior $\mathfrak{b}$ satisfies a timed formula $\mathsf{\Phi}$ if and only if $\mathfrak{b}_\psi \coloneqq (\breve{\psi}_0,t_0)(\breve{\psi}_1,t_1)\dots \models \mathsf{\Phi}$. 

We are now ready to state the problem addressed in this section.
\begin{problem} \label{problem 1 (icra19)}
	Consider a robot with dynamics governed by \eqref{eq:dynamics (icra19)}, operating in the workspace $\mathcal{W}$, with initial position $x(0) \in \mathcal{F}$. Given a timed formula $\mathsf{\Phi}$ over $\Psi$ and a labeling function $\widetilde{\mathcal{L}}$, develop a control strategy that results in a solution $x:[0,\infty)\to\mathcal{F}$, which achieves a timed behavior $\mathfrak{b}$ that yields the satisfaction of $\mathsf{\Phi}$.    
\end{problem}

\subsection{Problem Solution} \label{sec:main results (icra19)}

In this section we present the proposed solution, which consists of two layers: (i) a tuning-free continuous control law that guarantees the navigation of the agent to a desired point from \textit{all} obstacle-collision-free configurations, and (ii) a discrete time plan  over the regions of interest for the robot to follow, which employs formal verification and optimization techniques and is updated on-line.

\subsubsection{Motion Controller}
\label{subsec:motionC (icra19)}
The first part of the proposed solution is the design of a control protocol such that a transition to a region of interest is established, according to Def. \ref{def:transition (icra19)}. Assume, therefore, that $x(t_k)\in\mathcal{F}$, and more specifically, $x(t_k)\in\pi_k$ $(x(t_k)\in\pi_{\mathcal{W}})$ for some $t_k\in\mathbb{R}_{\geq 0}$ and $k\in\mathcal{K}_\mathcal{R}$. Given $\delta\in\mathbb{R}_{>0}$, we wish to find a time-varying state-feedback control law $u:\mathcal{F}\times[t_k,t_\ell]$, with $t_\ell \geq t_k+\delta$, such that $\pi_k \overset{\delta}{\longrightarrow} \pi_{\ell}$ ($\pi_{\mathcal{W}} \overset{\delta}{\longrightarrow} \pi_\ell$).
To that end, we first redefine the free space as 
\begin{align*}
\mathcal{F} \coloneqq \mathcal{W}\backslash \bigg( \bigcup\limits_{i\in\mathcal{J}} \mathcal{O}_i \cup \bigcup\limits_{m\in\mathcal{M}_s} \pi_m \Bigg),
\end{align*}
so that regions of interest that shall not be crossed during the transition are regarded as obstacles.

Following the previous work \cite{2018VrohRAL}, the tuple $\left(x(t_k), c_{\pi_l},  \delta\right)$ constitutes a well-defined instance of the \emph{Prescribed Time Scale Navigation Problem} \cite[Problem 1]{2018VrohRAL} in $\mathcal{F}$.
Theorem 2 of the aforementioned work suggests that the construction of the required feedback law $u:\mathcal{F}\times[t_k,t_\ell]\to\mathbb{R}^2$ reduces to the problem of smoothly transforming the free space $\mathcal{F}$ to a topologically equivalent, yet geometrically simpler, space. 

More specifically, we require a diffeomorphism $\mathbf{\mathrm{T}}: \mathcal{F} \rightarrow \mathcal{P}$ where $\mathcal{P}$ is a point world \cite{loizou2017navigation}; an open disk modulo a finite set with cardinality equal to the number of obstacles and regions of interest $|\mathcal{J}| + |\mathcal{M}_s|$. Under the prevailing Assumption \ref{ass:SphereWorld (icra19)}, \cite[Theorem 1]{vlantis2018robot} provides a computationally efficient method to determine the space $\mathcal{P}$ and the mapping $\mathbf{\mathrm{T}}$.

This allows us to apply the conclusions of \cite[Theorems 1, 2]{2018VrohRAL} which yield a feedback law $u:\mathcal{F}\times[t_k,t_\ell]\to\mathbb{R}^2$ such that the closed-loop system satisfies the properties of \autoref{def:transition (icra19)} therefore establishing the existence of the required timed transition.
More details can be found in \cite{2018VrohRAL}.

\subsubsection{High-Level Plan Generation} \label{subsec:High level (icra19)}
The second part of our solution is the derivation of a high-level timed plan over the regions of interest, which satisfies the given timed formula $\mathsf{\Phi}$. This plan will be generated using standard techniques from automata-based formal verification and optimization methodologies. Thanks to the proposed control law of the previous section that allows the transitions in the set $\widetilde{\Pi}$ in predefined time intervals, we can abstract the motion of the robotic agent as a finite transition system $\mathcal{T} \coloneqq \{\widetilde{\Pi}, \widetilde{\Pi}_0, \longrightarrow, \Psi, \widetilde{\mathcal{L}}, \gamma \}$, where $\widetilde{\Pi}$ is the set of states defined in Section \ref{sec:Problem-Formulation (icra19)}, $\widetilde{\Pi}_0\in \widetilde{\Pi}$ is the initial state, $\longrightarrow \coloneqq \widetilde{\Pi}\times\widetilde{\Pi}$ is a transition relation according to Def. \ref{def:transition (icra19)}, $\Psi$ and $\widetilde{\mathcal{L}}$ are the atomic propositions and the labeling function, respectively, as defined in Section \ref{sec:Problem-Formulation (icra19)}, and $\gamma:(\longrightarrow)\to\mathbb{R}_{> 0}$ is a cost associated with each transition. 
More specifically, we consider as cost the distance the agent has to cover from a region $\pi_k$ (or $\pi_{\mathcal{W}}$) to a region $\pi_\ell$. However, this cost is highly dependent on the initial robot configuration and the number and position of the obstacles between the initial and the goal regions, and cannot be computed explicitly. Therefore, we initially set $\gamma(\pi_k\to\pi_\ell) = \|c_k - c_\ell\|$, $\gamma(\pi_k\to\pi_k) = 0$, and $\gamma(\pi_\mathcal{W}\to\pi_k) = \gamma(\pi_k\to\pi_\mathcal{W}) = \|c_k - x(0)\|$, $\text{for all } k,\ell\in\mathcal{K}_\mathcal{R}$ with $k\neq \ell$, 
and proceed with the derivation of the timed plan as a timed sequence of regions in $\Pi$.


Firstly, the timed formula $\mathsf{\Phi}$ over the atomic propositions $\Psi$ is translated to the TBA $\mathcal{A}_t = (Q,Q_0,\mathsf{CL},\Psi,E,F)$ (see Appendix \ref{app:Logics} for more details) using off-the-shelf tools \cite{MightyL}. Secondly, we calculate the product B\"uchi Automaton $\mathcal{A}_\mathcal{P}$ as $\mathcal{A}_\mathcal{P} \coloneqq \mathcal{T}\otimes \mathcal{A}_t = (S, S_0, \longrightarrow_\mathcal{P}, F_\mathcal{P}, \gamma_\mathcal{P})$, where
\begin{itemize}
	\item $S = \widetilde{\Pi}\times Q$,
	\item $S_0 = \widetilde{\Pi}\times Q_0$,
	\item $\longrightarrow_\mathcal{P} \subset S\times \Phi(\mathsf{CL})\times 2^C \times S$ gives the set of edges; $e \coloneqq (s,g,R,s')\in \longrightarrow_\mathcal{P}$, with $s \coloneqq (\pi,q)$, $s'\coloneqq (\pi',q')\in S$ if and only if (i) $(q,g,R,\mathcal{L}(\pi),q')\in E$ and (ii)  $q = q'$, $(\pi,\pi')\in\longrightarrow$.
	\item $F_\mathcal{P}\subseteq \widetilde{\Pi}\times F$ with $s\coloneqq(\pi,q)\in F_\mathcal{P}$ if and only if $q\in F$ and $(s,\Phi(\mathsf{CL}),R,S)\in E$ for some state in $S$, \textit{i.e.,} there is always a transition from $s$,  for all the the possible valuations of the clocks $\mathsf{CL}$.
	\item $\gamma_\mathcal{P}: (\longrightarrow_\mathcal{P}^*) \to \mathbb{R}_{>0}$, with $\gamma_\mathcal{P}((s,g,R,s')) = \gamma(\pi\to\pi')$, where $(\longrightarrow_\mathcal{P}^*) \coloneqq \{ ((\pi,q),g,R,(\pi',q'))\in\longrightarrow_\mathcal{P} : \pi \neq \pi')\}$.		
\end{itemize}
We use the abbreviation $s \xrightarrow{{I}} s'$ for $(s,g,R,s')\in \longrightarrow_\mathcal{P}$, where ${I} \coloneqq \{g,R\}$.
Note that the product $\mathcal{A}_{\mathcal{P}}$ consists of a finite number of states, and therefore we can employ graph-search techniques to find the optimal timed path, with respect to the cost $\gamma_\mathcal{P}$, from the initial states $S_0$ to the accepting states $F_\mathcal{P}$, which will satisfy the given timed formula $\mathsf{\Phi}$ \cite{guo2013motion}. This path will contain a finite prefix --- a finite sequence of states to be visited --- and a infinite suffix --- a specific sequence of states to be visited infinitely many times \cite{guo2013motion,baier2008principles}. Moreover, note that the motion controller developed in Section \autoref{subsec:motionC (icra19)} can guarantee the safe navigation among two regions of interest in any predefined time interval.

By viewing $\mathcal{A}_\mathcal{P}$ as a graph, we can find a path that starts at the initial states $S_0$ and traverses an accepting state in $F_\mathcal{P}$ infinitely many times. Such a path has the form
\begin{multline*}
\bar{s}_{p_1} \xrightarrow{I_{1,2}} \bar{s}_{p_2} \xrightarrow{I_{2,3}}\ldots \xrightarrow{I_{L-1,L}} \bar{s}_{p_L} \xrightarrow{I_{L,L+1}} \\ 
\Big( \bar{s}_{p_{L+1}} \xrightarrow{I_{L+1,L+2}} \ldots \xrightarrow{I_{L+Z-1,L+Z}} \bar{s}_{p_{L+Z}}  \Big)^\omega
\end{multline*}
Here, $\bar{s}_{p_j}$, for $j\in\{1,\dots,L+Z\}$, denotes the sequence of states
\begin{equation*}
\bar{s}_{p_j} \coloneqq (\pi_{p_j}, q_{j_1}) \xrightarrow{I_{j_{1,2}}} \ldots \xrightarrow{I_{j_{(\ell_j-1),\ell_j}}} (\pi_{p_j}, q_{j_{\ell_j}}),
\end{equation*}
with $\pi_{p_j}\in \widetilde{\Pi}$, $q_{j_\iota} \in Q$, for $j\in\{1,\dots,L+Z\}$, $\iota \in \{1,\dots,\ell_j\}$, and $\ell_j\in\{1,\dots,|S|\}$. Moreover, $q_{(j+1)_1} = q_{j_{\ell_j}}$, $q_{(L+Z)_{\ell_{(L+Z)}}} = q_{(L+1)_1}$ and 
\begin{align*}
&I_{j,j+1} \coloneqq 
\Big\{g_{j,j+1}, R_{j,j+1}\Big\}, \  \  I_{j_{\iota,\iota+1}} \coloneqq \Big\{g_{j_{\iota,\iota+1}}, R_{j_{\iota,\iota+1}}\Big\},
\end{align*}
indicating the corresponding guards and reset maps, for $j\in\{1,1,\dots,L+Z-1\}, \iota \in \{1,\dots,\ell_j-1\}$. The transition set $I_{L+Z,L+1}$ is defined similarly.  Loosely speaking, the path consists of consecutive (at most $|S|$) transitions of the form $(\pi_j, q_{j_{\iota}})\xrightarrow{
	(\cdot)}(\pi_j, q_{j_{(\iota+1)}})$ among states in $\mathcal{A}_t$, where $\pi_j$ is fixed, and transitions of the form $(\pi_{p_j},q_{j_{\ell_j}})\xrightarrow{(\cdot)}(\pi_{p_{(j+1)}},q_{(j+1)_1})$ among the states of $\mathcal{T}$, where $q_{(j+1)_1} = q_{j_{\ell_j}}$ is fixed. 

Note that we have not yet associated any time intervals with the transitions $(\pi_{p_j},q_{j_{\ell_j}})\xrightarrow{(\cdot)}(\pi_{p_{(j+1)}},q_{(j+1)_1})$, which correspond to physical transitions among the regions of interest. We do that now by using the transition guards $g_{j,j+1}, g_{j_{\iota,\iota+1}}$. 
More specifically, consider the transitions 
\begin{equation*}
\small
\begin{split}
(\pi_{p_j}, q_{j_1}) \xrightarrow{I_{j_{1,2}}}  (\pi_{p_j}, q_{j_2}) \xrightarrow{I_{j_{2,3}}} \dots \xrightarrow{I_{j_{(\ell_j-1),\ell_j}}} (\pi_{p_j}, q_{j_{\ell_j}}) \xrightarrow{I_{j,j+1}} (\pi_{p_{j+1}}, q_{(j+1)_1}),
\end{split}
\end{equation*}
that encode the physical transition from $\pi_{p_j}$ to $\pi_{p_{j+1}}$ in $\mathcal{A}_\mathcal{P}$. The intersection of the respective guards $g_{j,j+1}, g_{j_{\iota,\iota+1}}$, $\iota\in\{1,\dots,\ell_{j-1}\}$, provides a time interval of the form $\mathcal{I}_{j,j+1}\in\{ [a,b], [a,b), (a,b], (a,b), [a,\infty), (a,\infty) \}$, with $a,b \in\mathbb{Q}_{> 0}$, $b>a$, such that,  $t_{j,j+1} \in \mathcal{I}_{j,j+1} \Rightarrow t_{j,j+1}\models g_{j,j+1}, t_{j,j+1}\models g_{j_{\iota,\iota+1}}$, for $\iota\in\{1,\dots,\ell_{j-1}\}$, where $t_{j,j+1}$ is the time duration of the navigation $\pi_j \xrightarrow{t_{j,j+1}} \pi_{j+1}$. Note that $\mathcal{I}_{j,j+1}$ might be a function of the previous transition duration $t_{j-1,j}$.

Since $\mathcal{I}_{j,j+1}$ is, in general, an infinite set, and we have, thus, infinitely many choices for $t_{j,j+1}$, we propose a procedure for assigning the time durations $t_{j,j+1}$, for each $j\in\{1,L+Z,-1\}$, and $t_{L+Z,L+1}$.
In particular, we formulate the transition times assignment as a convex optimization problem. To that end, let $t_p \coloneqq [t_{1,2},\dots,t_{L+Z-1,L+Z}, t_{L+Z,L+1}]^\top \in \mathbb{R}_{>0}^{L+Z+1}$ be the concatenation of the transition times constituting the variable of the following optimization problem:
\begin{subequations} \label{eq:optim (icra19)}
\begin{align}
&\underset{t_p}{\textrm{minimize}} \hspace*{-5pt}\sum_{j= 1}^{L+Z}\left(\frac{\gamma(\pi_{p_j}\rightarrow\pi_{p_{j+1}})}{t_{j,j+1}}\right) + \frac{\gamma(\pi_{p_{L+Z}}\rightarrow\pi_{p_{L+1}})}{t_{j,j+1}}, \\
&\textrm{subject to }
t_{j,j+1} \in \mathcal{I}_{j,j+1}, \text{ for all } j\in\{1,L+Z,-1\},\\
&\qquad \quad \quad ~ t_{L+Z,L+1}\in \mathcal{I}_{L+Z,L+1}.
\end{align}
\end{subequations}
Note that the objective function is a convex function of $t_p$ and the constraints can be expressed as linear inequalities on the problem variables. Thus, the above optimization problem is convex and can be efficiently solved using off-the-shelf software.
The choice of this particular cost function is motivated by the following two observations: (i) the time assigned to a transition is an increasing function of the transition cost, and (ii) brief transition times are penalized.
Furthermore, the imposed constraints guarantee the satisfaction of the formula provided that transitions are executed within the specified transition times. We also report empirical evidence from numerical simulations suggesting that reduction in control effort is achieved.

After solving the aforementioned optimization problem and obtaining the time durations $t_p$,
the robot performs the first transition using the motion controller presented in \autoref{subsec:motionC (icra19)} where $\delta$ is taken equal to the corresponding transition time.
Once the transition $\pi_{p_j}\xrightarrow{t_{j,j+1}}\pi_{p_{j+1}} $is completed, the corresponding transition cost $\gamma(\pi_j\rightarrow\pi_{j+1})$ is updated by being set equal to the length of the integral curve of the closed-loop system for the duration of the transition. The updated value of the transition cost is in some sense more accurate than the initial estimate based on the Euclidean distance since the existence of obstacles can potentially obstruct the straight line path between two regions of interest.

Having acquired this new information the associated optimization problem can be solved to acquire new values for the transition times. We note that after each transition the constraints of the optimization problem are altered. In particular, the TBA of the formula has to be shifted forward by an amount of time equal to the last performed transition which induces a change in the guards and, therefore, to the optimization problem constraints.
We assume that the time needed for solving the optimization problem is short enough so that satisfaction of the formula is not jeopardized. This assumption is reasonable enough primarily owing to the problem's low computational complexity and, secondarily, the fact that the previously computed values of the transition times are a good prior for initiating the numerical solver. Nevertheless, computational overhead can be accounted for in the constraints by allocating the required time, or optimization could be performed en route to the next region of interest with the transition times adjusted in an any-time fashion.

\subsection{Simulation Results}

To demonstrate the proposed scheme, we consider a task and motion planning problem for a robot operating in a planar office environment. In particular, we consider three points of interest and therefore have  $\Pi = \{c_{\pi_k}\}_{k\in\mathcal{K}_\mathcal{R}}$, where $\mathcal{K}_\mathcal{R} = \{1,2,3\}$. The corresponding regions of interest $\pi_k=\bar{\mathcal{B}}(c_{\pi_k}, r)$, where $r_{\pi_k}=0.2$ for $k\in\mathcal{K}_\mathcal{R}$, define the set ${\Pi} = \left\lbrace\pi_k\right\rbrace_{k\in\mathcal{K}_\mathcal{R}}$. The set of atomic prepositions is $\Psi = \tilde{\Pi}$ and the labeling function ${\mathcal{L}}:{\Pi}\rightarrow 2^\Psi$ is defined as $\pi_k\mapsto \{\pi_k\}$, $k\in\mathcal{K}_\mathcal{R}$. The scenario setting is illustrated in \autoref{fig:workspace (icra19)}.

We require that the robot \emph{``always visits each region of interest at least once every $120$ time units''} which is equivalent to the MITL formula
$
\mathsf{\Phi} = \bigwedge_{k\in\mathcal{K}_\mathcal{R}}\big(\square\lozenge_[0,120] \pi_k\big).$
The robot is initially located at $c_{\pi_1}\in \pi_1$ and, therefore,
the infinitely repeating cycle of transitions
$
\pi_1 \xrightarrow{t_{1,2}(1)} \pi_2 \xrightarrow{t_{2,3}(1)} \pi_3 \xrightarrow{t_{3,1}(1)}\pi_1\xrightarrow{t_{1,2}(2)} \ldots
$
with appropriately assigned transition times is an accepting run.
Let $t : {0,1,\dots}\rightarrow \mathbb{Q}_{> 0}^3$, $\kappa\mapsto [t_{1,2}(\kappa), t_{2,3}(\kappa), t_{3,1}(\kappa)]^\top $ which is defined recursively as follows: $t(0) \coloneqq [0,0,0]^\top$, then assuming $t(\kappa)$ is defined for some $\kappa\in\mathbb{N}_0$,
\[\tiny
t(\kappa+1) \coloneqq (U_3^\top)^{\kappa}\begin{bmatrix}
1&0&0\\
0&0&0\\
0&0&0
\end{bmatrix}U_3^{\kappa}\hat{t}(\kappa) +
(U_3^\top)^{\kappa}\begin{bmatrix}
0&0&0\\
0&1&0\\
0&0&1
\end{bmatrix}U_3^{\kappa}t(\kappa),
\]
where $U_3 \coloneqq \begin{bmatrix}
0&1&0\\
0&0&1\\
1&0&0
\end{bmatrix}$ is the upper shift matrix and $\hat{t}({\kappa})$ is the solution of the optimization problem \eqref{eq:optim (icra19)} under the following constraints:
\[
\begin{bmatrix}
1&0&0\\
1&1&0\\
1&1&1
\end{bmatrix}U_3^{\kappa}\hat{t}({\kappa})\leq 
\begin{bmatrix}
120\\
120\\
120\\
\end{bmatrix} -
\begin{bmatrix}
0&1&1\\
0&0&1\\
0&0&0
\end{bmatrix}U_3^{\kappa}t(\kappa).
\]
The motion controller results in collision-free trajectories (\autoref{fig:workspace (icra19)}), and by performing each transition time in the time derived from the optimization procedure results in a run that satisfies the formula $\mathsf{\Phi}$ (see bottom of \autoref{fig:schedule (icra19)}). Finally, it is worth noting that the transition times converge in just a few steps as illustrated on the top part of \autoref{fig:schedule (icra19)} and the overall control effort per suffix execution is reduced (\autoref{tbl:controlEffort (icra19)}). 
\begin{table}[h!]
	\caption{\textsc{Control Effort per Suffix Execution}}\label{tbl:controlEffort (icra19)}
	\centering
	\begin{tabularx}{\linewidth}{ l  X  X  X  X  X}
		\toprule
		Cycle & 1 & 2 & 3 & 4 & 5\\
		\midrule
		$\int \|u(x(\tau),\tau)\|^2\, d\tau$ & 8.06& 7.67 & 7.67 &7.65 & 7.66\\
		\bottomrule
	\end{tabularx}
\end{table}

\begin{figure}[h!]
	\centering
	\includegraphics[width=.75\textwidth]{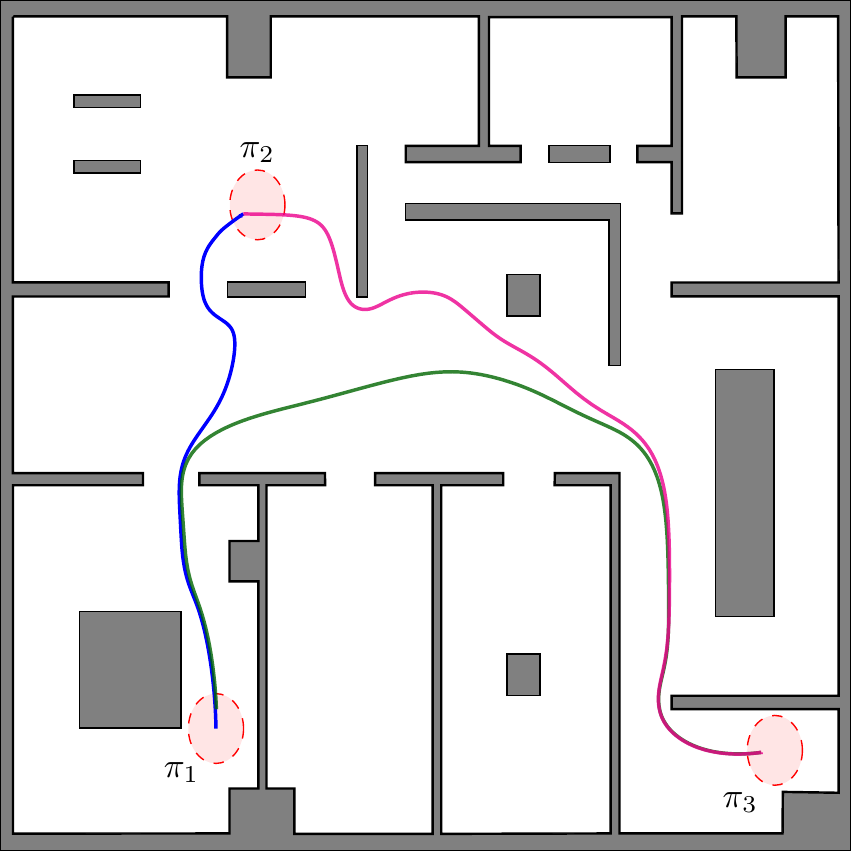}
	\caption{Workspace overview. The red discs correspond to the three regions of interest. The plotted paths are the resulting trajectories from the first out of the five executions of the suffix.}\label{fig:workspace (icra19)}
\end{figure}
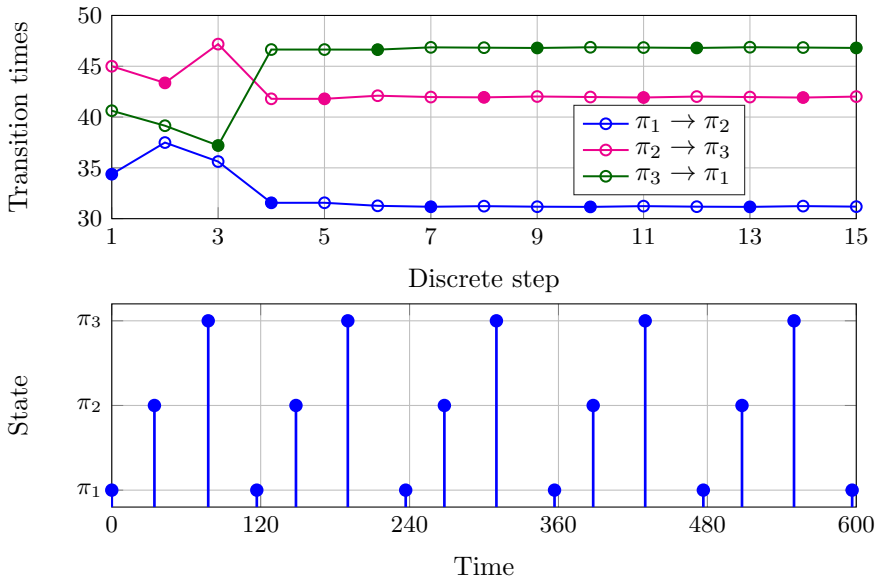
\begin{figure}[h!]
%
%
\begin{tikzpicture}
\pgfplotsset{every x tick label/.append style={font=\small, yshift=0ex}}
\pgfplotsset{every y tick label/.append style={font=\small, xshift=0ex}}
\begin{axis}[%
width=0.85*\linewidth,
height=0.55*1.91in,
at={(0 in,0in)},
scale only axis,
xmin=1,
xmax=15,
xlabel style={font=\color{black}},
xlabel={Discrete step},
ymin=30,
ymax=50,
ylabel style={font=\color{black}},
ylabel style={align=center},
ylabel=Transition times,
axis background/.style={fill=white},
xmajorgrids,
ymajorgrids,
legend style={at={(0.62,0.115)}, anchor=south west, legend cell align=left, align=left, draw=white!15!black},
xtick={1,3,5,7,9,11,13,15},
xtick pos=left,
ytick pos=left,
]
\addplot [color=blue, mark=o, mark options={solid, blue}, line width = 0.75pt]
  table[row sep=crcr]{%
1	34.3754129362276\\
2	37.4900046164788\\
3	35.6210895805993\\
4	31.5594599689439\\
5	31.5625429471231\\
6	31.2635148638612\\
7	31.1697725548059\\
8	31.2309380231369\\
9	31.1768745436276\\
10	31.1577197670838\\
11	31.2324131262107\\
12	31.1792429659967\\
13	31.1584112252896\\
14	31.2355941283686\\
15	31.1823910996003\\
};
\addlegendentry{$\pi_1\rightarrow \pi_2$}

\addplot [color=magenta, mark=o, mark options={solid, magenta}, line width = 0.75pt]
  table[row sep=crcr]{%
1	45.0010232585782\\
2	43.3623968124466\\
3	47.1800740765346\\
4	41.7954682910303\\
5	41.7940575436841\\
6	42.1009445146979\\
7	41.9688340222749\\
8	41.9397676062307\\
9	42.0230934085013\\
10	41.9724265358901\\
11	41.926333854544\\
12	42.0179990944535\\
13	41.9679520653167\\
14	41.9210454838386\\
15	42.0143181089489\\
};
\addlegendentry{$\pi_2\rightarrow \pi_3$}

\addplot [color=green!40!black, mark=o, mark options={solid, green!40!black}, line width = 0.75pt]
  table[row sep=crcr]{%
1	40.623419358921\\
2	39.1475737868774\\
3	37.1987150841393\\
4	46.6449765717485\\
5	46.643301140973\\
6	46.6354448125365\\
7	46.8612978252911\\
8	46.8291980504966\\
9	46.7999366838094\\
10	46.869757587883\\
11	46.8411566957965\\
12	46.8026625740851\\
13	46.8735406153388\\
14	46.8432640993204\\
15	46.8031954609839\\
};
\addlegendentry{$\pi_3\rightarrow \pi_1$}
\addplot[color=blue, mark=*, mark options={solid, blue}, only marks, forget plot] table[row sep=crcr] {%
1	34.3754129362276\\
4	31.5594599689439\\
7	31.1697725548059\\
10	31.1577197670838\\
13	31.1584112252896\\
};

\addplot[color=magenta, mark=*, mark options={solid, magenta}, only marks, forget plot] table[row sep=crcr] {%
2	43.3623968124466\\
5	41.7940575436841\\
8	41.9397676062307\\
11	41.926333854544\\
14	41.9210454838386\\
};

\addplot[color=green!40!black, mark=*, mark options={solid, green!40!black}, forget plot, only marks] table[row sep=crcr] {%
3	37.1987150841393\\
6	46.6354448125365\\
9	46.7999366838094\\
12	46.8026625740851\\
15	46.8031954609839\\
};
\end{axis}
\end{tikzpicture}%
%
%
\definecolor{mycolor1}{rgb}{0.00000,0.44700,0.74100}%
\begin{tikzpicture}
\pgfplotsset{every x tick label/.append style={font=\small, yshift=0ex}}
\pgfplotsset{every y tick label/.append style={font=\small, xshift=0ex}}
\begin{axis}[%
width=0.85*\linewidth,
height=0.55*1.91in,
at={(0,0)},
scale only axis,
xmin=0,
xmax=600,
ymin=0.8,
ymax=3.2,
axis background/.style={fill=white},
xmajorgrids,
ymajorgrids,
xlabel = \text{Time},
ylabel = \text{State},
xtick = {0,120,240,360,480,600},
ytick = {1, 2, 3},
yticklabels = {$\pi_1$, $\pi_2$, $\pi_3$},
]
\addplot[ycomb, color=blue, mark=*, mark options={solid, blue}, forget plot, line width = 1pt] table[row sep=crcr] {%
0	1\\
34.3754129362276	2\\
77.7378097486742	3\\
116.885383535552	1\\
148.444843504495	2\\
190.23890104818	3\\
236.882202189153	1\\
268.051974743959	2\\
309.991742350189	3\\
356.820940400686	1\\
387.97866016777	2\\
429.904994022314	3\\
476.74615071811	1\\
507.9045619434	2\\
549.825607427238	3\\
596.668871526559	1\\
};
\addplot[forget plot, color=white!15!black] table[row sep=crcr] {%
0	0\\
600	0\\
};
\end{axis}
\end{tikzpicture}%
	\caption{{(top)} The transition times calculated before each transition; filled marks correspond to actually used transition times. {(bottom)} The resulting timed run of the transition system.} 
	\label{fig:schedule (icra19)}
\end{figure}

\section{Sampling-based Motion Planning for Uncertain High-dimensional Systems via Adaptive Control} \label{sec:adaptive RRT}

We turn now our attention to the motion planning problem for uncertain high-dimensional systems, such as robotic manipulators. We integrate sampling-based motion planning techniques with intelligent adaptive control methodologies to tackle the problem of uncertain dynamics and collision-free navigation. 

\subsection{Problem Formulation}

Consider a robotic system with state $(q,\dot{q})\in \mathbb{T} \times \mathbb{R}^n \subset \mathbb{R}^{2n}$, $n\in\mathbb{N}$, {representing its positions and velocities}. Usual robotic structures (e.g., robotic manipulators) might consist of translational and rotational joints, which we define here as $q_{tr} \in \mathbb{R}^{n_{tr}}$ and $q_r \in [0,2\pi)^{n_r}$, respectively, with $n_{tr} + n_r = n$, and hence $\mathbb{T} \coloneqq \mathcal{W}_{tr} \times [0,2\pi)^{n_r}$, where $\mathcal{W}_{tr}$ is a closed subset of $\mathbb{R}^{n_{tr}}$. Without loss of generality, we assume that $q = [q_{tr}^\top,q_r^\top]^\top$. We consider that the equations of motion of the robot obey {the} standard $2$nd-order Lagrangian dynamics \eqref{eq:manipulator joint_dynamics (TCST_coop_manip)}
\begin{equation}
B(q)\ddot{q} + C_q(q,\dot{q})\dot{q} + g_q(q) + d_q(q,\dot{q},t) = \tau, \label{eq:dynamics (WAFR)}
\end{equation}
with the various terms as in \eqref{eq:manipulator joint_dynamics (TCST_coop_manip)}. 
We assume here that $d(\cdot)$ is continuous and uniformly bounded by a \textit{known} bound $\bar{d}$ as $\|d(t)\| \leq \bar{d}$, $\forall t\geq 0$ 
{We remind the reader that the dynamical terms $B(q)$, $C(q,\dot{q})$, $g(q)$ of \eqref{eq:dynamics (WAFR)} depend on the dynamical parameters of the robot, i.e., its mass and moment of inertia. These parameters are assumed to be \textit{unknown}, and hence they cannot be used in the planning and control modules. The same applies to the function $d(\cdot)$. Nevertheless, as will be shown later, having satisfying estimates for these terms renders the planning module for the robot less conservative in terms of collision checking. } 

We consider that the robot operates in a workspace $\mathcal{W} \subset \mathbb{R}^3$ filled with obstacles occupying a closed set $\mathcal{O} \subset \mathbb{R}^3$. We denote the set of points that consist the volume of the robot at configuration $q$ as $\mathcal{A}(q) \subset \mathbb{R}^3$. The collision-free space is defined as the open set $\mathcal{A}_\text{free} \coloneqq \{ q\in\mathbb{T} : \mathcal{A}(q) \cap \mathcal{O} = \emptyset \}$. Our goal is to achieve safe navigation of the robot to a predefined goal region $Q_g \subset \mathcal{A}_\text{free}$ from an initial configuration $q(0) \in \mathcal{A}_\text{free}$ via a path ${\boldsymbol{q}_\text{p}:[0,\sigma] \to \mathcal{A}_\text{free}}$ satisfying $\boldsymbol{q}_\text{p}(0) = q(0)$ and $\boldsymbol{q}_\text{p}(\sigma) \in Q_g$, for some positive $\sigma$.
	
	The problem we consider is the following:
	
	\begin{problem} \label{prob:1 (WAFR)}	
		Given $q(0))\in \mathcal{A}_{\text{free}}$ and $Q_g \subset \mathcal{A}_\text{free}$, respectively, design a control trajectory $u:[0,t_f] \to \mathbb{R}^{n}$, for some {finite} $t_f > 0$, such that the solution $q^\ast(t)$ of \eqref{eq:dynamics (WAFR)} satisfies $q^\ast(t) \in \mathcal{A}_\text{free}$, $\forall t\in[0,t_f]$, and $q^\ast(t_f) \in Q_g$.	
	\end{problem}
	
	The feasibility of Problem \ref{prob:1 (WAFR)} is established in the following assumption.
	
	\begin{assumption} \label{ass:path ass (WAFR)}
		There exists a (at least twice differentiable) path $\boldsymbol{q}_\textup{p}:[0,\sigma]\to \mathcal{A}_\textup{free}$ such that $\boldsymbol{q}_\textup{p}(0) = q(0)$ and $\boldsymbol{q}_\textup{p}(\sigma) \in Q_g$.
	\end{assumption}

	\subsection{Problem Solution}

	We present here the proposed solution for Problem \ref{prob:1 (WAFR)}. Our methodology follows a two-layer approach, consisting of a robust trajectory-tracking control design and a higher-level sampling-based motion planner.
	Firstly, we use an adaptive control protocol that compensates for the uncertain dynamical parameters of the {robot} and forces the system to evolve in a funnel around a desired trajectory, whose size depends on the initial estimates of the dynamical parameters and the bound of the external disturbances. Secondly, we develop a geometric sampling-based motion planner that uses this funnel to find a collision free trajectory from the initial to the goal configuration. Intuitively, the robust control design {helps} the motion planner procedure, which does not have to take into account the complete dynamics \eqref{eq:dynamics (WAFR)}.
	%

	\subsubsection{Control Design} \label{sec:adaptive controller (WAFR)}	
	
	We first recap the  dynamics linear parameterization with respect to the aforementioned unknown parameters, denoted by $\theta \in \mathbb{R}^l$, $l\in\mathbb{N}$. More specifically, similarly to Chapter \ref{chapter:cooperative manip}, it holds that		
	\begin{equation}
	B(a)d + C(a,b)c + g(a) = Y(a,b,c,d)\theta,  \label{eq:linear parameterization (WAFR)}
	\end{equation}
	$\forall a\in\mathbb{T}, b,c,d\in\mathbb{R}^{3n}$, {where $Y(\cdot)$ is a matrix independent of $\theta$. 
	}Let $q_\text{d}\coloneqq [q_{\text{d},tr}^\top, q_{\text{d},r}^\top]^\top :[0,t_f] \to \mathbb{T}$ be a reference trajectory, with $q_{\text{d},tr}\in\mathbb{R}^{n_{tr}}$ and $q_{\text{d},r}\in[0,2\pi)^{n_r}$ {being} its translational and rotational parts, respectively. Such a trajectory will be the output of the sampling-based motion planning algorithm that will be developed in the next section. {We wish to design the control input $\tau$ of \eqref{eq:dynamics (WAFR)}  such that $q(t)$ converges close to $q_\text{d}(t)$, despite the uncertainty in $\theta$. 
	We show how such a design can be used in the motion planning of a robotic system with dynamical uncertainties by developing a suitable variant of a standard adaptive control scheme. }
	
	{We start by defining the appropriate error metric between $q=[q_{tr}^\top,q_r^\top]^\top$ and $q_\text{d} = [q_{\text{d},tr}^\top, q_{\text{d},r}^\top]^\top$, which represents their distance. Regarding the translational part, we define the standard Euclidean error $e_{tr} \coloneqq q_{tr} - q_{\text{d},tr}$. For the rotation part, however, the same error $e_r \coloneqq q_r - q_{\text{d},r}$ does not represent the minimum distance metric, since $q_r$ evolves on the $n_r$-dimensional sphere, and its use might cause conservative or infeasible results in the planning layer. Hence, unlike standard adaptive control schemes for robotic manipulators, which drive the Euclidean difference $e_r(t)$ to zero (e.g., \cite{tomei1999robust,slotine1987adaptive}), we use the chordal metric $d_C(x,y) \coloneqq 1 - \cos(x,y) \in [0,2]$, $\forall x,y\in[0,2\pi)$, or $\bar{d}_C(x,y) \coloneqq \sum_{j\in\{1,\dots,\ell\}}d_C(x_j,y_j)$ for vectors $x = [x_1,\dots,x_\ell], y = [y_1,\dots,y_\ell] \in [0,2\pi)^\ell$. 
		Nevertheless, note that rotational joints subject to upper and/or lower mechanical limits evolve in $\mathbb{R}$ rather than the unit circle and should hence be included in $q_{tr}$ instead of $q_r$.} 
	
	We are now ready to define a suitable distance metric for $\mathbb{T}$ as follows: for $x\coloneqq [x_{tr}^\top,x_r^\top]^\top$, $y\coloneqq [y_{tr}^\top,y_r^\top]^\top$ $\in \mathbb{T}$ we define  $d_\mathbb{T}$ as 
	\begin{align*} 
	d_\mathbb{T}(x,y) \coloneqq \|x_{tr} - y_{tr}\|^2 + \bar{d}_C(x_r,y_r).
	\end{align*}
	Note, however, that the chordal metric induces a limitation with respect to tracking on the unit sphere. Consider $d_C(q_{r_j},q_{\text{d},r_j}) = 1 - \cos(e_{r_j})$, where we further define $e_{r_j} \coloneqq q_{r_j} - q_{\text{d},r_j}$ as the $j$th element of $e_r$, $j\in\{1,\dots,n_r\}$. Differentiation yields 
	\begin{equation*}
	\dot{d}_C(q_{r_j},q_{\text{d},r_j}) =\sin(e_{r_j})\dot{e}_{r_j},  \ \forall j\in\{1,\dots,n_r\},
	\end{equation*}
	which is zero when $e_{r_j} = 0$ or $e_{r_j} = \pi$. The second case is an undesired equilibrium, which  implies that the point $e_{r_j} = 0$ cannot be stabilized from \textit{all} initial conditions using a continuous controller. This is an inherent property of dynamics on the unit sphere due to topological obstructions (\cite{bhat2000topological}). {In the following, we devise a control scheme that, except for driving $q(t)$ to $q_\text{d}(t)$, guarantees that $e_{r_j}(t) \neq \pi$, $\forall t\in (0,t_f]$, provided that $e_{r_j}(0) \neq 0$, $\forall j\in\{1,\dots,n_r\}$. To do that, we define the mapping 
		\begin{equation} \label{eq:H metric (WAFR)}
		\mathsf{H}(x,y) \coloneqq \left[\tan\left(\frac{x_1-y_1}{2}\right),\dots,\tan\left(\frac{x_{n_r}-y_{n_r}}{2}\right)\right]^\top \in \left(-\frac{\pi}{2},\frac{\pi}{2}\right)^{n_r}
		\end{equation}
		for vectors $x = [x_1,\dots,x_{n_r}]^\top$, $y = [y_1,\dots,y_{n_r}]^\top \in [0,2\pi)^{n_r}$, as well as the signal $\eta_r \coloneqq \mathsf{H}(q_r,q_{\text{d},r})$. Note that $\eta_r$ is not defined when $e_{r_j} = \pi$ for some $j\in\{1,\dots,n_r\}$, which we exploit in the control design.}
	
		We define first the reference signals for  $\dot{q}_{tr}, \dot{q}_r$ as $\alpha_q \coloneqq [\alpha_{tr}^\top, \alpha_r^\top]^\top$, with 
		\begin{subequations} \label{eq:des velocity (WAFR)}
		\begin{align}	
		&\alpha_{tr}  \coloneqq -K_{tr} e_{tr} + \dot{q}_{\text{d},tr}, \\
		&\alpha_r  \coloneqq 
		\begin{bmatrix}
		\alpha_{r_1} \\ \vdots \\
		\alpha_{r_{n_r}} 
		\end{bmatrix}
		\coloneqq 
		\begin{bmatrix}
		\dot{q}_{\text{d},r_1} - k_{r_1} \cos\left(\frac{e_{r_1}}{2}\right) \sin\left(\frac{e_{r_1}}{2}\right)  \\ \vdots \\
		\dot{q}_{\text{d},r_{n_r}} - k_{r_{n_r}} \cos\left(\frac{e_{r_{n_r}}}{2}\right) \sin\left(\frac{e_{r_{n_r}}}{2}\right)
		\end{bmatrix}, \label{eq:des velocity r (WAFR)}
		\end{align} 
		\end{subequations}
		where ${K}_{tr} \in \mathbb{R}^{n_{tr}\times n_{tr}}$ is a symmetric positive definite gain matrix, and $k_{r_j} > 0$ are positive gain constants, $\forall j\in\{1,\dots,n_r\}$. Define also the associated velocity error 
		\begin{equation*}
			e_{v_q} \coloneqq \dot{q} - \alpha_q,
		\end{equation*}
		and the estimate $\hat{\theta} \in\mathbb{R}^l$ of $\theta$, as well as the error $e_\theta \coloneqq \hat{\theta} - \theta \in\mathbb{R}^l$.}
		
		Let now $\mathcal{R}_j$ be defined as 
		\begin{equation*}
		\mathcal{R}_j \coloneqq 
		\begin{cases}
		[0,\pi) & \text{ if } e_{r_j}(0) \in [0,\pi), \\ 
		(\pi,2\pi] & \text{ if } e_{r_j}(0) \in (\pi,2\pi),
		\end{cases}
		\end{equation*}
	and design
	the control law as $\tau: \mathcal{W}_{tr} \times \mathcal{R}_1\times\dots\mathcal{R}_{n_r}\times \mathbb{R}^6 \times \mathbb{R}^l \to \mathbb{R}^n$, with
	\begin{align} \label{eq:control law (WAFR)}
	\tau = \tau(e_x,e_{v_q},\hat{\theta}) \coloneqq Y_\alpha\hat{\theta} - K_v e_{v_q} - e_x,
	\end{align}
	where $Y_\alpha \coloneqq Y(q,\dot{q},\alpha_q,\dot{\alpha_q})$, {with $Y(\cdot)$ as given in \eqref{eq:linear parameterization (WAFR)},} the signal $e_x$ is defined as
	\begin{align*}
	e_x \coloneqq \begin{bmatrix}
	e_{tr},
	\frac{ \tan\left(\frac{e_{r_1}}{2}\right)  }{\cos\left(\frac{e_{r_1}}{2}\right)^2},
	\dots,
	\frac{ \tan\left(\frac{e_{r_{n_r}}}{2}\right)  }{\cos\left(\frac{e_{r_{n_r}}}{2}\right)^2}
	\end{bmatrix}^\top
	\end{align*}
	and $K_v\in\mathbb{R}^{n\times n}$ is a symmetric and positive definite gain matrix. Moreover, design the evolution of $\hat{\theta}$ as 
	\begin{equation} \label{eq:adaptation law (WAFR)}
	\dot{\hat{\theta}} = - \Gamma\left(\frac{1}{\underline{k}_v}Y_\alpha^\top e_{v_q} + \sigma_\theta \hat{\theta}\right),
	\end{equation}
	with any initial condition $\hat{\theta}(0)\in\mathbb{R}^l$, $\underline{k}_v\coloneqq \lambda_{\min}(K_v)$, and $\Gamma \in \mathbb{R}^{\ell \times \ell}$ being a symmetric positive definite gain matrix, {and $\sigma_\theta$ a positive constant}. 
{Note that the control law \eqref{eq:control law (WAFR)} is well-defined when $e_x \in \mathcal{R}_j$, since $e_{r_j} \neq \pi$, $\forall j\in\{1,\dots,n_r\}$. }
		
	The correctness of the aforementioned control scheme is proven in the subsequent theorem. 
	
			\begin{theorem} \label{th:control theorem (WAFR)}
				Consider the dynamics \eqref{eq:dynamics (WAFR)}, a reference trajectory $q_\textup{d}:[0,t_f]\to\mathbb{T}$, as well as the constant 
				\begin{align*} 
				V_0 \coloneqq \frac{1}{2}\|e_{tr}(0) \|^2 + \|\eta_r(0)\|^2  + \frac{1}{2\underline{k}_v}e_{v_q}(0)^\top B(q(0)) e_{v_q}(0) + \frac{1}{2}e_\theta(0)^\top \Gamma^{-1} e_\theta(0),
				\end{align*}
				Then, if $e_{r_j}(0) \neq \pi$, $\forall j\in\{1,\dots,n_r\}$, the control protocol \eqref{eq:control law (WAFR)}-\eqref{eq:adaptation law (WAFR)}  guarantees that 
				\begin{subequations} \label{eq:theorem bounds (WAFR)}
				\begin{align}
				\|e_{tr}(t)\| \leq \bar{e}_{tr} \coloneqq \max\left\{2V_0, \sqrt{\frac{d_x}{\underline{k}_{tr}}} \right\}, \ \
				\|\eta_r(t)\| \leq \bar{\eta}_r \coloneqq \max\left\{V_0, \sqrt{\frac{d_x}{\underline{k}_r}}  \right\}, 
				\end{align}
				\end{subequations}
				$\forall t\in[0,t_f]$, where $d_x$ is a positive constant satisfying $d_x \geq \frac{\bar{d}^2}{2\underline{k}_{v}^2} + \frac{\sigma_\theta}{2}\|\theta\|^2$, and $\underline{k}_{tr} \coloneqq \lambda_{\min}(K_{tr})$, $\underline{k}_r \coloneqq \min\{k_{r_1},\dots,k_{r_{n_r}}\}$. Moreover, it holds that 
				$e_{r_j}(t) \neq \pi$, $\forall j\in\{1,\dots,n_r\}$, and all closed-loop signals remain bounded, for all $t\in[0,t_f]$.
			\end{theorem}}
			\begin{proof}
				Let $x_R\coloneqq [e_{tr}^\top, \eta_r^\top, e_{v_q}^\top,e_\theta]^\top$ and consider the candidate Lyapunov function 
				\begin{align*}
				V(x_R) \coloneqq \frac{1}{2}\|e_{tr} \|^2 + \|\eta_r\|^2  + \frac{1}{2\underline{k}_v}e_{v_q}^\top B(q) e_{v_q} + \frac{1}{2}e_\theta^\top \Gamma^{-1} e_\theta,
				\end{align*}			
				Since $e_{r_j}(0) \neq \pi$, $V(x_R(0))$ is bounded by a constant $V(x_R(0)) \leq \bar{V}$. 
				
				Differentiation of $V$ yields 
				\begin{align*}
				\dot{V} =& e_{tr}^\top ( \dot{q}_{tr} - \dot{q}_{\text{d},tr}) + \sum_{j\in\{1,\dots,n_r\}}\frac{\tan\left(\frac{e_{r_j}}{2}\right)}{\cos\left(\frac{e_{r_j}}{2}\right)^2}(\dot{q}_{r_j} - \dot{q}_{\text{d},r_j})  + \frac{1}{\underline{k}_v}e_{v_q}^\top (\tau - C_q\dot{q} \\ & - g_q - d) 
				 + \frac{1}{2\underline{k}_v}e_{v_q}^\top \dot{B}e_{v_q} - \frac{1}{\underline{k}_v}e_{v_q}^\top B \dot{\alpha}_q
				+ e_\theta^\top \Gamma^{-1} \dot{\hat{\theta}},
				\end{align*}
				which, by substituting $\dot{q} =  e_{v_q} + \alpha_q$, becomes 
				\begin{align*}
				\dot{V} =& -e_{tr}^\top K_{tr} e_{tr} - \eta_r^\top K_r \eta_r + \frac{1}{2\underline{k}_v}e_{v_q}^\top \dot{B} e_{v_q} - \frac{1}{\underline{k}_v}e_{v_q}^\top C_q e_{v_q} + \\
				&\frac{1}{\underline{k}_v}e_{v_q}^\top(\tau + e_x - B\dot{\alpha_q} - C_q\alpha_q - g_q - d ) + e_\theta^\top \Gamma^{-1} \dot{\hat{\theta}},
				\end{align*}
				where $K_r \coloneqq \text{diag}\{k_{r_1},\dots, k_{r_{n_r}} \}$.
				By using the skew symmetric property of $\dot{B} - 2C$ and the {dynamics' linear parameterization \eqref{eq:linear parameterization (WAFR)}}, we obtain
				\begin{align*}
				\dot{V} =& -e_{tr}^\top K_{tr} e_{tr} -
				\eta_r^\top K_r \eta_r + \frac{1}{\underline{k}_v} e_{v_q}^\top(u + e_x - Y_\alpha \theta - d) + e_\theta^\top \Gamma^{-1} \dot{\hat{\theta}},
				\end{align*}
				and by substituting $\tau$ and $\dot{\hat{\theta}}$, as well as using $\|d(\cdot)\| \leq \bar{d}$, 
				\begin{align*}
				\dot{V} \leq & -e_{tr}^\top K_{tr} e_{tr} - 
				\eta_r^\top K_r \eta_r - \frac{1}{\underline{k}_v}e_{v_q}^\top K_v e_{v_q} + \frac{1}{\underline{k}_v}e_{v_q}^\top Y_\alpha e_\theta + \frac{1}{\underline{k}_v}\|e_{v_q}\| \bar{d} - \\
				&\frac{1}{\underline{k}_v} e_\theta^\top Y_\alpha^\top e_{v_q}   - \sigma_\theta e_\theta^\top \hat{\theta} \\
				\leq & -e_{tr}^\top K_{tr} e_{tr} -
				\eta_r^\top K_r \eta_r - \frac{1}{\underline{k}_v}e_{v_q}^\top K_v e_{v_q} + \frac{1}{\underline{k}_v}\|e_{v_q}\| \bar{d} - \sigma_\theta e_\theta^\top \hat{\theta} \\
				\leq & -\underline{k}_{tr} \|e_{tr}\|^2 -
				\underline{k}_r \|\eta_r\|^2 -\|e_{v_q}\|^2 + \frac{1}{\underline{k}_v}\|e_{v_q}\| \bar{d} - \sigma_\theta \|e_\theta\|^2 - \sigma_\theta e_\theta^\top \theta.
				\end{align*}
				Next, by using the identity $\alpha \beta \leq \frac{1}{2}\alpha^2 + \frac{1}{2}\beta^2$, $\forall \alpha, \beta \in \mathbb{R}$, we obtain 
				\begin{align*}
				\dot{V} \leq& -\underline{k}_{tr} \|e_{tr}\|^2 - \underline{k}_r\|\eta_r\|^2 -  \frac{1}{2} \|e_{v_q}\|^2  - \frac{\sigma_\theta}{2} \|e_\theta\|^2  +  \frac{\bar{d}^2}{2\underline{k}_v^2} + \frac{\sigma_\theta}{2}\|\theta\|^2 \\
				\leq & -\underline{k}_{tr} \|e_{tr}\|^2 - \underline{k}_r\|\eta_r\|^2 - \frac{1}{2} \|e_{v_q}\|^2  - \frac{\sigma_\theta}{2} \|e_\theta\|^2  + d_x.
				\end{align*}
				{Therefore, $\dot{V}$ is negative when $\|e_{tr}\| \geq \sqrt{\frac{d_x}{\underline{k}_{tr}}}$, or $\|\eta_r\| \geq \sqrt{\frac{d_x}{\underline{k}_r}}$, or $\|e_{v_q}\| \geq \sqrt{2d_x}$, or $\|e_{\theta}\| \geq \sqrt{\frac{2d_x}{\sigma_\theta}}$ and hence we conclude that there exists a finite $T$ such that the state is ultimately bounded as 
					\begin{align*}
					x_R(t) \in \Omega_x \coloneqq \Bigg\{x_R \in \mathbb{R}^{2n+\ell}: \  & \|e_{tr}(t)\| \leq \sqrt{\frac{d_x}{\underline{k}_{tr}}}, \ \  \|\eta_r(t)\| \leq \sqrt{\frac{d_x}{\underline{k}_r}},  \\
					& \hspace{10mm} \|e_{v_q}(t)\| \leq \sqrt{2d_x}, \ \ \|e_{\theta}(t)\| \leq \sqrt{\frac{2d_x}{\sigma_\theta}} \Bigg\}
					\end{align*}
					for all $t \geq T$}.	
				Since, outside $\Omega_x$ it holds that $\dot{V} < 0$, we obtain that $x \notin \Omega_x \Rightarrow V(x(t)) \leq V_0 \coloneqq V(x(0))$, $\forall t\geq 0$, i.e., $\|e_{tr}(t)\| \leq 2V_0$, $\|\eta_r(t)\| \leq V_0$, $\forall t\geq 0$. Therefore, we conclude that $\|e_{tr}(t)\| \leq \max\{ 2V_0, \sqrt{d_x/\underline{k}_{tr}} \} $, $\|\eta_{tr}(t)\| \leq \max\{ 2V_0, \sqrt{d_x/\underline{k}_{r}} \}$, $\forall t\geq 0$. Finally, {since $V(t)$ remains bounded $\forall t\in[0,t_f]$}, we conclude that all the closed loop signals remain bounded and $\cos(e_{r_j}(t)) \neq \pi$, $\forall t\in[0,t_f]$, $j\in\{1,\dots,n_r\}$. 			
			\end{proof}

			Note that the disturbance term $d(\cdot)$ prohibits the system from achieving asymptotic convergence, i.e., $\lim_{t\to\infty}(q(t) - q_\text{d}(t)) = 0$. Nevertheless, Theorem \ref{th:control theorem (WAFR)} establishes a funnel around the 
			desired trajectory $q_\text{d}$ where the state $q(t)$ will evolve in. This funnel will be used as clearance in the motion planner of the subsequent section to derive a collision-free path to the goal region. 			
			Note however, that this funnel cannot be accurately known by the user/designer, 
			since $V_0$ cannot be accurately known (the terms $B(q(0))$ and $e_\theta(0)$ contain the unknown terms $\theta$). Lower and upper bounds of $\theta$ can be obtained, however, since these involve mass and moments of inertia, which can be estimated by the geometry and the material of the links/motors. Hence, one can obtain an upper bound on $V_0$. {On the same note, a conservative estimate of $d_x$, appearing in \eqref{eq:theorem bounds (WAFR)}, can be obtained by estimating an upper bound of $d(\cdot)$ (e.g., by testing suitable trajectories on the robot) and using the aforementioned upper bound of $\theta$. Therefore, we can obtain an overestimate of the bounds in \eqref{eq:theorem bounds (WAFR)}, which will be used in the motion planner of the next section. These bounds can be tightened by appropriate tuning of the gain constants, as elaborated in the next remark. }

			\begin{remark} \label{rem:gains and free space (WAFR)}
				The collision-free geometric trajectory $q_\text{d}$ of the motion planner will connect the initial condition $q(0)$ to the goal and hence it is reasonable to enforce $q_\textup{d}(0) = q(0)$. By also reasonably assuming that $\dot{q}(0) = 0$,  $V_0$ from Theorem \ref{th:control theorem (WAFR)} becomes $V_0 = \frac{1}{2\underline{k}_v}\dot{q}_\textup{d}(0) B(q(0))\dot{q}_\textup{d}(0) + \frac{1}{2}e_\theta(0)^\top \Gamma^{-1} e_\theta(0)$, which can be rendered arbitrarily small by choosing large values for the control gains $\underline{k}_v$ and $\Gamma$. {In} the same vein, choosing large values for $\underline{k}_v$, $\underline{k}_{tr}$, and $\underline{k}_r$ shrinks the constants $\sqrt{\frac{d_x}{\underline{k}_{tr}}}$ and $\sqrt{\frac{d_x}{\underline{k}_r}}$, respectively. Therefore, the size of the funnel dictated by \eqref{eq:theorem bounds (WAFR)}
				can become smaller by appropriate gain tuning. This will lead to 
				less conservative solutions for the motion planner of the next section, as will be clarified in the next subsection. Nevertheless, it should be noted that too large gains might result in excessive control inputs that cannot be realized by the actuators in realistic systems. {Finally, note that the incorporation of $\underline{k}_v$ in the adaptation law \eqref{eq:adaptation law (WAFR)}, which is not common in standard adaptive control techniques, has been included to create an extra degree of freedom for reducing the value of $V_0$. This, along with the tracking using the chordal metric $d_C$ for the rotation part, constitute the differences of the proposed control scheme with respect to  standard adaptive control for uncertain robotic systems.}
			\end{remark}

			\subsection{Motion Planner} \label{sec:motion planner (WAFR)}
			
			We describe here the construction of the sampling-based motion planner, {referred to as {Bounded-RRT or B-RRT}}, that {drives} the robot from an initial {state} to the goal, which follows similar steps as the standard geometric RRT algorithm. 
			{Before presenting the algorithm, we define the extended-free space, which will be used to integrate the results from the feedback control of the previous subsection. In order to do that, we define first the open polyhedron as 
				\begin{equation}\label{eq:polyhedron def (WAFR)}
				\mathcal{P}(q,\delta) \coloneqq \{ y = [y_{tr}^\top,y_r^\top]^\top\in\mathbb{T} : \|y_{tr} - q_{tr}\| < \delta_{tr}, \|\mathsf{H}(y_r,q_r)\| < \delta_r \},
				\end{equation}
				for $q = [q_{tr}^\top,q_r^\top]^\top\in\mathbb{T}$ and $\delta=(\delta_{tr},\delta_r)\in\mathbb{R}^2$,
				where $\mathsf{H}(\cdot)$ is the metric introduced in \eqref{eq:H metric (WAFR)}. We define now the $\delta$-extended free space $\bar{\mathcal{A}}_{\text{free}}(\delta) \coloneqq \{q\in\mathbb{T}:\bar{\mathcal{A}}(q,\delta) \cap \mathcal{O} = \emptyset \}$, where $\bar{\mathcal{A}}(q,\delta) \coloneqq \bigcup_{x\in \mathcal{P}(q,\delta)} \mathcal{A}(x)$.
				Note that $\bar{\mathcal{A}}_{\text{free}}((\delta_{1_{tr}},\delta_{1_r})) \subseteq \bar{\mathcal{A}}_{\text{free}}((\delta_{2_{tr}},\delta_{2_r}))$ if $\delta_{1_{tr}} \geq \delta_{2_{tr}}$ and/or $\delta_{1_r} \geq \delta_{2_r}$. }
			
			{\begin{remark} \label{rem:path clearance (WAFR)}
					Since ${\mathcal{A}}_{\text{free}}$ is open, there exist positive constants $\delta_{tr}$, $\delta_r$ such that $Q_g \subset \bar{\mathcal{A}}_{\text{free}}((\delta_{tr},\delta_r))$ and the feasible path $\boldsymbol{q}_\text{p}$ from  Assumption \ref{ass:path ass (WAFR)} satisfies $\boldsymbol{q}_\text{p}(\nu) \in \bar{\mathcal{A}}_{\text{free}}((\delta_{tr},\delta_r))$, $\forall \nu\in[0,\sigma]$.
				\end{remark} } 
				
				{The control scheme of the previous subsection guarantees that the robot can track a trajectory within the bounds \eqref{eq:theorem bounds (WAFR)}. In other words, given a desired trajectory signal $q_\text{d}:[t_0,t_f]\to\mathbb{T}$, the control algorithm 
				\eqref{eq:control law (WAFR)} - \eqref{eq:adaptation law (WAFR)} guarantees that $q(t) \in \bar{\mathcal{A}}_\text{free}((\bar{e}_{tr},\bar{\eta}_r))$, $\forall t\in[t_0,t_f]$, with $\bar{e}_{tr}$, $\bar{\eta}_r$ as defined in \eqref{eq:theorem bounds (WAFR)}.
					Hence, the motion planner developed here takes that into account by producing trajectories that belong to the extended free space $\bar{\mathcal{A}}_\text{free}((\bar{e}_{tr},\bar{\eta}_r))$\footnote{{We keep the same notation $(\bar{e}_{tr},\bar{\eta}_r)$, although only upper bounds of these values can be actually estimated and hence used by the planner}}. }
				The respective algorithm is presented in Algorithm \ref{alg:main tree adaptive (WAFR)}. It  is a variant of the standard RRT algorithm. {The main difference,
					which constitutes the key point of the algorithm, is the procedure that aims to find a collision-free trajectory from a node on the tree towards the sampled point. In particular, the sampling of new nodes-points as well as the collision checker of the path between two nodes are carried out with respect to the extended free space $\bar{\mathcal{A}}_\text{free}((\bar{e}_{tr},\bar{\eta}_r))$.}
				Moreover, the motion planner does not need to integrate the system dynamics \eqref{eq:dynamics (WAFR)} with sampled input values in order {to} design a feasible and collision-free robot trajectory. Instead, we use the established evolution funnel to design a collision-free trajectory for the robot, without involving the dynamics. {This implies that the motion planner is purely geometrical.}
				
				\begin{algorithm}
					\caption{{B}-RRT}\label{alg:main tree adaptive (WAFR)}
					\begin{algorithmic}[1]
						\Procedure{TREE}{}					
						\State $\mathcal{V} \leftarrow \{q(0)\}$; $\mathcal{E} \leftarrow \emptyset$; $i \leftarrow 0$ 
						\While {$i < N_s$} 
						\State $\mathcal{G} \leftarrow (\mathcal{V},\mathcal{E})$;
						\State $q_\text{rand} \leftarrow \mathsf{Sample}(i)$; $i \leftarrow i+1$;			
						\State $q_\text{nearest} \leftarrow \mathsf{Nearest}(\mathcal{G},q_\text{rand})$;					
						\State $q_\text{new} \leftarrow \mathsf{Steer}(q_\text{nearest},q_\text{rand})$;
						\If {$\mathsf{ObstacleFree}(q_\text{nearest},q_\text{new})$}
						\State $V \leftarrow V\cup \{q_\text{new}\}$;	$\mathcal{E} \leftarrow \mathcal{E} \cup\{(q_\text{nearest},q_\text{new})\}$;						
						\EndIf
						\EndWhile
						\EndProcedure
					\end{algorithmic}
				\end{algorithm}
				
				The functions that appear in Algorithm \ref{alg:main tree adaptive (WAFR)} are the following:
				\begin{itemize}
					\item $\mathsf{Sample}(i)$: Samples $q_\text{rand}$ from a uniform distribution in the extended free space $\bar{\mathcal{A}}_{\text{free}}((\bar{e}_{tr},\bar{\eta}_r))$, where $\bar{e}_{tr},\bar{\eta}_r$ are the constants {from \eqref{eq:theorem bounds (WAFR)} that define the funnel polyhedron $\mathcal{P}(q_\text{d}(t),(\bar{e}_{tr},\bar{\eta}_r))$ (see \eqref{eq:polyhedron def (WAFR)}) around a reference trajectory $q_\text{d}(t)$} that  $q(t)$ can evolve in.
					
					\item $\mathsf{Nearest}(\mathcal{G},q)$: Finds the node $q_\text{nearest}$ in the tree such that $d_\mathbb{T}(q_\text{nearest},q)$ $=$ $\min_{z\in\mathcal{V}}d_\mathbb{T}(z,q).$		
					\item $\mathsf{Steer}(q,z)$: Computes a point $q_\text{new}$ lying on the straight line from $z$ to $q$ such that $d_\mathbb{T}(q,q_\text{new}) = \epsilon$, where $\epsilon$ is a {tuning constant that represents the incremental distance from $q$ to $q_\text{new}$}.
					\item $\mathsf{ObstacleFree}(q,z)$: Checks whether the path {$X_\text{Line}:[0,\sigma]\to \mathbb{T}$, for some positive $\sigma$,} from $q$ to $z$ is collision free with respect to the extended free space, i.e., check whether $q' \in \bar{\mathcal{A}}_{\text{free}}((\bar{e}_{tr},\bar{\eta}_r))$, $\forall q' \in X_\text{Line}$.					
				\end{itemize}

				{The difference hence of B-RRT with respect to the standard RRT algorithm is the use of the extended free space $\bar{\mathcal{A}}_\textup{free}((\bar{e}_{tr},\bar{\eta}_r))$ in the procedures of sampling new points (function $\mathsf{Sample}$) and checking collisions of the path between two nodes (function $\mathsf{ObstacleFree}$). {As stated before}, this stems from the control design of the previous section, which guarantees that the robot trajectory will evolve in $\bar{\mathcal{A}}_\textup{free}((\bar{e}_{tr},\bar{\eta}_r))$ with respect to a desired trajectory $q_\text{d}$.}
				
				{We briefly describe now the B-RRT algorithm. The tree $\mathcal{G}=(\mathcal{V},\mathcal{E})$ to be constructed is initialized in line 2, with the node set $\mathcal{V}$ initialized to the system's initial configuration $q(0)$, and the respective edge set $\mathcal{E}$ to the empty set.}
				The algorithm samples then a point $q_\text{rand}$ in the extended free space $\bar{\mathcal{A}}_{\text{free}}((\bar{e}_{tr},\bar{\eta}_r))$. 
				Then the nearest neighbor $q_\text{nearest}$, in terms of $d_\mathbb{T}$, is found in the tree (line 7), and a new point $q_\text{new}$ on the line between $q_\text{nearest}$ and $q_\text{rand}$ is computed; $q_\text{new}$ can be chosen such that $d_\mathbb{T}(q_\text{nearest},q_\text{new}) = \epsilon$, according to a predefined incremental distance $\epsilon$ (\cite{kuffner2000rrt}). If the line segment between $q_\text{rand}$ and $q_\text{new}$ belongs to the extended free space $\bar{\mathcal{A}}_{\text{free}}((\bar{e}_{tr},\bar{\eta}_r))$, then the respective {node $q_\text{new}$ and edge $\{q_\text{nearest},q_\text{new}\}$} are added to the tree (lines 9-11).
				After the execution of the algorithm, a standard search algorithm can be employed to find the sequence of edges that lead from $q(0)$ to $Q_g$.
				Moreover, note that the bounds in \eqref{eq:theorem bounds (WAFR)} concern 
				(at least twice) continuously differentiable trajectories. Therefore, the resulting solution path, which is formed by the concatenation of the respective edges in the tree, has to be converted to a such a trajectory. This procedure 
				might modify the initial path that was checked for collisions, and hence the smooth version should be re-checked for collisions in  $\bar{\mathcal{A}}_{\text{free}}((\bar{e}_{tr},\bar{\eta}_r))$. 			
				Subsequently, the resulting smooth (at least twice cont. different.) path is endowed with time constraints to derive a timed trajectory $q_\text{d}:[0,{t_f}] \to \bar{\mathcal{A}}_{\text{free}}((\bar{e}_{tr},\bar{\eta}_r))$, for some ${t_f} >0$, which is given as the desired trajectory input to the control protocol designed in the previous section.  The actual trajectory of the system $q(t)$ is guaranteed to track $q_\text{d}(t)$ in the funnel defined by $\bar{e}_{tr}, \bar{\eta}_r$. Since these bounds are taken into account in the design of the trajectory $q_\text{d}$ by Algorithm \ref{alg:main tree adaptive (WAFR)}, the system will remain collision free. {Note also that $t_f$ and hence the velocity of the formed trajectory $q_\text{d}$ is chosen by the user. Therefore, the robot can execute the respective path in a predefined time interval.}
				
				The probabilistic completeness of the algorithm is stated in the next theorem.
				
				\begin{theorem} \label{th:motion planner (WAFR)}
					Under Assumption \ref{ass:path ass (WAFR)} and for sufficiently high gains $\underline{k}_v$, $\underline{k}_{tr}$, $\underline{k}_r$, as introduced in eq. \eqref{eq:des velocity (WAFR)}, \eqref{eq:control law (WAFR)} and \eqref{eq:adaptation law (WAFR)}, Algorithm \ref{alg:main tree adaptive (WAFR)} is probabilistically complete.
				\end{theorem}
				
				\begin{proof}
						Assumption \ref{ass:path ass (WAFR)} and Remark \ref{rem:path clearance (WAFR)} imply that there exist positive $\delta_{tr}$ and $\delta_r$ and (at least) one twice differentiable path $\boldsymbol{q}_\textup{p}:[0,\sigma] \to \bar{\mathcal{A}}_{\text{free}}((\delta_{tr},\delta_r))$ connecting $q_0$ and $Q_g$. As stated in Remark \ref{rem:gains and free space (WAFR)}, by increasing the values of the control gains $\underline{k}_v$, $\underline{k}_{tr}$, $\underline{k}_r$, {one} can decrease the constants $\bar{e}_{tr}$, $\bar{\eta}_{tr}$ from eq. \eqref{eq:theorem bounds (WAFR)} such that $\bar{e}_{tr} < \delta_{tr}$, $\bar{\eta}_r < \delta_r$. Hence, $Q_g$ satisfies $Q_g \subset \bar{\mathcal{A}}_{\text{free}}((\delta_{tr},\delta_r)) \subset \bar{\mathcal{A}}_{\text{free}}((\bar{e}_{tr},\bar{\eta}_r))$ and the feasible path satisfies $\boldsymbol{q}_\textup{p}(\nu) \in  \bar{\mathcal{A}}_{\text{free}}((\delta_{tr},\delta_r)) \subset \bar{\mathcal{A}}_{\text{free}}((\bar{e}_{tr},\bar{\eta}_r))$, $\forall \nu \in [0,\sigma]$, which guarantees the feasibility of Algorithm \ref{alg:main tree adaptive (WAFR)}. 
						
						Next, by following similar arguments with Lemma 2 of \cite{kuffner2000rrt}, one can prove that for any $q\in\bar{\mathcal{A}}_\textup{free}((\bar{e}_{tr},\bar{\eta}_r))$ and $\epsilon >0$, it holds that $\lim_{i\to \infty} \mathbb{P}( D_{i,q} < \epsilon)$, where $D_{i,q}$ is the random variable associated with the minimum distance of the tree $\mathcal{G}$ to the point $q$ (in terms of $d_\mathbb{T}$) after iteration $i$, and $\mathbb{P}$ denotes the probability. Hence, the vertices $\mathcal{V}$ of $\mathcal{G}$ converge to the sampling distribution in $\bar{\mathcal{A}}_\textup{free}((\bar{e}_{tr},\bar{\eta}_r))$, which is assumed to be uniform.
						{Therefore, since $\boldsymbol{q}_\text{p}$ lies in $\bar{\mathcal{A}}_\textup{free}((\bar{e}_{tr},\bar{\eta}_r))$, a subset of $\mathcal{V}$  converges to it and the proof follows.}	
				\end{proof}
				
				\subsection{Collision Checking in $\bar{\mathcal{A}}_\text{free}(\bar{e}_{tr},\bar{\eta}_r)$} \label{sec:ext col check (WAFR)}
				
				Collision checking for points $q \in \mathbb{T}$ in variants of the standard RRT algorithm 
				is performed by checking whether $q$ belongs to $\mathcal{A}_\text{free}$ or not. For a line segment $X_\text{Line}$ {connecting two nodes of $\mathcal{V}$}, the latter is usually discretized into a finite set of points, which are checked separately for collision {(e.g., \cite{kavraki1996probabilistic,karaman2010incremental})}. Another approach is to consider an over-approximation of the convex hull of the points that form $X_\text{Line}$ (\cite{schulman2014motion}). In our case, however, we are interested in checking for collisions in the \textit{extended} free space $\bar{\mathcal{A}}_\text{free}(\bar{e}_{tr},\bar{\eta}_r)$. 
				Recall that the proposed feedback control scheme guarantees that $q(t) \in \mathcal{P}( q_\text{d}(t), (\bar{e}_{tr},\bar{\eta}_{r}))$ for any trajectory $q_\text{d}(t)$, formed by the {several line segments $X_\text{Line}$ that connect the nodes in $\mathcal{V}$ sampled in {Algorithm} \ref{alg:main tree adaptive (WAFR)}}. 
				Therefore, checking whether the points $q_s \in X_\text{Line}$ belong to $\mathcal{A}_\text{free}$ is not sufficient. That is, for each such point $q_s \in X_\text{Line}$, one must check whether $z \in \mathcal{A}_\text{free}$, $\forall z \in \mathcal{P}( q_s, (\bar{e}_{tr},\bar{\eta}_{r}))$, which is equivalent to checking if $q_s \in \bar{\mathcal{A}}_\text{free}(\bar{e}_{tr},\bar{\eta}_r)$. There are two procedures that one can use for that. Firstly, for each $q_s$, a finite number of points $z$ can be sampled from a uniform distribution in $\mathcal{P}( q_s, (\bar{e}_{tr},\bar{\eta}_{r}))$ and separately checked for collision. Then, for a sufficiently high number of such samples, and assuming a certain {``fat"-structure of the workspace obstacles (e.g., there are no long and skinny obstacles such as wires, cables and tree branches, etc., see (\cite{van1993complexity}) for more details)}, this approach can be considered to be complete, i.e., the resulting path will belong to the extended free space $\mathcal{A}_\text{free}$.  
				Secondly, we calculate the limit poses of each link of the robot, based on the lower and upper bounds by the joints that affect it, as defined by $(\bar{e}_{tr},\bar{\eta}_r)$. Subsequently, we compute the convex hull of these limit poses, which is expanded by an appropriate constant to yield an over-approximation of the swept volume of the potential motion of the link, as described in \cite{schulman2014motion}. The resulting shape is then checked for collisions for each link separately.

				\subsection{{Experimental} Results}
\begin{figure}[!ht]
					\centering
					{\includegraphics[width = 0.46\textwidth]{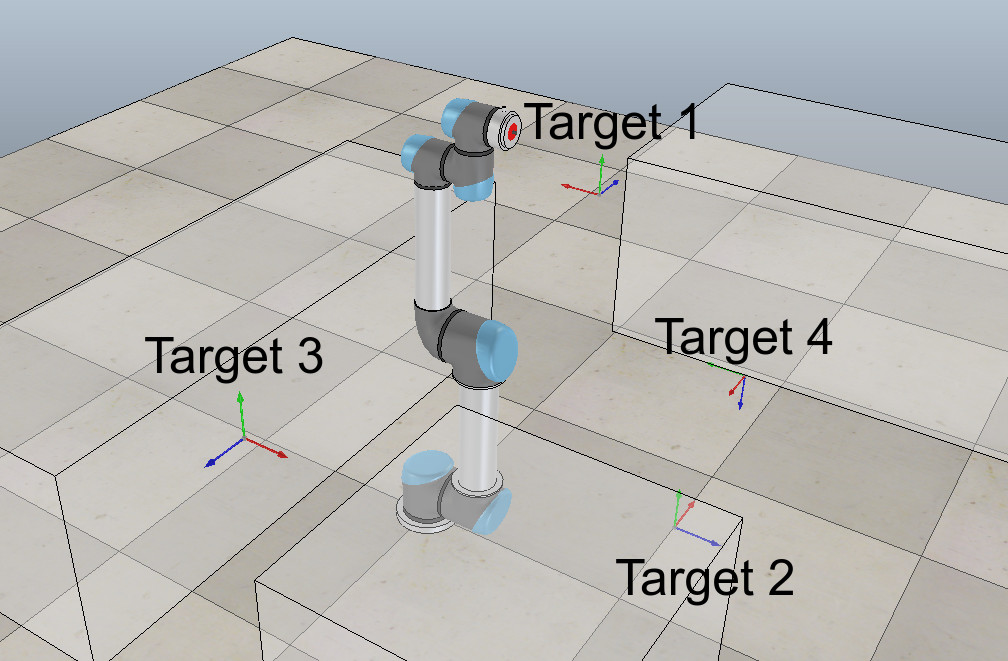}}				
					\caption{A UR5 robotic arm in an obstacle-cluttered environment with $4$ targets.} \label{fig:vrep_initial (WAFR)}
				\end{figure}

				This section presents {experimental} results for a UR5 robot, which consists of $6$ rotational degrees of freedom (see Fig. \ref{fig:vrep_initial (WAFR)}), using the V-REP environment (\cite{Vrep}).  We assume that the first joint is free to move on the unit circle, i.e., $q_{r_1} \in [0,2\pi)$, whereas the rest of the joints are restricted to $[-\pi,\pi]$ to avoid problematic configurations. 
				We consider that the robot end-effector has to sequentially navigate from its initial configuration $q_0 = [0,0,0,0,0,0]^\top$ rad to the following four target points (depicted in Fig. \ref{fig:vrep_initial (WAFR)}): 
				\begin{itemize}
					\item Target 1: $T_1 =[-0.15,-0.475,0.675]^\top$ m and orientation $[\frac{\pi}{2},0,0]^\top$ rad, which yields the configuration $q_1 = [-0.07, -1.05, 0.45, 2.3, 1.37, -1.33]^\top$ rad.
					\item Target 2: $T_2 = [-0.6,0,2.5]^\top$ m and orientation $[0,-\frac{\pi}{2},-\frac{\pi}{2}]^\top$ rad, which yields the configuration $q_2 = 
					[1.28, 0.35, 1.75, 0.03, 0.1, -1.22]^\top$ rad.
					\item Target 3: $T_3 = [-0.025,0.595,0.6]^\top$ m and orientation $[-\frac{\pi}{2},0,\pi]^\top$ rad, which yields the configuration $q_3 = [-0.08$, $0.85$, $-0.23$, $2.58$, $2.09$, $-2,36]^\top$ rad.
					\item Target 4: $T_4 = [-0.525,-0.55,0.28]^\top$ m and orientation $[\pi,0,-\frac{\pi}{2}]^\top$ rad, which yields the configuration $q_4 = [-0.7$, $-0.76$, $-1.05$, $-0.05$, $-3.08$, $2.37]\top$ rad.
				\end{itemize} 
				
\begin{figure}[!ht]
					\centering
					{\includegraphics[width = 0.45\textwidth]{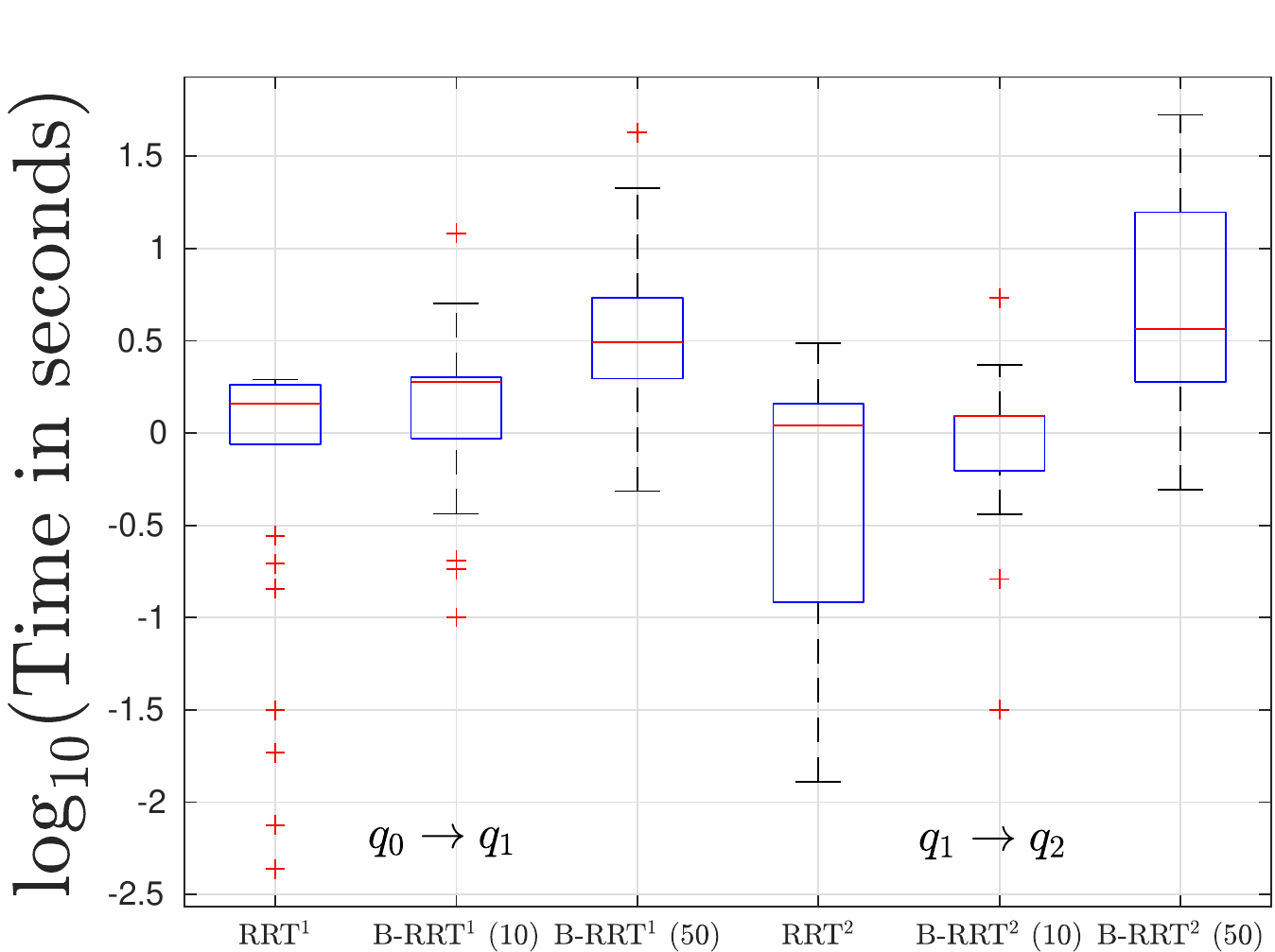}}
					{\includegraphics[width = 0.45\textwidth]{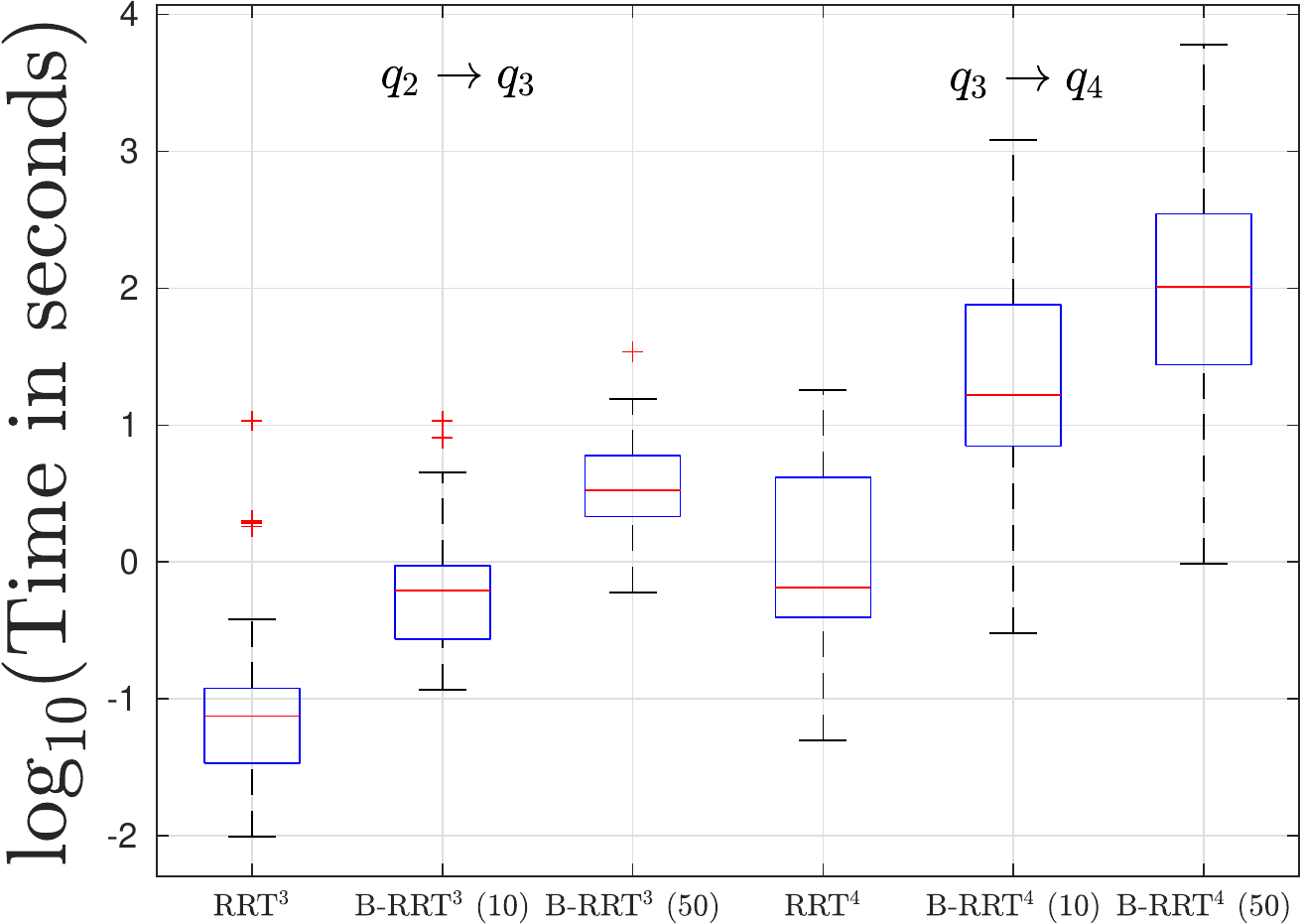}}
					{\includegraphics[width = 0.45\textwidth]{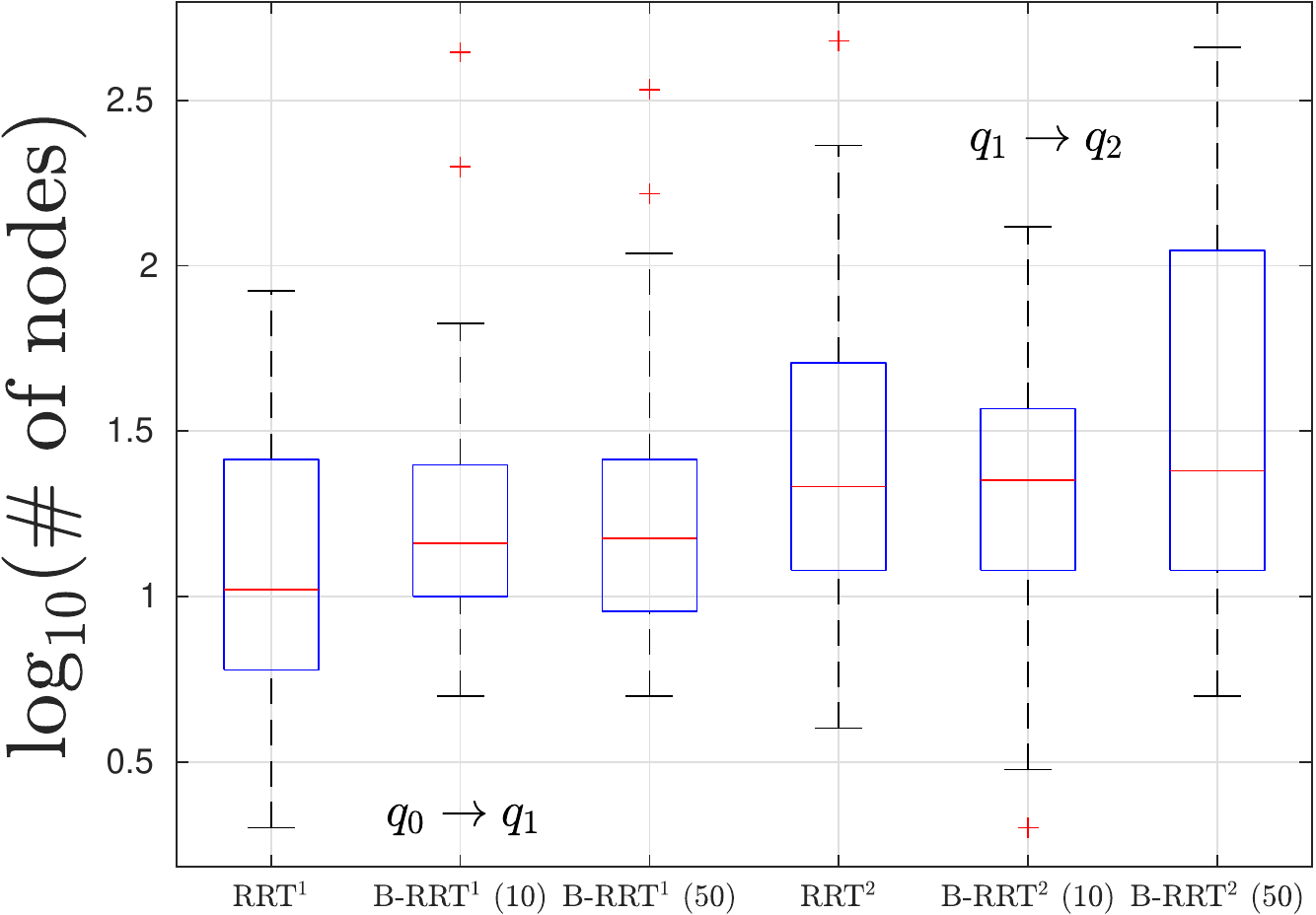}}
					{\includegraphics[width = 0.45\textwidth]{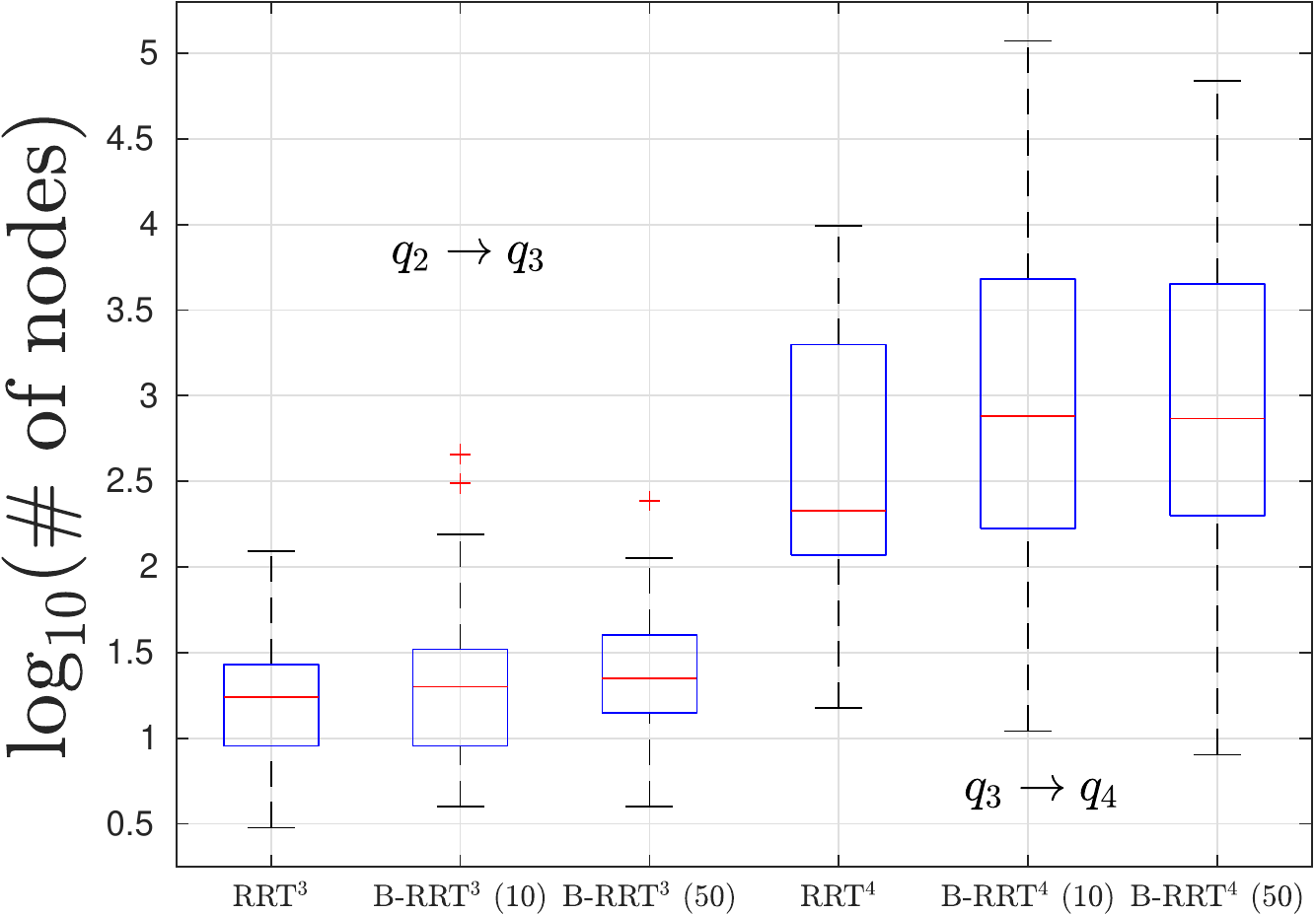}}
					\caption{Box plots showing the execution time (top) and the nodes (bottom) created of the three algorithms (in logarithmic scale) for the four paths (organized in two groups of two (left and right)); ``+' indicate the outliers. } \label{fig:basic_times_nodes (WAFR)}
				\end{figure}

				Regarding the collision checking in $\bar{\mathcal{A}}_\text{free}(\bar{e}_{tr},\bar{\eta}_r)$ of the B-RRT algorithm, {we check a finite number of samples around each point of the resulting trajectory $q_\text{d}$ for collision.}
				We run B-RRT with $10$ and $50$ such samples and we compared the results to a standard {geometric} RRT algorithm in terms of time per number of nodes. The results for {$30$ runs} of the algorithms are given in Fig. \ref{fig:basic_times_nodes (WAFR)} for the four paths, in logarithmic scale. One can notice that the average nodes created do not differ significantly among the different algorithms. As expected, however, B-RRT requires more time than the standard {geometric} RRT algorithm, since it checks the extra samples in $\bar{\mathcal{A}}_\text{free}(\bar{e}_{tr},\bar{\eta}_r)$ for collision. One can also notice that the time increases with the number of samples. However, more samples imply greater coverage of $\bar{\mathcal{A}}_\text{free}(\bar{e}_{tr},\bar{\eta}_r)$ and {hence the respective solutions are more likely to be complete with respect to collisions.}
				
\begin{figure}[!ht]
					\centering
					\subcaptionbox{}
					{\includegraphics[width = 0.425\textwidth,height=0.175\textheight]{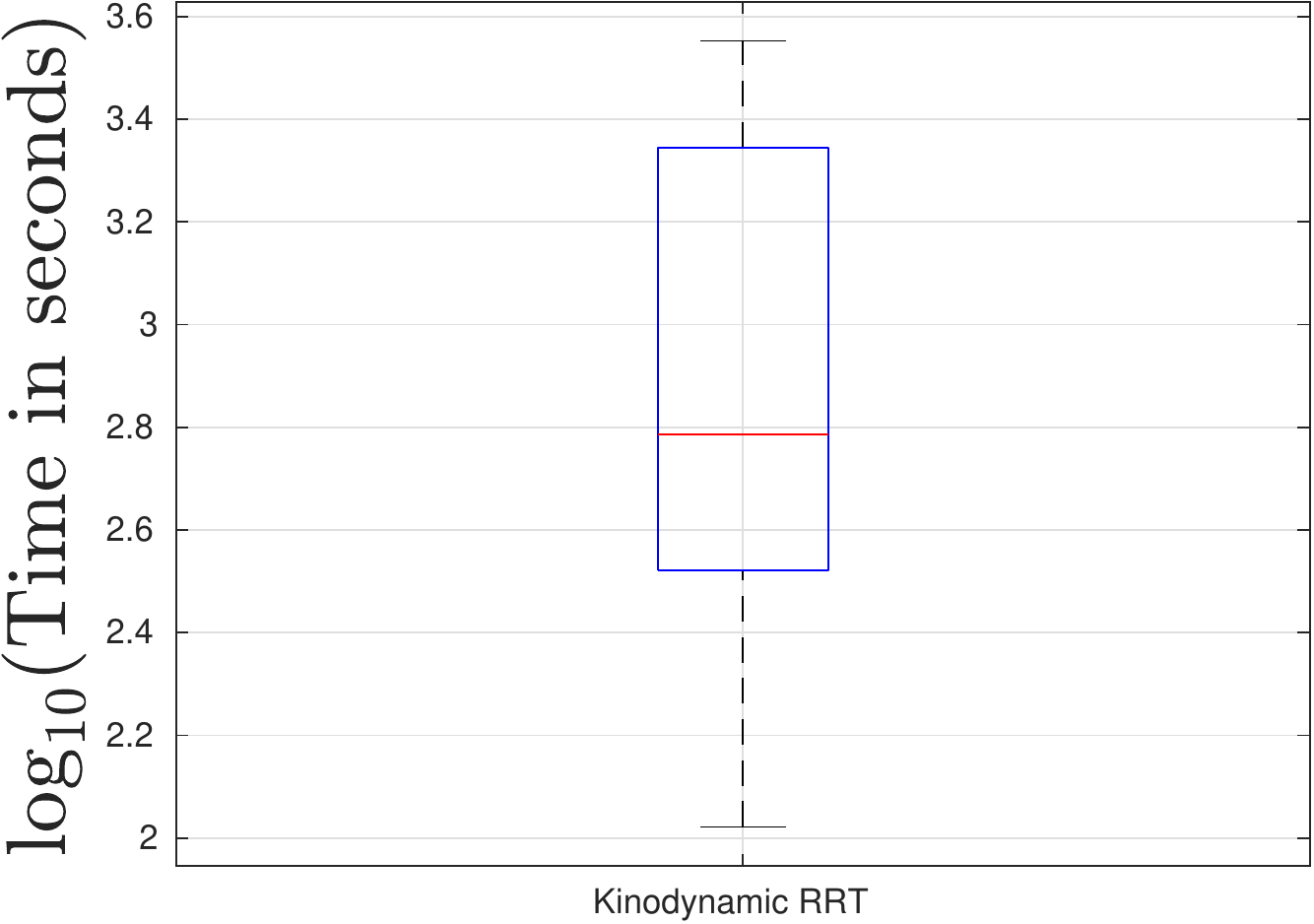}}
					\subcaptionbox{}
					{\includegraphics[width = 0.425\textwidth,height=0.175\textheight]{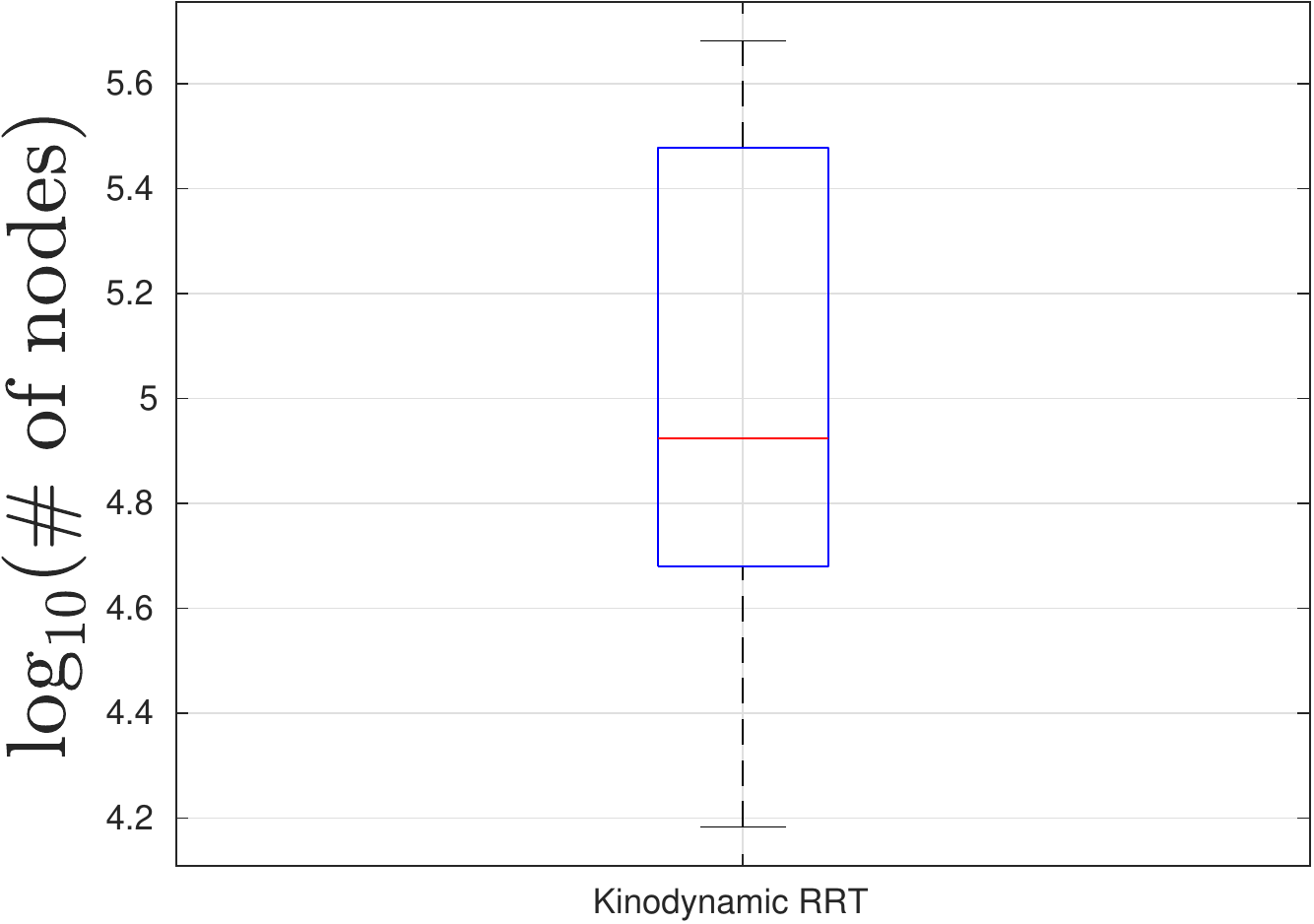}}
					\caption{Box plots showing the execution time (a) and number of nodes (b) created for the kinodynamic RRT in logarithmic scale (for the first two joints and the path $(0,0) \to (-\frac{\pi}{18},\frac{\pi}{4})$). } \label{fig:kinod_time_nodes (WAFR)}
				\end{figure}

				{Since, in contrast to the standard geometric RRT, B-RRT implicitly takes into account the robot dynamics through the  designed tracking control scheme and the respective extended free space $\bar{\mathcal{A}}_\text{free}(\bar{e}_{tr},\bar{\eta}_r)$, we compare the results to a standard kinodynamic RRT algorithm that simulates forward the robot dynamics, assuming {known} dynamical parameters. In particular, we run the algorithm only for the first two joints, with initial and goal configurations at $(0,0)$ and $(-\frac{\pi}{18},\frac{\pi}{4})$ rad, respectively, and keep the other joints fixed at $0$. For the forward simulation of the respective dynamics we choose a sampling step of $10^{-3}$ sec and total simulation time $30$ sec for each constant control input. The termination threshold distance is set to $0.25$ (with respect to the distance $d_\mathbb{T}$), i.e., the algorithm terminates when the forward simulation reached a configuration closer than $0.25$ units to the goal configuration. 	
					{The results for 10 runs of the algorithm are depicted in Fig. \ref{fig:kinod_time_nodes (WAFR)}, which provides the execution time and number of nodes created in logarithmic scale. Note that, even for this simple case (planning for only two joints), the execution time is comparable to the B-RRT case of $50$ samples in the fourth path scenario $q_3 \to q_4$.}
					As pointed out before, this is justified by the fact that the inputs are randomized as well as the complex dynamics of the considered robotic system. Hence, one concludes the necessity of an efficient technique that still takes into the robotic dynamics, which is given by our two-layer framework, combining an appropriately designed RRT planner with an ``intelligent" feedback control algorithm that also compensates for the uncertain dynamics. }
				
\begin{figure}[!ht]
					\centering
					\subcaptionbox{}
					{\includegraphics[width = 0.475\textwidth]{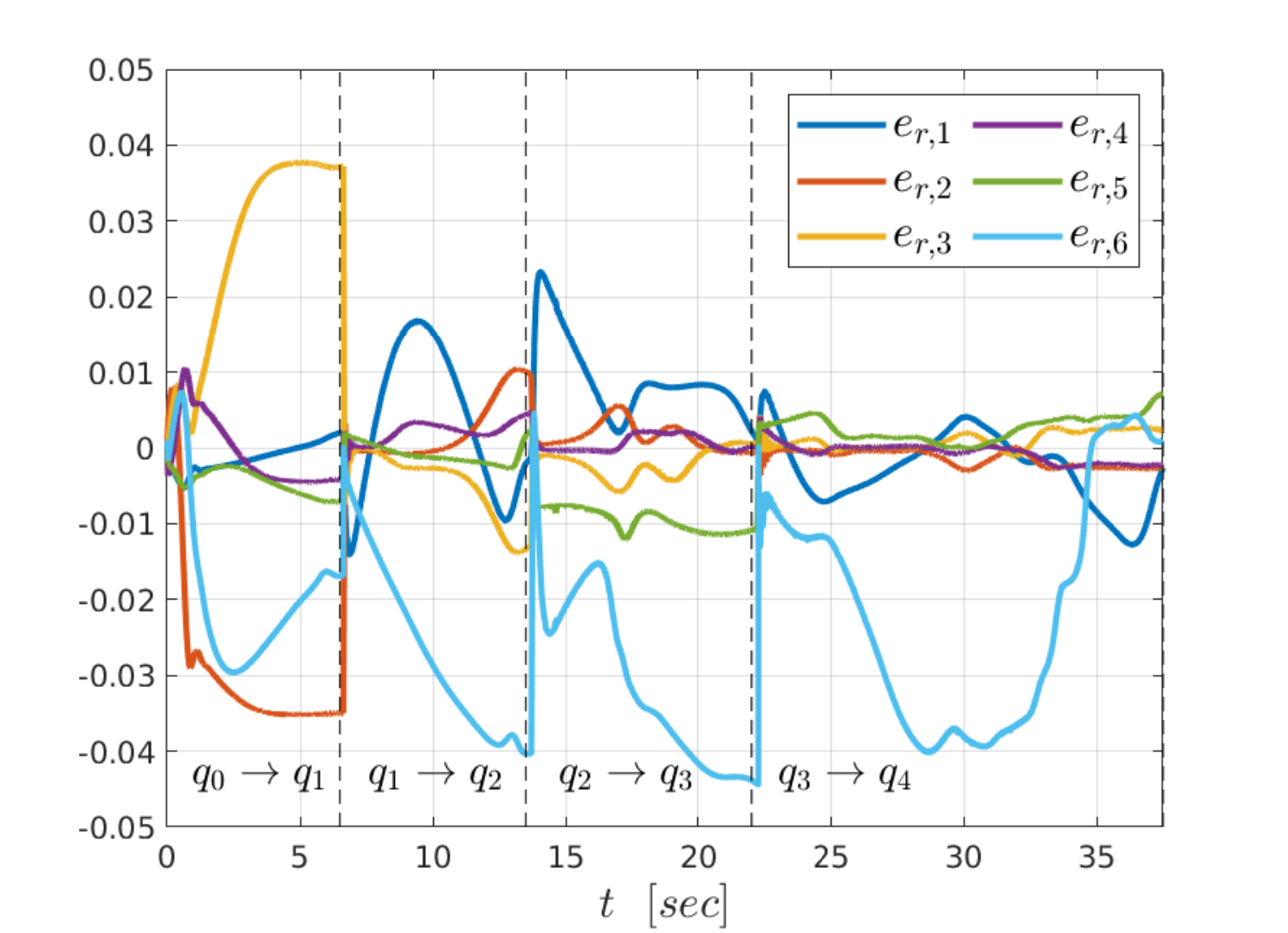}}
					\subcaptionbox{}
					{\includegraphics[width = 0.475\textwidth]{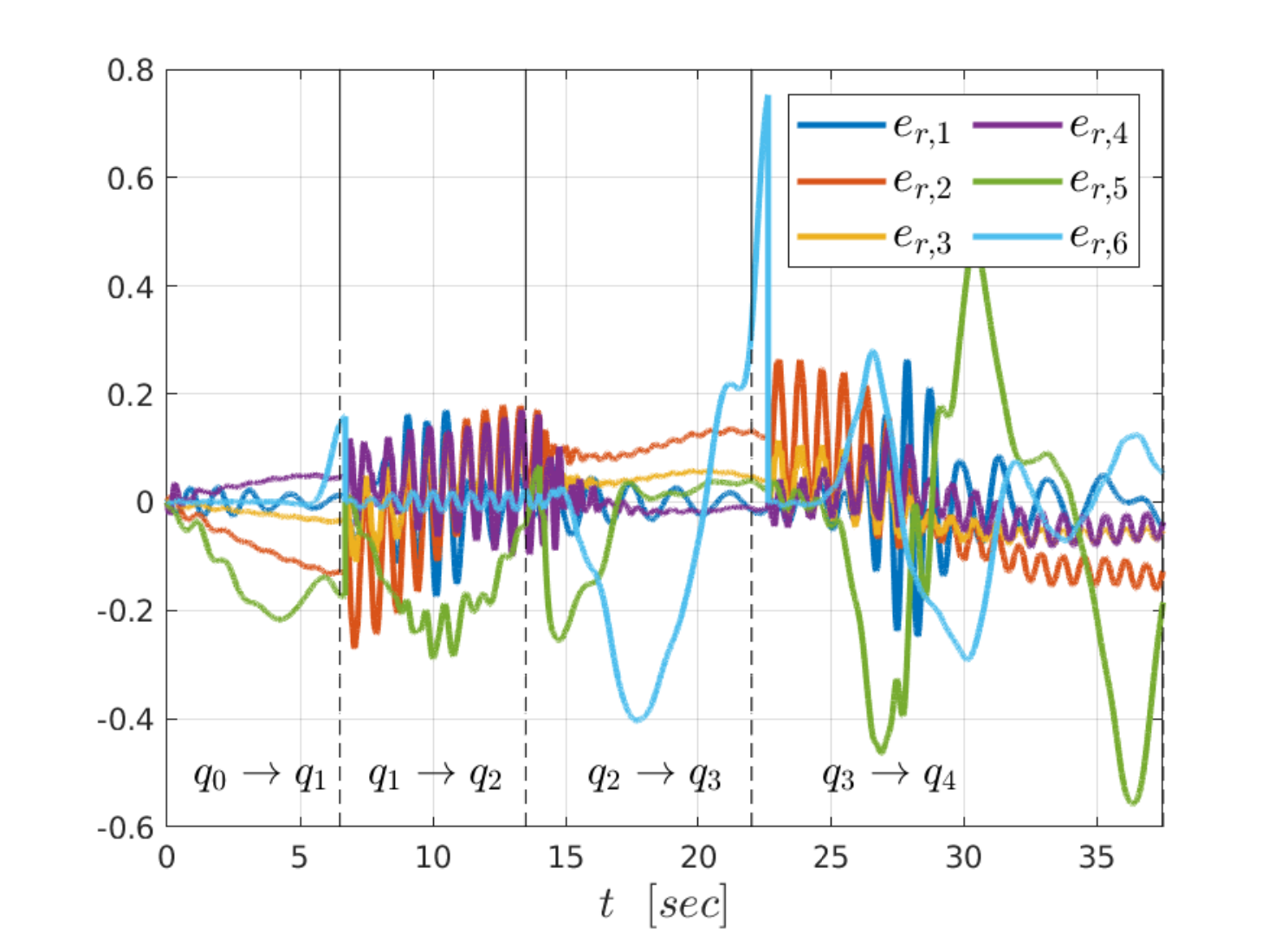}}
					\setlength{\belowcaptionskip}{-10pt}
					\caption{The error values $e_{r_j}(t) = e_{r_j}(t) - q^i_{\text{d},r_j}(t)$ for the adaptive controller (a) and the PID one (b). } \label{fig:errors (WAFR)}
				\end{figure}

				Next, we illustrate the motion of the robot through the four target points via the control design of Section \ref{sec:adaptive controller (WAFR)}. For each sub-path ($q_i \to q_{i+1}$, $\forall i\in\{0,1,2,3\}$) we fit a smooth timed trajectory $q^i_{\text{d}}(t)$, $\forall i\in\{0,\dots,3\}$ on the 
				generated nodes, whose total time duration depends on the distance between successive nodes.
				The estimates of the masses and inertias of the robot links and rotors, composing $\hat{\theta}$, were initialized at $60\%$ of the actual values. Morever, in view of \eqref{eq:theorem bounds (WAFR)}, we aim to impose an upper bound of $0.1$ rad for each $|q_{r_j} - q^i_{\text{d},j}|$, $\forall j \in\{1,\dots,6\}, i\in\{0,\dots,3\}$. To that end, we choose the control gains as $k_{r_1} = \dots = k_{r_6} = 0.005$, $K_v = \text{diag}\{[35,65,45,20,10,0.5]\}$, and $\Gamma = 50\text{diag}\{ \hat{\theta}(0)\}$. 
				The results are depicted in Fig. \ref{fig:errors (WAFR)} (a), which shows the error values $e_{r_j}(t) = e_{r_j}(t) - q^i_{\text{d},r_j}(t)$, $\forall j\in\{1,\dots,6\}$, and all paths $i\in\{0,\dots,3\}$. One can verify that the error values stay always bounded in the region $(-0.1,0.1)$ rad, achieving thus the desired performance. 
				{For comparison purposes, we also simulate a PID controller of the form $$\tau = -K_1 e_x - K_2 ( \dot{q} - \dot{q}_\text{d}) - K_3 \int e_x(\nu) d \nu,$$			
				where $K_1 = \text{diag}\{100,1000,1000,100,1,1\}$, and $K_2 = K_3 = I_6$ are positive definite gain matrices. The errors $e_{r_j}(t) = e_{r_j}(t) - q^i_{\text{d},r_j}(t)$, $\forall j\in\{1,\dots,6\}$, for the four paths $i\in\{0,\dots,3\}$ are shown in Fig. \ref{fig:errors (WAFR)} (b). Note that they exceed the interval $(-0.1,0.1)$, which defined the clearance in the B-RRT algorithm, jeopardizing hence the actual trajectory of the robot. A video illustrating the robot trajectory using the two control laws can be found here: \href{https://youtu.be/y7bCoUoTlPA}{https://youtu.be/y7bCoUoTlPA}.

\section{Asymptotic Tracking of Nonsmooth Feedback Stabilizable Unknown Systems with Prescribed Transient Response} \label{sec:TAC}

Finally, inspired by funnel-based techniques and the PPC methodology \cite{bechlioulis2008robust}, we propose in this section a novel control scheme that achieves asymptotic stability while complying with funnel constraints, for a $2$nd-order control affine uncertain and possibly non-smooth system.

\subsection{Problem Formulation} \label{sec:PF (TAC)}

We consider the asymptotic tracking control problem subject to transient constraints imposed by a predefined funnel. The consider systems are MIMO systems of the form 	
{
	\begin{subequations} \label{eq:system (TAC)}
		\begin{align} 
		\dot{x}_1 &= x_2, \\ 
		\dot{z} &= F_z(x,z,t), \\
		\dot{x}_2 &= F(x,z,t) + G(x,z,t)u, \ \ y = x_1  
		\end{align}
	\end{subequations}
	where $z \in\mathbb{R}^{n_z}$, $x\coloneqq {[x_1^\top,x_2^\top]^\top} \in \mathbb{R}^{2n}$, with $x_j \coloneqq [x_{j_1},\dots,x_{j_n}]^\top \in \mathbb{R}^n$, $\forall {j\in\{1,2\}}$, {are} the system's states, $y\coloneqq [y_1,\dots,y_n]^\top\in\mathbb{R}^n$ is the system's output, which is {required} to track a desired trajectory $y_\text{d}(t)$, and $F:{\mathbb{R}^{2n+n_z}}\times [t_0,\infty) \to \mathbb{R}^n$, $F_z : \mathbb{R}^{2n+n_z}\times[t_0,\infty) \to \mathbb{R}^{n_z} $  $G:{\mathbb{R}^{2n+n_z}} \times [t_0,\infty) \to \mathbb{R}^{n\times n}$ are unknown vector fields, not necessarily continuous everywhere. We assume that $x$ is available for measurement, whereas $z$ is not. In fact, the dynamics governing $z$ is called dynamic uncertainty and represents unmodeled dynamic phenomena that potentially affect the closed{-}loop response. 
	The {assumptions on} the system dynamics are restricted to local essential boundedness and measurability as well as controllability conditions on $G$ and internal stability of $z$, without considering 
	any uniform boundedness/growth condition or model approximation:
	{\begin{assumption} \label{ass:f(x,t) cont + bounded (TAC)}
			The maps $(x,z) \mapsto F(x,z,t) : \mathbb{R}^{2n+n_z}\to\mathbb{R}^n$, $(x,z) \mapsto G(x,z,t) : \mathbb{R}^{2n+n_z}\to\mathbb{R}^n$, $(x,z) \mapsto F_z(x,z,t) : \mathbb{R}^{2n+n_z}\to\mathbb{R}^{n_z}$ are Lebesgue measurable and locally  essentially bounded for each fixed $t\in[t_0,\infty)$, uniformly in $t$, and the maps $t\mapsto F(x,z,t):[t_0,\infty) \to\mathbb{R}^{n}$ and $t\mapsto G(x,z,t):[t_0,\infty) \to\mathbb{R}^{n}$ are Lebesgue measurable and uniformly bounded for each fixed $(x,z)\in \mathbb{R}^{2n+n_z}$, by unknown bounds.   
		\end{assumption}
		\begin{assumption} \label{ass:g(x,t) pd (TAC)}
			The matrix 
			\begin{equation*}
			\widetilde{G}(x,z,t) \coloneqq G(x,z,t) + G(x,z,t)^\top 
			\end{equation*}
			is positive definite, $\forall (x,z,t) \in \mathbb{R}^{2n+n_z} \times [t_0,\infty)$, i.e., $\lambda_\text{min}( \widetilde{G}(x,z,t))$ $>$ $0$, where $\lambda_\text{min}(\widetilde{G}(x,z,t))$ is {its \textit{unknown} minimum eigenvalue}.
		\end{assumption}
		\begin{assumption} \label{ass:internal dynamics (TAC)}	
			There exists a sufficiently smooth function $U_z:\mathbb{R}^{n_z}\to\mathbb{R}_{\geq 0}$ and class $\mathcal{K}_\infty$ functions $\underline{\gamma}_z(\cdot)$, $\bar{\gamma}_z(\cdot)$, $\gamma_z(\cdot)$ such that
			$\underline{\gamma}_z(\|z\|) \leq {U_z}(z) \leq \bar{\gamma}_z(\|z\|)$, and
			\begin{equation*}
			{\left(\frac{\partial U_z}{\partial z}\right)^\top F_z(x,z,t)} \leq -\gamma_z(\|z\|) + \pi_z(x,t), 
			\end{equation*}
			where $x\mapsto \pi_z(x,t) :\mathbb{R}^{2n} \to \mathbb{R}_{\geq 0}$ is continuous and class $\mathcal{K}_\infty$ for each fixed $t\in[t_0,\infty)$, and $t \mapsto \pi_z(x,t):[t_0,\infty)  \to \mathbb{R}_{\geq 0}$ is uniformly bounded for each fixed $x\in \mathbb{R}^{2n}$.
		\end{assumption}
	}
	\begin{assumption} \label{ass:state feedb (TAC)}
		The state $x$ is available for measurement.
	\end{assumption}
	\begin{assumption} \label{ass:y_d (TAC)}
		The desired trajectory and its derivatives are bounded by finite and unknown {constants {$\bar{y}_{\text{d},0}, \bar{y}_{\text{d},1} > 0$, i.e., 
				$\| y_d(t) \| < \bar{y}_{\text{d},0} \leq \bar{y}_\text{d}$, $\| y_d(t) \| < \bar{y}_{\text{d},1} \leq \bar{y}_\text{d}$,	
				${\forall} t\in[t_0,\infty)$, where $\bar{y}_\text{d}\coloneqq \max\{\bar{y}_{\text{d},0},\bar{y}_{\text{d},1}\}$}.}
	\end{assumption}
	Note that Assumption \ref{ass:g(x,t) pd (TAC)} is a sufficient controllability condition and Assumption \ref{ass:internal dynamics (TAC)} suggests that $z$ is input-to-state practically stable with respect to $x,t$ implying stable zero (internal) dynamics {\cite{bechlioulis2011robust}}. 

\subsection{{Problem Solution}} \label{sec:2nd order (TAC)}

The control objective is the \textit{asymptotic} output tracking of a desired bounded trajectory $y_\text{d}\coloneqq [y_{1,\text{d}},\dots,y_{n,\text{d}}]: [t_0,\infty) \to \mathbb{R}^n$, with bounded derivatives, as stated in Assumption \ref{ass:y_d (TAC)}. Moreover, as discussed before, we aim at imposing a certain predefined behavior for the transient response of the system. More specifically, motivated by funnel control techniques  (Appendix \ref{app:PPC}, \cite{ilchmann2007tracking,bechlioulis2010prescribed,berger2018funnel}), given $n$ predefined funnels, described by the smooth functions $\rho_{p_i}: [t_0,\infty)  \to [\underline{\rho}_{p_i},\bar{\rho}_{p_i}] \subset \mathbb{R}_{>0}$, where $\underline{\rho}_{p_i}$, $\bar{\rho}_{p_i} \in \mathbb{R}_{>0}$ are positive lower and upper bounds, respectively, we aim at guaranteeing that\footnote{The analysis can be extended to non-symmetric funnels.} $-\rho_{p_i}(t) > y_i(t) - y_{i,\text{d}}(t) > \rho_{p_i}(t)$, $\forall t\in[t_0,\infty)$, given that $-\rho_{p_i}(t_0) > y_i(t_0) - y_{i,\text{d}}(t_0) > \rho_{p_i}(t_0)$, $\forall i\in\{1,\dots,n\}$. 
These functions can encode maximum overshoot or convergence rate properties.
Note that, compared to the majority of the related works on funnel control (e.g., \cite{bechlioulis2011robust,ilchmann2007tracking,berger2018funnel},\cite{lee2019asymptotic}), we do not require arbitrarily small final values $\lim_{t\to \infty} \rho_{p_i}(t)$, which would achieve convergence of $y_i(t)-y_{i,\text{d}}(t)$ arbitrarily close to zero,
since one of the objectives is actual \textit{asymptotic stability}. 
In this section, the problem statement is as follows: 

\begin{problem} \label{problem:1 (TAC)}
	Consider the system \eqref{eq:system (TAC)} and let a desired trajectory $y_\text{d}:[t_0,\infty)\to\mathbb{R}^n$ as well as $n$ prescribed funnels, described by $\rho_{p_i}:[t_0,\infty)\to [\underline{\rho}_{p_i},\bar{\rho}_{p_i}]$, $\forall i\in\{1,\dots,n\}$. 
	Design a control protocol $u\in\mathbb{R}^n$ such that
	\begin{enumerate}
		\item $\lim_{t\to\infty}(y_i(t) - y_{i,\text{d}}(t)) = 0$, $\forall i\in\{1,\dots,n\}$
		\item $-\rho_{p_i}(t) > y_i(t) - y_{i,\text{d}}(t) > \rho_{p_i}(t)$, $\forall i\in\{1,\dots,n\}$, $t\in[t_0,\infty)$, 
	\end{enumerate} 
	and all closed loop signals remain bounded.
\end{problem}
Our solution to {Problem} \ref{problem:1 (TAC)} is based on 
the PPC error transformation, which converts the constrained error behavior $-\rho_{p_i}(t) > y_i(t) - y_{i,\text{d}}(t) > \rho_{p_i}(t)$ to an unconstrained one. More specifically, we define the errors
\begin{equation} \label{eq:e_p (TAC)}
e_p  \coloneqq \begin{bmatrix}
e_{p_1}, \dots, e_{p_n}
\end{bmatrix}^\top \coloneqq y - y_\text{d},
\end{equation}
as well as the error transformations $\varepsilon_{p_i}\in\mathbb{R}$ according to: 
\begin{equation} \label{eq:error transformation (TAC)}
e_{p_i} = \rho_{p_i} T(\varepsilon_{p_i}), \ \forall i\in\{1,\dots,n\},
\end{equation}
where $T:\mathbb{R} \to (-1,1)$ is a smooth, strictly increasing analytic function, with $T(0) = 0$.
Since $T$ is increasing, the inverse mapping $T^{-1}: (-1,1) \to \mathbb{R}$ is well-defined, and it holds that 
\begin{subequations} \label{eq:T as x to infty (TAC)}
	\begin{align}
	\lim_{\zeta \to -\infty}	T(\zeta) = -1, \ \ 
	\lim_{\zeta \to +\infty}	T(\zeta) = 1
	\end{align}
\end{subequations}
and hence, if $\varepsilon_{p_i}$ remains bounded in a compact set, the desired funnel objective $-\rho_{p_i}(t) < e_{p_i}(t) < \rho_{p_i}(t)$ is achieved, $\forall i\in\{1,\dots,n\}$. We further require that 
\begin{equation} \label{eq:T property (TAC)}
|\zeta| < \left|\frac{\partial T^{-1}(\zeta)}{\partial \zeta} T^{-1}(\zeta) \right|, \ \ \forall \zeta\in(-1,1).
\end{equation}
A possible choice that satisfies the aforementioned specifications is $T(\zeta) = \frac{\exp(\zeta) - 1}{\exp(\zeta)+1}$.

From \eqref{eq:error transformation (TAC)}, we obtain
\begin{equation} \label{eq:epsilon p (TAC)}
\varepsilon_{p_i} = T^{-1}\left( \frac{e_{p_i}}{\rho_{p_i}} \right),
\end{equation}
which, after differentiation, becomes
\begin{equation*}
\dot{\varepsilon}_{p_i} = \frac{r_{p_i}}{\rho_{p_i}}\left( x_{2_i} - \dot{y}_{i,\text{d}} - \frac{\dot{\rho}_{p_i}e_{p_i}}{\rho_{p_i}} \right),
\end{equation*}
or, in {stack} vector form,
\begin{equation} \label{eq:epsilon p dot (TAC)}
\dot{\varepsilon}_p = r_p \rho_p^{-1}\left(x_2 - \dot{y}_{\text{d}} - \dot{\rho}_p \rho_p^{-1}e_p \right),
\end{equation} 
where $\varepsilon_p \coloneqq [\varepsilon_{p_1},\dots,\varepsilon_{p_n}]^\top$, $r_{p_i} \coloneqq \frac{\partial T^{-1}(\zeta)}{\partial \zeta} \big|_{\zeta=\frac{e_{p_i}}{\rho_{p_i}}}$, $r_p \coloneqq \text{diag}\{r_{p_1},\dots,r_{p_n}\}$, and $\rho_p \coloneqq \text{diag}\{\rho_{p_1},\dots,\rho_{p_n}\}$. Due to the increasing property of $T(\cdot)$, it holds that $r_p$ is positive definite, and thus {in order to render $\dot{\varepsilon}_p$ negative} a straightforward choice for a desired value for $x_2$ is 
\begin{equation} \label{eq:x_v_des (TAC)}
x_{2,\text{d}} \coloneqq \dot{y}_\text{d} + \dot{\rho}_p\rho_p^{-1}e_p - k_p r_p\varepsilon_p,
\end{equation} 
where $k_p\in\mathbb{R}_{>0}$ is a positive and constant scalar gain. 
Since, however, $x_2$ is not the system's input, we follow a backstepping-like methodology and define the error 
\begin{equation} \label{eq:e_v (TAC)}
e_v \coloneqq \begin{bmatrix}
e_{v_1}, \dots,  e_{v_n}
\end{bmatrix}^\top \coloneqq x_2 - x_{2,\text{d}}.
\end{equation}
Next, we proceed in a similar manner and define a funnel for each $e_{v_i}$, $i\in\{1,\dots,n\}$, described by the functions $\rho_{v_i}:[t_0,\infty)\to [\underline{\rho}_{v_i},\bar{\rho}_{v_i}] \subset \mathbb{R}_{>0}$, where $\underline{\rho}_{v_i},\bar{\rho}_{v_i}\in\mathbb{R}_{>0}$ are the positive lower and upper bounds, respectively, with the constraint $\rho_{v_i}(t_0) > |e_{v_i}(t_0)|$, $i\in\{1,\dots,n\}$. {Note that $e_{v_i}(t_0) = x_2(t_0) - x_{2,\text{d}}(t_0)$ can be calculated at $t=t_0$ since it is a function of the state, the funnel functions and the desired trajectory profile}. Then, we define the open set
\small
\begin{align}  \label{eq:D_u,t (TAC)}
\mathcal{D}_{u,t} \coloneqq &\{(x,t)\in \mathbb{R}^{2n}\times [t_0,\infty) : \rho_p(t)^{-1}e_p \in (-1,1)^n,  \rho_v(t)^{-1} e_v \in (-1,1)^n\},
\end{align}  
\normalsize
and 
design the control law  $u:\mathcal{D}_{u,t} \to \mathbb{R}^n$ as 
{
	\begin{align} \label{eq:control law (TAC)}
	u = & - k_{v_2}\rho_v^{-1} \left(k_{v_3} \| r_p \varepsilon_p \| + k_{v_4} \hat{d}  \right)s_v 
	-k_{v_1} \rho_v^{-1} r_v \varepsilon_v 
	\end{align} 
	where 
	\begin{equation*}
	s_v \coloneqq \begin{cases}
	\frac{r_v \varepsilon_v}{\|r_v \varepsilon_v\|}, & \text{ if } \|r_v \varepsilon_v\| \neq 0, \\
	0, & \text{ otherwise, }
	\end{cases}	
	\end{equation*}
}$\rho_v \coloneqq \text{diag}\{\rho_{v_1},\dots,\rho_{v_n}\}$, $\varepsilon_v \coloneqq [\varepsilon_{v_1},\dots,\varepsilon_{v_n}]^\top$, $\varepsilon_{v_i} \coloneqq T^{-1}\left(\frac{e_{v_i}}{\rho_{v_i}}\right)$,  $r_v \coloneqq \text{diag}\{r_{v_1},\dots,r_{v_n}\}$, $r_{v_i} \coloneqq \frac{\partial T^{-1}(\zeta)}{\partial \zeta} \big|_{\zeta=\frac{e_{v_i}}{\rho_{v_i}}}$, $k_{v_i}\in\mathbb{R}_{> 0},i\in\{1,\dots,4\}$ are positive constant scalar gains, and $\hat{d}$ is an adaptive variable gain, subject to the constraint $\hat{d}(t_0) \geq 0$, and dynamics

\begin{equation} \label{eq:adaptation law (TAC)}
{\dot{\hat{d}} = \gamma_d \| r_v \varepsilon_v \|,}
\end{equation}
where $\gamma_d\in\mathbb{R}_{> 0}$ is a positive constant gain.

{\begin{remark}
		The control design procedure follows closely the prescribed performance backstepping-like methodology of Section \ref{subsec:PPC Controller (TCST_coop_manip)},  \ref{sec:formation control},  introduced in \cite{bechlioulis2014low}. {The desired signals and control laws there consist only of proportional terms with respect to the transformed errors $\varepsilon_p$, $\varepsilon_v$, i.e., $-k_pr_p\varepsilon_p$ and $-k_{v_1}\rho^{-1}_{v}r_v\varepsilon_v$ in \eqref{eq:x_v_des (TAC)} and \eqref{eq:control law (TAC)}, respectively, which are guaranteed to be ultimately bounded. In this work, we incorporate (a) the extra terms in \eqref{eq:x_v_des (TAC)} that would render \eqref{eq:epsilon p dot (TAC)} exponentially stable, and (b) the discontinuous term in \eqref{eq:control law (TAC)}, which, as will be {shown} in the sequel, {enforces} convergence of the transformed errors to zero, guaranteeing thus asymptotic stability. This is achieved without requiring the funnel functions to converge to zero. However, one can still set the prescribed funnel to converge arbitrarily close to zero, achieving thus a predefined convergence rate.}
	\end{remark}}
	
	\begin{remark}
		Note that no information regarding the dynamic model is incorporated in the control protocol \eqref{eq:e_p (TAC)}-\eqref{eq:adaptation law (TAC)}. All the necessary signals consist of the funnel terms $\rho_p$, $\rho_v$ and of known functions of the state and the desired trajectory $y_\text{d}$. Furthermore, no a-priori gain tuning is needed and, as the next theorem states, the solution of Problem \ref{problem:1 (TAC)} is guaranteed from \textit{all} initial conditions {that satisfy} $-\rho_{p_i}(t_0) > y_i(t_0) - y_{i,\text{d}}(t_0) > \rho_{p_i}(t_0)$, $\forall i\in\{1,\dots,n\}$. As will be revealed subsequently, the adaptive gain $\hat{d}$ compensates the unknown dynamic terms, which are proven to be bounded due to the confinement of the state in the prescribed funnels.	 
	\end{remark}
	
	
	The correctness of the control protocol \eqref{eq:e_p (TAC)}-\eqref{eq:adaptation law (TAC)} is shown in the next theorem.
	\begin{theorem} \label{th:theorem 1 (TAC)}
		Consider a system subject to the dynamics \eqref{eq:system (TAC)}, {Assumptions} \ref{ass:f(x,t) cont + bounded (TAC)}-\ref{ass:y_d (TAC)}, as well as a desired trajectory $y_\text{d}$ and funnels as described in Problem \ref{problem:1 (TAC)} satisfying $-\rho_{p_i}(t_0) > y_i(t_0) - y_{i,\text{d}}(t_0) > \rho_{p_i}(t_0)$, $\forall i\in\{1,\dots,n\}$.
		Then the control protocol \eqref{eq:e_p (TAC)}-\eqref{eq:adaptation law (TAC)} guarantees {the existence of at least one local Filippov solution  of the closed-loop system \eqref{eq:system (TAC)}-\eqref{eq:control law (TAC)} that solves Problem \ref{problem:1 (TAC)}. Moreover, every such local solution can be extended to a global solution
			and all closed-loop signals remain bounded, for all $t\geq t_0$.}
	\end{theorem}
	\begin{proof}
		The intuition of the subsequent proof is as follows: We first show the existence of at least one Filippov solution of the closed loop system in $\mathcal{D}_{u,t}$ for a time interval $I_t \subseteq [t_0,\infty)$. Next, we prove that for any of these solutions, the state remains bounded in $I$ by bounds independent of the endpoint of $I$. Hence, the dynamic terms of \eqref{eq:system (TAC)} are also upper bounded by a term, which we aim to compensate via the adaptation gain $\hat{d}$.
		
		We start by defining some terms that will be used in the subsequent analysis: 	
		\begin{align*}
			&M_p \coloneqq \max_{i\in\{1,\dots,n\}}\{ \bar{\rho}_{p_i}\} &			m_p \coloneqq \min_{i\in\{1,\dots,n\}}\{\underline{\rho}_{p_i}\}	\\
& M_{\dot{p}} \coloneqq \max_{i\in\{1,\dots,n\}}\{ \sup_{t\geq t_0} \{ |\dot{\rho}_{p_i}| \}\}
&M_v \coloneqq \max_{i\in\{1,\dots,n\}}\{ \bar{\rho}_{v_i}\} \\
&m_v \coloneqq \min_{i\in\{1,\dots,n\}}\{\underline{\rho}_{v_i}\} 
& \underline{\lambda} \coloneqq {\lambda_{\min}\left(\rho_v^{-1}\widetilde{G}(x,z,t)\rho_v^{-1}\right)} \\
& \bar{\beta} \coloneqq \left(k_{v_2}k_{v_4} \underline{\lambda} \right)^{-1}
&\underline{r}_p \coloneqq \inf_{\zeta\in(-1,1)} \frac{\partial T^{-1}(\zeta)}{\partial \zeta}.
		\end{align*}		
		Note that all the aforementioned terms are strictly positive. In particular, $\underline{\lambda}$ is strictly positive due to the definition of the funnels $\rho_v$ and Assumption \ref{ass:g(x,t) pd (TAC)}, and $\underline{r}_p$ is strictly positive due to the strictly increasing property of $T(\cdot)$ and hence of $T^{-1}(\cdot)$. Moreover, in view of \eqref{eq:T as x to infty (TAC)}, it holds that $\arg \inf_{\zeta\in(-1,1)} \frac{\partial T^{-1}(\zeta)}{\partial \zeta} \in(-1,1)$.

		By employing \eqref{eq:control law (TAC)}, \eqref{eq:adaptation law (TAC)}, we can write the closed loop system
		\begin{subequations} \label{eq:closed loop system (TAC)}
			\begin{align}				
			&\dot{x}_1 = x_2, \label{eq:closed loop system p (TAC)}\\
			&\dot{z} \in \mathsf{K}[F_z](x,z,t), \label{eq:closed loop system z (TAC)}\\
			&\dot{x}_2 \in \mathsf{K}[F](x,z,t) + \mathsf{K}[G](x,z,t) \mathsf{K}[u](x,t), \label{eq:closed loop system v (TAC)} \\
			&\dot{\hat{d}} = \gamma_d {\|r_v \varepsilon_v\|},
			\end{align}
		\end{subequations}
		{where $\mathsf{K}[F](x,z,t)$, $\mathsf{K}[G](x,z,t)$, $\mathsf{K}[u](x,t)$ 
			are the Filippov regularizations (see \eqref{eq:Filippov regular. (App_dynamical_systems)}) of the respective terms. For $u$ specifically, $\mathsf{K}[u](x,t)$ is formed by substituting the term $s_v$ with its reguralized term, which is $\mathsf{S}_v = \frac{r_v\varepsilon_v}{\|r_v\varepsilon_v\|}$ if $\|r_v\varepsilon_v\| \neq 0$, and $\mathsf{S}_v \in (-1,1)^n$ otherwise. Note that, in any case, it holds that $(r_v\varepsilon_v)^\top \mathsf{S}_v = \|r_v\varepsilon_v\|$.}
		Define now $\widetilde{x} \coloneqq [x^\top, z^\top, \hat{d}] \in \mathbb{R}^{2n+n_z+1}$ and
		consider the open set $\mathcal{D}_c \coloneqq \{ (\widetilde{x},t)\in \mathbb{R}^{2n+n_z+1}\times [t_0,\infty): (x,t)\in\mathcal{D}_{u,t}\}$.
		Since $\rho_{p_i}(t_0) > |e_{p_i}(t_0)|$ and $\rho_{v_i}(t_0) > |e_{v_i}(t_0)|$, $\forall i\in\{1\dots,n\}$, the set $\mathcal{D}_c$ is nonempty. Moreover, since $T(\cdot)$, and hence its derivative, are analytic, their zero sets have zero measure \cite{zeroSetAnalyticF} and thus the right hand-side of \eqref{eq:closed loop system (TAC)} is Lebesgue measurable and locally essentially bounded in $\widetilde{x}$ over the set $\{\widetilde{x}:(\widetilde{x},t)\in\mathcal{D}_c\}$, and Lebesgue measurable in $t$ over the set $\{t: (\widetilde{x},t)\in \mathcal{D}_c\}$. Hence, according to
		Prop. \ref{prop:Filippov exist (app_dynamical_systems)} of Appendix \ref{app:dynamical systems}, for each initial condition $(\widetilde{x}(t_0),t_0) \in \mathcal{D}_c$, there exists at least one Filippov solution $\widetilde{x}(t)$  of \eqref{eq:closed loop system (TAC)}, defined in $I_t \coloneqq [t_0, t_{\max})$, where $t_{\max} > t_0$ such that $(\widetilde{x}(t),t)\in\mathcal{D}_c$, $\forall t \in I_t$. By  applying \eqref{eq:epsilon p (TAC)}, we conclude the existence of the respective Filippov solutions $\varepsilon_p(t), \varepsilon_v(t)\in\mathbb{R}^n$, $\forall t\in I_t$. {Let now $\widetilde{x}(t_0)$ denote the initial condition of the system \eqref{eq:closed loop system (TAC)} satisfying $(\widetilde{x}(t_0),t_0)\in\mathcal{D}_c$} and consider the family of Filippov solutions starting from $\widetilde{x}(t_0)$ denoted by the set $\mathfrak{X}$.
		Note that, although not explicitly stated, $t_{\max}$ and $I_t$ might be different for each solution in $\mathfrak{X}$.
		We aim to prove that all $\varepsilon_p(t), \varepsilon_v(t)$ are bounded and converge to zero, for all $\widetilde{x}(t)\in\mathfrak{X}$.
		
		In view of the definition of $\mathcal{D}_c$ (see also \eqref{eq:D_u,t (TAC)}), for all $\widetilde{x}(t)\in\mathfrak{X}$ it holds that
		\begin{subequations} \label{eq:e_p e_v bounded tmax (TAC)}
		\begin{align}
		&		|e_{p_i}(t)| < \bar{\rho}_{p_i},  \\
		& |e_{v_i}(t)| < \bar{\rho}_{v_i},	\label{eq:e_v bounded tmax (TAC)}
		\end{align}
		\end{subequations} 	
		$\forall t\in I_t$, where $\bar{\rho}_{p_i}$ and $\bar{\rho}_{v_i}$ are the upper bounds of $\rho_{p_i}(t)$ and $\rho_{v_i}(t)$, respectively, $\forall i\in\{1,\dots,n\}$. Consider now the Lyapunov function $V_p \coloneqq \frac{1}{2}\|\varepsilon_p\|^2$, for which it holds, in view of \eqref{eq:epsilon p dot (TAC)}, \eqref{eq:x_v_des (TAC)}, \eqref{eq:e_v (TAC)}, and \eqref{eq:e_v bounded tmax (TAC)}
		\begin{align*}
		\dot{V}_p =& \varepsilon_p^\top r_p \rho_p^{-1} (x_2 - \dot{y}_\text{d} - \dot{\rho}_p \rho_p^{-1}e_p)  \\
		=& -k_p \varepsilon_p^\top r_p\rho_p^{-1} r_p\varepsilon_p + \varepsilon_p^\top r_p \rho_p^{-1} e_v  
		<  -\frac{k_p}{M_p}\|r_p \varepsilon_p\|^2 + \frac{M_v}{m_p}\|r_p \varepsilon_p \|,
		\end{align*}	 
		$\forall t\in I_t$.
		Hence, we conclude that $\dot{V}_p < 0$ when $\|r_p\varepsilon_p\| > \frac{M_v M_p}{k_p m_p}$. Since $r_{p_i}$ is positive definite, $\forall i\in\{1,\dots,n\}$, the latter is equivalent to $\|\varepsilon_p\| > \frac{M_v M_p}{k_p m_p \underline{r}_p}  \Rightarrow \dot{V}_p < 0$. Hence, we conclude that all $\widetilde{x}(t)\in\mathfrak{X}$ satisfy 
		\begin{equation*}
		\|\varepsilon_p(t)\| \leq \bar{\varepsilon}_p \coloneqq \max\left\{ \|\varepsilon_p(t_0)\|, \frac{M_vM_p}{k_p m	_p \underline{r}_p} \right\}.
		\end{equation*}
		Since $\bar{\varepsilon}_p$ is finite, it holds that $T(\bar{\varepsilon}_p) < 1$ and hence $|T(\varepsilon_{p_i}(t))| \leq T(\bar{\varepsilon}_p) < 1$, $\forall i\in\{1,\dots,n\}, t\in I_t$. Moreover, since $T(\cdot)$ and $T^{-1}(\cdot)$ are smooth, the derivative $\frac{\partial T^{-1}(\zeta)}{\partial \zeta}$ approaches infinity only when $\zeta \to \pm 1$. Therefore, in view of the definition of $r_{p_i}$ in \eqref{eq:epsilon p dot (TAC)}, we conclude the existence of a finite $\bar{r}_p > 0$ such that $\|r_p(t)\| \leq \bar{r}_p$, $\forall t\in I_t$. Next, 	
		\eqref{eq:error transformation (TAC)} implies that $\|e_p(t)\| \leq \bar{e}_p \coloneqq M_p T(\bar{\varepsilon}_p)\sqrt{n}$, $\forall t\in I_t$. 
		Hence, we conclude that $\|x_{2,\text{d}}(t)\| \leq \bar{x}_{2,\text{d}} \coloneqq \bar{y}_{\text{d}} + \frac{M_{\dot{p}}}{m_p} \bar{e}_p + k_p \bar{r}_p \bar{\varepsilon}_p$, $\forall t\in I_t$, where $\bar{y}_\text{d}$ is the uniform bound of the desired trajectory, introduced in Assumption \ref{ass:y_d (TAC)}. We also conclude that $\|x_1(t)\| \leq \bar{x}_1 \coloneqq \bar{e}_p + \bar{y}_\text{d}$, $\forall t\in I_t$. In addition, by employing $x_2 = e_v + x_{2,\text{d}}$ and \eqref{eq:e_v bounded tmax (TAC)},  we conclude that $\|x_2(t)\| < \widetilde{x}_2 \coloneqq M_v\sqrt{n} + \bar{x}_{2,\text{d}}$, $\forall t\in I_t$. 
		Finally, by differentiating $x_{2,\text{d}}$, employing the smoothness and boundedness of $\rho_p$ and its derivatives, the smoothness of $T(\cdot)$, the boundedness of $\ddot{y}_{\text{d}}(t)$ as well as the aforementioned bounds, we can conclude the existence of a bound $\bar{v}_\text{d}$ such that ${\|\dot{x}_{2,\text{d}}(t) \|} \leq \bar{v}_\text{d}$, $\forall t\in I_t$.
		
		{Furthermore, {the boundedness of $x(t)$ and Assumption \ref{ass:internal dynamics (TAC)} imply the existence of a positive finite constant $\bar{z}$ such that $\|z(t)\|\leq \bar{z}$, $\forall t\in I_t$.}
			Hence, since $F(x,z,t)$ is Lebesgue measurable and  locally essentially bounded in $\mathbb{R}^{2n+n_z}$ and $\|x_1(t)\| \leq \bar{x}_1 < \infty$, $\|x_2(t)\| < \widetilde{x}_2< \infty$, $\|z(t)\| \leq \bar{z}$, $\forall t\in I_t$,
			there exists some positive $\bar{F}$, such that ${\|F(x(t),z(t),t)\|} \overset{a.e.}{\leq} \bar{F}$, $\forall t\in I_t$, and hence, for each $(x,z)$, {since $\mathsf{K}[F]$ is formed by the convex closure of $F$}, it holds that $\max_{\zeta\in\mathsf{K}[F](x(t),z(t),t)} \{ \zeta \} \leq \bar{F}$, $\forall t\in I_t$ and $\widetilde{x}(t)\in\mathfrak{X}$}. Note that, in view of the aforementioned discussion, $\bar{F}$ depends solely on the initial conditions and the parameters of the funnel functions. Define now the finite constant term $d_b \in \mathbb{R}_{> 0}$ as
		\begin{equation} \label{eq:d (TAC)}
		d_b \coloneqq \frac{\bar{\beta}}{m_v} \left( \bar{F} + \bar{v}_\text{d} + M_{\dot{p}}\sqrt{n} \right).
		\end{equation} 	
		Note that the term in the {parenthesis} of \eqref{eq:d (TAC)} is an upper bound for the term ${\| F(x(t),z(t),t) - \dot{x}_{2,\text{d}}(t) - \dot{\rho}_v(t) \rho_v(t)^{-1}e_v(t) \|}$, for all $\widetilde{x}(t)\in\mathfrak{X}$ and almost all $t\in I_t$.
		
		Define also the signal $\widetilde{d} \coloneqq \hat{d} - d_b$, where $ \hat{d}$ is the adaptive gain introduced in \eqref{eq:control law (TAC)}.	
		Consider now the function 
		\begin{align*}
		V_v(\widetilde{\varepsilon})  \coloneqq& \bar{\alpha} V_p + \frac{\bar{\beta}}{2}\|\varepsilon_v\|^2 + \frac{1}{2\gamma_{\text{d}}}\widetilde{d}^2, 
		\end{align*}
		where $\widetilde{\varepsilon}\coloneqq [\varepsilon_p^\top, \varepsilon_v^\top, \widetilde{d}]^\top$, and $\bar{\alpha} > 0$ is a positive constant to be defined; $V_v(\widetilde{\varepsilon})$ satisfies $W_1(\widetilde{\varepsilon}) \leq V_v(\widetilde{\varepsilon}) \leq W_2(\widetilde{\varepsilon})$, for $W_1(\widetilde{\varepsilon}) \coloneqq \min\left\{\frac{\bar{\alpha}}{2},\frac{\bar{\beta}}{2},\frac{1}{2\gamma_\text{d}}\right\} \|\widetilde{\varepsilon}\|^2$ and $W_2(\widetilde{\varepsilon}) \coloneqq \max\left\{\frac{\bar{\alpha}}{2},\frac{\bar{\beta}}{2},\frac{1}{2\gamma_\text{d}}\right\} \|\widetilde{\varepsilon}\|^2$. Then, according to Lemma \ref{lem:Chain rule (App_dynamical_systems)} of Appendix \ref{app:dynamical systems}, $\dot{V}_v(\widetilde{\varepsilon}(t)) \overset{a.e.}{\in} \dot{\widetilde{V}}_v(\widetilde{\varepsilon}(t))$ with $$\dot{\widetilde{V}}_v \coloneqq \bigcap_{\xi \in \partial V_v(\widetilde{\varepsilon})} \xi^\top \mathsf{K}\begin{bmatrix}
		\dot{\widetilde{\varepsilon}} 
		\end{bmatrix}$$
		Since $V_v(\widetilde{\varepsilon})$ is continuously differentiable, its generalized gradient reduces to the standard gradient and thus it holds that $\dot{\widetilde{V}}_v =\nabla V_v^\top \mathsf{K}\begin{bmatrix}
		\dot{\widetilde{\varepsilon}}
		\end{bmatrix}$,
		where $\nabla V_v = [ \bar{\alpha} \varepsilon_p^\top, \bar{\beta}  \varepsilon_v^\top,  \frac{1}{\gamma_\text{d}}\widetilde{d} ]^\top$.
		After using \eqref{eq:system (TAC)}, \eqref{eq:control law (TAC)}, \eqref{eq:adaptation law (TAC)}, and $x_2 = x_{2,\text{d}} + e_v$, one obtains
		\small
		\begin{align*}
		\dot{\widetilde{V}}_v \subset \widetilde{W}_s \coloneqq &-\bar{\alpha} k_p \varepsilon_p^\top r_p \rho_p^{-1} r_p \varepsilon_p + \bar{\alpha}\varepsilon_p^\top r_p \rho_p^{-1} e_v - \bar{\beta} k_{v_1}\varepsilon_v^\top r_v \rho_v^{-1} \mathsf{K}[G](x,t) \rho_v^{-1}r_v\varepsilon_v \\
		& + \widetilde{d}{\|r_v\varepsilon_v\|} 
		 + \bar{\beta}\varepsilon_v^\top r_v \rho_v^{-1}\left( \mathsf{K}[F](x,t) - \dot{x}_{2,\text{d}} - \dot{\rho}_v\rho_v^{-1}e_v \right)  \\ 
		&- \bar{\beta} k_{v_2}\varepsilon_v^\top r_v \rho_v^{-1} \mathsf{K}[G](x,t) \rho_v^{-1} {\mathsf{S}_v\bigg(k_{v_3}  \|r_p\varepsilon_p\|} 
		+k_{v_4}\hat{d} \bigg).
		\end{align*}
		\normalsize
		Note that, since $\hat{d}(t_0) \geq 0$, \eqref{eq:adaptation law (TAC)} implies that $\hat{d}(t) \geq 0$, $\forall t\in I_t$. {Moreover, since the Filippov regularization \eqref{eq:Filippov regular. (App_dynamical_systems)} is defined as a closed set and $\dot{\widetilde{V}}_v \subset \widetilde{W}_s$, it holds that $\max_{\zeta\in \dot{\widetilde{V}}_v } \{\zeta\}  \leq  \max_{\zeta \in \widetilde{W}_s} \{ \zeta \}$.}	
		{By substituting $G = \frac{G+G^\top}{2} + \frac{G - G^\top}{2}$ and employing the skew-symmetry of the second term, we obtain in view of Assumption \ref{ass:g(x,t) pd (TAC)} and the definition of $d_b$ in \eqref{eq:d (TAC)}}:
		\begin{align*}
		 {\max_{\zeta\in \dot{\widetilde{V}}_v } \{\zeta\}  \leq  \max_{\zeta \in \widetilde{W}_s} \{ \zeta \}} \leq &- \bar{\alpha} \frac{k_p}{M_p} \|r_p \varepsilon_p\|^2 - k_{v_1} \bar{\beta} \underline{\lambda} \|r_v\varepsilon_v\|^2 - k_{v_2} k_{v_4}\bar{\beta} \underline{\lambda} {\|r_v\varepsilon_v\|}   \hat{d} - \\
		&\hspace{-20mm} k_{v_2} k_{v_3} \bar{\beta} \underline{\lambda} {\|r_v\varepsilon_v\| \|r_p \varepsilon_p\|}  + \widetilde{d}{\|r_v\varepsilon_v\|} + {\| r_v \varepsilon_v \|} d + {\bar{\alpha}\|\varepsilon_p^\top r_p \rho_p^{-1} e_v\|},
		\end{align*}		 
		for all solutions $\widetilde{x}(t)\in\mathfrak{X}$.
		By setting $\zeta=T(\varepsilon_{v_i})$ in \eqref{eq:T property (TAC)}, we obtain $|T(\varepsilon_{v_i})| \leq |r_{v_i} \varepsilon_{v_i}|$ and hence by employing $e_{v_i} = \rho_{v_i}T(\varepsilon_{v_i})$, $i\in\{1,\dots,n\}$, we obtain that 
		\begin{equation*}
		\bar{\alpha} {\|\varepsilon_p^\top r_p \rho_p^{-1} e_v\|} \leq \bar{\alpha} \frac{M_v}{m_p} {\|r_p\varepsilon_p\| \|r_v \varepsilon_v\|}.
		\end{equation*}
		Therefore, by setting $\bar{\alpha} \coloneqq \frac{k_{v_2}k_{v_3}m_p \bar{\beta} \underline{\lambda}}{M_v}$, employing $d_b = \hat{d} - \widetilde{d}$, and in view of the fact that $\bar{\beta} = (k_{v_2}k_{v_4}\underline{\lambda})^{-1}$, we obtain
		\begin{align*}
		\max_{\zeta {\in} \dot{\widetilde{V}}_v} \{ \zeta \} \leq & - \bar{\alpha} \frac{k_p}{M_p} \|r_p \varepsilon_p\|^2 - k_{v_1} \bar{\beta} \underline{\lambda} \|r_v\varepsilon_v\|^2  
		=: -W(\widetilde{\varepsilon}), 
		\end{align*}
		$\forall t\in I_t$, $\widetilde{x}(t)\in\mathfrak{X}$, where $W$ is continuous and positive semi-definite on $\mathbb{R}^{2n+1}$, since $r_v$ and $r_p$ are positive definite. 
		Hence, we conclude that $\zeta \leq-W(\widetilde{\varepsilon})$, $\forall \zeta\in\dot{\widetilde{V}}_v(\widetilde{\varepsilon}(t))$, 
		$\forall t\in I_t$ and all $\widetilde{x}(t)\in \mathfrak{X}$. Choose now any finite $r_a > 0$ and let $c_a < \min_{\|\widetilde{\varepsilon}\|=r_a}W_1(\widetilde{\varepsilon})$. Note that all the conditions of Theorem \ref{th:nonsmooth LaSalle (App_dynamical_systems)} in Appendix \ref{app:dynamical systems} are satisfied and hence, all Filippov solutions starting from $\widetilde{\varepsilon}(t_0)\in \Omega_f \coloneqq \{\widetilde{\varepsilon} \in \mathcal{B}(0,r_a) : W_2(\widetilde{\varepsilon}) \leq c_a\}$ are bounded and remain in $\Omega_f$, $\forall t\in I_t$. Moreover, $t_{\max} = \infty$, implying that $I_t = [t_0,\infty)$ and it also holds that $\lim_{t\to\infty} \|\varepsilon_p(t)\| = 0$ and $\lim_{t\to\infty} \|\varepsilon_v(t)\| = 0$, which, in view of the increasing property of $T(\cdot)$ and the fact that $T(0) = 0$, implies that $\lim_{t\to\infty} \|e_p(t)\| = 0$ and $\lim_{t\to\infty} \|e_v(t)\| = 0$. {Notice that Zeno behavior is avoided since $t_{\max} = \infty$.} 		
		
		Note that  $r_a$, and hence $c_a$, can be arbitrarily large allowing any finite initial condition $\widetilde{\varepsilon}$, which implies any $(\widetilde{x}(t_0),t_0) \in \mathcal{D}_c$. 
		In addition, it holds that $\|\widetilde{\varepsilon}\|^2 \leq \widetilde{c}\coloneqq (\max\{ \frac{\bar{\alpha}}{2},\frac{\bar{\beta}}{2},\frac{1}{2\gamma_\text{d}} \})^{-1}c_a$, which implies the boundedness of $\|\varepsilon_p\|$, $\|\varepsilon_v\|$ and $\widetilde{d}$ by $\sqrt{\widetilde{c}}$. Therefore, we conclude that $\|\hat{d}(t)\| \leq \bar{d}\coloneqq d_b + \sqrt{\widetilde{c}}$, $\forall t\in I_t$. Moreover, 	
		by employing \eqref{eq:error transformation (TAC)}, we conclude that  $|\rho_{v_i}(t)^{-1}e_{v_i}(t)| \leq T(\sqrt{\widetilde{c}}) < 1$, and hence $|e_{v_i}(t)| \leq M_v T(\sqrt{\widetilde{c}}) \Rightarrow \|x_2(t)\|  \leq \bar{x}_2 \coloneqq M_v T(\sqrt{\widetilde{c}}) \sqrt{n} + \bar{x}_{2,\text{d}}$,	
		$\forall t\in I_t$. Therefore, we conclude that all solutions are bounded in compact sets $\forall t\in I_t$, which means that
		$u$, and $\dot{\hat{d}}$, as designed in \eqref{eq:control law (TAC)} and \eqref{eq:adaptation law (TAC)}, respectively, remain also bounded, $\forall t\in I_t$.	
	\end{proof}

	{	
		\begin{remark}
			{
				Note that no boundedness assumptions or growth conditions are needed for the vector fields $F(x,z,t)$ and $G(x,z,t)$. In particular, the effect of $F(x,z,t)$ is canceled by the introduced {adaptive signal $\hat{d}$, which increases according to \eqref{eq:adaptation law (TAC)}. It is proved, {nevertheless}, that this adaptive signal remains bounded}. 
				Moreover, the response of the system is solely determined by the funnel functions $\rho_{p_i}$ and $\rho_{v_i}$, isolated from the system dynamics and the control gains selection. Nevertheless, we note that appropriate gain tuning might be needed to suppress chattering in real life scenarios.
				Similarly, note that the region of attraction (initial conditions) of $(\varepsilon_p,\varepsilon_v) = (0,0)$ is {independent} from the system dynamics and the control gain selection and depends only on the choice of the funnel functions $\rho_{p_i}$, $\forall i\in\{1,\dots,n\}$. In particular, if $\rho_{p_i}(t_0)$ are design parameters, we can always choose them such that $-\rho_{p_i}(t_0) < e_{p_i}(t_0) < \rho_{p_i}(t_0)$, $\forall i\in\{1,\dots,n\}$, which renders the result global. In fact, the choice $\lim_{t\to t_0^{+}}\frac{1}{\rho_{p_i}(t_0)} = 0$, $\forall i\in\{1,\dots,n\}$ \cite{berger2018funnel} is not excluded from {our} control scheme and does not restrict the initial condition $y(t_0)$. Moreover, noise can be taken into account  in the measurement of $x_2$, i.e., consider that $x_2 + \mathsf{n}(x,t)$ is available for measurement, where $\mathsf{n}(x,t)$ is an unknown noise signal with appropriate continuity and boundedness properties. By redefining $e_v = x_2 + \mathsf{n}(x,t) - x_{2,\text{d}}$ and including the time derivative of $\mathsf{n}(x,t)$ in \eqref{eq:d (TAC)}, the analysis still holds. Note, however, that {in this case it can only be deduced that} $\lim_{t\to\infty}(x_2(t) + \mathsf{n}(x,t) - x_{2,\text{d}}(t)) = 0$ and hence $x_2(t)$ does not necessarily converge to $x_{2,\text{d}}(t)$.}   
		\end{remark}
		\begin{remark}
			{Since funnel control traditionally guarantees confinement of the state in the desired funnel, a common practice is to tune the funnel to converge to arbitrarily small values, achieving thus ``practical stability", i.e., the state converging arbitrarily close to zero. Note that, in our case, the  funnel functions $\rho_{p_i}$, $\rho_{v_i}$, are not required to decrease to values arbitrarily close to zero, {yet} \textit{asymptotic stability} is still achieved. In fact, the proposed control schemes can be used to achieve {merely} asymptotic stability results without any funnel constraints, if the latter is not required. More specifically, given the initial errors $e_{p_i}(t_0)$, we can use the proposed control protocols by employing any \textit{constant} values $\rho_{p_i} > |e_{p_i}(t_0)|$, $\forall i\in\{1,\dots,n\}$. Finally, the proposed control scheme can be extended to systems of the form $\dot{x}_i =  \dot{x}_{i+1}, i\in\{1,\dots,k-1\},  \dot{x}_k = F(x,z,t) + G(x,z,t)u$ for some $k>0$, where the funnel constraints are set for the combined signal $
				\sum_{j\in\{1,\dots,k-1\}}e_p^{(k)}$. }
		\end{remark}
	}
	
	\subsection{Simulation Results} \label{sec:Simulation results (TAC)}
	
	{We consider here the simulation of two inverted pendulum connected by a spring and a damper \cite{bechlioulis2011robust}, with dynamics:		
		\begin{align*}
		&J_1\ddot{x}_{1_1} = g_s  \sin(x_{1_1}) - \frac{1}{4} F_s \cos(x_{1_1} - \theta_f) - T_{f_1} + u_1 \\
		&J_2\ddot{x}_{1_2} = 1.25 g_s \sin(x_{1_2}) + \frac{1}{4} F_s \cos(x_{1_2} - \theta_f) - T_{f_2} + \sigma_f(t) u_2,
		\end{align*}		
		where $F_s\coloneqq 150(d_s - \frac{1}{2}) + \dot{d}_s$ is the force between the connection points of the spring and damper at the pendulums, and \small$$d_s \coloneqq \sqrt{\frac{1}{4} + \frac{1}{4}(\sin(x_{1_1} - x_{1_2})) + \frac{1}{8}(1 - \cos(x_{1_2} - x_{1_1}))}$$\normalsize is the distance between these connection points; $\theta_f$ is defined as $$\theta_f\coloneqq \tan^{-1}\left( \frac{\frac{1}{4}(\cos(x_{1_2}) - \cos(x_{1_1})) }{\frac{1}{2} + \frac{1}{4}( \sin(x_1) - \sin(x_2) )} \right)$$ and $T_{f_1}$, $T_{f_2}$ are friction terms on the motors evolving according to $T_{f_i} \coloneqq \tau_{f_i} + \dot{\tau}_{f_i} + \dot{x}_{1_i}$, with
		\small
		\begin{align*}	 
		\dot{\tau}_{f_i} = \dot{x}_{1_i} - \frac{|\dot{x}_{1_i}|}{1 + \exp\left(-\left|\frac{\dot{x}_{1_i}}{0.1}\right|^2\right)}
		\end{align*}
		\normalsize
		The time varying signal $\sigma_f(t)$ is taken as:		
		$$\sigma_f(t) = \begin{cases}
		1 & \text{ if } t\in[0,3)\cup[3.5,\infty), \\
		0.5 & \text{ if } t\in[3,3.5)
		\end{cases}$$		
		modeling a loss of effectiveness of the second motor when $t\in[3,3.5)$. We also choose $g_s = 9.81$ as the gravity constant and $J_1 = 0.5$, $J_2 = 0.625$. The initial conditions are $t_0 = 0$, $x(0) = [0,0,0,0]^\top$ (rad, rad/s), $\tau_{f_1}(0)= \tau_{f_2}(0)=0$ and the desired trajectory $y_\text{d} = [2\cos(t), \frac{\pi}{2}-2\sin(t)]^\top$ rad. The prescribed funnel functions are chosen as $\rho_{p_i}(t) = 2.5\exp(-0.1t) + 2.5$, $\forall i\in\{1,2\}$, which converge to $2.5$. We also choose $\rho_{v_i}(t) = (\|e_v(0)\|_1-2)\exp(-0.1t) + 2.5$, as well as the gains $k_p = 10$, $k_{v_1} = 2\cdot 10^3$, $k_{v_2} = 0.1$, $k_{v_3} = 0.025$, $k_{v_4} = 0.05$, and $\gamma_\text{d} = 50$. The simulation {results are} depicted in Figs. \ref{fig:errors (TAC)}-\ref{fig:dhat (TAC)} for $t\in[0,60]$ $\sec$. More specifically, Fig. \ref{fig:errors (TAC)} depicts the errors $e_p(t), e_v(t)$ along with the performance functions $\rho_p(t)$, $\rho_v(t)$. One can conclude that $e_p(t)$ and $e_v(t)$ not only respect their imposed funnels but also converge asymptotically to zero, without the need of arbitrarily small values for $\lim_{t\to\infty}\rho_p(t)$ and $\lim_{t\to\infty}\rho_v(t)$. This can be verified also by Fig. \ref{fig:transf_errors (TAC)}, which depicts the evolution of the transformed errors $\varepsilon_p(t)$, $\varepsilon_v(t)$, $\forall t\in[0,60] \sec$, and shows their asymptotic convergence to zero. Finally, Figs. \ref{fig:inputs (TAC)} and \ref{fig:dhat (TAC)} illustrate the inputs $u(t)$ as well as the adaptation signal $\hat{d}(t)$, $\forall t\in[0,60] \sec$. {One can conclude the convergence of $\hat{d}(t)$ to a constant value as well as the boundedness of the control input $u(t)$, as was proved in the theoretical analysis.} }

	\begin{figure}
		\centering
		\includegraphics[scale = 0.65]{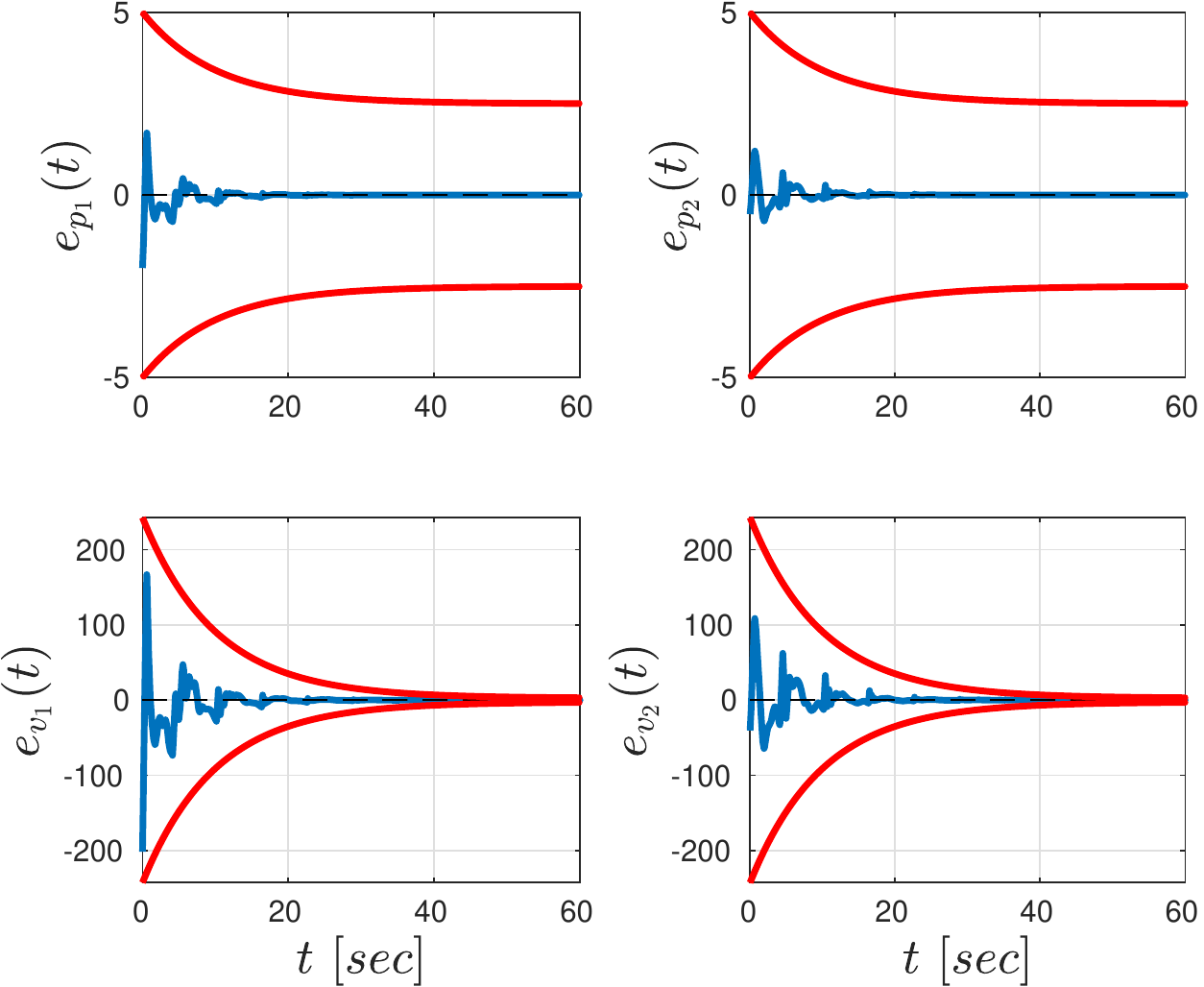}
		\caption{The evolution of the errors $e_p(t)$ (top), $e_v(t)$ (bottom), depicted with blue, along with the performance functions $\rho_p(t)$, $\rho_v(t)$, depicted with red, $\forall t\in[0,60] \ \sec$.}\label{fig:errors (TAC)}
	\end{figure}
	
	\begin{figure}
		\centering
		\includegraphics[scale = 0.55]{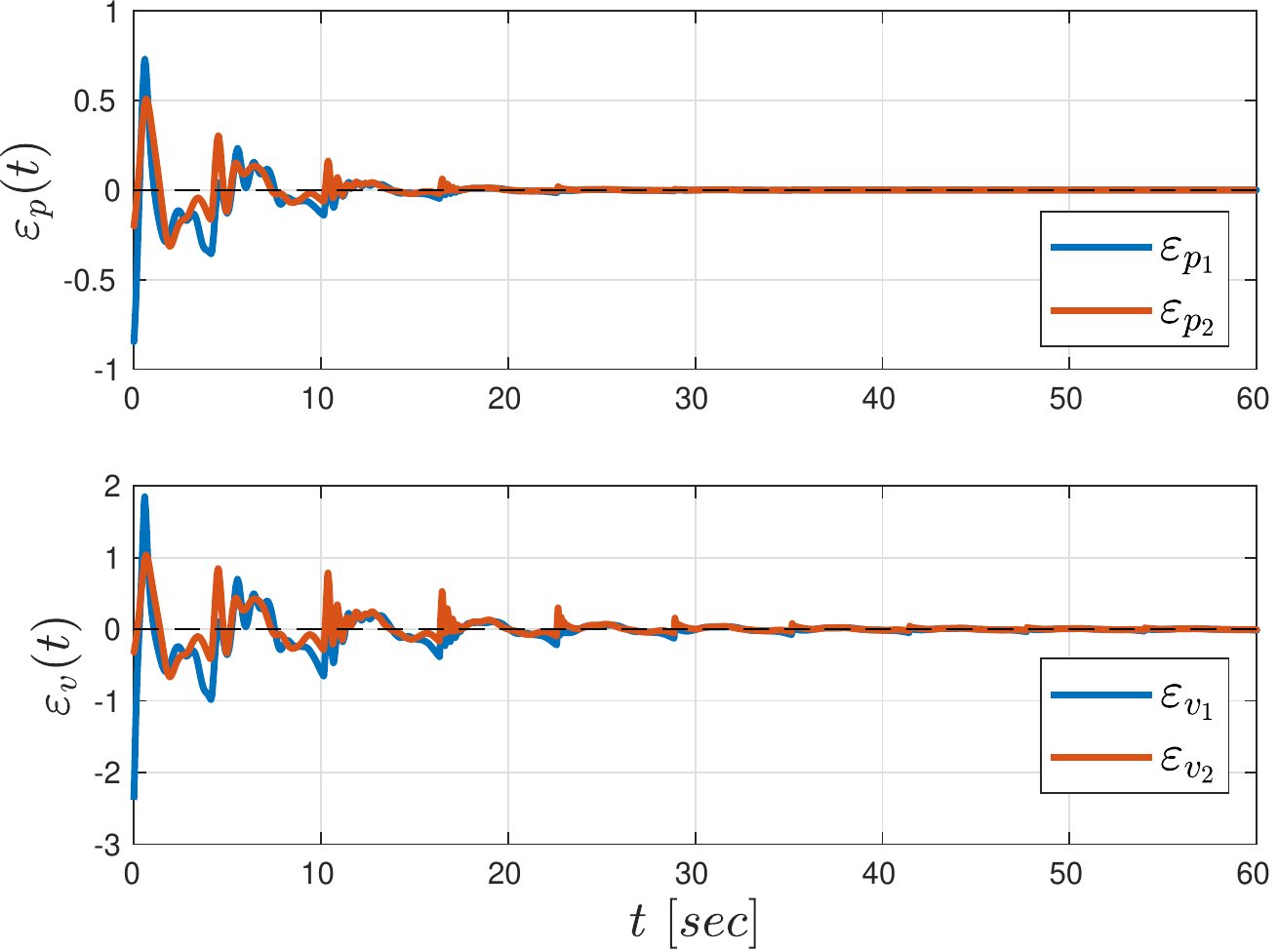}
		\caption{The evolution of the transformed errors $\varepsilon_p(t)$, $\varepsilon_v(t)$, $\forall t\in[0,60] \ \sec$.}\label{fig:transf_errors (TAC)}
	\end{figure}
	
	\begin{figure}
		\centering
		\includegraphics[scale = 0.5]{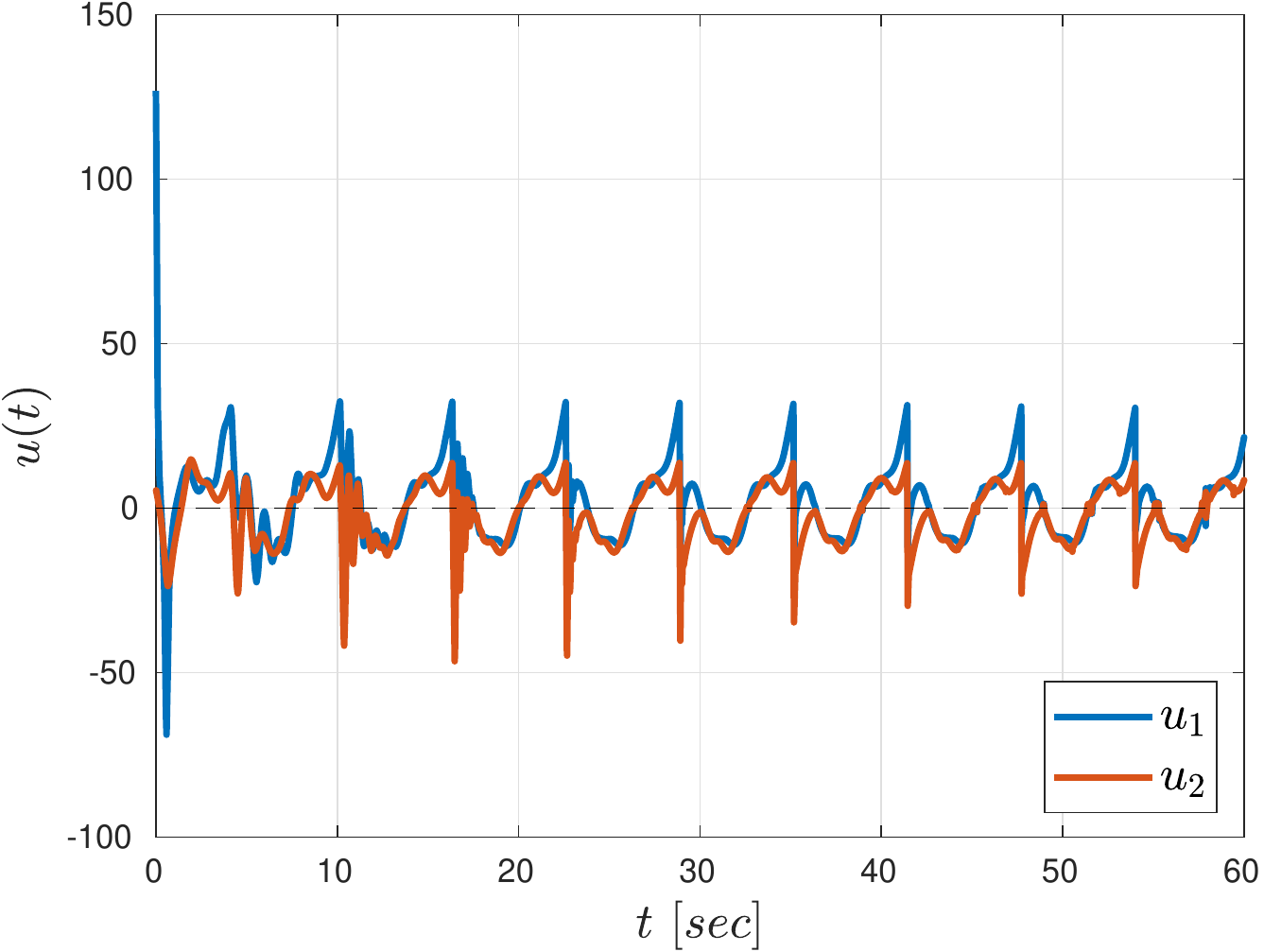}
		\caption{The evolution of the control inputs $u(t) = [u_1(t),u_2(t)]^\top$, $\forall t\in[0,60] \ \sec$.}\label{fig:inputs (TAC)}
	\end{figure}
	
	\begin{figure}
		\centering
		\includegraphics[scale = 0.5]{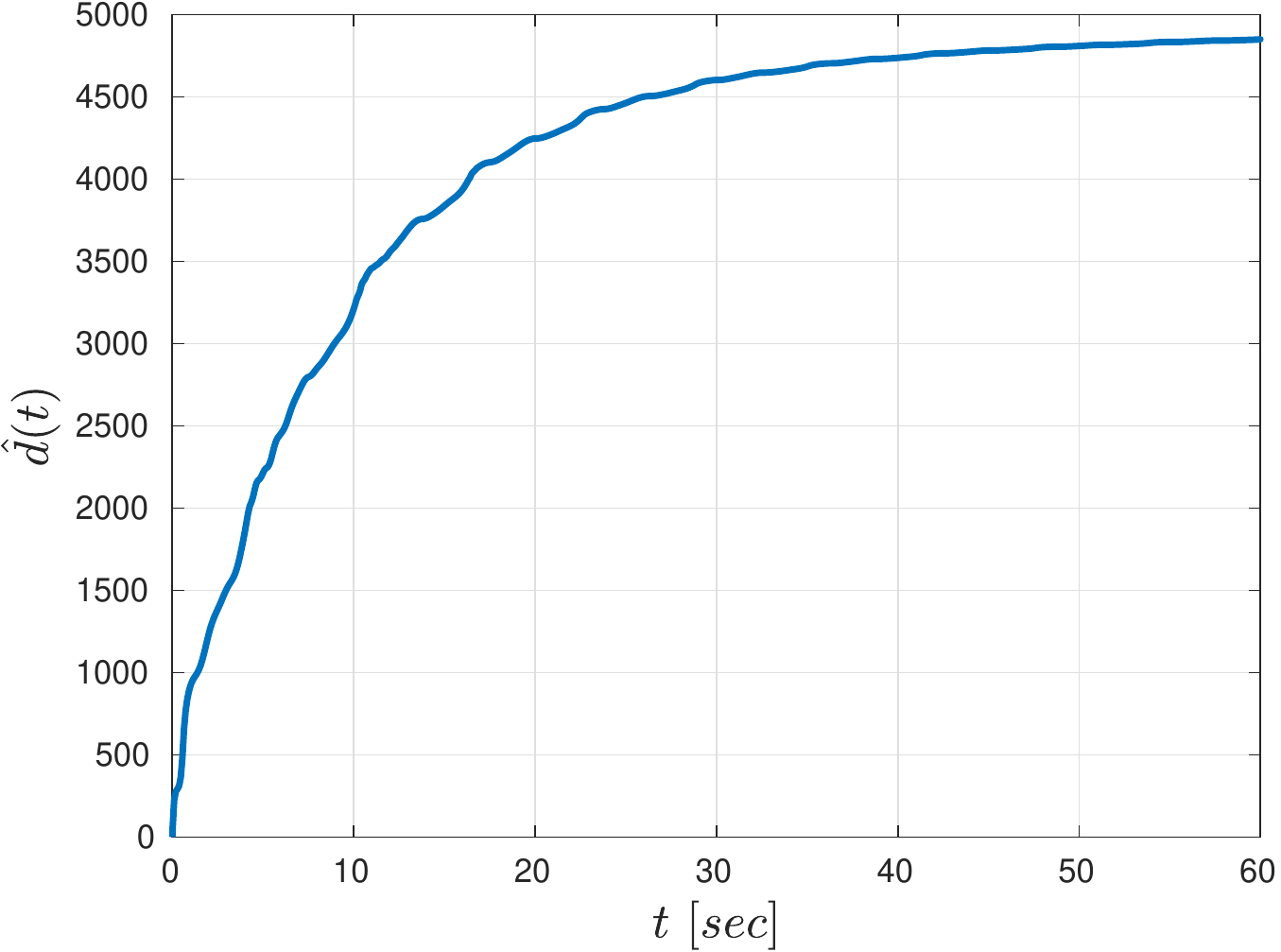}
		\caption{The evolution of the adaptation signal $\hat{d}(t)$, $\forall t\in[0,60] \ \sec$.}\label{fig:dhat (TAC)}
	\end{figure}

\section{Conclusion}

This chapter presented planning and control algorithms for single-agent systems. Firstly, we developed a hybrid algorithm for the planning of a robotic system under timed temporal logic formulas in an obstacle-cluttered workspace. By using previous results, we guaranteed the collision-free timed navigation leading to a timed abstraction of the system. A high-level planner and a novel optimization technique provided the timed path that satisfy the specification and is asymptotically optimal via reconfiguration. Secondly, we turned our attention to the motion planning of high-dimensional uncertain systems (e.g., robotic manipulators). We developed a two-layer framework by integrating adaptive control techniques and sampling-based motion planning. The closed-loop system provably navigated to a desired goal while avoiding collisions and compensating for the dynamic uncertainties. Finally, we developed a control scheme that guarantees asymptotic stability for an uncertain $2$nd-order system, while complying to funnel constraints, by integrating adaptive and discontinuous control methodologies.

\chapter{Summary and Future Research Directions} \label{chapter:conclusion}
This chapter summarizes the content of the thesis and provides potential future research directions. 

\section{Summary}

\vspace{2mm}

In Chapter \ref{chapter:cooperative manip}, we proposed a set of control algorithms for the cooperative manipulation of rigid objects. We tackled first the  case of rigid grasping contacts, and developed closed-form adaptive control algorithms, compensating for uncertainty in the dynamic parameters of the object and the agents, and Nonlinear Model Predictive Control schemes, taking into account constraints such as obstacle avoidance and input saturation. A Prescribed Performance Control methodology has been also developed to achieve prescribed transient and steady-state response for the object. 
Secondly, we considered the case of rolling contacts, for which we developed novel centralized and 	{decentralized} control algorithms that guarantee agents-object contact maintenance, along with object reference tracking. 

In Chapter \ref{chapter:formation}, we tackled the problem of multi-agent formation and its relation to rigid cooperative manipulation. Firstly, we developed a robust model-free decentralized control scheme for the formation control of a tree-graph multi-agent system with prescribed transient and steady-state response, subject to collision and connectivity constraints. Secondly, we associated rigid cooperative manipulation schemes to multi-agent rigidity theory. We related the grasp matrix of the former to the rigidity matrix of the latter and we used that to derive novel conditions for the internal force-free cooperative manipulation. 

In Chapter \ref{chapter:synthesis} we considered the problems of multi-agent navigation and leader-follower coordination subject to collision and/or connectivity constraints as well as uncertain dynamics. We first developed an adaptive control protocol for the problem of single-agent navigation in an obstacle-cluttered environment with uncertain dynamics under almost all initial conditions. This was extended to a prioritization-based decentralized scheme for multi-agent systems. Secondly, we proposed a novel adaptive control scheme for the leader-follower coordination, that is, navigation of a leader agent to a predefined pose, while guaranteeing collision avoidance and connectivity maintenance.  Finally, we designed an adaptive control protocol for the collision avoidance among ellipsoidal agents using a novel distance metric for $3$D ellipsoids.

In Chapter \ref{chapter:abstraction} we used previous continuous control schemes to derive appropriate discrete abstractions and synthesize controllers for the satisfaction of complex tasks expressed as temporal logic formulas. We first considered local tasks for multi-agent systems, such as UAVs and robotic manipulators, and then we focused on cases where unactuated objects have their own specifications. We considered discretizations both using predefined regions of interest as well as a full workspace partition. Collision avoidance was taken into account to define safe transitions among the discrete states.
Linear and Metric Interval Temporal Logic formulas were applied and control synthesis was performed using standard automata-based formal verification techniques.

In Chapter \ref{chapter:single agent} we developed extension algorithms for single-agent problems. Firstly, we considered the motion planning problem of a single agent in an obstacle-cluttered environment under timed temporal tasks. By using previous results on safe timed navigation, we developed an algorithm that guarantees the satisfaction of the timed specification as well as asymptotically optimal performance in terms of energy efficiency. Secondly, we addressed the problem of motion planning of high-dimensional complex systems in obstacle-cluttered environments with uncertain dynamics. We integrated adaptive control techniques with sampling-based motion planning algorithms to develop a two-layer framework that guarantees the safe navigation of the system to its goal. Finally, we developed a novel model-free adaptive control scheme that guarantees asymptotic stability of a class of nonlinear systems while respecting predefined funnel constraints.

\section{Future Research Directions}

Regarding the cooperative manipulation schemes of Chapter \ref{chapter:cooperative manip}, a strong assumption is that the agents  operate away from kinematic singularities (except for the NMPC frameworks). 
Future directions can aim at addressing this issue and guarantee singularity avoidance. Moreover, grasp reconfiguration that allows more modular schemes is a promising direction, as well as fault-tolerant extensions of the current schemes. Finally, real-time experiments with robotic agents with soft fingertip-type end-effectors should be attempted.  

Regarding Chapter \ref{chapter:formation}, future efforts can aim at extending the developed formation control scheme to more general multi-agent graph structures as well as taking into account collisions among all the agents. Moreover, one can notice that the control law for provably achieving zero internal forces in the rigid cooperative manipulation scheme, depending on the multi-agent rigidity matrix, is centralized. Therefore, one might aim at extending the proposed algorithm to a \textit{decentralized} scheme as well as considering compliant grasping contacts. Real-time experiments should be also conducted to showcase the validity of the proposed conditions. 

Regarding the multi-agent coordination of Chapter \ref{chapter:synthesis}, future works can address less conservative solutions for the multi-agent collision-free navigation. This can be also attempted by using the proposed leader-follower scheme and appropriate prioritization. Sampled inter-agent communication should be also used to resemble more realistic scenarios. 
Regarding the proposed ellipsoidal collision avoidance problem, one drawback that is required to be tackled in future works is the fact 
that the scheme interferes with the main assigned tasks, potentially causing them to fail (local minima scenarios). 

Regarding Chapter \ref{chapter:abstraction}, future directions are needed towards the generalization of the proposed multi-agent schemes to incorporate timed temporal specifications. Regarding the multi-agent-object hybrid scheme that is based on region-of-interest discretization, a decentralized extension must be considered. Moreover, failure of transition executions can be taken into account via
plan reconfiguration. Finally, future works should also focus on real-time experiments to further validate the proposed frameworks.

Finally, regarding Chapter \ref{chapter:single agent}, future works are required to focus mainly at extending the proposed frameworks to multi-agent schemes, as well as incorporation of input saturation constraints. 

As a final remark, another interesting topic of research is the consideration of delays in the multi-agent communication, as well as the sampled feedback and control realization, which have been neglected in this thesis. 
		
\appendix
\renewcommand{\chaptername}{Appendix}
	
\chapter{Dynamical Systems} \label{app:dynamical systems}
This Appendix provides preliminary background on the theory of dynamical systems. 
We consider both smooth and non-smooth systems. 

\section{Lipschitz Continuous Systems}

We start with defining standard results on the existence of solutions of ODEs.

Consider the initial value problem: 
\begin{equation}
\dot{{x}} = {h}({x},t), {x}(t_0)\in\Omega, \label{eq:initial value pr (App_dynamical_systems)}
\end{equation}
with ${h}:\Omega\times[t_0,\infty)\rightarrow\mathbb{R}^n$ where $\Omega\subset\mathbb{R}^n$ is a non-empty open set containing the origin, and $t_0\in \mathbb{R}_{\geq 0}$.
\begin{definition}
	\cite{sontag2013mathematical} A solution ${x}(t)$ of the initial value problem \eqref{eq:initial value pr (App_dynamical_systems)} is maximal if it has no proper right extension that is also a solution of \eqref{eq:initial value pr (App_dynamical_systems)}.
\end{definition}

\begin{theorem} \label{thm:ode solution (App_dynamical_systems)} \cite{bressan2007introduction}
	Let ${h}:\bar{\Omega}\coloneqq\Omega\times[t_0,\infty)\rightarrow\mathbb{R}^n$ from \eqref{eq:initial value pr (App_dynamical_systems)} satisfy the following conditions:
	\begin{enumerate}
		\item For every ${x}\in\mathbb{R}^n$, the function $t\to {h}({x},t)$ defined on $\Omega_x \coloneqq \{t : ({x},t)\in\bar{\Omega}\}$ is measurable. For every $t\in\mathbb{R}_{\geq 0}$, the function ${x}\to {h}({x},t)$ defined on $\Omega_t \coloneqq \{{x} : ({x},t)\in\bar{\Omega}\}$ is continuous.
		\item For every compact $K\subset \bar{\Omega}$, there exist constants $C_K, L_K$ such that 
		\begin{align*}
		&\|{h}({x},t) \| \leq C_K, \\
		&\|{h}({x},t)-{h}({y},t) \| \leq L_K \|{x}-{y} \|, 
		\end{align*}
		$\forall ({x},t),({y},t) \in K$.
	\end{enumerate}
	Then the initial value problem \eqref{eq:initial value pr (App_dynamical_systems)} with ${h}:\bar{\Omega}\rightarrow\mathbb{R}^n$ and
	some $x_0\in{\Omega}$, has a unique and maximal solution ${x}:[t_0,t_{\max})\to\mathbb{R}^n$, with $t_{\max} > t_0$ and $({x}(t),t)\in\bar{\Omega}, \forall t\in[t_0,t_{\max})$.
\end{theorem}

Note that the second condition imposed in the aforementioned theorem is a locally Lipschitz condition.

\begin{theorem} \label{thm:forward_completeness (App_dynamical_systems)} \cite{bressan2007introduction}
	Let the conditions of Theorem \ref{thm:ode solution (App_dynamical_systems)} hold in $\bar{\Omega}$ and let ${x}(t), t\in[t_0,t_{\max})$ be a maximal solution of the initial value problem \eqref{eq:initial value pr (App_dynamical_systems)}. Then, either $t_{\max} = \infty$ or 
	\begin{equation*}
	\lim\limits_{t\to t^{-}_{\max}}\Big(\| {x}(t) \| + \frac{1}{d_\mathcal{S}(({x}(t),t),\partial \bar{\Omega})} \Big) = \infty,
	\end{equation*}
	where $d_{\mathcal{S}}: \mathbb{R}^n\times2^{\mathbb{R}^n}$ is the distance of a point ${x}\in\mathbb{R}^n$ to a set $A$, defined as $d_{\mathcal{S}}({x},A) \coloneqq \inf\limits_{{y}\in A}\{\|{x}-{y}\|\}$.
\end{theorem}

\begin{definition} 
	The origin $x=0$ is the equilibrium point for \eqref{eq:initial value pr (App_dynamical_systems)} if 
	\begin{equation*}
		h(0,t) = 0, \forall t \in [t_0,\infty) 
	\end{equation*}
\end{definition}

We next provide the comparison function definitions, necessary for the stability classification of the equilibrium point.

\begin{definition} \label{def:class_K (App_dynamical_systems)}
	(\cite{khalil_nonlinear_systems,krstic1995nonlinear}) A continuous function $\alpha: [0, a) \to \mathbb{R}_{\ge 0}$ is said to belong to \emph{class} $\mathcal{K}$, if it is strictly increasing and $\alpha(0) = 0$. It is said to belong to class $\mathcal{K}_{\infty}$ if $a = \infty$ and $\lim_{r\to\infty}\alpha(r) = \infty$.
\end{definition}

\begin{definition} \label{def:class_KL (App_dynamical_systems)}
	(\cite{khalil_nonlinear_systems,krstic1995nonlinear}) A continuous function $\beta: [0, a) \times \mathbb{R}_{\ge 0} \to \mathbb{R}_{\ge 0}$ is said to belong to \emph{class} $\mathcal{KL}$, if:
	\begin{itemize}
		\item For each fixed $s$, $\beta(r, s) \in \mathcal{K}$ with respect to $r$.
		\item For each fixed $r$, $\beta(r, s)$ is decreasing with respect to $s$ and $\lim_{s\to\infty}\beta(r, s) = 0$.
	\end{itemize}
	It is said to belong to \emph{class} $\mathcal{KL}_\infty$ if, in addition, for each fixed $s$, the mapping $\beta(r,s)$ belongs to class $\mathcal{K}_\infty$ with respect to r.
\end{definition}

Now we can characterize the equilibrium point of \eqref{eq:initial value pr (App_dynamical_systems)} with respect to its stability.

\begin{definition} (\cite{khalil_nonlinear_systems,krstic1995nonlinear})
	The equilibrium point $x=0$ of \eqref{eq:initial value pr (App_dynamical_systems)} is 
	\begin{itemize}
		\item uniformly stable, if there exists a class $\mathcal{K}$ function $\gamma(\cdot)$ and a positive constant $c$ independent of $t_0$, such that 
		\begin{equation} \label{eq:uniformly stable (App_dynamical_systems)}
			\|x(t)\| \leq \gamma(\|x(t_0)\|), \forall t\geq t_0,  \| x(t_0) \| < c,
		\end{equation}
		\item uniformly asymptotically stable, if there exists a class $\mathcal{KL}$ function $\beta(\cdot,\cdot)$ and a positive constant $c$ independent of $t_0$, such that 
		\begin{equation} \label{eq:uniformly as stable (App_dynamical_systems)}
			\|x(t)\| \leq \beta(\|x(t_0)\|,t-t_0), \forall t\geq t_0, \|x(t_0)\| < c,
		\end{equation}
		\item exponentially stable, if \eqref{eq:uniformly as stable (App_dynamical_systems)} is satisfied with $\beta(r,s) = kr \exp(-\alpha s)$, $k,\alpha \in \mathbb{R}_{\geq 0}$,
		\item globally uniformly stable, if \eqref{eq:uniformly stable (App_dynamical_systems)} is satisfied with $\gamma \in \mathcal{K}_\infty$ for any initial state $x(t_0)$ and $\Omega = \mathbb{R}^n$,
		\item globally uniformly asymptotically stable, if \eqref{eq:uniformly as stable (App_dynamical_systems)} is satisfied with $\beta \in \mathcal{KL}_\infty$ for any initial state $x(t_0)$,
		\item globally exponentially stable, if \eqref{eq:uniformly as stable (App_dynamical_systems)} is satisfied for any initial state $x(t_0)$ and with $\beta(r,s) = kr \exp(-\alpha s)$, $k,\alpha \in \mathbb{R}_{\geq 0}$.
	\end{itemize}
\end{definition}

The main Lyapunov stability theorem is then formulated as follows:
\begin{theorem} (\cite{khalil_nonlinear_systems,krstic1995nonlinear})
	Let $x=0$ be an equilibrium point of \eqref{eq:initial value pr (App_dynamical_systems)}. Let $V: \Omega \times [t_0,\infty) \to \mathbb{R}_{\geq 0} $ be a continuously differentiable function such that, $\forall t \geq t_0$, $x\in \Omega$, 
	\begin{align*}
		& \gamma_1(\|x\|) \leq V(x,t) \leq \gamma_2(\|x\|), \\
		& \frac{\partial V}{\partial t} + \frac{\partial V}{\partial x} h(x,t) \leq -\gamma_3(\|x\|).
	\end{align*}
	Let $r \in\mathbb{R}$ such that $\mathcal{B}(0,r) \subset \Omega$. Then, the equilibrium point $x=0$ of \eqref{eq:initial value pr (App_dynamical_systems)} is 
	\begin{itemize}
		\item uniformly stable, if $\gamma_1$ and $\gamma_2$ are class $\mathcal{K}$ functions on $[0,r)$ and $\gamma_3(\cdot) \geq 0$ on $[0,r)$,
		\item uniformly asymptotically stable, if $\gamma_1$, $\gamma_2$, and $\gamma_3$ are class $\mathcal{K}$ functions on $[0,r)$,
		\item exponentially stable, if $\gamma_i(\rho) = k_i \rho^\alpha$ on $[0,r)$, $k_i$ $\alpha \in\mathbb{R}_{>0}$, $\forall i\in\{1,2,3\}$,
		\item globally uniformly stable, if $\Omega = \mathbb{R}^n$, $\gamma_1$ and $\gamma_2$ are class $\mathcal{K}_\infty$ functions, and $\gamma_3(\cdot) \geq 0$ on $\mathbb{R}_{\geq 0}$,
		\item globally uniformly asymptotically stable if $\Omega = \mathbb{R}^n$, $\gamma_1$ and $\gamma_2$ are class $\mathcal{K}_\infty$ functions, and $\gamma_3$ is a class $\mathcal{K}$ function on $\mathbb{R}_{\geq 0}$,
		\item globally exponentially stable, if $\Omega = \mathbb{R}^n$, $\gamma_i(\rho) = k_i \rho^\alpha$ on $\mathbb{R}_{\geq 0}$, $k_i$ $\alpha \in\mathbb{R}_{>0}$, $\forall i\in\{1,2,3\}$,
	\end{itemize}
\end{theorem}

We provide next standard invariance results for time-invariant and time-varying systems.


\begin{theorem}(LaSalle \cite{khalil_nonlinear_systems,krstic1995nonlinear}) \label{theorem:LaSalle (App_dynamical_systems)}
	Let $\Omega \subset \mathbb{R}^n$ be a positive invariant non-empty set of the time-invariant ODE $\dot{x} = h_I(x)$, where $h_I:\Omega\to\mathbb{R}^n$ is continuous and satisfies condition 2 of Theorem \eqref{thm:ode solution (App_dynamical_systems)}.  Let $V : \Omega \to \mathbb{R}_{\geq 0}$ be a continuously differentiable function $V(x)$ such that $\dot{V}(x) \leq 0$, $\forall x \in \Omega$. Let $E\coloneqq \{ x\in\Omega : \dot{V}(x) = 0 \}$, and let $M$ be the largest invariant set contained in $E$. Then, every bounded solution $x(t)$ starting in $\Omega$ converges to $M$ as $t\to \infty$.
\end{theorem}

\begin{lemma}(Barbalat \cite{khalil_nonlinear_systems,krstic1995nonlinear}) \label{lemma:barbalat (App_dynamical_systems)}
	Let $\phi:\mathbb{R}\to\mathbb{R}$ be a uniformly continuous function on $[0,\infty)$. Suppose that $\lim_{t\to\infty}\int_0^t \phi(\tau) d\tau$ exists and is finite. Then,
	\begin{equation*}
		\lim_{t\to\infty} \phi(t) = 0.
	\end{equation*}
\end{lemma}

We conclude the results for smooth systems with the standard ultimate boundedness theorem.

\begin{theorem} \label{th:uub_khalil (App_dynamical_systems)}
	(\cite{khalil_nonlinear_systems,krstic1995nonlinear})
	Let $x=0$ be an equilibrium point of \eqref{eq:initial value pr (App_dynamical_systems)}. Let $V: \Omega \times [t_0,\infty) \to \mathbb{R}$ be a continuously differentiable function such that 
	\begin{align*}
		& \gamma_1(\|x\|) \le V(x) \le \gamma_2(\|x\|) \\ 
		& \frac{\partial V}{\partial t} + \frac{\partial V}{
			\partial x}h(x,t) \le - W(x), \forall \|x\| \ge \mu > 0,
	\end{align*}
	$\forall t \ge 0$, $x \in \Omega$, where $\gamma_1$, $\gamma_2$ are class $\mathcal{K}$ functions and $W$ is a continuous positive definite function. Take $r > 0$ such that $\mathcal{B}(0,r) \subseteq \Omega$ and suppose that $\mu < \gamma_2^{-1}(\gamma_1(r))$. Then, there exist a class $\mathcal{K}_{\infty}$ function $
	\gamma_3$ and for every initial state $x(t_0)$ satisfying $\|x(t_0)\| \le \gamma_2^{-1}(\gamma_1(r))$, there exists a $T \ge 0$ such that 
	\begin{align}
	& \lVert x(t) \rVert \leq \gamma_3(\lVert x(t_0) \rVert), \forall \ t_0 \le t \leq T, \notag \\
	& \lVert x(t) \rVert \leq \gamma^{-1}_1(\gamma_2(\mu)), \forall t > T. \notag
	\end{align}
	Moreover, if $\Omega = \mathbb{R}^n$ and $\gamma_1$ belongs to class $\mathcal{K}_\infty$, then the aforementioned result holds for any initial state $x(t_0)$, with no restriction on how large $\mu$ is.
\end{theorem} 

Note that the aforementioned results also apply for the case where $x$ evolves in a manifold, by changing the $\|\cdot\|$ metric to the respective manifold one.

\section{Systems with Discontinuous Right-Hand-Side}

This section provides some equivalent results for non-smooth systems. 

Consider the following differential equation with a discontinuous
right-hand side:
\begin{equation} \label{eq:discont diff eq (App_dynamical_systems)}
\dot{x} = h(x,t),
\end{equation}
{where $h:\Omega \times [t_0,\infty) \to \mathbb{R}^n$, $\Omega\subset \mathbb{R}^n$, is Lebesgue measurable and locally essentially bounded, uniformly in $t$.	The Filippov regularization of $f$ is defined as \cite{paden1987calculus}}
\begin{equation} \label{eq:Filippov regular. (App_dynamical_systems)}
\mathsf{K}[f](x,t) \coloneqq \bigcap_{\delta > 0}\bigcap_{\mu(\bar{N})=0} \overline{\text{co}}(f(\mathcal{B}(x,\delta) \backslash \bar{N}),t),
\end{equation}
where $\bigcap_{\mu(\bar{N})=0}$ is the intersection over all sets $\bar{N}$ of Lebesgue measure zero, and $\overline{\text{co}}(E)$ is the convex closure of a set $E$. We are interested in the Filippov solutions of \eqref{eq:discont diff eq (App_dynamical_systems)}: 

\begin{definition}[\cite{fischer2013lasalle}] \label{def:Filipp sol (App_dynamical_systems)}	
	A function $x:[t_0,t_1)\to\mathbb{R}^n$, with $t_1 > t_0$, is called a Filippov solution of \eqref{eq:discont diff eq (App_dynamical_systems)} on $[t_0,t_1)$ if $x(t)$ is absolutely continuous and if, for almost all $t\in[t_0,t_1)$, it satisfies $\dot{x} \in \mathsf{K}[h](x,t)$, where $\mathsf{K}[h](x,t)$ is the Filippov regularization of $h(x,t)$. 
\end{definition}

The existence of Filippov solutions is given next.
\begin{proposition}[\cite{cortes2008discontinuous}] \label{prop:Filippov exist (app_dynamical_systems)}
	Let $\dot{x} \in \mathsf{K}[h](x,t)$, where $\mathsf{K}[h](x,t)$ is the Filippov regularization of $h(x,t)$. Let also $h(x,t)$ be measurable and locally essentially bounded in $x$ over $\Omega$, and measurable in $t$ over $[t_0,\infty)$. Then, there exists a Fillipov solution $x:[t_0,t_1) \to \mathbb{R}^n$ of \eqref{eq:discont diff eq (App_dynamical_systems)}. 	
\end{proposition}

We next provide the definitions for regular functions and generalized gradients.

\begin{definition}[\cite{fischer2013lasalle}] \label{def:Directional derivative (App_dynamical_systems)}	
	Given a function $h :\mathbb{R}^m \to \mathbb{R}^n$, the right directional derivative of $h$ at $x\in\mathbb{R}^m$ in the direction of $v\in\mathbb{R}^m$ is defined as 
	\begin{equation*}
		f'(x,v) \coloneqq \lim_{t\to 0^+} \frac{f(x+tv)-f(x)}{t}.
	\end{equation*}
	Additionally, the generalized directional derivative of $h$ at $x$ in the direction of $v$ is defined as 
	\begin{equation*}
	f^o(x,v) \coloneqq \lim_{y\to x} \sup_{t\to 0^+} \frac{f(x+tv)-f(y)}{t}.
	\end{equation*}
\end{definition}

\begin{definition}[\cite{fischer2013lasalle}] \label{def:Regular function (App_dynamical_systems)}	
	A function $h:\mathbb{R}^m \to \mathbb{R}^n$ is said to be regular at $x\in\mathbb{R}^m$ if for all $v\in\mathbb{R}^m$, the right directional derivative of $h$ at $x$ in the direction of $v$ exists and $f'(x,v) = f^o(x,v)$.
\end{definition}

\begin{definition}[\cite{fischer2013lasalle}] \label{def:Clarkes gen grad (App_dynamical_systems)}	
	For a function $V:\mathbb{R}^n \times [t_0,\infty) \to \mathbb{R}$ that is locally Lipschitz in $(x,t)$, define the generalized gradient of $V$ at $(x,t)$ by 
	\begin{equation*}
		\partial V(x,t) \coloneqq \overline{\text{co}}\left\{ \lim \nabla V(x,t) : (x_i,t_i) \to (x,t), (x_i,t_i) \notin \Omega_V \right\},
	\end{equation*}
	where $\Omega_V$ is the set of measure zero where the gradient of $V$ is not defined.
\end{definition}

\begin{lemma}[\cite{fischer2013lasalle}] \label{lem:Chain rule (App_dynamical_systems)}
	Let $x(t)$ be a Filippov solution of \eqref{eq:discont diff eq (App_dynamical_systems)} and $V:\Omega\times [t_0,t_1)\to\mathbb{R}$ be a locally Lipschitz, regular function. Then $V(x(t),t)$ is absolutely continuous, $\dot{V}(x(t),t)=\frac{\partial}{\partial t}V(x(t),t)$ exists almost everywhere (a.e.), i.e., for almost all $t\in [t_0,t_1)$, and $\dot{V}(x(t),t) \overset{\text{a.e}}{\in} \dot{\widetilde{V}}(x(t),t)$, where
	\begin{equation*}
	\dot{\widetilde{V}} \coloneqq \bigcap_{\xi\in\partial V(x,t)} \xi^\top \begin{bmatrix}
	\mathsf{K}[f](x,t) \\ 1
	\end{bmatrix}.
	\end{equation*} 
\end{lemma}

Finally, we provide the main invariance and stability result for the non-smooth type \eqref{eq:discont diff eq (App_dynamical_systems)}.

\begin{theorem}[\cite{fischer2013lasalle}] \label{th:nonsmooth LaSalle (App_dynamical_systems)}
	For the system given in \eqref{eq:discont diff eq (App_dynamical_systems)}, let $\Omega\subset\mathbb{R}^n$ be an open and connected set containing $x=0$ and suppose that $f$ is Lebesgue measurable and $x\mapsto f(x,t)$ is essentially locally bounded, uniformly in $t$. Let $V:\Omega\times [t_0,t_1) \to\mathbb{R}$ be locally Lipschitz and regular such that $W_1(x) \leq V(x,t) \leq W_2(x)$, $\forall t\in [t_0,t_1)$, $x\in\Omega$, and
	\begin{equation*}
	z \leq -W(x(t)), \ \ \forall z\in \dot{\widetilde{V}}(x(t),t), \ t\in [t_0,t_1), \ x \in \Omega,
	\end{equation*}
	where $W_1$ and $W_2$ are continuous positive definite functions and $W$ is a continuous positive semi-definite  on $\Omega$. Choose $r>0$ and $c>0$ such that $\bar{\mathcal{B}}(0,r) \subset \Omega$ and $c<\min_{\|x\|=r} W_1(x)$. Then for all Filippov solutions $x:[t_0,t_1)\to\mathbb{R}^n$ of \eqref{eq:discont diff eq (App_dynamical_systems)}, with $x(t_0)\in {\mathbb{D}} \coloneqq\{x\in \bar{\mathcal{B}}(0,r) : W_2(x) \leq c\}$, it holds that $t_1 = \infty$, $x(t)\in {\Omega}$, $\forall t\in[t_0,\infty)$, and  $\lim_{t\to\infty} W(x(t)) = 0$.
\end{theorem}

\section{Reduction Principle}

Consider now the autonomous time-invariant version of \eqref{eq:initial value pr (App_dynamical_systems)} 
\begin{equation} \label{eq:ode autonomous systems (App_dynamical_systems)}
	\dot{x} = h_I(x),
\end{equation}
where $h_I : \Omega \to \mathbb{R}^n$ is sufficiently smooth, and $x = 0$ is an equilibrium point, i.e., $h_I(0) = 0$. Let $A \coloneqq \left.\frac{\partial h_I}{\partial x}\right|_{x = 0}$ be the respective Jacobian matrix with $n_{+}$ eigenvalues with positive real part, $n_{-}$ eigenvalues with negative real part, and $n_0$ eigenvalues with zero real part. Let $T^c$ denote the linear (generalized) eigenspace of $A$ corresponding to the union of the $n_0$ eigenvalues on the imaginary axis. The next theorem provides the center manifold existence around the equilibrium point.

\begin{theorem}[\cite{kuznetsov2013elements}] 
	There is a locally defined smooth $n_0$-dimensional invariant manifold $W^c(0)$ of \eqref{eq:ode autonomous systems (App_dynamical_systems)} that is tangent to $T^c$ at $x=0$. 
\end{theorem}

System \eqref{eq:ode autonomous systems (App_dynamical_systems)} can be written as 
\begin{subequations} \label{eq:eigenbasis system (App_dynamical_systems)}
\begin{align}
	\dot{u} = B u + g(u,v), \\
	\dot{v} = C v  + h(u,v), 
\end{align}
\end{subequations}
where $u \in \mathbb{R}^{n_0}$, $v\in\mathbb{R}^{n_+ + n}$, $B \in \mathbb{R}^{n_0\times n_0}$ is a matrix with all its $n_0$ eigenvalues on the imaginary axis, while $C \in \mathbb{R}^{(n_+ + n_i)\times(n_+ + n_i)}$ has no eigenvalue on the imaginary axis. The functions $g$ and $h$ have Taylor expansions starting with at least quadratic terms. The center manifold $W^c$ of system \eqref{eq:eigenbasis system (App_dynamical_systems)} can be locally represented as a graph of a smooth function $V(\cdot)$ as $W^c = \{(u,v): v = V(u)\}$. 

\begin{theorem}[\cite{kuznetsov2013elements}] \label{theorem:reduction pr (App_dynamical_systems)}
	System \eqref{eq:eigenbasis system (App_dynamical_systems)} is locally topologically equivalent near the origin to the system 
	\begin{align*}
		\dot{u} &= Bu + g(u,V(u)),\\ 
		\dot{v} &= Cv
	\end{align*}
\end{theorem}


\chapter{Funnel Control} \label{app:PPC}
This Appendix provides preliminary background on funnel control and in particular, Prescribed Performance Control (PPC).

Funnel control describes the behavior where the output of the system is confined to a predefined funnel, as depicted in Fig. \ref{fig:funnel control (App_PPC)}. A special instance of funnel control, which this thesis focuses on, is Prescribed Performance Control, proposed in \cite{bechlioulis2008robust}, and describes the behavior where a tracking error $e:[t_0,\infty)\rightarrow\mathbb{R}$, with $t_0\in\mathbb{R}_{\geq 0}$, evolves strictly within a predefined region that is bounded by certain functions of time, achieving prescribed transient and steady-state performance.
The mathematical expression of prescribed performance is given by the inequalities:
\begin{equation}
-\rho_L(t) < e(t) < \rho_U(t),\ \ \forall t\geq t_0, \notag
\end{equation} 
where $\rho_L(t),\rho_U(t)$ are smooth and bounded functions of time satisfying $\lim\limits_{t\rightarrow\infty}\rho_L(t) > 0$ and $\lim\limits_{t\rightarrow\infty}\rho_U(t) > 0$, called performance functions. 

Although the functions $\rho_L(t),\rho_U(t)$ can be any bounded functions, it is usually preferred to be decaying functions. A particular interesting instance is the choice of 
the exponential performance functions $\rho_i(t) \coloneqq (\rho_{i, \scriptscriptstyle 0}-\rho_{i,\scriptscriptstyle \infty})\exp(-l_it)+\rho_{i,\scriptscriptstyle \infty}$, with $\rho_{i,\scriptscriptstyle 0},\rho_{i,\scriptscriptstyle \infty}, l_i\in\mathbb{R}_{>0}, i\in\{U,L\}$, appropriately chosen constants, the terms $\rho_{L,\scriptscriptstyle 0} \coloneqq\rho_L(0),\rho_{U,\scriptscriptstyle 0}\coloneqq\rho_U(0)$ are selected such that $\rho_{U,\scriptscriptstyle 0} > e(0) > \rho_{L,\scriptscriptstyle 0}$ and the terms $\rho_{L,\scriptscriptstyle \infty} \coloneqq\lim\limits_{t\rightarrow\infty}\rho_L(t),\rho_{U,\scriptscriptstyle \infty}\coloneqq\lim\limits_{t\rightarrow\infty}\rho_U(t)$ represent the maximum allowable size of the tracking error $e(t)$ at steady state, which may be set arbitrarily small to a value reflecting the resolution of the measurement device, thus achieving practical convergence of $e(t)$ to zero. Moreover, the decreasing rate of $\rho_L(t),\rho_U(t)$, which is affected by the constants $l_L, l_U$ in this case, introduces a lower bound on the required speed of convergence of $e(t)$. Therefore, the appropriate selection of the performance functions $\rho_L(t),\rho_U(t)$ imposes performance characteristics on the tracking error $e(t)$. 

The systems considered are general nonlinear systems of the form 
\begin{equation*}
	\dot{x} = f(x,t) + g(x,t)u
\end{equation*}
where $f,g: \mathbb{R}^n \times [t_0,\infty) \to \mathbb{R}^n$ are unknown functions satisfying certain continuity and controllability properties. 
We describe briefly the control design as well as the intuition behind it.
Firstly, given a desired signal $x_\text{d}$ and an error $e = x - x_\text{d}\in\mathbb{R}^n$, as well as a choice of performance functions $\rho_j(t)$, for each component $j\in\{1,\dots,n\}$, we define the normalized error 
\begin{equation*}
	\xi \coloneqq [\xi_1,\dots,\xi_n]^\top \coloneqq P^{-1} e,
\end{equation*}
where $P \coloneqq \text{diag}\{ [\rho_i]_{i\in\{1,\dots,n\}}\}$, as well as the transformation $\varepsilon:(-1,1)^n \to \mathbb{R}^n$, with
\begin{equation*}
	\varepsilon \coloneqq \begin{bmatrix}
	\ln\left(\frac{1 + \xi_{1}}{1 - \xi_{1}}\right), \dots, \ln\left(\frac{1 + \xi_n}{1 - \xi_n}\right)
	\end{bmatrix}^\top.
\end{equation*}
Intuitively, in order to guarantee $\xi \in (-1,1)^n$, we need to guarantee that $\varepsilon$ is bounded. This is attempted via the choice of control law 
\begin{equation*}
	u = -k P^{-1} \frac{\partial \varepsilon}{\partial \xi} \varepsilon,
\end{equation*} 
which acts as a barrier function, i.e., it increases to infinity as $\xi$ approaches the boundary of $(-1,1)^n$. 
Local existence of solutions guarantees that there exists a maximal time $\tau > t_0$ such that $\xi(t) \in (-1,1)^n$, $\forall t\in[t_0,\tau)$. 
Differentiating thus the well-defined (for $t\in[t_0,\tau)$) Lyapunov function candidate $V \coloneqq \frac{1}{2}\|\varepsilon\|^2$ yields
\begin{align*}
 	\dot{V} =& \varepsilon^\top \frac{\partial \varepsilon}{\partial \xi}^\top P^{-1}(f(x,t) - kg(x,t)P^{-1}\frac{\partial \varepsilon}{\partial \xi}\varepsilon - \dot{x}_\text{d} - \dot{P}\xi)  \\
 	\leq & -k \varepsilon^\top  \frac{\partial \varepsilon}{\partial \xi}^\top P^{-1} g(x,t) P^{-1}\frac{\partial \varepsilon}{\partial \xi}  \varepsilon + \left\| P^{-1}\frac{\partial \varepsilon}{\partial \xi}\varepsilon \right\| \left\| f(x,t) - \dot{x}_\text{d} - \dot{P}\xi  \right\|.
\end{align*}
By assuming Lipschitz $f(\cdot,t)$, uniform bounded $f(x,\cdot)$, and positive definite $g(x,t)$, we use the boundedness of $\xi(t) \in (-1,1)^n$, $\forall t\in[t_0,\tau)$ to conclude that the second term above is bounded by a constant $\bar{F}$, $\forall t\in[t_0,\tau)$. Hence it turns out that $\dot{V} < 0$ when $\left\| \frac{\partial \varepsilon}{\partial \xi} \varepsilon \right\| > \frac{F}{k\lambda_{\min}(g(x,t))}$, from which we can conclude the ultimate boundedness of $\varepsilon(t)$ in a compact set, and hence that $\tau = \infty$.

\begin{figure}
	\centering
	\begin{tikzpicture}
	\begin{axis}[xmax=15,xmin=0,ymin=-5,ymax= 5,xlabel={$t$},
	samples=50]
	\addplot[domain=0:15, red, thick] ({\x}, {4*exp(-\x)+0.4 + 1.5*exp(-(\x-10)^2/5) }) ;
	\addplot[domain=0:15, teal, thick] ({\x}, {-4*exp(-\x)-0.4 - 1.5*exp(-(\x-10)^2/5) }) ;
	\addplot[domain=0:15, blue, thick]  ({\x}, {2*exp(-\x) - 0.015*\x^2 + 0.1*x +  .1*rand + 1.5*exp(-(\x-15)^2/5) + 1.25*exp(-(\x-11.5)^2/5)}      );
	\addplot[domain=0:15, black, dashed] ({\x},{0});
	\legend{$\rho_U(t)$,$\rho_L(t)$,$e(t) $}
	\end{axis}
	\end{tikzpicture}
	\caption{Illustration of funnel control, where the error $e(t)$ is confined in the prescribed funnel defined by the functions $\rho_L(t)$, $\rho_U(t)$. \label{fig:funnel control (App_PPC)}}
\end{figure}
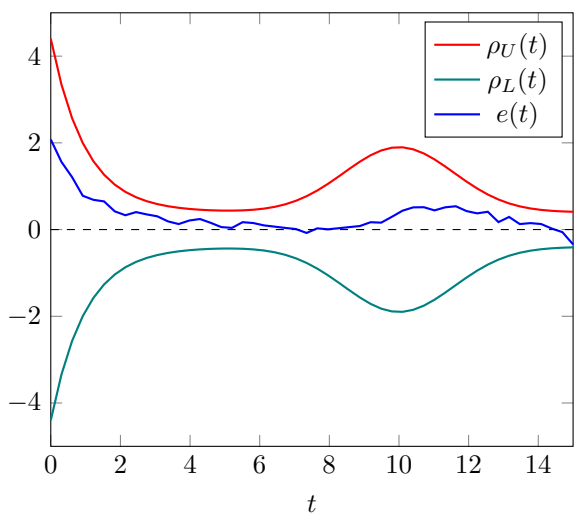

\chapter{Navigation Functions} \label{app:NF}
This Appendix provides preliminary background on navigation functions.

Navigation functions, initially proposed in \cite{koditschek1992robot} for single-point-sized robot navigation, are real-valued maps realized through cost functions, whose negated gradient field is attractive towards the goal configuration (referred to as the good or desirable set) and repulsive with respect to the obstacles set (referred to as the bad set which we want to avoid). We provide here a brief overview of the multi-agent versions introduced in \cite{Loizou2006} and \cite{dimarogonas2007decentralized}.

\section{Multirobot Navigation Functions (MRNFs)} \label{subsec:MAS NF (App:NF)}
Consider $N\in\mathbb{N}$ spherical robots, with center ${q}_i\in\mathbb{R}^n$, $n\in\mathbb{N}$, and radius $r_i \in\mathbb{R}_{>0}$, i.e., $\bar{\mathcal{B}}({q}_i,r_i)$, $i\in\mathcal{N}$, operating in an open spherical workspace $\mathcal{W}\coloneqq {\mathcal{B}}({0},r_0)$ of radius $r_0 \in\mathbb{R}_{>0}$. Each robot has a destination point ${q}_{\text{d}_i}\in\mathbb{R}^n, i\in\mathcal{N}$, and ${q}_{\text{d}} \coloneqq [{q}^\top_{\text{d}_1},\dots, {q}^\top_{\text{d}_N}]^\top$.
Let $\mathcal{F}\subset \mathbb{R}^n$ be a compact connected analytic manifold with boundary. A map $\varphi: \mathcal{F}\to[0,1]$ is a Multirobot Navigation Function (MRNF) if 
\begin{enumerate}
	\item It is analytic on $\mathcal{F}$,
	\item It has only one minimum at ${q}_\text{d}\in \text{Int}(F)$,
	\item Its Hessian at all critical points is full rank,
	\item $\lim\limits_{{q}\to \partial \mathcal{F}} = 1 > \varphi({q}')$, $\forall {q}'\in \text{Int}(F)$,
\end{enumerate}
where ${q}\coloneqq [{q}^\top_1,\dots,{q}^\top_N]^\top\in\mathbb{R}^{Nn}$. The class of MRNFs has the form
\small
\begin{equation*}
\varphi({q}) \coloneqq \frac{\gamma({q})}{\Big(\gamma({q})^{\kappa} + G({q}) \Big)^{\tfrac{1}{\kappa}}},
\end{equation*}
\normalsize
where $\gamma({q}) \coloneqq \| {q} - {q}_{\text{d}} \|^2$ is the goal function, $G({q})$ is the obstacle function, and $\kappa$ is a tunable gain; $\gamma^{-1}(0)$ denotes the desirable set and $G^{-1}(0)$ the set we want to avoid. Next we provide the procedure for the construction of the function $G$.

A robot proximity function, a measure for the distance between two robots $i,l\in\mathcal{N}$, is defined as $\beta_{i,l}({q}_i,{q}_l) \coloneqq \|{q}_i - {q}_l\|^2 - (r_i + r_l)^2$, $\forall i,l\in\mathcal{N}, i\neq l$. The term \textit{relation} is used to describe the possible collision schemes that can be defined in a multirobot team, possibly including obstacles.
The \textit{set of relations} between the members of the team can be defined as the set of all possible collision schemes between the members of the team. A binary relation is a relation between two robots. Any relation can be expressed as a set of binary relations. A \textit{relation tree} is the set of robot/obstacles that form a linked team. Each relation may consist of more than one relation tree. The number of binary  relations in  a  relation is called \textit{relation level}. Illustrative examples can be found in \cite{Loizou2006}. A \textit{relation proximity function} (RPF) provides a measure of the distance between the robots involved in a relation. Each relation has its own RPF. A RPF is the sum of the robot proximity functions of a relation.
It assumes the value of zero whenever the related robots collide (since the involved robot proximity functions will be zero) and increases with respect to the distance of the related robots. The RPF of relation $j$ at level $k$ is given by $(b_{R_j})_k \coloneqq \sum\limits_{(i,m)\in(R_j)_k}\beta_{i,m}$, where we omit the arguments ${q}_i,{q}_k$ for notational brevity.  A \textit{relation verification function} (RVF) is defined as 
\small
\begin{equation*}
g_{R_j} \coloneqq (b_{R_j})_k + \lambda \frac{ (b_{R_j})_k}{(b_{R_j})_k + ( B_{(R^C_j)_k} )^{\tfrac{1}{h}}}, 
\end{equation*}
\normalsize
where $\lambda, h > 0$, and $R^C_j$ is the complementary to $R_j$ set of relations in the same level $k$, $j$ is an index number defining the relation in level $k$, and $B_{R^C_j} \coloneqq \prod\limits_{m\in R^C_j} b_m$. The RVF serves as an analytic switch, which goes to zero only when the relation it represents is realized. By further introducing the workspace boundary obstacle functions as $G_0 \coloneqq \prod_{i\in\mathcal{N}}\Big\{ (r_0 -r_i)^2 - \|{q}_i\|^2 \Big\}$, we can define 
$G \coloneqq G_0\prod_{L=1}^{n_L}\prod_{j=1}^{n_{R,L}}(g_{R_j})_L,$
where $n_L$ is the number of levels and $n_{R,L}$ the number of relations in level $L$. It has been proved that, by choosing the parameter $\kappa$ large enough, the negated gradient field $-\nabla_{{q}}\varphi({q})$ leads to the destination configuration ${q}_{\text{d}}$, from almost all initial conditions \cite{Loizou2006}.

\section{Decentralized Navigation Functions (DNFs)} \label{subsec: DNF (App:NF)}

Consider now the class of \textit{decentralized navigation functions}, which has the form $\varphi_i:\mathcal{F}_i\rightarrow[0,1]$, with $\varphi_i({q}) \coloneqq \dfrac{\gamma_i({q_i}) + f_i(G_i)}{ (\gamma_i({q_i})^{\lambda_i} + G_i({q}))^{1/\kappa_i} }$. The key difference in this case is the term $G_i:\mathbb{R}^{3N}\rightarrow\mathbb{R}$ that is associated with the collision avoidance property of agent $i$ with the rest of the team and is based on the inter-agent \textit{decentralized} distance function \cite{dimarogonas2007decentralized}: $\beta_{ij}:\mathbb{R}^3\times\mathbb{R}^3\rightarrow\mathbb{R}$ with 
\begin{equation}
\beta_{ij}({p_i},{p_j}) \coloneqq \left\{ \begin{array}{ll} \lVert {p_i}-{p_j} \rVert^2-(r_i+r_j)^2, & \text{if } j\in\mathcal{N}_i\\ 
d^2_{s_i} -(r_i+r_j)^2, & \text{if } j\notin\mathcal{N}_i, \end{array} \right. \nonumber
\end{equation}
that represents the distance between agents $i$ and $j\in\mathcal{N}_i$. The term $f_i:\mathbb{R}\rightarrow\mathbb{R}$ is used in order to avoid inter-agent collisions in case one or more agents that take part in a collision scheme are very close to their goals. Note that in that case, the classical form of $\varphi_i$ would yield values very close to $0$, since agent $i$ is very close to its goal, without actively taking part in avoiding potential collisions. The term $f_i$, therefore, forces agent $i$
to avoid potential collisions.
Analytic expressions for $G_i$ and $f_i$ can be found in \cite{dimarogonas2007decentralized}. With the aforementioned tools, the control law for agent $i$ is ${u_i} = -k_i\dfrac{\partial \varphi_i({p})}{\partial {p_i}}$, which, as shown in \cite{dimarogonas2007decentralized}, drives all agents to their goal positions and guarantees inter-agent collision-avoidance. 	

\chapter{Nonlinear Model Predictive Control} \label{app:NMPC}
This Appendix provides preliminary background on Nonlinear Model Predictive Control (NMPC). 

Nonlinear Model Predictive Control has become an attractive feedback control strategy in applications where, except for stabilization or trajectory tracking of a system, certain state and input constraints must be also satisfied \cite{findeisen2002introduction,frank_1998_quasi_infinite,frank_2003_nmpc_bible}. In general, the NMPC is formulated as solving at each sampling time step an online Finite Horizon Optimal Control Problem (FHOCP) subject to system dynamics and state and input constraints. Based on measurements obtained at each sampling time step, the controller predicts the dynamic behavior of the system over a predictive horizon in the future and determines the input such that a predetermined open-loop performance objective is minimized. In order to incorporate feedback, the optimal open-loop input is implemented only until the next sampling time step. Using the new system state at the next sampling time step, the whole procedure (prediction and optimization) is repeated, moving the control and prediction horizon forward. Summarizing, a standard NMPC scheme works as follows 
\begin{enumerate}
	\item Obtain estimates of the states of the system.
	\item Calculate a constraint-conforming optimal input minimizing the desired cost function over the prediction horizon using the system model and the current state estimate for prediction.
	\item Implement the first part of the optimal input until the next sampling time step. 
	\item Go to step 1. 
\end{enumerate}

More technically, consider the stabilization problem for a class of systems described by the following nonlinear dynamical system:
\begin{equation} \label{eq:ode (App_NMPC)}
	\dot{x} = f(x,u),
\end{equation}
with $f:\mathbb{R}^n \times \mathbb{R}^m \to \mathbb{R}^n$ is locally Lipschitz continuous and satisfies $f(0,0) = 0$. The system is subject to the following input and state constraints:
\begin{equation*}
	u \in \mathcal{U} \subset \mathbb{R}^m, x \in \mathcal{X} \subset \mathbb{R}^n,
\end{equation*}
respectively, where $\mathcal{U}$ is compact and $\mathcal{X}$ is connected, and $(0,0)\in\mathcal{X}\times\mathcal{U}$. Denote by $h\in\mathbb{R}_{>0}$, $T_p > h$ the sampling step and the finite prediction horizon, respectively. Consider a sequence of sampling times $\{t_j\}, j\in\mathbb{N}$. Then, at every sampling time step $t_j$, the following FHOCP is solved:
\begin{subequations} \label{eq:fhocp (App_NMPC)}
\begin{align}
& \min_{\hat{u}(\cdot)}\left\{  E(\hat{x}(t_j+T_p)) + \int_{t_j}^{t_j + T_p}  F(\hat{x}(s),\hat{u}(s))\right\}  \\
& \text{subject to:}\\
& \dot{\hat{x}}(s) = f(\hat{x}(s),\hat{u}(s)), \ \hat{x}(t_j) = x(t_j), \\
& \hat{x}(s) \in \mathcal{X}, \ \hat{u}(s) \in \mathcal{U}, \ s \in [t_j,t_j + T_p],\\
& \hat{x}(t_j + T_p) \in \mathcal{E},
\end{align}
\end{subequations}
where $F$ and $E$ are running and terminal costs, respectively, usually chosen as 
\begin{align*}
	F(x,u) \coloneqq & x^\top Q x + u^\top R u \\
	E(x) \coloneqq & x^\top P x,
\end{align*}
?where $P\in\mathbb{R}^{n\times n}$ and $R\in\mathbb{R}^{m\times m}$ are positive definite matrix and $Q\in\mathbb{R}^{n\times n}$ a positive semi-definite weight matrix. The hat $\hat{\cdot}$ denotes predicted variables (internal to the controller), i.e., $\hat{x}(\cdot)$ is the solution of \eqref{eq:fhocp (App_NMPC)} driven by the input $\hat{u}(\cdot):[t_j,t_j+T_p]\to\mathcal{U}$ with the initial condition $x(t_j)$. Due to the fact that a finite prediction horizon is used, the actual closed-loop input and state trajectories will differ from the predicted open-loop trajectories, even if no model plant mismatch and no disturbances are present. This is the key difference between standard control strategies, where the feedback control law is obtained a priori, and NMPC, where the feedback control law is obtained online.  Since a finite horizon approach is used, the terminal set $\mathcal{E}$ is introduced and is appropriately designed as described in \cite{findeisen2002introduction,frank_1998_quasi_infinite} in order to guarantee the stability of  the  closed-loop  system.  The  solution  to  FHOCP \eqref{eq:fhocp (App_NMPC)} is  denoted by $\hat{u}^\star(\cdot,x(t_j))$. It defines the open-loop input that is applied to the system until the next sampling time step $t_{j+1}$ as 
\begin{equation*}
	u(s;x(t_j)) = \hat{u}^\star(s;x(t_j)), \ s\in[t_j,t_{j+1}).
\end{equation*}
The control input $u(s;x(t_j))$ is a feedback control law, since it is re-calculated at each sampling instant using the new state information. The solution of \eqref{eq:ode (App_NMPC)} from an initial state $x(t_j)$, applying an input $u:[t_j,t_{j+1}] \to \mathbb{R}^m$ is denoted by $x(s;u(\cdot),x(t_j))$, $s\in[t_j,t_{j+1}]$.
We next define the admissible control input.

\begin{definition} \label{def:admissible_control_input (App_NMPC)}
	A control input $u: [0, T_p] \to \mathbb{R}^m$ for a state $x_0$ is called \emph{admissible}, if 
	\begin{enumerate}
		\item it is piecewise continuous;
		\item $u(s) \in \mathcal{U}, \forall \ s \in [0, T_p]$;
		\item $x(s;u(\cdot), x_0) \in \mathcal{X}, \forall \ s \in [0, T_p]$;
		\item $x(T_p; u(\cdot),x_0) \in \mathcal{E}$;
	\end{enumerate}
\end{definition}

The following theorem states the stability results for NMPC.
\begin{theorem}[\cite{frank_2003_nmpc_bible}] \label{th:main theorem (App_NMPC)}
	Consider system \eqref{eq:ode (App_NMPC)} and suppose that 
	\begin{itemize}
		\item the NMPC open-loop optimal control problem \eqref{eq:fhocp (App_NMPC)} is feasible at $t=0$
		\item the terminal set $\mathcal{E} \subseteq \mathcal{X}$ is closed with $0\in\mathcal{E}$ and the terminal cost $E(x)$ is cont. differentiable and positive definite
		\item the terminal set and terminal cost are chosen such that, $\forall x\in\mathcal{E}$, there exists an (admissible) input $u_\mathcal{E}:[0,\delta] \to \mathcal{U}$ such that $x(s)\in\mathcal{E}$, $\forall s\in[0,\delta]$, and
		\begin{equation*}
			\frac{\partial E}{\partial x}f(x(s),u_\mathcal{E}(s)) + F(x(s),u_\mathcal{E}(s)) \leq 0, \forall s\in [0,\delta]
		\end{equation*}		
	\end{itemize}
	Then, the closed-loop system is asymptotically stable with the region of attraction being the sets of states for which the an admissible input exists.
\end{theorem}  

\chapter{Graph Theory and Rigid Frameworks} \label{app:Rigidity}
This Appendix provides preliminary background on graph theory and rigid frameworks. 

\section{Graph Theory} \label{sec:graph_theory (app_ridigity)}

An \textit{undirected graph} $\mathcal{G}$ is a pair $(\mathcal{N}, \mathcal{E})$, where $\mathcal{N}\coloneqq\{1,\dots,N\}$ is a finite set of $N\in\mathbb{N}$ nodes, representing a team of agents, and $\mathcal{E} \subseteq \{ (i,j) \in \mathcal{N}^2 : i < j \}$, with $K = |\mathcal{E}|$, is the set of edges that model the communication capabilities between neighboring agents. 
For each agent, its neighboring set $\mathcal{N}_i$ is defined as $\mathcal{N}_i \coloneqq \{ j \in \mathcal{N} \text{ s.t. } (i,j) \} \in \mathcal{E}\}$. A \textit{directed} graph is formed by the edges being $\mathcal{E} \subseteq \{ (i,j) \in \mathcal{N}^2 : i \neq j \}$, i.e., $(i,j)\in\mathcal{E}$ does not necessarily imply $(j,i)\in\mathcal{E}$. The complete graph of $N$ nodes, $N>1$, is denoted by $\mathcal{K}_N$.

If there is an edge $(i,j) \in \mathcal{E}$, then $i, j$ are called \textit{adjacent}. A \textit{path} of length $r$ from vertex $i$ to vertex $j$ is a sequence of $r+1$ distinct vertices, starting with $i$ and ending with $j$, such that consecutive vertices are adjacent. For $i = j$, the path is called a \text{cycle}. If there is a path between any two vertices of the graph $\mathcal{G}$, then $\mathcal{G}$ is called \textit{connected}. A connected graph is called a \text{tree} if it contains no cycles.

Consider an arbitrary orientation of $\mathcal{G}$, which assigns to each edge $(i,j) \in \mathcal{E}$ precisely one of the ordered pairs $(i, j)$ or $(j, i)$. When selecting the pair $(i, j)$, we say that $i$ is the tail and $j$ is the head of the edge $(i,j)$. By considering a numbering $k \in \mathcal{K} \coloneqq \{1, \dots , K\}$ of the graph's edge set, we define the \textit{incidence matrix} $D(\mathcal{G}) \coloneqq [d_{ik}] \in \mathbb{R}^{N\times M}$ as  
\begin{equation*}
	d_{ik} = \begin{cases}
	1, & \text{ if } i \text{ is the head of edge } k, \\
	-1, & \text{ if } i \text{ is the tail of edge } k, \\
	0, & \text{ otherwise}.\\
	\end{cases}
\end{equation*}

Moreover, the incidence matrix satisfies the following important property. 
\begin{property} \label{prop:D (App_dynamical_systems)}
	Consider the incidence matrix $D(\mathcal{G})$ of an arbitrary graph $\mathcal{G}$ as $D(\mathcal{G}) = [d_1, \dots,d_N]^\top\in\mathbb{R}^{N\times M}$, where $d^\top_i\in \mathbb{R}^M$ are the rows of $D$. Then it holds that $\sum_{i\in\{1,\dots,N\}}d_i = 0$, i.e., the rows of the 
	incidence matrix  sum up to zero.  
\end{property}


\begin{lemma} \label{lemma:tree formation (App_Rigidity)} \cite[Section III]{dimos_kalle_2008}
	Assume that the graph $\mathcal{G}$ is a connected tree. Then, $D(\mathcal{G})^\top \Delta D(\mathcal{G})$ is positive definite for any positive definite matrix $\Delta\in\mathbb{R}^{N\times N}$.
\end{lemma}

\section{Rigidity Theory} \label{sec:Rigidity theory (App_Rigidity)}

We review here some necessary notions on distance and bearing rigidity. 
\subsection{Distance Rigidity in $\mathbb{R}^n$}
Let an undirected graph $(\mathcal{G},\mathcal{E})$ and $p_i \in \mathbb{R}^n$ be the position of node $i \in\mathcal{N}$. Let also $p \coloneqq [p_1^\top,\dots,p_N^\top]^\top \in\mathbb{R}^{nN}$. A \textit{framework} in $\mathbb{R}^n$ is the tuple $(\mathcal{G},p)$. We assume that the nodes are not collinear in $\mathbb{R}^n$  or occupy the exact same position.
 
\begin{definition}
	Two frameworks $(\mathcal{G},p)$ and $(\mathcal{G},p')$ are \textit{distance equivalent} if $\|p_i - p_j\| = \|p'_i - p'_j \|$, $\forall (i,j) \in \mathcal{E}$.
\end{definition}

\begin{definition}
	Two frameworks $(\mathcal{G},p)$ and $(\mathcal{G},p')$ are \textit{distance congruent} if $\|p_i - p_j\| = \|p'_i - p'_j \|$, $\forall i,j \in \mathcal{N}$.
\end{definition}

\begin{definition}
	A framework $(\mathcal{G},p)$ is \textit{distance rigid} if there exists a constant $\epsilon \in\mathbb{R}_{>0}$ such that any framework $(\mathcal{G},p')$ that is distance equivalent to $(\mathcal{G},p)$ and satisfies $\|p'-p\| < \epsilon $ is also distance congruent to it.	
\end{definition}

\begin{definition}
	A framework $(\mathcal{G},p)$ is \textit{globally distance rigid} if an arbitrary framework that is distance equivalent to $(\mathcal{G},p)$ is also distance congruent to it.	
\end{definition}

Consider an orientation assigned to $\mathcal{G}$, as described in the previous section, as well as a numbering $\mathcal{K} \coloneqq \{1,\dots,K\}$, $K\coloneqq |\mathcal{E}|$. Let the interneighbor distances be expressed by $p_k \coloneqq p_i - p_j$, $\forall k=(i,j) \in \mathcal{E}$. Define the distance function $F_D : \mathbb{R}^{Nn} \to \mathbb{R}^{Kn}$ as 
\begin{equation*}
	F_D(p) \coloneqq \frac{1}{2}\begin{bmatrix}
	\|p_1\|^2, \dots, \|p_K\|
	\end{bmatrix} ^\top
\end{equation*}

The \textit{distance rigidity matrix} is defined as the respective Jacobian 
\begin{equation*}
	\mathcal{R}_D(p) \coloneqq \frac{\partial F_D(p)}{\partial p} \in\mathbb{R}^{K\times Nn}
\end{equation*}

Let now $\delta p \in \mathbb{R}^Nn$ be a variation of the configuration $p$. If $\mathcal{R}_D \delta p = 0$, then $\delta p $ is an \textit{infinitesimal distance motion} of $(\mathcal{G},p)$. An infinitesimal distance motion is \textit{trivial} if it corresponds only to a translation and a rotation of the entire framework. 

\begin{definition} \label{def:infinit. rigidity (App_Rigidity)}
	A framework is \textit{infinitesimally rigid} if all the infinitesimal distance motions are trivial. 
\end{definition}

\begin{lemma}
	A framework  $(\mathcal{G},p)$ is infinitesimally rigid if 
	\begin{itemize}
		\item $\textup{rank}(\mathcal{R}_D(p)) \geq 2N-3$, if $n=2$
		\item $\textup{rank}(\mathcal{R}_D(p)) \geq 3N-6$, if $n=3$
	\end{itemize}
\end{lemma}

\subsection{Bearing Rigidity in $\mathbb{SE}(3)$} 

Let a directed graph $(\mathcal{G},\mathcal{E})$ and now $x_i \coloneqq (p_i,R_i) \in \mathbb{SE}(3)$ be the \textit{pose} of node $i \in\mathcal{N}$. Let also $x \coloneqq (x_i,\dots,x_N) \in\mathbb{SE}^{3N}$. A \textit{framework} in $\mathbb{SE}(3)$ is the tuple $(\mathcal{G},x)$. We assume that the nodes are not collinear in $\mathbb{R}^n$  or occupy the exact same position. 

Let $b_k : \mathbb{SE}(3)^2\to \mathbb{S}^2$ be the relative bearings, with
$$ b_k(x_i,x_j) \coloneqq R_i^\top \frac{p_i - p_j}{\|p_i - p_j\|}, \forall k = (i,j)\in\mathcal{E}$$
as well as $b_{\mathcal{G}}(x) \coloneqq [b_1^\top,\dots,d_K^\top]^\top \in\mathbb{S}^{2N}$. 

\begin{definition}
	Two frameworks $(\mathcal{G},x)$ and $(\mathcal{G},x')$ are bearing equivalent if 
	\begin{equation*}
		R_i^\top \frac{p_i - p_j}{\|p_i - p_j\|} = R_i^{'\top} \frac{p'_i - p'_j}{\|p'_i - p'_j\|}, \forall (i,j)\in\mathcal{E}
	\end{equation*}
\end{definition}

\begin{definition}
	Two frameworks $(\mathcal{G},x)$ and $(\mathcal{G},x')$ are bearing congruent if 
	\begin{equation*}
	R_i^\top \frac{p_i - p_j}{\|p_i - p_j\|} = R_i^{'\top} \frac{p'_i - p'_j}{\|p'_i - p'_j\|}, \forall i,j \in\mathcal{N}
	\end{equation*}
\end{definition}

\begin{definition}
	A framework $(\mathcal{G},x)$ is globally bearing rigid in $\mathbb{SE}(3)$ if every framework which is equivalent to it is also congruent to it.
\end{definition}

\begin{definition}
	A framework $(\mathcal{G},x)$ is bearing rigid in $\mathbb{SE}(3)$ if there exists a neighborhood $\mathsf{S} \subset \mathbb{SE}(3)^N$ of $x$ such that
	\begin{equation*}
		b^{-1}_{\mathcal{K}_N}( b_{\mathcal{K}_N}(x)) \cap \mathsf{S} = b^{-1}_{\mathcal{G}}(b_\mathcal{G}(x)) \cap \mathsf{S}
	\end{equation*}
\end{definition}

Similar to distance rigidity, the respective rigidity matrix is defined as $\mathcal{R}_\mathcal{G}(x) \coloneqq \frac{\partial b_\mathcal{G}(x)}{\partial x}$, whose null-space describes the infinitesimal bearing motions of $(\mathcal{G},x)$ in $\mathbb{SE}(3)$. An infinitesimal bearing motion in $\mathbb{SE}(3)$ is trivial if it corresponds only to a translation, rotation, or dilation of the framework. The infinitesimal bearing rigidity in $\mathbb{SE}(3)$ follows from Definition \ref{def:infinit. rigidity (App_Rigidity)}.

\begin{lemma}
	A framework $(\mathcal{G},x)$ is infinitesimally bearing rigid in $\mathbb{SE}(3)$ if and only if $\textup{null}(\mathcal{R}_\mathcal{G}(x)) = \textup{null}(\mathcal{R}_{\mathcal{K}_N}(x))$, or, equivalently, $\textup{rank}(\mathcal{R}_\mathcal{G}(x)) = 6N-7$. 
\end{lemma}

\chapter{Temporal Logics-based Task Specification} \label{app:Logics}
This Appendix provides preliminary background on specifications expressed as temporal logic formulas. 

\section{Task Specification in LTL} \label{subsec:LTL (App_logics)}

\begin{definition}
	A transition system ($\mathcal{TS}$) is a tuple $\mathcal{T} \coloneqq (\Pi, \Pi_0, \to, \Psi, \mathcal{L}$), where $\Pi$ is a discrete finite set of states, $\Pi_0$ is a discrete finite set of initial states, $\to \subseteq \Pi\times\Pi$ is a transition relation, $\Psi$ is a discrete set of atomic propositions\footnote{boolean variables that are either true or false in a given state}, and $\mathcal{L}:\Pi\to 2^{\Psi}$ is a labeling function that assigns to each state the atomic propositions that are true in that state.
\end{definition}

\begin{definition}
	A \textit{run} of a $\mathcal{TS}$ is an infinite sequence 
	\begin{equation*}
	r_{\scriptscriptstyle \mathcal{TS}} \coloneqq \pi_1\pi_2\pi_3\dots,
	\end{equation*}
	with $\pi_1\in\Pi_0$, $\pi_i\in\Pi$, $\forall i\in\mathbb{N}$.
\end{definition}

\begin{definition}
	A \textit{word} $w_{\scriptscriptstyle \mathcal{TS}}$ of a run $r_{\scriptscriptstyle \mathcal{TS}}$ is the infinite sequence
	\begin{equation*}
	w_{\scriptscriptstyle \mathcal{TS}}(r_{\scriptscriptstyle \mathcal{TS}}) = w_1w_2w_3,\dots, 
	\end{equation*}
	where $w_i\in 2^{\Psi}, w_i=\mathcal{L}(\pi_i), \forall i\in\mathbb{N}$.
\end{definition}

We focus on the task specification $\mathsf{\Phi}$ given as a Linear Temporal Logic (LTL) formula. The basic ingredients of a LTL formula are a set of atomic propositions $\Psi$ and several boolean and temporal operators. LTL formulas are formed according to the following grammar \cite{baier2008principles}: $\mathsf{\Phi} ::= \mathsf{true}\: |\:a\: |\: \mathsf{\Phi}_{1} \land  \mathsf{\Phi}_{2}\: |\: \neg \mathsf{\Phi}\: |\:\bigcirc \mathsf{\Phi}\:|\:\mathsf{\Phi}_{1}\cup \mathsf{\Phi}_{2} $, where $a\in \Psi$, $\mathsf{\Phi}_1$ and $\mathsf{\Phi}_2$ are LTL formulas and $\bigcirc$, $\cup$ are the \textit{next} and \textit{until} operators, respectively. Definitions of other useful operators like $\square$ (\textit{always}), $\lozenge$ (\textit{eventually}) and $\Rightarrow$ (\textit{implication}) are omitted and can be found at \cite{baier2008principles}.
The semantics of LTL are defined over infinite words over $2^{\Psi}$. Intuitively, an atomic proposition $\psi\in \Psi$ is satisfied on a word $w_{\scriptscriptstyle \mathcal{TS}}=w_1w_2w_3\dots$ if it holds at its first position $w_1$, i.e. $\psi\in w_1$, denoted as $w_{\scriptscriptstyle \mathcal{TS}} \models \mathsf{\Phi}$. Formula $\bigcirc \mathsf{\Phi}$ holds true if $\mathsf{\Phi}$ is satisfied on the word suffix that begins in the next position $w_2$, whereas $\mathsf{\Phi}_1\cup \mathsf{\Phi}_2$ states that $\mathsf{\Phi}_1$ has to be true until $\mathsf{\Phi}_2$ becomes true. Finally, $\lozenge \mathsf{\Phi}$ and  $\square \mathsf{\Phi}$ holds on $w_{\scriptscriptstyle \mathcal{TS}}$ eventually and always, respectively. For a full definition of the LTL semantics, the reader is referred to \cite{baier2008principles}.

A LTL formula $\mathsf{\Phi}$ over a set of atomic propositions $\Psi$ can be translated to a B\"uchi Automaton $\mathcal{A}_\mathsf{\Phi}$ \cite{baier2008principles}. Then, by calculating the product of the transition system $\mathcal{TS} = (\Pi,\Pi_0,\to,\Psi,\mathcal{L})$ with $\mathcal{A}_\mathsf{\Phi}$ as $\widetilde{\mathcal{TS}} \coloneqq \mathcal{TS}\otimes \mathcal{A}_\mathsf{\Phi}$, we can find the runs of $\mathcal{TS}$ that satisfy the formula $\mathsf{\Phi}$. These runs can then be projected back to $\mathcal{TS}$, providing paths over $\Pi$ that satisfy $\mathsf{\Phi}$. More details regarding the technique can be found in \cite{baier2008principles}.

\section{Task Specification in MITL} 
\label{sec:MITL (App_logics)}

\begin{definition} \label{def:WTS (AR)}
	A \textit{Weighted Transition System ($\mathcal{WTS}$)} is a tuple 
	\begin{equation*}
	\mathcal{WTS} \coloneqq (\Pi,\Pi_0, \to,\Psi,\mathcal{L},\gamma),
	\end{equation*}
	where $\Pi$ is a discrete finite  set of states, $S_0\subseteq S$ is a discrete finite set of initial states,	$\to\subseteq \Pi\times \Pi$ is a transition relation, $\Psi$ is a finite set of atomic propositions, $\mathcal{L}:\Pi\rightarrow 2^{\Psi}$ is a labeling function and $\gamma:(\to)\rightarrow \mathbb{R}_{\geq 0}$ is a map that assigns a positive weight to each transition.	
\end{definition}

\begin{definition} \label{def:time sequence (App_logics)}
	\cite{alur1994theory} The time sequence $t_1t_2t_3\dots$ is an infinite sequence of time values $t_j\in\mathbb{R}_{\geq 0}, \forall j\in\mathbb{N}$, satisfying the following constraints:
	\begin{itemize}
		\item Monotonicity: $t_j < t_{j+1}, \forall j\in\mathbb{N}$.
		\item Progress: $\forall t'\in\mathbb{R}_{\geq 0}, \exists j\geq 1$ such that $t_j \geq t'$.
	\end{itemize}
\end{definition}

\begin{definition} \label{def:timed word (App_logics)}
	Let $\Psi$ be a finite set of atomic propositions. A timed word $w$ over $\Psi$ is an infinite sequence $$w = (w_1,t_1)(w_2,t_2),\dots,$$ where $w_1w_2\dots$ is an infinite word over $2^{\Psi}$ and $t_1t_2\dots$ is a time sequence according to Definition \ref{def:time sequence (App_logics)}.
\end{definition}

\begin{definition} \label{def:timed run of WTS (App_logics)} 
	A \textit{timed run} of a $\mathcal{WTS}$ is an infinite sequence $$r_{\scriptscriptstyle \mathcal{WTS}} = (r_1,t_1)(r_2,t_2)\dots,$$ such that $r_1\in \Pi_0$, and $r_j\in \Pi, (r_j,r_{j+1})\in$ $\to, \forall j\in\mathbb{N}$. The time stamps $t_j$ are inductively defined as
	\begin{enumerate}
		\item $t_1  = 0$,
		\item $t_{j+1} = t_j + \gamma(r_j, r_{j+1}), \forall j\in\mathbb{N}$.
	\end{enumerate} 
	The timed run $r_{\scriptscriptstyle \mathcal{WTS}}$ generates the timed word $$w_{\scriptscriptstyle \mathcal{WTS}}(r_{\scriptscriptstyle \mathcal{WTS}}) = w_1(r_1)w_2(r_2) \dots=(\mathcal{L}(r_1),t_1)(\mathcal{L}(r_2),t_2)\dots$$ over the set $2^{\Psi}$, where $\mathcal{L}(r_j)$ is the subset of atomic propositions that are true at state $r_j$ at time $t_j$, $\forall j\in\mathbb{N}$.
\end{definition}

The syntax of \textit{Metric Interval Temporal Logic (MITL)} over a set of atomic propositions $\Psi$ is defined by the grammar 
\begin{equation*}
\mathsf{\Phi} := p \:|\: \neg \mathsf{\Phi} \: |\: \mathsf{\Phi}_1 \land \mathsf{\Phi}_2  |\: \bigcirc_I \mathsf{\Phi} \:|\: \lozenge_I \mathsf{\Phi} \:|\: \square_I \mathsf{\Phi} \:|\:\mathsf{\Phi}_1 \mathcal{U}_I \mathsf{\Phi}_2,
\end{equation*}
where $p\in\Psi$, and $\bigcirc,\lozenge,\square$ and $\mathcal{U}$ are the next, future, always and until operators, respectively; $I$ is a nonempty time interval in one of the following forms: $[i_1,i_2],[i_1,i_2),(i_1,i_2],$ $(i_1,i_2),[i_1,\infty),(i_1,\infty)$ with $i_1,i_2\in\mathbb{R}_{\geq 0}, i_2>i_1$. MITL can be interpreted either in continuous or point-wise semantics. We utilize the latter and interpret MITL formulas over timed runs such as the ones produced by a $\mathcal{WTS}$. 

\begin{definition} \label{def:MITL (AR)}
	\cite{d2007expressiveness,ouaknine2005decidability}
	Given a run $r_{\scriptscriptstyle \mathcal{WTS}} = (r_1,t_1)(r_2,t_2)\dots$ of a WTS and a MITL formula $\mathsf{\Phi}$, we define $(r_{\scriptscriptstyle \mathcal{WTS}},j)\models\phi, j\in\mathbb{N}$ ($r_{\scriptscriptstyle \mathcal{WTS}}$ satisfies $\mathsf{\Phi}$ at $j$) as follows:
	\begin{align*} 
	(r_{\scriptscriptstyle \mathcal{WTS}},j) \models& p \Leftrightarrow p\in \mathcal{L}(r_j), \notag \\
	(r_{\scriptscriptstyle \mathcal{WTS}},j) \models& \neg \mathsf{\Phi} \Leftrightarrow (r_{\scriptscriptstyle \mathcal{WTS}},j) \not\models \mathsf{\Phi} \notag \\
	(r_{\scriptscriptstyle \mathcal{WTS}},j) \models&  \mathsf{\Phi}_1\land \mathsf{\Phi}_2 \Leftrightarrow (r_{\scriptscriptstyle \mathcal{WTS}},j) \models \mathsf{\Phi}_1 \text{ and } (r_{\scriptscriptstyle \mathcal{WTS}},j) \models \mathsf{\Phi}_2 \notag \\
	(r_{\scriptscriptstyle \mathcal{WTS}},j) \models& \bigcirc_I \mathsf{\Phi} \Leftrightarrow (r_{\scriptscriptstyle \mathcal{WTS}},j+1)\models \mathsf{\Phi} \text{ and } t_{j+1}-t_j\in I \notag \\	
	(r_{\scriptscriptstyle \mathcal{WTS}},j) \models&  \mathsf{\Phi}_1\mathcal{U}_I \mathsf{\Phi}_2 \Leftrightarrow \exists k,j, \text{ with } j\leq k, \text{s.t. } (r_{\scriptscriptstyle \mathcal{WTS}},k) \models \mathsf{\Phi}_2, t_k - t_j \in I \\ & \text{ and } (r_{\scriptscriptstyle \mathcal{WTS}},m)\models \mathsf{\Phi}_1, \notag  \forall m\in\{j,\dots,k\} 
	\end{align*}
	Also, $\lozenge_I \mathsf{\Phi} = \top\mathcal{U}_I \mathsf{\Phi}$ and $\square_I\mathsf{\Phi} = \neg\lozenge_I\neg \mathsf{\Phi}$. The sequence $r_{\scriptscriptstyle \mathcal{WTS}}$ satisfies $\mathsf{\Phi}$, denoted as $r_{\scriptscriptstyle \mathcal{WTS}}\models\mathsf{\Phi}$, if and only if $(r_{\scriptscriptstyle \mathcal{WTS}},1)\models \mathsf{\Phi}$.
\end{definition}

We provide next a description of \textit{Timed B\"uchi Automata (TBA)}, originally proposed in \cite{alur1994theory}. Let $\mathsf{CL} \coloneqq \{\mathsf{cl}_1,\dots,\mathsf{cl}_{|\mathsf{CL}|}\}$ be a finite set of \textit{clocks}. The set of \textit{clock constraints} $\Phi(\mathsf{CL})$ is defined by the grammar: 
\begin{equation*}
\phi \Coloneqq \top \ | \ \neg \phi \ | \ \phi_1 \ \land \ \phi_2 \ | \ \mathsf{cl} \bowtie \psi,
\end{equation*}
where $\mathsf{cl}\in \mathsf{CL}$ is a clock, $\psi \in \mathbb{Q}$ is a clock constraint, and $\bowtie \in \{<,>,\geq,\leq,=\}$. A \textit{clock valuation} is a mapping $v:\mathsf{CL} \to \mathbb{R}$ that assigns a value to each clock. A clock $\mathsf{cl}_i$ has valuation $v_i$ for $i\in\{1,\dots,|\mathsf{CL}|\}$. Given $v \coloneqq (v_1,\dots,v_{|\mathsf{CL}|})$ and $t\in\mathbb{R}_{\geq0}$, we denote by $v \models \phi$ and $t\models \phi$ the fact that the valuation $v$ and the time instant $t$, respectively, satisfy the clock constraint $\phi$.

\begin{definition}
	A \textit{Timed B\"uchi Automaton} is a tuple $$\mathcal{A}_t \coloneqq (Q, Q_0, \mathsf{CL}, \Psi, E, F),$$ where $Q$ is a finite set of locations, $Q_0 \subseteq Q$ is the set of initial locations, $\mathsf{CL}$ is a finite set of clocks, $\Psi$ is a finite set of atomic propositions that defines the input alphabet $2^{\Psi}$, $E \subset Q\times \Phi(\mathsf{CL})\times 2^\mathsf{CL} \times 2^{\Psi}\times Q$ gives the set of edges of the form $e = (q,g,R,\alpha,q')$, where $q$, $q'$ are the source and target locations, $g$ is the guard of edge, $R$ is a set of clocks to be reset upon executing the edge, and $\alpha$ is an input string; finally, $F\subseteq Q$ is a set of accepting locations.	 
\end{definition}

A state of $\mathcal{A}_t$ is a pair $(q,v)\in\mathbb{R}\times\mathbb{R}^{|\mathsf{CL}|}$. The initial state of $\mathcal{A}_t$ is $(q_0, 0_{|\mathsf{CL}|})$, with $q_0\in Q_{0}$. Given two states $(q,v)$, $(q',v')$, and an edge $e = (q,g,R,\alpha,q')$, there exists a \textit{discrete transition} $(q,v)\overset{e}{\longrightarrow} (q',v')$ if $v\models g$. Moreover, $v'_i = 0$, $\forall \mathsf{cl}_i\in R$, and $v'_i = v_i$, $\forall \mathsf{cl}_i \notin R$. Given $\delta \in \mathbb{R}$, there exists a \textit{time transition} $(q,v) \overset{\delta}{\longrightarrow} (q',v')$ if $q = q'$ and $v' = v + \delta$ (component-wise summation). We write $(q,v) \overset{\delta}{\longrightarrow}\overset{e}{\longrightarrow}(q',v')$ if there exists $q'',v''$ such that $(q,v)\overset{\delta}\longrightarrow(q'',v'')$ and $(q'',v'')\overset{e}{\longrightarrow}(q',v')$, with $q''=q$.

An infinite run of $\mathcal{A}_t$ starting at a state $(q_1,v_1)$ is an infinite sequence of time and discrete transitions $(q_1,v_1)\overset{\delta_1}{\longrightarrow}(q_1',v_1')\overset{e_1}{\longrightarrow}(q_2,v_2)\overset{\delta_2}{\longrightarrow}(q_2',v_2')\dots$, where $e_i = (q_i,g_i,R_i,\sigma_i,q'_i)$, $\forall i\in\mathbb{N}$. This run corresponds to the timed word $\mathsf{w}_t = (\sigma_1,\tau_1)(\sigma_2,\tau_2)$, with $\tau_{i+1} = \tau_i + \delta_i$, $\forall i\in\mathbb{N}$. The run is called accepting if $q_j\in F$ for infinitely many $j\in\mathbb{N}$. A timed word is called \textit{accepting} if there exists an accepting run associated with it. The problem of deciding the language emptiness of a given TBA is PSPACE-complete \cite{alur1994theory}. In other words, an accepting run of a given TBA can be synthesized, if one exists. Any timed formula $\mathsf{\Phi}$ over $\Psi$ originating from the decidable fragment of timed logics (\textit{e.g.,} MITL, discrete-time MTL, finite MTL\footnote{In this case, the generated Timed Automaton will have finite accepting runs.}, coFlat-MTL, Bounded-MTL  \cite{ouaknine2005decidability,bouyer07}) can be algorithmically translated into a TBA with input alphabet $2^{\Psi}$, such that the language of timed words that satisfy $\varphi$ is the language of timed words produced by the TBA. 

\chapter{Useful Properties} \label{app:useful_prop}
This appendix provides some additional technical properties that are useful throughout the thesis. 

\begin{proposition} \label{prop: f(x) (app_useful_prop)} 
	Let $f:\mathbb{R}_{\geq 0}\to \mathbb{R}$, with $f(x) \coloneqq \exp(x)(\exp(x)-1) - x^2$. Then it holds that $f(x) \geq 0$, $\forall x\in\mathbb{R}_{\geq 0}$.
\end{proposition}
\begin{proof}
	It holds that $\tfrac{\partial f(x)}{\partial x} = 2 \exp(2x) -\exp(x) - 2x > 0,
	\forall x\in\mathbb{R}_{\geq 0}$. Hence, 
	$f(x)\geq f(0)=0$, $\forall x\in\mathbb{R}_{\geq 0}$.
\end{proof}
\begin{proposition} \label{prop:e_R frobenious (app_useful_prop)} \cite{lee2017attitude}
	Let $R_1, R_2 \in  \mathbb{SO}(3)$, and $e_R \coloneqq S^{-1}( R^\top_1 R_2 - R^\top_2 R_1)$. Then $\|e_R \|^2  \coloneqq \|R_1 - R_2 \|^2_\text{F}\Big(1 - \tfrac{1}{8}\|R_1 - R_2 \|^2_\text{F} \Big)$.
\end{proposition}

\begin{proposition}  \label{prop:R trace (app_useful_prop)}
	Let $R_1, R_2 \in  \mathbb{SO}(3) $. Then, for the rotation matrix $R_2^\top R_1 \in  \mathbb{SO}(3) $ it holds that $-1 \le \text{tr}(R_2^\top R_1)\le 3$; $\text{tr}(R_2^\top R_1) = 3$ if and only if $R_2^\top R_1= I_3 \Leftrightarrow R_1 = R_2$; $\text{tr}[R_2^\top R_1] = -1$ when $R_1 = R_2 \exp(\pi \hat{s})$, for every $\hat{s}$ in the unit sphere.
\end{proposition}

Let $x, y\in\mathbb{R}^3$, $R \in \mathbb{SO}(3)$, and $A \in \mathbb{R}^{3\times 3}$. Then the following  hold \cite{lee10control}:
\begin{itemize}	
	\item $x^\top S(y) x = 0$;	
	\item $S(R x) = R S(x) R^\top$;	
	\item $-\frac{1}{2} \text{tr} \left[ S(x) S(y) \right] = x^\top y$;
	\item $\text{tr} \left[A S(x) \right] = \frac{1}{2} \text{tr} \left[ S(x) (A - A^\top) \right] = - x^\top S^{-1}(A-A^\top)$.
\end{itemize}

We provide next some useful properties of linear algebra.

\begin{definition} \label{def:left equiv. (app_useful_prop)}
	A matrix $A \in \mathbb{R}^{n\times m}$ is \emph{left equivalent} (or \emph{row equivalent}) to a matrix $B\in\mathbb{R}^{n\times m}$ if and only if there exists an invertible matrix $P\in\mathbb{R}^{n\times n}$ such that $A= PB$.
\end{definition}
	
The following propositions can be proved:
	
\begin{proposition} \label{prop:left equiv. and nullspace (app_useful_prop)}
	Let $A, B \in \mathbb{R}^{n\times m}$ Then $A$ and $B$ are left equivalent if and only if they have the same nullspace $\textup{null}(A) = \textup{null}(B)$, i.e., $Ax = 0 \Leftrightarrow Bx= 0$, for any $x\in\mathbb{R}^m$.
\end{proposition}
	
\begin{proposition} \label{prop: pseudoinv. of A, KA (app_useful_prop)}
	Let $A \in \mathbb{R}^{n\times m}$, and $B \coloneqq K A$, where $K\in\mathbb{R}^{n\times n}$ is an invertible matrix. Then it holds that 
	\begin{align*}
	A^\dagger A = B^\dagger B.
	\end{align*}
\end{proposition}

\begin{proposition} \label{prop: pseudo-inverse property identity (app_useful_prop)}
	Let $A, B \in \mathbb{R}^{n\times m}$ such that $\textup{range}(A^\top) = \textup{null}(B)$. Then it holds that 
	\begin{equation*}
	A^\dagger A + B^\dagger B = I_m.
	\end{equation*}
\end{proposition}



\begin{proposition} \label{prop:cubic (app_useful_prop)}
	Consider the cubic equation $f(\lambda) = c_3\lambda^3+c_2\lambda^2+c_1\lambda + c_0 = 0$ with $c_\ell\in\mathbb{R},\forall \ell\in\{0,\dots,3\}$ and roots $\lambda_1,\lambda_2,\lambda_3$, with $f(\lambda_1)=f(\lambda_2)=f(\lambda_3)=0$. Then, given its discriminant $\Delta \coloneqq (c_3)^4\prod_{\substack{i\in\{1,2\}\\\substack{j\in\{i+1,\dots,3\} } }}(\lambda_i-\lambda_j)^2$, the following hold: 
	\begin{enumerate} [(i)]
		\item $\Delta = 0 \Leftrightarrow \exists i,j\in\{1,2,3\}$, with $i\neq j$, such that $\lambda_i=\lambda_j$, i.e., at least two roots are equal, 
		\item $\Delta > 0 \Leftrightarrow \lambda_i\in\mathbb{R},\forall i\in\{1,2,3\}$, and $\lambda_i\neq\lambda_j, \forall i,j\in\{1,2,3\}$, with $i\neq j$, i.e., all roots are real and distinct.
	\end{enumerate}
\end{proposition}

\begin{proposition} \cite{choi2006continuousTRO} \label{prop:ellipsoids (app_useful_prop)}
	Consider two planar ellipsoids $$\mathcal{A} = \{z\in\mathbb{R}^3 \text{ s.t. } z^\top A(t) z \leq 0\}, \ \mathcal{B} = \{z\in\mathbb{R}^3 \text{ s.t. } z^\top B(t) z \leq 0  \},$$ with $z=[p^\top 1]^\top$, $p\in\mathbb{R}^2$, and $A, B:\mathbb{R}_{\geq 0}\to\mathbb{R}^{3\times3}$ terms that describe their motion in $2$D space.
	Given their characteristic polynomial  $f(\lambda) = \det(\lambda A - B)$, which has degree $3$, the following hold:
	\begin{enumerate}[(i)]
		\item  $\exists \lambda^* > 0 \text{ s.t. } f(\lambda^*)=0$, i.e,
		the polynomial $f(\lambda)$ has always one positive real root,
		\item $\mathcal{A}\cap\mathcal{B} = \emptyset$ if and only if the characteristic equation $f(\lambda) = 0$ has two distinct negative roots, i.e., $\exists \lambda^*_1, \lambda^*_2 <0$, with $\lambda^*_1\neq\lambda^*_2$, and $f(\lambda^*_1)=f(\lambda^*_2)=0$.
		\item {$\mathcal{A}\cap\mathcal{B} \neq \emptyset$ and $\textup{Int}{\mathcal{A}}\cap \textup{Int}(\mathcal{B})=\emptyset$, i.e., $\mathcal{A}$ and $\mathcal{B}$ touch externally, if and only if and only if $f(\lambda) = 0$ has a negative root with multiplicity $2$.}
	\end{enumerate}
\end{proposition}

\printindex
	
\bibliographystyle{unsrt}
\bibliography{./references_Chris}
	\end{document}